\documentclass[aps,pra,10pt,a4paper,superscriptaddress,onecolumn,preprintnumbers,floatfix,nofootinbib]{revtex4-2}

\usepackage{graphicx}
\usepackage[utf8]{inputenc}
\usepackage[T1]{fontenc}
\usepackage{float}
\usepackage{amsmath,amssymb,amsfonts,amstext}
\usepackage{natbib}
\usepackage{epsfig,color}
\usepackage{dsfont}
\usepackage{accents}
\usepackage{esint}
\usepackage[svgnames]{xcolor}
\usepackage{multirow}
\usepackage{subcaption}
\usepackage{journal_names}
\usepackage{enumitem}
\usepackage{adjustbox,pifont}
\usepackage{tabularx}
\usepackage{booktabs}
\usepackage{siunitx}
\usepackage{mathtools}
\usepackage{algorithm}
\usepackage{algpseudocode}
\usepackage[citecolor=blue,urlcolor=blue,unicode=true,pdfusetitle, bookmarks=true,bookmarksnumbered=true,bookmarksopen=true,bookmarksopenlevel=3, breaklinks=false,pdfborder={0 0 1},backref=false,colorlinks=true, linkcolor=blue]{hyperref}
\allowdisplaybreaks
\usepackage{tikz}
\usepackage[normalem]{ulem}
\usetikzlibrary{arrows,arrows.meta,shapes,positioning,shadows,trees,calc,intersections,external,intersections}
\usepackage{wrapfig}
\usepackage{comment} 
\usetikzlibrary{mindmap}
\usepackage{caption}
\usepackage[compact]{titlesec}

\makeatletter

\renewcommand*{\p@subsection}{}

\renewcommand*{\p@subsubsection}{}

\renewcommand*{\p@paragraph}{}
\makeatother

\numberwithin{equation}{section}
\newcommand{\kms}{$\,{\rm km\,s}^{-1}{\rm\,Mpc}^{-1}$}
\newcommand{\element}[2]{\ensuremath{{}^{#1}\mathrm{#2}}}
\newcommand{\aEM}{\alpha_{\rm EM}}
\newcommand{\me}{m_{\rm e}}
\newcommand{\dme}{\Delta \me/\me}
\newcommand{\lcdm}{$\Lambda$CDM}
\newcommand{\code}[1]{\texttt{#1}}

\newcommand{\dd}{{\rm d}}%Operator d, i.e. non-italic

\usepackage[utf8]{inputenc}
\inputencoding{latin1}
\inputencoding{utf8}
\usepackage[printonlyused]{acronym} 

\makeatletter
\AtBeginDocument{%
  \renewcommand*{\AC@hyperlink}[2]{%
    \begingroup
      \hypersetup{hidelinks}%
      \hyperlink{#1}{#2}%
    \endgroup
  }%
}
\makeatother

\begin{document}

\title{The CosmoVerse White Paper: Addressing observational tensions in cosmology with systematics and fundamental physics}

\let\mymaketitle\maketitle
\let\myauthor\author
\let\myaffiliation\affiliation
\author{The CosmoVerse Network}

\date[\relax]{\today}

\noaffiliation

\begin{abstract}
The standard model of cosmology has provided a good phenomenological description of a wide range of observations both at astrophysical and cosmological scales for several decades. This concordance model is constructed by a universal cosmological constant and supported by a matter sector described by the standard model of particle physics and a cold dark matter contribution, as well as very early-time inflationary physics, and underpinned by gravitation through general relativity. There have always been open questions about the soundness of the foundations of the standard model. However, recent years have shown that there may also be questions from the observational sector with the emergence of differences between certain cosmological probes. In this White Paper, we identify the key objectives that need to be addressed over the coming decade together with the core science projects that aim to meet these challenges. These discordances primarily rest on the divergence in the measurement of core cosmological parameters with varying levels of statistical confidence. These possible statistical tensions may be partially accounted for by systematics in various measurements or cosmological probes but there is also a growing indication of potential new physics beyond the standard model. After reviewing the principal probes used in the measurement of cosmological parameters, as well as potential systematics, we discuss the most promising array of potential new physics that may be observable in upcoming surveys. We also discuss the growing set of novel data analysis approaches that go beyond traditional methods to test physical models. These new methods will become increasingly important in the coming years as the volume of survey data continues to increase, and as the degeneracy between predictions of different physical models grows. There are several perspectives on the divergences between the values of cosmological parameters, such as the model-independent probes in the late Universe and model-dependent measurements in the early Universe, which we cover at length. The White Paper closes with a number of recommendations for the community to focus on for the upcoming decade of observational cosmology, statistical data analysis, and fundamental physics developments.
\end{abstract}

\maketitle

\begin{figure}[htbp]
    \centering
    \includegraphics[scale = 1.00]{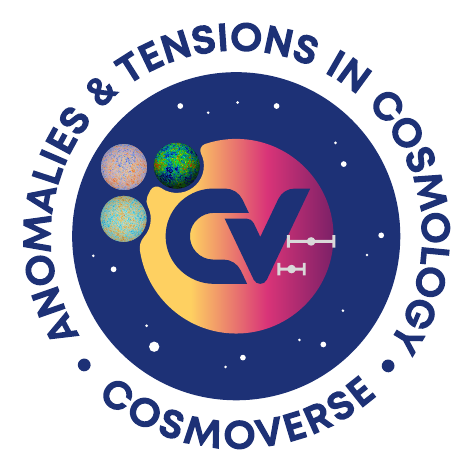}
\end{figure}

\newpage
% Author contribution information
\noindent \textbf{Editors\footnote{Detailed author information at the end of the document and lists alphabetized by first name}:} Eleonora Di Valentino\footnote{Email: \href{mailto:e.divalentino@sheffield.ac.uk}{e.divalentino@sheffield.ac.uk}}, Jackson Levi Said\footnote{Email: \href{mailto:jackson.said@um.edu.mt}{jackson.said@um.edu.mt}}\\

\noindent \textbf{Forward Writers:} Adam Riess, Agnieszka Pollo, and Vivian Poulin\\

\noindent \textbf{Section Coordinators:} Adrià Gómez-Valent, Amanda Weltman, Antonella Palmese, Caroline D. Huang, Carsten van de Bruck, Chandra Shekhar Saraf, Cheng-Yu Kuo, Cora Uhlemann, Daniela Grandón, Dante Paz, Dominique Eckert, Elsa M. Teixeira, Emmanuel N. Saridakis, Eoin \'O Colg\'ain, Florian Beutler, Florian Niedermann, Francesco Bajardi, Gabriela Barenboim, Giulia Gubitosi, Ilaria Musella, Indranil Banik, Istvan Szapudi, Jack Singal, Jaume Haro Cases, Jens Chluba, Jesús Torrado, Jurgen Mifsud, Karsten Jedamzik, Khaled Said, Konstantinos Dialektopoulos, Laura Herold, Leandros Perivolaropoulos, Lei Zu, Lluís Galbany, Louise Breuval, Luca Visinelli, Luis A. Escamilla, Luis A. Anchordoqui, M.M. Sheikh-Jabbari, Margherita Lembo, Maria Giovanna Dainotti, Maria Vincenzi, Marika Asgari, Martina Gerbino, Matteo Forconi, Michele Cantiello, Michele Moresco, Micol Benetti, Nils Sch\"oneberg, \"{O}zg\"{u}r Akarsu, Rafael C. Nunes, Reginald Christian Bernardo, Ricardo Chávez, Richard I. Anderson, Richard Watkins, Salvatore Capozziello, Siyang Li, Sunny Vagnozzi, Supriya Pan, Tommaso Treu, Vid Irsic, Will Handley, William Giar\`{e}, and Yukei Murakami\\

\noindent \textbf{Section Contributors:} Abdolali Banihashemi, Ad\`{e}le Poudou, Alan Heavens, Alan Kogut, Alba Domi, Aleksander \L{}ukasz Lenart, Alessandro Melchiorri, Alessandro Vadal\`{a}, Alexandra Amon, Alexander Bonilla, Alexander Reeves, Alexander Zhuk, Alfio Bonanno, Ali \"Ovg{\"u}n, Alice Pisani, Alireza Talebian, Amare Abebe, Amin Aboubrahim, Ana Luisa Gonz\'alez Mor\'an, Andr\'as Kov\'acs, Andreas Papatriantafyllou, Andrew R. Liddle, Andreas Lymperis, Andronikos Paliathanasis, Andrzej Borowiec, Anil Kumar Yadav, Anita Yadav, Anjan Ananda Sen, Anjitha John William, Anne Christine Davis, Anowar J. Shajib, Anthony Walters, Anto Idicherian Lonappan, Anton Chudaykin, Antonio Capodagli, Antonio da Silva, Antonio De Felice, Antonio Racioppi, Araceli Soler Oficial, Ariadna Montiel, Arianna Favale, Armando Bernui, Arrianne Crystal Velasco, Asta Heinesen, Athanasios Bakopoulos, Athanasios Chatzistavrakidis, Bahman Khanpour, Bangalore S. Sathyaprakash, Bartek Zgirski, Benjamin L'Huillier, Benoit Famaey, Bhuvnesh Jain, Bing Zhang, Biswajit Karmakar, Branko Dragovich, Brooks Thomas, Carlos Correa, Carlos G. Boiza, Catarina Marques, Celia Escamilla-Rivera, Charalampos Tzerefos, Chi Zhang, Chiara De Leo, Christian Pfeifer, Christine Lee, Christo Venter, Cláudio Gomes, Clecio Roque De bom, Cristian Moreno-Pulido, Damianos Iosifidis, Dan Grin, Daniel Blixt, Dan Scolnic, Daniele Oriti, Daria Dobrycheva, Dario Bettoni, David Benisty, David Fern\'andez-Arenas, David L. Wiltshire, David Sanchez Cid, David Tamayo, David Valls-Gabaud, Davide Pedrotti, Deng Wang, Denitsa Staicova, Despoina Totolou, Diego Rubiera-Garcia, Dinko Milakovi{\'c}, Dom Pesce, Dominique Sluse, Du\v{s}ko Borka, Ebrahim Yusofi, Elena Giusarma, Elena Terlevich, Elena Tomasetti, Elias C. Vagenas, Elisa Fazzari, Elisa G. M. Ferreira, Elvis Barakovic, Emanuela Dimastrogiovanni, Emil Brinch Holm, Emil Mottola, Emre \"{O}z\"{u}lker, Enrico Specogna, Enzo Brocato, Erik Jensko, Erika Antonette Enriquez, Esha Bhatia, Fabio Bresolin, Felipe Avila, Filippo Bouch\`{e}, Flavio Bombacigno, Fotios K. Anagnostopoulos, Francesco Pace, Francesco Sorrenti, Francisco S. N. Lobo, Fr\'ed\'eric Courbin, Frode K. Hansen, Greg Sloan, Gabriel Farrugia, Gabriel Lynch, Gabriela Garcia-Arroyo, Gabriella Raimondo, Gaetano Lambiase, Gagandeep S.~Anand, Gaspard Poulot, Genly Leon, Gerasimos Kouniatalis, Germano Nardini, G\'{e}za Cs\"{o}rnyei, Giacomo Galloni, Giada Bargiacchi, Giannis Papagiannopoulos, Giovanni Montani, Giovanni Otalora, Giulia De Somma, Giuliana Fiorentino, Giuseppe Fanizza, Giuseppe Gaetano Luciano, Giuseppe Sarracino, Gonzalo J. Olmo, Goran S. Djordjević, Guadalupe Ca\~nas-Herrera, Hanyu Cheng, Harry Desmond, Hassan Abdalla, Houzun Chen, Hsu-Wen Chiang, Hume A. Feldman, Hussain Gohar, Ido Ben-Dayan, Ignacio Sevilla-Noarbe, Ignatios Antoniadis, Ilim Cimdiker, In\^es S.~Albuquerque, Ioannis D. Gialamas, Ippocratis Saltas, Iryna Vavilova, Isidro G\'{o}mez-Vargas, Ismael Ayuso, Ismailov Nariman Zeynalabdi, Ivan De Martino, Ivonne Zavala, J. Alberto V\'{a}zquez, Jacobo Asorey, Janusz Gluza, Javier Rubio, Jenny G. Sorce, Jenny Wagner, Jeremy Sakstein, Jessica Santiago, Jim Braatz, Joan Sol\`a Peracaula, John Blakeslee, John Webb, Jose A. R. Cembranos, Jos\'e Pedro Mimoso, Joseph Jensen, Juan Garc\'ia-Bellido, Judit Prat, Kathleen Sammut, Kay Lehnert, Keith R.~Dienes, Kishan Deka, Konrad Kuijken, Krishna Naidoo, L\'aszl\'o \'Arp\'ad Gergely, Laur J\"arv, Laura Mersini-Houghton, Leila L. Graef, L\'eo Vacher, Levon Pogosian, Lilia Anguelova, Lindita Hamolli, Lu Yin, Luca Caloni, Luca Izzo, Lucas Macri , Luis E. Padilla, Luz \'Angela Garc\'ia, Maciej Bilicki, Mahdi Najafi, Manolis Plionis, Manuel Gonzalez-Espinoza, Manuel Hohmann, Marcel A. van der Westhuizen, Marcella Marconi, Marcin Postolak, Marco de Cesare, Marco Regis, Marek Biesiada, Maret Einasto, Margus Saal, Maria Caruana, Maria Petronikolou, Mariam Bouhmadi-L\'opez, Mariana Melo, Mariaveronica De Angelis, Marie-No\"elle C\'el\'erier, Marina Cort\^es, Mark Reid, Markus Michael Rau, Martin S. Sloth, Martti Raidal, Masahiro Takada, Masoume Reyhani, Massimiliano Romanello, Massimo Marengo, Mathias Garny, Mat\'{\i}as Leizerovich, Matteo Martinelli, Matteo Tagliazucchi, Mehmet Demirci, Miguel A. S. Pinto, Miguel A. Sabogal, Miguel A. Garc\'{i}a-Aspeitia , Milan Milo\v{s}evi\'{c}, Mina Ghodsi, Mustapha Ishak, Nelson J. Nunes, Nick Samaras, Nico Hamaus, Nico Schuster, Nicola Borghi, Nicola Deiosso, Nicola Tamanini, Nicolao Fornengo, Nihan Kat{\i}rc{\i}, Nikolaos E. Mavromatos, Nicholas Petropoulos, Nikolina \v{S}ar\v{c}evi\'c, Nils A. Nilsson, Nima Khosravi, Noemi Frusciante, Octavian Postavaru, Oem Trivedi, Oleksii Sokoliuk, Olga Mena, Paloma Morilla, Paolo Campeti, Paolo Salucci, Paula Boubel, Pawe\l{} Bielewicz, Pekka Heinämäki, Petar Suman, Petros Asimakis, Pierros Ntelis, Pilar Ruiz-Lapuente, Pran Nath, Predrag Jovanovi\'{c}, Purba Mukherjee, Rados{\l}aw Wojtak, Rafaela Gsponer, Rafid H. Dejrah, Rahul Shah, Rasmi Hajjar, Rebecca Briffa, Rebecca Habas, Reggie C. Pantig, Renier Mendoza, Riccardo Della Monica, Richard Stiskalek, Rishav Roshan, Rita B. Neves, Roberto Molinaro, Roberto Terlevich, Rocco D'Agostino, Rodrigo Sandoval-Orozco, Ronaldo C. Batista, Ruchika, Ruth Lazkoz, Saeed Rastgoo, Sahar Mohammadi, Salvatore Samuele Sirletti, Sandeep Haridasu, Sanjay Mandal, Saurya Das, Sebastian Bahamonde, Sebastian Grandis, Sebastian Trojanowski, Sergei D. Odintsov, Sergij Mazurenko, Shahab Joudaki, Sherry H. Suyu, Shouvik Roy Choudhury, Shruti Bhatporia, Shun-Sheng Li, Simeon Bird, Simon Birrer, Simone Paradiso, Simony Santos da Costa, Sofia Contarini, Sophie Henrot-Versill\'e, Spyros Basilakos, Stefano Casertano, Stefano Gariazzo, Stylianos A. Tsilioukas, Surajit Kalita, Suresh Kumar, Susana J. Landau, Sveva Castello, Swayamtrupta Panda, Tanja Petrushevska, Thanasis Karakasis, Thejs Brinckmann, Tiago B. Gon\c{c}alves, Tiziano Schiavone, Tom Abel, Tomi Koivisto, Torsten Bringmann, Umut Demirbozan, Utkarsh Kumar, Valerio Marra, Maurice H.P.M. van Putten, Vasileios Kalaitzidis, Vasiliki A. Mitsou, Vasilios Zarikas, Vedad Pasic, Venus Keus , Ver\'onica Motta, Vesna Borka Jovanovi\'{c}, V\'ictor H. C\'ardenas, Vincenzo Ripepi, Vincenzo Salzano, Violetta Impellizzeri, Vitor da Fonseca, Vittorio Ghirardini, Vladas Vansevi\v{c}ius, Weiqiang Yang, Wojciech Hellwing, Xin Ren, Yu-Min Hu, and Yuejia Zhai\\

\noindent \textbf{Endorsers:} Abdul Malik Sultan, Abdurakhmon Nosirov, Adrienn Pataki, Alessandro Santoni, Aliya Batool, Amlan Chakraborty, Aneta Wojnar, Arman Tursunov, Avik De, Ayush Hazarika, Baojiu Li, Benjamin Bose, Bivudutta Mishra, Bobomurat Ahmedov, Chandra Shekhar Saraf, Claudia Sc\'occola, Crescenzo Tortora, D'Arcy Kenworthy, Daniel E. Holz, David F. Mota, David S. Pereira, Devon M. Williams, Dillon Brout, Dong Ha Lee, Eduardo Guendelman, Edward Olex, Emanuelly Silva, Emre Onur Kahya, Enzo Brocato, Eva-Maria Mueller, Felipe Andrade-Oliveira, Feven Markos Hunde, F. R. Joaquim, Florian Pacaud, Francis-Yan Cyr-Racine, Pozo Nu\~nez, F, G\'abor R\'acz, Gene Carlo Belinario, Geraint F. Lewis, Gergely D\'alya, Giorgio Laverda, Guido Risaliti, Guillermo Franco-Abell\'an, Hayden Zammit, Hayley Camilleri, Helene M. Courtois, Hooman Moradpour, Igor de Oliveira Cardoso Pedreira, Il\'\i dio Lopes, Istv\'an Csabai, James W. Rohlf, J. Bogdanoska, Javier de Cruz P\'erez, Joan Bachs-Esteban, Joseph Sultana , Julien Lesgourgues, Jun-Qian Jiang, Karem Pe\~nal\'o Castillo, Kimet Jusufi, Lavinia Heisenberg, Laxmipriya Pati, L\'eon V.E. Koopmans, Lokesh kumar Duchaniya, Lucas Lombriser, Mahdieh Gol Bashmani Moghadam, Mar\'ia P\'erez Garrote, Mariano Dom\'{i}nguez, Marine Samsonyan, Mark Pace, Martin Kr\v{s}\v{s}\'ak, Masroor C. Pookkillath, Matteo Peronaci, Matteo Piani, Matthildi Raftogianni , Meet J. Vyas, Melina Michalopoulou, Merab Gogberashvili, Michael Klasen, Michele Cicoli, Miguel Quartin, Miguel Zumalac\'arregui, Milan S. Dimitrijevi\'c, Milos Dordevic, Mindaugas Kar\v{c}iauskas, Morgan Le~Delliou, Nastassia Grimm, Nicol\'as Augusto Kozameh, Nicoleta Voicu, Nicolina Pop, Nikos Chatzifotis, Odil Yunusov, Oliver Fabio Piattella, Paula Boubel, Pedro da Silveira Ferreira, P\'eter Raffai, Peter Schupp, Pierros Ntelis, Pradyumn Kumar Sahoo, Roberto V. Maluf, Ruth Durrer, S. A. Kadam, Sabino Matarrese, Samuel Brieden, Santiago Gonz\'alez-Gait\'an, Santosh V. Lohakare, Scott Watson, Shao-Jiang Wang, Simão Marques Nunes, Soumya Chakrabarti, Subinoy Das, Suvodip Mukherjee, Tajron Juri\'{c}, Tessa Baker, Theodoros Nakas, Tiago Barreiro, Upala Mukhopadhyay, Veljko Vuj\v{c}i\'{c}, Violetta Sagun, Vladimir A. Sre\'ckovi\'c, Wangzheng Zhang, Yo Toda, Yun-Song Piao, and Zahra Davari
\newpage

\setcounter{tocdepth}{3}
\tableofcontents
\clearpage

\section*{Executive Summary}
\addcontentsline{toc}{section}{Executive Summary}

The CosmoVerse network is born out of the CosmoVerse COST Action (formally CA21136 - Addressing observational tensions in cosmology with systematics and fundamental physics \cite{cosmoverseecost,cosmoversecost}), which traces its origins to the SNOWMASS 2021 effort in the cosmic tensions sector \cite{Abdalla:2022yfr,DiValentino:2020vhf,DiValentino:2020zio,DiValentino:2020vvd,DiValentino:2020srs}. This will be one of the key deliverables of the COST Action and one of its lasting legacies. The aim of this network is to establish synergy among researchers working across the disparate themes of observational cosmology, novel techniques of data analysis, and fundamental physics. This White Paper demonstrates how this effort has been successful, while also laying out a roadmap to sustain those efforts and expand the interdisciplinary nature of the topic across different areas of the community. Another core aspect of this effort is to harmonize approaches between the constituent subcommunities and to create a common language in which to discuss the topic of cosmic tensions.

In the White Paper, the accomplishment of these goals is demonstrated through the interwoven connections between the key communities of the network and the granular topics. This was achieved through a diverse set of actions, including the CosmoVerse Seminar Series led by Eleonora Di Valentino \cite{cosmoversecost_seminars}, the CosmoVerse conferences and the discussions and presentations involved in these events \cite{cosmoversecost_conferences}, the CosmoVerse School held in Corfu in 2024 \cite{cosmoversecost_school}, the CosmoVerse Training Series, which involved a significant effort to bridge \cite{cosmoversecost_training}, as well as the CosmoVerse Journal Club, led by Enrico Specogna and Mahdi Najafi \cite{cosmoversecost_jour_club}.

The CosmoVerse White Paper, edited by Jackson Levi Said and Eleonora Di Valentino, is a collective effort from the community to review the state of the art and identify opportunities to address tensions in cosmology over the coming years. This includes progress in observational cosmology and the exhaustive treatment of potential systematics, the development of new data analysis approaches for upcoming surveys and potential new physics models, as well as the construction of cosmological models that appropriately address the areas where the concordance model exhibits tensions. These topics are organized as follows:

\begin{itemize}
    \item Section \ref{sec:Intro} - Introduction and state-of-the-art. 
    \item Section \ref{sec:obs} - Observational cosmology. \textit{Coordinators: Adam Riess, Amanda Weltman, Antonella Palmese, Caroline D. Huang, Chandra Shekhar Saraf, Cheng-Yu Kuo, Cora Uhlemann, Dan Scolnic, Daniela Grandón, Dante Paz, Dominique Eckert, Florian Beutler, Gabriela Barenboim, Ilaria Musella, Istvan Szapudi, Jack Singal, Khaled Said, Leandros Perivolaropoulos, Lluís Galbany, Louise Breuval, Louise Breuval, Maria Giovanna Dainotti, Maria Vincenzi, Marika Asgari, Martina Gerbino, Margherita Lembo, Matteo Forconi, Michele Cantiello, Michele Moresco, Nils Schöneberg, Ricardo Chávez Murillo, Richard I. Anderson, Rick Watkins, Shahin Sheikh-Jabbari, Siyang Li, Tommaso Treu, Vid Iršič, Will Handley, William Giarè and Yukei Murakami}.
    \item Section \ref{sec:data_ana} - Novel data analysis techniques. \textit{Coordinators: Agnieszka Pollo, Adrià Gómez-Valent, Daniela Grandón, Jesus Torrado, Jurgen Mifsud, Lei Zu, Luis Escamilla, and Reginald Christian Bernardo}.
    \item Section \ref{sec:fun_phys} - Fundamental physics. \textit{Coordinators: Vivian Poulin, Carsten van de Bruck, Elsa Teixeira, Emmanuel N. Saridakis, Eoin O Colgain, Florian Niedermann, Francesco Bajardi, Giulia Gubitosi, Indranil Banik, Jaime Haro Cases, Jens Chluba, Karsten Jedamzik, Konstantinos F. Dialektopoulos, Laura Herold, Leandros Perivolaropoulos, Luca Visinelli, Luis Anchordoqui, Micol Benetti, Özgür Akarsu, Rafael Nunes, Sunny Vagnozzi, Supriya Pan}.
    \item Section \ref{sec:conclusion} - Discussion, White Paper recommendations, and future prospects.
\end{itemize}

In every section, the coordinators and contributing writers are identified at the beginning of each contribution. The preparation of the CosmoVerse White Paper involved a substantial number of people, with $\sim 65$ coordinators, $\sim 350$ contributing writers, and $\sim 130$ endorsers. The project was reviewed by Alba Domi, Alexandra Amon, Anjitha John William Mini Latha, Anton Chudaykin, Bivudutta Mishra, Emil Mottola, Emmanuel N. Saridakis, Florian Pacaud, Germano Nardini, Marcin Postolak, Mariano Dom\'{i}nguez, Miguel A. Garc\'ia-Aspeitia, Nelson J. Nunes, Oem Trivedi, Oliver Fabio Piattella, \"{O}zg\"{u}r Akarsu, Paolo Salucci, Pilar Ruiz-Lapuente, Rafid H. Dejrah, and Supriya Pan, and the bibliographic information was organized by Luca Visinelli. The notational conventions of the work are defined in Sec.~\ref{sec:conventions}. This is followed by a list of the glossary abbreviations in Sec.~\ref{sec:acronyms}. The CosmoVerse Action and White Paper were aided by the administrative contributions of Hiba Wazaz. \newpage

 \newpage
%%%%%%%%%%%%Section_1:
\section{Introduction} \label{sec:Intro}

Cosmology has entered an era of unprecedented precision in and volume of observational measurements, with large-scale surveys providing high-quality data across multiple redshifts and cosmic epochs. This wealth of observational data allows for a deeper understanding of the Universe's composition, dynamics, and structure formation processes of the Universe. However, these advancements have also revealed significant tensions between early- and late-time cosmological measurements, challenging the standard cosmological model. These discrepancies not only question the consistency of this framework but also suggest the possibility of unrecognized systematic errors or the need for new physics beyond the standard model of cosmology.

The standard model of cosmology, or \lcdm\, describes the Universe using a cosmological constant ($\Lambda$) and \ac{cdm}, and gravity through Einstein's \ac{gr}. It provides an excellent fit for a range of cosmological datasets, including the \ac{cmb} and large-scale structure surveys. Nevertheless, the emergence of tensions in key cosmological parameters has become increasingly difficult to ignore. Among these, the most prominent include discrepancies in the measurements of the Hubble constant ($H_0$), the amplitude of matter fluctuations ($S_8$), and the sound horizon at the epoch of \ac{bao}.

These tensions raise profound questions about our understanding of the Universe's expansion history, structure formation, and the fundamental nature of \ac{dm} and \ac{de}. If not due to systematic errors, such tensions may indicate the need for modifications of the standard model or additional components in the cosmic inventory. Addressing these issues requires a careful consideration of both observational and theoretical aspects, as well as a comprehensive approach that combines multiple cosmological probes.

\bigskip

\subsection{State of the art status of cosmological tensions \label{sec:state_of_the_art}}

The most statistically significant tension in cosmology is the $H_0$ tension, which refers to a significant and persistent discrepancy between measurements of the Hubble constant obtained from early- and late-time cosmological probes, challenging the completeness of the standard \lcdm\ model and suggesting the possible need for new physics. Early Universe constraints, primarily from the \textit{Planck} satellite (Sec.~\ref{sec:CMB}), which maps \ac{cmb} anisotropies, provide a precise estimate of $H_0 = 67.4 \pm 0.5$\kms. This result relies on the angular scale of the acoustic peaks in the \ac{cmb} power spectrum, under the assumptions of a flat \lcdm\ model with standard radiation content. Consistency with Planck's result has also been demonstrated by ground-based experiments such as \ac{act} and \ac{spt}, both of which measure the damping tail and lensing-induced smoothing of the \ac{cmb} power spectrum with high precision, reinforcing the lower $H_0$ value.

\ac{bao} (see Sec.~\ref{sec:BAO}) provide an additional, independent probe of the expansion history by measuring the characteristic scale imprinted by sound waves in the early Universe, observable in the large-scale distribution of galaxies. This standard ruler, linked to the sound horizon at the epoch of baryon drag, enables precise distance measurements at various redshifts. Surveys such as \ac{boss}, \ac{eboss}, and \ac{desi} have reported $H_0$ values consistent with \ac{cmb} constraints, further supporting the lower early Universe based estimate. Notably, \ac{desi}'s latest \ac{bao} results, based on over six million galaxies across multiple redshift bins, when calibrated by the Planck + \lcdm\ constraint on the sound horizon measured $H_0 = 68.5 \pm 0.6$\kms.

Late-time measurements of the Hubble constant $H_0$ using the distance ladder approach (Sec.~\ref{sec:dist_ladd}) suggest a higher value than early Universe constraints, contributing to the persistent Hubble tension. The most precise distance ladder method involves three steps: (1) geometric distance measurements to calibrate the luminosities of Cepheid variables using Gaia parallaxes, detached eclipsing binaries, and water masers; (2) using these calibrated Cepheids to standardize \ac{sn1} observed with \ac{hst}; and (3) measuring $H_0$ from \ac{sn1} distances in the Hubble flow, where cosmic expansion dominates.

Cepheid variables (Sec.~\ref{sec:Cepheids}) serve as primary standard candles due to their well-defined Period-Luminosity (P-L) relation, where the pulsation period correlates with intrinsic brightness. Systematic uncertainties, such as metallicity effects, crowding, and dust extinction, are mitigated through near-infrared photometry and consistent datasets from \ac{hst} and \ac{jwst}. The SH0ES collaboration recently measured $H_0 = 73.17 \pm 0.86$\kms using this approach, indicating a $5-6\sigma$ tension with Planck. \ac{jwst}'s independent observations, particularly in Cepheid-rich fields, have validated and refined these results by reducing crowding biases and confirming the reliability of the Cepheid calibration, strengthening the significance of the Hubble tension.

\ac{sn1} (Sec.~\ref{sec:SNeIa}) serves as the most common, far-field rung of the distance ladder, extending measurements into the Hubble flow. The Pantheon+ dataset provides precise $H_0$ measurements consistent with local results but in tension with early Universe estimates.

While the above tools remain central to the Hubble tension, offering the strongest leverage, alternative standard candles such as the \ac{trgb}, Mira variables, \ac{jagb} stars, \ac{sn2}, \ac{sbf}, and the \ac{btfr} provide valuable cross-checks. These independent methods, while varying in calibration techniques and stellar populations, consistently yield higher values of $H_0$ than early Universe constraints from the \ac{cmb}, emphasizing the robustness of the tension and fueling interest in theoretical refinements.

The \ac{trgb} method (Sec.~\ref{sec:TRGB}) measures distances using the sharp luminosity cutoff where \ac{rgb} stars ignite helium as a standard candle. Calibrated with galaxies like the Magellanic Clouds and NGC 4258, \ac{trgb} provides precise distance estimates. For state-of-the-art measurements, it yields consistent distance measures to \ac{sn1} hosts as Cepheids. \ac{trgb} remains a valuable cross-check for the Hubble tension, with minimal sensitivity to metallicity (if measured in the $I$-band) and dust.

Mira variables (Sec~\ref{sec:mira_stars}) are long-period pulsating stars used as standard candles for measuring $H_0$. Their period-luminosity relation, especially in the near-infrared, provides reliable distances. Calibrated using galaxies like the \ac{lmc} and NGC 4258, Miras offer an independent cross-check on $H_0$ with minimal sensitivity to metallicity and crowding.

\ac{jagb} stars (Sec.~\ref{sec:JAGB}) are standard candles used for measuring $H_0$ based on their narrow luminosity range in the near-infrared. They are carbon-rich \ac{agb} stars in an advanced stellar phase, providing distance estimates at long range.
Calibrated using galaxies like NGC 4258, \ac{jagb} measurements offer a route to calibrate \ac{sn1}. Though newer and less tested than Cepheids or \ac{trgb}, they contribute to cross-validation efforts.

\ac{sn2} (Sec.~\ref{sec:type_2_sn}) measure $H_0$ using the correlation between their luminosity and decline rate during the plateau phase. They offer an alternative to \ac{sn1}, aiding cross-checks in the Hubble tension.

The \ac{sbf} method (Sec.~\ref{sec:SBF}) measures $H_0$ using pixel-to-pixel luminosity variations in elliptical galaxies. Calibrated with nearby galaxies, \ac{sbf} provides $H_0$ estimates which are independent of \ac{sn1}.

The \ac{btfr} (Sec.~\ref{sec:btf_rel}) estimates $H_0$ using the correlation between a galaxy's rotational velocity and baryonic mass. Recent calibrations, particularly addressing zero-point uncertainties between different standard candles like Cepheids and \ac{trgb}, have refined the measurement to $H_0 = 76.3 \pm 2.1$(stat) $\pm 1.5$(sys)\kms, yielding another \ac{sn1}-independent measurement and further emphasizing the tension with early Universe estimates.

As an alternative to the distance ladder measurements, masers offer independent insights into the Hubble constant $H_0$. The maser technique (Sec.~\ref{sec:maser_drivers}) uses 22 GHz H$_2$O maser emissions from rotating disks around supermassive black holes to directly measure distances. Each of 5 masers (leaving NGC 4258 aside as its often used to calibrate the distance ladder)  yields a geometric distance and  $H_0 = 73.9 \pm 3.0$ \kms. The Megamaser Cosmology Project (MCP) has extended the precision of this method, producing results consistent with late-time measurements. Although rare, water masers provide a model-independent check, reinforcing the higher $H_0$ values obtained from local probes.

Strong gravitational lensing with time-delay measurements (Sec.~\ref{sec:strng_lens}) offers another independent method for estimating $H_0$. Multiple images of a background source, produced by a massive foreground lens, create time delays due to differences in the light paths. These delays depend on the lensing geometry and the Universe's expansion rate. Collaborations such as H0LiCOW and TDCOSMO have applied this technique, yielding $H_0 \approx 74.2 \pm 2.6$\kms, consistent with other late-time estimates.

\ac{gw} events with electromagnetic counterparts, known as standard sirens (Sec.~\ref{sec:GWs}), provide an independent estimate of $H_0$ by measuring the luminosity distance from \ac{gw}s emitted during compact object mergers. When an electromagnetic counterpart identifies the host galaxy, the redshift can be measured directly. Recent standard siren measurements yield $H_0 \approx 70.0 \pm 3.0$\kms, offering a model-independent constraint on cosmic expansion and contributing to the Hubble tension debate. However, this approach is still severely limited by the rarity of \ac{gw} events with EM counterparts, with only a single event which is too close for a reliable measurement of $H_0$.

\ac{cc} (Sec.~\ref{sec:CC}) estimate $H_0$ using the differential ages of passively evolving galaxies. By dating stellar populations, this method infers the expansion rate without relying on standard candles. However, unlike the use of distances, this measure is not empirical, depending on astrophysical modeling of aging stellar populations, their metallicities, star formation histories, and initial mass functions, with an uncertainty dominated by the modeling effort.  Recent measurements suggest $H_0 \approx 70.6 \pm 6.7$\kms, still consistent with both early and most late-time measurements. As a distance-ladder-independent probe, \ac{cc} offer a different dimension of study of the Hubble tension.

HII galaxies (Sec.~\ref{sec:HII_gal}) measure $H_0$ using the correlation between their luminosity and emission line flux from ionized gas. The strong emission lines, primarily from young massive stars, allow for distance estimates based on the luminosity-size relation. Calibrated using nearby galaxies, HII galaxy measurements provide independent constraints on $H_0$ consistent with other local probes.

\ac{desi} has provided a novel avenue for measuring $H_0$ through the Fundamental Plane (FP) relation of early-type galaxies, with a calibration tied to the distance to the Coma cluster (Sec.~\ref{sec:COMA}). By leveraging \ac{desi}'s extensive sample of over 4,000 early-type galaxies, the FP relation was calibrated to provide precise distance estimates. For a distance to the Coma cluster of $98.5 \pm 2.2$ Mpc based on \ac{sn1}, \ac{desi} yields a local value of $H_0 = 76.5 \pm 2.2$\kms. This result is in significant tension with the value inferred from Planck measurements where \lcdm\ is assumed, and which predicts a distance to Coma exceeding 110 Mpc. Historically, the distance to the Coma cluster has ranged from 90-100 Mpc, highlighting the discrepancy.  By extending the Hubble diagram with FP-calibrated distances, \ac{desi} highlights a persistent conflict between local measurements and early Universe predictions.

Extended \ac{qso} cosmologies (Sec.~\ref{sec:QSOs}) constrain $H_0$ using correlations between \ac{qso} luminosity and variability timescales. Their ability to probe higher redshifts than standard candles makes them valuable for testing cosmic expansion over extended epochs. Similarly, \ac{grb} observations (see Sec.~\ref{sec:GRBs}) use correlations between their luminosity and spectral properties to estimate $H_0$, serving as additional high-redshift probes relevant to the Hubble tension.

Despite significant methodological diversity and substantial precision improvements, the tension between early- and late-time measurements of the Hubble constant $H_0$ persists at a statistically significant level, exceeding $5.9\sigma$ (Planck 2018 vs. SH0ES, alone, but higher when combining measures). This discrepancy challenges the completeness of the standard \lcdm\ model and raises the possibility of new physics.

Related to the $H_0$ tension there is the sound horizon tension, which refers to a discrepancy in the inferred comoving sound horizon scale at the end of the baryon drag epoch, $r_s^{\rm drag}$, derived from early- and late-time cosmological measurements. This standard ruler, crucial for calibrating cosmological distances, is determined by the physics of the early Universe, specifically the acoustic oscillations in the photon-baryon plasma before recombination. It is primarily constrained by measurements of the \ac{cmb} power spectrum, where \textit{Planck} data assuming \lcdm\ suggest $r_s^{\rm drag} \approx 147.09 \pm 0.26$ Mpc. However, late-time measurements, such as \ac{bao} data combined with local distance ladder determinations of $H_0$, suggest a lower sound horizon, with \ac{bao}-based estimates yielding approximately 137 Mpc, a difference of about 7\% and a tension of $2.6\sigma$ significance. 

This discrepancy arises because $r_s^{\rm drag}$ is directly connected to the expansion rate around recombination, meaning changes to the early Universe physics could shift its value. Solutions involving early-time modifications often require an enhanced expansion rate before recombination, which can reduce $r_s^{\rm drag}$. Examples include models introducing additional relativistic species or \ac{ede} components. However, these scenarios face tight constraints from \ac{cmb} observations, as the acoustic peak structure is highly sensitive to changes in photon diffusion and gravitational driving effects. 

Late-time measurements rely on the \ac{bao} feature imprinted at the drag epoch, calibrated through local $H_0$ measurements, such as the SH0ES collaboration results. Since $r_s^{\rm drag}$ anchors the \ac{bao} scale, discrepancies in $H_0$ estimates propagate into inferred sound horizon measurements. This tension therefore highlights the need for consistent cross-calibration between early- and late-time probes and motivates further investigation into both systematics and extensions to the standard model of cosmology.

\begin{figure}[htbp]
    \centering
    \includegraphics[scale=1, trim=80 0 0 0, clip]{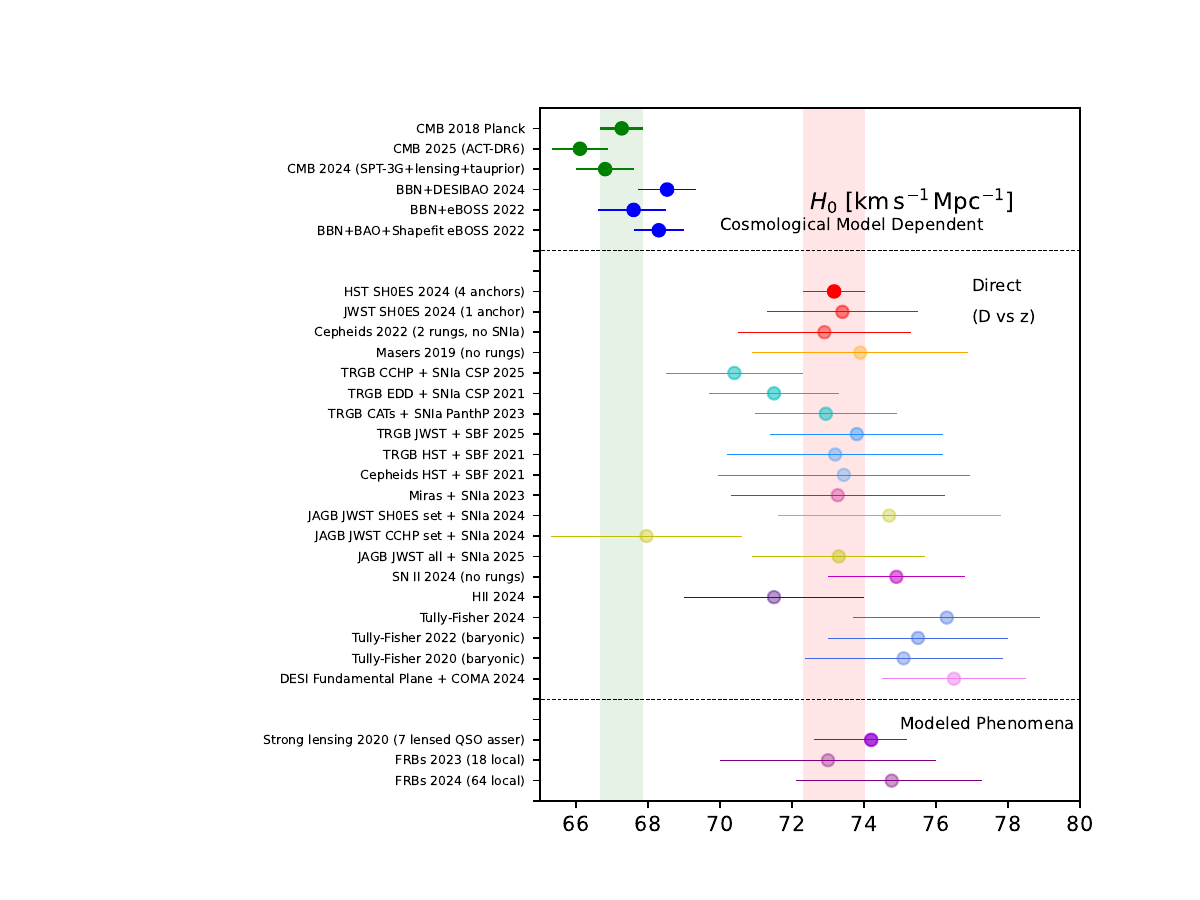}
    \caption{Summary of $H_0$ estimates from different cosmological probes with error bars smaller than $3.5$\kms.}
    \label{fig:Intro_H0_whisk}
\end{figure}

The combination of the estimates resulting from these, and other, approaches to estimating the value of $H_0$ points to a significant tension in the value of this parameter. This tension in the expansion rate of the Universe at current times is further detailed in Sec.~\ref{sec:obs_H0}, with the most prominent estimates shown in Fig.~\ref{fig:Intro_H0_whisk}.

\begin{figure}[htbp]
    \centering
    \includegraphics[scale=1, trim=80 0 0 0, clip]{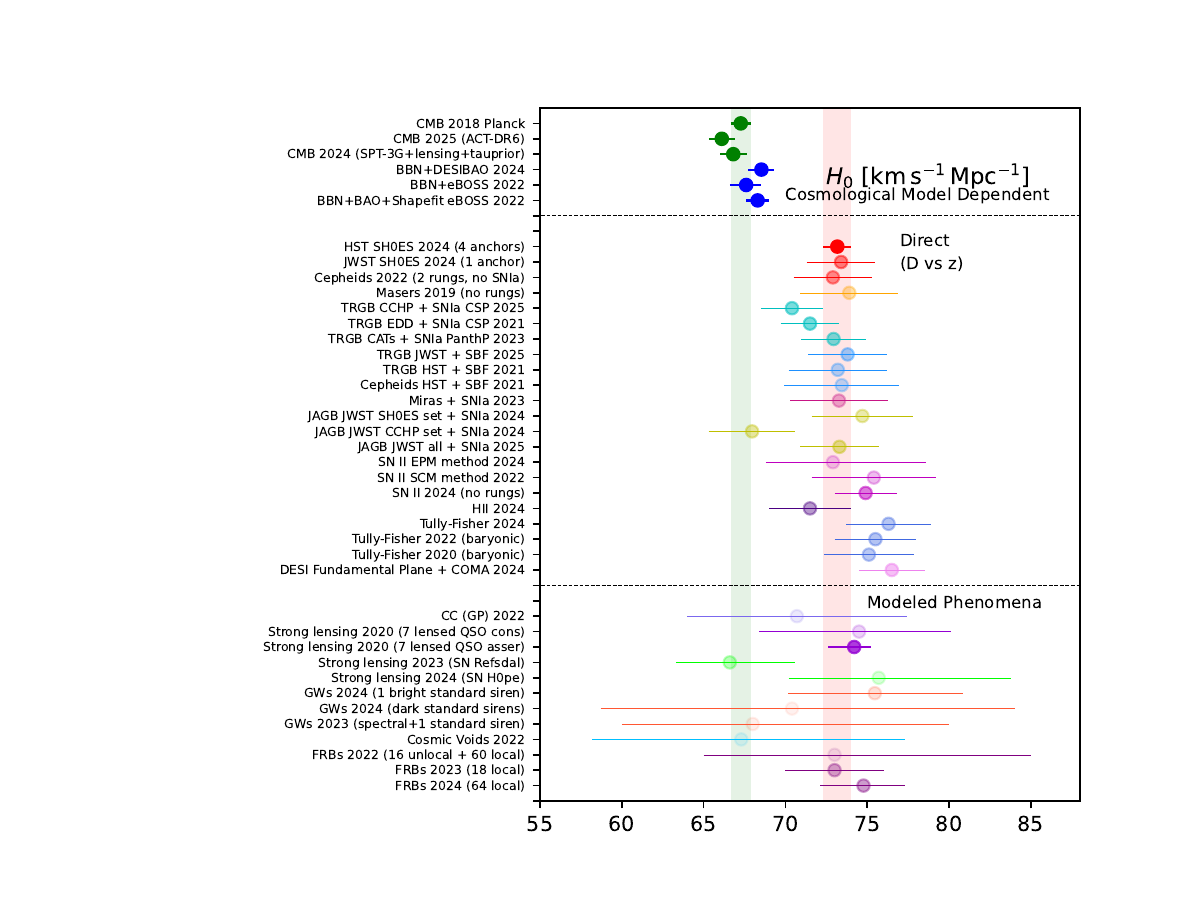}
    \caption{Summary of $H_0$ estimates from all the cosmological probes in this White Paper following the order of the different sections.}
    \label{fig:Intro_H0_whisk_full}
\end{figure}

\begin{figure}[htbp]
    \centering
    \includegraphics[scale=0.7]{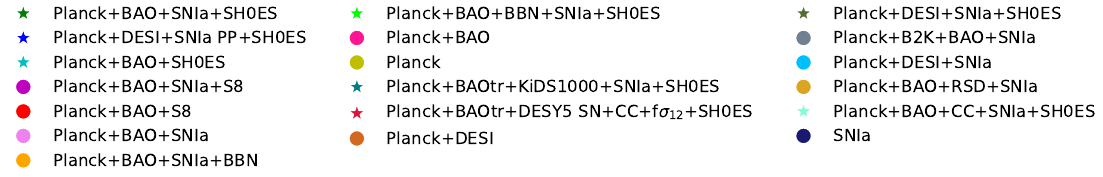}
    \includegraphics[scale=1, trim=80 0 0 30, clip]{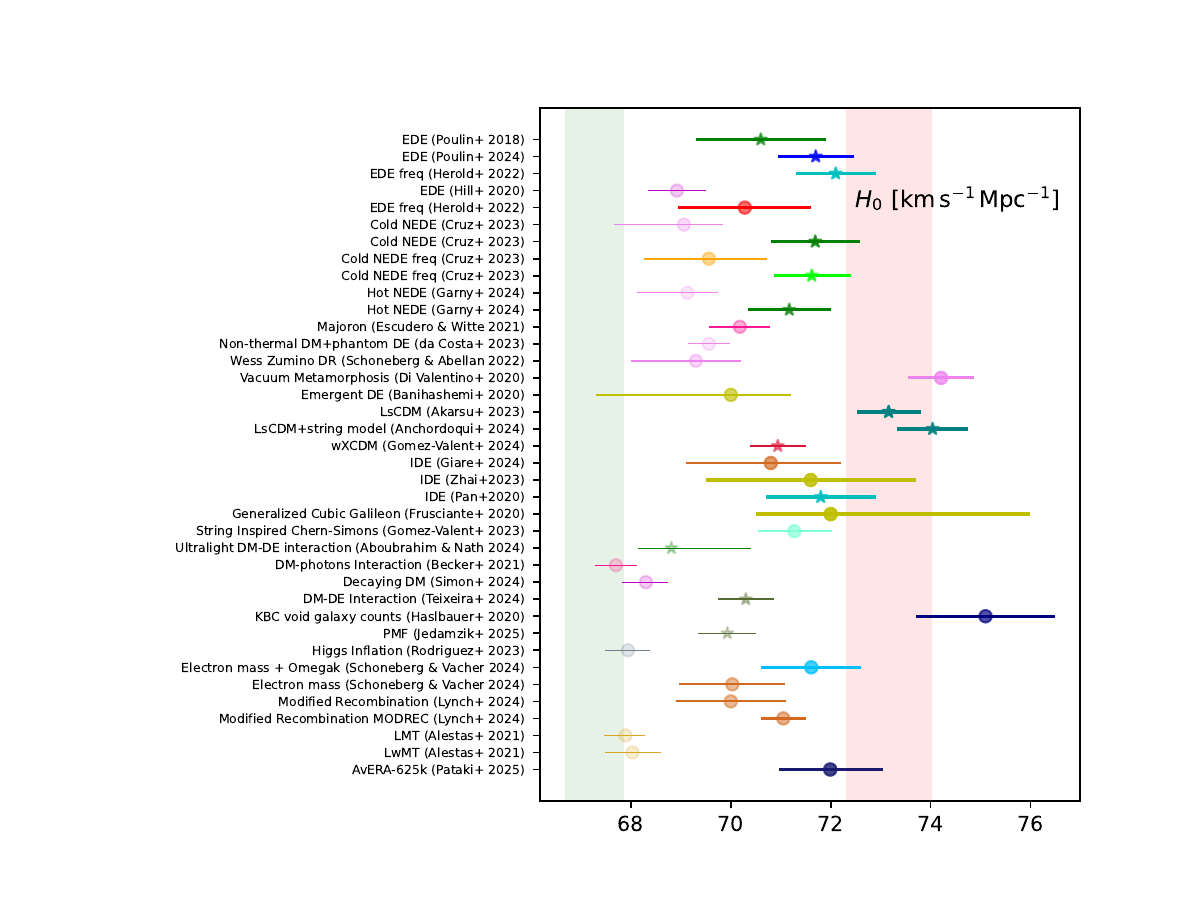}
    \caption{Summary of models proposed to solve the $H_0$ tension in this White Paper following the order of the different sections.}
    \label{fig:Intro_H0_whisk_solutions}
\end{figure}

Another interesting tension is that related to $S_8$, which highlights a persistent discrepancy between measurements of the amplitude of matter fluctuations on cosmological scales inferred from early and late Universe observations. The parameter $S_8$, defined as $S_8 \equiv \sigma_8 \sqrt{\Omega_m / 0.3}$, combines the clustering amplitude $\sigma_8$ (the root-mean-square of matter fluctuations on scales of $8 \, h^{-1}$ Mpc) with the present-day matter density parameter $\Omega_m$. It provides a key measure of the growth of cosmic structures, making it a crucial probe of the standard cosmological model.

Current measurements from early-time data, such as the \textit{Planck} 2018 results, which constrain the \ac{cmb} temperature and polarization anisotropies (Sec~\ref{sec:CMB_S8}), yield a high precision estimate of $S_8 = 0.834 \pm 0.016$ assuming the \lcdm\ model. However, late-time measurements derived from weak gravitational lensing (Sec.~\ref{sec:weak_lensing}) and galaxy clustering surveys (Sec.~\ref{sec:Gal_Clus}), including \ac{kids}, \ac{des}, and \ac{hsc}, consistently report lower values of $S_8$. For example, \ac{des} Year 3 results obtained from combined galaxy clustering and \ac{wl} measurements suggest $S_8 = 0.772 \pm 0.017$, while \ac{kids}-1000 and \ac{hsc} report similarly low values around $S_8 \approx 0.76$ with comparable uncertainties. This tension, at the $2-3\sigma$ level, persists across multiple independent data sets, indicating a possible systematic deviation between early and late-time probes of structure formation.

The origin of this tension remains under debate. On the one hand, it could be driven by systematic uncertainties in the analysis of \ac{wl} data. These systematics include shear calibration biases, uncertainties in photometric redshift estimates, and baryonic feedback effects that suppress the small-scale power spectrum due to processes like \ac{agn} feedback and gas ejection from galaxies. \ac{des} and \ac{kids} collaborations have both extensively explored the role of such systematics, yet the tension persists even after conservative scale cuts and improved modeling of non-linear clustering.

On the other hand, the $S_8$ tension might hint at the need for extensions to the standard \lcdm\ model. One potential modification involves evolving \ac{de} models, where a time-dependent equation of state parameter $w(z)$ could alter structure growth rates. Another possibility is modifications to the theory of gravity itself, such as $f(R)$ models or scalar-tensor theories, which could modify the relationship between the matter distribution and lensing potential. Additionally, the presence of massive neutrinos, which suppress structure formation at late-times due to their free-streaming behavior, could also contribute to lowering the observed $S_8$ value if the neutrino mass is larger than currently assumed in the base \lcdm\ model.

Notably, cross-correlation measurements between \ac{wl} and other probes, such as \ac{cmb} lensing (e.g., from Planck, \ac{act}, and \ac{spt}) and galaxy clustering from \ac{bao} surveys, have shown consistency with both early and late-time datasets, complicating the overall picture. Furthermore, analyses that vary only the normalization of the power spectrum, such as the $\sigma_8$ parameter itself, have not fully resolved the tension, suggesting a more complex interplay between cosmic structure formation and expansion history.

\begin{figure}[htbp]
    \centering
    \includegraphics[scale=1, trim=80 0 0 0, clip]{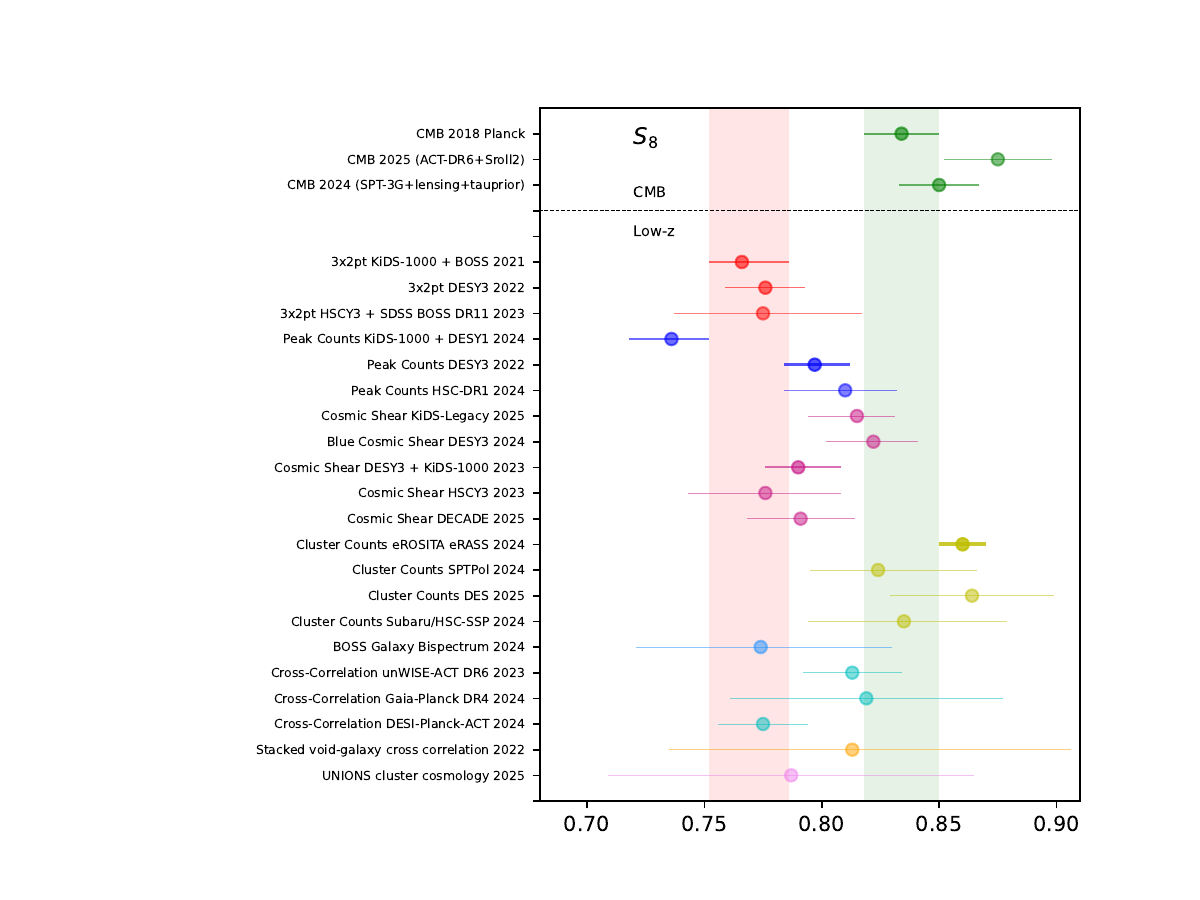}
    \caption{Summary of $S_8$ estimates from different cosmological probes.}
\label{fig:Intro_S8_whisk}
\end{figure}

The tension in the $S_8$ parameter may be milder but there is a growing effort to improve the constraints in observational estimates of the amplitude of matter fluctuations. These are detailed in Sec.~\ref{sec:S8tension_2.2}, with the most prominent estimates also shown in Fig.~\ref{fig:Intro_S8_whisk}.

\begin{figure}[htbp]
    \centering
    \includegraphics[scale=0.7]{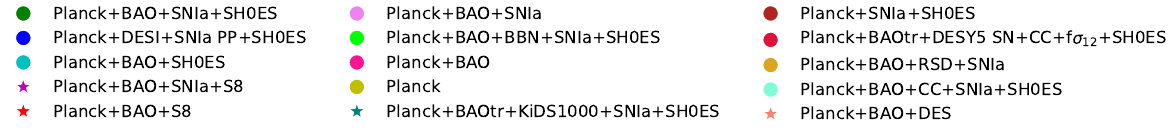}
    \includegraphics[scale=1, trim=80 0 0 30, clip]{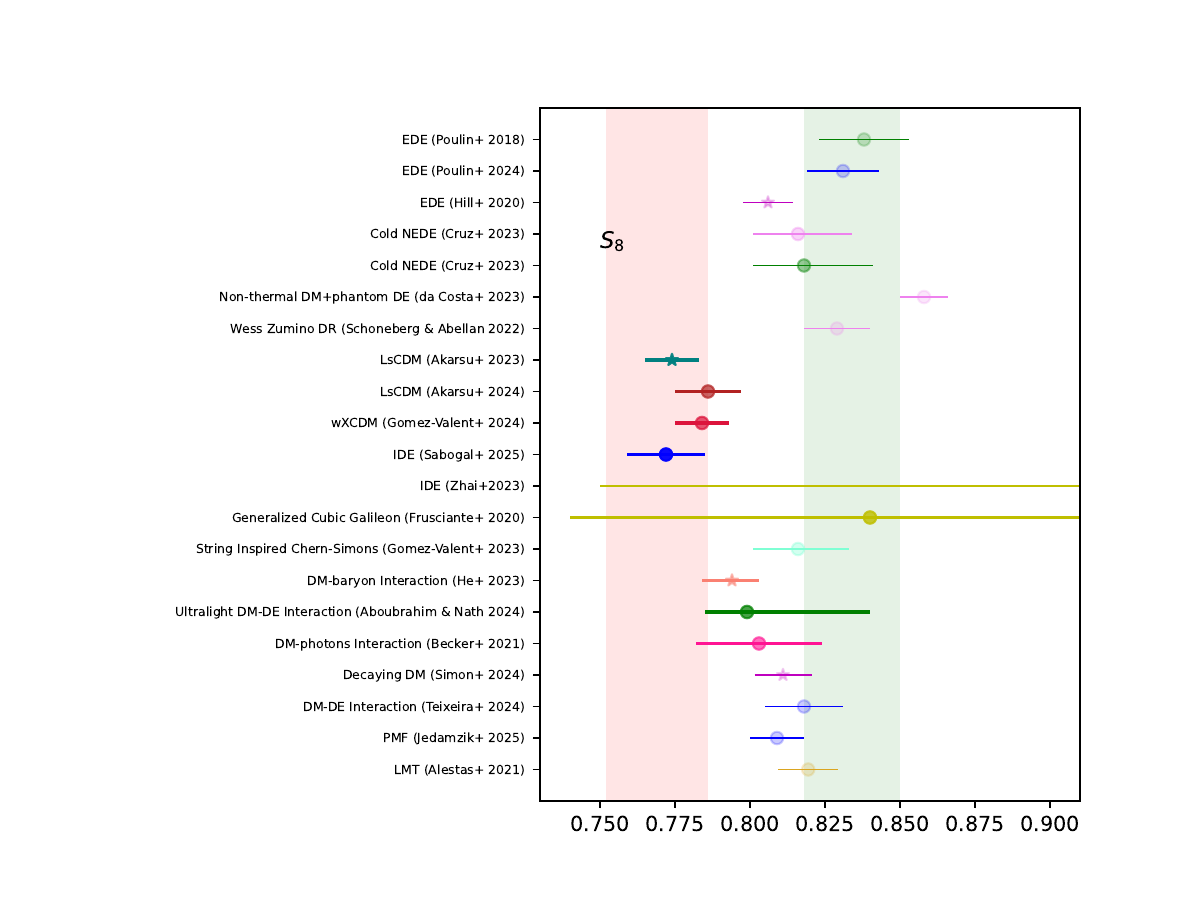}
    \caption{Summary of models proposed to solve the $S_8$ tension in this White Paper following the order of the different sections.}
    \label{fig:Intro_S8_whisk_solutions}
\end{figure}

\bigskip
\subsection{Early vs local measurements and efforts for a solution \label{sec:early_vs_late}}

\begin{figure}
\scalebox{0.95}{
\begin{tikzpicture}[mindmap, grow cyclic, every node/.style=concept, concept color=orange!40, 
	level 1/.append style={level distance=6cm,sibling angle=360/8},
	level 2/.append style={level distance=3cm,sibling angle=45}]

\node{Fundamental Physics Solutions}
child[concept color=blue!30, minimum size=1cm]{ node {Early-time solutions}
	child { node {Early dark energy}}
	child { node {New early dark energy}}
	child { node {Extra degrees of freedom (neutrino physics)}}
}
child[concept color=blue!30, minimum size=1cm]{ node {Varying fundamental constants}
}
child[concept color=yellow!30, minimum size=1cm]{ node {Late-time solutions}
	child { node {Late dark energy}}
	child { node {Dark energy transition models}}
	child { node {Interacting dark energy}}
}
child[concept color=gray!30, minimum size=1cm]{ node {Deviations from the Cosmological Principle}
}
child[concept color=teal!40, minimum size=1cm]{ node {Modified gravity}
	child { node {Higher order models}}
	child { node {Non-local models}}
	child { node {Higher dimensional models}}
	child { node {Non-Riemanian geometries}}
}
child[concept color=brown!30, minimum size=1cm]{ node {Quantum gravity phenomenology}
	child { node {Quantum gravity modified dynamics}}
	child { node {Quantum gravity effects on particles and fields}}
}
child[concept color=purple!50, minimum size=1cm]{ node {Matter sector solutions}
	child { node {cold dark matter}}
	child { node {Warm dark matter}}
	child { node {Interacting and decaying dark matter}}
}
child [concept color=violet!40, minimum size=1cm]{ node {Other solutions}
            child { node {Local voids} }
            child { node {Primordial magnetic fields} }
            child { node {Inflationary models} }
            };
\end{tikzpicture}
}

\caption{Fundamental physics provides a number of potential solutions to address the challenge of tensions in cosmology.}
\label{fig:IntroFunSols}
\end{figure}
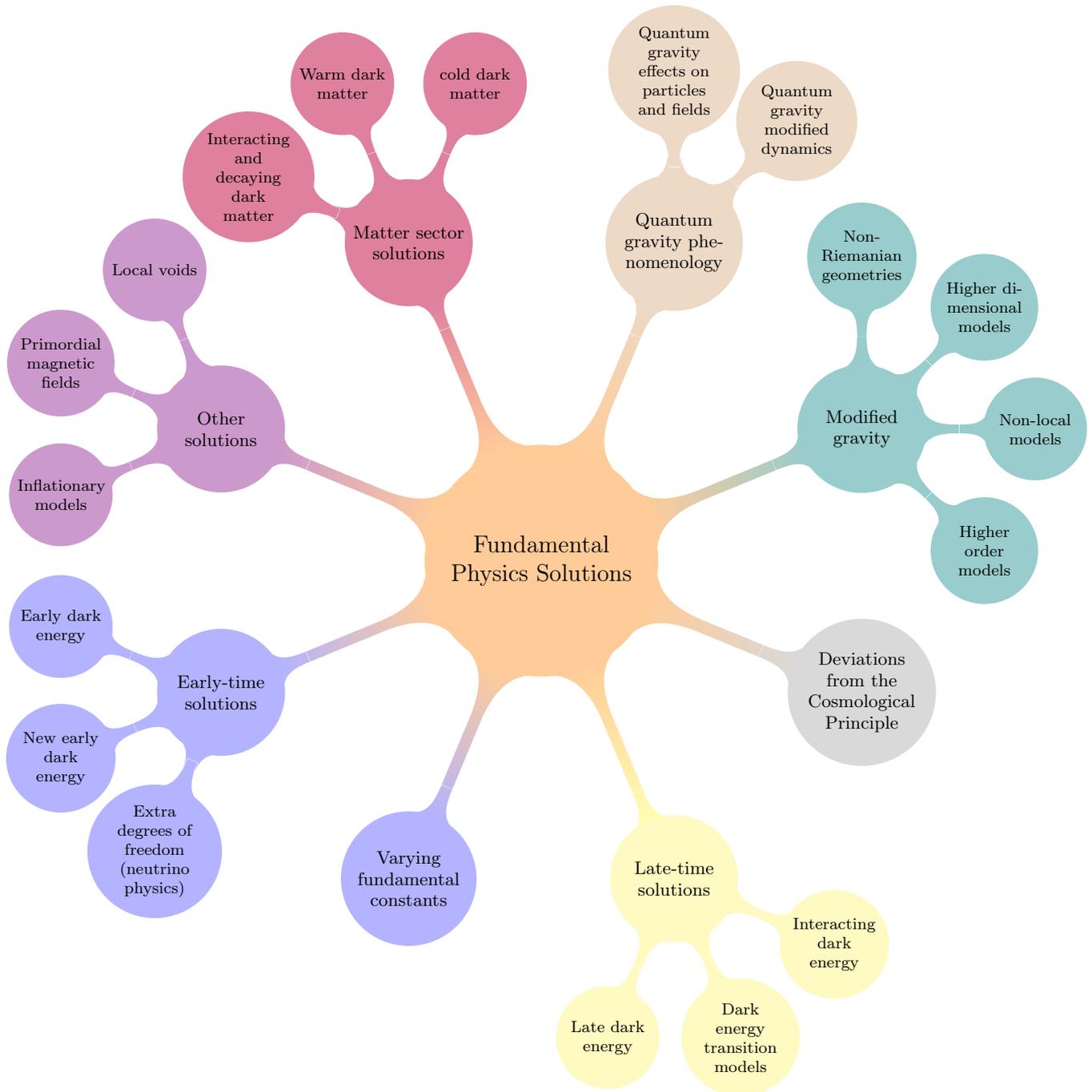

The persistence of the $H_0$ tension across independent datasets suggests either unresolved systematic errors or the need for extensions to the \lcdm\ framework. Numerous theoretical solutions have been proposed, aiming to reconcile the observed discrepancy while preserving the success of the standard cosmological model in describing early and late-time observables. The most prominent of these efforts for a solution are shown in Fig.~\ref{fig:IntroFunSols}, which are also highlighted below, and detailed in Sec.~\ref{sec:fun_phys}:

\begin{itemize}
    \item \textbf{Early Dark Energy (EDE)}: \ac{ede} models (Sec.~\ref{sec:EDE}) propose a transient \ac{de} component that briefly dominates the cosmic energy budget before recombination, reducing the size of the sound horizon and allowing for a higher inferred $H_0$ from \ac{cmb} data. This mechanism involves a scalar field with a potential that activates temporarily and dilutes before significantly affecting late-time cosmology. Several variants of \ac{ede} have been explored in the literature. Oscillatory \ac{ede} models, often motivated by axion-like fields, involve a scalar field oscillating around the recombination epoch, injecting energy into the expansion rate. \ac{nede} (see Sec.~\ref{sec:NEDE}) refines this idea by introducing a phase transition where the scalar field rapidly decays, creating a sharper and more controlled impact on the expansion rate. Adiabatic Fluctuation \ac{ede} models, meanwhile, adjust the evolution of the scalar potential to balance early- and late-time cosmological constraints, though these often introduce parameter degeneracies. While \ac{ede} models have shown promise in alleviating the $H_0$ tension by increasing the inferred value from early-time data, they also introduce challenges. The models can exacerbate tensions in other cosmological parameters, particularly the amplitude of matter fluctuations $S_8$ and the matter density $\Omega_m$. Depending on the specific data set, \ac{ede} can reduce the $H_0$ tension to $2-3\sigma$, however, recent \ac{cmb} and \ac{lss} data sets show no evidence for this model and disfavor large fractions of \ac{ede}. 

    \item \textbf{Late Dark Energy (LDE)}: \ac{lde} models (Sec.~\ref{sec:Late_DE}) attempt to address the Hubble tension by modifying the expansion history at low redshifts ($z \lesssim 1$). These models alter the equation of state parameter of \ac{de}, $w(z)$, away from the cosmological constant value $w = -1$. Some \ac{lde} scenarios involve a rapid transition in $w(z)$, where \ac{de} density evolves from a negative to a positive contribution, effectively accelerating the late-time expansion rate. Although \ac{lde} models can slightly raise $H_0$ without modifying the early Universe physics, they often face challenges in matching \ac{bao} and Type Ia supernova data simultaneously, as the shift in the expansion rate can introduce tension with large-scale structure constraints. 

    \item \textbf{Rapid transitions in the late-Universe}:
    Rapid transitions in the \ac{de} from negative to positive values in the late Universe (Sec.~\ref{sec:Grad_DE}) exhibit a sign-switching action in the vacuum energy while leaving the \ac{de} magnitude unchanged. These models allow for the possibility of simultaneously increasing the value of $H_0$ while also suppressing changes in $S_8$. The transition point is established primarily by \ac{bao} data sets, while also being supported by other contributions. Most models express an abrupt transition with a discontinuity in the vacuum energy density.
    
    \item \textbf{Interacting Dark Matter and Dark Energy (IDE) Models}: Interacting Dark Matter and \ac{ide} models (Sec.~\ref{sec:IDE}) propose a non-trivial energy exchange between \ac{dm} and \ac{de}, modifying both the cosmic expansion rate and structure growth. This interaction, described by a coupling function $Q$, allows energy to transfer from one component to the other. Depending on the direction and strength of the coupling, \ac{ide} can either slow down or accelerate the expansion rate. \ac{ide} models can increase $H_0$ estimates while modifying the growth of structure, offering a way to reduce the $S_8$ tension simultaneously. However, strong constraints from \ac{cmb} lensing and \ac{bao} measurements limit the parameter space of \ac{ide} models, and they often require fine-tuning to remain consistent with multiple datasets.

    \item \textbf{Modified Gravity Theories}: \ac{mg} models (Sec.~\ref{sec:MoG} and Sec.~\ref{sec:Early_mod_grav}) propose extensions to \ac{gr} by altering the Einstein-Hilbert action, introducing additional scalar degrees of freedom, altering the underlying geometry itself, or changing the spacetime dimensionality, among other alternatives, which can affect both the cosmic expansion rate and the growth of large-scale structure. These modifications can impact the lensing potential and structure formation, leading to changes in the inferred values of $H_0$ and $S_8$. Several frameworks have been explored to address the Hubble tension within \ac{mg}. $f(R)$ gravity, for example, generalizes \ac{gr} by replacing the Ricci scalar with a nonlinear function of $R$, introducing an additional scalar mode that modifies both the background expansion and structure growth. Horndeski theories, which include non-minimally coupled scalar fields with derivative interactions, allow modifications to both the expansion history and lensing effects but often face constraints from \ac{cmb} lensing and large-scale structure data. Other models, such as massive gravity and bimetric gravity, alter the graviton's properties, leading to modified cosmic acceleration and structure formation patterns. While some variants can reduce both the $H_0$ and $S_8$ tensions, they often require fine-tuning or introduce instabilities in the late Universe. Teleparallel gravity approaches, including $f(T)$ and $f(Q)$ theories, replace curvature with torsion or non-metricity as the primary geometrical property driving cosmic evolution. These models have been shown to affect the late-time expansion rate but remain under scrutiny for consistency with both early- and late-time datasets. Emerging curvature theories, such as AvERA \cite{Racz:2016rss,Beck:2018owr,Racz:2016rss} modify the late expansion history compared to the concordance model (Sec. ~\ref{sec:ISW}) to solve the $H_0$ and ISW puzzles. While \ac{mg} models offer a rich theoretical landscape to explore, many face difficulties in simultaneously resolving the $H_0$ and $S_8$ tensions while remaining consistent with solar system tests, \ac{cmb} lensing, and \ac{bao} constraints.
    
    \item \textbf{Exotic Scenarios and Non-Standard Dark Matter Models}: Exotic scenarios (Sec.~\ref{sec:CDM}, Sec.~\ref{sec:WDM}, and Sec.~\ref{sec:I_D_DM}) explore extensions to the standard cosmological model involving non-standard physics in both the \ac{dm} and \ac{de} sectors, which can influence the cosmic expansion history and structure formation. Among these are decaying dark matter (DDM) models, where \ac{dm} particles decay into lighter states. If this decay occurs before or during the recombination epoch, it can alter the expansion rate and reduce the sound horizon, leading to a higher inferred $H_0$. However, constraints from \ac{cmb} lensing and large-scale structure limit the viability of DDM as a complete solution. Ultralight scalar fields, such as axion-like particles or fuzzy dark matter, have also been proposed. These fields can modify the standard expansion rate through their contributions to the cosmic energy density, potentially mimicking an additional relativistic species or altering structure formation on small scales. Nevertheless, precise measurements of the \ac{cmb} power spectrum and the Lyman-alpha forest have placed stringent bounds on their properties. Primordial black holes and non-\ac{cdm} models represent further exotic avenues, with the latter deviating from the standard cold, collisionless \ac{dm} paradigm, often leading to a suppression of small-scale structure growth. While these models offer novel mechanisms to address cosmological tensions, most current models are tightly constrained by precision data from Planck, \ac{desi}, and the Pantheon+ supernova sample. However, they remain of theoretical interest, especially with upcoming data from \ac{cmb} Stage-4 and \ac{jwst} capable of probing extreme scenarios further.
    
    \item \textbf{Extra Relativistic Species and Neutrino Physics}: Extra relativistic species (Sec.~\ref{sec:Extra_DoF}) represent another class of proposed extensions aimed at resolving the Hubble tension. An increased number of relativistic degrees of freedom, commonly parametrized by $N_{\rm eff}$, leads to a higher early Universe expansion rate and a reduced sound horizon, potentially increasing the inferred $H_0$. This scenario can arise from new light particles, such as sterile neutrinos or dark radiation. Sterile neutrinos, in particular, can act as an additional relativistic species if they decouple before standard neutrinos while self-interacting or secret neutrino interactions can modify the thermal history of the Universe and delay neutrino decoupling, effectively altering $N_{\rm eff}$. However, current constraints from Planck, \ac{bao}, and \ac{cmb} lensing data have limited viable deviations to $\Delta N_{\rm eff} \approx 0.2$. While insufficient to fully resolve the Hubble tension, such scenarios remain relevant in the context of early Universe physics and potential beyond-the-Standard-Model extensions.

    \item \textbf{Local Void Hypothesis:} The local void hypothesis (Sec.~\ref{sec:Local_voids}) posits that the \ac{mw} resides in a large, underdense region of the Universe, potentially biasing measurements of the Hubble constant due to a locally faster expansion rate compared to the cosmic average. The idea suggests that a void could lead to a larger measured $H_0$ locally while being lower on cosmological scales. However, the size and depth of such a void required to explain the entire Hubble tension would be inconsistent with current observations. Studies using CosmicFlows-4 and Pantheon+ supernova data, as well as analyses of bulk flows, have not identified an underdensity significant enough to fully explain the observed discrepancy assuming \lcdm.
    
    \item \textbf{Primordial Magnetic Fields (PMFs):} \ac{pmf} (Sec.~\ref{sec:Pri_mag_fields}) contributions are generated before recombination could influence the early Universe's expansion history and the formation of large-scale structures. \ac{pmf} contributions can modify the photon-baryon plasma dynamics, alter the acoustic peaks in the \ac{cmb} power spectrum, and shift the inferred sound horizon, all of which can affect the measurement of $H_0$. However, current constraints from the \ac{cmb}, including non-detections of Faraday rotation and the damping tail suppression, limit the strength of \ac{pmf}s, making it unlikely that they can fully account for the observed tension.
    
    \item \textbf{Inflationary Models:} Certain inflationary scenarios (Sec.~\ref{sec:inflation}) propose modifications to the early Universe's physics that could indirectly affect the $H_0$ measurement. Models involving non-standard reheating phases, features in the inflationary potential, or a modified primordial power spectrum have been investigated in this context. These scenarios can alter the early expansion history or the acoustic peak structure, impacting the inferred value of $H_0$ from \ac{cmb} measurements. However, most inflationary modifications struggle to generate a significant enough shift in $H_0$ while remaining consistent with current \ac{cmb} and large-scale structure data.
    
    \item \textbf{Varying Fundamental Constants:} The idea of varying fundamental constants (Sec.~\ref{sec:Var_fun_const}) explores the possibility that parameters such as the fine-structure constant $\alpha$, the electron mass $m_e$, or the proton-to-electron mass ratio $\mu$ could change over cosmic time or with spatial position. If fundamental constants were to evolve, they could impact key cosmological observables, such as the sound horizon and the physics of recombination, altering the inferred values of $H_0$. Scalar fields coupled to the electromagnetic sector, as in models like varying-$\alpha$ theories, could drive such changes. However, stringent constraints from \ac{cmb}, \ac{bbn}, and \ac{qso} absorption spectra have significantly limited the variation of some of these constants and their ability to resolve the Hubble tension, while the others face issues of finding a coherent theoretical framework that can explain their variation.
    
    \item \textbf{Local Physics Solutions:} Local physics solutions (Sec.~\ref{sec:Local_sols}) suggest that the $H_0$ tension could arise from new physical effects specific to the local Universe rather than requiring modifications to the global cosmological model. One possibility is a local modification to the gravitational constant, $G_{\rm eff}$, which could alter the calibration of standard candles like Cepheid variables or \ac{sn1}, leading to a biased measurement of $H_0$. Another proposed local effect involves variations in the transparency of the \ac{ism} or modified extinction laws that could influence supernova observations. While some of these scenarios can explain a portion of the tension, they generally fail to provide a complete resolution and are often tightly constrained by measurements from cosmic bulk flows, galaxy clustering, and large-scale structure surveys.

    \item \textbf{Systematic Uncertainties and Calibration Issues}: A significant portion of the community continues to explore whether systematic errors could account for the observed tension. Potential sources include Cepheid calibration errors, host galaxy dust corrections, and selection effects in both early and late-time measurements. However, rigorous cross-calibrations between \ac{hst}, \ac{jwst}, and ground-based surveys like \ac{desi} and SH0ES have yet to identify a significant bias capable of resolving the tension completely.

\end{itemize}

\bigskip
\subsection{Additional curiosities and anomalies \label{sec:cur_and_ano}}

Other cosmological anomalies beyond the primary $H_0$ and $S_8$ tensions have emerged in recent years, revealing potential challenges to the standard \lcdm\ model. These discrepancies span a wide range of physical scales and cosmological probes, motivating deeper investigations into both systematic effects and new physics. Key anomalies include:

\begin{itemize}
    \item \textbf{The $A_{\rm lens}$ anomaly:} The $A_{\rm lens}$ parameter, introduced as a phenomenological extension to the standard \ac{cmb} analysis, quantifies the amplitude of lensing-induced smoothing of acoustic peaks in the \ac{cmb} power spectrum. Planck data suggest a value for $A_{\rm lens}$ slightly higher than expected under \lcdm, with a significance of $2-3\sigma$, potentially pointing to unmodeled effects in \ac{cmb} lensing or physics beyond standard cosmology (Sec.~\ref{sec:syst_and_A_lens}).

    \item \textbf{Evidence for nonzero curvature:} While the standard model assumes a flat universe, Planck's \ac{cmb} data, when analyzed without external priors, have shown mild evidence for a closed geometry with $\Omega_k < 0$. However, this claim remains controversial, as \ac{bao} and lensing measurements are consistent with a flat universe within $1\sigma$ (Sec.~\ref{sec:nonzero_curv}).

    \item \textbf{\ac{cmb} anisotropic anomalies:} Several statistically significant features have been identified in \ac{cmb} data, including the hemispherical power asymmetry, alignments in low multipole moments, and the cold spot anomaly. While most of these could be attributed to cosmic variance, their persistence in multiple datasets (e.g., Planck, \ac{wmap}) raises the question of whether they hint at new physics or residual systematics (Sec.~\ref{sec:Anomalies_CMB}).

    \item \textbf{The $w_0w_a$ tension:} Discrepancies have emerged in constraints on the \ac{de} equation of state parameters, $w_0$ and $w_a$, particularly when comparing results from \ac{cmb} data with late-time probes such as \ac{sn} and \ac{desi} \ac{bao}. Some results suggest a deviation from the cosmological constant value of $w_0 = -1$, which could indicate evolving \ac{de} or unresolved systematics (Sec.~\ref{sec:w0wa}).

    \item \textbf{Neutrino Tension:} The sum of neutrino masses, $\sum m_\nu$, inferred from cosmological data, such as \ac{cmb} and \ac{bao} observations, has shown tension with the lower bounds set by terrestrial neutrino oscillation experiments, which require $\sum m_\nu \gtrsim 0.06$ eV (normal ordering, NO) and $\sum m_\nu \gtrsim 0.1$ eV (inverted ordering, IO). Recent \ac{desi} \ac{bao} data combined with Planck and \ac{act} \ac{cmb} constraints yield tight upper limits, such as $\sum m_\nu < 0.05$ eV ($2\sigma$), challenging the inverted hierarchy and preferring the NO with a Bayes factor exceeding 46.5 in some datasets. Interestingly, some analyses allowing for extended models have reported a preference for \emph{negative} neutrino masses, linked to enhanced \ac{cmb} lensing signals in Planck and \ac{act} data. This tension could stem from unaccounted systematics in lensing measurements or may indicate new physics beyond minimal \lcdm, such as modified neutrino properties or evolving \ac{de} models (Sec.~\ref{sec:Neutri_Ten}).

    \item \textbf{Cosmic dipole anomalies:} Tensions exist between the inferred cosmic dipole from the \ac{cmb} and measurements from radio galaxies and \ac{qso}s. Separately, statistically significant directional variations in $H_0$ have been reported in both \ac{cmb} and the local Universe, providing an alternative perspective on \ac{cmb} anisotropic anomalies and bulk flow anomalies. These discrepancies could point to non-standard cosmic expansion or incomplete modeling of large-scale structure effects (Sec.~\ref{sec:dipoles}).

    \item \textbf{Big Bang Nucleosynthesis (BBN) anomalies:} Discrepancies exist between deuterium and helium abundances inferred from \ac{bbn} and those predicted by \ac{cmb}-based baryon density estimates, suggesting either measurement systematics or small deviations from standard physics during the early Universe (Sec.~\ref{sec:BBN}).

    \item \textbf{Lyman-$\alpha$ anomalies:} High-redshift Lyman-$\alpha$ forest measurements show mild tension with low-redshift constraints on the matter power spectrum, raising questions about the evolution of structure and thermal history of the \ac{igm} (Sec.~\ref{sec:Anomalies_lyman_alpha}).

    \item \textbf{Integrated Sachs-Wolfe (ISW) and cosmic superstructures:} Measurements of the \ac{isw} effect, which probes late-time structure growth through correlations between \ac{cmb} and large-scale structure data, have shown excess correlations on large scales, potentially linked to cosmic superstructures (Sec.~\ref{sec:ISW}).

    \item \textbf{Cosmic void anomalies:} Observations of cosmic voids have revealed unexpected properties in their size distribution and their contribution to the \ac{isw} effect, potentially suggesting deviations from the standard growth of structure (Sec.~\ref{sec:voids}).

    \item \textbf{Fast Radio Burst (FRB) probes:} \ac{frb} observations have recently emerged as a powerful probe of cosmic tensions, offering a unique way to constrain the expansion history and large-scale bulk flows. Preliminary results suggest some inconsistencies with early universe predictions. It is crucial to emphasize that in estimating $H_0$ from FRBs, a prior assumption of the underlying cosmological model is required. Specifically, the $H_0$ values presented in the associated Whisker plot are computed under the assumption of a $\Lambda$CDM cosmology. Deviations from the standard $\Lambda$CDM framework are expected to yield different $H_0$ estimates, reflecting the model dependence inherent in such measurements. Furthermore, the constraints on $H_0$ have progressively tightened over time, primarily due to the increasing number of well-localized \ac{frb}s and improvements in the statistical methodologies applied to their analysis (Sec.~\ref{sec:FRBs}).

    \item \textbf{Radio background excess:} Several observations of the diffuse radio sky have reported an unexplained excess in both the surface brightness and the anisotropy power of the radio background, which could be linked to new diffuse or low flux radio sources, \ac{pmf}s, exotic decays, or residual systematics across a range of radio observations (Sec.~\ref{sec:radio_background_excess}).

    \item \textbf{Bulk flow anomalies:} Peculiar velocity measurements and bulk flows inferred from galaxy catalogs show a higher amplitude than predicted by \lcdm, potentially pointing to the presence of large-scale structures beyond the standard model (Sec.~\ref{sec:bulk_flow}).

    \item \textbf{Ultra Long Period Cepheids:} These variable stars have been proposed as potential standard candles, but their properties appear inconsistent with standard stellar evolution models, raising questions about distance ladder calibrations (Sec.~\ref{sec:ULPC}).
\end{itemize}

Ongoing and upcoming surveys such as Euclid, The Rubin Observatory's \ac{lsst}, and the Roman Space Telescope will play a pivotal role in clarifying these anomalies. Additionally, methodological advances, including cross-correlations between multiple probes and the application of \ac{ml} techniques (Sec.~\ref{sec:data_ana}, Sec.~\ref{sec:ML_inference}--\ref{sec:Inference_cosmic_sim}), will help assess whether these tensions are genuine physical discoveries or the result of residual systematics. A deeper understanding of these anomalies is essential to determine whether extensions to the standard \lcdm\ framework are required and to uncover the fundamental nature of \ac{de}, \ac{dm}, and cosmic structure formation.

\bigskip
\subsection{Data analysis: How to tackle the problem \label{sec:data}}

The analysis of cosmological data has seen remarkable advancements, leveraging a wide range of statistical, computational, and interdisciplinary approaches. These methodologies have not only refined parameter constraints but also opened new avenues to tackle cosmological tensions such as the $H_0$ and $S_8$ discrepancies. These tensions challenge the completeness of the $\Lambda$CDM model, suggesting the possibility of new physics or unresolved systematics.

Cosmology simulators and \ac{mcmc} techniques (Sec~\ref{sec:MCMC}) are foundational tools for exploring high-dimensional parameter spaces. They allow for robust testing of cosmological models against observational data while accounting for degeneracies and systematics. Advances in \ac{mcmc} methods, such as nested sampling and Hamiltonian Monte Carlo, have improved the efficiency and accuracy of parameter estimation, particularly when combining data from the \ac{cmb}, \ac{bao}, and \ac{sn} datasets.

\ac{ml} approaches (Sec.~\ref{sec:ML_inference}) have emerged as powerful tools in cosmological analysis. Neural networks, \ac{gp}, and decision trees have demonstrated their utility in extracting patterns, accelerating data analysis, and refining model predictions. For example, \ac{ml} techniques have been used to process large-scale structure data and improve constraints on \ac{de} models. Hybrid approaches, combining \ac{ml} with traditional statistical methods, have further enhanced the reliability of these analyses by integrating the strengths of both paradigms.

Reconstruction methods (Sec.~\ref{sec:Recon_tech}) have also played a critical role, enabling the inference of the Universe's expansion history from observational data. Techniques such as \ac{gp} regression have been applied to reconstruct the Hubble parameter $H(z)$ and the growth of cosmic structures without assuming a specific cosmological model. These methods reduce systematic biases and provide model-independent insights into cosmological tensions.

Bio-inspired algorithms, particularly \ac{ga} (Sec.~\ref{sec:GA_selection}), have introduced novel ways to optimize model selection and parameter estimation in cosmology. These algorithms, inspired by evolutionary processes in nature, use mechanisms such as selection, mutation, and crossover to identify optimal solutions in complex and high-dimensional parameter spaces.
\ac{ga}s are particularly well-suited for exploring cosmological models where traditional gradient-based methods may struggle, such as those involving nonlinear dynamics or multimodal likelihood functions. By iteratively refining a population of candidate solutions, \ac{ga}s can efficiently locate regions of interest in the parameter space, even in the presence of degeneracies or non-Gaussian distributions.
Applications of \ac{ga}s in cosmology have been diverse. They have been employed to address key tensions, such as the $H_0$ and $S_8$ discrepancies, by testing alternative models including \ac{ede}, \ac{idm}-\ac{de} scenarios, and \ac{mg} theories. For instance, \ac{ga}s have been used to explore the parameter space of \ac{ide} models, identifying regions that minimize tensions with both early- and late-time observations. Their flexibility allows for the incorporation of priors from \ac{ml} or other inference techniques, further enhancing their robustness.
Moreover, hybrid approaches that combine \ac{ga}s with \ac{ml} have shown promise in improving the convergence and accuracy of solutions. These methods leverage the pattern recognition capabilities of neural networks or decision trees to guide the evolutionary search process, thereby reducing computational overhead while maintaining accuracy.
In addition to model testing, \ac{ga}s have proven effective in enhancing observational strategies. By simulating survey configurations and optimizing the allocation of observational resources, they help maximize the scientific output of upcoming missions. This includes optimizing survey strategies for \ac{wl}, galaxy clustering, and other probes of large-scale structure.
The adaptability of \ac{ga}s also extends to their role in reconstructing the initial conditions of the Universe from observed data. By evolving populations of initial conditions, these algorithms can identify those that best reproduce observed structures, offering insights into the fundamental physics of the early Universe.
Overall, bio-inspired algorithms like \ac{ga}s represent a powerful addition to the cosmological toolkit, enabling the exploration of complex models and the refinement of observational strategies. Their continued integration with other advanced methods, such as \ac{ml} and high-performance simulations, holds great potential for addressing the most pressing tensions in cosmology and uncovering new physics.

Cosmological simulations (Sec.~\ref{sec:Inference_cosmic_sim}) have become indispensable for interpreting observational data and testing theoretical models. High-resolution $N$-body simulations, hydrodynamic models, and semi-analytic methods provide a detailed understanding of the large-scale structure formation and the interplay of \ac{dm} and \ac{de}. These simulations enable the calibration of observables such as \ac{wl} signals, galaxy clustering, and \ac{bao} measurements, thereby improving the accuracy of cosmological parameter constraints.

Statistical tools, including profile likelihoods, have gained prominence in cosmology for estimating confidence intervals and assessing parameter significance. Unlike traditional Bayesian methods, profile likelihoods provide a frequentist alternative (Sec.~\ref{sec:frequen_approach}) that is less sensitive to prior assumptions. This makes them particularly useful in scenarios where prior information is limited or where strong assumptions may bias the results.
Profile likelihoods excel in analyzing models with complex parameter spaces, including those with non-Gaussian distributions or significant degeneracies between parameters. Such complexities often arise in cosmological contexts, for instance, when exploring \ac{ede} models, \ac{ide} scenarios, or the inclusion of additional relativistic species. These models frequently introduce additional degrees of freedom, leading to correlations that can obscure parameter constraints when traditional Bayesian approaches are employed.
A key advantage of the profile likelihood approach is its ability to disentangle degeneracies and provide robust confidence intervals without relying on the full posterior distribution. This is achieved by profiling the nuisance parameters, effectively marginalizing them without requiring explicit integration. As a result, the profile likelihood method can highlight the true parameter dependencies and offer more transparent interpretations of the data.
Applications of profile likelihoods in cosmology have included constraining the equation of state of \ac{de}, testing deviations from the standard $\Lambda$CDM model, and evaluating the statistical significance of cosmological tensions, such as those involving $H_0$ and $S_8$. For example, in the context of \ac{ede}, profile likelihoods have been used to assess whether parameter shifts can resolve these tensions or whether they indicate systematic effects in the data.
The computational efficiency of profile likelihoods also makes them a valuable tool in high-dimensional analyses, particularly for next-generation cosmological surveys. Their ability to isolate relevant subspaces of parameter space enables efficient exploration of hypotheses while minimizing the computational cost associated with evaluating full posterior distributions.
Furthermore, recent advances in computational methods have improved the applicability of profile likelihoods. Techniques such as adaptive mesh refinement and \ac{ml}-assisted likelihood evaluations have reduced the computational demands of high-precision analyses. These innovations make profile likelihoods increasingly attractive for analyzing the large datasets expected from upcoming missions like the Simons Observatory, Euclid, and the Roman Space Telescope.

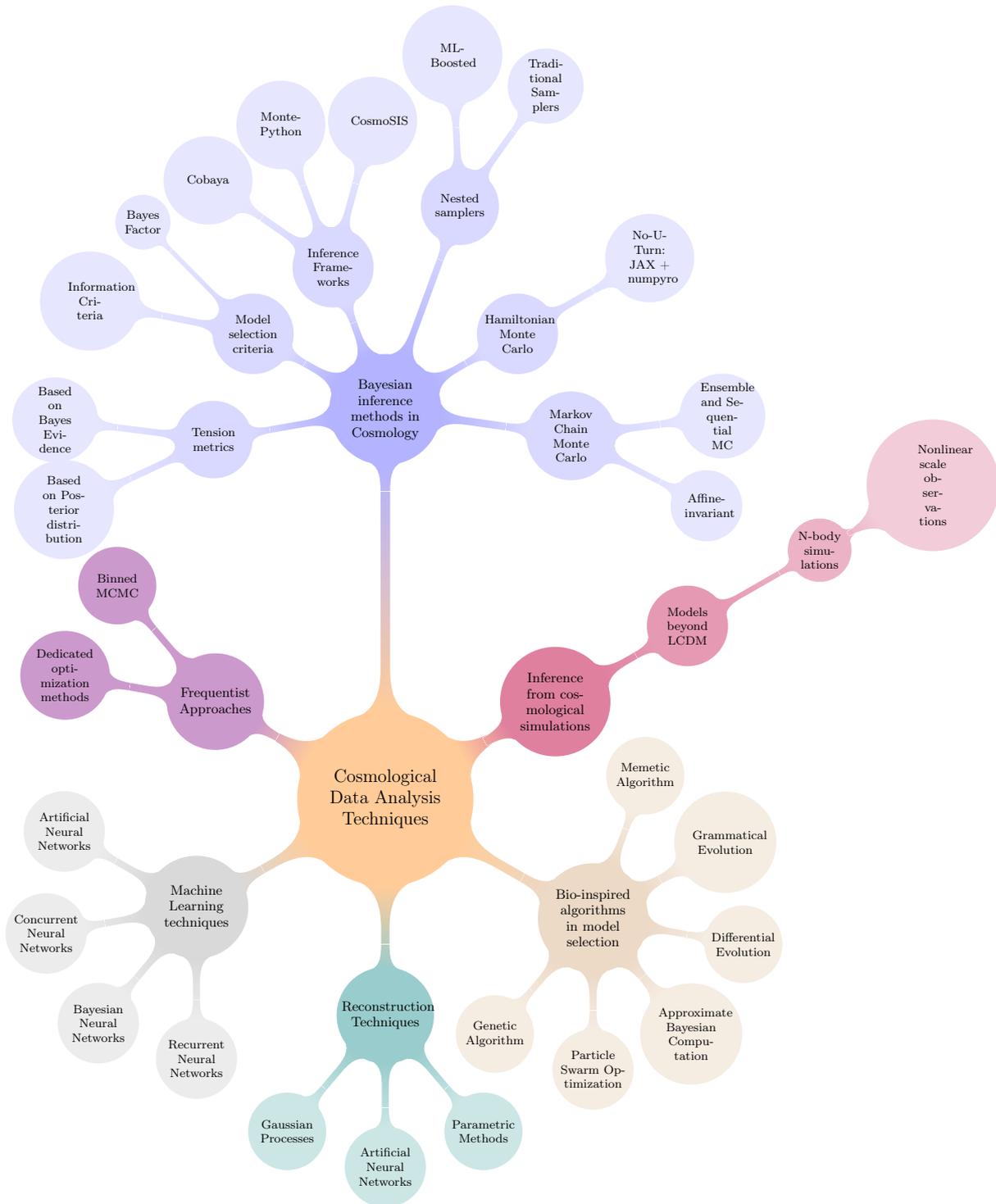
\begin{figure}
\scalebox{0.7}{
\begin{tikzpicture}[mindmap, grow cyclic, every node/.style=concept, concept color=orange!40, 
	level 1/.append style={level distance=6.5cm,sibling angle=360/6},
	level 2/.append style={level distance=3.5cm,sibling angle=40}, 
    level 3/.append style={level distance=3.5cm,sibling angle=35}, 
    level 4/.append style={level distance=3cm,sibling angle=0}]

\node{Cosmological Data Analysis Techniques}
child[level distance=5cm, concept color=gray!30, minimum size=1cm]{ node {Machine Learning techniques}
    child[concept color=gray!15]{ node {Artificial Neural Networks}}
	child[concept color=gray!15]{ node {Concurrent Neural Networks}}
	child[concept color=gray!15]{ node {Bayesian Neural Networks}}
	child[concept color=gray!15]{ node {Recurrent Neural Networks}}
}
child[level distance=5cm, concept color=teal!40, minimum size=1cm]{ node {Reconstruction Techniques}
	child[concept color=teal!20]{ node {Gaussian Processes}}
	child[concept color=teal!20]{ node {Artificial Neural Networks}}
	child[concept color=teal!20]{ node {Parametric Methods}}
}
child[level distance=5.5cm, concept color=brown!30, minimum size=1cm]{ node {Bio-inspired algorithms in model selection}
	child[concept color=brown!15]{ node {Genetic Algorithm}}
	child[concept color=brown!15]{ node {Particle Swarm Optimization}}
	child[concept color=brown!15]{ node[minimum size=2.3cm, draw]{Approximate Bayesian Computation}}
	child[concept color=brown!15]{ node {Differential Evolution}}
	child[concept color=brown!15]{ node[minimum size=2.3cm, draw]{Grammatical Evolution}}
	child[concept color=brown!15]{ node {Memetic Algorithm}}
}
child[level distance=4.5cm, concept color=purple!50, minimum size=1cm]{ node {Inference from cosmological simulations}
	child[concept color=purple!40]{ node {Models beyond LCDM}
	child[concept color=purple!30]{ node[minimum size=1cm, draw, font=\footnotesize]{N-body simulations}
	child[concept color=purple!20]{ node[minimum size=3cm, draw, font=\footnotesize]{Nonlinear scale observations}}}}
}
child[level distance=9cm, concept color=blue!30, minimum size=1cm]{ node {Bayesian inference methods in Cosmology}
	child[concept color=blue!15]{ node[minimum size=1cm, draw, xshift=2.5em]{Markov Chain Monte Carlo}
    child[concept color=blue!10]{ node[minimum size=1.6cm, draw, font=\footnotesize]{Affine-\\invariant}}
    child[concept color=blue!10]{ node[minimum size=1cm, draw, font=\footnotesize]{Ensemble and Sequential MC}}
    }
	child[concept color=blue!15]{ node {Hamiltonian Monte Carlo}
    child[concept color=blue!10]{ node[minimum size=2cm, draw, font=\footnotesize]{No-U-Turn: JAX + numpyro}}
    }
	child[level distance=5cm, concept color=blue!15]{ node {Nested samplers}
    child[concept color=blue!10]{ node[minimum size=1cm, draw, font=\footnotesize, xshift=-0.5em]{Tradi-\\tional Samplers}}
    child[concept color=blue!10]{ node[minimum size=2.25cm, draw, font=\footnotesize, xshift=-1em]{ML-Boosted}}
    }
    child[concept color=blue!15]{ node {Inference Frameworks}
    child[concept color=blue!10]{ node[minimum size=2cm, draw, font=\footnotesize]{CosmoSIS}}
    child[concept color=blue!10]{ node[minimum size=2cm, draw, font=\footnotesize]{Monte-\\Python}}
    child[concept color=blue!10]{ node[minimum size=2cm, draw, font=\footnotesize]{Cobaya}}
    }
    child[concept color=blue!15]{ node {Model selection criteria}
    child[concept color=blue!10]{ node[minimum size=1cm, draw, font=\footnotesize, xshift=-0.5em]{Bayes Factor}}
    child[concept color=blue!10]{ node[minimum size=2.25cm, draw, font=\footnotesize, xshift=-1em]{Information Criteria}}
    }
    child[level distance=4cm, concept color=blue!15]{ node {Tension metrics}
    child[concept color=blue!10]{ node[minimum size=1cm, draw, font=\footnotesize, xshift=-0.5em]{Based on Bayes Evidence}}
    child[concept color=blue!10]{ node[minimum size=2.25cm, draw, font=\footnotesize, xshift=-1em]{Based on Posterior distribution}}
    }
}
child [level distance=4.5cm, concept color=violet!40, minimum size=1cm]{ node {Frequentist Approaches}
            child { node {Binned MCMC} }
            child { node {Dedicated optimization methods} }
            };
\end{tikzpicture}
}

\caption{Schematic of the different data analysis approaches used in the cosmology community.}
\label{fig:IntroNewMethods}
\end{figure}

These methodologies, often used in combination, reflect a collaborative effort across theoretical, computational, and observational domains. For example, the integration of \ac{mcmc} methods with \ac{ml}-enhanced simulations and reconstruction techniques has provided a comprehensive framework for addressing persistent cosmological tensions. This synergy has also enabled the exploration of exotic scenarios, such as decaying \ac{dm}, \ac{pmf}s, and varying fundamental constants, which aim to explain anomalies in the current cosmological framework.

The ongoing advancements in data analysis tools and methodologies are pivotal for the next generation of cosmological surveys, including the Simons Observatory, \ac{cmbs4}, Euclid, and the Roman Space Telescope. These surveys will generate unprecedented datasets that require state-of-the-art techniques to extract meaningful insights. A schematic is shown in Fig.~\ref{fig:IntroNewMethods} of these methods. By embracing these innovations, the cosmological community continues to push the boundaries of our understanding of the Universe, addressing existing tensions and paving the way for new discoveries in fundamental physics.

\bigskip \newpage
%%%%%%%%%%%%Section_2:
\section{Observational cosmology and systematics} \label{sec:obs}

\subsection{H0 Tension: Measurements and systematics}\label{sec:obs_H0}

\noindent \textbf{Introduction:} Adam Riess\\

It is cliché to say we live in the era of precision cosmology—but it is true. Over the past several decades, the type, scope, and precision of cosmological measurements have grown enormously. Two of the most powerful tools, \ac{sn1}, and \ac{bao}s, joined the cadre of first-rank indicators during this time. Other observables—such as the \ac{cmb}, gravitational lensing, primordial abundances, and components of the distance ladder—have been refined and matured, sharpening our view of the Universe. It is an exhilarating time to be a cosmologist.
The surprising and non-intuitive composition of the Universe demands the full use of our observational toolkit. Whether the correct cosmological model is the simplest form of \lcdm\ or one with additional complexities will ultimately be determined by the quality of our measurements. In the following sections, you will read how the astronomical zoo of objects and features has been employed to produce precision tests of the Universe, with a level of control over systematic errors once known only to particle physics. To be fair, not every tool and technique has reached the same level of robustness (and some may never do so), but this is something you can judge for yourself as you explore the state of each art.
Perhaps the biggest question for observational cosmology in the 2020s is what to make of the growing evidence of tensions. The first appearance of any tension or anomaly is usually attributed to experimental error or systematics—an assumption that makes sense when playing the odds. However, signals that may herald new physics will be treated the same unless critical thinking and extensive analysis follow. At present, one or more tensions have surpassed thresholds of statistical significance, reproducibility, and independent cross-checks, earning them the continued attention of the field.
In the end, as Einstein once said, the Universe (or Lord) is subtle but not malicious. Our goal is to measure these subtleties.
\subsubsection{The distance ladder \label{sec:dist_ladd}}

\noindent \textbf{Coordinator:} Louise Breuval\\
\noindent \textbf{Contributors:} Adam Riess, Giulia De Somma, Leandros Perivolaropoulos, Lluís Galbany, Lucas Macri, Richard I. Anderson, Siyang Li, and Vladas Vansevicius
\\

The Hubble constant measures the present expansion rate of the Universe. This cosmological parameter represents the slope of the redshift-distance relation, $c z = H_0 D$, in the limit of $z \sim 0$. Only distant galaxies are sensitive to the Universe's expansion, unlike nearby systems which are dominated by local gravitational interactions. The best method to reach galaxies in this regime, called the \textit{Hubble Flow} ($0.02 < z < 0.20$), is to build a distance ladder based on a succession of distance indicators. The most widely used and best calibrated distance ladder is based on three rungs. First, geometric distances are used to calibrate the Period-Luminosity relation of Cepheid variables in nearby galaxies. This law is, in turn, adopted to calibrate \ac{sn1} on the second rung, which is limited to the volume where both Cepheids and \ac{sn1} are observable. Finally, on the third rung, distances and redshifts of \ac{sn1} in the \textit{Hubble Flow} directly measure $H_0$. Although Cepheids provide the most homogeneous and reliable set of distances to many nearby galaxies, they can be substituted by alternative standard candles such as \ac{trgb} stars \cite{Freedman:2019jwv, Anand:2024nim}, \ac{jagb} \cite{Lee:2023vku, Li:2024yoe}, or Mira variables \cite{Huang:2019yhh, Huang:2023frr}. On the second rung, \ac{sn1} can be replaced by \ac{sbf} \cite{Blakeslee:2021rqi, Anand:2024lbl}, \ac{sn2} \cite{deJaeger:2022lit, Csornyei:2023rpw} or the Tully-Fisher relation \cite{Kourkchi:2022ifq}. These methods provide valuable independent checks of the distance ladder.

\begin{figure}[ht!]
    \includegraphics[width=0.66\textwidth]{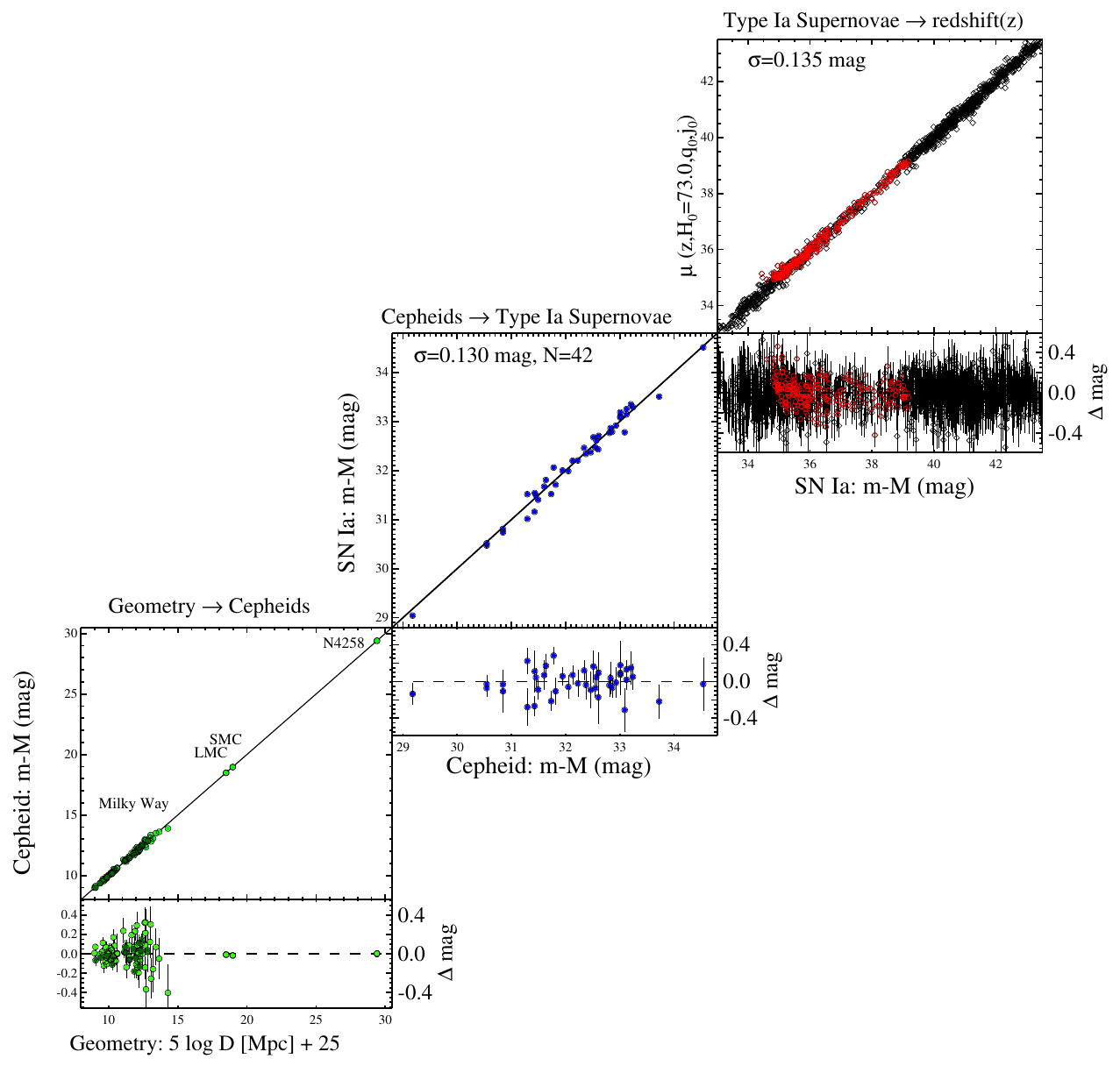}
    \caption{The Cepheid-\ac{sn1} distance ladder \cite{Riess:2024vfa}. The left-hand panel (first rung) shows the four anchor galaxies with geometric distances which are adopted to calibrate Cepheids. In the middle (second rung), 42 \ac{sn1} are calibrated with Cepheids. On the right-hand panel (third rung), redshifts and distances of \ac{sn1} in the Hubble Flow directly measure the expansion rate of the Universe. }  
    \label{fig:distance_ladder}
\end{figure}

\begin{figure}[ht!]
    \includegraphics[width=0.7\textwidth]{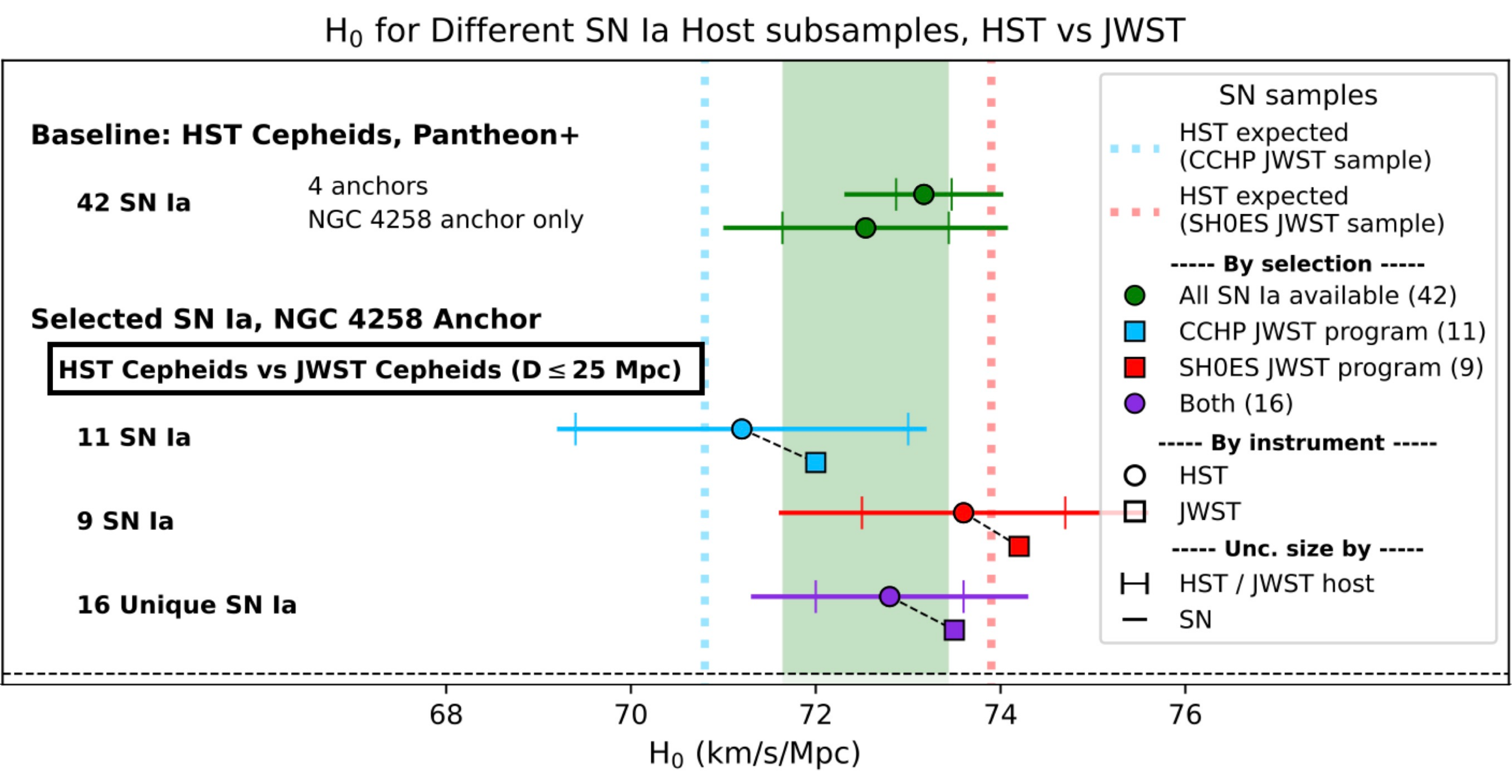}
    \caption{ Comparisons of $H_0$ between \ac{hst} Cepheids (baseline, green circles) and \ac{jwst} Cepheids (squares) for different \ac{sn1} host subsamples. The CCHP (blue) and SH0ES (red) subsamples selected for \ac{jwst} observations produce a difference of $3-4$\kms in $H_0$ owing to selection. The \ac{hst} and \ac{jwst} distance measurements themselves are in good agreement \cite{Riess:2024vfa}. \label{fig:HST_vs_JWST}}
\end{figure}

A Cepheid-\ac{sn1} distance ladder was included in the \ac{hst} Key Project on the Extragalactic Distance Scale \cite{HST:2000azd} result of $H_0 = 72 \pm 8$\kms (which calibrated multiple secondary distance indicators using Cepheids). Since then, this measurement was significantly improved using state-of-the-art data from later \ac{hst} instruments, homogeneous calibration, and near-IR Cepheid observations (Fig.~\ref{fig:distance_ladder}). The first rung of the distance ladder is currently supported by four geometric calibrators, each providing a direct absolute determination of the Cepheid luminosity: \textit{Gaia} DR3 parallaxes of Cepheids and host clusters in the \ac{mw} \cite{2023A&A...674A...1G}, detached eclipsing binaries distances in the Large and \ac{smc} \cite{Pietrzynski:2019cuz, 2020ApJ...904...13G}, and the water maser distance to NGC$\,$4258 \cite{Reid:2019tiq}. The second rung now comprises a total of 42 \ac{sn1} in 37 galaxies, where all hosts are observed with the same instrument (Wide Field Camera 3 (WFC3)) on the same \ac{hst}, an investment of more than $1000$ orbits of observing time. On the last rung, the Pantheon+ team provides $\sim300$ \ac{sn1} in the \textit{Hubble Flow} with the highest quality data and calibration, standardized across many surveys \cite{Scolnic:2021amr, Brout:2022vxf}. These recent developments provided the most precise $H_0$ measurement from a simultaneous fit of the three rungs with $73.17 \pm 0.86$\kms \cite{Riess:2021jrx, Breuval:2024lsv}. This \ac{hst}-based measurement has recently been confirmed with observations of a subsample of \ac{sn1} host galaxies with \ac{jwst}, which show excellent consistency between both telescopes \cite{Riess:2024vfa} (Fig.~\ref{fig:HST_vs_JWST}). Regardless of the method, essential elements for a precise $H_0$ measurement in the late Universe include the use of near-infrared photometry to minimize the impact of dust \cite{Galbany:2022zir} and consistent data to cancel flux calibration errors between rungs.

\bigskip
\subsubsection{Cepheid variables as standard candles \label{sec:Cepheids}}

\noindent \textbf{Coordinator:} Louise Breuval\\
\noindent \textbf{Contributors:} Adam Riess, Giulia De Somma, Leandros Perivolaropoulos, Lluís Galbany, Lucas Macri, Richard I. Anderson, Siyang Li, and Vladas Vansevicius
\\

%\begin{wrapfigure}{R}{0.45\textwidth}
 %   \centering
  %  \vspace{-10pt} 
   % \hspace{-60pt} 
\begin{figure}[h]
    \includegraphics[width=0.6\linewidth]{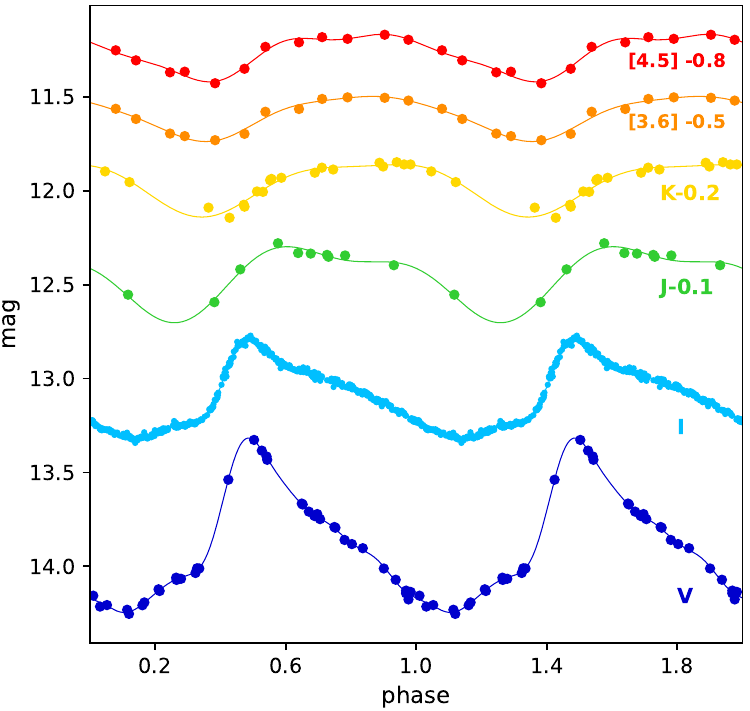}
    %\vspace{-5pt} 
    %\hspace{-50pt} 
    \caption{Multi-band light curves for Cepheid OGLE-SMC-CEP-4697 based on a compilation of data from Refs.~\cite{2016ApJ...816...49S, 2015AcA....65..297S, 2017MNRAS.472..808R}. 
    \label{fig:light_curves}}
\end{figure}

\paragraph{Cepheid variables as standard candles}

Classical Cepheids are luminous evolved stars located in the instability strip region of the Hertzsprung–Russell diagram. These stars undergo regular radial pulsations driven by two mechanisms: the $\kappa$ (opacity) and $\gamma$ (adiabatic exponent, $\Gamma_{3} - 1$) mechanisms, which operate in the partial ionization zones of abundant elements such as H, He, and He+ \cite{Zhevakin:1963dz, 1980tsp..book.....C}. Cepheids are regarded as the ``gold standard'' for distance measurements due to their tight Period-Luminosity (P-L) relation \cite{Leavitt:1912zz}, expressed as $M = a \log P + b$, where $M$ is the absolute magnitude, $P$ is the pulsation period, and $a$ and $b$ are constants. By measuring a Cepheid’s $P$, its $M$ can be determined, and its distance $D$ (in Mpc) can be inferred using the distance modulus: $m - M = 5 \log_{10} (D) + 25$, where $m$ is the apparent magnitude. From a theoretical perspective, the P-L relation arises from fundamental principles of stellar astrophysics \cite{1919ApJ....49...24S}. The pulsation period is related to the mean density $\rho$ of the star by $P \propto \rho^{-1/2}$, where $\rho \propto \mathcal{M}/R^3$ ($\mathcal{M}$ and $R$ are the stellar mass and radius, respectively). The Stefan-Boltzmann law, $L \propto R^2T_{\rm eff}^4$, and the mass-luminosity relation, $L \propto \mathcal{M}^{3.5}$, link the star’s luminosity ($L$) and effective temperature ($T_{\rm eff}$) to its pulsation period, providing a direct foundation for the P-L relation.

Cepheids play a key role in the distance ladder \cite{Riess:2021jrx}. Their P-L relation is locally calibrated with geometric distances and is used, in turn, to calibrate \ac{sn1} on the second rung. Cepheids are easy to identify due to their periodicity and are bright enough to be observed up to about 50 Mpc with the current generation of space telescopes. They are generally identified at optical wavelengths \cite{Hoffmann:2016nvl}, where their pulsation amplitudes peak ($\sim$ 1 mag in the $V$ band), allowing precise determination of their pulsation periods (Fig.~\ref{fig:light_curves}). Follow-up observations in the near-infrared minimize the impact of interstellar dust on their magnitudes. P-L relations are typically constructed using intensity-averaged magnitudes over the pulsation cycle. These averages are derived either through extensive photometric sampling or by fitting random-phase measurements with template light curves \cite{2015A&A...576A..30I, Breuval:2023rkw}. Using mean magnitudes significantly reduces the P-L dispersion by a factor of $\sim$2 compared to random-phase measurements.  

However, systematic uncertainties in Cepheid distances may arise due to differences between nearby calibrators and distant Cepheids observed in \ac{sn1} host galaxies. These include chemical composition, period range, crowding, dust properties, and potential nonlinearity of the P-L relation (see Ref.~\cite{2024arXiv240302801A} for a comprehensive review). In the following sections, we outline the most significant systematics affecting Cepheid distances, and we propose solutions to mitigate them.

\paragraph{Improvements in reducing the systematics}

\subparagraph{Photometric systems} 
Combining multiple photometric systems to observe Cepheids in the first and second rung introduces a $1.4-1.8\%$ systematic error in distance measurements, as described in Ref.~\cite{Riess:2019cxk}. To mitigate this, it is advantageous to use the same telescope and instrument, thereby minimizing flux-calibration errors. Ground-based telescopes, while offering dedicated facilities with high availability \cite{2015AcA....65..297S}, are impacted by atmospheric absorption and scattering. In contrast, space-based observatories like the \ac{hst} and \ac{jwst} provide superior image quality, are unaffected by weather, exhibit excellent long-term stability, and are not constrained by the day-night cycle. Prior to the launch of \ac{jwst}, WFC3 on \ac{hst} was the only instrument capable of resolving Cepheids in the most distant \ac{sn1} host galaxies at optical and near-infrared wavelengths \cite{Riess:2021jrx}. While WFC3 is well-suited for observing both faint and bright targets, it suffers from a slight non-linearity in the infrared. Specifically, photons from faint stars (low count rates) are measured less efficiently than those from bright stars (high count rates), introducing a Count Rate Non-Linearity (CRNL). Though small, this bias is critical for achieving percent-level precision and amounts to $0.0077 \pm 0.0006$ mag/dex \cite{2019wfc..rept....1R}. A CRNL correction is typically applied to bright Cepheids in anchor galaxies to ensure consistent calibration \cite{Riess:2021jrx, Breuval:2024lsv}.

\begin{figure}[h]
    \includegraphics[width=0.75\linewidth]{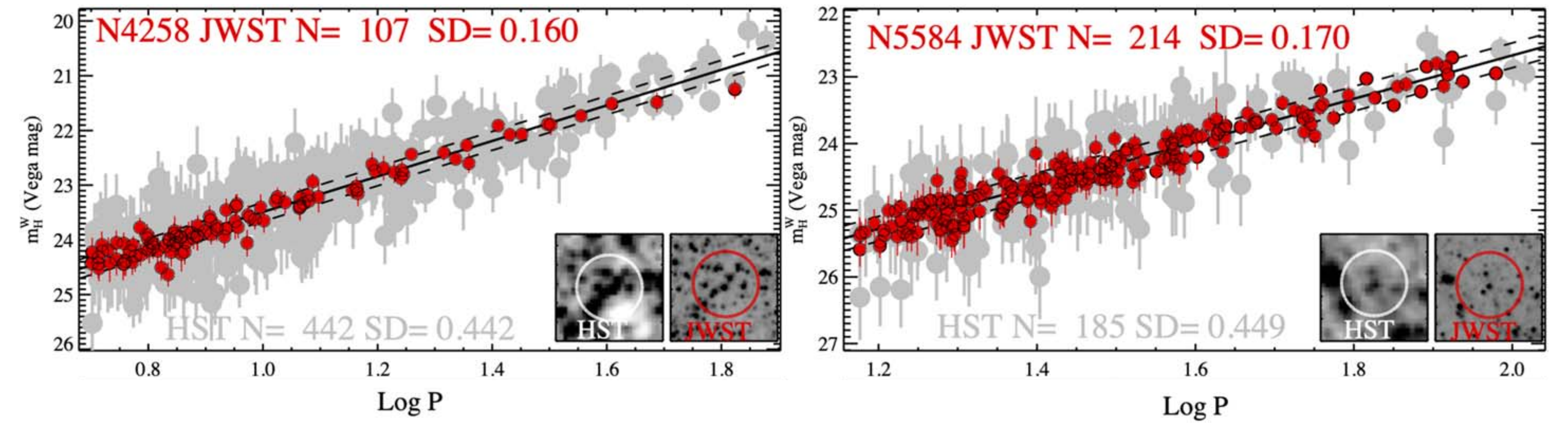}
    \caption{Cepheid P-L relations in the Wesenheit index obtained from \ac{hst}$F160W$ (grey) and \ac{jwst}$F150W$ (red), showing the 2.5 reduction in scatter in NGC 4258 (left) and NGC 5584 (right), figure taken from Ref.~\cite{Riess:2024ohe}. \label{fig:JWST_PL}}
\end{figure}

\subparagraph{Crowding} 
Separating Cepheids from their surrounding stellar populations is one of the most significant challenges in measuring Cepheids in \ac{sn1} host galaxies \cite{Freedman:2019jwv}. Contamination from redder stars, such as \ac{rgb} and \ac{agb} stars along the line of sight, limits the precision of Cepheid measurements, especially in the near-infrared. A common approach to address crowding is to add artificial stars of known brightness at random positions near Cepheids and recover their magnitudes using the same photometric methods applied to real stars \cite{Hoffmann:2016nvl}. This allows for a statistical correction for crowding (although such corrections remove the bias, but not the added scatter or crowding noise). With the advent of \ac{jwst}, the ability to separate Cepheids from background stars has improved significantly, resulting in a substantial reduction in crowding noise. Recent \ac{jwst} observations of Cepheids in \ac{sn1} host galaxies including the anchor galaxy NGC 4258 \cite{Riess:2023bfx, Riess:2024ohe} have demonstrated excellent agreement with prior \ac{hst} measurements, achieving a mean difference of only 0.01 mag. Furthermore, \ac{jwst} data reduced the scatter in the P-L relation by a factor of 2.5 (Fig.~\ref{fig:JWST_PL}). This result decisively rules out Cepheid crowding from \ac{hst} photometry as the cause of the Hubble tension, with a  $8.2 \sigma$ CL.

\subparagraph{Metallicity} 
The luminosity of Cepheids at a given period is known to correlate with their chemical composition, but the sign and magnitude of this dependence ($\gamma$) have historically been difficult to constrain \cite{Kennicutt:1997dm, Sakai:2004ys, Macri:2006wm}. Accounting for the metallicity dependence is essential in the distance ladder, particularly when calibrating Cepheids in the Magellanic Clouds, which are more metal-poor than typical \ac{sn1} host galaxies \cite{Riess:2019cxk, Breuval:2024lsv}. However, Cepheids in NGC$\,$4258 and the \ac{mw} resemble those in large spiral \ac{sn1} hosts, making metallicity differences negligible for determining $H_0$. Recent calibrations of the metallicity effect, leveraging accurate distances and expanded Cepheid samples, have converged on a consensus. These calibrations include studies using \textit{Gaia} parallaxes and individual spectroscopic abundances of \ac{mw} Cepheids \cite{2021MNRAS.508.4047R, Bhardwaj:2023mau, 2024A&A...681A..65T} and combined analyses of \ac{mw} and Magellanic Cloud Cepheids with geometric distances \cite{2022ApJ...939...89B, Breuval:2024lsv}. The derived metallicity dependence ($\gamma$) lies between $-0.2$ and $-0.3$ mag/dex, with the negative sign indicating that metal-rich Cepheids are intrinsically brighter than metal-poor ones. Recent theoretical studies have corroborated both the sign and magnitude of this effect \cite{Anderson:2016txx, 2022ApJS..262...25D}, particularly for the Wesenheit index (a dereddened magnitude combining multiple filters \cite{1982ApJ...253..575M}) employed in the SH0ES distance ladder \cite{Riess:2021jrx}. For nearby galaxies, direct [Fe/H] abundances can be determined from high-resolution spectroscopy of Cepheids \cite{Romaniello:2021vht, Bhardwaj:2023mau}. In more distant galaxies, metallicities are inferred from the [O/H] gradient measured via HII region spectroscopy \cite{Zaritsky:1994ht, Riess:2005zi, Riess:2009pv} or optical Integral Field Spectroscopy (IFS) of calibrator galaxies \cite{Galbany:2016cqf}. These methods are consistent in the \ac{mw}, where Cepheid spectroscopy agrees with HII region gradients to within $1\sigma \sim 0.05$ dex \cite{Riess:2021jrx}. Because the metallicities of Cepheids in \ac{sn1} host galaxies closely match those in the anchor galaxies, the Hubble constant is only weakly affected by the metallicity effect. Neglecting the correction would change $H_0$ by just 0.5\kms (in the higher $H_0$ direction) \cite{Riess:2021jrx}. Although the impact on $H_0$ is minimal, the metallicity correction plays a crucial role in ensuring consistency among anchor galaxies while independently matching the radial dependence with \textit{Gaia} parallaxes in the \ac{mw}.

\subparagraph{Binaries} 
While crowding can be mitigated using artificial star measurements, unresolved companion stars physically associated with Cepheids can bias their measured flux, thereby affecting inferred distances. A significant fraction of Cepheids reside in binary or multiple systems \cite{2019A&A...623A.116K}, and unresolved companions typically increase the measured brightness of Cepheids. 
However, because unresolved binarity is equally present in both \ac{sn1} hosts and anchor galaxies, its effect largely cancels out in calibrated distances. In the Wesenheit index adopted in the SH0ES distance ladder, \cite{2023ApJ...950..182K} used synthetic populations of binary Cepheids to estimate the impact of binarity, finding it to be a minor effect, contributing only 0.004 mag in distance modulus, or a 0.1\% change in $H_0$. Similarly, \cite{2018ApJ...861...36A} determined that the primary source of flux contamination arises from Cepheids located in open clusters, leading to an overestimate of $H_0$ by approximately 0.23\%. However, the fraction of Cepheids in open clusters is relatively small (see also Refs.~\cite{Mochejska:1999nd, Breuval:2023rkw}). Thus, while binarity and clustering introduce minor biases, their overall impact on $H_0$ remains negligible at the current level of precision. \\

\paragraph{Perspectives}

\subparagraph{Dust laws} 
Dust has long been a critical systematic uncertainty in the distance ladder \cite{Follin:2017ljs}. To correct for dust, one can either subtract $R_{\lambda} \times E(V-I)$ to Cepheid apparent magnitudes or adopt a reddening-free ``Wesenheit'' magnitude (see Ref.~\cite{1982ApJ...253..575M, Riess:2021jrx} for details). Additionally, to the uncertainties in the reddening $E(V-I)$ itself, the value of the $R_V$ parameter is not independently known in each \ac{sn1} host galaxy and at each Cepheid position, but rather is assumed to match the \ac{mw} \cite{Riess:2021jrx, Perivolaropoulos:2021bds}. However, the use of near-infrared observations greatly limits the impact of reddening, reducing the size of $R_{\lambda}$ to $\sim$ 0.4 and variations due to these different values of $R_V$ to $\pm 0.03$. Varying the reddening law across different hosts first requires subtracting the intrinsic color of Cepheids in order to separate the component of the color that results from dust reddening (see Sect.~6.3 and Appendix D in Ref.~\cite{Riess:2021jrx}). A complete sampling of the reddening curve at long and short wavelengths in each host as well as an improved characterization of the dependence of $R_V$ with host properties such as mass, star formation rate, and color \cite{2022ApJ...926..122H} can further reduce this source of uncertainty in Cepheid measurements.

\subparagraph{Cepheid parallaxes} 
Early measurements of trigonometric parallaxes were obtained with the Fine Guidance Sensor (FGS) on \ac{hst} \cite{Benedict:2006cp} and later by spatial scanning with \ac{hst}/WFC3 \cite{Casertano:2015dso, Riess:2018uxu} for $\sim$10 \ac{mw} Cepheids. The \ac{esa} \textit{Gaia} mission now provides parallaxes for thousands of Cepheids with individual precision  at the $\sim 5\%$ level \cite{2023A&A...674A...1G}. The main systematic uncertainty in this method arises from the parallax zero-point, a small corrective term estimated with distant \ac{qso}s, \ac{lmc} stars and binaries. The Early Data Release 3 constrained this term and characterized its dependence on magnitude, color, and position \cite{2021A&A...649A...4L}. However, \ac{mw} Cepheids are bright, resulting in limited sampling of the offset term for these stars. Fortunately, the parallax zero-point offset is an additive term, while the distance scale derived from parallaxes is multiplicative. This distinction allows for separation of the two terms, provided a sufficient range of Cepheid distances is available. The analysis of 75 \ac{mw} Cepheids with high S/N and \ac{hst}/WFC3 photometry in Ref.~\cite{Riess:2020fzl} yielded a residual parallax offset of $-14\,$\textmu as for Cepheid-like bright stars, which aligns well with the majority of independent studies for stars in this magnitude range \cite{Li:2022aho}. \ac{mw} Cepheids thus remain the strongest anchor of the distance ladder.

Greater individual precision can be achieved with Cepheids in open clusters by using the average parallax of cluster members \cite{Anderson:2012dv, Breuval:2020trd}. Cluster members, being fainter than Cepheids, allow for a more accurate calibration of the parallax offset. In this magnitude range, studies by Refs.~\cite{Riess:2022mme, Reyes:2022boz} find a zero-point consistent with zero, demonstrating good agreement between P-L relations derived from individual Cepheids and cluster members. These results further confirm the robust accuracy of \textit{Gaia} parallaxes across a broad range of magnitudes.

\subparagraph{New anchors for the cepheid distance ladder} 
The distance ladder is currently supported by geometric distances in four galaxies: \textit{Gaia} parallaxes in the \ac{mw} \cite{Riess:2020fzl, Riess:2022mme}, detached eclipsing binary (DEB) distances in the Large \cite{Riess:2019cxk, Pietrzynski:2019cuz} and Small \cite{2020ApJ...904...13G, Breuval:2024lsv} Magellanic Clouds, and the water maser distance to NGC 4258 \cite{Reid:2019tiq, Yuan:2022kxa}. Adding new anchors to the Cepheid calibration would reduce uncertainties and strengthen the first rung of the distance ladder. A galaxy can serve as an anchor if it meets two conditions: (1) having Cepheid photometry measured in the same system (e.g., \ac{hst}/WFC3) as the rest of the distance ladder and (2) having a sufficiently precise geometric distance. The two nearby galaxies, M31 and M33, are excellent candidates for this role, as they contain large Cepheid samples observed with \ac{hst}/WFC3 \cite{Li:2021qkc, Breuval:2023rkw}. However, they remain beyond the reach of late-type DEBs, which are too faint at these distances or suffer from imprecise geometric distances obtained with other methods \cite{Argon:2004re}. On the other hand, early-type DEBs are brighter and could enable geometric distance measurements for M31 and M33 \cite{Bonanos:2006jd}. This method relies on model atmosphere theory, in contrast to the 1\% precision surface brightness color relation used for late-type DEBs in the Magellanic Clouds. Significant advancements in these methods are anticipated \cite{2020ApJ...890..137T, 2021A&A...652A..26S}, and they should soon provide additional robust anchors for the Cepheid distance ladder.

\subparagraph{Future prospects} 
Ongoing surveys (e.g., \ac{ztf}) and the next generation of ground-based (Rubin, \ac{elt}) and space telescopes {\it (Roman)} will provide new insights into the Cepheid distance ladder \cite{2022AJ....164..154N, 2023ApJ...953...14N}. These advancements will enable the discovery of new Cepheids through deep optical time-series observations of distant galaxies, extending the volume of the local Universe where Cepheids can be studied. High-precision photometry for bright stars in the \ac{mw} and Magellanic Clouds will be measured by the PhotSat mission in optical bandpasses, which will also help characterize the extinction law in different environments. Direct abundance measurements obtained with ground-based \ac{elt} will complement current datasets and extend the range of distances over which different metallicity tracers -- such as Cepheids, blue supergiants, or HII regions -- can be compared \cite{Zaritsky:1994ht, Bresolin:2009bv, 2022ApJ...940...32B, Romaniello:2021vht}. Expanding Cepheid samples in both nearby and distant galaxies will allow further tests of the linearity and universality of the P-L relation, a subject of ongoing debate \cite{Ngeow:2005qc, Sandage:2008rq, Kodric:2013pz}. In \ac{sn1} host galaxies, long-period Cepheids dominate as they are the brightest, while in nearby anchors, they may saturate or have invalid parallaxes. Conversely, P-L relations in nearby galaxies often include mostly short-period Cepheids. This period disparity could influence the inferred distances in the event of a P-L break. Recent studies \cite{2016MNRAS.457.1644B, Riess:2021jrx, Kushnir:2024spm} show no evidence for a non-linear P-L relation and demonstrate that allowing for different slopes does not improve the P-L fit. \cite{Riess:2021jrx} also finds the P-L slope consistent with a single value within $1\sigma$ across the current sample of host galaxies. The significance of this effect will be confirmed with larger samples at greater distances. Additionally, the metallicity dependence of the P-L slope \cite{Anderson:2016txx, 2021MNRAS.508.4047R} will be clarified through observations in different environments. 

\bigskip
\subsubsection{Maser driven constraints \label{sec:maser_drivers}}

\noindent \textbf{Coordinator:} Cheng-Yu Kuo\\
\noindent \textbf{Contributors:} Dom Pesce, Jim Braatz, Mark Reid, and Violetta Impellizzeri
\\

For the purpose of measuring $H_0$ using accurate distances external galaxies, the so-called {\it H$_{2}$O maser technique}, e.g., see Ref.~\cite{Kuo:2012hg, Kuo:2014bqa} provides a novel tool that allows one to by-pass the traditional extragalactic distance ladders and measure the {\it angular-diameter distance} to a galaxy in a single step without relying on the \ac{cmb}. This method involves sub-milliarcsecond resolution imaging and single-dish monitoring of the 22 GHz H$_{2}$O maser emission from sub-parsec circumnuclear disks at the center of active galaxies. These {\it disk maser} emissions, which arise from the $J_{K_{-}K_{+}}=6_{16}-5_{23}$ transition\footnote{Here, $J$ is the total angular momentum of the H$_{2}$O molecule, with $K_{-}$ and $K_{+}$ representing the projections of $J$ on two molecular axes, e.g., see Ref.~\cite{1985ApJ...295..175C}.} of the ortho-H$_{2}$O molecule, usually have extremely high surface brightnesses, permitting mapping with Very Long Baseline interferometry (VLBI) that allows for a unique probe of the gas distribution and kinematics on sub-parsec scales at the centers of distant galaxies \cite{2005ARA&A..43..625L}.

\begin{figure}
    \includegraphics[width=9cm]{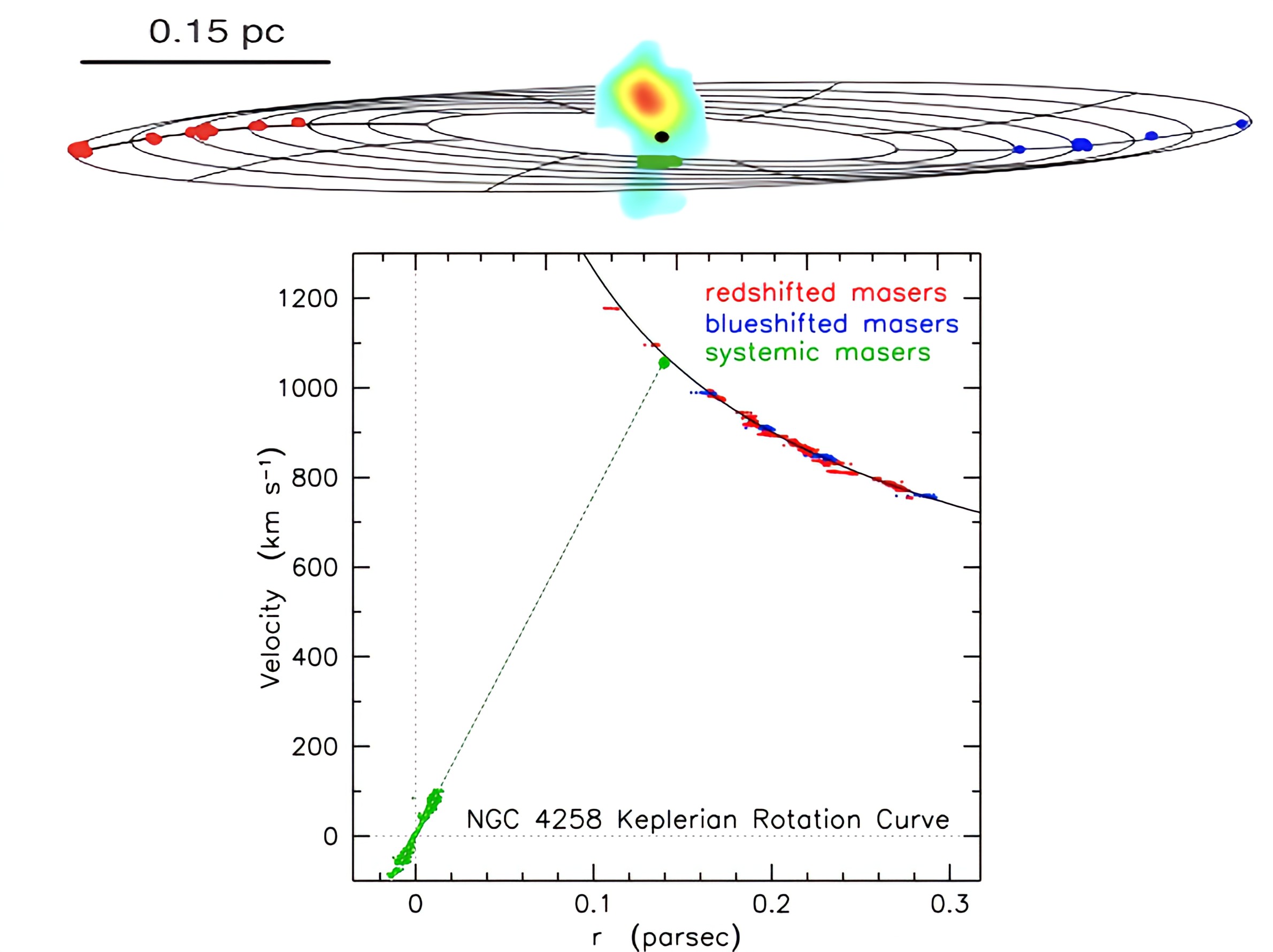}
    \caption{The distribution and Keplerian rotation curve \cite{Argon:2007ry, Kormendy:2013dxa} of the 22 GHz H$_{2}$O masers in NGC 4258. The red, blue, and green dots in the plot represent the redshifted, blueshifted, and systemic maser components, respectively.}
    \label{fig:ngc4258}
\end{figure}

As revealed in the prototypical maser galaxy NGC 4258, e.g., see Ref.~\cite{Argon:2007ry, Herrnstein:1999kd}, the masing gas in a disk maser system often resides in a nearly edge-on thin disk around the central supermassive black hole (BH). The rotation curve of the gas can be nearly perfectly traced by Keplerian rotation, enabling black hole (BH) mass measurements to percent-level accuracy, e.g., see Ref.~\cite{Kuo:2010uy, 2017ApJ...834...52G}. Given the assumption that the gas dynamics is dominated by the gravity of the central BH, the simplicity of disk geometry and kinematics allow one to use the orbital radii $r_{\rm sys}=v_{\rm sys}^{2}/a_{\rm sys}$ of the {\it systemic} maser components, see Fig.~\ref{fig:ngc4258}, e.g., see Ref.~\cite{Kuo:2010uy}, as a standard ruler for measuring the angular-diameter distance $D_{\rm A}$ of the galaxy, e.g., see Ref.~\cite{Herrnstein:1999kd}, where $v_{\rm sys}$ and $a_{\rm sys}$ standing for the orbital velocity and centripetal acceleration of a systemic maser component. By using $r_{\rm sys}$ as the standard ruler, one can easily show that the angular-diameter distance of a maser galaxy can be expressed as $D_{\rm A}=v_{0}^{2}{\rm sin}(i)/a\Delta\theta$, where $i$ and $\Delta\theta$ indicate the maser disk inclination and the angular radius of the systemic masers, respectively. As long as the masing components of the gas follow circular orbits, one can determine an accurate galaxy distance by measuring the four disk parameters including $a_{\rm sys}$, $v_{\rm sys}$, $i$, and $\Delta\theta$, where $a_{0}$ can be measured from single-dish monitoring of the maser lines, e.g., see Ref.~\cite{Kuo:2014bqa} and the rest of the three parameters can be obtained by modeling the maser disk in three dimensions, e.g., see Ref.~\cite{Reid:2012hm, Gao:2015tqd}.

\begin{figure}
    \includegraphics[width=11 cm]{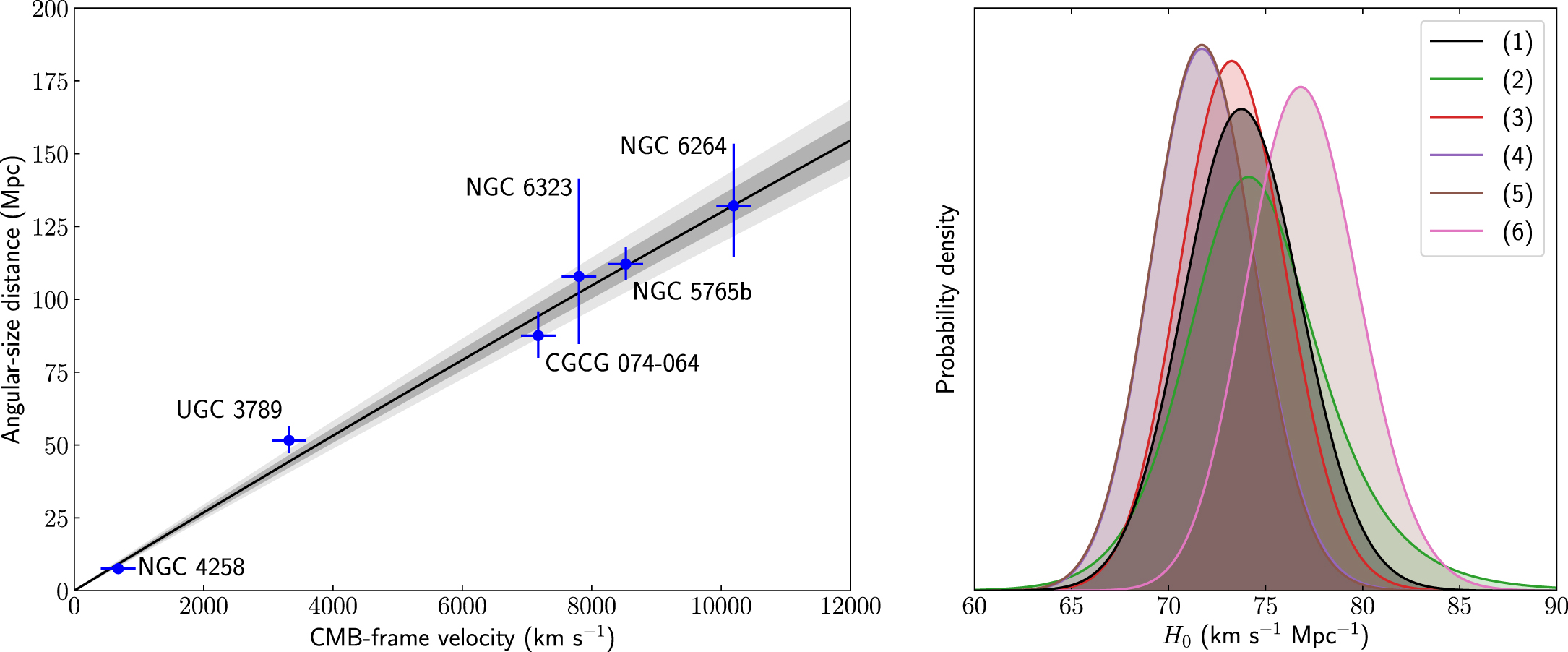}
    \caption{Left panel: Hubble diagram for the six maser galaxies considered in the MCP $H_0$ measurement \cite{Pesce:2020xfe}. Assuming a fixed velocity uncertainty of 250\,km~s$^{-1}$ associated with peculiar motions, the MCP constrains the Hubble constant to be $H_0 = (73.9 \pm 3.0)$\kms, independent of distance ladders and the \ac{cmb}. Right panel: posterior distributions for $H_{0}$ from the five different peculiar velocity treatments considered in Ref.~\cite{Pesce:2020xfe}, with the {\it fiducial} treatment plotted in black. }
    \label{fig:MCP_hubble_diagram}
\end{figure}

To apply this maser technique to galaxies in the local Universe, one has to first search for maser disks similar to NGC 4258 that allow for robust disk modeling. To this end, the Megamaser Cosmology Project (MCP; \cite{Reid:2008nm, Braatz:2010sg}) has carried out an extensive survey of H$_{2}$O megamaser emission from $>$4800 \ac{agn}s \cite{2018ApJ...860..169K,2020ApJ...892...18K} within redshift $z\lesssim 0.05$, resulting in the detection of $\gtrsim$30 candidate disk masers \cite{Pesce:2015tga}. The follow-up imaging of these candidates has increased the number of H$_{2}$O maser disks with high precision VLBI maps by a factor of $\gtrsim$4 over the past decade. The resultant Hubble constant (see Fig.~\ref{fig:MCP_hubble_diagram}) from this effort is $H_0 = 73.9 \pm 3.0$\kms \cite{Pesce:2020xfe}, a $\sim$4\% uncertain $H_0$ estimate determined based on six ``clean'' disk maser systems that have the required spectral qualities for reliable disk modeling. This maser-based $H_0$ measurement is well consistent with the majority of direct, {\it late Universe} measurements of Hubble constant. Its uncertainty is currently dominated by measurement errors in the maser position obtained from VLBI observation and in the determination of the maser acceleration with the single-dish monitoring. The systematic uncertainties, which could result from non-circular orbits of the masing gas, e.g., see Ref.~\cite{Reid:2012hm} or from the impacts of non-gravitational forces such as shocks in the maser disk, e.g., see Ref.~\cite{Bragg:2000in, Humphreys:2007ir, Pesce:2015tga}, are currently negligible in comparison with measurement errors. It is expected that the accuracy of the maser-based $H_0$ measurement can be further improved to $\sim$1\%, the long-term goal of the observational cosmology community, by measuring distances to $\gtrsim$50 maser galaxies, with $\sim$7\% accuracy per measurement \cite{2019BAAS...51c.446B}. It is promising that this goal can be realized after the full array operation of the next-generation Very Large Array (ngVLA), which will bring about an order of magnitude improvement in sensitivity, permitting efficient detection of disk maser systems in a $\sim$30 times larger volume compared with the MCP \cite{2019BAAS...51c.446B} as well as significant improvement in VLBI maser position measurement.
\bigskip
\subsubsection{On the tip of the red giant branch method \label{sec:TRGB}}

\noindent \textbf{Coordinator:} Richard I. Anderson\\
\noindent \textbf{Contributors:} Adam Riess, Gagandeep S.~Anand, Giulia de Somma, Ippocratis Saltas, Louise Breuval, Siyang Li, and Vladas Vansevicius
\\

\begin{figure}[ht!]
    \centering
    \includegraphics[width=0.6\linewidth]{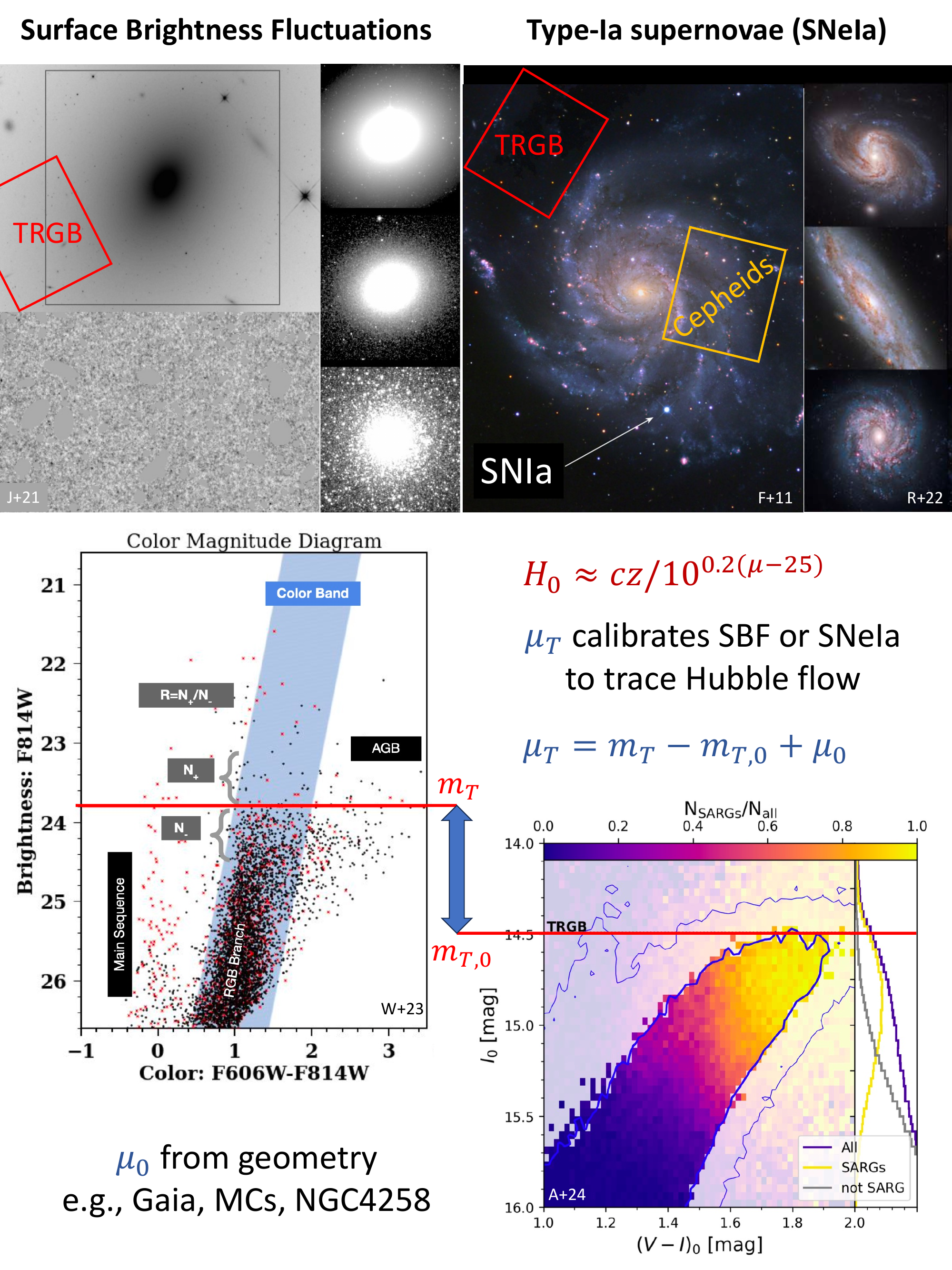}
    \caption{Illustrations of extragalactic distance ladders built on the \ac{trgb} method. The scales considered increase upwards. {\it Bottom right:} The \ac{trgb} feature is calibrated based on color-magnitude diagrams of stars whose distances ($\mu_0$) are known from geometric measurements, such as {\it Gaia} parallaxes, the Magellanic Clouds, or NGC4258. The choice and systematics of (the combination of) geometric anchors directly impact the Hubble constant by setting $\mu_0$. {\it Center left:} The apparent magnitude of the \ac{trgb} feature is determined in other galaxies.  Differences in the absolute magnitude between \ac{trgb} populations, either compared to the calibrating set or as field-to-field variations, must be mitigated by standardization. {\it Upper part, right:} The measured \ac{trgb} distances calibrate the fiducial absolute magnitude of \ac{sn1} in any type of galaxy. Cross-checks between multiple stellar standard candles can primarily be obtained in galaxies hosting young, intermediate-age, and old stellar populations at the same time. Note that \ac{trgb} fields should be placed on galaxy halos populated by old, metal-poor stars, whereas classical Cepheid fields target young stellar populations in supernova host galaxies. {\it Upper part, left:} Alternatively, the \ac{trgb} is used to calibrate the absolute scale of the \ac{sbf} method for distance determination to elliptical galaxies that trace the Hubble flow. \ac{sbf} exploits the fact that the variance of \ac{sbf} decreases as $d^2$.  \newline
    Figure credit: R.I.~Anderson based on images as labeled from: J+21 \cite{Jensen:2021ooi}; F+11 B.~J. Fulton, Las Cumbres Observatory; R+22 \cite{Riess:2021jrx}; W+23 \cite{Wu:2022hxf}, A+24 \cite{Anderson:2023aga}.  \label{Fig:TRGB_DL}}
\end{figure}

The \ac{trgb} method provides the most precise stellar alternative to classical Cepheids (Sec.~\ref{sec:Cepheids}) on the first and second rungs of the extragalactic distance ladder used to measure the Hubble constant \cite{Lee:1993jb}. Comparison between \ac{trgb} and Cepheid distances in supernova hosts shows very good agreement between both methods, Fig.~11 in Refs.~\cite{Anand:2024nim} or \cite{Riess:2024vfa}. Fig.~\ref{Fig:TRGB_DL} illustrates two paths to $H_0$ in which the \ac{trgb} calibrates either \ac{sbf} (Sec.~\ref{sec:SBF}) \cite{Anand:2024lbl,Jensen:2025aai} measured in elliptical galaxies or \ac{sn1} (Sec.~\ref{sec:SNeIa}), e.g., see Ref.~\cite{Freedman:2019jwv,Anand:2021sum,Scolnic:2023mrv} measured in any type of galaxy for tracing the Universe's isotropic expansion in the Hubble flow. In both cases, the \ac{trgb} serves as an intermediary to translate the relative apparent magnitude differences of \ac{sbf} or \ac{sn1} as a function of redshift to an absolute scale anchored to geometrically measured distances. As part of the past decade's unfurling Hubble tension, the \ac{trgb} method has seen major improvements and inspired deep investigations into systematics of late-Universe $H_0$ measurements. Here, we briefly review the astrophysical basis of the \ac{trgb} as a standard candle to determine luminosity distances, present methodological considerations relevant for determining \ac{trgb} distances to better than $\sim 5\%$, and consider likely future developments relevant for the $H_0$ tension in light of catalysts provided by new observational facilities and upcoming space missions. \cite{Li:2024gib} provides further background and in-depth discussion.

\paragraph{The astrophysical basis}

The \ac{trgb} as a standard candle is a robust empirical concept that is usefully supported by a solid understanding of the evolution of \ac{rgb} stars. Empirically, the magnitude $m_{\rm T}$ is measured as the inflection point of a \ac{rgb} \ac{lf} \cite{Hatt:2017rxl,Anderson:2023aga}. Measuring $m_{\rm T}$ therefore requires a large number of stars to avoid stochastic effects and, in particular, cannot be done for each star separately. This characteristic is shared with the J-region \ac{agb} method \cite{Nikolaev:2000tb,Madore:2020yqv} (Sec.~\ref{sec:JAGB}) and distinguishes the \ac{trgb} as a {\it statistical} standard candle from pulsating stars, such as classical Cepheids, whose luminosity can be measured and calibrated {\it individually} \cite{2024arXiv240302801A,Anderson:2024twinkle}. 

\begin{figure}{h}
\centering 
\includegraphics[width=0.5\textwidth]{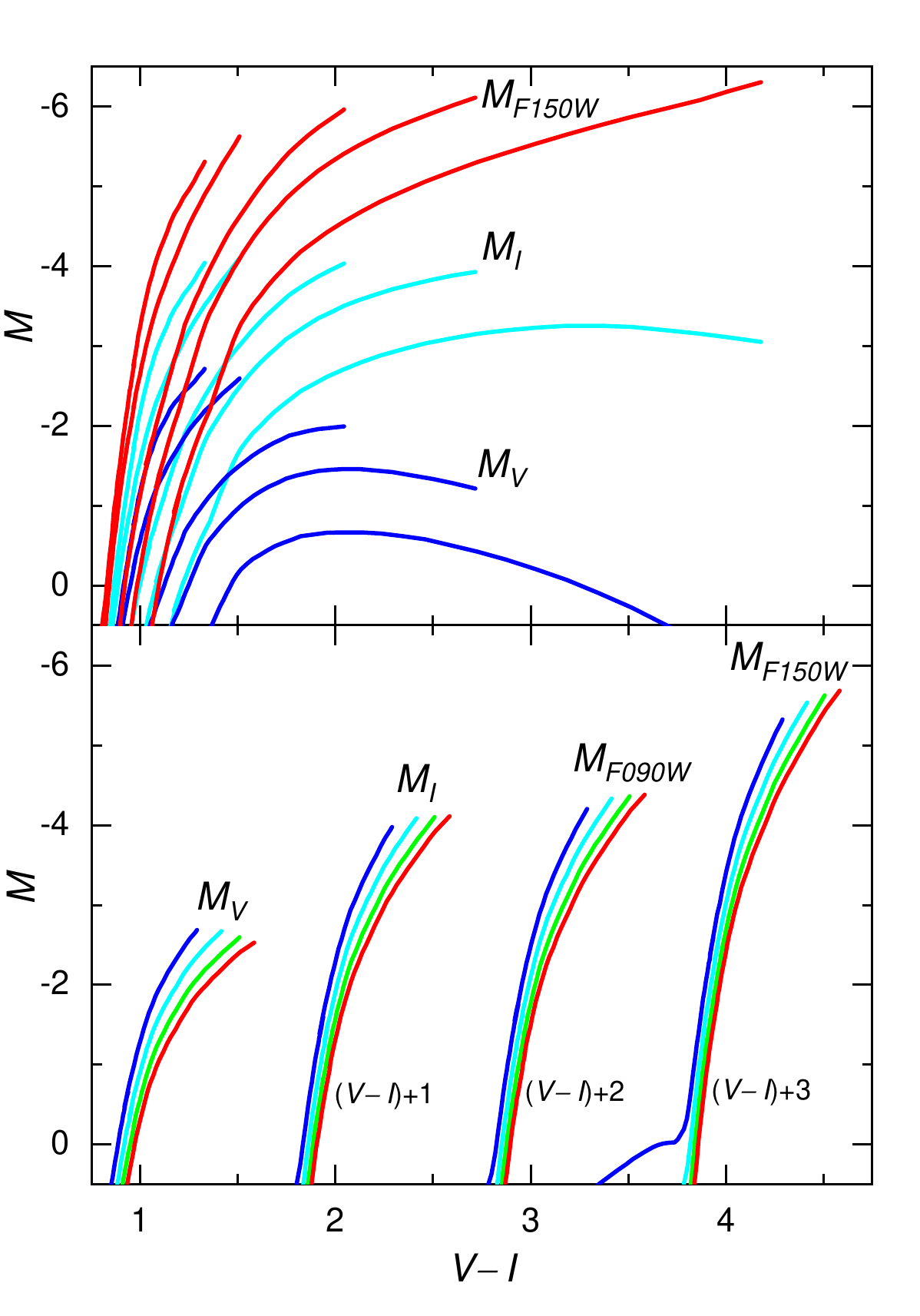}
\caption{Dependence of $M_{\rm T}$ on chemical composition and age based on PARSEC v.1.2S isochrones \cite{2014MNRAS.445.4287T,2015MNRAS.452.1068C}. {\it Top:}  $8$\,Gyr isochrones for increasing metallicity (from left to right: [M/H]$=-2.2, -1.5, -0.7, -0.4, 0.0$) in Johnson $V$ (dark blue), Cousins $I$ (cyan), and \ac{jwst}/NIRCAM F150W (red) passbands. {\it Bottom:} [M/H]$=-1.5$ isochrones for increasing ages (left to right) of 2 Gyr (blue), 4 Gyr (cyan), 8 Gyr (green), and 14 Gyr (red) in the Johnson $V$, Cousins $I$, \ac{jwst}/NIRCAM F090W and F150W passbands, successively offset in color for visibility.}
\label{Fig:TRGB_models}
\end{figure}

The \ac{trgb} feature is astrophysically explained by the very rapid thermonuclear ignition of degenerate helium cores of low-mass first-ascent \ac{rgb} stars, the so-called helium flash (HeF), see e.g., \cite{1997MNRAS.289..406S,Salaris:2002xf,2005essp.book.....S,2013osp..book.....C,2017A&A...606A..33S}. In particular, old \ac{rgb} with masses $\lesssim 1.6\,M_\odot$ develop helium cores of nearly equal mass ($\sim 0.5\,M_\odot$), resulting in a nearly equal luminosity of the HeF that can be exploited as a standard candle. Younger, higher-mass giants reach lower luminosity due to lower electron degeneracy at higher core temperatures and lower densities. This mass limit corresponds to an age-effect: giants older than $\sim 4$\,Gyr have nearly constant luminosity in $I-$band. Differences in chemical composition (metallicity) significantly affect \ac{trgb} luminosity and temperature due to line blanketing: the bolometric magnitude behaves as $M_{T,\mathrm{Bol}}\propto -0.19 \mathrm{[Fe/H]}$ \cite{2013osp..book.....C}. At the same time, higher metallicity results in cooler effective temperature. Hence, more metal-poor giants are brighter and bluer.  Thankfully, the dependence on metallicity is partially mitigated in Cousins $I-$band where the bolometric correction ($M_I = M_{\mathrm{Bol}} - BC_I$) nearly compensates the change in luminosity for low-metallicity ($\mathrm{[Fe/H]} \in [-0.7,-2.0]$) giants. Color-based metallicity calibrations can further mitigate this issue, e.g., see Ref.~\cite{Rizzi:2007ni,2017ApJ...835...28J}. Hence, the \ac{trgb} is a theoretically well-supported standard candle when low-metallicity ($\mathrm{[Fe/H]} < -0.7$) old ($> 4\,$Gyr) \ac{rgb} stars are observed in $I-$band. However, observations at infrared wavelengths require caution and additional study with respect to the aforementioned effects, in particular if the observed \ac{rgb} populations exhibit diversity in age and/or chemical composition, see Fig.~\ref{Fig:TRGB_models}. 

Theoretical predictions of the \ac{trgb} luminosity rely on stellar evolution models that require various assumptions and simplified treatment of stellar physics, including the modeling of opacities, diffusion, convection (mixing-length theory) and overshooting, mass-loss, electron screening, neutrino losses \cite{Farag:2023xid}, nuclear reaction rates, among others \cite{2017EPJWC.16004002C,2017A&A...606A..33S}. In particular, the dominant uncertainties related to radiative opacity introduce systematics of up to $1.6\%$. Simultaneously varying radiative and conductive opacities, nuclear reaction rates (e.g., triple-alpha reaction), and neutrino losses affect the bolometric Tip luminosity by $\delta L/L \sim 10^{-3}$, depending on mass and metallicity \cite{Saltas:2022aua}. Comparisons with observations are subject to larger uncertainties, notably related to the translation of bolometric luminosity to magnitude, which require stellar atmosphere models that are usually considered separately from the evolutionary models \cite{2017A&A...606A..33S}.

\paragraph{Methodological considerations and HST-based results}

\subparagraph{The absolute calibration of the \ac{trgb} method} requires measuring $m_{\rm T}$ in \ac{rgb} populations of known distance to obtain $M_{\rm T} = m_{\rm T} - \mu_0$. The best available, geometrically measured, distances (in ascending order) are: trigonometric parallaxes from the \ac{esa} {\it Gaia} mission \cite{Gaia:2016zol}, the distances to the \ac{lmc} and \ac{smc} measured by detached eclipsing binary stars \cite{Pietrzynski:2019cuz,2020ApJ...904...13G}, and the megamaser distance to NGC\,4258 (M106) \cite{Reid:2019tiq}. Importantly, the distance moduli $\mu_{0,i}$ of the \ac{lmc}, \ac{smc}, and NGC\,4258 are frequently used in distance ladders calibrated using stellar standard candles and hence their systematics are often shared among different $H_0$ measurements, e.g., see Ref.~\cite{Freedman:2021ahq,Riess:2021jrx,Madore:2023voh,Breuval:2024lsv}. Comparing distances to other galaxies measured using an absolute calibration based on the same anchor, such as NGC\,4258, thus yields the strongest test of distance systematics \citep[e.g.,][]{Csornyei:2023enu,Riess:2024vfa}. As implied in Fig.~\ref{Fig:TRGB_DL}, uncertainties in $\mu_{0}$ propagate directly into $H_0$ measurements, as they set the absolute scale of the secondary tracers (\ac{sn1}, \ac{sbf}) via the primary standard candle (here: \ac{trgb}). Basing an absolute \ac{trgb} calibration on the broadest possible set of ``anchors'' reduces possible bias and allows one to determine appropriate standardization procedures, e.g., to account for metallicity differences \cite{Rizzi:2007ni,Anand:2021sum}. At present, the most accurate \ac{trgb} calibrations to date are obtained in the Magellanic Clouds \cite{Anderson:2023aga,Koblischke:2024hft} and NGC\,4258 \cite{Anand:2024nim} using passbands similar to $I-$band, such as \ac{hst}'s ACS/F814W or \ac{jwst}'s NIRCAM/F090W. \ac{trgb} calibration based on {\it Gaia} parallaxes \cite{Li:2023pmo} has been achieved based on EDR3 parallaxes, which require correction for complex systematics, e.g., see Ref.~\cite{2021A&A...649A...4L,2021A&A...649A..13M,Riess:2020fzl,2021ApJ...910L...5H,2021ApJ...911L..20R,Reyes:2022boz,2023A&A...680A.105K}. Furthermore, a parallax-based \ac{trgb} calibration has been obtained for $\omega~$Centauri \cite{Soltis:2020gpl}. The period-color relation of stars near the \ac{trgb} may provide a useful avenue to deal with (differential) reddening when calibrating the \ac{trgb} based on field RGs \cite{Koblischke:2024hft}.

A critical element of determining accurate distances using standard candles is to ensure consistency between the absolute calibration and the standardized application in the target environments. However, there are measurable differences even between the $I-$band absolute magnitudes in the \ac{lmc} and the \ac{smc}, and they reflect noticeably in the variability periods of the small-amplitude \ac{rgb} stars that make up the \ac{rgb} population at the Tip \cite{Anderson:2023aga}. These ubiquitous small amplitude pulsations near the \ac{rgb} Tip allow us to probe the intrinsic diversity (e.g., in age and metallicity) of\ac{rgb} populations that is not typically known a priori and thus constitute a difficult-to-control astrophysical systematic at the level of a couple of percent \cite{Koblischke:2024hft}. Additionally, a common feature of \ac{rgb} populations in other galaxies is contamination by other stars, notably at higher luminosity (typically \ac{agb} stars). Unfortunately, the specifics and degree of such contamination differ from environment to environment (cf. CATs below) and are not reliably known a priori. The exception to this rule is globular clusters (GC), which rarely contain stars brighter than the \ac{trgb}, e.g., see Ref.~\cite{2021ApJ...920..129J}, so that measuring $m_{\rm T}$ conceptually corresponds to searching for the brightest cluster star without knowing how much it differs from the HeF luminosity. As a result, GC-based values of $m_{\rm T}$ represent lower limits to the desired measurement of $m_{\rm T}$ and are furthermore also subject to variations in metallicity and age, depending on the photometric band. Thus, GC-based $m_{\rm T}$ values differ conceptually from the values of $m_{\rm T}$ measured in \ac{rgb} populations of mixed age and metallicity. Further study is needed to assess to what degree such effects impact $M_{\rm T}$, and hence, $H_0$.

\subparagraph{Extinction} corrections are routinely applied for \ac{trgb} calibration in the Magellanic Clouds \cite{2021ApJS..252...23S}. However, dust corrections to \ac{trgb} measurements in other galaxies typically, e.g., see Ref.~\cite{Freedman:2021ahq} rely on all-sky dust extinction maps  \cite{Schlegel:1997yv,Schlafly:2010dz} that account only for \ac{mw} foreground dust and whose accuracy is limited in the vicinity of resolved galaxies. Recently, it has been pointed out \cite{Anderson:2021fsp} that small, albeit non-zero, extinction in galaxy halos leads to underestimated $H_0$ values and that dust extinction estimates based on background quasars \cite{2015ApJ...813....7P} provide a possibility for correcting this one-sided systematic. Statistical corrections for local extinction have since been applied \cite{Scolnic:2023mrv,Anand:2024nim}. However, further study of extinction effects and their variation across circumgalactic media would be useful to provide improved corrections.

\subparagraph{The TRGB magnitude ($m_{\rm T}$)} is measured either via an edge detection algorithm \cite{Lee:1993jb,Hatt:2017rxl}, such as a Sobel filter, or via a maximum likelihood fit of the \ac{lf} \cite{Mendez:2002ye,Makarov:2006wc,Li:2022aho}. When Sobel filters are used for edge detection, \ac{lf}s are typically smoothed to some degree to reduce noise \cite{Hatt:2017rxl}, and different weighting schemes have been considered in the literature \cite{Madore:2008tj, Wu:2022hxf}. Recently, it has been pointed out that both the \ac{lf} smoothing and the Sobel response weighting introduce biases that depend on the shape of the observed \ac{lf}s \cite{Anderson:2023aga}. In particular, Sobel response weighting was shown to introduce the tip-contrast relation determined independently using observations \cite{Wu:2022hxf}. Both issues can easily bias the measured value of $m_{\rm T}$ by $0.06$\,mag ($\sim 3\%$ in distance), and it is therefore crucial to apply the same \ac{trgb} measurement method in all target \ac{rgb} populations to avoid bias. Similarly, observer choices, such as color cuts (notably near the Tip), can affect the measured $m_{\rm T}$ and should be decided based on objective criteria \cite{Scolnic:2023mrv}.

\subparagraph{Field-to-field variations of $m_{\rm T}$} have been recently reported within the same galaxies. This led to the development of an unsupervised \ac{trgb} detection algorithm, called CATs (Comparative Analysis of \ac{trgb}s), which sought to reduce the impact of subjective observer choices on the measurement of $m_{\rm T}$ \cite{Wu:2022hxf,Scolnic:2023mrv}. In the process, a tip-contrast relation was determined, which has since been shown to result from the use of weighted Sobel response curves \cite{Anderson:2023aga}. Bias of \ac{trgb} distances can thus be avoided if either a) unweighted Sobel filters are used to measure $m_{\rm T}$ \cite{Anderson:2023aga}, or b) if an appropriate tip-contrast relation is used to standardize $m_{\rm T}$ a posteriori when weighted Sobel response curves are used \cite{Scolnic:2023mrv}. Given the simpler algorithm and the non-uniqueness of the tip contrast, which depends on the measured $m_{\rm T}$, it appears prudent to avoid Sobel filter weighting. First results from an $I-$band (F090W) \ac{trgb} calibration using the \ac{jwst} also support this approach \cite{Anand:2024nim}.

In observations based on the {\it Hubble} Space Telescope or \ac{jwst}, the placement of \ac{trgb} fields in target galaxies is crucial to ensure that the targeted \ac{rgb} population is similar or standardizable to the \ac{rgb} population that provides the absolute calibration \cite{Beaton:2018fyo}, for an example, e.g., see Ref.~\cite{Csornyei:2023enu}. Thanks to very large fields-of-view, observations with the \ac{esa} {\it Euclid} mission \cite{Euclid:2024yrr} or the future Nancy Grace Roman telescope \footnote{\url{https://roman.gsfc.nasa.gov}} will allow us to integrate field-to-field variation analysis into \ac{trgb} algorithms, rendering them much more robust. However, the existence of field-to-field variations within a single galaxy highlights the need for \ac{trgb} standardization to account for differences between \ac{rgb} populations in different galaxies. 

\subparagraph{The choice of photometric system} 
The \ac{trgb} has been most commonly measured in the $I-$band where it is nearly flat and relatively insensitive to metallicity and age effects  \cite{Lee:1993jb,Madore:2023voh}, cf. Fig.~\ref{Fig:TRGB_models}. There has been growing interest in NIR \ac{trgb} measurements due to the capabilities of the \ac{jwst} and the intrinsically higher ($1-2$\,mag) luminosity in the NIR \cite{2024ApJ...966..175N,Newman:2024fkx}, and several $I-$band and NIR observing programs are currently being analyzed with \ac{jwst} (e.g., GO-1685, 1995, 2875, 3055). However, the (color) slope of the \ac{trgb} at NIR wavelengths requires additional consideration. Both empirical (see Refs.~\cite{Valenti:2004jm, Dalcanton:2011si, Wu:2014zxa, Madore:2018qvn, Hoyt:2018ghz, 2020ApJ...898...57D, 2024ApJ...966..175N, Newman:2024fkx}) and theoretical approaches \cite{2017A&A...606A..33S, 2019ApJ...880...63M} have been considered for the time being, and further study is required to determine competitive and accurate \ac{trgb} distances at infrared wavelengths. Furthermore, $K-$corrections are expected to exceed $1\%$ in distance for single-band \ac{jwst} \ac{trgb} observations at distances above $\sim 70$\,Mpc for the F150W passband and shorter wavelengths \cite{Anderson:2021fsp}.

\paragraph{Implications for $H_0$ and the Hubble constant tension}
Several studies have reported $H_0$ measurements involving the \ac{trgb} as a standard candle, with rather significant differences. To understand these differences, it is crucial to consider the differing assumptions and methods underlying each analysis. In particular, Refs.~\cite{Freedman:2021ahq} and \cite{Anand:2021sum} applied different absolute \ac{trgb} calibrations and \ac{trgb} measurement methodologies to the same set of \ac{hst} $I-$band \ac{trgb} observations in conjunction with the CSP \ac{sn1} dataset in the Hubble flow and measured $H_0 = 69.8\pm0.6\pm1.6$\kms and $71.5 \pm 1.8$\kms, respectively. In an update to Ref.~\cite{Freedman:2021ahq}, Ref.~\cite{Freedman:2024eph}(v3) employed \ac{jwst} $I-$band (F090W) \ac{trgb} measurements together with CSP \ac{sn1} and obtained $H_0 = 70.4\pm 1.2 \pm 1.3 \pm 0.7$\,\kms. Ref.~\cite{Scolnic:2023mrv} corrected the tip-contrast \ac{trgb} systematic and replaced the CSP \ac{sn1} with Pantheon+ to ensure a consistent calibration between \ac{sn1} in host galaxies and the Hubble flow and found $72.94 \pm 1.98$\kms. Replacing \ac{hst} entirely with \ac{jwst} $I-$band \ac{trgb} observations of 17 unique \ac{sn1} yields $72.1\pm2.2\pm1.2$\kms \cite{Riess:2024vfa}. Recently, Ref.~\cite{Jensen:2025aai} replaced \ac{sn1} entirely with \ac{sbf} calibrated by \ac{jwst} $I-$band \ac{trgb} observations \cite{Anand:2024nim} and found $73.8 \pm 0.7 \pm 2.3$\kms, confirming the conclusion from Ref.~\cite{Scolnic:2023mrv} that at least half the difference between the $H_0$ values reported by Ref.~\cite{Freedman:2021ahq} and Ref.~\cite{Riess:2022mme} can be attributed to \ac{sn1} alone. 

The prospects for further improvements of the \ac{trgb} method and, in turn, to increase the accuracy in $H_0$ are very promising. New wide-field space-based telescopes (Euclid, Roman) will allow us to comprehensively measure $m_{\rm T}$ in a very large number of galaxies, while \ac{jwst} will allow to push the limits of \ac{trgb} distances. Potentially, {\it Roman} could allow to directly calibrate $m_{\rm T}$ in the \ac{lmc} and \ac{smc}, circumventing differences among photometric systems. However, the target magnitudes are close to saturation ($\sim 6$s at AB $15$\,mag in F087\footnote{\url{https://roman.gsfc.nasa.gov/science/apttables2021/table-timetosaturation.html}}). An \ac{sbf} calibration based on the \ac{trgb} method will yield an accurate $H_0$ measurement fully independent of the Cepheid-\ac{sn1} distance ladder \cite{Anand:2024lbl,Jensen:2025aai}, if different geometric distances are used to calibrate the two types of standard candles. Future improvements in the {\it Gaia} astrometric solution will play an important role in this endeavor, as will synthetic {\it Gaia} photometry based on $B_p$ and $R_p$ spectra \cite{2023A&A...674A..33G}.
Pushing the \ac{trgb} to the limit using \ac{jwst} may motivate infrared observations owing to the combination of \ac{rgb} stars being brighter in the near-IR and \ac{jwst} providing optimal sensitivity there, although it has been already shown that I-band \ac{jwst} \ac{trgb} measurements are feasible out to at least 50~Mpc \cite{2023ApJS..268...15W,Anand:2024lbl}. 

In order to ensure that stellar standard candles provide the clearest and most accurate picture of cosmic tensions, notably the Hubble constant tension, it is crucial to pursue the most direct and simple assessments of systematics. To this end, direct and detailed comparisons of distances measured using the \ac{trgb} method, classical Cepheids, and the \ac{jagb} method are preferred, cf. \cite{Riess:2024vfa}. In contrast, trying to understand \ac{trgb} or Cepheid-based systematics from published values of $H_0$ is complicated by several other possible differences between distance ladder set ups, e.g., with respect to the treatment of peculiar motions and supernova standardization \cite{Scolnic:2023mrv}. Cross-checks based on nearby galaxies, e.g., M31 \cite{Li:2021qkc}, M33 \cite{Lee:2022akw,Breuval:2023rkw}, and other nearby galaxies (e.g., \ac{hst} GO-17520), will be particularly insightful to this end as they provide the greatest possible precision for each of the standard candles and an optimal ability to investigate causes of systematic differences, e.g., using spectroscopy on future 30m-class telescopes, such as the \ac{elt}. 

\bigskip
\subsubsection{The surface brightness fluctuations method \label{sec:SBF}}

\noindent \textbf{Coordinator:} Michele Cantiello\\
\noindent \textbf{Contributors:} Enzo Brocato, Gabriella Raimondo, John Blakeslee, Joseph Jensen, and Rebecca Habas
\\

\paragraph{Introduction \& state of the art}

The \ac{sbf} method is a powerful technique used to measure distances of elliptical galaxies out to $\sim150$ Mpc, and may reach distances as far as $\sim300$ Mpc with telescopes like the \ac{jwst} \cite{Anand:2024lbl}. Introduced in the late 1980s, the \ac{sbf} method relies on the fact that the surface brightness of a galaxy exhibits small-scale variations due to the statistical distribution of stars \cite{1988AJ.....96..807T,1990AJ....100.1416T,Blakeslee:2009tc,2015ApJ...808...91J,Moresco:2022phi}. When imaging a galaxy, the number of stars that fall within each detector pixel varies, causing pixel-to-pixel variations in the surface brightness. The amplitude of these fluctuations is inversely proportional to the square of the distance to the galaxy \cite{Cantiello:2023obe}, allowing one to estimate distances from imaging alone. 

In principle, the \ac{sbf} method is straightforward. The only difficulty lies in isolating the signal fluctuations of the underlying Population II stars from contaminants such as globular clusters, star forming regions, and background galaxies. Once this is done, the absolute \ac{sbf} magnitude can be calibrated in a given passband, $\overline{M}_{\xi}$, using galaxies at well-known distances \cite{Tonry:2000aa} or through stellar populations numerical modeling \cite{Raimondo:2009qs}. The distance modulus $\mu_0\equiv (m{-}M)=5 \log10 (D)+25$ can then be inferred as usual:
$\mu_0 = \overline{m}_{\xi}-\overline{M}_{\xi}$~where $\overline{m}_{\xi}$ is the $\xi$-band observed fluctuation magnitude and  $D$ is the distance in Mpc. 

\begin{figure}
    \centering
    \includegraphics[width=0.8\columnwidth]{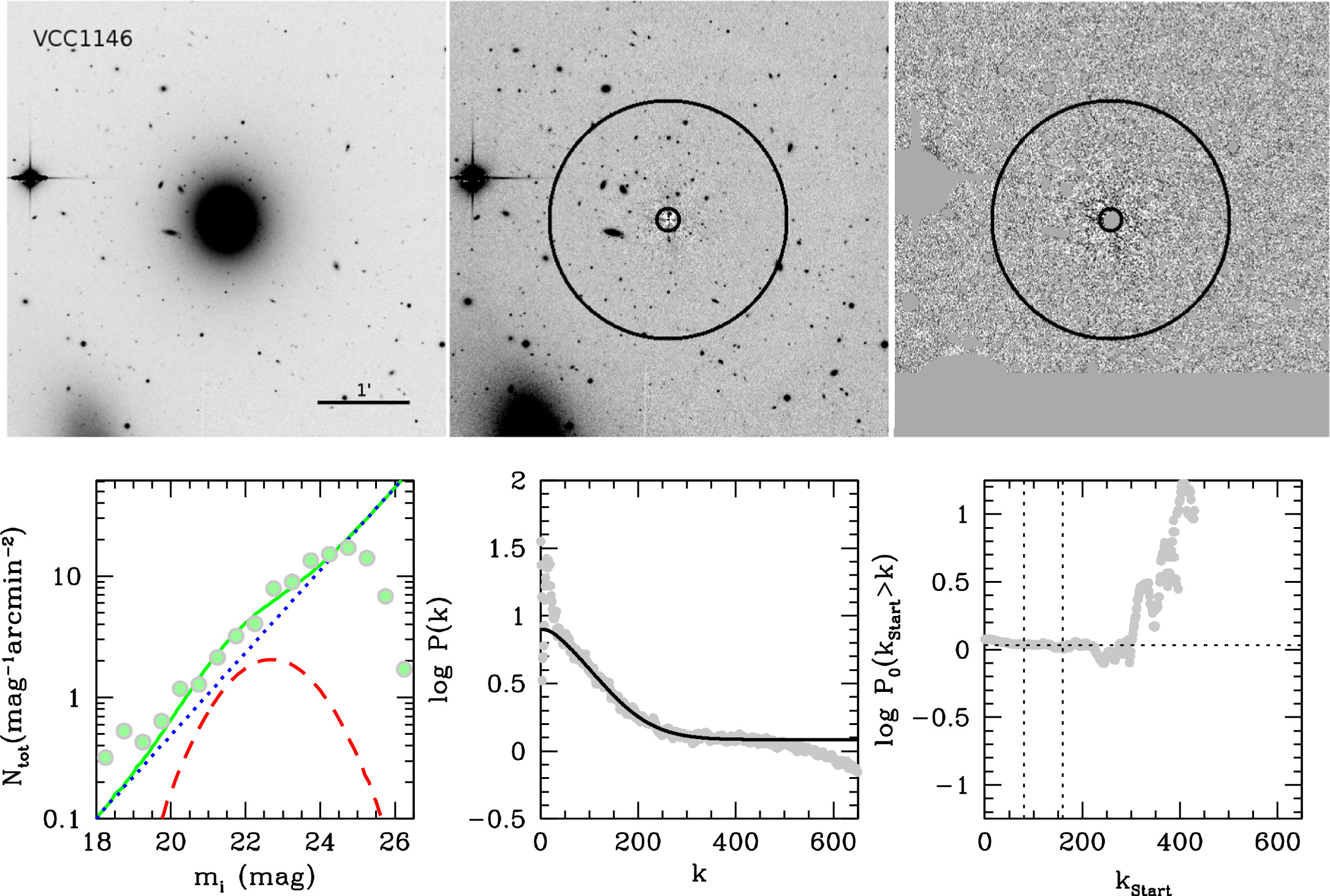}
\caption{\ac{sbf} analysis for the galaxy VCC\,1146. \textbf{Top row:} $i$-band image, residual image, and masked residual image (left to right). Black annuli in the second and third panels mark the inner and outer radii used for \ac{sbf} measurements. \textbf{Bottom left:} Fitted \ac{lf} for external sources. Green circles are observational data with a faint-end downturn from incompleteness. The solid green line shows the best fit, corrected for incompleteness, with blue dotted and red dashed curves representing the background galaxy and GC \ac{lf}s. \textbf{Bottom middle:} Azimuthal average of the residual power spectrum (gray circles) and the fit (solid black line). \textbf{Bottom right:} Fitted \(P_0\) values vs. \(k_{\text{start}}\), with \(k_{\text{end}} = 450\). Vertical lines at \(k_{\text{start}} = 80\) and \(160\) show the range over which the fit is stable. The median \(P_0\) in this range (dotted horizontal line) is adopted. Figure adapted from Ref.~\cite{2024ApJ...966..145C}.}    
\label{fig:sbf}
\end{figure}

In practical terms, the \ac{sbf} distance measurement involves several steps, summarized briefly here and illustrated in Fig.~\ref{fig:sbf}: \(i)\) determining and subtracting the sky background; \(ii)\) modeling and subtracting the galaxy (upper middle panel in Fig.~\ref{fig:sbf}); \(iii)\) masking all potential sources of contamination to the fluctuation signal (e.g., globular clusters within the galaxy, background galaxies, dust patches, etc.; upper right panel); \(iv)\) modeling the \ac{lf} of sources in the frame to estimate the spurious fluctuation contribution from unexcised sources (lower left panel); and \(v)\)  power spectrum analysis of the residual masked frame, to measure the amplitude of the \ac{sbf} signal (lower middle and right panels). Additional details on this procedure are available in the references cited in this review.

In recent years, the application of the \ac{sbf} method has seen significant advancements, particularly with the contributions from the high-resolution space-based data from the \ac{hst}. Recently, Refs.~\cite{Jensen:2021ooi} and~\cite{Blakeslee:2021rqi} used \ac{hst}/WFC3 data to measure highly accurate distances for a sample of $\sim60$ elliptical galaxies, taking particular care to analyze the $\overline{M}_{F110W}$ calibration, and succeeded in obtaining distances with median statistical uncertainties $\leq4\%$. 

From the ground, the Next Generation Virgo Cluster Survey (NGVS), a project based on 104 degrees$^2$ deep optical imaging from the CFHT \cite{2012ApJS..200....4F}, has provided a dataset ideal for \ac{sbf} measurements in the Virgo Cluster. The NGVS has enabled detailed studies of galaxy distances in this cornerstone galaxy cluster \cite{2024ApJ...966..145C}, and represents a crucial survey, serving as a precursor for future large ground- and space-based surveys.

Most of the focus for the use of \ac{sbf} has been on bright galaxies, where the method provides the lowest uncertainty, reaching levels $\leq5\%$ on single targets in well-designed observations. The advent of large area deep surveys has, however, led to the discovery of an increasingly large number of dwarf galaxies which play a fundamental role in Cosmology \cite{Sales:2022ich}. Several authors \cite{2018RNAAS...2..146B,2019ApJ...879...13C,2021ApJ...923..152K} have applied the \ac{sbf} technique to smaller and fainter systems. These studies have demonstrated that $\overline{m}$ can be used to estimate distances to dwarf galaxies, although with relatively large errors compared to massive galaxies, providing a typical uncertainty of $\delta D/D \sim 15\%$. Despite the relatively larger error, these elusive galaxies are often challenging targets for distance measurement due to their low surface brightness and complex stellar populations. This has opened new avenues for using \ac{sbf} to probe the Local Universe and improve the cosmic distance ladder.

\paragraph{The Hubble constant tension}
In addition to providing distance estimates for individual galaxies, \ac{sbf} can also be used to probe tensions in the measurement of the Hubble constant that have arisen between the \ac{cmb} observations \cite{Planck:2018vyg} and the value obtained from the late-Universe \cite{Riess:2021jrx}.

The \ac{sbf} method, with its ability to provide distance measurements to galaxies beyond the 50 Mpc range, offers a unique perspective on this tension, independent of the classical Cepheid/\ac{sn1} route. This topic has been explored by Ref.~\cite{Blakeslee:2021rqi} for a sample of bright galaxies between 20-100 Mpc, with $\overline{M}_{\rm F110W}$ calibrated using both Cepheids and \ac{trgb} distances. For both calibrations, the authors derived a value of $H_0$ that is fully consistent with the direct distance estimates in the local Universe. The work by Ref.~\cite{Blakeslee:2021rqi}  was based on a small sample, and this result should be corroborated with larger studies. That study, as well as some of the newly programmed ones \cite{Anand:2024lbl}, used \ac{sbf} in near-IR passbands where the amplitude of fluctuations is ``enhanced'' because the brightest stars contributing to the signal are brighter compared to the optical bands. This enhancement is key for the recent results on $H_0$ and will be crucial for future observations, as it allows for reliable measurements for galaxies at larger distances.

A complementary approach to the $H_0$ tension uses \ac{sbf} indirectly, as a calibration for \ac{sn1} distances. Rather than the Cepheid-\ac{sn1} route to Cosmology, one can use a Cepheid-\ac{sbf}-\ac{sn1} route. This method has larger systematic and statistical uncertainties from the extra rung in the distance ladder, but offers the advantage of having a larger calibration sample of \ac{sn1} from the local Universe because of the larger overlap in galaxies that host \ac{sn1} and have \ac{sbf} measurements. This approach was adopted by Ref.~\cite{Khetan:2020hmh}, who used a heterogeneous collection of \ac{sn1} and \ac{sbf} measurements from existing literature, obtaining $H_0=71.1 \pm 2.4 {\rm (stat)} \pm 3.4 {\rm (sys)}$\kms. More recently, Ref.~\cite{Garnavich:2022hef} used a very homogeneous sample of \ac{sbf} measurements from Ref.~\cite{Jensen:2021ooi}, and found that the Hubble-Lema\^{i}tre parameter derived from the revised SALT2 \cite{Scolnic:2021amr} parameter fit and calibrated with near-IR \ac{sbf} distances is $H_0=73.3 \pm 1.0 {\rm (stat)} \pm 2.7 {\rm (sys)}$\kms.
Note that the \ac{sn1}-independent estimate from Ref.~\cite{Blakeslee:2021rqi}, based on Cepheid zero-point calibration, yielded $H_0 = 73.3 \pm 0.7 {\rm (stat)} \pm 2.4 {\rm (sys)}$\kms, whereas the most recent recalibration, which relies on \ac{trgb} distances from \ac{jwst} observations of Virgo and Fornax galaxies by Ref.~\cite{Jensen:2025aai}, provided $H_0 = 73.8 \pm 0.7 {\rm (stat)} \pm 2.3 {\rm (sys)}$\kms.

\paragraph{Future prospects and projects}

The \ac{sbf} method is expected to become even more relevant in the future, taking advantage of the fact that most of the forthcoming facilities will operate in the near-IR regime, where the \ac{sbf} signal is much stronger. The \ac{jwst}, with its superior resolution and sensitivity, is expected to enable detailed \ac{sbf} studies of distant galaxies, potentially reaching $\sim300$ Mpc \cite{Anand:2024lbl}. The large sky surveys from the Vera Rubin Observatory's \ac{lsst} and from the Euclid mission, will further enhance the capabilities of \ac{sbf} by providing an enormous dataset for \ac{sbf}, largely dominated by the dwarfs. This will allow astronomers to apply the \ac{sbf} method to a much larger and more diverse sample of galaxies, improving the statistical robustness of the 3D mapping of the Universe within 40-70 Mpc.

On longer time scales, the Nancy Grace Roman Space Telescope (to be launched in the mid-late 2020s), will also play a crucial role in the advancement of \ac{sbf}, due to its combination of wide-field, high-resolution, and infrared wavelength coverage. Roman will enable high-precision \ac{sbf} measurements over large areas of the sky, potentially reaching the depth of the \ac{jwst}, but over a wider area. The future 30-40 m class ground based telescopes, with their adaptive optics (AO), may also be more useful for \ac{sbf} than the \ac{jwst}. Correcting for atmospheric distortions using the AO systems could provide better FWHM resolutions than the \ac{jwst} in certain passbands (e.g., $K$-band) which, combined with the huge telescope collecting area, would allow \ac{sbf} to be applied to more and more distant systems. Although time-consuming, because of the specific requests of AO observations of extended objects and the needs for AO activation loops, observing a handful of well-chosen targets with one of these massive telescopes could allow one to discriminate between different cosmological models, independently from \ac{sn1}. 

Beyond improved telescopes, \ac{sbf} observations also benefit from improved numerical stellar population studies. Accurate models of stellar populations are essential for interpreting \ac{sbf} measurements, reducing systematic uncertainties and deriving reliable $k$-corrections which cannot be ignored beyond $\sim150$ Mpc. Additionally, the application of \ac{sbf} to blue, low-mass galaxies requires the calibration of $\overline{M}$ across a broader range of stellar populations and environments. Stellar population modeling will offer a reference to test and prove the reliability of the \ac{sbf} method over a wide interval of population properties.

Theoretical efforts over the past 20 years have focused on incorporating stars in peculiar evolutionary stages, such as thermally pulsing \ac{agb} stars and hot HB stars in old-intermediate populations (e.g., see Ref.~\cite{Raimondo:2009qs,2020ApJS..250...33C}). These efforts have also included the impact of $\alpha$-elements and He-abundance enhancements. Advancements in this area will benefit from the insights gained by applying theories of late stellar evolution ---dominated by poorly understood physics such as mass loss processes--- to the observations in mid-infrared instruments like MIRI on the \ac{jwst}. This will result in more refined stellar population models for distance estimates and a better understanding of these crucial stages of stellar evolution.

Future \ac{sbf} measurements can also benefit from the adoption of \ac{ml} techniques. \ac{ml} methods require training on large samples that are not yet available, but when they are, \ac{ml} algorithms can optimize the measurement of \ac{sbf}, which is currently slow and heavily reliant on human intervention. \ac{ml} can further identify patterns and correlations in large datasets, improving the efficiency of \ac{sbf} measurements and reducing measurement time. Sufficient training datasets will be available after the first years of Euclid or \ac{lsst} public data releases.

In conclusion, the \ac{sbf} method currently stands as a robust and versatile tool in extragalactic distance measurements. To date, the method’s range of usefulness allows it to be adopted for tests on the $H_0$ tension, providing results that agree with the late-time/direct Hubble constant estimates. With advancements in telescope technology, data processing, and collaborative efforts, the \ac{sbf} method is well-positioned to make significant contributions to resolving cosmic tensions, possibly even beyond the $H_0$ parameter, and enhancing our understanding of the Universe. By implementing the proposed future developments, the astronomical community can further refine the \ac{sbf} technique and extend its applicability to new frontiers in extragalactic astronomy.
\bigskip
\subsubsection{Mira variables \label{sec:mira_stars}}

\noindent \textbf{Coordinator:} Caroline Huang\\
\noindent \textbf{Contributors:} Antonio Capodagli, Lucas Macri, and Massimo Marengo
\\

\noindent Mira variables are fundamental-mode, thermally-pulsing \ac{agb} stars with periods ranging from $\sim 100-1000$ days or longer. They fall into a broader category of variables known as Long-Period Variables (LPVs), which includes semi-regular variables, and OGLE small-amplitude \ac{rgb} stars, which are overtone or irregular pulsators. The pulsation of Miras is likely driven by a $\kappa$-mechanism similar to that found in Cepheids. As highly-evolved stars, they contribute to the chemical enrichment of the \ac{ism} through stellar mass loss, have low effective temperatures ($T_{\text{eff}} < 3500$ K) with spectral intensity peaking between 1-2 \textmu m, and are often among the brightest stars in an intermediate-to-old population ($L \sim 10^4 L_\odot$). They are also ubiquitous as nearly all stars ($0.8 M_{\odot} < M < 8M_{\odot}$) will experience a Mira phase in evolution. 

Like other \ac{agb} stars, they may be classified into two main spectral types based on their photospheric carbon-to-oxygen ratio (C/O ratio). The exact boundaries are somewhat fluid, but typical C/O values are: C/O ratio $\gtrsim 1$ are classified as C-rich (C-type), C/O ratio $\lesssim 0.5$ are O-rich (M-type), and C/O ratio $\sim 0.5 - 1.0$ are intermediate, or S-type. While the identification of carbonaceous or silicate molecular features in stellar spectra is the gold standard for classification, in practice, spectra are difficult or expensive to obtain, and colors, or color-color diagrams, are typically used as a proxy for spectral type. 

\begin{figure}[h]
    \centering
    \includegraphics[width=0.8\linewidth]{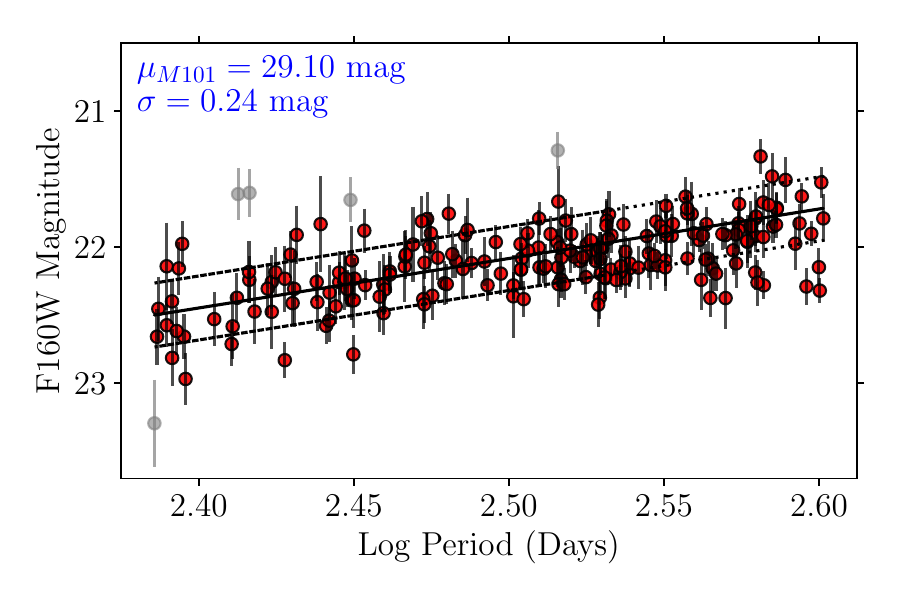} 
    \caption{Period-Luminosity Relation for short-period, presumed O-rich Miras in the \ac{sn1} host galaxy M101 from Ref.~\cite{Huang:2023frr}.}
    \label{fig:M101_PLR}
\end{figure}

The first Period-Luminosity Relations (PLRs) derived for Miras used bolometric magnitudes \cite{1981Natur.291..303G, 1989MNRAS.241..375F}. Over the past several decades, Miras have been observed in bands ranging from optical to far-infrared have been used to measure distances to many Local Group galaxies (see reviews by Refs.~\cite{Whitelock:2012yk, 2019IAUS..343..275W, Huang:2024exg} and references therein) and map the structure of the \ac{mw} \cite{2023ApJS..264...20I}. Studies of nearby dwarf galaxies have suggested that metallicity does not appear to have an observable effect on the pulsational properties of these stars \cite{2017ApJ...851..152B, 2019ApJ...877...49G}, although possible metallicity effects have been suggested by theoretical models. Recently, nonlinear stellar pulsational models have shown significant improvements in matching theoretical predictions of long-period variable PLRs with observations \cite{2021MNRAS.500.1575T} which may shed light on this apparent discrepancy in the future. 

\paragraph{Application}
For the purposes of this review, we will focus on the O-rich Miras, which are more commonly used as distance indicators because they follow PLRs with lower scatter in near-infrared wavelengths ($\sim$ 0.12 mag) compared to their C-rich counterparts. In order to create the PLRs, time-series observations --- often with either an uneven, power-law spacing or monthly sampling --- are also required in order to determine periods. 

The first rung of Mira distance ladder uses \emph{Hubble} Space Telescope observations of Miras in the water megamaser host galaxy NGC 4258 and ground-based near-infrared observations of Miras in the \ac{lmc} \cite{2017AJ....154..149Y, 2018AJ....156..112Y} as ``anchor'' galaxies to obtain an absolute magnitude calibration \cite{Huang:2019yhh}. In order to tie the ground-based observations to the \ac{hst} photometric system, Ref.~\cite{Huang:2018dbn} used a photometric transformation derived from O-rich Mira spectra. Unlike the Cepheid distance ladder, there is currently no precise, parallax-based absolute magnitude calibration. This is due to the fact that Miras (as well as \ac{agb} stars in general, and other evolved stars such as red supergiants) exist at the intersection of many difficulties for \emph{Gaia} --- they are highly luminous, red, have angular sizes larger than their parallaxes, and are known to have photocenter variations, all of which can contribute to underestimated uncertainties and bias in the inversion of their parallaxes \cite{2018A&A...617L...1C, 2020A&A...640A..23C, 2022A&A...667A..74A, 2023A&A...669A..49A}.

The second rung of the distance ladder is currently built on observations of Miras in two nearby Type Ia Supernova host galaxies --- NGC 1559 and M101 (hosts of SN 2005df and SN 2011fe respectively, with PL relation for M101 shown in Fig.~\ref{fig:M101_PLR}) \cite{Huang:2019yhh, Huang:2023frr}. Thus far, these are the only two \ac{sn1} host galaxies with Mira-based distances. The statistical uncertainty in the \ac{sn1} calibration is also the dominant source of error in the Mira-H$_0$ measurement (even after standardization, the uncertainty in \ac{sn1} peak magnitude is $\sim 0.1$ mag). Thus the greatest reduction in uncertainty will be obtained from obtaining more observations of Miras in \ac{sn1} host galaxies. 

\paragraph{Implications for Hubble constant and Hubble constant tension} The current most precise Mira-based measurement yields H$_0 = 72.37 \pm 2.97$\kms ($\sim$ 4\% total uncertainty, including both statistical and systematic components) \cite{Huang:2023frr}. While it is not yet as precise as more established distance indicators, as an independent measurement, this result does support the hypothesis that the local measurement of $H_0$ is greater than the early-Universe measurement with $\sim 95$\% confidence. In addition, it provides an important cross-check to distances made with other more established distance indicators such as Cepheids and \ac{trgb}. 

\paragraph{Prospects} The next decade should be particularly exciting for using Miras and other LPVs as distance indicators. In the NIR/MIR bands accessible with \ac{hst}, \ac{jwst} and \textit{Roman}, Miras have PLRs with smaller scatter than in the optical, nearly-sinusoidal light curves, and are relatively easy to identify as some of the brightest stars in resolved stellar populations. The Vera C.~Rubin Observatory's \ac{lsst} is expected to have sufficient depth and number of epochs to enable the detection of Miras at optical wavelengths ({\it griz}) \cite{2024MNRAS.531..110K} in galaxies out to $D\sim 10$~Mpc \cite{Yuan_2017_PhD}.

\bigskip
\subsubsection{Type Ia supernovae \label{sec:SNeIa}}

\noindent \textbf{Coordinator:} Maria Vincenzi\\
\noindent \textbf{Contributors:} Jenny Wagner, Lluís Galbany, Luca Izzo, M. Pilar Ruiz Lapuente, and Sanjay Mandal
\\

\noindent \ac{sn1} are thermonuclear explosions of carbon oxygen white dwarfs in close binary systems. Their peak brightness after various empirical corrections is highly homogeneous (r.m.s.\ $<$0.15 mag in $B$-band), hence their usage in cosmology as standardizable candles. \ac{sn1} have played and still play a crucial role in measurements of the Hubble constant, and they are typically used in two out of three rungs in the typical distance ladder.  In the second rung, their luminosities are calibrated with stellar distance indicators, e.g., Cepheids or \ac{trgb} or \ac{jagb}, in the same galaxy.  In the third rung, their brightness values calibrate the Hubble-Lema\^itre relation into what is now deemed the `Hubble flow' ($\sim$100 to 600 Mpc, or $z<0.15$), where the cosmological expansion dominates over the peculiar motions of galaxies. Ideally, one could remove the intermediate step (second rung), and go straight from geometric calibration to \ac{sn}. Unfortunately, the rate of \ac{sn} in the local Universe is not nearly frequent enough to provide multiple \ac{sn} in the few galaxies used for geometric anchors (the \ac{mw}, the \ac{lmc}, or the mega-maser NGC 4258). Even with Cepheids/\ac{trgb}/\ac{jagb} as a go-between, the low rate of \ac{sn} in the nearby Universe (roughly one per galaxy per 100 years) is the limiting component of the precision of $H_0$ measurements, for instance Ref.~\cite{Riess:2021jrx} utilizes every \ac{sn1} that pass cosmological quality requirements and Cepheid suitability within $40$ Mpc. The most comprehensive three-rung ladder $H_0$ measurements using Cepheids are shown in Fig.~\ref{fig:distance_ladder}.
 
\paragraph{SN Ia standardization and derivation of $H_0$}
We review here the formalism for deriving the Hubble constant with \ac{sn1} in the local distance ladder, as presented by Ref.~\cite{Riess:2021jrx}. Using a set of Cepheids or \ac{trgb} distance moduli ($\mu_0$) calibrated with geometric measurements such as parallax or megamasers (first rung in the distance ladder) and comparing these distances to brightnesses $m_{\rm X}$ of \ac{sn1} exploding in the same galaxies, we can estimate the single offset $M_{\rm B}$, which describes the absolute magnitude of an \ac{sn1} (second rung).
Following the SALT modeling framework \cite{SNLS:2007cqk,Kenworthy:2021azy},\footnote{The SALT2 and SALT3 frameworks have been used in all most recent cosmological analyses. Other frameworks are also available, i.e., SnooPy \cite{Burns:2010ka}, BayeSN \cite{Mandel:2010xj}.} \ac{sn1} standardized brightness, $m_{\rm X}$,\footnote{We call the standardized brightness $m_{\rm X}$ instead of $m_{\rm B}$ as in Ref.~\cite{Riess:2021jrx} to be clear that the brightness is standardized.} is generally measured with the Tripp formula \cite{1998A&A...331..815T} such that:
\begin{equation}\label{eq:SN_Tripp}
    m_{\rm X}=m_{\rm B}+\alpha x_1 - \beta c - \delta_{\rm Bias} + \delta_{\rm Host}\,,
\end{equation}
where $m_{\rm B}$, $x_1$ and $c$ are all independent properties of each light curve derived using the SALT light-curve model, $\alpha$ and $\beta$ are correlation coefficients that help standardize the brightness, $\delta_{\rm Bias}$ is a correction due to selection effects and other biases as predicted by simulations, and $\delta_{\rm Host}$ is a final correction due to residual correlations with host galaxy properties.

For a \ac{sn1} in the $i$-th Cepheid host, 
\begin{equation} 
    m_{\rm X,i}=\mu_{\rm 0,i}\!+\!M_{\rm B}\,, \label{eq:snmagalt} 
\end{equation}
where $M_{\rm B}$ is the fiducial \ac{sn1} absolute magnitude (assumed to be the same across the whole sample), and $\mu_{0,i}$ is the distance modulus derived from Cepheid measurements for the $i$th galaxy. The ladder is completed with a set of \ac{sn1} that measure the expansion rate quantified as the intercept, $a_{\rm B}$, of the distance (or magnitude)--redshift relation.  For an arbitrary expansion history and for $z>0$ as
\begin{equation} 
    a_{\rm B}=\log\,cz \left\{ 1 + {\frac{1}{2}}\left[1-q_0\right] {z} -{\frac{1}{6}}\left[1-q_0-3q_0^2+j_0 \right] z^2 + \mathcal{O}(z^3) \right\} - 0.2m_{\rm X}^{\rm HF}\,, \label{eq:aB} 
\end{equation}

\noindent measured from a set of \ac{sn1} ($z, m_{\rm X}^{\rm HF}$) in the Hubble flow,  where $z$ is the redshift due to expansion, $q_0$ is the deceleration parameter, and $j_0$ is the jerk parameter. Typically, for \lcdm, $j_0$ is set to 1.  The determination of $H_0$ follows from
\begin{equation} 
    \log\, H_0={0.2 M_{\rm B}\!+\!a_{\rm B}\!+\!5}\,. \label{eq:h0alt} 
\end{equation}

Covariances between rungs are taken into account following the approach in Refs.~\cite{SNLS:2011lii} and~\cite{Dhawan:2020xmp}. Finally, $q_0$ (and $j_0$) can only be constrained from \ac{sn1} data, without the requirement of any additional information.

\paragraph{Systematics on the path to $H_0$ with SN Ia}
$H_0$ measurements using \ac{sn1} are affected by various sources of systematic uncertainties. 
\begin{figure}
    \centering
\includegraphics[width=1\textwidth]{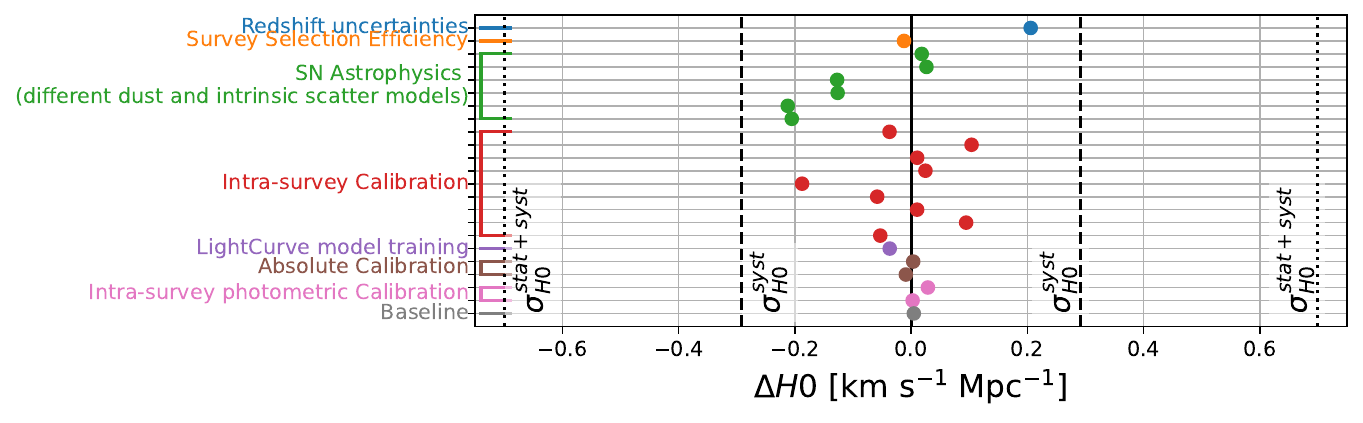}
\caption{Adapted from Ref.~\cite{Brout:2022vxf}, the impact on recovery of $H_0$ of the various systematic uncertainties tabulated. The units of these measurements are \kms.  The dashed lines are given at $\Delta {\rm H}_0$ of 0.7, which is the entire contribution of the uncertainty in Ref.~\cite{Riess:2021jrx} from \ac{sn} measurements. We labeled the different categories of systematic uncertainties.}
\label{fig:sysh0}
\vspace{.1in}
\end{figure}

\begin{table}[ht]
\centering
     \caption{A summary of the various cross-checks and systematics on the supernova component of the distance ladder. If two uncertainties are given, the first one is the statistical uncertainty and the second one is the systematic uncertainty.} \label{tab:table1}
    \begin{tabular}{|p{3cm}|p{11.5cm}|p{2.6cm}|}
    \hline
     \textbf{Reference} &     \textbf{Notes on the specific systematics check implemented}&                                          \textbf{Result} (\kms) \\
    \hline
 \textbf{Optical \ac{sn1}:} & ~ & ~ \\
    \cite{Murakami:2023xuy} &  Uses spectral feature twinning process to improve standardization; checks dust modeling, intrinsic scatter modeling; & $73.01 \pm 0.92$ \\
    \cite{Peterson:2021hel} &  Checks different models of peculiar velocities / bulk flows & $\sigma_{H_0}<0.2$ \\                
  \cite{Brownsberger:2021uue} &  Checks \ac{sn1} Calibration by allowing individual \ac{sn} survey offsets &                   $\sigma_{H_0}<0.2$ \\  
         \cite{Jones:2018vbn}  &    Checks impact of mass step, global vs.~local correlations &                        $\sigma_{H_0}<0.15$  \\
       \cite{CSP:2018rag} &     Checks light-curve fitting method; also does NIR fits &                $73 \pm 2$ \\
   \cite{Garnavich:2022hef} & Uses 4-rung distance ladder, checks \ac{sn1} host demographic systematic   &       $74.6 \pm 0.9  \pm 2.7$  \\
      \cite{Dhawan:2022yws} &   Uses \ac{ztf} data alone, check on \ac{sn1} calibration   &                    $76.94 \pm 6.4$  \\
   \textbf{NIR \ac{sn1}:} & ~ & ~ \\
   \cite{Dhawan:2017ywl} &    Uses literature NIR \ac{sn} (restframe $J$) and peak fitting   &              $72.8 \pm 2.8 $ \\
    \cite{Galbany:2022zir} &  Uses literature NIR \ac{sn} (restframe $J$ and $H$ band)  &        $72.3 \pm 1.4 \pm 1.4 $ \\
    \cite{Jones:2022mvo} & Uses  RAISIN$\backslash$+literature NIR \ac{sn} (restframe $Y$ band) and SNooPy fitting, check on dust & $75.9\pm2.2$  \\
    \cite{Dhawan:2022gac} &   Uses literature NIR \ac{sn} and BayesN fitting &           $74.82 \pm 0.97  \pm 0.84 $  \\
\hline
\end{tabular}
\end{table}

Ref.~\cite{Brout:2022vxf} gives a comprehensive overview of many of these systematics and how they may affect measurements of $H_0$. In Fig.~\ref{fig:sysh0}, adapted from Ref.~\cite{Brout:2022vxf}, we show the impact on $H_0$ of applying $1\sigma$ shift of each systematic.
We group the systematics into various categories: Redshifts, \ac{sn} Physics, Selection, Calibration, and \ac{mw} Dust.
Before delving into each category of systematics in detail, it's crucial to understand which systematics $H_0$ is most sensitive to, and conversely, which have minimal impact due to being mitigated. The distance ladder is structured so that the same probe is used in two of its three rungs (e.g., Cepheids in the first and second rungs, and \ac{sn1} in the second and third rungs). If a systematic error disproportionately affects \ac{sn1} in the second and third rungs, or introduces significant differences in the \ac{sn1} populations between these rungs, it will have a substantial impact on $H_0$. Conversely, if a systematic introduces a consistent offset that affects all \ac{sn1} uniformly, the impact of that systematic will be negligible due to the cancellation effect inherent in this formalism. In other words, consistent offsets will cancel out, mitigating the effect of the systematic error. \\
\noindent \textbf{Calibration:} Since many of the same surveys measure \ac{sn} in both the second and third rungs of the distance ladder, the effects of calibration systematics on $H_0$ is mitigated by the effect discussed above. As illustrated in Ref.~\cite{Brownsberger:2021uue} in the context of the latest SH0ES analysis, per-survey ``gray'' calibration offsets affect the uncertainty by $0.2$\kms at most. However, when different surveys are used for the second and third rungs, as shown in Ref.~\cite{CSP:2018rag,Freedman:2019jwv}, this cancellation does not occur and the effects of calibration can be larger.\\
 \noindent \textbf{Redshifts:} $H_0$ measurements are more sensitive to systematics related to redshift since redshift information is used only in the third rung of the distance ladder, with no systematic mitigation. Ref.~\cite{Peterson:2021hel} explore various bulk flow models in the nearby Universe, finding that changes in $H_0$ could be up to $0.2$\kms. 
 They also note that including or excluding peculiar velocity corrections could lead to $\Delta H_0 \sim 0.5$. Additionally, Refs.~\cite{Carr:2021lcj} and~\cite{2022yCat..19020014S} find redshift measurement biases causing uncertainties around $0.1$\kms.
 
\noindent \textbf{SN Physics and Selection:} 
\ac{sn} intrinsic astrophysics and the role of \ac{sn} dust remain one of the most poorly understood aspects of \ac{sn1} cosmology and can affect $H_0$ measurements \cite{Murakami:2023xuy,Popovic:2021yuo}. Most analyses have found that these effects have a small impact on $H_0$ because the differences between \ac{sn} sub-populations selected in the second and third rung are not expected to be significantly different.
For example, Ref.~\cite{Rigault:2014kaa} showed evidence for a correlation between standardized brightness and the age of the host galaxy (quantified estimating the specific star-formation at the \ac{sn} location). In earlier measurements like Ref.~\cite{Riess:2011yx}, the third rung of \ac{sn} had no galaxy-based selection applied. Only the second rung had this selection, tied to Cepheid discovery, which favored star-forming host galaxies. This would potentially lead to a bias in the recovery of $H_0$. The size of the bias would depend on the relative differential fraction of host-galaxy demographics between the second and third rung multiplied by the size of the correlation. Subsequent analyses \cite{Jones:2018vbn} showed that this effect would likely be insufficient to explain the Hubble tension. Still, in the most recent SH0ES analysis \cite{Riess:2021jrx}, the selection of \ac{sn} in the third rung of the distance ladder was done to be as similar as possible as the second rung. Only \ac{sn} found in star-forming galaxies are selected, which thereby removed the sensitivity to this systematic. Yet, the impact of this change was less than the statistical uncertainty from the supernova component of the distance ladder. Similarly, significant differences in dust extinction and/or color-related effects between \ac{sn} in the second and third rung could potentially bias $H_0$ measurements \cite{Wojtak:2024mgg}. \ac{sn} dust extinction and color-dependent corrections are encapsulated in the nuisance parameter $\beta$ (see Eq.~\eqref{eq:SN_Tripp}). As a cross-check, the $\beta$ parameter was fitted separately in \ac{sn} used in the second and third rung, and it was found to be consistent \cite{Brout:2022vxf}. This test, together with various NIR \ac{sn} $H_0$ measurements, see Table~\ref{tab:table1} \cite{Dhawan:2017ywl,Galbany:2022zir,Jones:2022mvo,Dhawan:2022gac}, suggests that dust or color-dependent effects are not expected to significantly bias $H_0$, or to be a significantly underestimated systematic in current $H_0$ measurements.\\
\noindent \textbf{Other systematics:}
Additional systematic tests, such as changing the light-curve fitter or adding spectroscopic information, have shown consistent results, within $\approx 0.3$\kms, e.g., see Ref.~\cite{CSP:2018rag,Murakami:2023xuy}. Even when isolating data to a single survey, as in Ref.~\cite{Dhawan:2022gac}, the results remain consistent, though with larger uncertainties due to the smaller sample size.

\paragraph{Variants on the path to $H_0$ with Supernovae Ia}
Some of the main cross-checks on the \ac{sn1} used for these analyses is varying the wavelength regime in which light curves are measured (i.e., optical to NIR) or the dataset used (i.e., the survey used to measure the light curves).  As $H_0$ constraints are limited by the number of \ac{sn} found within 40 Mpc, there is considerable overlap in the data between these various studies.  The most popular path to check and improve the distance ladder with \ac{sn1} is by measuring NIR light curves.  We list these papers in Table~\ref{tab:table1}.  Overall, even though the rest-frame band in which light curves are measured varies between these analyses, and the fitting method varies between these methods, there is generally very good agreement in the recovered values of $H_0$.  One challenge multiple of these studies have found (e.g., \cite{Dhawan:2017ywl, Jones:2022mvo}) is larger calibration offsets between samples than those found for optical studies.  A benefit of this type of study is the possibility of improved precision of distance measurements from NIR data, but the quality of older light curves has not typically been good enough to evaluate this possibility.  

An additional path is creating a ``4-rung distance ladder'', as done in Ref.~\cite{Garnavich:2022hef}. \ac{sn1} used in the SH0ES distance ladder are those found in late-type galaxies. To avoid this specific sub-sample, one can add another rung in the distance ladder from \ac{sbf} between \ac{trgb}/Cepheids and \ac{sn}.  The analysis of Ref.~\cite{Garnavich:2022hef} improves on that of Ref.~\cite{Khetan:2020hmh} because the latter follows a similar method, but uses an inhomogeneous set of \ac{sbf} measurements.
The inhomogeneous set significantly increases the scatter of the tie between \ac{sbf} and \ac{sn} measurements and appears to bias $H_0$ to lower values. Ref.~\cite{Garnavich:2022hef} find a value of $H_0=74.6\pm2.8$\kms, in good agreement with the SH0ES value.

\paragraph{Inverse distance ladder to $H_0$ with supernova}
The same set of \ac{sn1} used in the SH0ES distance ladder can also serve as uncalibrated relative distance indicators to constrain $H_0$ when combined with other probes like \ac{bao}. \ac{bao} constrain the expansion history $H(z)$ and extrapolate $H_0$. But this approach assumes the sound horizon from \ac{cmb} constraints and is model-dependent, relying on \lcdm\ to infer $H(z=0)$ from \ac{bao} data at $z\sim0.5$. However, \ac{sn1} can address the latter issue. Instead of calibrating \ac{sn1} to the distance ladder, they can be calibrated to the \ac{bao} distance scale at typical \ac{bao} redshifts ($z\sim0.5$), allowing \ac{sn1} to constrain the expansion history at later times ($z<0.1$) without assuming \lcdm. Studies such as Refs.~\cite{DES:2018rjw,Feeney:2018mkj,DES:2024ywx} using this technique found $H_0=68.57\pm0.9$\kms, consistent with \ac{cmb} under \lcdm. Since \ac{bao}s obtain the physical scale from the sound speed in the early Universe, these low $H_0$ values have prompted discussions about the impact of the sound horizon value.

\paragraph{Improving measurements of $H_0$ in the future}
Improving constraints on $H_0$ using \ac{sn1} and the distance ladder is challenging. The limiting factor in the past decade has been the fact that \ac{hst} can discover Cepheids/\ac{trgb} within a radius of 40 Mpc, and there is one to three \ac{sn1} per year exploding within this volume. However, new telescopes like the \ac{jwst} (or future instruments like the Nancy Grace Roman Space Telescope and the \ac{elt}) are already showing impressive improvements (both in quality and depth) in Cepheids, \ac{trgb} and \ac{jagb} measurements\footnote{Even a $25\%$ increase in the Cepheid distance would allow a doubling in the number of usable \ac{sn1} in the second rung of the distance ladder (the volume of discovered \ac{sn} will increase with distance cubed).} as shown by Refs.~\cite{Riess:2024vfa,Freedman:2024eph}, with exciting hints of shifts to lower $H_0$ values presented by Ref.~\cite{Riess:2024vfa}, even though the statistics is still too low to draw firm conclusions.

The other path towards improving the constraint from \ac{sn1} is to improve the precision of the measurements.  This is the path followed by papers like Refs.~\cite{Murakami:2023xuy,Ruiz-Lapuente:2023gow} which tried using spectral features to improve the standardization, or the large number of papers that measure \ac{sn1} in the NIR as better standard candles. The main challenge is that these new standardization approaches depend on the amount of data available for its application in past literature measurements. New types of measurements can be made for nearby \ac{sn} in the future, but we can not re-measure past \ac{sn1}.
\bigskip
\subsubsection{J-regions of the asymptotic giant branch methods \label{sec:JAGB}}

\noindent \textbf{Coordinator:} Siyang Li\\
\noindent \textbf{Contributors:} Adam Riess, Bartek Zgirski, Caroline Huang, Dan Scolnic, Gagandeep S.~Anand, Greg Sloan, Louise Breuval, Richard I. Anderson, and Stefano Casertano
\\

\begin{wrapfigure}{R}{0.4\textwidth}
\centering
\vspace{-10pt} 
\hspace{-60pt} 
\includegraphics[width=0.8\linewidth]{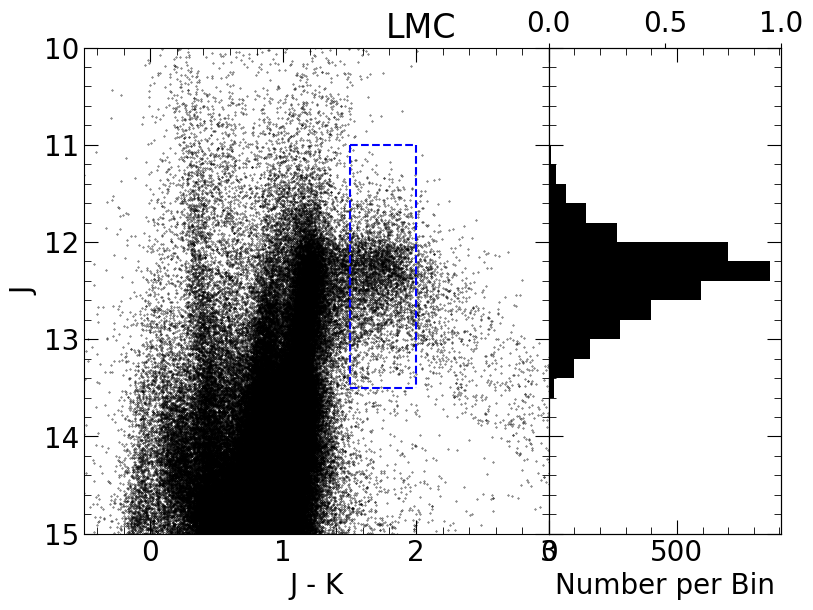}
\vspace{-10pt} 
\hspace{-50pt} 
\caption{J vs. J - K color-magnitude diagram of stars in the \ac{lmc} using the sample from Ref.~\cite{Macri:2014xpa}.}
\vspace{-20pt} 
\label{fig:LMC_CMD}
\end{wrapfigure}

The \ac{jagb} refers to a group of stars in a NIR color magnitude diagram (see Fig.~\ref{fig:LMC_CMD}) that contains thermally pulsating carbon-rich \ac{agb} stars with photospheric C/O greater than 1. The potential for using carbon stars (CS) as distance indicators was first realized in the 1980s \cite{1981ApJ...243..744R, 1984ApJ...287..138R, 1985ApJ...298..240R, 1987ApJ...323...79P, 1986ApJ...305..634C}, and the method was later revived in the 2000s \cite{2005A&A...442..159B, 2020MNRAS.495.2858R, Madore:2020yqv, Freedman:2020mho, 2021ApJ...916...19Z, Lee:2022akw}. These pioneering studies, among others, have proposed that the mean, median, mode, or model fit of a near-infrared J-region \ac{lf} can be used as a standard candle to measure extragalactic distances. The basis for doing so originates from the expectation that only oxygen-rich \ac{agb} stars that have masses falling within a relatively narrow range can evolve into carbon-rich \ac{agb} stars, constrained by hot bottom burning and the efficiency of the 3\textsuperscript{rd} dredge-up events.

\paragraph{Application}

The \ac{jagb} method relies on a measurement of the CS infrared magnitude distribution in a color range between weakly pulsating CS that produce little dust and strongly pulsating CS (i.e., Mira variables) that produce significant amounts of amorphous carbon dust. The \ac{jagb} is typically measured in the ``outer disk'' of a galaxy where the stellar population is young enough to contain a substantial amount of carbon \ac{agb} stars and far enough from the center to minimize the effects of crowding and internal reddening \cite{Lee:2023zsq}. The region is also typically chosen to be close enough to avoid sparser fields which increases the statistical uncertainty. In addition, the red background of galaxies can be similar to that of CS and contaminate the CS sample \cite{2021ApJ...916...19Z}. Consistent selection of fields is also important in maintaining consistent stellar populations.

Several methods have been proposed to measure the \ac{jagb} reference magnitude in the outer disk; these are typically designed to measure some variation of the peak of the J-region \ac{lf}. For instance, Refs.~\cite{Madore:2020yqv, Freedman:2020mho} use the mean and median, Ref.~\cite{Lee:2022akw} smooth a binned J-region \ac{lf} and take the mode, while Refs.~\cite{2020MNRAS.495.2858R, Parada:2020rpk} introduce maximum likelihood and binned versions of medians and calibrate their \ac{jagb} measurements using the skew parameter in a Lorentzian fit. Ref.~\cite{2021ApJ...916...19Z} fit a Gaussian+Quadratic model to the \ac{lf}. There is an important need for standardization of the technique to ensure accuracy.

\paragraph{Prospects}

The \ac{jagb} method has the potential to independently support or refute local measurements of the Hubble constant from the second rung of the distance ladder. Recent studies have found \ac{jagb}-based Hubble constants of 74.7 $\pm$ 2.1 (stat) $\pm$ 2.3 (sys)\kms, with a full range spanning $71$ to $78$\kms depending on the measurement method used \cite{Li:2024yoe}, and 67.96 $\pm$ 1.85 (stat) $\pm$ 1.90 (sys)\kms \cite{Lee:2024qzr, Freedman:2024eph}, noting that the lower value here originates from galaxy subsample selection, see Ref.~\cite{Riess:2024vfa}. However, it is important to be aware that this method is still much less mature than other standard candles, such as Cepheids. For this method to be robust, several aspects will need to be better understood, such as the empirical effects of metallicity, molecular atmospheric diversity effects on photometry, and the shift of CS from weak to strong pulsations in the context of population diversity.

In addition, past literature has found evidence of non-uniform asymmetry in the J-region \ac{lf} \cite{2020MNRAS.495.2858R, Parada:2020rpk, Parada:2023wyt, Li:2022aho}. This can produce methodological variations (i.e., the mean will be different from the median, mode, and model fit). Non-uniformity of the asymmetry can also result in a mismatch in the degree of methodological variations across the distance ladder, thus increasing uncertainties of the method. It will be important to standardize this effect.

The \ac{jagb} method introduces observational flexibility for measuring $H_0$. \ac{jagb} only requires a single epoch while Cepheids need multiple epochs to measure their periods and mean magnitudes. CS are ubiquitous in most galaxies, unlike Cepheids which are generally observed in face-on spiral galaxies. In the near-infrared, \ac{jagb} stars are as bright as long-period Cepheids, making them competitive for reaching large distances. On the other hand, the metallicity dependence of Cepheids is very well calibrated \cite{2022ApJ...939...89B}, which is not yet the case for CS. The \ac{jagb} feature is also brighter than \ac{trgb}. However, \ac{jagb} stars are an intermediate-age (300 Myr - 1 Gyr \cite{Madore:2020yqv}) population, and are therefore not as ubiquitous as \ac{trgb} stars, which populate essentially every galaxy.

The \ac{jagb} is primarily measured in the NIR, typically in the J- and H-bands (or space-based equivalents: \ac{hst} \emph{F110W} \cite{Lee:2023zsq}, \ac{jwst}, \emph{F115W}, \cite{Lee:2023vku} $\&$ \emph{F150W} \cite{Li:2024yoe}). It is important to understand whether either of these bands necessarily offers more consistency in color dependence over the other and whether either is sloped as a function of color. Multiple studies have found evidence of asymmetric \ac{lf} and non-flat color dependence in both the J- and H- bands \cite{2020MNRAS.495.2858R, Parada:2020rpk, Parada:2023wyt, Li:2022aho}. 

Increasing the \ac{sn1} sample size decreases fluctuations in H0 arising from cosmic variance and facilitates closer reversion to the mean (see Ref.~\cite{Riess:2024vfa}). Recently, Ref.~\cite{Li:2025ife} augmented the \ac{jagb} sample used to measure H0 by combining all available \ac{jagb} distances to \ac{sn1} host galaxies in the literature, as well as measuring new distances to galaxies from \ac{jwst} Cycles 1 \& 2 for a total of 15 galaxies hosting 18 \ac{sn1}. As in Ref.~\cite{Li:2024yoe}, they find methodical variations consequent of non-uniform asymmetry in the J-region \ac{lf}; taking the middle measurement variant (as described in Ref.~\cite{Li:2024yoe}) yields $H_0 = 73.3 \pm 1.4 {\rm (stat)} \pm 2.0 {\rm (sys)}$\kms.

A powerful way to characterize, and potentially better standardize, variations in the \ac{jagb} method is to conduct a field-to-field comparison of the \ac{jagb}. This approach has been implemented for the \ac{trgb} by the Comparative Analysis for \ac{trgb}s (CATs) team \cite{Wu:2022hxf, Scolnic:2023mrv, Li:2023utj}. They compared multiple fields in a given galaxy and developed a standardization procedure via a contrast ratio. Future missions, such as Roman, and more observations with \ac{jwst} can obtain a larger field coverage that can be used to measure the \ac{jagb} and make a similar analysis possible.

\bigskip
\subsubsection{The Hubble constant from Type II supernovae \label{sec:type_2_sn}}

\noindent \textbf{Coordinator:} Lluís Galbany\\
\noindent \textbf{Contributors:} Anil Kumar Yadav, Anto Idicherian Lonappan, David Benisty, G\'{e}za Csörnyei, Ismailov Nariman Zeynalabdi, and Vladas Vansevičius
\\

\begin{figure}[t!]
    \includegraphics[width=0.5\linewidth]{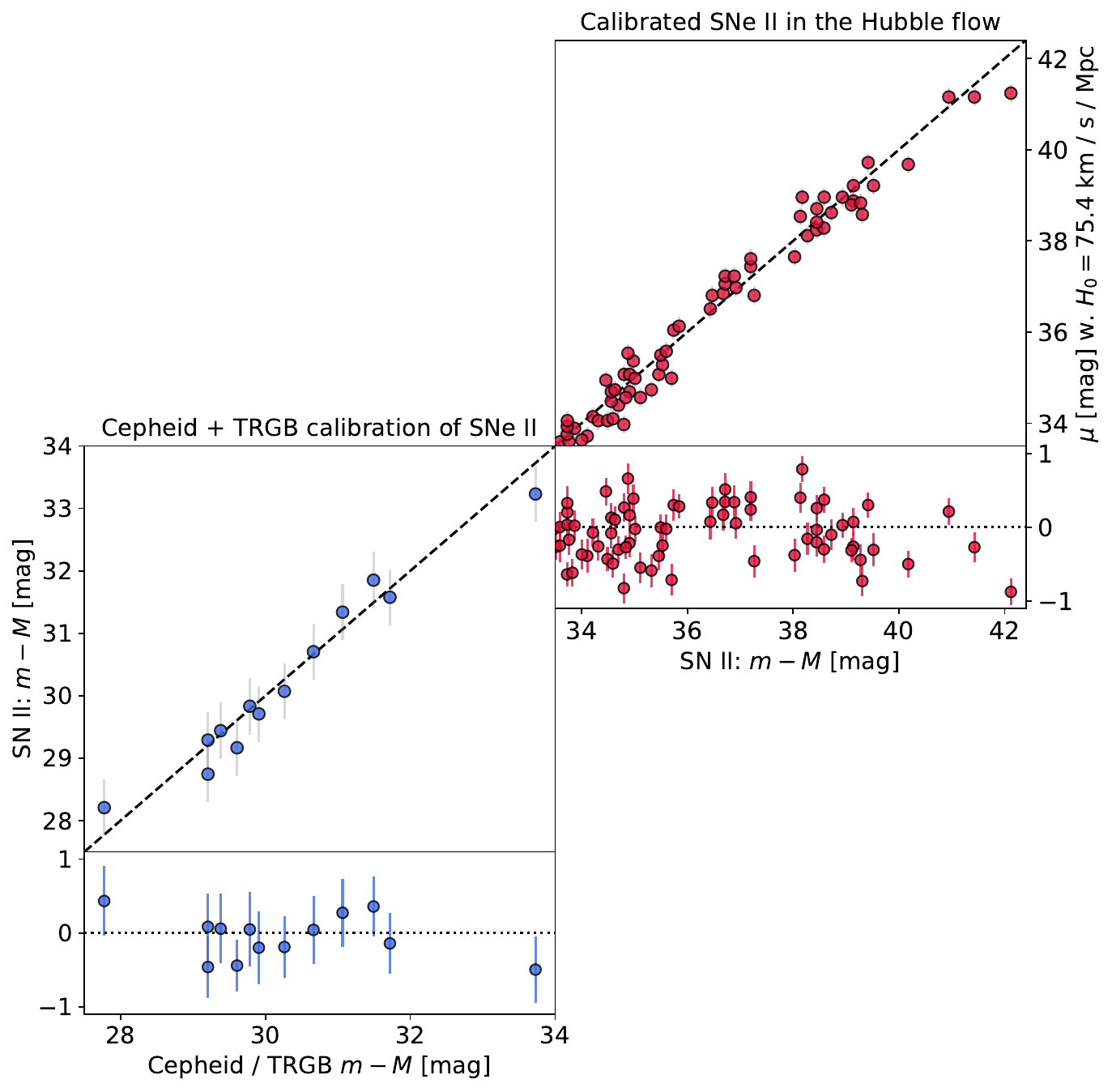}
    \raisebox{1.25cm}{\includegraphics[width=0.45\linewidth]{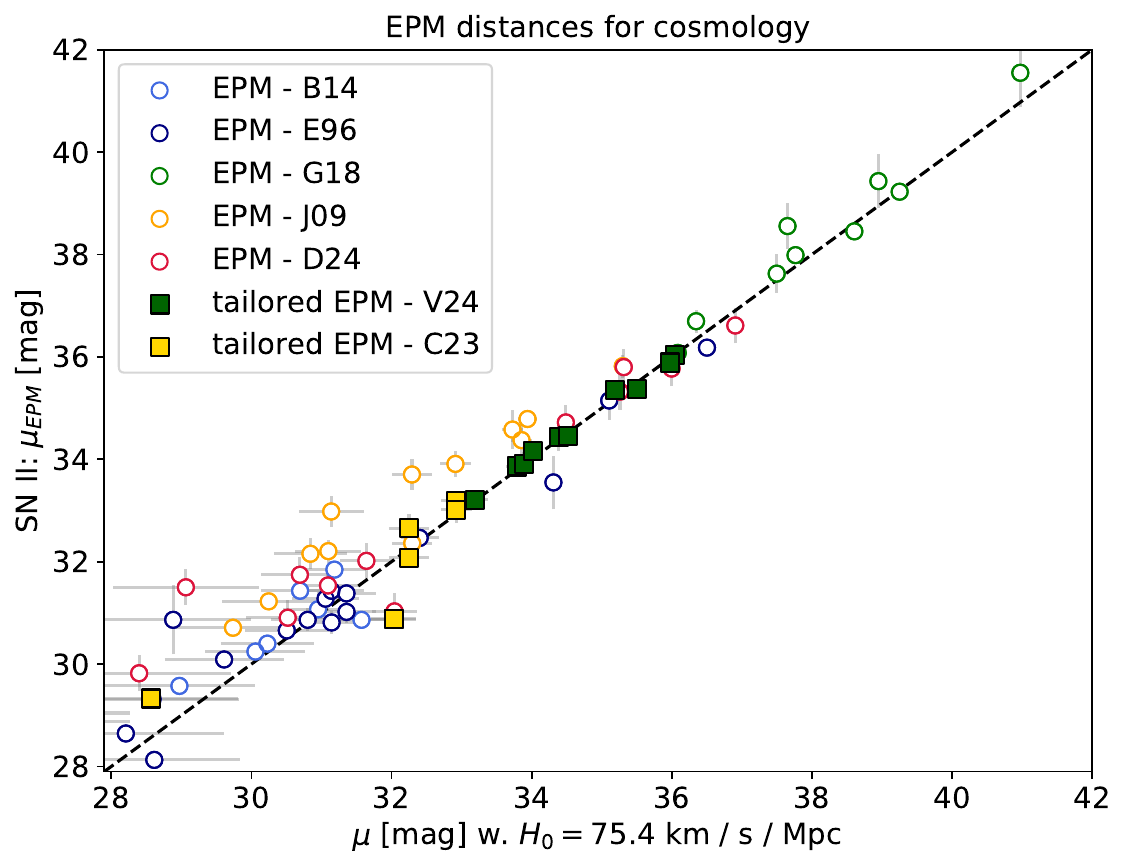}}
    \caption{\ac{sn2} for Hubble parameter estimation. \textbf{Left:} The distance ladder formalism applied on \ac{sn2}, as presented by Ref.~\cite{deJaeger:2022lit}. The bottom panels with the blue dots depict the intermediate rung of the distance ladder, where the \ac{sn2} brightnesses are calibrated through Cepheids and \ac{trgb}, along with the residuals after the standardization. The top panels with the red dots show the final rung of the distance ladder, where this standardization is applied on \ac{sn2} in the Hubble flow which allow for a Hubble constant estimation, while the residuals are shown on the panel below. \textbf{Right:} \ac{sn2} Hubble diagram using EPM distances, as updated from Ref.~\cite{deJaeger:2023vkm}. EPM distances do not require external calibration, hence they provide a single step measurement of $H_0$. The empty circles denote the EPM distances that employ dilution factors, while the filled squares mark the tailored EPM estimates that make use of radiative transfer modeling. The references for the distance estimates are as follows: B14 -- \cite{Bose:2014sza}, E96 -- \cite{1996ApJ...466..911E}, G18 -- \cite{Gall:2017gva}, J09 -- \cite{Jones:2008tz}, D24 -- \cite{Dhungana:2023dep}, V24 -- \cite{Vogl:2024bum} and C23 -- \cite{Csornyei:2023rpw}.}
    \label{fig:SNII-HD}
\end{figure}

With the currently ongoing tension between local and distant measurements of the Hubble constant, it is crucial to test and employ independent methods that adhere to a separate set of systematic uncertainties. \ac{sn2} offer such independent routes with sufficient accuracy. \ac{sn2} are the explosions of red supergiant stars, with multiple direct progenitor detections to date (e.g., see Refs.~\cite{Smartt:2009zr, Kilpatrick:2023pse}). Given the well constrained progenitor type along with their relatively simple composition (with the red supergiant retaining its hydrogen envelope and being made up mostly of H and He before the explosion) and the recent advancements in the understanding of the explosion mechanism, multiple theoretically well-founded distance estimation techniques exist for \ac{sn2}. The spectra of \ac{sn2} are characterized by the presence of broad P-Cygni profiles, which allow constraining the photospheric properties. To date, three techniques have been used to measure the Hubble constant: the expanding photosphere method (EPM, \cite{Kirshner:1974ghm}), the spectral modeling based techniques (either the spectral expanding atmosphere method SEAM \cite{Baron:2004wb} or the tailored EPM \cite{Dessart:2005gg}) and the standardizable candle method (SCM, \cite{Hamuy:2002tj}). Of the three, SCM is an empirical technique, which employs the relation present between the luminosity and the expansion velocity of the supernova. On the other hand, EPM and SEAM both aim to constrain the physical parameters and the luminosity based on the spectra directly, providing a single step measurement tool. Both Refs.~\cite{Gall:2017gva} and ~\cite{deJaeger:2022lit} demonstrated that SCM and EPM distance measurements up to redshifts of $z\approx 0.34$ are feasible, providing important alternative paths for cosmology, as depicted in Fig.~\ref{fig:SNII-HD}.

\paragraph{Methods and approaches to measure $H_0$ using SNe II}

The use of EPM for \ac{sn2}, which essentially consists of estimating the size of the photosphere, then comparing it to the observed flux by assuming a blackbody spectrum, was first demonstrated by Ref.~\cite{Kirshner:1974ghm}. This initial exploration yielded $H_0 = 65 \pm 15$\kms. However, it was highlighted later by Ref.~\cite{1981ApJ...250L..65W}, \ac{sn2} radiate with a much smaller surface flux than a blackbody of the same color temperature (the photosphere flux appears diluted), owing to the scattering dominated atmosphere of \ac{sn2} and the fact that the thermalization layer from which the blackbody radiation is generated is deeper than the photosphere. To take this into account, the EPM was refined by incorporating a multiplicative $\xi$ dilution factor, which relates the position of the thermalization layer to the photosphere, which was shown to depend largely only on the color temperature (see Refs.~\cite{1996ApJ...466..911E, Dessart:2005ax, Vogl:2018ckb}). Systematic discrepancies between the different sets of dilution factors can explain differences on the scale of 20\% in the inferred distance \cite{Jones:2008tz,Gall:2016qvq}. The EPM is to date one of the most frequently used distance estimation technique for \ac{sn2}, with the latest $H_0$ estimate being ${72.9}_{-4.3}^{+5.7}$\kms based on 12 \ac{sn2} following Ref.~\cite{Dhungana:2023dep}, using the dilution factors from Ref.~\cite{Dessart:2005ax}. To rule out the possible systematic effects introduced by dilution factors, one has to carry out radiative transfer based spectral modeling of observations, which yields self-consistent results for the physical parameters.

This is achieved in both the customized EPM and in the SEAM, which incorporate radiative transfer modeling of spectra into the distance estimation process. In both techniques, the modeling of the complete spectral time series is carried out, which yields the relevant physical parameters at each epoch in a self-consistent way, avoiding the use of dilution factors. However, estimating a single radiative transfer model takes hours to days, hence estimating a distance this way is a time-consuming process. Due to this, until recent years, this method was only used for a handful of \ac{sn} \cite{Baron:2004wb, Dessart:2005gg, Dessart:2007rt}. Recent advances in \ac{ml} allowed for the faster estimation of radiative transfer models, hence significantly speeding up this modeling process, as described in Ref.~\cite{Vogl:2019fhc}. This method was shown to provide internally consistent results
\cite{Csornyei:2023rpw} at a competitive precision
\cite{Csornyei:2023enu} and yielded a Hubble constant of $H_0 =
74.9^{+1.9}_{-1.9}$\kms based on the analysis of literature \ac{sn}
\cite{Vogl:2024bum}. However, this method is currently largely limited to local redshifts: to date only a few \ac{sn2} were observed far enough out into the Hubble flow in sufficient detail for the application of either tailored EPM or SEAM.

To date, the most widely used method for estimating $H_0$ using \ac{sn2} is the SCM \cite{Hamuy:2002tj,deJaeger:2020zpb,deJaeger:2022lit}. The method is based on an empirically found correlation between the luminosity and the photospheric expansion velocity of the objects \cite{2001PhDT.......173H,Hamuy:2002qx}, and in terms of philosophy is similar to the method employed for \ac{sn1} (e.g., see Ref.~\cite{Riess:2021jrx}). In contrast to the Ia's, however, to standardize \ac{sn2}, spectroscopic information is required in the form of line velocity measurements. Over the years, the standardization has been further extended with color term accounting for the varying level of extinction present for different \ac{sn} \cite{SNLS:2006mwe}. Recently, Ref.~\cite{deJaeger:2022lit} applied this method with nine calibrator \ac{sn} to estimate the Hubble constant by using \ac{sn2} as the final rung of the distance ladder, obtaining $H_0 = 75.4^{+3.8}_{-3.7}$\kms. The method is still under further development, with the aim of increasing the currently available set of calibrators with high precision cases, and also with attempts to investigate the need for additional standardization terms based on spectroscopic information.

These results show that \ac{sn2} indeed provides important and independent distance measures that can serve as the tools for a sanity check for multiple rungs of the distance ladder. With the steady increase in the number of objects with detailed observations, the coming years will see multiple \ac{sn2}-based Hubble constant estimates with competitive precision, which will allow for an alternative look at the Hubble tension.

\bigskip
\subsubsection{HII galaxy distance indicators \label{sec:HII_gal}}

\noindent \textbf{Coordinator:} Ricardo Ch\'avez\\
\noindent \textbf{Contributors:} Ana Luisa Gonz\'alez Mor\'an, David Fernández-Arenas, David Valls-Gabaud, Elena Terlevich, Fabio Bresolin, Iryna Vavilova, Ismailov Nariman Zeynalabdi, Manolis Plionis, Roberto Terlevich, Rodrigo Sandoval-Orozco, Spyros Basilakos, and Vladas Vansevičius
\\

HII galaxies (HIIGs) are characterized by intense, compact episodes of star formation predominantly occurring within dwarf irregular galaxies, significantly enhancing their luminosity. These galaxies are spectroscopically identified based on the prominent equivalent width of their Balmer emission lines, specifically \(EW(H\beta) > 50\)~\AA, which indicates their young age (less than 5 Myr). Similarly, Giant Extragalactic HII Regions (GEHRs) exhibit vigorous star formation but are typically found in the outer disks of late-type galaxies. The rest-frame optical spectra of both HIIGs and GEHRs display pronounced emission lines, indicative of gas ionization by massive Young Stellar Clusters (YSC) or Super Star Clusters (SSC), leading to similar spectral features \cite{1972ApJ...173...25S, 1977ApJ...211...62B, 1981MNRAS.195..839T, Kunth:1999af, Chavez:2014ria}.

Numerous studies have confirmed a consistent correlation, found by Ref.~\cite{1981MNRAS.195..839T}, between the luminosity of Balmer lines, such as \(L(H\beta)\), and the ionized gas velocity dispersion, \(\sigma\), measured through these emission lines, in both HIIGs and GEHRs. This relationship, known as the \(L-\sigma\) relation \cite{1988MNRAS.235..297M, Bordalo:2011yc, Chavez:2014ria}, is recognized as a potent cosmological distance indicator \cite{2011MNRAS.416.2981P, 2016MNRAS.462.2431C, Gonzalez-Moran:2019uij, Chavez:2024twa}, where GEHRs and nearby HIIGs are used as the “anchor” sample because their distances can be independently estimated from Cepheid
variables or \ac{trgb} measurements. As a result, the \(L-\sigma\) relation provides an exceptional method for utilizing this distance metric to investigate the Hubble flow over an extensive range of redshifts (\(z\)).

\begin{figure}
    \centering
    \includegraphics[width=0.8\columnwidth]{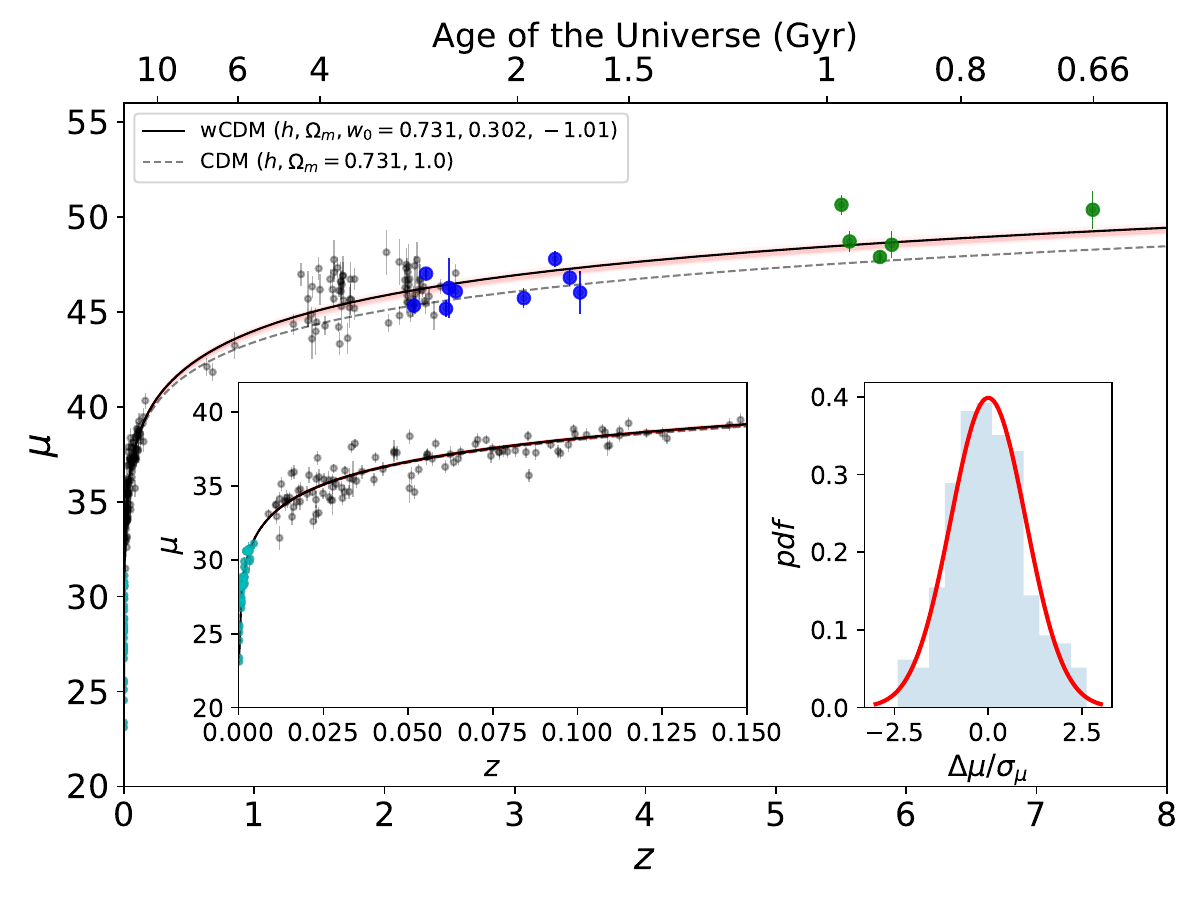}   
    \caption{Hubble diagram for GEHRs and HIIGs, here $z$ is the redshift and $\mu$ is the distance modulus. In cyan we present the ``anchor'' sample of 36 GEHRs which have been analyzed in Ref.~\cite{FernandezArenas:2017dux}, in black we present the full sample of 181 HIIGs which have been analyzed in Ref.~\cite{Chavez:2024twa}, while in blue we present the 9 new HIIGs from Ref.~\cite{2023A&A...676A..53L} and in green  5 new HIIGs studied with \ac{jwst} by Ref.~\cite{2024A&A...684A..87D}. The black line is the cosmological model that best fits the data with the red shaded area representing the 1$\sigma$ uncertainties to the model, while the gray dashed line is a flat cosmological model without \ac{de}. The inset at the left shows a close-up of the Hubble diagram for $z \leq 0.15$. The inset at the right presents the pulls \ac{pdf} of the entire sample of GEHRs and HIIGs and the red line shows the best Gaussian fit to the \ac{pdf}. Adapted from Ref.~\cite{Chavez:2024twa}.
    }
    \label{fig:hubdiag}
\end{figure}

The characteristic strong emission lines within the rest-frame optical spectra of GEHRs and HIIGs render them effective tools for exploring nascent star formation at high redshifts. Using instruments such as NIRSpec \cite{2016A&A...592A.113D} on board the \ac{jwst} \cite{Gardner:2006ky}, it is possible to study these regions up to \(z \sim 6.5\) through the H$\alpha$ emission line, or even up to \(z \sim 9\) using the H$\beta$ and [\text{OIII}]$\lambda\lambda4959,5007$ \AA\ emission lines. This capability enables the observation of luminous HIIGs that date back to the \ac{eor} as shown in Fig.~\ref{fig:hubdiag}.

Using a dataset of 231 GEHRs and HIIGs, some observed with NIRSpec on the \ac{jwst} up to \(z \sim 7.5\), a study by Ref.~\cite{Chavez:2024twa} employs the MultiNest Bayesian inference algorithm \cite{Feroz:2007kg, Feroz:2008xx, Feroz:2013hea} to derive constraints for various cosmological models. They used uniform, non-informative priors across all parameters for unbiased estimations. The derived constraints are detailed in Table~\ref{tab:const}, showing marginalized best-fit values and \(1\sigma\) uncertainties for each parameter, with some parameters held constant during the analysis. The study explores a generalized parameter space \(\theta = \{\alpha, \beta, h, \Omega_{\rm m}, w_0, w_a\}\), where \(\theta_n = \{\alpha, \beta\}\) represents nuisance parameters of the \(L-\sigma\) relation for GEHRs and HIIGs with \(\alpha\) as the intercept and \(\beta\) as the slope. For the flat \lcdm\ model, \(\theta_c = \{h, \Omega_{\rm m}, -1, 0\}\) sets \(h\) as the reduced Hubble constant and \(\Omega_{\rm m}\) as the total matter density, fixing the first two \ac{de} equation of state (\ac{de} \ac{eos}) parameters at \(w_0 = -1\) and \(w_a = 0\), corresponding to a cosmological constant (\(\Lambda\)). Adjusting \(w_0\) allows for models with evolving \ac{de} \ac{eos}, akin to quintessence \cite{Ratra:1987rm, Wetterich:1987fm}, while including \(w_a\) aligns with the \ac{cpl} model \cite{Chevallier:2000qy, Linder:2002et, Peebles:2002gy}.

\begin{table}[ht]
\caption{Marginalized best-fit parameter values and associated $1\sigma$ uncertainties for the HIIGs and anchor samples. Values enclosed in parentheses indicate parameters that were held constant during the analysis.} \label{tab:const}
\begin{tabular}{@{}lccccccc@{}}
\toprule
	Data Set       & $\alpha$ & $\beta$ & $h$ & $\Omega_{\rm m}$ & $w_0$ & $w_a$  & N\\
\midrule
	HIIG & --- & ($5.022\pm 0.058$) & --- & $0.282^{+0.037}_{-0.045}$ & (-1.0) & (0.0) & 195\\
	HIIG & --- & ($5.022\pm 0.058$) & --- & $0.278^{+0.092}_{-0.051}$ & $-1.21^{+0.45}_{-0.40} $ & (0.0) & 195\\
\midrule
	HIIG & ($33.268\pm 0.083$) & ($5.022\pm 0.058$) & $0.715\pm 0.018$ & $0.267^{+0.038}_{-0.048}$ & (-1.0) & (0.0) & 195\\
	HIIG & ($33.268\pm 0.083$) & ($5.022\pm 0.058$) & $0.718\pm 0.020$ & $0.278^{+0.091}_{-0.050}$ & $-1.22^{+0.46}_{-0.40}$ & (0.0) & 195\\
\midrule
    Anchor+HIIG & $33.28\pm 0.11$ & $4.997\pm 0.089$ & $0.730\pm 0.038$ & (0.3) & (-1.0) & (0.0) & 231\\
	Anchor+HIIG & $33.28\pm 0.14$ & $5.00\pm 0.11$ & $0.730\pm 0.040$ & $0.335^{+0.044}_{-0.055}$ & (-1.0) & (0.0) & 231 \\ 
	Anchor+HIIG & $33.29\pm 0.14$ & $4.99\pm 0.11$ & $0.731\pm 0.039$ & $0.302^{+0.12}_{-0.069}$ & $-1.01^{+0.52}_{-0.29}$ & (0.0) & 231\\ 
    Anchor+HIIG & $33.29\pm 0.14$ & $4.99\pm 0.11$ & $0.730\pm 0.039$ & $0.321^{+0.10}_{-0.063}$ & $-0.91^{+0.55}_{-0.33}$ & $-0.71^{+0.65}_{-1.2}$ & 231\\ 
\botrule
\end{tabular}
\end{table}

The constraints derived on cosmological parameters, specifically \( \lbrace h, \Omega_{\rm m}, w_0 \rbrace = \lbrace 0.731 \pm 0.039, 0.302^{+0.12}_{-0.069}, -1.01^{+0.52}_{-0.29} \rbrace \) (stat) from GHIIR and HIIG data, align closely with recent Pantheon+ results from 1550 \ac{sn1}, which produces \( \lbrace h, \Omega_{\rm m}, w_0 \rbrace = \lbrace 0.735 \pm 0.011, 0.334 \pm 0.018, -0.90 \pm 0.14 \rbrace \) \cite{Brout:2022vxf}. This agreement emphasizes the validity and importance of HIIGs as distance indicators within the context of current cosmological studies.

The enduring validity of the \(L-\sigma\) relation at high redshifts (\(z > 3\)), reaching into the \ac{eor}, suggests remarkable uniformity in HIIGs properties over vast cosmic timescales. This not only confirms the reliability of the \(L-\sigma\) relation as a cosmological tool but also illuminates the fundamental processes underlying the formation and evolution of early Universe galaxies.

In our efforts to improve the accuracy of cosmological parameters using HIIGs, the addition of data from \ac{jwst}, extending to \(z \sim 9\) and beyond, is proving crucial. The \ac{jwst}'s exceptional sensitivity and resolution enable detailed observations of HIIGs at these higher redshifts, offering a unique view of the early Universe. This data range is essential for analyzing the dynamics of the Universe's expansion across various cosmological scenarios.

\bigskip
\subsubsection{The baryonic Tully-Fisher relation approach \label{sec:btf_rel}}

\noindent \textbf{Coordinator:} Khaled Said\\
\noindent \textbf{Contributors:} Benoit Famaey, Cláudio Gomes, Du\v{s}ko Borka, Esha Bhatia, Jenny G. Sorce, Maurice H.P.M. van Putten, Milan Milo\v{s}evi\'{c}, Paolo Salucci, Paula Boubel, Predrag Jovanovi\'{c}, and Vesna Borka Jovanovi\'{c}
\\

The \ac{btfr} (\cite{McGaugh:2000sr,Bell:2000jt,Gurovich:2004vd,McGaugh:2005qe,Pfenniger:2004ib,Begum:2008gn,Trachternach:2009fb, Stark:2009ch, Gurovich:2010jx, Zaritsky:2014dca}) extends the classic Tully-Fisher (TF; \cite{Tully:1977fu,1988ApJ...330..579P,1995ApL&C..31..263T,Giovanelli:1996zw,Giovanelli:1996zx,Courteau:1997ap,Brent:1999uv,Karachentsev:2002rn,Bedregal:2006xq,Noordermeer:2007pa,Springob:2007vb, Williams:2010ug, Mocz:2012gk, Tully:2012ze, Rawle:2013lpa, Sorce:2013wt, Sorce:2014iva, Torres-Flores:2013jha, Said:2014jwa, Said:2016voa, Kourkchi:2020iyz,Kourkchi:2022ifq, Bell:2022kpt, Boubel:2023mfe}) relation by incorporating both the stellar and gas masses of galaxies, correlating a galaxy's baryonic mass ($M_{\rm b}$) with its rotational velocity ($V_\text{{rot}}$). This relation provides critical insights into galaxy formation and dynamics and is crucial in testing and constraining models of galaxy evolution within both the \lcdm\ framework and alternative theories \cite{Eisenstein:1994ni, McGaugh:1998tq, McGaugh:1998tn, Mo:1997vb, Steinmetz:1998gr, vandenBosch:1999by, Mayer:2003fs, Gnedin:2006zb, Governato:2006cq, Avila-Reese:2008irk, Dutton:2012jh, McGaugh:2011ac, Aumer:2013gpa, Desmond:2015nja}. 

The \ac{btfr} is defined as: 
    $M_{\rm b} \propto V_{\text{rot}}^x$,
where $M_{\rm b}$ is the sum of stellar and gas masses,
    $M_b = M_{*} + M_{\text{gas}}$
 and $x$ is typically around 4, suggesting a deep connection between the visible matter and the dynamics of galaxies \cite{McGaugh:2000sr,Lelli:2015rba,1993MNRAS.262..392S,Lapi:2018nuq}.

\paragraph{Historical context and development}
The original TF relation, proposed by Ref.~\cite{Tully:1977fu}, established a link between the intrinsic brightness of spiral galaxies and their maximum rotational velocities. This relation has been pivotal in determining distances to galaxies and measuring the Hubble constant \cite{Said:2023jur}. For example, Ref.~\cite{Tully:1977fu} applied their TF relation to derive distances to the Virgo and Ursa Major clusters, estimating a Hubble constant of approximately $H_0 \approx 84$\kms for Virgo and $H_0 \approx 75$\kms for Ursa Major. The extension to the \ac{btfr} was motivated by the need to include galaxies with significant gaseous components, particularly dwarf galaxies \cite{McGaugh:2000sr}. Over the years, numerous studies have validated and expanded the \ac{btfr}, demonstrating its robustness and utility in various cosmological and astrophysical applications \cite{McGaugh:2011nv,McGaugh:2011ac,Zaritsky:2014dca}. A crucial step towards the understanding of the physics of the TF and \ac{btfr} relationships has been the finding, in disk systems,  of a family of {\it local} Tully-Fisher-like relationships \cite{Yegorova:2006wv,Haridasu:2024ask,2020F}, holding at various specific galactocentric radii and thus dubbed as the Radial Tully Fisher (RTF: for a review see Ref.~\cite{Salucci:2018hqu}).

\begin{figure}[htbp]
    \centering
    \includegraphics[scale = 0.9]{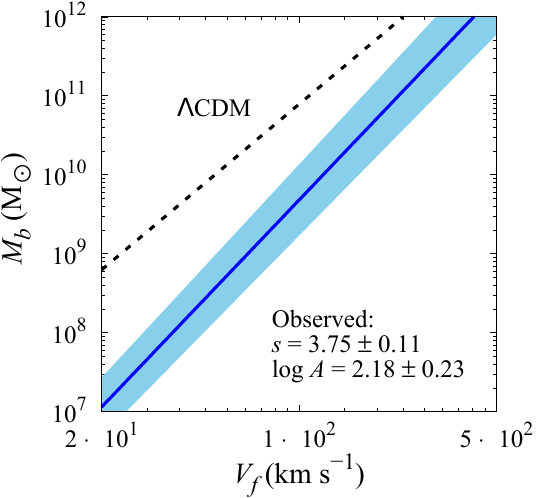}
    \caption[Comparison between the observed and predicted BTFR]{Comparison between the \ac{btfr} obtained from error-weighted fits of galaxies with accurate distances from Ref.~\protect\cite{Lelli:2015wst} to the \ac{btfr} predicted by \lcdm\ cosmology. The observed \ac{btfr} (blue solid line) includes a light blue band representing the intrinsic scatter of ~0.1 dex. The \lcdm\ prediction is shown as a black dashed line. Both relations are plotted using equations from Ref.~\protect\cite{Lelli:2015wst}: Eq. (6) for the observed \ac{btfr} and Eq. (8) for the \lcdm\ prediction. [Reprinted under CC BY 4.0. from: V. Borka Jovanović, D. Borka, and P. Jovanović, Contrib. Astron. Obs. Skalnate Pleso 55/2, 24 - 33 (2025)].}
\label{fig:BTFR_observed_vs_LCDM}
\end{figure}

\paragraph{Theoretical models and their implications}
\subparagraph{The standard \lcdm\ model}
In the \lcdm, the TF and \ac{btfr} are seen as consequences of the gravitational dynamics dictated by \ac{dm} halos and baryonic matter. The rotational velocities of galaxies are thought to be influenced by both visible and \ac{dm}, with the \ac{btfr} providing a means to probe these interactions \cite{McGaugh:2011ac}. However, as illustrated in Fig.~\ref{fig:BTFR_observed_vs_LCDM}, the observed slope and intrinsic scatter of the \ac{btfr} often differ from \lcdm\ predictions, which typically forecast a lower slope and higher scatter \cite{Lelli:2015wst}. This discrepancy poses challenges to the \lcdm\ model and prompts further investigation into galaxy dynamics and mass distribution.

\subparagraph{Modified gravity models}
Some studies propose that the \ac{btfr} can be explained without \ac{dm} through modified theories like $f(R)$ theories of gravity \cite{BorkaJovanovic:2016low,Capozziello:2017rvz,Capozziello:2018uht,Capozziello:2020dvd}. Alternative gravitational theories, such as \ac{mond} and non-minimal matter-curvature coupling models, offer different perspectives on the \ac{btfr}. \ac{mond}, proposed by \cite{Milgrom:1983ca}, adjusts Newtonian dynamics at low accelerations, predicting a \ac{btfr} with a slope exactly equal to 4, consistent with many observations \cite{McGaugh:2011nv,McGaugh:2011ac}. \ac{mond} has made several successful predictions regarding the detailed shapes of rotation curves, galaxy dynamics, and galaxy lensing  \cite{Famaey:2011kh, Banik:2021woo, Famaey:2025rma}.

Another approach, non-minimal matter-curvature coupling model, generalizes the pure gravity sector by introducing a generic function of the curvature scalar, $f_1(R)$, and a non-minimal coupling between the matter Lagrangian and another generic function of the curvature scalar, $f_2(R)$ \cite{Bertolami:2007gv},
\begin{equation}
    S=\int d^4x \sqrt{-g} \left(f_1(R)+f_2(R)\mathcal{L}\right)\,.
\end{equation}
 
This model leads to an extra force term in the geodesics for a perfect fluid. In three dimensions and in the Newtonian limit, assuming that the total acceleration $\vec{a}$ is collinear with the 3-force, $f$, and in the limit of very small gravitational accelerations, the Newtonian acceleration, $\vec{a}_{\rm N}$, is $\vec{a}_{\rm N}\approx \frac{a}{a_{\rm E}}\vec{a}$, where $a_{\rm E}^{-1}=(2f)^{-1}\left(1-\frac{f^2}{a^2}\right)$, which is remarkably similar to \ac{mond}'s result. Furthermore, $a\approx \sqrt{a_{\rm E} GM}/r=v^2/r$, hence a Tully-Fisher relation appears with a luminosity of the form $L\sim v^4$ with $v^4=a_{\rm E} GM$. Thus $a_{\rm E}=10^{-10} m~s^{-2}$ as in \ac{mond} or $a_{\rm E}=(8.5\pm 1.3)\times 10^{-10}{\rm ms}^{-2}$ \cite{Bertolami:2007gv}. This model has been shown to be consistent with several observations ranging from the cosmic version of the virial theorem at Abell 586 cluster \cite{Bertolami:2014hpa}, to Jeans instability in Bok globules \cite{Gomes:2020qhl, Gomes:2022jcs}, or to \ac{sn} distance data and the \ac{bao} data allowing for an attempt to solve the Hubble tension \cite{BarrosoVarela:2024htf}.

A finite sensitivity of weak gravitation in galaxy dynamics to background cosmology may be detectable below the de Sitter acceleration scale \cite{vanPutten:2017} using SPARC \cite{vanPutten:2018} and large galaxy surveys such as MaNGA \cite{Lee:2025xcd}. Redshift dependence would point to a potentially unified picture, linking \ac{btfr} to \ac{jwst} observations of ultra-high redshift galaxies at cosmic dawn \cite{vanPutten:2024a,vanPutten:2024b}. This approach may also place novel observational constraints on cosmological parameters, notably $H_0$  and the deceleration parameter $q_0$ \cite{Lee:2025xcd}.

\paragraph{Observations and cosmic tensions}
Various studies have extensively explored the \ac{btfr} using data from large galaxy surveys. The Spitzer Photometry and Accurate Rotation Curves (SPARC), for example, includes measurements of 175 rotationally supported galaxies in the near-IR, minimizing the effect of star-halo degeneracy and providing precise baryonic mass and rotational velocity data \cite{Lelli:2016uea,Lelli:2016zqa}. Such robust datasets enable accurate fits to the \ac{btfr} and facilitate comparisons between observed galaxy properties and theoretical predictions. The \ac{btfr} for the SPARC sample aligns with a scale of $x=4$. Research by Ref.~\cite{Schombert:2020pxm} using the \ac{btfr} with a sample of 95 independent galaxies also challenges the Hubble constant value $H_0 < 70$\kms with 95\% confidence level. A More recent study by Ref.~\cite{Kourkchi:2022ifq} demonstrated that using the \ac{btfr} allowed them to determine a Hubble constant of $H_0 = 75.5\pm2.5$\kms, which again challenges the standard cosmological model that suggests lower $H_0$ values \cite{Planck:2018vyg}. Additionally, studies have shown variations in \ac{btfr} parameters across different distance bins, which could contribute to the understanding of the Hubble tension \cite{Puech:2009nt, Miller:2011nt, 2016A&A...594A..77D, Shivaei:2015jdr, Straatman:2017ivb, Alestas:2021nmi}.

The RTF relationship, applied to a large sample of 843 local galaxies extending out to $z=0.03$, has recently been used as a distance indicator to determine the maximum allowed variance of the $H(z)/H_0$  parameter \cite{Haridasu:2024ask}. They found that the maximum allowed local ‘radial’ variation is $1\%$, which is not enough to resolve the $H_0$ tension. 
 In the near future, with the PROBES sample \cite{Stone:2022xxu} (3000 objects) we will be able to use the RTF to test the hypothesis of a Giant Local Void at greater distances and investigate the anisotropy of the expansion of the Universe (see Ref.~\cite{Haridasu:2024ask} for the present situation). Finally, the RTF can be applied at $z \sim 1-2$ to investigate a major part of the expansion history of the Universe \cite{Sharma:2024thv}.

 The original TF relation is also useful for tracing the distance-redshift relation, provided that the relation is carefully calibrated. Unlike the \ac{btfr}, the TF relation is not universally linear, so studies typically introduce more complex models to account for the curvature and varying intrinsic scatter \cite{Kourkchi:2020iyz, Boubel:2024cqw}. The Cosmicflows-4 (CF4) Tully-Fisher catalog \cite{Kourkchi:2022ifq} of $\sim$10,000 spiral galaxies is currently the largest TF dataset, combining HI line widths with photometric magnitudes. There have been several published $H_0$ measurements using this dataset. For the full CF4 TF catalog, Ref.~\cite{Kourkchi:2022ifq} derived $H_0$\,=\,75.5\,$\pm$\,2.5\kms with an estimated systematic error of $\pm$\,3\kms. Using an improved measurement methodology \cite{Boubel:2024cqw} and updated primary distance calibration \cite{Scolnic:2024oth}, this was re-measured as $H_0$\,=\,76.3\,$\pm$\,2.1\,(stat)\,$\pm$\,1.5\,(sys)\kms, where the statistical error reflects the relatively small number of primary calibrators in the sample ($\sim$ 50 objects with CPLR and/or \ac{trgb} distances). These are in agreement with other recent $H_0$ measurements using the TF relation as the final rung of the distance ladder, which consistently return $H_0>$ 70\kms (see Ref.~\cite{Said:2023jur} for a review). The next generation of Tully-Fisher datasets resulting from large surveys such as WALLABY \cite{Koribalski:2020ngl}, \ac{desi} \cite{Saulder:2023oqm}, and FAST \cite{2024SCPMA..6719511Z} will increase the precision of TF-derived measurements of the Hubble constant. 
 
\bigskip
\subsubsection{The Hubble tension in our Backyard: DESI and the nearness of the Coma cluster} \label{sec:COMA}

\noindent \textbf{Coordinator:} Yukei Murakami\\
\noindent \textbf{Contributors:} Daniel Scolnic\\

The recent work by \ac{desi}~\cite{Said:2024pwm} analyzed their first samples of early-type galaxies in the \ac{desi} peculiar velocity survey to study the fundamental plane (FP). FP is an empirical three-dimensional relation that links elliptical galaxies' radii, surface brightness, and the velocity dispersions~\cite{Djorgovski:1987vx,Dressler:1986rv}\footnote{See Sec.~\ref{sec:btf_rel} for the Tully-Fisher relation, a late-type galaxy counterpart}. This tight correlation serves as a standardization to measure distances to the galaxies over a large redshift range once the absolute distance scale is anchored to an independent, well-measured distance. In the \ac{desi} work~\cite{Said:2024pwm}, the Hubble Constant $H_0$ is measured with FP using an \ac{sbf} measurement to the Coma cluster to calibrate the distance scale (i.e., distance ladder). Alternatively, recent study~\cite{Scolnic:2024hbh} showed that one can calibrate the FP in the early universe using Planck+\lcdm\ inference of $H_0$ to measure the distance to the Coma cluster (inverse distance ladder).
The inverse distance ladder with Planck+\lcdm\--calibration of FP places Coma at $\sim10\%$ further than the direct measurement of distance by \ac{sbf}. In addition, the new \ac{sn1} distance, as well as other independent distance measurements to Coma collected over the past few decades, are all consistent with the \ac{sbf}, creating a tension against Planck+\lcdm\--prediction. This discrepancy provides a unique opportunity to shed light on the implications of the Hubble Tension in our local universe.

\paragraph{DESI FP and inverse distance ladder}
\ac{desi} has provided an extensive FP measurement and a Hubble diagram over the redshift range of $0.01 < z < 0.1$ using a sample of 4191 early-type galaxies in the Hubble flow and 226 galaxies in Coma~\cite{Said:2024pwm}. The FP relation, which correlates a galaxy’s velocity dispersion, surface brightness, and physical size, serves as a secondary distance indicator when calibrated with independent distance anchors. When the FP relation is calibrated using a near-infrared \ac{sbf} distance of $D_{\text{Coma}} = 99.1 \pm 5.8 $ Mpc~\cite{Jensen:2021ooi}, \ac{desi} derives a value of $ H_0 = 76.05 \pm 1.3 $\kms. This result relates the Hubble constant and the Coma distance as $H_0 = (76.05 \pm 1.3) \times \left( \frac{99.1\ \text{Mpc}}{D_\text{Coma}} \right)$\kms. This relation can be used to obtain the distance to Coma predicted by Planck+\lcdm, using the Planck-inferred value of the Hubble constant $H_0 = 67.4 $ \kms. The implied distance to Coma shifts to $ D_{\text{Coma}} = 111.8 \pm 1.8 $ Mpc, a distance larger than any of the historical measurements that we discuss later.

\paragraph{Latest Coma distance with ATLAS SNe Ia}
A recent analysis~\cite{Scolnic:2024hbh} compiled a sample of 10 spectroscopically confirmed \ac{sn1} within the Coma Cluster observed by the ATLAS survey~\cite{Rest:2018amw} and YSE~\cite{YoungSupernovaExperiment:2020dcd}, as well as two additional \ac{sn} previously included in the Pantheon+ dataset\cite{Scolnic:2021amr}. Applying the standardization and calibration methods established in the Pantheon+ analysis, the study derived a mean standardized peak brightness of $ m_B^0 = 15.712 \pm 0.041 $ mag for \ac{sn} in Coma cluster. Using the absolute magnitude calibration of \ac{sn1} from the \ac{hst} Cepheids~\cite{Riess:2021jrx}, this results in a distance modulus of $ \mu = 34.97 \pm 0.05 $ mag, corresponding to $ D_{\text{Coma}} = 98.5 \pm 2.2 $ Mpc. This measurement is in excellent agreement with the \ac{sbf} distance used to calibrate \ac{desi}, with a significantly reduced uncertainty. The measured distance is at 4.6$\sigma$ tension from the Planck-\lcdm prediction.

\begin{figure}
    \centering
    \includegraphics[scale = 0.40]{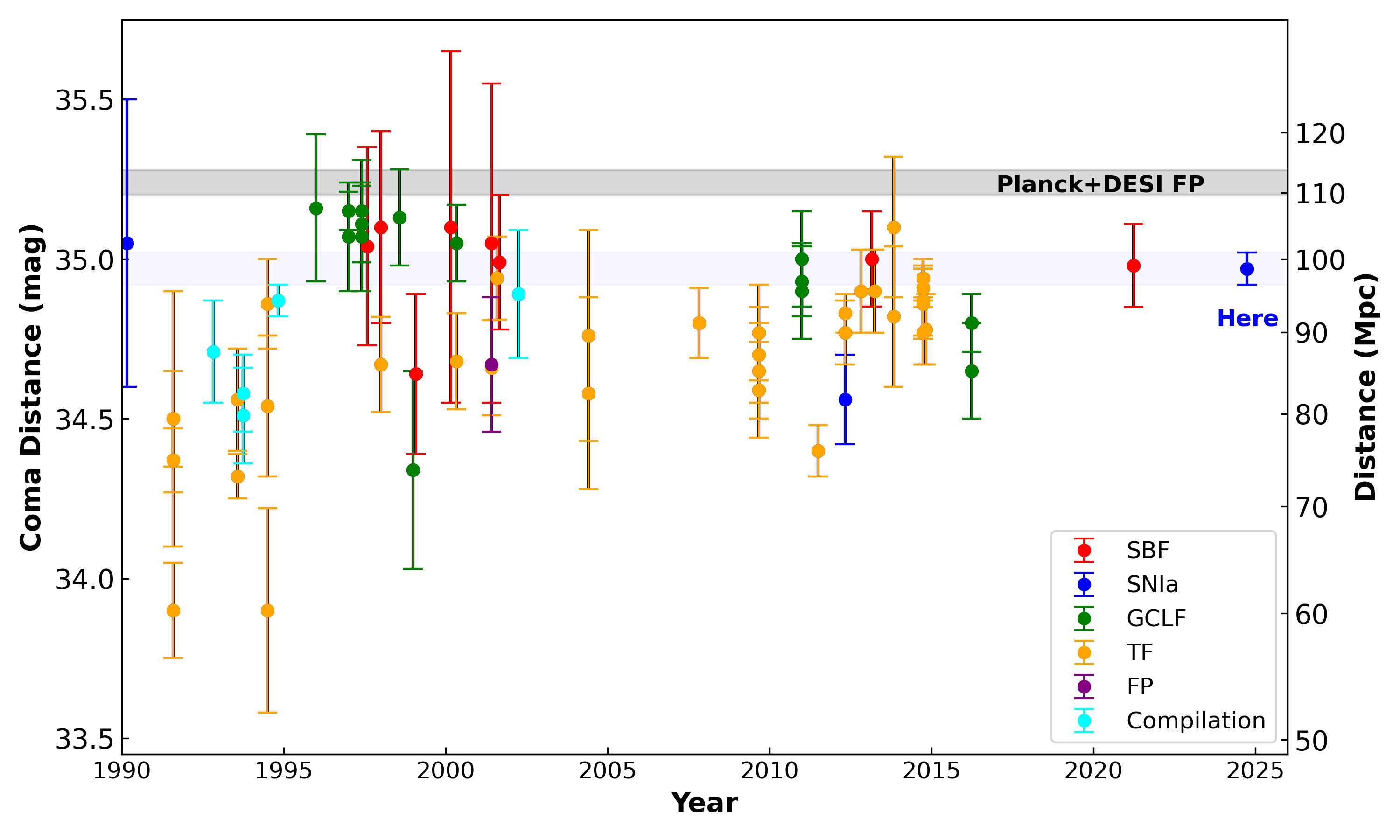}
    \caption{Historical (1990 onward) distance modulus measurements of the Coma cluster (as reviewed in~\cite{deGrijs:2020yfe}).  Only distance measurements that do not depend on redshift and $H_0$ are included. Figure taken from~\cite{Scolnic:2024hbh}, and the rightmost point represents the \ac{sn1} distance measured in their work.}
    \label{fig:coma_hist}
\end{figure}

\paragraph{New \& historical measurements in tension with \lcdm\ }
Historical measurements, as well as additional distance indicators and calibration methods further support a distance to Coma around 100 Mpc (see Fig.~\ref{fig:coma_hist}). A compilation of historical measurements summarized in~\cite{Carter:2008jg}, including  I-band Tully-Fisher~\cite{Brent:1999uv}, K-band \ac{sbf}~\cite{Jensen:1998dq}, I-band
\ac{sbf}~\cite{Thomsen:1997vz}, $D_n - \sigma$~\cite{Gregg:1997nd}, FP
\cite{Hjorth:1997mi} and Globular Cluster \ac{lf} (GCLF) \cite{Kavelaars:1999ec}, produce an average of $D_\text{Coma} = 95.1 \pm 3.1$ Mpc.
The \ac{hst} Key Project (KP) calibrated the FP relation using Cepheid variable stars, and its results, accounted for the recent updates, measure $ D_{\text{Coma}} = 85 \pm 8 $ Mpc~\cite{deGrijs:2020yfe}. 
A more recent \ac{jwst} measurement of \ac{trgb} and \ac{sbf}, allows to re-calibrate the FP relation by \ac{hst}-KP, which places Coma at $ 90 \pm 9 $ Mpc. When combined with the \ac{sbf} distance used in the \ac{desi} work and the new \ac{sn1} distance, these independent approaches yield a consensus distance of $ 98.0 \pm 2.0 $ Mpc.

The tension between local and Planck-calibrated distances to Coma reflects a broader challenge in reconciling the local and early-universe distance scales. If Coma were truly at $ D_{\text{Coma}} > 110 $ Mpc as \lcdm-Planck calibration of \ac{desi} FP predicts, it would be inconsistent with decades of direct distance measurements based on a wide range of methodologies. 
The upcoming year 1 data release for \ac{desi} FP is expected to be significantly larger than the current study, and the targeted \ac{jwst} observation of Coma galaxies (PI Jensen, GO 5989) is underway. When combined, these data will significantly reduce the statistical uncertainties in the FP-based $H_0$ or distance measurements, and the increased sky coverage in \ac{desi} data will allow FP to be calibrated with more than one galaxy clusters. Additionally, a few \ac{sn1} are being discovered in Coma every year, and the \ac{sn1}--based distance measurements are expected to be more robust in the next few years. Coma distance serves as an alternative or additional perspective on the issues surrounding the cosmological distance scales, and it showcases a new type of distance ladder analysis that is expected to develop rapidly in the coming years.

\bigskip
\subsubsection{Cosmic chronometers \label{sec:CC}}

\noindent \textbf{Coordinator:} Michele Moresco\\
\noindent \textbf{Contributors:} Adri\`a G\'omez-Valent, Anto Idicherian Lonappan, Arianna Favale, David Benisty, David Valls-Gabaud, Dinko Milakovic, Elena Tomasetti, Marek Biesiada, Rodrigo Sandoval-Orozco, Ruth Lazkoz, Sanjay Mandal, Swayamtrupta Panda, and Vavilova Iryna
\\

\begin{figure}[t!]
    \centering
    \includegraphics[width=0.95\textwidth]{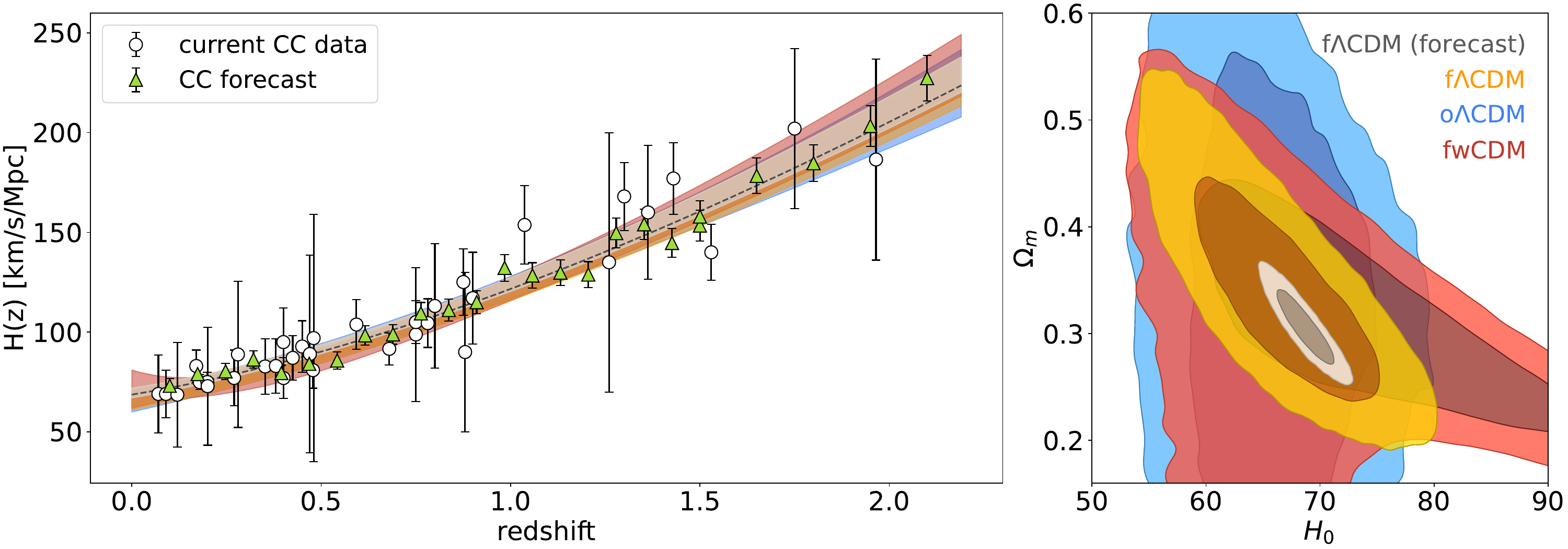}
    \caption{Left plot. H(z) measurements obtained with the \ac{cc} method from the literature (white points) and from future surveys  (green points, from Refs.~\cite{Moresco:2022phi, WST:2024rai}). Right plot. Cosmological constraints in the $\Omega_m$-$H_0$ plane; the different colors show the 68\% CL results (also reported in the left panel) obtained by fitting current data with a flat \lcdm\ (in blue), open \lcdm\ (in yellow), and flat $w$CDM (in red) cosmological model, and future \ac{cc} data with a flat \lcdm\ model (in grey).}
    \label{fig:CC}
\end{figure}

\ac{cc} have emerged in the last few years \cite{Moresco:2022phi} as one of the most promising cosmological probes that can provide direct measurements of the expansion history of the Universe without the need to rely on any assumed cosmological model. The method \cite{Jimenez:2001gg}, which is only based on the validity of the Cosmological Principle and on the assumption that General Relativity and standard physics hold in the environment of the galactic stars, proposes to derive the Hubble parameter $H(z)$ directly from the differential age evolution of the Universe ($dt$) in a redshift interval ($dz$), as:
\begin{equation}
    H(z)=-\frac{1}{(1+z)}\frac{dz}{dt}\,.
\end{equation}
To map the age evolution of the Universe, the best tracers that have been found are very massive ($\log_{10}(M/M_\odot)>11$) and passively evolving galaxies since they have been proven to be extremely homogeneous systems, with synchronized formation, representing at each redshift the oldest population of galaxies \cite{Moresco:2022phi}. Several authors have applied this method with different approaches to derive the differential age d$t$ \cite{Simon:2004tf, Stern:2009ep, Moresco:2015cya, Moresco:2016mzx, Ratsimbazafy:2017vga, Borghi:2021zsr, Jiao:2022aep, Tomasetti:2023kek, Jimenez:2023flo}. It currently counts more than 30 $H(z)$ measurements in the range $0<z<2.1$, with accuracy from $\sim$10-15\% at the higher redshifts down to $\sim$5\% at the lower redshifts (see Fig.~\ref{fig:CC}). See Refs.~\cite{Moresco:2022phi, Moresco:2023zys} for extensive reviews on the topic.

Since it provides a measurement of the Hubble parameter without assuming any cosmological model, recently, \ac{cc}s have been widely employed in discussing the Hubble tension. To fully take advantage of the potential of these data, several different cosmology-independent techniques have been explored, from \ac{gp} \cite{Busti:2014dua,Cao:2017ivt,Yu:2017iju,Gomez-Valent:2018hwc,Haridasu:2018gqm,Gomez-Valent:2021hda,Bonilla:2020wbn,OColgain:2021pyh,Liu:2022lqw, Yang:2022jkf,Favale:2023lnp,Favale:2024lgp,Yang:2024epu}, to  non-parametric smoothing \cite{Rani:2016wff, Li:2015nta}, Weighted Polynomial Regression method \cite{Gomez-Valent:2018hwc}, Pad{\'e} approximations \cite{Capozziello:2020ctn,Liu:2022lqw}, and  \ac{ann} architectures \cite{Yang:2024epu}; all these analysis provide a reconstruction of $H(z)$ independent of any cosmological model that can, in turn, be extrapolated to $z=0$ for a determination of the Hubble constant $H_0$. The other alternative is, instead, to assume a cosmological model and derive cosmological constraints from the parametric fit of the data \cite{Moresco:2012by,Moresco:2016nqq,Moresco:2022phi}. Both these approaches, considering all the sources of statistical and systematic errors, give comparable results, yielding a constraint on the Hubble constant of $H_0=70.7\pm6.7$\kms~\cite{Favale:2023lnp}. Future measurements, however, leveraging on much larger statistics (like from the \ac{esa} mission Euclid \cite{EUCLID:2011zbd} or from future spectrographs like WST \cite{WST:2024rai}) and better modeling of the data, promise to significantly improve these constraints, improving the accuracy from $\sim$8\% to $\sim$2\%. An example is given in Fig.~\ref{fig:CC}, where current and future \ac{cc} data are fitted with different cosmological models (flat \lcdm, open \lcdm, and $w$CDM) to provide constraints in the $\Omega_{\rm M}$-$H_0$ plane (for more details, see Refs.~\cite{Moresco:2022phi,WST:2024rai}).

A difference in $H_0$ can be reinterpreted also as a tension in the calibrators of the cosmic distance ladders, i.e., the absolute magnitude $M$ of standard candles such as \ac{sn1} and the standard ruler represented by the comoving sound horizon at the baryon-drag epoch, $r_d$. Assuming \ac{cc} as reliable \ac{cc}s, it is also possible to use them to measure these distance calibrators independently from the \ac{cmb} and the first rungs of the direct distance ladder. Thus, it is clear why \ac{cc} plays an important role in the discussion. So far, the uncertainties for $M$ and $r_d$ derived by applying these calibrations are still too large to arbitrate the tension \cite{Gomez-Valent:2021hda,Favale:2023lnp,Cogato:2023atm}, but also this is bound to improve in the near future with the advent of upcoming surveys and data. Euclid \cite{EUCLID:2011zbd}, for instance, is expected to increase by 2 orders of magnitude the currently available statistics \cite{Moresco:2022phi}. \ac{cc}s have been used to calibrate baryon acoustic oscillations and \ac{sn1} \cite{Heavens:2014rja, Verde:2016ccp,Gomez-Valent:2018hwc,Haridasu:2018gqm,Gomez-Valent:2021hda,Favale:2023lnp,Guo:2024pkx} and also other standardizable objects like ultra-compact radio \ac{qso}s \cite{Cao:2017ivt}, \ac{grb}s or \ac{qso}s \cite{Amati:2018tso,Wei:2019uss, Favale:2024lgp,Cogato:2023atm} which can be exploited to extend the ladder beyond the \ac{sn1} redshifts, $z>2$. 

Beyond the $H_0$ tension, \ac{cc}s are also the ideal data to test and compare with different cosmological models, from the simplest to the more exotic ones. Among the litmus tests of the \lcdm\ model, the $Om(z)$ diagnostic and its two-point version $Om(z_1,z_2)$ have a distinguishing feature \cite{Sahni:2014ooa} that depends only upon the expansion rate $H(z)$. Hence, \ac{cc}s have been used \cite{Zheng:2016jlq,Zheng:2018sxp} to perform tests of scenarios assuming either a cosmological constant or an evolving \ac{de} equation of state.
\bigskip
\subsubsection{Strong lensing and time delay measurements \label{sec:strng_lens}}

\noindent \textbf{Coordinator:} Tommaso Treu\\
\noindent \textbf{Contributors:} Alba Domi, Alexander Bonilla Rivera, Ali \"Ovg{\"u}n, Anowar J. Shajib, Clecio Roque De bom, David Valls-Gabaud, Dominique Sluse, Eoin O Colgain, Fr\'ed\'eric Courbin, Jenny Wagner, Lindita Hamolli, Predrag Jovanović, Sherry H. Suyu, Simon Birrer, Tanja Petrushevska, and Veronica Motta
\\

Time variable sources that are multiply imaged -- i.e., strongly lensed -- provide an opportunity to measure absolute distances, and therefore cosmological parameters including the Hubble constant $H_0$. The main advantages of this so-called time delay cosmography (TDC) with respect to other cosmological probes described in this white paper are: i) it is a direct measurement of distances well into the Hubble flow; ii) it is independent of all other types of distance measurements; iii) it relies on well understood fundamental physics such as general relativity. In this section, we provide a brief summary of TDC, the current state of the art, and future prospects. The reader is referred to dedicated reviews for more extensive discussion, e.g., see Ref.~\cite{Treu:2016ljm,Suyu:2018vqs,Treu:2022aqp,Suyu:2023jue,Birrer:2022chj}.

The method was originally proposed for lensed \ac{sn1} by Refsdal in Ref.~\cite{Refsdal:1964blz}, well before the discovery of any actual strong gravitational lens. Since the mid-eighties it has been applied to lensed \ac{qso}s \cite{Treu:2016ljm}, and only recently -- with the discovery of multiply imaged \ac{sn} \cite{XXX:2014xxi,Kelly:2014mwa} -- it has been possible to fulfill Refsdal's dream \cite{Kelly:2023mgv,Grillo:2024rhi}.

TDC measures angular diameter distances, typically between $z=0.5-3$, where the majority of deflector galaxies and lensed sources lie. More precisely, it measures the so-called time delay distance $D_{\rm \Delta t} = (1+z_{\rm d}) \frac {D_{\rm d} D_{\rm s}}{{D_{\rm ds}}}$, where $D$ indicates angular diameter distance, and the subscripts s and d stand for source and deflector, respectively. In combination with stellar kinematics, it also measures D$_{\rm d}$ \cite{Jee:2015yra}.

TDC is thus primarily sensitive to $H_0$. It is, however, also sensitive to other cosmological parameters, such as curvature or the \ac{de} equation of state, in a complementary way to probes such as the \ac{cmb} and \ac{sn1}, breaking some of the degeneracies of those probes \cite{Linder:2011dr,Suyu:2013kha,Collett:2019hrr,Shajib:2017omw}.

\begin{figure}[htbp]
    \centering
    \includegraphics[scale = 0.6]{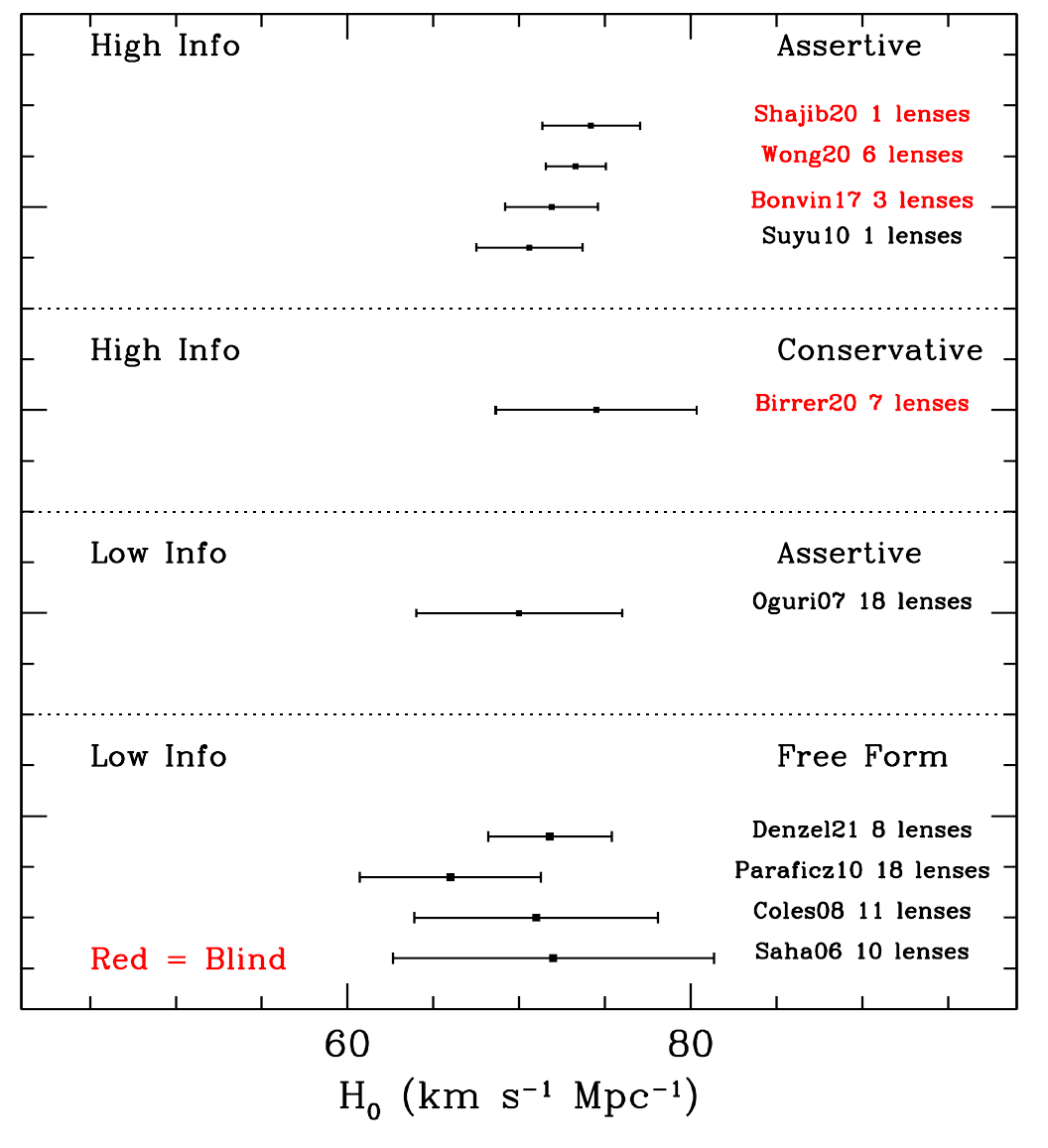}
    \caption[Comparison of inferences of H$_0$ based on TDC, in \lcdm\ cosmology, as of 2022]{The measurements are categorized according to i) assumptions on the mass distribution of the main deflector, ``assertive'' and ``conservative'' for simply parametrized models or ``free form'' for pixilated models; ii) to the amount of information used per lens, ``low info'' utilizes \ac{qso} positions and time delays, "high info" adds the extended surface brightness distribution of the multiple images of the \ac{qso} host galaxy (or ``Einstein Ring''), stellar kinematics of the main deflector, and number counts or \ac{wl} to estimate the line-of-sight convergence. For each measurement, the reference and the number of time delay lenses are given. The measurements shown in red were performed with blinding to prevent experimenter bias. The figure is taken from Ref.~\cite{Treu:2022aqp}, see therein for details.}
\label{fig:TDC}
\end{figure}

Several representative measurements using lensed \ac{qso}s prior to 2022 are summarized in Fig.~\ref{fig:TDC}. Already in 2020, just 7 lenses yielded $H_0$ at 2\% precision \cite{DES:2019fny, Millon:2019slk}, under standard assumptions about the mass density profile of the deflector. That measurement is in agreement with, and completely independent, of late Universe probes. If those assumptions are relaxed, and a parametrization of the mass density profile that is maximally degenerate with the Hubble constant via the mass-sheet degeneracy \cite{1985ApJ...289L...1F,Wagner:2018rae} is considered, the uncertainty increases to 8\% for the same 7 lenses \cite{Birrer:2020tax}. 
In 2023, the first lensed supernova \cite{Kelly:2023mgv} yielded an $H_0$ measurement with $\sim$7\% precision. Whereas this precision is very encouraging for a single system, especially given the complexity of modeling a cluster of galaxies, see also Ref.~\cite{Grillo:2024rhi}, it is not sufficient to distinguish between the early and late-time measurements and resolve the ``Hubble tension''. As more lensed \ac{sn} are being discovered and analyzed, e.g., see Ref.~\cite{Goobar:2016uuf,Goobar:2022wan,Rodney:2021keu,2022Natur.611..256C,2022TNSAN.169....1K,2024ApJ...961..171F,Pierel:2024frl,Pierel:2024rjr}, the precision is expected to improve rapidly, especially once the Roman and Rubin telescopes come on line \cite{Pierel:2020tav}. 

In order to increase the precision on $H_0$ to the desired level of 1\% under these weaker assumptions, three complementary approaches can be pursued \cite{Birrer:2020jyr}. The first is to increase the sample size. The Euclid, Roman, and Rubin Telescopes are expected to discover thousands of strong lenses, including lensed \ac{sn} \cite{Oguri:2010ns,Petrushevska:2020wmc,LSSTDarkEnrgyScience:2023atc,Collett:2015roa}.
There is little doubt that large numbers of lenses will be discovered. The challenge will be to transform them into cosmological probes by obtaining the necessary additional data and computing the necessary cosmography grade models. Additional data include high-resolution images, time delays, deflector and source redshift, deflector kinematics, and supernova classification if relevant. Some of these data will be provided by the surveys themselves, e.g., high-resolution images by Roman and Euclid \cite{Meng:2015qia,Pierel:2020tav} and some time delays by Rubin \cite{Liao:2014cka}, but others will require concerted follow-up effort.
A coordinated follow-up for measuring precise time delays is critical. \ac{qso} time delays require high precision lightcurves minimally affected by microlensing \cite{Jovanovic:2008ay, Millon:2020ugy}, 
while \ac{sn} require dedicated multi-color or spectroscopic follow-up \cite{Goldstein:2017bny,Bayer:2021ugw}. The effort will be substantial, and it will require, e.g., repurposing a 4m class telescope for time delays or building a dedicated satellite. The advent of \ac{elt}s will simplify matters, especially for high-resolution imaging and kinematics.

Modeling hundreds of lenses with precision sufficient for cosmography hundreds of lenses would be prohibitive at the moment, both in terms of human power and computing power. Fortunately, efforts are underway to develop automated pipelines \cite{Etherington:2022nzt,DES:2022dvw,Legin:2022ovl,2021MNRAS.503.2380S}, in some cases with the help of \ac{ml} algorithms \cite{Adam:2023xff,Park:2020eat,Schuldt:2022msj}, to speed things up. The first results are promising, in the sense that models from independent automated pipelines are found to agree in their predicted time delays within the uncertainties, when the data contain sufficient information models that agree to within the uncertainties in predicted time delay can be found \cite{Ertl:2022rqx}.

The second approach consists of increasing the precision per lens. In order to break the mass sheet degeneracy, non-lensing information is needed, for example, stellar kinematics of the deflector \cite{Treu:2002cb,Yildirim:2019vlv,TDCOSMO:2023hni,Shajib:2017omw}, the absolute luminosity of a lensed supernova Ia \cite{Birrer:2021use}, or \ac{wl} \cite{Khadka:2024bmw}.  If the quality of the kinematics is sufficient, in principle, each lens can deliver distances to 2-4\% precision, limited by systematic errors in stellar velocity dispersion measurements \cite{Yildirim:2021wdd}.

The third approach consists of utilizing the much more abundant non-time delay lenses  (a.k.a. ``external samples'') to learn about the internal structure of time delay lenses. This method has been shown to increase the precision of TDC significantly \cite{Birrer:2020tax}. The key to using external samples without introducing biases is to understand the selection function of both the time delay and non-time delay lenses sufficiently well. At the moment, this is not a limiting factor \cite{2023A&A...678A...4S} - statistical errors are larger - but the selection function will have to be properly modeled and understood in order to achieve 1\% precision and accuracy on $H_0$, sufficient to solve the Hubble tension.

As more lensed SN are being discovered and analyzed, the precision is expected to improve rapidly, especially once the Roman and Rubin telescopes come online \cite{Pascale:2024qjr}.

\bigskip
\subsubsection{Extended QSO cosmologies \label{sec:QSOs}}

\noindent \textbf{Coordinator:} Maria Giovanna Dainotti\\
\noindent \textbf{Contributors:} Aleksander \L{}ukasz Lenart, Catarina Marques, Celia Escamilla-Rivera, Giada Bargiacchi, Giovanni Montani, Rodrigo Sandoval-Orozco, and Swayamtrupta Panda
\\

\ac{qso}s are one of the most luminous non-transient energy sources, whose use at cosmological scales allows us to study the Universe at higher redshifts, e.g., up to $z \sim 7$ \cite{Lusso:2020pdb,2021ApJ...907L...1W}. At this redshift range, it is possible to reveal some interesting aspects and deviations from different cosmological models that can be indistinguishable at low $z$. In this path, some examples of the \ac{qso} cosmological use are: \textit{(i)} the reverberation mapping technique \cite{Watson:2011um,Haas2011A&A...535A..73H}, and \textit{(ii)} the relation between variability in X-ray amplitude and black hole mass scatter \cite{LaFranca:2014eba}. However, there is a lack of a clear definition of \ac{qso}s to order the diversity of \ac{agn} objects \cite{2018A&A...618A.179M}, although recent studies have begun to streamline the various \ac{qso} categories \cite{2018FrASS...5....6M, Panda:2019dok} with particular emphasis on sources accreting at or above the Eddington limit \cite{Panda:2022ncv}. There is a need to homogenize the modeling of broadband \ac{qso} spectra and examine and assess the evolution of \ac{qso} properties (Black Hole mass, Eddington ratio, viewing angle to the source) over a broad range of redshifts \cite{Prince:2021iuo, Dainotti:2022rfz}. Under such techniques, several \ac{qso} catalogs have been used to find cosmological constraints under certain theoretical assumptions, e.g., consistencies or inconsistencies between time delays from reverberation mapping of prominent emission lines extending beyond the cosmic noon \cite{Cao:2022pdv} and impact of the inherent heterogeneity in the data set on cosmological parameter constraints \cite{Cao:2023fpp}, constraints using \ac{grb}s with \ac{qso} cosmology analysis for future landscapes \cite{Dainotti:2022bzg}. 

Furthermore, studies including \ac{qso} catalogs combined with \ac{grb} observations achieve small uncertainties on cosmological parameters \cite{Bargiacchi:2023jse,Dainotti:2023bwq}. Moreover, combining the Risaliti-Lusso relation for \ac{qso} with \ac{sn1} can allow us to obtain corrections on the redshift evolution as a function of cosmology \cite{Lenart:2022nip}. Recent studies have been going through the Risaliti-Lusso relation, which has been validated in Ref.~\cite{Dainotti:2022rfz}, where $\Omega_{\rm m,0}$ value seems larger when only it is considered \ac{qso} baselines \cite{Yang:2019vgk, Khadka:2020whe,Bargiacchi:2021hdp,Colgain:2022nlb,Lenart:2022nip}, and some systematics in the \ac{qso}s measurements should be expected in future new cosmological probes \cite{Zajacek:2023qjm}. 
Despite the systematics, this relation is a promising cosmological probe since its observed dispersion on average is $\sim$0.20-0.25~dex, and
becomes significantly smaller when only the sources with the highest quality X-ray observations are considered 
\cite{Sacchi:2022ofz}, suggesting that intrinsically the X-ray to UV relation has a scatter lower than 0.1~dex, 
pointing to a tight, universal physical process regulating the energy flow between the accretion disk and the X-ray emitting region in \ac{qso}s. 
In this scheme, several studies related to extended theories of gravity have been performed and cosmologically constrained using \ac{qso} baselines, e.g., $f(T)$ cosmologies have been constrained through cosmography by considering non-flat and flat geometries \cite{Shabani:2023xfn} with \ac{grb} observables, using \ac{qso} objects \cite{Sabiee:2022iyo} detected through high-quality UV and X-ray fluxes up to $z\sim 5.1$ \cite{Lusso:2019akb}.  
In view of possible discrepancies among measurements of the matter critical parameter via sources with different $z$-ranges, it has been proposed a promising model in which
$\Omega_{0,m}$  becomes a dynamical quantity. The proposed scenario relies on a metric $f(R)$-gravity in the so-called Jordan frame and dominant linear contribution is retained in the potential term of the non-minimally coupled scalar field (a small deviation is included too). It is such a dominant contribution that is responsible for a non-standard critical matter parameter \cite{Dainotti:2024aha}.
From the cosmographic point of view, other analyses have highlighted a significant tension between the standard cosmological model and the best-fit cosmographic model, when using \ac{qso}s, \ac{sn1}, and \ac{grb}s \cite{Lusso:2019akb}, \ac{qso}s and \ac{sn1} \cite{Bargiacchi:2021fow}, and \ac{qso}s combined with \ac{grb}s, \ac{bao}, and \ac{sn1} \cite{Bargiacchi:2023rfd}. Also, some studies use \ac{qso}s as standard rulers \cite{Sabiee:2022iyo} by their angular size–luminosity using very-long-baseline interferometry \cite{1985AJ.....90.1599P}. In addition, the joint analysis of spectroastrometry and reverberation mapping can measure \ac{agn} distances and provide a new way to measure $H_0$ \cite{Wang:2019gaq}. Furthermore, recent results have considered \ac{qso} physics from UV, X-ray, and optical plane techniques behind the local observations as cosmological probes to study the $H_0$ tension \cite{Sandoval-Orozco:2023pit}.
Another application of \ac{qso}s in cosmology is the Sandage test of the cosmological redshift drift, i.e., a small dynamic change in the redshift of objects following the Hubble flow \cite{1962ApJ...136..319S}, which requires not only high instrumental resolution but also very bright targets.
One of the main objectives of the QUBRICS (QUasars as BRIght beacons for Cosmology in the Southern Hemisphere) survey \cite{G:2019dar} is to provide a sample of bright targets for the Sandage test.
The two brightest \ac{qso}s in this Golden Sample have started to be analyzed in the pilot program ``An ESPRESSO Redshift Drift Experiment'', which will be a complete end-to-end proof of concept for this experiment at ANDES in \ac{elt}, where the redshift drift signal is expected to be detected \cite{Liske:2008ph}.
Although the Sandage test is a direct and real-time model-independent mapping of the expansion rate of the Universe, recent theoretical studies such as Refs.~\cite{Marques:2023zuv} and~\cite{ANDES:2023cif} show synergies with other cosmological probes, in particular regarding the characterization of the physical properties of \ac{de}.

\begin{figure}
    \centering
         \includegraphics[scale=0.35]{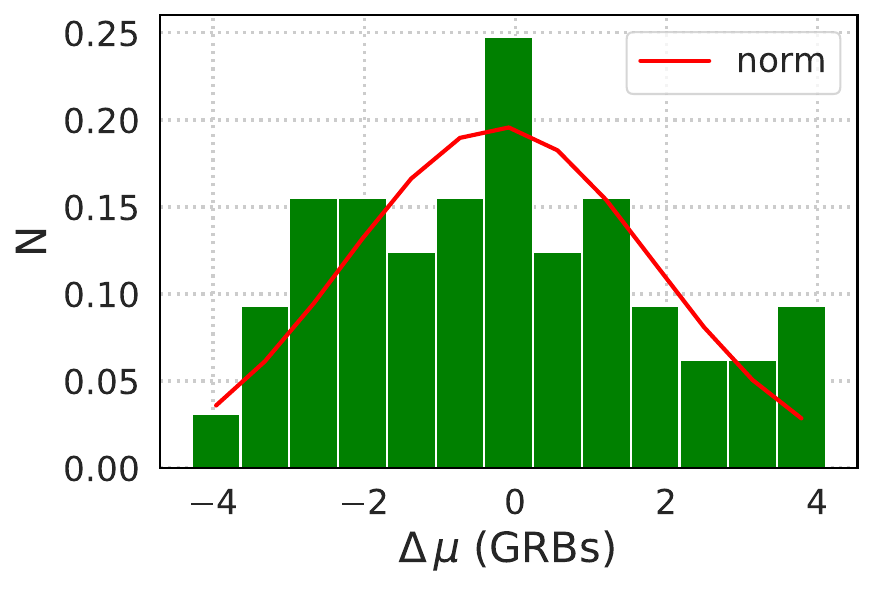}
     \includegraphics[scale=0.35]{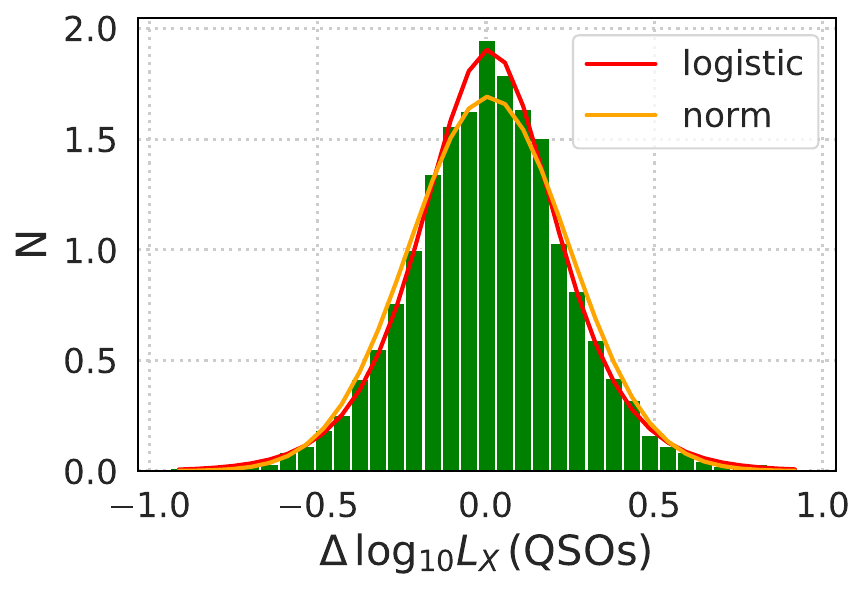}
     \includegraphics[scale=0.35]{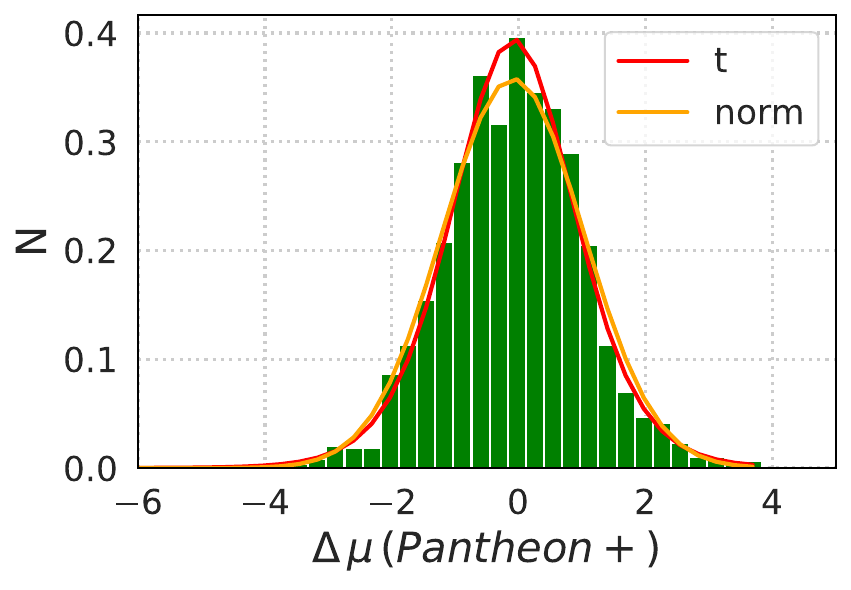}
     \caption{Normalized residuals $\Delta\mu$ histogram for \ac{grb}s (left panel), \ac{qso}s (middle panel), and \ac{sn1} from Pantheon + samples (right panel). The red curve is the best-fit distribution, while the orange one is the Gaussian distribution. In the case of \ac{grb}s, the two coincide since the Gaussian is the best-fit distribution.}
     \label{fig:hist}
 \end{figure}

Regarding the efforts to reach a higher precision in the determination of cosmological parameters, also a purely statistical approach has been employed. This is based on the fact that the normalized residuals of \ac{qso} logarithmic luminosities of the Risaliti-Lusso relation are not normally distributed. Hence, the Gaussian is not the most appropriate distribution to be used as cosmological likelihood. Indeed, Ref.~\cite{Bargiacchi:2023jse} prove that the \ac{qso} have the logistic distribution as a best-fit. This is shown in the right upper panel of Fig.~\ref{fig:hist} along with the same investigation for \ac{grb}s (left upper panel) and \ac{sn1} (lower panels). Remarkably, the employment of this proper likelihood for \ac{qso}s, compared to the standard Gaussian one, reduces the uncertainty on $H_0$ up to 35\%, on $\Omega_{\rm m,0}$ up to 27\%, on $\Omega_k$ up to 32\%, and on $w$ up to 31\%, when non-flat and $w$CDM models are investigated by combining \ac{qso}s with \ac{grb}s, \ac{bao}s, and \ac{sn1}, where every single probe is fitted with its own best-fit cosmological likelihood \cite{Dainotti:2023bwq}.
Another approach aiming at inferring cosmological parameters with lower uncertainties consists of identifying a \ac{qso} sub-sample composed only of sources that better follow the Risaliti-Lusso relation, i.e., are closer to the best-fit relation. Indeed, this method allows us to reduce the intrinsic dispersion of the correlation, thus yielding a better precision of the cosmological parameters. Different techniques have been applied to select such a sub-sample \cite{Dainotti:2023cpn,Dainotti:2024bth,Dainotti:2024aha} which have promoted \ac{qso}s as standalone cosmological probes, even with the same precision as \ac{sn1}.

\bigskip

\subsubsection{GRBs as cosmological standard candles \label{sec:GRBs}}

\noindent \textbf{Coordinator:} Maria Giovanna Dainotti\\
\noindent \textbf{Contributors:} Aleksander \L{}ukasz Lenart, Arianna Favale, Denitsa Staicova, Giada Bargiacchi, Hassan Abdalla, and Rados{\l}aw Wojtak
\\

Long-duration \ac{grb}s (LGRBs) could be used as standard candles, extending the capabilities of the Hubble diagram to measure distances further than currently possible with \ac{sn1}, thereby helping to constrain cosmological parameters. Phenomenological relations derived from spectral modeling, e.g., the Amati relation \cite{Amati:2002ny, Amati:2006ky, Amati:2009ts} can be used for this purpose. This relation connects the source-frame energy (E$_{i,p}$), where the gamma-ray spectral energy peaks, with the isotropic-equivalent bolometric energy (E$_{iso}$) released during the prompt phase \cite{Dirirsa:2019fcs}. Another significant empirical correlation in the prompt emission is the Yonetoku relation \cite{Yonetoku:2003gi} between E$_{peak}$ with the intrinsic peak luminosity L$_{iso}$. This correlation has been validated for both LGRBs and short-duration \ac{grb}s (SGRBs) \cite{Yonetoku:2003gi, Aldowma:2024nlp}.
Given the large variability in the feature of the prompt emission, a more promising approach is obtained with the use of correlations involving the plateau emission \cite{Dainotti:2008vw}, a relatively flat segment in the light curve (LC). The Dainotti relation relates the rest-frame time at the end of the plateau ($T^{*}_{a}$) and its corresponding luminosity ($L_a$). 
This relation succeeds in standardizing \ac{grb}s by utilizing the shape of their LCs. Indeed, around 500 \ac{grb}s observed by Swift exhibit a distinct, plateau phase in X-rays. The great advantage of this relation, compared to the ones in the prompt emission, is that the plateau phase can be attributed more straightforwardly to various theoretical models, including magnetar spin-down \cite{Metzger:2010pp,Rowlinson:2013ue,Rowlinson:2014dja,Stratta:2018xza} and forward/reverse shock mechanisms \cite{Uhm:2007nc,Uhm:2012yg,Hascoet:2014ira}.

This correlation was later updated in Refs.~\cite{Dainotti:2010ki,Dainotti:2011yz,Dainotti:2013fra}, and subsequently extended to a 3D correlation by incorporating the peak luminosity ($L_{peak}$) \cite{Dainotti:2010ki,Dainotti:2015gva,Dainotti:2016iqn,Dainotti:2017fem}, the so-called 3D Dainotti relation. \cite{Dainotti:2013fra,Dainotti:2015gva} demonstrated that this correlation is devoid of biases and it can be corrected for the redshift evolution.

Further studies by Refs.~\cite{Dainotti:2016iqn, Dainotti:2016yxl} confirmed the robustness of this correlation for high-quality LGRBs, which show well-defined plateau properties that obey the fundamental plane relation and constitute the so-called Gold sample. An improvement of this sample has been classified as the Platinum sample \cite{Dainotti:2020azn}. As a result, this correlation has been successfully applied as a cosmological tool \cite{Cardone:2009mr, Dainotti:2013cta,Dainotti:2022wli,Dainotti:2022ked,Dainotti:2023bwq}, providing insights into the high-redshift Universe. \ac{grb}s, being at high-$z$, could be one of the important tools to differentiate between competing cosmological models. Indeed, Ref.~\cite{Bargiacchi:2023jse} pinpointed traces of rising tension between flat \lcdm\ and the equivalent model with $\Omega_k\neq 0$. There is a trend that reveals $\Omega_k<0$, marking the importance of further studies in this direction at high-$z$.

\begin{figure}
    \centering
    \includegraphics[scale=0.09]
    {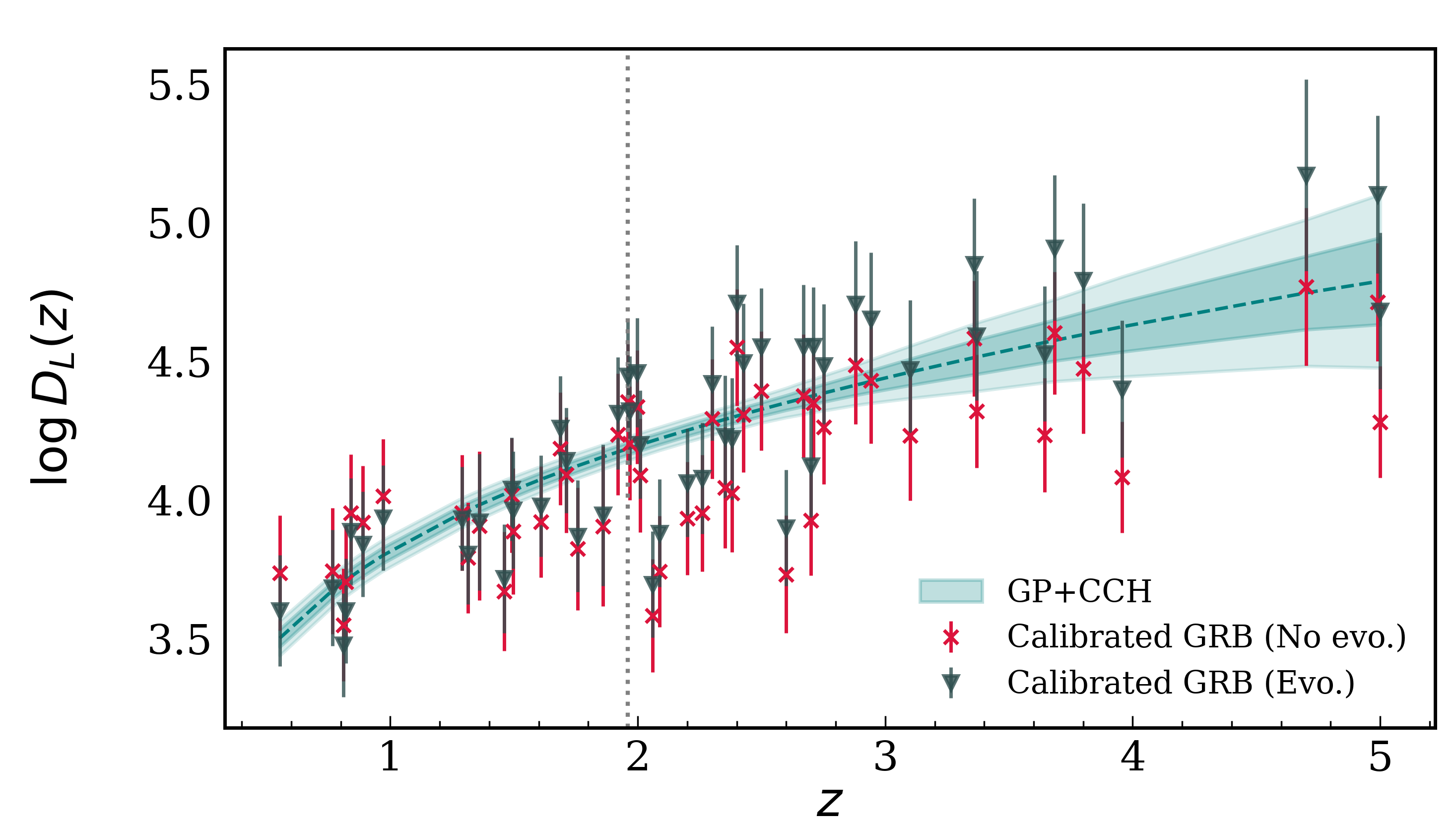}
    \caption{The distance luminosity vs redshift when a calibration with \ac{cc} is considered. The figure has been taken by Ref.~\cite{Favale:2024lgp}.}
    \label{fig:Hubblediag-OmegakGRB}
\end{figure}

The future cosmological role of \ac{grb}s has also been investigated in Ref.~\cite{Dainotti:2022wli} where it has been determined how many \ac{grb}s are needed as stand-alone probes to achieve a comparable precision on $\Omega_{m,0}$ to the one obtained by \ac{sn1} only. They obtained the same error measurements derived using \ac{sn1} in 2011 and 2014 with 142 and 284 simulated optical \ac{grb}s, respectively, considering the error bars on the variables halved. These error limits will be reached in 2038 and in 2047, respectively. Using a doubled sample, obtained by the current \ac{ml} approaches \cite{Dainotti:2024zhh,Dainotti:2024scc} allowing a LC reconstruction \cite{Dainotti:2023kwj} and the estimates of \ac{grb} redshifts, with error bars halved, the same precision as \ac{sn1} in 2011 and 2014, is reached now and in 2026, respectively. If we need to reach the current precision of \ac{sn1}, it would require 18 years from now, but this estimate is very conservative since it does not consider the new redshifts from Euclid and \ac{jwst}.

Indeed, in light of the existing cosmological tensions, such as the one on the Hubble constant, $H_0$, it has become fundamental to obtain unbiased cosmological distances via intermediate redshift probes. Thus, the calibration procedure of the \ac{grb} correlations must not be subject to strong model-dependent assumptions. Works have shown that this can be done by employing low-$z$ probes, such as \ac{sn1} from the Pantheon Plus sample, a collection of 1701 \ac{sn1}, discussed already in the previous section, in combination with advanced statistical techniques such as \ac{gp} \cite{Liang:2022smf} or cosmographical analyses \cite{Luongo:2020aqw, Mu:2023bsf}. On the other hand, data on the $H(z)$ obtained from \ac{cc} can play the role of calibrating the ladders \cite{Favale:2023lnp} since these data only rely on General Relativity and the validity of the standard physics in the environment of the galaxy stars. Ref.~\cite{Favale:2024lgp} found that current data on \ac{cc} can identify a subset of \ac{grb}s in the Platinum sample in $0.553\leq z\leq 1.96$ that reveals a tight fundamental plane relation, with one of the lowest intrinsic scatter observed to date, $\sigma_{int} = 0.20_{-0.05}^{+0.03}$. This result is obtained considering a relation corrected for evolutionary effects and with parameters compatible with the magnetar model \cite{Rowlinson:2013ue,Rowlinson:2014dja,Rea:2015gna,Stratta:2018xza}.
The \ac{grb} sub-sample pinpointed by this model-independent calibration can represent a valuable set of standardizable candles that is used to extend the cosmic distance ladder by providing unbiased luminosity distances up to $z=5$.
For \ac{qso}, calibrating with \ac{sn1}, varying only $H_0$ in the cases without and with correction for evolution, $H_0=73.76 \pm 2.18$\kms and $H_0=69.82 \pm 2.27$\kms, respectively \cite{Lenart:2022nip}. For \ac{grb}s, using the Gaussian priors, varying only $H_0$ in the cases without and with correction for evolution, $H_0=73.23 \pm 3.31$\kms and $H_0=72.87 \pm 2.92$\kms, respectively \cite{Dainotti:2022ked}.

Additionally, the 3D X-ray Dainotti relation has also been recently employed in cosmographic analyses to investigate the reliability of the standard cosmological model in a cosmology-independent way. In this concern, as mentioned in the previous section, \ac{grb}s combined with \ac{sn1}, \ac{qso}s, and \ac{bao} point toward a statistically significant discrepancy between the prediction of the concordance model and the observational data Ref.~\cite{Bargiacchi:2023rfd}. From another point of view, \cite{Alfano:2024ukk} constrained the transition epoch between a matter-dominated and a \ac{de}-dominated universe by using two cosmographic approaches and the combination of \ac{grb}s, \ac{sn1}, and \ac{bao} obtaining results compatible with the concordance model.
Concerning instead the statistical analyses already mentioned for \ac{qso}s, the residuals of the distance moduli of \ac{grb}s remarkably prove to be effectively Gaussian, as reported in the upper left panel of Fig.~\ref{fig:hist}, differently from the cases of \ac{qso}s and \ac{sn1}.

Given the importance of the 3D Dainotti relation, it is crucial to explore the existence of the same correlations but with different associated classes (not only the Platinum sample). The LGRBs associated with Supernovae Type Ib/c (\ac{grb}-\ac{sn}) are interesting since they have been discussed in the literature as possible standard candles \cite{Cano:2014oca,Dainotti:2020azn}. Ref.~\cite{Dainotti:2022mto} investigate the existence of probable correlations among \ac{sn} parameters and \ac{grb} prompt and afterglow features considering the largest compilation of \ac{grb}-v possible associations observed from 1997 up to 2021 and find a possible correlation between the \ac{grb} optical luminosity at the end of the plateau ($\log_{10} L_{a,opt}$) and the $\log_{10}$ of \ac{sn} rest-frame peak time ($\log_{10} t^{*}_{p}$). The correlation can be expressed as $\log_{10} L_{a,opt} = (9.43 \pm 1.47)\log_{10} t^{*}_{p} - (13.60 \pm 11.89)$. The uncertainties on the fitting parameters (16\%-87\%) are too high to allow using this correlation for standardizing \ac{grb}s, thus new observations are needed to validate it. If confirmed, then it will represent a crucial support for cosmological analysis.

Another interesting application of \ac{grb} physics in cosmology comes from the search for quantum gravity models. Some such models predict an energy-dependent speed of light, which can be observed as simultaneously emitted high and low-energy photons arriving at different times. Such a tiny quantum effect is expected to be amplified by very high energies of the photons and cosmological distances, making \ac{grb}s a promising probe. In Refs.~\cite{Staicova:2023vln, Staicova:2024ljn}, the authors explored the use of different \ac{grb} time-delay datasets in combination with the Pantheon/Pantheon+ \ac{sn1} datasets, \ac{bao}s, and the \ac{cmb} distance priors under different approximations for the intrinsic lag to constrain cosmological parameters and investigate the impact on the $H_0$ tension. The analysis revealed that the inclusion of such time-delay datasets still leads to a deviation in the parameter $c/H_0 r_d$, where $c$ is the speed of light and $r_d$ is the comoving sound horizon at the drag epoch. Such deviation can be interpreted as an artifact of the $H_0$ tension on the remaining quantities, demonstrating again that the tension is spread to the whole $H_0-r_d-\Omega_{\rm m,0}$ plane. Such a dataset is not directly affected by the $E_p-E_{iso}$ correlation, but relies critically on the assumption of Lorentz Invariance Violation (LIV), which so far has not been detected. In the case where LIV is not zero, \ac{grb} time delays would be a new independent cosmological probe. 

In addition, the constraints on the cosmological parameters can be obtained with the interaction of the gamma rays with their surrounding medium or other radiations. Gamma rays with energies above around 10 GeV are attenuated via interactions with the extragalactic background light (EBL) photons, resulting in electron-positron pair production. The attenuation effect was systematically measured in spectra of blazars and \ac{grb}s, using a wide range of instruments and techniques, e.g., see Ref.~\cite{Fermi-LAT:2018lqt,MAGIC:2019ozu}. Since the attenuation scales with the comoving distance, these measurements can be used to constrain cosmological parameters, in particular $H_0$, given a model of the EBL based on integrating light from galaxies in deep cosmological surveys \cite{Dominguez:2019jqc}. Recent joint modeling of the EBL and $\gamma$-ray attenuation yields $H_{0}=65.1^{+6.0}_{-4.9}$\kms, and $H_{0}=66.5^{+2.2}_{-2.1}$\kms when combined with the \ac{bao} observations and the \ac{bbn} prior \cite{Dominguez:2023rxa}. The $H_{0}$ measurements are independent of any external distance calibration, and the best-fit values are in close agreement with the \ac{cmb}-based Planck value \cite{Planck:2018vyg}.

\bigskip
\subsubsection{Gravitational wave constraints \label{sec:GWs}}

\noindent \textbf{Coordinator:} Antonella Palmese\\
\noindent \textbf{Contributors:} Bangalore S. Sathyaprakash, Ivan de Martino, Matteo Tagliazucchi, Nicola Borghi, and Nicola Tamanini
\\

The field of \ac{gw} astronomy has recently facilitated novel measurements of the Universe's expansion rate by exploiting unique properties of compact binary mergers. By providing a distinct perspective on the Universe, in contrast to traditional astronomical and cosmological probes (such as electromagnetic radiation, cosmic rays, and neutrinos), \ac{gw}s offer a complementary approach to addressing cosmological tensions.

Since 2015, a ground-based \ac{gw} detector network, including the \ac{ligo} \cite{LIGOScientific:2014pky} and later Virgo \cite{VIRGO:2014yos} and KAGRA \cite{KAGRA:2020agh}, has been detecting coalescences of compact object binaries, which so far include binary neutron star (BNS) \cite{LIGOScientific:2017vwq}, neutron star-black hole (NSBH) \cite{LIGOScientific:2021qlt}, and binary black hole (BBH) \cite{LIGOScientific:2016aoc} mergers. It is anticipated that \ac{ligo}-Virgo-KAGRA (LVK) will complete their fourth and fifth observing runs (O4 and O5) by the early 2030s \cite{KAGRA:2013rdx}, after which a significantly upgraded detector network, called A$^\#$ \cite{T2200287, VIRGO:2023elp} is expected to become operational. This will be followed by the deployment of next-generation (XG) observatories, such as the Cosmic Explorer \cite{Reitze:2019iox, Evans:2021gyd, Evans:2023euw} and \ac{et} \cite{ETDesign2020}. Additionally, the \ac{lisa} \cite{LISA:2017pwj} is scheduled for launch in the next decade and will be sensitive to \ac{gw}s from the inspiral and merger of binaries within the 0.1 mHz to 1 Hz frequency range, including massive BBHs and extreme mass ratio inspirals (EMRIs). \ac{lisa} and XG observatories will observe mergers throughout the cosmos from an epoch before the first stars formed.

\ac{gw} events can be used as ``standard sirens'' (StS; \cite{Schutz:1986gp,Holz:2005df}), as they can act as absolute distance indicators. The \ac{gw} signal of a compact binary merger is directly sensitive to the luminosity distance of the source and can therefore be combined with redshift information to act as a probe of the expansion of the Universe. Depending on where the redshift information is derived from, StSs are typically divided into different classes, with bright and dark StSs being the two main classes. In what follows, we briefly describe each method and the state of the field, including the latest measurements and systematics studies, and end with prospects for future measurements, focusing on $H_0$ constraints as a pathway to shed light on the Hubble tension.

\paragraph{Bright standard sirens}

For bright StSs, an electromagnetic (EM) counterpart is identified, and the redshift is derived from its host galaxy, assuming that both are unique. So far, there exists only one EM counterpart that is confidently associated with a \ac{gw} event, GW170817 \cite{LIGOScientific:2017ync}. This association has been used to derive the first StS measurement \cite{LIGOScientific:2017adf}, finding $H_0=70^{+12}_{-8}$\kms, a $\sim14\%$ precision measurement. Subsequent analyses have taken a more in-depth approach to estimate the peculiar velocity of the host galaxy \cite{Nicolaou:2019cip,Mukherjee:2019qmm,Howlett:2019mdh}. Moreover, various works attempted to take advantage of the EM observations to constrain the viewing angle of the binary and break the distance-inclination angle degeneracy which exists when estimating these parameters from the \ac{gw} data \cite{Guidorzi:2017ogy,Hotokezaka:2018dfi,Dhawan:2019phb,Palmese:2023beh}. The latest of these measurements reached a $\sim 7$\% precision and they are all consistent with both early and late-time measurements of $H_0$.

It is worth noting some of the major systematics that may affect future bright StS measurements. A StS analysis only using the \ac{gw} data and the host galaxy spectroscopic redshift of a high-confidence EM counterpart provides an $H_0$ measurement which is not expected to first order to be affected by major systematic uncertainties, assuming that the peculiar velocities measurements are well constrained (also considering that their uncertainties will become less important as we move towards larger \ac{gw} detector distance horizons \cite{Palmese:2019ehe}) and detector calibration errors are kept under control ($\lesssim 2\%$ \cite{Chen:2020dyt}). Although \ac{gw} selection effects are well understood and can be more easily modeled, selection effects from EM searches also need to be taken into account and may be coupled to binary parameters such as source luminosity distance and viewing angle \cite{Chen:2023dgw}. Beyond selection effects, if including EM estimates of the viewing angle, $H_0$ measurements may also be affected by modeling uncertainties of the jet structure \cite{Gianfagna:2023cgk}, kilonova geometry \cite{Heinzel:2020qlt}, or even a possible jet misalignment with the binary angular momentum \cite{Muller:2024wzl}.

Future bright StSs are most likely expected to arise from binary neutron star (BNS) and neutron star black hole (NSBH; which would prove valuable StSs \cite{Vitale:2018wlg}) mergers, but candidate EM counterparts to BBH mergers also exist \cite{Graham:2020gwr,Graham:2022xxu,Cabrera:2024wrk}. Although the BBH EM counterparts association is currently uncertain \cite{Ashton:2020kyr,Palmese:2021wcv}, they may in the future be used to infer the Hubble constant along with other cosmological parameters \cite{Mukherjee:2020kki,Gayathri:2020mra,Chen:2020gek,Bom:2023zgw,Alves:2020wnl}.

\paragraph{Dark standard sirens}

Most mergers detected by the LVK (currently in the order of $\sim 200$ when including significant detection candidates from the ongoing observing run) do not have an associated EM counterpart and are mostly comprised of BBH mergers. In that case, the redshift information necessary for an StS analysis may be taken from galaxy catalogs (e.g., see Refs.~\cite{Gair:2022zsa,Gray:2019ksv}), from the redshifted mass measured with the \ac{gw} data \cite{Farr:2019twy,Ezquiaga:2022zkx}, or a combination of the two \cite{Mastrogiovanni:2023emh, Gray:2023wgj, Borghi:2023opd}.

In the first case, the redshift distribution of galaxies along the line of sight of a \ac{gw} event's localization can be used to infer cosmological parameters, including $H_0$, from the distance-redshift relation. Various works have produced such measurements so far \cite{DES:2019ccw,LIGOScientific:2019zcs,LIGOScientific:2020zkf,LIGOScientific:2021aug,Palmese:2021mjm,DESI:2023fij,Alfradique:2023giv,Bom:2024afj,Finke:2021aom}, in general finding a precision on $H_0$ down to $\sim 20$\% following the third LVK observing run, again in agreement with both \emph{Planck} and SH0ES constraints. Even when combined with the bright StS available, the precision reaches $\sim10$\% (when ignoring EM viewing angle constraints) and does not yet allow us to distinguish between the two leading $H_0$ measurements.

Golden dark sirens are a subclass of dark sirens when the catalog contains, or follow-up observations reveal, a single galaxy in the 3D localization volume. Although rare, they may provide exquisite few percent-level measurements of $H_0$ \cite{Borhanian:2020vyr}.  
It is also possible to cross-correlate \ac{gw} events with galaxy catalogs to infer cosmological parameters (e.g., see Ref.~\cite{Diaz:2021pem}). In all cases, the \ac{gw} localization precision and the availability of extended spectroscopic galaxy catalogs \cite{Borghi:2023opd} will be crucial to establish dark StS as a cosmological probe competitive with bright StSs.

In the second case, the source redshift is reconstructed at the statistical level by taking advantage of the presence of features in the distributions of the BBH population properties, such as the mass gap that may be explained by the theory of pair-instability supernova \cite{Bond:1984sn, Zevin:2020gbd}, or the overdensity peaks in the mass distribution observed in LVK data \cite{KAGRA:2021duu,MaganaHernandez:2024qkz}.
This ``spectral sirens'' approach has allowed constraints on $H_0$ at the $\sim60\%$ level \cite{LIGOScientific:2021aug} and on modified \ac{gw} propagation \cite{Mancarella:2021ecn,Leyde:2022orh}. These constraints are expected to significantly improve with XG detectors, as all BBH events can be used as spectral sirens. However, this poses a computational challenge, since the time required for a full cosmological and population analysis scales linearly with the number of events.

The major sources of systematics for dark sirens methods are expected from incorrect assumptions on the mass function \cite{LIGOScientific:2021aug}; while inferring cosmological parameters in conjunction may mitigate this effect \cite{Mastrogiovanni:2023emh}, it is still crucial to correctly model the mass function including its potential evolution \cite{Ezquiaga:2022zkx}. Other systematics may arise when using photometric redshifts \cite{DES:2020nay,Turski:2023lxq}, and host galaxy weighting or galaxy assumptions that do not match the true underlying distribution of merger hosts \cite{Perna:2024lod,Hanselman:2024hqy}.

\begin{figure}[htbp]
    \centering
    \includegraphics[scale = 0.70]{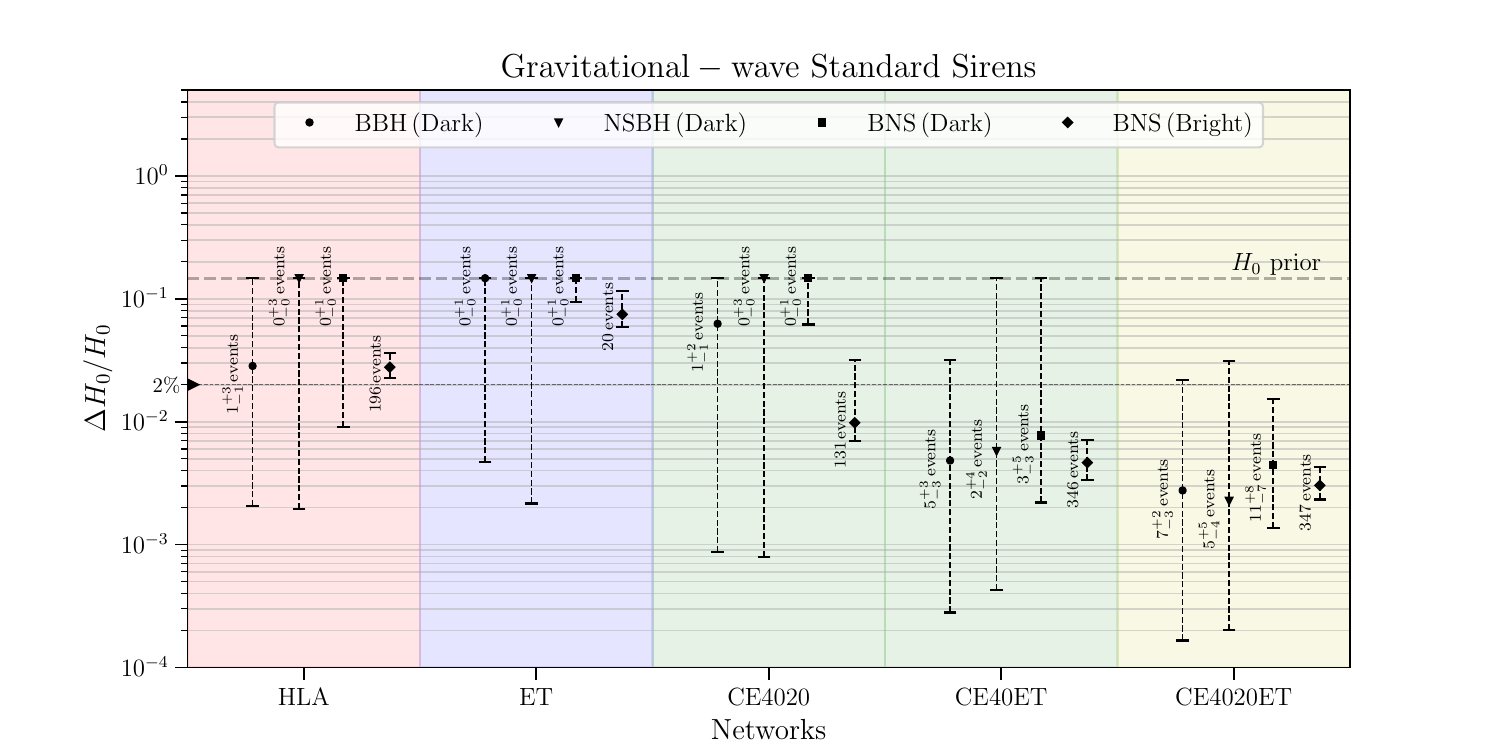}
    \caption[Short caption]{Precision with which $H_0$ would be constrained by standard sirens with the HLV network upgraded to A$^\#$ sensitivity and networks containing one or more XG observatories (see Sec.~\ref{sec:gw ground} text for details). Figure taken from Ref.~\cite{Chen:2024gdn}.}
\label{fig:GW_trinity}
\end{figure}

\paragraph{Prospects for ground-based detectors}
\label{sec:gw ground}
Fig.~\ref{fig:GW_trinity} in Ref.~\cite{Chen:2024gdn} summarizes the prospect of terrestrial \ac{gw} observatories in resolving the Hubble tension. The $x$-axis lists \ac{gw} detector networks that are expected to become operational or newly built over the next decade or more. For each network, the $y$-axis shows the precision with which the Hubble constant would be measured, for four classes of binary merger events: BBH, NSBH and BNS as golden dark sirens, as well as BNS mergers with associated EM counterpart. 
For each network the figure also shows the number of events expected to be detected in each source class. 
A target precision of 2\% could be achieved by the HLV network with dark sirens but is not guaranteed since the number of events expected is ${\cal O}(1).$ If the current median BNS merger rate holds, then the HLV network might observe $\sim 130$ bright sirens and accomplish the 2\% target. Assuming one \ac{et} detector and its synergy with Transient High Energy Sources and Early Universe Surveyor (THESEUS) \cite{THESEUS:2021uox} for bright StSs, one may achieve accuracy on $H_0$ of only $0.40$\kms in the case of a non-flat \lcdm\ model after five years of observations (which corresponds to $\sim 166$ bright StSs detections with redshift below $z\sim 4.3$) \cite{Califano:2022cmo}. \ac{et} alone, or a network of only two CE detectors,  will not be able to accomplish the 2\% target with dark sirens, although a network of two CE detectors could do so with bright sirens. A network consisting of an \ac{et} and at least one CE will determine $H_0$ to sub-percent precision with dark sirens of all source classes, as well as with hundreds of bright sirens.  
In the case of \ac{de} models, the accuracy on the Hubble constant may achieve 1\% after 10 years of observations due to the correlations between $H_0$ and the \ac{de} parameters of the specific model under consideration \cite{Califano:2022syd}.
It is also worth noting that bright sirens with XG will reach a percent level precision in distance in the local Universe, thus enabling precision measurements of $\sigma_8$ \cite{Palmese:2020kxn} which will inform our understanding of the $S_8$ tension, should it persist. With XG observatories we will enter the era of precision cosmology with \ac{gw} observations.

\paragraph{LISA perspective}

The \ac{lisa} \cite{LISA:2024hlh} will push \ac{gw} cosmology at high redshift ($z\gtrsim3$), mapping the expansion of the Universe in a still poorly charted cosmic epoch \cite{LISACosmologyWorkingGroup:2022jok}. \ac{lisa} will observe both bright sirens, in the form of massive black hole binary mergers with an identified EM counterpart \cite{Klein:2015hvg,Mangiagli:2022niy}, and dark sirens, in the form of extreme mass ratio inspirals (EMRIs) \cite{Babak:2017tow}. Independently these two populations of standard sirens are expected to deliver percent constraints on the Hubble constant \cite{Tamanini:2016zlh,LISACosmologyWorkingGroup:2019mwx,Mangiagli:2023ize,MacLeod:2007jd,Laghi:2021pqk,Toscani:2023gdf}. Combined together, however, they should provide a one-percent, or even better, constraint on $H_0$ \cite{Tamanini:2016uin}, making \ac{lisa} a rightful competitor to solve the Hubble tension. The large redshift of \ac{lisa} standard sirens will furthermore provide original tests of \lcdm\ and of general relativity, delivering new potential insights on the nature of \ac{de} \cite{LISACosmologyWorkingGroup:2019mwx,Caprini:2016qxs,Cai:2017yww,Corman:2021avn,Speri:2020hwc,Liu:2023onj}.

\bigskip
\subsubsection{The cosmic microwave background radiation \label{sec:CMB}}

\noindent \textbf{Coordinator:} Margherita Lembo and Martina Gerbino\\
\noindent \textbf{Contributors:} Anto Idicherian Lonappan, Chandra Shekhar Saraf, Enrico Specogna, Luca Caloni, \"{O}zg\"{u}r Akarsu, Nils A. Nilsson, Pawe\l{} Bielewicz, Simone Paradiso, and Venus Keus
\\

\paragraph{Review of CMB basics and recent observations} \label{sec:CMB_intro}
The \ac{cmb} plays a crucial role in shaping our understanding of modern physics and remains a powerful tool for advancing knowledge in cosmology and particle physics. 
The \ac{cmb} dates back to roughly 380000 years after the Big Bang, at a redshift of $z_* \simeq 1100$, when \ac{cmb} photons decoupled from electrons, close to the moment in the thermal history of the Universe when electrons and protons combined into neutral hydrogen atoms. Since then, \ac{cmb} photons free stream with a nearby blackbody distribution with a temperature today of approximately $T \simeq 2.7$ K~\cite{Fixsen:2009ug}. The \ac{cmb} spectrum peaks around 100 GHz, within the microwave frequency range of the electromagnetic spectrum. Primordial density perturbations at the last scattering surface gave rise to temperature anisotropies in the \ac{cmb} field of the order of $\Delta T/T \sim 10^{-5}$~\cite{COBE:1992syq}. The \ac{cmb} radiation, because of its quadrupolar anisotropy, is indeed also partially linearly polarized due to Thompson scattering, with an amplitude of $\sim 10\%$ of temperature anisotropies. 

The polarization pattern is usually decomposed into so-called (curl-free) E-modes and (divergence-free) B-modes. On small angular scales, both the temperature and the polarization fields undergo a tiny distortion when they pass close to large distributions of matter. The weak gravitational lensing effect (\ac{cmb} lensing) is one of the most important mechanisms that can generate secondary anisotropies in the \ac{cmb}. In particular, it smears out anisotropies at small angular scales in temperature and E-mode polarization, and it gives rise to an additional B-mode contribution sourced by the lensed E-mode pattern.

The first full sky mapping of the \ac{cmb} anisotropies was made by the COBE satellite~\cite{Mather:1990tfx}, refined by WMAP~\cite{WMAP:2003ivt} and finally by \textit{Planck}~\cite{Planck:2018vyg}, that delivered the state-of-the-art for full-sky \ac{cmb} measurements both in temperature and polarization. At smaller (arcmin) angular scales in temperature and E-mode polarization, observations have been dominated in sensitivity by the ground-based experiments \ac{act} and SPT; the search for B modes has been led so far by ground-based telescopes observing degree angular scales, i.e., BICEP/KECK, POLARBEAR/Simons Array.

Next-generation \ac{cmb} experiments, both space-borne (e.g., LiteBIRD~\cite{LiteBIRD:2022cnt}) and ground-based (e.g., Simons Observatory~\cite{SimonsObservatory:2018koc} and \ac{cmbs4}~\cite{CMB-S4:2016ple}), aim to achieve precise measurements of \ac{cmb} polarization. A clear science goal is to detect the imprint on \ac{cmb} pattern of primordial \ac{gw}s, which are the smoking gun of an inflationary phase in the early Universe~\cite{Kamionkowski:2015yta,LiteBIRD:2022cnt,Guzzetti:2016mkm}. Moreover, improved measurements of \ac{cmb} polarization would provide insights on potential new physics beyond the standard model~\cite{Gerbino:2016mqb,BICEP2:2017lpa,Pogosian:2018vfr,Bartolo:2018elp,Minami:2020odp,Namikawa:2020ffr,Choi:2021aze,Greco:2022ufo,Komatsu:2022nvu}. \ac{cmb} lensing is one of the most relevant observable of near-future \ac{cmb} experiments, as it provides an invaluable tool to reconstruct the integrated distribution of matter in the Universe and to place stringent constraints on physical properties and effects
mostly related to the late-time phases of the Universe, such as the growth of
structures and neutrino masses (see e.g., \textit{Planck}~\cite{Planck:2018lbu}, \ac{act}~\cite{ACT:2023kun}, \ac{spt}~\cite{SPT-3G:2021wgf}).

\paragraph{CMB-driven constraints in the context of cosmological tensions}
\label{sec:CMB_tensions}
\ac{cmb} observations allowed us to test the predictions of the standard cosmological model and to shape our knowledge of the Universe, its history, and composition.
Despite its phenomenological success, the \lcdm\ model is still incomplete in that, for example, fails to address the fundamental nature of the most abundant dark components. In addition, it has been facing challenges due to statistical tensions between results obtained with recent cosmological and astrophysical surveys of increased accuracy. 
Most notably, there is a severe ($>5\sigma$) discrepancy in the value of the Hubble constant $H_0$ estimated with \ac{cmb} observations with respect to the result obtained with local distance ladder measurements~\cite{Planck:2018vyg,Riess:2021jrx}. Furthermore, \ac{cmb} and \ac{lss} measurements show a less severe albeit still intriguing disagreement on the amplitude of matter perturbations quantified via the $\sigma_8$ parameter~\cite{Joudaki:2016kym,Asgari:2019fkq,Planck:2018vyg,DESI:2024mwx}.
These discrepancies and their implications will be explored in more detail in the following subsections from the perspective of \ac{cmb} observations.

\paragraph{Constraints on $H_0$ from CMB measurements}
\label{sec:CMB_constraints}

The Hubble tension is one of the major unresolved issues in modern cosmology. This tension refers to the discrepancy between local measurements of the Hubble constant, $H_0$, and the value inferred from early Universe observations assuming the \lcdm\ model. Early Universe estimates of $H_0$ are driven by \ac{cmb} observations made by the \textit{Planck} satellite, which yields the value $H_0 = 67.27 \pm 0.60$\kms at 68\% CL, using a combination of TT,TE,EE$+$lowE data~\cite{Planck:2018vyg} (see also Refs.~\cite{Rosenberg:2022sdy,Tristram:2023haj} for re-analyses of {\it Planck} data, with no significant deviations of $H_0$ from the {\it Planck} result). On the other hand, the most sensitive local measurement of $H_0$ is obtained by the SH$_0$ES collaboration, which finds $H_0 = 73.04 \pm 1.04$\kms exploiting type \ac{sn1} calibrated with Cepheids~\cite{Riess:2021jrx}. These two values are in 5$\sigma$ tension. 

Alternative \ac{cmb} datasets, such as \ac{wmap}~\cite{WMAP:2003ivt}, \ac{act}~\cite{ACT:2020gnv,ACT:2025fju}, and \ac{spt}~\cite{SPT-3G:2022hvq}, or \ac{cmb}-independent probes, such as combinations of \ac{bbn} with \ac{bao} measurements~\cite{Schoneberg:2019wmt}, yield $H_0$ values that align with those obtained from {\it Planck}. The
analysis of \ac{act}-DR6 data~\cite{ACT:2025fju} yields a 68\% CL value of
$H_0 = 66.11 \pm 0.79$\kms (\ac{act} TTTEEE+Planck-Sroll2 EE at large scales to
constrain tau) consistent with measurements obtained by \textit{Planck} satellite
and lower than local measurements. Similarly, the combination of \ac{act}-DR6
and \ac{wmap} data results in a value of $H_0 = 66.78 \pm 0.68$\kms at 68\%
CL~\cite{ACT:2025fju,ACT:2025tim}. Additionally, the combination of \ac{act} and \ac{wmap} data results in a value of $H_0 = 67.6 \pm 1.1$\kms at 68\% CL~\cite{ACT:2020gnv}. The same collaboration also reported an estimate of $H_0 = 68.3\pm1.1$\kms from measurements of \ac{cmb} lensing in combination with \ac{bao} data (6dF and SDSS)~\cite{ACT:2023kun}. The \ac{spt} collaboration finds a compatible result, with a value of  $H_0 = 68.3 \pm 1.5$\kms at 68\% CL~\cite{SPT-3G:2022hvq}, improved to $66.81\pm0.81$\kms with the use of the latest unlensed EE and \ac{cmb} lensing data in combination with a prior on the optical depth $\tau$~\cite{SPT-3G:2024atg}. It is relevant to emphasize that all the results above are independent of \textit{Planck} data.
On the other hand, late-time measurements different from SH$_0$ES, including additional methods for calibrating \ac{sn1} at large distances (see Sec.~\ref{sec:SNeIa}) or observations independent of \ac{sn1} (see e.g., Sec.~\ref{sec:type_2_sn}, Sec.~\ref{sec:CC}, Sec.~\ref{sec:HII_gal}, Sec.~\ref{sec:GWs}), are all suggesting higher values of the Hubble parameter.

\begin{figure}
    \centering
    \includegraphics[width=0.95\linewidth]{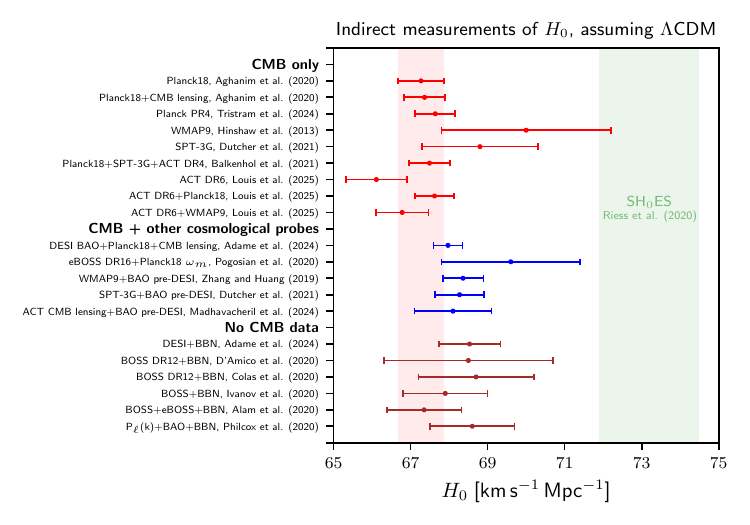}
    \caption{Whisker plot with 68\% CL constraints on the Hubble constant $H_0$ through different indirect measurements performed over the years. The constraints shown in this plot have been derived assuming \lcdm\ model. If not specified, ``BAO'' refers to a combination of \ac{bao} data prior to \ac{desi} (see the corresponding paper for more details). The green vertical band corresponds to the $H_0$ value from SH$_0$ES team~\cite{Riess:2021jrx}, while the red one corresponds to the value from \textit{Planck} 2018~\cite{Planck:2018vyg}. This figure has been adapted from Fig.~1 of Ref.~\cite{DiValentino:2021izs}.}
    \label{fig:H0whisker}
\end{figure}

Depending on the combination of measurements, the tension ranges roughly from 4$\sigma$ to 6$\sigma$. Fig.~\ref{fig:H0whisker} shows a comprehensive list of \ac{cmb} measurements of the Hubble parameter. It is important to emphasize that these results were derived under the assumption of the \lcdm\ model. For example, as shown in Ref.~\cite{SPT-3G:2024qkd}, the uncertainty on $H_0$ increases when considering extensions to the \lcdm\ framework.

Assuming a different cosmological model may lead to different estimates of $H_0$, either in the mean value or in the width (i.e., in the errorbar) of the \ac{pdf} (or both) \cite{DiValentino:2021izs, Abdalla:2022yfr, Schoneberg:2021qvd}.

Since the value of $H_0$ inferred from \ac{cmb} observations depends on the underlying cosmological model, this discrepancy with the local measurements of the Hubble parameter might be an indication of physics beyond the \lcdm\ model. An alternative possibility, which will be discussed in Sec.~\ref{sec:systematics}, is that this tension is due to unknown systematics that affect either early-time or late-time observations (or both). However, given the vast array of independent probes of $H_0$, it seems quite unlikely that systematic effects may be the only responsible for this discrepancy. This further motivates the search for an explanation in terms of new physics. The landscape of theoretical scenarios to address the $H_0$ tension is broadly discussed in Sec.~\ref{sec:fun_phys}. Here, we just recall that the easiest extensions to \lcdm\ are unable to solve the tension. More exotic scenarios, like those involving new BSM interactions between the components of the stress-energy tensor, the existence of new components or variation of fundamental constants, usually do not properly alleviate the tension, but rather yield estimates of $H_0$ with a larger uncertainty than the \lcdm-based inference. Some models that gained visibility in the past, like those with strongly self-interacting neutrinos, require new physics that is at odds with complementary constraints from laboratory probes~\cite{Berryman:2022hds}. The search for new physics is still ongoing.

In order to understand the possible physical origin of this tension, we need first to discuss how \ac{cmb} observations constrain $H_0$ (see Refs.~\cite{Bernal:2016gxb,Knox:2019rjx}). The key quantity from which the Hubble constant is inferred is the angular scale of the sound horizon at recombination, which we assume to happen instantaneously at redshift $z_*\simeq 1100$. This is defined as 
\begin{equation}
    \label{eq:acoustic_scale}
    \theta_s \equiv \frac{r_s(z_*)}{D_{\rm A}(z_*)} \, ,
\end{equation}
where $r_s(z_*)$ and $D_{\rm A}(z_*)$ are the comoving sound horizon at recombination and the comoving angular diameter distance to recombination, respectively:
\begin{equation}
\label{eq:sound_horizon}
    r_{s}(z_{*}) = S_{k}\bigg(\int_{z_*}^{\infty} \frac{c_s(z)}{H(z)} dz\bigg)\,, \qquad D_{A}(z_{*}) = S_{k}\bigg(\int_0^{z_*} \frac{1}{H(z)} dz\bigg)\,,
\end{equation}
with
\begin{equation}
    S_k(x) = 
        \begin{cases}
            \frac{1}{\sqrt{|k|}}\sinh (\sqrt{|k|}x) \,, & \text{ for } k < 0 \\
            x \,, & \text{ for } k = 0\\
            \frac{1}{\sqrt{|k|}}\sin (\sqrt{|k|}x) \,, & \text{ for } k > 0 \\
        \end{cases}\,.
\end{equation}
Here, $c_s(z) = 1/\sqrt{3[1+R(z)]}$ is the sound speed of the acoustic waves in the baryon-photon plasma, where $R(z) = 3\rho_{\rm b}(z)/(4\rho_\gamma(z)) = 3\omega_{\rm b}/(4\omega_\gamma) (1+z)^{-1}$ denotes the ratio between the baryon and photon energy densities, and $k$ is the spatial curvature of the Universe. Because of the so called ``geometric degeneracy''~\cite{Efstathiou:1998xx}, \ac{cmb} anisotropy measurements alone do not constrain the curvature of the Universe. To a large extent, the degeneracy can be broken using \ac{cmb} lensing effect. Tighter constraints can be imposed by analyzing \ac{cmb} data in combination with other cosmological probes (see Sec.~\ref{sec_cmb:other_probes}). The Hubble parameter, $H(z)$, is given in terms of the physical densities of \ac{de}, matter, photons, neutrinos and curvature as 
\begin{equation}
    \label{eq:Hubble_evolution}
    H(z) = (100 \; {\rm km} \; {\rm s}^{-1}  \; {\rm Mpc}^{-1}) \sqrt{\omega_{\rm DE}(z) + \omega_{\rm m,0}(1+z)^3 + \omega_\gamma(1+z)^4 + \omega_\nu(z)  + \omega_k (1+z)^2} \,.
\end{equation}
Within the \lcdm\ model, $\omega_{\rm DE}(z) = \omega_\Lambda$ and $\omega_k = 0$. The only unknown quantities in Eqs.~(\ref{eq:sound_horizon}-\ref{eq:Hubble_evolution}) are $\omega_{\rm b}$, $\omega_{\rm m,0}$ and $\omega_\Lambda$. The radiation density $\omega_\gamma$ is fixed by the \ac{cmb} temperature obtained from measurements of the \ac{cmb} blackbody spectrum~\cite{Fixsen:2009ug} and the neutrino density can be obtained at any time by properly integrating the neutrino distribution function\footnote{The parameter of the neutrino equation of state is not constant. Therefore, it is not possible to generally parametrize $\omega_\nu$ via a simple scaling of the redshift, unless one considers the limiting cases of ultra-relativistic and non-relativistic neutrinos.}.

Then, $H_0$ is inferred from \ac{cmb} data as follows (see Ref.~\cite{Knox:2019rjx} for more details). First, we determine the baryon and matter densities from their impact on \ac{cmb} power spectra~\cite{Planck:2016tof}. The baryon density affects the relative heights of the acoustic peaks and the diffusion damping in the \ac{cmb} power spectrum, while the matter density is mostly determined via the ``potential envelope'' effect, i.e., the increase in power for modes that re-entered the horizon before matter-radiation equality. The characteristic scale of this power boost corresponds to the comoving size of the horizon at matter-radiation equality projected on the last scattering surface, which depends on the ratio between the matter and radiation densities. Since $\omega_\Lambda$ is negligible at redshifts $z \ge z_*$, we have everything we need to determine $r_s(z_*)$ via Eq.~\eqref{eq:sound_horizon}.

Then, we obtain the acoustic scale $\theta_s$ from the spacing between the acoustic peaks, and, using Eq.~\eqref{eq:acoustic_scale} we derive $D_{\rm A}(z_*)$. As a last step, we adjust $\omega_\Lambda$ in order to match the value of $D_{\rm A}(z_*)$ calculated via Eq.~\eqref{eq:sound_horizon} with the one inferred as described above. Having determined the physical density of each component, we can finally reconstruct $H(z)$ via Eq.~\eqref{eq:Hubble_evolution}. For $z=0$, this provides us with the value of $H_0$.

Note that $\theta_s$ is very tightly constrained by observations. In particular, it is measured to 0.03\% accuracy by \textit{Planck} data, which gives $100\theta_s = 1.04109 \pm 0.00030$  using a combination of TT,TE,EE$+$lowE data~\cite{Planck:2018vyg}. Other \ac{cmb} probes measure $\theta_s$ to be consistent with Planck, such as $10^4 \theta_s= 104.056 \pm 0.031$ from \ac{act}-DR6 (TTTEEE+Planck-Sroll2 EE~\cite{ACT:2025fju}) and $100\theta_s = 1.04016 \pm 0.00067$ (\ac{spt}+\ac{wmap}~\cite{SPT-3G:2022hvq}).

\paragraph{CMB in combination with other cosmological probes} \label{sec_cmb:other_probes}

The same acoustic oscillations that we observe in the \ac{cmb} are also left imprinted in the galaxy power spectra in the form of \ac{bao}. 
The characteristic scale of \ac{bao} is the sound horizon at the drag epoch, $z_d \simeq 1059$~\cite{Planck:2018vyg}, i.e., the time when baryons were released from the drag of \ac{cmb} photons. This characteristic scale provides us with a cosmological standard ruler that serves as an independent way to measure the expansion rate of the Universe and improve our bounds on the cosmological parameters.
Notably, the inclusion of \ac{bao} data allows us to better constrain the spatial curvature of the Universe and break the aforementioned geometric degeneracy~\cite{Efstathiou:1998xx,Eisenstein:2006nj}. We refer to Sec.~\ref{sec:BAO} for a more detailed discussion of \ac{bao} measurements of $H_0$. In what follows, we report an incomplete list of the most recent estimates of $H_0$ obtained with the combination of \ac{cmb} and \ac{lss} data, to give a sample of the complementarity and constraining power of these combinations.

A recent analysis of \ac{desi} \ac{bao} data in combination with Planck PR4 primary anisotropies (using small, commander and CamSpec likelihoods), Planck and \ac{act}-DR6 \ac{cmb} lensing, leads to $H_0 = 68.17 \pm 0.28$\kms~\cite{DESI:2025zgx}. A combined analysis of \textit{Planck} CMB lensing, \ac{boss} DR12 galaxy power spectra, and the PANTHEON+supernova constraints yields $H_{0} = 64.8^{+2.2}_{-2.5}$\kms~\cite{Philcox:2022sgj}, by imposing a \ac{bbn} prior on physical baryon density and assuming \lcdm\ cosmology. A multi-tracer full-shape analysis of luminous red galaxy (LRG) and emission line galaxy (ELG) samples from the \ac{eboss} DR16 measures $H_{0} = 70.0\pm{2.3}$\kms in the \lcdm\ framework when combining the \ac{bbn} prior and fixing the spectral tilt $n_{s}$ to \textit{Planck} value~\cite{Zhao:2023ebp}. Another analysis of the \ac{boss} DR12 data by fixing $n_s$ and the baryon-to-\ac{dm} ratio $\Omega_{b}/\Omega_{\rm DM}$ to \textit{Planck} value gives $H_{0} = 68.5\pm{2.2}$\kms~\cite{DAmico:2019fhj}. Adopting a forward modeling approach for the \ac{boss} DR12 bispectrum monopole, with \ac{bbn} prior on $\Omega_{b}h^{2}$,~\cite{SimBIG:2023nol} found $H_{0} = 67.6^{+2.2}_{-1.8}$\kms. Another study of the \ac{boss} galaxy power spectrum with bispectrum monopole and quadrupole estimated $H_{0} = 69.2\pm{1.1}$\kms~\cite{DAmico:2022osl} within the \lcdm\ model, using \ac{bbn} and fixing spectral tilt to \textit{Planck} value. 

It is crucial to highlight that these results assume a \lcdm\ cosmology. A combination of \textit{Planck} TT,TE,EE+lowE, \textit{Planck} lensing, \ac{bao} measurements from \ac{boss} and \ac{eboss}, and Pantheon+ \ac{sn1} data, assuming a non-flat \lcdm\ model yielded $H_{0} = 68.24\pm{0.54}$\kms with $\Omega_{k} = 0.0004\pm 0.0017$~\cite{deCruzPerez:2024shj}. A similar constraint of $H_{0} = 68.53\pm{0.56}$\kms was found when treating $A_{\rm lens}$ as a free parameter (for more details, see the discussion on the $A_{\rm lens}$ anomaly in Sec.~\ref{sec:systematics}). Extending the analysis to non-flat $w$CDM model resulted in $H_{0} = 67.95\pm{0.66}$\kms and $\Omega_{k} = 0.0016\pm 0.0019$ (see Ref.~\cite{deCruzPerez:2024shj} and references therein). To conclude this list of constraints, the take-home message is that the combination of \ac{cmb} and \ac{lss} measurements provide estimates of $H_0$ which are in agreement with the (lower) value inferred from Planck alone. This is true also in case of combinations that do not include Planck anisotropies - or do not include Planck data at all - in their fit.

\paragraph{The role of potential systematic effects in CMB-based estimates of $H_0$}
\label{sec:systematics}
Unresolved systematic effects (hereafter systematics) in \ac{cmb} measurements could potentially bias the cosmological constraints. This possibility has motivated a collective effort to examine potential sources of systematics within the \ac{cmb} dataset. 

Features in the \textit{Planck} spectra (especially temperature and at small scales) have been also interpreted as hints to unsolved systematics. 
In this context, the $A_{\rm lens}$ anomaly in Planck data is noteworthy. First introduced in Ref.~\cite{Calabrese:2008rt}, $A_{\rm lens}$ is a phenomenological (somehow ``unphysical'') parameter used to rescale the effects of gravitational lensing on the \ac{cmb} angular power spectra. Therefore, the expected value is $A_{\rm lens}=1$. Interestingly, \textit{Planck} data (primary anisotropies) show a preference for $A_{\rm lens}>1$ at more than $2\sigma$, increasing to more than $3\sigma$ when combined with \ac{bao} data~\cite{Planck:2018vyg}. Theoretical explanations for such values are challenging, as it would require either a closed universe, posing conflicts with other datasets and simple inflationary models, or more exotic solutions such as modifications to General Relativity. Furthermore, the lensing anomaly does not appear in \textit{Planck} trispectrum data (i.e., in the lensing power spectrum), which offers an independent and complementary measurement of \ac{cmb} lensing. If not indicative of new physics or not interpreted as a statistical fluke, the $A_{\rm lens}$ anomaly might arise from an undetected systematics error in \textit{Planck} data and this systematic could potentially bias the measurement of the $H_0$ parameter. 
However, when the effect of $A_{\rm lens}$ is marginalized over in the analyses, the \textit{Planck} and \textit{Planck} + \ac{bao} constraints on $H_0$ shift only slightly towards higher values:
$H_0 =68.28\pm0.72$\kms and  $H_0 =68.23\pm0.49$\kms at $68\%$ CL, respectively~\cite{Planck:2018vyg}, not enough to provide a satisfying solution to the $H_0$ tension. If the $A_{\rm lens}$ anomaly is due to systematics which may impact the final $H_0$ constraints, it raises questions whether the same systematics can be fully captured by $A_{\rm lens}$ alone or if further modeling is needed. Moreover, recent analyses by \ac{act} and \ac{spt}, which also find a lower value of $H_0$ than local measurements, find no deviation from the standard lensing effect predicted by the \lcdm\ model~\cite{SPT-3G:2021eoc,ACT:2020gnv,ACT:2025fju}, supporting the idea that the $A_{\rm lens}$ anomaly might not be the right solution to the tension. 

The consistency of cosmological parameters estimated from different multipole ranges of Planck data has also been investigated extensively in the past years \cite{Planck:2015bpv, Couchot:2015eea, Addison:2015wyg}, sometimes claiming evidence for internal inconsistencies which may impact the $H_0$ estimates. However, it has been showed \cite{Planck:2016tof} that the shifts in cosmological parameters recovered from different ranges - mostly due to the combined effects in temperature of non-lensing-related residuals at high multipoles and power deficit at large scales ($\ell<30$) - are not very significant and consistent with sample variance.

In the context of potential systematics in \ac{cmb} data, it is also worth mentioning that the mild discrepancy seen between \ac{act}-DR4 and both \ac{wmap} and Planck has been solved with the final data release of \ac{act}. Indeed, \ac{act}-DR6 finds very good agreement with both Planck Legacy data (PR3) and Planck PR4 (NPIPE) already at the power spectrum level~\cite{ACT:2025fju}. In terms of cosmological parameters, the difference between the best fit values in \lcdm obtained with \ac{act}-DR6 and Planck is estimated to be within $1.6\sigma$ (for Planck Legacy) and $2.5\sigma$ (for Planck PR4), see~\cite{ACT:2025fju,ACT:2025tim} for details.

As far as \ac{spt} results are concerned, good agreement has been found~\cite{SPT-3G:2022hvq} between \ac{spt} and Planck, both at the power spectrum level (in TT over angular scales that are signal-dominated in both experiments) and at the cosmological parameter level (assuming \lcdm). As noted by the \ac{spt} collaboration, this excellent agreement between two effectively independent experiments (given the negligible overlap between observed sky fractions and different sensitivity to angular scales) is a strong argument in favor of the robustness of the results and of the consistency of \lcdm\ across different scales and spectra. A good agreement has been found also between \ac{spt} and \ac{wmap}. The agreement between \ac{spt} and \ac{act} is acceptable~\cite{SPT-3G:2022hvq}, with the largest shift in recovered \lcdm\ parameters being on $\theta$ ($2\sigma$ larger in \ac{act} than in \ac{spt}). However, a multi-dimensional statistical test results in no significant deviations to be noted and agree with an explanation in terms of statistical fluctuations. 

Another crucial aspect of the \ac{cmb} data analysis to be explored in the context of $H_0$ estimation is the possible impact of instrumental systematic effects. In \ac{cmb} data analysis, particular care is devoted to the quantification of residual systematics, which are corrected for through the pipeline and/or propagated in the likelihood analysis either by incorporating it in the noise model or modeling specific residuals in the data vector. In this context, the use of end-to-end simulations is key. Among the different sources of systematics, beam characterization and instrument calibration (especially in polarization) are particularly worrisome. As an example, the determination of polarization efficiencies is considered one of the main limitations of the Planck data products~\cite{Planck:2019nip}. Nevertheless, the use of a large suite of null tests and cross-checks on several data splits, and the comparison with realistic end-to-end simulations allow us to show, for all the data products released by the main \ac{cmb} experiments, that possible biases of cosmological parameters - included $H_0$ - induced by residual systematics are small-to-negligible~\cite{Planck:2018bsf,Planck:2018lkk,Planck:2019nip,ACT:2020frw,ACT:2020gnv,SPT-3G:2022hvq,SPT-3G:2024atg}. Similar analyses on dedicated simulations for next generation \ac{cmb} experiments have recently shown that, even in the presence of unaccounted-for systematics (i.e., effects which have not been corrected for in the analysis pipeline) due to, e.g., beam chromaticity, bandpass mismatch, polarization angle miscalibration and incorrect calibration in polarization, the bias on $H_0$ is at the level of a small fraction of $\sigma$~\cite{Giardiello:2024rew,Giardiello:2024uzz}. Therefore, we conclude that instrumental systematic effects cannot be the (only) source of discrepancy between the \ac{cmb}-based estimate of $H_0$ and local measurements.

Finally, the interplay between instrumental systematics and foreground contamination can have non-linear effects on the final results, potentially biasing cosmological parameter estimates, including the Hubble constant. The presence of galactic and extragalactic foregrounds, such as dust emission and radio sources, complicates the separation of the \ac{cmb} signal from other astrophysical contributions. This complexity underscores the need for precise modeling and mitigation strategies to ensure accurate parameter estimation, but it is unlikely the source of the $H_0$ tension.

\paragraph{$S_8$ tension}
\label{sec:CMB_S8}

The amplitude of the \ac{cmb} power spectrum, especially \ac{cmb} lensing, places stringent constraints on the matter density and hence on $\sigma_8$. Here, we briefly address the tension related to the amplitude of matter clustering in the late Universe, described by the parameter $S_8 \equiv \sigma_8 (\Omega_{\rm m,0}/0.3)^{0.5}$. For a more detailed discussion, refer to Sec.~\ref{sec:S8tension_2.2}.

\ac{cmb}-based estimates of $S_8$ are consistently larger than those derived from galaxy-based measurements~\cite{Planck:2018vyg,Rosenberg:2022sdy,Tristram:2023haj, ACT:2020gnv,ACT:2023kun,SPT-3G:2022hvq,SPT-3G:2024atg}. Observations of weak gravitational lensing at low redshifts ($z \lesssim 0.5 - 1$) indeed suggest weaker matter clustering than predicted by the \lcdm\ model with parameters inferred from \ac{cmb} data. Simply put, the distribution of galaxies and matter in the late Universe appears smoother than expected based on the evolution of the fluctuations seen in the \ac{cmb}. The \ac{cmb} estimate of $S_8$ is model-dependent and can shift with other parameters that affect the growth and amplitude of matter fluctuations. Notably, the optical depth to reionization ($\tau$), which carries the largest uncertainty among the \lcdm\ parameters, and the sum of neutrino masses ($\sum m_\nu$), which is a derived parameter and is usually fixed. Uncertainties in these parameters affect the derived $S_8$ constraints. Additionally, \textit{Planck}’s excess lensing anomaly, as discussed in Ref.~\cite{DiValentino:2018gcu}, can mimic a larger $S_8$. However, the \ac{cmb}-\ac{lss} tension persists even without \textit{Planck}'s lensing excess, as shown by the latest \ac{act}+\ac{wmap} analysis~\cite{ACT:2025fju,ACT:2025tim}, which reports a high value of $S_8 = 0.857 \pm 0.020$ without an anomalous lensing amplitude, and \ac{spt3g} analysis~\cite{SPT-3G:2024atg}, which is independent of temperature data and reports $S_8 = 0.850\pm0.017$.

On the theoretical side, efforts to resolve the $S_8$ tension involve modifying the matter or gravity sectors of the \lcdm\ model, leading to a range of alternative cosmological scenarios (see e.g., Sec.~\ref{sec:ETP_4.1} and Sec.~\ref{sec:LTP_4.2} for more details). Although some models alleviate the $S_8$ tension, most fail to provide consistent solutions when all cosmological probes are considered. In particular, due to the correlation between $H_0$ and $S_8$, models that resolve the $S_8$ tension often worsen the $H_0$ tension, and vice versa~\cite{Planck:2015lwi, Planck:2015koh, SPT:2016izt}. For example, late-time dark sector transitions, which prefer a higher $H_0$ value, often result in lower $\Omega_{\rm m,0}$ to preserve the measured value of $\Omega_{\rm m,0} h^2$, leading to changes in structure growth and \ac{cmb} anisotropies and typically yielding a higher $\sigma_8$ than \lcdm\ due to extended matter domination. Similarly, \ac{ede} models that address the $H_0$ tension tend to increase $\sigma_8$ because they require a higher initial curvature perturbation amplitude to counterbalance the unclustered component. It is therefore crucial to perform joint analyses fitting a full array of multiple datasets, without fixing any of the parameters of the model. Simultaneously any solution to the $S_8$ tension must be consistent with the growth history (typically studied through the parameter $f \sigma_8(z)$), probed by e.g., \ac{bao}, galaxy power spectrum, and void measurements. 
\bigskip
\subsubsection{Baryonic acoustic oscillations \label{sec:BAO}}

\noindent \textbf{Coordinator:} Florian Beutler\\
\noindent \textbf{Contributors:} Armando Bernui, David Benisty, Denitsa Staicova, Felipe Avila, Ignacio Sevilla-Noarbe, \"{O}zg\"{u}r Akarsu, Maret Einasto, Mustapha Ishak, Nicola Deiosso, Rafael C. Nunes, Ruchika, Samuel Brieden, and Sveva Castello
\\

\noindent \ac{bao} represent a characteristic scale in the distribution of galaxies, or more generally of any tracer of the matter density field~\cite{Peebles:1970ag, Sunyaev:1970bma}. This scale, known as the sound horizon $r_d$, originated from sound waves traveling through the early ($z \gtrsim 1060$) Universe before baryons and photons decoupled (see Refs.~\cite{Eisenstein:1997ik,Bassett:2009mm, Weinberg:2013agg} for reviews of the subject). 
We can employ this scale as a standard ruler, which allows us to measure the expansion history of the Universe. 
In practice, the \ac{bao} features can be extracted from the two-point correlation function of galaxies, where it appears as a peak~\cite{SDSS:2005xqv, DESI:2024lzq}, or from the Fourier-space equivalent, the power spectrum, where it is manifested as a series of oscillations~\cite{2dFGRS:2005yhx, SDSS:2009ocz, BOSS:2016hvq}. In the following, we will review the different methods to extract the \ac{bao} scale from the galaxy distribution. We will then discuss the most recent measurements from \ac{desi}.

\paragraph{Methods}\label{sec:methods}

The 3D position of a galaxy can be determined by measuring its location on the sky (in right ascension and declination) and its redshift. For this reason, measuring the \ac{bao} scale in the perpendicular direction to the line-of-sight and along the line-of-sight provides constraints on slightly different quantities. The angular component constrains the comoving angular diameter distance $D_\textrm{M}/r_d = 1/\Delta \theta$, where $\Delta \theta$ is the angular separation of the pair of galaxies, while the line-of-sight component constrains the comoving Hubble distance $D_\textrm{H}/r_d = 1/\Delta z$, where $\Delta z$ is the redshift separation of the galaxy pair. In the \ac{flrw} metric, the comoving angular diameter distance $D_\textrm{M}$, the related $D_{\rm A}$ and the Hubble distance $D_\textrm{H}$ are given by
\begin{equation}
    D_\textrm{M} =\frac{c}{ H_0 \sqrt{|\Omega_{K}|}  } \textrm{sinn}\left[|\Omega_{K}|^{1/2}\int_0^z \frac {dz'} {E(z')}\right]\,, \quad D_\textrm{A}=\frac{D_{\rm M}}{1+z}\,,\quad D_\textrm{H} = \frac{c}{H_0 E(z)} \,, 
    \label{eq:DA}
\end{equation}
where $\textrm{sinn}(x) \equiv \textrm{sin}(x)$, $x$, $\textrm{sinh}(x)$ for $\Omega_{K}<0$, $\Omega_{K}=0$, $\Omega_{K}>0$ respectively, and $E(z)$ is the normalized Hubble function. Rather than separating $D_\textrm{M}$ and $D_\textrm{H}$, often measurements are reported as a combination of the two given by 
$D_\textrm{V} \equiv \left[z D_\textrm{M}(z)^2 D_\textrm{H}(z)\right]^{1/3}$. \ac{bao} constraints from photometric galaxy surveys usually have significant redshift uncertainties that only allow for constraints on $D_\textrm{M}$. One important point to note from the equations above is that \ac{bao} measurements always constrain distances relative to the sound horizon scale. This introduces degeneracies between cosmological parameters and the sound horizon scale $r_d$. In \lcdm\ for example, \ac{bao} can only constrain $H_0r_d$, and breaking this degeneracy is crucial to obtain measurements of $H_0$ and provide key information in light of the Hubble tension. We will discuss this further in Sec.~\ref{sec:Hubble}.

The density field sources a velocity field that will displace galaxies away from their initial position, effectively leading to a redshift-dependent smoothing of the localized \ac{bao} feature. This effect can reduce the signal-to-noise by up to a factor of two at low redshift (e.g., see Ref~\cite{BOSS:2013rlg}) and also results in a sub-percent level systematic biasing of the \ac{bao} scale~\cite{Mehta:2011xf,Padmanabhan:2012hf}. Given that the velocity field is sourced by the density, one can estimate the displacement of galaxies assuming standard gravity and correct for this effect~\cite{Eisenstein:2006nk,Padmanabhan:2012hf}. Such techniques are known as density field reconstruction and are routinely applied to most spectroscopic \ac{bao} measurements. 

The standard \ac{bao} analysis relies on converting the angular position and the redshift into Cartesian coordinates, from which the 3D clustering statistics can be calculated. This step requires adopting a fiducial cosmology, which typically is taken to be a flat \lcdm\ model with parameters based on the \ac{cmb}. The \ac{bao} analysis accounts for these assumptions by including geometric scaling parameters and so far studies with alternative models (e.g., \ac{ede} or \ac{mg}) have not found that the fiducial cosmology assumptions inflict any significant bias in the standard \ac{bao} analysis~\cite{Bernal:2020vbb,Sanz-Wuhl:2024uvi,Pan:2023zgb,BOSS:2016sne}. Through density field reconstruction, the standard analysis also makes assumptions about the connection between the measured galaxy density field, the matter density field, and the large-scale velocity field. However, just like with the fiducial cosmology assumptions, the impact on the measurements has been shown to be negligible (e.g., see Refs.~\cite{Carter:2019ulk,Chen:2024tfp}). Assumptions about the number of relativistic particles, $N_{\rm eff}$ at high redshift have been shown to impact the \ac{bao} phase and could bias \ac{bao} measurements~\cite{Baumann:2017lmt, Baumann:2019keh}. The standard analysis usually fixes this parameter to the standard model value. While a Planck prior on $N_{\rm eff}$ (within \lcdm) reduces any potential bias to well below the uncertainties of current measurements, one still should be aware of these prior assumptions.

\begin{figure}
    \centering
    \includegraphics[width=\textwidth]{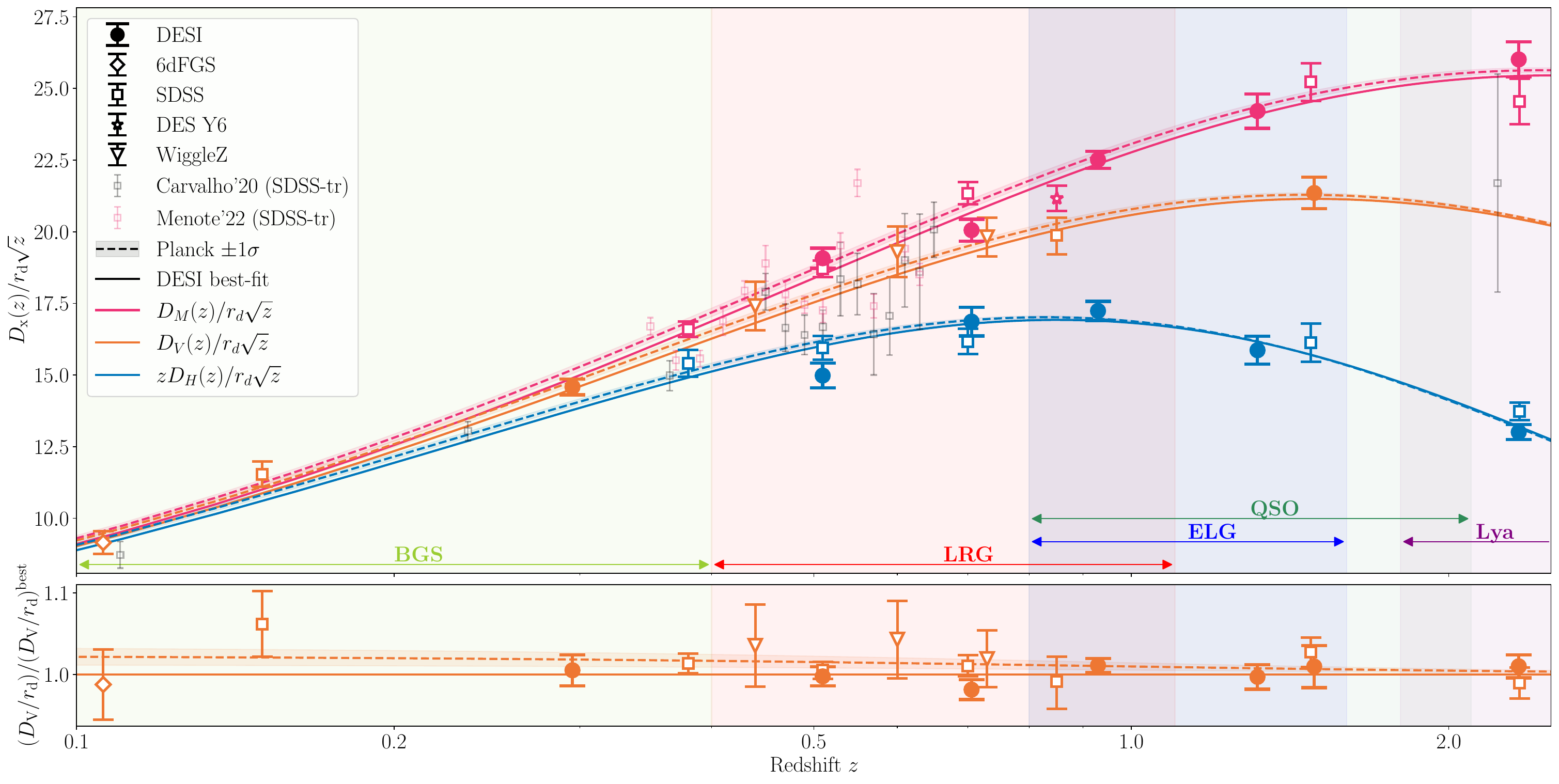} 
    \caption{\ac{desi} \ac{bao} measurements (filled circles) compared to prior \ac{bao} measurements (empty symbols) from spectroscopic (6dFGS, WiggleZ, SDSS) and photometric (\ac{des} Y6) surveys, where SDSS includes the MGS, \ac{boss} DR12 and \ac{eboss} DR16 galaxy and Lyman-$\alpha$ Samples.
    Additionally, we show the transverse \ac{bao} measured from SDSS catalogs (SDSS-tr) in thin redshift shells, neglecting the longitudinal modes. Note that these measurements are not official SDSS products, but were instead obtained independently in Refs.~\cite{Carvalho:2015ica,Alcaniz:2016ryy,Carvalho:2017tuu,
    deCarvalho:2017xye,deCarvalho:2021azj} (small grey squares) and Ref.\cite{Menote:2021jaq} (small pink squares).
    Horizontal arrows and colored regions indicate the redshift range spanned by each \ac{desi} tracer type. Solid lines show \ac{bao} distances (scaled by $\sqrt{z}$ for improved visibility) as a function of redshift $z$ for the standard flat \lcdm\ \ac{desi} best-fit, and dashed lines indicate the Planck best-fit with shaded region denoting the $\pm 1\sigma$ regime. \textbf{Upper panel:} Comparison between transverse (pink), radial (blue), and isotropic (orange) \ac{bao} measurements, where the latter is only displayed for data that does not exhibit an anisotropic measurement due to low signal-to-noise. \textbf{Lower panel:} Isotropic \ac{bao} distance residuals where the anisotropic \ac{bao} measurements from the upper panel (apart from \ac{des}-Y6 and SDSS-tr, which do not measure radial \ac{bao}) have been combined into an isotropic measurement (not shown in the upper panel to avoid redundancy).}
    \label{fig:1}
\end{figure}

Given the priors implicit in the standard \ac{bao} analysis, alternative analysis methods have been developed that can (at least partly) avoid the assumption of a fiducial cosmology~\cite{Sanchez:2010zg}. This technique often runs under the name \textit{transverse \ac{bao}}~\cite{Menote:2021jaq,Carvalho:2017tuu}, which should not be confused with the many angular \ac{bao} measurements in the literature which follow the standard analysis technique described above (see e.g., \ac{des} Y6 in Fig.~\ref{fig:1}). The basic idea of such methods is to bin the data into redshift bins and measure the angular \ac{bao} signature in those redshift bins through two-point angular-clustering statistics, avoiding the need to convert redshifts into distances. However, to account for the redshift evolution within the redshift bin one still needs to use a fiducial cosmology, and density field reconstruction is generally not applied in such cases. For this reason, such methods usually have much larger measurement uncertainties (see e.g., see Refs.~\cite{Camarena:2019rmj, Nunes:2020hzy,Nunes:2020uex,Arjona:2021hmg,Bernui:2023byc,Gomez-Valent:2023uof,Benetti:2017juy}).

\paragraph{BAO measurements}

The \ac{bao} feature has been measured both with photometric and spectroscopic surveys. Photometric surveys allow for a direct identification of a huge number of galaxies, but generally only provide poor photometric redshift information. Even for measurements of the angular \ac{bao} scale, the redshift information is necessary to split the sample into redshift bins, to avoid the smearing of the \ac{bao} signal due to its redshift evolution. Although the photometric redshift quality can be improved with an adequate subselection, this usually also reduces the number of galaxies in the sample~\cite{DES:2024wym}. To date, the best photometric \ac{bao} measurement comes from \ac{des} year-6 analysis using about $16\,000\,000$ galaxies and yielding a $2.1\%$ constraint on $D_\textrm{M}/r_d$~\cite{DES:2024pwq}. Observational systematics for such measurements are dominated by uncertainties in the redshift distribution, which are nevertheless still below the statistical error budget.

Compared to photometric surveys, spectroscopic surveys have the advantage of providing precise redshift measurements and detailed spectral information. This is achieved through the identification of specific spectral lines or features, such as the H$\alpha$, H$\beta$, OII lines or the $4000\,\mathring{A}$ break for galaxies, and broad emission lines for \ac{qso}s. The Lyman-$\alpha$ (Ly$\alpha$) forest represents a series of absorption features in the spectra of distant \ac{qso}s that can be used to map the distribution of intergalactic hydrogen gas and provides additional \ac{bao} measurements at higher redshift~\cite{McDonald:2006qs}.

The first convincing detections of the \ac{bao} signal in the distribution of galaxies came from the 2-degree Field Galaxy Redshift Survey (2dFGRS) and the Sloan Digital Sky Survey (SDSS) in the early and mid-2000s~\cite{2dFGRS:2001csf,2dFGRS:2005yhx,SDSS:2005xqv}, while the first detection in the Ly$\alpha$ forest was made by the \ac{boss}~\cite{BOSS:2013ola}. There have been many more subsequent detections in other galaxy surveys~\cite{Blake:2011en,Beutler:2011hx,BOSS:2013kxl,BOSS:2013igd,BOSS:2013rlg,BOSS:2016hvq,BOSS:2016apd, eBOSS:2020uxp, eBOSS:2020fvk}. The best constraints to date come from \ac{desi}, which measured the \ac{bao} signal in 8 independent redshift bins with a detection significance ranging from $3.3$ to $9.1\sigma$~\cite{DESI:2024uvr,DESI:2024lzq}. A comparison between these \ac{desi} measurements and previous \ac{bao} measurements is shown in Fig.~\ref{fig:1}.

Observational and instrumental systematics, such as fiber collisions~\cite{Bianchi:2018rhn}, distortions due to peculiar velocities (\ac{rsd}), non-linear matter and galaxy clustering, atmospheric dispersion, and spectrograph calibration are critical factors in any analysis based on spectroscopic surveys datasets. Since the \ac{bao} signal is located on very large scales ($r_d \sim 150\,$Mpc) and represents a distinguishable feature unlikely to be mimicked by any instrumental or physical process, \ac{bao} measurements have proven to be very robust (see e.g., see Ref.~\cite{Chen:2024tfp}). The galaxy clustering analysis of \ac{desi} identified the galaxy-halo connection (HOD) as the dominant systematic for their \ac{bao} measurement. However, all combined systematics still only added $5\%$ to the final statistical error budget even for the highest precision \ac{bao} measurement (see table 13 and 15 in Ref.~\cite{DESI:2024uvr}).

The \ac{bao} signal is most commonly measured as a mixture of radial and transversal modes. Alternatively, the \ac{bao} signature can be extracted in a thin redshift bin using the two-point angular-clustering statistics (where the angular separation of pairs instead of the comoving distances are computed, avoiding converting redshifts into distances using a fiducial cosmology), providing a measurement of the transverse \ac{bao}, $\Delta\theta_{\text{BAO}}(z)$. The finer the redshift bin, the purer this transverse contribution. If $r_d$ is known, $\Delta\theta_{\text{BAO}}(z)$ determines $D_{\rm M}(z)$, as explained in Sec.~\ref{sec:methods}. 
Fig.~\ref{fig:1} shows such transverse \ac{bao} measurements from SDSS data (labelled SDSS-tr) derived in Refs.~\cite{Menote:2021jaq} and~\cite{Carvalho:2015ica,Alcaniz:2016ryy,Carvalho:2017tuu,deCarvalho:2017xye,deCarvalho:2021azj}. However, for the remainder of this section, we will focus on the most recent \ac{desi} analysis.

\begin{figure}
 	\centering
    \includegraphics[width=0.95\textwidth]{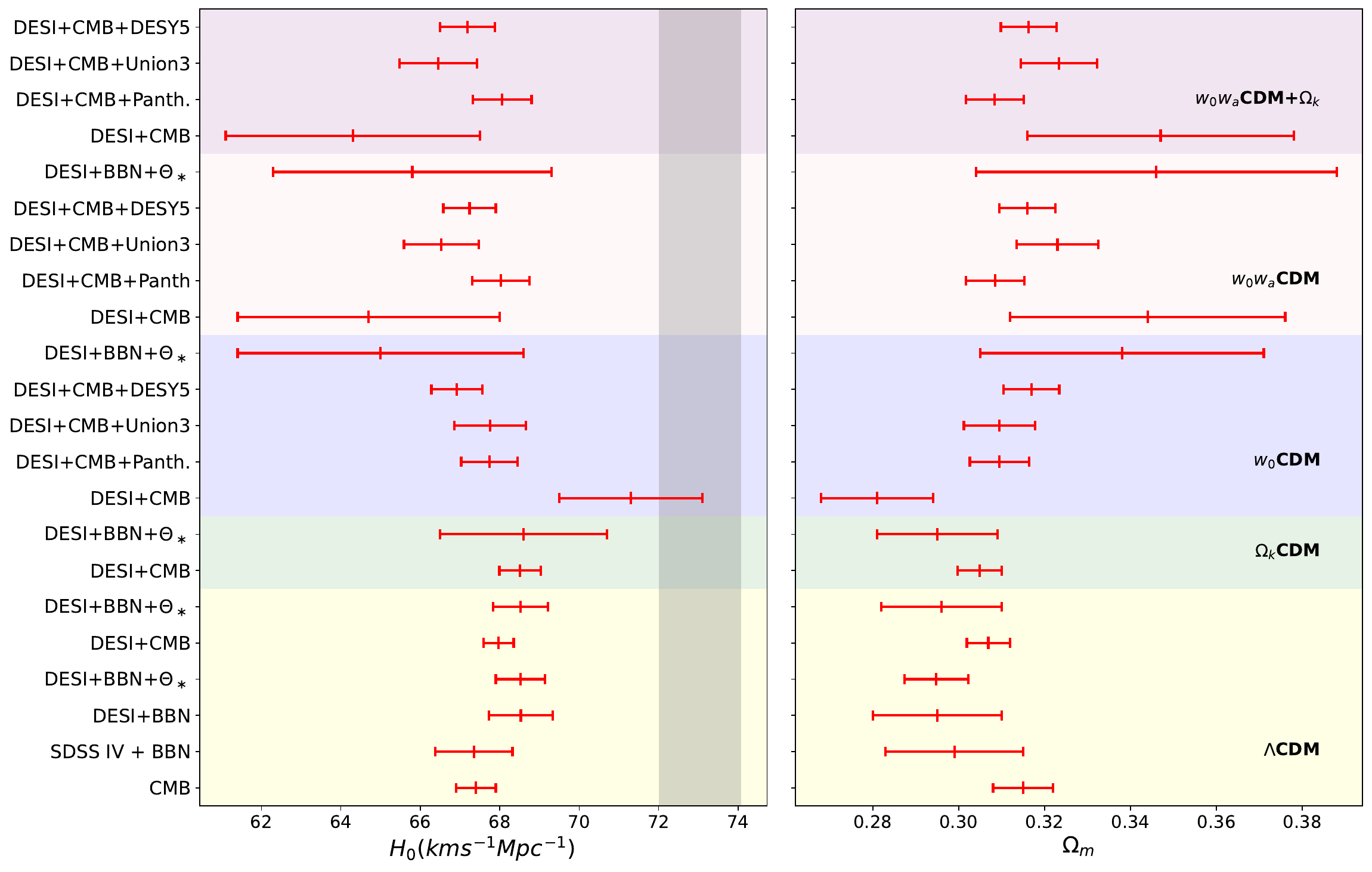}
    \caption{68\% credible-interval constraints on the Hubble constant $H_0$ (left panel) and the matter density $\Omega_{\rm m,0}$ (right panel), assuming different cosmological models and different combinations of datasets (on the plot, \ac{desi}, Planck 2018, Pantheon Plus, \ac{des}Y5, Union3) and the \ac{bbn} and $\theta_{\ast}$ priors. The gray band indicates the SH0ES measurement for $H_0$.} \label{fig:2}
\end{figure}

\paragraph{BAO and the $H_0$ tension}  
\label{sec:Hubble}

\begin{figure}
 	\centering
    \includegraphics[width=0.49\textwidth]{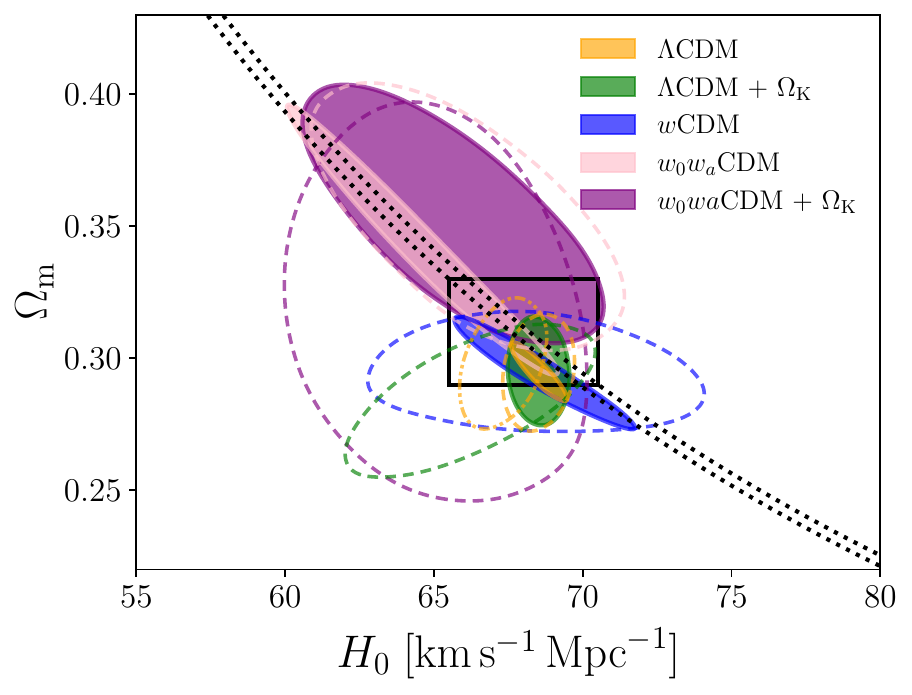}
    \includegraphics[width=0.49\textwidth]{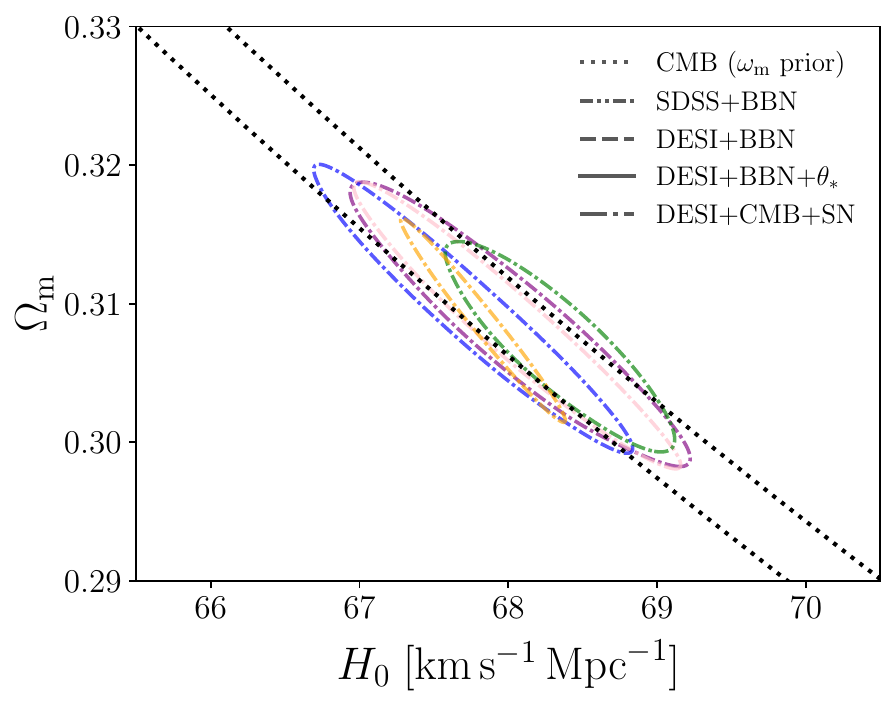}
    \caption{Illustrations of the degeneracy between the Hubble constant and the matter density parameter ($\Omega_{\rm m,0}$) constrained by different data combinations (indicated by line style) for different models (indicated by color) at 1$\sigma$. ``CMB'' refers to Planck + \ac{act} lensing data and ``SN'' to the Pantheon+ \ac{sn1} sample. The parameter region constrained by \ac{cmb} alone is shown as a prior on $\omega_\mathrm{m}$ (black dotted contours) that is independent of the late-time model extensions considered here. The most constraining data combination (\ac{desi}+\ac{cmb}+\ac{sn}) is shown on the right as a zoom into the black rectangle on the left for better visibility. For conciseness, the SDSS \ac{bao} + \ac{bbn} constraint (dash-dot-dotted) is shown for \lcdm\ (orange) only. Note that the legends shown in the left and right panels apply to both simultaneously. 
}
\label{fig:3}
\end{figure}

One of the challenges in the use of \ac{bao} measurements in cosmology is the degeneracy between the sound horizon $r_d$ and Hubble constant $H_0$. As mentioned in Sec.~\ref{sec:methods}, this arises because all measurable quantities in the \ac{bao} data depend on the product $c/H_0 r_d$. The only way to decouple them is by using additional information beyond the \ac{bao} 
scale~\cite{BOSS:2014hhw,Farren:2021grl,Philcox:2020xbv,Baxter:2020qlr,Brieden:2022heh, Philcox:2022sgj, Krolewski:2024jwj,Arendse:2019hev,Knox:2019rjx,Jiang:2025ylr} or imposing an external prior on $r_d$ from the \ac{cmb} or \ac{bbn}~\cite{Knox:2019rjx,eBOSS:2020yzd,DESI:2024mwx}. The 2024 \ac{desi} \ac{bao} results combined with a \ac{bbn} prior from Ref.~\cite{Schoneberg:2024ifp} yield a constraint of $H_0=68.53\pm0.80$\kms~\cite{DESI:2024uvr}. This measurement is independent of the \ac{cmb} and solely based on \ac{desi} \ac{bao} + \ac{bbn}. While this value of the Hubble constant is slightly higher than the one preferred by Planck~\cite{Planck:2018vyg} as well as the one reported in previous data from SDSS~\cite{eBOSS:2020yzd}, it is still in 3.4$\sigma$ tension with the $SH0ES$ result of Ref.~\cite{Riess:2020fzl}. Fig.~\ref{fig:2} shows the values for $H_0$ and $\Omega_{\rm m,0}$ inferred from the \ac{desi} data, along with the $SH0ES$ results on $H_0$ as a shaded band. One sees that only the $w_0$CDM model comes close to resolving the Hubble tension for the \ac{desi}+\ac{cmb} dataset to the price of much lower matter density. A further look at the \ac{desi} results is presented in Fig.~\ref{fig:3}, where we show the $1 \sigma$ posteriors for some of the models presented in Fig.~\ref{fig:2}. 
In this figure, the degeneracy between $H_0$ and $\Omega_{\rm m,0}$ manifests as the covariance of the inferred data, which is particularly visible on the right panel of Fig.~\ref{fig:3} zooming in on the \ac{desi}+\ac{bbn} models. This leads to the known issue that, across different models, increasing $H_0$ generally decreases $\Omega_{\rm m,0}$, and vice versa. On the plots, $r_d$ is missing since it is not a free parameter for all tested models.

\ac{bao} measurements rely on a model-dependent extrapolation to redshift zero when constraining the Hubble constant. However, combining \ac{bao} with \ac{sn1} datasets and the \ac{cmb} does constrain the redshift evolution and makes it difficult to find extensions to \lcdm\ that could resolve the Hubble tension~\cite{Percival:2007yw,BOSS:2014hhw,Giare:2024smz,Akarsu:2022typ,Akarsu:2024eoo,DES:2024ywx}.

Ref.~\cite{Smith:2022iax} has found that marginalizing over $r_d$ can lead to a degeneracy between $H_0$ and the primordial power spectrum and to a model-dependent $H_0$ measurement. A similar observation has been published in Ref.~\cite{Jedamzik:2020zmd} showing that models that only reduce $r_d$ cannot resolve the Hubble tension while at the same time remaining consistent with other cosmological datasets. A possible solution to this consists of working with the sound horizon as a free parameter (thus not making any assumption concerning the recombination physics) and using other quantities to break the degeneracy \cite{Pogosian:2020ded, Benisty:2020otr,Staicova:2022zuh}. In these works, it was shown that while this approach can produce interesting results when examining \lcdm\ alternatives, it does not solve the Hubble tension. Another approach consists in marginalizing over both $r_d$ and $H_0$ \cite{Staicova:2021ntm}, which integrates the quantity $c/H_0 r_d$ out of the likelihood ($\chi^2$) and thus removes entirely the dependence on these parameters from the theoretical predictions. The results obtained in this way show a slight preference for \ac{dde} models but they do not solve the Hubble tension. 

While early \ac{boss} and \ac{eboss} Ly-$\alpha$ \ac{bao} results showed a 1.5 to 2.5$\sigma$ tension with Planck-\lcdm~\cite{BOSS:2014hhw,eBOSS:2020yzd}, the most recent Ly-$\alpha$ analysis in \ac{desi}~\cite{DESI:2024lzq} shows no tension anymore. However, \ac{de} density reconstructions using \ac{desi} \ac{bao} still suggest the possibility of zero or negative \ac{de} densities for $z \gtrsim 1.5-2$ in some dataset combinations~\cite{DESI:2024aqx}. Additionally, a shift of the Ly-$\alpha$ \ac{bao} peak at the 2.2$\sigma$ and 3.5$\sigma$ levels in real and redshift space, respectively, was recently suggested in Ref.~\cite{Sinigaglia:2024kub}. The reasons behind these findings are yet to be understood, warranting further investigation. Since the comoving angular diameter distance to last scattering, $D_{\rm M}(z_*)$, is strictly constrained by \ac{cmb} data almost model-independently, a lower $H(z)$ for $z \gtrsim 1.5-2$ due to vanishing or negative \ac{de} density should be compensated by a higher $H(z)$ for $z \lesssim 1.5-2$, provided that the pre-recombination Universe is not modified~\cite{Akarsu:2021fol,Akarsu:2022typ,Akarsu:2023mfb,Adil:2023exv,Akarsu:2024eoo}. This results in a higher $H_0$ and correspondingly smaller $\Omega_{\rm m}$, which can reduce the $S_8$ value and address the most discussed tensions in the \lcdm\ model~\cite{Akarsu:2021fol,Akarsu:2022typ,Akarsu:2023mfb,Adil:2023exv,Akarsu:2024eoo}.

The most recent \ac{desi} data release, \ac{desi} DR2, improves on the DR1 findings, analyzing over 14 million galaxies and \ac{qso}s from three years of observations (surpassing SDSS in the effective volume of all tracers). Compared to DR1, the precision of \ac{bao} measurements has increased by approximately a factor of two, with uncertainties reduced to about 0.24\% for galaxy/\ac{qso} measurements and 0.65\% at high redshift from the Lyman-$\alpha$ forest (which is now twice the DR1 sample). The \ac{desi} DR2 combined with a \ac{bbn} prior yield $H_0= 68.51 \pm 0.58$\kms - 28\% more precise than DR1. The cosmological analysis shows stronger evidence for evolving \ac{de}, with statistical significance amounting to $2.8\sigma, 3.8\sigma$ and $4.2\sigma$ depending on which datasets are combined with \ac{desi} (Pantheon +, Union3 or \ac{des} Y5) and $3.1\sigma$ for \ac{desi}+\ac{cmb}. The results show a preference for a \ac{de} equation of state with $w_0 > -1$ and $w_a < 0$ and a possible phantom crossing in the past. While the degeneracy direction for the constraints in the $w_0-w_a$ plane approximately points towards the $\Lambda$CDM solution, the statistical evidence consistently challenges the standard $\Lambda$CDM across multiple analysis approaches and dataset combinations \cite{DESI:2025zgx}.

\paragraph{Outlook}

The \ac{desi} experiment is expected to publish its Y3 \ac{bao} analysis in 2025, while the final Y5 analysis should be made public in 2026/27~\cite{DESI:2016fyo}. The final dataset will be three times larger than the current Y1 catalog, and proposals for a \ac{desi} extension until 2029 and a second phase (\ac{desi}-II) well into the 2030s are under consideration. At the same time, the Euclid mission has also started, with the first cosmology analysis expected in 2025/26~\cite{Euclid:2024yrr}, and the Vera Rubin Observatory will see first light in 2025.

On a longer timescale, there have been several proposals for stage-V spectroscopic experiments with a start in the early to mid-2030s. These experiments increase the telescope aperture and the spectroscopic multiplexing capabilities with the aim of collecting hundreds of millions of galaxy redshifts (e.g., see Refs.~\cite{Schlegel:2022vrv,Schlegel:2019eqc,Ellis:2019gnt,WST:2024rai}). After the \ac{desi} mission is completed, the \ac{bao} signal in the low redshift Universe ($z\lesssim 1$) will have been measured up to the sample variance limit over almost the entire sky available for cosmological measurements ($\sim 15\,000$ to $18\,000\, \deg^2$), see figure 16 in Ref.~\cite{DESI:2023dwi}. Future experiments therefore focus on the high redshift regime, where significant gains are still possible.
\bigskip
\subsection{S8 Tension: measurements and systematics}\label{sec:S8tension_2.2}
\subsubsection{Weak lensing \label{sec:weak_lensing}}

\noindent \textbf{Coordinator:} Marika Asgari, Daniela Grandón, Cora Uhlemann\\
\noindent \textbf{Contributors:} David Sanchez Cid, Ignacio Sevilla, Judit Prat, Konrad Kuijken, Maciej Bilicki, Markus Michael Rau, Masahiro Takada, Nikolina \v{S}ar\v{c}evi\'c, and Shun-Sheng Li
\\

\noindent Cosmic shear is one of the key probes of the \ac{lss}. A cosmic shear analysis measures the weak gravitational lensing signal \cite{Jain:1996st, Kaiser:1991qi, Kaiser:1999ew, 1991ApJ...380....1M,Blandford:1991edc,VanWaerbeke:1998tu} imprinted in the observed galaxy shapes by \ac{lss} (see Ref.~\cite{Bartelmann:2016dvf, Kilbinger:2014cea, Prat:2025ucy} for a review and references therein). The strength of this signal depends directly on the total matter distribution between the source galaxies and the observer. Therefore, cosmic shear provides a precise mapping of the projected matter density field, capturing its statistical properties and enabling constraints on cosmological parameters. 

To perform a cosmic shear analysis, we need a galaxy catalog that comprises
two measurements: galaxy shapes (Sec.~\ref{subsection:WLshear}) and the redshift information (Sec.~\ref{subsection:WLphotoz}). 
The redshifts are used to divide source galaxies into tomographic bins.
The redshift distribution in each bin determines the strength of the lensing effect and serves as an input for the theoretical predictions of the cosmic shear signal, which require prescriptions for baryonic effects (Sec.~\ref{subsection:WLbaryons}) and the intrinsic alignment of galaxies (Sec.~~\ref{subsection:WLIA}).
In addition, the tomographic analysis allows us to extract information about the evolution of structures and to disentangle the astrophysical systematics from the cosmological information. 

The most commonly used estimator for the cosmic shear signal is the two-point statistic. 
In real space, the two-point shear correlation functions are measured as the average correlation between the shapes of galaxy pairs in parallel and orthogonal directions as a function of their angular separation \cite{Schneider:2002jd}.
Both of those correlations can be related to one angular convergence power spectrum, which is a line-of-sight projection of the matter power spectrum, and thus sensitive to nonlinear effects on small scales and low redshifts.
However, because the lensing field is a non-Gaussian random field, cosmic shear surveys have also implemented estimators that capture information beyond the traditional two-point statistics \cite{Euclid:2023uha}.
These higher-order or non-Gaussian statistics, such as the three-point shear correlation function \cite{Porth:2023dzx, Takada:2002qq,Heydenreich:2022lsa,Halder:2021itp}, the bispectrum \cite{Coulton:2018ebd,Kayo:2012nm}, the lensing convergence \ac{pdf} \cite{Castiblanco:2024xnd, Thiele:2023gqr,Giblin:2022ucn,Liu:2018dsw} or its moments \cite{DES:2021lsy}, and peak statistics \cite{Harnois-Deraps:2024ucb, Marques:2023bnr,Martinet:2015wza,Liu:2014fzc}, among others \cite{DES:2017eav, Cheng:2024kjv, Grandon:2024pek,Armijo:2024ujo}, have proven effective in enhancing the constraining power and promise to be valuable in Stage IV analysis \cite{Euclid:2023uha,Grandon:2024tud}. 

Cosmic shear analysis is most sensitive to the $S_8$ parameter and the strength of the intrinsic alignment signal of galaxies.
In fact, with the available data, $S_8$ is the only cosmological parameter that cosmic shear analysis can robustly constrain from two-point statistics.
A joint analysis of two-point function and higher-order statistics however can improve constraints in $\Omega_{\rm m,0}$, as well as self-calibrate systematics and break degeneracies between cosmological and nuisance parameters. These surveys consistently infer $S_8$ values that are $1-3\sigma$ lower than what is expected from the CMB anisotropies.

\begin{figure}[htbp]
\vskip\baselineskip\hspace{-14ex}
    \centering
    \includegraphics[scale = 0.70]{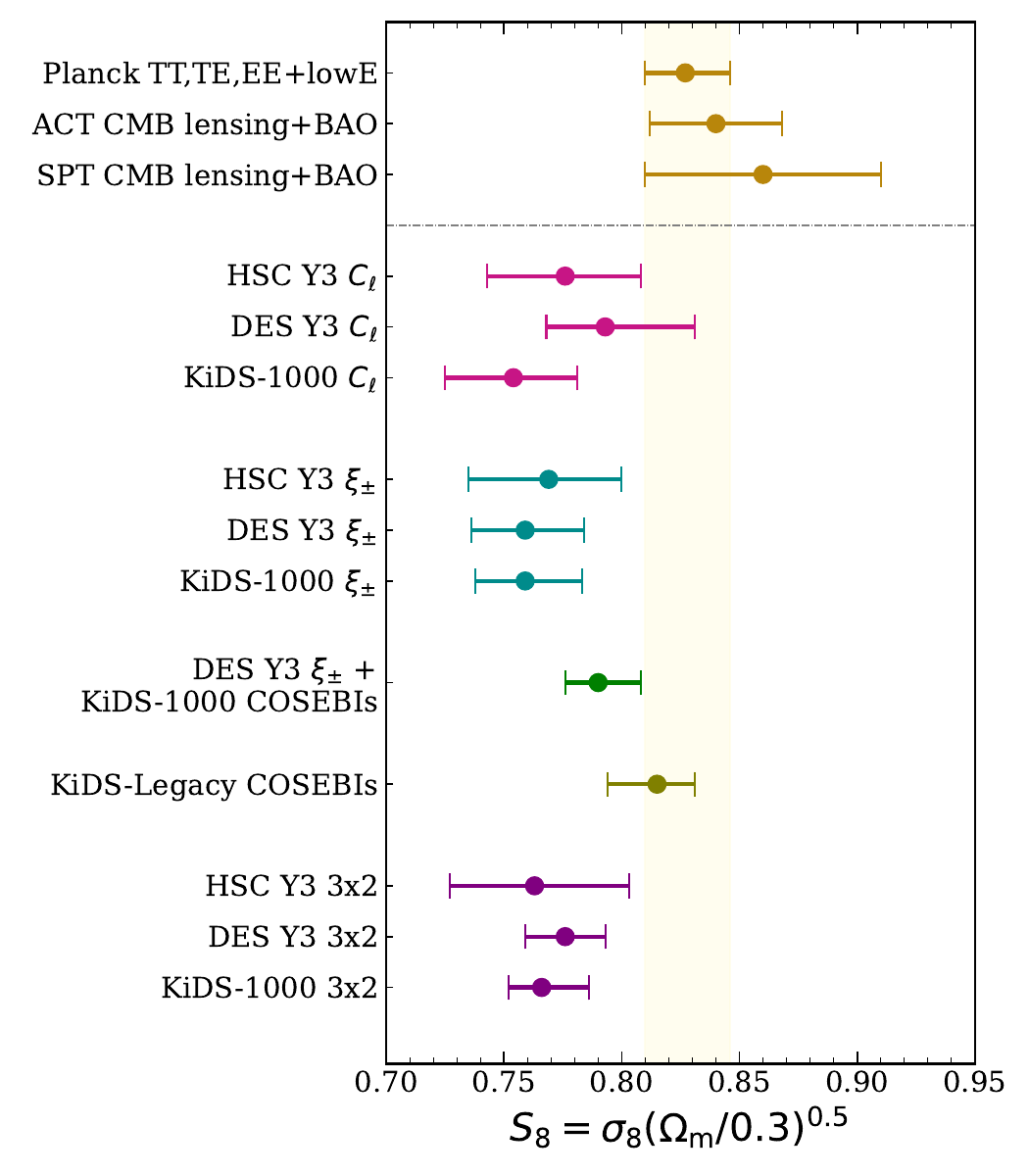}
    \caption[$S_8$ constraints from Stage III \ac{wl} surveys]{The $S_8$ constraints derived from the Stage III \ac{wl} surveys \ac{des} Y3 \cite{DES:2021vln,DES:2021bvc,DES:2021wwk,DES:2022qpf}, \ac{hsc} Y3 \cite{Dalal:2023olq,Li:2023tui,Sugiyama:2023fzm,Miyatake:2023njf}, \ac{kids}-1000 \cite{KiDS:2020suj,Heymans:2020gsg,KiDS:2021opn} and \ac{kids}-Legacy \cite{Wright:2025xka} from two-point shear statistics in harmonic space using $C_l$ or in angular space using $\xi_\pm$, COSEBIS and 3x2-point statistics including galaxy clustering. Also shown is the joint constraint from \ac{des} Y3+\ac{kids}-1000 \cite{Kilo-DegreeSurvey:2023gfr}. We include the CMB measurements from the Planck satellite \cite{Planck:2018lbu}, \ac{act} \cite{ACT:2023kun}, and \ac{spt} \cite{SPT:2019fqo} on top for comparison. The error bars denote 1-$\sigma$ uncertainties}
\label{fig:s8_2pt}
\end{figure}

The current generation of Stage III cosmic shear surveys—such as the \ac{kids} (\cite{Kuijken:2015vca,Kuijken:2019gsa}), \ac{des} (\cite{DES:2020ekd, DES:2018gui}), the Subaru \ac{hsc} Survey (\cite{Aihara:2017paw,HSC:2018mrq, Hamana:2019etx}) and the Dark Energy Camera All Data Everywhere (DECADE; \cite{Anbajagane:2025hlf}) —have conducted multiple cosmic shear analyses from at least two separate data releases. 
These surveys consistently infer $S_8$ values that are $2-3\sigma$ lower than what is expected from the \ac{cmb} anisotropies and \ac{cmb} lensing measurements by \textit{Planck} and \ac{act} \cite{Hildebrandt:2016iqg, Hildebrandt:2018yau, KiDS:2020suj,DES:2021vln,DES:2021bvc,DES:2017qwj, HSC:2018mrq, Hamana:2019etx,Dalal:2023olq}. A summary of these results is presented in Fig.~\ref{fig:s8_2pt}. This disagreement in the $S_8$ value between early- and late-time probes, often referred to as the $S_8$ tension, was first observed between the tomographic analysis of the Canada France Hawaii Telescope Lensing Survey (CFHTLenS) \cite{2013MNRAS.433.2545E} and the first \textit{Planck} cosmology results \cite{Planck:2013pxb}, at the level of $2.3\sigma$. More recently, the \ac{des} and \ac{kids} teams collaborated on a unified analysis pipeline which resulted in a joint constraint with a slightly higher value of $S_8$, translating to a 1.7$\sigma$ difference with \textit{Planck} \cite{Kilo-DegreeSurvey:2023gfr}. The most recent \ac{kids} results (\ac{kids}-Legacy) shows consistent results with Planck at the level of $1\sigma$. The most recent KiDS results (KiDS-Legacy) show consistent results with Planck at the level of $1\sigma$ \cite{Wright:2025xka}.
 
The ``3$\times$2pt'' analysis has emerged in the past decade as a mature cosmological probe that extends cosmic shear by incorporating two additional two-point correlations: galaxy-galaxy lensing (the cross-correlation of lens galaxy positions and source galaxy shapes) and photometric galaxy clustering (the autocorrelation of lens galaxy positions). The primary advantage of this combination is its ability to break the degeneracy between $\Omega_{\rm m,0}$ and $\sigma_8$, enabling robust constraints on $\Omega_{\rm m,0}$ in addition to $S_8$, as well as improving the self-calibration of the nuisance parameters, especially of the intrinsic alignment and redshift parameters. 
The three major current-generation surveys have conducted this kind of analysis, with their results presented in Fig.~\ref{fig:s8_2pt}. Generally, the $S_8$ values measured by  3$\times$2pt  have been found to  be $\sim1-2\sigma$ below the \textit{Planck} value predicted assuming \lcdm\ cosmology, thus with a slightly smaller tension with respect to the cosmic shear results. 

A summary of the $S_8$ constraints from low-redshift probes, including lensing surveys, is shown in Fig.~\ref{fig:Intro_S8_whisk}. In particular, results from peak count analyses—the most studied higher-order statistic—are presented for \ac{kids}-1000 \cite{Harnois-Deraps:2024ucb}, \ac{des} Y3 \cite{DES:2021epj}, and \ac{hsc} Y1 \cite{Marques:2023bnr}. The most precise constraints on $S_8$ come from the \ac{des} Y3 peak counts analysis, yielding $S_8 = 0.797^{+0.015}_{-0.013}$, which is in $1.5\sigma$ tension with Planck 2018 \cite{Planck:2018vyg}. The recently published \ac{kids}-Legacy COSEBIs results report $S_8 = 0.815^{+0.016}_{-0.021}$, showing less than $1\sigma$ disagreement with Planck \cite{Wright:2025xka}. On the other hand, a shear two-point correlation function analysis using only blue (star-forming) galaxies in \ac{des} Y1 yields $S_8 = 0.822^{+0.019}_{-0.020}$ \cite{DES:2024xvm}, indicating no tension with Planck \ac{cmb} results. These findings highlight the importance of mitigating astrophysical systematics—such as intrinsic alignments—to better understand the nature of the $s_8$ discrepancy.

The constraints on $S_8$ and other cosmological parameters can also be affected by the adopted statistical inference pipeline.
Most cosmic shear analyses use a Gaussian likelihood, which is defined using the data, the theoretical predictions, and the data covariance matrix (for exceptions employing simulation-based inference (SBI), e.g., see Ref.~\cite{DES:2023qwe,vonWietersheim-Kramsta:2024cks,Novaes:2024dyh}).
Studies have shown that an accurate estimation of the covariance matrix is crucial to avoid biases in cosmological constraints \cite{Reischke:2024fvk,Sellentin:2015waz}.
Additionally, properly accounting for likelihood asymmetry in two-point functions is essential for robust parameter inference \cite{Sellentin:2017koa}. 

In this section, we focus on weak gravitational lensing and the associated systematics that influence cosmic shear cosmological constraints.
These systematics can be broadly categorized as observational or theoretical.  
Observational systematics affect either the accuracy of shape measurements or the redshift distributions.  
On the theoretical side, achieving percent-level precision in predicting the matter power spectrum is required across a wide range of scales \cite{Chisari:2018prw}, determined by the survey depth and analysis choices. 
Since these scales extend into the nonlinear regime, an accurate nonlinear power spectrum prescription is necessary.
Additionally, two key astrophysical systematics—intrinsic galaxy alignments and baryonic feedback—remain among the least understood aspects of theoretical modeling for this signal.
Various modeling approaches have been developed to address these issues, which can generally be categorized into emulators based on $N$-body simulations \cite{Euclid:2020rfv,Arico:2020lhq}, and the more widely used halo model-based approaches, see Ref.~\cite{Asgari:2023mej} for a recent review, such as {\tt halofit} \cite{Smith:2002dz,Takahashi:2012em} and {\tt HMcode} \cite{Mead:2020vgs,Mead:2016zqy}. 
While emulators may offer greater accuracy within the specific parameter range they are trained on, they lack the flexibility of halo model approaches, which are capable of extrapolating beyond this range.
As a result, most cosmic shear analyses to date, including all of the flagship publications by current surveys have relied on the latter.
To limit the contributions from small scales, suitable linear combinations of the signal can achieve a nulling of the effective \ac{wl} kernel \cite{Bernardeau:2013rda}.
This can be used to perform scale cuts in physical rather than angular space and render predictions more robust to small-scale physics, e.g., see Ref.~\cite{Taylor:2018snp,Barthelemy:2019ciu}.
While this discussion focuses on current cosmic shear surveys, the next generation of cosmic shear observations (Stage IV surveys) is approaching \cite{EUCLID:2011zbd,Spergel:2015sza, LSST:2008ijt}.
As constraints tighten, more precise modeling and calibration will be required.
Consequently, the interplay of these systematic effects will become increasingly important in future analyses.

\paragraph{Observational effects: Shape measurement and calibration}\label{subsection:WLshear}

Extracting \ac{wl} signals from observed galaxy images is a non-trivial task. 
The cosmic shear signal is typically orders of magnitude smaller than the intrinsic galaxy ellipticities, necessitating robust statistical measurements of a large number of galaxies. Furthermore, systematic biases introduced by various instrumental and measurement effects further complicate the process.
While the statistical limitations are readily overcome with the ever-growing galaxy samples in ongoing and upcoming \ac{wl} surveys; addressing the systematic biases remains a challenging task.
This challenge is further exacerbated by increasingly stringent requirements driven by the growing statistical power of these surveys.
For upcoming Stage IV surveys, the overall residual shear biases must be reduced to the order of $10^{-3}$, compared to the current percent-level requirements of Stage III surveys, e.g., see Ref.~\cite{Massey:2012cz}. 
Achieving this stringent requirement demands continuous development of shear measurement and calibration techniques, with particular attention to subtle effects arising from object detection, blending, selection, and redshift estimation.

\subparagraph{Shape measurement and data-based calibration}

Given the critical importance of accurate shear measurement, the \ac{wl} community has dedicated significant early efforts to developing robust shear measurement algorithms.
A number of methods have now achieved an overall bias at the percent level.
Broadly, these shear measurement methods can be categorized into two main classes: moment-based methods, e.g., see Ref.~\cite{Kaiser:1994jb} and model-fitting methods, e.g., see Ref.~\cite{Miller:2007an,Bernstein:2013qsa}.
Moment-based methods estimate galaxy ellipticities from the second moments of observed galaxy images after correcting for the point-spread function (PSF) caused by instrumental and observational effects.
Conversely, model-fitting methods employ forward modeling with PSF-convolved parametric galaxy profiles. 
These techniques are susceptible to specific biases. 
Moment-based methods are affected by missing pixels, and contamination by light from detected and undetected neighboring galaxies, whereas model-fitting approaches suffer from ``model bias''. 
Additionally, both are affected by common biases such as ``noise bias'' and selection effects, see Ref.~\cite{Mandelbaum:2017jpr} for a recent review.  
These biases need to be calibrated and corrected for
shear measurement methods to meet the requirement of modern \ac{wl} surveys.

A recent significant advancement in shear measurement algorithms is the introduction of self-calibration or meta-calibration procedures that serve as a first-order correction to the initial shear measurement, e.g., see Ref.~\cite{FenechConti:2016oun,Huff:2017qxu}.
These techniques, which can rely on model profiles, observed galaxy images, or priors from deep observations, have demonstrated the ability to control the residual shear bias to a percent or sub-percent level, nearly meeting the requirements of current Stage III surveys, e.g., see Ref.~\cite{DES:2020lsz,Li:2021mvq,Li:2022pof}.
However, these data-based calibrations cannot correct biases introduced at the detection stage as required for the accuracy of Stage IV surveys, e.g., see Ref.~\cite{Hoekstra:2020zyl}.

\subparagraph{Detection biases and simulation-based calibration}

Detection biases arise primarily from blending, where the light of neighboring galaxies overlaps in the image plane. Pixel noise and PSF convolution further exacerbate these effects, introducing shear-dependent detection biases, e.g., see Ref.~\cite{Kaiser:1999ay,Bernstein:2001nz,Hirata:2003cv,Hartlap:2010ge}. Because these biases arise before shear measurement and are already imprinted in the data, they are difficult to calibrate using observations alone. Meta-calibration methods attempt to correct detection biases by introducing an additional detection step, known as metadetection, but this becomes challenging when accounting for redshift estimation~\cite{Sheldon:2019uxq}. Furthermore, the shear of galaxies varies with their environment and redshift, resulting in redshift-shear interplay effects that cannot be calibrated with observational data where the true shear is unknown, e.g., see Ref.~\cite{DES:2020lsz,Li:2022pof}.
Thus, simulation-based calibration is essential for shear measurement methods to meet the stringent requirements of Stage IV surveys. This approach originated from a series of community-wide blind challenges, where mock images with realistic galaxy properties and real data features were used to assess shear measurement algorithms~\cite{Heymans:2005rv,Massey:2006ha,Bridle:2009gg,Kitching:2012tu,Mandelbaum:2013esa}. With simulated images, where the ground truth is known, shear biases can be directly measured for any algorithm.

If the simulated images accurately replicate a survey’s properties (e.g., resolution, PSF, signal-to-noise SNR, size and ellipticity distributions, source density, and clustering), the measured shear biases can be used to calibrate residual biases in real observations.
This approach has been adopted by all Stage III surveys, e.g., see Ref.~\cite{DES:2020lsz,Li:2021mvq,Li:2022pof} and will remain vital for the Stage IV surveys.

\subparagraph{Impact on $S_8$ tension and future direction}

Decades of dedicated work on shear measurement and calibration have significantly improved the accuracy of shear measurement.
The latest \ac{kids} cosmic shear analysis demonstrates that residual shear biases, primarily stemming from uncertainties in galaxy profiles, introduce only sub-percent uncertainties in $S_8$ estimates~\cite{Li:2023azi}.
Shear measurement uncertainties are thus unlikely to be the main driver of the current $S_8$ tension.

Since shear measurement is closely tied to redshift estimation in real observations, developing a consistent, joint calibration of shear and redshift estimates using multi-band image simulations remains essential and will be a key focus for future simulation-based calibration, e.g., see Ref.~\cite{DES:2020lsz,Li:2022pof}.
Furthermore, due to the inherent limitations of image simulations—especially in capturing galaxy profile details—advancing shear measurement algorithms to reduce sensitivity to these intricacies remains a crucial goal.

\paragraph{Observational effects: Redshift Measurement and Calibration}\label{subsection:WLphotoz}
 
\ac{wl} measurements rely on two-dimensional (projected) quantities, meaning individual source redshifts are not strictly required.
However, redshift estimates are essential for cosmic shear tomography, which enhances the signal and probes the time evolution of the growth of structures.  
Historically, the requirements for photometric redshifts of individual lensing sources have been relatively lenient, but upcoming surveys such as \textit{Euclid} \cite{Euclid:2024yrr} and \ac{lsst} \cite{LSST:2008ijt} demand higher precision, e.g., see Ref.~\cite{Newman:2022rbn} for a review.

\subparagraph{Photometric redshift estimation}

Photometric redshift estimation is crucial for extracting cosmological information in optical surveys like \ac{lsst} and \textit{Euclid}.
In contrast to spectroscopic redshift measurement, photometric redshift inference uses image, or photometric information. 
Broad band optical surveys like \ac{lsst} and \textit{Euclid} take images in $100-200$ nanometer wide optical filter bands, which does not allow the recovery of atomic lines.
This limitation means that redshift estimation must rely on the overall shape and characteristic features of a galaxy’s spectral energy distribution (SED), for a review see Ref.~\cite{2019NatAs...3..212S,Newman:2022rbn}.
Two primary approaches exist: \ac{ml} and template fitting.
\ac{ml} methods \cite{2003LNCS.2859..226T, 2004PASP..116..345C, 2010ApJ...715..823G, 2013MNRAS.432.1483C,  2015MNRAS.449.1043B, 2015MNRAS.452.3710R, 2016A&C....16...34H} 
construct a direct mapping from observed photometry of individual galaxies to their redshift based on training data.
While this training data does not need to spatially overlap, it has to be representative and complete in the color-redshift mapping being learned. 
Template fitting methods, e.g., see Ref.~\cite{1999MNRAS.310..540A, Benitez:1998br,2006A&A...457..841I, 2006MNRAS.372..565F, 2015MNRAS.451.1848G, 2016MNRAS.460.4258L,Malz:2020epd} match observed photometry to pre-existing SED models.
It relies on accurate SED modeling and requires validation using similar calibration datasets.
Although complementary, both techniques face challenges such as epistemic uncertainty in the SED and selection function models and incompleteness of accurate reference data at the faint end of the color-magnitude space, e.g., see Ref.~\cite{2014MNRAS.444..129C, 2015APh....63...81N, 2017ApJ...841..111M, 2019ApJ...877...81M, 2020MNRAS.496.4769H}.

\subparagraph{Redshift distribution calibration for cosmic shear}

Beyond individual redshift estimates for cosmic shear sources, accurately calibrating their \textit{redshift distributions} is crucial for interpreting observed signals with underlying cosmological models. 
For a given set of sources -- usually selected in tomographic redshift bins -- it is essential to recover their underlying \textit{true} redshift distribution as closely as possible \cite{2006ApJ...636...21M}.
This is necessary to relate the observed signal to theoretical predictions, as the redshift distribution of sources determines the observed shape correlations via the matter power spectrum. Calibrating the redshift distribution in current \ac{wl} surveys is challenging due to the faint nature of most cosmic shear sources and the limited availability of spectroscopic measurements.
Consequently, a range of calibration methods have been developed, and efforts are ongoing to extend spectroscopic samples to better match the needs of cosmic shear surveys \cite{2015APh....63...81N, 2015ApJ...813...53M}.

One of the earliest approaches for redshift distribution calibration, employed by e.g., the Deep Lens Survey (DLS) \cite{Jee:2012hr} or CFHTLenS \cite{2013MNRAS.431.1547B}, involved stacking photo-$z$ posterior \ac{pdf}s, derived with Bayesian Photometric Redshifts (BPZ) \cite{BPZ2011_software}, to estimate the population distribution $\mathrm{d}N/\mathrm{d}z$. As Stage III surveys increased in statistical power, more reliable calibration methods were developed. These methods generally rely on either mapping the relation between colors and redshifts, or on the cross-correlation (clustering) approach. The former employs spectroscopic or narrow-filter multiband photometric calibration data to re-weight $\mathrm{d}N/\mathrm{d}z_\mathrm{phot}$, either directly in magnitude space \cite{2008MNRAS.390..118L,Hildebrandt:2016iqg,Hildebrandt:2018yau},  by incorporating additional properties, e.g., see Ref.~\cite{DES:2017ndt}, or  via self-organizing maps (SOM) \cite{Wright:2019fwm}. Two key ingredients are required for these calibration methods: spectroscopic data that closely match the cosmic shear sample (in magnitude, color, and redshift range) and a sufficient number of multi-band filters to break degeneracies in the color-redshift mapping. This is most readily achievable in \ac{kids}, which covers nine bands and includes multiple deep calibration fields \cite{2021A&A...647A.124H,2024A&A...686A.170W}. The \ac{des} redshift inference and calibration utilizes a mapping from deep multi-band to wide four-band data that is combined with spatial cross-correlations \cite{2021MNRAS.505.4249M, 2022MNRAS.513.5517C, 2022MNRAS.510.1223G}. This choice is motivated by studies \cite{2020MNRAS.496.4769H} that found that the photometric redshift biases from direct calibration strategies induced by the incompleteness in spectroscopic calibration samples are unacceptably high for \ac{des}. Similar to \ac{des}, the \ac{hsc} calibration \cite{Rau:2022wrq} relies on a combination of spectroscopic and narrow filter multiband calibration data as well as spatial cross-correlation measurements described in the following section. We note that the choice of photometric redshift calibration method and the incorporation of systematic effects is dictated by the survey specification and cannot be generalized across survey designs. However it can be concluded that the treatment of selection functions of calibration sources poses a major challenge for \ac{des}, \ac{hsc}, and \ac{kids}.

The cross-correlation method, also known as clustering redshifts, leverages the spatial correlation of galaxy positions to estimate $\mathrm{d}N/\mathrm{d}z$ for a photometric sample by measuring its angular cross-correlation with a reference sample that has known redshifts, e.g., see Ref.~\cite{2006ApJ...651...14S,2008ApJ...684...88N}. 
Either spectroscopic or high-fidelity photometric redshifts can serve as reference samples in this method.
A major limitation of the clustering redshift approach is the availability of wide-angle calibration samples that extend to sufficiently high redshifts. 
As a result, its application in current cosmic shear surveys has been restricted primarily to the calibration of low-redshift bins \cite{2021A&A...647A.124H,2021MNRAS.505.4249M,Rau:2022wrq}.

\subparagraph{Impact on \texorpdfstring{$S_8$}{} tension and future direction}

Ref.~\cite{Busch:2022pcx} showed that an improved redshift calibration of the full sample or calibrations using different subsets change the $S_8$ obtained from the fiducial \ac{kids}-1000 analysis at most at the level of $0.5\sigma$. For upcoming surveys, photometric redshift estimation based on \ac{ml} methods could benefit from training sample augmentation with simulated galaxies possessing otherwise unrepresented features \cite{Moskowitz:2024deh}. The complex selection function of spectroscopic surveys presents a major challenge in the calibration of photometric redshifts, especially at the faint end of the color-magnitude space \cite{2020MNRAS.496.4769H}.

Computationally efficient approaches for the marginalization over photometric redshift uncertainties are crucial to tackling the challenge of running inference in high-dimensional parameter spaces. Strategies can involve simplified $n(z)$ parametrizations, such as in terms of shifts of a mean redshift $\Delta z$ per bin, or a resampling approach using multiple \ac{mcmc} chains at a fixed $n(z)$ sampled from the uncertainties. Ref.~\cite{Zhang:2022xmt} showed that for the forecasted precision for \ac{hsc} Y3, these methods recover statistically consistent error bars. However, when the constraining power of the full \ac{hsc} survey (or other surveys of comparable or greater statistical power) is considered, the choice of the marginalization method may modify the $1\sigma$ uncertainties on $\Omega_{\rm m,0}-S_8$ constraints by a few percent.

\paragraph{Astrophysical effects: Baryonic feedback}\label{subsection:WLbaryons}

With the increasing precision of modern cosmological surveys, small-scale astrophysical processes—commonly referred to as baryonic feedback—have become increasingly relevant in \ac{wl} analyses.
Baryonic feedback, driven by galaxy formation, supernova explosions, and \ac{agn}, suppresses the total matter power spectrum on scales of $k \sim 1-10 \,h \,\text{Mpc}^{-1}$ by redistributing gas within and beyond halos and influencing through back-reaction \cite{vanDaalen:2011xb}. 
Unlike gravity, which acts as a long-range force, baryonic effects follow an inside-out pattern: halos, stars, galaxies, and central supermassive black holes form on small scales through hierarchical structure formation, while supernova and \ac{agn} feedback modify the baryon distribution by, for example, expelling intrahalo gas into the \ac{igm} \cite{Sunseri:2022txp, 2019JCAP...03..020S,2020MNRAS.495.4800A,2020JCAP...04..019S,2021JCAP...11..026S}. 
In particular, hydrodynamical simulations \cite{2014MNRAS.441.1270L,2018MNRAS.476.2999M,2017MNRAS.467.4739K,vanDaalen:2019pst,2023MNRAS.526.4978S} indicate that \ac{agn} feedback is one of the most significant effects in suppressing the matter power spectrum at small scales.
However, the spatial extent of the \ac{agn} ejected gas remains an open question—whether it affects a few Mpc or extends to tens of Mpc \cite{Chisari:2019tus,vanDaalen:2019pst}. 
Below this maximum scale, the total matter distribution remains unchanged due to the conservation of mass and momentum.
On smaller scales of $k > 10 \,h\,\text{Mpc}^{-1}$, baryonic effects such as gas cooling and star formation become efficient, leading to an upturn in the matter power spectrum \cite{Chisari:2018prw}. Since these effects alter the total matter distribution, cosmic shear summary statistics—such as two-point correlation functions — are significantly affected compared to a purely collisionless scenario.

\subparagraph{Mitigation strategies}

Despite the importance of baryonic feedback, modeling it from first principles remains challenging due to the complex and nonlinear nature of these processes. This introduces significant theoretical uncertainties in \ac{wl} analyses. A common mitigation strategy is to apply scale cuts to the summary statistics, excluding small scale data where hydrodynamical simulations suggest feedback effects dominate \cite{KiDS:2020suj,DES:2021vln,DES:2021bvc, Li:2023tui,Dalal:2023olq}. While this method preserves the robustness of cosmological constraints,  it comes at the cost of discarding high SNR scales that are rich in cosmological information, thereby limiting the constraining power of \ac{wl} surveys. 
To fully exploit the potential of upcoming Stage IV surveys, it is crucial to incorporate baryonic effects into cosmological analyses rather than simply removing affected scales.

\begin{figure}[t]
    \centering
    \includegraphics[scale = 0.70]{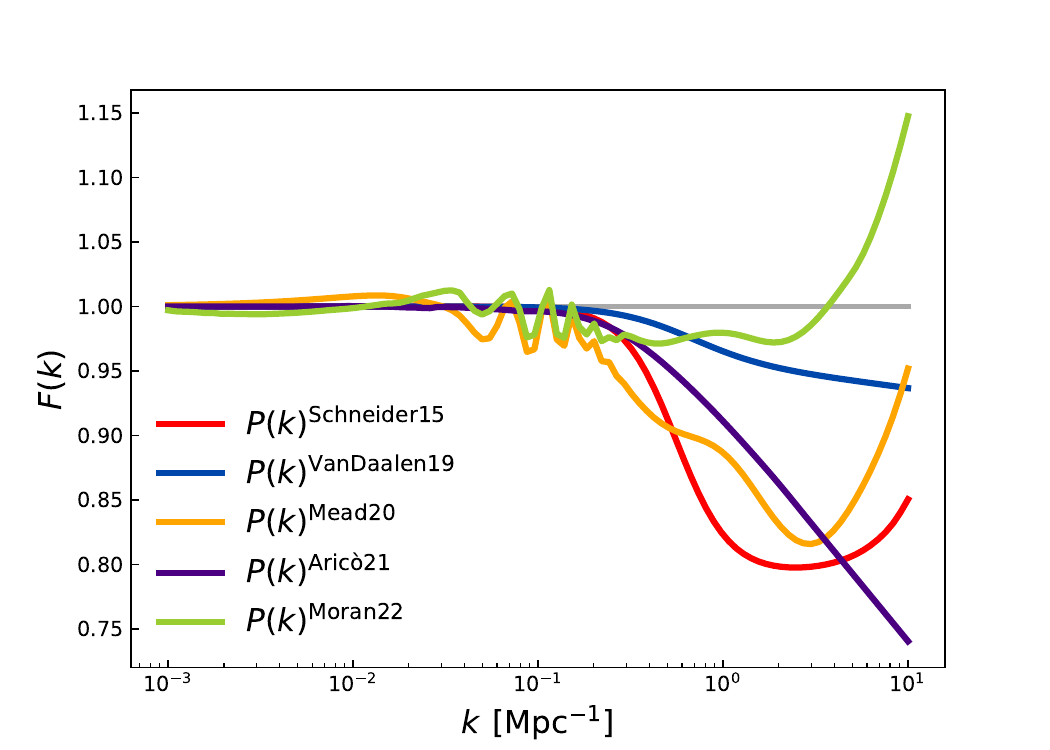}
    \caption[Impact of baryons on the power spectrum]{Suppression of the total matter power spectrum on small scales due to baryonic effects for various baryon models \cite{2015JCAP...12..049S,vanDaalen:2011xb, Arico:2020lhq,Mead:2020vgs,Moran:2022iwe}.
    All spectra were computed for a flat \lcdm\ cosmology with $\Omega_\mathrm{c} = 0.25$, $\Omega_\mathrm{b} = 0.05$, $\Omega_k = 0.0$, $\sigma_8 = 0.81$, $n_s = 0.96$, and $h = 0.67$, assuming no massive neutrinos.
    The $y$-axis shows the ratio $F(k)$ of each model including modeling of baryons, to the nonlinear matter power spectrum without baryonic effects.
    The nonlinear power spectrum was computed using {\tt pyccl.nonlin\_matter\_power} method, while baryonic suppression was modeled with the baryonic modules in {\tt pyCCL} \cite{LSSTDarkEnergyScience:2018yem}, all evaluated at $a = 1$ ($z = 0$) over the same $k$-range. 
    (Adapted from Ref.~\cite{sarcevic_2024_14602273}.)}
\label{fig:baryons_suppresion}
\end{figure}
Several modeling strategies have been developed to address this issue by incorporating baryonic feedback into the total matter power spectrum.
One widely used approach is the baryon correction model (BCM), which modifies the position of particles in $N$-body simulations using parametric prescriptions that approximate the impact of feedback on the \ac{dm} halo profiles \cite{2015JCAP...12..049S,Arico:2020lhq,Arico:2023ocu, 2019JCAP...03..020S}.

Another method integrates baryonic effects within the halo model framework by modifying density profiles and gas expulsion mechanisms \cite{2014JCAP...04..028F,Mead:2020vgs,Mead:2016zqy}. Some of these models and their impact on the matter power spectrum are summarized in Fig.~\ref{fig:baryons_suppresion}.
Principal Component Analysis (PCA) and alternative basis function approaches provide a more data-driven way to capture a broad range of feedback scenarios \cite{Huang:2018wpy}.
The most detailed recipe comes from cosmological hydrodynamical simulations, which explicitly simulate baryonic physics using sub-grid prescriptions for unresolved processes such as supernova and \ac{agn} feedback \cite{Chisari:2018prw}.
However, discrepancies arise due to differences in sub-grid implementations, calibration strategies, and simulation methods.
These variations lead to disagreements in both the amplitude and scale dependence of the suppression of the total matter power spectrum \cite{Chisari:2019tus}. 
Consequently, an effective model for \ac{wl} must be flexible enough to encompass the full range of plausible baryonic effects. 
Moreover, other nonlinear effects—including intrinsic alignments, nonlinear clustering, and reduced shear corrections—must be accounted for to ensure robust cosmological constraints.  
The aforementioned strategies provide a way to marginalize over baryonic feedback uncertainties while preserving small-scale cosmological information.
However, degeneracies between baryonic feedback and cosmological parameters may degrade the constraints on $S_8$, highlighting the need of estimators beyond the two-point functions that can disentangle these effects while maximizing the constraining power of next-generation surveys.

Higher-order statistics of the \ac{wl} field, such as bispectrum, peak counts \cite{Broxterman:2023zfn,Grandon:2024tud} and the one-point lensing \ac{pdf} \cite{Castiblanco:2024xnd}, are more sensitive to non-Gaussian and smaller-scale structures than two-point correlations \cite{Foreman:2019ahr, Grandon:2024pek, Martinet:2020omp,Lu:2021uvr}. Thus, combining the two-point correlations with the higher-order statistics would allow us to obtain tighter constraints on cosmological parameters \cite{Cheng:2024kjv, Marques:2023bnr}, while self-calibrating the baryonic effects \cite{Semboloni:2012yh}.

Finally, multi-probe approaches—combining \ac{wl} with X-ray, measurements of the \ac{sz} effect (thermal tSZ and kinetic kSZ), and \ac{frb} observations—offer complementary constraints on the distribution of baryons in large-scale structures \cite{Elbers:2024dad, DES:2024iny,Khrykin:2024lmd, DES:2024iny, 2023MNRAS.518..477A, 2022ApJ...928....9L}. 
These techniques collectively help preserve cosmological information while minimizing biases from astrophysical processes.

\subparagraph{Impact on cosmic tension and \(S_8\)}

To explore the impact of baryons on the inferred $S_8$, Stage III surveys have conducted parameter inference with two-point functions down to small scales. They found that moderate to small baryonic effects are present in the data \cite{Arico:2023ocu,Garcia-Garcia:2024gzy,Terasawa:2024agq, Amon:2022azi}. In particular, Ref.~\cite{Garcia-Garcia:2024gzy} performed a joint cosmic shear analysis of the Stage III surveys \ac{des}-Y3, \ac{kids}-1000 and \ac{hsc}-DR1 using small scale data in harmonic space. This reanalysis implements a single pipeline that extends the cosmic shear angular power spectrum to scales up to $\ell < 4500$, using $\tt{BACCOemu}$ to model the non-linear regime and baryonic effects in the total matter power spectrum. Their resulting $S_8$ constraint is 1.8$\sigma$ lower than \textit{Planck}, however, \cite{Garcia-Garcia:2024gzy} found a $\Omega_{\rm m,0}$ tension between both data sets of 3.5$\sigma$. When analyzing the parameter space in terms of variables without an implicit dependence on the Hubble constant $H_0$, ({$S_{12}, w_m$}), the authors find no tension with \textit{Planck}. On the other hand, Ref.~\cite{Terasawa:2024agq} explore signatures of baryonic effects in the \ac{hsc} Y3 two-point correlation functions $\xi_{\pm}$ down to small scales of 0.28 arcminutes (up to $k \sim 20 h $ Mpc$^{-1}$). The theoretical modeling is implemented by means of a non-linear \ac{dm}-only matter power spectrum emulator (similar to \textsc{DarkEmulator} \cite{Nishimichi:2018etk}). The authors found no significant shift in the inferred $S_8$ when including the small scales. Moreover, the \ac{dm}-only theory fits the observations within the statistical error of the survey, meaning baryons are not playing an important role in these measurements. Similar results were obtained for \ac{hsc} Y1 data using multiple higher-order statistics at small scales \cite{Grandon:2024pek}. These studies of cosmic shear data, together with other analyses from the Sunyaev-Zel’dovich observations, indicate that strong feedback scenarios are needed to reconcile the $S_8$ inferred from cosmic shear surveys with the value derived from \textit{Planck} \lcdm\ \cite{Amodeo:2020mmu}. This translates into a large suppression of the matter power spectrum, of the order of $\sim$25\% at $k \sim 1 h$ Mpc$^{-1}$. The latter, however, is larger than the suppression found from hydrodynamical simulations, and contradicts the constraints imposed by X-ray observations of the baryon
mass fraction for cluster-scale halos \cite{Amon:2022azi, Terasawa:2024agq}. Therefore, it is possible that baryons in combination with other non-linear effects and unmodeled systematics may explain the discrepancy between cosmic shear results and $\textit{Planck}$ \lcdm\ cosmology, or extensions of \lcdm\ and new physics are required.

\paragraph{Astrophysical effects: Intrinsic alignments of galaxies}\label{subsection:WLIA}

Intrinsic alignments (IA) of galaxies represent a significant astrophysical effect that can systematically affect the measurement of cosmic shear and, consequently, the inference of cosmological parameters (see Ref.~\cite{Lamman:2023hsj} for a comprehensive review). 
These alignments arise due to gravitational interactions and the tidal forces exerted by the large-scale structure of the Universe, causing the shapes of nearby galaxies to correlate with each other and with the surrounding matter density field. 

For cosmic shear two-point correlation functions, intrinsic alignments can be categorized primarily into two types: intrinsic-intrinsic (II) alignments and gravitational-intrinsic (GI) and intrinsic-gravitational (IG) alignments.
II alignments occur when the shapes of physically close galaxies are directly influenced by the same gravitational tidal field, leading them to align with each other.
GI and IG alignments describe the correlation between intrinsic galaxy shapes and the shear induced by gravitational lensing.
GI refers to the correlation between the intrinsic shape of a foreground galaxy and the shear of a background galaxy, while IG represents the reverse case \cite{Troxel:2014dba}.
These terms are related through symmetry properties of the shear field, with IG often being treated as the transpose of GI in theoretical and numerical treatments.

Galaxy-galaxy lensing (GGL) probe, or the correlation between galaxy shapes and galaxy positions, is also sensitive to intrinsic alignments when the shape and the position catalog overlap in redshift. 
GGL is sensitive only to GI types of alignments and through a different kernel than for cosmic shear (see equations in e.g., Ref.~\cite{LSST:2022sql}).
Because of this, 3$\times$2pt analyses significantly help to constrain the intrinsic alignment parameters. Recently, GGL shear-ratios have been used to fold in prior IA information to the cosmic shear analysis \cite{DES:2021jzg,DES:2021bvc,DES:2021vln}. 
Recently, inverse galaxy-galaxy lensing has also been proposed as an additional probe to constrain IA parameters \cite{Cross:2024vrg}.

\subparagraph{Modeling intrinsic alignments}

Intrinsic alignment (IA) models vary in complexity and scale-dependent effectiveness. The Nonlinear Alignment Model (NLA) \cite{Bridle:2007ft} has been the standard approach to model red (elliptical) galaxies. This model assumes a linear coupling between the nonlinear tidal field and galaxy shapes, typically parametrized by an amplitude parameter alongside an optional redshift evolution parameter. In this framework, blue (spiral) galaxies are assumed to have negligible intrinsic alignments. However, the linear tidal shear assumption breaks down on small scales, necessitating more sophisticated models.

The Tidal Alignment and Tidal Torquing (TATT) model \cite{Blazek:2017wbz} represents an important extension, introducing additional terms to capture the alignments of spiral galaxies where angular momentum plays a critical role. 
This model incorporates a separate amplitude parameter for the tidal torquing effect, along with its own redshift scaling parameter.
In total, the basic TATT model typically employs four parameters: $A_1$ (tidal alignment amplitude), $A_2$ (tidal torquing amplitude), $\eta_1$ (alignment redshift scaling), and $\eta_2$ (torquing redshift scaling). 
A fifth parameter, $b_{\mathrm{TA}}$ (the linear bias of source galaxies), is frequently included to account for the clustered distribution of source galaxies in observational data.
Many studies such as Refs.~\cite{Samuroff:2020gpm,DES:2022vuu,DES:2022aeh} demonstrate that while NLA fits measurements well above 6$-$8 Mpc/$h$, TATT extends this range down to 1$-$2 Mpc/$h$.
These findings emerge from both simulation-based tests ($N$-body with semi-analytic IA components or hydrodynamic simulations) and direct fits to observational data.

For modeling at even smaller scales, the halo model describes alignments using one-halo and two-halo terms, based on galaxy correlations within and between \ac{dm} halos \cite{Fortuna:2020vsz,Lamman:2023hsj} while the Effective Field Theory (EFT) framework treats alignments as a tensor field and incorporates small-scale physics through free parameters.
Ref. \cite{Bakx:2023mld} shows that EFT describes \ac{dm} halo intrinsic alignments up to $k_\mathrm{max}$ = 0.30 $h$/Mpc at $z = 0$, significantly outperforming both NLA and TATT which they find to only reach $k_\mathrm{max}$ = 0.05 $h$/Mpc. 

Despite these advances in model complexity, practical applications raise important questions about necessary sophistication.
The \ac{des} Y3 cosmic shear analysis \cite{DES:2021vln} found that NLA was sufficient to fit their data, even though TATT had been chosen for fiducial results in a blinded fashion. 
Similarly, Ref.~\cite{Samuroff:2020gpm} found no significant evidence for non-zero tidal torquing amplitude ($A_2$) in Illustris TNG hydrodynamic simulations, challenging the need for TATT's full 5-parameter framework in typical applications.
Moreover, it is important to note that increasing model complexity does not necessarily reduce biases. 
The interplay between IA and other uncertainties (e.g., from photometric redshift errors \cite{Leonard:2024nnw}) can introduce additional complications.
Recent work \cite{Sarcevic:2024tdr} shows that jointly modeling IA and source redshift distributions via a shared \ac{lf} can help mitigate such biases while maintaining consistency in \ac{wl} analyses.
Careful calibration and validation remain crucial to ensure these models accurately represent physical processes without introducing unintended distortions in cosmological analyses. 
To navigate these modeling challenges, researchers have developed strategic approaches for IA model selection.
Ref. \cite{Campos:2022gtg} recently introduced a data-driven methodology to objectively determine which IA model best suits a particular analysis, providing a more systematic framework for model choice. 
In parallel, many cosmological studies have adopted a multi-model approach, presenting results across several IA formulations to demonstrate robustness and identify potential model-dependent biases.
Below, we examine how these modeling decisions have shaped recent cosmological parameter constraints and what they reveal about the relative importance of IA modeling in contemporary \ac{wl} analyses.

\paragraph*{Modeling IA for higher order statistics} 
While IA modeling has primarily focused on two-point statistics, recent efforts have extended these frameworks beyond 2-point statistics.
Analytical advances have provided foundations for modeling IA in bispectra \cite{Merkel:2014wma,Burger:2023qef} and the lensing \ac{pdf} \cite{Barthelemy:2023mer}, while Ref.~\cite{Harnois-Deraps:2021krd} demonstrated that high signal-to-noise peaks in aperture mass maps can experience deviations up to 30\% for \textit{Euclid}-like surveys due to IA effects.
Many recent analyses employ simulation-based inference, which relies on map-level IA modeling and presents unique challenges.
For example, Ref.~\cite{DES:2023ycm} showed that source clustering impacts non-Gaussian statistics significantly more than two-point functions \citep{KiDS-1000:2024rdr}.
Current approaches address this either by implementing additional clustering terms in analytical predictions \cite{Barthelemy:2023mer} or by directly incorporating clustering effects in simulations \cite{DES:2023qwe}.
The latter approach extends beyond traditional NLA implementation by using simulation-based clustering rather than tree-level perturbation theory, showing that modeling IA at the map level can also offer advantages for various summary statistics.

\subparagraph{Impact on cosmic tension and \(S_8\)}

\begin{figure}[htbp]
    \centering
    \includegraphics[scale = 0.80]{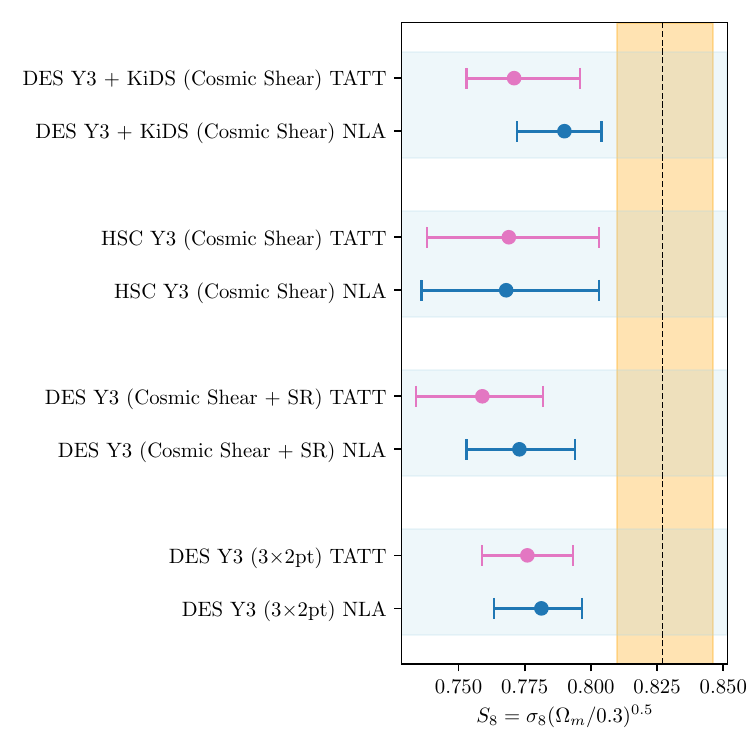}
    \caption[Short caption]{Impact of IA model choice to $S_8$ in recent \ac{wl} analyses assuming the \lcdm\ model.
    The \ac{des} Y3 + \ac{kids} results are from Ref.~\cite{Kilo-DegreeSurvey:2023gfr}, \ac{hsc} Y3 from Ref.~\cite{Dalal:2023olq}, \ac{des} Y3 cosmic shear (COSEBIS and $\xi_{\pm}$)+ SR (shear-ratio) from Ref.~\cite{DES:2021bvc,DES:2021vln} and \ac{des} Y3 3$\times$2pt from Ref.~\cite{DES:2021wwk}. The vertical dashed line is the \textit{Planck} value with its corresponding uncertainty shaded in orange. 
    All error bars correspond to 1-$\sigma$ uncertainties. }
\label{fig:s8_comparison_IA}
\end{figure}

Misestimations in intrinsic alignments can lead to significant biases in the structure growth parameter \(S_8\), thereby affecting our understanding of cosmic tensions. Recent cosmic shear analyses, as illustrated in Fig.~\ref{fig:s8_comparison_IA}, demonstrate that among various analysis choices, the selection of IA model often produces shifts in the $S_8$ constraints, which will become substantial for Stage IV surveys \cite{Krause:2015jqa}. The TATT model (pink points) tends to yield somewhat lower $S_8$ values compared to the NLA model (blue points) across different analysis combinations. 
This difference, while modest, may be relevant when considering potential tensions between early and late-Universe probes, as indicated by the comparison with the reference line in the figure.
The largest change comes from the joint analysis of data from the \ac{des} and \ac{kids}, which revealed that transitioning from the NLA model to the more complex TATT model can lead to heightened tension in \(S_8\) measurements. Quantitatively, the difference between the redshift-dependent NLA and TATT analysis results was 0.9\(\sigma\) in the \(S_8\) marginalized posterior, but reached 1.3\(\sigma\) when comparing the maximum a posteriori (MAP) values in the full parameter space.
This distinction is particularly significant as it indicates that the discrepancy arises from genuine differences in model physics rather than merely from differences in prior volumes or projection effects.
These results highlight the importance of carefully considering IA modeling choices when interpreting cosmological constraints from \ac{wl} surveys. 3x2pt analyses in \ac{des} Y3 were less susceptible to the IA model compared to statistics based on cosmic shear alone.

\subparagraph{Future directions}

As \ac{wl} analyses continue to advance, refining IA models and developing more realistic cosmological simulations will be essential for providing robust validation tests. 
These improvements will work synergistically with upcoming surveys such as \ac{lsst} by the Vera C. Rubin Observatory, which will deliver high-quality imaging across unprecedented sky areas, enabling studies of IA with greater statistical power and tightening constraints on cosmic shear measurements.

Exciting new approaches are also already emerging from recent observations.
Measurements have consistently found IA amplitudes lower than theoretical expectations, with blue galaxies showing negligible alignment signals \cite{DES:2018gxz}. 
Building on these findings, \cite{DES:2024xvm} proposed an innovative strategy of conducting cosmic shear analysis exclusively with blue galaxies -- potentially eliminating the need for complex IA modeling entirely.
While this approach sacrifices approximately half the galaxy sample, reducing statistical power, it could significantly mitigate systematic uncertainties.
The viability of this or other compromises will become clearer as theoretical understanding advances and next-generation survey data becomes available.

\paragraph{Confirmation bias and blinding}

Modern cosmological analyses in surveys like \ac{des}, \ac{kids}, and \ac{hsc} are often highly complex and require the seamless interaction of large teams responsible for different parts of the analysis.
Sources of epistemic error are difficult to control, which can lead to a dependence of cosmological parameter constraints on analysis choices.
At the same time, complementary cosmological constraints like from the \ac{cmb} might induce a subconscious bias in a select parameter subspace, like $\sigma_8$ and $\Omega_{m,0}$.
In cases where the value of the cosmological parameters are revealed during the analysis, this might lead to `confirmation bias', where analysis assumptions are subconsciously tuned towards a certain scientific narrative. 
The complexity of cosmological analysis combined with the significant dependency on analysis choices, therefore necessitates strict strategies to avoid this confirmation bias.
The blinding strategies differ across surveys but generally consist of catalog-level blinding where a set of synthetic and encrypted multiplicative shear values are artificially added to obfuscate the true shear field and analysis-level blinding where the inferred parameter posteriors are altered to avoid a ready comparison with prior work \cite{DES:2019zlw}. 
\bigskip
\subsubsection{Galaxy cluster counts \label{sec:Gal_Clus}}

\noindent \textbf{Coordinator:} Dominique Eckert\\
\noindent \textbf{Contributors:} Antonio da Silva, Filippo Bouch\`{e}, Francesco Pace, Iryna Vavilova, Jenny G. Sorce, Massimiliano Romanello, Rados{\l}aw Wojtak, Sebastian Grandis, Shahab Joudaki, and Vittorio Ghirardini
\\

\paragraph{Methodology}

Galaxy clusters are the most massive gravitationally bound structures in the present Universe and originate from the largest fluctuations of the primordial matter distribution. Their mass and number density strongly depends on the underlying cosmological parameters \cite{Allen:2011zs}, such that galaxy cluster surveys are powerful probes of cosmological parameters. In this section, we describe the main steps that are needed to extract cosmological information from galaxy cluster surveys. We identify the major bottlenecks that need to be addressed to advance our cosmological knowledge with the use of this technique, and summarize the main recent results in the field.

\subparagraph{The halo mass function and its cosmological dependency}

In the hierarchical structure formation scenario, small structures formed at high redshifts progressively merge under the influence of gravity to form the massive systems we see today. As such, the number density of halos and its time evolution trace the growth of structures in the Universe, which depends on the underlying cosmological model. The halo mass function describes the number of collapsed halos at a given time per unit volume and halo mass. Its shape, normalization, and redshift evolution are highly sensitive to cosmological parameters. 

In mathematical terms, the halo mass function can be expressed in the following way:
\begin{equation}
    \frac{\mathrm{dn}}{\mathrm{d}\ln{M}} = \frac{\rho_{\rm m,0}}{M}\,f(\sigma)\left|\frac{\mathrm{d}\ln{\sigma}}{\mathrm{d}\ln{M}}\right|\,,
    \label{eq:hmf}
\end{equation}
with $\rho_{\rm m,0}$ the mean matter density of the Universe at present day, $\sigma$ the mass variance on a mass scale $M$ and $f(\sigma)$ a function characterizing different fitting (or theoretical) expressions. Often, $f(\sigma)$ is written in as $\nu f(\nu)$, where the peak height $\nu=\delta_{\rm c}/\sigma$ is a function of mass and time. Here, $\delta_{\rm c}$ is the linearly extrapolated overdensity and can be evaluated within the framework of the spherical collapse model. Many different expressions exist for $f(\sigma)$ \cite{Press:1973iz,Sheth:1999mn,Jenkins:2000bv,Tinker:2008ff,Watson:2012mt,Despali:2015yla}.

\begin{figure}
 \centering
 \includegraphics[scale=0.48]{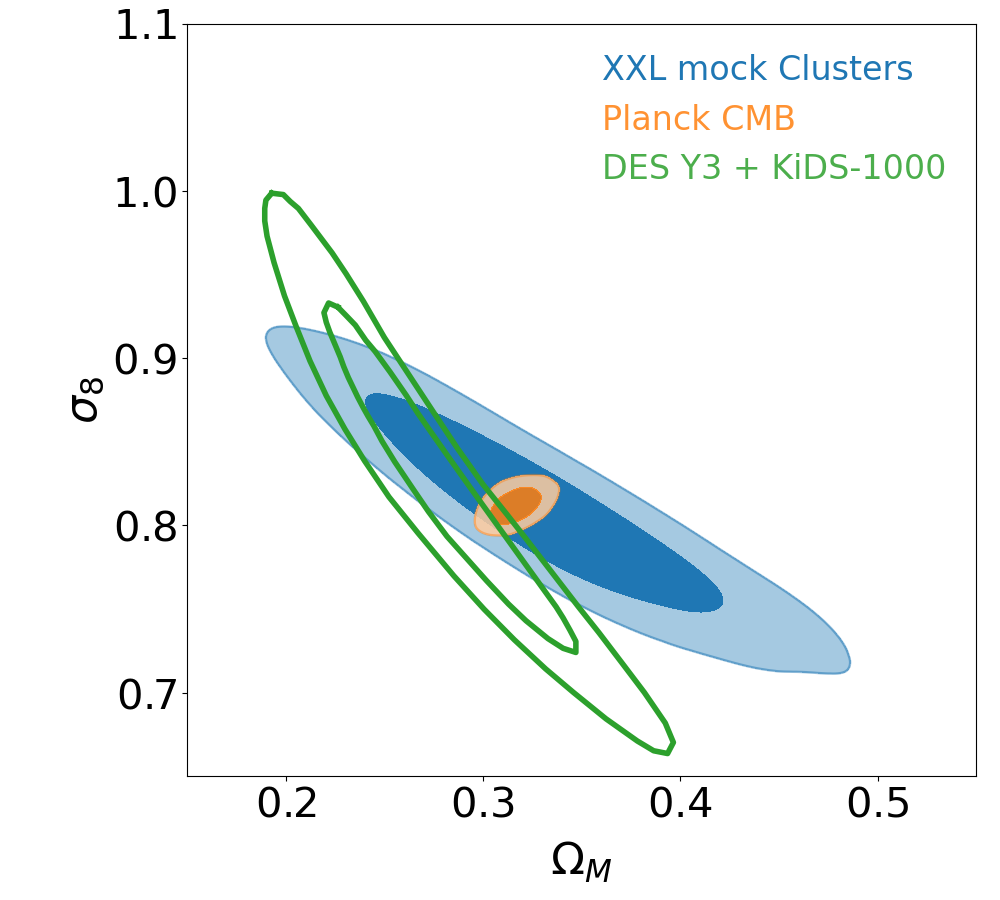}
 \caption{$\Omega_{\rm m,0}-\sigma_8$ degeneracy in typical cosmic shear and cluster counts experiments. The blue curve and shaded area show the degeneracy of the expected $\Omega_{\rm m,0}-\sigma_8$ contours in a mock 50 deg$^2$ X-ray cluster count survey, whereas the green contours show the same degeneracy in the combined \ac{des} Y3 + \ac{kids}-1000 cosmic shear estimate. We can see that the degeneracy is shallower in cluster counts experiments ($\sigma_8\propto \Omega_{\rm m,0}^{-0.3}$) compared to cosmic shear ($\sigma_8\propto \Omega_{\rm m,0}^{-0.5}$), such that any definition of $S_8$ that is optimal for either experiment will not perfectly follow the degeneracy of the other. For comparison, we add the \emph{Planck} \ac{cmb} $\Omega_{\rm m,0}-\sigma_8$ contours in orange.}
 \label{fig:Gal_Clus_C_Degen}
\end{figure}

\begin{figure}
 \centering
 \includegraphics[scale=0.48]{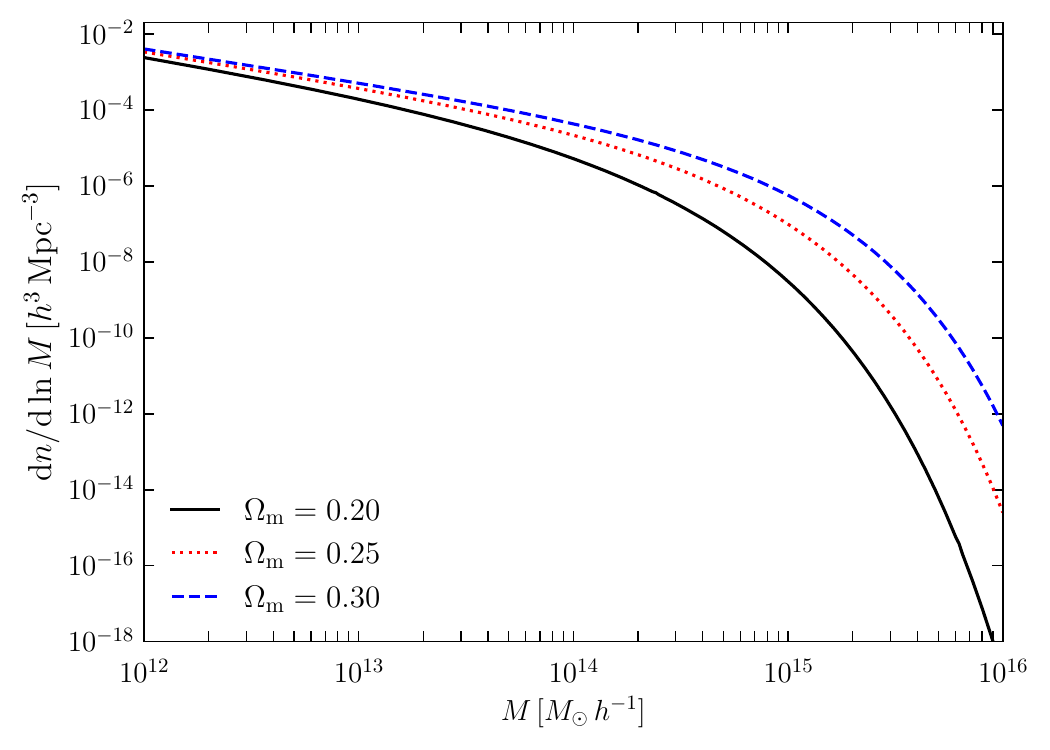}
 \includegraphics[scale=0.48]{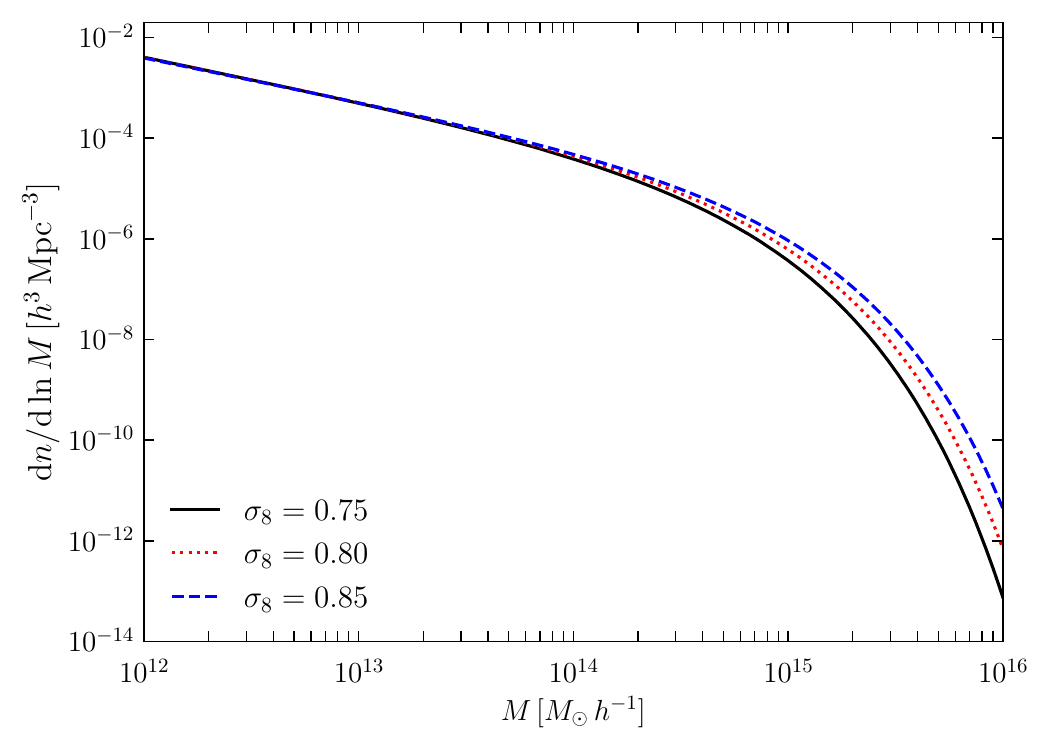}
 \caption{Dependence of the Tinker mass function on matter density $\Omega_{\rm m}$ (left panel) and on the matter spectrum normalization $\sigma_8$ (right panel) at $z=0$.}
 \label{fig:hmf}
\end{figure}

In Fig.~\ref{fig:hmf} we present the dependence of the Tinker halo mass function \cite{Tinker:2008ff} on the matter power spectrum normalization $\sigma_8$ (left panel) and on the matter density $\Omega_{\rm m,0}$. We see that the effect on the halo mass function is that of increasing the number of structures. A higher value of $\Omega_{\rm m,0}$ implies also a higher $\sigma_8$ value. The two parameters are highly degenerate at the high-mass end, as shown by Fig.~\ref{fig:Gal_Clus_C_Degen} where a $50\,\text{deg}^2$ mock cluster counts survey is compared with \ac{des}+\ac{kids} \ac{wl} and the latest Planck collaboration data. The degeneracy between $\Omega_,$ and $\sigma_8$ for galaxy cluster counts is clearly visible. 

It is important to note that halo masses are usually defined within a spherical overdensity $\Delta$ with respect to either the critical density $\rho_{\rm c}(z)$ or the mean density of the Universe $\rho_{\rm m}(z)$. For a given overdensity $\Delta_{\rm c}$ or $\Delta_{\rm m}$, the corresponding mass $M_\Delta$ and radius $R_\Delta$ are defined as the mass and radius within which the condition 
\begin{equation}
    \frac{M_{\Delta}}{4/3\pi R_\Delta^3} = \Delta_{\rm c/m} ~\rho_{\rm c/m}(z)\,,
\end{equation}
is satisfied. The choice of the halo mass function and of the observational mass definition must therefore be adapted to the chosen spherical overdensity. Common choices for the definition of the overdensity are $500c$, $200c$, and $200m$.

\subparagraph{Mass-observable scaling relations and the scatter around these relations}
\label{sec:scaling}

While the halo mass function is highly sensitive to cosmological parameters, constraining it from galaxy cluster surveys is a challenging task. The main underlying issue is that halo mass is not a directly observable quantity. Samples of galaxy clusters are extracted from surveys of the baryonic component that fills these structures, whether it is in the form of hot gas (X-ray, Sunyaev-Zel'dovich effect) or stars (optical/IR). Galaxy cluster detection techniques and their primary observables are described in more detail in Sec.~\ref{sec:surveys}. In all cases, the primary observable of the survey is not directly the halo mass but rather a quantity that depends on it. The relation between survey observable and the halo mass, usually referred as a \emph{scaling relation}, needs to be properly understood to turn the detected cluster samples into constraints on the halo mass function. 

If galaxy clusters grow and evolve in a self-similar way~\cite{Kaiser:1986ske}, the relations between observable quantities $\mathcal{O}$ and spherical overdensity halo mass $M_\Delta$ are expected to be power laws. While astrophysical effects such as mergers, cooling, and \ac{agn} feedback can cause deviations from the simple self-similar scenario, evidence suggests \cite{Truong:2016egq,Pop:2022rms,Pellissier:2023beo,Braspenning:2023bmq} that these relations can be well described by power laws with a log-normal intrinsic scatter $\sigma_{\log \mathcal{O}}$,
\begin{equation}
    \log \mathcal{O} = A_{\mathcal{O}M} + B_{\mathcal{O} M}\left(\frac{M_\Delta}{M_{pivot}}\right) + \gamma_{\mathcal{O}M}E(z)\pm \sigma_{\log \mathcal{O}}\,,
    \label{eq:scaling}
\end{equation}
with $E(z)=H(z)/H_0$. Here the parameters $A_{\mathcal{O}M}$ and $B_{\mathcal{O}M}$ are the normalization and the slope of the mass-observable scaling relation and $\gamma_{\mathcal{O}M}$ governs its redshift evolution. The scaling relation allows us to construct the probability of detecting a system with a measured observable $\mathcal{O}$ given the source's mass and redshift, $P(\mathcal{O}|M,z)$. A precise knowledge of these parameters is required to make accurate cosmological inferences from galaxy surveys. We discuss the associated issues in more detail in Sec.~\ref{sec:syst_mass}.

\subparagraph{Selection function, forward modeling, likelihood function}

The cosmological analysis of galaxy cluster catalogs follows a Bayesian inference approach. A generative model for the data given unknown model parameters has to be postulated, and the \ac{pdf} of the actual data as a function of the model parameters, the likelihood, has to be computed. The generative model assumed for a cluster catalog with observable $\hat{\mathcal{O}}$ and redshift $z$ at a sky position $\theta$ is a Poisson realization of the density field
\begin{equation}\label{eq:halo_number_field}
    \frac{\text{d}\tilde n}{\text{d}\hat{\mathcal{O}}} = \frac{\text{d}n}{\text{d}\hat{\mathcal{O}}} \left( 1 + b_\text{eff} \delta(\theta, z)\right)\,,
\end{equation}
with $b_\text{eff}$ the effective bias, $\delta(\theta, z)$ the matter density contrast, and 
\begin{equation}
     \frac{\text{d}n}{\text{d}\hat{\mathcal{O}}} = \int \text{d} \mathcal{O}\int \text{d} M \frac{\text{d}n}{\text{d} M} P(\mathcal{O} | M, z)  P(\hat{\mathcal{O}}| \mathcal{O}, z, \theta) P(\text{sel}| \hat{\mathcal{O}}, \mathcal{O}, z, \theta)\,,   
\end{equation}
where $\frac{\text{d}n}{\text{d}\ln M}$ is the halo mass function (Eq.~\eqref{eq:hmf}), $ P(\mathcal{O} | M, z) $ the observable mass relation and its scatter (Eq.~\eqref{eq:scaling}), $P(\hat{\mathcal{O}}| \mathcal{O}, z, \theta) $ the effect of instrumental noise of the intrinsic observable $\mathcal{O}$, and $ P(\text{sel}| \hat{\mathcal{O}}, \mathcal{O}, z, \theta)$ the selection function. The latter takes values between 0 and 1 based on the probability of including the cluster with properties  $(\hat{\mathcal{O}}, \mathcal{O}, z, \theta)$.

The second term in Eq.~\eqref{eq:halo_number_field} can be ignored for sufficiently high-mass and wide survey area samples, yielding the Poisson likelihood used by Refs.~\cite{Mantz:2014paa, Planck:2015lwi, SPT:2014wkb, SPT:2018njh, Chiu:2022qgb, DES:2023muu, DES:2024zpp, Ghirardini:2024yni}. For smaller area surveys and lower mass surveys, the second term contributes significantly via the variance of the matter field. In such cases, a composite Gaussian-Poisson likelihood has to be used \cite{XXL:2018ryw, DES:2018phx, DES:2020ahh, DES:2020mlx, Garrel:2021sgq, Lesci:2020qpk, Park:2021sbd, Sunayama:2023hfm}. The resulting likelihood depends on the cosmological parameters via the halo mass function and the cosmological volume. It also depends on the nuisance parameters describing the observable mass relation and, potentially, the selection function. Marginalizing over sufficiently flexible observable mass relation parameterizations is crucial to accurately represent the cluster population, especially with regard to the halo mass.

\paragraph{Galaxy cluster surveys}
\label{sec:surveys}

\subparagraph{X-ray}
The vast majority of the baryonic content of galaxy clusters is in the form of a hot ($>10^7$ [K]), highly ionized plasma that fills the system's gravitational potential well. Given the high temperatures involved, the intracluster medium (ICM) shines predominantly in the X-ray range through thermal bremsstrahlung and line emission. As such, galaxy clusters are luminous ($L_X=10^{44}-10^{45}$ [erg/s]) X-ray sources spanning a diameter of several arcmin. The overwhelming majority of extended X-ray sources in the extragalactic sky are galaxy clusters, such that X-ray surveys are powerful cluster detection machines. In the early 1990s, R\"ontgensatellit (ROSAT) \cite{Voges:1999ju, Borgani:1999ek, Piffaretti:2010my} performed an all-sky X-ray survey and several deeper but small area surveys, e.g., ROSAT Deep Cluster Survey (RDCS) \cite{Borgani:2001ir} which allowed for the detection of several thousand clusters. This spurred the build-up of several sub-samples, either volume-limited (REXCESS) \cite{Bohringer:2007vs}, or flux-limited (HIFLUGCS) \cite{Reiprich:2001zv, Schellenberger:2017wdw} with a well-understood selection function.

Deep follow-up observations of ROSAT-selected clusters with \emph{XMM-Newton} and \emph{Chandra} allowed to constrain cosmological parameters with cluster number counts by measuring the mass of these systems with better precision~\cite{Vikhlinin:2008ym}. \emph{XMM-Newton}, thanks to its larger field-of-view, enabled to scan deeply (30-40 ks) several tens of square degrees, enabling the serendipitous detection of the fainter galaxy group population, extending the mass range used in the cosmological analysis. The XXL survey~\cite{Pierre:2015cqe} detected $\sim450$ clusters over an area of 50 $\mathrm{deg}^2$ \cite{XXL:2018ryw, Garrel:2021sgq}. More recently, \emph{eROSITA} undertook several all-sky surveys, with the ultimate goal of detecting about $100,000$ galaxy clusters. The results from the first of these all-sky surveys were recently released \cite{Ghirardini:2024yni} and provided the detection of $\sim12,000$ clusters over half the sky.

\subparagraph{Sunyaev-Zel'dovich effect}

The thermal \ac{sz} effect \citep[tSZ;][]{Sunyaev:1972eq} describes the inverse-Compton scattering of \ac{cmb} photons off the hot electrons of the ICM. Photons crossing the path of a hot electron cloud occasionally encounter hot electrons and gain energy, which induces a modification of the spectral shape of the \ac{cmb}, see Ref.~\cite{Mroczkowski:2018nrv} for a review. At the peak frequency of the \ac{cmb}, the tSZ effect takes the form of a \emph{decrement} in the temperature of the \ac{cmb}, as a fraction of the photons are upscattered to higher frequencies. At first order, the spectral signal of the tSZ depends only on the Compton parameter, $y$, which describes the average energy gain over the line of sight. The total integrated signal over the area of a cluster, $Y_{SZ}$, is proportional to the integrated thermal energy of the ICM \cite{Kravtsov:2006db}. 

The advantage of the tSZ effect as a cluster detection technique is twofold. First, the signal is essentially independent of redshift, which makes it very efficient to detect high-redshift systems. Second, the primary tSZ observable, $Y_{SZ}$, is believed to be an excellent proxy of cluster mass \cite{Nagai:2007mt,Arnaud:2007br,Marrone:2011us,Planck:2011abm}. In the past decade, the \emph{Planck} survey yielded the detection of $>2,000$ galaxy clusters with this technique extending out to $z\sim1$ \cite{Planck:2013shx,Planck:2015koh}. While \emph{Planck} was very sensitive to the tSZ thanks to its wide frequency range and all-sky coverage, the detection efficiency of high-redshift sources was hampered by the poor \emph{Planck} beam, such that the bulk of the detected systems are at redshift $z<0.5$. Conversely, ground-based experiments have larger collecting areas and better angular resolution. \ac{spt} \cite{Williamson:2011jz,SPT:2014wbo,SPT:2023tib} and \ac{act} \cite{Hasselfield:2013wf,ACT:2020lcv} were now able to gather catalogs of several thousand galaxy clusters out to $z\sim2$. 

\subparagraph{Optical/near-IR photometry}

While in the past couple of decades, a great deal of attention was devoted to the selection of galaxy clusters through the ICM, there is now renewed interest in the detection of galaxy clusters as overdensities of galaxies thanks to the new generation of optical and near-IR surveys like \emph{Euclid} and \emph{Rubin}. As their name suggests, galaxy clusters were originally defined as such based on the detection of overdensities of galaxies in the same region of the sky. Based on visual inspection of photometric plates, Abell (1989) \cite{Abell:1989mu} constructed a catalog of more than 4,000 clusters that has been used for more than two decades as a reference of the local large-scale structure. Since that time, sophisticated galaxy cluster detection algorithms have been developed which allowed to extract large samples of clusters from galaxy surveys. Modern detection algorithms are mainly split into two classes: red sequence and photometric redshifts. Red sequence cluster finders \cite{SDSS:2007mmh,SDSS:2013jmz,Oguri:2014eba} take advantage of the predominance of passive red galaxies in clusters with respect to the field to search for overdensities of red galaxies. This method reduces spurious detections thanks to the use of the red sequence, which is known to trace rich environments \cite{Haines:2015zfa,Lopes:2013wfa}. Alternatively, methods based on photometric redshifts \cite{DAmico:2019fhj,Sarron:2017xjw} make use of galaxy SEDs to search for galaxy overdensities in three-dimensional space. These techniques rely on the availability of high-quality photometric redshifts but make no assumption on the existence of a red sequence. For a detailed comparison of various cluster finding algorithm, we refer to Ref.~\cite{Euclid:2019bue}. 

The main advantage of optical cluster catalogs is their high sensitivity, with the deepest current catalogs containing more than 100,000 clusters (e.g., see Refs.~\cite{DES:2018crd,Wen:2022vih,Maturi:2018fge,Oguri:2017khw,Wen:2024gho}). Modern algorithms also provide an estimate of the cluster redshift as a direct byproduct. The primary observable of optical cluster finders is the number of member galaxies, or \emph{richness} $N$. A standardized richness estimate is computed iteratively to include cluster members above a given absolute magnitude. Richness is known to correlate with halo mass \cite{Simet:2016mzg,DES:2018kma}, although with a large intrinsic scatter induced by projection effects and miscentering \cite{Ge:2018ktu,DES:2021zex}. 

\subparagraph{Shear selection}

The techniques discussed above all rely on detection through the baryonic content of clusters, whether it is in the form of gas or galaxies, whereas the vast majority of the mass is made of \ac{dm}. In the past decade, direct detection of galaxy clusters as peaks in projected \ac{wl} mass maps has become possible \cite{Miyazaki:2007jv,Miyazaki:2018dpu,Li:2021mvq,Chen:2024fpx,Leroy:2023rjx}. This technique allows to detect clusters directly through their total mass content rather than through a proxy that is related to the halo mass in a complex way. Thanks to the depth of the Subaru/\ac{hsc}-SSP survey ($\sim30$ galaxies per square arcmin), Ref.~\cite{Miyazaki:2018dpu} presented a sample of 65 shear-selected clusters within a $\sim160$ deg$^2$ area with a S/N greater than 4.7. All the most significant peaks were associated with galaxy overdensities or X-ray signals \cite{Ramos-Ceja:2021ujr}. More recently, the first cosmological constraints from shear-selected cluster samples were extracted from an area of $\sim500$ deg$^2$ \cite{Chen:2024fpx,Chiu:2024ptq}. While at the present day shear-selected cluster catalogs are much smaller than catalogs obtained from other techniques, upcoming lensing experiments like \emph{Euclid} will yield much larger samples directly extracted from the projected mass distribution. The primary issue associated with shear selection is projection associated with correlated large-scale structure and halo orientation, such that the masses of the detected systems are expected to be over-estimated by $\sim50\%$ \cite{Chen:2019cjt}. Future works will need to model projection effects in detail to make accurate cosmological inferences from large samples of shear-selected systems.

\paragraph{Mass calibration and systematics}

\subparagraph{Mass calibration techniques}

The number of galaxy clusters as an observable displays a direct degeneracy between the mean mass scale of the sample and the cosmological parameters, which dictate the number density of halos via the mass function and the cosmological volume. Number counts of galaxy clusters can thus only constrain cosmological parameters if the relation between halo mass and cluster observables is determined, a problem called mass calibration. Our knowledge of subgrid astrophysical processes limits the direct calibration of this relation on hydrodynamical simulation. It is thus very challenging to derive the degree of uncertainty on the mass calibration from pure simulation-based studies. The latter is, however, the main contributor to the final cosmological constraining power of cluster number counts experiments and should, therefore, be assessed in a reliable, empirical fashion. 

In the context of a \ac{cdm} cosmology, most of the halo mass is composed of \ac{dm}. Empirical mass calibration can thus only proceed via measurements of the cluster's gravitational potential. Three techniques have been established to observe the cluster gravitational potential: \emph{i)} hydrostatic equilibrium; \emph{ii)} cluster galaxy dynamics; \emph{iii)} \ac{wl}. Owing to its independence on the dynamical state, \ac{wl} has become in recent years the method of choice to calibrate galaxy cluster mass, see Ref.~ \cite{Umetsu:2020wlf} for a review. 

While \ac{wl} is expected to be close to unbiased on average, the signal-to-noise of individual cluster mass measurement is usually very low. Mass calibration is thus integrated into the likelihood framework discussed above by considering the marginal \ac{pdf} of the observable directly linked to halos mass $\hat{\mathcal{O}}_{\rm M}$ conditional on the observable one seeks to calibrate $\hat{\mathcal{O}}$ \cite{SPT:2014wkb, SPT:2018njh, DES:2023muu, DES:2024zpp}, reading
\begin{equation}
    p(\mathcal{O}_{\rm M} | \hat{\mathcal{O}}, z, \theta) = \left( \frac{\text{d}n}{\text{d}\hat{\mathcal{O}}}\right)^{-1} \frac{\text{d}n}{\text{d}\hat{\mathcal{O}}\text{d}\hat{\mathcal{O}}_{\rm M}}\,,
\end{equation}
with 
\begin{equation}
    \frac{\text{d}n}{\text{d}\hat{\mathcal{O}}\text{d}\hat{\mathcal{O}}_{\rm M}} = \int \text{d} \mathcal{O}\int \text{d} \mathcal{O}_{\rm M}\int \text{d} M \frac{\text{d}n}{\text{d} M} P(\mathcal{O}, \mathcal{O}_{\rm M} | M, z)  P(\hat{\mathcal{O}}| \mathcal{O}, z, \theta)P(\hat{\mathcal{O}}_{\rm M}| \mathcal{O}_{\rm M}, z) P(\text{sel}| \hat{\mathcal{O}}, \mathcal{O}, z, \theta)\,,
\end{equation}
where $P(\mathcal{O}, \mathcal{O}_{\rm M} | M, z)$ is the joint scaling relation between mass calibration observable and primary observable, and $P(\hat{\mathcal{O}}_{\rm M}| \mathcal{O}_{\rm M}, z)$ parameterized the measurement uncertainty on the mass calibration observable. Note that the correlation between the observable mass relation is crucial to account for physical selection effects \cite{DES:2022qkn, DES:2024gzt}. The marginal relation $P(\mathcal{O}_{\rm M} | M, z)$ needs to be anchored, and its parameters tightly constraint within prior, such that the overall likelihood of the mass calibration depends only on the parameters of the primary observable mass relation, thus constraining them.

\subparagraph{Mass calibration uncertainties}
\label{sec:syst_mass}

For the sake of brevity, we shall focus on the systematic uncertainty affecting \ac{wl} mass calibration. Given the advent of wide and deep photometric surveys, \ac{wl} mass calibration has emerged as the principal way to calibrate cluster number count. Three sources of statistical uncertainty impact the \ac{wl} signal of galaxy clusters: the shape noise due to the intrinsic ellipticity of source galaxies, the intrinsic heterogeneity of cluster mass profiles at the same mass (due to scatter around the concentration mass relation, varying degrees of substructure, orientation, triaxiality, etc.), and the statistical fluctuations in the line-of-sight integrated matter field \cite{Hoekstra:2002cq, Gruen:2015xxa}. Deeper and wider \ac{wl} surveys can only reduce the first of these three noise sources.

\ac{wl} mass calibration shares the systematics of the shear measurement and photometric redshift estimation inherent in the source catalogs from wide and deep photometric surveys. Furthermore, the displacement between the observed and true cluster position washes out the \ac{wl} signal (mis-centering). As clusters are significant overdensities in the galaxy field, depending on the source background selection, a fraction of unlensed cluster members might contaminate the source sample and dilute the signal (cluster member contamination). Finally, baryonic effects do not only alter the total halo mass but also the halo mass profile that sources the lensing. The combined impact of these effects is summarized by establishing the relation between the \ac{wl}L mass (that results from fitting the \ac{wl} signal) and the halo mass on realistic \ac{wl} simulations \cite{Applegate:2012kr, SPT:2017pbk, Grandis:2021aad}. \ac{wl} masses are biased low by 5-10\% depending on the fitting formula used and scatter around the true halo mass \cite{Becker:2010xj, Bahe:2011cb, Euclid:2023wwb}. The systematic uncertainties of the aforementioned effects are distilled into an uncertainty on the bias of the \ac{wl} mass \cite{Chiu:2021wst, DES:2023muu, DES:2024gzt, Kleinebreil:2024vfh}. Current understanding of baryon feedback puts a 2\% lower limit on this uncertainty \cite{Grandis:2021aad}. Depending on the \ac{wl} survey, photometric source redshift uncertainties increase this to 10\% at high cluster redshifts. The other effects have a quantifiably smaller impact, as, for instance, mis-centering and cluster member contamination are partially controlled empirically \cite{Bellagamba:2018gec, DES:2018kma, DES:2023muu, DES:2024gzt}. 

\subparagraph{Halo mass function}

The halo mass function described above only considers structures made by \ac{cdm}. However, halos contain a certain amount of baryons. Neglecting this contribution leads to a systematic bias in determining the halo mass function. This can be easily understood by considering that baryons will change the total mass of the halo.

Hydrodynamical simulations found that the effects of baryons are different whether, in the simulations, radiative effects are taken into account or gas heating is only due to gravitational effects. For instance, models with \ac{agn} feedback decrease the abundance of clusters compared to \ac{dm}-only simulations \cite{SPT:2014wkb}. The impact of baryons appears to depend on the halo mass. Considering an overdensity of 1500 times the critical density, baryons contribute up to 6\%--7\% to the total mass, while for 200 times the critical density, their contribution reduces to only about 1\%.

In terms of number counts, \cite{Cui:2011xc,Bocquet:2015pva} found that the number density of halos decreases up to 15\% at low masses and redshifts, while being in broad agreement with that of \ac{dm}-only simulations in the other cases. In addition, to better match X-ray and \ac{sz} observational studies, \cite{Bocquet:2015pva} proposed fits to estimate the baryon contribution when moving from a halo defined as 200 times the critical density to one found using 500 times this density. The retrieved halo masses in the presence of baryons can also be corrected to match the expected halo masses in gravity-only simulations \cite{Castro:2020yes,Grandis:2021aad}.

\paragraph{Results}

\subparagraph{Cosmological parameters measurements and tensions}

Cosmological results on the mean matter density at present time, $\Omega_\mathrm{m}$, from cluster number counts agree with \ac{cmb} results that it should be around 0.3. Recent results from eROSITA ($0.29 \pm 0.01$ \cite{Ghirardini:2024yni}), \ac{spt} ($0.29 \pm 0.03$ \cite{DES:2024zpp}), WtG ($0.26 \pm 0.03$ \cite{Mantz:2014paa}), XXL ($0.31 \pm 0.03$ \cite{Garrel:2021sgq}), \emph{Planck} \ac{sz} ($0.33 \pm 0.03$ \cite{Planck:2015lwi}) and \ac{des} ($0.32^{+0.08}_{-0.07}$ \cite{DES:2020cbm}), and UNIONS ($0.29\pm0.05$ \cite{Mpetha:2025bla}) are all consistent with the results from \emph{Planck} \ac{cmb} \cite{Planck:2018vyg} of $0.315 \pm 0.007$.

Conversely, there is currently some disagreement between the values of $\sigma_8$ from cluster number counts obtained in various studies. While some studies find $\sigma_8$ in agreement with the \ac{cmb} value of $0.811 \pm 0.006$ (eROSITA $0.88 \pm 0.02$ \cite{Ghirardini:2024yni};
\ac{spt} $0.82 \pm 0.03$ \cite{DES:2024zpp}; WtG $0.83 \pm 0.04$ \cite{Mantz:2014paa}; XXL $0.84 \pm 0.04$ \cite{Garrel:2021sgq}), \emph{Planck} \ac{sz} found a value of $0.76 \pm 0.03$ \cite{Planck:2015lwi} which is in tension with the \ac{cmb} at the level of $\sim2.5\sigma$.

Similar to the case of cosmic shear, and as highlighted in Fig.~\ref{fig:hmf}, consistency with \ac{cmb} should be assessed by considering the entire multi-dimensional posterior distribution. For this reason, it is customary to consider the quantity, $S_8 = \sigma_8 \left( \frac{\Omega_{\rm m,0}}{0.3} \right)^\alpha$, with $\alpha = 0.5$. We note, however, that the degeneracy between $\Omega_{\rm m,0}$ and $\sigma_8$ in cluster count experiments follows a slightly different slope from cosmic shear, with an index in the range 0.2 (e.g., \ac{spt} \cite{DES:2024zpp}) to 0.4 (e.g., eROSITA, \cite{Ghirardini:2024yni}).

Considering only experiments including an internally calibrated \ac{wl} mass-observable scaling relation, i.e., eROSITA, \ac{spt}, WtG, and XXL, the tension between $\Omega_\mathrm{m}$ and \ac{cmb} measurements is at most 1.2-$\sigma$, thus showing a high level of consistency between experiments. The same exercise applied to $\sigma_8$ indicates a slightly higher level of tension (1.7 $\sigma$), such that potential systematics still remain at this level of precision.

\subparagraph{Constraints on dark energy}
Cluster number counts are useful in providing constraints on the \ac{de} equation of state $w$. \ac{de} affects the growth of structures in the Universe at redshifts below $\sim1$, thereby reducing the growth of massive halos. Most current studies on the \ac{de} equation of state from cluster counts assume a constant equation of state $w$ and a flat universe. All the results published thus far are consistent with $w=-1$ (\emph{Chandra} $w = -1.14 \pm 0.21$ \cite{Vikhlinin:2008ym}; WtG $w = -0.98 \pm 0.15$ \cite{Mantz:2014paa}; \ac{spt} $w = -1.45 \pm 0.31$ \cite{DES:2024zpp}; eROSITA $w=-1.12 \pm 0.12$ \cite{Ghirardini:2024yni}). Similar results were obtained constraining the gas mass fraction to be constant with time ($w=-1.13^{+0.17}_{-0.20}$ \cite{SPT:2021vsu}).

\subparagraph{Alternative models}

The halo mass function, as discussed above, is a great tool to investigate the underlying cosmological model. For this reason, several authors studied its behavior in \ac{mg} models, such as the normal branch of Dvali-Gabadadze-Porrati theory (nDGP) \cite{Schmidt:2009sv} and $f(R)$ \cite{Kopp:2013lea} simulations. To model the HMF in \ac{mg} scenarios, $N$-body simulations are required to calibrate the difference between the HMF in the \ac{mg} simulations with respect to \lcdm. Ref.~\cite{Schmidt:2009sv} studied the evolution of perturbations for the nDGP model, which is characterized by a time-varying gravitational constant. The numerical nDGP mass function predicts up to three times more objects at very high masses, for strong deviations from \lcdm, explained by the fact that the halo mass function is very sensitive to $\sigma_8$.

A similar approach, but which combines the benefits of the previous ones is that used by Refs.~\cite{Kopp:2013lea,Hagstotz:2018onp} to study the halo mass function in $f(R)$ models. The authors parameterize the modifications induced by $f(R)$ models both in the spherical collapse model and in the HMF, calibrating these modifications on $N$-body simulations. These modifications are a function of mass, and redshift. Finally, Ref.~\cite{Gupta:2021pdy} showed that the halo mass function is universal when expressed in terms of $\ln{(\sigma^{-1})}$ rather than the halo mass $M$.

\paragraph{Other cosmological probes from galaxy clusters: cluster clustering}
The spatial distribution of galaxy clusters represents a valuable cosmological probe, depending on the composition of the Universe and on the nature of gravitational interaction. In particular, cluster clustering traces the growth rate of density perturbations, therefore it has been widely employed for inferring fundamental cosmological parameters, such as the matter density, $\Omega_{\rm m,0}$, and the amplitude of mass fluctuations on scales of $8\,h ^{-1}$ Mpc, $\sigma_8$ \cite{Sereno:2014eea, Marulli:2020uyy, Lesci:2022owx, Romanello:2023obk}, as well as the neutrino properties \cite{Cerbolini:2013uya} and deviations from the \lcdm\ model, possibly described by the \ac{dde} parameters, i.e., $w_0$ and $w_a$ \cite{Sartoris:2015aga}. Moreover, it can be used to significantly improve the precision on the constraints from cluster weak gravitational lensing \cite{Sereno:2014eea} and number counts \cite{Sartoris:2015aga}, when analyzed in combination with them.\\

Despite the limit in the exploitation of the clustering properties due to the difficulty in collecting large homogeneous cluster samples, which leads to poorer statistics with respect to galaxy samples, cosmology with galaxy clusters presents a series of relevant advantages. First, galaxy clusters, being placed at the nodes of the filamentary structure of the cosmic web, are hosted by the latest and most massive virialized haloes formed by the hierarchical growth of cosmic structures. Therefore they are highly biased tracers, more clustered than galaxies \cite{ Moscardini:2000iw, Sheth:1999su, Allen:2011zs}. While, in general, the galaxy bias is difficult to model properly, especially on small physical scales, and it is usually seen as a nuisance parameter in cosmological analyses based on galaxy statistics, the effective bias of galaxy cluster samples can be theoretically predicted \cite{Branchini:2016glc, Paech:2016hod, Lesci:2022owx, Romanello:2023obk}, depending on the mass and the redshift of the hosting haloes \cite{Tinker:2010my}. In turn, the cluster mass, mainly consisting of \ac{dm}, can be linked to directly observable mass proxies, through the so-called mass-observable scaling relation \cite{Okabe:2010mw, Allen:2011zs, Giodini:2013zqa, Bellagamba:2018gec}. However, the uncertainties on scaling relation parameters and their redshift evolution affect our ability to recover the true cluster mass and represent one of the limits of cosmological inference from galaxy clusters. This highlights the importance of having a robust \ac{wl} mass calibration, capable of assessing the systematics caused by halo orientation, selection, and projection effects \cite{DES:2018kma, DES:2020mlx}. A second advantage is that galaxy clusters present lower peculiar velocities with respect to their host galaxies, reducing the small-scale distortions in the clustering signal, namely the Fingers of God effect, caused by nonlinear dynamics and incoherent motion within virialized structures \cite{Veropalumbo:2013cua, Marulli:2015jga}. \\

\indent Cluster clustering is traditionally investigated through the analysis of the two-point correlation function, $\xi(r)$, which measures the excess probability of finding a pair of objects in the volume elements $\delta V_1$ and $\delta V_2$, at the comoving separation $r$, relative to that expected from a random distribution. Increasing attention is being paid to higher-order statistics, and in particular, to the three-point correlation function, $\zeta(r_{12}, r_{13}, r_{23})$, which provides the probability of finding triplets of objects at comoving separations $r_{12}$, $r_{13}$, and $r_{23}$, and has been successfully applied for the detection of the \ac{bao} peak of galaxy clusters \cite{Moresco:2020quj}. These probes require the assumption of a fiducial cosmology to convert the observed angular and redshift coordinates into distances, which might be different from the true one, resulting in geometric distortions \cite{Alcock:1979mp}, which have to be taken into account in the modeling of the clustering signal \cite{Marulli:2012na, BOSS:2013uda}. In recent years, the two-point redshift-space correlation function of a homogeneous sample of X-ray selected galaxy clusters from the XXL survey, carried out by the XMM-Newton satellite, has been used to measure a value of $\Omega_{\rm m,0}$ in full agreement with the \lcdm\ model \cite{Marulli:2018owk}. On the other hand, optically selected clusters from the Sloan Digital Sky Survey (SDSS) have been employed to get an estimate of $f\sigma_8$, which was found consistent with General Relativity predictions \cite{Marulli:2020uyy}, to determine the distance–redshift relation, $\Omega_{\rm m,0}$ and $H_0$, from the \ac{bao} peak \cite{Veropalumbo:2013cua, Veropalumbo:2015dpi} and, in combination to stacked gravitational lensing, to constrain $\sigma_8$ \cite{Sereno:2014eea}. Moreover, from the 3D clustering of the Planck \ac{sz} selected galaxy clusters, the Planck mass bias, $b_\mathrm{SZ}$, and $\Omega_{\rm m,0}$, have been derived, while $\sigma_8$ was not constrained \cite{Lesci:2023blg}. The tomographic cluster autocorrelation from the Constrain Dark Energy with X-ray (CODEX) sample, after redshift-dependent richness selection, has been employed to explore the $\Omega_{\rm m,0}-\sigma_8$ degeneracy, putting constraints on the structure growth parameter, $S_8$ \cite{Lindholm:2020gsd}. Since cluster clustering is particularly sensitive to these parameters, independent constraints on $S_8$ have been put by measuring the two-point correlation function of the photometric clusters detected in the third data release of \ac{kids}-DR3 \cite{Lesci:2022owx}. An equivalent approach consists in the study of the angular correlation function, $w(\theta)$, and its spherical harmonic counterpart, the angular power spectrum, $C_\ell$ \cite{1973ApJ...185..413P}. In particular, a first measurement of the auto-correlation cluster power spectrum and of the galaxy-galaxy cluster cross-spectrum has been performed on the SDSS-DR8 photometric sample \cite{Paech:2016hod}. Moreover, from the tomographic study of $w(\theta)$ and $C_\ell$ of the same \ac{kids}-DR3 catalog, competitive constraints on $S_8$ have been found \cite{Romanello:2023obk}. These results, being based on angular positions alone, are not affected by geometric distortions, highlighting the potential of angular cluster clustering as a cosmological probe in future photometric redshift surveys.
\bigskip
\subsubsection{Galaxy Clustering -- Other probes \label{sec:other_probes}}

\noindent \textbf{Coordinator:} Chandra Shekhar Saraf\\
\noindent \textbf{Contributors:} Kishan Deka, and Pawe\l{} Bielewicz \\

\paragraph{Cross-correlations}

\begin{figure}[ht]
    \centering
    \includegraphics[scale = 0.40]{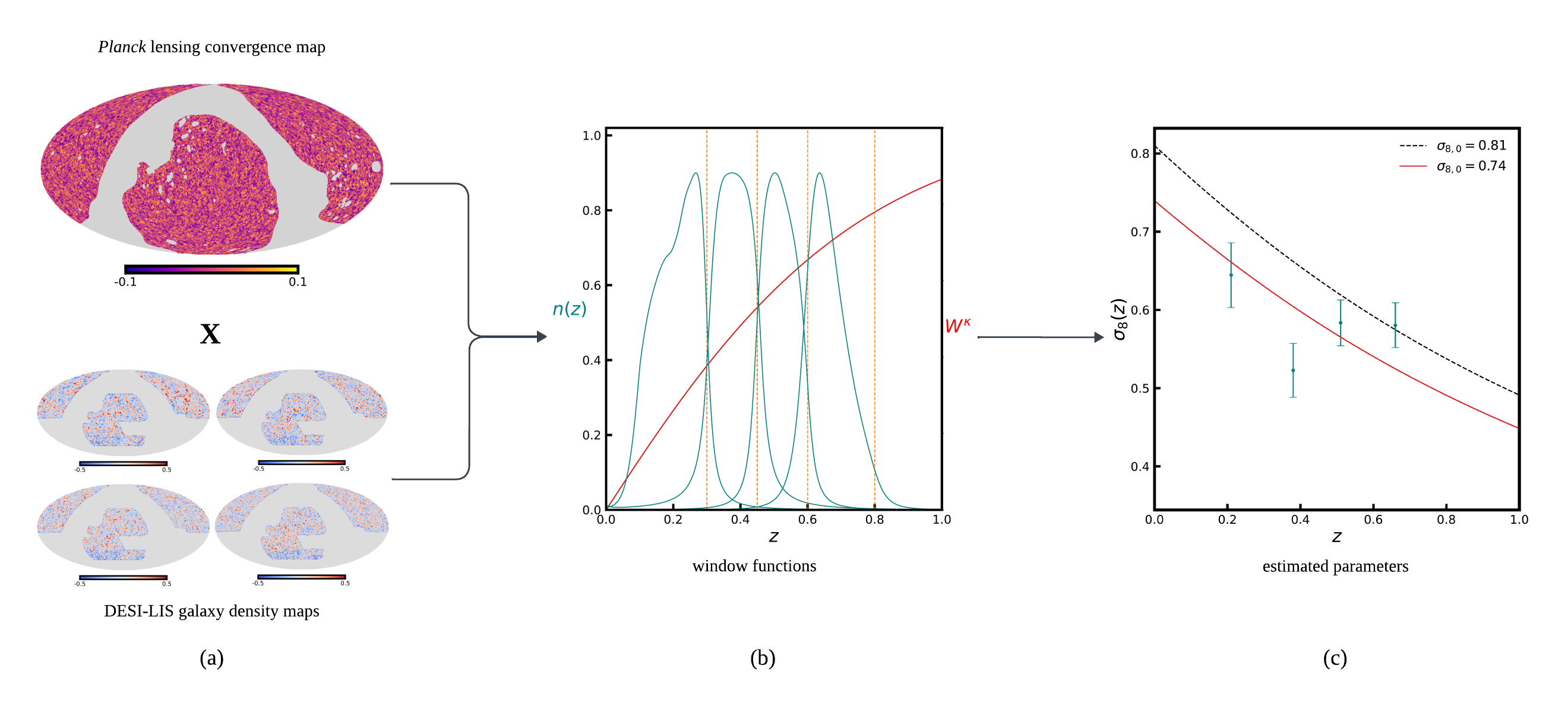}
    \caption[Cross-correlation schematic]{An illustration of estimating $\sigma_{8}$ parameter from tomographic cross-correlation measurements between \ac{desi}-LIS galaxy survey and \textit{Planck} lensing convergence map, assuming \lcdm\ model \cite{Saraf:2023bjo}. (a) The \textit{Planck} lensing convergence map and \ac{desi}-LIS galaxy density maps in four tomographic bins, smoothed with a Gaussian beam of $60'$ FWHM (only to better show the large scale distribution). (b) The galaxy redshift distribution for the four redshift bins (solid blue lines) and the \ac{cmb} lensing kernel (solid red line). The vertical dashed orange lines mark the boundaries of redshift bins. (c) The $\sigma_{8}$ parameter estimated from cross-correlation measurements of four tomographic bins. The red line is the evolution of $\sigma_{8}$ parameter as estimated from cross-correlation data. The dashed black line is the evolution of $\sigma_{8}$ parameter measured from the \textit{Planck} \ac{cmb} data alone.}
\label{fig:cross_correlation_schematic}
\end{figure}

The $S_{8}$ parameter can be measured from the redshift space clustering of galaxies by estimating the two- and three-point correlation functions or their Fourier counterparts, the galaxy power spectrum, and bispectrum. These statistics can be used to gain insights into the growth rate of cosmic structures, $f\sigma_{8}(z)$, from \ac{rsd} at the effective redshift of the galaxy sample \cite{eBOSS:2020qek,Li:2016bis,Alam:2015qta}. The galaxy clustering data can be combined with \ac{cmb} surveys to yield constraints on $\sigma_{8}$ and $S_{8}$.
Several studies \cite{ACT:2023oei, Piccirilli:2024xgo, Alonso:2023guh, BOSS:2016wmc} have reported values of $\sigma_{8}$ consistent with the \ac{cmb}-only analyses. An illustrative schematic of estimating $\sigma_{8}$ parameter from cross-correlation measurements between \textit{Planck} \ac{cmb} lensing convergence map and \ac{desi}'s Legacy Imaging Survey (\ac{desi}-LIS; \cite{DESI:2018ymu}) is shown in Fig.~\ref{fig:cross_correlation_schematic} (see Ref.~\cite{Saraf:2024log} for details on cross-correlation analysis). A cross-correlation between unWISE galaxies and \ac{act} DR6 \ac{cmb} lensing measurements found $S_{8} = 0.813\pm 0.021$, and $S_{8} = 0.810\pm 0.015$ with a combination of Planck and \ac{act} cross-correlations \cite{Alonso:2023guh}. These measurements are fully consistent with the predictions from primary \ac{cmb} measurements within our standard cosmology. Another study correlating Gaia–unWISE \ac{qso} catalog with \textit{Planck} PR4 \ac{cmb} lensing found $\sigma_{8} = 0.7766\pm 0.034$ and $\Omega_{\rm m,0} = 0.343^{+0.017}_{-0.019}$, which translates to $S_{8} = 0.819\pm 0.058$ \cite{Piccirilli:2024xgo}. However, many cross-correlation analyses between \ac{cmb} datasets and galaxy samples divided into narrow redshift bins \cite{Nakoneczny:2023nlt, DES:2022xxr, White:2021yvw, Sun:2021rhp, Krolewski:2021yqy, Hang:2020gwn, Peacock:2018xlz, DES:2015eqk} find low values of $\sigma_{8}$ and $S_{8}$ parameter (when $\Omega_{\rm m,0}$ is generally assumed to be consistent with the well constrained values from \ac{sn} and \ac{cmb} analyses), resulting in mild tension with \textit{Planck} analysis. A tomographic cross-correlation of \ac{des}-Y3 data and \ac{cmb} lensing from \ac{spt} and Planck estimated $S_{8} = 0.734^{+0.035}_{-0.028}$ \cite{DES:2022xxr}. Another cross-correlation of \ac{desi} Legacy Imaging survey LRGs and \ac{cmb} lensing from \textit{Planck} found $S_{8} = 0.765\pm 0.023$ and $S_{8} = 0.790^{+0.024}_{-0.027}$ with \ac{act} DR6 \cite{Sailer:2024coh}.

\paragraph{Full shape analyses}

The $\sigma_{8}$ parameter can also be measured by fitting the full shape of the observed power spectrum from spectroscopic surveys as well as the galaxy bispectrum \cite{Simon:2022csv, Kobayashi:2021oud, Chen:2021wdi, SimBIG:2023nol, Ivanov:2023qzb, Philcox:2021kcw}. Furthermore, the $\sigma_{8}$ parameter can be translated into $S_{8}$ parameter with additional measurements of $\Omega_{\rm m,0}$ (from power spectrum shape, \ac{sn} or \ac{cmb}). The measurements of $S_{8}$ parameters employing full shape analysis of the galaxy power spectrum \cite{Philcox:2021kcw,Simon:2022csv} are in tension with \textit{Planck} measurements \cite{Planck:2018vyg}, but with typically less than $3\sigma$ statistical significance. Recent constraints on the $S_{8}$ parameter from the \ac{boss} galaxy bispectrum multipoles found $S_{8} = 0.774^{+0.056}_{-0.053}$ \cite{SimBIG:2023nol} and $S_{8} = 0.77\pm 0.04$ \cite{Ivanov:2023qzb}, which are statistically consistent with both \ac{cmb} and \ac{wl} measurements.

\paragraph{$3\times 2$pt analyses}

The combined analysis of two-point statistics corresponding to galaxy clustering, galaxy-galaxy lensing, and cosmic shear is commonly known as the ``$3\times 2$pt'' analysis. The joint analyses provide us with an improved control of the systematic uncertainties such as galaxy bias and intrinsic alignments. The first $3\times 2$pt analysis were published with the datasets from \ac{kids}-450 $\times$ \{2dFLenS + \ac{boss}\} \cite{Joudaki:2017zdt}, \ac{kids}-450 $\times$ GAMA \cite{vanUitert:2017ieu} and \ac{des}-Y1 \cite{DES:2017myr}. The \ac{des}-Y1 analysis used photometric galaxies as both lens and sources, finding $S_{8} = 0.773^{+0.026}_{-0.020}$. The \ac{kids} collaboration, on the other hand, used overlapping imaging and spectroscopic surveys, resulting in $S_{8} = 0.742\pm 0.035$ when combined with 2dFLenS + \ac{boss}, and $S_{8} = 0.800^{+0.029}_{-0.027}$ with GAMA survey.

The updated constraints on $S_{8}$ with $3\times 2$pt analyses combining \ac{kids}-1000 with \{2dFLenS + \ac{boss}\} datasets gives $S_{8} = 0.766^{+0.020}_{-0.014}$ \cite{Heymans:2020gsg}, at $3.1\sigma$ tension with \textit{Planck} constraints within the \lcdm\ model. The recent measurements from the \ac{des}-Y3 datasets resulted in $S_{8} = 0.776\pm 0.017$ in the \lcdm\ cosmology and $S_{8} = 0.775^{+0.026}_{-0.024}$ assuming a $w$CDM model \cite{DES:2021wwk}. However, a joint analysis of \ac{des}-Y3 $3\times 2$pt data, \textit{Planck} \ac{cmb} anisotropy data (without lensing), \ac{eboss} \ac{bao} and \ac{rsd} measurements, and \ac{des} \ac{sn1} data provides $S_{8} = 0.812\pm 0.008$ in the \lcdm\ model. A joint cosmic shear analysis of \ac{kids}-1000 and \ac{des} Y3 data has obtained stringent constrain of $S_{8}= 0.790^{+0.018}_{-0.014}$ \cite{Kilo-DegreeSurvey:2023gfr}. Another $3\times 2$pt analyses performed with the \ac{hsc}-Y3 imaging data and SDSS \ac{boss} DR11 spectroscopic galaxies yielded $S_{8} = 0.775^{+0.043}_{-0.038}$ for the \lcdm\ model \cite{Sugiyama:2023fzm}. The \ac{hsc}-Y3 analysis is highly consistent with both \ac{kids}-1000 and \ac{des}-Y3 measurements of $S_{8}$ within one standard deviation.

\paragraph{Systematics}

The measurements of cosmological parameters from large-scale structure evolution at low redshifts require state-of-the-art analysis pipelines to correct for a large number of systematic effects. The observation sector can include effects such as atmospheric and extragalactic extinction, stellar contamination, airmass and blurring \cite{BOSS:2012coo}, fibre collisions \cite{Hahn:2016kiy}, angular and radial modes systematics \cite{deMattia:2019vdg}.

For the cross-correlation between galaxy clustering and \ac{cmb}, galaxy survey systematics such as photometric calibration errors, foreground contamination and catastrophic errors need to be modeled and mitigated for unbiased inferences \cite{Saraf:2021okc, Pullen:2015vtb, Krolewski:2021yqy, Joudaki:2016mvz, DES:2015eqk}. In order to allow tomographic cross-correlations with photometric galaxy surveys, both precise individual galaxy photometric redshifts and accurate redshift distributions for the ensemble, require techniques for redshift determination. Typically, a subset of spectroscopically observed sources are used to train, validate and calibrate photometric redshifts and redshift distributions. The accuracy of the redshift distribution directly impacts the accuracy of estimated cosmological parameters. Biases in the mean redshifts of the tomographic bins will cause a bias in $S_{8}$. Recently, \cite{Saraf:2024log,Saraf:2023bjo} have shown the importance of precise modeling of redshift error distributions and redshift bin mismatch of objects, when estimating $\sigma_{8}$ and $S_{8}$ parameters.

In the theory sector, the largest uncertainty comes from the accuracy of the models for non-linear clustering of galaxies. The cross-correlation measurements, full shape analyses, and the bispectrum multipoles, all depend on the accurate modeling of the non-linear regime in the redshift-space galaxy clustering. This uncertainty has been improved after the significant progress from the effective field theory of the large-scale structure which provides accurate description of galaxy clustering at large scales \cite{Ivanov:2019pdj,Zhao:2023ebp,Gsponer:2023wpm,DAmico:2022osl,Chen:2020zjt,DESI:2024lms,Maus:2024sbb,Baumann:2010tm,Carrasco:2012cv,Porto:2013qua,Lewandowski:2015ziq}.
\bigskip
\subsection{Other challenges}
\subsubsection{Systematics and the \texorpdfstring{$A_{\rm lens}$}{} parameter \label{sec:syst_and_A_lens}}

\noindent \textbf{Coordinator:} William Giar\`{e}\\
\noindent \textbf{Contributors:} Alessandro Melchiorri, Anto Idicherian Lonappan, Deng Wang, Elsa M. Teixeira, Enrico Specogna, Giulia Gubitosi, Ido Ben-Dayan, Matteo Forconi, \"{O}zg\"{u}r Akarsu, and Rita B. Neves
\\

On scales smaller than ten arcminutes, the interaction between \ac{cmb} photons and the Universe's large-scale structure becomes significant, giving rise to second-order anisotropies. One of the most important contributions is the gravitational deflection, or lensing, experienced by \ac{cmb} photons along their paths~\cite{Lewis:2006fu}. At arcminute scales, lensing deflections distort the observed image of the \ac{cmb} fluctuations, imprinting a distinctive non-Gaussian four-point correlation function (or trispectrum) in both the temperature and polarization anisotropies~\cite{Lewis:2006fu}. This signal can be extracted with a high signal-to-noise ratio by correlating power in different directions on the sky, leading to a direct reconstruction of the power spectrum of the \ac{cmb} lensing field, now measured at $\sim 40\sigma$~\cite{Planck:2018lbu,Carron:2022eyg,ACT:2023dou,ACT:2023kun}. 

Although gravitational lensing does not alter the overall distribution of primary \ac{cmb} anisotropies, it leaves distinctive signatures in the spectra of temperature and polarization anisotropies. The most relevant effects are the conversion of E-mode polarization to B-modes, the transfer of power to the damping tail, and the lensing-induced smoothing of the acoustic peaks and troughs in the TT, TE, and EE spectra. The degree of smoothing is directly tied to the amplitude of the power spectrum of the lensing potential, which is derived from the six \lcdm\ parameters. As a result, interesting consistency tests have been proposed to verify whether the amplitude of lensing inferred through the high-$\ell$ smoothing of acoustic peaks matches the theoretical predictions of the standard cosmological model. 

In Ref.~\cite{Calabrese:2008rt}, the phenomenological parameter $A_{\rm lens}$ was introduced to encompass various physical mechanisms that might affect the lensing amplitude. This parameter scales the amplitude of the lensing trispectrum, effectively accounting for the associated smoothing effects in the \ac{cmb} temperature and polarization spectra. $A_{\rm lens}$ is defined such that $A_{\rm lens} = 1$ matches the standard \lcdm\ prediction, whereas $A_{\rm lens} = 0$ corresponds to a scenario where \ac{cmb} lensing is disregarded altogether. By allowing $A_{\rm lens}$ to remain a free parameter in theoretical models, its value can be directly constrained by data, potentially confirming or deviating from the \lcdm\ predictions.

Focusing on the Planck-2018 data release (PR3), the analysis of the Planck \texttt{plik} likelihood for the TT, TE, and EE spectra at $\ell > 30$, combined with the \texttt{Commander} likelihood for the TT spectrum at $2 \leq \ell \leq 30$ and the \texttt{SimAll} likelihood for the EE spectrum at $2 \leq \ell \leq 30$, yields $A_{\rm lens} = 1.180 \pm 0.065$~\cite{Planck:2018vyg}. This finding suggests an excess lensing signal at approximately $2.8\sigma$, resulting from a significant improvement in $\chi^2$ of approximately $\Delta \chi^2 \sim 9.7$. This improvement primarily originates from high-$\ell$ temperature and polarization data, particularly within the multipole range $600 < \ell < 1500$. As shown in Fig. 24 of Ref.~\cite{Planck:2018vyg}, there is a visible preference for increased lensing smoothing in the oscillatory residuals in the TT spectrum at $1100 < \ell < 2000$. 

Since 2018, the Planck data has undergone substantial reanalyses. The new Planck PR4 (\texttt{NPIPE}) \ac{cmb} maps incorporate significant improvements, such as increased coverage of sky area at high frequencies, improved processing of time-ordered data, and approximately 8\% more data accounted for in the lensing trispectrum reconstruction. Updated likelihoods for temperature and polarization spectra have been released following these developments. The latest versions of \texttt{CamSpec}~\cite{Rosenberg:2022sdy} and \texttt{HiLLiPoP}~\cite{Tristram:2023haj} -- two likelihoods already employed in various studies by the Planck collaboration -- are now based on the Planck PR4 (\texttt{NPIPE}) maps, reducing small-scale noise compared to \texttt{plik} and enhancing the constraints on cosmological parameters by up to 10\%. Both likelihoods indicate a shift towards $A_{\rm lens} = 1$. For \texttt{CamSpec}, the constraints on the lensing amplitude inferred from high-$\ell$ smoothing of acoustic peaks are summarized in Table~6 of Ref.~\cite{Rosenberg:2022sdy}. Temperature and polarization data reduce the preference for $A_{\rm lens} > 1$ to less than $1.7\sigma$ while focusing solely on the TT spectrum increases the lensing anomaly to $2.3\sigma$. Constraints on $A_{\rm lens}$ resulting from the new \texttt{HiLLiPoP} likelihood are summarized in Table~6 of Ref.~\cite{Tristram:2023haj}. They consistently agree with \lcdm\ at $1\sigma$. 

These new reanalyses suggest that the preference for excess power smoothing has decreased in the \texttt{NPIPE} maps, lending weight to interpreting the lensing anomaly as systematics in the Planck PR3 data. The latter interpretation finds support in \ac{cmb} experiments other than Planck. \ac{act} and the \ac{spt} have released precise small-scale measurements of the spectra of temperature and polarization anisotropies~\cite{ACT:2020frw,SPT-3G:2021eoc,SPT-3G:2022hvq}, as well as precise reconstructions of the lensing trispectrum~\cite{ACT:2023dou,ACT:2023kun,SPT:2023jql}. The Data Release 4 of the \ac{act} constrains $A_{\rm lens} = 1.01 \pm 0.11$ in remarkable agreement with the baseline value (see Fig. 16 of Ref.~\cite{ACT:2020gnv}). Similarly, the \ac{spt} TT, TE, and EE spectra analysis gives $A_{\rm lens} = 0.87 \pm 0.11$, consistent with \lcdm\ at $1.2\sigma$. In this case, we refer to Table~V of Ref.~\cite{SPT-3G:2022hvq} and Table~VII of Ref.~\cite{SPT-3G:2021eoc} for earlier analyses involving only the EE and TE spectra (yielding $A_{\rm lens} = 0.98 \pm 0.12$). 

\begin{figure}
    \centering
    \includegraphics[width=0.45\textwidth]{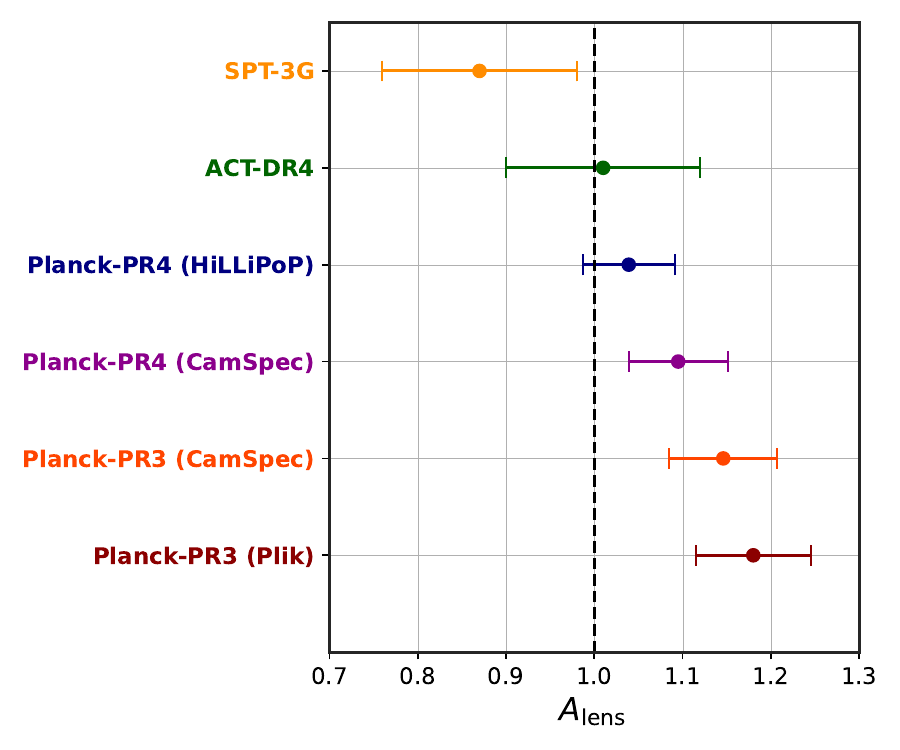}
    \caption{The whisker plot summarizes the 68\% CL intervals for the parameter $A_{\rm lens}$, inferred over the years from the analysis of lensing-induced smoothing of the acoustic peaks and troughs in the TT, TE, and EE spectra measured by various experiments with different likelihoods. Constraints based on the Planck-PR3 measurements indicate an excess of lensing, which is reduced in the recent Planck-PR4 updated likelihoods. Small-scale temperature and polarization spectra from \ac{act}-DR4 and \ac{spt3g} are in overall agreement with $A_{\rm lens} = 1$, consistently recovered within one standard deviation (or slightly more).
}
\label{fig:Alens_WP}
\end{figure}

Fig.~\ref{fig:Alens_WP} summarizes the 68\% CL intervals for the parameter $A_{\rm lens}$, inferred over the years from the analysis of lensing-induced smoothing of the acoustic peaks and troughs in the TT, TE, and EE spectra measured by various experiments and likelihoods discussed so far.

\bigskip
\subsubsection{Evidence for a nonvanishing \texorpdfstring{$\Omega_k$}{} \label{sec:nonzero_curv}}

\noindent \textbf{Coordinator:} Will Handley\\

\noindent The spatial curvature of the Universe, parameterized by $\Omega_k$, is a fundamental parameter in cosmology, intimately tied to the geometry and fate of the Universe. While the inflationary paradigm predicts a value extremely close to zero, it is important to observationally test this prediction. In the standard cosmological model, $\Omega_k$ is degenerate with other parameters when using \ac{cmb} data alone, such that \ac{cmb} data do not provide strong constraints on curvature. However, the addition of external data sets, such as \ac{bao} measurements, can break this degeneracy.

The Planck 2018 data release presented an intriguing hint of a possible departure from a flat universe. Analyzing the TT,TE,EE+lowE data, the Planck team found a preference for negative values of $\Omega_k$ in the range $-0.095<\Omega_k<-0.007$ at 99\% confidence level, with $\Delta\chi^2_{\rm eff}=-11$ compared to the baseline \lcdm\ model~\cite{Planck:2018vyg}. This preference was attributed to a combination of volume effects and better fits both the high-$\ell$ and low-$\ell$ data in closed models. Notably, closed models predict a higher lensing amplitude ($A_{\rm L}$), which can explain the preference for $A_{\rm L}>1$ in Planck data, and also a better fit to the low-$\ell$ temperature power spectrum. Adding Planck lensing data to TT,TE,EE+lowE, the constraint on $\Omega_k$ shifts to $\Omega_k=-0.0106 \pm 0.0065$ at 68\% confidence level, reducing the preference for closed models to less than $2\sigma$. Finally, including \ac{bao} data results in $\Omega_k = 0.0007 \pm 0.0019$, consistent with a flat universe at $1\sigma$~\cite{Planck:2018vyg}.

However, this conclusion has been challenged in the literature. Ref.~\cite{DiValentino:2019qzk} and Ref.~\cite{Handley:2019tkm} show that there is Bayesian evidence for a closed universe, albeit not quantified in the Planck paper, and that the Planck lensing reconstruction is in tension with TT,TE,EE+lowE data at $2.5\sigma$. In particular, the tension is driven by the preference for a higher value of $A_{\rm L}$ in closed models, which is not supported by the Planck lensing reconstruction. They argue that \ac{bao} data are also in tension with TT,TE,EE+lowE data, and that combining these datasets is not justified. Ref.~\cite{Efstathiou:2020wem} conversely shows that the preference for $\Omega_k < 0$ is reduced when using the CamSpec likelihood, which uses more data than the Planck baseline likelihood, and argue that \ac{bao} data strongly prefer flat universes.

More recently, Ref.~\cite{Glanville:2022xes} showed that the \ac{bao} scale measurements implicitly assume a fiducial \lcdm\ cosmology, and that the preference for a flat universe can be weakened when relaxing these assumptions. They find that using the full shape of the galaxy power spectra, the tension between \ac{bao} and Planck data in the $\Omega_k$ parameter is reduced to $\sim1.5\sigma$.

The tension between Planck \ac{cmb} data and \ac{bao} measurements in the context of non-vanishing curvature has motivated numerous works that explore its possible origin and its implications for cosmology. For example, Ref.~\cite{Vagnozzi:2020dfn} used \ac{cc} as an alternative dataset to break the geometric degeneracy, finding $\Omega_k = -0.0054 \pm 0.0055$, consistent with a flat universe. However, the tension between different $H_0$ measurements persists even in the context of non-flat models~\cite{Vagnozzi:2020rcz, Park:2017xbl, Park:2018tgj}.

The curvature tension therefore remains an open problem in cosmology. Future observations, such as those from \ac{desi}, Euclid, and the Vera Rubin Observatory will provide more precise \ac{bao} measurements, which will allow us to test the robustness of the flat Universe assumption and potentially shed light on the origin of this discrepancy.
\bigskip
\subsubsection{Anisotropic anomalies in the cosmic microwave background radiation \label{sec:Anomalies_CMB}}

\noindent \textbf{Coordinator:} Leandros Perivolaropoulos\\
\noindent \textbf{Contributors:} Andr\'as Kov\'acs, Anto Idicherian Lonappan, Emanuela Dimastrogiovanni, Eoin \'O Colg\'ain, Frode K. Hansen, Giulia Gubitosi, Laura Mersini-Houghton, Marina Cort\^es, Nils A. Nilsson, Shahin Sheikh-Jabbari, and Venus Keus
\\

The \ac{cmb}, a relic from the early Universe, provides crucial insights into cosmology. Observations by COBE, \ac{wmap}, and \textit{Planck} have revealed several anisotropic anomalies that may challenge the prevailing \lcdm\ cosmological model and the cosmological principle, prompting a reevaluation of our understanding of the early Universe's structure and the inflationary paradigm \cite{WMAP:2012fli,Planck:2015mrs}.

One such anomaly is the unusually low quadrupole moment ($\ell = 2$) in the \ac{cmb} power spectrum, which deviates from the predictions of cosmic variance and the \lcdm\ model \cite{WMAP:2003zzr,Planck:2015bpv}. This anomaly has implications for inflationary cosmology and may require theoretical adjustments. The low quadrupole moment was first observed by COBE and later confirmed by \ac{wmap} and \textit{Planck}, with increasing precision. The statistical significance of this anomaly has been thoroughly investigated, and its persistence across multiple observations challenges our understanding of the primordial power spectrum and the inflationary paradigm. Theoretical efforts to explain the low quadrupole moment have included modifications to the inflation model, such as a running spectral index or a cutoff in the primordial power spectrum at large scales \cite{Contaldi:2003zv,Cline:2003ve}.

Another striking feature is the alignment of the quadrupole and octopole moments, which appears to be oriented with the ecliptic plane and the motion of the Solar System \cite{Schwarz:2004gk,Copi:2005ff}. This alignment violates statistical isotropy and suggests potential cosmological or local explanations. The quadrupole-octopole alignment was first reported using \ac{wmap} data and later confirmed by \textit{Planck}, with both missions providing strong evidence for its existence. The alignment has been studied extensively, with some proposing that it could be a signature of cosmic topology \cite{deOliveira-Costa:2006sst}, while others have investigated the possibility of local foreground contamination or systematic effects \cite{Slosar:2004xj,Hajian:2007pi}. The statistical significance of this alignment and its potential origins remain active areas of research.

\begin{figure}[htbp]
\centering
\includegraphics[width=0.5\textwidth]{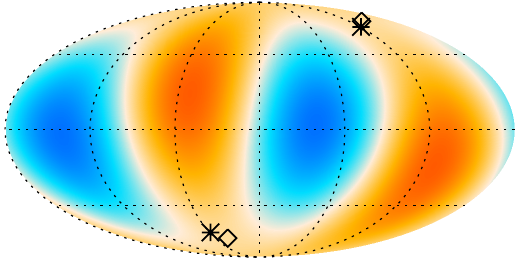}
\includegraphics[width=0.5\textwidth]{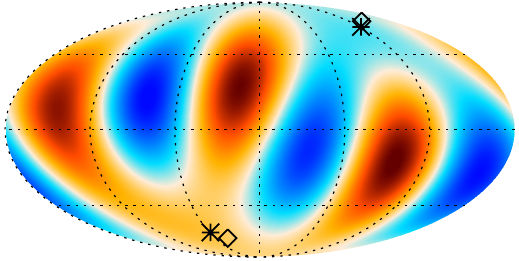}
\caption{Upper Panel: Derived quadrupole moment from the Planck SMICA map, showing the preferred axis of the quadrupole. Lower Panel: Derived octopole moment, indicating the alignment with the quadrupole. These alignments suggest a significant anomaly in the \ac{cmb}, challenging the assumption of statistical isotropy. The plus and star symbols indicate the axes of the
quadrupole and octopole, respectively, around which the angular
momentum dispersion is maximized. The diamond symbols
correspond to the quadrupole axes after correction for the kinematic
quadrupole. From Ref.~\cite{Planck:2013lks}. }
\label{fig:CMB_alignment}
\end{figure}

The upper and lower panels of Fig.~ \ref{fig:CMB_alignment} illustrate the derived quadrupole and octopole moments, respectively. The alignment of these moments with the ecliptic plane highlights one of the key anisotropic anomalies in the \ac{cmb}. This alignment suggests a possible violation of statistical isotropy and indicates that there may be underlying cosmological or local causes for this observed pattern. Understanding the significance and origins of these alignments is crucial for interpreting the implications for the standard cosmological model and exploring potential new physics.

The \ac{cmb} also exhibits a hemispherical power asymmetry, known as the dipolar power distribution, where one half of the celestial sphere has significantly more temperature fluctuations than the other \cite{Eriksen:2003db,Planck:2015igc}. The scale and statistical significance of this asymmetry raise questions about the density variations in the early Universe. The hemispherical power asymmetry was first detected in \ac{wmap} data and later confirmed by \textit{Planck}. The hemispherical asymmetry can be considered as two different anomalies as described in detail in Ref.~\cite{Planck:2015igc}. For larger scales, $\ell<100$, the asymmetry can be modeled as a dipolar modulation of an isotropic field. The $p$-value of this modulation asymmetry has been reported as $0.1-1$\%, but considerably less significant when correcting for the choice of $\ell_\mathrm{max}$. The second hemispherical asymmetry, the angular clustering asymmetry, extends to much smaller scales. By estimating the \ac{cmb} power spectrum locally in different parts of the sky, maps of the spatial distribution of the power spectrum can be estimated for different multipole ranges. When estimating a dipole for these local power spectrum maps at different scales, these dipoles are found to cluster. This clustering is found to persist at least to $\ell=1000$ and only $1/1000$ simulations show similar clustering. The clustering direction is close to the direction of the large scale dipolar asymmetry. The origin of this asymmetry remains unclear, with some proposing that it could be a signature of non-Gaussianity in the primordial perturbations \cite{Erickcek:2008sm,Schmidt:2012ky,Hansen:2018pgg}, while others have explored the possibility of a super-horizon scale mode modulating the primordial power spectrum \cite{Gordon:2004ez}.

Additionally, there is a parity asymmetry in the \ac{cmb}, with a preference for odd-parity modes over even-parity modes, challenging the scale invariance of primordial fluctuations \cite{Land:2005ad,Kim:2010st}. The parity asymmetry was first reported using \ac{wmap} data and later confirmed by \textit{Planck}, with both missions providing evidence for a statistically significant difference between the power in odd and even multipoles. This asymmetry has been studied in the context of the primordial tensor-to-scalar ratio, with some suggesting that it could be a signature of chiral gravity \cite{Lue:1998mq,Alexander:2006mt}. The physical origin of this asymmetry and its implications for the inflationary model remain open questions.

The Cold Spot is another intriguing anomaly – a large, unusually cold region in the \ac{cmb}, surrounded by a hot ring \cite{Vielva:2003et,Cruz:2006fy,Nadathur:2014tfa}. Both the \ac{wmap} and \textit{Planck} missions identified it as a significant deviation from the Gaussian random field expectation ($\sim3\sigma$), and its region has been studied extensively. To explain it, novel cosmological mechanisms \cite{Inoue:2006rd,Holman:2006an}, a cosmic texture \cite{Cruz:2006sv}, a large void in the line-of-sight \cite{Rudnick:2007kw}, and a systematic effect \cite{Vielva:2003et, Lambas:2023gzy} have all been investigated. Importantly, the \textit{Eridanus supervoid} ($R\approx200 h^{-1} {\rm Mpc}$, $\delta\approx-0.2$) was detected at $z<0.3$ aligned with the Cold Spot, based on galaxy counts from multiple surveys \cite{Szapudi:2014zha,Mackenzie:2017ioh}. This evidence is supported by reconstructions of the cosmic velocity field \cite{Courtois:2017mrq} and \ac{wl} analyses by the \ac{des} team \cite{DES:2021cge}, suggesting a causal relation between these individually rare objects in the \ac{cmb} and in the cosmic web. Yet, the statistical significance of the Cold Spot and its potential origins remain active areas of research.

Another very significant \ac{cmb} anomaly, which potentially could give rise to many of the above mentioned anomalies, is the cooling of \ac{cmb} photons around nearby galaxies. First discovered in Ref.~\cite{Luparello:2022kqb}, the stacking of \ac{cmb} temperatures in areas around the halos in nearby $z<0.015$ late type spiral galaxies showed lower \ac{cmb} temperatures compared to elsewhere on the sky. This was followed up in Ref.~\cite{Hansen:2023gra} where it was shown that an attempt at modeling the discovered cooling of \ac{cmb} photons around galactic halos gives rise to several of the observed anomalies. In particular, the hemispherical asymmetries and the cold spot arise naturally in such a scenario with the correct directions on the sky. Correlations are found between the largest scales of the \ac{cmb} and the nearby galaxy distribution obtained from the 2MRS galaxy catalog \cite{Huchra:2011ii}. In Ref.~\cite{Lambas:2023gzy} a detailed study of the galaxies in the cold spot areas showed that the shape and size of the cold spot to a large degree could be explained by cooling of the galaxies in this area. In particular, the nearby large Eridanus group of galaxies is located at this position. In Ref.~\cite{Cruz:2024xbh}, the cooling was found to correlate to the nearby cosmic density field. In Ref.~\cite{Hansen:2024vgs}, the cooling is detected at the $5.7\sigma$ significance level when looking at galactic halos in the most massive nearby cosmic filaments. Even when correcting for the look-elsewhere-effect by allowing for different choices of galaxy properties, the detection is stronger than in any of 10.000 simulated \ac{cmb} skies. Finally, Ref.~\cite{Toscano:2024xhc} shows that by masking the affected area on the sky, the estimate of the cosmological parameters is not significantly altered. The origin of the cooling of \ac{cmb} photons in galactic halos is unknown, but in Ref.~\cite{Hansen:2024vgs}, it is speculated whether \ac{dm} could be involved.

The statistical independence and potential interconnections between these anomalies are crucial areas of investigation. For example, the relationship between the low quadrupole moment and the lack of large-angle correlations in the \ac{cmb} is of particular interest \cite{Copi:2008hw,Sarkar:2010yj}. The lack of large-angle correlations was first reported using COBE data and later confirmed by \ac{wmap} and \textit{Planck}, with all three missions providing evidence for a significant deficit of correlations on angular scales greater than $\sim60^\circ$. This anomaly has been studied in the context of cosmic topology \cite{Luminet:2003dx,Aurich:2007yx}, with some proposing that it could be a signature of a non-trivial topology of the Universe. The relationship between the lack of large-angle correlations and the low quadrupole moment has also been investigated, with some suggesting that they may share a common origin \cite{Copi:2006tu}.

Collectively, these anomalies pose a significant challenge to the standard model of cosmology. While some of these anomalies may be the result of foreground contamination or systematic effects, their persistence across multiple observations and their statistical significance suggests that they may have a cosmological origin. The standard \lcdm\ model, which assumes a flat, homogeneous, and isotropic Universe, struggles to explain these anomalies, and their existence may require modifications to the model or the development of new theoretical frameworks.

Upcoming observational missions, such as the \textit{Euclid} space telescope and next-generation ground-based \ac{cmb} experiments, have the potential to shed new light on these anomalies \cite{EUCLID:2011zbd, CMB-S4:2016ple}. These missions will provide unprecedented sensitivity and angular resolution, allowing for a more detailed study of the \ac{cmb} and its anomalies. In particular, the \textit{Euclid} mission will provide detailed measurements of the large-scale structure of the Universe, which can be used to test models of the early Universe and the origin of the \ac{cmb} anomalies. Next-generation \ac{cmb} experiments, such as \ac{cmbs4}, will provide a significant improvement in sensitivity and angular resolution, allowing for a more detailed study of the \ac{cmb} power spectrum and its anomalies.

Several theoretical frameworks and models have been proposed to explain these anomalies, such as the role of topology in large-angle \ac{cmb} correlations \cite{Aurich:2007yx}, the impact of a cosmological \ac{gw} background \cite{Hogan:2006we}, and the effects of superhorizon isocurvature \ac{de} \cite{Gordon:2004ez}. Other models, including anisotropic k-essence \cite{Chimento:2005ua}, loop quantum cosmology \cite{Tsujikawa:2004dm}, non-canonical anisotropic inflation \cite{Watanabe:2009ct}, and unexpected topology of temperature fluctuations \cite{Luminet:2003dx}, have also been explored. These models aim to provide a theoretical framework that can account for the observed anomalies while maintaining consistency with other cosmological observations.

The development of new theoretical models and advanced simulations will also be crucial in understanding the early Universe's complexities. Theoretical efforts to explain the \ac{cmb} anomalies will require a deep understanding of the physics of the early Universe, including the dynamics of inflation, the generation of primordial perturbations, and the evolution of the Universe in the presence of \ac{de} and \ac{dm}. Advanced simulations, such as those performed with the \textit{Planck} Sky Model \cite{Delabrouille:2012ye}, will be essential in understanding the impact of foreground contamination and systematic effects on the observed \ac{cmb} anomalies.

In the early 2000's in the program of investigation of the selection of the initial conditions of our Universe \cite{Mersini-Houghton:2006phg, Mersini-Houghton:2005axz, Kobakhidze:2004gm, Holman:2005ei}, authors proposed to allow the wave-function of the Universe to propagate on the landscape of string theory and used quantum cosmology to derive the probability of our origin. This was the first work where the answer was derived from an underlying fundamental theory. It showed that in contrast to previous beliefs, the most likely Universe to spontaneously come into existence is the one that starts at very high energy. The authors proposed to use the quantum entanglement between branches of the wave function as a way to test the theory and, for the first time to have a handle to test the existence of the quantum multiverse \cite{Holman:2006an, Holman:2006ny}. Traces of earlier entanglement are imprinted as anomalies in our sky. The authors calculated and predicted a series of seven anomalies, including the Cold Spot, power asymmetry, alignments, and suppression of power in lowest multipoles, etc., all of which are by now observed. Status of the predicted anomalies against observations showed perfect agreement \cite{DiValentino:2016nni, DiValentino:2016ziq, DiValentino:2018wum, Mersini-Houghton:2016jno}.

As cosmology continues to evolve, the anisotropic anomalies in the \ac{cmb} serve as a reminder of the importance of critically examining our models and assumptions, driving us to deepen our understanding of the cosmos. The existence of these anomalies suggests that our current understanding of the early Universe and the origin of cosmic structure may be incomplete, and that new theoretical frameworks and observational efforts may be necessary to fully explain the observed features of the \ac{cmb}. As we continue to study these anomalies and their implications for cosmology, we may uncover new insights into the fundamental nature of the Universe and the physical processes that shaped its evolution.

\bigskip

\subsubsection{Hints of dynamical dark energy in DESI baryon acoustic oscillations and beyond \label{sec:w0wa}}

\noindent \textbf{Coordinator:} William Giar\`{e}\\
\noindent \textbf{Contributors:} Anton Chudaykin
\\

\noindent \ac{desi} has recently released data from its first (DR1)\cite{DESI:2024mwx,DESI:2024uvr,DESI:2024kob,DESI:2024lzq,DESI:2024aqx} and second (DR2)\cite{DESI:2025zgx,DESI:2025zpo,DESI:2025jet,Brodzeller:2025bbw} year of \ac{bao} measurements, based on observations of tens of millions of extragalactic objects, including galaxies, \ac{qso}s, and Lyman-$\alpha$ forest tracers. These \ac{desi} \ac{bao} observations provide precise constraints on the transverse comoving distance, the Hubble rate, and their combination (all relative to the sound horizon at the drag epoch) across seven redshift bins in the range $0.1 < z < 4.2$.

One of the most notable results from \ac{desi} \ac{bao} observations concerns the nature of \ac{de}. As initially highlighted by the \ac{desi} collaboration’s DR1 results~\cite{DESI:2024mwx,DESI:2024kob} and recently confirmed by DR2 results~\cite{DESI:2025zgx,DESI:2025kuo}, the combination of \ac{desi} \ac{bao} with \ac{cmb} data from the \textit{Planck} satellite and \ac{sn1} distance moduli measurements from three independent \ac{sn1} samples (i.e., the Pantheon-plus catalog~\cite{Scolnic:2021amr, Brout:2022vxf}, the Union3 compilation~\cite{Rubin:2023ovl}, and five-year observations from \ac{des}y5~\cite{DES:2024jxu,DES:2024upw,DES:2024hip} ) provides moderate to strong evidence for a time-evolving \ac{de} component, commonly referred to as \ac{dde}. One important aspect of this preference is that the data indicates a phantom crossing scenario, in which the dark energy equation of state crosses the $w=-1$~\cite{DESI:2025fii}.
While this behavior presents theoretical challenges for most simple scalar-field models, it can be realized in multi-field scenarios, alternative models of gravity~\cite{Vikman:2004dc,Carroll:2003st,Hu:2004kh,Creminelli:2008wc}.
The DESI preference for an evolving \ac{de} has been explored in the context of general scalar-tensor Horndeski theories~\cite{Chudaykin:2024gol,Ye:2024ywg,Ishak:2024jhs,Chudaykin:2025gdn} and in modified gravity models with non-minimal coupling \cite{Ye:2024ywg,Wolf:2024stt}.

Given the potential impact of these results on our understanding of the Universe, caution is essential, and thoroughly testing the robustness of these findings is of paramount importance. Unsurprisingly, a significant portion of the cosmology and high-energy physics community has actively engaged with this issue, clarifying and/or bringing up several aspects and concerns surrounding this preference towards \ac{dde}. In the following, we review some of the key results that have emerged from the consistency checks conducted over the past few months, highlighting both the strengths and weaknesses of the observed preference, while stressing aspects that warrant further clarification.

\begin{itemize}
\item \textbf{Parameterization of the Equation of State --} Barring, for the moment, any potential systematic issues in the different datasets involved in the analyses, a first key aspect that has undergone significant cross-checking~\cite{DESI:2024aqx,Hernandez-Almada:2024ost,Malekjani:2024bgi,Ramadan:2024kmn,Carloni:2024zpl,Berghaus:2024kra,Qu:2024lpx,Notari:2024rti,Adolf:2024twn,Wolf:2024stt,Jiang:2024xnu,Dinda:2024ktd,Wolf:2025jlc,Sousa-Neto:2025gpj} is the parameterization used to describe the \ac{de} \ac{eos}. Originally, the \ac{desi} collaboration parameterized the time evolution of the \ac{eos} using the linear \ac{cpl} form, $w(a) = w_0 + w_a (1-a)$, arguing that various combinations of data indicate a consistent preference for a present-day quintessence-like \ac{eos} ($w_0 > -1$) that crosses the phantom barrier ($w_a < 0$). Specifically, within this parameterization, \ac{desi} DR1 \ac{bao} measurements, when combined with Planck \ac{cmb} and \ac{des}y5 \ac{sn1}, yield a preference for \ac{dde} at a significance level of $\sim 3.9\sigma$. This preference decreases to $\sim 2.5\sigma$ ($\sim 3.5\sigma$) when replacing \ac{des}y5 with Pantheon-plus (Union3)~\cite{DESI:2024mwx}. Notably, within the same \ac{cpl} parameterization, these results remain stable when substituting \ac{desi} DR1 \ac{bao} with the latest DR2 \ac{desi} measurements. The latter not only confirms the preference for \ac{dde} but also consistently strengthens it by approximately $\sim 0.3\sigma$ across all dataset combinations with \ac{bao} and \ac{sn1}. Although the \ac{cpl} parameterization has been shown to match the background evolution of distances arising from the exact \ac{de} equations of motion with about 0.1\% accuracy for viable cosmologies across a broad range of physics (including scalar fields, \ac{mg}, and phase transitions see, e.g., Refs.~\cite{Linder:2002et,dePutter:2008wt}) alternative parameterizations that deviate from \ac{cpl} at both $z \ll 1$ and $z \gtrsim 1$ remain consistent with current observations. In Ref.~\cite{Giare:2024gpk}, it was argued that assuming the \ac{cpl} parameterization is not the primary driver of this preference. When combining \ac{cmb}, \ac{desi} \ac{bao}, and \ac{sn1} measurements within various \ac{eos} parameterizations, $w_0$ consistently remains in the quintessence regime, while the constraints on $w_a$ indicate a preference for a dynamical evolution crossing into the phantom regime (see also Ref.~\cite{Wolf:2025jlc}). This result holds for both \ac{desi} DR1 and the latest \ac{desi} DR2 \ac{bao} measurements, as confirmed by the \ac{desi} collaboration in Ref.~\cite{DESI:2025kuo}. In this sense, the preference is to be considered robust against different models.

\begin{figure}[ht!]
    \centering
    \includegraphics[width=0.9\linewidth]{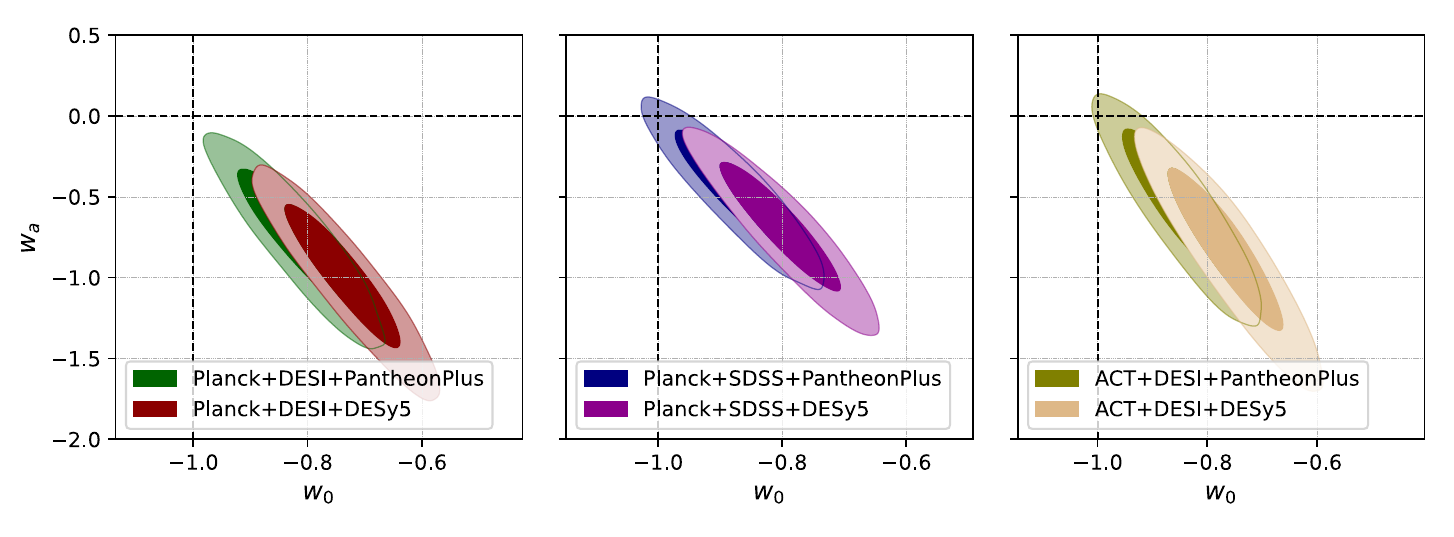}
    \caption{Observational constraints in the $w_0$-$w_a$ plane (at 68\% and 95\% CL) assuming a \ac{cpl} parametrization for \ac{de} across different datasets. The left panel shows constraints from a combination of the Planck \texttt{Plik} likelihood for high-$\ell$ TT, TE, and EE PR3 spectra, the \texttt{Commander} and \texttt{SimAll} likelihoods for low-$\ell$ temperature and polarization (TT and EE), and Planck PR4 \texttt{NPIPE} combined with \ac{act}-DR6 lensing likelihoods, along with \ac{desi} DR1 \ac{bao} and Pantheon-plus / \ac{des}y5 \ac{sn1} data. The middle panel examines the impact of replacing \ac{desi} \ac{bao} with SDSS \ac{bao}, while the right panel explores how the constraints shift when Planck is replaced with \ac{act}-DR4 for temperature and polarization spectra and \ac{act}-DR6 for lensing.}
    \label{fig:DDE_comparison}
\end{figure}

\item \textbf{DESI BAO measurements (and beyond)--} Since the preference for \ac{dde} was first highlighted by the \ac{desi} collaboration, the impact of \ac{desi} \ac{bao} DR1 data on this preference has been carefully scrutinized, with systematic effects in \ac{desi} \ac{bao} measurements remaining a topic of debate, see, e.g., Refs.~\cite{Wang:2024pui,Colgain:2024xqj,Naredo-Tuero:2024sgf,Sapone:2024ltl,Chudaykin:2024gol}. Notably, the \ac{desi} \ac{bao} measurement at $z = 0.71$ exhibits a $\sim3\sigma$ tension with predictions from the Planck best-fit \lcdm\ cosmology, making it a key driver of several hints for new physics, including (part of) the preference for \ac{dde}. However, it is crucial to emphasize that \textit{(i)} the latest \ac{desi} DR2 \ac{bao} measurements not only confirm the preference for \ac{dde} but also consistently strengthen it by approximately $\sim 0.3\sigma$ across all dataset combinations with \ac{bao} and \ac{sn1}. While this increase in preference from \ac{desi} DR1 to \ac{desi} DR2 is not highly statistically significant, it definitively reinforces the signal, suggesting that the initial preference for \ac{dde} observed in \ac{desi} DR1 may not be merely a statistical fluctuation resulting from a preliminary one-year observation. \textit{(ii)} In combined analyses of \ac{cmb}, \ac{bao}, and \ac{sn1} data, the preference for \ac{dde} is not exclusively driven by \ac{desi} \ac{bao}, whether from DR1 or DR2. In fact, as reported in Ref.~\cite{Giare:2025pzu}, a similar shift toward \ac{dde} is observed when replacing \ac{desi} \ac{bao} measurements with SDSS \ac{bao}. Planck \ac{cmb} combined with SDSS \ac{bao} and \ac{des}y5 \ac{sn1} indicates a preference for \ac{dde} at a statistical significance exceeding $2.5\sigma$, see also Fig.~\ref{fig:DDE_comparison}. Likewise, Planck \ac{cmb} combined with SDSS \ac{bao} and Union 3 points to a \ac{dde} component at  $\gtrsim 2\sigma$ significance. The only combination where this preference is notably weakened below $2\sigma$ is when SDSS \ac{bao} and Pantheon-plus \ac{sn1} are analyzed together with Planck \ac{cmb}. Importantly, across all independent datasets -- each showing hints of deviation from a cosmological constant at varying levels of statistical significance -- a consistent shift in parameter space is observed in the same direction, namely towards a quintessence-like equation of state in the present epoch that transitions into a phantom-like regime in the past. Overall, while \ac{desi} \ac{bao} data undoubtedly strengthen the preference for \ac{dde}, the other datasets involved in the analysis -- most prominently \ac{sn1} distance modulus measurements -- play an equally significant role in shaping this result. For further discussions on \ac{bao} data, we refer to Sec.~\ref{sec:BAO}.

\item \textbf{\ac{sn1}--} It quickly became clear that \ac{sn1} data play a major role in driving the preference for \ac{dde}. Excluding \ac{desi} \ac{bao} and considering data combinations involving Planck \ac{cmb} and either \ac{des}y5 or Union3 (i.e., without including any \ac{bao} surveys) already leads to a preference for \ac{dde} at $\sim 2-2.5\sigma$ (see, e.g., Refs.~\cite{Giare:2025pzu,DES:2025bxy}). Given the crucial role of \ac{sn1} measurements in this trend, the potential impact of systematic effects has been widely investigated \cite{Colgain:2022nlb,Colgain:2022rxy,Malekjani:2023ple,Colgain:2024ksa, Colgain:2024mtg,Notari:2024zmi}, particularly for the \ac{des}y5 sample, which exhibits the strongest shift toward \ac{dde}. For instance, Ref.~\cite{Efstathiou:2024xcq} highlighted the influence of \ac{sn1} measurements, reporting a cross-correlation between the PantheonPlus and \ac{des}y5 supernova samples that suggested a calibration difference of $\sim 0.04$ mag between low and high redshifts. Correcting for this offset brings the \ac{des}y5 sample into closer agreement with Planck’s \lcdm\ cosmology. Since the parameter space favored by the uncorrected \ac{des}y5 sample diverges from many other cosmological datasets, it has been suggested that the apparent evidence for \ac{dde} might primarily arise from systematic effects in \ac{des}y5 \ac{sn1}. In response to these concerns, the \ac{des} collaboration in Ref.~\cite{DES:2025tir} showed that the debated $\sim 0.04$ mag offset between Pantheon+ and \ac{des}y5 at low and high redshift is partly due to improvements in the modeling of supernova intrinsic scatter and host galaxy properties in \ac{des}y5 (which account for up to $\sim 43\%$ of the offset) and partly ($\sim 38\%$) due to a misleading comparison. The latter arises because different selection functions characterize the \ac{des} subsets included in Pantheon+ and \ac{des}-SN5YR, leading to differences in individual supernova distance measurements due to distinct bias corrections. Therefore, while \ac{sn1} data do play a pivotal role in shaping the preference for \ac{dde}, these findings might warrant additional tests to carefully assess potential systematics before drawing definitive conclusions. For further discussions on \ac{sn1} data, we refer to Sec.~\ref{sec:SNeIa}.

\item \textbf{Planck CMB data (and beyond):} The final dataset considered in these analyses is the \ac{cmb} angular power spectra, measured by the Planck Collaboration. While a \ac{dde} component primarily affects the late-time evolution of the Universe, it also has a non-negligible impact on the \ac{cmb} spectra. The most significant effect of \ac{de} dynamics on the \ac{cmb} angular power spectrum is observed in the amplitude of the \ac{isw} plateau at very large angular scales. This amplitude is mainly determined by primordial inflationary parameters ($A_s$ and $n_s$) and contributions from the late-time \ac{isw} effect, which is sensitive to \ac{de} dynamics (see also Sec.~\ref{sec:ISW}). Specifically, a phantom (quintessence) \ac{de} component suppresses (enhances) the decay of gravitational potentials that source the late \ac{isw} effect. Therefore, assessing the impact of Planck measurements on the preference for \ac{dde} is certainly worthwhile. As discussed in Ref.~\cite{Giare:2024ocw}, Planck data -- particularly the large-scale temperature and E-mode polarization measurements -- play a crucial role in reinforcing the preference for \ac{dde} reported by the \ac{desi} collaboration. Several studies in the literature~\cite{Planck:2018vyg, Escamilla:2023oce, Giare:2023ejv, Ben-Dayan:2024uvx, Peng:2025nez, Chan-GyungPark:2025cri, Chan-GyungPark:2024brx} have highlighted that temperature and polarization data at large angular scales ($\ell \leq 30$) drive a number of mild anomalies, including a latent preference for a phantom-like \ac{de} component~\cite{Planck:2018vyg, Escamilla:2023oce, Giare:2023ejv}. As discussed in Ref.~\cite{Giare:2024ocw}, temperature and E-mode polarization anisotropy measurements at $\ell \lesssim 30$ are precisely the subsets of Planck data strengthening the shift toward \ac{dde}, as well. Notably, when Planck's large-scale data are excluded from the analysis or when considering alternative \ac{cmb} experiments that are independent of Planck, the preference for \ac{dde} diminishes significantly. In these cases, no strong preference for \ac{dde} is observed when combining \ac{desi} with Pantheon-plus \ac{sn1} data, and the \lcdm\ model remains consistent within (or close to) the 95\% confidence level results. For instance, no convincing preference for \ac{dde} is found in any combination involving \ac{spt} data, whether using \ac{desi} \ac{bao} together with Pantheon-plus or \ac{des}y5, or even when including Planck/\ac{wmap} large-scale temperature and polarization measurements. Similarly, in analyses involving \ac{act} data, a preference for \ac{dde} is observed only when combining \ac{act} with \ac{desi} \ac{bao} and \ac{des}y5 \ac{sn1}, while it gets diluted when replacing \ac{des}y5 with Pantheon-plus, see also Fig.~\ref{fig:DDE_comparison}. However, when Planck large-scale temperature and polarization measurements are combined with \ac{act} small-scale temperature and polarization data, a preference for \ac{dde} (re-)emerges. Long story short, in the \ac{cmb} front of the analysis, the shift toward \ac{dde} is primarily strengthened by Planck’s large-scale data, while other \ac{cmb} experiments generally weaken the case for \ac{dde}. Taking a conservative stance, this adds to broader concerns about potential systematic issues raised in other datasets discussed earlier.

\end{itemize}

\bigskip
\subsubsection{Neutrino tensions \label{sec:Neutri_Ten}}

\noindent \textbf{Coordinator:} Gabriela Barenboim\\
\noindent \textbf{Contributors:} Ad\`{e}le Poudou, Janusz Gluza, Mariana Melo, Matteo Forconi, Olga Mena, Rasmi Hajjar, Rishav Roshan, and Stefano Gariazzo
\\

Neutrino physics is in its Golden Age era. Neutrinos possess unique characteristics that set them apart. Firstly, they are significantly lighter than other fermions by several orders of magnitude. Remarkably, no direct mass measurement has yet provided evidence for a non-zero neutrino mass; for all detected neutrinos, $E=pc$ within errors. Secondly, they have their own cosmological background. Neutrinos are often described as ``elusive'' due to their nature, yet they are incredibly abundant throughout the Universe. Their near absence of interaction is why we seldom detect their presence. This relic neutrino background has never been detected directly. Nevertheless, neutrinos are hot thermal relics, and their masses, even if tiny, affect the Universe's evolution, having its largest impact on the growth of structure at small scales due to the neutrino free steaming nature. Therefore, albeit indirectly, we can set an upper bound on them by cosmological observations.
Although official Planck \ac{cmb} results report $\sum m_i<0.24$~eV~\cite{Planck:2018vyg}, the addition of \ac{lss} data in the form of \ac{bao} allows reaching $\sum m_i<0.09$~eV~\cite{DiValentino:2021hoh} or even $\sum m_i<0.072$~eV from the very recent \ac{desi} \ac{bao} survey observations~\cite{DESI:2024mwx,Jiang:2024viw}, all at 95\% CL. Similar bounds are obtained when different likelihood approaches are adopted for analyzing Planck data \cite{Tristram:2023haj}, while the limits are relaxed when high-multipole \ac{cmb} data are considered instead of Planck \cite{DiValentino:2023fei}.

Nevertheless, the tightest limits to date are those reported in Ref.~\cite{Wang:2024hen} after the addition to Planck and \ac{desi} observations of other background probes, such as \ac{cc}, galaxy clusters angular diameter distances, and \ac{grb}s distance moduli.  For instance, the combination of \ac{cmb} with \ac{grb}s and \ac{desi} \ac{bao} provides $\sum m_\nu <0.049$ eV. The most constraining bound is obtained when \ac{cmb}, \ac{sn1} luminosity distances, \ac{desi} \ac{bao}, and all background probes are combined: this limit is $0.043$ eV at $95\%$~CL. Background probes are therefore able to provide strong bounds on the neutrino mass due to both the preferred higher mean value of the Hubble constant and the smaller errors on both $H_0$ and the matter mass-energy density $\Omega_{\rm m}$. Consequently, cosmological limits currently possess the highest constraining power on neutrino masses if the standard scenario is assumed for neutrino decoupling and if neutrinos are massive stable particles. 

In the past, neutrinos were supposed to provide a possible explanation to the $H_0$ and $\sigma_8$ tensions, see e.g.,~\cite{Hamann:2013iba}.
More recently, however, it has been shown that an increase in $H_0$ due to neutrino physics is normally related to an increase also in $\sigma_8$, see e.g.,~Fig.~34 in Ref.~\cite{Planck:2018vyg}. A solution to both tensions thanks to neutrinos seems therefore unlikely.

On the other hand, neutrino oscillations measured at terrestrial experiments while being insensitive to the overall neutrino mass scale, determine the value of two squared mass differences, the atmospheric $|\Delta m^2_{31}| \approx 2.55\cdot 10^{-3}$~eV$^2$ and the solar $\Delta m^2_{21} \approx 7.5\cdot 10^{-5}$~eV$^2$  splittings~\cite{deSalas:2020pgw,Esteban:2020cvm}. 
Since the sign of $|\Delta m^2_{31}|$ is unknown, two mass orderings are possible, the \emph{normal} (NO) and the \emph{inverted} (IO)  orderings: 
\begin{equation}
    \sum m_i =
    \left\{
    \begin{array}{ll}
        \mbox{NO:} &
        m_\nu^2 + \sqrt{ m_\nu^2 + \Delta m_{21}^2} +  \sqrt{ m_\nu^2 + \Delta m_{31}^2 }
        \geq0.058 \mbox{ eV}\,, \\    
        \mbox{IO:} &
        m_\nu^2
        + \sqrt{|\Delta m_{31}^2|+m_\nu^2}
        + \sqrt{|\Delta m_{31}^2|+\Delta m_{21}^2 + m_\nu^2}
        \geq0.101 \mbox{ eV}\,.
    \end{array}
    \right.
\end{equation}
Notice that not only some of the above-mentioned current cosmological limits are \emph{below} the minimum sum of the neutrino masses allowed in the inverted hierarchical scenario, but also some of them are \emph{below} the minimum mass required from neutrino oscillation probes. Very interestingly, several of the possible data combinations imply a neutrino mass limit at $2\sigma$ smaller than the minimum expected from oscillation experiments, i.e., $\sum m_\nu \lesssim 0.06$ eV, pointing towards a clear tension between cosmological and oscillation neutrino mass limits, see Ref.~\cite{Wang:2024hen}. Therefore, despite the fact that such tension has so far not been present~\cite{Gariazzo:2023joe}, not only one but several sets of cosmological measurements indicate a clear problem between these cosmological and terrestrial searches of neutrino masses. 
Cosmological constraints, therefore, at present, reflect some \emph{tension regarding neutrino masses}:

\begin{figure}[ht]
    \centering
    \includegraphics[width=0.5\linewidth]{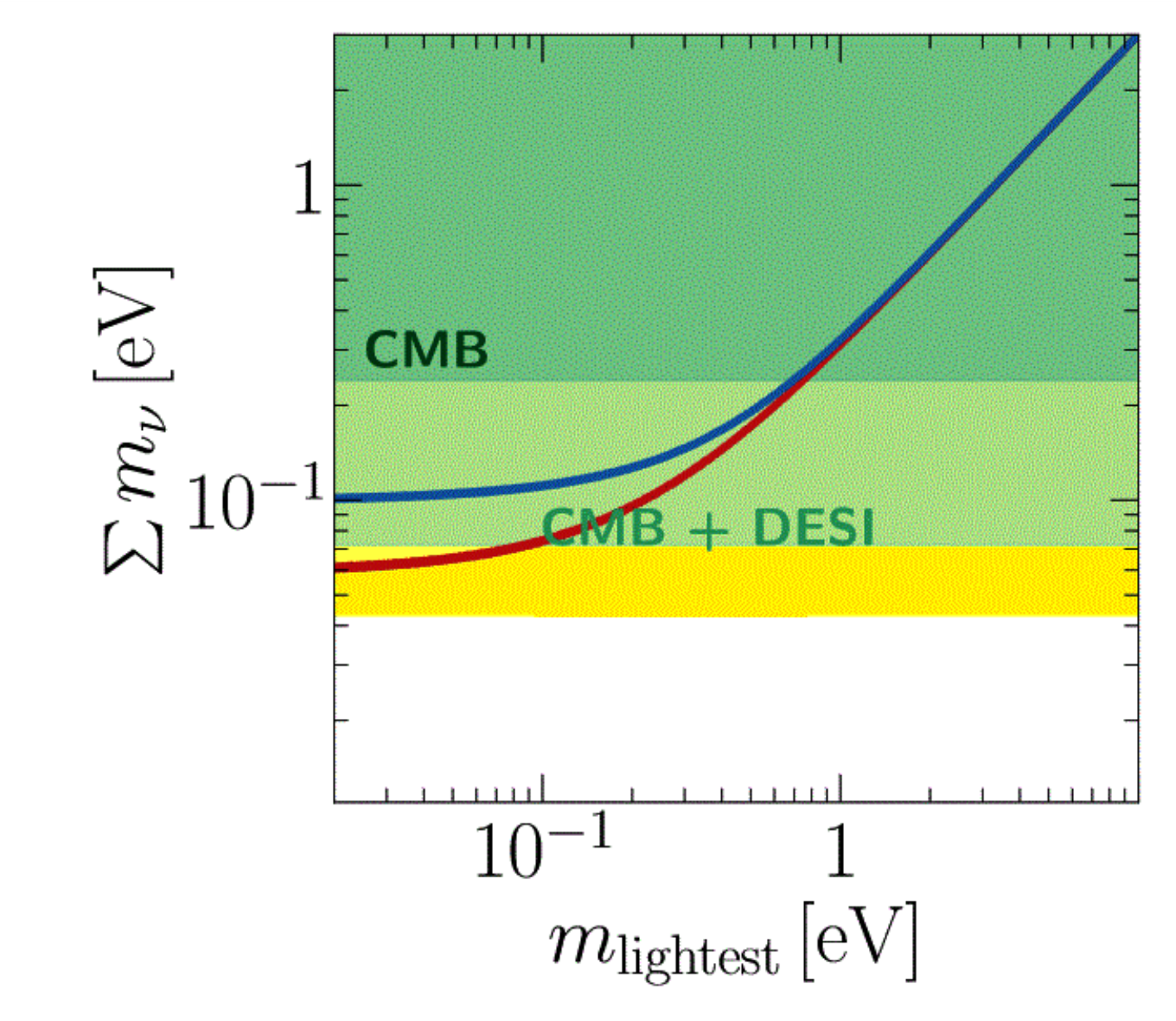}
    \caption{The sum of the neutrino masses as a function of the
lightest neutrino mass for normal (red) and inverted (blue) orderings. Cosmological constraints from \ac{cmb}, \ac{cmb} plus \ac{desi}, and including measurements from \ac{cc}, distance moduli from \ac{grb}s, and angular diameter distances from galaxy clusters are also depicted by the shaded green and yellow regions.}
    \label{fig:neutrinos}
\end{figure}

\begin{itemize}
\item Through cosmological observations, we are about to rule out the minimum value of $\sum m_i$ allowed by neutrino oscillations. It is important to remember that cosmology does not directly measure neutrino masses but rather constrains the neutrino energy density, which is proportional to neutrino masses in the standard scenario. This relation could be altered by the presence of new physics: current cosmological bounds may point to the existence of very exotic cosmological scenarios (possibly related to dark sector physics) and/or non-standard neutrino physics since these bounds are extremely robust within simple extensions of the \lcdm\ model; see, e.g.,~\cite{diValentino:2022njd}.
Possible scenarios to relax the cosmological neutrino mass bounds include time-varying neutrino masses~\cite{Lorenz:2018fzb}, decaying neutrinos~\cite{Escudero:2020ped,Chacko:2020hmh,Chacko:2019nej,FrancoAbellan:2021hdb}, strongly-interacting neutrinos~\cite{Kreisch:2019yzn,Camarena:2024daj}, long-range neutrino forces~\cite{Esteban:2021ozz}, among others~\cite{Oldengott:2019lke,Dvali:2016uhn,Barenboim:2024wek,Craig:2024tky,Yadav:2024duq}.

\item Using Planck 2018, \ac{boss} Lyman-alpha, and \ac{desi} data, if negative neutrino masses are allowed in the fitting pipeline, a slightly better fit is obtained with $ \sum m_i <0$ than with $\sum m_i \geq0$~\cite{Craig:2024tky}. It should be noted, however, that the absolute sign of the mass of a fermion is a phase (i.e., is unobservable) and therefore this preference for negative masses should be taken as an indication of an incomplete model rather than an actual physical measurement.
\end{itemize}
These tensions will need to be addressed in the coming years, and upcoming cosmological measurements, such as those from e.g., future observations by \ac{desi} or \ac{cmbs4}, will be able to test all the current tensions (and possibly additional ones; see below) while sharpening the cosmological neutrino mass limits.

Last but not least,  the Karlsruhe Tritium Neutrino (KATRIN) experiment, a tritium beta decay experiment, may directly discover the electron neutrino mass, finding a positive signal.\footnote{The current limit is $m_{\beta}<0.45$~eV (90\% CL)~\cite{KATRIN:2024cdt}, and the expected sensitivity is 0.2 eV (90\% CL)~\cite{Drexlin:2013lha}} Such a putative neutrino mass detection by KATRIN will show an additional clear \emph{tension} with current cosmological mass limits. As previously stated, possible inconsistencies among laboratory and cosmological searches would definitely point to a much richer neutrino sector, to a departure from the standard model of cosmology, or to a combination of both. Such a situation would offer exciting possibilities to discover new physics. Neutrino physics has brought the first departure from Standard Model Physics and may also be the first one to falsify the current standard cosmological scenarios where they are hot thermal stable relics interacting exclusively via weak interactions. 

\bigskip
\subsubsection{Cosmic dipoles \label{sec:dipoles}}

\noindent \textbf{Coordinator:} M.M. Sheikh-Jabbari\\
\noindent \textbf{Contributors:} Dinko Milakovi{\'c}, Eoin \'O Colg\'ain, Francesco Sorrenti, Iryna Vavilova, Jenny Wagner, Jos\'e Pedro Mimoso, Laura Mersini-Houghton, Leandros Perivolaropoulos, Lu Yin, Maciej Bilicki, and Manolis Plionis
\\

Cosmological principle, the \textit{assumption} that the Universe around us at cosmological scales is (statistically or on the average) homogeneous and isotropic, is the cornerstone of the modern cosmology, see Ref.~\cite{Aluri:2022hzs} for a recent review. \ac{cmb} temperature fluctuations \cite{Planck:2018vyg} have been regarded as a very precise verification of cosmic isotropy at the surface of last scattering at around $z\sim 1100$. However, the existence of a very small deviation from isotropy in the \ac{cmb} is not ruled out theoretically \cite{ellis2012relativistic} and references therein or observationally \cite{Schwarz:2015cma, Fleury:2016fda, Aluri:2022hzs, Jones:2023ncn, Fosalba:2020gls, Yeung:2022smn, Euclid:2024wly} and references therein. Such small anisotropies, if they exist and are not due to systematics, in principle can grow in time through the large structure formation, especially when we enter a nonlinear growth regime. It is therefore, prudent to thoroughly explore evidence for traces of anisotropy especially in late cosmology. 

Any physical observable may be expanded in terms of spherical harmonics on the celestial sphere. The lowest mode is $\ell=0$ which preserves isotropy. The next mode is $\ell=1$, corresponding to dipoles of the physical observable in question. Given that deviations from spherical symmetry are expected to be small, one expects higher multipoles to have smaller contributions and hence harder to detect. On the other hand, the dipole component, which is the largest mode beyond isotropy, is observer dependent and can be a kinematical effect. For example, it is well established that the dipole in the \ac{cmb}, which corresponds to $\sim 10^{-4}-10^{-3}$ fluctuation in the \ac{cmb} temperature (compared to usual $10^{-5}$ fluctuations),  can be attributed to peculiar motion of the observer compared to the ``\ac{cmb} frame'' that moves with the Hubble flow. So, exploring the dipoles in cosmic observables one should always note this frame dependence. The very local Universe is obviously not isotropic around us. One can ask how far one should look to be confident in the  Hubble/\ac{cmb} frame. It is established that this distance is larger than $\sim 100$ Mpc, as indicated also by the convergence scale of the acceleration dipole due  to matter fluctuations traced by galaxies and clusters of galaxies distributions, first at
Ref.~\cite{1988MNRAS.234..401P,1991MNRAS.249...46P, 1991ApJ...376L...1S}
and later at Ref.~\cite{Branchini:1995tk,Plionis:1997wj,Rowan-Robinson:1999zsn,2005ASPC..329...89K}. The ``bulk flow'' in the Local Universe is a collective phenomenon due to the peculiar motions of matter structures \cite{Hoffman:2017ako} that instead of moving in random directions, appear to follow an approximate dipole velocity flow, see Ref.~\cite{Lopes:2024vfz} and references therein.  The existence of bulk flows that are too large for \lcdm\ expectations may be taken as a sign of departure from isotropy that shows itself as a dipole in various cosmological observables. 

\paragraph{Non-comoving cosmology, tilted cosmology} 
The cosmological dipole is a frame dependent notion. To distinguish kinematical and a non-kinematical (not removable by the choice of frame) one needs to study usual \ac{flrw} cosmological models in a non-comoving frame, see Ref.~\cite{Cembranos:2019plq,Tsagas:2021dsl, Tsagas:2021tqa, Asvesta:2022fts, Santiago:2022yjt}. A related, but conceptually different, notion is the ``tilt'' introduced by King and Ellis \cite{King:1972td}. Suppose we choose a comoving frame, a frame in which metric does not have time-space off-diagonal elements, $g_{ti}=0$. In this frame cosmic fluids may exhibit a momentum flow, i.e., the time-space component of energy momentum tensor $T^t_i$ may be nonzero. Tilt may be different for different components of cosmic fluid, e.g., (pressureless) matter or radiation may have different tilts and hence there does not exist a Hubble or \ac{cmb} frame.  Tilt is a dynamical variable in cosmology, like the shear or Hubble parameter, e.g., see Ref.~\cite{Tsagas:2007yx, ellis2012relativistic,  Krishnan:2022qbv, Krishnan:2022uar, Ebrahimian:2023svi, Allahyari:2023kfm} and references therein. This is in contrast with the non-comoving cosmology in which the dipole components are not dynamic ones. The existence of tilts yields dipole anisotropy in various cosmological observables. 

\paragraph{Observational hints for cosmic dipoles} 
We start by recalling that typical cosmological observables are redshift (as a measure of distance), look-back time, Hubble expansion rate, deceleration/acceleration parameter, luminosity distance, angular diameter distance, and number counts. In an anisotropic cosmology, these quantities besides the redshift $z$ also depend on line-of-sight. Redshift, the ratio of the frequency of the photon emitted by a source and the one observed by the observer minus one, receives a Doppler shift if there is a bulk flow of velocity $v$:
\begin{equation}
    1+z=(1+z_0) \Delta\,, \qquad \Delta :=(1-v/c \cos\theta_{\text{\tiny{LoS}}})/\sqrt{1-v^2/c^2}\,, 
\end{equation}
where $1+z_0=1/a$ is redshift when the source and observer have zero relative velocity and $\theta_{\text{\tiny{LoS}}}$ denotes the line of sight angle. For more formal and detailed derivations and analysis see Refs.~\cite{Tsagas:2015mua, Heinesen:2020sre, Heinesen:2020bej, Heinesen:2021azp, Guandalin:2022tyl, Maartens:2023tib, Ebrahimian:2023svi}.

\paragraph{Ellis-Baldwin test} 
A standard technique for searching for a dipole component in the distribution of sources in the sky is the seminal Ellis-Baldwin test \cite{1984MNRAS.206..377E}. The number $N$ per solid angle of a local source in the sky (say radio galaxies or \ac{qso}) in two frames moving with a relative velocity $v$ and angle  $\cos\theta_{\text{\tiny{LoS}}}$ is given by
\begin{equation}
    \frac{\text{d}N}{\text{d}\Omega}\big{|}_{\text{obs}}=\frac{\text{d}N}{\text{d}\Omega}\big{|}_{\text{rest}}\ \Delta^{2+k}\,,
\end{equation}
where $k$ is a constant parameterizing astrophysical characteristics of the source population.

\paragraph{Cosmic microwave background dipole} 
A dipole component in the \ac{cmb} is customarily treated as purely kinematical, due to the motion of the observer. This may be viewed as the definition of ``\ac{cmb} frame'', a frame in which the \ac{cmb} has no dipole component. While one can always find such a frame, only in an isotropic universe all other cosmological observables are necessarily dipole free in the \ac{cmb} frame. The heliocentric frame, in which our cosmological observations are made, is moving with respect to the \ac{cmb} frame with velocity $v = (369.82 \pm 0.11)\,{\rm km s}^{-1}$ along the direction (in degrees) $(l, b) = (264.02 \pm 0.0085, 48.253 \pm 0.004)$ \cite{Planck:2018vyg}. While the \ac{cmb} dipole may be treated as kinematical, there have been persistent \ac{cmb} quadrupole and octopole components which may be hints of deviations from the cosmological principle, see Refs.~\cite{Schwarz:2015cma, Perivolaropoulos:2021jda, Aluri:2022hzs} for reviews.

The Planck \ac{cmb} data in the \ac{cmb} frame may still show traces of a dipole once one makes a hemispherical split analysis (HSA) of the data. This method involves splitting the celestial sphere into two hemispheres and fitting \lcdm\ to \ac{cmb} data in each hemisphere. One then reads the values of $H_0$ and $\Omega_{\rm m,0}$ in each hemisphere, rotates the hemisphere separation plane in the sky, and records the ratio of the difference between the two values of $H_0$ in each hemisphere, $\Delta H_0$, over the mean $H_0$ value over the whole sky, $\bar{H}_0$. This analysis has yielded a maximal value of $\Delta H_0 / \bar{H}_0 \simeq 10\%$, roughly along $45^\circ$ to the direction of the \ac{cmb} kinematic dipole \cite{Yeung:2022smn}, {a direction close to the axis of the hemispherical power asymmetry, a well-recognized \ac{cmb} anomaly \cite{Schwarz:2015cma}.} Intriguingly, this is the same level as the discrepancy in Hubble tension; however see Ref.~\cite{Akarsu:2021max}.

\paragraph{Maximum Temperature Asymmetry (MTA)} is another dipolar anomaly in the \ac{cmb} which is identified by examining the temperature differences between opposite pixels in the \ac{cmb} sky map after subtracting the dipole component. MTA selects a preferred axis by maximizing the temperature difference between these opposite pixels and indicates a preferred direction in the \ac{cmb} temperature fluctuations, signature of a dipole. See Ref.~\cite{Mariano:2012ia} for MTA analysis in the \ac{wmap} 7-year data. The proximity of the MTA axis to the $\alpha$ dipole, \ac{de} dipole, and dark flow directions (see below) further strengthens the case for a potential underlying physical cause for these anisotropies.

\paragraph{Fine structure constant dipole} Initially reported by Ref.~\cite{Webb:2010hc}, was identified through the analysis of \ac{qso} absorption spectra using the Very Large Telescope (VLT) and the Keck Observatory. The dipole suggests a spatial variation in the fine structure constant $\alpha$ across the sky, with a dipole axis pointing towards $(l, b) = (331^\circ, -14^\circ)$, and an amplitude $A = (0.97 \pm 0.21) \times 10^{-5}$. However, recent studies have reduced the significance of the $\alpha$ dipole to less than 4$\sigma$ \cite{Wilczynska:2020rxx, Perivolaropoulos:2021jda}.

\paragraph{Type Ia supernovae dipole} 
\ac{sn1} are widely accepted as standard candles with a fairly good distribution over the sky [JLA, Pantheon(+), Union3, \ac{des}-5Y, \ac{ztf}]. The \ac{sn1} observations are of course made in the heliocentric frame, correcting for the presumed (peculiar) motion w.r.t. the \ac{cmb} frame, assuming that a Hubble flow frame exists and that the far enough (farther than $\sim 100$ Mpc \ac{sn1}) are already in the Hubble flow frame w.r.t us. This may obscure possible dipole components in the data in the heliocentric frame \cite{Mohayaee:2021jzi}. \ac{sn1} data  may be explored for a possible cosmological dipole component in 3 different kind of analyses:

(I) \textit{Ellis-Baldwin test for SNIa} We still do not have enough observed \ac{sn1} to carry out this test. It may be feasible in some years.

(II) \textit{HSA using a cosmographic expansion, dipole fitting method.} For  relatively low redshift data, $z\lesssim 1$, one can expand Hubble parameter and luminosity distance in powers of $z$, 
\begin{equation}
    H = H_0\left[1+  (1+q_0) z+ {\cal O}(z^2)\right]\,, \qquad  D_{\rm L}(z)=\frac{cz}{H_0}\left[1+ \frac12(1-q_0) z+ {\cal O}(z^2)\right]\,.
\end{equation}
One may now consider a dipole in $H_0$, i.e., $H_0(\theta)=H_0(1+{\cal D}_H\cos\theta)$ or similarly in acceleration, $q_0(\theta)=q_0(1+{\cal D}_q\cos\theta)$ and perform HSA. This test has been carried out for JLA \cite{Colin:2019opb,Mohayaee:2020wxf} and Pantheon(+) \cite{Mohayaee:2021jzi, Singal:2021crs, Horstmann:2021jjg, Asvesta:2022fts, Cowell:2022ehf} datasets claiming for a small dipole in $H_0$ and a notable dipole for $q_0$ roughly along the \ac{cmb} dipole in the heliocentric frame. See also, \cite{McConville:2023xav, Lopes:2024vfz}. There have been analyses that claim not seeing a dipole see e.g., \cite{Andrade:2017iam, Bengaly:2019ibu, Rahman:2021mti,  Salehi:2020hek, Hu:2020mzd, Brout:2022vxf,Dhawan:2022lze, Sapone:2020wwz,Bengaly:2024ree}. We need to wait for more \ac{sn1} data, e.g, the \ac{des} data and \ac{ztf} to confirm or refute these claims. 

(III) \textit{HSA and dipole fitting of \lcdm\ into SNIa data}. This analysis started in 2010 in Ref.~\cite{Antoniou:2010gw} with the Union2 \ac{sn1} dataset and identified a significant anisotropy in the accelerating expansion rate in direction $(l, b) = (309^\circ {^{+23^\circ}_{-3^\circ}}, 18^\circ {^{+11^\circ}_{-10^\circ}})$. The anisotropy level $\Delta \Omega_{0m, \text{max}} = 0.43 \pm 0.06$ was consistent with statistical isotropy in about 30\% of simulations. The alignment with other observations like bulk velocity flows and \ac{cmb} low multipole moments, fine structure constant, and \ac{de} dipole obtained through MTA method \cite{Mariano:2012wx} suggested a potential underlying physical cause.

Similar recent approaches use the luminosity distance and distance ladder relation $\mu=5\log D_{\rm L}(z)+25$ with $D_{\rm L}(z)=\frac{c}{H_0}\int^z \text{d}z'/\sqrt{1 - \Omega_{\rm m,0}+(1+z')^3\Omega_{\rm m,0}}$ and perform HSA. This has been carried out yielding a small variation $\Delta H_0\lesssim$ few \kms over the sky \cite{Krishnan:2021jmh}, see also Refs.~\cite{Bahr-Kalus:2012yjc, Zhao:2019azy, Kalbouneh:2022tfw, McConville:2023xav, Tang:2023kzs, Hu:2023eyf} and Refs.~\cite{Sorrenti:2022zat, Horstmann:2021jjg, McConville:2023xav} for further analysis. It would be instructive to repeat a similar analysis with {other datasets like \ac{des} 5Y \cite{McConville:2023xav, Tang:2023kzs} or \ac{ztf} data release. Alternatively, one may repeat a similar analysis with absolute magnitudes of \ac{sn1} (instead of $H_0$ and $\Omega_{\rm m,0}$) \cite{Perivolaropoulos:2023tdt}. Analyzing the latest \ac{ztf} data release, particularly at low redshifts, may yield further valuable information. Ultimately, the combination of these different approaches and datasets will enhance our understanding of cosmic anisotropies and contribute to a more comprehensive picture of the Universe's large-scale structure.

\paragraph{Quasar (QSO) dipoles} 
\ac{qso} catalogs contain about 1000 times more data points compared to \ac{sn1} datasets, nonetheless, unlike the \ac{sn1} their standardizability \cite{Risaliti:2015zla, Risaliti:2018reu, Moresco:2022phi} is still a matter of debate, see Refs.~\cite{Khadka:2021xcc,  Khadka:2022aeg, Zajacek:2023qjm, Colgain:2024clf} and references therein. The 3 anisotropy/dipole searches mentioned above using \ac{sn1} data, can be repeated for the \ac{qso} samples. \textit{Ellis-Baldwin test for \ac{qso}} has been carried out yielding a significant (up to $4.9\sigma$) departure from isotropy \cite{Secrest:2020has}: \ac{qso} of redshifts $z\gtrsim 1$ are not in Hubble/\ac{cmb} frame, while in the heliocentric frame they move in the same direction as the \ac{cmb} dipole, their peculiar velocity is few orders of magnitude bigger than the \ac{cmb} velocity $370\,{\rm km s}^{-1}$, see also Refs.~\cite{Singal:2014wra,Dam:2022wwh}. This result is still under debate e.g., see Refs.~\cite{Kothari:2022bjr, Abghari:2024eja} and references therein. 

The HSA for \ac{qso} data with the dipole fitting method has also been carried out, in a cosmographic expansion and/or using Risaliti-Lusso \cite{Risaliti:2018reu} X-ray/UV flux relation. This has also led to a $2-3\sigma$ level result for a dipole component in $H_0$ \cite{Luongo:2021nqh}.

\paragraph{Radio Galaxy dipole} 
Radio galaxies are typically located at cosmological distances and hence within the standard cosmological principle paradigm, they are expected to be moving with the Hubble flow; they should not exhibit a flow in the \ac{cmb} frame and \ac{cmb} frame should be their rest frame. This was first put to the test in 2002 \cite{Blake:2002gx}, confirming the expectation. However, the same observation has been repeated with better accuracy, refuting the earlier result in about $3\sigma$ level \cite{Colin:2017juj}, finding radio galaxies moving with speed 4 times bigger than the \ac{cmb} velocity in the heliocentric frame, and roughly in the same direction. Similar results have been reported in Refs.~\cite{Singal:2011dy, Rubart:2013tx, Bengaly:2018ykb, Singal:2019pqq, Siewert:2020krp, Secrest:2022uvx, Wagenveld:2023kvi, Singal:2023lqm}, see however Refs.~\cite{Gibelyou:2012ri, Bengaly:2019zhr, Murray:2021frz, Cheng:2023eba, daSilveiraFerreira:2024ddn, Oayda:2024hnu}.

\paragraph{Gamma-ray burst dipole} 
\ac{grb}s may also be used to perform similar dipole tests as in \ac{qso} or \ac{sn1}. However, a relatively low number of data points in \ac{grb} samples and their still debated standardizability \cite{Amati:2002ny, Schaefer:2006pa,  Basilakos:2008tp, Liang:2008kx} (see Ref.~\cite{Dainotti:2022bzg} for a recent review) puts limits on using \ac{grb}s as reliable cosmological observables. See Refs.~\cite{Luongo:2021nqh, Cao:2021irf} for the search of a dipole in $H_0$ through an HSA within \lcdm\ and \ac{grb} data, reporting $2-3\sigma$ deviation from isotropy.

\paragraph{Cosmicflow-4 (CF4) bulk flow}  
A measure for a possible (non)kinematic dipole may be inferred from peculiar velocities of large catalogs of distant galaxies with known distances (which typically means cosmologically close $\lesssim 500$ Mpc). Such large distance, non-random and coherent in direction velocity fields, if they exist, are called bulk flows. Bulk flows may be compared to the expected Hubble flow. CF4 (superseded CF2 and CF3) \cite{Tully:2022rbj} complies $\sim 56000$ galaxies to this end, while reporting $H_0 = 74.6\pm 0.8\pm 3(Sys)$ \kms also reports their peculiar velocities.\footnote{It is notable that this value for $H_0$ has independent systematics from those based on cosmic distance ladder.} The amplitude and alignment of the inferred velocity field from the CF4 data is at $\sim 2-3 \sigma$ discrepancy with respect to the \lcdm\ model \cite{Hoffman:2023pac}. While closer galaxies at around $50-100$ Mpc seem to show no excess or deficit in their velocity fields compared to expected Hubble flow, farther ones show sizable excesses (by more than 3 times) \cite{Watkins:2023rll, Whitford:2023oww} and in a direction compatible with the \ac{cmb} dipole, confirming similar anomalies seen in CF3 \cite{Peery:2018wie}. Similar bulk flows have also been reported from the low redshift part of Pantheon+ compilation ($0.015\leq z \leq 0.06$); these flows are toward the Shapley supercluster \cite{Lopes:2024vfz} (see also Ref.~\cite{McConville:2023xav}).

A specific kind of bulk flow (as defined above) has been given a different name, \textit{dark flows}.  Dark flows are associated with peculiar velocities of galaxy clusters that can be measured through fluctuations in \ac{cmb} generated by the \ac{cmb} photons off  X-ray emitting gas inside clusters \cite{Kashlinsky:2008ut}. Dark flows are found in scales larger than $300$ Mpc and the dark flow dipole is a dipole found at the position of galaxy clusters in filtered maps of \ac{cmb} temperature anisotropies \cite{Atrio-Barandela:2014nda} see also Refs.~\cite{Kashlinsky:2009dw, Kashlinsky:2010ur, Atrio-Barandela:2013ywa}, see however, Ref.~\cite{Planck:2018nkj}. It is hard to accommodate dark flow dipoles in the \lcdm\ cosmology.

\paragraph{Cosmological models accommodating the presence of cosmic dipoles} 
As discussed, dipoles can appear in various cosmological observables, such as the distribution of \ac{qso} and \ac{sn1}, distances, acceleration parameter, $H_0$, the fine-structure constant $\alpha$. One can construct models that accommodate dipoles in some of these observables and/or introduce mechanisms to yield large-scale inhomogeneities or anisotropies in the Universe. Below, we discuss some of the prominent theoretical models that could explain the observed cosmic dipoles.

\textit{Off-Center observers in spherical cosmological scale inhomogeneities.}
Off-center observers situated in a spherically symmetric inhomogeneous universe can see a dipolar anisotropy due to the observer's position relative to the center of the inhomogeneity. This idea is explored in models where large-scale structures or voids introduce anisotropies due to the observer's peculiar location within the cosmic structure \cite{Grande:2011hm, Alnes:2005nq}.

\textit{Primordial dipolar horizon scale perturbations.} Primordial perturbations generated during the inflationary epoch can also have dipole anisotropy that can be imprinted on the largest scales and re-enter the horizon at later times, leading to observable dipolar anisotropies in the \ac{cmb} and other cosmological observables. Such mechanisms can be related to specific inflationary models that predict large-scale anisotropies \cite{Erickcek:2008sm, Ackerman:2007nb}.

\textit{Nontrivial topology of the cosmos.}
The spatial topology of the Universe may play a role in creating cosmic dipoles. Nontrivial topologies, such as those involving multiply connected spaces, can introduce preferred directions and anisotropies on large scales which may be detectable in the distribution of galaxies, the \ac{cmb}, and other cosmological datasets \cite{Luminet:2003dx, Aurich:2004fq}. (Non-trivial topology may induce anisotropy through parity-violation without involving a dipole \cite{COMPACT:2022gbl, Jones:2023ncn}.)

\textit{Hubble scale topological defects.} Hubble scale topological defect, such as a global monopole created by the \ac{de} scalar field at recent cosmological times, known as topological quintessence, can be another source of dipolar anisotropy in various observable (from the off-center observer viewpoint) \cite{Mariano:2012wx, Perivolaropoulos:2021jda, BuenoSanchez:2011wr, Perivolaropoulos:2012mca}. This scenario may also explain the alignment of various observed dipoles, such as the \ac{qso} dipole, the \ac{sn1} dipole, and the $\alpha$ dipole.

\textit{Other physical mechanisms.}
Other physical mechanisms that can induce cosmic dipoles include:\\
- \textit{Anisotropic Dark Energy}: Models  with an anisotropic equation of state for \ac{de} can lead to directional dependencies in the expansion rate of the Universe, thereby creating dipolar anisotropies \cite{Koivisto:2005mm, Battye:2009ze,Barrow:1997sy}.\\
- \textit{Large-Scale magnetic fields}  can introduce anisotropies in the \ac{cmb} and other observables, potentially explaining some of the observed dipoles \cite{Barrow:1997mj, Campanelli:2009tk}.\\
- \textit{Vector field models}: Inflationary models involving vector fields can naturally generate anisotropic perturbations \cite{Maleknejad:2012fw}, leading to observable dipoles in various cosmological datasets \cite{Dimopoulos:2009am, Karciauskas:2008bc}.\\
- \textit{Tilted models}: Tilted cosmological models, where observers have peculiar velocities relative to the cosmic rest frame, can lead to locally observed acceleration and dipole-like anisotropies in the deceleration parameter \cite{Tsagas:2011wq, Anton:2023icm, Pasten:2023rpc, Tsagas:2021tqa, Tsagas:2021ldz, Colin:2019opb, Tsagas:2015mua, Kothari:2022bjr, Krishnan:2022qbv, Krishnan:2022uar, Ebrahimian:2023svi, Allahyari:2023kfm}.

These theoretical models provide a rich framework for understanding the observed cosmic dipoles. Future observations and more refined data will be crucial in testing these models and determining the true nature of the large-scale anisotropies in our Universe.

\paragraph{Conclusion and future directions} 
We briefly discussed dipoles in various cosmological observables, from the dipole in the distribution of astrophysical sources like \ac{sn1}, \ac{qso}, \ac{grb} and radio galaxies over the sky, to dipole in the coherent flows of matter (clusters and superclusters), to dipoles in the cosmological model observables like $H_0, \Omega_{\rm m,0}$. Besides the dipoles we discussed here, there have been reports on an intrinsic (non-kinematical) dipole in the \ac{cmb} \cite{Roldan:2016ayx, Yasini:2016dnd,Ferreira:2020aqa,Khan:2022lpx,Kester:2023qmm}, the dipole in the fine structure constant \cite{Webb:2010hc, King:2012id, Mariano:2012wx} and in the distribution of baryon matter in the Universe using \ac{frb} \cite{Qiang:2019zrs, Lin:2021syj}. Moreover, if the \ac{cmb} dipole is interpreted as our departure from the Hubble flow it will induce ‘time dilation dipole’, a directionally-dependent time dilation over the sky. Detection of time dilation dipole can provide a new assessment of the cosmological principle \cite{Oayda:2023gxk, Mittal:2023xub}. 

These dipoles, if of a cosmological origin and if not kinematical, hint at a breakdown of the cosmological principle, the very pillar of modern cosmology. Most of the dipoles discussed here may be individually of not a large statistical significance, nonetheless, the synergy between them suggests that they may not be dismissed. Future data with better sky coverage will be instrumental in establishing/refuting true cosmological dipoles. If established, one should take more serious steps in accommodating a cosmological dipole in the cosmological models, such first steps have been taken in \cite{Krishnan:2022qbv, Ebrahimian:2023svi}.
\bigskip
\subsubsection{Big bang nucleosynthesis \label{sec:BBN}}

\noindent \textbf{Coordinator:} Nils Sch\"oneberg\\
\noindent \textbf{Contributors:} Dinko Milakovi{\'c}, Ismailov Nariman Zeynalabdi, John Webb, Luca Izzo, and Venus Keus
\\

\ac{bbn} describes the process of generating the atomic nuclei of light elements shortly after the big bang as a result of the Universe cooling down beyond the point at which photo-disintegration of these nuclei becomes inefficient. Since the heavier elements need to be built up from lighter components,\footnote{The reason for this is that the direct $n$-particle fusion is suppressed by a power of $\eta_b^n$ where $\eta_b \sim 6\cdot 10^{-10}$ is the tiny baryon-to-photon ratio. As such, subsequent progressive 2-nuclei fusion processes are more relevant for the fusion of heavier elements, which makes the isobar stability gaps crucially important.} the process of the nucleosynthesis only begins when the Deuterium photo-disintegration becomes inefficient, which occurs when the photon temperature is around $T \sim 100\mathrm{keV}$ (corresponding to redshift $z \sim 5 \cdot 10^8$). For reviews see Refs.~\cite{Iocco:2008va,Cyburt:2015mya,Grohs:2023voo,ParticleDataGroup:2022pth}. Not all stable elements are readily produced from \ac{bbn} due to the instability of isobars with 5 or 8 baryons. In particular, the stable elements produced in significant proportion from \ac{bbn} are hydrogen (\element{1}{H} and \element{2}{H}), Helium (\element{3}{He} and \element{4}{He}), and Lithium (\element{6}{Li} and \element{7}{Li}), with the remaining elements produced only in tiny proportions compared to later stellar and cosmic-ray induced generation. As such, the primordial abundances of these heavier elements are almost impossible to measure, given that already for the small abundance of Lithium the post-\ac{bbn} generation presents a significant systematic, see Sec.~\ref{ssec:lithium}. Despite generally showing an excellent agreement between predicted and observed abundances, we discuss hints at possible tensions for these different elements in Sec.~\ref{ssec:lithium}--Sec.~\ref{ssec:deuterium}. Furthermore, \ac{bbn} can play a critical role in other tensions, see Sec.~\ref{ssec:impact}.

\paragraph{Lithium tension}\label{ssec:lithium}

The two stable isotopes of Lithium (\element{6}{Li} and \element{7}{Li}) are produced only in very subdominant amounts, with \element{7}{Li} at a level $10^{-9}$ times below that of hydrogen (\element{1}{H}). At face value, the prediction of the \element{7}{Li} abundance from standard \ac{bbn} is about 2-4 times higher than that measured in the atmospheres of metal-poor stars in the galactic halo, for example, see Ref.~\cite{Yeh:2020mgl} computes \element{7}{Li}/H=$(4.94\pm0.72)\cdot 10^{-10}$ (for a \ac{cmb}-motivated baryon density, see also Refs.~\cite{Coc:2011az,Consiglio:2017pot,Pitrou:2020etk,Singh:2023ugh,Gariazzo:2021iiu} for other such high computations) while Ref.~\cite{Sbordone:2010zi} measured \element{7}{Li}/H=$(1.58^{+0.35}_{-0.28})\cdot 10^{-10}$ (see Ref.~\cite{Howk:2012rb} for a higher measurement in the \ac{smc}, possibly related 
to nova enrichment \cite{2022MNRAS.510.5302I}). See Ref.~\cite{Fields:2011zzb} for a review. Na\"ively this appears to be a large issue for standard \ac{bbn}. However, there are a few caveats to take into account.

The observations of the \element{7}{Li} abundances are extrapolated to zero metallicity (iron abundance), but the plateau originally observed (the so-called Spite plateau) Ref.~\cite{Spite:1982dd,1988A&A...192..192R} appears to not persist at lower metallicities (increased scatter falling on average below the plateau), see for example Refs.~\cite{Fields:2011zzb,2021MNRAS.505..200M,Makki:2024sjq}. The issue is that Lithium can be burnt up in stellar environments, possibly even in metal-poor stars. In particular, convective mixing in metal-poor stars could bring the Lithium down to layers with a temperature above $3 \cdot 10^{6}$ K sufficient for further burning, see for example Ref.~\cite{2015MNRAS.452.3256F}. This stellar depletion argument is still under investigation, for a recent discussion see Ref.~\cite{Fields:2022mpw} -- 
In this case \element{6}{Li} would be burnt up and thus its abundance is used as a useful argument on whether such burning is expected to have taken place. However, the observation of \element{6}{Li} is even more difficult due to its even smaller abundance (and its ready generation through cosmic rays). Initial observations indicating a plateau at \element{6}{Li}/\element{7}{Li}$\sim$0.05  (\ac{bbn} predictions are at $\sim$$10^{-4}$ Refs.~\cite{Coc:2011az,Cyburt:2015mya}) would indicate that stellar burning could not have taken place~\cite{Smith:1993zzg,1994ApJ...428L..25H,1997ApJ...491..772H,1998ApJ...506..405S,1999ApJ...523..797H,Cayrel:1999kx,Asplund:2005yt}, but newer observations currently only put upper bounds \cite{Steffen:2009yr,Perez:2009ax,Lind:2013iza,Wang:2021urb}, overall making stellar depletion a likely explanation. See Ref.~\cite{Korn:2024gel} for another argument towards stellar processes and Ref.~\cite{GALAH:2020aun} for possible observations of a truly primordial plateau.

Another argument could be made on the aspect of the nuclear reaction rates, though the rates appear to be consistent, see Ref.~\cite{Fields:2022mpw} for a discussion. As such, while nuclear rates are unlikely to be the cause of the Lithium tension, stellar depletion could very well be an entirely sufficient explanation. Given this unclear footing of the Lithium tension, any claims of a ``required'' modification of standard \ac{bbn} should be seen with great caution. Despite this, a range of non-standard \ac{bbn} models have been proposed as possible solutions to the Lithium tension, such as for example Refs.~\cite{Olive:2012xf,Coc:2014gia,Jedamzik:2004ip,Jedamzik:2004er,Goudelis:2015wpa,Hou:2017uap,Nakamura:2017qtu,Yamazaki:2017uvc,Makki:2019zem,Makki:2019xem,Makki:2024sjq,Burns:2024ods,Huang:2021dba,Deppisch:2020sqh}, and Ref.~\cite{Mathews:2019hbi} for a review. 

\paragraph{Helium anomaly}\label{ssec:helium}

Primordial \element{3}{He} is remarkably difficult to measure \cite{ParticleDataGroup:2022pth} (see also Ref.~\cite{Bania:2002yj}) since \element{3}{He} is generally readily produced and depleted in stars. It is thus rarely used to constrain \ac{bbn}. Conversely, \element{4}{He} is a crucial part of the \ac{bbn} predictions that can be readily measured in ionized metal poor extragalactic HII regions. Since \element{4}{He} is produced in stellar cores, the primordial Helium mass fraction $Y_P$ is usually found by extrapolating the ratio of \element{4}{He}/\element{1}{H} from the measured systems as a function of metallicity to zero metallicity, which is presumed to represent the primordial value. These determinations from 214 systems \cite{Aver:2020fon,Valerdi:2019beb,Fernandez:2019hds,Kurichin:2021ppm,Hsyu:2020uqb,2021MNRAS.505.3624V} (which roughly average to $Y_P = 0.245 \pm 0.003$ \cite{ParticleDataGroup:2022pth}) generally agree extremely well with the value predicted from baryon abundance and neutrino abundance determinations from the \textit{Planck} satellite under the standard \ac{bbn} assumptions (giving $Y_P = 0.241\pm 0.025$ \cite{Planck:2018vyg}). The recent determination from the EMPRESS survey on the Subaru telescope by Ref.~\cite{Matsumoto:2022tlr} returns a much lower value ($Y_P = 0.2370 \pm 0.0033$) compared to all previous results, including that of Ref.~\cite{Hsyu:2020uqb} (which gets $Y_P = 0.2436 \pm 0.0040$). This is interesting because Ref.~\cite{Matsumoto:2022tlr} uses the entire sample 1 of Ref.~\cite{Hsyu:2020uqb} and the only difference is the addition of five additional extremely metal-poor galaxies (J1631+4426, J0133+1342, J0825+3532, J0125+0759, and J0935-0115). Four out of five of these galaxies lie significantly below the expected Helium mass fraction (from the fit of Ref.~\cite{Hsyu:2020uqb}) given their metallicity, see Fig.~\ref{fig:data}. Taking the measurements at face value, one would conclude a non-zero lepton asymmetry \cite{Takahashi:2022cpn,Escudero:2022okz} and a slight preference for additional relativistic degrees of freedom \cite{Matsumoto:2022tlr,Takahashi:2022cpn}. However, such an analysis is yet to be independently confirmed.We also note that Ref.~\cite{Izotov:2014fga} finds a higher-than-mean value.

\paragraph{Deuterium -- A new tension?}\label{ssec:deuterium}

The final element that is produced in decent quantity from \ac{bbn} is Deuterium (\element{2}{H}, alternatively D). While the Deuterium measurements are largely consistent, see Fig.~\ref{fig:data}, 
its abundance can also be seen to be in tension with the baryon abundance determined from \ac{cmb} data, depending on which modeling of the nuclear rates is performed \cite{Pitrou:2020etk,Pitrou:2021vqr,Pisanti:2020efz}, thus requiring new measurements of nuclear rates to clear up the posited tension. In particular, the crucial rates for Deuterium are
\begin{equation}
    \element{2}{H} + \element{2}{H} \to \element{3}{H} + p^+ ~\mathrm{(ddp)}\,, \qquad \qquad \element{2}{H} + \element{2}{H} \to \element{3}{He} + n ~\mathrm{(ddn)}\,.
\end{equation}
Conversely, to probe the consistency with the \ac{cmb} one can also check what nuclear rates would be preferred given the baryon abundance and neutrino number from the \ac{cmb} (see Sec.~\ref{sec:CMB}) \cite{DiValentino:2014cta,Pitrou:2021vqr}.

\paragraph{Impact of BBN on other tensions}\label{ssec:impact}

While \ac{bbn} is extremely self-consistent, it can also be used to support other tensions. For example, the combination of \ac{bao}+\ac{bbn} data provides one of the tightest non-Planck determinations of a low $H_0$ value, see for example Refs.~\cite{Addison:2013haa, BOSS:2014hhw,Addison:2017fdm,eBOSS:2019qwo,Cuceu:2019for,Schoneberg:2019wmt,Schoneberg:2022ggi,DESI:2024mwx} (from DESI DR1 \cite{DESI:2024mwx}, $68.51 \pm 0.58$km s$^{-1}$ Mpc$^{-1}$ from DESI DR2, or $67.6^{+0.9}_{-1.0}$km s$^{-1}$ Mpc$^{-1}$ from eBOSS (and $68.3 \pm 0.7$km s$^{-1}$ Mpc$^{-1}$ when additionally including eBOSS ShapeFit \cite{Addison:2017fdm}), see also Sec.~\ref{sec:BAO}), recently reaching a precision of $H_0 = 68.52 \pm 0.62$\kms \cite{DESI:2024mwx}, which puts a significant strain on solutions of the Hubble tension that focus solely on the \ac{cmb}. See also in particular Sec.~\ref{sec:BAO}. \ac{bbn} data also aids in constraining dark radiation or varying constant solutions, see Sec.~\ref{sec:Extra_DoF} and Sec.~\ref{sec:Var_fun_const} (although the latter is not tightly enough constrained to exclude it as a solution to the Hubble tension, see Ref.~\cite{Schoneberg:2024ynd}).

\begin{figure}
    \centering
    \includegraphics[width=0.45\textwidth]{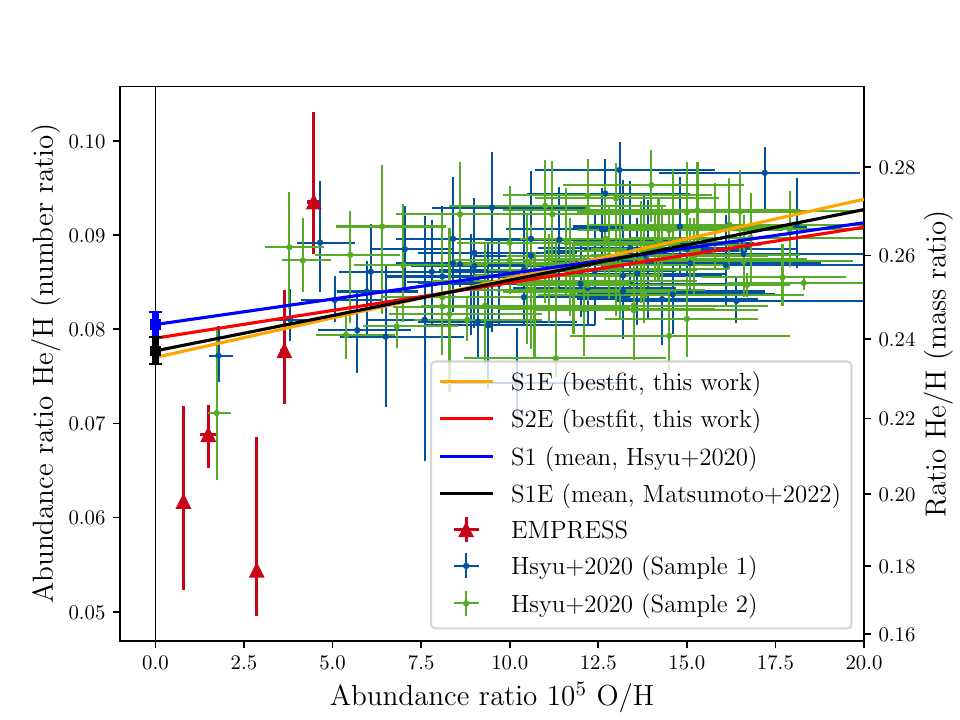}
    \includegraphics[width=0.45\textwidth]{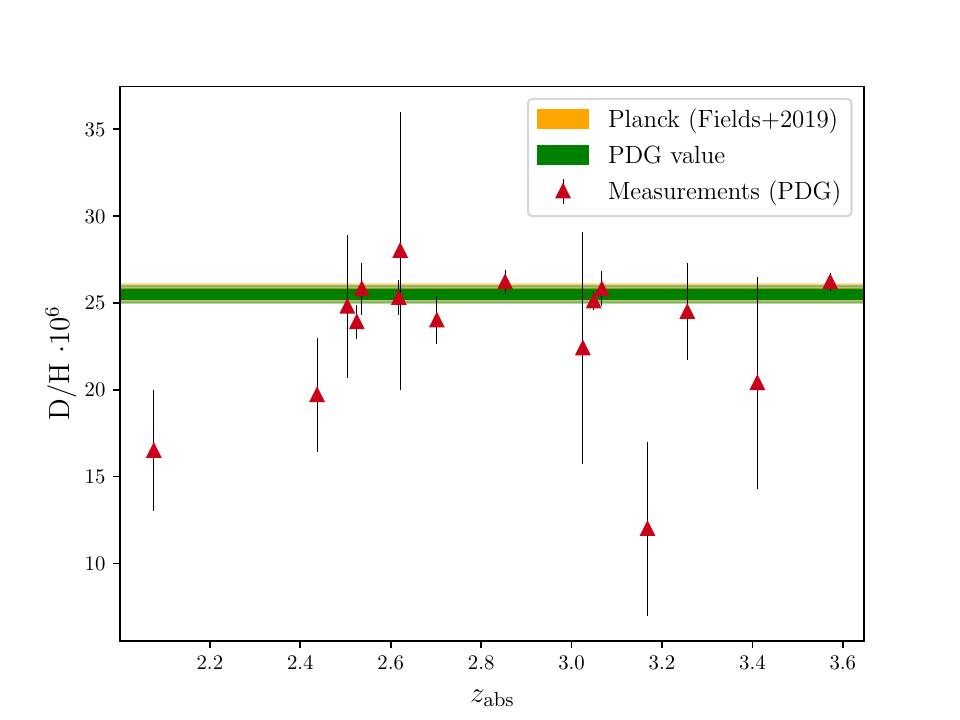}
    \caption{Left: A comparison of the Helium abundance measurements from Ref.~\cite{Hsyu:2020uqb} (Hysu+2020) to some recent fits of Ref.~\cite{Matsumoto:2022tlr} (Mastumoto+2022). For the fits, S1 is sample 1, S2 is sample 2, and E denotes the EMPRESS data. Right: A comparison of Deuterium abundance measurements from Ref.~\cite{ParticleDataGroup:2022pth}(PDG) compared to the corresponding summary value and the Planck prediction according to Ref.~\cite{Fields:2019pfx} (Fields+2019).}
    \label{fig:data}
\end{figure}

\bigskip
\subsubsection{Anomalies with Lyman-\texorpdfstring{$\alpha$}{} measurements \label{sec:Anomalies_lyman_alpha}}

\noindent \textbf{Coordinator:} Vid Ir\v{s}i\v{c}\\
\noindent \textbf{Contributors:} Simeon Bird\\

\noindent Similarly to how galaxies trace the matter distribution at lower redshifts, the \ac{igm} has been used to trace the matter distribution at the time when first galaxies are forming ($z>2$) (e.g., Ref.~\cite{McQuinn:2015icp}). The most well-studied observable of the \ac{igm} is the Lyman-$\alpha$ (Ly$\alpha$) forest -- a series of absorption lines in the spectra of \ac{qso}s, that arise due to the scattering of \ac{qso} light on the intervening neutral hydrogen atoms in the ground state. In the last two decades the Ly$\alpha$ forest has been measured in large spectroscopic surveys to study the precise position of the \ac{bao} \cite{eBOSS:2020tmo,DESI:2024lzq}, full-shape \ac{rsd} \cite{Cuceu:2021hlk,deBelsunce:2024knf}, as well as small-scale 1D clustering statistics \cite{DESI:2023xwh,Karacayli:2023afs}. The latter relies on the estimation of the 1D power spectrum along each \ac{qso} spectrum and then averaged over all the \ac{qso} spectra in the data sample \cite{Narayanan:2000tp,Seljak:2006bg}. This methodology has also been applied to smaller samples, with 10-100's of high-resolution and high signal-to-noise \ac{qso} spectra \cite{Viel:2012sd,Irsic:2017sop,Boera:2018vzq,Karacayli:2021jeg}. The highest precision constraints from the Ly$\alpha$ forest on the amplitude and shape of the matter power spectrum come from the measurements of the 1D flux power spectra.

The cosmological parameter inference from the 1D flux power spectra relies on state-of-the-art cosmological simulations \cite{Villasenor:2022aiy,Puchwein:2022wvk,Doughty:2023kko,Bird:2023evb} as well as a data compression technique \cite{SDSS:2004aee}. This method assumes that at $2<z<4$ and small scales ($<10\;\mathrm{Mpc}/h$) the Ly$\alpha$ forest is constraining only two effective parameters -- the amplitude and slope of the linear matter power spectrum at a pivot wavenumber and redshift, with the exact pivot point depending on the survey. This data compression has been shown to be valid also for some of the extensions of the \lcdm\ model, such as cosmology with massive neutrinos or running of the spectral index \cite{Pedersen:2022anu}.

However, this methodology is typically valid when limited to scales $>1\;\mathrm{Mpc}/h$ or cosmological models that do not drastically change the behavior of the small-scale linear matter power spectrum (such as light \ac{dm} models). In those cases, at least one more parameter can be extracted from the data corresponding, for example, to the scale of suppression of the matter power spectrum due to free-streaming nature of \ac{dm} \cite{Viel:2005qj}. This has been successfully used to produce bounds on various types of alternative \ac{dm} models \cite{Irsic:2017yje,Armengaud:2017nkf,Palanque-Delabrouille:2019iyz,Rogers:2020cup,Irsic:2023equ}.

While the ongoing \ac{desi} survey will provide updated constraints, the current most precise measurements of the 1D flux power spectrum at $0.1-1.0\;h\;\mathrm{Mpc^{-1}}$ come from the SDSS-IV/\ac{eboss} spectroscopic survey \cite{eBOSS:2018qyj}. A subsequent cosmological inference analysis \cite{Palanque-Delabrouille:2019iyz} reported a tension with the \ac{cmb} inferred cosmological parameters, suggesting a hint of new physics in the form of a running of the spectral index. No non-statistical effects in the data analysis were identified that could explain the signal and significant effort was invested towards mitigation strategies of systematic effects. A later re-analysis of the \ac{eboss} DR14 measurements used improved cosmological simulations and sampling techniques \cite{Fernandez:2023grg} and pointed to possible internal tension in the data between lower redshift ($z<2.5$) and higher redshift ($z>2.5$) Ly$\alpha$ data. The analysis also proposed that the previous tension driving the constraints on the running of the spectral index can be completely alleviated at the expense of the increased tension in the amplitude of the matter clustering $\sigma_8$. The S8 tension between the amplitude of structure expected by the \ac{cmb} and the amplitude of structure detected by \ac{wl} at lower redshifts has also been detected at $z=2$ by the Ly$\alpha$ analysis of Ref.~\cite{Fernandez:2023grg}, with agreement between the measurements of the Ly$\alpha$ forest and \ac{wl}.

Similarly, several analyses of high-resolution data pointed to a small tension between the \ac{cmb} and Ly$\alpha$ forest from the high-resolution observations of \ac{qso} spectra \cite{Irsic:2017ixq,Yeche:2017upn}. A $2\sigma$ tension was reported between Planck \ac{cmb} analysis and high redshift ($3<z<5.5$) Lyman-$\alpha$ forest data from XQ-100/MIKE/HIRES in Ref.~\cite{Esposito:2022plo}. The study argued that several systematic effects in the data, such as \ac{qso} continuum mis-estimation, residual high column density systems not masked in the data selection, and astrophysical uncertainties in the mean transmission of the high redshift Lyman-$\alpha$ scattering could potentially alleviate the tension. A further study combining both high redshift XQ-100/MIKE/HIRES and low redshift \ac{eboss} DR14 data \cite{Goldstein:2023gnw} showcased that the internal tension is strongest in the compressed parameter space of amplitude and slope of the linear matter power spectrum at pivot scale ($k_p\sim 1 \;h\;\mathrm{Mpc^{-1}}$) and redshift ($z\sim2.5$) as probed by the Lyman-$\alpha$ forest data. Both studies \cite{Esposito:2022plo,Goldstein:2023gnw} further proposed that systematics in the data would pull in the direction of higher slope and lower amplitude (or vice-versa). Study of Ref.~\cite{Goldstein:2023gnw} then argued that for that reason \ac{ede} models are severely penalized in this part of the parameter space and can thus be robustly excluded. A recent study by Ref.~\cite{Rogers:2023upm} summarized these findings and proposed a new physics model of mixed ultra-light axions to solve the tension between \ac{eboss} and Planck \ac{cmb} data. While successful in explaining the results of the \ac{eboss} survey, such a model would induce heavy suppression of matter power spectrum on smaller scales $<1\;h^{-1}\;\mathrm{Mpc}$, that would be detected high redshift data \cite{Irsic:2017sop,Boera:2018vzq}. These data sets access smaller scales and put stringent constraints on such on \ac{dm} suppression, including mixed \ac{dm} models \cite{Kobayashi:2017jcf,Baur:2017stq}.

The dependence of the Ly$\alpha$ 1D analysis on the cosmological simulations makes independent tests costly. Moreover, the cosmological parameter inference is not immune to astrophysical effects. While certain physical properties of the \ac{igm} are becoming well understood, and can be measured independently, such as the thermal history evolution \cite{Gaikwad:2020art}; the impact of other effects, such as inhomogeneous helium reionization, on the 1D flux power spectrum is not yet fully understood \cite{Montero-Camacho:2019ucp,Molaro:2023sys}. Aside from the uncertainty of the astrophysics of the \ac{igm}, several observational systematic effects can potentially explain the differences between the current measurements. 

Nevertheless, these anomalies in the Ly$\alpha$ forest data could offer a fascinating hint of new physics that can be tested and explored in a unique range of redshifts and clustering scales probed. 
\bigskip
\subsubsection{Cosmic superstructures and the ISW anomalies \label{sec:ISW}}

\noindent \textbf{Coordinator:} Istvan Szapudi\\
\noindent \textbf{Contributors:} Andr\'as Kov\'acs, Christine Lee, Deng Wang, Maret Einasto, Mina Ghodsi, and Pekka Heinämäki
\\

%\subsubsection{The largest structures in the cosmic web and the ISW Puzzle}

\paragraph{Galaxy superclusters}
\label{sect:sclintro} 

The largest structures in the cosmic web are superclusters of galaxies - overdensity regions that embed the richest galaxy clusters connected by filaments of poor clusters and galaxy groups \cite{2014dmcw.book.....E,Einasto:2023ymm, Sankhyayan:2023tii}. Galaxies and galaxy systems form due to initial density perturbations of different scales. Perturbations of a scale of about $100$$h^{-1}$ Mpc give rise to the largest superclusters, the largest coherent systems in the Universe \cite{Einasto:2019jyr}. The sizes of the richest and largest superclusters are up to almost $200$$h^{-1}$ Mpc \cite{1994MNRAS.269..301E,Lietzen:2016thc,Sankhyayan:2023tii}. With their galaxies, groups, clusters, filaments, and gas, superclusters can be considered as miniature versions of universes. Therefore, they are ideal laboratories to study the properties and evolution of various elements of the cosmic web \cite{Einasto:2021qgx,Aghanim:2024qsp}.

Superclusters as the connected overdensity regions in the cosmic web have been defined based on individual objects (typically optical or X-ray groups and clusters of galaxies) or luminosity-density or velocity fields \cite{1994MNRAS.269..301E, Liivamagi:2010jg,Tully:2014gfa,Dupuy:2023ffz,Sankhyayan:2023tii}.

The extreme cases of observed objects usually provide the most stringent tests for theories; this motivates the need for a detailed understanding of various properties of the richest superclusters \cite{Einasto:2011qf}. Various properties of superclusters can be used to test cosmological models, such as their abundance, masses, sizes, shapes, morphology, the properties of their high-density cores, and so on \cite{Einasto:2021qgx,Zuniga:2024wau}. Recent observational results (e.g., \ac{desi} Collaboration \cite{DESI:2024mwx}) have shown a tension in the Hubble value based on different observational probes, challenging the concordance model of the \ac{de} equation of state. Superclusters, representing the grand finale of hierarchical merging events, evolve at the forefront under the influence of two competing forces driven by \ac{de} and \ac{dm}. They may provide a useful probe through their superior mass and extent. For example, the abundance of rich superclusters at a given epoch and the stacked signal through the \ac{isw} effect generated by large superclusters can be useful to constrain the \ac{de} equation of state \cite{Lim:2012gv, Nadathur:2011iu,Granett:2008ju,Gialamas:2024lyw,Reyhani:2024cnr}. In what follows, we focus on supercluster imprint on the \ac{cmb}, and on the properties of supercluster planes, shells, and on the regularity in the supercluster distribution.

\paragraph{The ISW puzzle}
\label{sect:iswintro} 

When photons cross the decaying potential well of a supercluster, they become hotter; in voids, they become colder. This phenomenon is the \ac{isw} effect \cite{Sachs:1967er}. The \ac{isw} puzzle consists of anomalous signal levels detected when stacking the \ac{cmb} aligned with supervoids and superclusters. The effect has persisted since 2008 \cite{Granett:2008ju}. 
 
The linear contribution to the late-time \ac{isw} \ac{cmb} $T_{\mathrm{ISW}}$ is integrated over the line-of-sight from the present time ($z=0$) to the surface of last scattering ($z_\mathrm{LS}$) \cite{Sachs:1967er},
\begin{equation}
    \label{eq:ISW_definition}
    \frac{\Delta T_\mathrm{ISW}}{\overline{T}}(\boldsymbol{\hat n}) = 2\int_0^{z_\mathrm{LS}} \frac{a}{H(z)}\dot\Phi\left(\boldsymbol{\hat n},\chi(z)\right)\,\mathrm{d}z\ = -2\int_0^{z_\mathrm{LS}} a\left(1-f(z)\right)\Phi\left(\boldsymbol{\hat n},z\right)\,\mathrm{d}z\,,
\end{equation}
where $a$ is the scale factor, $\chi$ is the comoving distance, $\Phi$ is the gravitational potential, and $H(z)$ is the Hubble parameter. The second equality reveals the connection with the logarithmic growth function $f(\tau) \equiv d \ln D / d \ln a $ \cite{Beck:2018owr}. While the analogous non-linear Rees-Sciama effect~\cite{Rees:1968zza} happens in all models, the \ac{isw} effect gauges the divergence of growth history from the Einstein-de Sitter (EdS) model where it is zero.

The \ac{cmb} stacked towards superclusters and supervoids creates hot and cold spots, respectively (c.f., Fig.~\ref{fig:granett2008}). The \ac{isw} amplitude relative to concordance expectations, $A_\mathrm{ISW}\equiv \Delta T_\mathrm{obs}/\Delta T_\mathrm{\Lambda\text{CDM}}$  quantifies any anomalies when $A_\mathrm{ISW}\neq 1$. The first measurement to stack superstructures \cite{Granett:2008ju} found $A_\mathrm{ISW}\simeq 4$ with the correct signs for supervoids and superclusters. 

\begin{figure}[htbp]
    \centering
    \includegraphics[scale = 0.65]{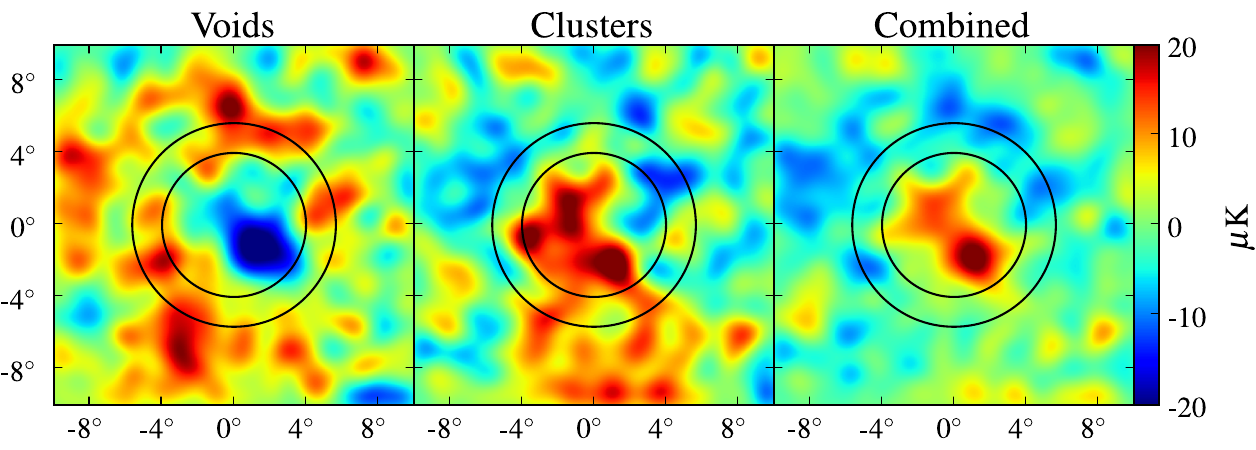}
    \caption[]{The first stacking measurement of \ac{wmap} towards SDSS LRGs \cite{Granett:2008ju} from 2008. The significance exceeded 4.4$\sigma$ with  $A\simeq 4$. The anomalous signal has been reproduced with Planck \ac{cmb} maps and in most wide-angle \ac{lss} maps over different areas of the sky.}
\label{fig:granett2008}
\end{figure}

\paragraph{Summary of observations}
Later stacking measurements, mainly based on voids, are consistent with $A_{\rm ISW}\approx4-5$ \cite{Nadathur:2011iu,Flender:2012wu,Hernandez-Monteagudo:2012xre,Aiola:2014cna,Soltan:2018upx}. Cross-correlating the \ac{cmb} with large-scale structure is typically consistent with the concordance model \cite{Planck:2015fcm,Hang:2020gwn}, although with less statistical power from a given data set. Stacking superstructures focuses on the highest signal-to-noise data
in contrast with the unweighted average of the two-point statistics.

The \ac{boss} redshift survey
updated the void stacking measurements of the SDSS. Void finding techniques solving the ``void-in-void'' problem by combining voids into larger structures confirmed the earlier excess \cite{Cai:2016rdk,Kovacs:2017hxj}. Algorithms with no merging identified no excess \cite{Nadathur:2016hky}. Stacking \cite{DES:2016zxh,DES:2018nlb} \ac{des} Year-1 and Year-3 photometric redshift data sets further corroborated earlier results by Ref.~\cite{Granett:2008ju}. Since \ac{des} covers a different part of the sky than the original SDSS, it lessens the likelihood of a statistical fluke. The combined \ac{des} and \ac{boss} data sets \cite{Kovacs:2017hxj} yield the ultimate result $A_\mathrm{ISW}\approx5.2\pm1.6$ in the redshift range of $0.2<z<0.9$.
Fig.~\ref{fig:ISW_tension} summarizes the principal observations of the \ac{isw} tension. 

\paragraph{The cold spot}

The stacking measurements draw attention to the largest anomaly of the \ac{cmb}, the Cold Spot (CS), an exceptionally cold of approximately $70$ \textmu K area centered on Galactic coordinates $(l,b) \simeq (209^\circ,-57^\circ)$ \cite{WMAP:2012fli,Vielva:2003et,Cruz:2004ce}. Could it be due to the \ac{isw} effect? The Eridanus void, found by Refs.~\cite{Szapudi:2014zha,Finelli:2014yha} targeting an extended underdensity aligned with the CS, is the likely cause. While the full CS signal with substructures taken into account predicted by the concordance model $A_\mathrm{ISW}=1$ \cite{Mackenzie:2017ioh} is not enough to explain observations, the excess $A_\mathrm{ISW}\approx 4-5$ would suffice. Bayesian hypothesis testing overwhelmingly favors the hypothesis of the Eridanus void causing the CS over chance alignment \cite{Szapudi:2014zha}.

\paragraph{AvERA and the ISW and Hubble puzzles}

The Average Expansion Rate Approximation (AvERA) model \cite{Racz:2016rss,Beck:2018owr,Racz:2016rss} simultaneously explains the excess \ac{isw} signal and Hubble constant anomaly, see Ref.~\cite{Pataki:2025ubv}, where they fit $H_0 = 71.99^{ +1.05}_{-1.03}$\kms to \ac{sn1}. AvERA approximates emerging curvature models \cite{Rasanen:2011ki,Buchert:2015iva,Wiltshire:2024mph} with a modified $N$-body simulation. AvERA reverses the order of averaging and solving the Friedman equation; a minor but surprisingly consequential modification of the standard $N$-body algorithm. Its motivation is the Separate Universe Hypothesis (SUH), stating that an under- or overdense region evolves like a universe with modified cosmological parameters. Recent general relativistic simulations \cite{Williams:2024vlv} confirmed that emerging curvature in voids dominates the late cosmic expansion history as predicted by AvERA.

The final Hubble constant in AvERA depends slightly on the coarse-graining parameter. Between the extreme scales of the box size (no effect) and resolution (SUH breaks down), the final results are only mildly sensitive to coarse-graining: an order of magnitude change will alter the results by a few percent. The best fit, expressed as a mass scale of $\simeq 1-2 \times 10^{11} M_\odot$, reproduces the locally observed higher Hubble constant \cite{Racz:2016rss}.

The AvERA expansion history is similar to concordance expectations with minor differences in detail. By definition, it starts on the same trajectory as a \lcdm\ or EdS model at high $z$. Around $z\simeq 4$, the initial collapse of high-density regions accelerates growth compared to the concordance model. At lower redshifts, $z\simeq 1.5$, the voids start dominating. Their expansion stunts growth more effectively than $\Lambda$, thus the ``void effect'' explains the local higher Hubble constant. The higher derivative also produces an excess \ac{isw} effect \cite{Beck:2018owr} comparable to observations. Thus, the AvERA \emph{explains the Hubble and \ac{isw} puzzles simultaneously}. In AvERA, the coarse-graining scale drives the local Hubble constant. While boosting the coarse-graining scale by a factor of two induces only a few percent change in $H_0$, a reasonable choice of a few times $10^{11}M_\odot$ (Lagrangian) achieves perfect consistency with the local measurements. The \emph{late complexity} of AvERA expansion history affects $S_8$ as well, although this has not been investigated in detail. 

AvERA has a robust and surprising prediction: a sign change of the \ac{isw} effect above $z \simeq 1.5$ \cite{Beck:2018owr}. In contrast, the concordance \ac{de} models predict an unmeasurably small \ac{isw} signal at the same redshift.

\begin{figure}[htbp]
\begin{center}
\includegraphics[scale=0.35]{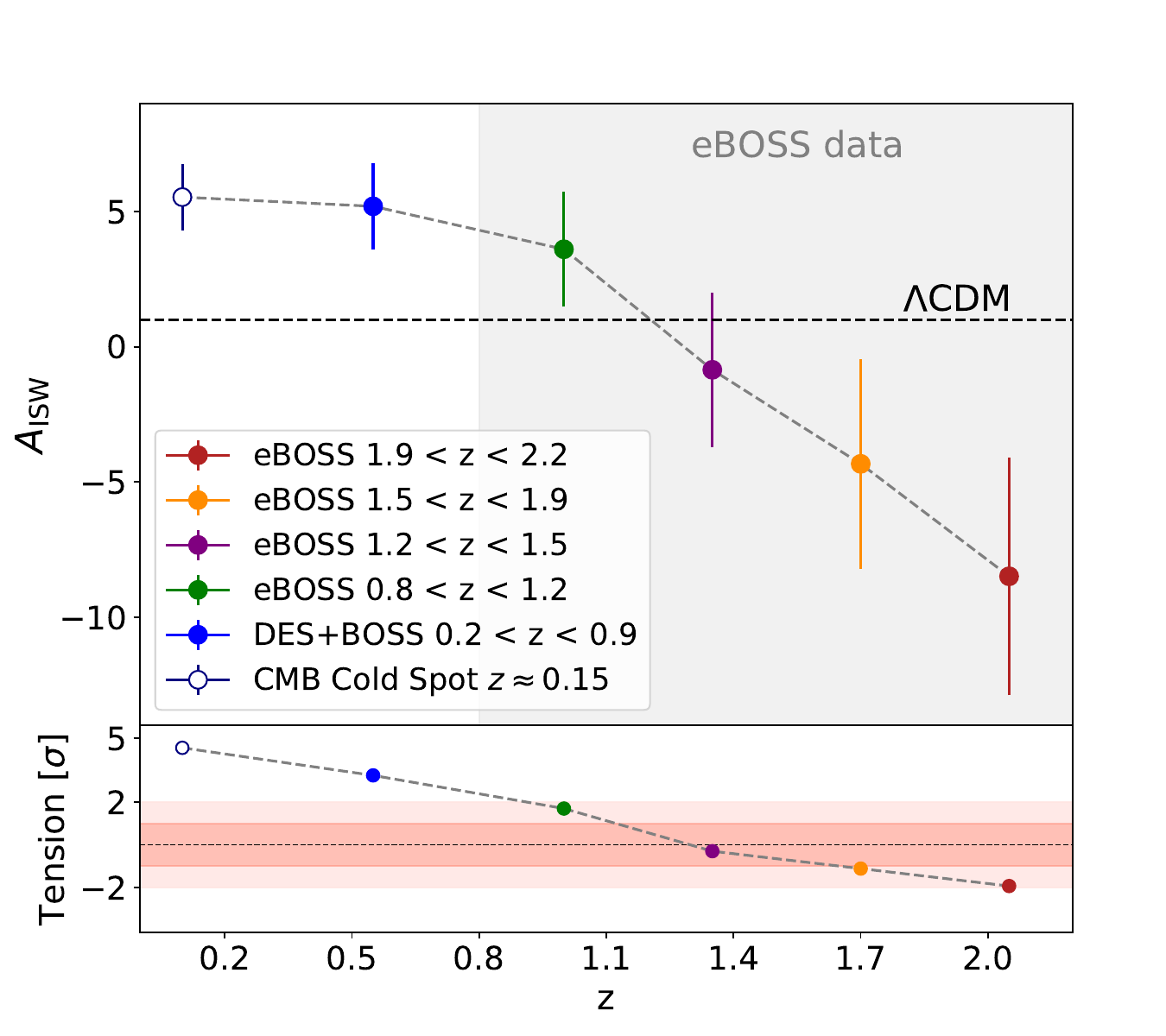}
\includegraphics[scale=0.35]{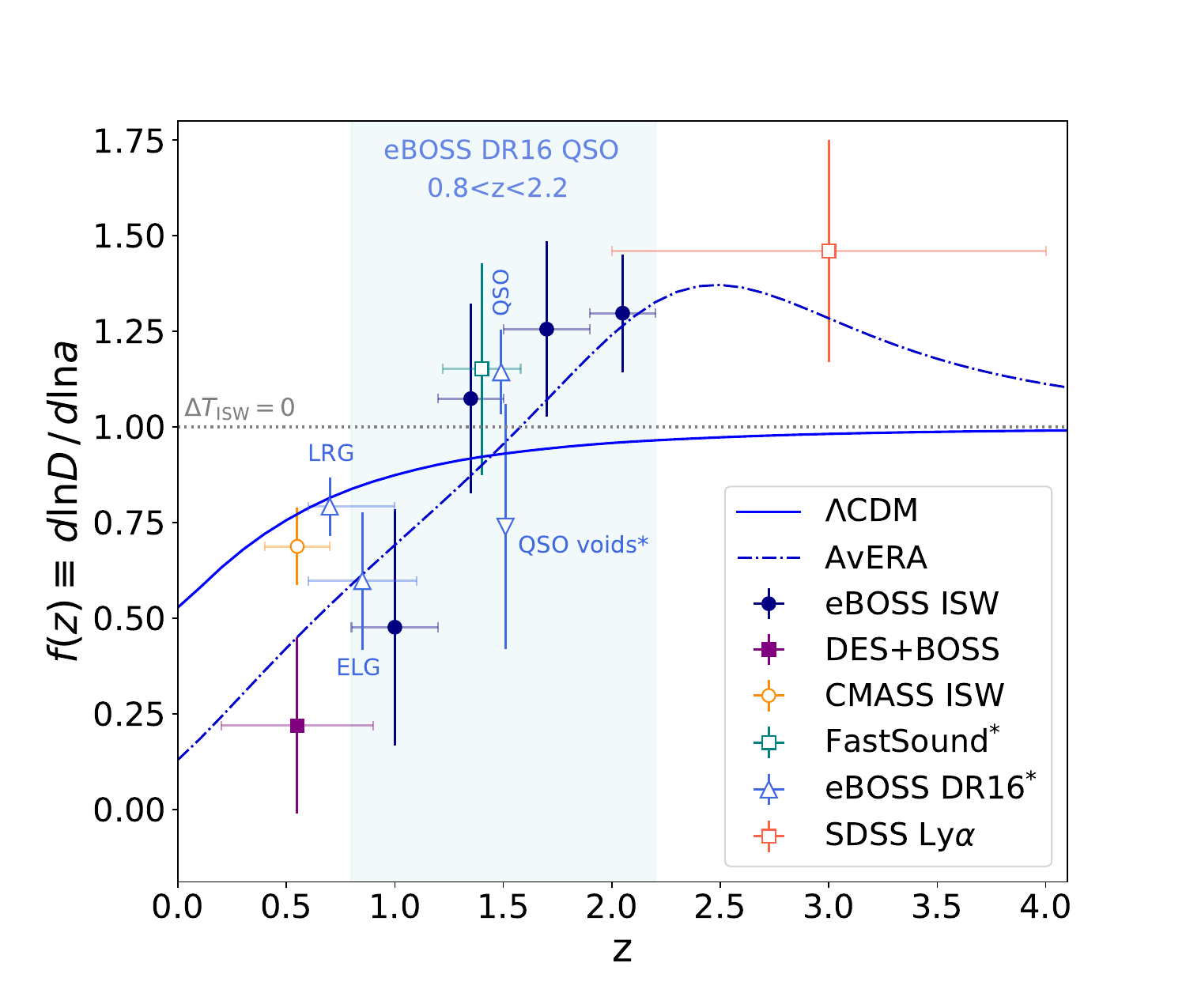}
\caption{\label{fig:ISW_tension} 
These figures from Ref.~\cite{Kovacs:2021mnf} sum up the \ac{isw} tension as of today.
Left: Observed \ac{isw} amplitudes in different redshift bins. The excess \ac{isw} signal transitions around $z\lesssim1.5$ and changes sign. The bottom panel displays the significance of the deviations compared to \lcdm\ predictions ($1\sigma$ and $2\sigma$ correspond to shaded bands).
Right: Solid and dashed lines show the cosmic growth history in\ \emph{Planck} 2018 \lcdm\ cosmology and in AvERA as used by Ref.~\cite{Hang:2020gwn}. Points with an asterisk display $f\sigma_{8}(z)$ constraints divided by the \emph{Planck} $\sigma_{8}(z)$ value. 
The \ac{des}, \ac{boss}, and \ac{eboss} \ac{isw} anomalies have a consistent trend in terms of the re-scaled \lcdm\ growth rate values ($A_{\rm ISW}(z)\times[1-f^{\Lambda {\rm CDM}}(z)]$). The CMASS \cite{Nadathur:2016hky} and \ac{eboss} LRGs \ac{isw} amplitude is not significantly anomalous. Nevertheless, the \ac{eboss} ELG, \ac{qso}, and high-$z$ constraints from the FastSound and SDSS Ly$\alpha$ lean toward the \emph{late complexity} trend predicted by the AvERA.}
\end{center}
\end{figure}

\paragraph{Opposite sign ISW effect}

The \ac{eboss} DR16 \ac{qso}s \cite{eBOSS:2015zvd} enabled the first look for \ac{isw} sign reversal \cite{Kovacs:2021mnf}. The data cover a crucial redshift range $0.8<z<2.2$ where \lcdm\ and AvERA diverge: the first tends to zero while the latter predicts a sign change around $z\simeq 1.5$.

The concordance model signal was estimated with a \ac{qso} mock catalog from the Millennium-XXL for comparison with stacking 800 supervoids of the \ac{eboss} DR16 \ac{qso} catalog. The excess signal $A_\mathrm{ ISW}\approx3.6\pm2.1$ in the redshift range of $0.8<z<1.2$ is comparable with earlier observations. At $1.5<z<2.2$, the AvERA-predicted opposite-sign \ac{isw} signal emerged in $2.7\sigma$ tension with the \lcdm. Despite the moderate significance of these measurements, taken together with the excess at low redshift and the \ac{cmb} CS, it suggests a more complex growth history than any variations of the concordance model. The observed \emph{late complexity} (c.f., Fig.~\ref{fig:ISW_tension}) consistent with AvERA implies that alternative models, such as emerging curvature, will explain the late expansion history of the Universe.

\paragraph{Supercluster complexes and planes}
\label{sect:planes} 

In several cases, the richest superclusters form complexes in which several very rich superclusters are almost connected. These are, for example, the Sloan Great Wall at redshift $z = 0.08$ and the \ac{boss} Great Wall at redshift $z = 0.47$ with a length of over $250$$h^{-1}$ Mpc \cite{Gott:2003pf,Lietzen:2016thc,Einasto:2016rhe,Einasto:2022zqp}. It has been questioned whether the presence of several such rich complexes in the local Universe is in agreement with the standard cosmological model \cite{Einasto:2006es,Granett:2008ju,Sheth:2011gi,Park:2012dn}.

Also, in the local Universe galaxy superclusters are arranged on huge planes which span many hundreds of Mpc in space. Among these are the Local Supercluster Plane, discovered by Ref.~\cite{1983IAUS..104..405E} in the distribution of superclusters. Rich optical and X-ray clusters, as well as luminous galaxies and radio galaxies in the local Universe are located on this plane \cite{1991AuJPh..44..759S, 1994MNRAS.269..301E, Boehringer:2021pvv,Peebles:2023jzn}. Perpendicular to the Local supercluster plane, \cite{Einasto:1996zd} discovered in the distribution of rich superclusters another plane - the Dominant supercluster plane. It is not yet clear whether the presence of such planes are in agreement with our standard cosmological model. For example, using constrained simulations of the local Universe in the $500$$h^{-1}$ Mpc box \cite{Dolag:2023xds} found that a plane with the extent of approximately $100$$h^{-1}$ Mpc can be recovered in simulations. However,  0.28\% of random realizations of simulations only matched both underdense and overdense regions of the local Universe, and their simulation box was not large enough to test the presence of a supercluster plane hundreds of Mpc long. Also, they did not analyze the probability of finding two perpendicular supercluster planes. Refs.~\cite{Peebles:2021gou,Peebles:2023jzn} speculate that such planes may be a signature of long, nearly straight strings.

\paragraph{Regularity in the distribution of rich superclusters and supercluster shells}
\label{sect:regularity} 

Ref.~\cite{1994MNRAS.269..301E} analyzed for the first time the spatial distribution of superclusters based on the all-sky catalog of superclusters using various methods. In their study \cite{1994MNRAS.269..301E} discovered that local rich superclusters are arranged in an almost regular pattern with the characteristic distance between superclusters $120 - 140$$h^{-1}$Mpc, see also Ref.~\cite{Toomet:1999yw} on the regularity periodogram of the local structures. The power spectrum of rich clusters has a bump at this scale, and the correlation function has a series of wiggles \cite{Einasto:1997md, Einasto:1997wc, Tago:2000ia}. Rich superclusters are separated by giant voids or supervoids \cite{Lindner:1995ce,Kovacs:2015hew}. Superclusters are not arranged randomly in the cosmic web, and this makes projections smaller. As a result, supervoids are in some cases connected and form very large voids, as the Eridanus supervoid. Such huge structures can leave their signature to the \ac{cmb}. These signatures are discussed in Sec.~\ref{sect:iswintro}. In addition, Ref.~\cite{Einasto:2014coa} detected a hint of the regularity with the scale of about $400$$h^{-1}$Mpc in the distribution of \ac{qso} systems at redshift range  $1 < z < 1.8$.

The Sloan Digital Sky Survey covers a part of the regular patterns of superclusters. Ref.~\cite{Einasto:2015vea} showed that rich superclusters in the sky region, covered by the SDSS, are arranged in six intertwined shell-like structures with the same characteristic radius as found for the whole pattern of superclusters, approximately $120 - 140$ $h^{-1}$ Mpc \cite{Einasto:2015vea}. The centers and walls of these shells are marked by rich galaxy clusters in superclusters such as the Bootes supercluster, the Sloan Great Wall, the Corona Borealis supercluster, the Ursa Major supercluster, and other superclusters, see also Ref.~\cite{Einasto:2011zc}. The most prominent of these shells was recently determined also using the CosmicFlows 4 data, and named ``Ho'oleilana''~\cite{Tully:2023epf}. 

The huge sizes of these shells tell us that the origin of these shells, and regular pattern in the supercluster distribution should come from the very early Universe. Ref.~\cite{Einasto:2015vea} concluded that the process behind these shells is still unknown. Ref.~\cite{Tully:2023epf} interpreted the shell which they named as Ho´oleilana as \ac{bao} shell. However, in the \ac{bao} shells the central mass is always much higher than the mass in shell walls \cite{Arnalte-Mur:2011nke}. Analysis of the luminosities and masses of superclusters have shown that rich superclusters typically have mass-to-light  ratios $M/L \approx 300$ \cite{Einasto:2015fma,Einasto:2016rhe, Heinamaki:2022ccr}. Therefore, we can estimate masses of superclusters in the Ho´oleilana using their luminosities provided by Ref.~\cite{Einasto:2015vea} and this $M/L$ value. The mass of the central cluster in the Bootes supercluster, Abell cluster A1795 is $M \approx 6.6-11.2 \times10^{14} M_\odot$ \cite{Vikhlinin:2004uu, Tempel:2014cya}. The mass of the  Bootes supercluster itself is the lowest among superclusters in this shell, $M_{Boo} \approx 0.1 \times10^{16} M_\odot$. Masses of superclusters in the walls of this structure are more than a hundred times higher than the mass of the central cluster, A1795, of the order of $M_{scl} \approx 0.2 - 3.5 \times10^{16} M_\odot$, and their total mass is at least $M_{tot} \approx 25 \times10^{16} M_\odot$.

Therefore, masses of superclusters contradict the interpretation of Ho'oleilana as \ac{bao} shell, as proposed in Ref.~\cite{Tully:2023epf}. Ref.~\cite{Einasto:2015vea} already provided arguments against \ac{bao} interpretation. Namely, the radius of the shell is larger than $\approx 109$ $h^{-1}$ Mpc, the \ac{bao} scale, and shell is wide, being in the interval of $120 - 140$ $h^{-1}$ Mpc. Ref.~\cite{Tully:2023epf} proposed that for the interpretation that this structure represents a \ac{bao} shell, the value of the Hubble constant should be approximately $77$\kms, see Ref.~\cite{Tully:2023epf} for details. However, so high value of H0 is very unlikely. Therefore it is still, even thirty years after discovery, an open question whether the presence of such structures can be explained within the \lcdm\ cosmological model, and which processes in the very early Universe are behind this regularity.

\bigskip
\subsubsection{JWST anomalies \label{sec:JWST_anomalies}}

\noindent \textbf{Coordinator:} Matteo Forconi\\
\noindent \textbf{Contributors:} William Giar\`{e}\\

\noindent One of the major challenges in testing the \lcdm\ model is the lack of direct observations at high redshift ($z\gtrsim10$). Current probes, such as \ac{bao} and \ac{sn1}, do not reach this regime, making it difficult to study the accuracy of the \lcdm\ model around these epochs; this is particularly important considering that the large-scale structures we observe today largely emerged in the early stages of the Universe's evolution. The \ac{jwst}, however, has been able to achieve this goal, offering a glimpse into these distant epochs of galaxy evolution history. Because of the expansion of our Universe, photons emitted by distant sources experience a non-negligible redshifting, such that UV light falls into the infrared spectrum today. Moreover, the farther away the source (i.e., the earlier in cosmic time), the fainter the observed signal. To address these observational challenges, \ac{jwst} Near Infrared Camera (NIRCam)~\cite{2005SPIE.5904....1R} covers a wavelength range of $[0.6-5]$\textmu m, improving the coverage of its predecessor, \ac{hst}, that was optimized for visible wavelengths. Coupled with a primary mirror which is nearly three times larger than that of \ac{hst}, \ac{jwst} significantly increases our ability to detect and resolve faint and very distant targets.

Recent \ac{jwst} observations have identified a large population of photometric galaxy candidates~\cite{2023ApJ...942L..27S, 2022ApJ...938L..15C, Finkelstein_CEERS_I1, 2022ApJ...940L..14N, Treu:2022iti, Harikane:2022rqt, 2023ApJ...946L..16P, 2023ApJ...949L..18P, 2023MNRAS.523.1036B, 2023MNRAS.518.4755A}. Many of these galaxies are very bright~\cite{2022ApJ...940L..14N,2022ApJ...938L..15C,Bouwens:2022gqg}, with UV \ac{lf}s that differ from previous theoretical and observational predictions~\cite{Harikane:2022rqt,2024ApJ...969L...2F,2024MNRAS.527.5004M,2024MNRAS.527.5929Y,2024ApJ...975..285W,Sabti:2023xwo}. They are found at remarkably high redshifts ($z\geq 12$)~\cite{2023MNRAS.518.6011D,2023MNRAS.519.3064F,Harikane:2022rqt,2023ApJ...943L...9Z,Yan:2022sxd} and some exhibit extremely large stellar masses ( $M\geq 10^{10.5}M_\odot$)~\cite{Labbe:2022ahb,2024Natur.635..311X,2024ApJ...965...98C,2024MNRAS.533.1808W,2024AJ....168..113C}. The unveiling of these extreme objects is difficult to reconcile within the standard galaxy formation models and, by extension, the assumed underlying cosmology (\lcdm)~\cite{Lovell:2022bhx,Boylan-Kolchin:2022kae,Forconi:2023izg}. Moreover, \ac{jwst}'s ability to discover new \ac{sn} candidates~\cite{Lu:2022utg,Pierel:2024ewl,Vinko:2024jqp,Coulter:2025nqt,Moriya:2025coz,Moriya:2023jfq} introduces additional prospects for tackling tension in cosmology~\cite{Dainotti:2025qxz,Alonso:2023oro}.

It is important to emphasize that these preliminary findings are from photometric data which, alone, are not enough to draw definitive conclusions. A major source of uncertainty is the difficulty in differentiating early star-forming galaxies from quiescent galaxies at lower redshifts~\cite{2024ApJ...968...34W}, as well as the spectral energy distribution templates used to interpret photometric bands may be unsuitable for very massive galaxies in the early Universe - though updated templates have been suggested, e.g., see Refs.~\cite{2023ApJ...951L..40S,2023MNRAS.524.2312E}. Nonetheless, comparisons of photometric and spectroscopic redshifts in overlapping samples of galaxies with both measurements further validate these observations~\cite{2023ApJ...951L..22A,2023ApJ...957L..34W,Robertson:2022gdk,2023NatAs...7..622C,Fujimoto:2023orx,2024Natur.633..318C,2024ApJ...972..143C,Gimenez-Arteaga:2022ubw}.

A way to alleviate this emerging tension is to reconsider the physics and assumptions behind the structure evolution and star formation process at high redshift~\cite{2024A&A...684A.207F,Mason:2022tiy,Kravtsov:2024tmk,Hegde:2024kph,2024ApJ...975..192G,Desprez:2023pif,2023MNRAS.523.3201D,Ferrara:2022dqw,2024ApJ...963...74W,2024ApJ...961...73N,2025ApJ...978L..42C,Haslbauer:2022vnq}. Because measured stellar masses strongly depend on the Initial Mass Function (IMF), adopting a top-heavy IMF~\cite{Ventura:2024mbp,2024ApJ...963...74W,2024MNRAS.529.3563T} instead of the typical standard assumption of a Salpeter IMF~\cite{Salpeter:1955it} could shift the inferred stellar masses. Additionally, increasing the gas temperatures can lead to a greater contribution to the brightness of early galaxies and consequently lowering the stellar mass estimates, although feedback effects might compensate this mechanism~\cite{Cueto:2023ayd}. Another critical factor is the star formation history and star formation efficiency $\epsilon$. The results are highly sensitive to the model adopted~\cite{2022ApJ...927..170T,2023MNRAS.519.5859W,Iocco:2024rez,Pallottini:2023yqg}; a higher $\epsilon$ can explain large early galaxies by accelerating the reionization process ~\cite{2024A&A...689A.244C,2025MNRAS.537.1826T,2025ApJ...978...89H,2023MNRAS.526.2542C,Wang:2023gla} though this must be reconciled with constraints from lower-redshift data. Indeed, extremely high $\epsilon$ with minimal feedback is required to explain bright $z\sim16$ \ac{jwst} candidates~\cite{Qin:2023rtf,2023MNRAS.519.1201A}. 

These possible explanations deal with the complex physics relating star formation to the evolution of \ac{dm} halos at high redshifts. However, emerging evidence suggests the issue might lie  within our cosmological model. While it is premature to draw any definitive conclusions from these preliminary observations, if neither of the aforementioned possibilities can reconcile theory with the \ac{jwst} data, it may be necessary to rethink the assumptions of the underlying cosmological model itself. 

A common way to quantify discrepancies with \ac{jwst} data is through the \textit{Cumulative Stellar Mass Density} (e.g., see Refs.~\cite{Labbe:2022ahb, 2023ApJ...942L..27S, 2024Natur.635..311X,Navarro-Carrera:2023ytd}), defined as
\begin{equation}
    \rho_\star(\bar{M})=\epsilon f_{\rm b}\int^{z_2}_{z_1}\int^{\infty}_{\bar{M}}\frac{dn_h}{dM}MdM\frac{dV}{V(z_1,z_2)}\,,
    \label{CSMD}
\end{equation}
where $f_{\rm b}=\Omega_{\rm b}/\Omega_{\rm m}$ is the cosmic baryon fraction, $\epsilon$ is the efficiency of converting baryons into stars, $V$ is the comoving volume of the Universe between redshift $z_2$ and $z_1$ and $dn_h/dM$ is the mass function. For computing the mass function, the halo multiplicity~\cite{Press:1973iz} is needed and one of the most used is the Sheth-Tormen alternative~\cite{Sheth:1999mn,Sheth:1999su} due to its theoretical motivations for collapsed halos~\cite{Maggiore:2009rw,Achitouv:2012ux} and multiple successful N-body tests~\cite{Despali:2015yla,Reed:2006rw} (even though other phenomenological fitting mass functions exist~\cite{Basilakos:2010fb, Bhattacharya:2010wy}). A further assumption often made is to use a Top-Hat window function for smoothing the density field. In focusing on the cosmological framework, one typically sets $\epsilon$ to a fixed value (for a conservative approach usually $\epsilon=0.2$ is enough). In principle, star formation efficiency can be a function of the halo mass~\cite{Tacchella:2018qny} and further adjustments to star formation physics might be needed for more precise computations~\cite{2024AJ....168..113C}. Nevertheless, for massive halos (e.g., CEERS observations~\cite{Labbe:2022ahb}), one can approximate $\epsilon$ as a smoothly varying power law. If the mass range is short, this approximation does not affect greatly the assumptions for the model. Lastly, the cosmic baryon fraction is usually considered, instead of computing the baryon evolution in different halos~\cite{Allen:2004cd,Vikhlinin:2005mp,Kravtsov:2005ab,Borgani:2009cd,Mantz:2014xba,Maio:2014qwa,Panchal:2024dcl}. If $f_{\rm b}$ is not a function of the halo mass, it plays a role of a multiplicative factor and a greater $f_{\rm b}$ can only push the theoretical predictions towards the observed data points. All these methodological choices and simplifications are widely used in the literature and allow to present conservative results that can be directly compared with similar works following the same approach.

One strategy for reconciling \ac{jwst} data is to boost the halo mass function at the tail, thereby producing enough massive galaxies at early-times. This can be achieved with massive primordial black holes~\cite{Huang:2024aog,Yuan:2023bvh} with a fractional contribution of $f_{\mathrm{PBH}}\sim 10^{-3}$, consistent with current constraints~\cite{Carr:2023tpt,Dolgov:2023ijt}. Super massive primordial black holes not only can accelerate the structure formation process but they could also explain the observed super massive black holes that, otherwise, would have insufficient time to form from Pop III stellar remnants~\cite{Trinca:2022jav,  Pacucci:2023oci, Maiolino:2023zdu, Mo:2023ioo, CEERSTeam:2023qgy, Bogdan:2023ilu, Schneider:2023xxr, Hai-LongHuang:2024vvz, Trinca:2024dgt}. Other approaches involve modifying the primordial fluctuations through, for instance, rare density fluctuations from the inflaton field~\cite{Kumar:2025gon}, a blue tilted~\cite{Parashari:2023cui,Hirano:2023auh} or modified~\cite{Padmanabhan:2023esp} primordial power spectrum, or also introducing non-Gaussianities in the initial condition, affecting the abundance of \ac{dm} halos~\cite{Biagetti:2022ode}. Barring self-interactions, the properties of \ac{dm} halos around galaxies imply that $m_a > 10^{-21} \,$eV i.e. that this particle is indistinguishable from a cold one on galactic scale \cite{DeLaurentis:2022nrv}.

Axion-related solutions offer a further possibility for understanding \ac{jwst}’s detection of unusually massive galaxies~\cite{Guo:2023hyp}. For example, axion miniclusters can boost early galaxy formation~\cite{Hutsi:2022fzw} though their required masses exceeded super-radiance constraints. Alternatively, axions with masses between  $10^{-22}\mathrm{eV}\lesssim m_a \lesssim 10^{-19}\mathrm{eV}$ with delayed oscillations, can produce efficient axion field fragmentation driving the formation of more massive galaxies~\cite{Bird:2023pkr}. Introducing a fuzzy \ac{dm} component, usually modeled with axions, in the structure formation process can lead to suppression of small halos and galaxies, thus potentially alleviating tension if paired with altered star formation efficiency~\cite{Gong:2022qjx}. Yet another axion-like solution is found in axion quark nuggets~\cite{Zhitnitsky:2023znn}.

These \ac{jwst} observations can also be interpreted by invoking \ac{pmf}~\cite{Zhang:2024yph} or cosmic string loops with a tension below the limit imposed by \ac{cmb} observations but above constraints from astrophysical probes (that can be avoided by adjusting the effective low radius cut-off)~\cite{Jiao:2023wcn,Jiao:2024rcr,Koehler:2024gzv}. Observed massive galaxies also allow us to revisit the proprieties of \ac{wdm} particles~\cite{Dayal:2023nwi,Maio:2022lzg,Gandolfi:2022bcm}. Explored in the context of \ac{ede}, the fit to \ac{jwst} observations is improved with respect to the one in the minimal \lcdm\ framework. In fact, \ac{ede} leads to an increased $n_s$ and simultaneously raises the fraction of \ac{ede}, that allows for a larger number of halos and massive galaxies, and, at the same time, achieving a non-negligible improvement toward easing the Hubble tension~\cite{Forconi:2023hsj,Klypin:2020tud,Shen:2024hpx,Jiang:2024tll}. Another possibility to increase the predicted cumulative stellar mass density in agreement with \ac{jwst} observations, is to employ alternatives in the \ac{de} sector such as phantom crossing in the \ac{de} equation of state~\cite{Menci:2022wia}, a negative cosmological constant~\cite{Menci:2024hop,Menci:2024rbq,Adil:2023ara} a sign-switching model~\cite{Paraskevas:2023itu} or a simple \ac{cpl} parametrization~\cite{Wang:2023ros}. Conversely, interactive dark energy appears to exacerbate the tension~\cite{Forconi:2023hsj}.
\bigskip
\subsubsection{Cosmic voids \label{sec:voids}}

\noindent \textbf{Coordinator:} Dante Paz\\
\noindent \textbf{Contributors:} Alice Pisani, Carlos Correa, Cora Uhlemann, Maret Einasto, Mina Ghodsi, Nico Hamaus, Nico Schuster, Sofia Contarini, and Umut Demirbozan
\\

\paragraph{Properties and identification of cosmic voids}

Cosmic voids are vast underdense regions of the Universe. Since their discovery \cite{1978MNRAS.185..357J,1978ApJ...222..784G,Kirshner:1981wz,deLapparent:1985umo}, they have been recognized as powerful cosmological laboratories \cite{Zeldovich:1982zz,Peebles:2001nv,Padilla:2014hea,Pisani:2019cvo}. Hence, they encode key information about the expansion history and geometry of the Universe \cite{Hamaus:2013qja,Pisani:2019cvo}. The statistical properties of voids depend on two factors: (i) the matter tracers used to map the large-scale structure, namely, galaxies, and (ii) the method used to identify them from the spatial distribution of these tracers. Regarding the first factor, the non-trivial biased relation between galaxies and matter distribution in low-density environments has been studied intensively in recent years \cite{Lindner:1995ce,Chan:2014qka,Pollina:2016gsi,Contarini:2019qwf,Einasto:2019mwi}.
Regarding (ii), there are different classes of void finders; for a comparison
see Ref.~\cite{Colberg:2008qg}. Some methods exploit the watershed transform
technique, see for instance Refs.~\cite{Neyrinck:2007gy,Sutter:2014haa}, others are based on dynamical properties \cite{Elyiv:2014pqa}, and some methods identify low
integrated density in spherical or free-shape regions, see for instance
Ref.~\cite{Paz:2022zug}. Despite the intrinsic differences between all the methods,
there is a consensus on the basic statistical properties of voids. Roughly
speaking, voids are underdense regions with densities as low as $10-20\%$ of the
average density of the Universe, their properties described here are almost
exclusively governed by gravity
\cite{Ruiz:2015tja, Lambas:2015afa, Schuster:2023pmt, Williams:2024vlv}, and the motion
of tracers around them is consistent with linear dynamics
\cite{Paz:2013sza, Schuster:2022ogh}. They can have diameters starting from few
Mpc, spanning several tens of Mpc and the sizes of the largest
voids may even exceed hundreds of Mpc
\cite{1994MNRAS.269..301E,Park:2012dn,Szapudi:2014eza}. Such huge voids are
delineated by the richest structures -  galaxy superclusters, connected by a
hierarchical pattern of filaments within these large voids \cite{Lindner:1995ce,
Park:2012dn}.

\paragraph{Void size function as cosmological test}

The evolution of underdense regions in the Universe, particularly isolated
spherically symmetric density perturbations, involves several key stages.
Initially, these regions expand until internal matter overtakes the outer
shells, transitioning from linear to non-linear growth. As voids expand, their
density decreases, forming spherical shapes with reverse top-hat density
profiles, leading to super-Hubble bubbles with suppressed structure growth and
boundary ridges. There are two modes of this void evolution: the void-in-void
mode, where regions expand at all scales, and the void-in-cloud mode, where
voids expand but surrounding matter density leads to contraction and eventual
collapse \cite{Sheth:2003py,Paz:2013sza}. The spherical expansion model,
combined with excursion set theory, can model the void size function (VSF)—the
number density of voids as a function of their comoving size
\cite{Sheth:2003py, Jennings:2013nsa, Verza:2024rbm}. The VSF depends on the power spectrum of fluctuations, the growth rate of structures, and the critical density threshold when expanding shells cross. Using the cosmological model, we can predict the abundance of voids of a given comoving size. By comparing this modeled function with observations of abundances at different redshift bins, it is possible to conduct cosmological tests.

\paragraph{The void-galaxy cross-correlation function}

The cross-correlation function for voids and galaxies can also be used to perform an Alcock-Paczynski test, similar to the galaxy correlation function at \ac{bao} scales. The one void term can be interpreted as the stacked mean void profile and serves as a standard sphere at different redshifts to test the Universe's geometry and the relations between the angular diameter and the comoving distances. This profile, in combination with a model for \ac{rsd}, can be used to test cosmological parameters and gravity models \cite{Lavaux:2011yh,Hamaus:2016wka,Cai:2016jek,Correa:2018vge}.

\paragraph{Redshift-space distortions in voids}

The assumption of spherical symmetry for the galaxy distribution around voids breaks down in observations due to \ac{rsd} acting along the line of sight. The general characteristics of \ac{rsd} include an elongation of inner correlation contours towards larger scales (due to void expansion) and a flattening of the correlation levels at large scales due to the infall and lower expansion rates at the void outskirts. These anisotropic patterns encode valuable information about void dynamics and the galaxy velocity field. The first model for \ac{rsd} in voids is the linear model, equivalent to the Kaiser factor in the galaxy autocorrelation \cite{Cai:2016jek}. The pairwise velocity distribution can also be described with the Gaussian streaming model \cite{Paz:2013sza}. For the impact of \ac{rsd} in VSF and void-galaxy cross correlation see for instance \cite{Correa:2020ddy,Correa:2021wqw}.

\begin{figure}
    \begin{center}
    \includegraphics[width = 0.9\textwidth]{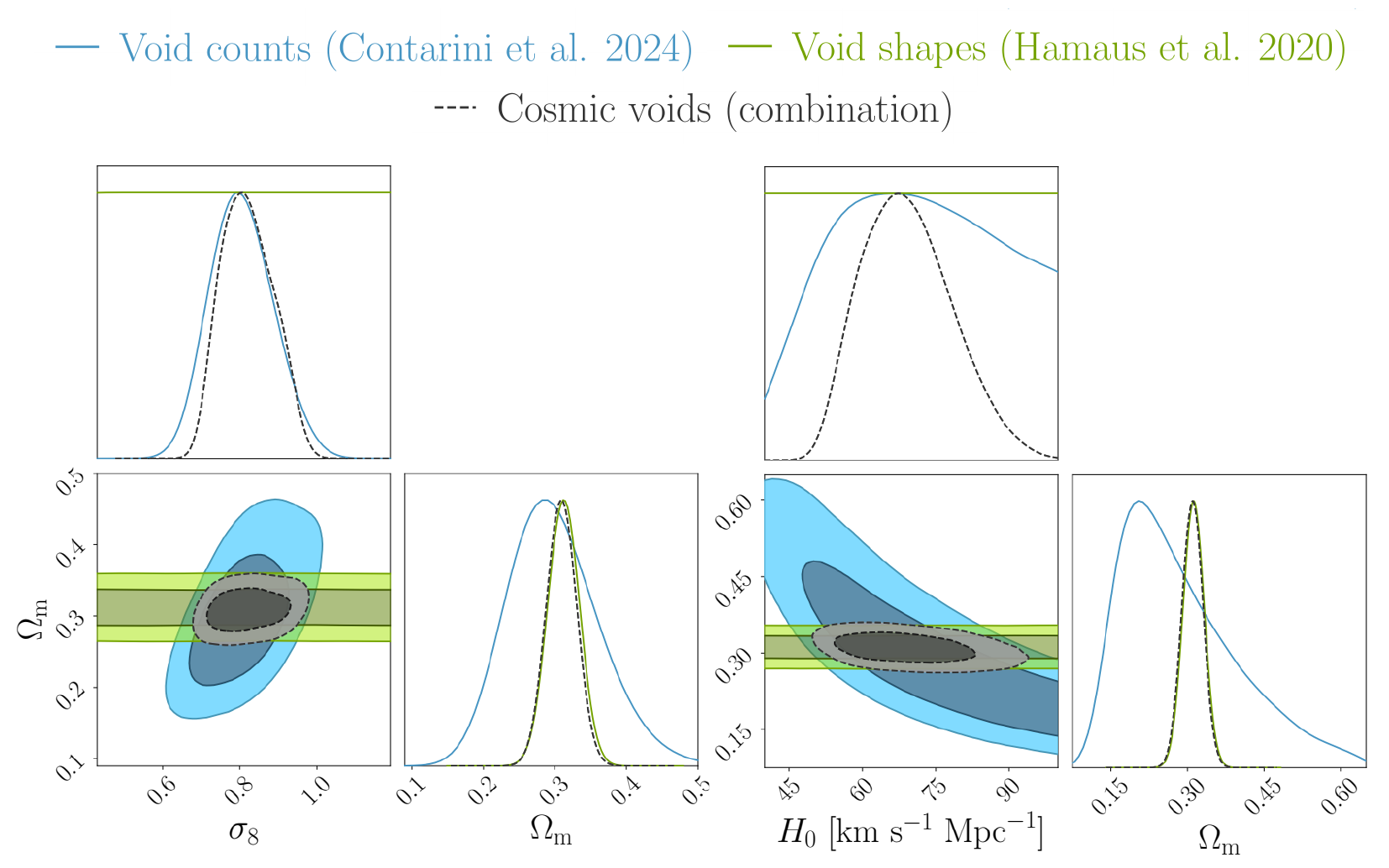}
    \caption[]{Confidence contours from void counts (light blue, \cite{Contarini:2022nvd}) and void shape distortions (green, \cite{Hamaus:2020cbu}) of the \ac{boss} DR12 voids, and their combination
(black) as independent constraints on $\Omega_m$, $H_0$ and $\sigma_8$. Adapted from \cite{Contarini:2022nvd}, licensed under Creative Commons Attribution 4.0 International (CC BY 4.0)}
\label{fig:contarini2024}
\end{center}
\end{figure}

\paragraph{An independent probe in the landscape of cosmic tensions}  

Recent studies have highlighted the potential to test the standard cosmological model using voids in upcoming redshift surveys, such as Euclid \cite{Euclid:2021xmh,Euclid:2022qtk,Euclid:2022hdx}. In a recent analysis presented in Ref.~\cite{Contarini:2022nvd}, the void size function in the \ac{boss} DR12 galaxies were examined. Through a combined analysis of void abundances and void-galaxy cross-correlations, these findings proved to be both competitive and compatible with cosmological constraints derived from other probes (see Fig.~\ref{fig:contarini2024}). Furthermore, recent studies of the \ac{cmb} lensing imprints of cosmic voids have become a valuable test for the cosmological model. In particular, the \ac{boss} survey reported up to 5.3$\sigma$ \cite{2020ApJ...890..168R}, \ac{des} reached up to 5.9$\sigma$ \cite{DES:2024bnq}, the WISE--PanSTARRS combination obtained 13.3$\sigma$ \cite{Camacho-Ciurana:2023afl}, and the \ac{desi} Legacy Survey DR9 achieved up to 17$\sigma$ \cite{Sartori:2024wam}, all in excellent agreement with \lcdm\ predictions. However, the achieved precision on the parameters does not yet allow for a decisive position in the context of current cosmological tensions. Nevertheless, these results emphasize the significance of cosmic voids as they provide an independent cosmological probe that comes at no extra costs in modern cosmological surveys.\bigskip
\subsubsection{Fast radio burst probes of cosmic tensions \label{sec:FRBs}}

\noindent \textbf{Coordinator:} Amanda Weltman\\
\noindent \textbf{Contributors:} Anthony Walters, Bing Zhang, Christo Venter, Shruti Bhatporia, and Surajit Kalita
\\

\paragraph{Introduction to fast radio burst cosmology}

\ac{frb} observations are relatively recently discovered bright transient events (typically observed for a few ms duration) detected in the radio spectrum \cite{Lorimer:2007qn}. Of particular use to us in cosmology, is the observation that they originate at cosmological distance, apparently from host galaxies at these distances \cite{Tendulkar:2017vuq}, and thus they provide us with potentially very powerful probes of cosmology \cite{Walters:2017afr,Deng:2013aga} as well as the \ac{igm} \cite{Walters:2019cie,Macquart:2020lln}.

To date, there have been approximately 800 observed \ac{frb}s with information available in the public domain. The majority of these have been detected by the Canadian Hydrogen Intensity Mapping Experiment~(CHIME).\footnote{\url{https://www.chime-frb.ca/catalog}} The primary open problem of \ac{frb} science is understanding their origins and the progenitor mechanism at work \cite{Platts:2018hiy}. Compelling evidence has accumulated for a magnetar engine, as exemplified by the only Galactic \ac{frb} 20200428D detected from a magnetar in the \ac{mw}. On the other hand, some other engines for cosmological \ac{frb}s are still possible (see Refs.~\cite{Platts:2018hiy,Zhang:2020qgp,Zhang:2022uzl} for detailed reviews on \ac{frb} progenitor mechanisms). Perhaps the greatest open puzzle is the curious observation that while some \ac{frb}s appear to repeat, not all of them do, and there does not appear to be any pattern or periodicity to so-called \ac{frb} repeaters. While there may be some selection effects at work, it is likely that there are simply different classes of bursts and thus even more to discover about \ac{frb}s and with \ac{frb}s. Hence, multi-wavelength observations or detections of \ac{gw}s \cite{Kalita:2022uyu} and neutrinos from \ac{frb} sites will be important to better understand their progenitor mechanisms.

Recent searches for persistent radio emission from one-off and repeating \ac{frb}s identified a nearly unresolved source for \ac{frb} 20190714A, with a peak brightness of 53 µJy/beam, marking the first detection of persistent radio emission potentially linked to a non-repeating \ac{frb}. Multi-wavelength follow-ups in ultraviolet, optical, X-ray, and gamma-ray bands set upper limits \cite{HESS:2021smp}, aiding the distinction between repeating and one-off \ac{frb}s. Persistent emission modeling can constrain central source energetics and test magnetar-driven emission scenarios. Follow-ups also refine host galaxy properties, localization, and redshift-dependent \ac{frb} characteristics, enhancing their use in cosmology.

\paragraph{Fundamental parameter inference from Fast Radio Bursts}

\begin{figure}
    \begin{center}
    \includegraphics[scale=0.35]{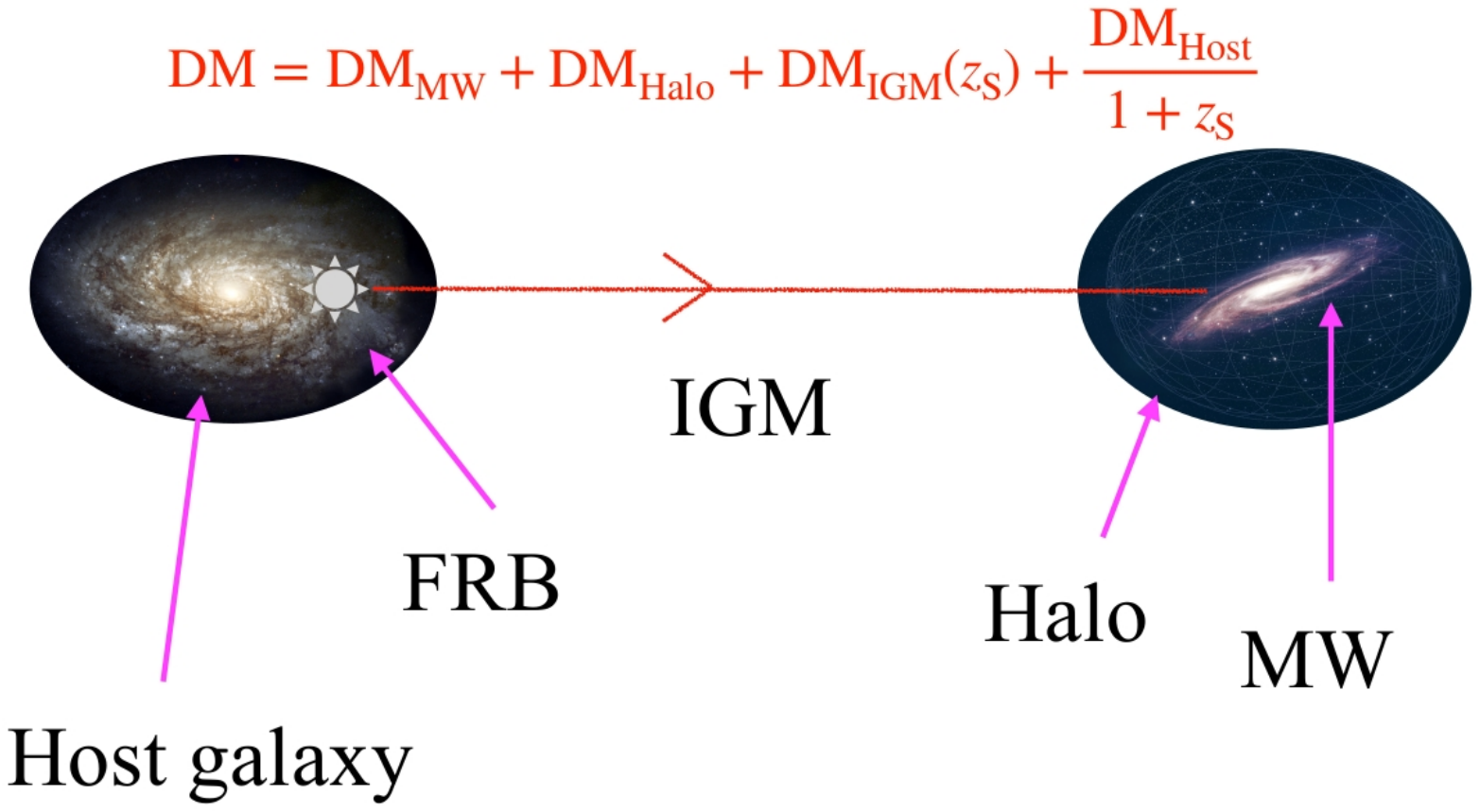}
    \caption{Illustrative diagram indicating key components of contributing to measured DM of \ac{frb}s.}
    \label{fig:FRB DM}
    \end{center}
\end{figure}

One of the key parameters in understanding \ac{frb}s is the dispersion measure (DM), which gives information about the ionized plasma along the path of the light ray. Broadly it gets contributions from each of the 4 environments along the line of sight, our Galaxy, its halo, the \ac{igm}, and the host galaxy as illustrated in Fig.~\ref{fig:FRB DM}. The $(1+z_\mathrm{S})$ factor in the denominator of the last term accounts for the observed value of redshifted $\mathrm{DM}_\mathrm{Host}$. In general, $\mathrm{DM}_\mathrm{IGM}$ contains the information of the underlying cosmology. The average $\mathrm{DM}_\mathrm{IGM}$ is given by the following equation (now known as the Macquart relation \cite{Macquart:2020lln})
\begin{align}\label{Eq: Macquart}
    \langle \mathrm{DM}_\mathrm{IGM}(z_\mathrm{S})\rangle &= \frac{3c \Omega_\mathrm{b} H_0^2}{8\pi G m_\mathrm{p}} \int_{0}^{z_\mathrm{S}} \frac{f_\mathrm{IGM}(z)\chi(z)(1 + z)}{H(z)} \dd{z},
\end{align}
where $\Omega_\mathrm{b}$ is the baryonic matter density, $m_\mathrm{p}$ is the proton mass, $f_\mathrm{IGM}$ is the baryon mass fraction in the \ac{igm}, and $\chi(z)$ is the ionisation fraction along the line of sight. As $\langle \mathrm{DM}_\mathrm{IGM}\rangle$ contains cosmological parameters, this equation can be used to constrain their values. For instance, utilizing a set of mock \ac{frb}s, \ac{de} equation of state for $w$CDM cosmology was constrained \cite{Zhou:2014yta,Gao:2014iva} and predicted that with $\mathcal{O}(10^4)$ localized \ac{frb}s, this constraint can be tighter than the same obtained from \ac{bao} measurements.

\ac{frb}s have been observed over a relatively narrow redshift range to date, 100\,MHz -- 8\,GHz \cite{Pleunis:2020vug,Gajjar:2018bth}, and partly this is most likely a result of observation bias, where we can more easily observe the bursts that are closest to us. This has limited their potential in constraining cosmological parameters directly \cite{Walters:2017afr}, though using cosmology as a prior allows for \ac{igm} constraints, specifically a potential solution to the missing baryon problem \cite{Walters:2019cie,Macquart:2020lln}. Moreover, using a specific case of \ac{frb}\,150418, earlier studies established a limit on the photon mass of $m_\gamma<1.8\times 10^{-14}\rm\,eV\,c^{-2}$ \cite{Bonetti:2016cpo} and this bound has been continuously strengthened with the inclusion of additional \ac{frb}s~\cite{Wang:2021nrl,Lin:2023jaq}. Furthermore, localized \ac{frb}s have been used to constrain the parameterized post-Newtonian (PPN) parameter \cite{Reischke:2023gjv} and other fundamental constants like the fine-structure constant and the proton-to-electron mass ratio \cite{Kalita:2024kyo,Lemos:2024jbl}.

\paragraph{Hubble constant estimations using Fast Radio Bursts}

Several recent studies have leveraged \ac{frb}s with measured redshifts to constrain $H_0$ within a Bayesian framework. This approach compares the measured $\mathrm{DM}_\mathrm{IGM}$ of localized \ac{frb}s with their theoretical predictions, yielding constraints on $H_0$ and other cosmological parameters. Ref.~\cite{Walters:2017afr} was among the first to investigate the utility of \ac{frb}s as cosmological probes through simulations. Their mock \ac{frb} catalog highlighted the challenge posed by DM variations due to \ac{igm} inhomogeneities. Notably, the most significant improvement was observed in the $\Omega_{\rm b,0} h^2$ parameter when combining \ac{frb}s with \ac{cmb}, \ac{bao}, \ac{sn}, and $H_0$ data. Additionally, they found that \ac{frb}s offered limited constraints on the \ac{de} equation of state, while the curvature parameter ($\Omega_k$) showed some improvement when combined with other data. Moving forward, Ref.~\cite{Macquart:2020lln} leveraged 8 localised \ac{frb}s to constrain $\Omega_{\rm b,0} = 0.051^{+0.021}_{-0.025}h_{70}^{-1}$ where $h_{70} = H_0/(70 \rm\,km\,s^{-1}\,Mpc^{-1})$, consistent with values derived from \ac{cmb} and \ac{bbn} data. Another interesting work using a strong lensing effect with 10 \ac{frb}s found $H_0 \approx 70$\kms \cite{Li:2017mek}.

More recently, as more \ac{frb} data are available, these constraints have been revised to a great extent. Ref.~\cite{Hagstotz:2021jzu} analyzed 9 localized \ac{frb}s and found $H_0 = 62.3\pm 9.1$\kms, assuming a homogeneous host galaxy contribution, which may limit the complete validation of the result. Subsequently, Ref.~\cite{Wu:2021jyk} classified 18 localized \ac{frb}s based on host galaxy morphology and employed the IllustrisTNG simulation to estimate individual host contributions. This approach yielded $H_0 = 68.81^{+4.99}_{-4.33}$\kms. However, their model-based host DM values were lower than the actual reported values for some \ac{frb}s. Moreover, Ref.~\cite{James:2022dcx} included 16 localized and 60 unlocalized \ac{frb}s detected by ASKAP, estimating $H_0 = 73^{+12}_{-8}$\kms. Their larger error bars, despite a substantial number of \ac{frb}s, likely reflect the inclusion of systematic uncertainties. Moving beyond, Ref.~\cite{Baptista:2023uqu} incorporated a scatter in the Macquart relation with 78 \ac{frb}s (21 localized) and obtained $H_0 = 85.3^{+9.4}_{-8.1}$\kms. Ref.~\cite{Liu:2022bmn} employed a model-independent method with 18 localized \ac{frb}s and found $H_0 = 71 \pm 3$\kms. Conversely, Ref.~\cite{Wei:2023avr} estimated $H_0 = 95.8^{+7.8}_{-9.2}$\kms using 24 localized \ac{frb}s with a wide flat prior on $\mathrm{DM}_\mathrm{Halo}$. Furthermore, Ref.~\cite{Zhao:2022yiv} combined 12 unlocalized \ac{frb}s with \ac{bbn} data, resulting in $H_0 = 80.4^{+24.1}_{-19.4}$\kms. Additionally, Ref.~\cite{Gao:2023izj} used a combination of 18 localized \ac{frb}s and the Pantheon dataset to estimate $H_0 = 65.5^{+6.4}_{-5.4}$\kms. Ref.~\cite{Fortunato:2024hfm} propose a novel method using \ac{ann} architectures for 23 localized \ac{frb}s to estimate $H_0 = 67.3 \pm 6.6$\kms. Furthermore, utilizing Bayesian analysis with different likelihood functions and distinct host distributions for 64 localized \ac{frb}s, Ref.~\cite{Kalita:2024xae} obtained $H_0$ values well above $70$\kms aligning with the late-Universe $H_0$ values with 1$\sigma$ error bars no longer overlap with those obtained from early-Universe measurements. More recently, accounting 98 localized \ac{frb}s, Ref.~\cite{Piratova-Moreno:2025cpc} showed multiple estimates of $H_0$ utilizing different methodologies.

As we see, \ac{frb}s do offer preliminary constraints on cosmological parameters, although they currently lack the power to definitively resolve existing tensions in cosmology. While not as stringent as constraints derived from established probes like \ac{cmb} and \ac{sn1}, \ac{frb}-derived constraints are significantly tighter than those obtained from \ac{gw} observations. A primary contributor to the discrepancies in $H_0$ values reported by different researchers lies in the choice of models for \ac{igm} and host DM contributions, which remain largely unconstrained from observations. Future surveys aiming to detect a larger number of \ac{frb}s, with a substantial fraction localized to their host galaxies, hold promise for significantly improved constraints. These advancements could potentially contribute to resolving the current Hubble tension.

\paragraph{Understanding nature of Dark Matter using Fast Radio Bursts}

Beyond constraining the Hubble constant and other cosmological parameters related to the evolution of the Universe, \ac{frb}s hold promise for constraining the fraction of primordial mass black holes made up of \ac{dm} ($f_\mathrm{PBH}$). In this context, strong gravitational lensing of \ac{frb}s plays a crucial role. When light rays from \ac{frb}s pass near a massive object with mass $M_\mathrm{L}$, they can be deflected, potentially producing multiple, time-delayed copies of the original burst. The time delays and magnification ($\mu$) of these lensed images provide an estimate of the source's optical depth ($\tau$), which is linked to $f_\mathrm{PBH}$ as follows
\begin{align}\label{Eq: optical depth}
    \tau(M_\mathrm{L},z_\mathrm{S}) = &\frac{3}{2}f_\mathrm{PBH} \Omega_\mathrm{c}\int_0^{z_\mathrm{S}} \dd{z_\mathrm{L}} \frac{H_0^2}{cH(z_\mathrm{L})} \frac{D_\mathrm{L} D_\mathrm{LS}}{D_\mathrm{S}} \left(1+z_\mathrm{L}\right)^2 \left[y_\mathrm{max}^2(\mu) - y_\mathrm{min}^2(M_\mathrm{L},z_\mathrm{L})\right]\,,
\end{align}
where $D_\mathrm{L}$ and $D_\mathrm{S}$ are respectively angular diameter distances to the lens object and the source from the observer, $D_\mathrm{LS}$ is the same between lens and source, $y_\mathrm{min}$ and $y_\mathrm{max}$ are minimum and maximum impact parameters, respectively. Ref.~\cite{Munoz:2016tmg} pioneered this method, setting an initial constraint of $f_\mathrm{PBH}<0.08$ assuming \ac{frb} lensing by black holes exceeding mass $M>20\,M_\odot$. Subsequent studies incorporated \ac{frb} microstructure \cite{Sammons:2020kyk} and extended mass functions \cite{Laha:2018zav} to refine these bounds. Notably, Ref.~\cite{Liao:2020wae} visualized using a null detection of lensed \ac{frb}s within a sample of 110 real \ac{frb} observations.  Further investigations explored the impact of intervening plasma, which mimics the lensing effects. Ref.~\cite{Leung:2022vcx} demonstrated this using 172 CHIME \ac{frb}s, highlighting the need to account for such decoherence or scattering screens. More recently, Ref.~\cite{Kalita:2023eeq} analyzed 636 \ac{frb}s from CHIME and the absence of confirmed lensing events suggesting that \ac{mg} might introduce a screening effect akin to the plasma scenario. In summary, \ac{frb}s offer constraints on $f_\mathrm{PBH}$ in the mass range of approximately $10^{-4}-10^4\,M_\odot$ and these constraints are comparable or even surpassing existing experiments like OGLE, EROS, Icarus, and MACHO.

The landscape of \ac{frb}-based constraints is constantly evolving as new and more powerful radio telescopes come online. Arrays such as Hydrogen Intensity and Real-time Analysis eXperiment~(HIRAX) \cite{Crichton:2021hlc}, Deep Synoptic Array~(DSA)-2000 \cite{Connor:2022bwl}, Bustling Universe Radio Survey Telescope in Taiwan~(BURSTT) \cite{Ho:2023feo} and \ac{ska} \cite{Weltman:2018zrl} promise to detect a significantly larger number of \ac{frb}s, including potentially localized ones. This will undoubtedly lead to further refinements in our understanding of the Universe.
\bigskip
\subsubsection{Radio background excess \label{sec:radio_background_excess}}

\noindent \textbf{Coordinator:} Jack Singal\\
\noindent \textbf{Contributors:} Alan Kogut, Marco Regis, and Nicolao Fornengo\\

\noindent Throughout most of the electromagnetic spectrum the observed level of surface brightness and anisotropy of the sky background radiation is roughly consistent with that expected from known emission mechanisms from astrophysical and cosmological sources.  This is the case for the backgrounds at infrared, e.g., see Ref.~\cite{Hauser:2001xs}, microwave, e.g., see Ref.~\cite{Planck:2018nkj}, optical/UV, e.g., see Ref.~\cite{Gilmore:2009zb}, X-ray, e.g., see Ref.~\cite{Cappelluti:2017miu}, and gamma-ray, e.g., see Ref.~\cite{Fermi-LAT:2015otn} wavelengths. However, in the radio region of the electromagnetic spectrum, both the observed surface brightness level of the photon background and the level of its anisotropy angular power are seemingly much larger than that which can be produced by known source classes, resulting in a tension that is, as of now, an anomaly.

As recently summarized in Ref.~\cite{Singal:2022jaf}, with the known  structure resulting from Galactic diffuse emission subtracted, the surface brightness of the level of diffuse background radiation on the sky, from at least $\sim$20~MHz to 3~GHz, as a function of frequency $\nu$ is, in radiometric temperature units
\begin{equation}
    T_\mathrm{BGND}(\nu) = 30.4 \pm 2.6 \mathrm{K} \, \left(\frac{\nu}{310\mathrm{~MHz}}\right)^{-2.66 \pm 0.04} \, + \, T_{\rm CMB}\,,
\label{T_B}
\end{equation}
where $T_{\rm CMB}$ is the frequency-independent contribution of 2.725~K due to the \ac{cmb}. This background level has been measured by the Absolute Radiometer for Cosmology, Astrophysics, and Diffuse Emission 2 (ARCADE~2) \cite{Fixsen:2009xn,Singal:2009xq} at the high frequency end, and several radio maps at lower frequencies from which an absolute zero level calibration can be determined, including recently the Long Wavelength Array (LWA) \cite{Dowell:2018mdb}.  Because the spectral index of $-2.6$ is consistent with synchrotron radiation, this has been referred to as the radio synchrotron background (RSB). The RSB surface brightness level is shown in radiometric temperature units and spectral energy density units in Fig.~\ref{fig:fig}.

\begin{figure}
    \begin{center}
    \includegraphics[width=0.4\textwidth,height=0.3\textwidth]{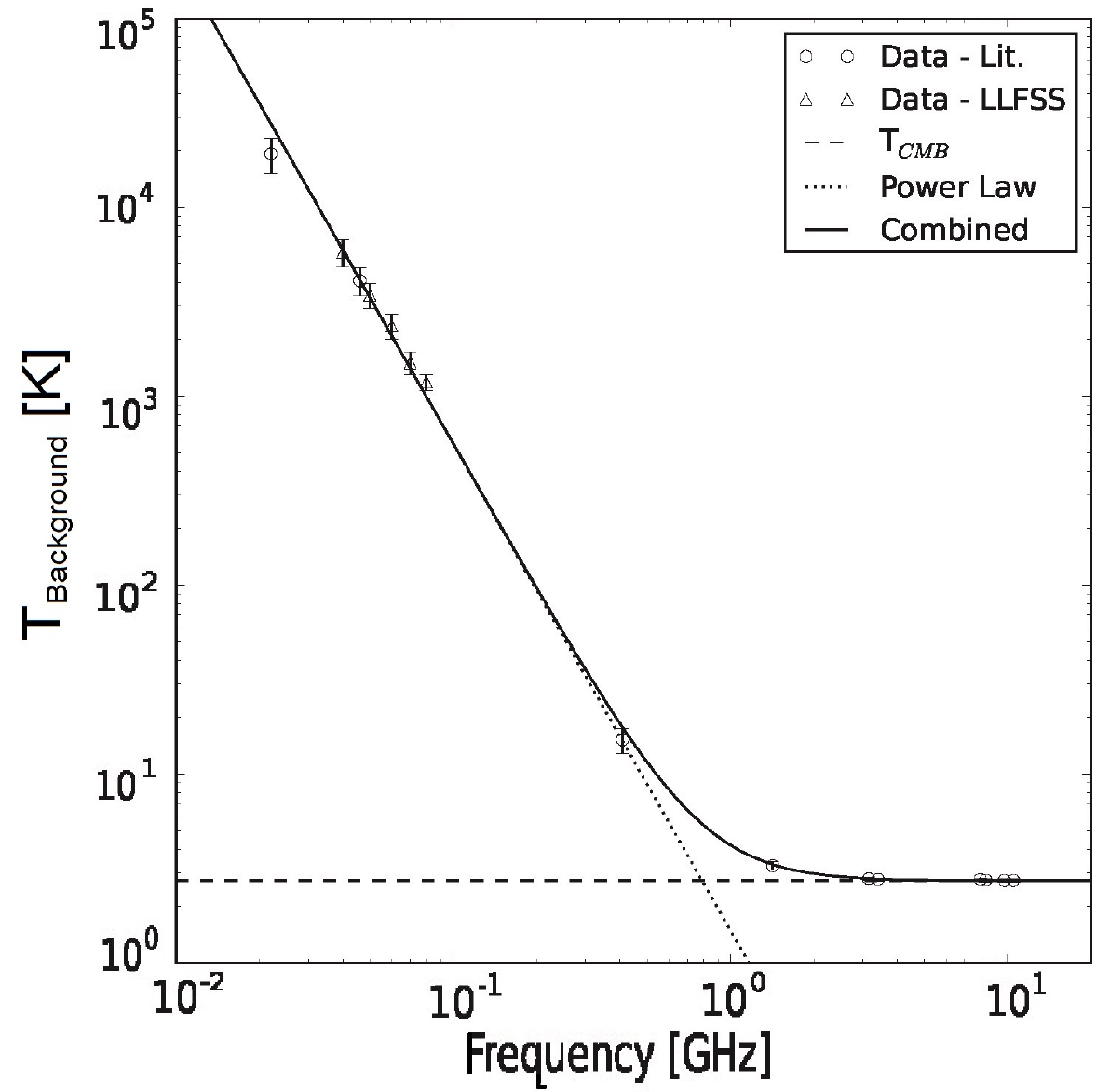}
    \includegraphics[width=0.45\textwidth]{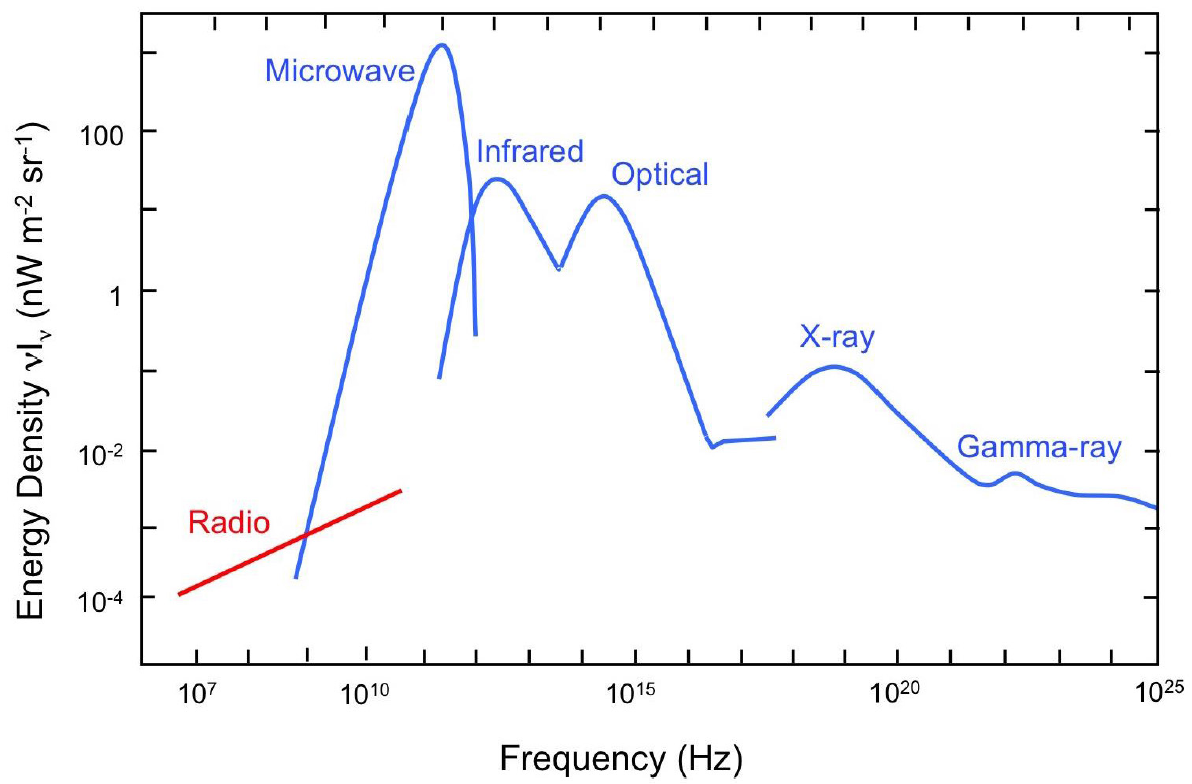}
    \caption{{\bf Left}: The radio sky zero level in radiometric temperature units, reproduced from Ref.~\cite{Dowell:2018mdb}, as measured by several different instruments or surveys reporting an absolute zero-level calibration. Results are shown for ARCADE~2 at 3--10 GHz \cite{Fixsen:2009xn,Singal:2009xq}, Reich \& Reich at 1.4 GHz, \cite{Johannesson:2021exs}, Ref.~\cite{Haslam:1982zz} at 408 MHz,  Ref.~\cite{1999A&AS..140..145M} at 45 MHz, and Ref.~\cite{Roger:1999jy} at 22 MHz, as well as several points reported by Ref.~\cite{Dowell:2018mdb}.  {\bf Right}: The photon backgrounds in the Universe in units of spectral energy surface brightness density.  Reproduced from Ref.~\cite{Singal:2017jlh}.}
    \label{fig:fig}
    \end{center}
\end{figure}

This level of surface brightness is several times higher than can be produced by known classes of radio sources in the Universe.  Studies of radio source counts show that the total emission from the known discrete extragalactic sources, particularly \ac{agn} and star-forming galaxies, is around a factor of five lower than the surface brightness level of the RSB \cite{Condon:2012ug,Hardcastle:2020dfj,2023MNRAS.520.2668H}.  Producing the surface brightness of the RSB with discrete extragalactic sources would require an enormous number of a new class of low-flux sources \cite{Condon:2012ug}.  

For lines of sight far away from the Galactic plane, the RSB is considerably brighter than the expected contribution from models of Galactic diffuse emission \cite{Kogut:2009xv}.  The possibility that the RSB originates from a larger spherical halo of radio emission surrounding our Galaxy is highly disfavored by several considerations, e.g., see Ref.~\cite{Singal:2022jaf,Kogut:2009xv} including that it would make our Galaxy highly atypical \cite{Singal:2015tta}.  Observational constraints on the polarization structure of the diffuse radio emission rule out the Local Bubble as an origin scenario \cite{Krause:2021xav}.

In a parallel situation to the surface brightness, the anisotropy level of the RSB, as measured at 140~MHz, is also higher than that which could result from models of known source classes \cite{Offringa:2021rwp,Cowie:2023svj}.  The measured angular power spectrum of the diffuse radio emission is a power law (in $K^2$ or ``$(\Delta T)_\ell^2$'' units) with increasing $\ell$ with index $\beta = 2.17 \pm 0,08$ from $700 \lesssim \ell \lesssim 4000$ (angular scales from $\sim$2' to $\sim$20'), indicative of unclustered point sources, diffuse sources, or a large number of faint, clustered point sources \cite{Cowie:2023svj}. However, its level of anisotropy power rules out the former, given that the density of high flux point sources is constrained by source counts observations such as e.g., \cite{Hardcastle:2020dfj}. This leaves very numerous but highly clustered point sources or diffuse sources as the source classes that can seemingly reproduce the level of the observed angular power spectrum of the RSB. 

There are other observational constraints on possible origin scenarios for the RSB.  Any possible source class and emission mechanism must not:

$\bullet$ overproduce the observed level of the background radiation at far-infrared wavelengths, given the known correlation between radio and far-infrared emission in galaxies, e.g., see Ref.~\cite{Ponente:2011se}.  Any origin scenario involving galaxies or their components would need this correlation to evolve with redshift \cite{Singal:2009dv,Todarello:2023nqd}.

$\bullet$ overproduce the observed level of the X-ray background through inverse-Compton upscattering of the \ac{cmb} and/or optical/UV background \cite{Singal:2009dv}. If the RSB indeed originates from a synchrotron process, this puts a lower limit on the strength of the magnetic field in the sources from which the RSB originates of around 1$\mu$G at redshift $z=0$.  At higher redshifts this lower limit increases by a factor of up to $(1+z)^4$ due to the increase in the surface brightness level of the \ac{cmb} at higher redshifts.

$\bullet$ overproduce the observed limits on the 21-cm absorption trough, due to the possible presence of the RSB at redshifts $z\sim 10$ increasing the temperature of the background relative to the temperature of the 21-cm transition, e.g., see Ref.~\cite{Fialkov:2019vnb}.

$\bullet$ overproduce the observed cross-correlation angular power spectrum between the RSB and optical source catalogs, indicating that the vast majority of its sources, if discrete, must be at $z \gtrsim 0.5$ \cite{Todarello:2023nqd}.

Some RSB origin scenarios that have been suggested include \ac{sn} of massive population~III stars \cite{Biermann:2014lna}, emission from Alfv\'{e}n reacceleration in merging galaxy clusters \cite{Fang:2015dga}, annihilating \ac{dm} in halos or filaments \cite{Fornengo:2011cn,Hooper:2012jc,Fang:2014joa,Fortes:2019jgo} or ultracompact halos \cite{Yang:2012qi}, ``dark'' stars in the early Universe \cite{Spolyar:2009nt, Rindler-Daller:2020yqe}, dense nuggets of quarks \cite{Lawson:2012zu}, accretion onto primordial black holes \cite{Cappelluti:2021usg,Mittal:2021dpe,Acharya:2022txp}, decays to dark photons \cite{Pospelov:2018kdh}, dark photon decays \cite{Caputo:2022keo,Acharya:2022vck}, and other injections of photons in the early Universe from processes \cite{Acharya:2023bln} including superconducting cosmic strings \cite{Cyr:2023yvj} and decays of relic neutrinos \cite{Dev:2023wel}.  Some of these models likely violate one or more of the observational constraints discussed here, while others contain parameter spaces where they do not.

Upcoming observatories, and observations possible with current facilities, have the potential to further constrain and test RSB origin scenarios.  A measurement, such as with the space-based LuSSE-Night \cite{2023arXiv230110345B} to determine if the spectrum hardens below 10~MHz would be useful in this regard.  The radio \ac{sz} effect \cite{Lee:2021nwv,Holder:2021qtm} when combined with the well-known \ac{cmb} \ac{sz} effect, would result in observed cluster radio emission having a null frequency at 
$700 \leq \nu \leq 800$~MHz, below which there would be an increment in the observed surface brightness in the direction of a cluster and above which there would be a decrement, each on the order of $\lesssim 1$~mk (both on top of the contribution from the background RSB). A detection or lack thereof at the required sensitivity in clusters of various redshifts would constrain the redshift(s) of the origin of the RSB. The radio background being far in excess, in both surface brightness and anisotropy angular power, of that which is expected to be produced by known source classes and processes in the Universe is an intriguing issue.

\bigskip
\subsubsection{Tension between the large scale bulk flow and the standard cosmological model \label{sec:bulk_flow}}

\noindent \textbf{Coordinator:} Richard Watkins\\
\noindent \textbf{Contributors:} Hume A. Feldman\\

\noindent The Cosmological Principle requires that the motions of galaxies averaged over a sphere of radius $R$ (bulk flow) should go to zero as $R$ becomes cosmologically large.  Thus the determination of the Large-Scale Bulk Flow is an important probe of this principle and of the Standard Cosmological Model more generally.  Estimating the bulk flow requires large catalogs of galaxy peculiar velocity measurements; in the last decade, the quantity and quality of this type of data has improved to where bulk flows can provide an important test of our understanding of the Universe.

\paragraph{Analyzing the bulk flow using the Cosmic Flows 4 catalog}

The Cosmic Flows 4 (CF4) peculiar velocity catalog \cite{Tully:2022rbj} contains velocities and their uncertainties for 38,057 groups and individual galaxies; this represents most of the peculiar velocity data in existence.  Analyzing the bulk flow using this data presents several significant challenges.  First, we can only measure the radial component of the peculiar velocity.  Second, peculiar velocities typically have large uncertainties; objects in the CF4 catalog typically have uncertainties that are around 15\% of the redshift.  This means that since distant galaxies have much larger absolute uncertainties, most of the information in the catalog is concentrated nearby.  Thus care must be taken to ensure that the bulk flow estimate is not dominated by objects near the center of the volume.  Finally, the CF4 survey has a very irregular distribution both radially and on the sky.  Thus analyzing the bulk flow using the CF4 requires a method that can ``even out'' the distribution of the information in the volume of interest both so that the volume is uniformly sampled and so that radial flows do not make spurious contributions.  The latter is particularly important given the current tension in the value of the Hubble constant.  The use of an incorrect value of the Hubble constant in calculating peculiar velocities can result in phantom radial flows that can contribute to the bulk flow if the volume of interest is not sampled in an isotropic manner.    

Ref.~\cite{Watkins:2023rll} deals with these challenges by analyzing the CF4 catalog using the Minimum Variance (MV) method \cite{Watkins:2008hf,Feldman:2009es,Peery:2018wie}. The MV method generates estimates of the bulk flow components, in a volume of radius R, that is as close as possible to those that would be measured from an ideal (isotropic and well sampled) survey; thus it balances the information in the catalog so that the volume is evenly sampled.  In addition, a constraint is put on the MV bulk flow estimates so that they are completely independent of the value of the Hubble constant.  It should be noted that even though the analysis uses only radial velocity measurements, if it is assumed that the velocity field is irrotational (as it should be if it were generated via gravitational instability), then the MV method estimates reflect averages of the full three-dimensional velocities.

\paragraph{The bulk flow tension}

Ref.~\cite{Watkins:2023rll} determined that the largest radius for which the bulk flow could be estimated with the CF4 catalog is around  200$h^{-1}$Mpc. Their bulk flow estimates as a function of radius are shown in Fig.~\ref{fig:bf} and displayed in Table~\ref{tab:prob}. The figure shows both the estimates and uncertainties for the components of the bulk flow (in galactic coordinates) as well as the magnitude. Included in the figure (in red) is the expected magnitude of the bulk flow as a function of radius calculated using the Cosmological Standard Model. Contrary to expectations, they find that the bulk flow magnitude increases with increasing radius.  In addition, they find that the bulk flow has a much larger magnitude than expected in the Cosmological Standard Model \cite{Planck:2018vyg}.  Indeed, they found the probability of obtaining a bulk flow as large or larger in the Standard Model for $R=200h^{-1}$Mpc to be $1.5\times 10^{-6}$, equivalent to about $4.8\sigma$. Ref.~\cite{Whitford:2023oww} find a similar magnitude for the bulk flow (also using the MV method) but estimate larger uncertainties using cosmological simulation data. Ref.~\cite{Hoffman:2023pac} also analyzes the bulk flow from the CF4 catalog using a very different method and obtains a result that is consistent with the Standard model. However, it is difficult to discern the sensitivity of their bulk flow to different scale motions. Thus the large-scale bulk flow potentially poses a significant challenge to the standard model, but determining the strength of this challenge will require resolving differences in analysis methods.  

\begin{table}[ht]
    \centering
    \caption{Summary of Bulk Flows for $R=150 h^{-1}$ Mpc and $R=200 h^{-1}$ Mpc.  The uncertainties include both the theoretical difference between the estimate and the bulk flow from an ideal survey and the measurement noise.} \label{tab:prob}
    \begin{tabular}{lccc}
    & $R=150 h^{-1}$ Mpc& $R=200 h^{-1}$ Mpc \\
    \hline
    Expectation (km/s) & 139 & 120\\
    Bulk Flow (km/s) & $395\pm 29$ & $427\pm 37$\\
    Direction & $l=297^\circ\pm 4^\circ$ &$l=298^\circ\pm   5^\circ$\\
    & $b=-4^\circ\pm 3^\circ$ & $b=-7^\circ\pm 4^\circ$ \\
    $\chi^2$ with 3 d.o.f.  &20.19 & 29.84\\
    Probability &$1.54\times10^{-4}$ & $1.49\times10^{-6}$\\
    \hline
    \end{tabular}
\end{table}

\begin{figure}[htbp]
    \begin{center}
    \includegraphics[scale = 0.8]{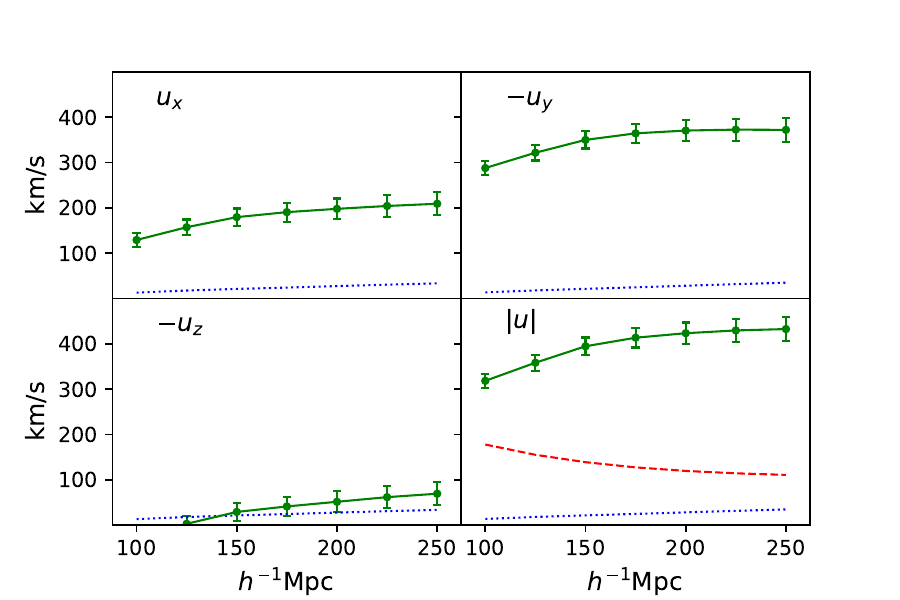}
    \caption[Short caption]{The green points with error bars show the bulk flow components and magnitude estimated from the CF4 catalog as a function of radius $R$.  The error bars indicate the uncertainty in the estimates due to measurement noise. The dotted blue lines show the theoretical standard deviation of the expected differences between the bulk flow estimates and the bulk flow from an ideal survey calculated using the cosmological standard model (not including measurement noise).  The red dashed line indicates the theoretical expectation for the magnitude of the bulk flow calculated using the Cosmological Standard Model. The figure is reproduced with permission from Ref.~\cite{Watkins:2023rll}.}
    \label{fig:bf}
    \end{center}
\end{figure}

\bigskip
\subsubsection{Ultra long period cepheids \label{sec:ULPC}}

\noindent \textbf{Coordinator:} Ilaria Musella\\
\noindent \textbf{Contributors:} Giuliana Fiorentino, Marcella Marconi, Roberto Molinaro, and Vincenzo Ripepi
\\

\noindent The Ultra Long Period Cepheids are pulsating stars characterized by periods longer than $\sim 80$ days with a mean absolute magnitude in the $I$ band $-9 < M_I < -7$ mag. Thanks to their significant brightness they are observable at very long distances, up to cosmologically interesting scales. Thus they could represent competitive standard candles, not requiring the combination with secondary distance indicators \cite{Bird:2008hn,Fiorentino:2010si, Marconi:2010wr, 2021MNRAS.501..866M,2022Univ....8..335M,2015AJ....149...66N,2024ApJS..275...26M} to reach the Hubble flow. On this basis, using  ULPs might reduce the possible effect of systematic errors on the calibration of the extragalactic distance scale and, in turn, on the local determination of $H_0$. Despite their expected important role in distance scale studies, their application as standard candles is still challenging due to the relatively small sample of ULPs with well-sampled light curves covering more than one cycle and some inconsistencies between theoretical evolutionary and pulsational predictions and the observed properties. Indeed, the number of known ULPs is 73 observed in different galaxies. In particular, only one has recently been discovered in our Galaxy in the Gaia DR3 catalog. Their position in the period-luminosity planes and the color-magnitude diagrams suggest that these objects represent the extension at higher luminosity and mass of the Classical Cepheids, but their mass-luminosity relation could be different according to their evolutionary stage. As recently suggested in Ref.~\cite{2024ApJS..275...26M}, the consistency between ULP and Classical Cepheids relations increases when the photometry accuracy improves, thus supporting the hypothesis that they are the same type of pulsating variables but with different mass and period ranges.

However, at such high luminosity and mass levels, stellar evolution does not predict the occurrence of the blue loop typical of the central helium-burning phase of intermediate-mass stars, so the population of the Instability Strip might correspond to the first crossing toward the \ac{rgb}. Moreover, the observed pulsation period provides a strong constraint on the stellar mass at fixed temperature and luminosity. Current evolutionary tracks matching the observed position of the ULPs do not always provide a combination of mass, luminosity, and effective temperature consistent with the observed periodicity. 

The agreement between the ULP and Classical Cepheid Wesenheit relations tends to confirm the Hubble constant value obtained based on Classical Cepheids, even if the remaining uncertainties on these objects' evolutionary and pulsational properties need to be solved and the dispersion of their Wesenheit relation has to be reduced to contribute to the understanding of the Hubble tension. To this aim, future theoretical and observational investigations will include different steps: to better understand the evolutionary status of the ULPs through a detailed comparison with updated sets of evolutionary tracks and isochrones, to produce an extended grid of nonlinear convective pulsation models covering the high mass and luminosity regime expected for these variable start to increase the number of ULPs and improve their photometric accuracy. In this respect, exploring the Gaia database to find new ULPs will be fundamental. Recently, the first \ac{mw} ULP was found among the variables classified as Long Period Variables in Gaia DR3 and there are likely other misclassified ULPs. In addition, the forthcoming Rubin-\ac{lsst} survey will also represent a fundamental opportunity to improve the photometry of the known Local Group ULPs and/or increase the sample. 
\bigskip \newpage
%%%%%%%%%%%%Section_3:
\section{Data analysis in cosmology} \label{sec:data_ana}

\noindent \textbf{Coordinator:} Agnieszka Pollo\\

The volume and complexity of data available for observational cosmology has increased greatly during the last years, and in the coming years, the amount of data generated by different ground-based and space observatories and experiments is expected to grow by orders of magnitude. 

For example, the Early Data Release of \ac{desi} comprising 2 million spectra of extragalactic sources amounts to 80 TB, and it is only 2\% of the expected full final \ac{desi} catalog \cite{DESI:2023ytc,DESI:2025fxa,DESI:2025ejh}. Imaging data are even more voluminous: a single exposure of the Dark Energy Camera (DECam) produces an image of the size of a gigabyte \cite{DES:2015wtr}; and the latest \ac{des} Data Release 2 \cite{LineaScienceServer:2021mgv} was built upon 76,217 such single-epoch images. Similarly, one night of observations with the Subaru \ac{hsc} camera results in several hundred gigabytes of data \cite{2018PASJ...70S...1M}, which translated into $\sim$ 70 TB database already with the first data release of the Hyper Suprime-Cam Subaru Strategic Program \cite{Aihara:2017tri}, and several hundreds of terabytes after the third data release \cite{Aihara:2021jwb}. 

All these numbers are expected to fade soon with the expected arrival of Euclid \cite{Euclid:2024yrr}, Vera Rubin Observatory \cite{LSST:2008ijt}, or \ac{ska} data.\footnote{\url{https://www.skao.int/}} In particular, Vera Rubin Observatory is expected to deliver more than 500 petabytes of imaging data per year, while for \ac{ska}the data volume is expected to exceed 700 petabytes per year.\footnote{\url{https://www.skao.int/sites/default/files/documents/SKA-TEL-SKO-0001818-01_DataProdSummary-signed_0.pdf}}

In addition to the expected multiwavelength and multimessenger observational or experimental data, there exists a whole additional realm of data useful for cosmology, i.e., simulations. Simulated catalogs are being used to test different cosmological scenarios, to probe mechanisms of baryonic interactions in the dark-matter-dominated Universe, especially at small, non-linear scales; and finally, to create mock realizations of real catalogs which can be used both for physical predictions and for tests of specific measurement techniques against a variety of biases introduced by observational strategy and technique. The creation of simulated catalogs is usually a multi-level process, requiring a lot of computational power and creating amounts of data much larger than the corresponding observational data themselves. 

Using all types of data expected in the near future, both observed and simulated, to provide reliable scientific results, requires developing new approaches to data handling and analysis at different levels. Firstly, all these data need to be properly stored and transferred. Ideally, they should also be made publicly accessible in formats that are easy to handle for users from different scientific communities. Datasets observed at different wavelengths or coming from different experiments need to be cross-correlated to facilitate multi-messenger studies. Finally, scientific analysis of these data results in the creation of derived data sets, in amounts often exceeding the original data. Storage, transfer, and subsequent analysis of large amounts of data is not only costly but also requires a well thought out technological and strategic approach. ASTRONET Roadmap for years 2022-2035\footnote{\url{https://www.astronet-eu.org/wp-content/uploads/2023/05/Astronet_RoadMap2022-2035_Interactive.pdf}} indicates near-future big data as one of the biggest challenges that astronomy in Europe will face in the coming decade and recommends the development of a ``Tiered'' approach for data infrastructure, to optimize the use of available resources.

With new big data, unavoidably, come new methods, ideally fully automated and requiring little human intervention (e.g., see Ref.~\cite{Moriwaki:2023sdh}). It is therefore not surprising that a majority of sections in this chapter deal with different aspects of \ac{ml}-based approaches to the data analysis, in order to speed up and optimize the science output, and to probe all possible parameter spaces. However, with the growing popularity of these methods, new challenges arise.

An old saying popular in \ac{ml} community says ``rubbish in, rubbish out''; it concisely summarizes the fact that the errors or biases that exist in the original data, also those unknown to the users, are unavoidably transferred to the results. Informed application of any machine-learning methods requires an in-depth understanding of the properties and limitations of the data. A related problem is a common ``black box'' approach to \ac{ml} applications. Insufficient understanding of algorithms used may result in users not being aware of their limitations and applicability to a given scientific problem and given data. The next big challenges are related to the interpretability and reproducibility of the results obtained with the aid of machine-learning-based methods. In particular, reproducibility strengthens the importance of keeping codes and data open to the scientific community so that all the scientific results obtained can be independently verified. However, it cannot be forgotten that storage - especially long term storage - and computing power come with a cost, both monetary, and environmental. The latter aspect draws more and more attention especially now, when commercial \ac{ml} and artificial intelligence projects fuel an increasing demand for resources.

Observational cosmology has already become an extremely data-intensive endeavor, and the future large, complex and interdependent data sets that will be generated by observatories, space missions, real and mock experiments, theoretical model simulations, and different types of numerical simulations will require new tools and approaches, and close co-operation between different fields of science. It is safe to say that cosmology and extragalactic astrophysics are now only at the beginning of this road.
\subsection{Bayesian inference methods in Cosmology \label{sec:MCMC}}

\noindent \textbf{Coordinator:} Jesús Torrado\\
\noindent \textbf{Contributors:} Alessandro Vadal\`{a}, Benjamin l'Huillier, Denitsa Staicova, Guadalupe Ca\~nas-Herrera, Jenny G. Sorce, Laura Herold, Mat\'{\i}as Leizerovich, Matteo Martinelli, Ruth Lazkoz, and Susana J. Landau
\\

\noindent Modern cosmology has entered an era of precision measurements, where extracting robust constraints on theoretical models from increasingly complex datasets demands sophisticated statistical approaches. Monte Carlo (MC) methods have emerged as the cornerstone of parameter inference in cosmology, allowing researchers to efficiently explore high-dimensional parameter spaces and quantify uncertainties in a fully Bayesian framework. This section presents an overview of the most commonly used MC techniques in cosmological analyses, their strengths and limitations, and practical considerations for their implementation.

The goal of Bayesian parameter inference is to determine the posterior $p(\boldsymbol{\theta}|\boldsymbol{d},\mathcal{M})$, which is the probability of the parameters of model $\mathcal{M}$ taking some value $\boldsymbol{\theta}$, given the data $\boldsymbol{d}$. The posterior can be related to the likelihood of the data $\boldsymbol{d}$ being a realization of the model with given parameter values, i.e.,  $\mathcal{L}(\boldsymbol{d}|\boldsymbol{\theta},\mathcal{M})$, via the Bayes theorem
\begin{equation}
    \label{eq:Bayes_theorem}
    p(\boldsymbol{\theta}|\boldsymbol{d},\mathcal{M}) = \frac{\mathcal{L}(\boldsymbol{d}|\boldsymbol{\theta},\mathcal{M}) \, \pi(\boldsymbol{\theta},\mathcal{M})}{\mathcal{Z}(\boldsymbol{d},\mathcal{M})}\,,
\end{equation}
where \( \pi(\boldsymbol{\theta},\mathcal{M}) \) is the prior and \( \mathcal{Z}(\mathbf{d},\mathcal{M}) \) the model evidence. Since one typically cares about the constraints imposed on the model parameters, which are independent of the overall normalization of the posterior, the model evidence is often neglected as it contributes only a constant factor. However, as we will discuss below, this evidence becomes important for model selection. In standard parameter estimation contexts, we usually drop the explicit conditioning on the model $\mathcal{M}$. When required, lower-dimensional constraints, e.g., Bayesian credible intervals, are easily obtained by marginalization, i.e., integration of the posterior over the remaining parameters. By construction, the model evidence is the \textit{marginal likelihood} over the full parameter space, $\mathcal{Z}(\boldsymbol{d})=\int\mathcal{L}(\boldsymbol{d}|\boldsymbol\theta)\pi(\boldsymbol\theta)\,\mathrm{d}\boldsymbol\theta$, determining the probability of the data having been realized under the given model, whatever the values of its parameters (hence their usefulness for model comparison).

There are various Monte Carlo (MC) techniques aimed at determining the posterior $p(\boldsymbol{\theta}|\mathbf{d})$ by means of obtaining a fair sample from it. Below we offer a short review of the most popular MC algorithms used by the astrophysical and cosmological community, which is illustrated in Fig.\ \ref{fig:MCMMC_flowchart}. These approaches have diverse strengths, but also limitations, which we refer to as ``sampling problems'' below.

Moreover, the choice of prior $\pi(\boldsymbol{\theta})$ can lead to unwanted (and sometimes unknown) effects like prior dependence or projection/prior volume effects, which we refer to as ``modeling problems'' and will be discussed below in the context of cosmological tensions.\footnote{Of course, the modeling of the likelihood $\mathcal{L}(d|\boldsymbol{\theta})$ is also often based on approximations and can lead to uncertainties. Since the likelihood is the basis of all parameter inference, not only MC approaches, we will not discuss those here.}

\begin{figure}[t!]
    \centering
\includegraphics[width=0.75\textwidth]{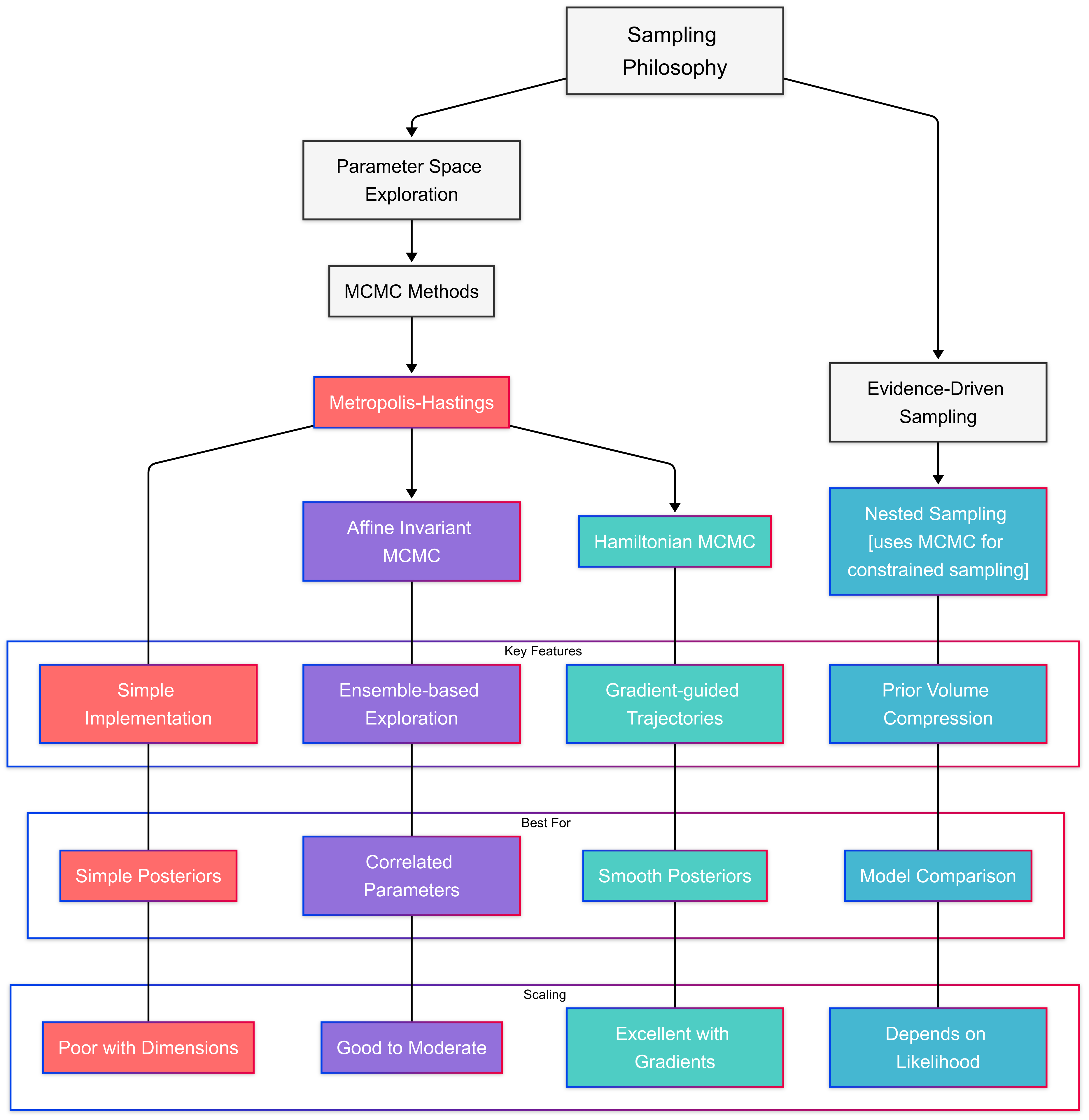}
 \caption{A flowchart summarizing the main differences between different samplers.}
    \label{fig:MCMMC_flowchart}
\end{figure}

\subsubsection{Markov chain Monte Carlo sampling algorithms}

\ac{mcmc} algorithms explore the parameter space through a proposal-acceptance mechanism, building a \textit{chain} of states, or points in parameter space, from which a (decorrelated) fair sample from the posterior can be extracted. Robust convergence tests (usually based on the stability of a single Markov chain or similarity between parallel ones), are crucial to avoid biased inferences and inaccurate error estimates, guaranteeing that the \ac{mcmc} samples reliably represent the target posterior distribution.

The family of \ac{mcmc} algorithms encompasses a large number of different approaches. They can be classified under different criteria: the strategy followed to propose new points, the acceptance-rejection test performed on the proposals, the distribution from which they effectively sample (not always the target distribution), whether they pre-condition the parameter space, and how.

In the simplest approach, at each step $\boldsymbol\theta$, a new state $\boldsymbol\theta^\prime$ is drawn from a proposal density $q(\boldsymbol\theta^\prime|\boldsymbol\theta)$ and accepted with probability
\begin{equation}
    \alpha = \min\left(1, \frac{\phi(\boldsymbol\theta^\prime)q(\boldsymbol\theta|\boldsymbol\theta^\prime)}{\phi(\boldsymbol\theta)q(\boldsymbol\theta^\prime|\boldsymbol\theta)}\right)\,,
\end{equation}
where $\phi(\boldsymbol\theta)$ is the target distribution, i.e., the posterior in a Bayesian framework. This test is called the \textit{Metropolis-Hastings} (MH) criterion.
Rejected proposals increase the \textit{weight} of the current state within the chain. A typical choice for the proposal distribution is a multivariate normal $\boldsymbol\theta^\prime\sim\mathcal{N}(\boldsymbol\theta, \boldsymbol\Sigma_T)$, with a variable dispersion covariance \( \boldsymbol\Sigma_T \) tailored to parameter values at each step.

While unsophisticated, this method can produce accurate inference for simple distributions in large dimensions, but its efficiency depends critically on the choice of an appropriate proposal distribution, and otherwise deteriorates quickly for highly correlated or complex distributions.

Efficiency can be increased by using the current chain (or parallel chains) to compute an affine transformation that would decorrelate the posterior locally (or globally if sufficiently Gaussian).
This affine transformation can be chosen so that proposals involve the re-calculation of only parts of the likelihood pipeline \cite{Lewis:2013hha} (as implemented in \code{MontePython} \cite{Audren:2012wb,Brinckmann:2018cvx} and \code{Cobaya} \cite{Torrado:2020dgo}).
Parallelization, together with these strategies, can make \ac{mcmc} well-suited for computationally expensive likelihood evaluations.

For more complicated posterior structures, such as curving degeneracies that would need a local transformation to be decorrelated, \textit{ensemble methods} such as \code{emcee}\footnote{\url{https://github.com/dfm/emcee}} \cite{Foreman-Mackey:2012any} can use the information on the position of multiple walkers to automatically adapt proposals.
More recently, the authors of \code{pocoMC}\footnote{\url{https://github.com/minaskar/pocomc}} \cite{Karamanis:2022alw} have proposed the use of Sequential Monte Carlo (i.e., sequentially sampling from a family of distributions that interpolate between the prior and the posterior) while learning a generalized transformation that decorrelates arbitrary parameter spaces, using Normalizing Flows \cite{JMLR:v22:19-1028}. \code{pocoMC} can compute model evidences along MC samples.

\textit{Slice sampling} \cite{Neal2003} is an alternative adaptive approach to propose new Markov chain states, that automatically adjusts step sizes to the local shape of the distribution. For each parameter update, it samples uniformly from a randomly-oriented ``slice'' under the \ac{pdf} curve, $S_y = \{\boldsymbol\theta: y < \phi(\boldsymbol\theta)\}$, where $y$ is uniformly drawn from $[0, \phi(\boldsymbol{\theta}_\text{current})]$. A proposal $\boldsymbol\theta^\prime$ is then uniformly sampled from this slice through a stepping-out and shrinking procedure. This adaptivity makes slice sampling particularly robust, requiring minimal tuning while maintaining good mixing properties. The algorithm naturally accommodates different scales in different regions of the parameter space, though its efficiency can decrease in higher dimensions.
A modern implementation combining slice sampling with an ensemble approach, with applications in cosmology and astronomy, is \code{zeus}\footnote{\url{https://github.com/minaskar/zeus}} \cite{Karamanis:2021tsx}. Slice sampling has also been used to increase the efficiency and robustness of nested samplers (see below).

\subsubsection{Hamiltonian Monte Carlo}

Hamiltonian Monte Carlo (HMC), also known as \textit{hybrid} Monte Carlo, is an \ac{mcmc} approach that incorporates gradient information through an augmented parameter space with momentum variables. The system evolves according to Hamiltonian dynamics
\begin{equation}
    \mathcal{H}(\boldsymbol \theta, \boldsymbol p) = -\ln\phi(\boldsymbol\theta) + \frac{1}{2}\boldsymbol p^T\boldsymbol M^{-1}\boldsymbol p\,,
\end{equation}
where $\boldsymbol M$ is a mass matrix. In this approach, proposals for accepted/rejected new states are obtained by letting the current state evolve according to an energy-conserving trajectory for a set amount of time -- in practice some number of steps of corresponding of small but finite duration. By following continuous trajectories determined by local dynamics, rather than diffusive random steps, HMC significantly reduces random-walk behavior, improving sampling efficiency for posterior with complicated geometries.

A key challenge in HMC is selecting an appropriate integration time for the dynamic evolution of the steps, i.e., the number and duration of finite steps. The No-U-Turn Sampler (NUTS) \cite{JMLR:v15:hoffman14a} refines this approach by adaptively determining the trajectory length, ensuring efficient exploration without manual tuning. By dynamically terminating paths upon noticing signs of retracing, NUTS avoids redundant computations while preserving detailed balance, thus enhancing the efficiency, particularly in high-dimensional spaces.

In order to exploit the advantages of HMC, cosmological codes and likelihoods need to be able to compute derivatives with respect to input parameters and intermediate observables, respectively. The most common approach is to build these codes using \textit{automatically-differentiable} numerical frameworks that are able to produce gradients for arbitrary inputs. The most capable and popular one at the time of writing this paper is the \code{Python} library \code{JAX} \cite{jax2018github}. There is a growing body of \code{JAX}-based
cosmological and astronomical libraries
(\code{JAX-COSMO}\footnote{\url{https://github.com/DifferentiableUniverseInitiative/jax\_cosmo}} \cite{Campagne:2023ter}, \code{DISCO-DJ}\footnote{\url{https://github.com/ohahn/DISCO-EB}} \cite{Hahn:2023nvb})
and likelihoods
(\code{candl}\footnote{\url{https://github.com/Lbalkenhol/candl}} \cite{Balkenhol:2024sbv}),
as well as machine-learning cosmological emulators
(\code{CosmoPower-JAX}\footnote{\url{https://github.com/dpiras/cosmopower-jax}} \cite{Piras:2023aub}).
Another popular programming framework that allows for automatic-differentiation is
\code{julia},\footnote{\url{https://github.com/JuliaLang/julia}} that has been employed for example in
\code{LimberJack.jl}\footnote{\url{https://github.com/jaimerzp/LimberJack.jl}} \cite{Ruiz-Zapatero:2023hdf}, \code{Capse.jl}\footnote{\url{https://github.com/CosmologicalEmulators/Capse.jl}} \cite{Bonici:2023xjk} and \code{Effort.jl}\footnote{\url{https://github.com/CosmologicalEmulators/Effort.jl}} \cite{Bonici:2025ltp}.

An implementation of NUTS that can be interfaced with \code{JAX}-based codes is \code{numpyro}\footnote{\url{https://github.com/pyro-ppl/numpyro}} \cite{2018arXiv181009538B}. Using \code{numpyro} and diverse \code{JAX}-based cosmological libraries and emulators, recent works \cite{Piras:2024dml,Mootoovaloo:2024lpv} have reported efficiency gains of up to orders of magnitude, especially for parameter spaces of large dimensionality that are expected when performing inference on data from next-generation surveys.

\subsubsection{Nested sampling}

Nested sampling \cite{Skilling:2006ns} transforms the evidence calculation into a one-dimensional integral over prior mass $X$ accumulated up to a given likelihood value
\begin{equation}
  \mathcal{Z} = \int_0^1 \mathcal{L}(X)\,\mathrm{d}X
  \qquad\text{with}\qquad
  X(\lambda) = \int_{\mathcal{L}(\boldsymbol\theta)>\lambda} \pi(\boldsymbol\theta)\,\mathrm{d}\boldsymbol\theta\,,
\end{equation}
where the likelihood is evaluated such that $\mathcal{L}(X(\lambda))=\lambda$ for some likelihood-cutoff value $\lambda$. Numerically, this integral is computed as a weighted sum $\mathcal{Z} \sim \sum_{i=1}^{m} w_{i} L_{i}$, with $w_{i} \sim \Delta X$. The algorithm maintains a set of \textit{live points} updated at every step: the one with the lowest likelihood is dropped (and referred to as a \textit{dead} point), and a new one is sampled restricted to having a likelihood larger than that of the recently-dead point. The prior volume shrinkage $\Delta X$ is computed probabilistically and assigned the likelihood value of the dead point. In addition to the evidence calculation, the pool of dead points may be used to construct a fair sample from the posterior.

The procedure to sample a point from the prior subject to a minimum likelihood, i.e., within an \textit{iso-likelihood} contour, is the main hurdle of the algorithm.
Some implementations, such as \code{MultiNest}\footnote{\url{https://github.com/farhanferoz/MultiNest}} \cite{Feroz:2008xx}, construct a minimal ellipsoid containing the set of live points, and sample uniformly from an enlarged version of it.
\code{UltraNest}\footnote{\url{https://github.com/JohannesBuchner/UltraNest}} \cite{Buchner2021}, implements a different sampling approach based on ellipsoids centered around the live points themselves, and includes more robust uncertainty estimation for the integrated evidence.
Others, like \code{PolyChord}\footnote{\url{https://github.com/PolyChord/PolyChordLite}} \cite{Handley:2015fda}, propose new points by running a short affine-invariant \ac{mcmc} chain (in particular using slice-sampling) from one of the current live points. \ac{mcmc}-based approaches have better dimensionality scaling, but tend to be slower for simpler problems.
Other modern implementations like \code{dynesty}\footnote{\url{https://github.com/joshspeagle/dynesty}} \cite{Speagle2020} dynamically vary the number of live points to improve efficiency. All these approaches allow for efficient sampling of multi-modal distributions by using clustering algorithms
to separate and evolve different subsets of live points in parallel.

In general, nested samplers are preferred when the geometry of the posterior distribution is complicated or exhibits multiple peaks. Due to its ability to compute the Bayesian evidence, nested sampling is also particularly valuable for model selection problems.
Nested sampling can however be computationally very expensive for highly dimensional parameter spaces, but it parallelizes very efficiently up to the number of live points.

There have been in recent years various attempts at accelerating nested sampling using \ac{ml}, focusing on efficiently sampling within iso-likelihood contours.
\code{nautilus}\footnote{\url{https://github.com/johannesulf/nautilus}} \cite{Lange:2023ydq} and \code{neuralike}\footnote{\url{https://github.com/igomezv/neuralike}} \cite{Gomez-Vargas:2024izm} train a neural network on the spatial dependence of the likelihood in the vicinity of the current live set (only on the live set for \code{neuralike}) to construct an estimate of the iso-likelihood contour, and reject points predicted to be outside it before the true likelihood is evaluated.
\code{nessai}\footnote{\url{https://github.com/mj-will/nessai}} \cite{Williams:2023ppp} takes a similar approach, training a \textit{normalizing flow} on the set of live points to find a transformation of the iso-likelihood contour into a much simpler hyper-sphere.

\subsubsection{Practical considerations for choosing an MC sampler}

The sampling methods discussed above differ in both their theoretical foundations and practical applications, each offering distinct advantages and limitations in cosmological parameter inference.

Traditional Metropolis-Hastings \ac{mcmc} uses local random-walk proposals, leading to $O(\sqrt{d})$ scaling in dimensionality $d$. While simple and reliable, these proposals often require many likelihood evaluations, making them computationally expensive, especially for high-dimensional parameter spaces or when the likelihood poorly constrains the parameters. Non-Gaussian posteriors, particularly multi-modal distributions, can further challenge efficient parameter space exploration.
A good dimensionality scaling can be retained for simple distributions using decorrelating affine-invariant transforms. For more complicated, highly non-Gaussian, and multi-modal distributions, ensemble and sequential MC approaches can help, especially when in combination with generalized decorrelating transforms such as NFs.

Slice sampling takes a geometric approach by sampling from level sets of the \ac{pdf}, automatically adjusting to the local structure of the posterior. This self-adaptive behavior results in $O(d)$ scaling without requiring manual tuning, making it effective for problems with widely varying parameter scales, such as fitting galaxy \ac{lf}s where parameters span multiple orders of magnitude. While computationally inexpensive in low dimensions, its efficiency can degrade in high-dimensional spaces.

HMC improves upon this by using gradient information to make physics-inspired proposals, achieving better $O(d^{1/4})$ scaling. This makes HMC particularly effective for problems with strong parameter correlations, such as hierarchical models in galaxy clustering. However, it requires differentiable likelihoods and careful tuning of integration parameters.

Nested sampling transforms the problem into a one-dimensional integration over nested likelihood contours. Despite its $O(d^3)$ scaling due to constrained sampling requirements
when using simple acceptance/rejection sampling, it offers important advantages: direct computation of the evidence $\mathcal{Z}$ (essential for model comparison in \ac{de} studies) and natural handling of multi-modal distributions. This makes it particularly valuable in cosmology, \ac{gw} physics, and gravitational lensing analyses.
A better dimensionality scaling can be achieved by using slice-sampling for constrained sampling, or performing acceptance/rejection sampling on a surrogate model.

The choice of sampler often depends on specific requirements like parameter space dimensionality, likelihood characteristics, and the need for evidence calculations \cite{Buchner:2021kpm, Colgain:2023bge,Albert:2024zsh, Staicova:2025huq}.
For problems in which the likelihood is extremely fast, $\mathcal{O}(<10^{-2}\,\mathrm{s})$, and the dimensionality is of orders or a few 10s, simple approaches should suffice, such as affine-invariant \ac{mcmc}, or a non-boosted nested sampler if a model evidence estimate is needed. For harder problems (very high dimensionality, strong non-Gaussianity, multi-modality), the use of the more advanced methods discussed above, such as preconditioned Sequential Monte Carlo and \ac{ml}-boosted nested samplers, can prove optimal. For extremely slow likelihoods, $\mathcal{O}(>10\,\mathrm{s})$, the approaches discussed in Sec.~\ref{sec:ML_inference} such as \textit{surrogate posterior} methods or \textit{simulation-based inference} may be the only viable option, at the cost of some statistical robustness.

An illustration of the problem of dimensionality and the complexity of the distribution can be seen Fig.~\ref{fig:MCMMC_runtime}. On it we show how the runtime and the accuracy scale with dimensions for a multi-dimensional Gaussian distribution and the Rosenbrock distribution. For the accuracy test,  we use the Wasserstein distance metric \cite{2024arXiv240503664R}, which measures the true distributional similarity between sampler outputs and reference distributions.
One can see that all samplers show excellent performance on Gaussian problems across dimensions, but exhibit a dramatic accuracy drop-off for Rosenbrock problems beyond 6-8 dimensions, reflecting the inherent challenge of sampling correctly from complex, highly correlated distributions in higher dimensions. 

\begin{figure}[t!]
    \centering
\includegraphics[width=1\textwidth]
{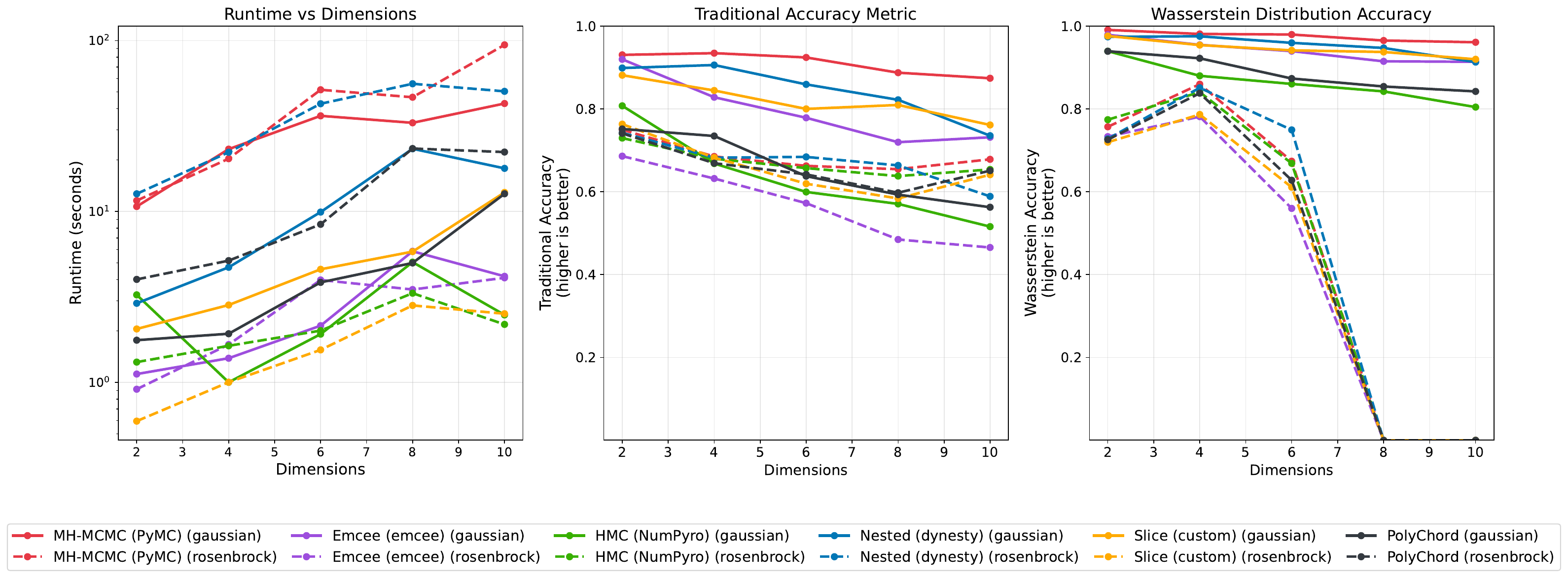}
 \caption{Illustration of how different sampling methods scale with dimensionality and target distribution complexity. The left panel shows runtime scaling with increasing dimensions, while the middle and right panels display sampler accuracy using two different metrics. Results are shown for both a simple multivariate Gaussian distribution (solid lines) and the challenging banana-shaped Rosenbrock distribution (dashed lines). The traditional accuracy metric combines deviations from mean and function evaluation errors. The Wasserstein accuracy metric uses optimal transport theory to measure distributional similarity between samples and theoretical distributions.
 } 
    \label{fig:MCMMC_runtime}
\end{figure}
\subsubsection{Cosmological inference frameworks}

There exist a number of public codes facilitating the integration of MC samplers such as the ones mentioned above with cosmological inference pipelines: numerical codes to calculate cosmological observables (Boltzmann codes such as \code{CAMB} \cite{Lewis:1999bs}, \code{CLASS} \cite{Blas:2011rf}, and \code{PyCosmo} \cite{Moser:2021rej} and their extensions, as well as emulators) and experimental likelihoods for cosmological and astrophysical surveys and data.

The most used ones in the last years are
\code{CosmoSIS}\footnote{\url{https://github.com/joezuntz/cosmosis?tab=readme-ov-file}} \cite{Zuntz:2014csq}
\code{MontePython}\footnote{\url{https://github.com/brinckmann/montepython\_public}} \cite{Audren:2012wb,Brinckmann:2018cvx}
and \code{Cobaya}\footnote{\url{https://github.com/CobayaSampler/cobaya}} \cite{Torrado:2020dgo} (which uses \code{GetDist}\footnote{\url{https://github.com/cmbant/getdist}} \cite{Lewis:2019xzd} for analyzing MC results). The choice for a particular one will be determined by the scientific problem at hand, since each provides a different combination of features and integrates different theoretical codes and experimental likelihoods.

Their use is thoroughly recommended since they facilitate not only the interfacing between the different parts of the inference pipeline, but also help in specifying the models and priors as well as the configuration of samplers and theoretical codes, and provide sane defaults for each. They usually either include in their source or automate the installation of cosmological codes and likelihoods, as well as MC samplers.

\subsubsection{Modeling problems: Prior dependence and projection effects}

Analyses of present-day data require a significant modeling effort in order to achieve accurate theoretical predictions that can be compared to data without biasing the results. This includes the modeling of complicated physics and systematic or non-linear effects, which require introducing additional parameters (nuisance parameters) that need to be kept free during the sampling of the corresponding posterior distribution. If the data is not sensitive to all parameters in such a high-dimensional spaces, this can lead to a flat likelihood surface $\mathcal{L}(\mathcal{D}|\boldsymbol\theta)$ in Eq.~\eqref{eq:Bayes_theorem}, which in turn can result in a sensitivity to the choice of prior $\pi(\boldsymbol\theta)$. If the prior distribution is not informed by past experiments or theory, the dependence of the results on $\pi(\boldsymbol\theta)$ may be unwanted and needs to be explored in a sensitivity analysis. A second -- more subtle -- dependence of the one-dimensional posteriors on the prior is introduced at marginalization. Since the marginalization procedure takes into account the posterior volume, for very flat or very non-Gaussian likelihood surfaces, there may be an (unwanted) up-weighting of regions with large (prior) volume. These so-called projection or prior volume effects can lead to shifts of the marginalized posterior away from the maximum-likelihood estimate or ``best fit'' point (see also Sec.~\ref{sec:frequen_approach}), and may bias the constraints on the physics of the problem if the choice of prior was not sufficiently informed.

An example of this in the context of the $\sigma_8$ tension is the \ac{eftoflss} (e.g., see Refs.~\cite{Baumann:2010tm,Carrasco:2012cv,Senatore:2014via,Senatore:2014eva,Senatore:2014vja}), a perturbation-theory based technique to describe the power spectrum of biased tracers up to mildly non-linear scales, that is applied to \ac{lss} measurements.
In such an approach, several new parameters are introduced, as they enter the perturbative expansion of the galaxy density field. These parameters will be free parameters in the context of an \ac{mcmc} analysis and, therefore, require a choice of prior. However, the arbitrary nature of the choice of such priors can become a problem for the analysis, since the inferred cosmological parameters can depend on the particular choice of prior. It has been shown how different prior choices, e.g., extending the range of allowed values to extreme intervals, can significantly affect the outcome of the analysis, with projection effects shifting some parameters, such as the primordial amplitude $A_{\rm s}$ and the amplitude of matter fluctuations, $\sigma_8$, with respect to their true values~\cite{Carrilho:2022mon, Simon:2022lde}.
It was further demonstrated that different -- albeit theoretically equivalent -- choices of the EFT parametrization lead to discrepant inferred cosmological parameters~\cite{Simon:2022lde, Maus:2023rtr}. Ref.~\cite{Holm:2023laa} confirms a significant impact of priors on the nuisance parameters on the inferred cosmological parameter using frequentist profile likelihoods. Approaches to choose well motivated priors on the EFT parameters include the use of a Jeffreys prior~\cite{Donald-McCann:2023kpx, Zhao:2023ebp} or simulation-inferred priors~\cite{Ivanov:2024xgb,Chudaykin:2024wlw,Zhang:2024thl}.
Such an effect hinders the robustness of the analysis results, as strong assumptions, only partially motivated by theoretical considerations, need to be done on the choice of priors for the extra parameters. 

A similar projection effect, shifting the values of parameters away from their true value, can be introduced also when considering priors on the baryon abundance  $\Omega_{\rm b,0}\,h^2$. Several nuisance effects modeled in \ac{lss} observables require the value of this parameter, which is however not strongly constrained by \ac{lss} observations. For such a reason the choice of prior for the abundance of baryons can be significant for the analysis, as switching from a uniform prior to a Gaussian constrain as provided by \ac{bbn} \cite{Schoneberg:2024ifp} can change the outcome in a significant way \cite{Simon:2022lde}.

In the context of the Hubble tension, an impact of prior effects on the inferred value of $H_0$ has been reported and explored for several extended models, for example: \ac{ede} (see Sec.~\ref{sec:EDE})~\cite{Smith:2023oop, Murgia:2020ryi,Smith:2020rxx, Herold:2021ksg, Herold:2022iib, Reeves:2022aoi, Efstathiou:2023fbn}, \ac{nede} (see Sec.~\ref{sec:NEDE})~\cite{Niedermann:2020dwg, Cruz:2023cxy}, number of relativistic species (see Sec.~\ref{sec:Extra_DoF})~\cite{Hamann:2007pi,Hamann:2011hu, Henrot-Versille:2018ujq}, Brans-Dicke model (see Sec.~\ref{sec:MoG})~\cite{Gomez-Valent:2022hkb} decaying \ac{dm} (see Sec.~\ref{sec:I_D_DM})~\cite{Holm:2022kkd} and more.

\subsubsection{Model comparison criteria}
\label{sec:3p1:modelselection}

The increasing complexity of cosmological models and datasets necessitates robust statistical frameworks for model comparison and consistency checks. Information criteria and Bayesian methods provide quantitative tools for selecting between competing models while accounting for model complexity and data support.

There exist several information criteria that can be used to compare models \cite{Liddle:2007fy}. Common criteria that balance goodness-of-fit of a model against its complexity are the Akaike Information Criterion (AIC) and the Bayesian Information Criterion
\begin{equation}
  \text{AIC} = -2\ln(\mathcal{L}_{\text{max}}) + 2k\,,
  \qquad\qquad
  \text{BIC} = -2\ln(\mathcal{L}_{\text{max}}) + k\ln(N)\,,
\end{equation}
where $\mathcal{L}_{\text{max}}$ is the maximum likelihood, $k$ is the number of free parameters, and $N$ is the effective number of data points on which the likelihood is defined.
When comparing a \textit{test} model versus a \textit{baseline} one (e.g., \lcdm) using ICs (AIC and BIC), we compute the difference in criterion values as
$$ \Delta \mathrm{IC}_{\text{test}} = \mathrm{IC}_\text{baseline} - \mathrm{IC}_{\text{test}}\,.$$
The model with the \textit{lowest IC} is preferred \cite{Jeffreys:1939xee}, meaning, in this sign convention, $\Delta \text{IC} > 0$ favors the test model, while $\Delta \text{IC} < 0$ favors the baseline model. The value of $|\Delta \text{IC}|$ indicates the strength of the preference:
$|\Delta \text{IC}| \geq 2$ (weak),
$|\Delta \text{IC}| \geq 6$ (medium),
$|\Delta \text{IC}| \geq 10$ (strong).

The Deviance Information Criterion (DIC) extends AIC to hierarchical models by accounting for parameter uncertainty:
\begin{equation}
    \text{DIC} = -2\ln(\mathcal{L}(\boldsymbol{\hat{\theta}})) + 2p_D\,,   \qquad\text{with}\qquad
  p_D = 2\left[\ln(\mathcal{L}(\boldsymbol{\hat{\theta}})) - \langle\ln(\mathcal{L}(\boldsymbol\theta))\rangle\right]\,,
\end{equation}
where $\boldsymbol{\hat{\theta}}$ is the posterior mean of the free parameters, $p_D$ the effective number of them, and $\langle\ln(\mathcal{L}(\boldsymbol\theta))\rangle$ the mean log-likelihood over the posterior samples. Unlike AIC and BIC, DIC uses the entire posterior distribution rather than just the maximum likelihood estimate, making it particularly suitable for Bayesian hierarchical models where the effective number of parameters may be less than the actual number due to prior constraints \cite{Spiegelhalter:2002yvw}.

The Bayes Factor (BF) provides a fully Bayesian approach to model comparison
\begin{equation}
    \text{BF}_{12} = \frac{\mathcal{Z}_1}{\mathcal{Z}_2} = \frac{\int \mathcal{L}_1(\boldsymbol\theta_1)\pi_1(\boldsymbol\theta_1)\,\mathrm{d}\boldsymbol\theta_1}{\int \mathcal{L}_2(\boldsymbol\theta_2)\pi_2(\boldsymbol\theta_2)\,\mathrm{d}\boldsymbol\theta_2}\,.
\end{equation}
Jeffreys' scale {is commonly used to interpret} $\ln(\text{BF}_{12})$ values: $0$--$1$ indicates inconclusive evidence, $1$--$3$ suggests positive evidence, $3$--$5$ shows strong evidence, and values larger than 5 represent very strong evidence in favor of model 1 \cite{Jeffreys:1939xee, Trotta:2008qt}. While AIC and BIC are easily computed from \ac{mcmc} chains, Bayes factors require accurate evidence estimation, typically obtained through nested sampling or thermodynamic integration methods.
As a fully Bayesian approach, the value of the Bayes Factor depends strongly on the choice of prior density, unlike for information criteria, so a prior sensitivity analysis is advisable.

\subsubsection{Tension metrics}

Tension metrics are statistical tools that are used to quantify the agreement (or lack thereof) between the estimation of cosmological parameters obtained with different data sets, beyond the widely applied \textit{``rule of thumb''}. According to this rule the distance between two posterior 1-d distributions can be quantified in 1-dimensional standard deviations by $N_\sigma=({\mu_{\rm \rm A} - \mu_{\rm B}})/{\sqrt{\sigma_{\rm A}^2+\sigma_{\rm B}^2}}$ where $\mu_{\rm A/B}$ and $\sigma_{\rm A/B}$ refer to the means and variances obtained with each data set.
 Most studied tension metrics can be grouped into two families: i) those based on Bayesian evidence \cite{Marshall:2006,Handley:2019wlz} and ii) those based on the posterior distribution \cite{Raveri:2019,Raveri:2020,Raveri:2021}. While group (i) answers the question about what hypothesis is preferred by the data under the assumed model, group (ii) intends to establish the statistical significance between the posteriors of data sets A and B within the parameter space of both experiments.
For each metric, an estimator is computed and a corresponding probability $P$ is defined that quantifies the agreement or disagreement between data sets in terms of its corresponding equivalent 1-dimensional $N_\sigma$ defined as: $P={\rm Erf} ({N_{\sigma}}/{\sqrt{2}})$. In most cases the probability $P$ is related to the \textit{probability-to-exceed} ($\mathrm{PTE}$) of an estimator $Q$ that follows a $\chi^2_d$ distribution, defined as
\begin{equation}
    {\rm PTE} = \int_{Q}^{\infty} \chi_d^2(x) \,\mathrm{d}x\,.
\end{equation}

\vskip 1ex \noindent
\textbf{i) Metrics defined on Bayesian evidence ($\mathcal{Z}$).}
The Bayesian ratio statistic $R$  corresponding to two datasets $A$ and $B$, and the combination $AB$ of both of them, can be written as \cite{Marshall:2006}
\begin{equation}
    R = \frac{\mathcal{Z}_{\rm AB}}{\mathcal{Z}_{\rm A} \mathcal{Z}_{\rm B}}\,.
\end{equation}
$R \gg 1$ is interpreted as both datasets being consistent, while $R \ll 1$ means that the datasets are inconsistent. However, it has been shown that the $R$ estimator depends on the choice of the prior and can therefore underestimate inconsistencies. This estimator has two primary contributions, one from the unlikeliness of two datasets being in agreement,  the \textit{information ratio} $I$, and another from their disagreement, the \textit{suspiciousness} $S$ \cite{Handley:2019wlz}
\begin{equation}
  \ln I = \mathcal{D}_{\rm A}+\mathcal{D}_{\rm B} - \mathcal{D}_{\rm AB}\,,
  \qquad\qquad
  \ln S = \ln R - \ln I\,.
\end{equation}
Here $\mathcal{D}$ refers to the Kullback-Leibler divergence between prior and posterior, quantifying the information gain/compression produced by the given data set. The suspiciousness estimator $S$ remains unaltered under a change of the prior widths as long as this change does not significantly alter the posterior. In addition, if the posterior is a $d$-dimensional Gaussian, the quantity $d-2\ln S$ follows a $\chi^2_d$ distribution. Under that approximation, the probability of two datasets being discordant by chance can be computed as the $\mathrm{PTE}$ of $d-2\ln S$.

\vskip 1ex \noindent
\textbf{ii) Metrics based on the posterior distribution.}
First, we discuss some metrics based on quadratic estimators that require Gaussian posteriors. For these, the probability ${\rm P}$ of the data sets agreeing is defined as ${\rm P}=1-\mathrm{PTE}$ where $\mathrm{PTE}$ is the probability-to-exceed defined above. Let us call $\boldsymbol{\hat{\theta}}_{\rm A}$ the mean of the  parameters inferred with dataset A  and $\boldsymbol{\hat{\theta}}_{\rm B}$ the one obtained with dataset B, while $\boldsymbol{\hat{\Sigma}}_{\rm A}$ and $ \boldsymbol{\hat{\Sigma}}_{\rm B}$ refer to their respective covariance matrices. The method called \textit{parameter differences in standard form} relies on the quadratic estimator \cite{Raveri:2019}
\begin{equation}
    Q_{\rm DM}= (\boldsymbol{\hat{\theta}}_{\rm B}-\boldsymbol{\hat{\theta}}_{\rm A})^{T} (\boldsymbol{\hat{\Sigma}}_{\rm B}+\boldsymbol{\hat{\Sigma}}_{\rm A})^{-1}(\boldsymbol{\hat{\theta}}_{\rm B}-\boldsymbol{\hat{\theta}}_{\rm A})\,,
    \label{eq: Q_DM}
\end{equation}
This estimator follows a $\chi^{2}_{\nu}$ distribution with $\nu=\rm{Rank}[\boldsymbol{\hat{\Sigma}}_{\rm B}+\boldsymbol{\hat{\Sigma}}_{\rm A}]$ and can be regarded as a generalization of the \textit{rule of thumb}. The method called \textit{parameter differences in updated form} measures how one data set updates the other and is based on \cite{Raveri:2019}
\begin{equation}
    Q_{\rm UDM}= (\boldsymbol{\hat{\theta}}_{\rm AB}-\boldsymbol{\hat{\theta}}_{\rm A})^{T} (\boldsymbol{\hat{\Sigma}}_{\rm AB}-\boldsymbol{\hat{\Sigma}}_{\rm A})^{-1}(\boldsymbol{\hat{\theta}}_{\rm AB}-\boldsymbol{\hat{\theta}}_{\rm A})\,,
    \label{eq: Q_UDM}
\end{equation}
Here $\boldsymbol{\hat{\theta}}_{\rm AB}$ are the inferred parameters obtained with the joint datasets A and B, while $\boldsymbol{\hat{\Sigma}}_{\rm AB}$ is the respective covariance matrix.  This estimator follows a $\chi^{2}_{\nu}$ distribution with $\rm \nu=rank[(\boldsymbol{\hat{\Sigma}}_{\rm AB}-\boldsymbol{\hat{\Sigma}}_{\rm A})]$ degrees of freedom. Next, we describe the \textit{Goodness-of-fit loss} which evaluates the likelihood function at the maximum of the posterior of the joint distribution and of each distribution separately \cite{Raveri:2019}
\begin{equation}
    Q_{\rm DMAP} = 2 \ln \mathcal{L}_{\rm A}(\boldsymbol\theta_{p,{\rm A}}) + 2 \ln \mathcal{L}_{B}(\boldsymbol\theta_{p,{\rm B}}) - 2 \ln \mathcal{L}_{\rm AB}(\boldsymbol\theta_{p,{\rm AB}})\, ,
    \label{eq: Q_DMAP}
\end{equation}
Here $\boldsymbol\theta_{p, {\rm A/B}}$ are the Maximum a posteriori (MAP) parameters considering the dataset A/B. The estimator $Q_{\rm DMAP}$ follows a $\chi^{2}$ distribution with $\Delta \nu = \nu^{\rm A} + \nu^{\rm B} - \nu^{\rm AB}\, $ degrees of freedom, where for a given data set/combination $\nu = N - \mathrm{tr}[\boldsymbol\Sigma_{\pi}^{-1}\,\boldsymbol\Sigma_{p}]$, with $N$ being the effective number of data points defining the likelihood, and $\boldsymbol\Sigma_{\pi}$, $\boldsymbol\Sigma_{p}$ the covariance matrices of the prior and the posterior, respectively. This metric is effective at evaluating how well the theoretical prediction fits the data, but does not evaluate directly the disagreement between inferred parameters.

\begin{figure}[h!]
    \centering
    \begin{subfigure}{0.475\linewidth}
	   \includegraphics[width=\linewidth]{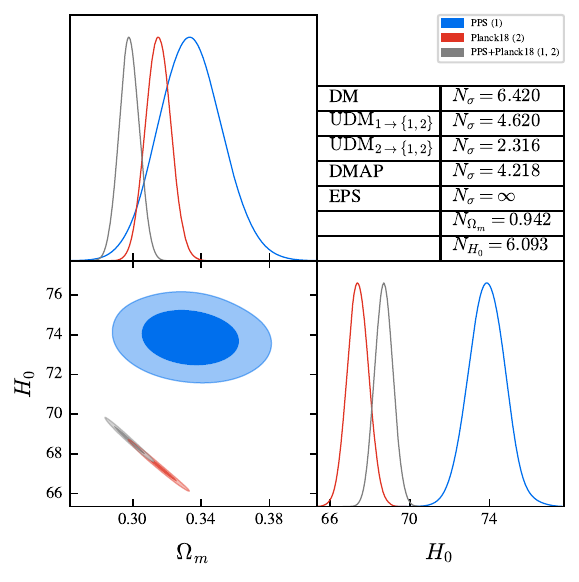}
    \end{subfigure}
    \begin{subfigure}{0.475\linewidth}
	   \includegraphics[width=\linewidth]{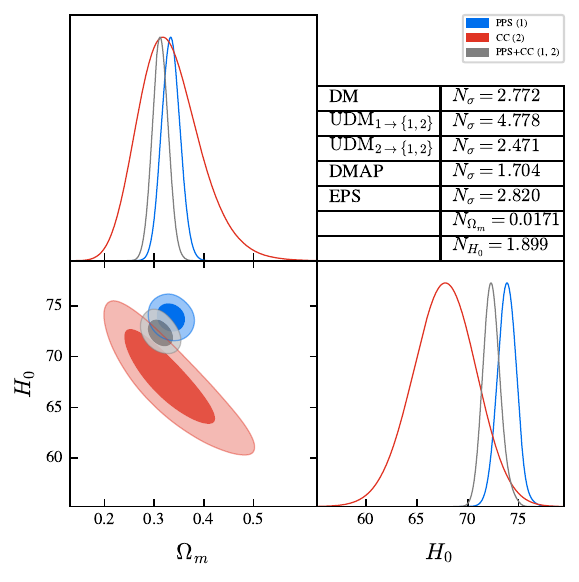}
    \end{subfigure}
    \caption{Tensions metrics applied to \textbf{a)} Planck 2018 vs Pantheon Plus + SH0ES (PPS), \textbf{b)} \ac{cc} vs Pantheon Plus + SH0ES.  In all cases, blue contours correspond to PPS and gray contours to the joint analysis. $N_\sigma$  refers to the 1-d standard deviations that correspond to a 1-d \ac{pdf} associated with each metric estimator that quantifies the disagreement between different data sets (see the beginning of this subsection). $Q_{\rm UDM}$ quantifies how the parameters inferred from one dataset are updated when another is incorporated into the analysis. $Q_{\rm DMAP}$ is related to the difference in goodness of fit to the different data of the underlying model. $Q_{\rm DM}$ and EPS quantify the difference in the value of the inferred parameters. $N_{\Omega_{m}}$ and $N_{H_0}$ quantify the tension in each inferred parameter applying the 1-d rule of thumb, adapted from Ref.~\cite{Leizerovich:2024}${}^{\dagger\dagger}$.}
    \label{fig:tensions}
\end{figure}

{\let\thefootnote\relax\footnotetext{${}^{\dagger\dagger}$Reprinted from Physics Letters B, Volume 855, id.138844 Leizerovich, M., Landau, S.J., Scóccola, C.G. Tensions in cosmology: A discussion of statistical tools to determine inconsistencies Copyright (2024), with permission from Elsevier. }}
 
The next method, called \textit{exact parameter shift} (EPS) does not require Gaussian posteriors and  is based on the parameter difference probability \cite{Raveri:2019,Raveri:2021}
\begin{equation}
    P(\Delta\boldsymbol\theta) = \int P_{\rm A}(\boldsymbol\theta)P_{\rm B}(\boldsymbol\theta-\Delta\boldsymbol\theta)\,\mathrm{d}\boldsymbol\theta\, ,
    \label{eq: Delta_theta}
\end{equation}
where $\Delta\boldsymbol\theta=\boldsymbol{\hat{\theta}}_{\rm A}-\boldsymbol{\hat{\theta}}_{\rm B}$ is the difference between the means of the posterior parameters that correspond to datasets A/B.
The probability to identify the tension  is calculated by summing over all the values of $P(\Delta\boldsymbol\theta)$ over the iso-contour corresponding to no difference $\Delta\boldsymbol\theta = 0$
\begin{equation}
    \Delta = \int_{P(\Delta\boldsymbol\theta)>P(\boldsymbol{0})} P(\Delta\boldsymbol\theta)\,\mathrm{d}\Delta\boldsymbol\theta\,.
    \label{eq: Delta}
\end{equation}
Usually, this integral is computed numerically from the posterior samples. In cases of strong disagreement, this estimator may be difficult to compute.\footnote{We point out that in the case of strong disagreement, care must be taken with the sampling of the tails of the distributions.}

Many authors use the tools presented here to discuss tensions in the inferred value of cosmological parameters with recent data sets \cite{Lemos:2021, DESI:2024mwx, Leizerovich:2024}.
It has been discussed \cite{Leizerovich:2024} that the different metrics answer different questions. For example, $Q_{\rm UDM}$ quantifies how a given data set updates the values of the inferred parameters of another data set, while $Q_{\rm DMAP}$ quantifies the difference of the goodness of fit between the individual and joint data sets to the underlying theoretical model. On the other hand, to quantify the tension between the value of inferred parameters, the appropriate metrics are $Q_{\rm DM}$ and EPS, emphasizing that the former requires posteriors to be sufficiently Gaussian, while the latter does not. Fig.~\ref{fig:tensions} shows two examples of 2-d contours and posteriors that correspond to different data sets and the corresponding $N_\sigma$ obtained with each metric, together with the ones computed with the 1-d \textit{rule of thumb} for each parameter. If EPS can be computed, usually the equivalent $N_\sigma$ that corresponds to \ac{dm} and EPS indicate similar values, while there is a difference with the ones obtained using the \textit{rule of thumb} \cite{Leizerovich:2024}. Note that the \textit{rule of thumb} only quantifies the tension in a given parameter, while the metrics presented above quantify the disagreement in the whole parameter space.

\bigskip
\subsection{Machine learning based inference techniques \label{sec:ML_inference}}

\noindent \textbf{Coordinator:} Jurgen Mifsud\\
\noindent \textbf{Contributors:} Alba Domi, Anto Idicherian Lonappan, Benjamin L'Huillier, Celia Escamilla-Rivera, Clecio Roque De bom, Daniela Grandón, Daria Dobrycheva, David Valls-Gabaud, Denitsa Staicova, Elena Giusarma, Filippo Bouch\`{e}, Germano Nardini, Ippocratis Saltas, Iryna Vavilova, J Alberto Vazquez, Jacobo Asorey, Jenny G. Sorce, Judit Prat, Konstantinos Dialektopoulos, Leandros Perivolaropoulos, Luca Izzo, Luis A. Escamilla, Matteo Martinelli, Purba Mukherjee, Rahul Shah, and Ruth Lazkoz
\\

\noindent In traditional approaches, analytical likelihood functions are employed to characterize the \ac{pdf}s of observed data, enabling parameter estimation through posterior inference. However, certain datasets exhibit errors that cannot be adequately described by simple analytical distributions, such as multivariate Gaussian distributions. Consequently, the true likelihood in such cases is often complex and lacks an explicit analytical form, making traditional parameter inference methods challenging to implement. To address this issue, various likelihood-free techniques have been developed to circumvent the direct computation of likelihood functions.

In standard Bayesian inference, posterior distributions are typically explored using methods such as \ac{mcmc} sampling, variational inference, or other Bayesian computational techniques. These conventional approaches generally rely on evaluating the likelihood function for the models and parameters under consideration. However, for highly complex and computationally intensive models, the likelihood function may become intractable, and simulations can demand significant time and resources, rendering parameter inference impractical. Therefore, parameter estimation techniques that mitigate or overcome these challenges are highly beneficial, particularly in cosmological studies. Likelihood-free inference has emerged as a promising paradigm for Bayesian inference in scenarios involving complex generative models, leveraging only forward simulations to perform analysis.

With the current generation of large-scale structure observational programs, such as the European Space Agency's Euclid mission \cite{EUCLID:2011zbd} and \ac{lsst} \cite{LSST:2008ijt}, sub-percent-level precision measurements are anticipated. The upcoming decade is thus expected to witness an unprecedented increase in the quantity, diversity, and quality of multi-wavelength astronomical observations of the large-scale structure. This surge in data will necessitate the development of highly sophisticated computational tools. At this stage, the limitations may shift from data quality or availability to the capabilities of statistical and data-driven methodologies. \ac{ml} techniques have demonstrated substantial potential in addressing the computational challenges associated with traditional statistical methods, positioning them as valuable tools for advancing cosmological analyses.

For instance, current \ac{ml} techniques were adopted to the scatter in cluster mass estimates \cite{Ho:2019zap} and to distinguish between standard and \ac{mg} theories from statistically similar \ac{wl} maps \cite{Peel:2018aei}. Such techniques have also been found to be useful for the next generation \ac{cmb} experiments \cite{Caldeira:2018ojb},
 N-body simulations \cite{He:2018ggn}, cosmological parameters inference \cite{Ravanbakhsh:2016xpe}, \ac{de} model comparison \cite{Escamilla-Rivera:2019hqt}, supernova classification \cite{ANTARES:2018uvq} and strong lensing probes \cite{Lanusse:2017vha}.

Cosmology is experiencing an unprecedented growth in the volume and complexity of astronomical datasets, driven by large-scale surveys like Euclid, \ac{lsst}, and \ac{desi}. These datasets are used to construct data vectors for cosmological parameter inference, with which we can shed light on current parameter tensions within \lcdm. To fully leverage the wealth of cosmological information encoded in these datasets, \ac{ml} methods have emerged as a powerful tool. 

In this section, we will discuss a number of \ac{ml} based inference techniques, with an emphasis on those techniques which were adapted to address cosmological tensions. Neural networks are becoming ever more widely
employed in physics, including in the field of cosmology. We refer the interested reader to a review of the core concepts surrounding the use of neural networks in Refs.~\cite{Carleo:2019ptp, Mehta:2018dln}. In this section we discuss the cosmological applications of artificial neural networks (Sec.~\ref{sec:ML_Inf_ANN}), convolutional neural networks (Sec.~\ref{sec:ML_Inf_CNN}), Bayesian neural networks (Sec.~\ref{sec:ML_Inf_BNN}), and deep learning (Sec.~\ref{sec:ML_Inf_ladder}).

\subsubsection{Artificial neural networks}
\label{sec:ML_Inf_ANN}

The method by which \ac{ann} architectures \cite{Lovell:2022bhx} are used to learn to mimic the iterative process by which an \ac{mcmc} analysis takes place is described in this section together with some details of the adopted methodology to optimize the \ac{ann} structure for comparative analyses. \ac{ann}s are non-parametric techniques in the sense that they do not contain the cosmological parameters themselves, unlike the likelihood functions. This may extend cosmological analyses to otherwise overly complex systems, or reduce the computational requirements of regular cosmological model analyses making the approach more accessible. Inspired by biological neural networks, an \ac{ann} is a setup composed of a collection of neurons that are organized into layers~\cite{aggarwal2018neural,Wang:2020sxl,Gomez-Vargas:2021zyl}. This family of algorithms offers a powerful base on which to apply complex numerical tasks. \ac{ann}s are optimized to work in computer systems with a large number of threads which has become a feasible prospect in recent years with the development of very powerful graphical processing units (GPUs), thus making the process of training \ac{ann}s competitive with traditional techniques. In recent years, \ac{ann}s have become very useful in meeting the accuracy and computational efficiency required for analyses in cosmology.
\ac{ann}s are constructed with input and output layers to represent the \ac{mcmc} parameter set that is being sampled and output likelihood respectively. These layers are connected to a series of internal, or hidden, layers that are structured in a way to optimize how closely the \ac{ann} can imitate the \ac{mcmc} iterations. Each layer in this network is composed of neurons, each of which is connected to the neurons of the layers preceding and after it, as illustrated in the left panel of Fig.~\ref{fig:ML_Inf_colfi-arch} for a single data set. 

For the task of inferring cosmological parameter constraints, the observational data is fed to the input layer, then the information of observational data passes through each hidden layer, and finally the cosmological parameters are outputted from the output layer. Specifically, each layer accepts a vector, the elements of which are called neurons, from the former layer as input, then applies a linear transformation and a nonlinear activation on the input, and finally propagates the current result to the next layer.

\begin{figure}
    \centering
    \includegraphics[width=0.45\linewidth]{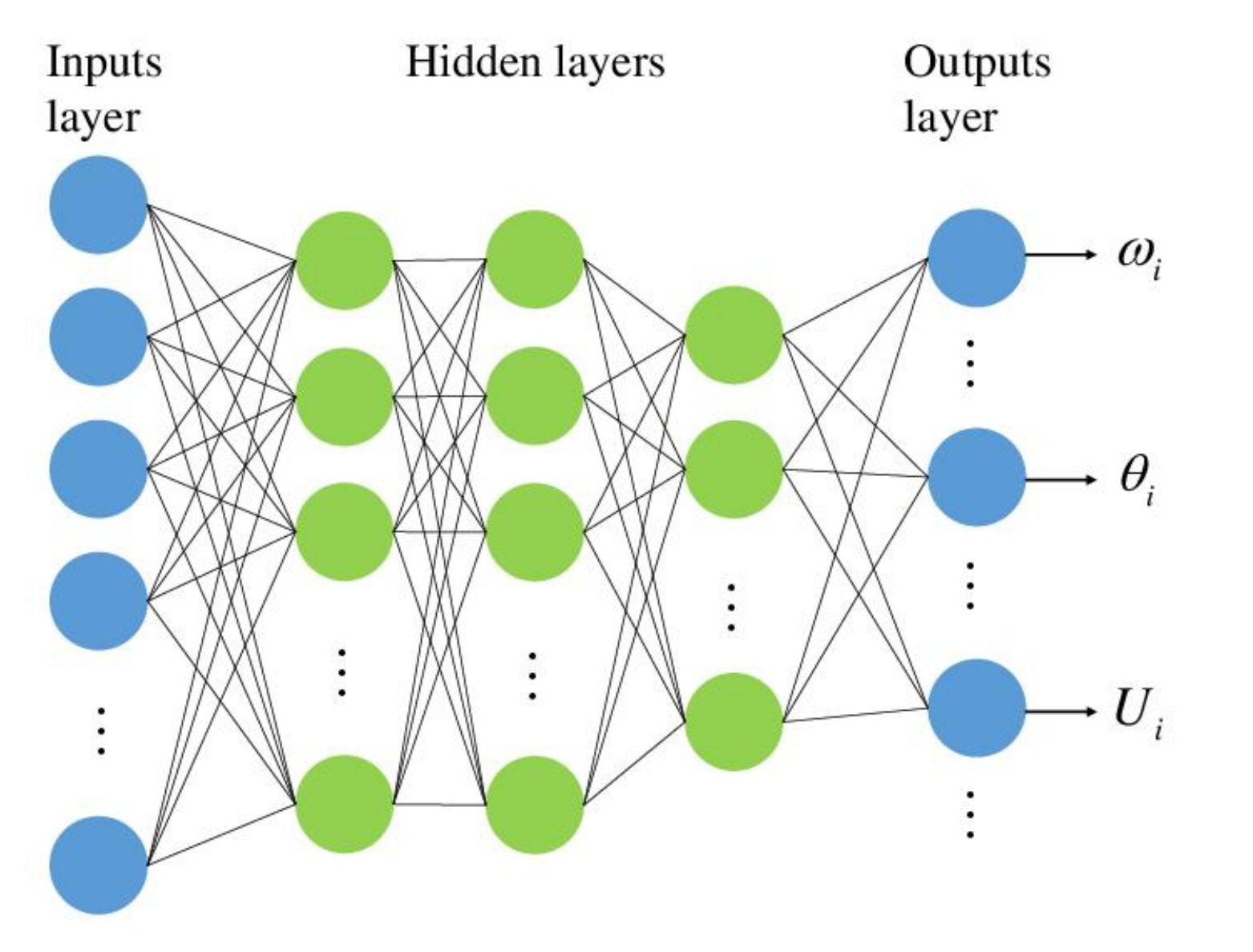}
    \includegraphics[width=0.45\linewidth]{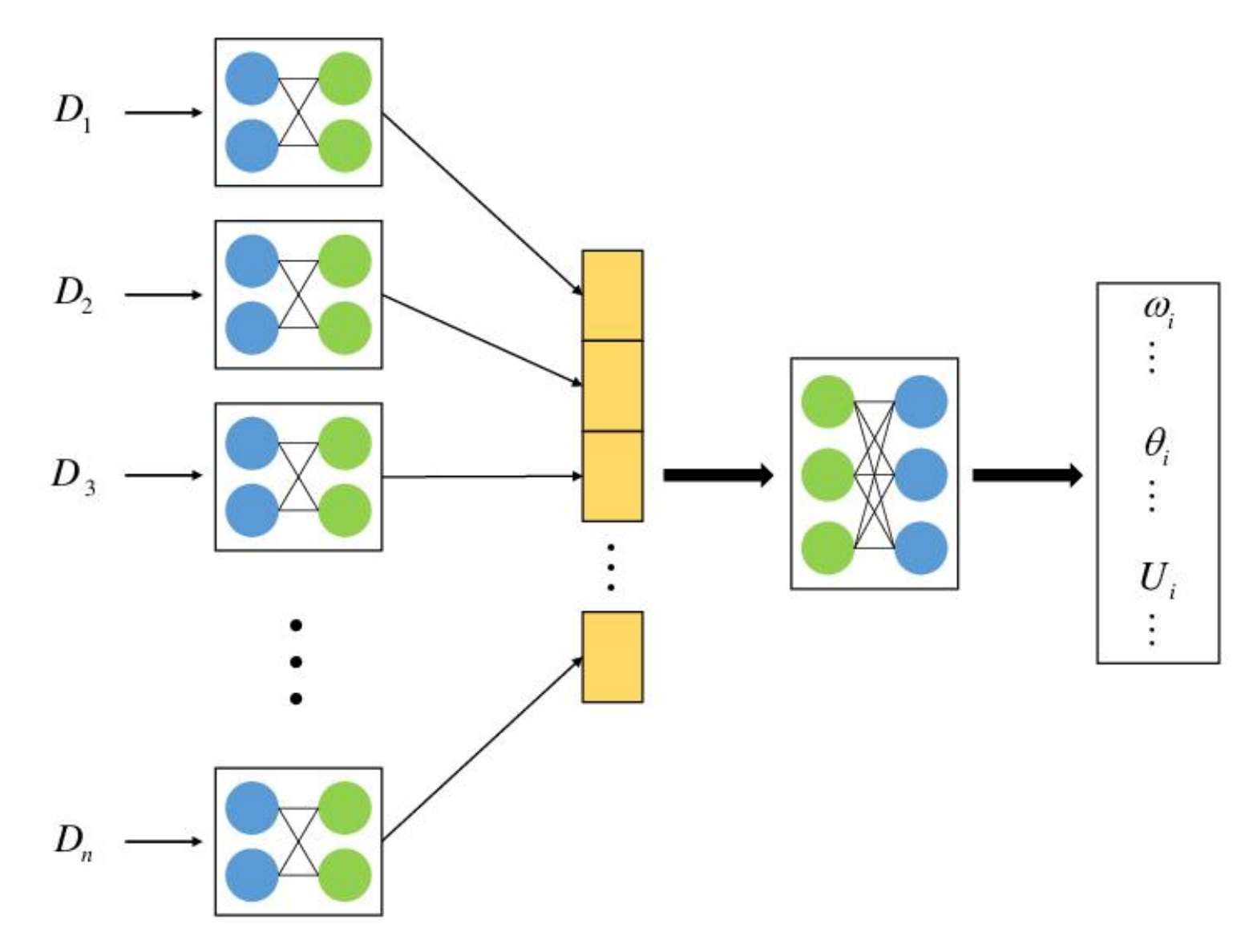}
    \caption{The network structure used for parameter estimation in Ref.~\cite{Wang:2023vej}. The structure of the left panel is for one dataset, while the multi-branch network of the right panel is for multiple datasets \{$D_1,D_2,D_3,...,D_n$\} that are from different experiments.}
    \label{fig:ML_Inf_colfi-arch}
\end{figure}

\ac{ann}s have been widely implemented as emulators of cosmological observables, such as the lensing power spectrum \citep{Manrique-Yus:2019hqc, SpurioMancini:2021ppk}, \ac{cmb} source functions \cite{Albers:2019rzt, Gunther:2022pto}, the \ac{cmb} temperature anisotropies power spectrum and the matter power spectrum \citep{Auld:2006pm, SpurioMancini:2021ppk}. The main purpose of these applications is to accelerate the statistical analysis, when expensive Boltzmann codes and big volumes of data prevent us from efficiently sampling the posterior distributions. In particular, CosmicNet I aims to remove bottlenecks from Einstein-Boltzmann solvers (such as CLASS or CAMB) by training neural network emulators that learn the mapping from four \lcdm\ parameters to source functions of \ac{cmb} anisotropies. The resulting trained \ac{ann} model is then injected into the CLASS code for public use. This work was later extended to include extensions of a flat \lcdm, such as \ac{dde}, spatial curvature. Emulators are crucial for summary statistics of cosmological probes that lack an analytical prescription that models their dependence on cosmology and other parameters. This is the case of higher-order statistics of cosmic fields, such as the peak counts (a distribution of local maxima in a field) and the one-point \ac{pdf}. As emulators are only approximations of the target observables, recent works have explored how to propagate the uncertainty derived from such approximation, in order to safeguard parameter inference against biases \cite{Grandon:2022gdr}.

In recent years, likelihood-free analyses relied on ensembles of neural networks as density estimators of the sampling distribution of the data. In Refs.~\cite{Wang:2020hmn, Wang:2023vej}, a cosmological likelihood-free inference technique has been developed. The left panel of Fig.~\ref{fig:ML_Inf_colfi_schematic} depicts the schematic diagram of the likelihood-free architecture adopted in Ref.~\cite{Wang:2023vej} containing the main processes of training and parameter estimation. First, a class object for the cosmological model that contains the simulation method of the measurements is developed, where the simulation method here is used to generate the training set. Then, the initial parameters are set without any constraints in their selection. Subsequently, the training set will be simulated automatically, where two sampling methods can be considered to generate cosmological parameters: sampling uniformly in a hypercube or a hyper-ellipsoid. It should be noted that the posterior distribution is unknown before the first estimation using the \ac{ann} model. Thus, the cosmological parameters here cannot be generated in a hyper-ellipsoid. Therefore, the cosmological parameters are uniformly generated in a hypercube using the set of initial parameters. Consequently, an \ac{ann} model will be constructed automatically based on the respective training set. At the same time, the training set is then preprocessed before training the network, while noise will be automatically generated based on the observation errors and added to the training set. The training set will then be normalized, and the training set will be fed to the \ac{ann} model, consisting of thousands of epochs.

Cosmological parameters are then estimated using the well-trained \ac{ann} model. In simple terms, the \ac{ann} model actually learns a mapping between the data space of the measurements and the parameter space of cosmological parameters, as depicted in the right panel of Fig.~\ref{fig:ML_Inf_colfi_schematic} via a mapping between the data space of measurements and the cosmological parameter space. Therefore, in order to infer the respective posterior distribution, the corresponding distribution of the measurements is fed to the \ac{ann} model. Indeed, a large number of data-like samples of the measurements using the observational data are generated and fed to the \ac{ann} model to obtain the corresponding \ac{ann} chain, similar to the concept of an \ac{mcmc} chain.

The \ac{ann} method in Refs.~\cite{Wang:2020hmn, Wang:2023vej} was shown to be very successful in estimating cosmological parameters with one data set, which was then extended to multiple data sets by utilizing a multi-branch network
The latter multi-branch network is illustrated in the right panel of Fig.~\ref{fig:ML_Inf_colfi-arch}, in which it was successfully tested with \ac{cmb}, \ac{sn1}, and \ac{bao} datasets in Refs.~\cite{Wang:2020hmn, Wang:2023vej}, as illustrated in the derived results of Fig.~\ref{fig:ML_Inf_ANN-MCMCl} with the Planck 2015 data set. Thus, the \ac{ann} method was shown to be capable of estimating parameters using data sets of multiple experiments and is a direct competitor of the customary \ac{mcmc} technique.

\begin{figure}
    \centering
    \includegraphics[width=0.45\linewidth]{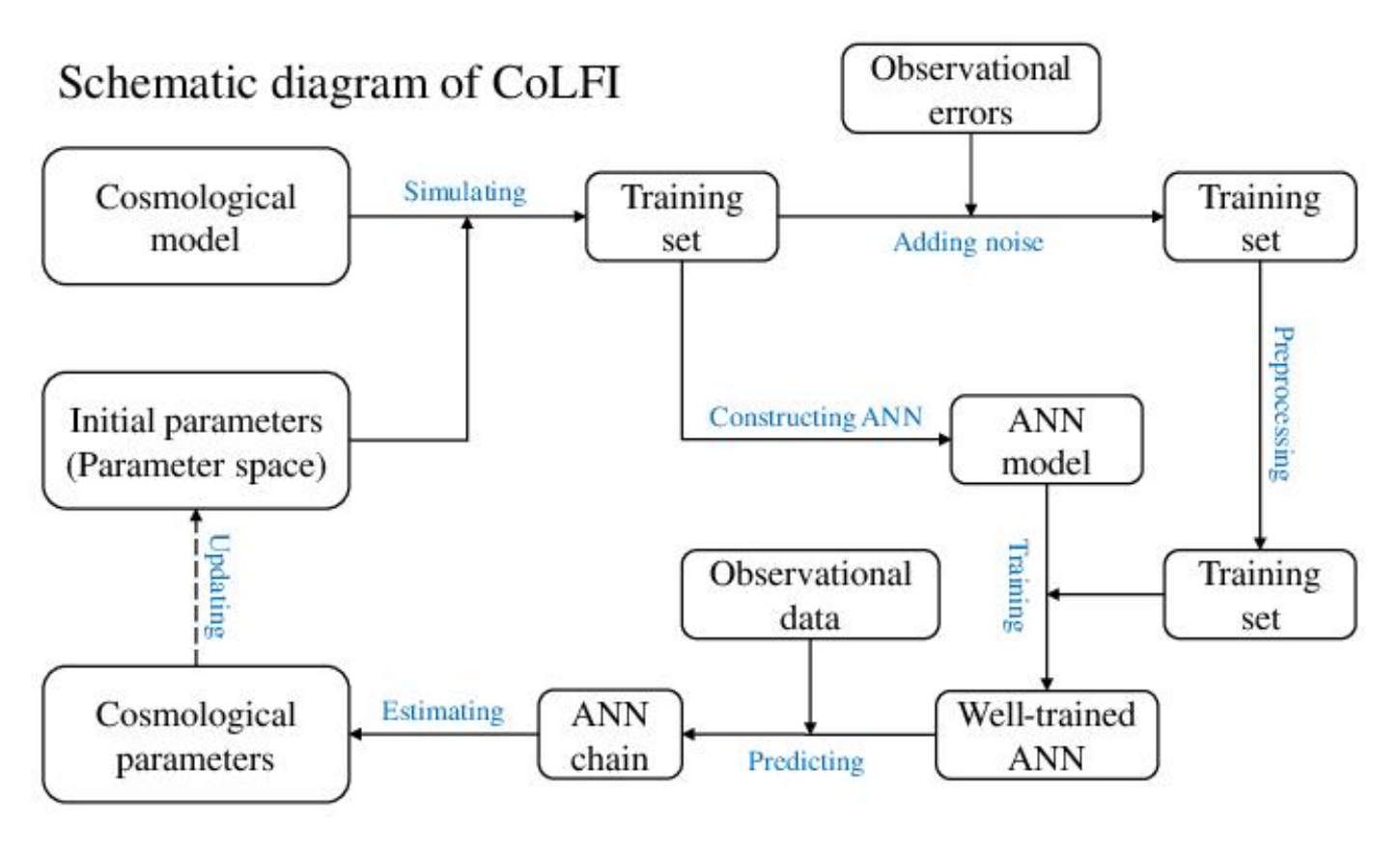}
        \includegraphics[width=0.45\linewidth]{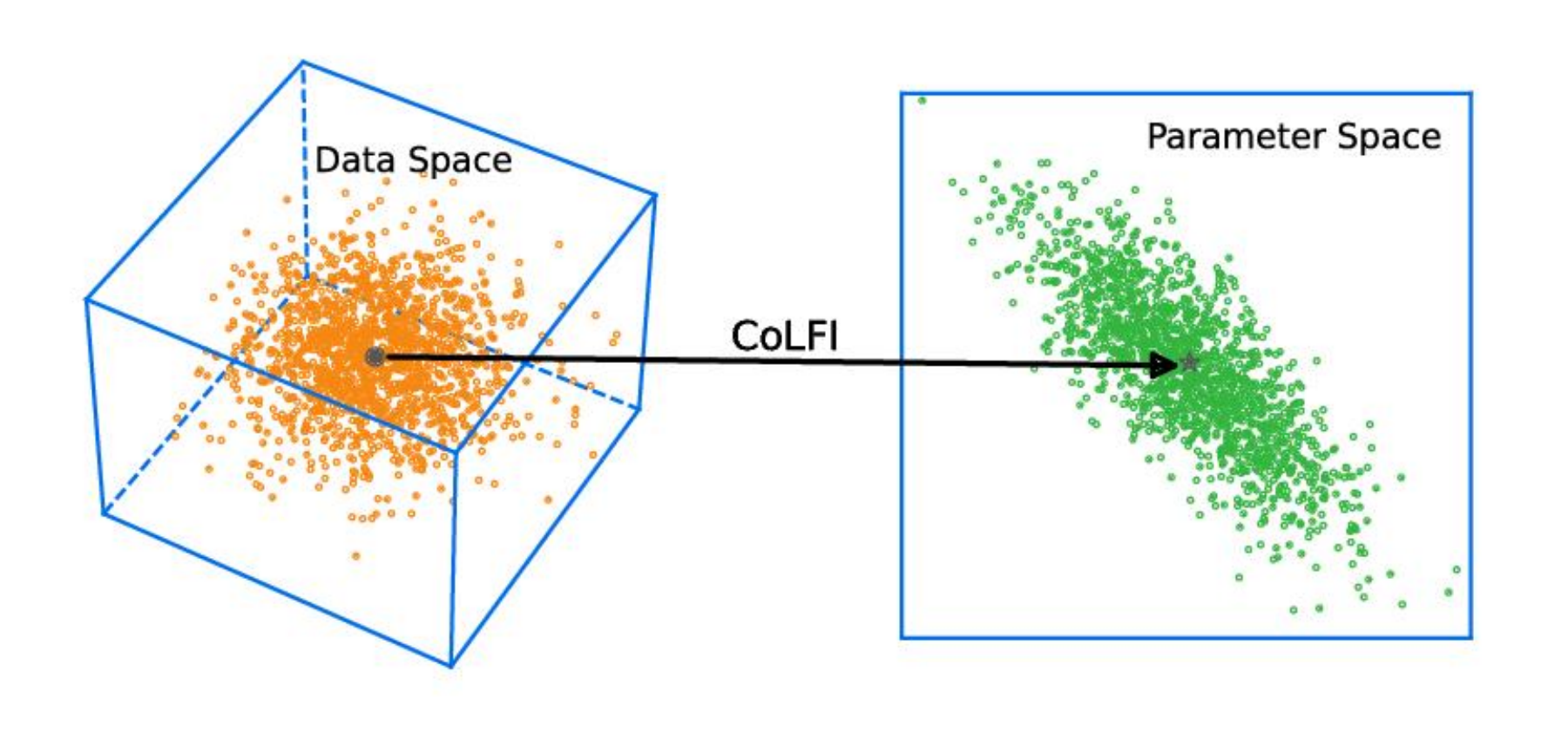}
    \caption{The left panel depicts the schematic diagram of the likelihood-free architecture adopted in Ref.~\cite{Wang:2023vej}. In the right panel, a visualization of a mapping between the data space of measurements and the cosmological parameter space is illustrated.}
    \label{fig:ML_Inf_colfi_schematic}
\end{figure}

\begin{figure}
    \centering
    \includegraphics[width=0.6\linewidth]{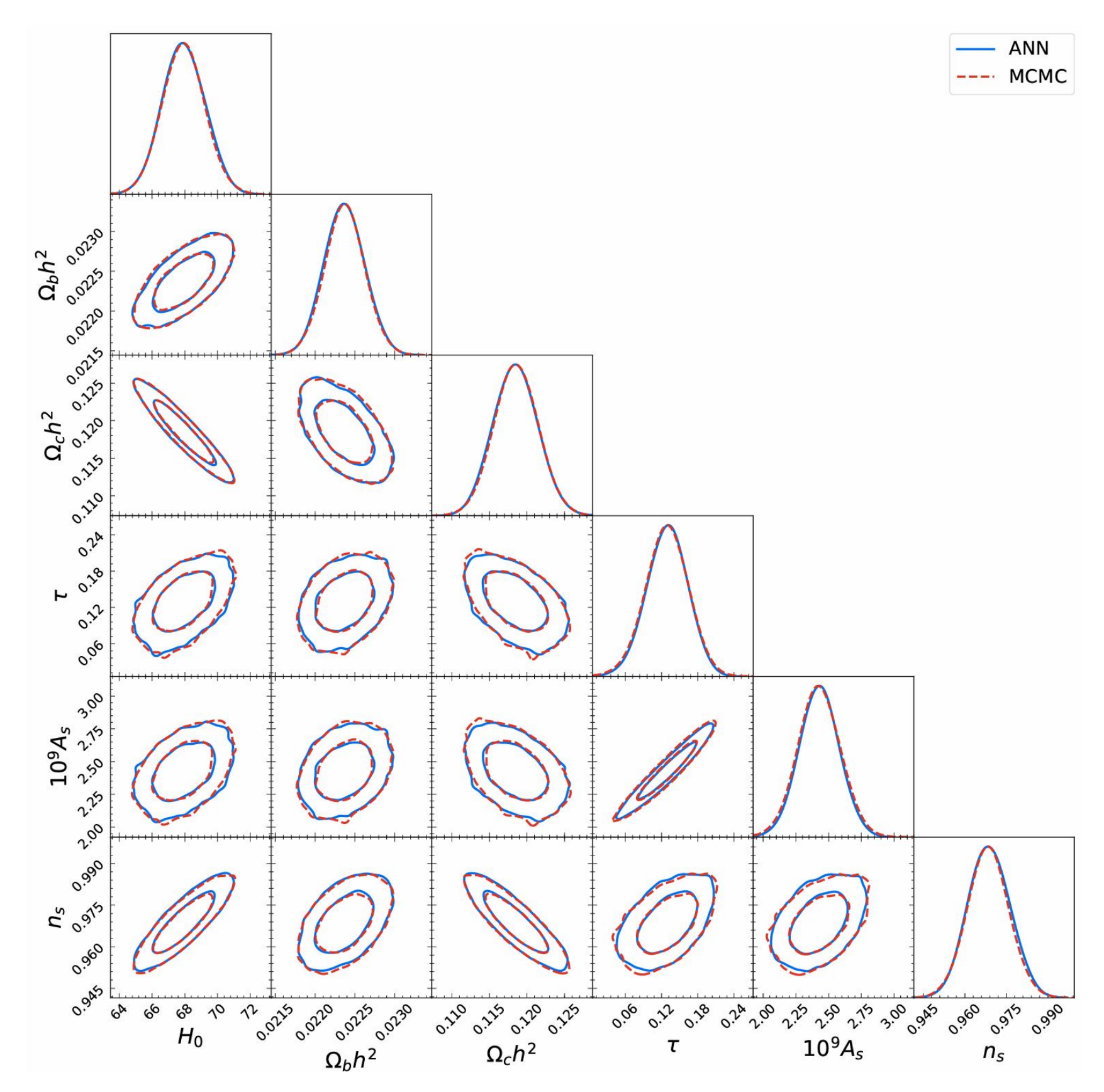}
    \caption{One-dimensional and two-dimensional marginalized distributions \cite{Wang:2023vej} constrained from Planck 2015 power spectrum. The blue solid lines are the results of the \ac{ann} method, the red dashed lines represent those of the \ac{mcmc} method, and the gray circles are the fiducial values of the cosmological parameters.}
    \label{fig:ML_Inf_ANN-MCMCl}
\end{figure}

\subsubsection{Convolutional neural networks}
\label{sec:ML_Inf_CNN}

With the advent of precision cosmology, we are reaching an era in which it is extremely important to avoid any potential biases in the cosmological analysis when extracting cosmological parameters. For example, most of the time large-scale structure analyses are done using summary statistical observables like 2-point correlation functions, after assuming a cosmology to go from angular coordinates to Cartesian ones. This type of analysis can bias the cosmological parameters and a tomographic analysis can avoid this \cite{Asorey:2012rd}.

One alternative to the use of summary statistics is the use of \ac{cnn} architectures. They can be used to make direct inferences on cosmological parameters from astrophysical images. For instance, Ref.~\cite{Sabiu:2021aea} used \ac{cnn}s to constrain axions as \ac{dm} from 21cm line intensity mapping probes.  In Fig.~\ref{fig:ML_Inf_CNN-schematic} we show a schematic of the adopted \ac{cnn} model architecture. It took as input a
 $64\times64\times128$ lightcone data cube where each voxel has dimension $4.68\mathrm{cMpc}\times4.68\mathrm{cMpc} \times1.48\mathrm{Mhz}$ (cMpc means co-moving Mpc).

\begin{figure}
    \centering
    \includegraphics[width=0.95\linewidth]{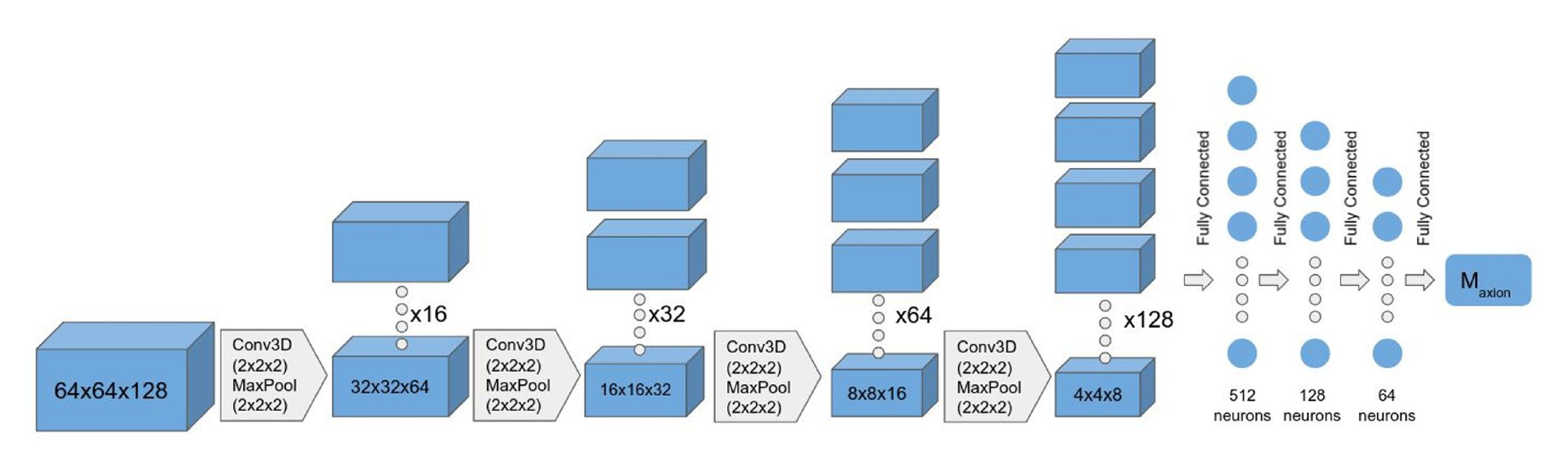}
    \caption{The architecture of the \ac{cnn} adopted in Ref.~\cite{Sabiu:2021aea}, with input data cube of $64\times64\times128$ voxels where each voxel has dimension $4.68\mathrm{cMpc}\times4.68\mathrm{cMpc} \times1.48\mathrm{Mhz}$ (cMpc means co-moving Mpc).}
    \label{fig:ML_Inf_CNN-schematic}
\end{figure}

\begin{figure}
    \centering
    \includegraphics[width=0.5\linewidth]{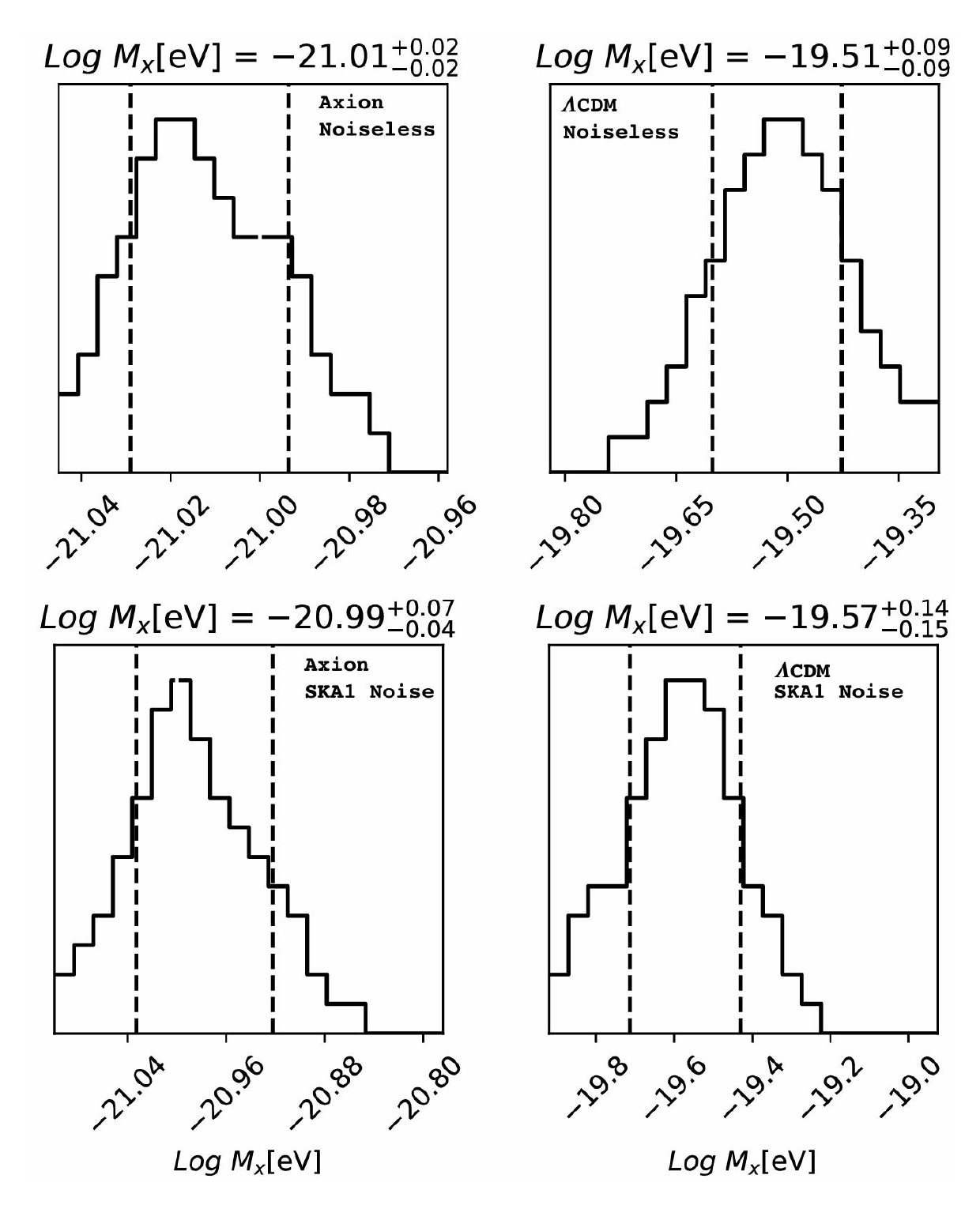}
    \caption{The distribution of the \ac{cnn} inferred predictions \cite{Sabiu:2021aea} on 100 noiseless simulations fixed axion mass of $10^{-21}$eV, along with the vertical dashed lines showing the 1-$\sigma$ (68\%) credible interval. We should remark that the \ac{cnn} could predict the axion mass because the \ac{cnn} is trained only with the data including axions.}
    \label{fig:ML_Inf_Sabiu-mass}
\end{figure}

If \ac{dm} were described by axion-like particles (ALPs), the small scale structure would change the value of $\Omega_{m, 0}$ leading to a change in $H_0$ and $\sigma_8$, potentially solving the potential tensions. To make a direct inference from \ac{cnn}s, we need a training set of simulations of the given astrophysical image for different sets of parameters (e.g., axion mass or $H_0$).  The input layer is basically the simulation with the angular maps and the frequency maps while the final output layer is the vector with the cosmological parameters inferred (in this case, the axion mass). The input layer is filtered and pooled down before being inputted as a vector to the connected neural network layers. The data is passed through the network a number of epochs updating the corresponding weights each time.

The inferred results, as depicted in Fig.~\ref{fig:ML_Inf_Sabiu-mass}, show that the \ac{cnn} successfully recovered the true values of the axion mass in the testing data with a precision of $\sim20\%$ across a broad range of masses. To evaluate the uncertainty in the recovered values, 100 additional independent simulations were generated for both an axion model and a standard \ac{cdm} model. For an axion mass of $M_X=10^{-21}$eV, a mass uncertainty of ${}^{+5.94}_{-4.80}\times10^{-21}$eV was inferred. It was further concluded that this method could potentially enable the detection of axion \ac{dm} using \ac{ska}1-Low with $68\%$ confidence if the axion mass is $M_X<1.86\times10^{-20}$eV. However, these findings depend on the Planck 2015 cosmological parameters and the specific design parameters of future \ac{ska}1-Low configurations.

A \ac{cnn} was also adopted to perform the typically computationally expensive task of estimating the parameters of \ac{gw} events. The considered \ac{cnn} in Ref.~\cite{Andres-Carcasona:2023rnk} is able to produce posterior distributions that in all cases are compatible with the already published results. The schematic of the respective \ac{cnn} is depicted in Fig.~\ref{fig:ML_Inf_CNN-GW}, whereas the inferred results for the event GW200224\_222234 are illustrated in Fig.~\ref{fig:ML_Inf_GW-event}. In the case of event GW200224\_222234, all the estimated parameters were found to be in accordance with those published in the GWTC-3 catalog. Having said that, the effective spin was the worst performing parameter. Furthermore, the uncertainty yielded by the \ac{cnn} on the sky position was found to be too large to accurately pinpoint the event, but it still is unbiased and is expected to produce a first alert for the instruments that look for electromagnetic counterparts to start pointing their instruments to a given patch in the sky. The latter would be more feasible with the upcoming more accurate pipelines which are expected to improve the precision of the sky localization. 
Overall, the \ac{gw} event results indicate a good response from \ac{cnn}s. One of the most important advantages is its computational efficiency, in which the \ac{cnn} model could provide several posterior samples in a fraction of the time required by Bayesian inference methods, thus facilitating multi-messenger astronomy. 

\begin{figure}
    \centering
    \includegraphics[width=0.85\linewidth]{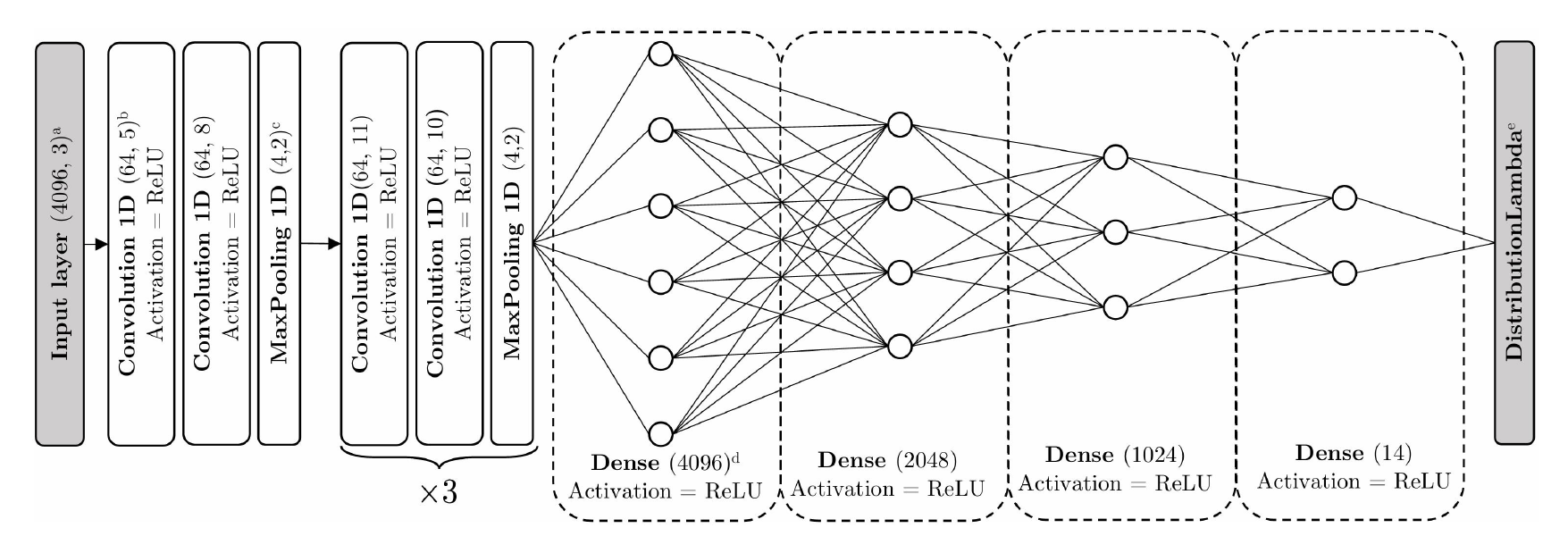}
    \caption{Architecture of the \ac{cnn} used in Ref. \cite{Andres-Carcasona:2023rnk}. The first set of layers are 1-dimensional convolutions and 1-dimensional maxpooling ones, followed by a set of fully connected dense layers. Between each of the dense layers, a dropout of 20\% is applied to prevent overfitting during the training stage.}
    \label{fig:ML_Inf_CNN-GW}
\end{figure}

\begin{figure}
    \centering
    \includegraphics[width=0.65\linewidth]{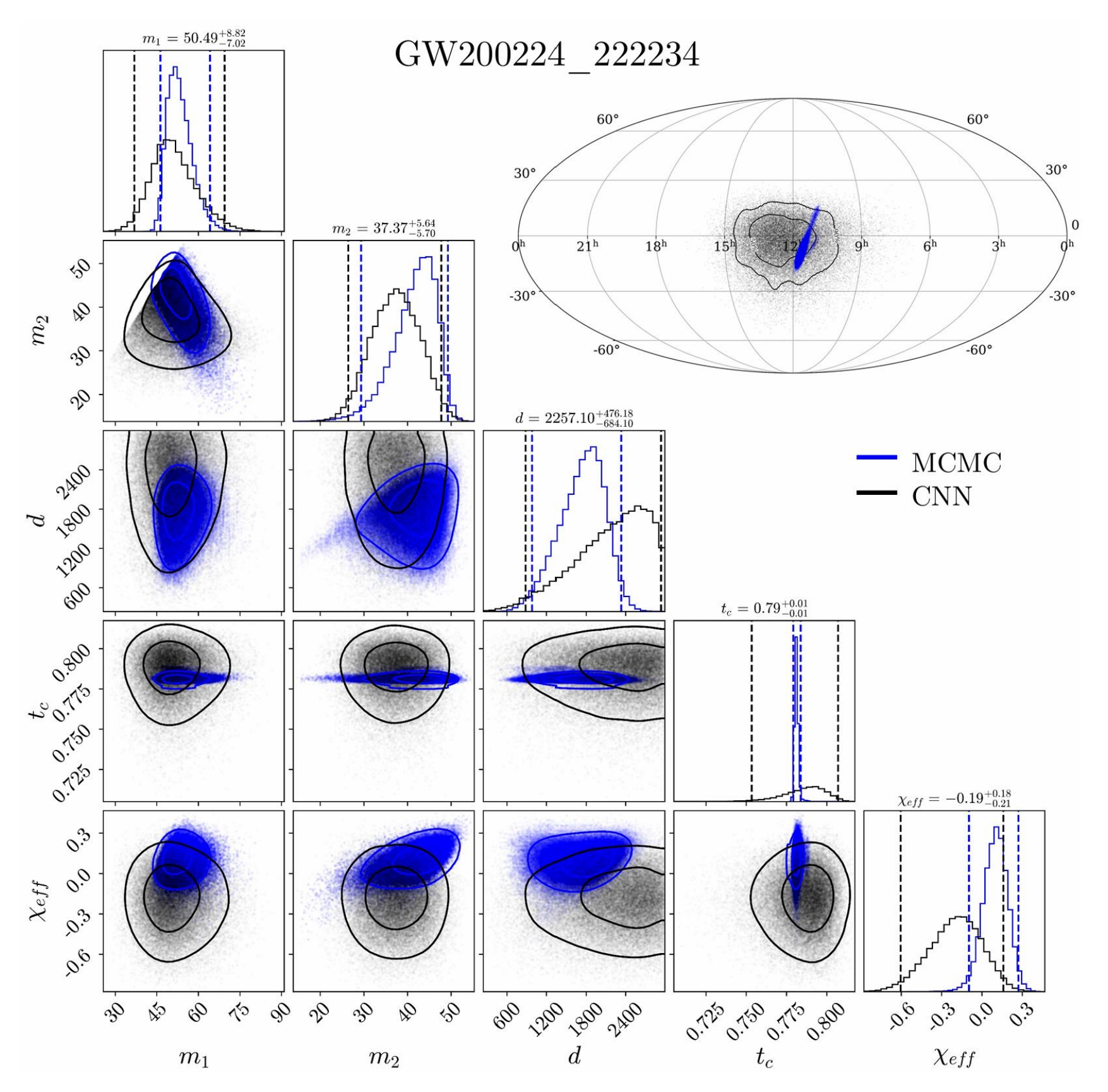}
    \caption{Full posterior distribution inferred \cite{Andres-Carcasona:2023rnk} from the \ac{cnn} and \ac{mcmc} approaches for the GW200224\_222234 event.}
    \label{fig:ML_Inf_GW-event}
\end{figure}

\subsubsection{Bayesian neural networks}
\label{sec:ML_Inf_BNN}

A \ac{bnn} that outputs Gaussian \ac{pdf}s was adopted in Ref.~\cite{Jones:2023xix}. The \ac{bnn} has five input nodes for the five-band \textit{grizy} photometry. A parameter grid search was performed in order to optimize for free parameters, including the number of epochs, number of layers, number of nodes per layer, learning rate, loss function, activation function, and optimizer. The \ac{bnn} has a final output node that produces a mean and standard deviation assuming a Gaussian distribution for each photo-$z$ prediction. 
 
A key attribute of the \ac{bnn} model is the production of photo-$z$ uncertainties, which are needed for using photo-$z$ results in cosmological analyses. It was also found that the \ac{bnn} produces accurate uncertainties. In Ref.~\cite{Jones:2023xix}, the \ac{hsc} Public Data Release 2~\cite{Aihara:2019xyr}, which is designed to reach similar depths as \ac{lsst} but over a smaller portion of the sky was adopted for the \ac{bnn}. In total, the data consists of 286,401 galaxies with broadband grizy photometry and known spectroscopic redshifts. The considered galaxy sample covers a redshift of $0.01 < z < 4$, however the majority of the sample lies between $0.01<z<2.5$. In the \ac{bnn} analysis of Ref.~\cite{Jones:2023xix}, 80\% of the galaxies were used for training, 10\% for validation, and 10\% for testing. The performance of the \ac{bnn} technique is clearly illustrated in Fig.~\ref{fig:ML_Inf_BNN-photoz}, in which the \ac{bnn} was reported to have superior
 photo-$z$ estimations with respect to other competing models.
 
\begin{figure}
    \centering
    \includegraphics[width=0.8\linewidth]{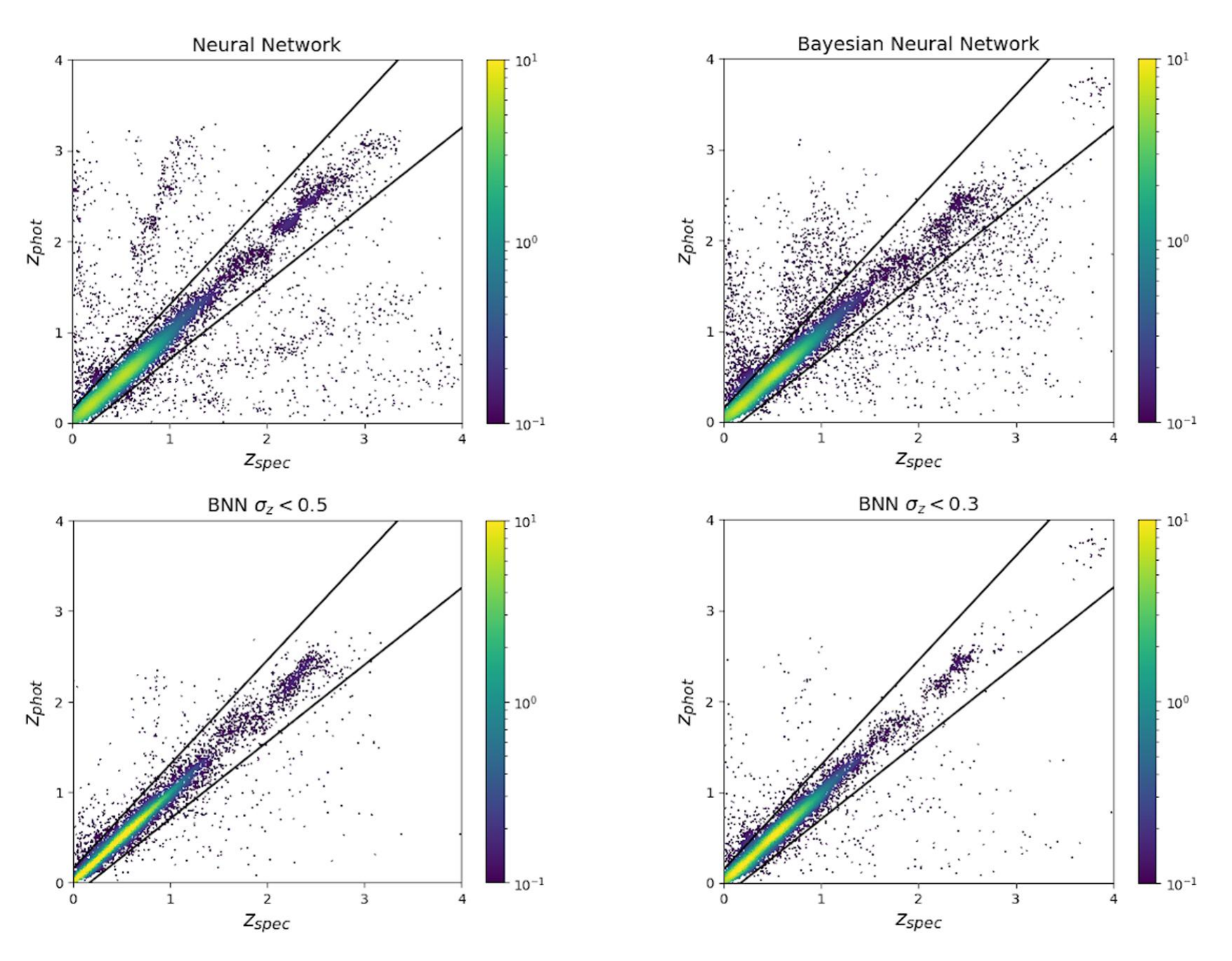}
    \caption{Visualization of the neural network (top left) and \ac{bnn} (top right) performance compared to the \ac{bnn}N with outlier removal criteria $\sigma_z<0.5$ (bottom left) and $\sigma_z<0.3$ (bottom right) \cite{Jones:2023xix}.}
    \label{fig:ML_Inf_BNN-photoz}
\end{figure}

The capability of \ac{bnn}s was confronted in Refs.~\cite{Mancarella:2020jyu,Thummel:2024nhv} on different avenues. The first is how effective
\ac{bnn}s can be in recognizing the distinct features in the power spectrum for a particular modification to the concordance model of cosmology, such as $f(R)$ or Dvali–Gabadadze–Porrati (DGP) \cite{Dvali:2000hr}. The second avenue that was exploited in Refs.~\cite{Mancarella:2020jyu,Thummel:2024nhv} is their ability to detect a deviation from \lcdm\ in the power spectrum
irrespective of the particular modification. Two \ac{bnn}s with the same architecture were trained in Refs.~\cite{Mancarella:2020jyu,Thummel:2024nhv}, the first for five labels divided between \lcdm\ and the considered four extensions, whilst the second trained to distinguish between the two labels of \lcdm\ and non-\lcdm. Due to the fact that only four redshift bins were considered in these works, it was beneficial to treat the data as four separate time series and use one-dimensional convolutional
layers.

The architecture of the network used to train both the five-label and two-label networks is depicted in the left panel of Fig.~\ref{fig:ML_Inf_BBN-arch-wf}. Initially, the adopted structure consisted of three 1D convolutional flip out layers with 8, 16 and 32 filters, kernel sizes of 10, 5, and 2 with strides of 2, 2, and 1, respectively. Each of the first two 1D convolutional layers was followed by a max pooling layer with a pool size of 2 and a pooling stride of 2 for the first max pooling layer and a pooling stride of 1 for
the second max pooling layer. After both of these max pooling layers there is a batch normalization layer. Following the final convolutional layer there is a global average pooling layer to reduce the filter size to one in order to pass it to a dense layer with 32 nodes. Finally, after a
further batch normalization there is a softmax layer consisting of five or two neurons for either the five- or two-label networks, respectively.

It was found that a five-label network trained to classify between \lcdm, $f(R)$ gravity, DGP gravity, $w$CDM and a ``random'' class provided more reliable predictions than a two-label network trained to distinguish simply between \lcdm\ and non-\lcdm. While generally being less sensitive to variations in the noise distribution, it can also determine whether a power spectrum does not belong to any class included in the training set. Since the selection of the correct model is crucial when performing conventional statistical analyses such as with \ac{mcmc}s, this ability could prove beneficial in indicating prospective models to consider. However, the network used in this work is currently limited to classification tasks while the notion of model selection on firm statistical grounds in the context of \ac{bnn}s remains an open problem.

It could be concluded that \ac{bnn}s may provide a powerful new means to search for hints of new physics in cosmological datasets. In particular, it is anticipated that they will serve as a powerful ``filter'', allowing us to narrow down the theory space before moving on to constrain model parameters with \ac{mcmc}s while perhaps even signaling the presence of new physics that does not belong to any known model.

\begin{figure}
    \centering
    \includegraphics[width=0.52\linewidth]{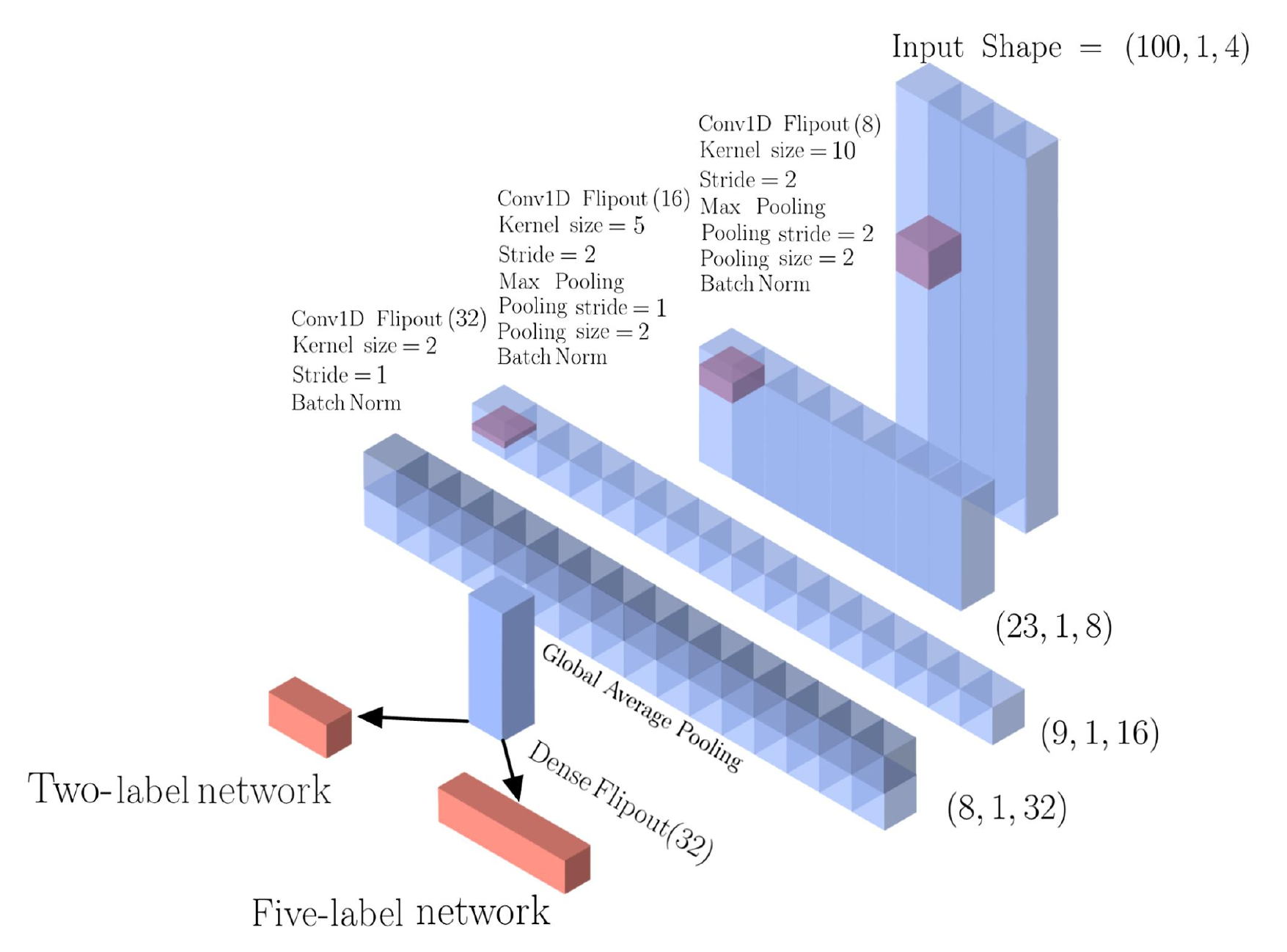}
    \includegraphics[width=0.37\linewidth]{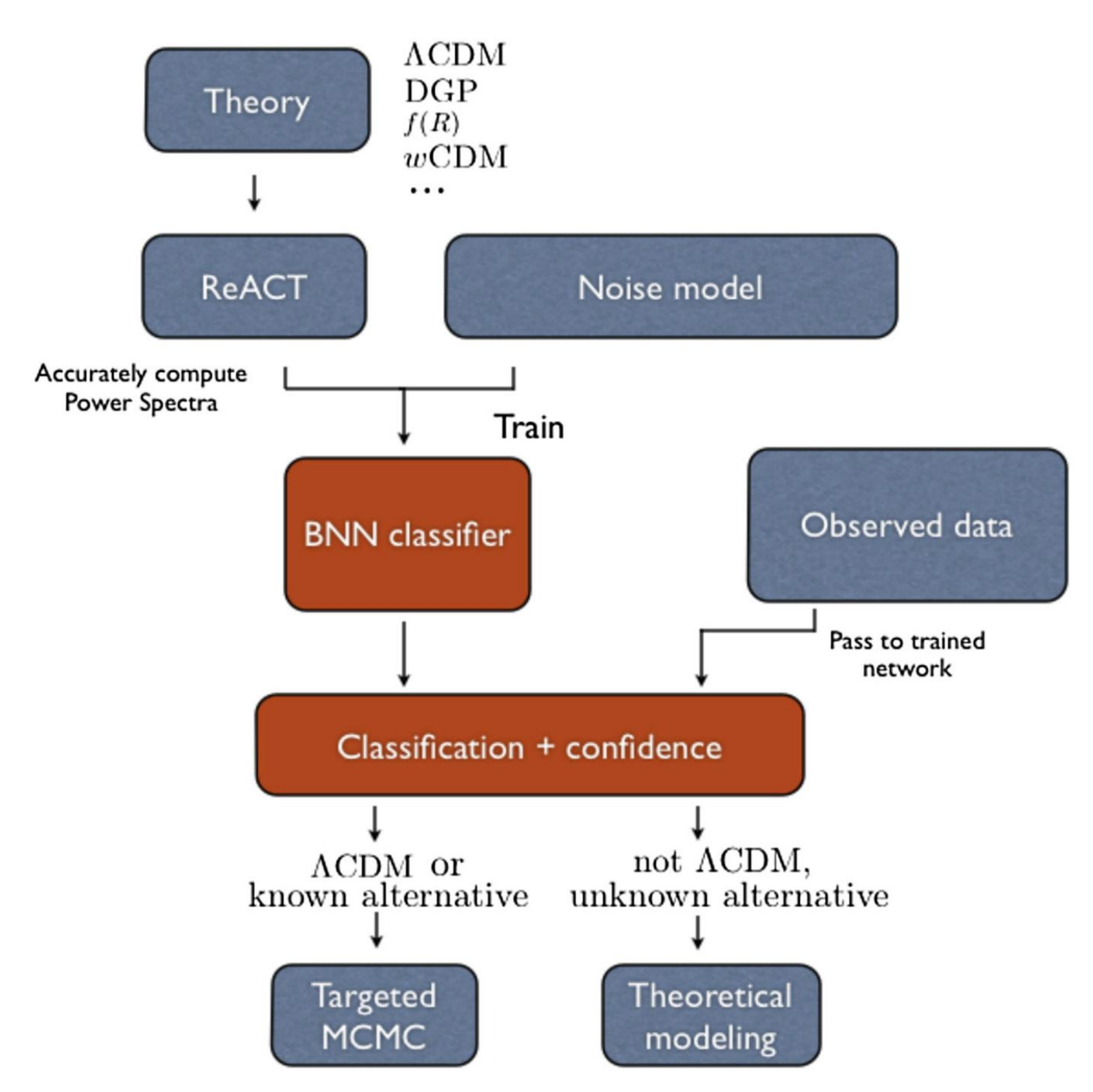}
    \caption{The right panel illustrates the \ac{bnn} workflow as adopted in Refs.~\cite{Mancarella:2020jyu,Thummel:2024nhv}, whilst the left panel depicts the corresponding \ac{bnn} architecture employed for both the five-label and two-label classification tasks. The height of each block illustrates the dimension size for each layer, while the number of blocks per layer corresponds to the number of filters. The dense blocks embedded in the first three transparent layers indicate the kernels for the first three one-dimensional convolutional layers scaled by their respective size. }
    \label{fig:ML_Inf_BBN-arch-wf}
\end{figure}

\subsubsection{Deep learning}
\label{sec:ML_Inf_ladder}

Neural networks have lately become ubiquitous in cosmology \cite{Mukherjee:2022yyq,Wang:2019vxv,Escamilla-Rivera:2021vyw,Olvera:2021jlq,Gomez-Vargas:2022bsm,Gomez-Vargas:2021zyl,Giambagli:2023ngt,Dialektopoulos:2023dhb,Dialektopoulos:2023jam,Mukherjee:2024akt,Zhang:2023xgr,Zhang:2023ucf,Xie:2023ydk,Tang:2020nmw,Wang:2020hmn,Liu:2023rrr}. In this regard, the \texttt{LADDER} - Learning Algorithm for Deep Distance Estimation and Reconstruction - suite is a novel deep learning algorithm designed to learn the cosmic distance ladder in a model-independent, non-parametric manner \cite{Shah:2024slr,Shah:2024slr}. The schematic overview of the training algorithm is depicted in Fig.~\ref{fig:ML_Inf_ladder}, whilst the corresponding algorithm is illustrated in Fig.~\ref{fig:ML_Inf_ladder-algorithm}. Trained on late-time datasets like the Pantheon \ac{sn1} dataset \cite{Pan-STARRS1:2017jku}, \texttt{LADDER} incorporates associated errors and complete covariance information. It interpolates from the joint distribution of a randomly chosen subset of the dataset to estimate target variables and errors simultaneously, effectively handling correlations and the sequential nature of the data. This leads to robust predictions resilient to input noise and outliers, even in data-sparse regions. Optimized by physically motivated metrics such as monotonicity and smoothness, \texttt{LADDER} leverages the Gaussian nature of the observations by employing the Kullback–Leibler (KL) divergence as the loss function during training. The LSTM network, found to perform best with the Pantheon dataset, is included in the \texttt{LADDER} suite, available publicly on \href{https://github.com/rahulshah1397/LADDER}{GitHub (https://github.com/rahulshah1397/LADDER)} under a MIT License and version 1.0 is archived in Zenodo \cite{Shah:2024slr}.

Having learned the cosmic distance ladder, \texttt{LADDER} is a powerful tool for cosmological applications. It can independently verify the consistency of \ac{sn1} datasets like Pantheon+ \cite{Scolnic:2021amr} and \ac{des} \cite{DES:2024jxu}, and serve as a pathology test for qualitatively different datasets, such as 2D vs. 3D \ac{bao}. Additionally, it acts as a model-independent calibrator for high-redshift datasets, including \ac{grb}s and \ac{qso}s. Moreover, previous \ac{ml} approaches in cosmology struggled with stable error prediction at higher redshifts, rendering them unsuitable for precision cosmological tests. \texttt{LADDER} overcomes these challenges and shows promise in extrapolating beyond available data, useful for simulating intermediate-redshift data or augmenting current data to higher redshifts.

In Ref.~\cite{Shah:2024gfu} \texttt{LADDER} is employed to recalibrate SDSS \ac{bao} and \ac{desi} \ac{bao} in a model-independent manner, which helps address the longstanding $H_0$ and $S_8$ tensions simultaneously. Traditionally, \ac{bao} distances are inferred using the sound horizon at the drag epoch ($r_d$), which is calibrated based on \ac{cmb} data under the assumption of the \lcdm\ model. In contrast, \texttt{LADDER}, trained on the Pantheon dataset, derives $r_d$ purely from observational data without relying on cosmological model assumptions, instead incorporating merely an astrophysical prior on the absolute magnitude of \ac{sn1} ($M_B$). When combined with Planck 2018 data, this recalibrated \ac{bao} framework, along with Pantheon, yields joint constraints on \lcdm\ that significantly alleviate the $H_0$ and $S_8$ tensions, due to an ``in-plane shift", despite their negative correlation. These results highlight the potential biases introduced by \ac{cmb}-based \ac{bao} calibration and demonstrate the robustness of data-driven approaches, offering a promising new avenue for precision cosmology.

\begin{figure}
    \centering
    \includegraphics[width=0.8\linewidth]{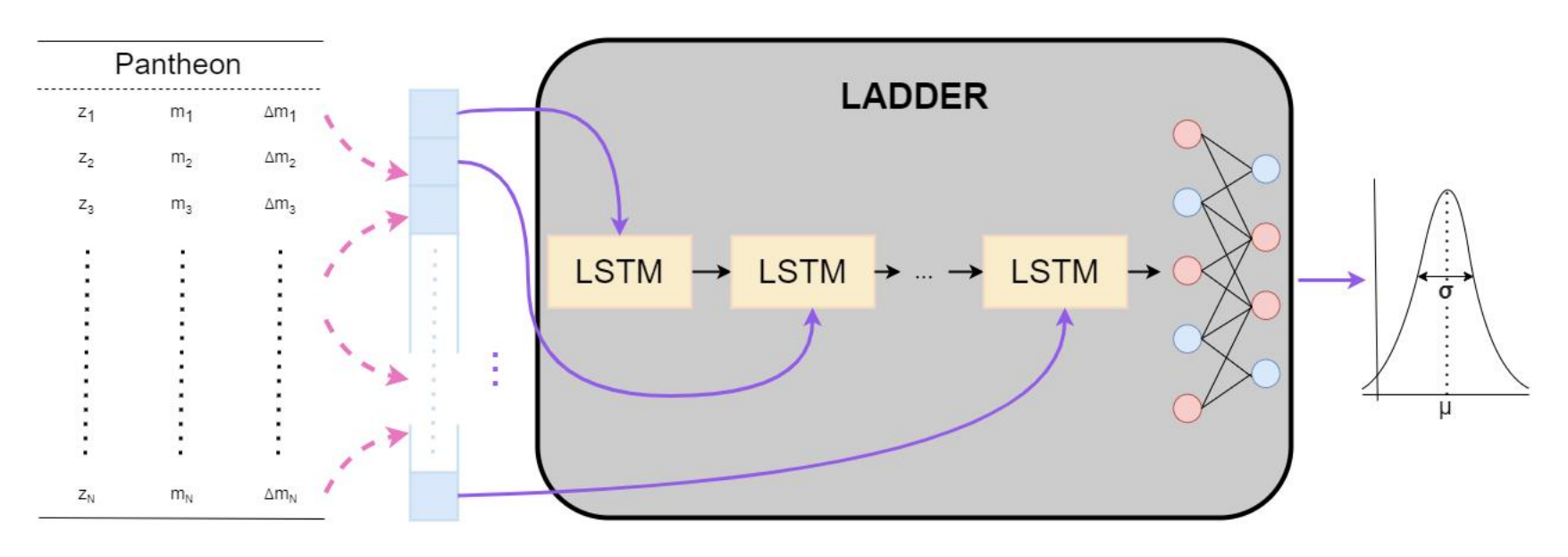}
    \caption{Schematic overview of the deep learning algorithm adopted in Ref.~\cite{Shah:2024slr}.}
    \label{fig:ML_Inf_ladder}
\end{figure}

\begin{figure}
    \centering
    \includegraphics[width=0.8\linewidth]{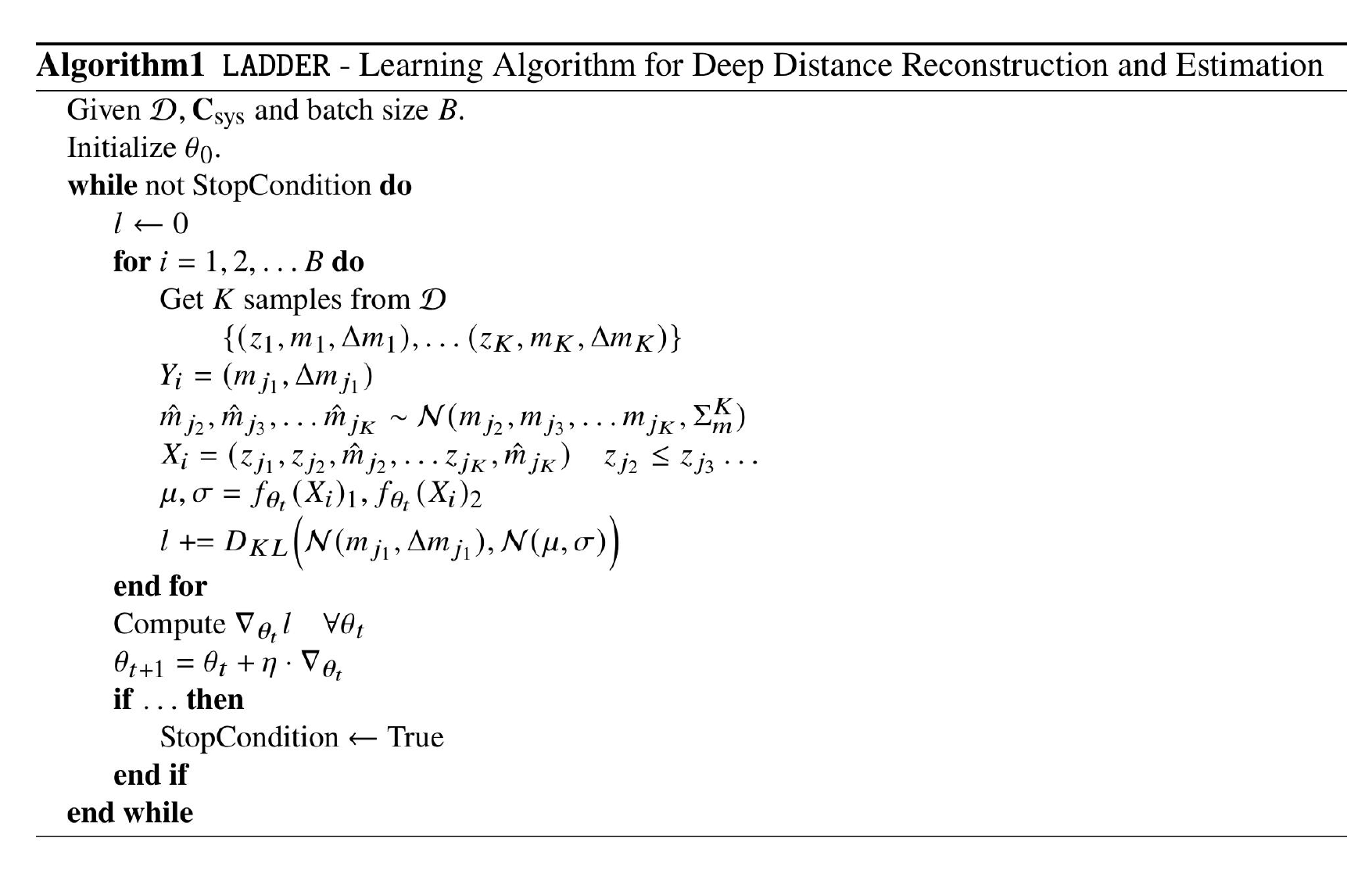}
    \caption{The deep learning algorithm adopted in Ref.~\cite{Shah:2024slr}.}
    \label{fig:ML_Inf_ladder-algorithm}
\end{figure}

Beyond conventional deep learning architectures, more advanced generative models have been developed to tackle complex challenges in cosmological inference. These methods, such as normalizing flows~\cite{2019arXiv191202762P} and diffusion models~\cite{2015arXiv150303585S,2020arXiv201002502S}, extend deep learning capabilities by efficiently learning high-dimensional \ac{pdf}s and accelerating likelihood-free inference. Unlike traditional \ac{mcmc} approaches, which often struggle with computational inefficiency, generative models can rapidly approximate complex posteriors while maintaining high accuracy.

Normalizing flows have shown promise in addressing cosmological tensions, such as discrepancies in the Hubble constant and matter fluctuation amplitude ($S_8$). By modeling joint \ac{pdf}s across datasets, they enhance uncertainty quantification and provide a robust statistical framework for analyzing these tensions~\cite{Bevins:2022qsc,Friedman:2022lds}. One example is the \texttt{EMUFLOW} framework~\cite{Mootoovaloo:2024sao}, which utilizes normalizing flows to model joint posterior distributions from multiple datasets, enabling efficient constraint combination without increasing parameter space dimensionality. Additionally, methods employing normalizing flows have been developed for non-Gaussian tension estimation, improving the quantification of agreement levels between different cosmological experiments and offering insights into discrepancies such as the $H_0$ tension ~\cite{Raveri:2021wfz}. Further developments in normalizing flow architectures have introduced approaches that incorporate symmetries of the universe, such as translation and rotation equivariance, optimizing parameter inference for cosmological fields~\cite{Dai:2022dso}. Other extensions~\cite{Dai:2023lcb} leverage hierarchical wavelet-based decompositions to improve robustness across spatial scales, while iterative flow-based techniques facilitate the transformation of complex \ac{pdf}s into tractable forms, improving efficiency in cosmological data analysis.

Diffusion models, leveraging stochastic iterative denoising processes, have also been applied successfully to cosmological parameter inference. These models excel in uncertainty quantification and high-fidelity generative sampling of cosmological fields, particularly in large-scale structure and \ac{wl} studies. A recent application integrates diffusion models with Hamiltonian Monte Carlo methods~\cite{Mudur:2023smm,Mudur:2024fkh}, enabling accurate posterior sampling and accelerated convergence in extracting key cosmological parameters such as $\Omega_{m,0}$ and $S_8$. The ability of diffusion models to reconstruct high-fidelity statistical distributions while remaining robust to observational noise makes them valuable tools for precision cosmology. As \ac{ml} techniques continue to evolve, integrating generative models into cosmological inference is expected to enhance the precision and reliability of parameter estimation, especially for upcoming surveys like \ac{lsst} and Euclid. These advancements underscore the growing role of deep learning-driven methodologies in shaping the future of cosmological analysis.

\subsubsection{Conclusion}

We reviewed the application of \ac{ml} methods for the inference of cosmological parameters, emphasizing their versatility and capacity to address the inherent complexity of astronomical datasets. By integrating \ac{ml} techniques into cosmological inference, we enable efficient and accurate predictions, facilitating faster convergence of traditional algorithms such as \ac{mcmc} and advancing implicit likelihood or likelihood-free inference using deep learning. These innovations are particularly valuable in handling the high-dimensional, noisy, and often incomplete nature of cosmological data.

Across the subsections of this review, it is evident that the \ac{ml} techniques considered provide powerful new tools to uncover potential signals of new physics within cosmological datasets. These methods act as a "filter," allowing researchers to effectively narrow down the theory space and refine subsequent analyses with standard techniques like \ac{mcmc}. Furthermore, \ac{ml} based approaches may even reveal evidence of new physics that does not fit within the framework of existing theoretical models, opening new avenues for exploration.

While alternative methodologies exist, we emphasize the transformative potential of advanced learning algorithms in optimizing the extraction of information from cosmological data. By demonstrating their utility, we aim to showcase how these novel approaches can complement or enhance traditional inference techniques. The innovative integration of \ac{ml} into the cosmological framework promises to address current challenges, such as improving parameter constraints and identifying hidden structures or patterns in data, thereby paving the way for more accurate reconstructions of cosmological phenomena.

We also highlight the possibility of yet unexplored \ac{ml} algorithms that may surpass the performance of those outlined in this review. This underscores the importance of continued exploration and innovation within the \ac{ml} and cosmology communities. By encouraging the adoption and further development of \ac{ml} based inference techniques, we aim to inspire new efforts to tackle unresolved questions in cosmology from fresh perspectives. Such endeavors would significantly improve the precision and reliability of cosmological analyses, contributing to a deeper understanding of the Universe's fundamental nature and addressing ongoing tensions in cosmological observations.

In conclusion, the integration of \ac{ml} methods into cosmological research represents a paradigm shift, offering not only computational efficiency but also novel insights into the underlying physics of the Universe. We urge the community to actively engage in the exploration of these advanced tools, as they hold the potential to revolutionize the study of cosmology in the years to come.

\bigskip
\subsection{Reconstruction techniques \label{sec:Recon_tech}}

\noindent \textbf{Coordinator:} Luis Escamilla, Daniela Grandón\\
\noindent \textbf{Contributors:} Adrià Gómez-Valent, Anil Kumar Yadav, Anjan Ananda Sen, Anto Idicherian Lonappan, Ariadna Montiel, Arianna Favale, Benjamin L'Huillier, Biesiada Marek, Celia Escamilla-Rivera, David Benisty, David Valls-Gabaud, Eoin \'O Colg\'ain, Filippo Bouch\`{e}, Iryna Vavilova, Isidro G\'{o}mez-Vargas, J. Alberto V\'{a}zquez, Jenny G. Sorce, Jenny Wagner, Jurgen Mifsud, Konstantinos Dialektopoulos, Luis E. Padilla, Matteo Martinelli, Miguel A. Sabogal, Purba Mukherjee, Rafael C. Nunes, Rahul Shah, Rocco D'Agostino, Ruth Lazkoz, V\'ictor H. C\'ardenas, and Wojciech Hellwing
\\

Reconstruction techniques have become crucial for understanding the phenomenology of \ac{de}, the expansion history, and the growth of large-scale structures in our Universe. They aim to capture specific features or general trends in the data, test a predefined physical model against observations, or explore a flexible, theory-agnostic framework to elucidate the phenomenology of cosmological functions. In general, reconstructions can be categorized into parametric and non-parametric approaches. They mainly differ in their underlying assumptions about a cosmological model and the functional form of the reconstructed quantity. The former imposes an explicit form on the function being reconstructed, thereby reducing uncertainties in the result. However, the a priori assumption of a specific model may limit the flexibility of the reconstruction and introduce biases regarding the properties of the cosmological function of interest. To mitigate this, some studies instead perform a non-parametric (``free-form'') reconstruction within a theory-agnostic framework. Non-parametric reconstructions are particularly useful for gaining insights into the dynamics of cosmological functions when the true underlying function is not directly observable but can be inferred from observed data.

Many works on the parametric and non-parametric approaches investigate the impact of $H_0$ priors, to see whether the $H_0$ tension is also imprinted in the derived cosmological reconstructions. Hence, these techniques can shed light on the cosmological tensions and advance our understanding of their effects on a broader context. Moreover, a joint analysis of parametric and non-parametric reconstructions using cosmological observations is valuable to robustly identify departures from the \lcdm\ model that emerge despite the differences of the methods used (e.g., see Ref.~\cite{Bernardo:2021cxi,Wang:2023ghk, Rani:2015lia} for examples). In summary, these methods serve as data-driven frameworks to decode the phenomenology of cosmological functions from complex datasets, and offer an alternative approach to distinguish between competing cosmological models.

In what follows we present some examples of reconstructions and how they can be classified according to their methodology.

\subsubsection{Parametric reconstructions}

In a parametric reconstruction, the desired physical quantity is modeled using an analytical function with a set of free parameters that describe its behavior across cosmic time or spatial scales. The free parameters of the postulated function, and the cosmological parameters, are hence jointly inferred in light of the observational data. Hence, parameterizations are a crucial tool for modeling unknown physical phenomena in a mathematically tractable way. As an example, in the absence of a fundamental and well-defined theory of \ac{de}, several functions, i.e., $w(z)$, $\rho(z)$, have been parameterized in a broad number of different ways. These parametrizations track the dynamics of a \ac{de} back in time simply extending the cosmological model with extra parameters. 
While parametric methods are computationally efficient and easy to interpret, they are inherently limited by the assumptions built into the chosen parameterization. A poorly chosen model may lead to biased results that fail to capture the true complexity of the underlying physics. As such, model selection criteria for various parametric candidates are critical in assessing the robustness of the results.

\subsubsection{Phenomenological parameterizations}

Phenomenological parameterizations focus on capturing specific behaviors, often without direct derivation from fundamental physics. Their primary goal is to test whether certain trends (such as evolving equations of state, deviations from General Relativity, or modified growth of structure) are compatible with observational data. These parameterizations provide a practical way to explore a broad range of possibilities without requiring a detailed theoretical foundation. Some well-known examples include:

\begin{itemize}
    \item \ac{cpl} parameterization \cite{Linder:2002et,Chevallier:2000qy}
    for the \ac{de} equation of state
    \begin{equation}
        w(z) = w_0 + w_a z/(1+z)\,,
    \end{equation}
    which is the general go-to parameterization used to test if any dynamic behavior can be present in \ac{de} (this parameterization is further explained in Sec.~\ref{sec:w0wa}). The \ac{desi} collaboration recently employed this parameterization in their analysis of year-one \ac{bao} data \cite{DESI:2024mwx} and found evidence suggesting a preference for a \ac{dde} model with a significance exceeding $2\sigma$. 
    \item The Jassal-Bagla-Padmanabhan (JBP) parameterization \citep{Jassal:2004ej}, also for the \ac{de} equation of state
    \begin{equation}
        w(z) = w_0 + w_a z/(1+z)^2\,,
    \end{equation}
    which serves as a modification of the \ac{cpl} designed to exhibit different behaviors at large redshifts.
    \item Interaction term between \ac{de} and \ac{dm} \citep{Wang:2016lxa, Amendola:1999er, Cardenas:2018nem, 2019GReGr..51...42G}
    \begin{equation}
        Q=\xi H\rho_{\rm{DM}}\,,
    \end{equation}
    with the parameter $\xi$ governing the strength of the interaction. It is proposed as a way to represent the interaction as a function of time ($H$) and energy density ($\rho_{\rm{DM}}$). Given its phenomenological nature, this approach can be extended further, as demonstrated in Ref.~\cite{Sun:2010vz}, where the interaction term was modified to depend not only on \ac{dm} but also on \ac{de}. In this case, the interaction takes the form $Q=\xi H(\rho_{\rm{DM}}-\alpha\rho_{\rm{DE}})$ with the new parameter $\alpha$ regulating the strength of the dependence on \ac{de}. Moreover, the authors in Ref.~\cite{Cardenas:2018nem} discuss the interaction between \ac{dm} and \ac{de} in the context of the second law of thermodynamics and its consequences on \ac{de} evolution. For a more complete discussion on \ac{de}-\ac{dm} interactions please refer to Sec.~\ref{sec:I_D_DM}.
    
    \item A parametric form for the \ac{de} density $X(z)$ was developed by Refs.~\cite{Wang:2001ht, Wang:2001da, Cardenas:2014jya} and further tested in Refs.~\cite{Grandon:2021nls, Bernardo:2021cxi}. The quadratic parametrization corresponds to 
    \begin{equation}
       X(z) \equiv \frac{\rho_{\rm{DE}}(z)}{\rho_{\rm{DE}}(0)} = 1+ \left(4x_1 -x_2-3\right) \left( \frac{z}{z_m} \right) - 2 \left(2 x_1-x_2-1 \right) \left( \frac{z}{z_m} \right)^2\,,
    \end{equation}
    where $z_m$ is the maximum redshift in the data set, $x_1$ and $x_2$ are the free parameters of the model to be constrained in light of the data. When $x_1=x_2=1$, the parametrization reduces to \lcdm\ where the \ac{de} density is constant $X(z)=1$. This approach aims to study evidence for \ac{de} evolution by allowing its energy density to change with redshift. Reconstructing $X(z)$, instead of $w(z)$, is advantageous since this quantity is more directly traced by the luminosity distance. A cubic parametrization was also explored in Ref.~\cite{Grandon:2021nls} to allow for an extra degree of freedom and to investigate whether further transitions with respect to $X(z)=1$ are found in the reconstruction.
    
    \item The growth rate parameterization \cite{Linder:2005in}
    \begin{equation}
        f(z) = \Omega_{m,0}(z)^\gamma\,,
    \end{equation}
    where $\gamma$ is named the growth index, which for \lcdm\ has a value of $\sim0.55$. A recent application of this parameterization can be found in Ref.~\cite{Nguyen:2023fip}, where an analysis using $f\sigma_8$ and Planck \ac{cmb} revealed a tension of $4.2\sigma$ with the concordance model's expected value of $\gamma=0.55$, suggesting a potential need for modifications to the way the standard model predicts structure growth. 
    \item The gravitational slip \cite{Caldwell:2007cw}
    \begin{equation}
        \eta = \frac{\Phi}{\Psi}\,,
    \end{equation}
    where $\Phi$ and $\Psi$ are the metric potentials in the perturbed Einstein equations. If $\eta$ deviates from $1$ there may be an indication for \ac{mg}. In Ref.~\cite{Amendola:2013qna} four different models for $\eta$ were analyzed using forecasts from \ac{wl}, galaxy clustering, and supernova data to evaluate how effectively future surveys could constrain the slip parameter. Furthermore, in Ref.~\cite{Ranjbar:2024bip} a deviation from \ac{gr} was found using \ac{gw} data along with the slip parameter. 
\end{itemize}

As we can see, phenomenological parameterizations allow for a broad exploration of deviations from the standard model without committing to a specific underlying theory.

\subsubsection{Physically motivated parameterizations}

Physically motivated parameterizations, in contrast, are derived from fundamental theories and ensure consistency with known physics. These models are often constructed from Lagrangian formulations, field equations, or extensions of Einstein’s General Relativity, ensuring that theoretical principles like energy conditions and stability constraints are satisfied. Some of these parameterizations which can be found in the literature are:

\begin{itemize}
    \item Scalar fields for \ac{de}. A well-known one being the Quintessence \citep{Ratra:1987rm}, where a scalar field $\phi$ with a potential $V(\phi)$ leads to an equation of state
    \begin{equation}
        w(z)=\frac{\dot{\phi}/2-V}{\dot{\phi}/2+V}\,,
    \end{equation}
    having the peculiarity of behaving as, in which the kinetic term has the wrong sign, $\dot{\phi}^2/2 \rightarrow -\dot{\phi}^2/2$, leading to $w(z)>-1$. Complementary to it we have the Phantom scalar field \cite{Caldwell:1999ew} where $w(z)<-1$. The combination of them, resulting in the interacting-two-scalar-field model named Quintom \cite{Elizalde:2004mq, Cai:2009zp, Guo:2004fq} has also been studied due to its peculiarity of being able to cross the Phantom-divide line ($w=-1$) \cite{Tot:2022dpr, Vazquez:2023kyx}. K-essence scalar field \cite{Armendariz-Picon:2000ulo}, which can be seen as an extension of Quintessence where non-canonical kinetic terms are included in the Lagrangian.
    
    \item Horndeski theories of \ac{mg} \cite{Horndeski:1974wa, Clifton:2011jh, Gleyzes:2014dya, LISACosmologyWorkingGroup:2019mwx}, which provide the most general scalar-tensor theories leading to second-order field equations.
    
    \item $f(R)$ theories \cite{Starobinsky:1980te} which modify the Ricci scalar in the Einstein-Hilbert action
    \begin{equation}
        S = \int d^4x\sqrt{-g} \bigg(\frac{1}{16\pi G}f(R) + \mathcal{L}_{\rm m} \bigg)\,,
    \end{equation}
    where the $f(R)$ can take many forms. In Ref.~\cite{Leizerovich:2021ksf} two different parameterizations for $f(R)$ were studied, them being $f(R) = R-2\Lambda(1-e^{-R/\Lambda b})$ and $f(R)=R-2\Lambda\big(1-\frac{1}{1+(R/\Lambda b)^n} \big)$, finding concordance with the standard model when using \ac{bao}, \ac{sn} and \ac{cc} data. Similarly, in Ref.~\cite{Pan:2021tpk} a preference for the standard model is recovered when the two models $f(R)= R-\beta/R^n$ and $f(R)=R-\alpha\ln R-\beta$ are used in tandem with simulated \ac{gw} data.
    \item For \ac{dm} models, we encounter \ac{wdm} \cite{Bode:2000gq, Liu:2024edl, Lin:2024fyw, Oman:2024kru, Rose:2023qbw}, where a nonzero velocity dispersion modifies the matter power spectrum. Another example is Self-Interacting \ac{dm} \cite{Spergel:1999mh, DES:2023bzs, Yang:2024uqb, Bringmann:2016din}, in which a cross-section $\sigma/m$ accounts for self-interactions, affecting structure formation at small scales while preserving the \ac{cdm} behavior on larger scales.
\end{itemize}

Both phenomenological and physically motivated parameterizations aim to describe the evolution of cosmic components, test deviations from \lcdm, and assess the viability of new models against observations. Phenomenological parameterizations provide a flexible way to analyze trends, while the physically motivated ones ensure consistency with fundamental physics. The choice between the two depends on whether the focus is on empirical data fitting or theoretical consistency.

\subsubsection{Model-independent parameterizations}

While both phenomenological and physically motivated parameterizations help constrain the behavior of cosmological functions, they inherently require assumptions about the functional form. 
However, these assumptions can limit the flexibility of the analysis, as the chosen parameterization may not fully capture the true underlying features of the function being reconstructed.
To overcome this limitation, one can employ more sophisticated parameterization methods that minimize the number of underlying assumptions. 
Some of these approaches, referred to as model-independent methods, remain parametric but offer significantly greater flexibility, allowing them to capture a wider range of possible features in the data.

\paragraph{Basis representation}

Basis-representation parameterizations provide a flexible and systematic way to reconstruct cosmological functions by expanding them in terms of a chosen set of basis functions. Common examples found in the literature include:

\begin{itemize}
    
    \item Given its simplicity and ease of use, the Taylor expansion has been applied to many different cosmological functions. As an example, in Ref.~\cite{Visser:2003vq} the \ac{de} equation of state, through its pressure, was expressed as a truncated Taylor series
    \begin{equation}
        p=p_0+\sum^{N-1}_{n=1}\frac{1}{n!}\frac{d^np}{d\rho^n}\bigg\vert_0(\rho-\rho_0)+\mathcal{O}[(\rho-\rho_0)^N]\,.
    \end{equation}
    This idea can be extended even further. By performing a Taylor expansion of the Hubble parameter around $z=0$, one arrives at the cosmographic approach \cite{Visser:2004bf,Cattoen:2008th,Dunsby:2015ers,Capozziello:2019cav}, which will be explored in more detail in a later subsection.

    \item In a manner similar to the Taylor series expansion, one can instead adopt a different basis, such as a Fourier series. In Ref.~\cite{Tamayo:2019gqj}, this approach was applied to the \ac{de} equation of state, revealing a preference for an oscillatory behavior at late-times. The equation of state was found to cross the phantom divide multiple times, which has been found to be a stable solution for the Kinetic Gravity Braiding model proposed in Ref.~\cite{Deffayet:2010qz}, a minimally coupled Horndeski model. This however does not hold for other single scalar field models, where gravitational instabilities appear when crossing the phantom divide.
    
    \item The Padé approximation could be seen as an extension to the Taylor series, being itself a rational of two polynomial series. It has the advantage of a faster convergence rate, but potentially using more new terms/parameters. Its uses are well varied in the literature, going from the luminosity distance \cite{Gruber:2013wua, Aviles:2014rma, Wei:2013jya, Capozziello:2020ctn, Capozziello:2023ccw}, the Hubble expansion rate \cite{Capozziello:2023ccw}, the \ac{de} \ac{eos} \cite{Wei:2013jya} and even $f(R)$ \cite{Capozziello:2017ddd}. In Ref.~\cite{Gruber:2013wua} the luminosity distance was expressed as $d_{\rm L}=\frac{a_1z}{1+b_1z+b_2z^2}$, which is usually called the $(1,2)$ Padé approximant for $d_{\rm L}$, finding some agreement with Planck's \ac{cmb} data at small redshifts but allowing \ac{de} to be dynamical. 
    
    \item Wavelet expansion. Wavelets work as a basis and have the peculiarity of being localized in frequency and configuration space, which is useful for capturing both global trends and local fluctuations. In Ref.~\cite{Hojjati:2009ab}, a wavelet-basis was used for the \ac{de} \ac{eos} in the next manner
    \begin{equation}
        w(z_j)+1 = \sum P_i\psi_i(z_j)\,,
    \end{equation}
    with $z_j$ being the redshift points at which $w$ is calculated, $P_i$ the coefficients of the expansion and $\psi_i$ the wavelet function. In this particular work the wavelets used were the Haar ones (where $\psi_i(x) = 1$ for $0\leq x\leq1/2$, $\psi_i(x) = -1$ for $1/2< x\leq1$ and $\psi_i(x) = 0$ otherwise) and, using \ac{sn1}, \ac{wmap} and \ac{bao} data, a hint for dynamical \ac{de} was found.

    \item Reconstructions using an orthonormal basis set of functions (ONB) has been used to reconstruct the normalized cosmic expansion function, $H(z)/H_0$, from the Pantheon dataset \cite{Wagner:2018jxp}, without assuming a specific cosmology. In this approach, the luminosity distance function $D_\mathrm{L}(a,\boldsymbol{c})$ was expanded into orthonormal basis functions $\phi_\alpha(a)$ as
    \begin{align}
        D_\mathrm{L}(a,\boldsymbol{c}) = \sum \limits_{\alpha=0}^{N_\mathrm{B}-1} c_\alpha \phi_\alpha(a) = \boldsymbol{c} \circ \Phi\,,
    \label{eq:series_expansion}
    \end{align}
    where $a$ is the scale factor. The $c_\alpha$ elements of $\boldsymbol{c} \in \mathbb{R}^{N_\mathrm{B}}$ denote the weights of the $N_\mathrm{B}$ basis functions. These terms are summed up in the short-hand notation of the right hand side. The choice of the ONB can be adapted to the problem at hand, as has been done in Ref.~\cite{Wagner:2018jxp}, but other general choices like Chebycheff polynomials can be used.
    Then the values of $c_\alpha$ are obtained by minimizing the $\chi^2$ function \cite{Mignone:2007tj}. Using a special ONB adapted to approximate a \lcdm\ universe with its first few basis functions, \cite{Wagner:2018jxp} showed that the reconstructed cosmic expansion function from the Pantheon set of \ac{sn1} still yields very broad confidence bounds, such that \lcdm\ and other competitive models of \ac{de} are still statistically consistent with the dataset.

\end{itemize}

\paragraph{Interpolation methods}

Another model-independent approach to parameterization, similar to the basis representation method, involves interpolation techniques. In this approach, a set of ``positions'' is defined and then connected in a specific manner, with the chosen connection method determining the type of interpolation used. These positions act as free parameters, whose optimal values are determined through Bayesian parameter inference algorithms, such as \ac{mcmc} or Nested Sampling.

An example of such a class of methods consists in binning the function one wants to reconstruct and treat each one of its binned values as a free parameter of the analysis, e.g., one could aim at constraining the value of $w(z)$ at a given number of redshift $z_i$. This completely removes assumptions made on the trend of the function, but significantly increases the number of free parameters to take into account. When using this approach the bins are usually connected with hyperbolic tangents to preserve continuity. As a function it can be represented as
\begin{equation}
    w(z)= w_1+\sum^{N-1}_{i=1}\frac{w_{i+1} - w_i}{2}\bigg(1+\tanh{\Big(\frac{z-z_i}{\xi}\Big)} \bigg)\,,
\label{bin_equation}
\end{equation}
where $N$ is the number of bins, $w_i$ the amplitude of the bin, $z_i$ the position where the bin begins in the $z$ axis and $\xi$ is a hyperparameter which governs how ``smooth'' is the transition from one bin to the other (a higher value represents a higher smoothness). Being one of the most utilized model-independent interpolation methods it has a wide range of applications such as in reconstructing: the neutrino mass \cite{Lorenz:2021alz}; the primordial power spectrum and inflaton potential \cite{Hazra:2013nca}; the \ac{de} energy density \cite{Escamilla:2021uoj, Wang:2018fng}; the \ac{de} \ac{eos} \cite{Zhao:2012aw, Escamilla:2021uoj, Zhao:2017cud}; the interaction kernel in an \ac{ide} model \cite{cai2010dark, Escamilla-Rivera:2021vyw}; a \ac{de} perfect-fluid model \cite{Moss:2021obd}. Another example of a binned analysis (or interpolation) is the reconstruction of the functions $\mu(z)$ and $\Sigma(z)$, which describe deviations from General Relativity (e.g., see Refs.~\cite{Amendola:2007rr,Bertschinger:2008zb,2010PhRvD..81j4023P}). In Refs.~\cite{Raveri:2021dbu,Pogosian:2021mcs}, joint constraints on these functions and on the \ac{de} density parameter $\Omega_{\rm DE}(z)$ were obtained combining \ac{cmb}, \ac{bao}, and \ac{sn} data. 
The results of this analysis show hints for deviations from General Relativity, mainly in the $\Sigma(z)$ function which encodes deviations from the gravitational lensing effect expected in the standard model, while also reducing both $H_0$ tension to $\sim2\sigma$ (when comparing the inferred value of $H_0=69.44\pm1.3$\kms with the SH0ES one) and the $S_8$ tension (by inferring a value of $S_8=0.780\pm0.033$ which is in line with \ac{des} Y3's measured value of $S_8=0.769\pm0.016$).

Another choice of interpolation is commonly referred to as ``nodal reconstruction'' \cite{Hee:2016nho}. Here the interpolations are done using linear, cubic, or higher order splines, to fill in the gaps between a certain number of nodes. The simplest example is the linear interpolation, that is, given two coordinates $(z_i, f_i)$ and $(z_{i+1}, f_{i+1})$, the function behaves as follows
\begin{equation}
    f(z) = f_i + \frac{f_{i+1}- f_i}{z_{i+1}-z_i}(z-z_i), \quad z \in [z_i, z_{i+1}]\,.
\label{linear_interp}
\end{equation}
This method has been used to reconstruct several important cosmological quantities, i.e., the primordial power spectrum \cite{Guo:2011re, Ravenni:2016vjd, Aslanyan:2014mqa, Handley:2019fll}, the \ac{de} energy density and the \ac{de} \ac{eos} \cite{Escamilla:2021uoj, AlbertoVazquez:2012ofj}, the expansion rate of the Universe \cite{Tutusaus:2018ulu} and the neutrino mass \cite{Lorenz:2021alz}; just to mention a few.

An unusual type of interpolation makes use of \ac{gp}. The mathematical formalism behind it was already presented in Section 3.1, and further details of its applications in reconstruction will be discussed in the following subsection. This interpolation works in a similar manner to the nodal reconstruction, but with the obvious distinction that the connection between the varying nodes is made through \ac{gp}. This approach can mitigate some issues that the other two approaches present, such as the choice of the bins' width and positions \cite{Gerardi:2019obr}, while also presenting the advantage of being infinitely differentiable. Given its relatively new implementation it has not been used widely. In Ref.~\cite{Gerardi:2019obr} was first implemented with the equation of state parameter, and in Ref.~\cite{Escamilla-Rivera:2019hqt} was used to reconstruct the interaction kernel of an \ac{ide} model.

\subsubsection{Cosmography}

The degeneracy among various theories proposed in recent years to explain \ac{de} has driven the investigation of methods that enable the study of cosmic expansion without relying on predefined cosmological models. This is the case of the well-known cosmographic approach \cite{Visser:2004bf,Cattoen:2008th,Dunsby:2015ers,Capozziello:2019cav}, which depends only on the cosmological principle and uses series expansions of the luminosity distance around the current time. 
The power of the cosmographic method arises from the fact that it 
involves observables that can be directly compared with data and guarantees independence from any assumed \ac{de} equation of state.
Thus, the cosmographic method has been extensively utilized to distinguish between various theoretical models that appear similar when compared to observations \cite{Aviles:2012ay,Bamba:2012cp,Capozziello:2017uam,Capozziello:2022jbw}.

However, the cosmographic method faces two main challenges that may limit its effectiveness as an accurate tool for describing cosmic expansion. The first challenge is the necessity for a substantial amount of data to distinguish between the cosmological constant and a \ac{dde}. 
The second issue concerns the use of high-redshift data to explore potential deviations from the standard cosmological model. The latter aspect conflicts with the core principle of the standard cosmographic technique, which relies on a Taylor expansion series centered around $z = 0$. As a result, large error propagations due to convergence issues often compromise the effectiveness of the method.
Over time, various alternatives to the standard cosmographic method have been explored to address these limitations. One approach involves using auxiliary variables and developing expansion series for cosmological observables based on re-parametrizations of the redshift variable converging as $z\gg 1$ \cite{Cattoen:2007sk,Risaliti:2018reu,Capozziello:2020ctn}.

An alternative way to tackle the aforementioned challenges is to utilize rational polynomials that can heal convergence problems typical of the standard Taylor-based cosmography. A prominent example of this technique is offered by Pad\`e polynomials, which can be calibrated to maximize the convergence radius, thereby enabling a more stable fitting process at high redshifts \cite{Gruber:2013wua,Wei:2013jya,Aviles:2014rma}. The benefits of Pad\`e polynomials have recently been exploited in various theoretical contexts to explore potential deviations from Einstein's gravity \cite{Capozziello:2022wgl,Capozziello:2022uak,Capozziello:2023ccw}.
Furthermore, the cosmographic method based on Pad\`e polynomials, as well as the auxiliary $y$-redhift parametrization, has been adopted to evaluate in a model-independent way the cosmological tensions in the $H_0$ and $\sigma_8$ measurements suggested by recent observations \cite{DAgostino:2023cgx}. 
Another strategy to extend the convergence radius of the cosmographic series and address the subjectivity in truncating the expansion, which may still be an issue with the Padé method, involves using Chebyshev polynomials. It has been shown that these polynomials can significantly reduce the uncertainties in higher-order terms of the cosmographic series, thus offering a precise description of the Universe's evolution at late-times \cite{Capozziello:2017nbu,Capozziello:2020awd}. 

The Weighted Function Regression Method, first employed in Ref.~\cite{Gomez-Valent:2018hwc}, addresses the subjectivity involved in choosing the truncation order of cosmographical expressions. It does so by considering all cosmographical orders when reconstructing cosmological functions, and it applies robust Bayesian weights to penalize the use of additional parameters. In practice, the contribution of higher orders can be neglected since their weights are significantly suppressed. This method has been used to reconstruct the Hubble and deceleration functions, as well as to estimate the Hubble and deceleration parameters in a model-independent way with low-redshift data, see Ref.~\cite{Gomez-Valent:2018hwc} and \cite{Gomez-Valent:2018gvm}, respectively.

\subsubsection{Non-parametric methods}

Non-parametric reconstructions are data-driven methods that do not rely on a specific functional form of the reconstructed function and hence strive to minimize model assumptions. In other words, the aim is to describe the data without imposing strong a priori constraints. Among the most common applications of non-parametric methods we find \ac{gp} regression, \ac{ann}s, the local regression smoothing method, and the iterative smoothing method. These methods vary in their flexibility to reconstruct highly non-linear relations between variables and how the uncertainties on the reconstruction are obtained. We proceed to describe them and their applications in cosmology.

\paragraph{Gaussian Processes}

Both powerful and versatile tools in \ac{ml} and statistical analysis \cite{rasmussen2006gaussian,williams2005gaussian}, \ac{gp} are a Bayesian technique that generalizes distributions over functions, extending Gaussian distributions into function space \cite{Seikel:2012uu}. This approach allows for the reconstruction of a function \( f(x) \) at every point \( x \) using an observational data set \(\left\{\left(x_{i},\, f(x_{i}) + \sigma_{i}\right) \mid i=1, \ldots, N\right\}\), without needing to assume a predetermined specific functional parameterization. The reconstructed function and its derivatives are Gaussian random variables with a mean \(\mu(x)\) and variance \( \text{cov}\left[ f(x), f(x) \right] \) at each data point \( x \). The functions at different points \( x \) and \( \tilde{x} \) are related by a covariance function \( K(x, \tilde{x}) \), commonly referred to as a kernel, which depends on a small set of hyperparameters. Although there is a wide range of covariance functions available in the literature \cite{williams2005gaussian, Sun:2021pbu}, the hyperparameters are generally constant, as their values characterize the function's smoothness rather than modeling such behavior. Most cosmological applications of \ac{gp} use a zero mean function and optimize the hyperparameters \cite{Dhawan:2021mel}. A strict Bayesian approach is to marginalize over the hyperparameters. As for the mean function, while stationary processes can be described by a constant mean function, non-stationary ones such as the cosmic distances or growth may not. The authors in Ref.~\cite{Hwang:2022hla} showed how marginalizing over a \emph{reasonable} family of mean functions and over the hyperparameters yields a robust and unbiased result. 
An application of this \emph{full marginalization} approach is to use \ac{gp} as a forward model. 
One can generate \emph{untrained} \ac{gp} by sampling the hyperparameters, and calculating the likelihood. This allows forward modeling of quantities such as the growth rate $f\sigma_8(z)$ and distances while combining several datasets \cite{Avila:2022xad,Joudaki:2017zhq,Keeley:2019hmw,Hwang:2022hla,Calderon:2022cfj,Calderon:2023msm, Reyes:2022exv}. 
In cosmology, \ac{gp} techniques have been successfully employed to reconstruct the dynamics of cosmological functions with minimal physical assumptions about the Universe's geometry or the nature of its main components. These include the background the expansion rate of the Universe \( H(z) \) \cite{Sabogal:2024qxs, Zhang:2018gjb, Briffa:2020qli, Banerjee:2023rvg, Favale:2023lnp, Wang:2023ghk, Banerjee:2023evd, Busti:2014dua, Gomez-Valent:2018hwc, Haridasu:2018gqm} as can be seen in Fig.~\ref{fig:neuralANNrrec}, the deceleration parameter $q(z)$ \cite{Bengaly:2019ibu, Mukherjee:2020vkx, 2012PhRvD..85l3530S, Zhang:2016tto}, the  distance duality relation \cite{Liao:2019qoc, Mukherjee:2021kcu, Renzi:2021xii}, and the cosmological jerk parameter \cite{Mukherjee:2020ytg}. To explore the dynamics of \ac{de}, examples include reconstructions of the \ac{de} energy density \cite{Bernardo:2021cxi, Grandon:2021nls, Calderon:2022cfj}, the scalar field dynamics \cite{Jesus:2021bxq}, the equation of state \( w(z) \) \cite{Bonilla:2020wbn, Wang:2017jdm, Holsclaw:2010sk},
and interaction between \ac{de} and \ac{dm} \cite{Aljaf:2020eqh, Bonilla:2021dql, Escamilla:2023shf, Mukherjee:2021ggf}.
These applications have served to study \ac{mg} models \cite{Briffa:2020qli,Pinho:2018unz,Briffa:2020qli,2023CQGra..40v5003M,Fortunato:2023ypc,2023PDU....4101240S,Oliveira:2023uid,Gadbail:2024rpp,Mu:2023zct,Sultana:2022qzn,2022EPJC...82..811E,Wang:2022xdw,2022PDU....3500926S,Ren:2022aeo,Cai:2019bdh}, cosmic curvature \cite{Liu:2020pfa, Mukherjee:2022ujw, Pan:2023omz, Qi:2023oxv, Wang:2022rvf,Wu:2022fmr,Yang:2020bpv,Wang:2020dbt}, and \ac{gw}s \cite{Belgacem:2019zzu, Zheng:2020tau, Cai:2017yww, Shah:2023rqb}, inflaton speed of sound's profile \cite{Canas-Herrera:2020mme}, and many other perspectives \cite{Dinda:2022jih, Favale:2024sdq, Favale:2024lgp, Gomez-Valent:2021hda,Rodrigues:2021wyk,Mukherjee:2023lqr,Li:2023pot,Mukherjee:2024ryz,Khurshudyan:2024gpn,vonMarttens:2020apn,Lorenz:2021alz,Mukherjee:2022lkt,Mukherjee:2023yxq}. In particular, the authors in Ref.~\cite{Canas-Herrera:2020mme} adopt a slightly different approach. They demonstrated how mildly-informative
physical priors can be imposed on a \ac{gp} reconstruction in a Bayesian way so that robust constraints on physical parameters can be extracted along with a non-parametric reconstruction.
In summary, all these applications have proven particularly valuable for testing the robustness of \lcdm\ and identifying potential deviations from our current understanding of the Universe.

\paragraph{Artificial neural networks}

\ac{ann}s have also seen a steep increase in their application to cosmology (for further information on this please refer to Sec.~\ref{sec:ML_Inf_ANN}, Sec.~\ref{sec:ML_Inf_CNN}, Sec.~\ref{sec:ML_Inf_BNN} and Sec.~\ref{sec:ML_Inf_ladder}). The two main advantages of \ac{ann}s over other non-parametric techniques are that they can capture (highly) non-linear relations in complex datasets, and in principle do not require the data to follow a specific statistical distribution. For this reason, the cosmological community has developed various neural network reconstructions of cosmological functions. Some cosmological functions reconstructed with \ac{ann}s include the distance modulus $\mu(z)$ from the Pantheon and Pantheon+ compilation \cite{Escamilla-Rivera:2019hqt, Wang:2019vxv, Dialektopoulos:2023dhb, Gomez-Vargas:2022bsm, Shah:2024slr, Shah:2024gfu} and the Hubble function $H(z)$ \cite{Wang:2019vxv, Dialektopoulos:2021wde, Gomez-Vargas:2021zyl, Dialektopoulos:2023jam,Sharma:2024mtq}. In particular, a novel approach was developed in Ref.~\cite{Escamilla-Rivera:2019hqt} where the authors train a recurrent neural network to learn the mapping from redshift $z$ to distance modulus $\mu(z)$ using the Pantheon \ac{sn1} sample. This model, combined with a \ac{bnn}, is also able to propagate the uncertainties into the predicted function $\hat{\mu}(z)$ and thereby into cosmological constraints. This approach was also implemented to calibrate \ac{grb}s at high redshift \cite{Escamilla-Rivera:2021vyw}. Other applications include \ac{ann}s reconstructions using \ac{lsst} simulated data \cite{Mitra:2024ahj}, 
\ac{cnn}s to reconstruct the \ac{bao} signal \cite{Mao:2020vdp}, \ac{ann}s for growth data $f\sigma_8$ \cite{Dialektopoulos:2021wde, Gomez-Vargas:2021zyl}, and velocities of rotation curves \cite{Garcia-Arroyo:2024rdj}. For visual reference, in Fig.~\ref{fig:neuralANNrrec} an \ac{ann} was used to reconstruct $H(z)$. Other approaches generate \ac{ann} models for small datasets and their errors using feedforward neural networks with Monte Carlo Dropout and hyperparameter grid optimization \cite{Gomez-Vargas:2021zyl} or \ac{ga}s \cite{Gomez-Vargas:2022bsm}. 

As with other non-parametric reconstruction methods, the goal is to generate an \ac{ann} model based on observational data and then compare the resulting reconstruction with theoretical predictions from parametric methods to allow for robust conclusions. In this direction, different cosmological models have been analyzed using neural network reconstructions. Examples include \lcdm~\cite{Wang:2019vxv, Wang:2020dbt, Gomez-Vargas:2021zyl}, the \ac{cpl} model \cite{Escamilla-Rivera:2019hqt, Wang:2019vxv, Gomez-Vargas:2021zyl, Mitra:2024ahj}, Chaplygin gas models \cite{Escamilla-Rivera:2019hqt}, and Horndeski gravity \cite{Dialektopoulos:2023jam}, among others.

\begin{figure}
    \centering
    \makebox[11cm][c]{
    \includegraphics[width=0.52\linewidth]{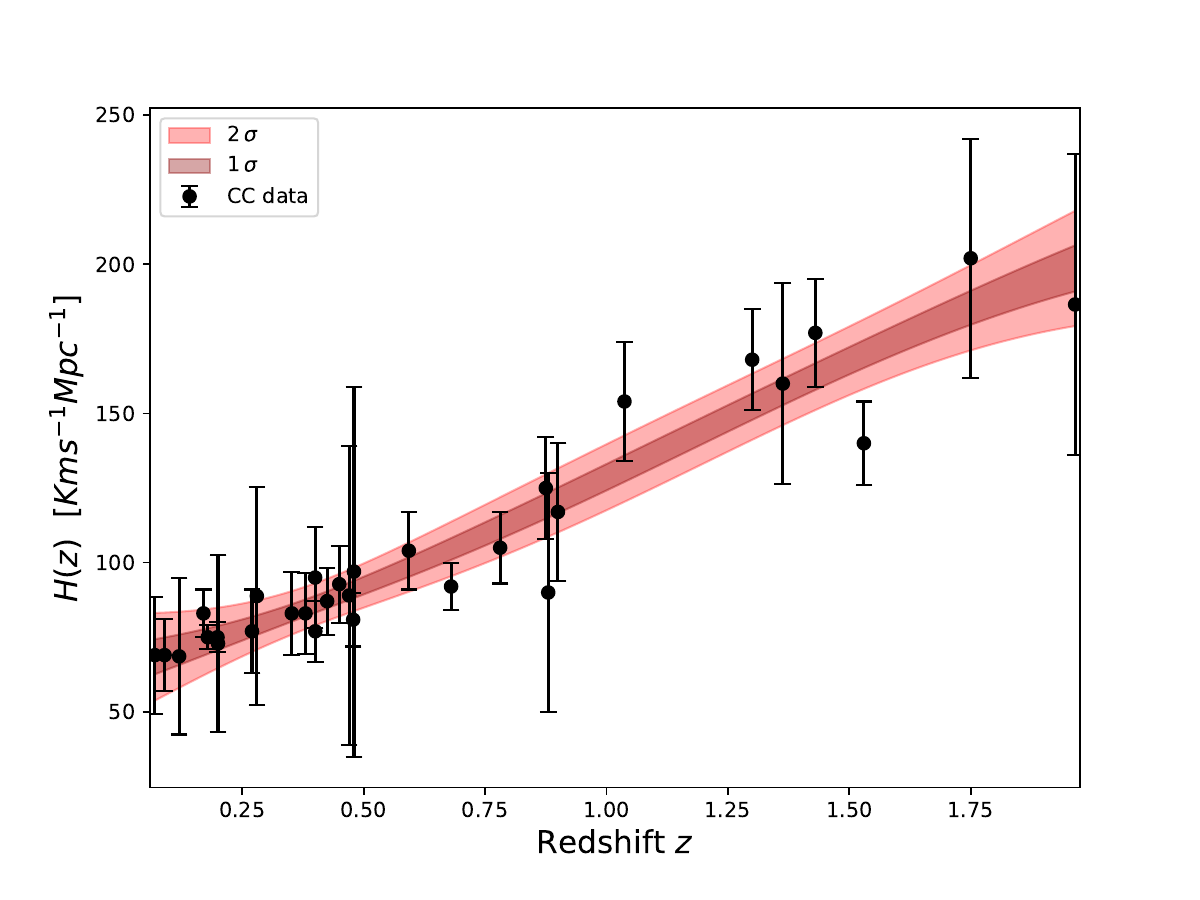}
    \includegraphics[width=0.5\linewidth]{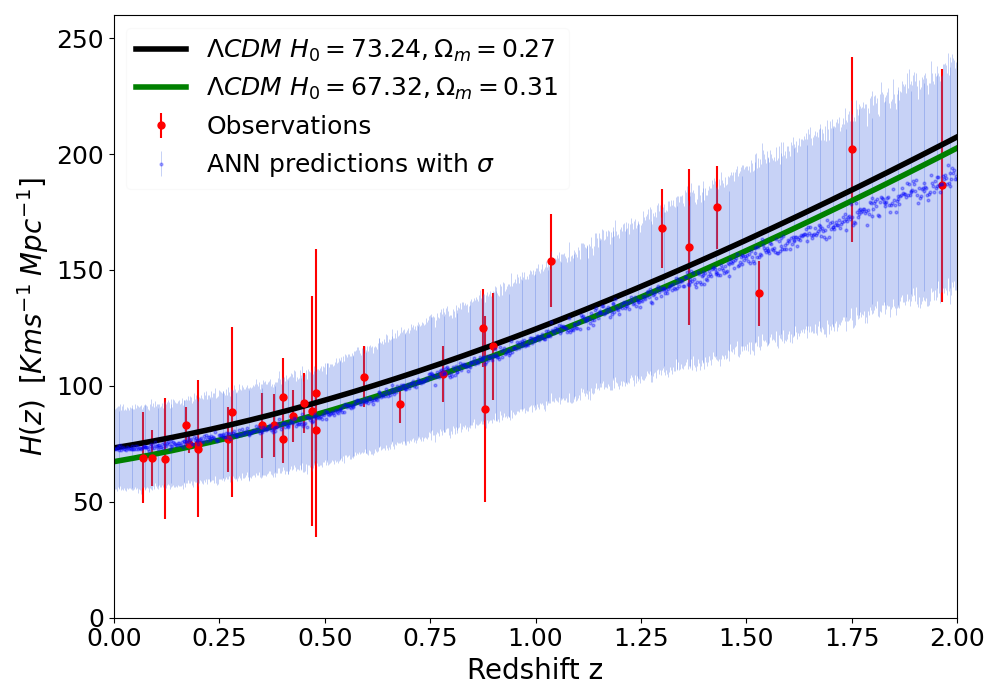}
    }
    \caption{Non-parametric reconstructions of the Hubble function $H(z)$ using \ac{cc}s. Two different approaches were used: \ac{gp} regression (left) and \ac{ann}s (right). The \ac{ann} reconstruction was made following the approach in Ref.~\cite{Gomez-Vargas:2021zyl}. The analysis models observational uncertainties using Monte Carlo Dropout, as also employed in Refs.~\cite{Wang:2020dbt, Mitra:2024ahj, Escamilla-Rivera:2019hqt}. The cosmic chronometer data from \cite{Stern:2009ep} and theoretical predictions from the \lcdm\ model are included for reference.}
    \label{fig:neuralANNrrec}
\end{figure}

\paragraph{Iterative smoothing method}

The authors in Ref.~\cite{Shafieloo:2005nd} used this non-parametric method to smooth supernova data using a Gaussian smoothing function. This method aims to reconstruct cosmological quantities such as the expansion rate, $H(z)$, and the equation of state of \ac{de}, $w(z)$, in a model-independent manner. The only assumptions made by this method are the smoothing scale and a guess background model for the quantity under study. However, it is necessary to use a bootstrapping method to determine the optimal guess model. In Ref.~\cite{Shafieloo:2005nd}, the authors used an iterative method to estimate the guess model. They started with a simple cosmological model, such as \lcdm, as the initial guess model and then the results were used as the next step in the iteration. With each iteration, it was expected the guess model to become more accurate, thus giving a result that is less and less biased towards the initial guess model used. Indeed, it was noted that using different models for the initial guess does not affect the final result as long as the process is iterated several times.

The method requires careful consideration of the smoothing scale. A very small smoothing scale gives an accurate but noisy guess model, therefore after a few iterations, the result will become too noisy to be of any use. It's better to use a larger smoothing scale for smoother results. On the other hand, the bias of the final result decreases with each iteration, since with each iteration one gets closer to the true model. The bias decreases non-linearly with the number of iterations. In Ref.~\cite{Shafieloo:2005nd}, points out that after about 10 iterations, for moderate values of the smoothing scale, the bias is acceptably small. Also, beyond this, the bias still decreases with the number of iterations but the decrease is negligible while the process takes more time and results in larger errors on the parameters. In Refs.~\cite{Shafieloo:2007cs,Shafieloo:2009hi,Shafieloo:2018gin} the method was improved to reconstruct the distance modulus in a model-independent way and then employed in Refs.~\cite{Shafieloo:2012yh,Koo:2020wro}, to name just a few examples.
In particular, this method was applied to reconstruct the cosmic expansion history and growth to test curvature, \ac{de}, and \ac{gr} \cite{LHuillier:2016mtc,LHuillier:2017ani,Shafieloo:2018gin,LHuillier:2024rmp}.

\paragraph{Local regression smoothing and simulation extrapolation (Loess+Simex)}

The LOcally wEighted Scatterplot Smoothing (Loess), also known as local polynomial regression, is a non-parametric method for data analysis that does not require specifying a predefined relationship between dependent and independent variables. It generalizes standard least-squares methods and is widely used for non-parametric simple regression in various disciplines. \textit{Loess} aims to depict the global trend of a dataset by fitting low-degree polynomials to subsets of data around each observation, giving more weight to points closer to the target observation using a Kernel-based weighting system. Typically, first or second-order polynomials are used since higher orders do not significantly enhance results and increase computational complexity.
The process is applied iteratively to cover the entire data range, resulting in a comprehensive trend depiction. Important aspects of the method include selecting the number of data points for each fit through a smoothing parameter called span, $s$, determining the polynomial degree, and choosing the weight function form. Additionally, confidence intervals are constructed by calculating the variance of the fitted values, assuming Gaussian distributed errors. Despite some bias in the estimation, the cross-validation technique provides accurate confidence intervals around the \textit{Loess} curve. To account for the observational errors on real data, additional methods are needed. To this end, the authors in Ref.~\cite{Montiel:2014fpa} proposed to combine the \textit{Loess} method with the simulation-extrapolation method, also called \textit{Simex}, which is a simulation-based method designed to minimize bias resulting from the inclusion of covariates that are prone to errors. Estimates are derived by introducing additional measurement errors, using a form of resampling. This resampling helps identify the pattern of measurement error. After estimating the pattern, final estimates are obtained by extrapolating back to the scenario where there is no measurement error.

\textit{Loess} and \textit{Simex} were first used together in cosmology in Ref.~\cite{Montiel:2014fpa} to reconstruct the expansion history $H(z)$. Later, this technique was also used to reconstruct galaxy rotation curves \cite{2019MNRAS.488.5127F}, the distance modulus \cite{Escamilla-Rivera:2021rbe}, and other cosmological quantities \cite{Rani:2015lia,Rana:2015feb,Escamilla-Rivera:2015odt}, with also accurate results.

\bigskip
\subsection{Bio-inspired algorithms in model selection \label{sec:GA_selection}}

\noindent \textbf{Coordinator:} Reginald Christian Bernardo\\
\noindent \textbf{Contributors:} Anto Idicherian Lonappan, Antonio da Silva, Arrianne Crystal Velasc, Celia Escamilla-Rivera, David Valls-Gabaud, Dinko Milakovic, Erika Antonette Enriquez, Filippo Bouche, Isidro G\'{o}mez-Vargas, J. Alberto V\'{a}zquez, Jenny G. Sorce, John K. Webb, Jurgen Mifsud, and Renier Mendoza\\

We briefly discuss \ac{ga}s (Secs.~\ref{subsubsec:genetic_algorithm}-\ref{subsubsec:machine_learning_and_variants}), their applications to cosmology (Sec.~\ref{subsubsec:ga_for_cosmology}) and potential relevance to the understanding of cosmological tensions (Sec.~\ref{subsubsec:ga_for_tensions}).

\subsubsection{Genetic algorithm}
\label{subsubsec:genetic_algorithm}

A \ac{ga} is a biology-inspired optimization strategy that mainly takes elements of natural evolution in order to single out one solution that is the fittest from a pool of similarly naturally selected individual solutions. \ac{ga}s are powerful optimization methods, called meta-heuristic because they do not use derivatives to find the optimum, and they guarantee to find the best solution under certain conditions, despite the challenges posed by local optimality \cite{rudolph1994convergence}. This has been used in a wide variety of scientific problems such as in high energy physics \cite{Akrami:2009hp} and \ac{gw} astronomy \cite{Crowder:2006wh}, and is known to be able to find global optimum and resolve tiny differences between seemingly degenerate solutions to a problem.
\ac{ga} is particularly well-suited to bypass issues related to complex, high-dimensional parameter spaces and multimodal functions. In cosmology, it was introduced as a means to overcome the bias in selecting a cosmological model in order to infer the properties of \ac{de} \cite{Bogdanos:2009ib}. This was then followed by Ref.~\cite{Nesseris:2010ep} and Ref.~\cite{Nesseris:2012tt}, which has further marketed \ac{ga} as an alternative tool for cosmological analysis through the estimation of uncertainty. An excellent recent introduction to \ac{ga} for cosmological parameter estimation is given in Ref.~\cite{Medel-Esquivel:2023nov}.

\ac{ga}s operate with a population of individuals (possible solutions), where each individual is characterized by a chromosome,
which in turn is described by a set of genes. These individuals make up the population that evolves over successive generations. The key ingredients of \ac{ga} are described in Table~\ref{tab:GA_ingredients}, while the evolutionary process is illustrated in Fig.~\ref{fig:GA_flowchart}.

\begin{table}[ht!]
\centering
\caption{Key ingredients of \ac{ga}s \cite{gad2023pygad}.}
\begin{tabular}{p{2.5cm}  p{13.5cm}}
\textit{Fitness-function} & Determines the fitness of individuals in a population. This function can be tailored to the specific problem, such as minimizing the Euclidean distance or optimizing the likelihood in cosmological applications. The fitness function is a predefined metric used to rank solutions. \\
\textit{Selection} & Defines the portion of the population that will advance to the next generation. A common method is the `roulette wheel' selection, where fitter individuals have higher chances of being selected for reproduction. \\
\textit{Elitism} &Ensures that the individuals that passed the selection process can be directly added to the next generation. Elitism guarantees that the best fitness value will not get worse after every generation. \\
\textit{Crossover} & Refers to the process of combining the genetic information of two parents to produce offspring. This mechanism allows for the exchange of genes between individuals, promoting the inheritance of favorable traits while introducing genetic diversity into the population. \\
\textit{Mutation} & Involves the random alteration of an individual's genes. This step is crucial for introducing new genetic variations, ensuring that the population does not stagnate and continues to evolve towards better solutions. This allows \ac{ga}s to escape local solutions and obtain global solutions. \\

\end{tabular}
\label{tab:GA_ingredients}
\end{table}

\begin{figure}[ht!]
\begin{center}
\begin{tikzpicture}[
    node distance = 1.0cm,
    every node/.style = {rectangle, draw, align=center, minimum width=2cm, minimum height=0.5cm}
]

\node (init) {Initialization};
\node (select) [below of=init] {Selection};
\node (crossover) [below of=select] {Crossover};
\node (elite) [left=1cm of crossover] {Elitism};
\node (mutation) [below of=crossover] {Mutation};
\node (newgen) [below of=mutation] {New Generation};

\draw[->, shorten >=2pt, shorten <=2pt] (init) -- (select);
\draw[->, shorten >=2pt, shorten <=2pt] (select) -- (crossover);
\draw[->, shorten >=2pt, shorten <=2pt] (crossover) -- (mutation);
\draw[->, shorten >=2pt, shorten <=2pt] (mutation) -- (newgen);
\draw[->, shorten >=2pt, shorten <=2pt] (newgen.east) -| ++(1.5cm,0) |- (init.east);
\draw[->, shorten >=2pt, shorten <=2pt] (select.west) -| (elite.north);
\draw[->, shorten >=2pt, shorten <=2pt] (elite.south) |- (newgen.west);

\end{tikzpicture}
\caption{Flowchart illustrating the key steps in \ac{ga}s \cite{gad2023pygad}.}
\label{fig:GA_flowchart}
\end{center}
\end{figure}
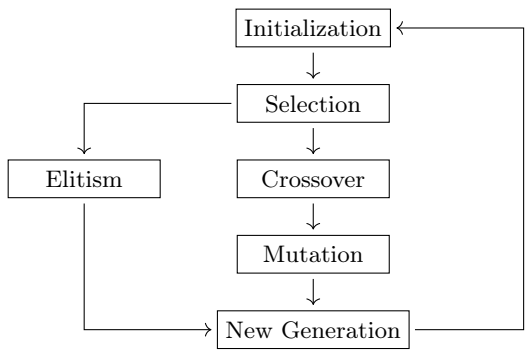

In essence, \ac{ga}s sample from a population of individuals ranked by a fitness scale. The fittest individuals are more likely to survive and reproduce through crossover and mutation, creating offspring with improved traits. The best or elite individuals are stored to guarantee that the population is improving through the generations. This iterative process continues until the optimal solution, or the fittest individual, emerges. In cosmology, the gene can be thought of as each cosmological parameter, such as $w_0$ and $w_a$ in \ac{cpl} cosmology, and the chromosome as the string of genes, e.g., $(H_0, \Omega_{\rm m,0}, w_0, w_a)$ \cite{Medel-Esquivel:2023nov}. A \ac{ga} can, of course, be applied in a variety of ways; notably, in the pioneering works \cite{Bogdanos:2009ib, Nesseris:2010ep, Nesseris:2012tt} the \ac{ga} has been used in the context of grammatical evolution (GE), where the elements and key ingredients of \ac{ga}s are utilized to search for solutions within function spaces, alleviating the arbitrariness of cosmological parametrization. More recently, \ac{ga}s have been used to eliminate the arbitrary choice of a kernel function in \ac{gp} regression \cite{Bernardo:2021mfs} and to optimize neural networks \cite{Gomez-Vargas:2022bsm}. It has also found applications in spectroscopic modeling \cite{2017MNRAS.468.1639B, 2017Univ....3...34B, Lee:2020lof}, and in conjunction with information criteria for model selection \cite{2021MNRAS.501.2268W}.

\subsubsection{Machine learning and GA variants}
\label{subsubsec:machine_learning_and_variants}

\begin{algorithm}[H]
    \caption{\ac{ga} pseudocode.}
    \textbf{Notation:} $f$: fitness function, $[b_1,b_2]$: population range, $D$: population dimension, $r$: mutation rate, $N$: population size,\\ $G$: max. number of generations, $m$: individual index, $t$: generation index, $M(t)$: population at generation $t$, $m^*$: best solution \\
    \textbf{Input:} $f$, $[b_1,b_2]$, $D$, $r$, $N$, $G$\\
    \textbf{Output:} $m^*$, $f(m^*)$
    \begin{algorithmic}[1]
        \State Define the maximum number $G$ of generations and the mutation rate $r$
        \State Set the generation counter $t \leftarrow 0$ 
        \State Generate an initial random population $M(0)$ consisting of $N$ individuals having dimension $D$ and range $[b_1,b_2]$
        \While {$t < G$}
            \State Compute and save the fitness $f(m)$ of each individual $m$ in the current population $M(t)$
            \State Sort $M(t)$ with respect to the fitness of each $m$
            \State Define selection probabilities $p(m)$ for each $m$ in $M(t)$
            \State Choose a proportion of $M(t)$ based on the selection probabilities $p(m)$
            \State Apply crossover on parent individuals to produce offspring 
            \State Apply mutation on offspring with a probability based on the mutation rate $r$
            \State Store the new generation $M(t + 1)$ of individuals    
            \State Select the elite individuals in $M(t + 1)$ to be preserved for the next generations
            \State $t \leftarrow t + 1$
        \EndWhile
        \State Return the best individual $m^*$ of the final generation and its fitness $f(m^*)$
    \end{algorithmic} \label{algo1}
\end{algorithm}

In this section, we dig a little further into the details on \ac{ga}s in order to get a better grip of its intricacies, and use it as a template to introduce some of its more familiar variants that have also been considered in the field. The basic steps of \ac{ga}s are summarized in the pseudocode shown in Algorithm \ref{algo1}, based on the notation in Ref.~\cite{Nesseris:2012tt} and Ref.~\cite{gad2023pygad}.

Algorithm \ref{algo1} highlights that \ac{ga}s are fundamentally an optimization strategy, inspired by the principles of natural selection and genetics. In cosmology, however, particularly through the lens of grammatical evolution, \ac{ga}s transcend traditional optimization by serving as an \ac{ml} method. The resulting combined method utilizes the adaptive search capabilities of \ac{ga} and GE to incrementally evolve a grammar that approximates cosmological functions, such as cosmic expansion and growth rates, from input data. By discovering functional forms that capture the underlying dynamics of cosmological phenomena, \ac{ga}s in this context demonstrate their dual role in both optimizing and learning, revealing intricate patterns that may be governing our Universe. This versatility is notably owed to the pioneering works of Refs.~\cite{Bogdanos:2009ib, Nesseris:2010ep, Nesseris:2012tt}, which fully fleshed out \ac{ga}'s flexibility for extracting insights on the elusive \ac{de} given late-time data. This opens up \ac{ga}s to the same wealth of applications \ac{ml} has been applied to in cosmology, such as in the calibration of very high redshift observables \cite{Escamilla-Rivera:2021vyw}, non-parametric cosmological reconstructions \cite{Escamilla-Rivera:2021rbe}, Bayesian deep learning for \ac{de} \cite{Escamilla-Rivera:2020fxq, Escamilla-Rivera:2020szs, Escamilla-Rivera:2019hqt}, and in testing the validity of routines such as the cosmographic approach \cite{Munoz:2020gok}.

Other than \ac{ml}, another broad scientific discipline where \ac{ga}s are prominent is optimization, specifically under the categories of meta-heuristic, nature-inspired, and evolutionary algorithms. Within this realm, \ac{ga}s have various variants, such as memetic algorithms, which are designed to prevent premature convergence by incorporating local search strategies \cite{Moscato2010}. Given their inherently distributed nature, \ac{ga}s stand to benefit significantly from the advent of quantum computers and their parallelism. Quantum \ac{ga}s exploit quantum mechanical phenomena such as superposition and entanglement to implement quantum evolutionary concepts and operators \cite{Ross:2020iux}: quantum chromosomes, entangled crossovers, quantum elitism, etc. The quantum nature of these algorithms introduces a non-zero probability of adding new genetic material to the population, thereby mitigating premature convergence \cite{Acampora:2022ker}.

In recent years, several quantum \ac{ga}s have been developed \cite{Acampora:2021dhd, Acampora:2022ker, Ibarrondo:2022psk, Ibarrondo:2022emj}, featuring either a fully quantum architecture or the integration of quantum operators within a classical \ac{ga} framework. Implementing these innovative \ac{ga}s on actual quantum devices has yielded enhanced performances compared to their classical counterparts, despite the limitations of today's noisy quantum hardware \cite{Acampora:2022ker, Ibarrondo:2022psk}. The first application of quantum \ac{ga}s in cosmological analysis is currently underway, aiming at minimizing the chi-squared function of different cosmological probes: \ac{sn1}, baryon acoustic oscillations, \ac{cmb} \cite{QGA_Authors_2024}. This development marks a significant step forward, combining the strengths of quantum computing and \ac{ga}s to tackle complex problems in cosmology.

One of the more familiar variants of \ac{ga}s in the astrophysics and cosmology community is particle swarm optimization (PSO). PSO is inspired by the social behavior of birds flocking or fish schooling, where each particle (or candidate solution) adjusts its position in the search space based on its own experience and the experience of neighboring particles. The algorithm involves iteratively updating the velocities and positions of the particles, guiding them towards the best solutions found so far. This method has been used in Ref.~\cite{Prasad:2011rd} to infer cosmological parameters given \ac{cmb} data and in Ref.~\cite{Ruiz:2013sla} to calibrate semi-analytic galaxy formation models in a fixed cosmological background. PSO has been employed even earlier in astrophysics, such as in Ref.~\cite{Skokos:2005ii} to search for periodic orbits in three-dimensional galactic potentials. Recently, the method has found application in \ac{gw} analysis as seen in Refs.~\cite{DarkMachinesHighDimensionalSamplingGroup:2021wkt}.

It is now becoming clearer that \ac{ga} and PSO approaches are only the tip of the iceberg, or perhaps an opening to Pandora's box, of mathematical optimization techniques available for astrophysical and cosmological research. The recent review by Refs.~\cite{DarkMachinesHighDimensionalSamplingGroup:2021wkt} discusses several optimization methods, including particle swarm optimization, in high energy and astrophysics applications. In the following sections, we shall introduce a Bayesian tool inspired by PSO, the Approximate Bayesian Computation-Sequential Monte Carlo (ABC-SMC) \cite{doi:10.1098/rsif.2008.0172, 10.1093/bioinformatics/btp619, 2009arXiv0910.4472T}, which has been applied recently in cosmological model selection and parameter estimation \cite{Bernardo:2022pyz, Bernardo:2022ggl}. Work on a comparative study of nature-inspired optimization methods applied to cosmological parameter estimation, including \ac{mcmc}, \ac{ga}, Improved Multi-Operator Differential Evolution \cite{9185577}, and the Philippine Eagle Optimization Algorithm \cite{2021arXiv211210318E}, is in progress, showing optimistic results for parameter estimation and cosmological reconstruction \cite{PEOA_cosmo_Bernardo_etal_2024}.

\subsubsection{GA for cosmology}
\label{subsubsec:ga_for_cosmology}

In cosmology, \ac{ga}s have mostly been used in the context of grammatical evolution \cite{Bogdanos:2009ib,Nesseris:2010ep,Nesseris:2012tt}. This was motivated by the fact that the Universe's most dominant component at late-times is dark, and a non-parametric reconstruction via \ac{ga}-GE seems particularly well suited for such a problem. In grammatical evolution, one deals with a set of grammar functions, often polynomials are considered as well as trigonometric, logarithm, and exponential functions, that are exposed to the \ac{ga} optimization scheme. The output of \ac{ga}-GE is thus a functional approximation of the ``true'' function that has spawned the observed data; this can then be characterized with a fundamental model, such as \lcdm\ or $w_0w_a$CDM \ac{cpl}, if the data were cosmological in nature. Uncertainty estimation in the context of \ac{ga}-GE was also studied in Refs.~\cite{Bogdanos:2009ib, Nesseris:2010ep, Nesseris:2012tt,Bernardo:2025flj,Bernardo:2025zbv}, suggesting various converging estimates by a Fisher matrix formalism, bootstrapping {\it a la} Monte Carlo, and a path integral approach.

Recently, \ac{ga}-GE and other \ac{ml} strategies were used in Ref.~\cite{Kim:2023unc} to draw insights from late-time expansion data. As with any \ac{ml} approach, however, the output of the reconstruction via \ac{ga}-GE can generally be quite dependent on the choice of the hyperparameters, for \ac{ga}-GE, this would be the grammar functions. This can be put to good use if there is prior knowledge of the functional behavior of the underlying fundamental model, such as in Ref.~\cite{Lodha:2023jru} where trigonometric grammar functions have been utilized to search for oscillating features in the primordial power spectrum. This turned out to be a promising avenue to address simultaneously the tensions in the Hubble constant and the matter power spectrum \cite{Antony:2022ert, Hazra:2022rdl}. Furthermore, the \ac{ga}-GE approach has been applied to \ac{sn}, baryon acoustic oscillations, \ac{cc}, \ac{rsd}, and normalized Hubble rate measurements to symbolically reconstruct the cosmological functions in Ref.~\cite{Alestas:2022gcg}, and was also used to forecast constraints on the \lcdm\ model in Refs.~\cite{EUCLID:2020syl, Euclid:2021cfn, Euclid:2021frk}.

Another way to use \ac{ga}s in cosmology that has gained some traction recently is to directly use it to estimate the cosmological parameters, e.g., phenomenological \ac{de} parameters $w_0$ and $w_a$, in a given model \cite{Medel-Esquivel:2023nov}. In this way, the uncertainty can similarly be estimated using a Fisher matrix approach or via bootstrapping, resulting in an approximation to the posterior of the cosmological parameters, that can be used with \ac{mcmc} to strengthen the robustness of the analysis. We illustrate this very briefly with a quick \ac{ga} computation (Fig.~\ref{fig:ga_constraints_cc_pantheon}), to show the constraints obtained by \ac{ga}s and \ac{mcmc} in spatially-curved \lcdm\ and flat \ac{cpl} given cosmic chronometer measurements \cite{Moresco:2020fbm} and \ac{sn} observations \cite{Brout:2021mpj, Brout:2022vxf, Scolnic:2021amr}. For the \ac{ga} parameters, we consider the likelihood ${\cal L}$ as a fitness function, a ``tournament'' type selection with a rate of $30\%$, an adaptive mutation rate of $(80\%, 20\%)$, and a ``scattered'' crossover type with a probability of $50\%$. In our notation for adaptive mutation, the first number in the tuple $(a, b)$ denotes the fraction $a$ of the genes in a chromosome that are going to be mutated for low quality solutions (ranked accordingly by fitness), and the second number corresponds to the fraction $b$ of chromosomes that are to be mutated for high quality ones.

\begin{figure}[ht!]
    \centering
    \begin{subfigure}{0.475\linewidth}
	   \includegraphics[width=\linewidth]{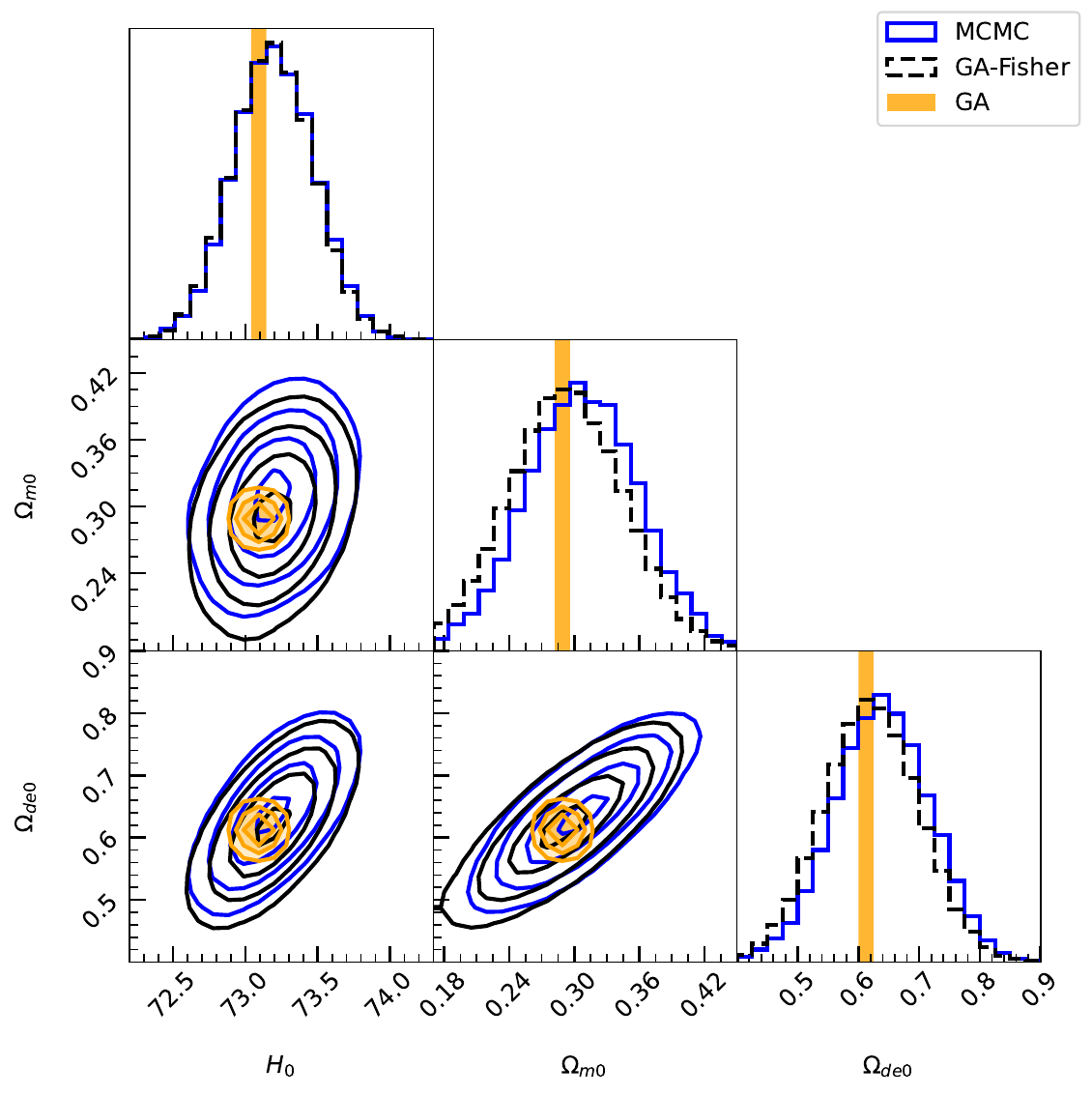}
	   \caption{Curved \lcdm}
    \end{subfigure}
    \begin{subfigure}{0.475\linewidth}
	   \includegraphics[width=\linewidth]{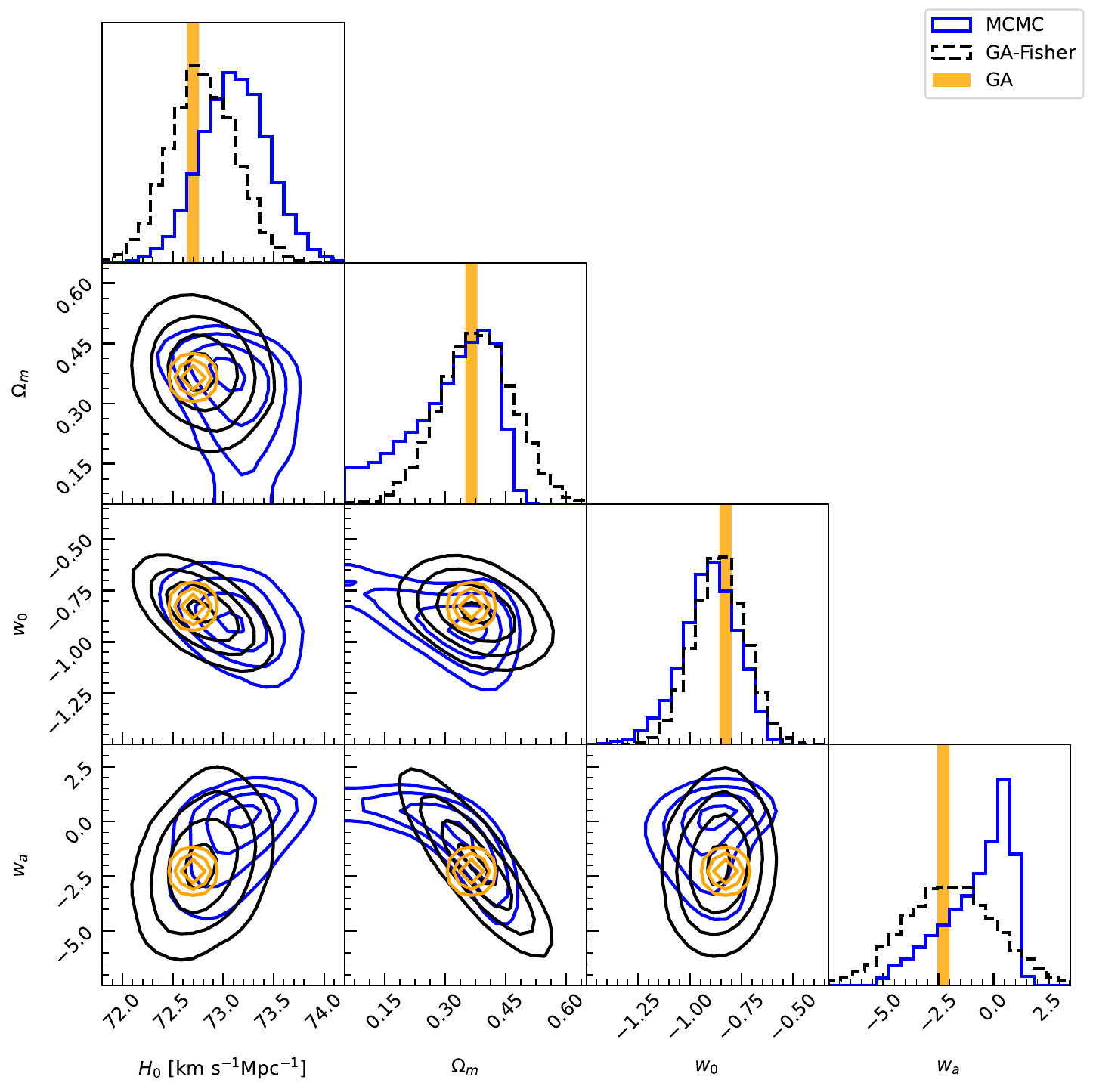}
	   \caption{\ac{cpl} ($w_0w_a$CDM)}
    \end{subfigure}
    \caption{Constraints on a spatially-curved \lcdm\ and a flat \ac{cpl} ($w_0w_a$CDM) models with \ac{cc}s \cite{Moresco:2020fbm} and \ac{sn} \cite{Brout:2021mpj, Brout:2022vxf, Scolnic:2021amr}; blue corresponds to \ac{mcmc} results; black to \ac{ga}-Fisher (\ac{ga} Fisher matrix uncertainty estimation hybrid); orange are the localized samples (trimmed outliers) in the \ac{ga} final population containing the best solution.}
    \label{fig:ga_constraints_cc_pantheon}
\end{figure}

In non-flat \lcdm, it can be seen that the \ac{ga} and \ac{mcmc} results agree with each other on the inferred cosmological parameter space to a high degree, including correlations \cite{GA_Demystified_Bernardo_Chen_2024}. The \ac{ga} output population can furthermore be used to chip in to the results; as shown in this case it is also consistent with \ac{mcmc}. However, in the more complex \ac{cpl} model, the analogous constraints reveal a slight hint of disassociation between the results of \ac{mcmc} and \ac{ga}, particularly with the \ac{de} parameters. The disagreement is not to be worried about regardless since the deviations are well within the inferred confidence regions of either method. This quick exercise nonetheless shows \ac{ga} in action in cosmological parameter estimation, and reinforces one of our messages that \ac{ga}s can act as a supporting tool to \ac{mcmc} for cosmological analysis. This is especially true for models that are more complex than \lcdm, despite the different nature of \ac{mcmc} methods that sample the posterior distribution, while \ac{ga}s maximize the likelihood.

\ac{ga}s and their variants have been considered in applications that assess improve the output of other methods, such as in optimizing Neural Networks \cite{Gomez-Vargas:2022bsm, Zhang:2023ucf} and speeding up Bayesian inference \cite{Gomez-Vargas:2024izm}. Similarly ABC-SMC and \ac{ga} have been applied to tackle the issue of kernel selection of \ac{gp} \cite{10.3389/fbuil.2017.00052} when it is applied for cosmological reconstruction \cite{Bernardo:2021mfs, Zhang:2023pis}; leading toward an \ac{ml} hybrid that no longer depends on any one arbitrary choice of a kernel function. Furthermore, ABC-SMC has shown to be quite useful for parametric reconstruction and model selection, as shown in Ref.~\cite{Bernardo:2022pyz}, in the end catering to its users with one model and its parameters that best suit the data. Another notable application of \ac{ga}s is given in Ref.~\cite{2024IJMPD..3350114O} to test the viability of an emergent universe scenario using Planck data and \ac{bbn}.

\subsubsection{GA for cosmological tensions}
\label{subsubsec:ga_for_tensions}

We emphasize that \ac{ga}s for cosmological parameter estimation are not meant as a replacement for \ac{mcmc}. \ac{ga}s are rather a support tool for the traditional \ac{mcmc}, offering unique advantages given its ability to navigate complex, non-linear, and high-dimensional parameter space, for example, to have faster parameter estimations and more information to define the priors of an \ac{mcmc}. Subsequently, \ac{ga}s can become very useful in tackling cosmological tensions and systematics, especially given the fact that biases might not be traceable by other methods. This has been teased out in recent results to be discussed.

Ref.~\cite{Lodha:2023jru} used \ac{ga}s to search for local features in the primordial power spectrum using Planck data. \ac{ga}s were applied as a reconstruction tool by using its main ingredients in a grammar space of functions, such that the overall implementation looks for the most suitable parametrization of the power spectrum all while constraining the parameters of each parametrization. It resulted in significant improvements to the fit, compared with prior Planck likelihoods \cite{Lodha:2023jru}. Such local features were shown to be able to reduce simultaneously both the Hubble and the $\sigma_8$ tensions, giving a possible unified fundamental solution to the cosmic tensions traceable to inflation \cite{Antony:2022ert, Hazra:2022rdl}. Notably, \ac{ga}'s capacity to break degeneracy between otherwise acceptable solutions has proven to be an invaluable lens for identifying local features in the primordial power spectrum.

On the other hand, Ref.~\cite{Aizpuru:2021vhd} used \ac{ga}s to constrain the comoving sound horizon at the baryon drag epoch. They also checked it against traditional recombination codes or numerical fitting via the Eisenstein-Hu approach. \ac{ga}s were again applied in an independent cosmological model fashion by working on grammar functions. Subsequently, \ac{ga} constraints to the sound horizon at baryon drag prevented usual cosmological parameter biases that affect traditional methods. Because the Hubble constant from \ac{bao} measurements relies on constraints to the sound horizon at baryon drag epoch, \ac{ga}s may well be able to help in understanding the cosmic tensions and the role of systematics. The observable is also directly influenced by the matter density, giving this a route for \ac{ga} to chip into the $\sigma_8$ tension.

Ref.~\cite{Arjona:2021mzf} furthermore used \ac{ga}s to test cosmology at the perturbative level. They used \ac{ga}s to constrain growth rate observations, based on a synthetic \lcdm\ cosmological model. \ac{ga}s led easily to the reconstruction of the underlying cosmology and ruled out other often considered phenomenological and \ac{mg} models including $w$CDM, designer $f(R)$, and the Hu-Sawicki model. This impressive application highlights another aspect of \ac{ga}s---the ability to see a signal through noise, even when the noise is as huge as in growth rate observations. \cite{Gangopadhyay:2023nli} used \ac{ga}s together with cosmological background and perturbations data, leading to a support for phantom \ac{de} behavior in the dark sector in order to alleviate the cosmic tensions.

Ref.~\cite{Bernardo:2022ggl} used a \ac{ga} variant ABC to pit together Hubble constant priors that are representative of the Hubble tension. This has seen the solution consistently evolve toward the direction of the Planck values, independent of the shape of the priors and the data sets (\ac{cc}s, \ac{sn}, and \ac{rsd}) used in combination. However, this is heavily influenced by the use of \ac{cc}s which were considered as a baseline data set in the analysis, since \ac{cc}s prefer low values of the Hubble constant. Nonetheless, this serves as another example of when \ac{ga}s can be used to choose which priors are more consistent for given data sets. An earlier work has used ABC in the same vein \cite{Bernardo:2022pyz}, but in a more extensive way by using the \ac{ga} variant to select between \lcdm\ and phenomenological parameterizations of \ac{de}, sometimes referred to as $X$CDM. The results of this work have been quite intriguing, all evolved toward $X$CDM with a preference for Hubble constant values consistent with the Planck constraint, regardless of the choice of priors, including Hubble constant priors, and data sets considered. However, this has similarly relied on \ac{cc}s as an anchor of the data, and the resulting low values of the Hubble constant may be traced to this. 

\begin{figure}[ht!]
    \centering
    \begin{subfigure}{0.475\linewidth}
	   \includegraphics[width=\linewidth]{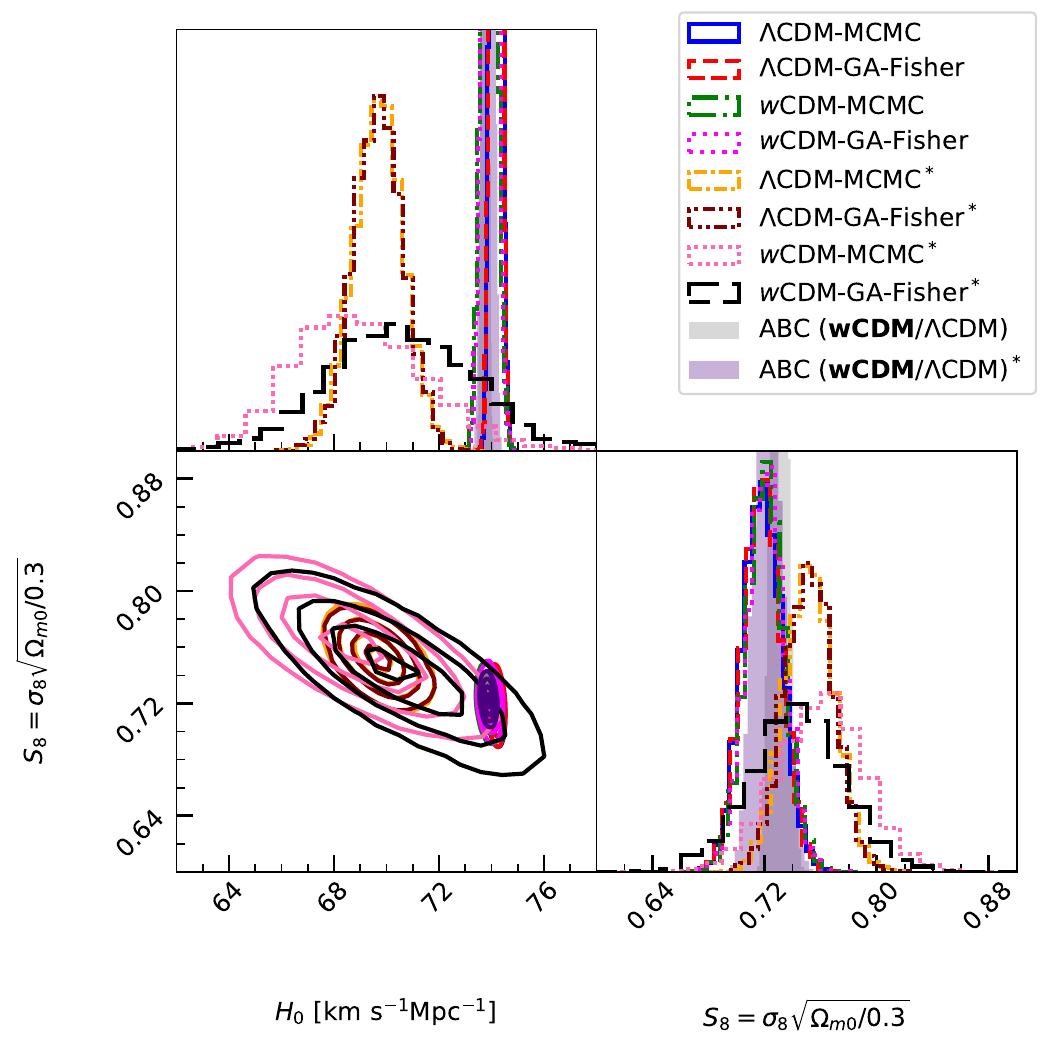}
    \end{subfigure}
    \begin{subfigure}{0.475\linewidth}
	   \includegraphics[width=\linewidth]{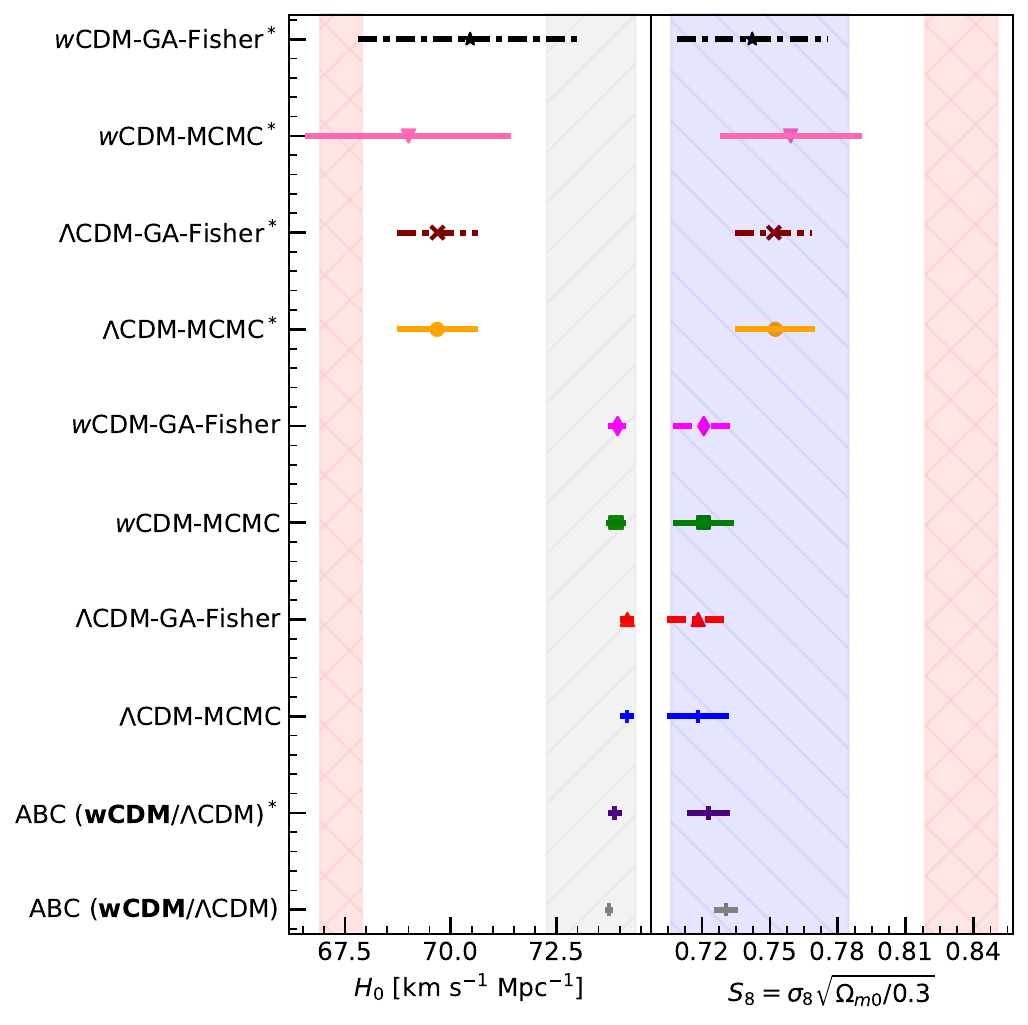}
    \end{subfigure}
    \caption{$H_0$ and $S_8=\sigma_8 \sqrt{ \Omega_{\rm m,0}/0.3 }$ constraints on a spatially-flat \lcdm\ and $w$CDM models using \ac{mcmc}, \ac{ga}, and ABC. Data sets used are \ac{cc}s \cite{Moresco:2020fbm}, standardized distances (Pantheon$+$/SH$0$ES \cite{Brout:2021mpj, Brout:2022vxf, Scolnic:2021amr}), growth rate measurements (compiled in Ref.~\cite{Kazantzidis:2018rnb}), and baryon acoustic oscillations (\ac{desi} year-1 \cite{DESI:2024mwx}). Red vertical bands are from Planck \cite{Planck:2018vyg}, blue from \ac{kids}-450 \cite{Kuijken:2015vca, Hildebrandt:2016iqg, FenechConti:2016oun, Joudaki:2016kym}, and gray from Ref.~\cite{Riess:2021jrx}. An asterisk in the superscript in the label denotes that the Pantheon$+$/SH$0$ES data was not considered. In the ABC label, the bold faced model is the preferred one.}
    \label{fig:ga_h0s8_constraints}
\end{figure}

Fig.~\ref{fig:ga_h0s8_constraints} presents a comparison of $H_0$ and $S_8=\sigma_8 \sqrt{ \Omega_{\rm m,0}/0.3}$ \cite{DiValentino:2020vvd} constraints from \ac{mcmc}, \ac{ga}, and ABC-SMC using a combination of late-time data. The results highlight the ability of \ac{ga}s, discussed extensively in this review, to address cosmological tensions by corroborating \ac{mcmc} outcomes and even providing its own distinctive results. All cases examined show remarkable consistency between \ac{ga}s and \ac{mcmc}, with \ac{ga}s successfully capturing correlations within the parameter space. Notably, all independent methods considered generally support a low $S_8$ and a high $H_0$, although this conclusion is contingent on the datasets used. Additionally, ABC, which compares \lcdm\ and $w$CDM while constraining their parameters, demonstrated narrower posterior distributions compared to \ac{mcmc}, particularly when the algorithm extends over numerous generations before selecting a preferred model \cite{Bernardo:2022pyz}. Nonetheless, ABC constraints remain consistent due to the method's capacity to mitigate prior dependence \cite{Bernardo:2022pyz, Bernardo:2022ggl}.

To sum up, \ac{ga}s offer several significant advantages as they are particularly effective for solving complex optimization problems where traditional methods falter. \ac{ga}s are robust and flexible, capable of finding optimal or near-optimal solutions in large, multi-dimensional search spaces. \ac{ga}s are also inherently parallel, allowing to explore multiple solutions simultaneously and increase the likelihood of discovering high quality solutions efficiently. Understandably, \ac{ga}s also have some disadvantages, such as computational efficiency, requiring significant processing power and time, especially for large-scale problems. Its performance also relies heavily on the choice of its hyperparameters such as population size, mutation rate, and crossover rate. Regardless, this is certainly one method that the community may find worth further investing into, for its flexibility and applicability to an incredible range of scenarios, embodied by the various ways \ac{ga}s and their variants have been applied in astrophysics and cosmology.

When looking for a needle in a haystack, one may turn to \ac{mcmc} to identify the most probable region where the needle could be. Reversely, one may turn to \ac{ga}s to identify this needle, systematically sifting through the haystack and evolving toward the solution with precision and persistence. This needle may well be the solution to the cosmic tensions, if not a huge hint to this cosmological conundrum.

\bigskip
\subsection{Inference from cosmological simulations \label{sec:Inference_cosmic_sim}}

\noindent \textbf{Coordinator:} Lei Zu\\
\noindent \textbf{Contributors:} Alan Heavens, Andrew Liddle, Anto Idicherian Lonappan, Antonio da Silva, Benjamin L'Huillier, Chi Zhang, Daniela Grandón, David Benisty, Elena Giusarma, Filippo Bouchè, Houzun Chen, Jenny G. Sorce, Jenny Wagner, Jurgen Mifsud, Luz \'Angela Garc\'ia, Marika Asgari, Nikolaos E. Mavromatos, Oleksii Sokoliuk, Ruth Lazkoz, Vasiliki A. Mitsou, and Wojciech Hellwing
\\

\noindent In this section, we explore cosmological simulation and their connection to cosmological tensions. We begin by introducing the role of cosmological simulations in modeling nonlinear scale physics and the relevant tools. Next, we examine how cosmological simulations serve as a powerful tool for investigating physics beyond the \lcdm. Finally, we discuss key observations and associated tensions at small scales, showing how cosmological simulations bridge theoretical predictions and observational data.

\subsubsection{Cosmological simulation}

When studying the large-scale structure evolution of the Universe, 
the density perturbation $\delta=\frac{\delta\rho_{\rm m}}{\rho_{\rm m}}$ becomes much greater than unity ($\delta \gg 1$) at small scales like $k \gtrsim 1 h/\rm{Mpc}$. In this nonlinear regime, linear perturbation theory fails to accurately describe the evolution of structures. To overcome this limitation, $N$-body simulations, a particle based method, are employed to model the behavior of phase space particles-representing a group of \ac{dm} particles within a given cosmological framework. Additionally, some cosmological simulations incorporate hydrodynamics to account for the effects of baryons. These simulations produce detailed predictions of structure formation, enabling direct comparisons with observational data. While computationally intensive, cosmological simulations are essential for understanding the Universe at scales dominated by nonlinear effects. They serve as a critical link between theoretical cosmological models and observations. By providing insights into nonlinear physics, cosmological simulations play a pivotal role in addressing tensions between the standard \lcdm\ model and observational data.

\subsubsection{Nonlinear effects}

$N$-body simulations describe structure formation at scales dominated by the nonlinear effects of gravity. In these simulations, ``particles'' represent \ac{dm}, characterized by properties such as mass, position and velocity. The initial conditions for these simulations are derived from theoretical models of the early Universe, typically set at high redshifts ($z \sim 100$) when the Universe was nearly homogeneous. On small scales like $k \gtrsim 10 h/\rm{Mpc}$, hydrodynamic effects become significant. The inclusion of hydrodynamics is necessary in simulations to accurately capture the influence of baryonic processes. \textbf{GADGET} (GAlaxies with \ac{dm} and Gas intEracT) is one of the most widely used $N$-body and hydrodynamics simulation codes~\cite{Springel:2000yr,Springel:2005mi,Springel:2020plp}. It combines $N$-body methods for gravitational interactions with Smoothed Particle Hydrodynamics (SPH) for modeling gas dynamics, enabling the study of both collisionless systems (e.g., \ac{dm}) and hydrodynamic phenomena. Building on \textbf{GADGET}, other advanced codes have been developed, such as \textbf{AREPO}~\cite{Springel:2009aa}, which uses a moving mesh approach instead of SPH, and \textbf{GIZMO}~\cite{Hopkins:2014qka}, which replaces modern SPH with a mesh-free hydrodynamics solver for improved accuracy. \textbf{SWIFT} is also a fully open-source highly-parallel, versatile, and modular coupled hydrodynamics, gravity, cosmology, and galaxy-formation code~\cite{SWIFT:2023dix}. In addition, Adaptive Mesh Refinement based codes such as \textbf{RAMSES} (Refined Adaptive Mesh with Static Expansion)~\cite{Teyssier:2001cp} are also extensively used for cosmological and astrophysical simulations. Due to the computational intensity of high-resolution simulations, several groups have made their results publicly available. Examples include the \textbf{Millennium Simulation}~\cite{Springel:2005nw}, \textbf{Horizon Run Simulations}~\cite{Kim:2008kf}, \textbf{Illustris and IllustrisTNG}~\cite{Vogelsberger:2014dza,Springel:2017tpz}, \textbf{ Bolshoi Simulation}~\cite{Klypin:2010qw}, \textbf{EAGLE}~\cite{Schaye:2014tpa,Crain:2015poa}, \textbf{Quijote Simulations}~\cite{Villaescusa-Navarro:2019bje}. Researchers can utilize these simulated results for their own science targets. 

\begin{figure}[t!]
\centering
\includegraphics[scale = 0.8]{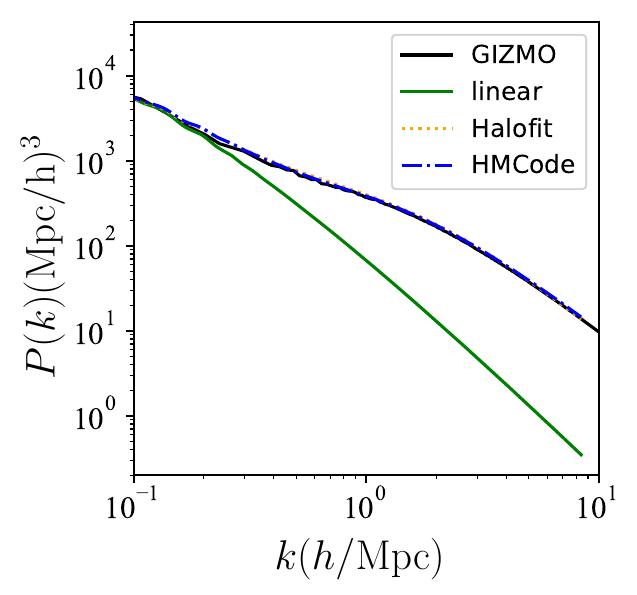}
\caption{The matter power spectrum with the same \lcdm\ cosmological parameters for $k \sim 0.1-10$ $h/\rm{Mpc}$ at $z=0$. The green solid line shows the linear results. The black solid line represents the $N$-body simulation results by using the publicly available code \textbf{GIZMO} while the orange dotted (blue dash-dotted) line is calculated by the nonlinear analytical code Halofit (HMCode).
}
\label{Fig:simulation}
\end{figure}

Using some $N$-body simulation results as benchmarks~\cite{Jenkins:2000bv,Kauffmann:1999fq}, 
several semi-analytic models are developed to predict the nonlinear matter power spectrum,
such as Halofit~\cite{Smith:2002dz,Takahashi:2012em} and HMCode~\cite{Mead:2015yca,Mead:2020vgs}.
These semi-analytic models significantly reduce computational time while remaining applicable across a wide range of cosmological parameters. 
Halofit employs simulation results to construct fitting formulas, while HMCode, a refined version of the halo model, introduces modifications that enhance its consistency with simulation outcomes. Additionally, \ac{ml} techniques have been increasingly applied as emulator to predict nonlinear structure formation. For example, the \textbf{Quijote Simulations} have been widely employed in \ac{ml} applications for cosmology, providing large-scale training datasets for deep learning models aimed at improving parameter inference and nonlinear structure formation predictions~\cite{Hortua:2021vvj,Lazanu:2021tdl}. 

Fig.~\ref{Fig:simulation} shows the matter power spectrum in the standard \lcdm\ model for $k \sim 0.1-10 h/\rm{Mpc}$ at $z=0$. The linear matter power spectrum (green solid line) significantly underestimates the power at nonlinear scales  $k>0.1h/\rm{Mpc}$. Nonlinear results from $N$-body simulations (black solid line, GIZMO) align closely with analytical models such as Halofit (orange dotted line) and HMCode (blue dash-dotted line). Within the \lcdm\ framework, both Halofit and HMCode achieve accuracy within 10\% compared to the $N$-body simulation. HMCode also exhibits superior performance when incorporating extensions beyond \lcdm, such as baryonic feedback, \ac{wdm}, massive neutrinos, \ac{dde}, and \ac{mg}~\cite{Mead:2015yca,Mead:2020vgs}. However, if the initial matter power spectrum or the late-time dynamical evolution significantly deviates from the \lcdm\ scenario, the reliability of HMCode predictions is doubtful. As shown in Fig.~\ref{Fig:compare}, when addressing new physics beyond \lcdm, at nonlinear scales, both HMCode and Halofit fail to provide accurate nonlinear corrections, see Ref.~\cite{Zhang:2024mmg} for details. In such cases, conducting cosmological simulations correctly becomes essential to study the structure evolution in the specific model.

\begin{figure}[t!]
\centering
\hspace{0.9cm}
\includegraphics[scale = 0.4]{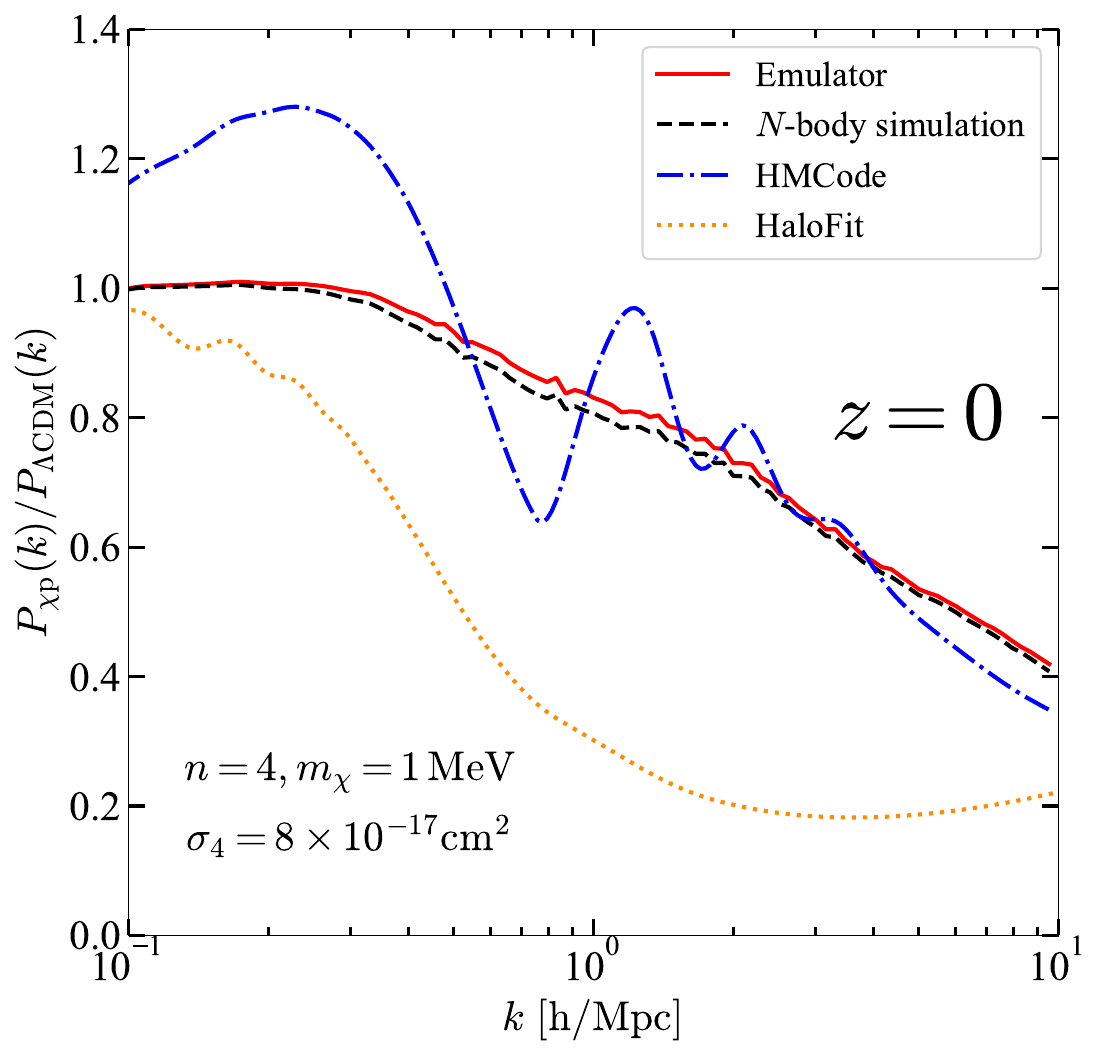}
\caption{The ratio of the nonlinear matter power spectrum
at $z=0$ in \ac{dm}-proton scattering scenario~\cite{Zhang:2024mmg}. The red solid lines denote the emulator developed in Ref.~\cite{Zhang:2024mmg}, the black dashed lines are the $N$-body simulation results from GIZMO, the blue dashed-dotted lines are the correction from HMCode and the orange dotted lines are the correction from HaloFit.}
\label{Fig:compare}
\end{figure}

\subsubsection{Cosmological simulations beyond \texorpdfstring{$\Lambda$}{lambda}CDM}

To explore new physics beyond the \lcdm\ framework at nonlinear scales, various groups have developed cosmological simulations in different cosmological scenarios.  One of the most studied extensions is the \ac{wdm} model. The free-streaming motion of \ac{wdm}, due to its thermal velocity, suppresses structure formation on small scales. Numerous cosmological simulations of \ac{wdm}~\cite{Lovell:2013ola,Bose:2015mga,Maio:2014qwa,Shtanov:2023xcs} have investigated the resulting mass distributions and halo properties. These studies reveal suppressed mass distributions below a scale determined by the mass of \ac{wdm} particle, and shallower density slopes systematically compared to \ac{cdm}.

Another critical area of interest is the total mass of neutrinos, which influences the evolution of the total energy density and introduces free-streaming effects. Several studies have incorporated massive neutrinos into cosmological simulation codes using additional particle sets or relativistic approaches~\cite{Villaescusa-Navarro:2013pva,Adamek:2017uiq, Liu:2017now}. A higher total neutrino mass results in a stronger suppression of the matter power spectrum at small scales. For instance, the suppression of the total matter power spectrum at $k \sim 1 h/\rm{Mpc}$ is approximately 5\% for $m_\nu=0.06\,\rm{eV}$, and increases to 25\% for $m_\nu=0.3\,\rm{eV}$, see Ref.~\cite{Adamek:2017uiq} for details.

In addition to neutrino effects, cosmological simulations have also been extended to study alternative gravity models and \ac{de} effects beyond \ac{cdm}. Various computational tools have been developed to simulate structure formation under \ac{mg} theories, including \textsc{ecosmog} \cite{Li:2011vk}, \textsc{mg-gadget} \cite{Puchwein:2013lza}, \textsc{me-gadget} \cite{Zhang:2018glx}, and \textsc{mg-arepo} \cite{Arnold:2019vpg}, \textsc{mg-pcola} \cite{Wright:2017dkw}, \textsc{mg-quijote} \cite{quijote_mg}. \textsc{ECOSMOG} incorporates \ac{mg} models, including $f(R)$ gravity, chameleon models, and symmetron theories, using an AMR framework to achieve high-resolution nonlinear simulations. \textsc{me-gadget} allows for an arbitrary phenomenological (with varying $H(a)$ and $G_{\rm N}(a)$) and \ac{idm} (with varying $m_{\rm DE}(a)$) models to be used in the gravity-only simulation setting. \textsc{mg-pcola} is a fast N-body solver extending the COLA method to \ac{mg}, balancing speed and accuracy for large-scale structure studies. Recent \textsc{mg-pcola} simulations incorporating massive neutrinos reveal that their inclusion leads to additional suppression of structure growth. In particular, in scalar-tensor models, this suppression enhances the damping of small-scale power beyond what is expected from standard neutrino free-streaming, reinforcing the interplay between neutrinos and \ac{mg}. \textsc{mg-quijote} is a large suite of \ac{mg} simulations tailored for \ac{ml} applications and parameter inference, covering various \ac{mg} scenarios and offering a benchmark dataset for cosmological studies.

Furthermore, if \ac{dm} has additional non-gravitational interactions—for instance, with baryons/electrons~\cite{Chen:2002yh,Nguyen:2021cnb,Zhang:2024mmg}, neutrinos~\cite{Serra:2009uu,Mangano:2006mp,Wilkinson:2014ksa}, or dark radiation~\cite{Cyr-Racine:2015ihg,Vogelsberger:2015gpr,Zu:2023rmc}—this will introduce additional pressure that suppresses small-scale structure formation. These interactions modify the Boltzmann equation in the early Universe by adding energy and momentum exchange terms, creating a pressure that counteracts gravitational collapse. This suppression occurs primarily during the early Universe, when matter densities are high enough to trigger collisions. As the Universe expands, the interaction rate falls below the Hubble rate, i.e., \ac{dm} decouples. After that, the interactions become negligible on galactic scales. 
Thus the primary deviations between \ac{idm} and \ac{cdm} arise during the early Universe, influencing the initial conditions for structure formation. A common approach involves calculating the linear evolution of the matter power spectrum first, including \ac{dm} interactions, during the early phase when density perturbations remain small and linear theory is applicable. These results are then used as initial conditions (at $z\sim 100$) for later cosmological simulations when \ac{dm} interactions can be neglected. Notably, interactions often imprint unique oscillation features in the initial power spectrum, similar to those observed in baryons. However, these oscillations diminish during nonlinear evolution as the Universe evolves. As a result, these distinctive oscillation structures are expected to be observable primarily at high redshifts~\cite{Bohr:2020yoe,Zu:2023rmc,Zhang:2024mmg}.

\subsubsection{Cosmological simulations in nonlinear observation and cosmic tension}

Cosmological simulations model particle behavior and predict \ac{dm} properties at nonlinear scales, providing a crucial tool for comparing theoretical predictions with observations such as \ac{wl}, Lyman-$\alpha$, halo density profile, upcoming cosmic 21cm, and so on. Various discrepancies between \lcdm\ model and observations have emerged at different scales, including the $S_8$ tension, core-cusp problem, too-big-to-fail problem, the diversity of dwarf galaxy rotation curves, missing satellites problem, and so on. Cosmological simulations serve as a powerful bridge between theory and observations, helping to address discrepancies and exploring the physics beyond standard \lcdm. In this section, we discuss key nonlinear observations and the related tensions between \lcdm\ and the measurements.

\paragraph{Weak lensing}

\ac{wl} provides a direct method for mapping the late-time large-scale structure of the Universe by statistically analyzing the shape distortions of numerous background galaxies induced by foreground matter fields. By calculating the two-point correlation functions of the shapes of galaxy pairs, \ac{wl} measurements have widely been used for exploring the matter distributions in our local Universe. \ac{wl} is sensitive to the scale down to $k \gtrsim 1~h/\rm{Mpc}$ at low redshift(typically $z<2$), where nonlinear gravitational evolution significantly affects the matter power spectrum~\cite{HSC:2018mrq,Hamana:2019etx,DES:2021bvc,DES:2021wwk,KiDS:2020suj}. On smaller scales ($k \gtrsim 10~h/\rm{Mpc}$), baryonic processes, such as gas cooling, feedback from star formation, and \ac{agn}, leading significantly systematic uncertainties to \ac{wl} measurement. To mitigate uncertainties in baryonic physics, many analyses mask or marginalize over these small-scale regions~\cite{DES:2021bvc,DES:2021wwk,Garcia-Garcia:2024gzy}. Several other approaches have been developed to address this uncertainty. For example, one method that preserves small-scale measurements is the Bayesian Model Averaging (BMA) framework presented in Ref.~\cite{Grandon:2024tud}. In this approach, BMA combines the individual posterior distributions of multiple competing baryonic feedback models derived from hydrodynamical simulations to provide robust cosmological constraints. Furthermore, the nonlinear evolution of cosmic structures at late-times introduces strong non-Gaussianities in the matter density field, requiring estimators beyond two-point functions to capture this additional information \cite{Euclid:2023uha}. These estimators, often referred to as higher-order or non-Gaussian statistics, are useful for breaking parameter degeneracies \cite{2020PhRvD.102j3531A, 2018A&A...619A..38P}, self-calibrating systematics \cite{Semboloni:2012yh}, and overall improving cosmological constraints compared to two-point functions \cite{Marques:2023bnr, Thiele:2023gqr, Grandon:2024pek, 2021A&A...645L..11A, 2021MNRAS.506.1623H, Liu:2014fzc}. The theoretical modeling of these estimators relies heavily on cosmological simulations, both to characterize their dependence on cosmological parameters and to build the likelihoods, where the covariance matrix is typically estimated from hundreds of simulations at a fixed cosmology \cite{Sellentin:2015waz}. An alternative to adding estimators of higher-order statistics is to use the full information from the field using Bayesian hierarchical modeling, e.g., see Ref.~\cite{Jasche:2012kq,Jasche:2018oym,Porqueres:2020wwf,Porqueres:2021clw,Porqueres:2023drp}, or Bayesian simulation-based inference based on neural compression of the field, e.g., see Ref.~\cite{Makinen:2024xph,DES:2024xij}, which avoid the need to estimate covariance matrices. Both these approaches rely heavily on simulations. We refer the reader to Sec.~\ref{sec:obs_H0} for a detailed discussion on \ac{wl}.  

In \lcdm\ framework, \ac{wl} surveys have reported an $S_8$ tension (see Sec.~\ref{sec:obs_H0} for details): the $S_8$ parameter measured from \ac{wl} survey is lower than the value inferred from Planck \ac{cmb} data, with a significance of 2-3$\sigma$~\cite{MacCrann:2014wfa,Hildebrandt:2018yau,DES:2021wwk}. Despite some uncertainties in observational measurements, various beyond \lcdm\ models have been proposed to address this issue, such as suppressed small-scale matter power due to interactions between \ac{dm} and dark radiation \cite{Rubira:2022xhb,Zu:2023rmc,Brinckmann:2022ajr,DES:2020mpv}, \ac{dm}-baryon interactions~\cite{He:2023dbn}, or \ac{dm}-neutrino scattering~\cite{DiValentino:2017oaw,Zu:2025lrk}. Cosmological simulations play a crucial role in addressing these issues. Upcoming more precise large scale surveys like Vera C. Rubin Observatory \cite{LSST:2008ijt}, Euclid \cite{EUCLID:2011zbd} and China Space Station Telescope (CSST) \cite{Zhan:2021} are expected to provide crucial insights and help resolve these tensions.

\paragraph{Lyman-$\alpha$ forest}

Lyman-$\alpha$ forest refers to a series of absorption lines observed in the spectra of distant \ac{qso}s. These lines are caused by the interaction of light from the \ac{qso} with intervening clouds of neutral hydrogen gas along the line of sight. As a result, the Lyman-$\alpha$ forest is widely used as a tracer of neutral hydrogen gas distribution and, by extension, the underlying matter distribution, assuming a correlation between mass and neutral gas. The Lyman-$\alpha$ forest probes the matter power spectrum in the mildly nonlinear regime over a broad range of redshifts, down to scales of $ 1 - 80\,h/ \rm{Mpc}$ (at $z = 2-6$ in ground-based observation) \cite{Croft:2000hs,Viel:2008ige,Viel:2013fqw}. Compared to \ac{wl} surveys, the Lyman-$\alpha$ forest is sensitive to smaller scale structures at higher redshifts. Since the Lyman-$\alpha$ forest mainly traces the \ac{igm} in the cosmic web, directly correlated with the neutral hydrogen gas distribution, its signal is highly influenced by the baryonic physics~\cite{Viel:2004bf}. High-precision hydrodynamical simulations that accurately account for baryonic physics are essential for studying the Lyman-$\alpha$ forest. Using these simulation results, researchers can investigate physics beyond the \lcdm, including \ac{wdm}~\cite{Viel:2005qj,Boyarsky:2008xj,Viel:2013fqw,Palanque-Delabrouille:2019iyz}, fuzzy \ac{dm}~\cite{Irsic:2017yje,Kobayashi:2017jcf,Rogers:2020ltq} and neutrino mass~\cite{Garny:2020rom,Palanque-Delabrouille:2015pga}. Recent simulations suggest that small tensions may exist between \lcdm\ cosmological simulation and Lyman-$\alpha$ observation (see Sec.~\ref{sec:Anomalies_lyman_alpha} for the anomalies in Lyman-$\alpha$ measurements), which can be explained by mechanisms such as the dark photon heating or \ac{dm}-neutrino interaction~\cite{Bolton:2022hpt,Hooper:2021rjc}. An improved understanding of baryonic physics is crucial to addressing these issues and refining our interpretation of Lyman-$\alpha$ observation.

\paragraph{Cosmic 21 cm}

The cosmic 21 cm signal refers to the hyperfine transition of neutral hydrogen atoms, which emits or absorbs radiation at a wavelength of 21 cm (frequency of 1.42 GHz). This signal is a powerful tool for probing the evolution of the Universe at high redshift, across a wide range of cosmic epochs, from the Dark Ages (prior to star formation) to the \ac{eor} and beyond \cite{Furlanetto:2006jb,Pritchard:2011xb}. The 21 cm signal traces the distribution of matter and can map the evolution of large-scale structures over cosmic time. Although the angular resolution of the 21 cm signal is limited to tracing ionized bubbles on scales of several Mpc, which corresponds to very large structures, it remains a powerful tool for exploring the nonlinear scale structure. The small scale structure significantly impacts the halo collapsing. In general, any model that suppresses or modifies the amplitude of \ac{dm} fluctuations on small scales could affect the 21 cm cosmic dawn signal. Several previous works have discussed using the future cosmic 21 cm data to probe the physics beyond \lcdm~\cite{Schneider:2018xba,Lopez-Honorez:2018ipk,Escudero:2018thh,Munoz:2019hjh,Yoshiura:2019zxq,Munoz:2020mue}. One key advantage of the cosmic 21 cm signal is its ability to trace the mass distribution at very high redshifts, where the original information is preserved due to limited impact from structure formation~\cite{Bohr:2020yoe,Lovell:2017eec}. By modeling the dynamics of \ac{dm} and the feedback of baryons, cosmological simulations play a pivotal role in understanding and interpreting the cosmic 21 cm signal~\cite{Hassan:2015aba,Bohr:2020yoe,Munoz:2020mue}. Upcoming cosmic 21 cm observations, such as the Low-Frequency Array (LOFAR) \cite{LOFAR:2013jil}, the Hydrogen Epoch of Reionization Array (HERA) \cite{DeBoer:2016tnn}, and the \ac{ska}~\cite{Koopmans:2015sua}, are expected to open a new window into the high redshift Universe and provide novel insights into the nature of \ac{dm}.

\paragraph{Dwarf galaxy scale puzzles}

At small scales, such as those of dwarf galaxies, significant tensions exist between predictions from $N$-body simulations and observational data~\cite{Bullock:2017xww}. Several key discrepancies are summarized below:

\begin{description}
\item[Core-cusp Problem] Gravity only $N$-body simulations of \ac{dm} in a \ac{cdm} Universe predict that \ac{dm} halos should have a ``cusp''—a sharp rise in density toward their center, known as Navarro-Frenk-White profile, or NFW~\cite{Navarro:1995iw}. However, observations of low-mass galaxies of different luminosity and morphology \cite{Salucci:2018hqu} indicate that their \ac{dm} halos instead have a core, a central region with nearly constant density~\cite{deBlok:2009sp,Gentile:2004tb}. 

\item[Diversity Problem for Rotation Curves] Simulations predict that galaxies with the same maximum circular velocity should have similar halo density profiles. However, observations reveal a significant variation in the central densities of these galaxies~\cite{Bullock:1999he,Oman:2015xda}.

\item[Missing Satellites Problem] Simulations predict a large number of small \ac{dm} halos, but observations show a much smaller number of dwarf satellite galaxies around larger galaxies, such as the \ac{mw}~\cite{Klypin:1999uc,Moore:1999nt}.

\item[Too-Big-to-Fail Problem] The number of massive \ac{dm} subhalos predicted by simulations is inconsistent with the number of observed bright satellite galaxies in systems like the \ac{mw}~\cite{Boylan-Kolchin:2011qkt,Boylan-Kolchin:2011lmk}.
\end{description}

These small-scale issues highlight a disconnect between $N$-body simulations and observations, particularly in our local group. They suggest that gravity-only \ac{cdm} $N$-body simulations are insufficient to fully describe the Universe on these scales. Several hypotheses have been proposed to resolve these discrepancies, involving both astrophysical processes and modifications to the standard \lcdm\ model. Stellar and supernova feedback can redistribute baryonic matter, influencing the gravitational potential and altering the density profile of the \ac{dm} and the halo mass function~\cite{Navarro:1996bv,Oh:2010mc}. The feedback process can also eject gas from massive subhalos, suppressing star formation to explain the too big to fail problem~\cite{Zolotov:2012xd,Arraki:2012bu,Brooks:2012vi,Brook:2014hda,Dutton:2015nvy}. Another way to solve this small-scale puzzle requires modifications to \lcdm. Several models have been proposed such as Self-Interacting \ac{dm} (SIDM)~\cite{Spergel:1999mh} and fuzzy DM~\cite{Hu:2000ke}. SIDM introduces \ac{dm}-\ac{dm} interactions that redistribute energy, forming cores in halos, and leading to more diverse halo density profiles and a different subhalo mass function. $N$-body simulations incorporating SIDM have been used to test these predictions and compare them with the observations. Many SIDM simulations were performed and used to discuss the \ac{dm} halo distribution after the first proposal~\cite{Moore:2000fp,Yoshida:2000bx,Burkert:2000di,Kochanek:2000pi,Yoshida:2000uw,Dave:2000ar}. Early studies constrained the SIDM cross section through observations, such as strong lensing~\cite{Miralda-Escude:2000tvu}. The results show that the parameter space for the SIDM model to solve the small scale problem have already been constrained. However, subsequent simulations with higher resolution and improved algorithms revealed that earlier constraints may have been overestimated~\cite{Rocha:2012jg,Peter:2012jh,Vogelsberger:2012ku,Elbert:2014bma}. For fuzzy \ac{dm}, the introduction of ultra-light bosons with kiloparsec-scale wavelength naturally suppresses structure formation at these scales. Unlike classical particles, fuzzy \ac{dm} behaves more like a wave, requiring the solution of quantum-mechanical wave equations in $N$-body simulations. Studies have shown that fuzzy \ac{dm} naturally forms solitonic cores at the centers of halos due to quantum pressure while simultaneously suppressing small-scale structure formation as a consequence of the uncertainty principle~\cite{Schive:2014dra,Mocz:2017wlg,Veltmaat:2018dfz}. This provides a compelling explanation for the core-cusp problem and the missing satellites problem. For now, resolving small-scale issues remains an ongoing challenge, as both baryonic feedback processes and alternative \ac{dm} properties beyond \ac{cdm} have their own strengths and limitations. High-resolution cosmological simulations, combined with advanced algorithms to incorporate baryonic physics and additional \ac{dm} interactions, provide a crucial tool for understanding these small-scale puzzles. These simulations become a bridge between the theoretical predictions from proposed models and the observational data. They help us better understand the existing cosmological tensions.
\bigskip
\subsection{Profile likelihoods in cosmology\label{sec:frequen_approach}} 

\noindent \textbf{Coordinator:} Adrià Gómez-Valent\\
\noindent \textbf{Contributors:} Elisa G. M. Ferreira, Emil Brinch Holm, Giacomo Galloni, Laura Herold, Paolo Campeti, Sophie Henrot-Versill\'e, Vivian Poulin, and William Giarè
\\

\subsubsection{Motivation}

In cosmology, it is very common to encounter high-dimensional parameter spaces. Efficient exploration of these spaces through the sampling
of their corresponding multivariate distributions is commonly achieved using Monte Carlo techniques. In these analyses, statistical information is contained in the resulting \ac{mcmc}, from which one can infer the posterior distribution and other derived Bayesian products. In parameter spaces of dimension greater than two one cannot visualize the posterior distribution in the original parameter space. One needs to apply the so-called marginalization technique to project the results into spaces of lower dimension and compute the marginalized posterior distributions of the individual parameters and their constraints at the desired confidence level. This is also done to obtain the contour plots in all the two-dimensional planes of interest. Marginalization is extremely efficient and is a very practical tool that eases the visualization and interpretation of the results in Bayesian analyses. However, when the original (non-marginalized) \ac{pdf} has non-Gaussian features, the marginalized posterior distributions obtained do not exclusively reflect how the parameters are distributed based on their ability to explain the data but also according to their integrated probability weight. This introduces volume (or marginalization) effects, which can potentially bias conclusions in a significant way. In other words, if deviations from Gaussianity are large, the marginalized posterior can hide points in parameter space that fit the data well. Another potential drawback of the marginalization method is the impact of the prior, which, even if assumed to be flat, is still informative in all cases. For instance, a flat prior on a variable $x$ or on its logarithm, $\log x$, might induce important changes in the constraints of a model, even if the boundaries of the priors are perfectly consistent. This is due to volume effects, again. Hence, volume effects can significantly impact the interpretation of results extracted from \ac{mcmc}s. Detecting these biases is of utmost importance. In the context of cosmology, accurately quantifying these biases is essential, e.g., for a correct assessment of cosmological tensions in the \lcdm\ model and beyond, as well as in data compression performed by galaxy and \ac{wl} surveys, see Sec.~\ref{sec:application} for details. 

A mismatch between the point in parameter space that maximizes the original posterior distribution and the point constructed with the values of the parameters that maximize the individual one-dimensional posteriors already hints at the non-negligible impact of marginalization effects. One can investigate the impact of marginalization effects using the profile likelihood (PL). The PL is a method from frequentist statistics to construct parameter intervals, and therefore, by construction inherently independent of a prior. It ensures, in particular, that the results (on the parameter inference) do not depend on the choice of the parameterization of the problem. This part of the work aims to explain the basics of PLs and is organized as follows. In Sec.~\ref{sec:CI} we describe how to build confidence intervals using PLs. In Sec.~\ref{sec:calculation} we describe several methods to compute them, both approximately from the \ac{mcmc}s or, more precisely, using advanced numerical optimization techniques with the help of concrete codes. Finally, in Sec.~\ref{sec:application} we give some examples of applications of the PLs in cosmology, providing a rich list of references.

\subsubsection{Confidence intervals from Profile Likelihoods}\label{sec:CI}
 
While Bayesian credible intervals are statements about the degree of belief of the value of the true parameter, frequentist confidence intervals are based on the \textit{coverage}: frequentist confidence intervals are said to \textit{cover} the true value of the parameter with a confidence level $\alpha$ if -- after repetition of the experiment -- a fraction $\alpha$ of the experiments contain the true value of the parameter (e.g., see Ref.~\cite{ParticleDataGroup:2022pth} for a review, or Ref.~\cite{Herold:2024enb} in the context of cosmology).  Confidence intervals with correct coverage can be constructed using the \textit{Neyman construction} \cite{Neyman:1937uhy}. This construction requires knowledge of the full \ac{pdf} $P(\hat{\boldsymbol{\mu}}|\boldsymbol{\mu}_\mathrm{true})$, where $\hat{\boldsymbol{\mu}}$ are the maximum likelihood estimates (MLE) of the model parameters $\boldsymbol{\mu}$ estimated from observations and $\boldsymbol{\mu}_\mathrm{true}$ the (hypothetical) true values of these parameters. Since this probability is typically not known a priori, it needs to be estimated by generating mock realizations of the data. This makes the Neyman construction computationally expensive. Fortunately, a theorem by Wilks \cite{Wilks:1938dza} helps in many practical applications, which states: In the limit of a large data set, the log-likelihood ratio 
\begin{equation}
    \label{eq:chi2_wilks}
   \Delta\chi^2 = -2\log R(\boldsymbol{x}, \boldsymbol{\mu})\,, \qquad R(\boldsymbol{x}, \boldsymbol{\mu}) = \frac{\mathcal{L}(\boldsymbol{x}|\boldsymbol{\mu})}{\mathcal{L}(\boldsymbol{x}|\hat{\boldsymbol{\mu}})}\,,
\end{equation}
follows a $\chi^2$-distribution. Here $\mathcal{L}$ is the likelihood function and $\boldsymbol{x}$ is the data vector. In this case, the Neyman construction can be replaced by a much simpler \textit{graphical construction} based on the profile likelihood. Wilks theorem holds trivially if the p.d.f.\ is Gaussian (for every possible value of the true parameters). We note, however, that Wilks' theorem holds for the full-dimensional likelihood (for any possible value of the true parameters) and not the profile likelihood. Although this effect vanishes in the large data limit, it is often very difficult to assess whether this limit is reached in practice~\cite{pawitan2001all}, and the confidence intervals are, therefore, only approximate (see Ref.~\cite{Herold:2024enb} for checks of the asymptotic assumptions in cosmology).

The profile likelihood for a parameter of interest $\mu$ (in one dimension) can be obtained by evaluating Eq.~\eqref{eq:chi2_wilks} for different values of $\mu$. In the presence of other cosmological parameters and nuisance parameters, $\boldsymbol{\nu}$, the profile likelihood is 
calculated by setting $\mu$ to various values within a range of interest and minimizing $\chi^2(\mu, \boldsymbol{\nu}) = -2\log\mathcal{L}(\mu, \boldsymbol{\nu})$ with respect to all parameters $\boldsymbol \nu$. The global MLE, or ``best-fit'', corresponds to the minimum value $\chi^2_{\rm min}$ by design.
A graphical frequentist confidence interval for $\mu$ can now be constructed using the difference $\Delta\chi^2(\mu) = \chi^2(\mu) - \chi^2_{\rm min}$. When $\mu$ is far from its physical boundary, a confidence interval at confidence level $\alpha$ is obtained by applying a fixed threshold $\Delta\chi^2_{\rm th}$ to $\Delta\chi^2(\mu)$. This threshold is chosen such that the cumulative distribution function of the $\chi^2$ distribution with one degree of freedom equals $\alpha$. Notable values are $\Delta\chi^2_{\rm th}=1$ for 68\% CL and $\Delta\chi^2_{\rm th}=3.84$ for 95\% CL \cite{Trotta:2017wnx}. This method is applicable to both parabolic (i.e., for a Gaussian-distributed parameter) and non-parabolic $\Delta\chi^2(\mu)$ because of the invariance of the MLE under reparameterization.

When the parameter estimate is near its physical boundary, the conventional graphical profile likelihood construction for frequentist confidence intervals can be inadequate. It may result in empty intervals and fail to maintain the frequentist coverage property if the choice between an upper limit or a two-sided interval is made based on the data (the so-called ``flip-flopping''). An adapted Neyman construction, also known as Feldman-Cousins (FC) prescription \cite{Feldman:1997qc} in the particle-physics literature, addresses these issues. To describe this method, we restrict the discussion to one-dimensional data $x$ and a single model parameter $\mu$ for clarity. For each value of $\mu$ (with an unknown true value) and each observable $x$, we compute the likelihood ratio $R(x,\mu)$, with $\mu$ (and its MLE $\hat{\mu}\equiv \mu_\mathrm{best}$) restricted to its physically allowed values. 
The \textit{confidence belt} at the desired CL $\alpha$  is constructed by selecting an \textit{acceptance interval} $[x_1, x_2]$ for each $\mu$ satisfying
\begin{equation}\label{eq:system}
    \begin{cases}
     R(x_1, \mu) = R(x_2, \mu)\,, \\
      \int_{x_1}^{x_2} P(x|\mu) \, dx = \alpha\,,
    \end{cases}
\end{equation}
where $P(x|\mu)$ is the p.d.f.\ of $x$ given $\mu$, and values of $x$ are added to the acceptance interval in order of decreasing likelihood ratio. The confidence belt is formed by the union of all acceptance intervals $[x_1(\mu), x_2(\mu)]$. Intersecting the confidence belt with a line at $x = x_0$, where $x_0$ is the value of $x$ minimizing $\chi^2$ (i.e., the value observed in an experiment), yields the confidence interval $[\mu_1, \mu_2]$ for $\mu$.
The adapted Neyman or FC prescription provides a method to determine the endpoints of confidence intervals, which smoothly transitions between an upper limit and a two-sided interval, ensuring exact frequentist coverage even for parameters with non-Gaussian p.d.f.'s (and if Wilks' theorem does not hold). This is in contrast to the conservatism (overcoverage) inherent in Bayesian limits in the same context \cite{Cousins:1994yw, Feldman:1997qc}. While overcoverage might not be as problematic as undercoverage, it reduces the ability to reject false hypotheses. This highlights the value of examining frequentist intervals.

If the p.d.f.\ of the parameter is Gaussian and the MLE of the parameter near a physical boundary at $\mu=0$, then $\mu_{\rm best}=\max(0,\,x)$ and  the likelihood ratio in simply becomes \cite{Feldman:1997qc}
\begin{equation}
    R(x, \mu) = \begin{cases} 
    \exp(-(x-\mu)^2/2), & \text{for } x \geq 0\,,\\
    \exp(x\mu-\mu^2/2), & \text{for } x < 0\,,
    \end{cases}
\end{equation}
where $x$ and $\mu$ are in units of $\sigma$, the width of the parabolic fit to $\Delta\chi^2(\mu)$. Then, the confidence interval can be derived by solving Eq.~\eqref{eq:system}, with $P(x|\mu)$ being a Gaussian with mean $\mu$ and unit variance. If, however, the p.d.f.\ of the parameter is non-Gaussian and Wilks' theorem doesn't hold, a full Neyman construction of the confidence intervals using simulations for several input values of the parameter of interest is necessary, e.g., see Ref.~\cite{LiteBIRD:2023zmo, SPIDER:2021ncy} for details.

With the necessary modifications, this entire treatment can be generalized to multiple dimensions. For example, under the assumption of Gaussianity, the iso-$\chi^2$ curves for a two-dimensional case will be ellipses on the parameter space, whose angle of the major axis depends on the correlation of the two parameters of interest. Furthermore, if no physical boundary is close to the absolute minimum of $\chi^2$ and if we consider the limit of a large data set, it is possible to adopt a graphical approach to derive the 68\% and 95\% confidence regions, as we mentioned for the one-dimensional case. This time, one must search for $\Delta\chi^2_{\rm th}=2.3$ for 68\% CL and $\Delta\chi^2_{\rm th}=5.99$ for 95\% CL regions.
Similarly to the one-dimensional case, if Wilks' theorem doesn't hold, one should adopt a full Neyman construction to extract the confidence regions.

\subsubsection{Calculation of profile likelihoods}\label{sec:calculation}

Computing a profile likelihood for a single parameter involves a series of $M$ optimizations, typically with $M\sim \mathcal{O}(10)$, in the $N-1$ dimensional parameter space. Furthermore, these cannot be reused for profiles in other parameters, unlike an \ac{mcmc}, which simultaneously produces parameter constraints for all varied parameters. Since the optimizations themselves are moderately difficult, constructing a full profile likelihood analysis of several parameters can be very computationally demanding. In this subsection, we present some of the strategies adopted in the literature to accommodate the often high computational cost of computing profile likelihoods.

\paragraph{Optimisation strategies}
A converged set of \ac{mcmc} chains of course contains information about the likelihood values at a large number of points in parameter space. Although \ac{mcmc} algorithms are poor optimisers~\cite{Hamann:2011hu}, they can be used to estimate profile likelihoods by binning their points along the parameter $\theta$ being profiled. The bin around a fixed value of $\theta=\theta_0$ can be defined, for example, by a homogeneous central binning, as is done in Refs.~\cite{Reid:2009nq,Gomez-Valent:2022hkb, Colgain:2023bge,Giare:2023qqn, Karwal:2024qpt}, or by taking the $n$ points of the \ac{mcmc} that have a value of $\theta$ closest to the fixed value $\theta_0$, as done in Ref.~\cite{Holm:2023uwa}, with $n$ a fixed fraction of the total amount of \ac{mcmc} points. Due to the finite sampling of the \ac{mcmc}, the largest likelihood value found in each bin will generally be smaller than its optimized value at $\theta_0$, which leads to noise in the profile likelihood curve. The great advantage of this approach, however, is that it has no computational cost if an \ac{mcmc} is already given, so it can be used as an inexpensive initial test to check for marginalization and prior effects~\cite{Gomez-Valent:2022hkb}, see also Ref.~\cite{Giare:2023qqn}. Furthermore, such a profile can be used as the starting point for explicit optimizations at each fixed value of the profiled parameter.

When the binned \ac{mcmc} is not adequately accurate, the profile likelihood can be computed by explicit optimization in the parameter subspace with the profiled parameter taking on a series of fixed values. Generally speaking, the most efficient optimization algorithms explore the parameter space for the highest likelihood value by using the gradient of the likelihood with respect to parameter space. Examples of such gradient-based optimizations algorithms often used in cosmology include the MIGRAD optimizer of the MINUIT package, which is a variable metric algorithm using the finite-difference estimates of the first derivative of the likelihood function~\cite{minuit}. References~\cite{Galloni:2024lre,Moretti:2023drg,Campeti:2022acx,Campeti:2022vom,Planck:2013nga}, amongst others, used MINUIT to construct profile likelihoods and found it to be computationally efficient.

In some cases, however, the cosmological likelihood functions are noisy, for example, due to approximation switching in the underlying theory code or insufficient precision settings, leading to inefficiency in the gradient-based algorithms. For example, Refs.~\cite{Karwal:2024qpt,Schoneberg:2021qvd,Reeves:2022aoi,Hannestad:2000wx} found that gradient-free methods outperformed the gradient-based ones. The most popular gradient-free method employed in the literature is simulated annealing~\cite{Kirkpatrick:1983zz}, which works by running an \ac{mcmc} using the modified likelihood $\Tilde{L} \equiv L^{1/T}$ with an iteratively decreasing temperature $T$. The decreasing temperature enhances the peak structures in the likelihoods and eventually traps the \ac{mcmc} close to the best fit. Simulated annealing has been used for profile likelihood construction in Refs.~\cite{Herold:2021ksg,Herold:2022iib,Reeves:2022aoi,Holm:2022kkd,Cruz:2023cxy,Holm:2023laa,Efstathiou:2023fbn,Holm:2023uwa,Karwal:2024qpt}, amongst others. Unfortunately, simulated annealing is very sensitive to its hyperparameters, which include the initial point of the optimization, the covariance matrix of the proposal distribution of the \ac{mcmc}, and the rate at which the temperature decreases. Broadly speaking, three approaches have been suggested to inform the initial point and covariance matrix:
\begin{itemize}
    \item \textit{Running fixed-parameter \ac{mcmc}s to obtain information.} Refs.~\cite{Herold:2021ksg,Reeves:2022aoi,Herold:2022iib} run individual \ac{mcmc}s for each fixed value of the profiled parameter and use the best fit and covariance matrices of their chains in the optimisation. This method produces great initial points and local covariance matrices, although with the disadvantage of the added computational cost of running the additional \ac{mcmc}s.

    \item \textit{Using information from an existing \ac{mcmc}.} The \textsc{CAMEL}~\cite{Henrot-Versille:2016htt} and \textsc{PROSPECT} codes~\cite{Holm:2023uwa} assume the user to have a converged \ac{mcmc} in the full parameter space before constructing the profile. Using this, it uses the binned \ac{mcmc} profile likelihood estimate discussed in the preceding subsection, and furthermore constructs individual covariance matrices from the points in each bin. This method usually gives great initial points and local covariances matrices, but can be sensitive to the quality of the initial \ac{mcmc}.

    \item \textit{Optimising the profile sequentially and reusing information from previous optimizations.} The \textsc{Procoli} code~\cite{Karwal:2024qpt} also assumes the availability of a converged \ac{mcmc} at the beginning and uses its maximum a posteriori point and covariance matrix to find the global best fit. \textsc{Procoli} then carries out optimizations at equidistant points along the profiled parameter direction sequentially, using the result of the previous iteration as the initial point of the next. This method yields great initial points, at the sacrifice of the parallelization of the optimizations.
\end{itemize}

\paragraph{Publicly available profile likelihood codes}
At the time of writing, there exist several publicly available tools for computing profile likelihoods in cosmology; here, we present them in order of the recency of their release.
\begin{itemize}    
    \item \textsc{cobaya fork}~\cite{Galloni:2024lre}: Can be used with \textsc{cobaya}~\cite{Torrado:2020dgo}. Initialises points from an existing \ac{mcmc} as described above. Usable as a sampler inside \textsc{cobaya}, making it easy to use if an \ac{mcmc} has already been run with \textsc{cobaya}\footnote{\url{https://github.com/ggalloni/cobaya/tree/profile_sampler}}.

    \item \textsc{Procoli}~\cite{Karwal:2024qpt}: Can be used with \textsc{MontePython}~\cite{Brinckmann:2018cvx,Audren:2012wb}. Uses the sequential strategy described above. Additionally allows for the extraction of individual $\chi^2$ values for each experiment used in the data\footnote{\url{https://github.com/tkarwal/procoli}}.

    \item \textsc{Prospect}~\cite{Holm:2023uwa}: Can be used with \textsc{MontePython} and \textsc{cobaya}. Uses an assumed precomputed \ac{mcmc} as initialization, as described in the last section. Introduces an automatic step size tuning scheme, and comes with a snapshot checkpoint scheme that enables efficient parallelization and inspections of the run\footnote{\url{https://github.com/AarhusCosmology/prospect_public}}.

    \item \textsc{Camel}~\cite{Henrot-Versille:2016htt}: Code written in C++. The minimization algorithm is based on MINUIT~\cite{minuit}\footnote{\url{http://camel.in2p3.fr/wiki/pmwiki.php}}.
\end{itemize}
All of the above codes (except \textsc{Camel}) work directly with the cosmology-specifying parameter files of \textsc{MontePython} or \textsc{cobaya}, respectively, that have been used, for example, in a complementary Bayesian analysis. They therefore also work with the theory codes already available in these, such as \textsc{CLASS}~\cite{Blas:2011rf} and, in the case of \textsc{cobaya}, \textsc{CAMB}~\cite{Lewis:1999bs}. They also employ the \textsc{GetDist} code~\cite{Lewis:2019xzd} for analysis. 

While the brute-force computation of full triangle plots of profile likelihoods, analogous to the most common product of the Bayesian analysis, is computationally unfeasible, future advances in gradient-based inference and emulators of likelihood codes are expected to make extensive profile likelihood analyses more broadly available. For example, already available is the neural network emulator \textsc{CONNECT}\footnote{\url{https://github.com/AarhusCosmology/connect_public}}~\cite{Nygaard:2022wri} has a run-mode that efficiently computes full triangle plots of profile likelihoods using gradient-based optimization, as explored in reference~\cite{Nygaard:2023cus}.

\subsubsection{Applications of profile likelihood in cosmology}\label{sec:application}

\begin{itemize}
\item {\it EDE \& NEDE:} Due to the complicated parameter structure of \ac{ede} (see Sec.~\ref{sec:EDE}), volume effects can have a strong impact on the constraints of this model. Using PLs, Refs.~\cite{Herold:2021ksg, Herold:2022iib} showed that tight upper limits on \ac{ede} from \ac{cmb} and \ac{lss} in Bayesian analyses, which questioned the ability of \ac{ede} to resolve the Hubble tension, are partially driven by prior volume effects, while in a frequentist analysis \ac{ede} presents a viable solution to the tension. Ref.~\cite{Reeves:2022aoi} uses PLs to explore the interplay between \ac{ede} and massive neutrinos. In Ref.~\cite{Efstathiou:2023fbn}, both a Bayesian and PL analysis under an updated \textit{Planck} \ac{cmb} pipeline give tight upper limits on the fraction of \ac{ede}. Moreover, Ref.~\cite{Cruz:2023cxy} constrained the \ac{nede} (see Sec.~\ref{sec:NEDE}) model and showed that the frequentist analysis gives greater evidence of the model due to volume effects in the Bayesian posterior associated with the \lcdm\ limit.
\item {\it EFTofLSS:} Ref.~\cite{Holm:2023laa} studied the impacts of priors on the nuisance parameters of the \ac{eftoflss} analysis method applied to full-shape power spectrum from galaxy surveys. They found that the priors usually employed to constrain the parameters to their physical regime were informative and had a significant effect on the constraining power of the analysis method.

\item {\it Decaying dark matter:} Ref.~\cite{Holm:2022kkd} studied a model where a fraction of the \ac{dm} is allowed to decay into dark radiation. With two new parameters, the abundance of the decaying part of the \ac{cdm} and its lifetime, series volume effects are found in the limits of parameter space that recover the \lcdm\ model; these include the case of infinite lifetime and vanishing abundance. In particular, whereas the Bayesian analyses prefer either of these limits,~\cite{Holm:2022kkd} shows that the best fit is associated with a narrow log-likelihood peak at a lifetime corresponding to decays just around recombination.

\item {\it Phenomenological transition of dark matter to dark radiation:} Ref.~\cite{Holm:2023uwa} tested the \textsc{PROSPECT} code on a model where a fraction of the \ac{dm} transitions to dark radiation, characterized by three parameters: The fraction of \ac{dm} that transitions, the central redshift of the transition, and a parameter that controls the shape of the transition~\cite{Bringmann:2018jpr}. The \lcdm\ limit is recovered in many limits of parameter space, leading to significant volume effects and parameterization dependencies that were diagnosed with profile likelihoods.

\item \textit{Lensing}: A complete study of the so-called $A_{\rm L}$ tension in the $\Lambda {\rm CDM}+A_{\rm L}$ has been done in Ref.~\cite{Couchot:2015eea}. 

\item \textit{Neutrino}: One of the prototypical applications of profile likelihood in cosmology is for neutrinos. Since neutrino mass constraints are close to the boundary $\sum m_\nu > 0$ and the non-Gaussian posteriors found \ac{mcmc} data analysis, profile likelihood was used to obtain the confidence interval on the sum of the neutrino mass. This has been done in the context of \ac{wmap}5, Sloan Digital Sky Survey (SDSS), and \ac{hst} in Ref.~\cite{Reid:2009nq} and for \textit{Planck} data in Ref.~\cite{Planck:2013nga}. In Ref.~\cite{Couchot:2017pvz} and Ref.~\cite{Gonzalez-Morales:2011tyq}, they study the effect of priors in constraining the number of neutrino species using cosmological data and no evidence for deviations from the standard number of neutrino using profile likelihood. In Ref.~\cite{Couchot:2017pvz} was also highlighted the link with $A_{\rm lens}$ on the current limit. Refs.~\cite{Naredo-Tuero:2024sgf, Herold:2024nvk} explore the preference for ``negative'' neutrino masses with profile likelihoods.

\item \textit{Number of relativistic species, $N_{\rm eff}$}: Profile likelihood was used to test the possibility that extra relativistic species are present beyond the standard model prediction. A discrepancy between the posterior and profile likelihood confidence interval was found in $N_{\rm eff}$ in Refs.~\cite{Hamann:2007pi,Hamann:2011hu}, indicating volume effects when using Wilkinson Microwave Anisotropy Probe (\ac{wmap}7) and \ac{hst} data. It was applied to Planck data in Ref.~\cite{Henrot-Versille:2018ujq}. 

\item \textit{Gravitational wave energy density}: Constraints on the primordial \ac{gw} density using \ac{cmb} data and a study of the impact on cosmic string models has also been derived in Ref.~\cite{Henrot-Versille:2014jua}. 

\item \textit{Tensor-to-scalar ratio}: Profile likelihood was used to analyse the tensor-to-scalar ratio, $r$, with current data. In Ref.~\cite{Campeti:2022vom} the most recent \textit{Planck} and BICEP/Keck \ac{cmb} data was used in order to gain a clearer perspective on the discrepancy between Bayesian and frequentist upper limits on $r$ previously found by the SPIDER collaboration \cite{SPIDER:2021ncy}.  A similar analysis was done in Ref.~\cite{Galloni:2024lre} combining \ac{cmb} data from Planck PR4 and BICEP/Keck, and \ac{gw} data from  \ac{ligo}-Virgo-KAGRA. With the goal of testing the impact of \ac{mg} theories on $r$, in Ref.~\cite{Capistrano:2024kuc} combining data from \ac{cmb} from \textit{Planck} DR4 and BICEP/Keck and \ac{bao} data from 6dF Galaxy Survey, SDSS DR7 and \ac{eboss}, the Bayesian and frequentist upper limits on $r$ were obtained.

\item \textit{Inflation}: The advantages of using a profile likelihood, and more in general, of constructing frequentist confidence intervals with the FC prescription, are particularly evident when trying to constrain inflationary models. This typically allows for more informative constraints and physics insights, especially on models with numerous additional parameters, which are typically degenerate and hard to constrain simultaneously, especially with current data. An example is provided by the latest constraints from \textit{Planck} and BICEP/Keck on axion-U(1) inflation \cite{Campeti:2022acx} and by forecasted constraints on axion-SU(2) inflation parameters for the future \ac{cmb} satellite survey \textit{LiteBIRD} \cite{LiteBIRD:2023zmo}. Another example is given by the attempt to constrain the spectral tilt of tensor primordial perturbations with the latest \ac{cmb} and \ac{gw} interferometers data \cite{Galloni:2024lre}.
Profile likelihoods were also used to search for features in the primordial power spectrum imprinted in \ac{cmb} and large-scale structure formation observables, using data from \textit{Planck} and WiggleZ \cite{Hu:2014hra}.

\item {\it Brans-Dicke model with cosmological constant:} The Brans-Dicke model with a constant positive vacuum energy density, the so-called BD-\lcdm\ model, has been confronted with observations
multiple times in the last years \cite{Avilez:2013dxa,deCruzPerez:2018cjx,SolaPeracaula:2019zsl,SolaPeracaula:2020vpg,Joudaki:2020shz}. In 
Ref.~\cite{SolaPeracaula:2020vpg} the authors showed that in the absence of the Planck 2018 high-$\ell$ \ac{cmb} polarization and lensing data, but still under a very rich dataset including the Planck 2018 full temperature and low-$\ell$ polarization likelihoods, together with the state-of-the-art data on \ac{sn1}, \ac{bao}, \ac{cc} and \ac{rsd}, it is possible to loosen the $H_0$ tension in BD-\lcdm, while keeping $\sigma_8\sim 0.790$. In Ref.~\cite{Gomez-Valent:2022hkb} the author tested the impact of volume effects in BD-\lcdm\ through the analysis of the profile distribution and showed that they do not play a major role in this model.

\item {\it Coupled quintessence:} Coupled quintessence was firstly studied in Refs.~\cite{Wetterich:1994bg,Amendola:1999er} considering a scalar-mediated 
interaction (fifth force) between \ac{dm} particles, with the scalar field playing the role of \ac{de}. Constraints on this model have been reported e.g., in Refs.~\cite{Pettorino:2012ts,Pettorino:2013oxa,Planck:2015bue,Barros:2018efl,Gomez-Valent:2020mqn,Gomez-Valent:2022bku,Goh:2022gxo}. Volume effects have been recently quantified in Ref.~\cite{Gomez-Valent:2022hkb} for the case of a Peebles-Ratra potential \cite{Peebles:1987ek,Ratra:1987rm} using the profile distributions built directly from the \ac{mcmc}. Their impact on the results are found to be small.

\item {\it Neutrino-Dark Matter Interactions:} Ref.~\cite{Giare:2023qqn} studied a model featuring scatter-like interactions between neutrinos and \ac{dm}. Allowing the interaction strength and the total neutrino mass to vary, the model can have up to two parameters more than \lcdm. A profile likelihood analysis was employed to test possible volume effects. In this case, no relevant volume effects were found. The profile likelihood supported the results derived by marginalizing over the parameter space, confirming a mild preference for non-vanishing interactions when small-scale \ac{cmb} data are considered~\cite{Brax:2023rrf,Brax:2023tvn}.

\item {\it Baryon Acoustic Oscillations:} Profile likelihood has been used, sometimes in combination with Bayesian methods, to quantify \ac{bao} errors in determining the fitting scale in surveys. This has been done in the context of many surveys, like the Sloan Digital Sky Survey (SDSS) \ac{boss} \cite{BOSS:2012tck} and the \ac{eboss} \cite{eBOSS:2017cqx}, or \ac{des} \cite{DES:2017rfo,DES:2018fiv}. In \cite{Ruggeri:2019kjl}, they extend this method to add noise and compare different approaches to determine the error in determining the \ac{bao} scale. 
Profile likelihood was also used to fit the Lyman-$\alpha$ (Ly$\alpha$) forest 3D correlation function \cite{Cuceu:2020dnl}. Using mocks, they found that profile likelihood is a good approximation for fitting the \ac{bao} peak from the Ly$\alpha$
forest correlation functions and is in full agreement with the Bayesian analysis (which is not the case for frequentist maximum likelihood estimators, which assume Gaussianity).

\end{itemize}

\bigskip \newpage
%%%%%%%%%%%%Section_4:
\section{Fundamental physics} \label{sec:fun_phys}

\noindent \textbf{Coordinator:} Vivian Poulin\\

Cosmology stands at a crossroads. The concordance \lcdm\ model, which had shown a great level of success in describing a wide variety of cosmological observations, is being challenged by high precision data as reviewed in Sec.~\ref{sec:obs}. However, the \lcdm\ model has raised profound questions about the nature of its dominant components --\ac{dm} and \ac{de}-- as well as the mechanism at the origin of the primordial fluctuations (usually described by inflation) that remains largely unknown. Barring the existence of unknown systematic errors, these observational issues thus present a crucial opportunity to explore new physics beyond the \lcdm\ paradigm. This section examines the collective efforts of the scientific community to interpret the implications of observations that challenge the \lcdm\ model, with a particular focus on what the $H_0$ and $S_8$  tensions may reveal about fundamental physics.

At the core of the \lcdm\ value of $H_0$ is the angular size of the sound horizon at recombination $\theta_s(z_*) = r_s(z_*)/D_{\rm A}(z_*)$, that \ac{cmb} data have determined at ${\cal O}\sim 0.1\%$ precision \cite{Planck:2018vyg}. In the flat \ac{flrw} metric, the angular diameter distance to recombination $D_{\rm A}(z_*)$ that carries information about $H_0$ can be computed as
\begin{equation}
   D_{\rm A}(z_*)=(1+z_*)^{-1}\int_0^{z_*}\frac{dz}{H(z)}\,,
\end{equation}
where under flat \lcdm, $H(z) = H_0\sqrt{\Omega_{\rm r}(1+z)^4 + \Omega_{\rm m,0}(1+z)^3 + \Omega_\Lambda}$. Extracting the angular diameter distance from $\theta_s(z_*)$ requires knowledge of the sound horizon
\begin{equation}
   r_s(z_*)=\int^\infty_{z_*}\frac{c_s(z)dz}{H(z)}\,,~~~{\rm with}~~~c_s(z)=\frac{1}{\sqrt{3(1+4\rho_{\rm b}(z)/3\rho_{\rm r}(z))}}\,.
\end{equation}
A value of $r_s(z_*)$ can be computed within a model, given knowledge of the cosmological parameters.  Within \lcdm, {\em Planck} data have allowed us to determine $r_s(z_*) = 144.39\pm0.3$ Mpc (TTTEEE+lowE) from the complex structure of peaks and troughs \cite{Planck:2018vyg}. The value of $H_0$ is then adjusted in order to match the corresponding value of $\theta_s(z_*)$, which yields $H_0=67.27\pm0.6$\kms. Similarly, the value of $S_8$ can be inferred from the linear matter power spectrum $P(k)$ given a set of cosmological parameters reconstructed from a fit to a given dataset as 
\begin{equation}
\label{eq:S8}
    S_8 \equiv \sqrt{\frac{\Omega_{\rm m,0}}{0.3}}\times\bigg\{\sigma_8\equiv\int_0^\infty \frac{k^3}{2\pi^3}P(k) W_8^2(k)d\ln k\bigg\}\,,
\end{equation}
where $W_8(k)$ is a window function describing a sphere of radius $R=8$Mpc$/h$ in Fourier space. Under the \lcdm\ model fit to {\it Planck} data, it yields $S_8=0.834\pm0.016$.

The solutions to the $H_0$ discrepancy generally fall into two categories based on the requirement to leave the precisely measured value of $\theta_s(z_*)$ unaffected \cite{Knox:2019rjx,DiValentino:2021izs,Schoneberg:2021qvd}. First, there are those that reduce the value of the sound horizon $r_s(z_*)$ such as to compensate for the shorter angular diameter distance implied by a larger $H_0$. Those typically correspond to early-time  (pre-recombination) modifications and include for instance include, new relativistic species, \ac{ede}, or modified recombination scenarios. Second, late-time solutions modify the expansion history after recombination such that the angular diameter distance to recombination remains unaffected. These include e.g., phantom \ac{de}, \ac{ide}, or decaying \ac{dm}. Both avenues offer advantages and disadvantages: the former scenarios may more easily adjust the shape of the expansion history that is constrained by \ac{bao} and \ac{sn1} (beyond the mere value of $H_0$), but will be strongly constrained by \ac{cmb} observations, while the latter scenarios can more easily leave the \ac{cmb} unaffected by exploiting the geometrical degeneracy, but are strongly constrained by late-time observations \cite{Schoneberg:2021qvd}.

Resolving the $S_8$ tension, on the other hand, requires suppressing structure growth, which can be done through new \ac{dm} interactions, evolving \ac{de} models, or modifications to general relativity. Yet, it remains to be established precisely what scales and what cosmic era must be affected in order to achieve concordance between all our \ac{lss} observations: is the tension solely appearing at small scales, that are potentially affected by large uncertainties related to non-linear and baryonic effects, or does it also appear on large-scales at late-times, where the physics is much more linear and under control? Answering these questions will drive the community effort to find a solution to the tension. Moreover, there is a clear interplay between the $H_0$ and $S_8$ tensions. First, simply because the definition of the physical scale $k$ involved in $S_8$ in Eq.~\eqref{eq:S8} is trivially influenced by the value of $h$: as the power spectrum as a strong dependence on the scale $k$ (falling like $k^{-3}$) tension in $S_8$ may appear due to different values of $h$ in the analysis, fooling us in comparing different physical scales \cite{Sanchez:2020vvb,Secco:2022kqg,Forconi:2025cwp}. However, even when corrected (e.g., through the use of $R=12$ Mpc as an absolute scale \cite{Sanchez:2020vvb}), the tension can be influenced by the new physics introduced to resolve the Hubble tension. This stems from the observation that distance indicators, that provide measurements of $\Omega_{\rm m,0}$ and $H_0$, necessarily imply a measurement of the physical matter density $\omega_{\rm m,0} \equiv \Omega_{\rm m,0}h^2$ \cite{Jedamzik:2020zmd,Poulin:2024ken,Pedrotti:2024kpn}. Hence, a larger value of $h$ will automatically imply a larger value of $\omega_{\rm m,0}$, leading to earlier matter domination and thus a larger amplitude of fluctuations today. 

Finding a solution to either or both tensions is an active area of research. As we illustrate within each section, understanding and resolving these tensions will not only improve the precision in measuring fundamental cosmological parameters but also offer a window into the fundamental physics governing the Universe \cite{DiValentino:2021izs}. Whether these tensions indicate systematic biases in data or genuine hints of new physics, they represent one of the most exciting frontiers in modern cosmology, potentially unlocking new insights into \ac{dm}, \ac{de}, and the laws of gravity.
\subsection{Early-time proposals}  \label{sec:ETP_4.1}
\subsubsection{Early dark energy and variants \label{sec:EDE}}

\noindent \textbf{Coordinator:} Laura Herold\\
\noindent \textbf{Contributors:} Adrià Gómez-Valent, Alexander Reeves, Alireza Talebian, David Benisty, Elisa G. M. Ferreira, Gabriela Garcia-Arroyo, Ivonne Zavala, Joan Solà Peracaula, Lu Yin, Luis Anchordoqui, Luz Ángela García, Matteo Forconi, Nikolaos Mavromatos, Rafaela Gsponer, and Vivian Poulin
\\

\paragraph{Introduction and motivation}

\ac{ede} models (for reviews see Refs.~\cite{Kamionkowski:2022pkx, Poulin:2023lkg, McDonough:2023qcu}) are a promising class of models proposed to resolve the Hubble tension with new physics. The central idea behind \ac{ede} is to introduce an additional component to the energy density of the Universe at early-times, which increases the expansion rate $H(z)$ before recombination $z^*$. This leads to a decrease of the (comoving) size of the sound horizon: $r_s(z^*) = \int_{z^*}^\infty c_s(z)/H(z)\mathrm{d}z$, where $c_s$ is the sound speed in the baryon-photon plasma. To understand why this increases $H_0$ at present times, note that the angular size of the sound horizon $\theta_s = r_s/D_{\rm A}$ is directly and precisely observed by the \ac{cmb}, e.g., see Ref.~\cite{Planck:2018vyg}. Hence, to maintain the same size of $\theta_s$, the (comoving) angular diameter distance reduces, $D_{\rm A}(z^*) = \int_0^{z^*} \mathrm{d}z/H(z)$, and leads to an increase of $H_0$. 

All \ac{ede}-type solutions to the Hubble tension have in common a \ac{de}-like rapid expansion with an equation of state $w \simeq -1$ around $z^*$, which peaks at some critical redshift $z_c$ and is followed by a phase in which the background energy density decays or dilutes (see Fig.~\ref{fig:EDE}, left). \ac{ede} models often take the form of a cosmological scalar, whose background dynamics follow the homogeneous Klein-Gordon equation: the field is initially frozen in its potential, such that the background energy density is constant. The fractional contribution of the field to the total energy density, $\rho_\mathrm{EDE}/\rho_\mathrm{Tot}$, increases over time in a manner similar to that of \ac{de}. Eventually, Hubble friction dropping below a critical value, or a phase transition changing the shape of the potential, releases the scalar, at which point the field becomes dynamical, and the background energy density dilutes away faster than matter. The contribution of \ac{ede} to the Hubble rate is therefore localized before recombination, where it efficiently acts to reduce the size of the sound horizon. 

\ac{ede}-like models have already been studied before the appearance of the Hubble tension, e.g., in the context of quintessence models \cite{Albrecht:1999rm, Doran:2000jt,Wetterich:2004pv,Doran:2006kp} or in the context of recurring \ac{de} \cite{Kamionkowski:2014zda}. The (axion-like) \ac{ede} model has gained significant attention as a proposed solution to the Hubble tension~\cite{Karwal:2016vyq, Poulin:2018dzj, Poulin:2018cxd}. To fully resolve the tension, a fractional \ac{ede} contribution of $\sim 10\%$ around the time of matter-radiation equality is required~\cite{Smith:2019ihp}, reflecting the percentage difference in $H_0$ estimates between \textit{Planck}~\cite{Planck:2018vyg} and SH0ES~\cite{Riess:2021jrx}.
Although \ac{ede} models present a promising class of solutions when compared to other proposed solutions \cite{Knox:2019rjx, Schoneberg:2021qvd, Khalife:2023qbu}, the improvement in observational constraints from the \ac{cmb} and \ac{lss} in the last few years poses challenges to this class of models. 

\begin{figure}[ht]
     \centering
          \begin{subfigure}[b]{0.42\textwidth}
         \centering
         \includegraphics[width=\textwidth]{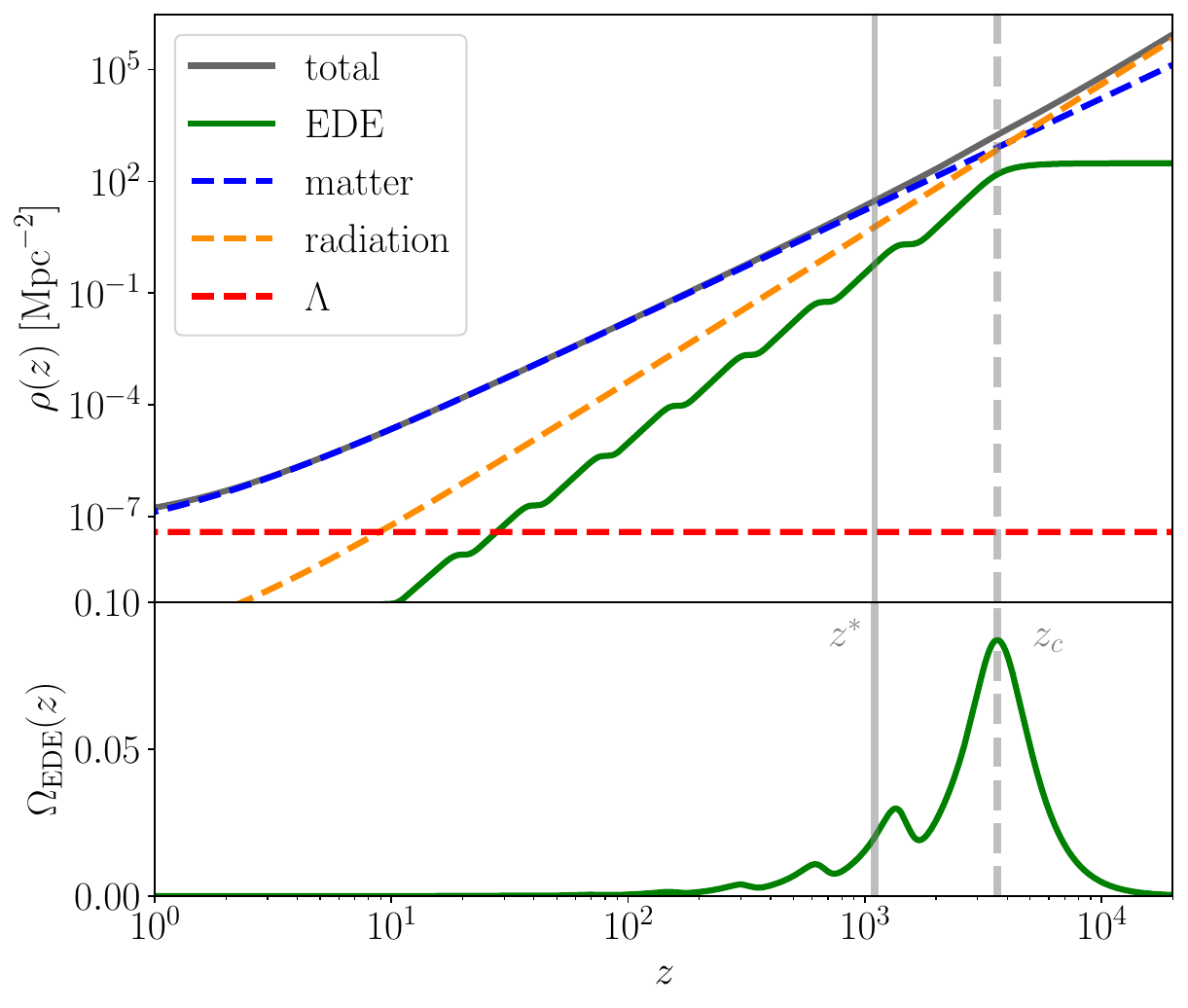}
     \end{subfigure}
     \hfill
     \begin{subfigure}[b]{0.53\textwidth}
         \centering
         \includegraphics[width=\textwidth]{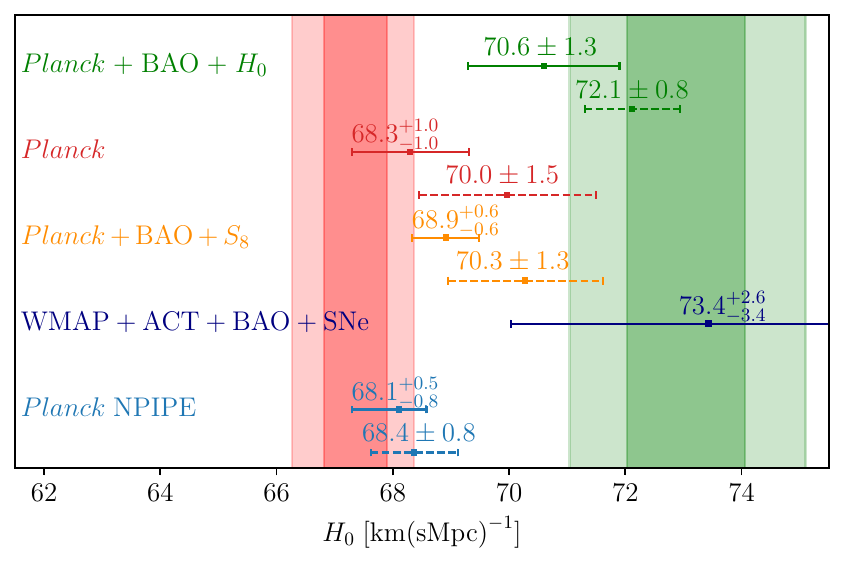}
     \end{subfigure}
     \hfill
         \caption{\textit{Left:} Energy densities, $\rho(z)$, of different components of the Universe as indicated in the legend. \ac{ede} (green) plays a subdominant rule at all redshifts, $z$. The fractional energy density $\Omega_\mathrm{EDE}(z)$ (bottom subplot) peaks at the critical redshift, $z_c$, and decays around recombination, $z^*$. \textit{Right:} A comparison of constraints on $H_0$ within the axion-like \ac{ede} model under different data sets as indicated in the figure, where solid (dashed) lines represent Bayesian (frequentist) constraints. The constraints are taken from Refs.~\cite{Poulin:2018cxd, Herold:2022iib, Hill:2020osr, Poulin:2021bjr, Efstathiou:2023fbn} (see text).}
        \label{fig:EDE}
        \vspace{-0.4cm}
\end{figure}

\paragraph{The axion-like EDE model}

The originally proposed \ac{ede} model \cite{Karwal:2016vyq, Poulin:2018dzj, Poulin:2018cxd} consists of a scalar field $\phi$ in a potential of the form 
\begin{equation}
    \label{eq:axEDE_potential}
    V(\phi) = V_0 [1-\cos(\phi/f)]^n\,,
\end{equation}
where $V_0 = m^2f^2$ with $m$ and $f$ being referred to as ``mass'' and ``decay constant'', respectively. The index $n$ is typically fixed to $n=3$ as this provides the best fit to the data \cite{Smith:2019ihp, Poulin:2023lkg}. The parameters $m$ and $f$, along with the initial value of the scalar field in the potential, $\theta_i = \phi_i/f$, need to be fit from the data. This ``particle-physics parameterization'' is often traded for the ``phenomenological parameterization'', which consists of $f_\mathrm{EDE}$, the maximum fraction of \ac{ede}, $f_{\rm EDE} \equiv \rho_{\rm EDE}(z_c)/\rho_{\rm Tot}(z_{c})$, at the critical redshift, $z_c$, which is determined by the onset of oscillations, along with the initial value of the scalar field $\theta_i$.

Predictions for \ac{ede} cosmologies have been found using the effective fluid theory approach \cite{Poulin:2018cxd} and by directly solving the linearized scalar field evolution equations \cite{Smith:2019ihp,McDonough:2021pdg}. Extensions of \texttt{CLASS} \cite{Blas:2011rf} using the latter approach are publicly available\footnote{\url{https://github.com/PoulinV/AxiCLASS}}\footnote{\url{https://github.com/mwt5345/class_ede}}.  \texttt{CAMB}~\cite{Lewis:1999bs} includes \ac{ede} using the effective fluid approximation\footnote{\url{https://github.com/cmbant/CAMB/tree/master}}. 

The axion-like \ac{ede} model faces some theoretical challenges: the desired degeneracy between $f_\mathrm{EDE}$ and $H_0$, and with that, the ability of \ac{ede} to solve the tension sensitively depends on the choice of the parameters $z_c$ and $\theta_i$. Since there is no a priori reason for these parameters to have specific values, this leads to a ``fine-tuning'' problem. Moreover, the shape of the potential with index $n=3$ is considered fine-tuned itself since it requires that lower-order terms in the axion-like potential are zero. Therefore, the axion-like \ac{ede} model is often seen as a phenomenological model of \ac{ede}. 
Nonetheless, there are approaches to derive the axion-like \ac{ede} model from fundamental theories such as string theory \cite{Alexander:2019rsc,Cicoli:2023qri,McDonough:2022pku} or to motivate $n=3$ by higher-order instanton corrections \cite{Kappl:2015esy}.

\paragraph{EDE variants}

From the axion-like \ac{ede} model, we learned valuable lessons about the challenges of \ac{ede} as a solution to the Hubble tension. In order to have the maximal possible leverage to resolve the tension, we need a specific form of potential, \ac{ede} needs to become relevant around matter-radiation equality and it needs to decay sufficiently fast before recombination. These theoretical challenges, or ``fine-tunings'', have inspired many variants of \ac{ede}, which will be presented here. 

Several \ac{ede} models consist of \textbf{slow-rolling scalar fields} minimally coupled to the metric, with different shapes of potentials~\cite{Yin:2020dwl,Ye:2020btb,Albrecht:1999rm, Adil:2022hkj}. Rock 'n' Roll \ac{ede} assumes a power law potential of the form $V(\phi) \sim \phi^{2n}$, which approximates the axion-like potential for small $\phi/f$ \cite{Agrawal:2019lmo}. Acoustic DE generalizes the \ac{ede} model by using a phenomenological fluid description of \ac{ede} parameterized by the sound speed $c_s^2$ and equation-of-state parameter $w_\mathrm{ADE}$ \cite{Lin:2019qug}. 

A class of axion-like \ac{ede} models arises in the \textbf{string-inspired Chern-Simons (CS) gravity}~\cite{Jackiw:2003pm}, which characterizes the low energy limit of string theory~\cite{Duncan:1992vz}. In the simplest of such models~\cite{Basilakos:2019acj, Basilakos:2020qmu, Mavromatos:2020kzj, Guendelman:2022cop}, it leads to the generic form of \ac{rvm} inflation \cite{Perico:2013mna, Lima:2013dmf, Sola:2015rra, SolaPeracaula:2019kfm}. In the specific context of string-inspired CS gravity, it is called Stringy \ac{rvm} (StRVM) \cite{Mavromatos:2020kzj, Mavromatos:2021urx}, in which there is a pre-\ac{rvm}-inflationary era dominated by a stiff-fluid of gravitational (Kalb-Ramond) axions that characterize the massless gravitational string multiplet. At the end of such eras, condensation of primordial chiral \ac{gw}s leads to a condensation of the gravitational anomaly CS term that couples the model to string axions that are either of Kalb-Ramond type or arise from compactification~\cite{Svrcek:2006yi}. The condensed CS term leads to linear terms in the axion potential~\cite{Dorlis:2024yqw,Mavromatos:2024pho}, responsible for \ac{rvm} inflation, while the nonperturbative instanton effects lead to additional periodic modulations. In the context of the StRVM~\cite{Basilakos:2019acj, Basilakos:2020qmu, Mavromatos:2020kzj}, the coefficient of the linear axion term is not a constant, but an \ac{rvm}-type functional of even powers of the Hubble parameter. This differentiates the model from other linear axion monodromy models that arise from brane compactification~\cite{McAllister:2008hb}. In Refs.~\cite{Dorlis:2024yqw,Mavromatos:2024pho} it was shown that the CS condensate leads to a {\it metastable} \ac{ede}, as there are imaginary parts in the effective action, pointing to a finite-lifetime of the respective vacua. Notably, different forms of the StRVM can also help alleviate, in modern eras, the Hubble and growth-of-structure tensions~\cite{Gomez-Valent:2023hov,Gomez-Valent:2024tdb}. In the framework of $\alpha$-attractors in inflation \cite{Kallosh:2013hoa, Kallosh:2013yoa, Galante:2014ifa}, a potential for \ac{ede} was proposed in Ref.~\cite{Braglia:2020bym} that can reproduce various cases based on its functional form.

\textbf{Unified models of EDE and late DE} have been proposed in Refs.~\cite{Brissenden:2023yko, Chowdhury:2023opo, Ramadan:2023ivw}. Ref.~\cite{Brissenden:2023yko} introduces a model where a scalar field can explain both \ac{ede} and late \ac{de} in a joined manner in the context of $\alpha$-attractors. In Ref.~\cite{Chowdhury:2023opo}, a unified model of \ac{ede} and late \ac{de} is introduced, which can also enhance the primordial \ac{gw} spectrum at PTA scales. Moreover, the \ac{ede} scalar potential can arise from a D-brane moving in the compact dimensions of warped compactifications in the context of D-brane models of cosmology. Ref.~\cite{Ramadan:2023ivw} used a dynamical system approach to analyze a unified model of \ac{ede} and late \ac{de}. Scaling \ac{ede} is characterized by a constant \ac{ede} fraction in both the radiation- and matter-dominated epochs. This behavior is naturally found in quintessence models with exponential potentials and dates back to the concept of scalar field \ac{de} \cite{Wetterich:1987fm}; see also Ref.~\cite{Copeland:1997et,Barreiro:1999zs}. The very tight constraints imposed by the \ac{cmb} disfavor this type of models as a solution to the tension \cite{Gomez-Valent:2021cbe, Doran:2000jt, Doran:2006kp, Pettorino:2013ia}, even when allowing for a coupling between \ac{ede} and \ac{dm} \cite{Gomez-Valent:2022bku}. In the case of tracker solutions, \cite{Copeland:2023zqz} studied \ac{ede} using quintessence and K-essence in the tracker regimes. 

The possibility of \textbf{\ac{ede} field coupling to other particles} was also explored. Refs.~\cite{Karwal:2021vpk, McDonough:2021pdg, Lin:2022phm} introduce a model with a coupling between \ac{ede} and \ac{dm}. The coupling modifies the scalar potential, including a matter contribution and a modulation of the \ac{dm} mass. In a similar spirit, the coupled scalar \ac{dm} to \ac{ede} has been studied in Refs.~\cite{Gomez-Valent:2022bku,Liu:2023kce,Garcia-Arroyo:2024tqq}. The waterfall \ac{ede} model was explored in Ref.~\cite{Talebian:2023lkk}, in which the \ac{ede} phase ends through spontaneous symmetry breaking. This mechanism has the potential to generate a dark-photon-\ac{dm} component. In Ref.~\cite{Berghaus:2019cls}, the \ac{ede} phase is realized and ended similarly to the warm inflation model. The possibility of \ac{ede} coupling to neutrinos was studied in Refs.~\citep{Sakstein:2019fmf,CarrilloGonzalez:2020oac} with the coupling to neutrinos being a natural trigger for the \ac{ede} field at around the matter-radiation equality when the neutrinos become non-relativistic. The possibility of these models not being able to solve the Hubble tension was raised in Ref.~\cite{deSouza:2023sqp} and later disputed in Ref.~\cite{CarrilloGonzalez:2023lma}. 
Emerging \ac{de} models consider a non-negligible contribution of the \ac{de} density during the radiation domination era. This effect is introduced through an evolving equation of state that, at early-times, mimics radiation and at late-times asymptotically reaches $\omega \rightarrow -1$ \cite{Li:2019yem,Garcia:2020sjl,Benaoum:2023ekz}. This phenomenological emergent \ac{de} model \cite{Li:2020ybr} shows promising results when confronted with data \cite{Staicova:2021ntm}.

Instead of using a specific \ac{ede} model, one can use a \textbf{model-independent reconstruction} of the \ac{ede} expansion history. Ref.~\cite{Gomez-Valent:2021cbe} uses a tomographic approach, in which $\rho_\mathrm{EDE}$ is grouped into redshift bins $z$ and constrained by confronting this model with data. They find that the tightest upper bound on $\rho_{\rm EDE}$ occurs around the \ac{cmb} decoupling, whereas much weaker constraints are obtained before and after recombination, leaving room for the \ac{ede} model to alleviate the Hubble tension. Ref.~\cite{Moss:2021obd} determine $\rho_\mathrm{EDE}$ in scale-factor bins, finding that the maximum $\rho_\mathrm{EDE}$ is reached at about $z\sim10^{5}$, which is different from the $z_c$ derived with the standard approach. Alternatively, a parametric equation of state can be fit \citep{Nojiri:2021dze}. Moreover, different microphysics of \ac{ede} could have important effects~\citep{Sabla:2022xzj}. In Ref.~\cite{Moshafi:2022mva}, a phenomenological model is proposed that can mimic the \ac{ede} phase through multiple transitions in the vacuum \ac{de} density.

\paragraph{Constraints on EDE from CMB, BAO and SN}

In the following sections, we will discuss observational constraints on the axion-like \ac{ede} model from different data sets (see Fig.~\ref{fig:EDE}, right, for a whisker plot). When analyzing \ac{ede} with \textit{Planck} primary and lensing \ac{cmb} \cite{Planck:2015bpv}, \ac{bao} \cite{Beutler:2011hx,Ross:2014qpa,Mueller:2016kpu}, \ac{sn1} \cite{Pan-STARRS1:2017jku}, and the $H_0$ measurement of the SH0ES collaboration \cite{Riess:2018byc, Riess:2019cxk}, these data sets generally prefer high values of $f_\mathrm{EDE}$ and high values of $H_0$ consistent with direct measurements.\footnote{All data combinations and model implementations in this section use the same \ac{sn1} sample and \ac{bao} data, but differ in $H_0$ and \textit{Planck} data.} The goodness of fit, i.e., the $\chi^2$, reduces compared to \lcdm, albeit with three extra parameters compared to \lcdm. This is mainly driven by the ability of \ac{ede} to accommodate higher $H_0$ values by reducing $r_s$, which in turn leads to higher values of $n_s$ and $\omega_\mathrm{cdm}$. Ref.~\cite{Poulin:2018cxd} uses the effective fluid description of the axion-like \ac{ede} model, Planck 2015 \cite{Planck:2015bpv} and SH0ES 2018 \cite{Riess:2018byc} and obtains $f_{\text{EDE}} \sim 5\% $ at $z_c \sim 5000$ and $H_0 = 70.6 \pm 1.3$\kms ($\Delta \chi^2 = -14.2$, first errorbar in Fig.~\ref{fig:EDE}), whereas in Ref.~\cite{Smith:2019ihp} using the scalar field description and updating to the SH0ES 2019 value \cite{Riess:2019cxk} results in $f_{\text{EDE}} \sim 10\%$ at $z_c \sim 3700$ and $H_0 = 71.49 \pm 1.20$\kms ($\Delta \chi^2 = -20.3$).\footnote{All central intervals in this section are quoted at $68\%$ CL while all upper limits are quoted at 95\% CL} Updating to Planck 2018 data, \cite{Murgia:2020ryi,McDonough:2021pdg} also use the scalar field description yielding $f_{\text{EDE}} \sim 14\%$ at $z_c \sim 3800$ and $H_0 = 71.2 \pm 1.1$\kms ($\Delta \chi^2 = -16.2$) while updating to D\ac{desi} \ac{bao} yields $H_0 = 71.7 \pm 0.76$\kms. Finally, those constraints were recently updated with Planck 2020 NPIPE data \cite{Efstathiou:2023fbn} yielding $H_0=71.24\pm0.77$\kms for $f_{\text{EDE}} =0.107\pm0.023$ at $\log_{10}(z_c) = 3.585^{+0.049}_{-0.15}$ ($\Delta \chi^2 = -28.0$).

\paragraph{EDE and LSS}

Confronting the \ac{ede} model with \ac{lss} data can impose stringent constraints on $f_\mathrm{EDE}$. This can be understood as \ac{lss} data typically prefer lower values of the amplitude of clustering $\sigma_8$ and the related parameter $S_8$ compared to \ac{cmb} data (the so-called ``$S_8$ discrepancy''). Ref.~\cite{Poulin:2018cxd} find $S_8 = 0.838 \pm 0.015$ for Planck, \ac{boss} \ac{bao}, Pantheon and SH0Es data, and more recently Ref.~\cite{Poulin:2024ken} find $S_8 = 0.831\pm 0.012$ for Planck, \ac{desi} \ac{bao}, Pantheon+ and SH0ES data. However, $f_\mathrm{EDE}$ typically correlates positively with the amount of late-time clustering due to increased $\omega_\mathrm{cdm}$ in models with higher $f_\mathrm{EDE}$ which compensates for the \ac{ede}-induced boost to the early Integrated Sachs-Wolfe (eISW) effect to maintain a good fit to the \ac{cmb}~\cite{Smith:2019ihp, Vagnozzi:2021gjh, Ye:2021nej, Pedreira:2023qqt}. Including \ac{lss} data thus weakens the evidence for \ac{ede} compared to using \ac{cmb} data alone by providing an independent constraint on the clustering amplitude~\cite{Hill:2020osr}. 

Several studies have shown that the combination of \ac{boss} full-shape and \ac{bao} data \cite{BOSS:2015npt,BOSS:2015fqm,Beutler:2011hx,Ross:2014qpa,eBOSS:2019ytm,eBOSS:2019qwo} (based on the effective field theory of \ac{lss}, e.g., see Refs.~\cite{Baumann:2010tm, Carrasco:2012cv}) alongside the \ac{cmb} give tight upper bounds on $f_\mathrm{EDE}$~\cite{Ivanov:2020ril, DAmico:2020ods, Gsponer:2023wpm} when not including any direct measurements of $H_0$. \ac{wl} data have also been explored in the context of the \ac{ede} model. For example, see Ref.~\cite{Hill:2020osr} where \ac{des}-YI 3x2pt data \cite{DES:2017myr}, as well as $S_8$ priors from \ac{kids} and \ac{hsc} \cite{Hildebrandt:2016iqg,Hildebrandt:2018yau,HSC:2018mrq}, were used to derive tight upper limits of $f_\mathrm{EDE}<0.060$ and $H_0=68.92^{+0.57}_{-0.59}$\kms while giving high values of $S_8 = 0.8060 \pm 0.0082$ when combined with \textit{Planck} \cite{Planck:2018vyg} leading to the conclusion that \ac{lss} data can rule out the \ac{ede} model (fifth errorbar in Fig.~\ref{fig:EDE}). In contrast, \cite{Murgia:2020ryi} examine the \ac{ede} model in the context of \ac{kids} \cite{Hildebrandt:2018yau} and DES \cite{Asgari:2019fkq} data and find that a one-parameter \ac{ede} model, which fixes $\{z_c,\,\theta_i \}$, does not lead to a significantly worsened $S_8$ discrepancy compared to \lcdm\ and advocate for caution when combining datasets that are statistically discrepant. Furthermore, \cite{Smith:2020rxx} argue that the stringent constraints found when combining \ac{cmb} data with \ac{boss} full-shape and \ac{bao} likelihoods can be traced to a potential tension in the power spectrum amplitude $A_s$ between \textit{Planck} and \ac{boss} which is present even in \lcdm. Thus, while it is clear that \ac{lss} data can play an important role in constraining the \ac{ede} model, the existence of subtle inconsistencies between \ac{cmb} and \ac{lss} data, present even within the \lcdm\ framework, complicates the interpretation.

It is also worth mentioning the important role that the use of $\sigma_8$ as a cosmological parameter might play in this discussion. The $\sigma_8$ parameter is computed as a convolution of the matter power spectrum and a window function at a characteristic scale $R_8\equiv 8h^{-1}$ Mpc. Thus, $\sigma_8$ provides information about the amplitude of the power spectrum at a scale that changes in models with different values of $H_0$. To address this issue, \cite{Sanchez:2020vvb} suggested using the parameter $\sigma_{12}$, which is computed at a fixed scale $R_{12}=12$ Mpc that does not depend on $h$. In Ref.~\cite{Gomez-Valent:2022hkb} it is shown that despite the positive correlation between $\sigma_8$ and $f_{\rm EDE}$ in fitting analyses with \ac{cmb}+\ac{bao}+\ac{sn1}, the correlation between $\sigma_{12}$ and $f_{\rm EDE}$ is negligible. 

Attempts have been made to address both the $H_0$ and $S_8$ tensions simultaneously by expanding the \ac{ede} model parameter space to include components that reduce the amplitude of clustering at late-times. For example, Refs.~\cite{Reeves:2022aoi, Gomez-Valent:2022bku} allow for a free neutrino mass in the \ac{ede} model finding that current data limits the effectiveness of massive neutrinos to suppress structure formation at small scales due to the stringent constraint placed on the sum of neutrino masses from the \textit{Planck} 2018 data which is not much weakened in the context of \ac{ede}. Other works have also considered a coupling between \ac{ede} and \ac{dm} in order to try to decrease the value of $\omega_{\rm cdm}$ and make the growth of large-scale structures less efficient at late-times \cite{Karwal:2021vpk,Gomez-Valent:2022bku,McDonough:2021pdg}.

\paragraph{Prior effects and frequentist constraints}

Due to the complicated parameter structure of the \ac{ede} model, the tight upper limits on \ac{ede} from \ac{cmb} and \ac{lss} are not only driven by physical effects but also affected by prior volume (or projection) effects in the \ac{mcmc} posterior. These prior effects arise due to the nested parameter structure of the model: while $f_\mathrm{EDE}$ controls the fraction of \ac{ede}, $z_c$ and $\theta_i$ are auxiliary parameters describing the model in more detail. When $f_\mathrm{EDE}$ approaches zero, the \lcdm\ limit is recovered, and $z_c$, $\theta_i$ becomes redundant and unconstrained. This leads to a larger prior volume at $f_\mathrm{EDE} \approx 0$ than at $f_\mathrm{EDE} > 0$ and a non-Gaussian posterior, which in turn can lead to a preference for the \lcdm\ limit in the marginalized posterior. In
Refs.~\cite{Murgia:2020ryi, Smith:2020rxx} (and Ref.~\cite{Niedermann:2020dwg} for \ac{nede}) the influence of prior effects was illustrated by fixing $z_c$, $\theta_i$ in the analysis and varying only $f_\mathrm{EDE}$ along with the \lcdm\ parameters, which leads to larger allowed values of $f_\mathrm{EDE} = 0.0523^{+0.026}_{-0.036}$ when using the same \ac{boss} full-shape data \cite{BOSS:2016wmc,Ivanov:2019pdj,DAmico:2019fhj} with the effective field theory of \ac{lss} \cite{Baumann:2010tm, Carrasco:2012cv} combined with \ac{bao} and \ac{cmb} \cite{Planck:2018nkj}. 

Since these effects are related to the prior, which is an inherently Bayesian quantity, Refs.~\cite{Herold:2021ksg, Herold:2022iib} constructed frequentist constraints based on the profile likelihood, which are independent of priors and only depend on the likelihood (top three dashed error bars in Fig.~\ref{fig:EDE}). Using \textit{Planck} \ac{cmb} \cite{Planck:2018nkj} and \ac{boss} full-shape power spectrum data \cite{BOSS:2016wmc, Ivanov:2019pdj, DAmico:2019fhj, Beutler:2021eqq} based on the effective field theory of \ac{lss}, the profile likelihood yields higher values of $f_\mathrm{EDE} = 0.087 \pm 0.037$ and $H_0 = 70.57 \pm 1.36\,$\kms ($\Delta\chi^2 = -5.6$) than the constraints from Bayesian posteriors. Additionally, including SH0ES data pushes the inferred $H_0 = 72.1 \pm 0.8$\kms even further up. Even when additionally including 3x2pt data from DES \cite{DES:2021wwk}, the profile likelihood yields moderately high values of $f_\mathrm{EDE} \approx 6\,\%$ and $H_0 = 70.28 \pm 1.33$\kms. This shows that the Bayesian constraints from \ac{cmb} and \ac{lss} are partially driven by prior effects and indicates that more data is necessary to fully rule out or detect \ac{ede}. 
These results agree with Ref.~\cite{Gomez-Valent:2022hkb}, which uses an approximate profile distribution obtained from \ac{mcmc} samples, which is computationally cheaper than computing a profile likelihood. In a Bayesian context, \cite{Gsponer:2023wpm} explores the impact of priors on constraints from the effective field theory of \ac{lss} on \ac{ede}. Using a Jeffreys prior on nuisance parameters mitigates projection effects from the effective field theory of \ac{lss}, resulting in smaller overall volume projection effects on \ac{ede} when combined with other datasets. Ref.~\cite{Piras:2025eip} utilizes a neural network to obtain a data-driven parameterization of \ac{ede}.

\paragraph{Ground-based CMB experiments and alternative CMB pipelines}

The \textit{Planck} experiment offers high-precision measurements of the large- to intermediate-scale \ac{cmb} temperature and polarization power spectra. Ground-based \ac{cmb} experiments such as \ac{act} and \ac{spt} can improve over \textit{Planck} data, especially at small scales ($\ell \gtrsim 2000$) and in polarization measurements~\cite{ACT:2020frw}. Constraining \ac{ede} with data from the \ac{act} Data Release 4 (TT, TE, and EE power spectra)~\cite{ACT:2020frw} combined with large-scale \textit{Planck} TT \cite{Planck:2018nkj, Planck:2018vyg,  Planck:2019nip}, \textit{Planck} \ac{cmb} lensing~\cite{Planck:2018lbu}, and \ac{bao} data \cite{Ross:2014qpa,Beutler:2011hx,BOSS:2016wmc} yields $f_\mathrm{EDE}=0.091^{+0.020}_{-0.036}$ and $H_0=70.9^{+1.0}_{-2.0}$\kms  \cite{Schoneberg:2021qvd,Hill:2021yec,Poulin:2021bjr}.  The combination of \ac{act}, \ac{wmap} \cite{WMAP:2012fli}, \ac{bao}, Pantheon \cite{Pan-STARRS1:2017jku} in a Bayesian analysis leads to $f_\mathrm{EDE}=0.158^{+0.015}_{-0.094}$ and $H_0=73.43^{+2.6}_{-3.4}$\kms~\cite{Poulin:2021bjr} (seventh errorbar in Fig.~\ref{fig:EDE}). The combination of data sets \ac{act} DR4, \ac{spt3g} TEEE 2018 \cite{SPT-3G:2021eoc, LaPosta:2021pgm}, and \textit{Planck} TT650TEEE \cite{Planck:2018vyg} gives a similar preference for \ac{ede} \cite{LaPosta:2021pgm,Smith:2022hwi,Jiang:2022uyg}. Hence, replacing \textit{Planck} data with \ac{act} data leads to a preference of high values of $f_\mathrm{EDE}$ and $H_0$ without the inclusion of any $H_0$ prior \cite{Benevento:2020fev, Camarena:2021jlr,Efstathiou:2021ocp}. However, the inclusion of \ac{spt} temperature and \textit{Planck} temperature TT at $\ell > 650$ strongly decreases this preference, with $f_\mathrm{EDE}<0.071$ in a Bayesian analysis \cite{Smith:2023oop}. 

This preference for \ac{ede} diminishes with recent \ac{act} DR6 data, finding $f_\mathrm{EDE} < 0.12$ for \ac{act} combined with \textit{Planck}, and even lower for \ac{act} alone \cite{ACT:2025fju, ACT:2025tim}. For \ac{act} DR6 alone, the \ac{ede} model does not show any improvement in fit compared to \lcdm, suggesting that the preference for \ac{ede} in the previous \ac{act} DR4 could be due to a statistical fluctuation in the EE and TE power spectra. 

The most commonly used \textit{Planck} likelihood for multipoles $\ell>30$ is the \texttt{Plik} likelihood \cite{Planck:2019nip}. The \texttt{CAMSPEC}~\cite{Efstathiou:2019mdh} and \texttt{Hillipop}~\cite{Tristram:2023haj} likelihoods are alternative likelihoods, which differ in data cuts and analysis choices. Ref.~\cite{Efstathiou:2023fbn} use the \texttt{CAMSPEC} likelihood with the \textit{Planck} PR4 NPIPE maps \cite{Planck:2020olo} along with \ac{bao} data from \ac{boss} \cite{BOSS:2016wmc}, SDSS \cite{Ross:2014qpa} and 6dFGS \cite{Beutler:2011hx} and supernova data from Pantheon+ \cite{Brout:2022vxf} to constrain the \ac{ede} model. They find low values of $H_0=68.11^{+0.47}_{-0.82}$\kms and $f_\mathrm{EDE}<0.061$ in a Bayesian analysis, and similarly $H_0=68.37 \pm 0.75$\kms and $f_\mathrm{EDE}<0.094$ in a frequentist analysis (bottom two errorbars in Fig.~\ref{fig:EDE}). Using the \texttt{Hillipop} likelihood, Ref.~\cite{McDonough:2023qcu} find similarly tight constraints on \ac{ede}, as well as Ref.~\cite{Jiang:2023bsz}. Hence, both alternative \textit{Planck} likelihoods pose a challenge to the axion-like \ac{ede} model's ability to solve the $H_0$ tension.

\paragraph{Complementary EDE observables}

If \ac{ede} existed in the early Universe, it is natural to expect evidence other than the $H_0$ tension of this deviation from \lcdm. In this section, we discuss possible such signatures in data sets different from the commonly used data discussed above. Galaxies and small-scale \ac{lss} probes could provide important insight into \ac{ede} models. For example, \ac{ede} cosmologies predict \textbf{more massive clusters and an increased abundance of galaxy-mass haloes} at higher redshifts \cite{Klypin:2020tud}. Refs.~\cite{Forconi:2023hsj, Liu:2024yan, Shen:2024hpx} propose that \ac{ede} cosmologies could accommodate the massive high-redshift \ac{jwst} galaxies. However, while the fit to \ac{jwst} data improves, \ac{ede} possibly worsens the fit to Hubble Space Telescope data. 
\textbf{Lyman-$\alpha$ forest} data from SDSS \ac{eboss} \cite{eBOSS:2018qyj} and MIKE/HIRAS \cite{Viel:2013fqw} give tight upper limits on \ac{ede} \cite{Goldstein:2023gnw} due to the preference of lower values of $n_s$ than \textit{Planck}. However, the Lyman-$\alpha$ forest data shows a discrepancy with \textit{Planck} data \cite{Rogers:2023upm}, which warrants further investigation. Furthermore, an expansion rate differing from \lcdm\ would affect the relation between redshift and age, which can be constrained by the \textbf{age of ancient objects} in the Universe \cite{Boylan-Kolchin:2021fvy}. Hence, improved age constraints will be able to constrain \ac{ede} cosmologies. 

Measurements of the \ac{cmb} anisotropies along with other \ac{cmb} observables could hold clues about deviations from \lcdm. For example, if \ac{ede} would couple to photons in a parity-violating manner, it could explain the tentative evidence of \textbf{cosmic birefringence} observed in \textit{Planck} data \cite{Minami:2020odp, Diego-Palazuelos:2022mcp, Komatsu:2022nvu}. The \ac{ede} field with the potential in Eq.~\eqref{eq:axEDE_potential} predicts a unique shape of the EB power spectra, which is not favored by \textit{Planck} data and hence constrains a possible parity-violating coupling of \ac{ede} to photons \cite{Choi:2021aze, Murai:2022zur, Eskilt:2023nxm, Yin:2023srb}. Moreover, \textbf{spectral distortions} are sensitive to physics in the (pre-) recombination era. Since \ac{ede} is dynamical during this epoch, spectral distortion measurements could be sensitive to distinct \ac{ede} signatures \cite{Hart:2022agu}. 

Theoretical considerations and the requirement of a consistent UV completion can place constraints on \ac{ede} models. For example, Ref.~\cite{Rudelius:2022gyu} employed the \textbf{axion weak gravity conjecture} \footnote{The axion weak gravity conjecture is part of a broader program known as the ``swampland''. This program aims to define the boundary between effective field theories that are consistent with quantum gravity and those that exist in the “swampland” of incompatible theories. The precise statement of the axion weak gravity conjecture remains a topic of debate \cite{Hebecker:2016dsw,Heidenreich:2015nta}.} to derive an upper bound on the axion decay constant. Specifically, for $n = 1$ in the potential Eq.~\eqref{eq:axEDE_potential}, this leads to $f < 0.008\, M_{\text{Pl}}$, which is not consistent with the typically assumed parameter values. Potential ways to circumvent or relax this constraint exist \cite{Rudelius:2022gyu}. Moreover, \ac{ede} cosmologies typically prefer \textbf{higher values of $\boldsymbol n_s$} close to unity, which would point to a scale invariant Harrison-Zeldovich spectrum~\cite{Ye:2021nej, Jiang:2022uyg, Jiang:2022qlj, Peng:2023bik, Wang:2024tjd} and would have important implications for inflation~\cite{Kallosh:2022ggf,Ye:2022efx}.

Hence, while \ac{ede} models provide a promising class of models to resolve the Hubble tension, theoretical considerations and new data provide challenges to these models. 

\bigskip
\subsubsection{New early dark energy and variants \label{sec:NEDE}}

\noindent \textbf{Coordinator:} Florian Niedermann\\
\noindent \textbf{Contributors:} Martin S. Sloth, and Mathias Garny\\

\ac{nede} is a theoretical framework to address the Hubble tension where a scalar field $\psi$ undergoes a phase transition between BBN and recombination on a time scale that is short compared to the Hubble time, such as, e.g., a first-order phase transition~\cite{Niedermann:2023ssr,Niedermann:2019olb}. In contrast to other phenomenological approaches to the Hubble tension, \ac{nede} is rooted in concrete microphysics, which is operative at the $\mathrm{eV}$ scale. Current \ac{cmb} temperature and polarization measurements are sensitive to the precise mechanism triggering the phase transition, and, therefore, the phenomenology differs depending on the concrete \ac{nede} realization. The two main examples are Cold and Hot \ac{nede}, where the trigger of the phase transition is a second scalar field or the temperature of a dark sector, respectively. In the case of Cold \ac{nede}, the phenomenology shares some features with other \ac{ede}-type models, but differs both at background and perturbation level, particularly after the phase transition. On the other hand, Hot \ac{nede} is more similar to strongly interacting dark radiation (SIDR) models,  and can be viewed as a UV completion of these types of models, explaining naturally the creation of extra radiation after BBN in a supercooled phase transition.  In a nutshell, the idea is summarized in Fig.~\ref{fig:NEDE}, where the left panel explains the field-theoretic idea and the right panel depicts the cosmological model.  Initially, the \ac{nede} field $\psi$ is stuck in a false minimum of its potential. It is separated from the true minimum by a large potential barrier. The (false) vacuum energy associated with $\psi$ constitutes an early \ac{de} component (orange line). However, as the trigger mechanism kicks in, the barrier shrinks, initiating a strong first-order phase transition. This corresponds to the nucleation of bubbles that separate the true from the false vacuum. Eventually, all of space is converted to the true vacuum. In the case of a fast phase transition, this can be described in a fluid model in terms of an abruptly decaying early \ac{de} component (red line). It leaves behind a residual vacuum energy, which can be identified with \ac{de} (green line). As a closely related idea, we mention ChainEDE, which relies on a series of a large number of first-order phase transitions to address the $H_0$ tension~\cite{Freese:2021rjq}.

\begin{figure}[ht]
     \centering
          \begin{subfigure}[b]{0.47\textwidth}
         \centering
         \includegraphics[width=\textwidth]{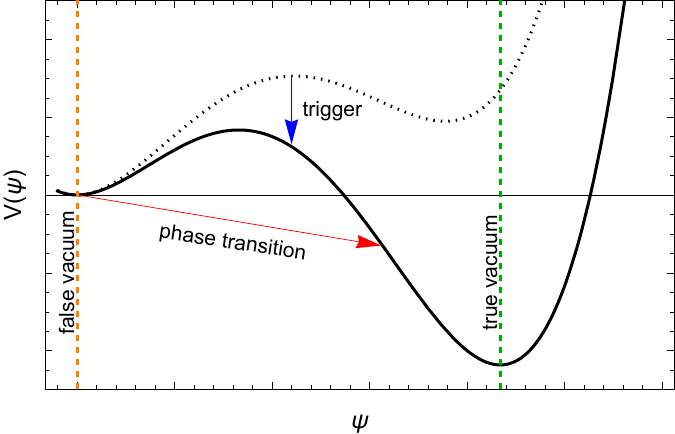}\vspace{0.2cm}
         \caption{Microscopic picture}
         \label{fig:three sin x}
     \end{subfigure}
     \hfill
     \begin{subfigure}[b]{0.47\textwidth}
         \centering
         \includegraphics[width=\textwidth]{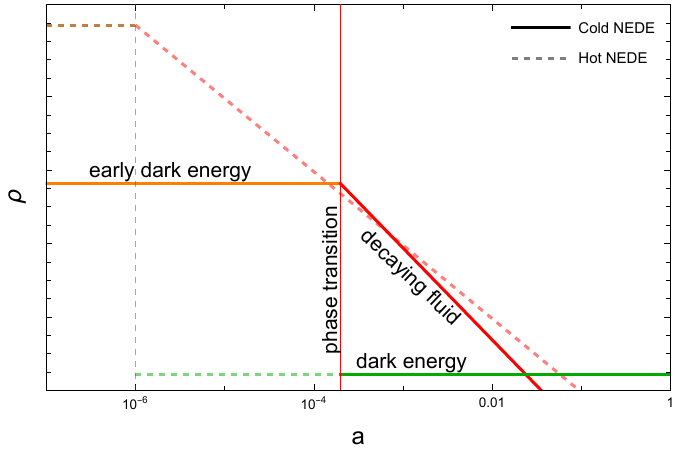}
         \caption{Cosmological fluid model}
         \label{fig:y equals x}
     \end{subfigure}
     \hfill
         \caption{Schematic representation of the \ac{nede} framework, explaining its microphysics and phenomenology. \underline{Left:} The \ac{nede} field decays in a triggered first-order phase transition, giving rise to an energy injection before recombination. \ac{ede} is identified with the field's vacuum energy before the transition, and \ac{de} with the true vacuum after the transition. \underline{Right:} On cosmological length scales, a fluid description can be applied. Cold and Hot \ac{nede} energy densities are represented as the solid and dashed lines, respectively.}
        \label{fig:NEDE}
        \vspace{-0.4cm}
\end{figure}

\paragraph{Cold New Early Dark Energy}
Cold \ac{nede} introduces an additional, ultralight scalar field $\phi$ to trigger the phase transition~\cite{Niedermann:2020dwg,Niedermann:2019olb}. Its effective potential valid at low energies is
\begin{equation}\label{cold_NEDE}
    V(\psi,\phi) =\frac{\lambda}{4}\psi^4+\frac{\beta}{2} M^2 \psi^2 -\frac{1}{3}\alpha M \psi^3 + \frac{1}{2}m^2\phi^2 +\frac{1}{2}\tilde\lambda \phi^2\psi^2  \ldots\,,
\end{equation}
where $\alpha$, $\beta$, $\lambda$, and $\tilde \lambda$ are dimensionless EFT parameters. The model introduces a hierarchy of scales, where the trigger field $\phi$ is an ultralight field with mass $m \sim 10^{-27} \mathrm{eV}$ and the tunneling field $\psi$ is much heavier with mass $M \sim \mathrm{eV}$. A possible high-energy completion has been argued to be possible within a multi-axion framework, where the small masses are protected through an approximate shift symmetry~\cite{Cruz:2023lmn}. In terms of its cosmological fluid description, Cold \ac{nede} introduces four additional parameters: the fluid's maximal energy fraction, $f_\mathrm{NEDE}$, the redshift $z_*$ of the phase transition, the equation of state parameter of the post-phase transition fluid, $w_\mathrm{NEDE}$, and the relic abundance of the trigger field today $\Omega_\phi$. 

The phenomenological model has been implemented in the Boltzmann code \texttt{TriggerCLASS}\footnote{\href{https://github.com/NEDE-Cosmo/TriggerCLASS.git}{https://github.com/NEDE-Cosmo/TriggerCLASS.git}} and tested extensively against cosmological data in the literature. Most work has been done in the limit where $\Omega_\phi \ll 1$, corresponding to a subdominant trigger field. For example, using Planck 2018~\cite{Planck:2018vyg}, Pantheon~\cite{Pan-STARRS1:2017jku}, and \ac{bao}~\cite{Beutler:2011hx,Ross:2014qpa,BOSS:2016wmc} data, the $H_0$ tension was found to be reduced to $2 \sigma$ when employing the $Q_\mathrm{DMAP}$ tension measure~\cite{Schoneberg:2021qvd} (with a value for $H_0$ taken from Ref.~\cite{Riess:2020fzl}). This result was supported later in Ref.~\cite{Cruz:2023cxy} in a profile likelihood approach, which gets around prior volume issues, that are otherwise are known to drive the posteriors for $f_\mathrm{NEDE}$ towards small values~\cite{Niedermann:2020dwg} (see also the discussion in Sec.~\ref{sec:EDE} for more details). Without including a prior on $H_0$, it was found that $H_0 = 69.56^{+1.16}_{-1.29}$ \kms and $f_\mathrm{NEDE} = 0.076^{+0.040}_{-0.035}$ at $ 68 \% $  CL, amounting to a $2 \sigma$ indication for a non-vanishing fraction of \ac{nede}. Including the SH0ES prior, $H_0 = 71.62^{+0.78}_{-0.76}$ \kms and $f_\mathrm{NEDE} = 0.136^{+0.024}_{-0.026}$ was obtained at $68 \%$ CL, which amounts to a $5 \sigma$ evidence for a non-vanishing fraction of \ac{nede}. The model has also been studied in the presence of additional \ac{lss}~\cite{Niedermann:2020qbw,Cruz:2023cxy} and ground-based \ac{cmb} data~\cite{Poulin:2021bjr,Haridasu:2022dyp,Cruz:2022oqk}, broadly matching the baseline results cited above, although it should be noted that axiEDE and Cold \ac{nede} respond differently to \ac{act} and \ac{spt} data~\cite{Poulin:2021bjr}.
Recently the assumption $\Omega_\phi \ll1$ was dropped in Ref.~\cite{Cruz:2023lmn}. In that case, the trigger field can make a sizable contribution to the energy budget today in the form of a fuzzy \ac{dm} component (for a phenomenologically similar model see also Ref.~\cite{Allali:2021azp}). With their non-vanishing pressure fluctuations, the trigger field's perturbations act against gravitational collapse on small scales as required for solving also the $S_8$ tension. As a result, the model was found to simultaneously reduce both tensions\footnote{This is another important difference at the phenomenological level between axiEDE (and other \ac{ede} type models) and \ac{nede}. While axiEDE is known to worsen the $S_8$ tension, \ac{nede} has the potential to resolve both the $H_0$ tension and the $S_8$ tension.} below $2 \sigma$ (with the above choice of datasets $S_8 = 0.818^{+0.023}_{-0.017}$ was obtained).

\begin{figure}
\begin{center}
    \includegraphics[width=0.5\textwidth]{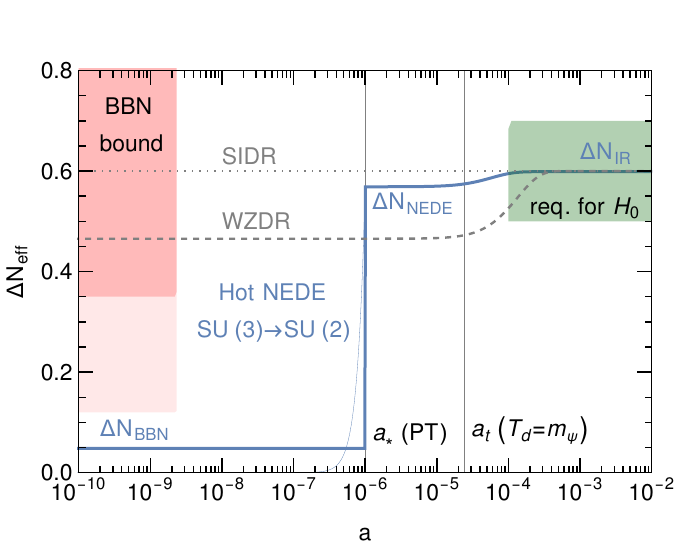}
\end{center}
    \caption{\label{fig:HotNEDEoverview}
    Evolution of $\Delta N_{\rm eff}$ in Hot \ac{nede} with a supercooled Coleman-Weinberg phase transition (PT) between the \ac{bbn} and recombination epochs, leading to a strong increase ($\Delta N_{\rm BBN}\to\Delta N_{\rm NEDE}$) due to the latent heat being converted into relativistic $SU(2)$ dark gauge and Higgs bosons during the PT, and a second slight increase due to the dark Higgs becoming non-relativistic ($\Delta N_{\rm NEDE}\to \Delta N_{\rm IR}$). This evolution is intimately linked to the underlying Coleman-Weinberg model. Also shown are bounds from \ac{bbn}, and the values required to solve the $H_0$ tension around recombination, as well as two models featuring extra dark radiation but no phase transition (SIDR and WZDR~\cite{Aloni:2021eaq}).}
\end{figure}

\paragraph{Hot New Early Dark Energy}

In Hot \ac{nede}~\cite{Niedermann:2021vgd,Niedermann:2021ijp}, the phase transition is triggered by the decreasing dark sector temperature $T_d$, and the sound horizon is lowered due to the dark radiation generated from the latent heat, i.e., the false vacuum energy released into the dark sector during the transition~\cite{Garny:2024ums} (for a comparison to Cold \ac{nede} see Table~\ref{tab:compare}). If the phase transition occurs between \ac{bbn} and recombination, this setup naturally reconciles SIDR-type solutions to the Hubble tension with \ac{bbn} bounds on extra radiation~\cite{Garny:2024ums} (see Fig.~\ref{fig:HotNEDEoverview}). It furthermore differs from SIDR at the perturbation level, and makes complementary predictions. In particular, the model predicts oscillatory features on small scales in the matter power spectrum alongside a \ac{sgwb} signal that can be looked for in Lyman-$\alpha$ forest data and pulsar timing arrays, respectively.

Hot \ac{nede} can be realized by the well-known Coleman-Weinberg mechanism of spontaneous symmetry breaking within a dark sector featuring a dark Higgs field $\psi$ that transforms under a dark gauge symmetry $SU(N)$. This setup naturally leads to the required supercooled first-order phase transition, breaking the initial $SU(N)$ to $SU(N-1)$, as well as to self-interactions described by the non-Abelian gauge symmetry. The dynamics is captured by the effective potential~\cite{Garny:2024ums}
\begin{align}\label{hot_NEDE}
    V({\psi}; T_d) =  B\psi^4\left(\ln\frac{\psi^2}{v^2}-\frac12\right)  -\frac{\mu_{\rm eff}^2}{2}{\psi}^2\left(1-\frac{\psi^2}{2v^2}\right)  + \Delta V_\mathrm{thermal}({\psi}; T_d) \,,
\end{align}
where the dimensionless parameter $B \sim g^4 $ depends on the gauge coupling $g$.  During the \ac{nede} phase transition, $\psi$ picks up a super-$\mathrm{eV}$ expectation value $v$. Due to an approximate conformal symmetry  (controlled by the soft breaking parameter $\mu_\mathrm{eff} \sim \mathrm{eV}$), the phase transition exhibits a significant amount of supercooling, where the latent heat dominates over the dark radiation plasma, giving rise to an \ac{ede} component. 
In a more fundamental picture, $\psi$ is the modulus of a dark sector Higgs multiplet, breaking an initial $SU(N)$ gauge symmetry to $SU(N-1)$. The corresponding massless gauge bosons alongside the light Higgs degrees of freedom are then populated during the phase transition, reheating the dark sector and introducing a sizable increase in the effective number of relativistic degrees of freedom, $\Delta N_\mathrm{eff}$. Although both Cold and Hot \ac{nede} feature an \ac{ede} component, their phenomenology differs significantly. In particular, the Hot \ac{nede} phase transition occurs before the \ac{cmb} epoch at redshifts $z_* \gtrsim 10^5$ (see the right panel in Fig.~\ref{fig:NEDE}). As a result, Hot \ac{nede} lowers the sound horizon $r_s$ (and thus increases $H_0$ as explained in Sec.~\ref{sec:EDE}) through the energy injection provided by the strongly interacting dark sector plasma \textit{after} the phase transition and, on a phenomenological level, is thus more akin to SIDR models (with mass threshold effects) such as Ref.~\cite{Aloni:2021eaq}. However, its main phenomenological advantage is that it gets around \ac{bbn} constraints on $\Delta N_\mathrm{eff}$, which otherwise would rule out SIDR models as solutions to the Hubble tension~\cite{Schoneberg:2022grr,Garny:2024ums}, see Fig.~\ref{fig:HotNEDEoverview} (for other ideas to avoid \ac{bbn} bounds see Refs.~\cite{Berlin:2017ftj,Berbig:2020wve,Escudero:2022gez,Aloni:2023tff,Fischler:2010xz,Hooper:2011aj,Bjaelde:2012wi,Choi:2012zna,Nygaard:2023gel}). A first test against cosmological data~\cite{Garny:2024ums} showed that the model reduces the $Q_\mathrm{DMAP}$ tension to $2.8 \sigma$. Moreover, it was found that $H_0 = 96.13^{+0.62}_{-1.00}$ \kms and $H_0=71.17 \pm 0.83$ \kms without and with including a Gaussian prior on $H_0$, respectively (datasets as for cold \ac{nede} except for the prior on $H_0$ taken from Ref.~\cite{Riess:2021jrx}).

\begin{table*}[t]
\small
\renewcommand{\arraystretch}{1.3}
	\centering
    \caption{\label{tab:compare} Comparison of Cold and Hot \ac{nede}, both relying on a fast-triggered phase transition before recombination to address the Hubble tension.}
	\begin{tabular}{|c||c|c|} \hline 
        \textbf{\ac{nede} model} & \textbf{Cold \ac{nede}} & \textbf{Hot \ac{nede}} \\ \hline \hline
        \textbf{model family} & \ac{ede} & (stepped) strongly interacting dark radiation\\ \hline
        \textbf{physics lowering $r_s$} & false vacuum energy &  latent heat released as SIDR \\ \hline
        \textbf{trigger} & ultralight field & dark sector temperature \\ \hline
        \textbf{redshift of phase transition }& $10^3  \lesssim z_* \lesssim 10^4$&$10^5  \lesssim z_* \lesssim 10^9$\\ \hline
        \textbf{equation of state parameter}&  $w_\mathrm{NEDE}>1/3$ ($\sim 2/3$)&$w_\mathrm{NEDE}=1/3$\\ \hline
        \textbf{microphysics}& axion-like particle & dark sector Higgs   \\ \hline
    \end{tabular}
\end{table*}
\bigskip
\subsubsection{Extra relativistic degrees of freedom \label{sec:Extra_DoF}}

\noindent \textbf{Coordinator:} Sunny Vagnozzi\\
\noindent \textbf{Contributors:} Ad\`{e}le Poudou, Alexander Bonilla Rivera, Biswajit Karmakar, Branko Dragovic, Davide Pedrotti, Diego Rubiera-Garcia, L\'aszl\'o \'Arp\'ad Gergely, Leila L. Graef, Luca Visinelli, Luis Anchordoqui, Marcin Postolak, Margus Saal, Mariana Melo, Matteo Forconi, \"Ozg\"ur Akarsu, Salvatore Capozziello, Sebastian Bahamonde, Shouvik Roy Choudhury, Simony Santos da Costa, Stefano Gariazzo, Thejs Brinckmann, Utkarsh Kumar, Vivian Poulin, William Giar\`e, and Wojciech Hellwing
\\

Introducing extra relativistic species (ERS, also referred to as ``dark radiation'') is one of the most well-motivated and straightforward extensions of the \lcdm\ model, but also one of the simplest ways of raising the inferred value of $H_0$~\cite{Gariazzo:2023hch}. The ESR energy density is most easily parametrized via the effective number of neutrino species $N_{\text{eff}}$~\cite{Steigman:1977kc}, controlling the relation between the total radiation energy density $\rho_r$ and the photon energy density $\rho_{\gamma}$~\cite{Archidiacono:2013fha,Archidiacono:2022ich}
\begin{eqnarray}
    \rho_r = \rho_{\gamma} \left [ 1+\frac{7}{8} \left ( \frac{4}{11} \right ) ^{\frac{4}{3}}N_{\text{eff}} \right ] \,.
    \label{eq:neff}
\end{eqnarray}
In the presence of only three standard neutrinos undergoing instantaneous decoupling, $N_{\text{eff}}=3$. However, in the standard cosmological model, $N_{\text{eff}}^{\text{SM}}$ is slightly larger than $3$, owing to the fact that the neutrino decoupling process is not instantaneous. The latest determinations yield $N_{\text{eff}}^{\text{SM}} \simeq 3.044$~\cite{Akita:2020szl,Froustey:2020mcq,Bennett:2020zkv,Drewes:2024wbw}. After electron-positron annihilation, the only relativistic particles in the standard cosmological model are neutrinos and photons. However, several well-motivated extensions of the Standard Model of particle physics predict the existence of additional particles which could be relativistic at epochs of cosmological interest, including but not limited to light sterile neutrinos~\cite{Garcia:2011fia,Anchordoqui:2011nh,Anchordoqui:2012qu,Jacques:2013xr,RoyChoudhury:2018bsd}, Goldstone bosons~\cite{Weinberg:2013kea, Lin:2022xbu}, axions, axion-like particles, and ultra-light \ac{dm}~\cite{Arias:2012az,Marsh:2015xka,Baumann:2016wac,Poulin:2018dzj,DEramo:2018vss,Ferreira:2020fam,Semertzidis:2021rxs,Chadha-Day:2021szb,Green:2021hjh,OHare:2024nmr,Poulin:2018dzj,Caloni:2022uya,DEramo:2022nvb}, light or massless dark photons~\cite{Cline:2012is,Fan:2013yva,Vogel:2013raa,Petraki:2014uza,Foot:2014uba,Foot:2014osa,Foot:2016wvj,Flambaum:2019cih,Anchordoqui:2019yzc}, and so on~\cite{Steigman:2013yua,Brust:2013ova}): such particles lead to a value of $\Delta N_{\text{eff}} = N_{\text{eff}}-N_{\text{eff}}^{\text{SM}} \neq 0$. It is worth noting that a non-zero value of $\Delta N_{\text{eff}}$ does not necessarily imply the existence of new particles. For instance, it could be associated with a non-standard distribution of standard neutrinos, or the presence of neutrino asymmetries (i.e., a non-zero chemical potential)~\cite{Nunes:2017xon,Bonilla:2018nau}.
In addition, a \ac{sgwb}, possibly at the origin of the signal detected in pulsar timing array data~\cite{NANOGrav:2020bcs,Goncharov:2021oub,EPTA:2021crs,Antoniadis:2022pcn}, could also provide a significant contribution to $N_{\text{eff}}$~\cite{Allen:1997ad,Smith:2006nka,Boyle:2007zx,Kuroyanagi:2014nba,Cabass:2015jwe,Ben-Dayan:2019gll,Aich:2019obd,Giare:2022wxq,NANOGrav:2023hvm,Benetti:2021uea,Vagnozzi:2023lwo}. Moreover, $\Delta N_{\text{eff}}$ can in principle be negative, for instance, in low-reheating scenarios following inflation, where the reheating temperature $T_{\text{rh}}$ can be as low as a few MeV, leading to incomplete thermalization of standard neutrinos~\cite{Kawasaki:1999na,Gelmini:2004ah,deSalas:2015glj,Gerbino:2016sgw}.

A non-zero value of $\Delta N_{\text{eff}}$ leads to a host of signatures at both the times of \ac{bbn} and recombination. ERS increases the physical radiation density and thereby raises the pre-recombination expansion rate $H(z)$ before recombination (as well as after, although by then the radiation component is sub-dominant). At the \ac{bbn} epoch, this pushes the freeze-out of nuclear interactions towards higher temperatures, raising the yield of light elements. At recombination, the presence of ERS leads to a variety of direct and indirect signatures on the \ac{cmb}: however, once the redshift of matter-radiation equality $z_{\text{eq}}$ and the acoustic angular scales $\theta_s$ are fixed, the only remaining effects of $\Delta N_{\text{eff}}$ on the \ac{cmb} power spectra consist of an increased (Silk) damping of the higher acoustic peaks and a phase-shift of the acoustic peaks towards larger scales (e.g., see Refs.~\cite{Hou:2011ec,Baumann:2018muz,Vagnozzi:2019utt}). All these effects indeed lead to some of the tightest cosmological constraints on $\Delta N_{\text{eff}}$: current \ac{bbn} and \ac{cmb} inferences both broadly indicate $\Delta N_{\text{eff}} \lesssim 0.2-0.3$, while next-generation cosmological surveys will be able to improve the sensitivity of $\Delta N_{\text{eff}}$ by almost an order of magnitude~\cite{CMB-S4:2016ple,SimonsObservatory:2018koc}.

\begin{figure}[ht]
\centering
\includegraphics[width=0.7\linewidth]{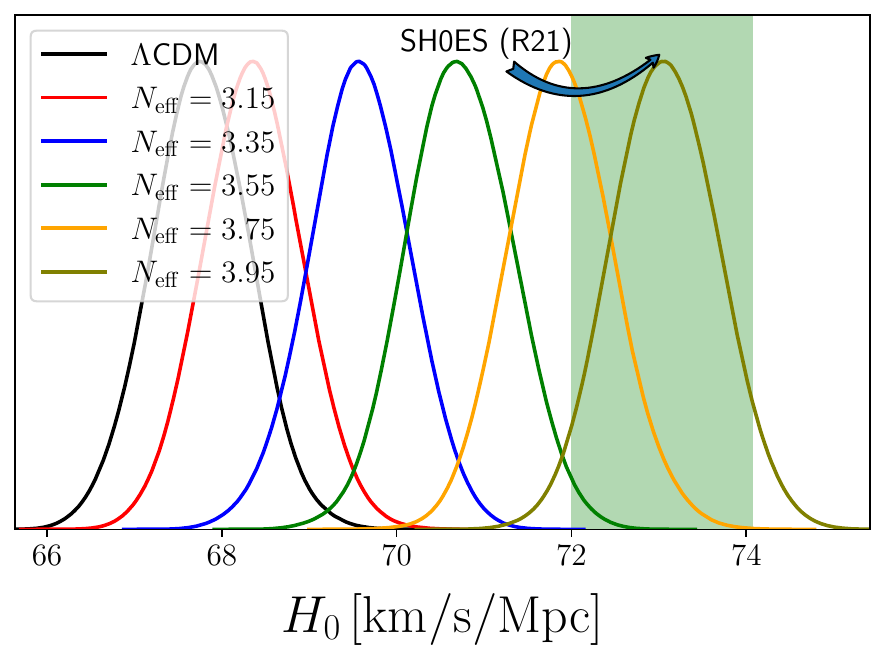}
\caption{Posterior distributions for $H_0$ inferred from a combination of \textit{Planck} 2018, SDSS \ac{bao}, and \textit{PantheonPlus} \ac{sn1} data for different fixed values of $N_{\text{eff}}$. The green shaded region corresponds to the \textit{R21} measurement $H_0=73.04\pm 1.04$\kms as inferred by the SH0ES team in Ref.~\cite{Riess:2021jrx}.}
\label{fig:h0neff}
\end{figure}

Increasing $N_{\text{eff}}$ while keeping the physical baryon and \ac{dm} densities $\omega_b$ and $\omega_c$ fixed has the effect of increasing the pre-recombination expansion rate, thus decreasing the comoving sound horizon at recombination $r_s$, which then requires a higher value of $H_0$ in order to keep $\theta_s$ fixed. This is the reason why ERS are one of the most economical ways of raising $H_0$ and are generally considered a benchmark model in this sense. These effects lead to a positive correlation between $N_{\text{eff}}$ and $H_0$~\cite{DiValentino:2016hlg,Bernal:2016gxb}, with an increase in the former leading to an increase in the latter, as captured by the following linear relation~\cite{Vagnozzi:2019ezj}, as can be seen in Fig.~\ref{fig:h0neff}
\begin{eqnarray}
    \Delta H_0 = H_0\vert_{\Delta N_{\text{eff}} \neq 0} - H_0\vert_{\Lambda{\text{CDM}}} \simeq 5.9\Delta N_{\text{eff}}\,,
    \label{eq:deltah0neff}
\end{eqnarray}
as calibrated to \textit{Planck} 2018 \ac{cmb}, \ac{bao}, and \textit{Pantheon} \ac{sn1} measurements in 2020. This relation, first obtained in Ref.~\cite{Vagnozzi:2019ezj}, was illustrated in~ Ref.\cite{Pedreira:2023qqt} in the $H_0$-$\sigma_8$ plane, together with the analogous relation for several tension-resolving candidate models.

From Eq.~\eqref{eq:deltah0neff} we see that increasing $H_0$ to a level that solves the Hubble tension requires $N_{\text{eff}} \gtrsim 4$, corresponding to the effect of an extra fully thermalized neutrino which, importantly, is completely excluded by \ac{bbn} and \ac{cmb} considerations. The main reason why the \ac{cmb} excludes high values of $N_{\text{eff}}$ is that they disproportionately alter the damping scale $r_d$, in turn altering the ratio $r_s/r_d$, which is tightly constrained especially by high-multipole polarization measurements to be close to its \lcdm\ value (see e.g., the discussion in Ref.~\cite{Knox:2019rjx}). Moreover, the neutrino drag/phase shift effect~\cite{Bashinsky:2003tk,Follin:2015hya} is also altered to an extent that is incompatible with the data. This is why the simplest, 7-parameter \lcdm+$N_{\text{eff}}$ model fails to solve the Hubble tension, despite constituting a useful benchmark~\cite{DiValentino:2019dzu,Seto:2021xua,DiValentino:2021izs}. Within the simplest 7-parameter \lcdm+$N_{\text{eff}}$ model one infers $N_{\text{eff}}=2.89 \pm 0.19$, $H_0=66.3 \pm 1.4$\kms, and $S_8=0.831 \pm 0.013$ from \textit{Planck} 2018 temperature, polarization, and lensing data, whereas further adding SDSS \ac{bao} and \textit{Pantheon} \ac{sn1} data change these numbers to $N_{\text{eff}}=3.00 \pm 0.17$, $H_0=67.4 \pm 1.1$\kms, and $S_8=0.823 \pm 0.011$ respectively (in what follows, the switch from \textit{Pantheon} to \textit{PantheonPlus} \ac{sn1} data is not expected to bring about significant changes).

The cosmological evolution of ERS characterized by Eq.~\eqref{eq:neff} is de facto equivalent to that of massless neutrinos, and for this reason we can refer to the \lcdm+$N_{\text{eff}}$ model as describing \textit{free-streaming} ERS. It is worth noting that the above considerations hold for some of the simplest dark radiation models, including light sterile neutrinos~\cite{Hagstotz:2020ukm}, models with lepton asymmetries/non-zero neutrino chemical potential~\cite{Barenboim:2016lxv,Seto:2021tad}, thermal axions~\cite{DEramo:2018vss}, and so on. Relaxing the assumption that ERS are free-streaming can potentially help address the above shortcomings. The simplest possibility for such a \textit{self-interacting dark radiation} is one where the ERS component is strongly self-coupled and therefore constitutes a perfect fluid~\cite{Baumann:2015rya}, whose Boltzmann hierarchy can be truncated to the first moment (with vanishing anisotropic stress). Nevertheless, when confronted with cosmological data, such a scenario fails to solve the Hubble tension~\cite{Schoneberg:2021qvd}, as it is unable to reduce the neutrino drag and Silk damping effects to an extent that is allowed by the data while raising $H_0$ sufficiently. When confronted with \textit{Planck} 2018 and SDSS \ac{bao} data, one finds $H_0= 68.67 \pm 0.84$\kms. Other natural extensions envisage the possibility of both free-streaming and self-interacting ERS, for instance within the ``dark sector equilibration'' model of Ref.~\cite{Blinov:2020hmc}, and the ``recoupling'' or ``instantaneous decoupling'' models (such as in atomic \ac{dm}~\cite{Kaplan:2009de,Cyr-Racine:2012tfp,Bansal:2022qbi} or twin Higgs~\cite{Chacko:2018vss}) of Ref.~\cite{Brinckmann:2022ajr}, which, however, can only partially alleviate the tension, leading to figures very similar to the previous ones. Yet another possibility to undo the unwanted effects of free-streaming ERS and improve the tension-solving ability of self-interacting dark radiation features scattering between the latter and the \ac{dm} component: however, this model can only partially alleviate the Hubble tension and is disfavored from a model-comparison perspective due to its higher statistical complexity~\cite{Schoneberg:2021qvd}. In this case, one finds $H_0= 68.55 \pm 0.92$\kms from a fit to \textit{Planck} 2018 and SDSS \ac{bao} data~\cite{Schoneberg:2021qvd}.

A class of models that has received considerable interest in this context features non-standard interactions that do not involve the dark radiation components, but the otherwise standard neutrino species. Non-standard neutrino interactions, for instance via a four-point contact interaction mediated by a sufficiently massive ($m \gtrsim 1\,{\text{keV}}$) mediator, as studied in \cite{Cyr-Racine:2013jua, Archidiacono:2013dua,Lancaster:2017ksf,Oldengott:2017fhy,Barenboim:2019tux,Das:2020xke,RoyChoudhury:2020dmd,Brinckmann:2020bcn,RoyChoudhury:2022rva,Brinckmann:2022ajr,Das:2023npl,Camarena:2023cku,He:2023oke,Bostan:2023ped,Camarena:2024daj} can suppress or delay neutrino free-streaming, and, with it, the free-streaming-induced phase shift. for fixed values of the sound horizon $r_s$ (which is mostly unaffected by these interactions), the position of the first acoustic peak would therefore be at higher multipoles. Reversing it to the observed position requires a higher value of $H_0$. It has been shown that a combination of self-interacting neutrinos, additional dark radiation components ($N_{\text{eff}} \sim 4$), and large neutrino masses ($M_{\nu} \sim 1\,{\text{eV}}$) can offset the unwanted effects of dark radiation on the damping tail of the \ac{cmb}, and allow for large values of $H_0$ while not spoiling the fit to \ac{cmb} temperature data~\cite{Kreisch:2019yzn}. A fit to the latter and \ac{bao} data reveals a bi-modal distribution for the interaction strength, with stronger values of the interaction strength ($\sim 10^9$ times that of the weak interaction strength) being the tension-solving ones. Nevertheless, this solution is strongly challenged from both the data and theoretical perspectives. From the data side, small-scale polarization data strongly disfavor these models with strong self-interaction strength~\cite{RoyChoudhury:2020dmd,Brinckmann:2020bcn}), which are also partially challenged by \ac{bao} data (due to the fact that $r_s$ is unchanged)~\cite{Camarena:2024daj}. In particular, when confronted against \textit{Planck} 2018, SDSS \ac{bao}, and \textit{PantheonPlus} \ac{sn1} data, one infers $H_0= 67.3^{+2.2}_{-2.1}$\kms and $H_0=66.7^{+2.2}_{-2.1}$\kms for the moderately interacting and strongly interacting modes respectively~\cite{RoyChoudhury:2020dmd,Brinckmann:2020bcn}. Finally, from the model-building side, a model in which all three neutrinos self-interact equally is excluded by laboratory constraints \cite{Blinov:2019gcj}, while constructing a phenomenologically viable model has proven challenging \cite{Bostan:2023ped}. EFT analyses of \ac{boss} data reveal support for a strongly interacting neutrino mode~\cite{Camarena:2023cku}, although an analysis combined with \ac{cmb} data still disfavors the strongly self-interacting neutrino model where all 3 neutrinos are interacting~\cite{Camarena:2024daj}. Interestingly, parameter degeneracies allow for a lower value of the scalar spectral index in the interacting neutrino model, which can be leveraged to allow for inflationary models that are otherwise ruled out within \lcdm, such as natural inflation and Coleman-Weinberg inflation~\cite{Barenboim:2019tux,RoyChoudhury:2022rva,Bostan:2023ped}. It is worth noticing that large values of $N_{\text{eff}}$ are in tension with \ac{bbn}, unless this extra dark radiation is generated after the \ac{bbn} epoch. This is the scenario studied in Ref.~\cite{daCosta:2023mow}, where a fraction of the \ac{dm} abundance results from the decay of an unstable and initially thermally decoupled heavy particle, which decays into a \ac{dm} particle (initially relativistic and therefore contributing to $N_{\text{eff}}$, before becoming cold once the Universe has expanded sufficiently) and a Standard Model particle such as neutrinos or photons. In this case, one finds in the best case $H_0=69.08 \pm 0.71$\kms and $S_8=0.850 \pm 0.008$ from a fit to \textit{Planck} 2018, SDSS \ac{bao}, and \textit{Pantheon} \ac{sn1} data.

Although the simplest self-interacting neutrino models appear to be ruled out, the framework remains a very interesting benchmark one and gives a good idea of possible ingredients one might want to consider in order to make non-minimal ERS models viable. Motivated by these considerations, various works have explored models of eV-scale Majorons, pseudo-Goldstone bosons associated with the spontaneous breaking of a global $U(1)$ lepton number symmetry. The Majoron enjoys weak couplings to neutrinos, although in the theoretically best motivated coupling limit, it cannot induce the four-point contact interactions mentioned earlier. The resulting phenomenology is instead dominated by inverse neutrino decays and Majoron decays. Overall, these effects still damp neutrino free-streaming, although with a different time dependence, which leads to a better performance of the model when confronted against cosmological data. It should be noted that Majoron decays also lead to a value of $\Delta N_{\text{eff}} \sim 0.11$~\cite{Escudero:2019gvw}. Nevertheless, although relatively successful, Majoron models fall short of completely solving the Hubble tension~\cite{Barenboim:2020dmg}, with $H_0=70.06^{+2.21}_{-2.31}$\kms from a fit to \textit{Planck} 2018, SDSS \ac{bao}, and \textit{Pantheon} \ac{sn1} data. A fit to \textit{Planck} 2018 and SDSS \ac{bao} data instead leads to $H_0 = 70.18 \pm 0.61\,$\kms, whereas the $S_8$ values inferred within the model are not available \cite{Escudero:2021rfi}.

The above are only examples of some of the ERS models considered in the literature, and it is beyond our scope to give a full list of examples. Other interesting possibilities which have been considered in the literature involve self-interacting sterile neutrinos~\cite{Archidiacono:2014nda,Forastieri:2017oma,Archidiacono:2020yey,Corona:2021qxl}, or the so-called stepped fluids~\cite{Aloni:2021eaq,Joseph:2022jsf,Bagherian:2024obh,Cho:2024lhp}. The latter are a particular class of self-interacting ERS models, consisting of a mix of massless and massive particles: when the temperature of the dark sector drops below the mass of a massive species, the latter deposits entropy in the lighter ones, leading to a ``step'' in the ERS energy density. Concrete Lagrangian realizations of these classes of models have been studied, for instance, within the context of ``Wess-Zumino dark radiation'', and have been found to be relatively promising in alleviating the Hubble tension, although current data sets do not show a strong preference for the model~\cite{Schoneberg:2022grr}. From a fit to \textit{Planck} 2018, SDSS \ac{bao}, and \textit{Pantheon} \ac{sn1} data, one finds $H_0=69.3^{+0.9}_{-1.3}$\kms and $S_8=0.829 \pm 0.011$ for the Wess-Zumino dark radiation model, whereas for a more general stepped dark radiation model these change to $H_0=68.8^{+1.3}_{-1.5}$\kms and $S_8=0.823 \pm 0.013$ respectively~\cite{Aloni:2021eaq}.

Overall, despite the fact that there is no consensus ERS model that solves the Hubble tension, many of these models possess interesting features that could guide the community in the right direction. It would therefore not be surprising if ERS were to play at least some role in the Hubble tension. These and related aspects are being actively investigated.

\bigskip
\subsection{Late-time proposals}  \label{sec:LTP_4.2}
\subsubsection{Late dark energy \label{sec:Late_DE}}

\noindent \textbf{Coordinator:} Rafael C. Nunes\\
\noindent \textbf{Contributors:} Abdolali Banihashemi, Adri\`a G\'omez-Valent, Anil Kumar Yadav, Anjan Ananda Sen, Anne Christine Davis, Arianna Favale, Benjamin L'Huillier, Branko Dragovic, Brooks Thomas, Carlos G. Boiza, Carsten van de Bruck, Celia Escamilla-Rivera, Cláudio Gomes, Cristian Moreno, Daniele Oriti, David Benisty, David Tamayo, Davide Pedrotti, Elsa Teixeira, Emmanuel N. Saridakis, Gaetano Lambiase, Giulia De Somma, Hsu-Wen Chiang, Hussain Gohar, Ilim Cimdiker, Jaume Haro, Javier Rubio, Joan Sol\`a Peracaula, Juan Garc\'ia-Bellido, Jurgen Mifsud, Kathleen Sammut, Keith R.~Dienes, Laur J\"arv, Laura Mersin, Leandros Perivolaropoulos, Leila L. Graef, Lilia Anguelova, Luis A. Escamilla, Luz \'Angela Garc\'ia, Mahdi Najafi, Margus Saal, Mariam Bouhmadi-L\'opez, Masoume Reyhani, Mina Ghodsi Yengejeh, Nikolaos E. Mavromatos, Nima Khosravi, Oem Trivedi, \"Ozg\"ur Akarsu, Paloma Morilla, Paolo Salucci, Purba Mukherjee, Rahul Shah, Ronaldo C. Batista, Ruth Lazkoz, Salvatore Capozziello, Sanjay Mandal, Sebastian Bahamonde, Simony Santos da Costa, Suresh Kumar, Utkarsh Kumar, V\'ictor H. C\'ardenas, Vasiliki A. Mitsou, Vincenzo Salzano, Vivian Poulin, and Wojciech Hellwing
\\

\paragraph{Current status of simplified DE model parameterizations circa 2024}

The key property of \ac{de} is its \ac{eos}, defined as $w \equiv \frac{P_{\rm x}}{\rho_{\rm x}}$. Extensions of the \lcdm\ model, where $w$ can either be a constant or a dynamical function of cosmic time, represent the simplest parametric approaches for testing deviations from the \lcdm\ framework. Fig.~\ref{fig:whisker_lateDE} presents the current state-of-the-art constraints on the \ac{eos}, $w$, at the 68\% CL, inferred from various dataset combinations under the assumption of the $w$CDM model  \cite{Yang:2021flj}. The figure is adapted from Ref.~\cite{Escamilla:2023oce}. As widely discussed in the literature, models with $w < -1$, when constrained using \ac{cmb} data alone, tend to predict higher values of $H_0$, due to a strong degeneracy between $w$ and $H_0$ \cite{Dong:2023jtk}. To achieve a more robust analysis, a joint dataset combination of \ac{cmb}+\ac{bao}+\ac{sn}+\ac{cc} yields $H_0 = 68.6^{+1.7}_{-1.5}$\kms, which is in good agreement with the \lcdm. 

Robust and up-to-date analyses of simple extensions to the well-known dynamic $w(z)$CDM models are presented in Refs.~\cite{Giare:2024gpk,Najafi:2024qzm}. The results include the most recent \ac{bao} measurements obtained by \ac{desi}. The preference for the $w(z)$CDM model remains robust, regardless of the parameterization used, but none of the parameterizations are able to resolve the $H_0$ tension. Similar or identical simple parametric models within the $w(z)$CDM class have been discussed in the literature. In general, while certain datasets may favor a dynamic parametrization of \ac{de}, these models are not capable of resolving the tension in $H_0$.

\begin{figure}[!t]
\centering
\includegraphics[width=0.6\textwidth]{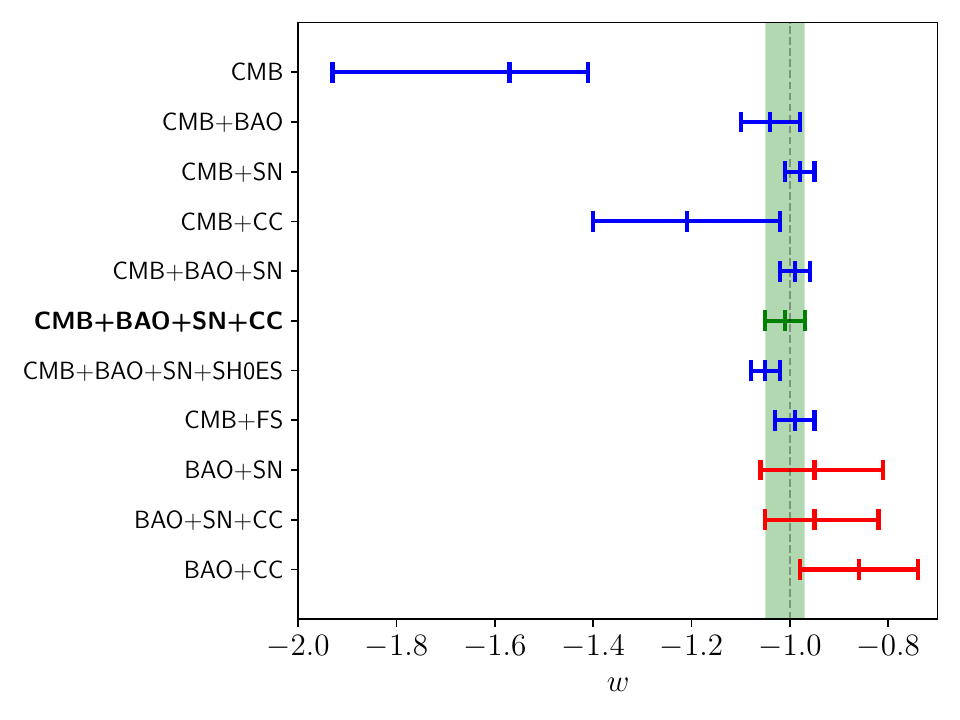}
\caption{The whisker plot summarizes the 68\% CL intervals for the \ac{de} \ac{eos} parameter, $w$, inferred from various dataset combinations under the assumption of the $w$CDM model. Results that incorporate \ac{cmb} data are represented by blue bars, except for the final consensus dataset combination (\ac{cmb}+\ac{bao}+\ac{sn}+\ac{cc}), which is shown in green. In contrast, results obtained without the \ac{cmb} data are depicted in red. The green band highlights the 68\% CL interval derived from the final consensus dataset combination, yielding $w = -1.013^{+0.038}_{-0.043}$. The grey vertical dashed line marks the cosmological constant value, $w = -1$. Figure taken from Ref.~\cite{Escamilla:2023oce}.}
\label{fig:whisker_lateDE}
\end{figure}

\paragraph{Challenges for late dark energy models from \ac{bao} and \ac{sn1} Data on Deformations of \texorpdfstring{$H(z)$}{H(z)} and the Ultra-late Physics Transition Approach}

\ac{lde} models aim to resolve the Hubble tension by introducing degrees of freedom that deform the Planck18 \lcdm\ form of $E(z) \equiv H(z)/H_0$, reconciling the early sound horizon scale with late-time Cepheid calibrators. However, these models face significant challenges in simultaneously fitting the SH0ES $H_0$ value, the sound horizon scale at recombination, and distances from \ac{sn1} and anisotropic \ac{bao} at $z \in [0.01, 2.5]$. Many models also exacerbate the growth tension \cite{Vagnozzi:2019kvw,Alestas:2020zol,Yang:2018qmz,Benevento:2020fev,Camarena:2021jlr,Perivolaropoulos:2021jda,Marra:2021fvf,Alestas:2021luu,Alestas:2021xes,Heisenberg:2022gqk,Adil:2022hkj,Abdalla:2022yfr,Frion:2023xwq,Gomez-Valent:2023uof,Escamilla:2023oce,Kumar:2023bqj,Efstathiou:2021ocp,Capozziello:2024stm,Bousis:2024rnb,Alestas:2020mvb,Briffa:2021nxg,LeviSaid:2021yat,Theodoropoulos:2021hkk}.

Recent analyses also propose a rapid transition in the effective gravitational constant, $G_{\text{eff}}$, at ultra-late times ($z_t \lesssim 0.01$) to address the tension \cite{Marra:2021fvf,Perivolaropoulos:2021bds,Alestas:2021nmi,Alestas:2021xes,Alestas:2022xxm,Perivolaropoulos:2022txg,Paraskevas:2024ytz,Paraskevas:2023aae,Perivolaropoulos:2022khd}. This is based on the mismatch in absolute magnitude ($M_{\rm B}$) between \ac{cmb} and local Universe observations. The luminosity distance-redshift relation
\[ \mu(z) = m_{\rm B}(z) - M_{\rm B} = 5 \log_{10} \left( \frac{d_{\rm L}(z)}{10 \, \text{pc}} \right)\,, \]
suggests \ac{sn1} luminosities at $z > z_t$ are consistent with constant $G_{\text{eff}}$ \cite{Perivolaropoulos:2021jda}. A sudden transition in $G_{\text{eff}}$ at $z_t$ could explain the $\Delta M_{\rm B} \approx -0.2 \, \text{mag}$ deviation in local \ac{sn} needed to reconcile high- and low-redshift $H_0$ values. This model addresses the $H_0$ tension and suggests reduced growth of density perturbations without affecting \lcdm\ background expansion \cite{Perivolaropoulos:2021jda,Alestas:2021nmi,Alestas:2022xxm,Perivolaropoulos:2022txg}. Future \ac{gw}, astrophysical, and cosmological perturbation analyses may test this hypothesis in the context of physical models \cite{Perivolaropoulos:2022txg,Paraskevas:2024ytz,Perivolaropoulos:2022vql}. The ultra-late $G_{\text{eff}}$ transition provides a compelling alternative framework warranting further investigation.

On the other hand, when the inverse distance ladder is built with angular (or 2D) \ac{bao}, instead of anisotropic (or 3D) \ac{bao}, the phenomenology required to solve the Hubble tension is completely different. Angular \ac{bao} data are claimed to be less subject to model dependencies, since in contrast to anisotropic \ac{bao} analyses, no fiducial model is used to convert redshifts and angles into distances of a tracer map. Despite being obtained from the same parent catalogs, 2D and 3D \ac{bao} data are known to be in tension \cite{Nunes:2020uex,Nunes:2020hzy,Camarena:2019rmj,Bernui:2023byc,Gomez-Valent:2023uof,Favale:2024sdq}, pointing to the existence of unaccounted-for systematic errors in one or both data sets or an underestimation of their uncertainties. When angular \ac{bao} is employed in the fitting analysis, it is possible to find a late-time solution to both the Hubble and growth tensions without violating the constancy of $M_{\rm B}$ \cite{Gomez-Valent:2023uof}. The latter requires the effective \ac{de} density to become negative at $z \gtrsim 1.5-2$ to compensate for the late-time increase of $H(z)$ and not spoil the description of the \ac{cmb}. This possibility has been recently realized in the context of a model with a sign-switching cosmological constant \cite{Akarsu:2021fol,Akarsu:2022typ,Akarsu:2024eoo,Akarsu:2019hmw,Akarsu:2023mfb,Anchordoqui:2024gfa,DeFelice:2023bwq} and even more effectively in the model presented in Ref.~\cite{Gomez-Valent:2024tdb}.

\paragraph{Holographic dark energy and gravity-thermodynamics correspondence models}

The holographic principle, rooted in quantum gravity, proposes that a system's entropy is determined by its surface area rather than its volume. Holographic dark energy (HDE) utilizes this principle to describe late-time \ac{de} as an infrared cutoff for specific HDE models. Various alternative HDE formulations exist, including the Tsallis and Barrow models, each with distinct energy density equations and parameters \cite{tHooft:1993dmi,Susskind:1994vu,Bousso:1999xy,Cohen:1998zx,Wang:2016och,Nojiri:2021iko,Nojiri:2020wmh,Trivedi:2024rhp,Tavayef:2018xwx,Saridakis:2020zol,Moradpour:2018ivi,Drepanou:2021jiv,Granda:2008dk, Mukherjee:2017oom, Adhikary:2021xym,Moradpour:2020dfm,SayahianJahromi:2018irq,Trivedi:2024inb,Moradpour:2023ayk}. Moreover, choices for the infrared cutoff can vary, ranging from simple options such as the Hubble horizon to more intricate ones like the Granda-Oliveros or Nojiri-Odintsov cutoffs. Beyond theoretical motivations, HDE models potentially resolve tensions in $H_0$ measurements. Fitting these models to comprehensive datasets yields values consistent with recent observations, possibly enhanced by incorporating sterile neutrinos \cite{Guo:2018ans}.

On the other hand, modified cosmology based on gravity-thermodynamics correspondence \cite{Jacobson:1995ab,Verlinde:2010hp,Padmanabhan:2003pk,Cai:2005ra} is another way to explore the Universe's evolution. The relationship between black hole quantities and conventional thermodynamics \cite{Bekenstein:1973ur,Hawking:1974rv} has been extended to cosmic horizons \cite{Gibbons:1977mu}, with a range of applications \cite{Jacobson:1995ab,Cai:2005ra,Padmanabhan:2003pk}. In particular, these concepts have been used for entropic force cosmological models \cite{Easson:2010av,Gohar:2023lta,Komatsu:2013qia,Nunes:2015xsa,Moradpour:2016tuw,Abreu:2017fhw}, in which entropic force terms account for the Universe's accelerated expansion. Furthermore, modifications to holographic dark energy models have been made using different notions of entropy \cite{Tsallis:2012js, Barrow:2020tzx, Cimidiker:2023kle} coming from statistical mechanics and thermodynamics. Indeed, the fundamental limit established by the Bekenstein entropy bound \cite{1981PhRvD..23..287B} has had important consequences for understanding fundamental aspects of geometry and possible ways of reducing the Hubble tension \cite{vanPutten:2024xwe} consistent with data from Refs.~\cite{Camarena:2019moy,DeSimone:2024lvy}.

The $H_0$ and $\sigma_8$ tension can be concurrently relieved by the modified cosmology employing Tsallis-Cirto nonextensive entropy \cite{Tsallis:2012js}, as demonstrated in Ref.~\cite{Basilakos:2023kvk}. The authors demonstrate how a phantom effective \ac{de}, which is recognized as one of the adequate processes that might relieve $H_0$ tension, can be obtained for specific Tsallis-Cirto parameter choices. Furthermore, given the same parameter choice, they find an enhanced friction term and an effective Newton's constant smaller than the conventional one, therefore solving the $\sigma_8$ tension. In Ref.~\cite{Asghari:2021lzu}, the $\sigma_8$ tension is examined using Jacobson's method \cite{Jacobson:1995ab} in the nonextensive setting. Additionally, Hubble tension can be relieved in entropic force cosmological models \cite{Gohar:2020bod, Gohar:2023lta} because the interaction terms naturally arise from reversible and irreversible processes across the horizons in the evolution equations. Nevertheless, further work needs to be done to develop these approaches. Moreover, a wide range of holographic dark energy models have been utilized to examine these cosmological tensions, for example, Tsallis holographic dark energy \cite{daSilva:2020bdc}. Addressing these cosmological challenges will require further research on gravity-thermodynamic relationship to cosmological applications.  

\paragraph{Modeling late dark energy: Examining the phantom, tracker, quintom, multifield descriptions and their clustering properties}

Phantom-scalar-field \ac{de} cosmologies are usually denoted by an \ac{eos} that satisfies $w<-1$ \cite{Melchiorri:2002ux, Caldwell:1999ew}. However, the current observations have shown a good agreement with the standard \lcdm\ model. Within these observational analyses, some suggestions invoke the presence of a phantom divide boundary. Furthermore, these phantom scalar field \ac{de} models have introduced scalar fields $\phi$ to study the dynamics of evolving \ac{de} through this phantom divided boundary, however, within this scheme, we experiment with issues like fine-tuning \cite{Melchiorri:2004bs}. To find a viable solution, tracker \ac{de} scenarios have been analyzed \cite{Ratra:1987rm, Steinhardt:1999nw}, where $\phi$ controls the energy density and it is possible to obtain attractor background solutions. Some parametrizations in this regard \cite{Roy:2018nce,LinaresCedeno:2021aqk,Roy:2022fif} are used to verify if this model can solve (or relax) the $H_0$  tension. 
To introduce a phantom scalar field term, we consider a gravitational action as
\begin{equation}
    \label{eq:action}
    S = \frac{1}{2\kappa^2} \int d^4x \sqrt{-g} \left( R + \mathcal{L}_\phi (\phi, \partial_\mu \phi) + \mathcal{L}_{\rm m} \right)\,,
\end{equation}
where $\mathcal{L}_\phi$ is the Lagrangian for a phantom scalar field. Considering a flat \ac{flrw} metric, we can vary Eq.~\eqref{eq:action} with respect to the metric and $\phi$ to obtain the gravitational field equations and the Klein-Gordon equation. In this scheme, it is standard to consider a new set of hyperbolic polar coordinates which can ease the numerical solutions \cite{LinaresCedeno:2021aqk}.

From the perspective of this ansatz, it was reported that this kind of phantom/tracker model does not address the Hubble tension when considering a compressed Planck likelihood, given a $H_0 = 69.1^{+0.5}_{-0.6}$\kms \cite{LinaresCedeno:2021aqk} value, which is in a $3\sigma$ CL tension with the latest local result \cite{Riess:2021jrx}. However, Ref.~\cite{Najera:2024dsy} explored the possibility of reconstructing the constraints using the full Planck \cite{Planck:2018vyg} likelihood in addition to model-independent \ac{cmb} baselines, e.g., ACTPol DR4 \cite{ACT:2020gnv}, the \ac{spt3g} \cite{SPT-3G:2021eoc} and the \ac{wmap}9 \cite{WMAP:2012nax} datasets, showing that phantom tracker scalar field cosmologies can reduce the statistical tension on $H_0$ to below 3$\sigma$.

However, quintom \ac{de}, featuring two scalar fields with distinct kinetic energies, blends quintessence and phantom characteristics, and has been shown to be a framework of interest. Model-independent techniques reveal a fluctuating \ac{eos}, crossing the phantom divide line ($w_{\rm DE}=-1$) multiple times. Quintom models offer a promising alternative to standard quintessence and phantom models, potentially providing a more accurate fit with fewer parameters. Several studies explore quintom's potential in resolving the $H_0$ tension, with analyses coupling the phantom component to \ac{dm} and examining various quintessence, phantom, and quintom models. These studies yield various estimates of the Hubble constant, suggesting the possibility of reducing the tension by up to 2.6 $\sigma$ \cite{Feng:2004ad, Guo:2004fq, Hu:2004kh, Cai:2009zp, Qiu:2010ux, Zhang:2018zwu, Leon:2018lnd, Panpanich:2019fxq, Zhao:2017cud, Wang:2019ufm, Tamayo:2019gqj, Escamilla:2021uoj, Escamilla:2024fzq, Vazquez:2020ani,Vazquez:2023kyx, Xia:2013dea, Fu:2023tlp, Roy:2023vxk}.

Multifield models of cosmic acceleration arise from the coupling of several scalar fields to gravity. The standard action that describes this system is
\begin{equation} \label{Action_gen}
S = \frac{1}{2\kappa^2} \int d^4x \sqrt{-g} \left[R - G_{IJ} (\{\phi^L\}) \partial_{\mu} \phi^I \partial^{\mu} \phi^J - V (\{ \phi^L \}) \right] \,,
\end{equation}
where $g_{\mu \nu}$ is the four-dimensional space-time metric and $G_{IJ}$ is the metric on the manifold parameterized by the scalars $\phi^I$ with $I = 1,2,...,n$\,. When the field-space metric $G_{IJ}$ has curvature, such multifield models can lead to novel effects, compared to single-scalar field models, in the context of both cosmological inflation and \ac{lde}. Specifically, for \ac{lde} models of this type, see Refs.~\cite{Cicoli:2020cfj,Cicoli:2020noz,Akrami:2020zfz,Eskilt:2022zky,Anguelova:2021jxu,Anguelova:2023dui}.

The qualitatively new features of the multifield case arise from solutions of the background equations of motion, whose field-space trajectories are (strongly) non-geodesic. In the case of two fields, that is, when $n = 2$, the deviation from a geodesic is measured by the turning rate of the background trajectory $(\phi^1_0 (t), \phi^2_0 (t))$ in field space, $\Omega = - N_I D_t T^I$, where $T^I$ and $N_I$ are unit vectors, respectively, tangent and normal to the trajectory; also $D_t T^I \equiv \dot{\phi}_0^J \nabla_J T^I$\,. Strongly non-geodesic trajectories are characterized by large turning rates. Multifield models of \ac{lde}, whose background solutions have such rapid-turning field-space trajectories, can have an equation of state parameter $w$ very close to $-1$ and, despite that, can be observationally distinct from a cosmological constant \cite{Akrami:2020zfz,Eskilt:2022zky,Anguelova:2021jxu,Anguelova:2023dui}. This is because the speed of sound $c_s$ of the dark-energy perturbations around the background solution can be (significantly) reduced compared to the speed of light.

\ac{de} perturbations around the exact background solution of Ref.~\cite{Anguelova:2021jxu} were studied in Ref.~\cite{Anguelova:2023dui}. It was shown there that the speed of sound of these perturbations is significantly reduced. Combining that with an equation of state parameter $w \approx -1$\,, as well as incorporating matter in the exact background solution, it was further argued in Ref.~\cite{Anguelova:2023dui} that this model of multifield \ac{de} is very promising for alleviating simultaneously the $\sigma_8$ and Hubble tensions.

More generally, a low sound speed of \ac{de} perturbations is a natural possibility of any k-essence model \cite{Armendariz-Picon:1999hyi,Garriga:1999vw,Hu:2004kh} and also phantom models \cite{Creminelli:2008wc}. In this case, \ac{de} perturbations can grow on small scales and impact the structure formation, see Ref.~\cite{Batista:2021uhb} for a review. Recent efforts in understanding the effects of \ac{de} perturbations in structure formation include Refs.~\cite{Dakin:2019vnj,Hassani:2019lmy,Blot:2022mnt,Batista:2022ixz}, but we still need an accurate determination of their impact on nonlinear observables.

With more freedom in the perturbative sector, clustering \ac{de} models could circumvent the usual trend seen in late \ac{de} models of worsening the growth tension. However, some explorations of this case were conducted only for non-phantom \ac{eos}. Recent analyses of this scenario indicate that \ac{de} with low sound speed worsens the $S_8$ tension \cite{Ben-Dayan:2023htq} and the very low $c_s$ values are not allowed by current data \cite{Dinda:2023mad}.

\paragraph{Running vacuum models}
\label{sec:RVM}

Quantum fluctuations in the expanding Universe induce the running of the vacuum energy density (VED). An example, motivated by dualities in string theory is Ref.~\cite{Bastero-Gil:2001smv} where the spacetime vacuum is such that its trans-Planckian modes are coupled and track the Hubble expansion rate \cite{Mersini-Houghton:2001cwp, Mersini-Houghton:2002eir, Bastero-Gil:2001rmj, Bastero-Gil:2002plx, Bastero-Gil:2003hfz}. The running character of the vacuum was originally motivated using renormalization group arguments in curved spacetime \cite{Shapiro:2000dz,Sola:2007sv,Shapiro:2009dh}, but more formal calculations have been recently performed in the context of QFT in curved spacetime using adiabatic regularization and renormalization techniques
\cite{Moreno-Pulido:2020anb,Moreno-Pulido:2022phq,Moreno-Pulido:2023ryo} as well as effective low-energy string theory \cite{Basilakos:2019acj,Basilakos:2020qmu,Mavromatos:2021urx}. See Refs.~\cite{Sola:2013gha,SolaPeracaula:2022hpd} for dedicated reviews of the \ac{rvm} within the QFT approach and \cite{Mavromatos:2020kzj} for the stringy version. The running VED takes the following form
\begin{equation}\label{eq:RVdens}
    \rho_{\rm vac}(H)=\frac{3}{8\pi G}(c_0+\nu H^2+\tilde{\nu}\dot{H})+\mathcal{O}(H^4)\,.
\end{equation}
In contrast to the rigid energy density $\rho_\Lambda=\Lambda/8\pi G$ associated with the cosmological constant in \lcdm, $\rho_{\rm vac}$ also receives dynamical contributions from the quantized matter fields. For $\nu,\tilde{\nu}\neq 0$ the VED becomes dynamical. Eq.~\eqref{eq:RVdens} lies at the core of the class of \ac{rvm}s. The higher-order terms $\mathcal{O}(H^4)$ can explain inflation with a graceful exit without the need of {\it ad hoc} scalar fields \cite{Lima:2013dmf,Perico:2013mna,Basilakos:2013xpa,Sola:2015rra,Sola:2015rra,Basilakos:2019zsf}. Here we just focus briefly on the low-energy terms appearing in Eq.~\eqref{eq:RVdens}, which are controlled by the parameters $\nu$ and $\tilde{\nu}$. Their values depend on the masses and non-minimal couplings (in the case of the scalar fields) of the various particle species of the theory, and they are expected to be $\lesssim \mathcal{O}(10^{-4}-10^{-2})$, although they are ultimately determined from observations. In spite of being small, the running of the vacuum can leave a non-negligible imprint on the post-inflationary stages of cosmic expansion both at the background and perturbation levels. The time evolution of $\rho_{\rm vac}$ may imply either the non-conservation of matter and/or the time-evolution of the gravitational coupling $G$ and other fundamental ``constants'' of nature, and this can lead to a variety of scenarios explored in the literature \cite{Fritzsch:2012qc,Gomez-Valent:2014fda,Gomez-Valent:2015pia,Fritzsch:2015lua,Fritzsch:2016ewd,SolaPeracaula:2023wqw}. The \ac{rvm}s are able to accommodate the wealth of cosmological data at our disposal and have a positive bearing on the cosmological tensions. The \ac{rvm}s were actually the first to point out significant hints of \ac{de} (vacuum) dynamics in the Universe using a large amount of cosmological data, almost one decade before the advent of \ac{desi} \cite{Sola:2015wwa,Sola:2016jky}, see also  Refs.~\cite{Sola:2015wwa,Sola:2016jky,SolaPeracaula:2016qlq,SolaPeracaula:2017esw,Sola:2017znb,Gomez-Valent:2017idt,Gomez-Valent:2018nib} and Refs.~\cite{Gomez-Valent:2014rxa,Geng:2017apd,Tsiapi:2018she,Asimakis:2021yct}. The most updated constraints on these models can be found in the recent works Refs.~\cite{SolaPeracaula:2021gxi,SolaPeracaula:2023swx}, where the authors analyze the \ac{rvm}s in light of a complete dataset under the simplified assumption $\tilde{\nu}=\nu/2$ and show that coupling between \ac{dm} and the running vacuum can significantly loosen the growth tension, especially when the vacuum dynamics is activated at $z\sim 1$, and also that a joint running of $\rho_{\rm vac}$ and $G$ can strongly mitigate both the growth and $H_0$ tensions. This is also true in the Brans-Dicke model \cite{deCruzPerez:2018cjx,SolaPeracaula:2019zsl,
SolaPeracaula:2020vpg,deCruzPerez:2023wzd}. 

The stringy version of the \ac{rvm} \cite{Basilakos:2019acj,Basilakos:2020qmu,Mavromatos:2021urx,Mavromatos:2020kzj} provides both early- and late-dark-energy models within string theory \cite{Mavromatos:2021sew,Mavromatos:2022xdo}. This model, a Chern-Simons modification of \ac{gr} \cite{Jackiw:2003pm}, results in a vacuum energy density with logarithmic corrections in the Hubble parameter, $H^2 \ln H$. These corrections, arising from quantum graviton effects in an expanding Universe \cite{Alexandre:2013nqa}, help alleviate the $H_0$ and growth-of-structure tensions \cite{Gomez-Valent:2023hov}, and can dominate logarithmic corrections from matter QFT effects \cite{Moreno-Pulido:2020anb,Moreno-Pulido:2022phq,Moreno-Pulido:2023ryo}.

\paragraph{Emergent dark energy}

Dynamic \ac{de} models are gaining attention over the simple cosmological constant, $\Lambda$, to reconcile discrepancies between \ac{cmb} and S$H_0$ES data \cite{Li:2019yem,Yang:2021eud}. One such model is Critically Emergent Dark Energy (CEDE), where \ac{de} emerges as a critical phenomenon \cite{Banihashemi:2018oxo,Banihashemi:2018has,Banihashemi:2020wtb}. In CEDE, \ac{de} density grows from zero in a phase transition, with an order parameter sensitive to photon thermal baths. Below a critical redshift, $z_c$, DE self-interaction dominates, leading to universal order. This model uses Ginzburg-Landau theory to describe phase transitions and derives \ac{de} density dependence below $z_c$ as \cite{Ginzburg:1950sr,Banihashemi:2020wtb} 
\begin{equation}
\label{GLTofDE}
    \kappa^2 \rho_{\rm DE}(z)/3 H_0^2=(1-\Omega_{\rm m,0}-\Omega_{\rm r,0})\sqrt{1 -z/z_c}\,.
\end{equation}
Confronting this phantom-like model with the \ac{cmb} data \cite{Planck:2019nip}, yields $H_0=70.0^{+1.2}_{-2.7}$\kms with less $\chi^2$ than \lcdm. Adding \ac{sn} data \cite{Pan-STARRS1:2017jku} and/or \ac{bao} data \cite{BOSS:2012tck,Beutler:2011hx,Blake:2011en,Ross:2014qpa}, pushes $z_c$ toward larger values where practically \lcdm\ is restored; and the $H_0$ posterior shrinks to the tension zone of Planck18. CEDE offers richer predictions beyond phantom-like \ac{de}. Considering the specific scale dependence of \ac{de} density fluctuations in this model \cite{Banihashemi:2022vfv}, it alleviates both low-$\ell$/high-$\ell$ inconsistencies in the \ac{cmb} angular power spectrum and the $A_{\rm L}$ anomaly reported by Planck \cite{Planck:2018vyg}. Assuming CEDE, $1\sigma$ posterior intervals of cosmological parameters are constrained with $\ell<800$ and $\ell>800$ overlap. Additionally, in this framework, $A_{\rm L}$ is consistent with unity well within the $1\sigma$ region.

Another Phenomenologically Emergent Dark Energy (PEDE) model developed to address the Hubble tension is \citep{Li:2019yem}
\begin{equation}
    \Omega_{\text{DE}}(z) = \Omega_{\text{DE},0} \left[1 - \tanh\left(\log_{10}(1 + z)\right)\right]\,.
\end{equation}

The lack of additional parameters makes the model especially intriguing. However, Ref.~\cite{Rezaei:2020mrj} argues that while the PEDE model fits background data well, it fails to match cluster-scale observations compared to \lcdm\ at the perturbation level. Various authors further analyzed this model statistically with diverse cosmological datasets and in extended frameworks \cite{Pan:2019hac, Mukherjee:2023lqr, Shah:2023rqb, Shah:2024rme, Hernandez-Almada:2020uyr, Yang:2020ope, Li:2020ybr, Liu:2020vgn, DiValentino:2021rjj, Alestas:2020mvb, Hernandez-Almada:2024ost,Garcia-Aspeitia:2022uxz,Pizzuti:2020tdl}. However, it offers a worse fit to the combined datasets in comparison to \lcdm\ \cite{Schoneberg:2021qvd}. As extensions of the PEDE model, include 
the generalized emergent dark energy (GEDE) \cite{Li:2020ybr,Yang:2021eud} and the modified emergent dark energy (MEDE) models \cite{Benaoum:2020qsi}. Both the MEDE \& GEDE frameworks can help in reducing the $H_0$ tension to $\lesssim 3\sigma$ CL.

\paragraph{Vacuum metamorphosis}

The Vacuum Metamorphosis (VM) model, proposed to explain the late-time accelerated expansion of the Universe, is based on non-perturbative quantum gravitational effects \cite{Parker:2000pr, Parker:2003as, Caldwell:2005xb}. It features a minimally coupled, ultra-light scalar field with mass \(m \sim 10^{-33}\) eV. The model's key aspect is a gravitational phase transition occurring when the Ricci scalar curvature \(R\) reaches the order of the scalar field's squared mass, \(m^2\), which is related to the current matter density parameter \(\Omega_{\rm m,0}\). In this framework, the gravitational phase transition occurs at the critical redshift  $z_t = -1 + \frac{3\Omega_{\rm m,0}}{4(1 - M)}$, where $M$ is a free parameter of the theory. 

This transition leads to an effective radiation component in the Universe post-transition, resulting in a De-Sitter phase, distinct from the \lcdm\ scenario. In the original VM scenario, there is no cosmological constant at high redshifts, and the matter density parameter \(\Omega_{\rm m,0}\) is given by \(\left[\frac{4}{3} \left(3M(1 - M - \Omega_k - \Omega_{\rm r,0})\right)^{3}\right]^{1/4}\), where \(\Omega_k = -k/H_0^2\) and \(\Omega_{\rm r,0}\) is the radiation density parameter. Joint analysis with \textit{Planck} 2018 + \ac{bao} + Pantheon yields $H_0 = 74.21\pm0.66$\kms \cite{DiValentino:2020kha}, which resolves the $H_0$ tension within $1\sigma$. An extension of the VM model, namely the VM-VEV model, includes a cosmological constant at high redshifts, represented by the vacuum expectation value of the massive scalar field. This extension requires additional conditions such as the transition occurring in the past ($z_t \geq 0$) and the \ac{de} density being non-negative post-transition i.e., \(\Omega_{\text{DE}}(z > z_{t}) \geq 0\), respectively. Joint analysis with \textit{Planck} 2018 + \ac{bao} + Pantheon leads to $H_0=73.26\pm0.32$\kms \cite{DiValentino:2020kha}. While the VM \& VM-VEV models can align \(H_0\) with the SH0ES value, they exhibit a poorer fit to low-redshift data, such as \ac{bao} and Pantheon, compared to \lcdm.

\paragraph{Others DE frameworks and their implications for the $H_0$ tension}

Rather than initially adopting a particular physical model for \ac{de}, given our lack of clear insight into its origin, we can opt for an exploratory approach utilizing a probing function to represent the \ac{de} density. The \ac{dde} model, generalized to $w(z) = \sum_{n} w_n x(z)$, where the parameters \( w_n \) are fixed by observations and \( x(z) \) is a function of redshift that can correspond to both phantom and non-phantom fields, has been shown to alleviate the \(\sigma_8\) tension \cite{Lambiase:2018ows} as well as the \( H_0 \) tension. It is worth noting that the \( H_0 \) tension improves further when the neutrino mass \(\sum_\nu m_\nu\) is taken into account \cite{Battye:2013xqa}. Another approach was initiated in Refs.~\cite{Wang:2001ht,Wang:2001da,Wang:2003cs}, where the authors used both linear and quadratic probe functions for $X(z) = \frac{\rho_{\rm DE}(z)}{\rho_{\rm DE}(0)}$. For \lcdm, \(X(z) = 1\), while \(X(z) \neq 1\) for \textit{any} redshift \(z\) indicates \ac{de} evolution. A quadratic probe function used in the literature is proposed in Ref.~\cite{Cardenas:2014jya}. This perspective shows that such an approach may prefer \(X(z)\) decreasing with \(z\), even taking negative values for \(z > 1\). Using the most recent \ac{bao} observations from the \ac{desi} \cite{DESI:2024mwx}, the same trend was found in Ref.~\cite{DESI:2024aqx}. This trend, including the possibility of a negative \(X(z)\), poses a challenge for \ac{de} modeling and may have profound implications in light of recent cosmological tensions and the foundations of standard cosmology. 

A pseudo-Rip \ac{de} scenario, leading to asymptotically de Sitter behavior, is proposed in Ref.~\cite{Lazkoz:2023oqc}. This model appears to be statistically favored over the consensus \lcdm\ model according to some Bayesian discriminators. The equation of state parameter, typically evaluated at the present time, is given by $w_0 = -1 - \frac{\eta \lambda \text{sech}(2\lambda)}{3 \arctan(\sinh(\lambda))}$, where $\eta$ is a free parameter of the model. It is concluded that the pseudo-Rip \lcdm\ model significantly improves the description of late-time ($z \leq 2.5$) data more decisively than that at higher redshifts.

Late-time \ac{de} transitions at redshifts \( z \ll 0.1 \) \cite{Benevento:2020fev} can make the predicted value of \( H_0 \) compatible with SH0ES measurements. Conversely, in Ref.~\cite{Huang:2024erq}, a model-independent constraint on late-time models shows strong evidence against homogeneous new physics over the \lcdm\ model. Surprisingly, despite the absence of \( H_0 \) and \( M_{\rm B} \) tensions in the local Universe, Ref.~\cite{Kitazawa:2023peg} argues that late-time data solutions to the \( H_0 \) tension require a smaller sound horizon at the recombination era. Additional discussions on \( r_d \) scales and \( H_0 \) tension can be found in Refs.~\cite{Gomez-Valent:2023uof,Ruchika:2024lgi}. The authors in Ref.~\cite{Keeley:2022ojz} argue that new physics at low redshifts cannot resolve the \( H_0 \) tension.

It is important to emphasize that simple scalar field quintessence models have been widely studied for their ability to address the coincidence problem through tracking behavior. A possibility is to use generalized axion-like models with inverse-cosine-like potential \cite{Hossain:2023lxs, Boiza:2024azh, Boiza:2024fmr,Hossain:2025grx}, which provides natural entrance and exit of the tracking regime within a single-field framework. While it is not ruled out, the Hubble tension can be slightly reduced \cite{Chiang:2025qxg}. Also, a different approach to tackle the late acceleration of the Universe is by invoking a modified Chaplygin gas \cite{Benaoum:2002zs,Stefancic:2004kb,Bouhmadi-Lopez:2015oxa,Yang:2019nhz}, being the simplest case when its equation of state deviates from that of a cosmological constant by a simple power law of its energy density. This was analyzed from a perturbation point of view in Ref.~\cite{Albarran:2016mdu}. 
%A full cosmological fit tackling the $H_0$ and $S_8$ tension has been done in Ref.~\cite{CGC2025Benetti}, showing a moderate preference to \lcdm\ in some cases \cite{CGC2025Benetti}.

We also highlight that the cosmological stasis framework \cite{Dienes:2021woi, Dienes:2023ziv,Dienes:2024wnu} is a phenomenon arising within many BSM cosmologies in which the abundances $\Omega_i$ of multiple cosmological energy components with different equation of state parameters $w_i$ remain constant across extended periods despite cosmic expansion.  This occurs due to dynamical feedback within the equations of motion for the corresponding energy densities, with the stasis state serving as a global dynamical attractor.  Such systems evolve toward stasis irrespective of initial conditions and remain in stasis until some underlying mechanism alters the dynamical feedback.  Most importantly, the effective equation of state parameter $w_{\rm eff} = \sum_i\Omega_i w_i$ for the Universe remains constant during stasis but nevertheless generically differs from the canonical values associated with, e.g., matter, radiation, or vacuum energy.  Stasis at either early or late-times (relative to recombination) can therefore potentially broaden the scope of possibilities for reconciling cosmological tensions. In a similar context, we can also mention the effects of the backreaction of super-Hubble cosmological fluctuations on the late-time accelerated expansion \cite{Abramo:1997hu}. Some works suggested that this mechanism could drive a dynamical self-regulating relaxation of the cosmological constant, as first speculated in Ref.~\cite{Brandenberger:1999su}. This behaviour could potentially result in an oscillatory effective \ac{de} \cite{Alvarez:2025kma}. A discussion on the tension problem within these scenarios can be found in Ref.~\cite{Alvarez:2025kma}, which also provides further references on these models. 

\bigskip
\subsubsection{Dark energy models  exhibiting a rapid density transition from negative to positive values in the late Universe \label{sec:Grad_DE}}

\noindent \textbf{Coordinator:} \"Ozg\"ur Akarsu, Rafael C. Nunes\\
\noindent \textbf{Contributors:} Adrià Gómez-Valent, Alexander Zhuk, Anil Kumar Yadav, Anjan Ananda Sen, Antonio De Felice, Branko Dragovic, Davide Pedrotti, Emmanuel N. Saridakis, Emre \"{O}z\"{u}lker, Eoin \'O Colg\'ain, Hanyu Cheng, J. Alberto V\'{a}zquez, Joan Sol\`a Peracaula, Jurgen Mifsud, Laur Järv, Leandros Perivolaropoulos, Leila L. Graef, Luca Visinelli, Luis Anchordoqui, M.M. Sheikh-Jabbari, Nihan Kat{\i}rc{\i}, Ruth Lazkoz, and Suresh Kumar
\\

\noindent The possible need for \ac{de} assuming negative density values at high redshifts was first highlighted by \ac{boss} collaboration~\cite{BOSS:2014hhw}, detecting \(\Lambda > 0\) for \(z < 1\) but favoring \(\rho_{\rm DE} < 0\) for \(z > 1.6\), particularly based on Ly-\(\alpha\) \ac{bao} measurements at \(z_{\rm eff} \approx 2.34\), which exhibited a \(2.5\sigma\) tension with Planck-\lcdm~\cite{Planck:2018vyg} and suggested a non-monotonic \(H(z)\) evolution that is difficult to reconcile with strictly \(\rho_{\rm DE}\geq 0\) in \ac{gr}. Independently, Ref.~\cite{Sahni:2014ooa} proposed that this discrepancy with Planck-\lcdm, known as the Ly-\(\alpha\) \ac{bao} anomaly, could be explained by an \ac{mg} model in which \(\Lambda > 0\) is dynamically screened, leading to an effective \(\rho_{\rm DE}\) passing below zero, accompanied by a singularity in its \ac{eos} parameter \(w_{\rm DE}\), at \(z \sim 2.3\). This discrepancy was later reduced to $\sim1.5\sigma$ with the completed \ac{eboss}~\cite{eBOSS:2020tmo}. However, model-independent/non-parametric reconstructions of \(\rho_{\rm DE}\), using the \ac{bao} data, consistently favor \(\rho_{\rm DE}\) crossing below zero or vanishing for \(z \gtrsim 1.5-2\). Moreover, such reconstructions of \(w_{\rm DE}\)---though inherently unable to capture a sign change in \ac{de} density---persistently favor \(w_{\rm DE} \sim -1\) for \(z \lesssim 1-1.5\) but trend toward large negative values below \(-1\) at \(z \sim 1.5-2\)---which is expected right after a \ac{de} density smoothly transitions from negative to positive as the Universe expands, therefore potentially indicating a sign change in \(\rho_{\rm DE}\). The recent \ac{desi} Ly-\(\alpha\) \ac{bao} data shows no tension with Planck-\lcdm~\cite{DESI:2024mwx}. However, while \ac{desi} \ac{bao} data---when analyzed using the \ac{cpl} parametrization---provides more than \(3\sigma\) evidence for dynamical \ac{de}~\cite{DESI:2024mwx}, a less noted but significant finding is that non-parametric reconstructions of \ac{de} density also indicate the possibility of vanishing or negative values for \(z \gtrsim 1.5-2\)~\cite{DESI:2024aqx,Escamilla:2024ahl}, consistent with trends observed in pre-\ac{desi} \ac{bao} data, e.g., SDSS \ac{bao}~\cite{Escamilla:2023shf,Sabogal:2024qxs,Escamilla:2024ahl}.

It was not immediately recognized that addressing the Ly-\(\alpha\) anomaly could naturally be linked to the \(H_0\) and \(S_8\) tensions, both of which emerged sometime after the Ly-\(\alpha\) anomaly was first identified. A brief explanation of this connection is as follows~\cite{Akarsu:2021fol,Akarsu:2022typ}: If a \ac{de} model leaves the pre-recombination Universe remains unaltered, as in the standard \lcdm\ model, then the comoving sound horizon at last scattering, \(r_* = \int_{z_*}^{\infty} c_{\mathrm{s}} H(z)^{-1} \mathrm{d}z\), is expected to remain effectively unchanged from its \lcdm\ counterpart. The Planck \ac{cmb} spectra provide precise, nearly model-independent measurements of the angular scale of the sound horizon, \(\theta_* = r_*/D_M(z_*)\), and the present-day physical matter density, \(\Omega_{\rm m,0} h^2\), derived from the peak structure and damping tail. Consequently, in \ac{de} models featuring negative \ac{de} densities at high redshifts, say, for \(z > z_\dagger\), both the comoving angular diameter distance to last scattering, \(D_M(z_*) = c \int_{0}^{z_*} H(z)^{-1} \mathrm{d}z\), and \(\Omega_{\rm m,0} h^2\) must remain consistent with their Planck-\lcdm\ inferred values. This requires that any suppression of \(H(z)\) at \(z > z_\dagger\), due to negative \ac{de} density, must be compensated by an enhancement of \(H(z)\) at \(z < z_\dagger\) to maintain consistency with the Planck-\lcdm-inferred \(D_M(z_*)\). As a result, this mechanism naturally increases \(H_0\) (implying a fainter \(M_{\rm B}\)) and decreases \(\Omega_{\rm m,0}\) relative to Planck-\lcdm. A later transition (i.e., smaller \(z_\dagger\)) leads to a prolonged phase of suppression, or similarly, more negative \ac{de} density values lead to more suppression; both effects amplify the enhancement in \(H_0\) and reduction in \(\Omega_{\rm m,0}\), provided the transition occurs before the negative \ac{de} density becomes dominant---beyond which expansion would halt and contraction would ensue. This framework also predicts higher \(\sigma_8\) values because the suppressed \(H(z)\) at \(z > z_\dagger\) reduces cosmic friction, thereby enhancing structure formation at high redshifts~\cite{Paraskevas:2024ytz,Akarsu:2025ijk}. However, the lower present-day matter density leads to reduced \(S_8 = \sigma_8\sqrt{\Omega_{\rm m,0}/0.3}\) values, potentially resolving the \(S_8\) tension. This reduced cosmic friction enhances structure formation for $z>z_\dagger$~\cite{Akarsu:2022typ}, potentially explaining the \ac{jwst} anomaly~\cite{Menci:2022wia, Biagetti:2022qnp, Wang:2022jvx, Forconi:2023hsj}---where deep-space observations at \(z \gtrsim 5\) indicate stronger structure growth than predicted by Planck-\lcdm.

In the rapidly growing literature, fitting within the framework outlined above, Refs.~\cite{Dutta:2018vmq} and \cite{Akarsu:2019hmw} are the earliest studies explicitly connecting the Ly-\(\alpha\) anomaly with the \(H_0\) tension. Ref.~\cite{Dutta:2018vmq} assumed that the Universe is consistent with Planck-\lcdm\ for \(z \gtrsim 4\) and reconstructed \(H(z)\) using low-redshift data, including Ly-\(\alpha\) \ac{bao}. They found that the \ac{de} density reaches a minimum within a certain redshift range and becomes negative for \(z \gtrsim 2\), accompanied by higher \(H_0\) values, and argued that this behavior can be most simply explained by an AdS-like cosmological constant ($\Lambda<0$) combined with an evolving \ac{de} component. Ref.~\cite{Akarsu:2019hmw} proposed that inertial mass density, \(\varrho \equiv \rho + p\), maybe more fundamental than energy density and introduced a \ac{de} parametrization with a minimal dynamical deviation from the usual vacuum energy/cosmological constant (\(\varrho = 0\)) in the form \(\varrho \propto \rho^{\lambda}\), referred to as graduated dark energy (gDE) (\({\varrho = \rm const}\), simple-gDE~\cite{Acquaviva:2021jov}). gDE exhibits a wide range of behaviors depending on \(\lambda\); notably, for large negative \(\lambda\), it serves as a phenomenological model for an AdS-to-dS-like transition in \ac{de}. It was shown via gDE that joint observational data, including but not limited to Planck-\ac{cmb} and Ly-\(\alpha\) \ac{bao}, suggest \(\lambda \sim -18\), indicating a behavior analogous to a rapid AdS-to-dS transition at \(z_\dagger \sim 2\), consequently alleviating both the \(H_0\) tension and the Ly-\(\alpha\) anomaly. With this constraint on \(\lambda\), gDE resembles a smooth step-like function, yielding \(\rho_{\rm gDE}/\rho_{\rm c0} \sim -0.7\) with $w_{\rm DE}\gtrsim-1$ for \(z_\dagger \gtrsim 2\) before rapidly switching sign at \(z_\dagger \sim 2\) and settling at \(\rho_{\rm gDE}/\rho_{\rm c0} \approx 0.7\) with $w_{\rm DE}\lesssim-1$ for \(0 \leq z \lesssim 2\).\footnote{For a minimally interacting \ac{de} with \(\rho_{\rm DE}\) transitioning from negative to positive at \(z_{\dagger}\) as the Universe expands (\(z\) decreases), physical consistency demands \(\rho_{\rm DE} < 0\) with \(w_{\rm DE} > -1\) just before the transition and \(\rho_{\rm DE} > 0\) with \(w_{\rm DE} < -1\) just after, ensuring a smooth crossing through \(\rho_{\rm DE} = 0\) while maintaining the continuity equation. Since \(p_{\rm DE}\) remains finite, \(w_{\rm DE}\) exhibits a singularity at \(z = z_\dagger\), with \(\lim_{z\to z_\dagger^{\pm}} w(z) = \pm\infty\). However, this is a safe singularity---\(\rho_{\rm DE}\) remains finite, and \(w_{\rm DE}\) diverges solely because \(\rho_{\rm DE}\) crosses zero while \(p_{\rm DE}\) remains finite and nonzero. Consequently, given that \(H^2 \propto \rho_m + \rho_{\rm DE}\), \(H\) and other kinematical parameters evolve smoothly; in scalar field realizations, the sound speed remains luminal, further reinforcing the regularity of the crossing~\cite{Akarsu:2025gwi}. By contrast, the alternative scenario of sign-changing \(\rho_{\rm DE}\)---where \(\rho_{\rm DE} < 0\) with \(w_{\rm DE} < -1\) just before the transition and \(\rho_{\rm DE} > 0\) with \(w_{\rm DE} > -1\) just afterward---leads to singularities in both $w_{\rm DE}$ and $\rho_{\rm DE}$, as \(\lim_{z\to z_\dagger^{\pm}} w_{\rm DE}(z) = \mp\infty\) and \(\lim_{z\to z_\dagger^{\pm}} \rho_{\rm DE}(z) = \mp\infty\), thereby leading to a breakdown of the spacetime metric. Thus, only the first scenario is physically viable, predicting a phantom-like phase (\(w_{\rm DE} < -1\)) following the transition. Consequently, the tendency of model-agnostic reconstructions of \(w_{\rm DE}(z)\) and various \ac{eos} parameterizations (e.g., \ac{cpl}) to favor large negative values beyond \(-1\) for \(z \sim 1.5-2\) may indicate negative or vanishing \ac{de} densities for \(z \gtrsim 1.5-2\). For further discussion, see Refs.~\cite{Ozulker:2022slu,Adil:2023exv,Paraskevas:2024ytz,Akarsu:2025gwi}} This led to the conjecture that around \(z_\dagger \sim 2\), the Universe underwent a period of rapid \textit{mirror} AdS-to-dS transition in vacuum energy---rapid sign-switching cosmological constant, \(\Lambda_{\rm s}\), from negative to positive while preserving the magnitude---or a similar phenomenon. 

The $\Lambda_{\rm s}$CDM framework~\cite{Akarsu:2021fol,Akarsu:2022typ,Akarsu:2023mfb,Yadav:2024duq,Akarsu:2024eoo,Akarsu:2024qsi} extends \lcdm\ by replacing the usual cosmological constant (\(\Lambda\)) with a dynamically evolving counterpart (\(\Lambda_{\rm s}\)) that undergoes a mirror AdS-to-dS transition in the late Universe, while leaving other standard cosmological components—such as \ac{cdm}, baryons, pre-recombination physics, \ac{bbn}, and inflation paradigm—unchanged. This transition can typically be described using sigmoid-like functions, such as the smooth approximation of the signum function, \({\rm sgn}\,x \approx \tanh(kx)\), where \(k > 1\) and \(x\) represents either redshift (\(z\)) or scale (\(a\)). An example is \(\Lambda_{\rm s}(z) = \Lambda_{\rm s0} \tanh[\nu(z_{\dagger}-z)]/\tanh[\nu z_\dagger]\), where \(\nu > 1\) controls the sharpness of the transition, \(\Lambda_{\rm s0} > 0\) is the present-day value, and \(z_{\dagger}\) denotes the transition redshift. For a rapid transition (e.g., \(\nu \gtrsim 10\)) occurring at \(z_{\dagger} \sim 2\), this function effectively behaves as \(\Lambda_{\rm s} \approx \Lambda_{\rm s0}\) for \(z \lesssim 2\) and \(\Lambda_{\rm s} \approx -\Lambda_{\rm s0}\) for \(z \gtrsim 2\). In the limiting case \(\nu \to \infty\), the transition becomes instantaneous, \(\Lambda_{\rm s}(z) \to \Lambda_{\rm s0} \, {\rm sgn}[z_{\dagger} - z]\), defining the \textit{abrupt} \(\Lambda_{\rm s}\)CDM model, which extends \lcdm\ by a single additional parameter, serving as an idealized representation of a rapid mirror AdS-to-dS transition. Abrupt \(\Lambda_{\rm s}\)CDM~\cite{Akarsu:2021fol,Akarsu:2022typ,Akarsu:2023mfb,Yadav:2024duq,Akarsu:2024eoo,Akarsu:2024qsi} has emerged as one of the most promising and economical extensions of \lcdm; introduces only a single additional parameter, $z_\dagger\sim2$ (estimated through robust observational analyses), beyond \lcdm, resolving major cosmological tensions---including those in \(H_0\), \(M_{\rm B}\), and \(S_8\)---as well as the \ac{bao} Ly-\(\alpha\) anomaly, while yielding an age of the Universe consistent with estimates from the oldest globular clusters. Additionally, when allowing variations in $m_\nu$ and $N_{\rm eff}$, it predicts values consistent with the standard model of particle physics~\cite{Yadav:2024duq}, suggesting that it may avoid the recently proposed anomaly in which cosmological data (such as \ac{desi} \ac{bao}), within the \lcdm\ framework, appears to favor $m_\nu<0$~\cite{Craig:2024tky,Wang:2024hen,Green:2024xbb,Herold:2024enb,Elbers:2024sha,Ge:2024kac} (see Sec.~\ref{sec:Neutri_Ten}). From a physical perspective, $\Lambda_{\rm s}$CDM is identical to \lcdm\ for $z < z_{\dagger}$, featuring a dS-like cosmological constant after the transition, but introduces a minimal modification by adopting an AdS-like cosmological constant for $z > z_{\dagger}$, i.e., for all redshifts prior to the transition. However, from a phenomenological perspective---viz., in terms of the Universe’s expansion dynamics and observational signatures---the impact of this modification is effectively confined to redshifts $z \lesssim z_{\dagger}\sim2$. Specifically, $\Lambda_{\rm s}$CDM replicates the $H(z)$ of \lcdm\ for $z < z_{\dagger}$---albeit with systematically larger values---introduces $H(z)$ deformation around $z \sim z_{\dagger}$, and becomes nearly indistinguishable from \lcdm\ at higher redshifts ($z \gtrsim 3$). Consequently, from a phenomenological standpoint, $\Lambda_{\rm s}$CDM is a post-recombination/late-time modification to \lcdm. Abrupt \(\Lambda_{\rm s}\)CDM, the simplest phenomenological realization of the \(\Lambda_{\rm s}\)CDM framework, has been extensively studied~\cite{Akarsu:2021fol,Akarsu:2022typ,Akarsu:2023mfb,Yadav:2024duq,Akarsu:2024eoo}. However, its abrupt transition at \(z = z_{\dagger}\) introduces a discontinuity, leading to a type II (sudden) singularity~\cite{Barrow:2004xh} at $z=z_\dagger$, though Ref.~\cite{Paraskevas:2024ytz} has shown this has negligible impact on cosmic structure formation and evolution. Nonetheless, this singularity suggests that the abrupt \(\Lambda_{\rm s}\)CDM model should be interpreted as an idealized approximation, effectively serving as a proxy for a rapid yet smooth transition.\footnote{A well-defined formulation necessitates a smooth transition, enabling a detailed study of perturbations but introducing theoretical challenges. A rapid AdS-to-dS transition can sharply increase \(\dot{H}\), inducing transient super-acceleration (\(\dot{H} > 0\)), which may affect cosmological observables and, within GR, is often linked to ghost instabilities and WEC violations. These impose constraints on the transition's rapidity in \ac{gr} but may be circumvented in type-II minimally \ac{mg} theories such as VCDM~\cite{DeFelice:2020eju,DeFelice:2020cpt}. For further discussion, see~\cite{Akarsu:2024qsi,Akarsu:2024eoo,Akarsu:2025gwi}.}

We refer readers to Refs.~\cite{Vazquez:2012ag,BOSS:2014hwf,Sahni:2002dx,Sahni:2014ooa,Bag:2021cqm,BOSS:2014hhw,DiValentino:2017rcr,Mortsell:2018mfj,Poulin:2018zxs,Capozziello:2018jya,Wang:2018fng,Banihashemi:2018oxo,Dutta:2018vmq,Banihashemi:2018has,Akarsu:2019ygx,Li:2019yem,Visinelli:2019qqu,Ye:2020btb,Perez:2020cwa,Akarsu:2020yqa,DeFelice:2020cpt,Calderon:2020hoc,Ye:2020oix,Paliathanasis:2020sfe,Bonilla:2020wbn,Acquaviva:2021jov,Bernardo:2021cxi,Escamilla:2021uoj,Akarsu:2022lhx,Bernardo:2022pyz,DiGennaro:2022ykp,Ong:2022wrs,Malekjani:2023ple,Alexandre:2023nmh,Gomez-Valent:2023uof,Medel-Esquivel:2023nov,Tiwari:2023jle,Anchordoqui:2023woo,Anchordoqui:2024gfa,Anchordoqui:2024dqc,Gomez-Valent:2024tdb,Bousis:2024rnb,Wang:2024hwd,Colgain:2024ksa,Yadav:2024duq,Toda:2024ncp,Akarsu:2024nas,Souza:2024qwd,Mukherjee:2025myk,Tyagi:2024cqp,Manoharan:2024thb,Gomez-Valent:2024ejh,Akarsu:2024qsi,Akarsu:2024eoo,Dwivedi:2024okk,Giare:2025pzu,Keeley:2025stf,Akarsu:2025gwi} for further theoretical and observational studies---including model agnostic reconstructions (see Sec.~\ref{sec:Recon_tech})---that explore \ac{de} with negative densities, often consistent with an AdS-like cosmological constant, at $z \gtrsim 1.5-2$, and aimed at addressing major cosmological tensions. Phantom \ac{de} models, which typically feature $\rho_{\rm DE}$ that decreases with redshift but are conventionally assumed to yield $\rho_{\rm DE}>0$, are known to mitigate the $H_0$ tension. Among these, the \textit{phantom crossing model}, proposed phenomenologically in Ref.~\cite{DiValentino:2020naf} (DMS20~\cite{Adil:2023exv}), stands out. A recent analysis, which considered this model as a particular example of the broader Omnipotent \ac{de} class~\cite{Adil:2023exv}, reaffirmed its success while also revealing that its ability to assume negative densities for \(z \gtrsim 2\)---mimicking an AdS-like cosmological constant beyond sufficiently high $z$---is central to its effectiveness. \ac{ide} models~\cite{Kumar:2017dnp,DiValentino:2017iww,Yang:2018uae,Pan:2019gop,Kumar:2019wfs,DiValentino:2019jae,DiValentino:2019ffd,Lucca:2020zjb,Gomez-Valent:2020mqn,Kumar:2021eev,Nunes:2022bhn,Bernui:2023byc,Giare:2024smz,Sabogal:2025mkp} (see Sec.~\ref{sec:IDE}) offer an alternative approach to resolving the \(H_0\) tension; however, model-independent reconstructions of the \ac{ide} kernel~\cite{Escamilla:2023shf} suggest that negative \ac{de} densities at \(z \gtrsim 2\) persist as a possibility.

While late-Universe rapid AdS-to-dS (or analogous) transitions in \ac{de}, as proposed by $\Lambda_{\rm s}$CDM, were initially viewed as challenging to reconcile with a robust physical mechanism, the remarkable phenomenological success of this approach---despite its simplicity---has prompted deeper theoretical inquiries. Even established frameworks, upon re-examination, have been found to accommodate such transitions within previously overlooked solution spaces, prompting researchers to adopt a fresh perspective on familiar theories. For instance, \(\Lambda_{\rm s}\)CDM\(^+\)~\cite{Anchordoqui:2023woo,Anchordoqui:2024gfa,Anchordoqui:2024dqc} proposes a stringy realization of \(\Lambda_{\rm s}\)CDM; it was shown that, despite the AdS swampland conjecture suggesting that a late-universe AdS-to-dS transition is unlikely due to the arbitrarily large separation between AdS and dS vacua in moduli space, such a transition can nonetheless be realized through the Casimir forces of fields inhabiting the bulk.\footnote{By combining swampland conjectures with observational data, it was proposed that the cosmological hierarchy problem---i.e., the smallness of \ac{de} density in Planck units---could be understood as an asymptotic field-space limit corresponding to the decompactification of a micron-sized extra (dark) dimension~\cite{Montero:2022prj}. Within this framework, Casimir forces from fields inhabiting the dark dimension can drive an AdS-to-dS transition~\cite{Anchordoqui:2023woo}, forming the basis of the \(\Lambda_{\rm s}\)CDM\(_{\pm}\) and \(\Lambda_{\rm s}\)CDM\(^+\). Specifically, a 5D Einstein-de Sitter gravity action compactified on a circle induces a runaway potential inherited from the 5D cosmological term~\cite{Anchordoqui:2023etp}. If the 5D cosmological constant is small, the quantum contribution of the lightest 5D modes---identified with Casimir energy~\cite{Arkani-Hamed:2007ryu}---becomes significant. A minimal setup requires a 5D mass spectrum comprising the graviton, three generations of light right-handed neutrinos, and a real scalar field \(\varphi\) with a potential featuring two local minima~\cite{Anchordoqui:2023woo}. At \(z_\dagger \sim 2\), \(\varphi\) undergoes quantum tunneling from the false to the true vacuum~\cite{Anchordoqui:2024dqc}, acquiring a larger mass and suppressing its Casimir energy contribution. This modifies the balance of fermionic and bosonic degrees of freedom, triggering the AdS-to-dS transition. The deep infrared fields of the dark sector contribute to the effective number of relativistic neutrino-like species.}  Building on the same theoretical framework, \(\Lambda_{\rm s}\)CDM\(_{\pm}\)~\cite{Soriano:2025gxd} extends \(\Lambda_{\rm s}\)CDM\(^+\) by allowing the AdS phase to have arbitrary depths, considering different curvature radii in the AdS and dS phases. It was demonstrated in Ref.~\cite{Alexandre:2023nmh} that, in various formulations of GR, a $\Lambda_{\rm s}$ can arise naturally through an overall signature change of the metric. \(\Lambda_{\rm s}\)VCDM~\cite{Akarsu:2024qsi,Akarsu:2024eoo} advances the \(\Lambda_{\rm s}\)CDM framework into a theoretically complete physical cosmology, offering a fully predictive description of the Universe, including the AdS-to-dS transition epoch itself. It was shown that the mirror AdS-to-dS transition can be effectively realized within a type-II minimally \ac{mg} framework called VCDM~\cite{DeFelice:2020eju,DeFelice:2020cpt}, through a specific Lagrangian incorporating an auxiliary scalar field with a smoothly sewed two-segmented linear potential. Ref.~\cite{Akarsu:2024nas} demonstrated that the teleparallel \( f(T) \) gravity, specifically its exponential infrared form~\cite{Awad:2017yod}, which has shown significant promise in addressing the \(H_0\) tension~\cite{Hashim:2020sez,Hashim:2021pkq} in its solution space giving phantom-like effective \ac{de}, admits previously overlooked solution spaces, which accommodate an alternative scenario where the effective \ac{de} transitions smoothly from negative to positive at \(z_\dagger \sim 1.5\), while remaining consistent with \ac{cmb} power spectra. Building on these insights, \( f(T) \)-\(\Lambda_{\rm s}\)CDM successfully maps the background dynamics of \(\Lambda_{\rm s}\)CDM into the \( f(T) \) gravity framework~\cite{Souza:2024qwd}, further establishing a theoretical foundation for AdS-to-dS-like transitions in the late Universe. Ph-\(\Lambda_{\rm s}\)CDM~\cite{Akarsu:2025gwi}, introduced a phantom \ac{de} model, which is a specific realization of a general scalar field with a hyperbolic tangent potential that induces smooth AdS-to-dS, 0-to-dS, and dS-to-dS transitions. Despite its negative kinetic term, the step-like potential prevents pathologies like unbounded energy growth, Big Rip, and WEC violations, ensuring smooth evolution of cosmological parameters. Notably, the AdS-to-dS transition in \ac{de} density does not precisely parallel that of the potential, persisting longer due to the kinetic term’s contribution. Even if different realizations of \(\Lambda_{\rm s}\)CDM yield identical background dynamics, they still exhibit differences. GR-based \(\Lambda_{\rm s}\)CDM models~\cite{Akarsu:2021fol, Akarsu:2022typ, Akarsu:2023mfb, Akarsu:2025gwi}, \(\Lambda_{\rm s}\)VCDM~\cite{Akarsu:2024qsi, Akarsu:2024eoo}, and \( f(T) \)-\(\Lambda_{\rm s}\)CDM differ in their predictions for linear perturbations, while the string-inspired \(\Lambda_{\rm s}\)CDM\(^+\)~\cite{Anchordoqui:2023woo, Anchordoqui:2024gfa, Anchordoqui:2024dqc, Soriano:2025gxd} predicts a modest excess in the total effective number of neutrino species, with \(N_{\rm eff} = 3.294\). Such distinguishing features are invaluable, as they provide a means to compare and ultimately discriminate between these alternative realizations using observational data. Other than the models explicitly realizing/resembling the background dynamics of \(\Lambda_{\rm s}\)CDM, there exist various other approaches that fit within the broader paradigm outlined earlier. These include brane-world models~\cite{Sahni:2002dx,Sahni:2014ooa,Bag:2021cqm}, multiple axion models~\cite{Cicoli:2018kdo,Ruchika:2020avj}, energy-momentum log gravity~\cite{Akarsu:2019ygx}, bimetric gravity~\cite{Dwivedi:2024okk}, Horndeski gravity~\cite{Tiwari:2023jle}, holographic \ac{de}~\cite{Tyagi:2024cqp}, Granda–Oliveros holographic DE~\cite{Manoharan:2024thb}, composite \ac{de} (\(w\)XCDM)~\cite{Gomez-Valent:2024tdb,Gomez-Valent:2024ejh}, inspired by the $\Lambda$XCDM framework~\cite{Grande:2006nn}---extends the abrupt \(\Lambda_{\rm s}\)CDM model by introducing two free parameters that allow \( w_{\rm DE} \) to take arbitrary constant values before and after the transition\footnote{The \(w\)XCDM model~\cite{Gomez-Valent:2024tdb,Gomez-Valent:2024ejh}, inspired by \(\Lambda\)XCDM~\cite{Grande:2006nn}, introduces a composite \ac{de} scenario with two components: \(X\) (for \(z > z_\dagger\)) and \(Y\) (for \(z < z_\dagger\)). \(Y\) behaves as running vacuum energy with a quintessence-like \ac{eos} (\(w_Y \gtrsim -1\)), while \(X\), termed ‘phantom matter’ (PM), has a phantom-like \ac{eos} (\(w_X \leq -1\)) but negative energy density, mimicking effective string action terms at low energies~\cite{Mavromatos:2020kzj}. The Kalb-Ramond axion and gravitational Chern-Simons term generate a ‘phantom vacuum’~\cite{Mavromatos:2021urx}, where PM enhances structure formation at high \(z\), potentially explaining the \ac{jwst} anomaly. The de Sitter vacuum is restored via gChS condensates, and data support \(w_Y > -1\), \(w_X < -1\), aligning with \ac{desi} results~\cite{DESI:2024mwx}.}---, the Omnipotent \ac{de} concept~\cite{Adil:2023exv}, with its specific realization DMS20~\cite{DiValentino:2020naf}, represents a class of \ac{de} models characterized by non-monotonic densities and transitions across $w_{\rm DE}=-1$, DE pressure parametrizations~\cite{Sen:2007gk, Kumar:2012dv}, which yield $w_{\rm DE}$ similar DMS20, the Lotka-Volterra model of two interacting fluids~\cite{Ong:2022wrs}, running Barrow entropy~\cite{DiGennaro:2022ykp}, multiple-transition vacuum \ac{de} models~\cite{Moshafi:2022mva}, and scenarios invoking a modification of the gravitational constant between super- and sub-horizon regimes, motivated by the Ho\v{r}ava–Lifshitz proposal or the Einstein-aether framework~\cite{Wen:2023wes}.

Thus, a rapidly growing body of literature has emerged within this class of models---particularly in recent years---with many having been tested against a variety of observational datasets, consistently revealing a strong correlation between the redshift of the sign change in the DE density, $z_\dagger$, and the model’s ability to address major cosmological tensions, especially those involving $H_0$ and $S_8$. To provide insight, we highlight several representative observational constraints. The abrupt $\Lambda_{\rm s}$CDM model, using the combined Planck+BAOtr+\ac{kids}1000+\ac{sn1}+SH0ES dataset, yields $z_\dagger = 1.72^{+0.09}_{-0.12}$, predicting $H_0 = 73.16 \pm 0.64\,{\rm km\,s^{-1}\,Mpc^{-1}}$ and $S_8 = 0.774 \pm 0.009$, thereby fully resolving both tensions~\cite{Akarsu:2023mfb}. When analyzed with the Planck+SNIa+SH0ES combination, the same model gives $z_\dagger = 1.83^{+0.11}_{-0.19}$, $H_0 = 72.07 \pm 0.88$, and $S_8 = 0.786 \pm 0.011$, still strongly alleviating both tensions, albeit trending slightly toward \lcdm\ values—as expected for higher $z_\dagger$, which reduces the impact of the AdS-like phase~\cite{Akarsu:2024eoo}. The $\Lambda_{\rm s}$CDM$^+$, the string-theory-inspired realization, analyzed with the same combined dataset as the first case, yields a higher transition redshift of $z_\dagger = 2.105 \pm 0.264$; yet, due to its enhanced $N_{\rm eff} = 3.249$, it predicts an even larger Hubble constant, $H_0 = 74.04 \pm 0.71$, thus also fully resolving the Hubble tension~\cite{Anchordoqui:2024gfa}. The $w$XCDM model, the two-parameter extension of the abrupt $\Lambda_{\rm s}$CDM, yields a slightly lower transition redshift $z_\dagger = 1.46 \pm 0.02$, predicting $H_0 = 70.94 \pm 0.56\,{\rm km\,s^{-1}\,Mpc^{-1}}$ and $S_8 = 0.784 \pm 0.009$ based on the Planck+BAOtr+DESY5 \ac{sn}+\ac{cc}+$f\sigma_{12}$+SH0ES dataset, thereby successfully addressing both tensions, while differing in that it features an EoS parameter crossing from $w_{\rm DE} < -1$ to $w_{\rm DE} > -1$ across the transition from negative to positive energy density~\cite{Gomez-Valent:2024ejh}. These examples show that, despite variations in dataset and model details, scenarios featuring AdS-to-dS or analogous transitions with $z_\dagger \sim 1.5\text{--}2$ consistently succeed in addressing both the $H_0$ and $S_8$ tensions. Further insights come from $\Lambda_{\rm s}$CDM$_{\pm}$, the extension of $\Lambda_{\rm s}$CDM$^+$ that allows arbitrary AdS depths, revealing a degeneracy whereby deeper AdS phases permit higher $z_\dagger$ values while still resolving the $H_0$ tension~\cite{Soriano:2025gxd}. Even the few observational findings highlighted here provide a compelling case for continued theoretical and observational investigation of this class of models.

If future observations continue to support these models, the implications for theoretical physics would be profound, as an AdS-like cosmological constant is a theoretical sweet spot, favored by the AdS/CFT correspondence~\cite{Maldacena:1997re} and string theory frameworks~\cite{Bousso:2000xa}. Recent work~\cite{Demirtas:2021ote} has demonstrated that a supersymmetric vacuum in string theory can naturally produce an AdS-like cosmological constant at present-day energy scales, motivating \ac{de} scenarios where the field evolves on a potential with an AdS minimum rather than the standard dS minimum. Such a negative cosmological constant could shape both the current cosmic acceleration and the Universe’s long-term evolution, offering a compelling link between fundamental theory and observations. In this direction, Ref.~\cite{Visinelli:2019qqu} examined a \ac{de} model consisting of an AdS-like cosmological constant and a \ac{de} component with an \ac{eos} parameter \( w_\phi \), finding no evidence for an AdS-like cosmological constant and a mild preference for an effective phantom \ac{de} component, though \lcdm\ remains favored. Ref.~\cite{Sen:2021wld} (see also~\cite{Calderon:2020hoc}) extends \(w\)CDM and \ac{cpl}-\ac{cdm} by introducing an AdS-like cosmological constant, demonstrating improvements in both data fits and alleviation of the \(H_0\) tension, while further studies~\cite{Adil:2023ara, Menci:2024rbq, Menci:2024hop} show that this model can also address the \ac{jwst} anomaly. Indeed, the models discussed in this section are expected to generally enhance structure formation at high $z$, making them natural candidates for explaining the \ac{jwst} anomaly~\cite{Menci:2022wia, Biagetti:2022qnp, Wang:2022jvx, Forconi:2023hsj}. However, their rigorous quantitative analysis of perturbation evolution and cosmic structuring is necessary. The evolution of cosmic structures and linear perturbations in the (abrupt) \(\Lambda_{\rm s}\)CDM model has been studied in Refs.~\cite{Paraskevas:2023itu,Paraskevas:2024ytz,Akarsu:2025ijk}. Specifically, Ref.~\cite{Paraskevas:2023itu} found that for \(\Omega_{\Lambda,0} = -0.7\) and \(\Omega_{\rm m,0} = 0.3\), \(\Lambda_{\rm s}\)CDM predicts up to an 80\% increase in cluster density for turnaround redshifts \(z_{\text{max}} \gtrsim 2\), suggesting a potential explanation to the \ac{jwst} anomaly. Ref.~\cite{Paraskevas:2024ytz} demonstrated that even in abrupt $\Lambda_{\rm s}$CDM (the most extreme case), the transition itself has no significant impact on bound structures, preserving model viability. Additionally, Ref.~\cite{Akarsu:2025ijk} showed that the growth index remains \(\gamma \sim 0.55\) as in \lcdm. However, Planck \ac{cmb} data predicts \(\Omega_{\rm m,0} = 0.28\) for \(\Lambda_{\rm s}\)CDM and \(\Omega_{\rm m,0} = 0.32\) for \lcdm, leading to growth rates of \(f = 0.49\) and \(f = 0.53\) at \(z = 0\), respectively. Notably, \(\Lambda_{\rm s}\)CDM predicts a value closer to \( f = 0.48 \), recently obtained from \ac{lss} data when \(\gamma\) is treated as a free parameter in \lcdm~\cite{Nguyen:2023fip}, indicating its potential to resolve the structure growth anomaly. \ac{desi} data (combination with other datasets) have further validated \ac{de} models incorporating an AdS-like cosmological constant~\cite{Wang:2024hwd,Mukherjee:2025myk}, while obviating the need for phantom \ac{de} in the \ac{de} sector~\cite{Mukherjee:2025myk}. The post-reionization HI~21-cm signal is explored as a probe for an AdS-like cosmological constant in the \ac{de} sector~\cite{Dash:2023scq}. In summary, a negative cosmological constant in the \ac{de} sector is physically motivated, consistent with current data, and may yield distinctive cosmological and astrophysical signatures—including the possibility of a bouncing future universe~\cite{Andrei:2022rhi}.

Ref.~\cite{Vagnozzi:2023nrq} suggested that a promising approach to resolving cosmological tensions may involve combining early- and late-time new physics to better fit the data. Ref.~\cite{Toda:2024ncp} explored a hybrid model integrating abrupt \(\Lambda_{\rm s}\)CDM with a varying electron mass mechanism, identifying it as a promising candidate. However, they found that this combination does not improve the tension, as the two models push \(\Omega_{\rm m,0}\) in opposite directions. Ref.~\cite{Yadav:2024duq} analyzed the \(\Lambda_{\rm s}\)CDM+\(N_{\rm eff}\)+\(\sum m_{\nu}\) and \lcdm\ + \(N_{\rm eff}\)+\(\sum m_{\nu}\) models, allowing variations in \(N_{\rm eff}\) and \(\sum m_\nu\) within abrupt \(\Lambda_{\rm s}\)CDM and \lcdm. They found that \(\Lambda_{\rm s}\)CDM+\(N_{\rm eff}\)+\(\sum m_{\nu}\) consistently fits the data better while preserving the success of \(\Lambda_{\rm s}\)CDM and predicting standard neutrino properties. In contrast, when \lcdm\ + \(N_{\rm eff}\)+\(\sum m_{\nu}\) yields high \(H_0\) values, it does so at the cost of large \(\Delta N_{\rm eff}\), whereas \(\Lambda_{\rm s}\)CDM+\(N_{\rm eff}\)+\(\sum m_{\nu}\) achieves similarly high \(H_0\) values while remaining consistent with the standard \(N_{\rm eff}\). This suggests that the \(\Lambda_{\rm s}\)CDM+\(N_{\rm eff}\)+\(\sum m_{\nu}\) alleviates the need for pre-recombination new physics, at least concerning neutrino properties.
\bigskip
\subsubsection{Interacting dark energy \label{sec:IDE}}

\noindent \textbf{Coordinator:} Carsten van de Bruck\\
\noindent \textbf{Contributors:} Amare Abebe, Dario Bettoni, David Benisty, David Tamayo, Denitsa Staicova, Diego Rubiera-Garcia, Emmanuel Saridakis, Emre \"Oz\"ulker, Kay Lehnert, Leila L. Graef, Lu Yin, Luis Anchordoqui, Marcel A. van der Westhuizen, Marco de Cesare, Nikolaos E. Mavromatos, Oem Trivedi, Oleksii Sokoliuk, Purba Mukherjee, Rahul Shah, Sveva Castello, Vitor da Fonseca, and Yuejia Zhai
\\

\paragraph{Introduction}

In this subsection, we discuss models with an interaction between \ac{de} and \ac{dm}. 
Such \ac{ide} models have been introduced to address problems of the \lcdm\ model, including the $H_0$ and $S_8$ tensions (for comprehensive reviews, e.g., see Ref.~\cite{Bolotin:2013jpa, DiValentino:2021izs, Wang:2024vmw}). Some \ac{ide} models can reduce the $H_0$ tension from 5$\sigma$ to 3.6$\sigma$ \cite{DiValentino:2021izs}. 

In a general model, the conservation equations for \ac{dm} and \ac{de} take the form
\begin{eqnarray}
    \nabla_\mu T^{\mu\nu}_{\rm(DM)} = Q^\nu\,,~~~{\rm and}~~~\nabla_\mu T^{\mu\nu}_{\rm(DE)} = -Q^\nu\,,
\end{eqnarray}
ensuring that the total energy-momentum tensor of \ac{dm} and \ac{de} remains conserved, namely, $\nabla_\mu (T^{\mu\nu}_{\rm(DM)}+T^{\mu\nu}_{\rm(DE)})=0$. Often, the resulting background equations are written in the form 
\begin{equation}\label{background}
    \dot\rho_{\rm DM} + 3H\rho_{\rm DM} = Q\,, ~~~{\rm and}~~~ \dot\rho_{\rm DE} + 3H(\rho_{\rm DE}+p_{\rm DE})  = - Q\,.
\end{equation} 

In order for \ac{ide} models to be viable candidates for addressing the $H_0$ and $\sigma_8$ tensions, special attention must be given to the physicality of the parameter space \cite{vanderWesthuizen:2023hcl}. In these models, there is not always a mechanism to halt the energy transfer when either the \ac{dm} or \ac{de} density becomes zero (i.e., $Q\neq 0$ in Eq.~\eqref{background} when $\rho_{\text{DE/DM}}=0$), which can lead to negative energies. The case of $Q>0$ corresponds to an energy transfer from \ac{de} to \ac{dm}. It has been reported that \ac{cmb} observations seem to favor an energy transfer from \ac{de} to \ac{dm}, \ac{wl} measurements and thermodynamical considerations suggest an energy transfer from \ac{dm} to \ac{de} \cite{Alcaniz:2005dg, Pavon:2007gt, Wang:2016lxa, daFonseca:2021imp}.

One well studied class of models is coupled quintessence, in which $Q^\nu = \beta \phi^{,\nu} T^\mu_{(\rm DM)\mu}$  and $Q=-\beta T^\mu_{(\rm DM)\mu} \dot\phi$, where $\phi$ is the \ac{de} scalar field and $T^\mu_{(\rm DM)\mu}$ denotes the trace of the \ac{dm} energy--momentum tensor \cite{Wetterich:1994bg,Amendola:1999er,CarrilloGonzalez:2017cll}. There also exists an extension of these models in which \ac{dm} and \ac{de} are disformally coupled \cite{Zumalacarregui:2010wj,vandeBruck:2015ida,Teixeira:2019hil,vandeBruck:2020fjo}. Such scalar field theories of \ac{ide} may alleviate the Hubble tension but do not fully solve it \cite{VanDeBruck:2017mua,Gomez-Valent:2020mqn,Goh:2022gxo}. 

Other, more phenomenological approaches propose a coupling of the form $Q^\mu= Q u^{\mu}_{\rm DM} / a $ or $Q^\mu = Q u^{\mu}_{\rm DE} / a$ \cite{Valiviita:2008iv,Gavela:2009cy} (in these expressions, $a$ is the scale factor and $u^\mu_{\rm DM}$ or $u^\mu_{\rm DE}$ are the four-velocity of \ac{dm} or \ac{de} fluid, respectively). The interaction $Q$ is usually written as $Q=\xi H \rho_{\rm DE}$ or $\xi H \rho_{\rm DM}$, where $\xi$ is a dimensionless coupling constant, determining the size and direction of energy/momentum flow. The case $Q=\xi H \rho_{\rm DE}$ was recently studied in Ref.~\cite{Zhai:2023yny} (see also~\cite{DiValentino:2017iww} for a previous study), assuming an equation of state of \ac{de} $w_{\rm DE} = -0.999$, using the \ac{cmb} data provided by Planck, \ac{wmap} and \ac{act}. Different dataset combinations resulted in consistent results, showing evidence for a non-zero coupling and a Hubble expansion rate $H_0$ consistent with local measurements. Using Planck alone, the analysis results in $H_0=71.6 \pm 2.1$\kms, \ac{act} alone results in $H_0 = 72.6^{+3.4}_{-2.6}$\kms and the combination of \ac{act} and Planck results in $H_0=71.4^{+2.5}_{-2.8}$\kms. In Ref.~\cite{Giare:2024ytc}, the assumption about the equation of state was relaxed, allowing for a dynamical evolution via the parametrization $w_{\rm DE}(a) = w_0 + w_a(1-a)$. It was found that models featuring a dynamical phantom equation of state ($w_0<-1$) perform worse than the non-dynamical case. On the other hand, models featuring a dynamical quintessence equation of state ($w(a)>-1$ at any redshift) perform better in attempting to increase the value of $H_0$ compared to the respective
non-dynamical case. However, when considering the joint analysis of \ac{cmb}, \ac{bao} SDSS, and \ac{sn} data, no significant increase in $H_0$ to solve the Hubble tension was found, while it is a promising solution considering \ac{desi} \ac{bao} data~\cite{Giare:2024smz}.

In the following, we present a number of other \ac{ide} models.

\paragraph{Models}

\subparagraph{Nonlinear interacting dark energy}

This class of models is an extension of the models discussed above. A general form of the nonlinear \ac{ide} models is given by $Q=3H \xi F(\rho_{\rm DE}, \rho_{\rm DM})$, where $F$ is a nonlinear function of the energy densities and $\xi$ is a free constant parameter. Generally, when $Q>0$, the $H_0$ tension worsens and the $S_8$ alleviates; conversely, when $Q<0$, the $H_0$ tension alleviates and the $S_8$ worsens \cite{vanderWesthuizen:2023hcl}. The authors of Ref.~\cite{Paliathanasis:2019hbi} investigate nonlinear \ac{ide} as an interaction between \ac{dm} and a scalar field, in Ref.~\cite{De-Santiago:2016oeu} a model is discussed that could be interpreted as a particular case of a running vacuum model $\Lambda(H)$.

Power laws \ac{ide} models with the general form $Q = 3H\xi \rho_{\rm DM}^p \rho_{\rm DE}^s (\rho_{\rm DM}+ \rho_{\rm DE})^r$, where $p$, $s$ and $r$ integers, were studied in Ref.~\cite{Arevalo:2011hh}, and observational constraints for specific cases provided in Ref.~\cite{Ebrahimi:2016rvi}. Observational constraints of $Q = 3H\xi\left(\rho_{\rm DM} + \rho_{\rm DE} +\frac{\rho_{\rm DM}\rho_{\rm DE}}{\rho_{\rm DM}+ \rho_{\rm DE}}\right)$ are discussed in Ref.~\cite{Khurshudyan:2017qtd}. The model $Q=\gamma\rho_{\rm DM}^\alpha\rho_{\rm DE}^\beta$, using \ac{cmb}, \ac{bao}, \ac{rsd}, and \ac{sn1} data, predicts lower values of $f(z)\sigma_8(z)$ at $z < 1$ comparing to \lcdm, which alleviates the tension of \lcdm\ with various \ac{rsd} data \cite{Cheng:2019bkh}. Other model types, such as exponential models $Q=3H\xi \rho_{\rm DE}\exp\left(\rho_{\rm DE}/\rho_{\rm DM} -1\right)$ \cite{Yang:2018pej}, and logarithmic models $Q = 3H\xi\rho_{\rm DE} \log(\rho_{\rm DE}/\rho_{\rm DM})$ and $Q =3H\xi\rho_{\rm DE}  \log(\rho_{\rm DM}/\rho_{\rm DE})$ \cite{Khurshudyan:2017kmf}, were studied but showed no significant promising results.

Models with the forms $Q=3H\xi \rho_{\rm DE}\sin(\rho_{\rm DE}/\rho_{\rm DM} -1)$ and $Q=3H\xi \rho_{\rm DE}[1 +\sin(\rho_{\rm DE}/\rho_{\rm DM} -1)]$ have gained special attention for their potential to alleviate the $H_0$ tension \cite{Pan:2020bur}. Analysis with Planck 2018 gives $H_0=72.67^{+5.43}_{-8.26}$\kms at 68\% CL, and Planck 2018+\ac{bao} gives $H_0=69.17^{+1.52}_{-1.71}$\kms at 68\% CL. Hence, for Planck 2018 and Planck 2018+\ac{bao}, the $H_0$ tension with respect to SH0ES ($H_0 = 73.0 \pm 1.0$\kms) is reduced down to 0.05$\sigma$, 2$\sigma$ respectively; notably, for Planck 2018 alone case the tension is completely solved. However, these results should be interpreted with caution due to potential biases and uncertainties. Additionally, a theoretical justification for the specific form of $Q$  is necessary to ensure that it is not merely an ad hoc model.

\subparagraph{Diffusion interactions} 

A diffusive interaction between \ac{de} and \ac{dm} was introduced in Refs.~\cite{Haba:2016swv,Koutsoumbas:2017fxp,Calogero:2013zba}. The diffusion of energy density between \ac{de} into \ac{dm} uses a non-conserved stress energy tensor $T^{\mu\nu}$ with a source current $j^\mu$, $\nabla_\mu T^{\mu\nu}_{(\text{m})}=\gamma^{2} j^\nu\,,$ where $\gamma^{2}$ is the coupling diffusion coefficient of the fluid. The current $j^\mu$ is a time-like covariant conserved vector field $\nabla_\mu j^{\mu} = 0$ which describes the conservation of the number of particles in the system. In a homogeneous expansion, the modified Friedman equations read
\begin{equation}
    \dot{\rho}_{\rm DM} + 3 H \rho_{\rm DM} = \frac{\gamma}{a^3} \,, \quad  \dot{\rho}_\Lambda  = -\frac{\gamma}{a^3}\,, 
\end{equation}
The contribution of the current goes as $\sim a^{-3}$ since the current is covariantly conserved. In this way, there is a compensation between \ac{de} and \ac{dm}. Ref.~\cite{PiratovaMoreno:2023bpg} introduces cases with a diffusion constant that could be represented for a scalar field $\phi$ or a perfect fluid, leading to late-time forms of \ac{de} and a \ac{de} density parameter that could be interpreted as a perturbation of the \lcdm\ model.

Another example of diffusion interactions is represented by models where the energy-momentum transfer vector comes from a potential, $Q_{\mu}=\nabla_{\mu} J$. This class of models finds a natural embedding within unimodular gravity. In fact, in unimodular gravity the total energy-momentum of matter fields need not be conserved in general \cite{Josset:2016vrq}. The energy density of \ac{de} is then identified with the potential $J$ up to an arbitrary integration constant, $\rho_{\rm DE}=J+\Lambda_0$, and the equation of state is $w_{\rm DE}=-1$ identically. Models of this kind have been proposed to describe effective diffusion processes due to an underlying discrete spacetime structure \cite{Perez:2017krv,Perez:2019gyd}. In Ref.~\cite{Perez:2020cwa} it was argued that such diffusive interactions ---with energy transfer from \ac{dm} to \ac{de}--- can alleviate the Hubble tension, see Refs.~\cite{Corral:2020lxt,LinaresCedeno:2020uxx,Landau:2022mhm} for a more detailed analysis. This class of models is also formally equivalent to ``interacting vacuum models'', which have been studied within the context of general relativity \cite{Wands:2012vg,Sebastianutti:2023dbt}. Embedding minimally \ac{ide} with $w_{\rm DE}=-1$ within UG has the following advantages: \textit{i}) the vacuum energy does not gravitate in UG removing the age-old problem regarding the puzzling absence of the vacuum energy density in cosmological observations~\cite{Weinberg:1988cp,Ellis:2010uc}, \textit{ii}) the well-known perturbative large-scale instability studied in Ref.~\cite{Valiviita:2008iv} does not exist in UG~\cite{deCesare:2021wmk}, \textit{iii}) since the \ac{de} is an effective source in UG rather than a physical one, $\rho_{\rm DE}$ can attain negative values opening up a large phenomenological landscape to address the cosmological tensions. Moreover, it opens up a path for quantum gravity motivated interaction terms; e.g., see Refs.~\cite{Josset:2016vrq,Perez:2017krv,Perez:2019gyd}.

An important aspect often forgotten in the study of such diffusive models is to study the effect of large-scale structure formation in such a fluid environment. Growth-rate analyses can shed light on whether such a model alleviates the $\sigma_8$ tension or not.

\subparagraph{Metastable dark energy}

Models of metastable \ac{de} have been analyzed in several works, see for instance Refs.~\cite{Shafieloo:2016bpk, deSouza:2024sfl, DiValentino:2021izs, Li:2019san, Yang:2020zuk, Urbanowski:2021waa, Urbanowski:2022iug, Abdalla:2012ug, Landim:2016isc, Landim:2017lyq, Stojkovic:2007dw, Greenwood:2008qp}. In particular, in Refs.~\cite{Shafieloo:2016bpk} and \cite{Li:2019san}, phenomenological models in this context were investigated, considering the case of a constant \ac{de} decay rate, depending only on the intrinsic properties of \ac{de} and the type of decay channel. The following cases of metastable \ac{de} decaying in three distinct ways were considered: (I) exponentially; (II) into \ac{dm}; (III) into dark radiation. Among these models, model II showed slightly better consistency \cite{Li:2019san}. This model is sometimes called metastable \ac{ide} \cite{DiValentino:2021izs}, since it is described by Eq.~\eqref{background} with $Q=\Gamma \rho_{\rm DE}$, where $\Gamma$ is the \ac{de} decay rate.

In Ref.~\cite{Li:2019san} it was shown that despite the fit for this model against Pantheon + \ac{bao} data providing higher values for $H_{0}$, when the \ac{cmb} distance priors from Planck 2018 are included, the Hubble tension is restored. Later on, in  Ref.~\cite{Yang:2020zuk}, a full analysis was performed using different combinations of data from Planck 2018, \ac{bao}, \ac{des}, R19 \cite{Planck:2018vyg, Riess:2019cxk}. A larger value of $H_{0}$ was then supported, solving the tension with R20 \cite{Riess:2020fzl} within $1\sigma$. Future data with higher precision must provide us with a clearer understanding of the performance of these models concerning the tension. Some discussion on motivations for metastable \ac{ide} arising from quantum mechanics can be found in Refs.~\cite{Urbanowski:2021waa, Urbanowski:2022iug}. We can also find some examples from field theory descriptions, as in Refs.~\cite{Abdalla:2012ug, Landim:2016isc, Landim:2017lyq, deSouza:2024sfl}, for instance.

Another model of metastable \ac{ede} is provided by the so-called Chern-Simons gravity inspired from string theory, studied in Refs.~\cite{Basilakos:2019acj,Mavromatos:2020kzj,Dorlis:2024yqw}. It has been shown that, under some conditions, condensation of chiral \ac{gw}s in early epochs of the Universe, can lead to a non-trivial condensate of the Chern-Simons gravitational anomaly term. The latter is characterized by the presence of imaginary parts, which are such that they lead to a metastable \ac{ede}, with a lifetime that can be in the phenomenologically right ballpark of larger than 50-60 e-foldings. The model, which also leads to \ac{rvm} type inflation, due to the non-linear $H^4$ dependence of the (real part of the) pertinent vacuum energy density, can also help alleviate the Hubble and growth-of-structure tensions in modern eras~\cite{Gomez-Valent:2023hov}.

\subparagraph{Elastic dark energy interactions}

In \ac{de}-\ac{dm} elastic interactions models the two species interact via velocity or momentum transfer. There are various ways of implementing these exchanges \cite{Simpson:2010vh,Pourtsidou:2013nha,Asghari:2019qld} but here, for the sake of concreteness, we focus on the particular realization presented in Ref.~\cite{Asghari:2019qld} and further developed in Refs.~\cite{Figueruelo:2021elm,BeltranJimenez:2021wbq,Poulin:2022sgp} (see Ref.~\cite{Jimenez:2024lmm} for a short review), which is based on the following phenomenological implementation of the coupling
\begin{equation}
    \nabla_\nu T^{\mu\nu}_{\rm DM} = \alpha\left(u^\mu_{\rm DM}-u^\mu_{\rm DE}\right)\,,\quad  \nabla_\nu T^{\mu\nu}_{\rm de} = -\alpha\left(u^\mu_{\rm DM}-u^\mu_{\rm DE}\right)\,,
\end{equation}
where $\alpha$ measures the strength of the interaction. It is clear that the interaction is active only when there is relative motion between the two species, i.e., at sub-Hubble scales while, at the largest scales, where the two fluids share a common reference frame, the interaction is absent. This means, in turn, that no modifications to the background evolution are produced. As a consequence, these models are not designed to address the $H_0$ tension but have an impact on relaxing the $\sigma_8$ one. In fact, at the level of linear perturbation, Euler equation receives a new contribution proportional to the relative velocity between the two species with a time dependent interaction rate $\Gamma_\alpha = \alpha a^4/\Omega_{\rm DM}$ in the case of \ac{dm} and further weighted by the relative abundance of \ac{dm} to \ac{de} $R = a^{3w}\Omega_{\rm DM}/(1+w)/\Omega_{\rm DE}$ in the case of \ac{de}. Since the interaction grows with the Universe expansion as $a^4$,   eventually, $\Gamma_\alpha \sim H$ and the interaction with \ac{de} becomes relevant starting to exert a drag on \ac{dm} which suppresses its clustering. This allows for the onset of a dark-coupling epoch at late-time as desired and as suggested by data that points towards a suppression in the clustering at $z\lesssim 2$. By considering Planck 2018 data releases, \ac{sn1} and Baryon Acoustic Oscillations data it has been shown  that the interaction can indeed reduce the value of $\sigma_8$  and does so without affecting $\Omega_m$ ($\sigma_8 = 0.753^{+0.011+0.022}_{
-0.011-0.021}$, $\Omega_m = 0.311^{+0.007+0.014}_{
-0.007-0.013}$) and that the presence of the interaction is strongly favored \cite{BeltranJimenez:2021wbq,Figueruelo:2021elm}. The fact that no correlations between the interacting parameter and background ones is introduced is a particularly welcome feature since it implies no worsening of the $H_0$ tension. Remarkably, the inclusion of $S_8$ data allows for a constraint at percent level of the interaction by surveys like JPAS, Euclid, and \ac{desi} \cite{Figueruelo:2021elm} while \ac{ska}-like experiments should have enough sensitivity to detect the interaction parameter \cite{BeltranJimenez:2022irm}. Finally, it is worth mentioning that the inclusion of massive neutrinos does not spoil the detection prospects for this model \cite{Jimenez:2024bhv} and that other implementations and analyses reach similar conclusions \cite{Baldi:2016zom, Pourtsidou:2016ico,Poulin:2022sgp}.

\subparagraph{Fading dark matter}

If the swampland conjecture for \ac{de} is correct, it would lead to the prediction that the \lcdm\ model, where \ac{de} is constant, cannot be veracious~\cite{Agrawal:2018own, Kinney:2018nny}. Quite independently of swampland considerations, there have been proposals for a rolling scalar field as the source of\ac{de}. If this were the case, the swampland distance conjecture suggests that the rolling field would lead at late-times to an exponentially light tower of states. Identifying this tower as residing in the dark sector yields a natural coupling of the scalar field to the \ac{dm}, leading to a continually reducing \ac{dm} mass as the scalar field rolls in the recent cosmological epoch. It has been shown that the way in which the tower of light states evolves over time could help to reduce (though not fully eliminate) the $H_0$
 tension~\cite{Agrawal:2019dlm}. More unambiguously, the coupling of the scalar field to the \ac{dm} leads to a reduction of mass, or fading of \ac{dm}, which is compensated by a bigger value of \ac{de}. The latter becomes more noticeable in the present accelerating epoch, leading to an increase in $H_0$. Salam-Sezgin 6-dimensional supergravity~\cite{Salam:1984cj} and its string realization of Cveti\v c-Gibbons-Pope~\cite{Cvetic:2003xr} can bring to fruition the fading \ac{dm} model~\cite{Anchordoqui:2019amx, Anchordoqui:2020sqo}.

\subparagraph{Late-Time interacting constant/dynamical dark energy}

Late-time dark energy interacting with \ac{dm} has been extensively explored in the context of the $H_0$ and $\sigma_8$ tensions. Numerous studies have been carried out in the context of the Hubble tension using different forms of the interaction term and $\Lambda$/constant \ac{eos} late-time dark energy, and by imposing various theoretical bounds on the parameter space to prevent instabilities \cite{Benisty:2024lmj,Hoerning:2023hks,Lucca:2020zjb,Vagnozzi:2023nrq,DiValentino:2017iww,DiValentino:2019ffd,Pan:2023mie,Wang:2024vmw,Yang:2018euj,Bhattacharyya:2018fwb,DiValentino:2019exe,Bernui:2023byc,Yao:2022kub,Gariazzo:2021qtg,Guo:2021rrz,Nunes:2021zzi,Zhao:2022ycr,Gao:2021xnk,Pan:2020bur,Amirhashchi:2020qep,Pan:2019gop,Pan:2019jqh,Yang:2018uae,Gao:2022ahg,Kumar:2019wfs,Kumar:2021eev,Yang:2021hxg,Yang:2019uog}. Similarly in the context of the clustering tension significant studies include \cite{DiValentino:2019ffd,DiValentino:2019jae,Lucca:2021dxo,Bhattacharyya:2018fwb,Gao:2022ahg,Kumar:2019wfs, Kumar:2021eev,An:2017crg,Mukherjee:2017oom, Sinha:2019axe, Sinha:2021tnr,Sinha:2022dze}. In Ref.~\cite{Giare:2024ytc} an interacting setup with interaction proportional to the \ac{de} density ($Q\propto\rho_\mathrm{de}$), was investigated in the light of late-time \ac{dde} assuming a \ac{cpl} parametrization in the context of the Hubble tension, and found no significant scope of alleviation of the tension with various combinations of datasets in either the phantom or quintessence regimes. In Ref.~\cite{Shah:2024rme}, using the same form of interaction, various other parametrizations of \ac{lde} (MEDE, \ac{cpl}, JBP) were considered. It was reported here that a quintessence \ac{eos} in such an interacting scenario does not help with the Hubble tension, and noticeably makes the $\sigma_8$ tension worse. However, a phantom \ac{eos}, although did not affect the Hubble tension, showed significant scope in alleviating the $\sigma_8$ tension (especially for the \ac{cpl} and JBP parametrizations). The inclusion of \ac{rsd} data in this scenario helped obtain tighter constraints on certain parameters, but resulted in a relatively less pronounced relaxation of the clustering tension, which might indicate \lcdm\ bias in \ac{rsd} data. A phenomenological parametrization, where a scalar field depends linearly on the number of e-folds, was used to test its interaction with \ac{dm} \cite{daFonseca:2021imp}. However, the constrained interacting dark energy (CIDER) model, with a background identical to the standard \lcdm\ model, appears to better address the $\sigma_8$ tension \cite{Barros:2018efl,Barros:2022bdv}.

\subparagraph{Late-time singularities}

Depending on the choice for the interacting term $Q$, late-time singularities, for which infinities in the Hubble factor (or on its derivatives) appear, may arise at finite future times \cite{Trivedi:2023zlf,deHaro:2023lbq,Nojiri:2005sx,Nojiri:2017ncd,Nojiri:2005sr,Bamba:2012cp,Bamba:2008ut,Nojiri:2004ip}. Since some of them represent possible ends for the Universe, this undermines the consistency of the theory. Furthermore, they have been shown to occur for both linear and non-linear forms of $Q$.

\paragraph{Modelling and Observational Aspects of IDE Models}
In this section, we present some other aspects of \ac{ide} models, such as $N$-body structure formation simulations of certain \ac{ide} models, and comparing \ac{ide} models to data.

\subparagraph{Marginalization approach to IDE}

A recent estimate of \ac{ide} contribution from Pantheon and Pantheon Plus, \ac{cc} and transversal \ac{bao} \cite{Benisty:2024lmj} has shown evidence for up to $2\sigma$ deviation from \lcdm. This study has been done in the framework of the marginalization approach which aims to circumvent the Hubble tension by integrating $H_0$ and $r_d$ out of the model, leaving only $\Omega_m$, $w$ and $\xi$ as parameters. The results show a strong dependence on the used \ac{sn1} dataset and whether it is Cepheids calibrated or not, with the latter remaining compatible with \lcdm. 

\subparagraph{$N$-body simulations of IDE}

Several simulations of the structure formation were run up to date under the \ac{ide} ansatz, namely works of Refs.~\cite{Zhang:2018glx,Zhao:2022ycr}, where they have considered several cases of \ac{ide} and derived such probes of the \ac{lss} as matter power spectrum, halo mass function, concentration-mass relation, and halo bias. The most interesting parameter is the concentration-mass relation, namely $c_{200}$ that can be related to the X-ray data and galaxy rotation curves, which via the maximum likelihood method imposes the redshift evolving constraints on the \ac{de}-\ac{dm} coupling parameter $\xi$, with the joint posterior distribution giving a best fit of $\xi= 0.071 \pm 0.034$.

\subparagraph{Distinguishing IDE from modified gravity}

A word of caution is required when constraining \ac{ide} models based on large-scale structure observables. Models involving a fifth force acting on \ac{dm} lead to a breaking of the weak equivalence principle for this component and thus a deviation in the Euler equation, affecting the growth of cosmic structure $f$. However, the resulting observational signatures on current and forecasted measurements of $f \sigma_8$ present a strong degeneracy with gravity modifications in the Poisson equation \cite{Castello:2022uuu, Castello:2023zjr}. \ac{wl}, which is sensitive to the sum of the two gravitational potentials describing the geometry of the Universe, is also generically unable to break this degeneracy. This indicates that standard large-scale structure analyses may not be able to distinguish \ac{ide} from gravity modifications, leading to a potential ambiguity among models proposed to address cosmological tensions.  

Luckily, the two scenarios can be disentangled by considering measurements of gravitational redshift \cite{Bonvin:2023ghp, Castello:2024jmq}, an observable accessible in two-point correlations by the coming generation of galaxy surveys \cite{Sobral-Blanco:2021cks, Sobral-Blanco:2022oel}. This effect is sensitive to the gravitational potential encoding the distortion of time and appearing in the Euler equation, thus providing a direct test of the weak equivalence principle and a clear way to discriminate between the two scenarios.

\subparagraph{Model-Independent Reconstructions of IDE Kernel}  

In Ref.~\cite{Escamilla:2023shf}, the interaction kernel $Q(z)$, which governs the energy exchange between \ac{dm} and \ac{de}, is reconstructed within a fully model-independent framework, without imposing any predefined functional form. Employing both binned step-function reconstructions and \ac{gp} interpolation, the authors infer the redshift evolution of $Q(z)$ directly from observational data, including \ac{bao}, \ac{cc}, and the Pantheon+ supernova sample. The reconstructed kernel displays nontrivial features—most notably, multiple sign changes and mild oscillatory behavior at the $1\sigma$ level—suggesting dynamic and potentially reversible energy transfer between the dark components. As a result of this evolving interaction, the effective \ac{de} density is found to cross zero and become negative at redshifts $z \gtrsim 2$, demonstrating that \ac{ide} scenarios do not necessarily exclude negative DE densities at high redshift (see Sec.~\ref{sec:Grad_DE}), despite the conventional assumption that DE remains positive in \ac{ide} models.

\subparagraph{Application of the parametrized post-Friedmann framework in IDE}

Understanding the interaction of \ac{de} necessitates examining both the expansion history of the Universe and the growth of cosmic structures \cite{Li:2015vla, Feng:2019mym}. It has been observed that in the \ac{ide} scenario, calculating perturbations often results in divergent curvature perturbations on superhorizon scales, posing a significant problem for \ac{ide} cosmology \cite{Li:2013bya, Feng:2016djj, Zhao:2017urm}.

To address this issue, the parametrized post-Friedmann (PPF) framework is an effective theory that treats \ac{de} perturbations based solely on the fundamental properties of \ac{de} \cite{Li:2014cee, Li:2014eha, Guo:2017hea, Zhang:2017ize, Feng:2018yew}. This approach extends the parametrized PPF framework, originally designed for uncoupled \ac{de}, and can mitigate the instability in \ac{ide} cosmology \cite{Li:2023fdk}. Consequently, the entire parameter space of \ac{ide} models can be explored using observational data \cite{Guo:2017deu, Li:2017usw, Guo:2018ans, Feng:2019jqa, Gao:2021xnk, Zhang:2021yof,Li:2024qso}. 

A generalized framework to study \ac{ide} models is also provided by the effective theory of \ac{ide} \cite{Gleyzes:2015pma}. This formalism relies on a model-independent approach at the level of the action and provides a description of linear cosmological perturbations in scalar-tensor theories of gravity with a homogeneous and isotropic background. This allows for a direct comparison with cosmological data to constrain a rich non-standard phenomenology \cite{Gleyzes:2015rua, Castello:2023zjr}.

\bigskip
\subsection{Modified gravity}
\subsubsection{Modified gravity in light of cosmic tensions \label{sec:MoG}}

\noindent \textbf{Coordinator:} Francesco Bajardi, Micol Benetti, Salvatore Capozziello\\
\noindent \textbf{Contributors:} Adri\`a G\'omez-Valent, Alessandro Vadalà, Ali \"Ovg\"un, Amare Abebe, Andronikos Paliathanasis, Anil Kumar Yadav, Araceli Soler Oficial, Christian Pfeifer, Daniel Blixt, David Benisty, Diego Rubiera-Garcia, Du\v{s}ko Borka, Emmanuel N. Saridakis, Erik Jensko, Francisco S. N. Lobo, Gabriel Farrugia, Gaetano Lambiase, Giovanni Montani, Giuseppe Sarracino, Hussain Gohar, Ilim Cimdiker, In\^es S.~Albuquerque, Ismael Ayuso, Joan Sol\`a Peracaula, Konstantinos F. Dialektopoulos, L\'aszl\'o  \'A. Gergely, Marcin Postolak, Marco de Cesare, Maria Caruana, Mariaveronica De Angelis, Nihan Kat{\i}rc{\i}, Nils A. Nilsson, Noemi Frusciante, Pierros Ntelis, Predrag Jovanovi\'{c}, Rebecca Briffa, Rocco D'Agostino, Saeed Rastgoo, Sanjay Mandal, Sergei D. Odintsov, Tiago B. Gon\c{c}alves, Tiziano Schiavone, Vesna Borka Jovanovi\'{c}, and Vincenzo Salzano
\\

\noindent The shortcomings exhibited by \ac{gr} on different energy scales question its validity as the best theory to describe the gravitational interaction \cite{Will:2014kxa}. Indeed, at the quantum level it shows problems with its extension to a quantum theory of gravity \cite{Addazi:2021xuf,AlvesBatista:2023wqm}, since it is non-renormalizable \cite{Goroff:1985th}, and the prediction of infinite gravitational tidal forces, by the existence of singularities \cite{Barack:2018yly}. In the large-scale scenario, the observed Universe's accelerating expansion \cite{Riess:2021jrx,HST:2000azd} cannot be predicted within the context of the standard model of cosmology without introducing the cosmological constant, whose theoretical value assessed by Quantum Field Theory differs from the value inferred by the Friedman equations \cite{Weinberg:1988cp,Carroll:2000fy}. The repulsive force driving the Universe expansion is dubbed ``Dark Energy'', whose existence, to date, is only based on indirect observations \cite{SupernovaSearchTeam:1998fmf,SupernovaCosmologyProject:1998vns}. In addition, on galactic scales, the dynamics of the farthest stars orbiting around the center of galaxies (and, generally, incompatibilities on the \ac{lss}) represents a further open issue \cite{Davis:1985rj, Bertone:2016nfn} currently addressed to a never-detected form of matter called ``Dark Matter'' \cite{Aprile:2009dv,Misiaszek:2023sxe}.  
 
These incompatibilities therefore open up the possibility of new (or rather, extended) scenarios, and here we will examine possible modifications to gravity.

Among the most important and widely considered approaches to address the aforementioned problems related to the unknown nature of \ac{de} and \ac{dm}, as well as cosmological tensions,  are modifications to \ac{gr} itself. These can broadly be split into two main categories, whose features will be considered in detail in the next sections. The first category, referred to as {\it extended theories of gravity}, includes extensions of the Einstein-Hilbert action by other curvature terms (see Ref.~\cite{Bajardi:2022ypn} for a review on the topic). The second category, referred to as {\it alternative theories of gravity} consists of models modifying the basic principles of \ac{gr}, such as the Equivalence principle or Lorentz invariance. The main purpose of modified theories of gravity is to address \ac{gr} shortcomings on different scales, providing a different view to describe gravity, potentially able to fix issues related to cosmology \cite{CANTATA:2021ktz} and the quantum formalism of the gravitational interaction \cite{Addazi:2021xuf}. Extended and alternative theories of gravity naturally introduce extra degrees of freedom with respect to \ac{gr}, which might potentially accommodate cosmic tensions. Several promising \ac{mg} models involve modifications to the Einstein-Hilbert gravitational action, such as incorporating higher-order curvature invariants \cite{Stelle:1976gc, Bajardi:2020mdp, Bajardi:2021hya}, establishing connections between geometry and scalar fields \cite{Halliwell:1986ja, Uzan:1999ch, Clifton:2011jh, Urban:2020lfk}, or introducing additional geometric features to spacetime beyond curvature, such as torsion (Poincar\'e gauge gravity and teleparallel gravity) \cite{Krssak:2018ywd,Bahamonde:2021gfp} or non-metricity (metric-affine gravity and symmetric teleparallel gravity) \cite{JimenezCano:2021rlu, Ayuso:2020dcu}, or non-linear connections (Finsler gravity) \cite{Hohmann:2021zbt,Pfeifer:2019wus}. Different classes of \ac{mg} theories are outlined in Fig.~\ref{fig:modgrav}.

\begin{figure}[!t]
\centering
\includegraphics[width=0.9\textwidth]{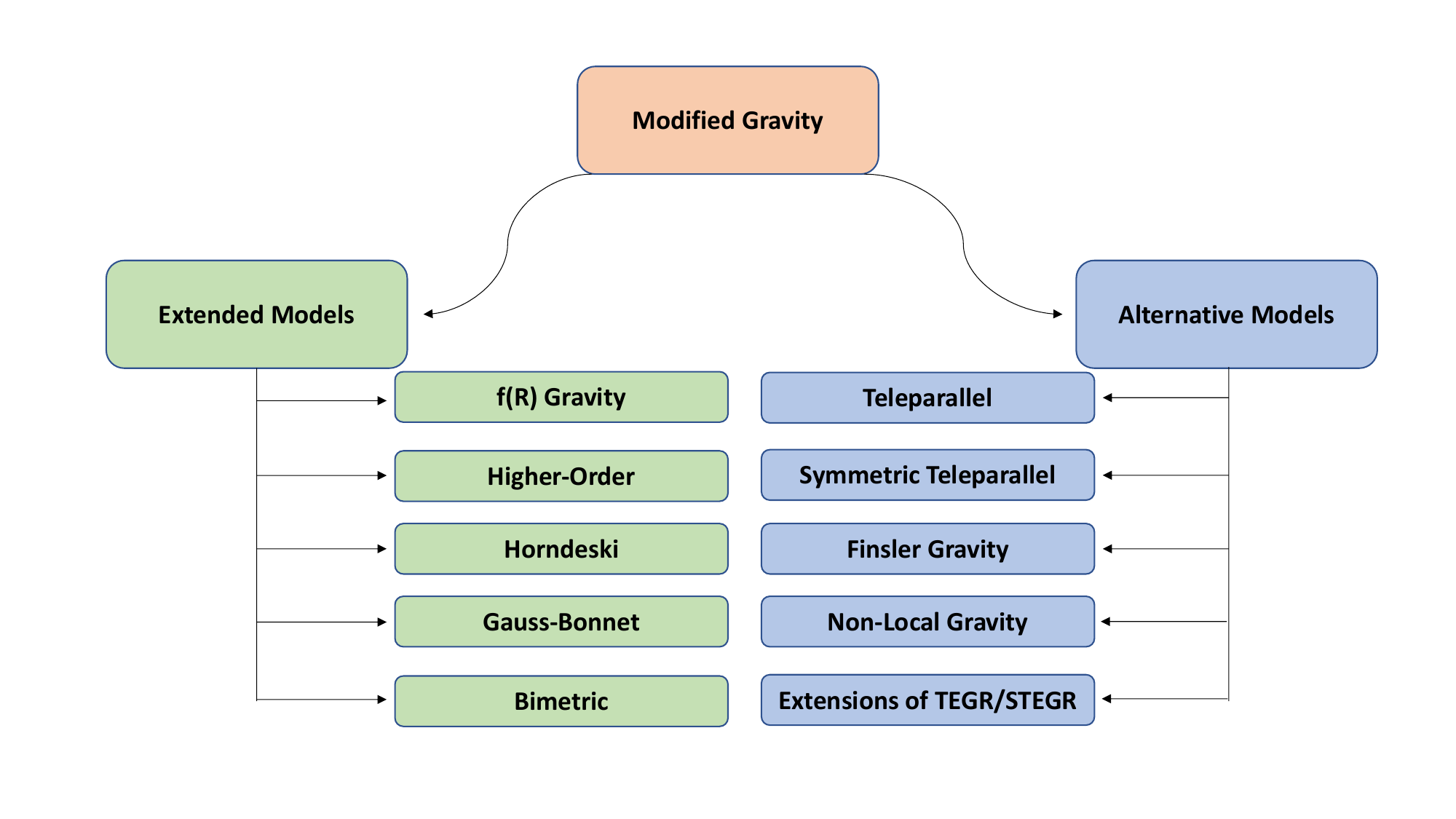}
\caption{A summary of the \ac{mg} roadmap, outlining the potential extensions of \ac{gr}.}
\label{fig:modgrav}
\end{figure}

\paragraph{Extended theories of gravity}

One of the famous examples of \ac{gr} extensions is the $f(R)$ gravity models, where the Ricci scalar $R$ in the Einstein-Hilbert action is replaced by a general function $f(R)$. This modification leads to fourth-order field equations, which can explain cosmic acceleration without invoking \ac{de} \cite{Barrow:1988xh,Sotiriou:2008rp, DeFelice:2010aj, Capozziello:2011et, Starobinsky:2007hu, Paliathanasis:2011jq, Nojiri:2006gh, Papagiannopoulos:2018mez, Bajardi:2022ocw, Nojiri:2010wj, Nojiri:2017ncd}. Various formulations of this theory have been explored, some of which can produce deviations in the Newtonian potential \cite{Capozziello:2021goa}, potentially addressing phenomena like the galaxy rotation curves without resorting to \ac{dm}. Notably, the Starobinsky model \cite{Starobinsky:1980te}, which introduces a quadratic term in the scalar curvature to explain cosmic inflation, has attracted significant attention within this framework. Extra degrees of freedom in $f(R)$ gravity affect the Hubble function $H(z)$, providing deviations from the \lcdm\ paradigm. In the equivalent scalar-tensor representation in the Jordan frame of $f(R)$ gravity, in Ref.~\cite{Schiavone:2022wvq} it is shown that the scalar field dynamics may lead to a definition of an effective Hubble constant $H_0^{\textrm{eff}}(z)$ that depends on the redshift $z$. In this regard, the non-minimal coupling between the scalar field and the metric plays a crucial role. Notably, $H_0^{\textrm{eff}}(z)$ may successfully address the Hubble tension since it could reconcile measurements and estimates of $H_0$ obtained at different redshifts, from local probes at $0\lesssim z\lesssim2$ to \ac{cmb} at $z\sim 1100$. Accordingly, the general scalar field potential is computed in Ref.~\cite{Schiavone:2022wvq}, as well as the respective $f(R)$ functional form in the low-redshift regime within $f(R)$ quadratic gravity. Furthermore, Ref.~\cite{Montani:2023xpd,Montani:2024xys} shows that a rescaling of the Universe’s expansion rate can emerge without presupposing a specific dependence on the redshift (as done in Ref.~\cite{Schiavone:2022wvq}). Indeed, it is just the non-minimally coupled scalar field of the scalar-tensor representation responsible for the Hubble constant scaling. Then, the additional \ac{dde} source was implemented in the $f(R)$ gravity to provide a smooth transition between a quintessence regime and phantom fluids, resulting in an effective Hubble constant that exhibits a plateau behavior for $z\gtrsim 5$ matching the Planck prediction.  For equivalent approaches, see also Refs.~\cite{Montani:2024ntj,Montani:2023ywn,Montani:2024pou,BarrosoVarela:2024ozs}. These scenarios with $H_0^{\textrm{eff}}(z)$ were motivated by binned analysis of the \ac{sn1} Pantheon sample and \ac{bao}s \cite{Dainotti:2021pqg,Dainotti:2022bzg,Dainotti:2025qxz}, pointing out a slowly decreasing trend of the Hubble constant. In Ref.~\cite{Nojiri:2022ski}, by exploiting the $f(R)$ gravity correspondence with a scalar-tensor theory, the authors provide a condition in which the $H_0$ tension is alleviated. Specifically, this condition is based on the existence of a metastable de Sitter point that occurs for redshifts near the recombination.

The Einstein-Hilbert action could be extended also by introducing dynamical scalar fields non-minimally coupled with the geometry \cite{Ayuso:2014kda}. Scalar-tensor theories are often taken into account to address the dynamics of the early Universe, in the framework of the inflationary paradigm  \cite{Barrow:1995fj}. In fact, inflation is usually conceived as generated by a scalar field $\phi$, called the inflaton, which is supposed to be responsible for the accelerating expansion of the early Universe. According to this picture, inflation should be led by some scalar field driving the cosmic acceleration between $10^{-34}$ and $10^{-35}$ s after the Big Bang, generating an isotropic and homogeneous Universe. The theory of inflation describes also the production of particles after the early-time accelerating expansion (reheating) \cite{Shtanov:1994ce, Allahverdi:2010xz, Greene:1999hw}. Moreover, to tackle the cosmological constant problem or the evolution of cosmological vacuum energy, new degrees of freedom for the gravitational field have to be considered.

In this scenario, Horndeski theory represents the most general scalar-tensor theory with second-order field equations in four dimensions \cite{Horndeski:1974wa,Kobayashi:2019hrl}. It encompasses a wide range of models and offers a comprehensive framework for modifying the gravitational interaction by incorporating an extra degree of freedom through a scalar field \cite{Deffayet:2011gz,Kase:2018aps,DAgostino:2019hvh}. This results in significant and intriguing consequences for both inflation and \ac{de} physics. The detection of the GW170817 event together with the electromagnetic counterpart \cite{LIGOScientific:2017vwq} provides one of the most stringent constraints on scalar-tensor theories that propose an unusual speed of \ac{gw}s. This imposes severe limits on Galileons and extends to other scalar-tensor theories, including the quartic and quintic sectors of Horndeski's theory \cite{Ezquiaga:2017ekz,Bonilla:2019mbm,Saltas:2022ybg,Babichev:2024kfo}. Nevertheless, it has been noted that since the energy scales observed at \ac{ligo} lie very close to the typical cutoff of these types of theories, the translation of the bound on the speed of \ac{gw}s to a cosmological setting might be non-trivial \cite{deRham:2018red}. 

Scalar-Tensor models have been taken into account also to address the $H_0$ tension problem. For instance, in Ref.~\cite{Ballardini:2020iws} Planck 2018 data are used to constrain the simplest models of scalar-tensor theories, accommodating a higher value for $H_0$ and therefore alleviating the tension between Planck/\ac{bao} and distance-ladder measurement from \ac{sn1} data. The capability of scalar-tensor and bi-scalar–tensor modified theories to alleviate the tension is also considered in Refs.~\cite{Petronikolou:2021shp,Petronikolou:2023cwu}, and other works \cite{Ballardini:2023mzm,Ferrari:2023qnh,Ferrari:2025egk}. 

Even when limiting the analysis to Horndeski models that respect the \ac{gw}s' speed bound, observational constraints on $H_0$ using Planck data alone have been shown to yield values that are consistent with local determinations at $2 \sigma$ for the Galileon Ghost Condensate \cite{Peirone:2019aua} and $1 \sigma$ for the Generalized Cubic Covariant Galileon \cite{Frusciante:2019puu, Dwivedi:2024okk}. More recent studies have also considered Horndeski gravity to specifically address cosmic tensions, showing that particular setups free of ghost instabilities could lead to interesting late-time features possibly alleviating the $H_0$ tension \cite{Banerjee:2022ynv,Tiwari:2023jle}. 

Cosmic tensions have also been addressed in the context of the time-honored  Brans-Dicke (BD) theory \cite{Brans:1961sx,Dicke:1961gz}, which is the prototype of the large family of scalar-tensor theories (see also Ref.~\cite{Clifton:2011jh}). BD gravity embodies the first historical attempt at incorporating a dynamical gravitational coupling to \ac{gr}. It also constitutes the simplest scalar-tensor theory which can be embedded within the much broader class of Horndeski models \cite{Horndeski:1974wa} and, in fact, many of these models reduce to BD at cosmological scales \cite{Avilez:2013dxa}. Apart from the metric tensor, BD gravity involves a new (scalar) degree of freedom, $\varphi$, which controls the strength (and the possible time evolution) of the gravitational interaction $G_{\varphi}=G/\varphi$, with $G$ being the (locally measured) Newton constant transforming the gravitational coupling constant into a new dynamic variable \cite{Ayuso:2019bnw}. The scalar field $\varphi$ is non-minimally coupled to gravity via the Ricci scalar and its dynamics is controlled by a dimensionless parameter $\omega_{\rm BD}$, the BD parameter (or equivalently $\epsilon_{\rm BD}\equiv 1/\omega_{\rm BD}$).

Let us denote by BD-\lcdm\ the Brans-Dicke counterpart of the standard \lcdm, which represents the BD model with a cosmological constant, $\Lambda$. The cosmological term $\Lambda$ is actually not present in the original BD theory, but in modern times $\Lambda$ is obviously needed to trigger the late-time observed acceleration of the Universe. The BD-\lcdm\ extension of the original BD model has been recently tested in the light of a large set of cosmological data of various sorts \cite{SolaPeracaula:2020vpg}, such as \ac{sn1}+\ac{bao}+$H(z)$+\ac{cmb}+\ac{lss}, where $H(z)$ may include or not the SH0ES prior on $H_0$. If the latter is included, one finds \cite{SolaPeracaula:2020vpg} that the BD-\lcdm\ model is remarkably successful in reducing the $H_0$ and $\sigma_8$ tensions to inconspicuous levels of $\sim 1.5 \sigma$ (when the high-$l$ polarization and lensing data from Planck 2018 are not considered) \cite{SolaPeracaula:2020vpg}. Notice that The \ac{rvm} is mimicked by the BD-\lcdm\ model, particularly the type-II \ac{rvm} version with variable $G_{\varphi}$  \cite{SolaPeracaula:2021gxi,SolaPeracaula:2023swx,deCruzPerez:2023wzd,deCruzPerez:2018cjx}. The BD-\lcdm\ turns out to fit the cosmological data significantly better than the standard \lcdm\ model, and this is confirmed by different statistical criteria (see Refs.~\cite{SolaPeracaula:2019zsl, SolaPeracaula:2020vpg, SolaPeracaula:2021gxi, deCruzPerez:2023wzd} for details). On the other hand, adding a constant potential for the scalar field does not correspond to the usual vacuum energy in BD theory. An explicit detailed theoretical and observational investigation of this BD extension of \lcdm\ model, studied in Ref.~\cite{Akarsu:2019pvi}, reveals that the model does not alleviate the tensions with no significant deviations from \lcdm.

Values of the effective cosmological gravitational coupling $G_{\varphi}$ about $\sim 7-9\%$ larger than $G$ are preferred at $\sim 3\sigma$ CL This is possible thanks to the fact that $\varphi$ remains below 1 throughout the entire cosmic evolution, while preserving the matter and radiation energy densities very close to the typical \lcdm\ values. As a result, one finds higher $H(z)$ values during all the stages of the cosmic history without changing dramatically the abundances of the matter species in the pre- and post-recombination epochs. The lowering of the sound horizon at the baryon-drag epoch, $r_d$, is accompanied by an increase in the Hubble parameter. This helps to alleviate the $H_0$ tension in good agreement with other relevant background datasets, such as e.g., \ac{bao} and \ac{sn1}. In order not to spoil the correct fit to the \ac{cmb} temperature data, the model is prone to yield values of the spectral index $n_s$ that are considerably larger (viz. closer to $1$ from below) than the standard \lcdm\ model. This causes, however,  no problem since the dynamics of $\varphi$ can compensate for the changes in the matter power spectrum introduced by this fact (cf. Sec. 3 of Ref.~\cite{SolaPeracaula:2020vpg}). Remarkably, the model is able to cut back the $\sigma_8$ tension at the same time. The relief of this tension can be accomplished by means of a negative value of $\epsilon_{\rm BD}$, or alternatively through a positive value, provided that sufficiently massive (light) neutrinos are allowed \cite{SolaPeracaula:2020vpg}. In both cases one finds $|\epsilon_{\rm BD}|\lesssim\mathcal{O}(10^{-3})$. The BD parameter has a direct impact on the \ac{lss} data through the linear perturbation equations. For instance,  for $\epsilon_{\rm BD}<0$ the friction term in the equation of the matter density contrast grows, and the source term decreases. Both effects contribute in sync to the lowering of $\sigma_8$, thus providing a physical explanation for a possible solution to the growth tension. It goes without saying that the compatibility between the larger cosmological value obtained for the gravitational coupling and the value measured locally, $G$, should be possible provided one can find an appropriate screening mechanism. Different studies show that these mechanisms are possible, although other works find that their implementation may not be so straightforward \cite{Gomez-Valent:2021joz}.

Another way to extend the \ac{gr} action is to introduce higher-order curvature invariants \cite{Barrow:1989nb,Cotsakis:1993zz,Cotsakis:1993ezo,Carloni:2018yoz,Amendola:1993bg,Cuzinatto:2018chu,Gottlober:1989ww,Paliathanasis:2024ozx,Wands:1993uu,Berkin:1990nu,Bajardi:2023shq}. This is the case of Gauss-Bonnet gravity, where the action includes a specific combination of curvature invariants known as the Gauss-Bonnet term. In this context, particularly intriguing are theories incorporating quadratic curvature terms, as they stem from \ac{gr} viewed as an Effective Field Theory aiming to construct effective models towards quantum gravity. Notably, certain combinations of higher-order contractions of the Riemann tensor yield a topological surface term in four dimensions, known as the Gauss-Bonnet topological scalar, $\mathcal{G}$. In four dimensions, this scalar equates to the Euler density, which, according to the generalized Gauss-Bonnet theorem, provides the Euler characteristic when integrated over the manifold. The Gauss-Bonnet invariant is often utilized for its topological properties to simplify dynamics. However, in four dimensions, to render its contribution non-trivial, it is typically coupled to a dynamical scalar field \cite{Carter:2005fu, Bajardi:2020xfj,Millano:2023czt,Millano:2023gkt,SantosDaCosta:2018bbw} or incorporated as a function in the gravitational action \cite{Li:2007jm,Cognola:2006eg, Nojiri:2005jg, Nojiri:2005vv}. This approach, where a function $f(\mathcal{G})$ is added to the scalar curvature, can mimic the behavior of a cosmological constant at late-times and, in the limit $f(\mathcal{G}) = 0$, \ac{gr} is restored. Another avenue involves considering higher dimensions where $\mathcal{G}$ is non-trivial and coupling the Gauss-Bonnet invariant to a constant diverging when $D = 4$. 
Gauss-Bonnet gravity has been studied to address the $H_0$-tension \cite{Wang:2021kuw, Benetti:2018zhv}, finding that can be greatly resolved within $2 \sigma$ level.

Bimetric gravity \cite{Hassan:2011zd} is a consistent theory of non-linearly interacting spin-2 fields, one massless and one massive, which allows for a broad range of cosmological expansion histories and is compatible with cosmological as well as local tests of gravity \cite{Hogas:2021fmr,Hogas:2021lns,Hogas:2022owf}. In its most general form, the theory features four additional parameters in addition to the standard \lcdm\ ones. There are no free functions, unlike several other \ac{mg} theories. Studies based on a restricted class of two-parameter models have shown that the value of $H_0$ inferred from the inverse distance ladder is only slightly increased compared to \lcdm\ \cite{Mortsell:2018mfj}; however, such models do not comply with the constraints required for a working screening mechanism \cite{Hogas:2021lns}. A more recent analysis \cite{Dwivedi:2024okk} based on more general three-parameter models has shown that, using (model-independent) transverse 2D \ac{bao} data in combination with \ac{sn} and \ac{cmb}, results in a value of $H_0$ which is closer to the SH0ES value. On the other hand, using 3D \ac{bao} data produces a lower $H_0$ value, suggesting a potential bias in the latter. A study of the $S_8$ tension has not been performed at present, since a framework for structure formation in bimetric gravity has not yet been developed.

Functors of actions theories (FAT), is part of the extensions of gravity theories, and they predict the existence of \textit{actionions}, i.e., the actionic fluctuations and field-particles, which are analogs of energetic fluctuations and field particles. In light of deviations from \ac{de} equations of state, $w_0 \simeq -1.1$, FAT predicts the existence of actionic fluctuations which are of the order of one tenth of the observed volume \cite{Ntelis:2020pak,Ntelis:2024fzh}.

\paragraph{Alternative theories of gravity}

The second class of modifications revises the basic foundations of \ac{gr}, such as the Equivalence Principle, metric compatibility, Lorentz invariance, \emph{etc.}. One such approach involves considering a more general geometric framework with an affine connection different from the Levi-Civita one, thereby introducing both torsion and non-metricity into spacetime. Torsion is associated with the antisymmetric part of the connection, while non-metricity arises from the non-vanishing covariant derivative of the metric tensor. Within this framework, the Riemann tensor and its contractions in \ac{gr} can be expressed in terms of torsion, non-metricity, or a combination of both. The gravity formulations characterized by curvature, torsion, and non-metricity are dynamically equivalent, differing only by a boundary term in their respective actions, reason for which they are often termed the ``geometric trinity of gravity'' \cite{BeltranJimenez:2019esp}. This classical equivalence involving boundary terms that do not affect the field equations (and are thus omitted from the action formulation), is viewed from the perspective of \ac{gr} as motivated by the possibility of formulating \ac{gr} as a gauge theory \cite{Capozziello:2022zzh,Krssak:2018ywd,Bahamonde:2021gfp,Hohmann:2017duq}.

In this framework, a consistent gravitational theory can be formulated by promoting torsion over curvature as the sole governing factor of spacetime, resulting in a theory that precisely mirrors the dynamics of \ac{gr}. This theory, known as the ``Teleparallel Equivalent of General Relativity'' (TEGR) \cite{Maluf:2013gaa}, has garnered significant attention in recent years, undergoing extensive analysis \cite{Xu:2012jf, Krssak:2018ywd, Bahamonde:2021gfp, Obukhov:2002tm, Geng:2011ka,Hohmann:2017duq}. TEGR presents a theoretical framework that interprets gravity as a consequence of torsion within the fabric of spacetime. In this context, gravitational interactions are described through a set of tetrad fields, also referred to as ``vierbeins'', which form the basis for depicting spacetime geometry. These tetrad fields define a torsion tensor, serving as the source of gravity in the theory and representing the antisymmetric component of the Christoffel connection.

An extensively studied model alongside TEGR is the ``Symmetric Teleparallel Equivalent of General Relativity'' (STEGR), which describes spacetime through non-metricity. Non-metricity allows for the consideration that spacetime may not adhere to the metric compatibility condition, a fundamental assumption in \ac{gr}. While torsion arises directly from the antisymmetry of the affine connection, non-metricity emerges when the covariant derivative of the metric tensor is non-zero, denoted as $\nabla_{\sigma} g_{\mu \nu} \neq 0$.
As consisting of field equations entirely equivalent to those of \ac{gr},  both TEGR and STEGR fall short of addressing the limitations \ac{gr} imposes on larger scales. Consequently, similar to $f(R)$ gravity within the metric formalism, modifications to the Lagrangian density of TEGR can be explored in various ways \cite{Bajardi:2021tul}, including introducing new torsion \cite{Hayashi:1979qx,Bahamonde:2017wwk} and non-metricity invariants \cite{Jarv:2018bgs}, coupling these to additional scalar or axion fields \cite{Hohmann:2020dgy} or introducing an arbitrary function of the torsion scalar, leading to $f(T)$ gravity \cite{Cai:2015emx, Li:2010cg}. The latter has been proposed as a potential solution to late-time cosmological issues, such as the Universe's accelerated expansion \cite{Ferraro:2006jd, Linder:2010py}. The characteristics of gravity theories including torsion and non-metricity, such as $f(T)$, $f(Q)$ (with $T$ being the torsion scalar and $Q$ the non-metricity scalar) gravity, new general relativity and many more, are currently under investigation, particularly concerning their applications in cosmology and astrophysics \cite{Chen:2010va, Bamba:2010wb, Kofinas:2014owa,Paliathanasis:2016vsw,Finch:2018gkh,Farrugia:2018gyz,Soudi:2018dhv,Farrugia:2020fcu,Dimakis:2023oje, Capozziello:2022wgl, Albuquerque:2022eac, Anagnostopoulos:2022gej, Capozziello:2022tvv, Bajardi:2020fxh, Banerjee:2021mqk, DAgostino:2022tdk,Hohmann:2022wrk,Khyllep:2021pcu,Capozziello:2023giq,Paliathanasis:2023nkb,Paliathanasis:2021uvd,Yang:2024kdo,Paliathanasis:2024yea,Wu:2024vcr,2016APS..APR.L1049C}. Specifically, in the context of cosmological tensions, Refs.~\cite{Cai:2019bdh,Ren:2022aeo,Aljaf:2022fbk} explores methods to address the $H_0$ tension within $f(T)$ models; in Refs.~\cite{Yan:2019gbw,Ren:2021tfi} it is shown how to alleviate both the $H_0$ and $\sigma_8$ tensions simultaneously within torsional gravity from the perspective of effective field theory;  in Ref.~\cite{Nunes:2018xbm} the evolution of scalar perturbations in $f(T)$ gravity and its effects on the \ac{cmb} anisotropy is evaluated to show that $f(T)$ models do not provide tension on the Hubble constant that prevails in the \lcdm\ cosmology; in Ref.~\cite{DAgostino:2020dhv} the authors demonstrate that the $f(T)$ models prefer a higher value of $H_0$ with respect to the Planck prediction, in better agreement with local estimates. Similar effects can be achieved in some classes of the beyond-generalized Proca model that, at the background level, resembles $f(T)$ theory \cite{deRham:2021efp}. Finally, in Refs.~\cite{Barros:2020bgg,Wang:2024eai,Sakr:2024eee} solutions to the tensions were proposed in the framework of $f(Q)$ gravity. 

In another class of these models, Finsler gravity \cite{Pfeifer:2011xi,Hohmann:2021zbt,Basilakos:2013ij,Basilakos:2013hua,Papagiannopoulos:2017whb,Papagiannopoulos:2020mmm} offers an intriguing new way to derive the gravitational field of many particle systems, when they are modeled as a kinetic gas, instead of as a (perfect) fluid \cite{Hohmann:2020yia,Hohmann:2019sni}. The dynamics of a kinetic gas is described by a 1-particle distribution function (1PDF) on the 1-particle phase space of spacetime, which takes the velocity distribution of the gas particles into account. Usually, when the gravitational field of a kinetic gas is derived, one obtains the energy-momentum tensor of the gas by averaging the 1PDF over the velocity distribution of the particles. By means of this averaging procedure, the information about the velocity distribution of the gas particles are lost and the gravitational field is derived from an effective fluid energy momentum-tensor as a source term in the Einstein equations. Finsler spacetime geometry describes gravity in terms of a curved 1-particle phase space of a curved spacetime (technically the tangent or co-tangent bundle). Here, the 1PDF directly sources the gravitational field of the kinetic gas without losing information through velocity averaging. The contribution of the velocity (kinetic energy) distribution of the gas to its gravitational field is fully taken into account in Finslerian gravity, rather than being averaged away \cite{Hohmann:2020yia,Hohmann:2019sni}. The first promising solutions of Finsler gravity in the cosmological context have been obtained \cite{Heefer:2023tgf}, which show a linearly expanding Universe as a vacuum background solution. Further studies are ongoing to demonstrate that the contribution of the kinetic energy of a many-particle system to its gravitational field can be the solution to the Hubble tension and our understanding of \ac{de}.

It is worth pointing out that some alternative theories propose that gravity is mediated by a graviton with a small non-zero mass $m_g$. Experimental limits on $m_g$ are set using models such as the Yukawa potential and dispersion relations, tested on astrophysical and cosmological scales. In addition to detectors like \ac{ligo}/Virgo or \ac{lisa}, graviton mass limits can be derived from electromagnetic observations of gravitational systems, such as constraints on S-stars orbits \cite{Zakharov:2016lzv, Jovanovic:2022twh}, or from Schwarzschild precession of S2 in Yukawa gravity \cite{Jovanovic:2023tcc}, consistent with \ac{ligo} data.

Some theories of gravity relax the assumption of nonminimal coupling between the geometry and matter sources. 
This happens, for instance, in $f(R,\mathcal{L}_{\rm m})$ \cite{Bertolami:2007gv, Harko:2010mv}, $f(R,\mathcal{T})$ \cite{Harko:2011kv}, $f(R,T_{\mu\nu}T^{\mu\nu})$ \cite{Katirci:2013okf, Roshan:2016mbt,Akarsu:2017ohj,Board:2017ign}, and $f(R,\mathcal{T},R_{\mu\nu}T^{\mu \nu})$ \cite{Haghani:2013oma,Ayuso:2014jda} theories of gravity, where $\mathcal{L}_{\rm m}$ is the matter Lagrangian density, and $\mathcal{T}$ the trace of the energy-momentum tensor $T_{\mu\nu}$ (in its standard definition). Additionally, one can construct similar nonminimal coupled theories in the torsional framework, such as in $f(T,\mathcal{L}_{\rm m})$ \cite{Harko:2014sja,Carloni:2015lsa} and in $f(T,\mathcal{T})$ \cite{Harko:2014aja} theories of gravity, where $T$ is the torsion scalar.

A characteristic of these theories with nonminimal geometry-matter coupling is that the covariant divergence of the matter energy-momentum tensor can in general be nonzero (though the standard continuity equation can be recovered in particular models). A new avenue has been recently opened by modifying the introduction of the material source in the usual EH action, such as matter-type modified theories of gravity with $f(\mathcal{L}_{\rm m},\mathcal{T}, T_{\mu\nu} T^{\mu\nu})$, since these theories are equivalent to nonminimal interaction models in \ac{gr} \cite{Akarsu:2023lre}. Depending on the form of interaction determined by $f$ function, some nontrivial dynamics easing cosmological tensions \cite{Escamilla:2023shf}, difficult to achieve via simple interaction kernels can be achieved, these \ac{de} models have effects on early and late-times of the Universe and hence may address the current tensions within \lcdm\ model. The extra two degrees of freedom have the potential to offer more room to accommodate cosmic tensions. However, this also provides a challenge, given that Einstein-Boltzmann codes are commonly prepared to deal with at most one extra degree of freedom, it is, at the moment, more challenging to compute the evolution of perturbations and constrain these theories.

So far, studies have focused on the simplest classes of models.
An example of studying dynamics of scalar perturbations, using the quasistatic approximation, is in Ref.~\cite{Alvarenga:2013syu} for $f(R,\mathcal{T})= R + f_2(\mathcal{T})$ models with $f_2(\mathcal{T})\propto \mathcal{T}^{1/2}$ to ensure the usual matter continuity equation. This work finds that there is a strong scale $k$ dependence of the matter perturbations, not supported by observations.

A more recent work Ref.~\cite{Asghari:2024obf} reports on the possibility of $f(R,\mathcal{T})= R + \lambda \mathcal{T}$ alleviating the $\sigma_8$ tension, while increasing the $H_0$ tension.
In the Hu–Sawicki model of $f(R)$ gravity the $\sigma_8$-tension  observations worsen \cite{Lambiase:2018ows}, while it might be alleviated in viscosity in the \ac{dm} model (in modified cosmological models, massive neutrinos suppress the matter power spectrum on the small length scales, which implies that the bounds on neutrino mass get modified too)
\cite{Anand:2017ktp,Anand:2017wsj, Parashari:2021qjg, Mohanty:2018ame}.

To conclude, it is worth noticing that most alternative/extended theories of gravity may be subject to screening mechanisms, such as the \textit{chameleon} mechanism \cite{Khoury:2003aq,Khoury:2003rn}, which can be encountered in several scalar-tensor theories \cite{Li:2010re,Paliathanasis:2023ttu} as well as in \ac{mg} models \cite{Farajollahi:2011ym,Paliathanasis:2021nqa,Paliathanasis:2024sle,Paliathanasis:2024gwp,Paliathanasis:2020plf,Gurzadyan:2019yir,Gurzadyan:2021jrw,Gurzadyan:2021hgh,Gurzadyan:2023ttf,Gurzadyan:2023hmg,Gurzadyan:2025ekn,Gurzadyan:2025epn,Erdem:2024vsr} that can be recast in terms of an additional scalar field (e.g., see Ref.~\cite{Borowiec:2023kmq}), like $f(R)$ gravity. In this case, the scalar field is coupled to the matter fields so that its mass is environment-dependent. As long as the matter density is high enough, the scalar field acquires a heavy mass around the potential minimum, strongly reducing the range of the fifth force it mediates, and making it essentially unobservable, see however \cite{Vagnozzi:2021quy, OShea:2024jjw}.

On the contrary, on cosmological scales, far from regions with higher densities, the scalar field has a lighter mass, effectively mediating the additional gravitational interaction. Another screening mechanism is the \textit{symmetron} screening \cite{Hinterbichler:2010es}, where the potential of the scalar field $V(\phi)$ breaks the parity symmetry in low-density regions, which instead is restored at smaller scales. In this mechanism, the scalar field has a vacuum expectation value (VEV) that depends on local matter density, becoming larger in low-density regions and smaller in high-density regions. Moreover, the coupling of the scalar field to matter is proportional to the VEV, so that the scalar couples with gravitational strength in the low-mass-density regions, while it is decoupled, and then screened, in the high-density ones. In Ref.~\cite{Hogas:2023vim} it has been concluded that symmetron screening cannot be used to alleviate the Hubble tension.

Finally, the Vainshtein mechanism \cite{Babichev:2013usa} is important for some \ac{mg} models, such as massive gravity, Dvali–Gabadadze–Porrati (DGP), or Galileon theory. The Vainshtein mechanism manifests itself at non-linear scales, below a certain radius known as the Vainshtein radius. Nearby massive bodies \ac{gr} are recovered through a strong kinetic self-coupling, weakening the interaction with matter and reducing the propagation of the extra degree of freedom. Recently, it was shown that bimetric gravity can alleviate the $H_0$ tension (refer to Sec.~\ref{sec:LTP_4.2}), and could potentially alleviate also the $S_8$ tension \cite{Dwivedi:2024okk}.\footnote{However, it must be noted that drawing accurate predictions for the formation of \ac{lss} in bimetric gravity is challenging, so at present the $S_8$ tension can only be addressed under some simplifying assumptions. Regarding the $H_0$ tension the alleviation comes with the drawback that more parameters are introduced and bimetric gravity is preferred by AIC \cite{Akaike:1974vps} but \lcdm\ is preferred by BIC \cite{Schwarz:1978tpv} with the latter giving a stronger penalty for introducing more parameters to the model in consideration.}

\noindent According to the latest observations, one of the most promising extended gravity models is the generalized cubic covariant Galileon model. Using only \ac{cmb} data, provides $H_0 = 72^{+8}_{-5}$\kms, $\sigma_8 = 0.88^{+0.07}_{-0.05}$ at a 95\% confidence level \cite{Frusciante:2019puu} and $S_8 = 0.84 \pm 0.10$. At the same time, the tensions are restored by combining the \ac{cmb} data with external datasets.

In the framework of alternative theories, particularly encouraging is the $\Delta_4$ model, which is a specific non-local gravity model providing non-local quantum corrections to standard \ac{gr}. This model predicts the values $H_0 = 70.3 \pm 0.9$\kms and $\sigma_8 = 0.82 \pm 0.02$ \cite{Belgacem:2017cqo}, using the \ac{cmb}, \ac{bao} and \ac{sn} datasets, which is higher than Planck predictions, in better agreement with the local estimates. The related $S_8$ value is $S_8 = 0.81 \pm 0.02$.

All in all, although there is not a universally accepted \ac{mg} model that resolves the cosmological tensions, many of these models offer promising aspects that could help guide future research. Most of the models considered in this section provide a higher value for $H_0$, though implications on $\sigma_8$ should be further explored. As cosmological data continue to improve, these theories will be subject to increasingly stringent tests, potentially leading to a deeper understanding of the fundamental nature of gravity and the Universe. Therefore, it is quite possible that extended/alternative theories could contribute, at least in part, to addressing the $H_0$ and $\sigma_8$ tensions. 

\bigskip
\subsubsection{Early modifications to gravity \label{sec:Early_mod_grav}}

\noindent \textbf{Coordinator:} Emmanuel N. Saridakis, Konstantinos F. Dialektopoulos\\
\noindent \textbf{Contributors:} Andrzej Borowiec, Alfio Maurizio Bonanno, Ali \"Ovg\"un, Andreas Lymperis, Andreas Papatriantafyllou, Anil Kumar Yadav, Antonio Racioppi, Athanasios Bakopoulos, Athanasios Chatzistavrakidis, Branko Dragovic, Carlos G. Boiza, Charalampos Tzerefos, Chiara De Leo, Christian Pfeifer, Cláudio Gomes, Damianos Iosifidis, Daniele Oriti, David Benisty, Davide Pedrotti, Despoina Totolou, Diego Rubiera-Garcia, Elias Vagenas, Elisa Fazzari, Elvis Baraković, Flavio Bombacigno, Fotios K. Anagnostopoulos, Francisco S. N. Lobo, Gaetano Lambiase, Genly Leon, Gerasimos Kouniatalis, Giannis Papagiannopoulos, Giovanni Otalora, Giulia De Somma, Giuseppe Gaetano Luciano, Gonzalo J. Olmo, Jos\'e Pedro Mimoso, Jurgen Mifsud, Kathleen Sammut, L\'aszl\'o \'Arp\'ad Gergely, Laur Järv, Manolis Plionis, Manuel Gonzalez-Espinoza, Manuel Hohmann, Margus Saal, Maria Petronikolou, Miguel A. S. Pinto, Nikolaos E. Mavromatos, Nicholas Petropoulos, Oem Trivedi, \"Ozg\"ur Akarsu, Petros Asimakis, Saeed Rastgoo, Salvatore Capozziello, Saurya Das, Sebastian Bahamonde, Simony Santos da Costa, Spyros Basilakos, Stylianos A. Tsilioukas, Supriya Pan, Thanasis Karakasis, Tiago B. Gon\c{c}alves, Tomi Koivisto, Vasilios Zarikas, Vedad Pasic, Xin Ren, Yuejia Zhai, and Yu-Min Hu
\\

\ac{mg} is a quite general framework that can offer alleviation to the tensions, since apart from late-time solutions (which were analyzed in the previous subsection) one can apply it at early-times too \cite{DiValentino:2022uvj}. In order to solve the Hubble tension, it is possible to introduce modifications to the standard cosmological model. The general idea is that new assumptions at pre- and post-recombination are needed. For this reason, changing some of these assumptions, i.e., changing the cosmological model before or after the recombination epoch, may be the solution to obtain a different value of $H_0$. One possible solution is to consider a \ac{mg} scenario before the recombination epoch. Since the \ac{bao} data constrain $H r_d$, the final goal is to obtain a lower value of the sound horizon at the drag epoch $r_d$ and thus a higher value of $H_0$. Some constraints from models adhering to this approach are shown in Table~\ref{tab:E_MoG_sample}.

The expression of the sound horizon is $r_d=\int_{z_d}^\infty dz c_s(z)/H(z)$, where at sufficiently high redshift the sound speed $c_s(z)$ can be considered almost constant and equal to $c/\sqrt{3}$. To decrease the value of $r_d$ we need that, in a certain redshift interval $z\in(z_1,z_2)$, the expansion of the Universe is higher with respect to the \lcdm\ case in the same value. An example of this approach can be found in Ref.~\cite{Braglia:2020auw}, where it is shown that the Hubble tension can be reduced by assuming a \ac{mg} model acting at early-times with a scalar field non-minimally coupled to the Ricci scalar. Furthermore, in Ref.~\cite{Benevento:2022cql}, they explore the possibility of obtaining a lower value of the sound horizon, and consequently a higher value of $H_0$, by using a \ac{mg} model that allows for a phenomenological shift in the effective Planck mass on cosmological scales before recombination, through a non-minimal coupling of a scalar field with gravity.

In Refs.~\cite{Basilakos:2019acj,Mavromatos:2020kzj,Mavromatos:2021urx} it was proposed that the dynamics of the early Universe is described by a Chern-Simons (CS) gravitational model, characterized by gravitational anomalies, which arises in the low-energy limit of string theory~\cite{Duncan:1992vz,Svrcek:2006yi}, after appropriate compactification to (3+1) spacetime dimensions. The model involves the graviton and Kalb-Ramond (KR), or gravitational, axion field (assuming a constant dilaton). The latter couples to the anomalous terms. Chiral primordial \ac{gw} can condense under some circumstances, leading to a non-trivial condensate of the CS terms~\cite{Alexander:2004us,Lyth:2005jf,Dorlis:2024yqw}, and hence a linear axion potential, resembling the axion monodromy situation in string-theory models~\cite{McAllister:2008hb}. Such linear-axion potentials lead to inflation~\cite{Mavromatos:2022xdo,Dorlis:2024yqw}, which however are of the so-called \ac{rvm} type~\cite{Lima:2013dmf}. This means that the corresponding vacuum energy density contains dominant terms of order $H^4$ (where $H$ is the Hubble parameter, which is almost constant during inflation).

It is such non-linear terms that drive the \ac{rvm} inflation. In Ref.~\cite{Dorlis:2024yqw}, it was shown that the CS condensates are \emph{metastable}, in the sense that they contain imaginary parts~\cite{Mavromatos:2024pho}, which can determine the duration of the \ac{rvm} inflation. The model contains a pre-\ac{rvm}-inflationary era characterized by stiff-KR-axion-matter dominance. It is during such an epoch that chiral \ac{gw}s are formed, by either non-spherically-symmetric mergers of primordial (rotating) black holes, or collapses of domain walls, the latter arising, for instance, from dynamical breaking of supergravity~\cite{Alexandre:2013iva,Alexandre:2014lla}, which can characterize the early epochs of such string-inspired models after the Big Bang~\cite{Mavromatos:2020kzj}).

The detailed transition from such an era to the \ac{rvm} inflation is discussed in Ref.~\cite{Dorlis:2024yqw}, using a dynamical system approach. The model can also lead to post-inflationary late-eras \ac{de} modifications, involving vacuum energy terms of the form $H^2  \mathrm{ln}(H)$, which are held responsible for the simultaneous alleviation of $H_0$ and growth-of-structure tensions~\cite{Gomez-Valent:2023hov}.

The simplest generalization of Einstein's theory are the so-called $ f(R) $ theories, which can be further generalized by including a non-minimal coupling between the matter Lagrangian and another generic function of the curvature \cite{Bertolami:2007gv}
$$
S=\int d^4x \sqrt{-g} \left(f_1(R)+f_2(R)\mathcal{L}\right)\,.
$$
This model gives rise to an extra force term in the geodesics for a perfect fluid. In fact, stemming from the Liouville theorem in phase space, the flux density of particles is shown to be covariantly conserved, and the Boltzmann H-theorem is still preserved as the entropy vector flux is a non-decreasing function of the spacetime coordinates likewise in general relativity. Despite the distribution function being formally equivalent to Einstein theory for a homogeneous and isotropic universe, some quantities differ, namely as the effects of the non-minimal coupling appear on the radiation density evolution and on the matter Lagrangian choice, which is no longer degenerate, for instance, Ref.~\cite{Bertolami:2020ldj}. In these theories, the baryon asymmetry is generated through an effective coupling between the Ricci scalar and the net baryon current, which dynamically breaks CPT invariance, thus matching the observed asymmetry for very small deviations from general relativity. Moreover, the resulting temperatures are compatible with the subsequent formation of the primordial abundances of light elements \cite{Ramos:2017cot}. Hence, among others, this scenario allows for a good agreement with \ac{sn} distance data and the \ac{bao} data for different exponents in a power-law type non-minimal coupling function, thus alleviating the Hubble tension \cite{BarrosoVarela:2024htf}.

A consistent gauge theory of gravity and spacetime based on the Lorentz symmetry was found rather recently \cite{Zlosnik:2018qvg,Nikjoo:2023flm,Gallagher:2023ghl}. Addressing the problem of time, the theory predicts that the space that emerges via spontaneous breaking of the Lorentz symmetry could be massive \cite{Zlosnik:2018qvg,Gallagher:2021tgx,Koivisto:2023epd}. This would be the minimal explanation of the missing mass in the Universe, and a slight extension of the scenario might alleviate the $ H_0 $ tension through a cosmic spin as has been discussed also in the context of Poincaré models of gravity \cite{Koivisto:2023epd,Poplawski:2019tub,Izaurieta:2020xpk,Akhshabi:2023xan}. An ultraviolet completion mentioned in Ref.~\cite{Koivisto:2022uvd} leads to modifications of early cosmology, whereas the relevance of a different description of nonperturbative effects to the cosmological tensions was speculated in Ref.~\cite{Iosifidis:2024ksa}. Ref.~\cite{Benisty:2018ufz} rewrites the higher curvature theories using a gauge theory of gravity.

As a subcase of metric-affine and gauge theories of gravity, teleparallel gravity has also been well-studied in the early Universe. The Teleparallel Equivalent of General Relativity (TEGR) describes gravitational interactions as the torsion of the connection of spacetime. Several teleparallel inflationary scenarios have been proposed in the literature, in order to change the early-Universe evolution and eventually alleviate the $ H_0 $ tension, and here we briefly summarize some of them. Born-Infeld inspired inflation originates from the work of Born and Infeld, who proposed a finite self-energy for point-like particles to avoid divergences in physics. This approach has influenced various gravity theories, including Eddington-inspired Born-Infeld gravity, which addresses singularities like black holes and the big bang. In teleparallel cosmology, Born-Infeld inspired models use an $ f(T) $ gravity Lagrangian, featuring a scaling parameter $ \lambda_{\rm BI} $ that induces inflation without an inflaton field \cite{Ferraro:2006jd,Ferraro:2008ey,Fiorini:2009ux}. This model aligns with the \ac{cmb} and transitions to standard \lcdm\ at later times, providing a comprehensive framework for early Universe dynamics \cite{Jana:2014aca,Nesseris:2004wj,Fiorini:2013kba,Bouhmadi-Lopez:2014tna,Fiorini:2015hob}.

In Loop Quantum Cosmology within the \ac{flrw} framework, the matter bounce scenario is described by a modified Friedmann equation, resulting in a non-singular bounce at $ t = 0 $. This model can be reformulated in terms of $f(T)$ gravity \cite{Wilson-Ewing:2012lmx,Cai:2011tc,Bamba:2012ka,Amoros:2013nxa,Casalino:2020kdr,Haro:2014wha,deHaro:2014tla,Haro:2014xtk,deHaro:2015wda}. The Lagrangian can be adjusted to fit observations by varying the critical density $ \rho_{\rm cr} $. Scalar fields, affecting spectral indices and power spectra, are crucial for matching \ac{cmb} observations and ensuring reheating. Perturbations reveal additional degrees of freedom, impacting the scalar and tensor power spectra, with the tensor-scalar ratio potentially exceeding observed bounds, dependent on scalar field dynamics.

Higgs inflation explores the coupling of the Higgs field with gravitational theories like GR, TEGR, and STEGR. The Higgs potential $ V(\phi) = \frac{\lambda}{4}(\phi^2 - v^2)^2 $ can be coupled with the torsion scalar $ T $ to investigate inflationary scenarios. In this context, a particular action is used, incorporating kinetic and potential functions \cite{Raatikainen:2019qey}. Some studies reveal issues, such as a high tensor-to-scalar ratio or failing to achieve inflationary models. Despite these challenges, alternative scalar field couplings and potential functions may yield more practical inflationary models, suggesting further exploration is necessary for natural inflation in teleparallel gravity.

When the canonical quantization techniques of Loop Quantum Gravity are applied to cosmological scenarios, one finds that the big bang singularity is generically resolved by a big bounce \cite{Ashtekar:2003hd,Ashtekar:2006rx,Ashtekar:2006wn,Ashtekar:2007em,Taveras:2008ke,Diener:2014hba}. This happens even when different quantization prescriptions are used, which leads to various nonsingular models of loop quantum cosmologies \cite{Yang:2009fp,Dapor:2017rwv,Li:2018opr,Li:2018fco,Assanioussi:2019iye} (known as LQC, mLQC-I, and mLQC-II \cite{Li:2018opr}) that differ in their concrete pre-bounce and post-bounce dynamics. In Ref.~\cite{Delhom:2023xxp}, it has been found that the effective dynamics of those three quantum-corrected models can be accurately approximated by a 3-parameter family of metric-affine $ f(R) $ theories of gravity \cite{Olmo:2011uz}, where two of the parameters can be set by constraints at the bounce and at low curvatures (\ac{gr} limit), while the third one can be determined (numerically) by fitting the background evolution. Interestingly, the non-perturbative dynamics of all three models are dominated by a logarithmic correction \cite{Olmo:2008nf}. Additionally, the best fit value of the free parameter can be well approximated by elementary combinations of the bounce density and the Barbero-Immirzi parameter. On the other hand, it is possible to recover regular big-bounce scenarios also in the presence of a dynamical Barbero-Immirzi field, when a purely torsional Nieh-Yan term on a \ac{flrw} background \cite{Bombacigno:2016siz,Bombacigno:2018tyw} and its non-metrical generalization in Bianchi-I cosmologies \cite{Bombacigno:2021bpk} are taken into account. This idea naturally leads to projective-invariant metric-affine formulations of Chern-Simons \ac{mg}, for which big bounce and de Sitter solutions arise \cite{Boudet:2022nub}. A peculiar aspect of these models is that they predict the coupling of metric tensor modes to torsion tensor components, leading to torsional birefringence and the possibility of distinguishing the usual metric version of Chern-Simons gravity from its metric-affine counterpart via the quasinormal mode emission of compact objects \cite{Boudet:2022wmb} (see also Ref.~\cite{Bombacigno:2022naf}).

Axions and Axion-Like Particles (ALPs) provide well-motivated candidates to address important problems in cosmology, maintaining at the same time a significant discovery potential in current and future experiments. Among their uses, they can serve as candidates for \ac{de} within the quintessence scenario \cite{Choi:1999xn,Kim:2002tq}. ALPs arise naturally in string-inspired models \cite{Svrcek:2006yi}, often not single but as a plethora of light pseudoscalar fields \cite{Arvanitaki:2009fg}. Their characteristic sinusoidal potential generates models of \ac{ede} that can potentially resolve the Hubble tension \cite{Poulin:2018cxd,Smith:2019ihp,Chakraborty:2021vcr}. Remarkably, although these models are loosely inspired by string theory, recent attempts highlight the potential of embedding them into stable type IIB string compactifications \cite{McDonough:2022pku,Cicoli:2023qri}, which would provide a fundamental starting point. The appearance of multiple ALPs in stringy scenarios allows for alignment mechanisms that combine them and can lead to models where at different stages of the cosmological evolution different ALPs act as inflaton, QCD axion, and quintessence \cite{Kim:2009cp,Chatzistavrakidis:2012bb}. Pseudoscalar fields were also considered in the context of teleparallel and symmetric teleparallel gravity, where they couple to the CP-violating quadratic terms that can be formed with the torsion tensor or/and the non-metricity tensor \cite{Hohmann:2020dgy,Chatzistavrakidis:2020wum,Zhang:2023scq}. Such additional fields could contribute to the cosmological dynamics in the early Universe \cite{Lattanzi:2009mg,Hohmann:2020dgy} and can also lead to \ac{gw} velocity birefringence \cite{Li:2020xjt,Chatzistavrakidis:2021oyp}, which can be tested using multi-messenger approaches \cite{Lagos:2024boe}.

\ac{dm} creation during or after inflation can be associated with irreversible thermodynamic processes \cite{Su:2017esf, Harko:2021gnz, Matei:2023ler}. Refs.~\cite{Prigogine:1988jax,Prigogine:1989zz} proposed an alternative cosmology based on the irreversible thermodynamics of open systems, in which the explanation for the production of macroscopic matter and entropy in the early Universe relies on a reinterpretation of the matter energy-momentum tensor that includes an irreversible creation term. However, this contrasts with the covariant conservation of the energy-momentum tensor in GR. 

In the last decade, several \ac{mg} theories that contain nonminimal couplings between geometry and matter have been proposed, such as $f(R,\mathcal{L}_{\rm m})$ \cite{Harko:2010mv}, $f(R,T)$ \cite{Harko:2011kv}, $f(R,T_{\mu\nu}T^{\mu\nu})$ \cite{Katirci:2013okf, Roshan:2016mbt, Akarsu:2017ohj, Board:2017ign} and $f(R,T,R_{\mu\nu}T^{\mu \nu})$ \cite{Haghani:2013oma} theories of gravity, where $R$ and $R_{\mu\nu}$ are the Ricci scalar and tensor, $\mathcal{L}_{\rm m}$ the matter Lagrangian density, and $T$ the trace of the energy-momentum tensor $T_{\mu\nu}$. A property of all these theories is that the matter energy-momentum tensor is not conserved. This feature allowed Harko~\cite{Harko:2014pqa} to physically interpret such non-conservation as an irreversible energy flow from the gravitational field to the matter sector that could result in particle creation, by using the thermodynamics of open systems approach. The effects and implications of the irreversible matter creation processes on the late cosmological evolution have been studied on some of these modified theories of gravity \cite{Harko:2014pqa, Harko:2015pma, Pinto:2022tlu, Cipriano:2023yhv} (see Ref.~\cite{Pinto:2023phl} for a review). It was shown in Ref.~\cite{Akarsu:2023lre} that if such theories include the Ricci scalar $R$ in the Einstein–Hilbert form, i.e., assuming minimal geometry-matter coupling—they are effectively equivalent to nonminimal interaction models within the framework of \ac{gr}. These \ac{mg} theories could be further explored in the context of the early Universe \ac{dm} creation that can eventually lead to alleviation of the $H_0$ tension.

\lcdm\ paradigm is very efficient in describing our Universe; however, it is broadly acknowledged that the model ought to break down and exhibit limitations in at least the two extreme phases of the Universe, namely its very early and very late phases. In the very early Universe, Planck scale or quantum gravity effects are expected to set in, resulting in potentially significant conclusions, some of which may be measurable. While proposals for quantum gravity abound, and one can in principle take any theory and examine its implications for the above stages of the Universe, here we will primarily be interested in a virtual model-independent prediction of theories of quantum gravity, namely the \ac{gup}. The \ac{gup} predicts significant changes to several thermodynamic quantities, while it leaves some others intact \cite{Basilakos:2010vs,Das:2021wxq,Das:2021nbq,Escamilla:2024xmz}. 

A generic and often-used version of \ac{gup} is 
\begin{gather}
   [x_i,p_j] = i\hbar\,\left[
    1 - \alpha \left( 
    p\delta_{ij} + \frac{p_i p_j}{p} \right)
    + \alpha^2 \left(p^2 \delta_{ij} + 3 p_i p_j \right)  
    \right] \label{gup1} \\
 \Delta x \Delta p \geq \frac{\hbar}{2} 
\left[ 1 - 2\alpha \langle p \rangle + 4\alpha^2 \langle p \rangle^2
\right]~\,,\label{gup2}
\end{gather}
where the dimensional parameter $\alpha$ can be traded off with 
a dimensional one $\alpha_0$, via
$    \alpha =  \alpha_0 \ell_{Pl}/\hbar$ ,
where $\ell_{Pl} \simeq 10^{-35}$\,m is the standard Planck length
and $1 \leq \alpha_0 \leq 10^{15}$, the upper bound dictated by the fact that no Planck scale effects have been observed at the LHC, which has a characteristic length scale of $10^{-20}$\,m $\simeq 10^{15}\, \ell_{Pl}$ (corresponding to $10$\,TeV of energy), which the 
``new length scale'' $\alpha_0 \ell_{Pl}$ should not exceed.  

It follows that quantum gravity effects can in principle take effect right from the electroweak to the Planck scale. Since our focus is in the very early Universe, we will be concerned near the Planck scale. As explained in Ref.~\cite{Basilakos:2010vs} the standard Heisenberg Uncertainty Principle and the \ac{gup} can be translated into a corresponding time-energy uncertainty relation. This, coupled with the first Friedmann equation, can in turn be used to estimate the Planck energy, mass, density, time, temperature, and effective number density. Since the \ac{gup} terms can differ significantly from the standard Heisenberg one near the Planck scale, it is not surprising that many of the quantities alluded to above undergo significant changes in the early Universe. Perhaps the most significant among them is the Planck energy density, whose ratio to the observed \ac{de} density increases by at least a factor of $4$ to about $10^{119}$. 

Additionally, the same uncertainty principle can in fact affect the Friedmann equations and cosmological predictions in the radiation dominated era, as shown in Refs.~\cite{Das:2021wxq,Das:2021nbq}. There,     an interaction term coupling baryon current and space-time was used to satisfy the first two Sakharov conditions, which along with the modified Friedmann equations break thermal equilibrium and thus satisfy the third and last Sakharov condition. The results can provide a rich phenomenology.

In the \ac{gup} approach, where a modification to the canonical algebra of the system leads to minimal uncertainties in the configuration and/or momenta, the evolution of the \ac{flrw} Universe with canonical variables $\left(c,p\right)$ in the presence of scalar matter field $\phi$ with its momentum $p_{\phi}$, can be described by the action of the \ac{gup} modified Wheeler-deWitt equation on the wave function of the Universe $\Psi$, $\left(\partial_{\phi}^{2}+\widehat{\Theta}_{\text{GUP}}\right)\Psi=0$, where $\widehat{\Theta}_{\text{GUP}}$ is a self-adjoint operator. In one approach~\cite{Battisti:2007jd}, which is equivalent to a cut-off in the length, $[\mathbf{q},\mathbf{p}]=i\left(1+\beta\mathbf{p}^{2}\right)$, using suitable wave-packets for the deep Planckian regime, it is found that the wave-packets do not fall in the classical singularity and the \ac{gup} Universe exhibit a stationary behavior in approaching the Planckian region. This implies that no bounce is present in this model. In another approach using \ac{gup}-modified Friedmann equations \cite{Kouwn:2018rmp}, it is found that the \ac{de} density scales as $\tilde{\alpha}H^{4}$, where $\tilde{\alpha}$ is a \ac{gup} parameter associated to the minimum length, and $H$ is the Hubble parameter. This means that this approach to \ac{gup} cannot explain the acceleration of the present Universe, since the energy density decreases very quickly. It is worth mentioning that \ac{gup} comes in various forms and a recent proposal based on a different modification of algebra (minimal momenta) \cite{Fragomeno:2024tlh} together with an improved prescription may be a better approach. The reason is that 1) the momenta of gravitational field are related to the metric or triads so it is more natural to have minimal values in those quantities, and 2) without an improved scheme, just like LQC, one might obtain incorrect classical or asymptotic limits and hence the theory might not also be correct in the Planckian regime.

The Asymptotic Safety approach to Quantum Gravity offers a potential solution to the problem of quantizing the gravitational field. This approach is grounded in the existence of an ultraviolet non-Gaussian fixed point for the couplings of the $\sqrt{g} R$ and $\sqrt{g}$ operators in the Euclidean theory \cite{Reuter:1996cp}. A direct consequence of this fixed point is the dynamic suppression of Newton constant $G$ at very high energies, which has significant implications for early cosmological models. In Ref.~\cite{Reuter:2005kb}, a comprehensive cosmological evolution from the inflationary phase to a late de Sitter phase is discussed. Additionally, \cite{Bonanno:2007wg} proposes the idea that the Universe originates from a state of zero entropy and evolves towards an accelerated expansion phase.

Following Ref.~\cite{Weinberg:2009wa}, one can also expect various modifications to the Effective Action at the inflationary scale. In particular, it has been shown in Ref.~\cite{Bonanno:2015fga} that an $f(R)$ Lagrangian of the form $L = R + \alpha R^{3/2}+ \frac{R^2}{6 m^2} -\Lambda $ represents a modification of the well-known Starobinsky quadratic Lagrangian. This modification arises due to the presence of higher curvature relevant operators generated by the renormalized flow near the non-Gaussian fixed point. The predicted scalar-to-tensor ratio $r$ is significantly higher than in the standard Starobinsky model, making it a potential target for testing in future \ac{cmb} missions like LiteBIRD.

The cosmological evolution of the early Universe is tested at temperature \(T \lesssim 1\) MeV through the prediction of \ac{bbn}. For temperatures \(T \gg 1\) MeV, it is not excluded that new physics occurred, offering the possibility that non-standard cosmologies might have played a relevant role in the interplay between particle physics and cosmology. The Ice Cube experiment, for example, has revealed a high-energy astrophysical neutrino flux, with energies \(\sim O(1)\) PeV \cite{IceCube:2015qii,IceCube:2014flw}. Among the various scenarios to explain this phenomenon, it has been proposed the minimal model of \ac{dm} decay \cite{Chianese:2016smc}. This interaction is also able to generate the correct abundance of \ac{dm} in the Universe \cite{Chianese:2016smc}. In the standard cosmological model, the rate of \ac{dm} decay, needed to obtain the correct \ac{dm} relic abundance, differs by several orders of magnitude as compared with that needed to explain the Ice Cube data, making the four-dimensional operator unsuitable. However, such a discrepancy can be reconciled in modified cosmologies, so that the Ice Cube neutrino rate and \ac{dm} relic density can be consistently explained \cite{Lambiase:2018yql,Jizba:2022bfz,Jizba:2023fkp}.

In Fig. \ref{fig:earlymodgrav} we summarize the various early modified gravity theories that have been proposed to alleviate the tensions, while some corresponding predictions for $H_0$ and $\sigma _8$ are summarized in Table~\ref{tab:E_MoG_sample}.

\begin{table}[ht!]
    \centering
    \caption{Early modifications to gravity: predictions.}
    \label{tab:E_MoG_sample}
    \begin{tabular}{|c|c|c|}
        \hline
        Model & $H_0$ & $\sigma _8$ \\
        \hline
        Early \ac{mg} and the sound horizon \cite{Poulin:2023lkg}   & $>70$   & $<0.8$   \\
        String-inspired Chern-Simons modifications \cite{Gomez-Valent:2023hov}   & $71.27 ^{+0.76}_{-0.73}$    & $ 0.816 ^{+0.017} _{-0.015}$   \\
        Non-minimally coupled models \cite{BarrosoVarela:2024htf}  & $70.59 ^{+0.52}_{-0.53}$   & -   \\
        Axions and Axion-like particles \cite{Lagos:2024boe}  & $70.6 \pm 1.3$  & -  \\
        \hline
    \end{tabular}
\end{table}

\begin{figure}
\centering
\includegraphics[width=0.9\textwidth]{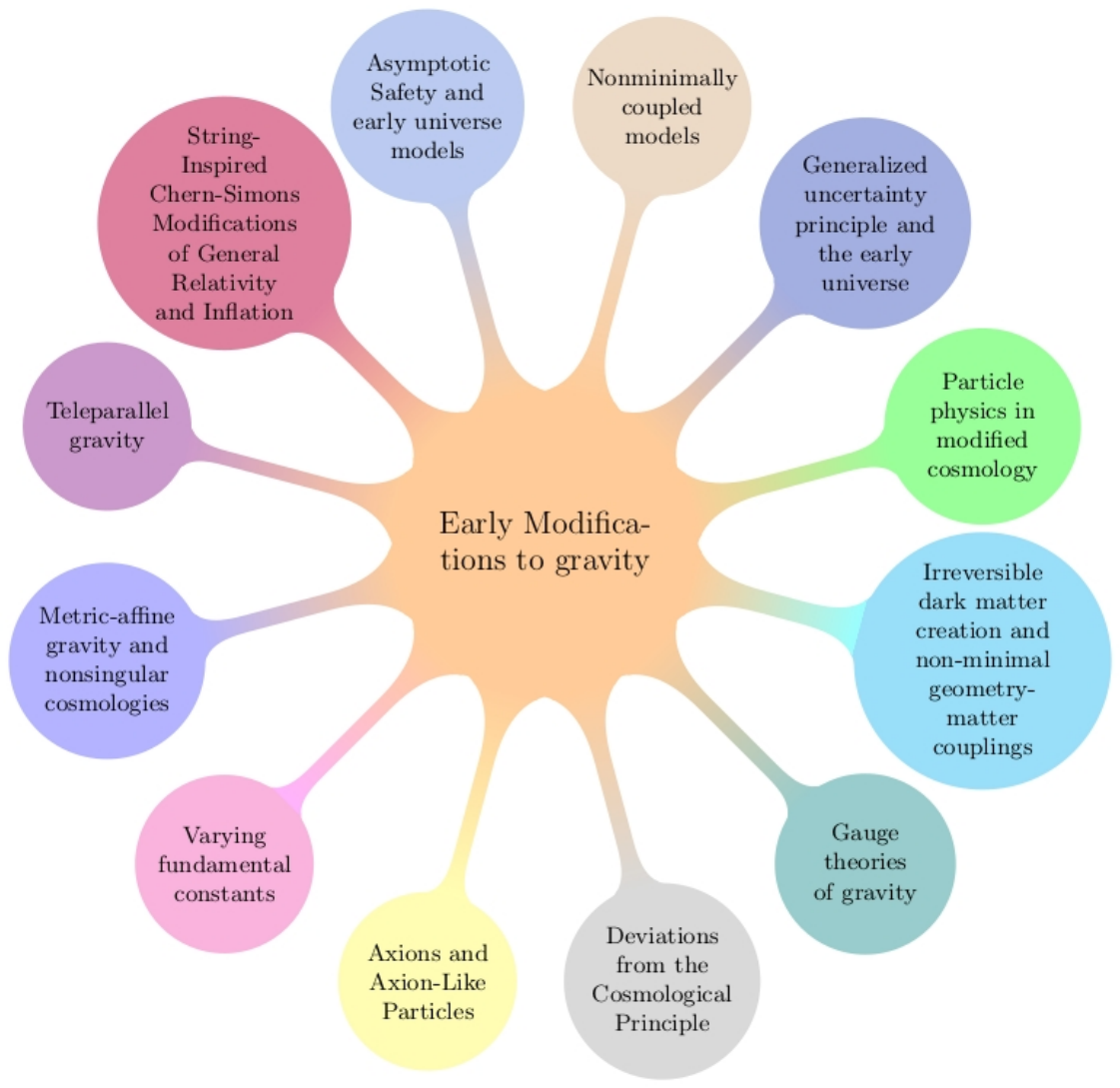}
\caption{Possible early modifications to gravity.}
\label{fig:earlymodgrav}
\end{figure}
\subsection{Matter sector solutions}
\subsubsection{Cold dark matter \label{sec:CDM}}

\noindent \textbf{Coordinator:} Luca Visinelli\\
\noindent \textbf{Contributors:} Andrzej Borowiec, Anil Kumar Yadav, Branko Dragovic, Davide Pedrotti, Goran S. Djordjevic, Ioannis D. Gialamas, Ippocratis Saltas, Ivan De Martino, Janusz Gluza, Jenny G. Sorce, Jenny Wagner, Krishna Naidoo, Marcin Postolak, Martti Raidal, Mehmet Demirci, Paolo Salucci, Pran Nath, Reggie C. Pantig, Riccardo Della Monica, Rishav Roshan, Sebastian Trojanowski, Torsten Bringmann, Venus Keus, and Wojciech Hellwing
\\

\paragraph{Motivation and evidences}

The circular velocity of stars in a spiral galaxy can be inferred from the measurements of the Doppler shift of atomic lines at different distances $R$ from the galactic center, leading to a velocity distribution $v=v(R)$ known as the {\it rotation curve}. The distribution of the luminous matter in virialized objects does not match that of the gravitational matter, demanding an additional non-luminous component.

Historically, this ``missing mass problem'' in the late Universe led to the proposition that there is a yet undetected form of matter~\cite{1930MeLuF.125....1L, Zwicky:1933gu, Zwicky:1937zza, Rubin:1970zza, Freeman:1970mx}. Subsequently, \ac{cmb} observations revealed a missing mass in the power spectrum already at redshift $z=1100$, which could also be attributed to the same form of matter we call \ac{dm} today. Evidence for the presence of \ac{dm} in the Universe spans from galactic (rotation curves of spiral galaxies) to extra-galactic (clusters of galaxies, weak gravitational lensing observations) to cosmological scales (\ac{lss}, acoustic peaks in the \ac{cmb}, \ac{bbn}, non-linear growth of inhomogeneities. See Ref.~\cite{deSwart:2017heh} for a historic reconstruction and Refs.~\cite{Jungman:1995df, Bertone:2004pz, Iocco:2015xga, Salucci:2018hqu, deMartino:2020gfi, Cirelli:2024ssz} for reviews. N-body simulations with \ac{dm} only reproduce well the \ac{lss}~\cite{Davis:1985rj, Navarro:1995iw, Vogelsberger:2019ynw} and are advancing to incorporate the essential \emph{gastrophysics} effects~\cite{Vogelsberger:2014dza, Dave:2019yyq}.

The role of \ac{dm} in addressing cosmic tensions is of profound interest from both theoretical and observational perspectives. If discovered, a \ac{dm} particle would be a relic from a period before \ac{bbn}, for which we currently have no direct data. The properties of \ac{dm}—such as its clustering power spectrum, free-streaming velocity, and particle characteristics—would enable us to better model the pre-\ac{bbn} epoch and differentiate between various ``early Universe'' proposals aimed at resolving these tensions. Furthermore, some theories that address the $H_0$ tension explicitly involve interactions between \ac{dm} and other components, including \ac{de}. Pinning down the properties of \ac{dm} is crucial for understanding its interactions with other particles, including those in exotic \ac{de} models.

\paragraph{Theory}

The evidence supporting the existence of non-baryonic \ac{dm} is compelling across all observed astrophysical scales. While alternative theories are still under consideration, the notion of \ac{cdm} has emerged as the dominant framework. \ac{dm} might exist as macroscopic objects or as a particle, either fundamental or composite, proposed in extensions beyond the SM framework. A non-exhaustive summary of the properties (scattering cross section and target mass) of some candidates is provided in Fig.~\ref{fig:DMtargetplot}, inspired by previous work in Refs.~\cite{Roszkowski:2004jc, Kim:2008hd, Baer:2014eja}.

\begin{figure}[!t]
\centering
\includegraphics[width=0.7\textwidth]{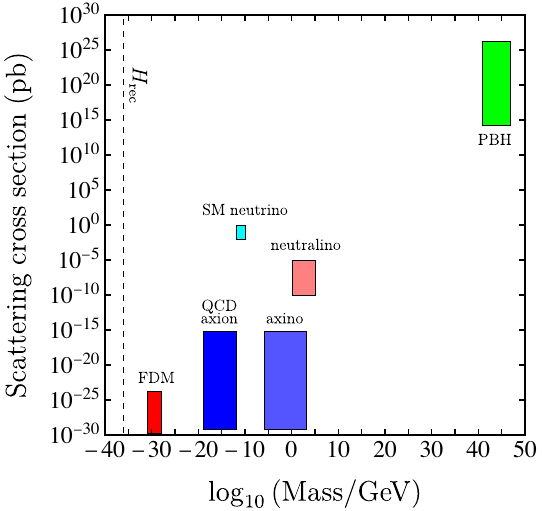}
\caption{An estimate of the scattering cross section for some \ac{dm} candidates, as a function of the \ac{dm} mass. The plot is inspired from Refs.~\cite{Roszkowski:2004jc, Kim:2008hd, Baer:2014eja}. As a comparison, the SM neutrino is also included.}
\label{fig:DMtargetplot}
\end{figure}

\subparagraph{Macroscopic DM} One of the earliest propositions on the nature of \ac{dm} is that it could appear in the form of known astrophysical objects of mass $M$ such as planets or stellar black holes~\cite{Paczynski:1985jf,Griest:1990vu}. This conjecture has been challenged by gravitational microlensing surveys, which collectively exclude these objects as the main \ac{dm} component over the mass range $10^{-11} \lesssim M/M_\odot \lesssim 10$~\cite{MACHO:2000qbb, EROS-2:2006ryy, Wyrzykowski:2010mh, Niikura:2017zjd}.

Along with known astrophysical objects, massive \ac{dm} candidates consist of primordial black holes which are produced in the early Universe by a plethora of mechanisms at different mass scales. Stable primordial black holes possess a mass $M \gtrsim 5\times 10^{14}\,$g so that their Page time~\cite{Page:1976df} exceeds the age of the Universe, making them suitable \ac{dm} candidates. A more stringent bound $M \gtrsim 10^{17}\,$g is obtained from the non-observation of Hawking emission in the extragalactic $\gamma$-ray flux and in X-ray detectors~\cite{Ricotti:2007au, Gaggero:2016dpq, Acharya:2020jbv, Korwar:2023kpy}. Lensing constraints also apply to primordial black holes, leaving an open window of opportunity where primordial black holes are the \ac{dm} in the mass range $M \in [10^{17} \textrm{-} 10^{22}]{\rm\,g}$ under the assumption of a monochromatic mass function.

\subparagraph{Particle DM} An intriguing possibility is that \ac{dm} results from a completely new sector of physics not comprised within the SM. This includes candidates such as sterile neutrinos, axions, supersymmetric particles (gravitinos, neutralinos, axinos), weakly/feebly/strongly interacting massive particles, Q-balls, or ultralight scalars. Generally speaking, \ac{dm} particles should be i) non-relativistic at recombination, ii) neutral or feebly charged, iii) stable over cosmological timescales, iv) feebly interacting with the SM. Interactions between the dark and visible sectors can help alleviate cosmological tensions. For example, a model in which dark matter interacts with the baryon sector yields $S_8 = 0.794\, (0.804)^{+0.009}_{-0.010}$ for the marginalized posterior (maximum of the full posterior), when fit to \textit{Planck} 2018, SDSS \ac{bao}, and \ac{des}-Y3 data at 1$\sigma$ CL~\cite{He:2023dbn}. In a scenario where dark matter feebly interacts with photons, a fit to \textit{Planck} 2018 + SDSS \ac{bao} leads to $S_8 = 0.803 \pm 0.021$ and $H_0 = 67.70 \pm 0.43\,$\kms at 1$\sigma$ CL~\cite{Becker:2020hzj}. While the upper limit for the mass $m$ of a particle candidate is the Planck mass, the lower limit for a fermionic species is provided by the Gunn-Tremaine condition~\cite{Tremaine:1979we}, which reads $m \gtrsim 100$\,eV from consideration on dwarf spheroidal galaxies~\cite{Alvey:2020xsk}. For a bosonic species, the lower mass limit is often considered as the Hubble rate at matter-radiation equality, $m \gtrsim 10^{-27}\,$eV, from the consideration that the ultralight field should oscillate by the time at which the matter component becomes relevant over radiation. Some particle models that have been explored include the following. Explaining the missing mass across cosmic times and scales by a single extension of the known cosmological model has motivated the search for further corroborations of its existence and led particle physicists to investigate potential particle candidates. 

\subparagraph{Flavor discrete symmetry models} The SM fails to explain neutrino masses, mixings, and \ac{dm} with the WIMP paradigm comprehensively. Standing at this juncture, it is certainly a tempting challenge to find a common origin of these two seemingly uncorrelated sectors and much effort goes beyond the SM of particle physics to explore scenarios that can accommodate a candidate of \ac{dm} and explain non-zero neutrino masses and mixing~\cite{Chauhan:2023faf}. Typically, models based on non-Abelian discrete flavor or modular symmetries are considered to explain observed neutrino mixing, and breaking of the same symmetry may lead to the residual symmetry, which at the same time stabilizes the \ac{dm}~\cite{Keus:2013hya, Chauhan:2023faf}. For example, in models based on the $A_4$ discrete group, light neutrino masses can be associated with tree-level type-I seesaw mechanism and the one-loop contributions accommodating viable \ac{dm} candidates responsible for observed relic abundance of \ac{dm}~\cite{CentellesChulia:2016rms}. In such models, constraints coming from the lepton flavor violating processes are also important, restricting \ac{dm} masses and leaving ranges of \ac{dm} parameters that can be probed in collider studies.

\subparagraph{Supersymmetry} Since $B$ and $L$-violating operators in the SM have dimension $d\geq5$, the renormalizable Lagrangian possesses an accidental global $B$-$L$ symmetry~\cite{Weinberg:1979sa, Wilczek:1979hc, Weinberg:1980bf, Weldon:1980gi}, which is imposed in supersymmetric extensions of the SM to avoid experimental consequences such as a large proton decay rate. Such a symmetry leads to an R-parity which is conserved in scattering and decay processes, leading to supersymmetric particles to be produced in pairs. The same R-parity leads to the stability of the lightest supersymmetric particle, which could be a \ac{dm} candidate if it is neutral. Supersymmetry, first postulated to address the hierarchy problem, has thus long provided us with a long list of \ac{dm} candidates~\cite{Ellis:1983ew}. These include neutralinos~\cite{Drees:1992am, Baer:1995nc, Barger:1997kb, Lahanas:1999uy, Cyburt:2009pg, Han:2014nba, Roszkowski:2017nbc}, sneutrinos~\cite{Falk:1994es, Arina:2007tm}, gravitinos~\cite{Weinberg:1982zq, Khlopov:1984pf, Ellis:1984eq, Bolz:2000fu, Kawasaki:2008qe, Pradler:2006qh, Rychkov:2007uq, Ellis:2015jpg, Dudas:2017rpa, Kaneta:2019zgw, Eberl:2020fml}, and axinos~\cite{Covi:1999ty, Covi:2001nw}, massive spin-2 particles~\cite{Aoki:2014cla, Aoki:2016zgp, Babichev:2016hir, Babichev:2016bxi, Kolb:2023dzp} resulting from modifications of gravity, and Kaluza-Klein excitations~\cite{Cheng:2002ej, Servant:2002aq, Hooper:2007qk} found in theories incorporating extra dimensions. However, some of the simplest versions of the theory are under stress due to the non-observation of any supersymmetric particles in collider experiments. The lightest neutralino, among others, is an example of a much broader category of particle \ac{dm} candidates dubbed weakly interacting massive particles (WIMPs), produced thermally in the early Universe.

\subparagraph{Weakly interacting massive particles (WIMPs)}

WIMPs, some of the most studied \ac{dm} candidates, are hypothesized to have been produced in the early Universe as they were initially in thermal equilibrium with the primordial bath. As the Universe expanded and cooled, the production of WIMPs ceased and they ``froze out'' of thermal equilibrium. The freeze-out temperature is defined as the point when the Hubble expansion rate exceeds the WIMP annihilation rate. The leftover density of WIMPs after freeze-out, i.e., their relic density, depends on their annihilation cross-section. The annihilation cross-section needed for WIMPs to achieve the observed \ac{dm} density today is naturally of the same order of magnitude as the weak nuclear force cross-section. This coincidence is known as the ``WIMP miracle'' because it suggests that WIMPs could naturally account for \ac{dm} without needing highly tuned parameters. Since the WIMP miracle provides a compelling argument for them as the leading \ac{dm} candidate, many direct and indirect detection experiments are built to probe the typical WIMP mass range of $\sim\mathcal{O}$(GeV) to $\sim\mathcal{O}$(100\,GeV)~\cite{Jungman:1995df, Bertone:2004pz}. Even though direct and indirect experiments continue to constrain the parameter space for the vanilla WIMPs scenario - as for most \ac{dm} models - exciting new directions in the WIMP paradigm, such as CP-violating \ac{dm}~\cite{Cordero-Cid:2016krd}, multi-component \ac{dm}~\cite{Hernandez-Sanchez:2020aop}, pseudo-Goldstone \ac{dm}~\cite{Alanne:2018zjm}, have been proposed which not only provide viable \ac{dm} candidates but also address other shortcomings of the SM~\cite{Keus:2019szx}. Even if the WIMP miracle is not realized, loopholes exist such as coannihilation~\cite{Griest:1990kh, Drees:1992am, Edsjo:1997bg}, annihilation to slightly heavier states~\cite{Griest:1990kh}, p-wave annihilation, resonances~\cite{Griest:1990kh, Gondolo:1990dk}, and Sommerfeld enhancement~\cite{Hisano:2004pv, Arkani-Hamed:2008hhe}.

\subparagraph{Sub-GeV thermal DM} The WIMP paradigm can be generalized to lower \ac{dm} masses and coupling constants, and various sub-GeV \ac{dm} candidates have been proposed that can also be produced thermally in the early Universe~\cite{Boehm:2003hm,Pospelov:2007mp,Feng:2008ya}. In this case, their mass is typically bounded from below to $m > \mathcal{O}(\textrm{MeV})$ by astrophysical constraints and \ac{bbn}, while even stronger indirect detection bounds are present for light \ac{dm} candidates annihilating to SM species via $s$-wave channels, see the recent review in Ref.~\cite{Zurek:2024qfm}. In other models, however, dedicated direct detection techniques are needed to probe them~\cite{Essig:2022dfa}, and multiple accelerator-based searches are ongoing and planned~\cite{Krnjaic:2022ozp}.

\subparagraph{Sterile neutrinos} A minimal extension of the SM with the addition of a Weyl fermion $N$  as a sterile neutrino could provide viable \ac{dm} candidates. This particle couples to the SM through a Yukawa term such as $\sim yNLH$, where $L$ is a left-handed lepton doublet and $H$ is the SM Higgs doublet, which could give rise to the small neutrino masses through the seesaw mechanism~\cite{Yanagida:1979gs, Yanagida:1980xy, Mohapatra:1979ia, Schechter:1980gr} and possible lead to baryogenesis via the leptogenesis mechanism~\cite{Fukugita:1986hr}. Sterile neutrinos of mass $m < m_e$ might decay into three active neutrinos, thus constraining the sterile mixing angle $\theta$ for a given $m$~\cite{Pal:1981rm}. The abundance of sterile neutrino \ac{dm} is fixed through a freeze-in mechanism in which active neutrinos oscillate into $N$ before decoupling~\cite{Dodelson:1993je}. Possible caveats include a finite lepton asymmetry~\cite{Shi:1998km} or the presence of self-interactions~\cite{Hansen:2017rxr, DeGouvea:2019wpf, Bringmann:2022aim}.

\subparagraph{The QCD axion} The QCD axion~\cite{Wilczek:1977pj, Weinberg:1977ma} is a light pseudo-scalar coupled with the SM gluons, introduced to solve the strong-CP problem through the Peccei-Quinn mechanism~\cite{Peccei:1977hh}. The interaction with the gluon leads to a mixing of the axion with the neutral pions and to a small mass $m_a \sim 6{\rm\,\mu eV}(10^{12}{\rm\,GeV}/f_a)$, where $f_a$ is the energy scale at which the infrared description breaks. Because of its interaction with the gluon that induces a coupling with the charged pions, this particle inevitably couples with the SM photon. Models that generalize the Peccei-Quinn mechanism leading to a very light axion and a prediction for the axion-photon coupling $g_{a\gamma\gamma}$ have been long studied and constitute the benchmark for laboratory axion searches, such as the KSVZ~\cite{Kim:1979if, Shifman:1979if} and the DFSZ axion~\cite{Zhitnitsky:1980tq, Dine:1981rt}. A nonrelativistic population of cosmic QCD axions is produced through the vacuum misalignment mechanism~\cite{Preskill:1982cy, Abbott:1982af, Dine:1982ah}, leading to a number density in axions with a large occupation number at the time $T_{\rm osc} \sim\,$GeV at which coherent oscillations in the field begin. This population of axions can possibly address the \ac{dm} abundance for given values of the mass $m_a$ and of the initial value of the axion field $\theta_i\,f_a$~\cite{Linde:1991km, Hertzberg:2008wr, Visinelli:2009zm}. The uncertainties in the assessment of the axion parameters can be divided into cosmological~\cite{Lazarides:1990xp, Visinelli:2009kt}, particle content~\cite{DiLuzio:2016sbl, DiLuzio:2017pfr}, and theoretical~\cite{Barr:1992qq}, see Refs.~\cite{Marsh:2015xka, DiLuzio:2020wdo} for reviews. A robust experimental search is currently ongoing worldwide to try and detect this particle~\cite{Irastorza:2018dyq, Chadha-Day:2021szb}.

\subparagraph{Ultralight bosons} 
Ultralight bosons or ``fuzzy'' \ac{dm} (FDM) with masses in the range $m \sim 10^{-23}$-$10^{-18}$\,eV would possess a de Broglie wavelength of astronomical size~\cite{Hu:2000ke, Amendola:2005ad, Marsh:2013ywa}. An ultralight scalar, naturally generated in string theory, is generally considered to be real or pseudo-scalar field minimally coupled to the metric, and acquires a small mass from various UV mechanisms~\cite{Arvanitaki:2009fg, Marsh:2013taa}. When the Universe cools below a certain critical temperature, the field starts to roll down and oscillates about the minimum of its potential~\cite{Arias:2012az, Hui:2016ltb, Visinelli:2017imh, Visinelli:2018utg}. N-body cosmological simulations have shown the formation of a \ac{dm} core within each virialized halo, which is supported against gravitational collapse by the internal quantum pressure due to the Heisenberg uncertainty principle on the de Broglie scale, surrounded by an interference pattern representing the oscillation in the density and velocity fields of the \ac{dm} particles~\cite{Schive:2014dra, Schive:2014hza, Nori:2018hud, Mocz:2019pyf, Veltmaat:2019hou, May:2022gus}. A lower bound on the boson mass is derived by considering the quantum pressure introduced at cosmological scales, which suppresses the power spectrum of linear inhomogeneities at large wavelengths and leads to a lower bound $m \gtrsim 10^{-21}$\,eV~\cite{Irsic:2017yje, Armengaud:2017nkf, Kobayashi:2017jcf, Zimmermann:2024xvd, Teodori:2025rul}. Stronger constraints on the mass and the abundance of ultralight scalars are placed from a variety of observations which involve detection strategies at astronomical or cosmological scales~\cite{DeMartino:2017qsa, Hlozek:2017zzf, Bar:2018acw, Bar:2019bqz, Pozo:2020fft, deMartino:2020gfi, Rogers:2020ltq, Chan:2020exg, Dalal:2022rmp, DeMartino:2023cgg, DellaMonica:2023dcw}, which could nevertheless depend on the details of cosmology and the astrophysics involved. An issue with FDM candidates lies in that such particles cannot produce the very large regions of about constant density observed at the center of large galaxies~\cite{Burkert:2020laq, DeLaurentis:2022nrv}. Ultralight bosons coupled to a dark energy fluid can help relieve both the $H_0$ and $S_8$ tensions. For instance, a model fit to \textit{Planck} 2018, lensing, Pantheon, SH0ES, and WiggleZ LSS data results in $H_0 = 68.81^{+1.60}_{-0.67}\,$\kms and $S_8 = 0.7993^{+0.0410}_{-0.0140}$ at 1$\sigma$ CL~\cite{Aboubrahim:2024spa}.

\paragraph{Experimental efforts}

\subparagraph{Laboratory searches} 
Given the vast landscape of particle \ac{dm} candidates, several techniques have been implemented for their searches. Here, we highlight the detection of WIMPs and axions in underground detectors.

If \ac{dm} is in the form of WIMPs with a small coupling to SM nucleons, it is possible to detect a small event rate $R$ based on \ac{dm}-nucleon elastic scattering, totaling $R \sim N_T n_{\rm DM}\sigma_Av$. Here, $N_T$ is the number of nucleon targets in the experiment, $n_{\rm DM}$ is the number density of \ac{dm} in the Solar system, $v \sim 220{\rm\,km/s}$ is the relative velocity, and $\sigma_A$ is the scattering cross section off a nucleon of atomic number $A$. For spin-independent (SI) interactions, the \ac{dm} scattering cross section is often quoted in terms of $\sigma_{\rm SI} \approx \sigma_A/A^4$~\cite{Jungman:1995df, Ullio:2000bv}. For spin-dependent (SD) scatterings, the enhancement with the target mass is more modest and reads $\sigma_A \approx \sigma_{\rm SD}\,A^2(J+1)/J$, where $J$ is the spin of the nucleus. The most sensitive experiments include liquid xenon-based detectors: XENONnT~\cite{XENON:2023cxc}, LUX-ZEPLIN (LZ)~\cite{LZ:2022lsv}, and PandaX-4T~\cite{PandaX:2022xas}, which all set a bound on the SI \ac{dm}-nucleon cross section $\sigma_{\rm SI} \lesssim 10^{-47}{\rm\,cm^2}$ for the \ac{dm} mass $m \sim 40\,$GeV. The argon-based detector DarkSide-50~\cite{DarkSide-50:2022qzh} is the most sensitive in targeting the SI cross section for WIMP masses below about $3\,$GeV, with proposed experiments including DarkSide-20k~\cite{DarkSide-20k:2017zyg} and ArDM~\cite{ArDM:2016zfm}. Note, that the results depend on several assumptions on: i) the distribution of \ac{dm} in the solar system and around the Earth, ii) the universal coupling between the WIMP and the nucleons, iii) the mass range of the mediator, iv) the knowledge of the nuclear form factors, v) the effective theory used to model the interaction, vi) the absence of self-interactions, vii) the scattering process being elastic. All of these caveats have been thoroughly investigated in the literature in relation to realistic \ac{dm} models.

The laboratory search for the QCD axion and other light bosons proceeds by means of resonant cavities in a ``haloscope''~\cite{Sikivie:1983ip, Sikivie:1985yu}, light shining through wall experiments~\cite{OSQAR:2007oyv, ALPS:2009des, Bahre:2013ywa, OSQAR:2015qdv}, dielectric haloscopes~\cite{Horns:2012jf}, and \ac{dm} radio searches~\cite{Sikivie:2013laa}. In a resonant cavity, a galactic axion converts into a microwave photon through the interaction with the virtual photon of an external magnetic field~\cite{ADMX:2001dbg, Brubaker:2016ktl, Barbieri:2016vwg, Alesini:2020vny, Adair:2022rtw, Alesini:2023qed, ArguedasCuendis:2019swy}. \ac{dm} radio searches are employed in DMRadio/ABRACADABRA~\cite{Ouellet:2018beu} and, for light axions not related to the QCD theory, with SuperMAG~\cite{Fedderke:2021aqo}. Scalar fields produced in the Sun are searched through helioscopes: CAST~\cite{CAST:2017uph} and IAXO~\cite{Armengaud:2014gea}.

\subparagraph{Indirect detection} 
Indirect detection of \ac{dm} involves observing the potential byproducts of \ac{dm} interactions rather than revealing a new particle. This method relies on \ac{dm} particles annihilating or decaying into SM particles, which are then detected in astronomical and terrestrial instruments. The byproducts of these interactions typically include (1) gamma rays~\cite{Gunn:1978gr, Stecker:1978du}, high-energy photons that can travel vast distances and are easier to trace back to their source; (2) Neutrinos~\cite{Krauss:1985ks, Freese:1985qw, Gaisser:1986ha}, which are nearly massless and interact very weakly with matter, making them hard to detect but potentially useful for indirect detection; and (3) Charged cosmic rays~\cite{Silk:1984zy, Stecker:1985jc, Ellis:1988qp} such as electrons, positrons, protons, and antiprotons. An excess of these particles in cosmic rays could indicate \ac{dm} interactions~\cite{Cirelli:2010xx}.

Different detectors and observatories are used to identify the potential signals from \ac{dm} interactions. Instruments like the Fermi Gamma-ray Space Telescope observe the sky for gamma-ray emissions, focusing on high \ac{dm} density regions like the Galactic Center or dwarf spheroidal and dwarf spirals~\cite{Pieri:2009je, Ackermann:2010rg, Fermi-LAT:2015att, Ajello:2015mfa, HAWC:2023vtl}. Facilities such as IceCube in Antarctica detect neutrinos by observing the Cherenkov radiation produced when neutrinos interact with ice~\cite{IceCube:2021xzo,IceCube:2023ies}. Instruments like the Alpha Magnetic Spectrometer (AMS-02) on the International Space Station measure the flux of cosmic rays, looking for anomalies that could indicate \ac{dm} annihilation or decay~\cite{AMS:2013fma}.

One key challenge for indirect detection is to distinguish potential \ac{dm} signals from astrophysical backgrounds. For instance, gamma rays can be produced by various astrophysical sources like pulsars, supernova remnants, or \ac{agn}, while neutrinos are produced in the Sun and in the atmosphere. Cosmic rays have numerous sources, including the Sun and \ac{sn}, making it difficult to attribute any observed excess to \ac{dm}. For this, fingerprints are searched through cross-correlation studies using data from different wavelengths and types of particles to identify coincident excesses that could point at \ac{dm}, and spectrum analysis where the energy spectrum of detected particles is analyzed to identify features characteristic of \ac{dm} annihilation or decay.

Currently, there is no definitive detection of \ac{dm} through indirect methods, but some intriguing signals and excesses warrant further investigation. For instance, Fermi has observed a gamma-ray excess in the Galactic Center which could be due to \ac{dm} annihilation~\cite{Fermi-LAT:2017opo}. Also, the AMS-02 has detected an excess of positrons in cosmic rays, which might indicate \ac{dm}~\cite{AMS:2013fma}, although astrophysical sources like pulsars are a competing explanation. Future instruments and missions, such as the Cherenkov Telescope Array (CTA)~\cite{CTAConsortium:2012fwj,CTA:2015yxo,CTAConsortium:2017dvg,Morselli:2023xeb,CherenkovTelescopeArray:2024osy} and the next-generation neutrino observatories will provide more sensitive measurements and potentially offer clearer insights into the nature of \ac{dm}.

Inferring the distribution and amount of \ac{dm} by its gravitational interactions is a complementary route to the detection of byproducts. Gravitational lensing is a very suitable probe because it only relies on light deflection, with no further need for the modeling of luminous matter effects. Since the first detection of a strongly lensed \ac{qso}~\cite{Walsh:1979nx, 1983Sci...219...54G}, strong gravitational lensing has successfully constrained masses of galaxies and galaxy clusters to about 10\%~\cite{Galan:2024gzo, Meneghetti:2016hcr}. While the initial lens modeling of the light-deflecting mass density profiles overestimated the \ac{dm} inferred from strong lensing, increasing complexity of the lens models has reduced the amount of \ac{dm} necessary to produce the observed, highly magnified and distorted images of background galaxies~\cite{Meneghetti:2006gh}. Yet, open questions still remain whether the amount of substructures observed in galaxy clusters agree with \lcdm~\cite{Wagner:2019orn, Griffiths:2021tim, Lin:2022mrn, Meneghetti:2023fug, Vegetti:2023mgp}.

\bigskip
\subsubsection{Warm dark matter \label{sec:WDM}}

\noindent \textbf{Coordinator:} Supriya Pan\\
\noindent \textbf{Contributors:} Biswajit Karmakar, Branko Dragovic, Cláudio Gomes, Cora Uhlemann, Davide Pedrotti, Emmanuel N. Saridakis, Ioannis D. Gialamas, Janusz Gluza, Massimiliano Romanello, Mehmet Demirci, Pran Nath, Reggie C. Pantig, Riccardo Della Monica, Torsten Bringmann, Venus Keus, Weiqiang Yang, and Wojciech Hellwing
\\

\noindent \paragraph{Constraints on WDM from observation of high-redshift galaxies and reionization}

The introduction of \ac{wdm} particles, with a mass of the order of a few keV, can alleviate some of the challenges of the \lcdm\ model at the kpc scales, namely the Missing Satellites problem or the  Too-Big-to-Fail problem~\cite{Bullock:2017xww}. Indeed, during the structure formation process, the higher thermal velocity of non-collisional \ac{wdm} particles determines a flow from overdense to underdense regions, canceling the cosmic perturbations with a physical size smaller than the so-called free streaming length, $\lambda_{\rm FS}$. This leads to a suppression in the matter power spectrum, $P(k)$, which becomes more important with decreasing \ac{wdm} particle mass, $m_{\chi}$, and has macroscopic consequences on the number density of low-mass haloes.

In addition to baryonic effects related to the star formation efficiency in \ac{dm} haloes, small-scale modifications in the halo mass function can also alter the high-$z$ number density of faint galaxies, resulting in a turn-over in the rest frame galaxy ultraviolet \ac{lf} (UV \ac{lf}) \cite{Castellano:2016hlk, Yue:2017hbz} and in a delay in the galaxy formation \cite{Dayal:2014nva, Menci:2018lis}. From the comparison between the abundance of faint galaxies and the number density of \ac{dm} haloes predicted in the context of \ac{wdm} cosmologies, it is possible to derive a lower limit on $m_{\chi}$, independent of the baryonic physics involved in the galaxy formation process \cite{Menci:2016eui}.

In turn, the progressive steepening of the UV \ac{lf} faint-end slope at high redshift confirms the fundamental role of faint galaxies during the \ac{eor}, when the intergalactic hydrogen passed from a neutral to an ionized state for the effect of energetic photons emitted by the
primeval sources of light, at $z\gtrsim 6$. Indeed, the UV \ac{lf} is directly linked to the star formation rate (SFR) \citep{Bouwens:2014fua} and so to the existence of a hot stellar population that acts as a source of ionizing photons responsible for the reheating of the \ac{igm} \citep{Bouwens:2015vha, Robertson:2015uda, Finkelstein:2019sbd}. In particular, their number density is given by $\dot{N}_{\rm ion} = f_{\rm esc} \xi_{\rm ion} \rho_{\rm UV}$, where the UV luminosity density, $\rho_{\rm UV}$, derived from the integral of the UV \ac{lf}, is multiplied by two quantities, namely the ionizing photon production efficiency, $\xi_{\rm ion}$, and the escape fraction, $f_{\rm esc}$. The former describes how efficiently it is possible to get ionizing photons from a UV continuum radiation field and depends on different astrophysical quantities, such as the initial mass function (IMF), the metallicity and the age of the stellar population. The latter is defined as the fraction of ionizing photons that escape from the emitting galaxies and contribute to the phase transition of the \ac{igm}. Due to the lack of information about the geometry of the \ac{ism} in high-$z$ galaxies, $f_{\rm esc}$ is not well constrained and it is difficult to model properly. Moreover, as a multiplicative factor of the total ionizing photons budget, it degenerates with the \ac{wdm} particles mass, thus 
summarizing most of our uncertainties about the reconstruction of the reionization history in both \lcdm\ and $\Lambda$WDM scenarios \cite{Carucci:2018tzf, Romanello:2021gnp}. Despite these astrophysical uncertainties, constraints on $m_{\chi}$ can be obtained through observations of the \ac{lf} \cite{Corasaniti:2016epp, Rudakovskyi:2021jyf, Lapi:2022aaq} and the SFR density \cite{Gandolfi:2022bcm} up to high redshift, and from the comparison of the expected electron scattering optical depth, $\tau_{\rm es}$, positive with an ionized \ac{igm}, with empirical data from Planck \citep{Lapi:2015zea}.

\paragraph{Candidates}

Traditionally, \ac{wdm} is defined as a fermion with 2 degrees of freedom that decouples relativistically and has a temperature to match the observed \ac{dm} relic density with zero chemical potential. This allows to express observational limits on the free-streaming length as 
constraints on the mass of the \ac{dm} particle. Lyman-$\alpha$ observations, for example, firmly exclude $m_{\rm WDM}>1.9 \, \mathrm{keV}$~\cite{Garzilli:2019qki} (though sometimes more aggressive bounds have been derived, reaching up to $m_\mathrm{WDM} > 5.3 \, \mathrm{keV}$~\cite{Palanque-Delabrouille:2019iyz}).

From the theoretical point of view, an excellent \ac{wdm} candidate consists of a fermion that is a singlet under the standard model gauge group, commonly referred to as sterile (or right-handed) neutrino~\cite{Drewes:2016upu,Dasgupta:2021ies}. The resulting mixing with active neutrinos, through the neutrino portal, is constrained to mixing angles too small for sterile neutrinos to fully thermalize in the early Universe. Instead, \ac{dm} production proceeds via oscillations, combined with neutral and charged current interactions of the active neutrinos with the SM heat bath~\cite{Dodelson:1993je}. Parameter regions for which this mechanism produces the correct \ac{dm} abundance in its simplest form are excluded~\cite{Abazajian:2017tcc}. 
These bounds can be evaded by enhancing the production through oscillations by introducing a large primordial lepton asymmetry~\cite{Shi:1998km}, new self-interactions of the SM~\cite{DeGouvea:2019wpf,Kelly:2020pcy}
or sterile neutrinos~\cite{Bringmann:2022aim,Astros:2023xhe}. Further alternative scenarios that remain viable include the decay of a new scalar state combined with subsequent entropy injection~\cite{Shaposhnikov:2006xi,Kusenko:2006rh,Petraki:2007gq} and thermal production via extended gauge sectors~\cite{Bezrukov:2009th,Kusenko:2010ik,Dror:2020jzy,Heikinheimo:2018luc}.
Common to these scenarios is that the \ac{wdm} mass bounds in general become model-dependent, and therefore have to be re-expressed by using model-independent bounds on the free-streaming length (for the case of self-interacting sterile neutrinos, notably, the bound on the sound-horizon due to late kinetic decoupling can be more stringent~\cite{Bringmann:2022aim}).

Fuzzy dark matter (FDM): Fuzzy \ac{dm} (FDM), also known as wave \ac{dm}, quantum wave dark matter, or ultra-light axion \ac{dm}, is an interesting area of research that bridges concepts from quantum mechanics, astrophysics, and cosmology, which offers potential solutions to some longstanding puzzles in our understanding of the Universe. It is a theoretical model proposing that \ac{dm} consists of ultra-light particles with boson masses on the order of $m_{\rm b} = 10^{-22}$\,eV \cite{Schive:2014dra,Hui:2016ltb}.

Unlike traditional particle \ac{dm} models, fuzzy \ac{dm} behaves more like a wave due to its extremely low mass. This wave-like nature leads to unique quantum mechanical effects on large scales~\cite{May:2021wwp,Gough:2024gim}. This wave-like behavior of FDM results in a quantum pressure that opposes gravitational collapse on small scales. This can suppress the formation of small-scale structures, potentially addressing the ``missing satellites problem'' observed in the \ac{mw} and other galaxies. In terms of halo structure, the density of \ac{dm} halos is predicted to have a central core with a soliton-like structure. This differs from the sharp cusps predicted by \ac{cdm} models and can help resolve some discrepancies between observations and simulations of galactic cores. While ultra-light, FDM behaves like a \ac{cdm} on cosmological scales, providing a good fit for large-scale structure formation and \ac{cmb} observations. Some works show hints that ultralight axions can improve consistency between \ac{cmb} and galaxy clustering data, potentially reducing the $S_8$ tension \cite{Rogers:2023ezo}. Tensions between the former and Lyman-alpha constraints could be alleviated with a component of ultralight or \ac{wdm} \cite{Rogers:2023upm}.
Regarding astrophysical tests, precise measurements of the distribution and movement of stars in galaxies, the structure of \ac{dm} halos, and gravitational lensing effects can be done. The low mass end for ultralight \ac{dm} has been ruled out by the dynamics of ultra-faint dwarf galaxies based on stellar kinematics in virialized halos \cite{Dalal:2022rmp} and the Lyman-alpha forest \cite{Irsic:2017yje} using hydrodynamical simulations, where uncertainties in the modeling remain.  Recently, constraints on the soliton mass using the Event Horizon Telescope (EHT) results for the black hole shadows of M87* and Sgr. A* was found \cite{Pantig:2022sjb}, confirming the theoretical model of FDM. It was also tested and constrained using the motion of the S2 star around Sgr. A* \cite{DellaMonica:2023dcw}. How the FDM soliton evolves while being accreted on a black hole helps determine the stages of accretion flow \cite{Cardoso:2022nzc}. Not only for black holes, the effect of solitonic quantum wave \ac{dm} was also explored through wormhole solutions \cite{Mustafa:2023qxf}. Indeed, these studies explored how the effects of FDM can manifest through astrophysical objects, acting as an alternative method for Earth-based laboratory detection.

Based on the existing literature, it is clear that \ac{wdm} models are potentially reached since some indications for alleviating the $S_8$ tension have been noted. However, the models in this category are not fully explored widely unlike other cosmological models. We anticipate that \ac{wdm} remains a potentially rich area that needs to be investigated widely aiming to understand how the tensions on $H_0$ and $S_8$ can be relieved. 
\bigskip
\subsubsection{Interacting and decaying dark matter \label{sec:I_D_DM}}

\noindent \textbf{Coordinator:} Elsa M. Teixeira\\
\noindent \textbf{Contributors:} Amare Abebe, Amin Aboubrahim, Andrzej Borowiec, Anil Kumar Yadav, Branko Dragovic, Brooks Thomas, Davide Pedrotti, Emmanuel N. Saridakis, Emre \"Oz\"ulker, Gaspard Poulot, Ioannis D. Gialamas, Ivonne Zavala Carrasco, Jose A. R. Cembranos, Keith R.~Dienes, Luis Anchordoqui, Marcin Postolak, Nikolaos E. Mavromatos, Paolo Salucci, Pran Nath, Rafael C. Nunes, Reggie C. Pantig, Riccardo Della Monica, Rishav Roshan, Simony Santos da Costa, Tomi Koivisto, Torsten Bringmann, V\'ictor H. C\'ardenas, Vasiliki A. Mitsou, Venus Keus, Vivian Poulin, William Giar\`e, and Wojciech Hellwing
\\

\paragraph{Interacting dark matter}

Models of \ac{idm} replace the \ac{cdm} component in \lcdm\ by posing either a self-interaction between \ac{dm} particles or interactions between \ac{dm} and the other species in the Universe. Consequently, the mass of the \ac{dm} particles bears some dependence on such interaction, defined in an \ac{flrw} background as 
\begin{equation}
    \dot{\rho}_{\rm IDM} + 3 H \rho_{\rm IDM} = Q_i\, ,
\end{equation}
where $Q_i$ embodies the interaction between \ac{dm} and the $i$-th species with energy density $ \rho_i$ (whose conservation equation will include a symmetric interacting term to ensure energy conservation). One key feature in favor of several models of \ac{idm} is that they can exhibit suppression of the matter power spectrum in relation to its \lcdm\ counterpart. This effect is relevant in terms of attempts to address the cosmological tensions, particularly the $S_8$ tension, even upon including constraints from the \ac{cmb} data. In the following, we present various types of \ac{idm} models and discuss their relevance for the resolution of cosmic tensions.

\subparagraph{Dark matter interactions with baryonic matter}

Theoretical models often consider \ac{dm}-baryonic matter (\ac{dm}-BM) scattering to modify the evolution of density perturbations. For instance, velocity-dependent cross-sections in these interactions can impact the growth rate of cosmic structures, affecting the matter power spectrum and, hence, the derived value for the $S_8$ parameter \cite{Boddy:2018wzy,Li:2022mdj,He:2023dbn}. These interactions may also modify the reionization history through changes in the ionization fraction, optical depth, and \ac{cmb} data-derived cosmological parameters \cite{Barkana:2018lgd}. Recent simulations suggest even weak \ac{dm}-BM interactions can leave observable effects on large-scale structures and the Lyman-$\alpha$ forest \cite{Gluscevic:2017ywp}. Screened \ac{dm} models propose local environment-dependent interaction strengths \cite{Borowiec:2023kmq} while models like matter bounce scenarios propose alternatives to the standard inflationary paradigm \cite{Postolak:2024xtm, Sakstein:2019qgn}. These \ac{dm}-BM interaction models must balance improving cosmological fits without violating constraints from other astrophysical observations, such as galaxy clustering and \ac{wl} surveys, thereby representing a promising but challenging research avenue \cite{Buen-Abad:2021mvc}. Further observational data from upcoming missions and theoretical refinement are needed to assess their viability as a potential solution to the cosmological discrepancies \cite{Dvorkin:2013cea}.

Recent astrophysical evidence shows discrepancies between observed mass distributions in galaxies and predictions from the collisionless \lcdm\ model \cite{Jungman:1995df}. 
Notably, a constant density region is often found in galaxies' innermost areas, contrary to the expected cuspy density profile \cite{Navarro:1995iw, Moore:1994yx, Navarro:1996bv, Oh:2008ww, Gentile:2004tb, Salucci:2018hqu}. 
Additionally, \ac{dm} halo parameters correlate with the structural parameters of the luminous component, such as the half-light radius $R_{50}$ and stellar mass $M_\star$ \cite{Kormendy:2004se, Donato:2004af, Donato:2009ab, Gentile:2009bw, DiPaolo:2019eib}. This entanglement between the dark and luminous components suggests some interaction between \ac{dm} and BM \cite{Sharma:2021jdw}, occurring only where the product of \ac{dm} and BM densities exceeds a threshold \cite{Salucci:2020eqo, Sharma:2021jdw}. Outside these regions, \ac{dm} halos retain their original NFW profiles. The \ac{dm}-BM interaction, observed over several Gyrs, does not affect the formation of galaxy dark halos but gradually modifies the halo density over time, erasing the initial cusp. This process may be observable in the recent mass distribution of high-redshift galaxies \cite{Sharma:2021jdw}. While the macroscopic effects of this interaction are clearer, the microscopic processes remain uncertain. Numerous candidates for \ac{idm} have been proposed (e.g., see Refs.~\cite{Shoji:2023pdg, Choi:2020pyy, Yengejeh:2022tpa, Wu:2022wzw}), and investigations continue. Observations suggest these interactions could occur in secluded regions of galaxies, such as stars, molecular clouds, neutron stars, black holes, and central supermassive black holes, aligning with various studies (e.g., see Refs.~\cite{Bell:2021fye, Leung:2022wcf}).

\subparagraph{Dark matter interactions with photons}

In Ref.~\cite{Bagherian:2024obh}, an extension of the \lcdm\ model is proposed where \ac{dm} couples with photons ($\gamma$), resulting in a nonconservation of particle numbers. This model suggests that \ac{dm} particles gradually decay into photons over time without significant scattering processes, slightly deviating from the standard cosmic evolution. It offers a potential solution to the cosmic tensions (see Fig.~\ref{fig:Intro_S8_whisk}). 
Based on the theoretical study of couplings between \ac{dm} and relativistic relics in Ref.~\cite{Akarsu:2018aro}, a lower limit on the \ac{dm} mass has been derived \cite{Zhou:2022ygn} for \ac{dm}-$\gamma$ interactions: $m_{\chi} > 8.7$ keV at 95\% CL (assuming it saturates 100\% of the observed relic density). Alternatively, one could derive a bound on the abundance of such \ac{dm} candidates depending on the mass~\cite{Heikinheimo:2018luc}. Observational constraints on \ac{dm}-$\gamma$ scattering cross-section were derived in Ref.~\cite{Stadler:2018jin}, with $\sigma_{{\rm DM}-\gamma} \leq 2.25 \times 10^{-6} \sigma_{\rm Th}, (m_{\rm DM}/\text{GeV})$ at 95\% CL. A multi-\ac{idm} framework, including interactions with photons, was explored in Ref.~\cite{Becker:2020hzj}, simultaneously addressing the $H_0$ and $S_8$ tensions. The prospects for testing \ac{dm}-$\gamma$ interactions through \ac{cmb} spectral distortions have been discussed in Ref.~\cite{Ali-Haimoud:2021lka}. The black hole shadow, a dark region observed against a bright background, is closely related to the photonsphere, the region where photons orbit the black hole. By performing backward tracing techniques, one can study how the shadow deviates from the classical prediction due to \ac{dm} effects. Studies like Refs.~\cite{Xu:2018wow,Konoplya:2022hbl} have shown that \ac{dm}-$\gamma$ interactions in the photonsphere can alter the shadow's shape, illustrating how astrophysical objects can be used as independent \ac{dm} probes. Numerous studies have since explored different \ac{dm} models and their effects on black hole properties \cite{Hou:2018avu,Hou:2018bar,Haroon:2018ryd,Xu:2020jpv,Jusufi:2020cpn,Xu:2021dkv,Nampalliwar:2021tyz,Jusufi:2020zln,Konoplya:2021ube,Saurabh:2020zqg,Pantig:2022toh,Pantig:2022whj,Atamurotov:2021hck,Jusufi:2022jxu,Pantig:2021zqe,Liu:2022ygf,Pantig:2022sjb,Anjum:2023axh,Ovgun:2023wmc,Errehymy:2023xpc,Qiao:2022nic,Zhou:2022eft,Capozziello:2023rfv,Capozziello:2023tbo,Liu:2023xtb, DellaMonica:2023dcw,Yang:2023tip,Gomez:2024ack,Qiao:2024ehj,Macedo:2024qky,Wu:2024hxr,Pantig:2024rmr}, concluding that significant \ac{dm} concentrations are necessary for observable effects on a black hole's shadow \cite{Konoplya:2019sns}. 

\subparagraph{Dark matter interaction with dark radiation}

The possibility that the multiple cosmic discrepancies are tied to the existence of a dark sector filled with a thermal bath of dark relativistic species, also known as dark radiation (DR), interacting with \ac{dm} has been widely explored (e.g., see Ref.~\cite{Buen-Abad:2015ova,Lesgourgues:2015wza,Chacko:2016kgg,Buen-Abad:2017gxg,Buen-Abad:2018mas,Aloni:2021eaq,Joseph:2022jsf,Buen-Abad:2022kgf,Schoneberg:2022grr,Schoneberg:2023rnx,Buen-Abad:2023uva,Bagherian:2024obh}). In the standard model, the effective number of neutrinos accounting for the three neutrino species, $N_{\rm eff}=3$. Still, the presence of DR can modify this value, reducing the sound horizon and adjusting the inverse distance ladder calibration of the \ac{bao} and \ac{sn1}. This is discussed extensively in Sec.~\ref{sec:Extra_DoF}.  

The DR-\ac{dm} interaction has two primary effects: it introduces a drag term in the \ac{dm} Euler equation, reducing the growth of \ac{dm} perturbations and potentially explaining the low $S_8$ values measured by \ac{wl} surveys; and it influences DR perturbation dynamics, reducing their free-streaming, which can result in a larger contribution to $N_{\rm eff}$ and align with $H_0$ measurements from the S$H_0$ES experiment. Various models embody these ideas, such as the ``non-Abelian \ac{dm} model'' \cite{Buen-Abad:2015ova,Lesgourgues:2015wza,Buen-Abad:2017gxg}, ``partially acoustic \ac{dm}" \cite{Chacko:2016kgg}, ``cannibal \ac{dm}'' \cite{Buen-Abad:2018mas}, stepped-dark radiation \cite{Aloni:2021eaq,Joseph:2022jsf,Schoneberg:2022grr,Schoneberg:2023rnx,Buen-Abad:2022kgf,Buen-Abad:2023uva}, and neutrino-\ac{dm} interaction~\cite{Zu:2025lrk}. Interestingly, these models may also address other \ac{cdm} issues, such as the cusp/core problem, the so-called ``too big to fail" (TBTF) issue, and the missing satellites problem (see Ref.~\cite{Bullock:2017xww} for a review). Indeed, Ref.~\cite{vandenAarssen:2012vpm} is the first particle-physics proposal to address all three problems simultaneously. The model in Ref.~\cite{Bringmann:2013vra} integrates this into DR as sterile neutrinos, introducing late-time effects that influence $H_0/\sigma_8$. Ref.~\cite{Bringmann:2016ilk} offers a comprehensive Lagrangian-level classification of \ac{dm}-DR interaction models, showing how \ac{cdm} can mimic a \ac{wdm} cut-off in the matter power spectrum. 
Alternatively, model-independent analyses can be conducted through the ``effective field theory of structure formation'' (ETHOS) \cite{Cyr-Racine:2015ihg, Vogelsberger:2015gpr}, which provides a mapping from Lagrangian parameters of \ac{dm}-DR models to phenomenological parameters of the linear power spectrum, with numerical simulations extending this to the full non-linear power spectrum.

\subparagraph{Dark matter interaction with dark energy}

Interactions within the dark sector may help resolve various tensions simultaneously by leveraging on the unknown nature of \ac{de} and \ac{dm} (see Sec.~\ref{sec:IDE} for a more thorough account of \ac{ide} models). For instance, in Refs.~\cite{Poulot:2024sex,vandeBruck:2022xbk,Teixeira:2024qmw}, the authors show that by applying fluid approximation methods for rapidly oscillating scalar field \ac{dm} coupled to scalar field \ac{de} through a Lagrangian approach, \ac{dm} can effectively be treated as a fluid on cosmological scales under the appropriate Hubble-averaged behavior. These models are expected to help alleviate cosmic tensions, namely on $S_8$, due to their contributions in suppressing the matter power spectrum on small scales, and on $H_0$ through the modified background dynamics.

A recent paper proposed a cosmological model of \ac{dm} and \ac{de} based on a particle physics Lagrangian \cite{Aboubrahim:2024spa} to address cosmic tensions. The model involves two interacting ultralight scalar fields, one for \ac{dm} and one for \ac{de}, with a coupling between them and was tested against various cosmological data sets, including Planck \cite{Planck:2018vyg, Planck:2018nkj, Planck:2019nip}, \ac{bao} \cite{Ross:2014qpa, BOSS:2012bus, BOSS:2016wmc, eBOSS:2020yzd, Howlett:2014opa, Beutler:2011hx}, and Pantheon and SH0ES data \cite{Brout:2022vxf, Riess:2021jrx}. It demonstrated an ability to alleviate the $H_0$ tension, reducing the discrepancy to about $\sim 2\sigma$, and also resolved the $S_8$ tension.

\paragraph{Decaying dark matter}

Another approach posits instead that a fraction of the \ac{dm} sector is undergoing a decay process into another species, falling under the broad heading of decaying \ac{dm} (DDM) models. This scenario is motivated by theoretical considerations, can explain specific experimental results, and addresses small-scale issues in the \ac{cdm} paradigm. Initially developed in the 1980s, the DDM model has gained renewed interest in light of cosmic tensions. Accordingly, such models conventionally contain two extra parameters: the lifetime of DDM $\tau$ plus the initial fraction of DDM to total \ac{dm} $f_{\rm DDM}$. The decay is expressed in \ac{flrw} as
\begin{equation}
    \dot{\rho}_{\rm DDM} + 3 H \rho_{\rm DDM} = - \Gamma_i\, \rho_{\rm DDM}\,,
\end{equation}
where $\Gamma_i$ is the decay width of DDM (related to its lifetime as $\Gamma = \tau^{-1}$) into the species $\rho_i$. There are several variations of the DDM hypothesis. However, these scenarios alone are unlikely to resolve the $H_0$ tension, which would require, for example, assuming an earlier decay of \ac{dm}.

\subparagraph{Dark matter decays into dark decay products}

Various scenarios involving the decay of an unstable component of multicomponent \ac{dm} into dark radiation have been proposed to address the $H_0$ and $S_8$ tensions \cite{Menestrina:2011mz,Hooper:2011aj,Gonzalez-Garcia:2012djt,Berezhiani:2015yta,Vattis:2019efj,Enqvist:2015ara,FrancoAbellan:2020xnr,FrancoAbellan:2021sxk,Liu:2021mkv}. For short-lived particles ($\Gamma \gtrsim 10^6~{\rm Gyr}^{-1}$), \ac{dm} decays into dark radiation at early-times ($t \ll t_{\rm LS} \sim 1.17 \times 10^{13}~{\rm s}$, where $t_{\rm LS}$ denotes the time of last scattering), increasing the expansion rate and reducing the sound horizon size $r_s$, thus increasing $H_0$ \cite{Menestrina:2011mz,Hooper:2011aj,Gonzalez-Garcia:2012djt}. 
Since the angular size of the sound horizon at the last scattering surface  $\theta_{\rm LS} \equiv r_{\rm LS}/D_M(t_{\rm LS})$ is a \ac{cmb} observable that must be kept fixed, a reduction of $r_{\rm LS}$ simultaneously decreases the comoving angular diameter distance from a present-day observer to the last scattering surface $D_M(t_{\rm LS})$, and increases $H_0$.
For long-lived particles, \ac{dm} is depleted into radiation after $t_{\rm LS}$, shifting matter-\ac{de} equality to earlier times and allowing a late-time increase in $H_0$ \cite{Berezhiani:2015yta,Vattis:2019efj}. Moreover, two-body decays that transfer energy from \ac{dm} to DR, reducing the late Universe matter content, help accommodate local $S_8$ measurements \cite{Enqvist:2015ara,FrancoAbellan:2020xnr,FrancoAbellan:2021sxk}. For $\Gamma \gtrsim H_0 \sim 0.7~{\rm Gyr}^{-1}$, most unstable \ac{dm} particles have decayed by redshift $z=3$, impacting observations like those from IceCube if sterile neutrinos play the role of DR \cite{Anchordoqui:2015lqa,Anchordoqui:2021dls}. If $\Gamma \lesssim H_0$, only a fraction of \ac{dm} particles have decayed. Recent \ac{cmb} data severely constrain the fraction of unstable \ac{dm} in these scenarios \cite{Chudaykin:2016yfk,Poulin:2016nat,Clark:2020miy,Chudaykin:2017ptd,Nygaard:2020sow,Anchordoqui:2020djl,Heikinheimo:2018luc}. Short-lived particles are constrained by \ac{cmb} polarization \cite{Nygaard:2020sow, Anchordoqui:2020djl}, and low redshift \ac{dm} depletion reduces \ac{cmb} lensing power, conflicting with Planck data \cite{Chudaykin:2016yfk, Poulin:2016nat, Clark:2020miy}. Including \ac{bao} measurements further tighten constraints on the fraction of long-lived particles \cite{Chudaykin:2017ptd,Nygaard:2020sow}. Current bounds make a decaying \ac{dm} solution to the $H_0$ tension unlikely, though a combination of scenarios with multiple decaying \ac{dm} particles (for example decays into a combination \ac{cdm} and \ac{wdm} \cite{Davari:2022uwd}) at different epochs may effectively ease this tension.

More promising is the role that DDM can play regarding the $S_8$ tension, especially in models where the (dark) decay products are massive particles. These products get a velocity kick $\epsilon \equiv v/c$ from the decay, suppressing structure growth on scales below their free-streaming length while behaving like regular \ac{cdm} at the background level, thus avoiding the strong constraints from \ac{bao} data. These scenarios have been tested against various cosmological data \cite{FrancoAbellan:2020xnr, FrancoAbellan:2021sxk}, including the full shape of \ac{boss} power spectrum \cite{Simon:2022ftd} and \ac{kids} \ac{wl} measurements \cite{Bucko:2023eix}, and shown to reduce the $S_8$ tension to $\sim 1.5\sigma$ for lifetimes $\Gamma^{-1} \simeq 120$ Gyr and velocity kick $\epsilon \simeq 1.2\%$ \cite{Simon:2022ftd}. In addition, depending on the lifetime and kinetic energy of the decay, DDM can address or mitigate various tensions related to small-scale structures and galaxy formation \cite{Sigurdson:2003vy, Cembranos:2005us, Kaplinghat:2005sy, Strigari:2006jf, Cembranos:2007fj} such as the missing satellites problem \cite{Klypin:1999uc, Moore:1999nt}, by estimating the power spectrum cut-off scale.

It is also worthwhile to move beyond the specific idea of decaying particle \ac{dm} and consider a more generic parameterization to describe the potential conversion of a \ac{dm} component to DR, with a transition rate that may not necessarily follow exponential decay~\cite{Bringmann:2018jpr, DES:2020mpv}. This approach includes traditional decaying \ac{dm} but also encompasses scenarios such as merging primordial black holes (PBHs) or late-time \ac{dm} self-annihilation in the presence of long-range forces. In these generalized scenarios, the $S_8$ and $H_0$ tensions can be mitigated in somewhat different ways, though a complete resolution of these tensions remains unlikely.

\subparagraph{Dynamical dark matter}

DDM models the decay of a \ac{dm} ensemble across epochs, balancing the collective lifetimes against cosmological abundances among various \ac{dm} components \cite{Dienes:2011ja}. One realization of such a DDM ensemble that has been extensively studied consists of a tower of dark Kaluza-Klein states associated with a large extra dimension;  such a tower then allows for the possibility of ``intra-ensemble'' decays from heavier to lighter dark states \cite{Dienes:2011sa,Dienes:2020bmn}.  Such decays can modify the \ac{dm} velocity distribution over time \cite{Dienes:2020bmn} and potentially address the $S_8$ tension \cite{Anchordoqui:2022svl, Obied:2023clp}. This idea has recently found new relevance within the Swampland program \cite{Vafa:2005ui}, which suggests an extra mesoscopic dimension (``dark dimension'') in the micron range \cite{Montero:2022prj}, leading to a tower of weakly interacting light \ac{dm} particles which are KK excitations of the graviton \cite{Gonzalo:2022jac}.  While DDM scenarios of this sort may potentially address the $S_8$ tension, they are also subject to constraints on the expansion rate at late-times from, e.g., \ac{sn1} \cite{Desai:2019pvs}.

\subparagraph{Dark sectors with varying equations of state}

In the case of soft \ac{dm} and soft \ac{de} \cite{Saridakis:2021qxb,Saridakis:2021xqy} one considers that effectively the dark sectors have a different equation of state at large scales, namely at scales entering the Friedmann equations, and a different one at intermediate scales, namely at scales entering the perturbation equations, features that are typical in soft materials \cite{2011RvMP...83.1367S}. In this case, the perturbative-level properties are slightly changed, and one can easily alleviate the $S_8$ tension. It has also been shown that such scenarios can be compatible with \ac{jwst} observations, offering a potential explanation for the unexpectedly rapid emergence of massive galaxies at high redshifts \cite{Davari:2023tam}.

\subparagraph{Some applications of decaying dark matter}

\ac{dm} decaying during or after recombination can alter the \ac{cmb} power spectrum by injecting energy that reionises the \ac{igm}, constraining the \ac{dm} lifetime to $\tau_\text{DM} \gtrsim 10^{25}$ s for photon or $e^+e^-$ decay products \cite{Poulin:2016anj, Slatyer:2016qyl, Planck:2018vyg, Heikinheimo:2018luc}. Additionally, null detections of diffuse X$/\gamma$-rays impose strict constraints on the \ac{dm} lifetime. In \ac{dm}-dominated galaxies or clusters, $e^+e^-$ decay products can produce radio waves through electromagnetic interactions, observable by radio telescopes like the \ac{ska} \cite{Colafrancesco:2015ola}, providing a better probe for decaying \ac{dm} compared to current gamma-ray observations. A recent study \cite{Dutta:2022wuc} found that \ac{dm} decay width $\Gamma_\text{DM} \gtrsim 10^{-30}~{\rm s}^{-1}$ is detectable with \ac{ska}. These constraints were used in Refs.~\cite{King:2023ayw, King:2023ztb} to understand the scale of Quantum Gravity, assuming that it breaks all approximate global symmetries, including \ac{dm} stability. Moreover, null detection of X-ray signals has ruled out minimal \ac{dm} models like $\nu$MSM \cite{Asaka:2005pn} for \ac{dm} produced from active-sterile oscillation and has severely constrained the \ac{dm} mass above the MeV scale for feebly interacting massive particles produced from gauge boson decays \cite{Datta:2021elq}.

\bigskip
\subsection{Other solutions}
\subsubsection{Estimates based on a possible local void \label{sec:Local_voids}}

\noindent \textbf{Coordinator:} Indranil Banik\\
\noindent \textbf{Contributors:} Alireza Talebian, Bahman Khanpour, Ebrahim Yusofi, Harry Desmond, Paolo Salucci, Richard Stiskalek, Sahar Mohammadi, Sandeep Haridasu, Sergij Mazurenko, Sveva Castello, and Vasileios Kalaitzidis
\\

\noindent Cosmic voids may help to alleviate the Hubble tension. A void effectively adds negative mass to a localized region of an otherwise homogeneous universe. The repulsive gravitational effect of this negative mass causes matter to flow away from the void, further deepening it and enhancing its repulsive effect. Recent studies in a void-dominated late universe \cite{1988MNRAS.235.1301R, Yusofi:2019sai, Mohammadi:2023idz, Moshafi:2024guo} indicate that cosmic voids achieve maximum entropy and thermal equilibrium \cite{Ahmadi:2024ono, Shahriar:2024vqx}. This hierarchical evolution may lead to the formation of large spherical cosmic voids through the merging of smaller sub-voids via the void-in-void process \cite{Sheth:2003py, vandeWeygaert:2009hr, Contarini:2022nvd}. Studies suggest that clusters and voids can exert positive pressure individually \cite{Mohammadi:2023idz, Shahriar:2024vqx}. Their coexistence in the Universe's web-like structure results in opposing pressures: overdensities generate positive pressure that pulls them closer together, while voids create negative expansion pressure that separates galaxies. The outflow from a local void could inflate observed redshifts and thus estimates of $H_0$, though this effect is not sufficient to solve the Hubble tension in \lcdm\ \cite{Wu:2017fpr, Camarena:2018nbr}.

Near-infrared observations covering 90\% of the sky suggest that we live inside the Keenan-Barger-Cowie (KBC) void \cite{Keenan:2013mfa}. The results are deep enough to cover most of the galaxy \ac{lf}, making them representative of the underlying matter distribution. The KBC void or Local Hole is evident throughout the electromagnetic spectrum, including in X-rays \cite{Bohringer:2014okf, Bohringer:2019tyj}, optical \cite{1990MNRAS.242P..43M, 1990IAUS..139..269S}, infrared \cite{Huang:1996zb, Busswell:2003ta, Frith:2003tb, Frith:2004wd, Frith:2005et, Keenan:2013mfa, Whitbourn:2013mwa, Whitbourn:2016irk, Wong:2021fvu}, and radio wavelengths \cite{Rubart:2013tx, Rubart:2014lia}. Comparison with the Millennium XXL simulation \cite{Angulo:2012ep} suggests that the KBC void is $6\sigma$ discrepant with \lcdm\ \cite{Haslbauer:2020xaa}. Since the underlying observations are galaxy number counts in redshift space, the results are unaffected by the assumed $H_0$.

Various proposals have been made arguing that gravitationally driven outflows from the KBC void could solve the Hubble tension \cite{Keenan:2014qwa, Shanks:2018rka, Shanks:2019inu, Ding:2019mmw, Martin:2021wvb}. In a homogeneously expanding universe, we would expect that $cz' = \dot{a} = H_0$, where $z' \equiv \dd z/\dd r$ is the rate at which $z$ rises with $r$. However, if we are located near the centre of a large void, the outflow velocity would rise from zero at the void centre to some maximum further out, before gradually decaying away. The initial rise with $r$ would inflate $cz'$. To estimate how much, we note that the KBC void has an underdensity in redshift space of $\delta_{\mathrm{obs}} = 46 \pm 6\%$ (see figure~11 of Ref.~\cite{Keenan:2013mfa}). The actual underdensity is about half as much because of \ac{rsd} \cite{Kaiser:1987qv}, the reduction in distance to any fixed redshift due to outflow from a local void inflating $cz'$. The reduced volume reduces the number counts, making the underdensity appear deeper in redshift space than it actually is. Accounting for this and assuming the required reduction in comoving density from the nearly homogeneous early Universe must be due to an increase in comoving volume, \cite{Haslbauer:2020xaa} argued in their equation~5 that
\begin{eqnarray}
    \frac{cz'}{H_0} ~=~ \left( 1 - \delta{_\mathrm{obs}} \right)^{-1/6}\,.
\end{eqnarray}
Thus, we would generally expect the locally estimated $H_0$ (actually $cz'$) to exceed $H_0$ estimated from the \ac{cmb} or other high-redshift probes by $11 \pm 2\%$. This is only a rough estimate -- the actual result depends on how the void has evolved over cosmic history. This must differ from the growth of voids in \lcdm\ given that density fluctuations on the KBC void's comoving scale of 300~Mpc are very well observed at the time of recombination and match \lcdm\ expectations \cite{Tristram:2023haj}, implying that density perturbations on these scales grow faster than in \lcdm. It is suggested that this is caused by a modification to gravity on length scales $\gtrsim 100$~Mpc, which would not affect the early Universe as its smaller age reduced the cosmic horizon, the distance that light and gravity could have traveled \cite{Haslbauer:2020xaa}.

A further argument for this scenario is the observed bulk flow curve out to $z = 0.083$ or about 350~Mpc \cite{Watkins:2023rll} based on the CF4 galaxy catalog \cite{Tully:2022rbj}. The bulk flow measures the average vector velocity of tracers within a spherical region centered on the Sun, albeit using only line of sight velocities. Ref.~\cite{Watkins:2023rll} argue that the observed bulk flow on the largest probed scales is roughly quadruple the \lcdm\ expectation and thus in $>5\sigma$ tension with it (see also Sec.~\ref{sec:bulk_flow}). The bulk flow curve in a full-sky survey is independent of the assumed $H_0$, which sets the monopole of the velocity field, while the bulk flow is a function of the orthogonal dipole. There are additional subtleties when the sky coverage is not complete, but since CF4 covers most of the sky, a slight adjustment to the estimator is sufficient to deal with the small unobserved regions of the sky, which mainly lie at low Galactic latitudes (see figure~2 of Ref.~\cite{Tully:2022rbj}). However, the sky coverage becomes less complete at $z \gtrsim 0.05$ as CF4 mostly relies on SDSS. There can be spurious effects even at lower $z$ because CF4 is reliant on combining surveys, each with their own zero-point and limited sky coverage.

The local void scenario was explored in great detail by Ref.~\cite{Haslbauer:2020xaa}, who constructed semi-analytic models of a small initial underdensity at $z = 9$ and evolved it to today. Those authors used Milgromian \ac{mond} (\cite{Milgrom:1983ca, Famaey:2011kh, Banik:2021woo}) to enhance the gravity from any given (under)density distribution, allowing the exploration of self-consistent models with enhanced structure growth on the relevant scales. A good joint fit was obtained to the density profile of the KBC void and the observed magnitude of the Hubble tension. An important conclusion was that our location in the void needs to be fine-tuned at the 2\% level, which is not a strong argument against the model \cite{Haslbauer:2020xaa}. The velocity field in their model was recently calculated in more detail \cite{Mazurenko:2023sex}. The predicted bulk flow curve was found to agree well with the observations of Ref.~\cite{Watkins:2023rll}, which are themselves in good agreement with the measurements reported by Ref.~\cite{Whitford:2023oww}. The rising bulk flow curve implies we are located fairly close to the void center, minimizing any impact on the \ac{cmb} quadrupole due to gravitational lensing \cite{Alnes:2006pf, Nistane:2019yzd}. Ongoing work aims to perform this peculiar velocity inference at the field level rather than relying on the bulk flow summary statistic.

The $H_0$ tension should disappear at high $z$ once the density returns to the cosmic mean and the void's impact gradually decays. Indeed, the ages of the oldest Galactic stars and \ac{cc}s (Sec.~\ref{sec:CC}) suggest a low $H_0$ consistent with the \ac{cmb} value and well below the local $cz'$ \cite{Cimatti:2023gil, Xiang:2024cve, Cogato:2023atm, Guo:2024pkx}. In fact, the Hubble tension is largely a mismatch between the local $cz'$ and $H_0$ estimated in other ways, typically from higher $z$ data \cite{Perivolaropoulos:2024yxv}. This can be done assuming \lcdm, leading to the concept of $H_0(z)$, the value of $H_0$ inferred from data in a narrow redshift range centered on $z$. Although \lcdm\ predicts a flat curve, there is evidence for a declining trend \cite{Krishnan:2020obg, Krishnan:2020vaf, Dainotti:2021pqg, Dainotti:2022bzg, Schiavone:2022wvq, Montani:2023xpd, Colgain:2022rxy}. The most recent analyses carefully minimize covariance between $H_0$ estimates from different redshift bins, finding $6\sigma$ evidence for a declining trend \cite{Jia:2022ycc, Jia:2024wix}. Similar techniques can be applied to construct simulated $H_0(z)$ curves that map the decay of the void's impact on $z$ \cite{2025MNRAS.536.3232M}. Their figure~3 shows good agreement with the observed $H_0(z)$ \cite{Jia:2022ycc, Jia:2024wix}.

\ac{bao} observables also deviate from \lcdm\ in the manner predicted by the local void scenario \cite{Banik:2025dlo}. These discrepancies (especially in \ac{desi} data; \cite{DESI:2024mwx}) have alternatively been interpreted in terms of the \ac{de} density varying with time \cite{Giare:2024gpk, Rezazadeh:2022lsf, Wang:2024pui}. Regardless of the interpretation, the Hubble tension seemingly becomes weaker at high redshift, and not merely due to larger uncertainties \cite{Bousis:2024rnb}. A recent study examined a variable \ac{de} fluid with a quadratic \ac{eos} in a late-time context \cite{Khanpour:2017das, Moshafi:2024guo, Shahriar:2024vqx}. Such an \ac{eos} may emerge from cluster/void mergers and could represent a redshift-dependent \ac{eos} parameter for cosmic acceleration \cite{Yusofi:2019sai, Mohammadi:2023idz, Moshafi:2024guo, Shahriar:2024vqx}.

The main issue facing the void model at the moment is that the Hubble tension does not clearly decay at high $z$ in the analysis of \ac{sn} \cite{Kenworthy:2019qwq}, though there are some hints of such a decay \cite{Jia:2022ycc, Dainotti:2021pqg, Dainotti:2022bzg}. While the \ac{bao} ruler has a fixed comoving size at $z \lesssim 1000$, \ac{sn} are not standard candles but merely standardizable \cite{Phillips:1993ng}. The intrinsic \ac{sn} luminosity depends on color and stretch, or how rapidly the light curve decays \cite{1998A&A...331..815T, Brout:2022vxf}. There are not enough \ac{sn} in host galaxies with Cepheid distances to reliably constrain these dependencies, which must be obtained jointly with the other cosmological parameters in the Hubble flow sample. This is usually done assuming some parametrized form for the relation between luminosity distance and redshift derived from an isotropic and homogeneous Friedmann cosmological model \cite{Friedman:1922kd, 1924ZPhy...21..326F}, though with an additional free parameter in the \ac{de} equation of state $w$ \cite{March:2011xa, Marriner:2011mf, Kessler:2016uwi}. \ac{sn} calibrated through the distance ladder, in combination with \ac{bao} data and the \emph{Planck} distance priors \cite{Planck:2018vyg}, suggest compatibility with a homogeneous cosmology and indicate that a potential local void would not significantly impact the measurement of $H_0$ from \ac{sn} \cite{Castello:2021uad, Camarena:2021mjr, Camarena:2022iae} -- but this would need to be confirmed by less model-dependent approaches as a cross-check \cite{Lane:2023ndt, Seifert:2024bqr}. A particular concern is that the typical stretch and color of \ac{sn} exhibit strong trends with $z$ \cite{DES:2021yis, Nicolas:2020lql, Wiseman:2023fcn}. Due to the need to calibrate how both affect the intrinsic \ac{sn} luminosity, these trends may mask the decaying away of the excess redshift induced by a local void \cite{2025MNRAS.536.3232M}. One might think that neglecting an actually present local void would cause the optimal homogeneous Friedmann model to provide a poor fit to the data, but the \ac{sn} magnitude uncertainties are artificially inflated such that the reduced $\chi^2 = 1$, completely hiding any such issue \cite{Keeley:2022iba}. The results also change slightly if we assume that it is the color rather than magnitude that has an extra unknown source of noise, for instance, due to dust \cite{Mandel:2016rks}. There is ongoing work on revising the \ac{sn} fitting procedure to allow a local void to inflate redshifts in the nearby Universe, jointly considering cosmological and standardization parameters.

While the KBC void could potentially solve the Hubble tension, it is not evident in the reconstructed local density field \cite{Jasche:2018oym} and its existence has not been validated kinematically \cite{Kenworthy:2019qwq, Lukovic:2019ryg}. A recent analysis of galaxies using the Radial Tully-Fisher (RTF) relation \cite{Yegorova:2006wv} provides a very promising way to investigate the possibility \cite{Haridasu:2024ask}. They demonstrated comprehensively that a local void capable of alleviating the Hubble tension is highly unlikely to reside at $z \lesssim 0.01$ (as suggested by Ref.~\cite{Whitbourn:2013mwa}) because it would have been evident in the analysis of the local cosmic expansion through the RTF distance indicator. This agrees with previous studies over the same redshift range using standard \ac{sn} distance indicators and datasets \cite{Kenworthy:2019qwq, Lukovic:2019ryg} -- though a plausible local void would inflate redshifts much further out \cite{2025MNRAS.536.3232M}. Remarkably, the RTF indicator could easily become the best distance indicator for $z \lesssim 0.15$, where it could be applied to a galaxy sample several times larger than \ac{sn} datasets. In each galaxy, the RTF method would provide multiple distance measurements, each as accurate as those obtained from \ac{sn}.

The KBC void is unique among solutions to the Hubble tension in that it was proposed long before the Hubble tension became known. Its large size and depth and the anomalous bulk flow curve all suggest that structure formation on scales of several hundred Mpc is more efficient than expected in \lcdm, as also suggested by the high redshift, mass, and collision velocity of the El Gordo interacting galaxy clusters \cite{Asencio:2020mqh, Asencio:2023yul}. If so, the Hubble tension might be resolved without adjusting the background cosmology, preserving its excellent fit to high-redshift datasets and the ages of the oldest stars \cite{Cimatti:2023gil, Grillo:2024rhi, Guo:2024pkx}. In the future, important tests will be provided by techniques that extend beyond the void radius. For instance, peculiar velocity reconstructions using galaxy scaling relations already constrain $cz'$ out to $z = 0.1$ \cite{Boubel:2024cqw, Said:2024pwm, Scolnic:2024oth, Scolnic:2024hbh}. Extending them to $z \gtrsim 0.3$ would test the predicted decaying of the void's effect at larger $z$ \cite{2025MNRAS.536.3232M}. \ac{bao} measurements suffer the opposite problem: they are most accurate at $z \gtrsim 0.5$, well beyond the void's influence \cite{DESI:2024mwx}. It is crucial to extend \ac{bao} measurements to $z \lesssim 0.2$, where a local void should have an appreciable impact \cite{Banik:2025dlo}. More detailed analyses of existing datasets would also help, for instance obtaining bulk flow measurements using \ac{sn} to check results obtained so far using galaxies and extend them further out \cite{Kalbouneh:2022tfw, Sorrenti:2022zat, Hu:2023eyf, Hu:2024big, Sah:2024csa, Huang:2024erq}. The inferred bulk flows can serve as a consistency check on measurements using the far more numerous galaxies.
\bigskip
\subsubsection{Primordial magnetic fields \label{sec:Pri_mag_fields}}

\noindent \textbf{Coordinator:} Karsten Jedamzik\\
\noindent \textbf{Contributors:} Alireza Talebian, Anto Idicherian Lonappan, Cláudio Gomes, Gaetano Lambiase, Iryna Vavilova, Levon Pogosian, and Tom Abel
\\

\noindent The possible existence of \ac{pmf}s in the early Universe has been a research topic for decades (for further reviews see Refs.~\cite{Durrer:2013pga,Subramanian:2015lua,Vachaspati:2020blt}). The underlying question is if magnetic fields observed in galaxies, clusters of galaxies, filaments, and in particular voids of the extragalactic medium \cite{Elyiv:2009bx,Neronov:1900zz,Tavecchio:2010mk,Tavecchio:2010ja,Taylor:2011bn,Vovk:2011aa,Dolag:2010ni} could be the result of a magnetogenesis process in the very early Universe which magnetized the Universe already well before recombination. The alternative would be that all/or most of the observed magnetic fields are due to astrophysical processes such as dynamos and outflows. The impact of a putative \ac{pmf}s on the \ac{cmb} has therefore been studied in great detail \cite{Subramanian:1998fn,Jedamzik:1999bm,Durrer:1999bk,Seshadri:2000ky,Mack:2001gc,Subramanian:2002nh,Subramanian:2003sh,Mollerach:2003nq,Lewis:2004kg,Scoccola:2004ke,Sethi:2004pe,Kosowsky:2004zh,Kahniashvili:2005xe,Brown:2005kr,Zizzo:2005az,Chen:2004nf,Lewis:2004ef,Tashiro:2005hc,Yamazaki:2006bq,Kahniashvili:2006hy,Giovannini:2007qn,Seshadri:2009sy,Caprini:2009vk,Cai:2010uw,Trivedi:2010gi,Brown:2010jd,Shiraishi:2010yk,Shiraishi:2011dh,Trivedi:2011vt,Yamazaki:2010nf,Paoletti:2010rx,Shaw:2010ea,Kunze:2010ys,Pogosian:2012jd,Paoletti:2012bb,Kunze:2013uja,Shiraishi:2013wua,Trivedi:2013wqa,Ballardini:2014jta,Kahniashvili:2014dfa,Kunze:2014eka,Ade:2015cva,Ganc:2014wia,Chluba:2015lpa,Ade:2015cva,Zucca:2016iur,Sutton:2017jgr,Pogosian:2018vfr,Minoda:2020bod}. Most proposed effects result in upper limits around comoving $\sim 1\,$nG, which is however a field strength about two orders of magnitude stronger than that required for cluster magnetic fields $\sim 1\mu$G to be explained entirely by a \ac{pmf}s \cite{Banerjee:2003xk}. It has been realized that the anisotropies in the \ac{cmb} are particularly sensitive to one effect, the creation of small-scale baryon inhomogeneities by \ac{pmf}s \cite{Jedamzik:2011cu,Jedamzik:2013gua}. 
 
A \ac{pmf} of comoving field strength $\sim 0.05\,$nG on comoving length scales $\sim\,$kpc produces slightly non-linear baryon inhomogeneities on such scales during recombination.  This is possible since kpc scales are well below the photon mean free path such that magnetic pressure is counter-acted only by the small baryon pressure and not the photon pressure. This process is often referred to as ``baryon clumping''.

Note that any \ac{pmf} is characterized by a continuous spectrum, containing initially magnetic power on a large range of scales, either with a scale-invariant spectrum due to inflationary magnetogenesis or a very blue Batchelor spectrum due to magnetogenesis during cosmic phase transitions. In the latter scenario the bulk of the \ac{pmf} is dissipated as the Universe expands, and a comoving $\sim 0.05\,$nG field left shortly before recombination results in a $\sim 0.01\,$nG field at the present epoch, approximately the value to explain cluster magnetic fields. The process of \ac{pmf}-induced baryon clumping has been studied numerically and allowed to set stringent upper limits on the final total present \ac{pmf} strength of $\sim 0.01\,$nG for a Batchelor spectrum \cite{Jedamzik:2018itu}. This upper limit may be subject to factor two changes when the newer Planck \ac{cmb} data and/or a more realistic treatment of the \ac{pmf} effects on the \ac{cmb} are employed.

Recombination in a clumpy baryon medium proceeds faster than in a homogeneous medium due to the non-linearity in the recombination rate (i.e., $\alpha_e \langle n_e n_p \rangle > \alpha_e \langle n_e \rangle\langle n_p \rangle$ where $\alpha_e$ is the recombination coefficient and $n_e$, $n_p$ are electron- and proton- density, respectively, with $n_e\approx n_p$). Recombination therefore occurs earlier when \ac{pmf}s are present, such that the sound horizon $r_{\star}$ is reduced \cite{Jedamzik:2020krr}. A reduction of the sound horizon compared to \lcdm\ is the ingredient utilized by essentially all early-time solutions for the Hubble tension (see Sec.~\ref{sec:EDE} for an explanation). The effect of baryon clumping on the \ac{cmb} has been analyzed by several groups \cite{Jedamzik:2020krr,Thiele:2021okz,Rashkovetskyi:2021rwg,Galli:2021mxk}. Here the clumping was treated in toy three-zone models, averaging the ionization fraction from three different and independent regions with the average density the same as in an homogeneous universe. Due to the absence of detailed knowledge of the \ac{pmf}s generated baryon \ac{pdf} function, i.e., the probability to find a baryon at a certain density, and it's evolution, non-evolutionary models with an ad hoc \ac{pdf} were chosen. In Ref.~\cite{Jedamzik:2020krr} it was found that when Planck data in combination with three late-time $H_0$ determinations was confronted to such clumping models, clumping was preferred at the $\sim 3\sigma$ CL with inferred $H_0 \approx 69.8 - 71\,$\kms depending on the chosen baryon \ac{pdf}. Though the fit to the \ac{cmb} data was slightly worse than in \lcdm\ it was still reasonably good (a PTE value of 0.17 compared to 0.20 for \lcdm\ without \ac{pmf}s). On the other hand, when other data is included, such as \ac{bao} and \ac{sn}, slightly lower $H_0 \approx 69.7-70.6\,$\kms were inferred. It is now known that the combination of Planck \ac{cmb} data and \ac{bao} data essentially disallows $H_0$ as high as the SH0ES value $H_0\approx 73\,$\kms in all early-type solutions to the Hubble tension if the physical matter density parameter $\Omega_mh^2$ has a value comparable to the one in the best-fit \lcdm\ model \cite{Jedamzik:2020zmd}. When the model was confronted only to \ac{cmb} data (Planck in combination with \ac{spt3g} and/or \ac{act}), but no late-time $H_0$ measurements were used, clumping was no further detected but only limited from above \cite{Thiele:2021okz,Rashkovetskyi:2021rwg,Galli:2021mxk}. It was argued that clumping is heavily constrained by the reduced Silk damping in conflict with the \ac{act} data \cite{Thiele:2021okz}, see below however.

\begin{figure}[htbp]
    \centering
    \includegraphics[width=0.5\textwidth]{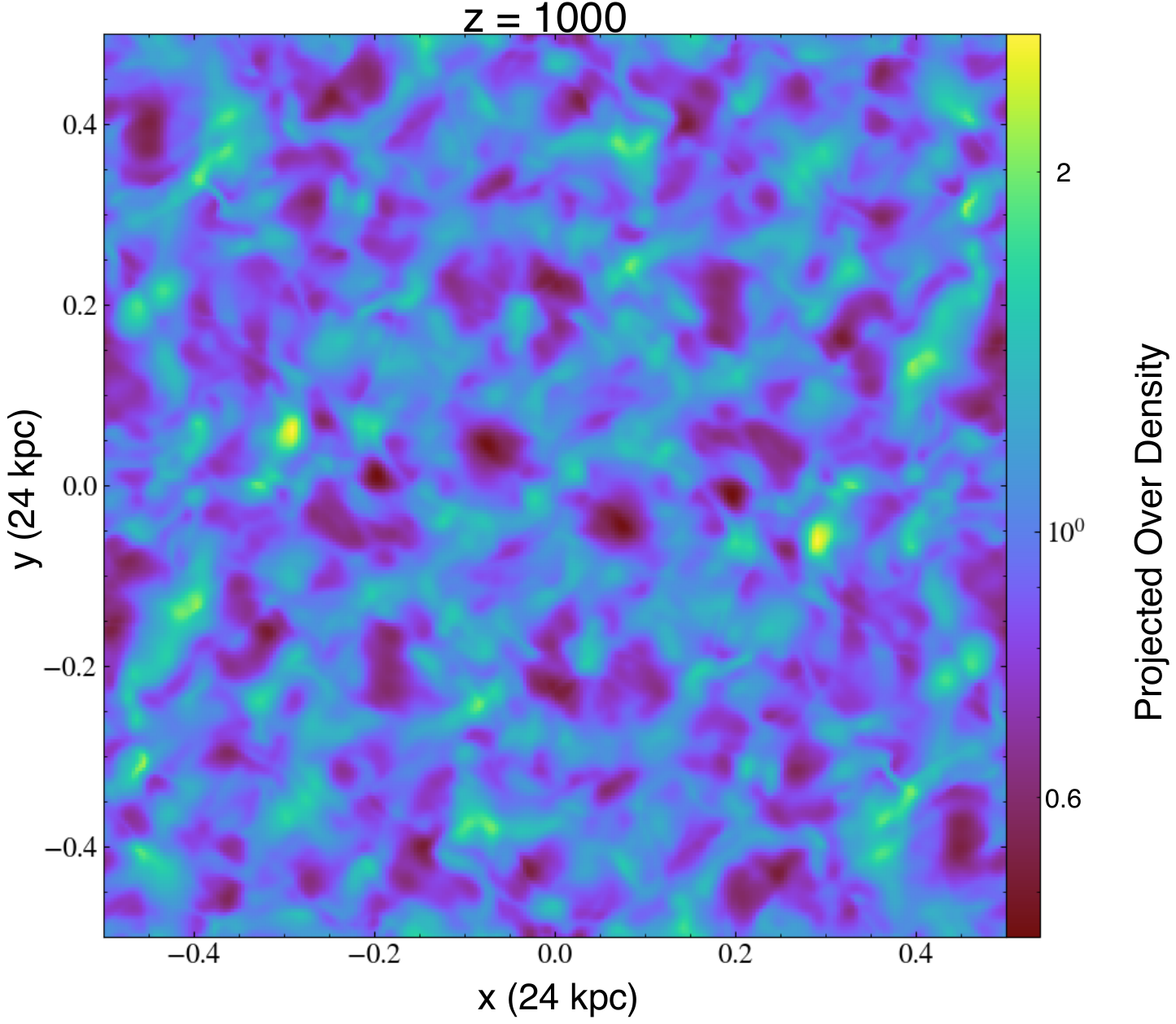}
    \includegraphics[width=0.45\textwidth]{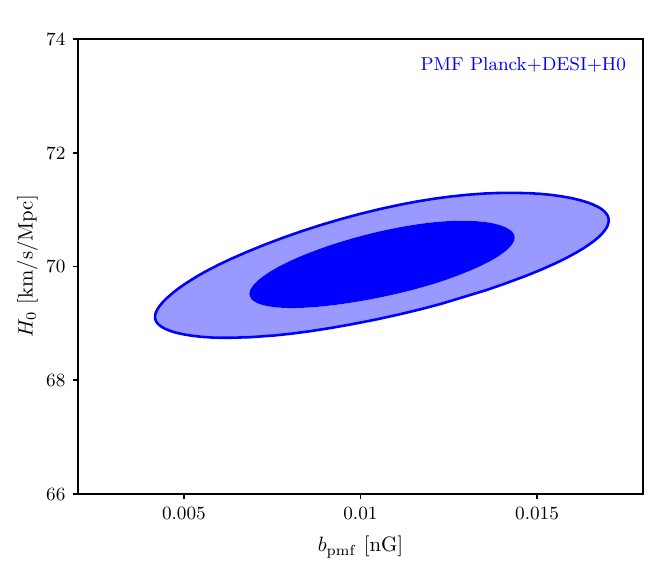}
    \caption{\emph{Left:} Results of an MHD simulation \cite{Jedamzik:2023rfd} starting at redshift $z = 4500$ with an initial comoving magnetic field of 0.526 nG with Batchelor spectrum and initially vanishing density perturbations and peculiar flows. The figure shows the (projected) density fluctuations generated at $z = 1000$ by the stochastic magnetic field.  \emph{Right:} Preliminary results of a \ac{mcmc} comparing the theoretical prediction of a \lcdm\ model with a \ac{pmf} with Batchelor spectrum to the combination of Planck \ac{cmb} data \cite{Planck:2018vyg}, \ac{desi}-1Y \ac{bao} data \cite{DESI:2024mwx}, and the SH0ES \cite{Riess:2021jrx} Hubble constant determination. It is seen that a final magnetic field of $B\approx 0.01\,$nG is preferred leading to Hubble constants of $H_0\approx 70\,$\kms, significantly larger than those predicted in  \lcdm\ without a \ac{pmf}. This field strength is close to that needed to explain observed cluster magnetic fields entirely by a \ac{pmf}. }
    \label{fig:figMHD}
\end{figure}

These promising trends tempted a recent much more sophisticated study of recombination in the presence of \ac{pmf}s \cite{Jedamzik:2023rfd}. Instead of using three-zone models, MHD simulations of \ac{pmf}s before, during, and after recombination were performed automatically leading to the correct baryon \ac{pdf} and its evolution. The simulations were performed down to redshift $z = 10$, providing a direct connection between the magnitude of clumping during recombination and the \ac{pmf} total strength at the onset of structure formation. A chemistry solver coupled to the MHD simulations was used to self-consistently compute the ionization fraction in each volume element. The average perturbation of the ionization fraction compared to a homogeneous no-\ac{pmf} model was computed and can be used to compute the anisotropies in the \ac{cmb} in a magnetized universe. The study revealed the possible importance of two further effects, namely the mixing of Lyman-alpha photons between different regions, and enhanced Lyman-alpha photon loss due to peculiar motions of the baryon fluid. Whereas the second effect was found to be subdominant for weak magnetic fields (final field $\sim 0.01\,$nG for Batchelor spectrum) the almost complete mixing of Lyman-alpha photons from regions at different densities leads to a further reduction of the ionization fraction during recombination. It also results in an enhancement of Silk damping such that magnetized models often have virtually identical Silk damping to non-magnetized ones.

The left panel of Fig.~\ref{fig:figMHD} shows the projected baryon overdensity, i.e., $(\rho -\langle\rho\rangle )/\langle\rho\rangle )$ of a magnetized universe at $z =1000$, with density clumps entirely created by the existing stochastic \ac{pmf} with Batchelor spectrum. There is currently no other than \ac{pmf} explanation for the putative existence of baryon isocurvature density fluctuations before recombination as other scenarios (pre-existing isocurvature fluctuations) are ruled out by \ac{bbn} or are not volume-filling enough (isocurvature fluctuations due to cosmic strings). The right panel of Fig.~\ref{fig:figMHD} shows the result if a magnetized \lcdm\ model is compared to Planck, \ac{desi}, and SH0ES data. It is found that a universe without \ac{pmf}s is essentially ruled out by this data combination. Moreover, the predicted Hubble constant is around $70\,$\kms. Though this value does not match the SH0ES value it is in accordance with other observational determinations of $H_0$ \cite{Lee:2024qzr,Freedman:2024eph}. It is intriguing to realize that the final present day magnetic field $B_0 \approx 0.01\,$nG is essentially that required to explain the origin of cluster magnetic fields. It furthermore is also an explanation of the putative observations of an extragalactic field by $\gamma$-ray observations. Further observational advances in small-scale \ac{cmb} anisotropy experiments (\ac{act}, \ac{spt3g}, S4) and $\gamma$-ray observations of a compact-size halo of powerful blazars (CTA with sensitivity less 0.1 TeV and sufficient angular resolution \cite{Korochkin:2020pvg}) have the potential to establish that the early Universe had already been magnetized.
\bigskip
\subsubsection{Feasibility of inflationary models to ameliorate the cosmic tensions \label{sec:Infla_models}}

\noindent \textbf{Coordinator:} Jaume de Haro, Luis A. Anchordoqui\\
\noindent \textbf{Contributors:} Ali \"Ovg\"un, Alireza Talebian, Andrew R. Liddle, Andrzej Borowiec, Anil Kumar Yadav, Anto Idicherian Lonappan, Antonio Racioppi, Benjamin L'Huillier, Carsten Van De Bruck, Cláudio Gomes, Cristian Moreno-Pulido, Dario Bettoni, David Benitsy, Deng Wang, Denitsa Staicova, Diego Rubiera-Garcia, Elias C. Vagenas, Emanuela Dimastrogiovanni, Emmanuel N. Saridakis, Giulia Gubitosi, Giuseppe Fanizza, Goran S. Djordjević, Ido Ben-Dayan, Ignatios Antoniadis, Ioannis D. Gialamas, Javier Rubio, Joan Sol\`a Peracaula, Juan Garcia-Bellido, Laura Mersini-Houghton, Leila L. Graef, M.M. Sheikh-Jabbari, Margus Saal, Marina Cort\^es, Micol Benetti, Milan Milo\v{s}evi\'{c}, Nikolaos E. Mavromatos, Nils A. Nilsson, Octavian Postavaru, Oem Trivedi, \"{O}zg\"{u}r Akarsu, Petar Suman, Rishav Roshan, Salvatore Samuele Sirletti, Saurya Das, Simony Santos Da Costa, and Venus Keus
\\

\paragraph{Modeling the inflationary expansion and its generalities in connection to the cosmic tensions} 

Inflation was first suggested to solve the flatness and horizon problems~\cite{Guth:1980zm}. The simplest inflationary model, known as ``vanilla'' or single-field inflation, describes the expansion of the Universe using a single scalar field $\phi$ with a suitable potential  $V(\phi)$. The dynamics of this field are governed by the conservation and Friedmann equations. The most popular approximation for the scalar field is the so-called \textbf{slow-roll approximation}, with the potential energy dominating the energy density of the Universe, i.e., $\dot{\phi}^2 \ll V(\phi)$ and  $\ddot{\phi} \ll 3 H \dot{\phi}$, where $H$ is the Hubble rate. This leads to an approximately exponential expansion, with the scale factor evolving as $a(t) \sim e^{Ht}$. It is possible to define the slow roll parameters: $\epsilon = M_{\rm Pl}^2 \ [V'(\phi)/V(\phi)]^2/2$  and $\eta = M_{\rm Pl}^2 \ [V''(\phi)/V(\phi)]$. Using them, one can relate inflationary observables, such as the scalar spectral index $n_s \approx 1 - 6 \epsilon + 2 \eta$ and the tensor-to-scalar ratio $r \approx 16 \epsilon$, to the shape of the inflaton potential $V(\phi)$~\cite{Liddle:1993nr}. A point worth noting at this juncture is that an approach to addressing the $H_0$ tension involves studying the slow-rolling dynamics of a self-interacting scalar field, which can lead to an emerging $H$ as a function of redshift; however, it should also be noted that while the scalar field could be responsible for a quintessence slow-rolling, it has not been associated thus far to inflationary models~\cite{Montani:2023ywn}.  

Vanilla inflation predicts a nearly scale-invariant spectrum of scalar perturbations with a slight red tilt, quantified by the scalar spectral index $n_s\approx0.96-0.98$. However, the slow-roll approximation is only one way to go, there are other examples such as fast-roll approximation \cite{Linde:2001ae}, constant-roll \cite{Motohashi:2014ppa,Guerrero:2020lng}, rapid-turn inflation \cite{Bjorkmo:2019qno,Anguelova:2022foz}, 
scalar field inflation in the presence of a non-minimal coupling between matter and curvature~\cite{Gomes:2016cwj}
symmergent inflation \cite{Cimdiker:2020enx, Bostan:2023hjp}, gauge-field inflation \cite{Maleknejad:2011sq, Adshead:2012kp, Maleknejad:2012fw, Garretson:1992vt, Anber:2006xt,Barnaby:2010vf,Cook:2011hg,Sorbo:2011rz, Anber:2012du,Adshead:2012kp,Dimastrogiovanni:2012ew,Adshead:2016omu,Dimastrogiovanni:2016fuu,Agrawal:2017awz,Caldwell:2017chz,Thorne:2017jft,Dimastrogiovanni:2018xnn,Fujita:2018vmv,Domcke:2018rvv,Lozanov:2018kpk,Watanabe:2020ctz,Holland:2020jdh,Domcke:2020zez,Iarygina:2023mtj,Ishiwata:2021yne,Durrer:2024ibi,Dimastrogiovanni:2024xvc,Gaztanaga:2024vtr}, $s$-dual inflation~\cite{Anchordoqui:2014uua,Anchordoqui:2021eox}, nonlinear electrodynamics inflation \cite{Ovgun:2016oit,Benaoum:2022uta,Otalora:2018bso,Ovgun:2017iwg,Kruglov:2015fbl,DeLorenci:2002mi,Novello:2003kh,Novello:2006ng,Vollick:2008dx}, and symmergent gravity theory \cite{Cimdiker:2020enx,Bostan:2023hjp}.

Many inflationary models predict primordial non-Gaussianities, which play a crucial role in understanding the early Universe's evolution and testing the inflationary paradigm. Non-Gaussianities are quantified by the non-linearity parameter $f_{\rm NL}$, which measures the amplitude of non-Gaussian fluctuations in the primordial curvature perturbations. These fluctuations require analysis of higher-order correlation functions, such as the bispectrum, beyond the two-point correlation function used for Gaussian fluctuations. Previous \ac{cmb} experiments, including \ac{wmap} and early releases of the Planck data, aimed to detect these non-Gaussianities. The most recent constraints from the Planck 2018 results provide $f_{\rm NL}$ values for the standard templates: $f_{\rm NL}^{\text{local}} = -0.9 \pm 5.1$, $f_{\rm NL}^{\text{equilateral}} = -26 \pm 47$, and $f_{\rm NL}^{\text{orthogonal}} = -38 \pm 24$, what indicates no statistically significant detection of primordial non-Gaussianities yet. However, it is important to keep in mind that multi-field or exotic single field inflationary models can lead to strongly scale-dependent non-Gaussianity~\cite{LoVerde:2007ri}, which could evade large-scale constraints while still significantly affecting small scales. Such scale-dependent primordial non-Gaussianities could actually alleviate the $S_8$ tension in the non-linear regime of structure formation, without affecting the linear regime \cite{Stahl:2024stz}. 

Future \ac{cmb} surveys, such as those planned with the Simons Observatory, \ac{cmbs4}, and the Euclid mission, promise significantly improved constraints on $f_{\rm NL}$. The Simons Observatory aims to achieve sensitivity levels that could reduce the uncertainties on $f_{\rm NL}$ to around $\Delta f_{\rm NL} \sim 1$. Similarly, \ac{cmbs4}, with its planned deep and wide \ac{cmb} observations, is expected to further tighten these constraints and provide unprecedented precision in measuring primordial non-Gaussianities. The Euclid mission, primarily designed for large-scale structure surveys, will complement \ac{cmb} observations by probing the distribution of galaxies and the large-scale structure of the Universe, thereby offering additional avenues to constrain $f_{\rm NL}$. These upcoming observations are expected to enhance our ability to test a broader range of inflationary models and deepen our understanding of models that could alleviate the $S_8$ tension.

\paragraph{Classes of inflationary models with ramifications for the cosmic tensions}\label{sec:inflation}

There are numerous extensions or alternatives to vanilla inflation, justified in various ways. In this subsection, we summarize the generalities of inflationary models that have a connection to the cosmic tensions and call attention to the role these specific models have in ameliorating the tensions.

\subparagraph{Higgs inflation} 

In the simplest Higgs inflation model \cite{Bezrukov:2007ep}, the Standard Model Higgs field is coupled non-minimally to gravity through a large dimensionless parameter, leading effectively to an almost flat Einstein-frame potential able to support inflation for a sufficient number of $e$-folds (for a review, see Ref.~\cite{Rubio:2018ogq}). One of the appealing aspects of this scenario is its minimalist approach, as it does not require the introduction of new fields or couplings beyond the Standard Model content. In particular, the reheating stage can be computed in detail, leading, in terms of the number of last $e$-folds $N_\star$,  to precise predictions for the spectral tilt $n_s = 1 -2/N_\star \,{\simeq}\, 0.966$ and the tensor-to-scalar ratio $r= 12/N_\star^2\, {\simeq} \,0.0033$ \cite{Bezrukov:2008ut,Garcia-Bellido:2008ycs,Repond:2016sol}. Additionally, the scenario offers a compelling connection with accelerator experiments, providing a detailed description of the power spectrum across a vast range of scales \cite{Bezrukov:2014ipa,Bezrukov:2017dyv}. The \ac{mcmc} analysis of current cosmological data confirms the model's observational viability and demonstrates that for the interval $N_\star \in [54.5,60]$, it can break down the $H_0 - \sigma_8$ correlation \cite{Rodrigues:2021txa}. Specifically, considering an instantaneous transition to radiation-domination, the $H_0$ tension is reduced to approximately $3\sigma$ \cite{Riess:2018uxu}, constraining $H_0= 67.94 \pm 0.45$\kms, while the value of $\sigma_8 = 0.793 \pm 0.003$ is in complete agreement with the \ac{kids}-1000 results \cite{Rodrigues:2023kiz}. The constrained $S_8=0.8019 \pm 0.0097$ is in 2$\sigma$ with Planck, \ac{des} and \ac{kids}-1000 data.

Several extensions of the minimal Higgs inflation scenario have been considered in the literature \cite{Shaposhnikov:2008xb,Garcia-Bellido:2011kqb,Germani:2010gm,Bauer:2010jg,Rasanen:2018ihz,Rubio:2019ypq,Shaposhnikov:2020gts,Piani:2022gon}. The so-called Higgs-Dilaton model addresses the current accelerated expansion of the Universe by introducing a unimodular constraint that manifests itself as the strength of a runaway Einstein-frame potential for a dilaton field \cite{Shaposhnikov:2008xb}. A significant feature of this scenario is the existence of specific consistency relations between the spectral tilt of primordial density perturbations, the tensor-to-scalar ratio, and the \ac{de} equation of state~\cite{Garcia-Bellido:2011kqb}. Predictions of these relations are in tension with direct measurements of $H_0$ using low-redshift \ac{sn}, but they align well with current \ac{cmb} observations and large-scale structure data, offering distinctive signatures that could be tested by future galaxy surveys \cite{Casas:2017wjh,Trashorras:2016azl,Casas:2018fum}.

\subparagraph{Starobinsky inflation}

Starobinsky cosmology is renowned as an excellent large-field slow-roll inflationary model~\cite{Starobinsky:1980te}, aligning very well with \ac{cmb} observations~\cite{Planck:2018jri, BICEP:2021xfz}. The key feature of the Starobinsky potential, essential for achieving slow-roll inflation, is its asymptotic behavior, where $V_s(\phi) \rightarrow \text{const.} $ as $\phi \rightarrow \infty $. Despite its celebrated success, the model does not seem to align well with proposals that could resolve the cosmic tensions. For example, it was noted in Ref.~\cite{Giare:2024akf} that Starobinsky inflation can hardly coexist with an \ac{ede} fraction $\sim  0.06$, which is only able to reduce the $H_0$ tension down to about $3\sigma$.

\subparagraph{Brane inflation} 

A theory that stems from the string theory, involves the interaction between branes in higher-dimensional space. Inflation occurs due to the motion of branes within a compactified extra-dimensional space, such as in the Klebanov-Strassler throat. The separation between branes acts as the inflaton, and their eventual collision ends inflation. This scenario can lead to observable levels of non-Gaussianity in the \ac{cmb} power spectrum and the distribution of large-scale structures in the Universe, as well as potentially detectable \ac{gw}s. The dynamics of the inflaton in brane inflation are governed by the Dirac--Born--Infeld (DBI) action, which is a type of non-canonical action~\cite{Silverstein:2003hf,Alishahiha:2004eh,Chen:2005ad,Chen:2005fe, Peiris:2007gz}. Matrix inflation (a.k.a. M-flation)~\cite{Ashoorioon:2009wa} may also be viewed as a multi-brane version of the brane-inflation scenarios. 

Tachyon cosmological models of inflation, based on the dynamics of a D3-brane in the bulk of the second Randall-Sundrum model, have been extensively considered \cite{Dimitrijevic:2018nlc}. Particle creation and reheating in a braneworld inflationary scenario with a tachyon field were extended to include matter in the bulk. It has been demonstrated how the interaction of the tachyon with the U(1) gauge field drives the cosmological creation of massless particles and where estimates have been found for the resulting reheating at the end of inflation \cite{Bilic:2017yll}. A holographic braneworld scenario with a D3-brane located at the holographic boundary of an asymptotic $AdS_5$ bulk \cite{Bilic:2016fgp} has shown significantly better agreement with Planck data \cite{Planck:2018jri}.

The cosmic tensions have been addressed in several models based on brane inflation. For example, in Ref.~\cite{Bag:2021cqm} it was shown that phantom braneworld prefers a
higher value of $H_0$ providing a much better fit to the local measurements. In addition, in Ref.~\cite{Lin:2022gbl} it was pointed out that the simplest D-term inflation can be consistent with the cosmic string bound provided by observations of gravitational. Consistency with observations requires a spectral index $n_s = 1$ in accordance with
the pre-recombination proposals to alleviate the $H_0$ tension (e.g., \ac{ede}).

\subparagraph{Quintessential inflation} 

After the discovery of the current cosmic acceleration \cite{Weinberg:1988cp,Lombriser:2019jia,Frieman:2008sn,Riess:2019cxk}, several theoretical mechanisms were developed to explain it. Quintessence models broadly encompass scenarios with one or more scalar fields with appropriate potential/dynamics that can drive the late-time cosmic acceleration, allowing for much richer possibilities beyond that simple case, in which one gets a dynamical equation of state parameter. An important feature of quintessential models is the need to examine ghosts and other non-physical features of their perturbations \cite{Creminelli:2008wc}. For recent applications in cosmology, see Refs.~\cite{Park:2021jmi,Goldstein:2022okd,Anchordoqui:2025fgz}. One of the simplest ways to do it is the so-called {\it quintessential inflation}~\cite{Peebles:1998qn,WaliHossain:2014usl,deHaro:2016hpl,deHaro:2016cdm,Haro:2018zdb,Haro:2019gsv,Haro:2019peq}, where the inflaton field is the only responsible for both accelerated phases. Several authors developed and improved the original Peebles--Vilenkin model~\cite{Dimopoulos:2001ix,Geng:2015fla,Dimopoulos:2017zvq,Geng:2017mic}, also embedding it in more fundamental frameworks \cite{Hossain:2014xha,Hossain:2014coa,Rubio:2017gty,Bettoni:2021qfs}. In particular, $\alpha$-attractors quintessential-inflation parameters can be constrained by stage IV surveys \cite{Akrami:2020zxw} and they have been tested using {\it Planck} and low-redshift data \cite{Giare:2024sdl} and scale invariant models \cite{Casas:2018fum}. Of particular interest here, the quintessential inflation, coming from the Lorentzian distribution introduced in Refs.~\cite{Benisty:2020xqm,Benisty:2020qta}, agrees with the recent observations and is in agreement with the SH0ES $H_0$ measurement~\cite{AresteSalo:2021lmp}. Indeed by confronting the model to cosmological observations, one obtains $H_0 = 72.25 \pm 0.74$\kms at 68\% CL~\cite{AresteSalo:2021lmp,AresteSalo:2021wgb}.

\subparagraph{K-essence, multifield inflation, and their generalizations} 

K-essence extends the standard model by including a non-canonical kinetic term in the Lagrangian, $P(\phi,X)$, where $X=\frac{1}{2} \partial_\mu \phi \partial^\mu \phi $~\cite{Armendariz-Picon:1999hyi,Armendariz-Picon:2000ulo, Rendall:2005fv, Langlois:2008mn,Romano:2020kmj}. K-essence models can help address cosmic tensions through their unique equation of state properties and field dynamics
affecting both early and late Universe evolution. In
Ref.~\cite{HosseiniMansoori:2024pdq} researchers demonstrate how k-essence
can modify the expansion history in ways that specifically target both
the Hubble and structure growth tensions. Similarly,
Ref.~\cite{Tian:2021omz} examines how an \ac{ede} model triggered by the
radiation-matter transition can be realized in k-essence. K-essence's
theoretical flexibility makes it particularly valuable for tension
resolution, as shown in Refs.~\cite{Jawad:2022dal, Hussain:2024qrd}, where
k-essence is shown to maintain consistency with both observational
constraints and theoretical requirements. The multifield dynamics have
been further studied in Ref.~\cite{Giare:2023kiv}. The generalized
k-essence can further contribute to models affecting the sound speed
of perturbations and the equation of state, which can modify
cosmological distance ladder and thus influence tension-related
parameters \cite{Staicova:2020wph,Dinda:2023mad,Staicova:2025lni}.

\subparagraph{Uniform five-dimensional inflation}

Compact extra dimensions can obtain large size by higher dimensional
inflation, relating the weakness of the actual gravitational force to
the size of the observable
Universe~\cite{Anchordoqui:2022svl,Anchordoqui:2023etp}. A solution to
the horizon problem implies that the fundamental scale of gravity must
be smaller than $10^{13}~{\rm GeV}$ which can be realized in a
brane-world framework for any number of extra dimensions. However, the
requirement of (approximate) flat power spectrum of primordial density
fluctuations consistent with present observations make this simple
proposal possible only for one extra dimension at around the micron
scale~\cite{Anchordoqui:2023etp,Antoniadis:2023sya}. Consistency with
2018 {\it Planck} \ac{cmb} data is supported by numerical simulations~\cite{Petretti:2024mjy,Anchordoqui:2024amx,Hirose:2025pzm}. After the end of
five-dimensional inflation, the radion modulus can be stabilized at a
vacuum with positive energy of the order of the present \ac{de}
scale. An attractive mechanism for radion stabilization is based on the contribution to the Casimir energy of fields
propagating in the bulk: the graviton, a real scalar field, and three
right-handed neutrinos~\cite{Anchordoqui:2023woo}.  The scalar field
has a potential holding two local minima with very small differences in
vacuum energy and bigger curvature (mass) of the lower one, and thus
when the false vacuum tunnels to its true vacuum state, the field
becomes more massive and its contribution to the Casimir energy
becomes exponentially suppressed~\cite{Anchordoqui:2024dqc}. The tunneling process then changes
the difference between the total number of fermionic and bosonic
degrees of freedom contributing to the quantum corrections of the
vacuum energy, yielding a sign-switching cosmological constant, $\Lambda_{\rm s}$, see Sec.~\ref{sec:Grad_DE}. Despite the fact that  when inflation comes to an end radion stabilization necessitates a dark sector with $\Delta N_{\rm eff} \sim  0.25$ (i.e., saturating {\it Planck}'s upper limit~\cite{Planck:2018vyg}), in the style of Ref.~\cite{Akarsu:2023mfb}, uniform five-dimensional inflation could
help simultaneously resolve the $H_0$, $S_8$, and $M_B$ tensions~\cite{Anchordoqui:2024gfa,Soriano:2025gxd}.

\subparagraph{Running vacuum models} 

The \ac{rvm} approach (see Refs.~\cite{SolaPeracaula:2022hpd,Sola:2013gha,Sola:2015rra} and references therein), is a framework for an effective description of cosmology, and it has implications for the time evolution of the fundamental constants \cite{Fritzsch:2012qc,SolaPeracaula:2023wqw}. The \ac{rvm} framework
\cite{Mavromatos:2020kzj} is based upon the idea that the cosmological vacuum energy densities and pressure are functions of even powers of the Hubble parameter $H(t)$ (and, under circumstances,  also of logarithmic non-polynomial corrections of $H(t)$). In the cosmic vacuum, which characterizes inflation, a de Sitter equation of state $p(H(t)) = - \rho(H(t))$, with
\begin{equation}
    \rho(H(t)) = \frac{\Lambda(H(t))}{\kappa^2} = \frac{3}{\kappa^2}\Big(c_0 + \nu \, H^2 + \alpha\, \frac{H^4}{H_I^2} + \dots \Big)\,,
\end{equation}
is satisfied, although, during each post-inflationary era, the total energy density and pressure is supplemented by the corresponding contributions of matter and/or radiation, as recent quantum-field-theory calculations in the \ac{rvm} context have shown~\cite{Moreno-Pulido:2020anb,Moreno-Pulido:2022phq,Moreno-Pulido:2022upl,Moreno-Pulido:2023ryo}. We remark that the coefficients $\nu$ and $\alpha$ are considered as constants during each epoch of the Universe's evolution; $H_I$ is the scale of inflation, which, according to data in Ref.~\cite{Planck:2018jri}, is currently set to $H_I \lesssim 10^{-5}\, M_{\rm Pl}$, in order of magnitude.
The important feature of the \ac{rvm} framework is that inflation does not require any external inflaton fields, but it is mainly due to the non-linear gravitational dynamics due to the $H^4$ and higher-order terms in the expression of $\rho(H(t))$~\cite{Lima:2013dmf,Perico:2013mna}, which dominate the early Universe eras. Note that the \ac{rvm} can provide a smooth evolution of the Universe, and can also account for its thermodynamic history and properties~\cite{Lima:2015mca,SolaPeracaula:2019kfm}.

This generic framework finds a concrete realisation in the context of Quantum Field Theory (QFT) in curved spacetime when the vacuum energy density is renormalized using the adiabatic renormalization method, in which a subtraction is performed at an off-shell scale $M$ subsequently identified with the Hubble rate $H(t)$ at each cosmic epoch. This eliminates the quartic mass contributions $\sim m^4$ (hence the need for fine tuning)  and induces a dynamics of the vacuum energy of the above form~\cite{Moreno-Pulido:2020anb,Moreno-Pulido:2022phq,Moreno-Pulido:2022upl,Moreno-Pulido:2023ryo}.  An alternative realisation, termed ``Stringy \ac{rvm}'' (StRVM), is found when \ac{rvm} is embedded in string theory cosmologies, under the assumption~\cite{Basilakos:2019acj,Mavromatos:2020kzj,Mavromatos:2021urx} that the early Universe consists of fields appearing in the bosonic massless ground-state multiplet of superstrings. The model assumes that the dilaton is stabilised to a constant value, through the minimization of an appropriate potential, possibly induced by string loops~\cite{Basilakos:2020qmu}. The pertinent string-inspired, low-energy gravitational theory is a Chern-Simons (CS) gravity~\cite{Jackiw:2003pm,Green:2012oqa}.  The CS nature of gravity arises from the linear coupling of the string-model independent axion~\cite{Duncan:1992vz,Svrcek:2006yi} to the CS gravitational anomaly, which is due to the Green--Schwarz counterterms~\cite{Green:1984sg} that modify the antisymmetric-tensor field strength as a consequence of the requirement of cancellation of gauge and gravitational anomalies in the extra dimensional string space. In the StRVM, the gravitational anomalies in the primordial Universe are not assumed to be canceled, although such a cancellation occurs at the exit from \ac{rvm} inflation, after the generation of chiral fermionic matter~\cite{Basilakos:2019acj,Mavromatos:2020kzj,Mavromatos:2021urx}, due to the decay of the \ac{rvm} vacuum. The \ac{rvm} inflation in the StRVM is induced by condensates of the gravitational CS term due to condensation of primordial \ac{gw}, which are formed during a pre-inflationary epoch~\cite{Mavromatos:2020kzj,Mavromatos:2021urx}. An estimate for the CS condensate can be provided within a second quantization of weak \ac{gw} formalism~\cite{Alexander:2004us,Lyth:2005jf, Dorlis:2024yqw}. It is found to be proportional to $H^4(t)$, which during inflation varies very slowly with the cosmic time. As shown in Ref.~\cite{Dorlis:2024yqw}, the formation of the CS condensate provides an approximately linear KR-axion monodromy potential, leading to inflation in the sense that the latter corresponds to a saddle point of the evolution of an appropriate dynamical system of variables. This result is general and refers to all scenarios with linear axion potentials, such as those obtained, for instance, from appropriate compactifications of string theories~\cite{McAllister:2008hb}. 

Finally, let us notice the intriguing potential implication of this framework on the alleviation of cosmological tensions~\cite{Gomez-Valent:2023hov,Gomez-Valent:2024tdb}. Indeed, in the context of StRVM, quantum graviton corrections induce in general corrections to the effective vacuum energy density of the non-polynomial form $H^{2n}\, {\rm ln}(H)$, $n \in \mathbb Z^+$. Such corrections have been computed explicitly in the context of dynamically broken supergravity models~\cite{Alexandre:2013nqa}, which can also be cast in an \ac{rvm} format~\cite{Basilakos:2015yoa} and can describe a pre-\ac{rvm} inflationary epoch of StRVM. In cases such corrections survive in modern epochs, they can provide a resolution of the Hubble and, under some circumstances, also of the growth-of-structure tension: $H_0=71.27^{+0.76}_{-0.73}$ and $\sigma_8=0.816^{+0.017}_{-0.015}$~\cite{Gomez-Valent:2023hov}. See also the ``phantom matter'' proposal of Ref.~\cite{Gomez-Valent:2024tdb}, inspired by the StRVM approach and providing a potential resolution of both tensions.
Subdominant logarithmic corrections on $H$ can also arise within the conventional \ac{rvm}, by integrating out matter fields in QFT of fermions and bosons in expanding spacetimes~\cite{Moreno-Pulido:2020anb,Moreno-Pulido:2022phq,Moreno-Pulido:2022upl,Moreno-Pulido:2023ryo}. 

In a related context to the \ac{rvm}, we can mention the works from Ref.~\cite{Brandenberger:2018fdd} and \cite{Comeau:2023euf} on the back-reaction effect of super-Hubble cosmological fluctuations. Ref.~\cite{Brandenberger:2018fdd} examines a scenario in which a large bare cosmological constant drives early accelerated expansion and shows that the backreaction effect of infrared fluctuations can reduce it. Such kind of models are expected to have an impact on the $H_{0}$ tension problem, as discussed in Ref. \cite{Alvarez:2025kma}.
\bigskip

\subsection{The cosmological principle \label{sec:Cosmo_prin}}

\noindent \textbf{Coordinator:} Eoin \'O Colg\'ain\\
\noindent \textbf{Contributors:} Alexander Bonilla Rivera, Andr\'as Kov\'acs, Anil Kumar Yadav, Antonio da Silva, Asta Heinesen, Daniele Oriti, Dario Bettoni, David Benisty, David L. Wiltshire, Diego Rubiera-Garcia, Dinko Milakovic, Du\v sko Borka, Ebrahim Yusofi, Elias C. Vagenas, Giuseppe Fanizza, Goran S. Djordjević, Hassan Abdalla, Ido Ben-Dayan, Iryna Vavilova, Jenny G. Sorce, Jenny Wagner, Jessica Santiago, John K. Webb, Jos\'e Pedro Mimoso, Juan Garcia-Bellido, Laura Mersini-Houghton, Leandros Perivolaropoulos, M.M. Sheikh-Jabbari, Maciej Bilicki, Marie-No\"elle C\'e l\'erier, Marina Cort\^es, Milan Milo\v{s}evi\'{c}, Nihan Kat{\i}rc{\i}, Nils A. Nilsson, \"Ozg\"ur Akarsu, Predrag Jovanovi\'c, Saurya Das, Tajron Juri\'{c}, V\'ictor H. C\'ardenas, Valerio Marra, Vesna Borka Jovanovi\'c, and Wojciech Hellwing
\\

\noindent 
In recent years the status of the cosmological principle (CP), a fundamental assumption in modern cosmology, has become less clear cut. However, what stymies debate is the absence of a consensus definition of the CP. The CP is commonly presented as the idea that the Universe is \textit{isotropic} (the same in every direction for a comoving observer who is at rest with respect to the
%who has no peculiar motion relative to the -- JSantiago: I would avoid mentioning '`peculiar motion'' here since the notion of a peculiar motion pre-requires the validity of the CP in order to make sense and it is better to keep a non-circular definition.
\emph{cosmic fluid} of the expanding Universe) and \textit{homogeneous} (the same at every point on a hypersurface corresponding to a given cosmic time) at suitably large scales. Homogeneity implies that there is no special location, such as a preferred observing position or center, while isotropy states that there is no special direction, such as an axis. These two statements do not automatically imply one another, and the possible departures from each of them are currently among the most prominent research topics in cosmology. 

Unfortunately, the terminology is loose and it is always possible to find a definition that is perfectly untestable. First, what is meant by large scales is rarely discussed. Secondly, given that we have only one vantage point in the Universe, testing homogeneity is difficult. In particular, depending on how one defines the latter, one could end up with a claimed ``large scale'' varying from $\sim 70 h^{-1}$ Mpc \cite{Hogg:2004vw, Yadav:2005vv, Scrimgeour:2012wt} up to $ \sim 260 h^{-1}$ Mpc \cite{Yadav:2010cc}. Thirdly, even when scientists argue that there is an anomaly with the CP, great care is required to separate \lcdm\ predictions from generic predictions of isotropic/homogeneous Universes. For example, $ \sim 260 h^{-1}$ Mpc \cite{Yadav:2010cc} is a claimed upper bound for the homogeneity scale in a \lcdm\ universe. Moreover, homogeneity is also challenged by large structures \cite{Gott:2003pf, Horvath:2014wga, Balazs:2015xsa, Lopez:2022kbz, Lopez:2024rzp}, but one inevitably must compare to the predictions of a specific \ac{flrw} model, i.e., \lcdm, and often, one either finds comparable structures in simulations or one can question the statistics \cite{Sheth:2011gi, Park:2012dn, Nadathur:2013mva, Ukwatta:2015rxa, Sawala:2025ajg}. Finally, a point very rarely discussed, and which may pass by as a minimal assumption, is the very existence of a ``cosmic fluid''. Besides how harmless those words might sound, what this actually assumes is the existence of a \emph{unique reference frame} which is at rest with respect to the average distribution of the \ac{cmb}, baryonic matter, \ac{dm} and \ac{de} simultaneously in different time periods of the history of the Universe. Looking closely at this definition, one then opens Pandora's box, in which to start, we have a lack of proper and robust definition of averages in GR, followed by questions regarding averaging scales and even if the averaging scales should be the same or not for different fluids.\footnote{Applying the CP for a void-dominated cosmology \cite{Yusofi:2019sai, Mohammadi:2023idz, Moshafi:2024guo, 1988MNRAS.235.1301R}, involving superclusters and cosmic voids, necessitates that cosmic voids tend towards achieving maximum symmetry. The result of such an evolution would be a vast spherical cosmic void, which is formed by the merging of smaller sub-voids (void-in-void process) \cite{Contarini:2022nvd, vandeWeygaert:2009hr, Sheth:2003py}. The density of the largest cosmic voids will be closer and closer to that of the Universe, making the CP more applicable to the present void-dominated Universe.} Admittedly, these are excellent theoretical questions, but if one approaches cosmology at a granular scale, constructing an observationally testable model is impossible. In cosmology one should not lose track of the \textit{grossly} simplified assumptions - justifiable on the relatively poor quality of astronomical data - being made.

\begin{figure}[ht]
     \centering
          \begin{subfigure}[b]{0.47\textwidth}
         \centering
         \includegraphics[width=\textwidth]{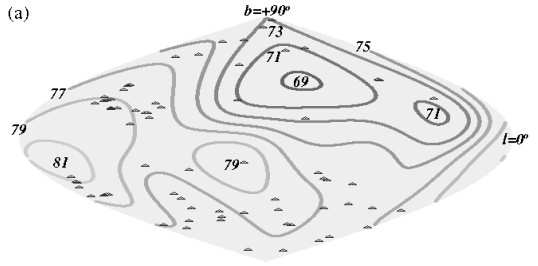}\vspace{0.2cm}
         \caption{\ac{hst} Key Project}
         \label{fig:McClure_H0}
     \end{subfigure}
     \hfill
     \begin{subfigure}[b]{0.47\textwidth}
         \centering
         \includegraphics[width=\textwidth]{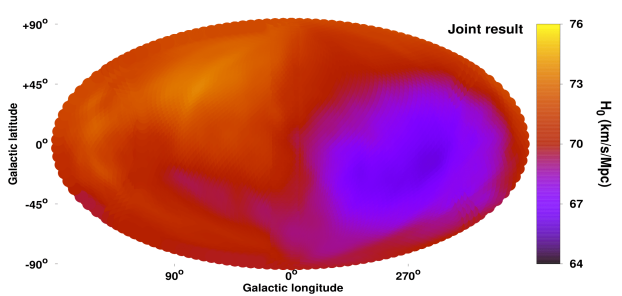}
         \caption{Galaxy cluster scaling}
         \label{fig:Migkas_H0}
     \end{subfigure}
     \hfill
         \caption{Left: Angular variations of $H_0$ on the sky in galactic coordinates in the aftermath of the \ac{hst} Key Project. Reproduced from Figure 1 (a) of Ref.~\cite{McClure:2007vv}. Right: Angular variations of $H_0$ on the sky in galactic coordinates from galaxy cluster scaling relations. Reproduced from Figure 7 of Ref.~\cite{Migkas:2021zdo}.}
        \label{fig:H0var_on_sky}
        \vspace{-0.4cm}
\end{figure}

The Universe is a gravitating system. Gravity is \ac{gr}, at least in a minimal setting. This necessitates a choice for a spacetime metric $g_{\mu \nu}$ to insert on the left hand side of Einstein's equation. Here the modus operandi is to start from the most symmetric possibility. This leads one to a maximally symmetric spacetime in 4D, namely Minkowski, anti-de Sitter, and de Sitter metrics. From these three, the latter is the only physically relevant possibility and appears in the asymptotic future of any universe with late-time accelerated expansion. 
The pursuit of symmetry also leads one to the \textit{perfect cosmological principle} (PCP), according to which the Universe is homogeneous and isotropic in space \textit{and} time. This means that the Universe in the large is also unchanging with time. The PCP was the basis of the steady-state cosmology, an alternative to the Big Bang theory \cite{Bondi:1948qk,Hoyle:1948zz}, but this required a physically contrived scenario whereby the density of matter remained unchanged due to a continuous creation of matter.

\subsubsection{FLRW spacetime}

In pursuit of a more physical model, one can always pursue classes of less restrictive spacetimes. The concession one makes is reducing maximal symmetry in the 4D spacetime to maximal symmetry in a 3D space. This leads one to the \ac{flrw} metric, a cornerstone of modern cosmology. The \ac{flrw} spacetime metric $g_{\mu \nu}$ is given by Ref.~\cite{Friedman:1922kd}
\begin{equation}
\label{eq:FLRW}
    d s^2 = g_{\mu \nu} d x^{\mu} d x^{\nu} = -c^2 dt^2 + a(t)^2 \left( \frac{dr^2}{1 - k r^2} + r^2 (d\theta^2 + \sin^2 \theta \, d\phi^2) \right)\,, 
\end{equation}
where \(a(t)\) is the scale factor, \(k\) represents the spatial curvature, and \((t, r, \theta, \phi)\) are the comoving coordinates. 
Combining \ac{flrw} with \ac{gr}, the Einstein equation reduces to the Friedmann equation, 
\footnote{The Friedmann equation can be derived by moving to the accelerated frame using the scale factor~\cite{Benisty:2019wpm,Guendelman:2020xfv}.} 
which has an important consequence that the Hubble constant $H_0 = H(z=0)$ arises as an integration constant \cite{Krishnan:2020vaf}. Note, Eq.~\eqref{eq:FLRW} simply specifies the spacetime and not the model. However, once one fully specifies the model, e.g., \lcdm, the onus is on the astronomy and cosmology communities to demonstrate that $H_0$ is a constant in cosmic time or redshift. If $H_0$ is not constant in a given model, one runs into a ``Hubble tension'' problem, e.g., see Refs.~\cite{Planck:2018vyg, Riess:2021jrx}.

This then brings us to the simplest way to test the metric in Eq.~\eqref{eq:FLRW}. $H_0$ can be determined model independently in the local Universe $z \lesssim 0.2$ using Cepheids, \ac{trgb}, or other possible primary distance indicators in order to calibrate \ac{sn1} \cite{Riess:2021jrx, Freedman:2021ahq}. 
Note that working locally, one can safely decouple the cosmological model \cite{Dhawan:2020xmp}. In short, one should make sure that there are no statistically significant angular variations of $H_0$ between different directions on the sky, because as discussed, $H_0$ must be a constant. Already, galaxy cluster scaling relations point to $\sim 9 \%$ changes in $H_0$ in the local Universe \cite{Migkas:2020fza, Migkas:2021zdo, Pandya:2024jqg}. Evidently, if the precision on $H_0$ is $\sim 5-10 \%$, this is no problem, but it may become problematic if one pursues $\sim 1\%$ precision. Since nobody builds the distance ladder on galaxy clusters, it is imperative to recover this result in \ac{sn1}. Preliminary results exist, but given the relatively small size of \ac{sn1} samples, which become worse when restricted to lower redshifts, angular $H_0$ variations at low redshifts are consistent with statistical fluctuations ($ < 2 \sigma$) \cite{McConville:2023xav, Lopes:2024vfz} (see also Refs.~\cite{McClure:2007vv, Krishnan:2021jmh, Zhai:2022zif, Dhawan:2022lze, Hu:2023eyf}).\footnote{One can also fit a dipole to \ac{sn1} \cite{Cooke:2009ws,Sorrenti:2022zat, Sorrenti:2024ztg, Sorrenti:2024ugq}. One may find \cite{Sorrenti:2024ugq} that \ac{sn1} and \ac{cmb} are not in the same frame, but that the difference can be interpreted as a bulk peculiar velocity $v_{\textrm{bulk}}$ consistent with \lcdm\ expectations. Note, these analyses are typically biased towards the lowest redshift \ac{sn1}, so one needs to study the redshift dependence of $v_{\textrm{bulk}}$ in shells, where overlap is expected with the cosmicflows program \cite{Watkins:2023rll}.} Given the better statistics in cosmicflows (CF) data, a data set comprising \ac{sn1} alongside other distance indicators, the observation of an anisotropic $H_0$ has recently been pushed to $3.9 \sigma$ \cite{Boubel:2024cmh}. It should be stressed that there is no guarantee that one can get to a $>3 \sigma$ variation in $H_0$ from \ac{sn1} alone, but the test can and should be done. Note that a statistically significant angular variation in $H_0$ would make both the Hubble constant and the Hubble tension ill-defined. Since, $S_8$ is a parameter within the \lcdm\ model, and the model builds upon \ac{flrw}, a breakdown of \ac{flrw} would also make $S_8$ tension ill-defined \footnote{As an aside, since an anisotropy makes a contribution to the energy budget of the Universe scaling as $a^{-6}$, which is relevant in the early Universe, it is plausible a new anisotropic cosmological model can be found that alleviates the tension with larger local $H_0$ determinations, e.g., see Refs.~\cite{Akarsu:2021max, Yadav:2023yyb}.}.  

In addition, locally, there are also claims from the CF program that there are anisotropic bulk flows \cite{Hoffman:2017ako}, which may no longer converge to \lcdm\ expectations \cite{Watkins:2023rll} (see also Refs.~\cite{Whitford:2023oww, Hoffman:2023pac}), thereby echoing a kinematic \ac{sz} anomaly, one disputed by the {\it Planck} collaboration \cite{Planck:2013rgv}, from a decade previously \cite{Kashlinsky:2008ut, Kashlinsky:2008us, Kashlinsky:2009dw, Kashlinsky:2010ur}. An upgrade of CF data to include WALLABY \cite{Koribalski:2020ngl, 2022PASA...39...58W, 2024PASA...41...88M}, FAST \cite{2024SCPMA..6719511Z} and \ac{desi} data  \cite{Said:2024pwm} recently found that the homogeneity scale is not yet recovered at $300 h^{-1}$ Mpc \cite{Courtois:2025xcs}, thereby strictly violating the $260 h^{-1}$ bound \cite{Yadav:2010cc}. In tandem, the inferred bulk flow was larger than \lcdm\ predictions, but remained consistent with the model. Note, in Ref.~\cite{Watkins:2023rll} an isotropic expansion or constant $H_0$ is assumed, but the bulk flow becomes anomalously large. CF data makes use of a host of distance indicators, including \ac{sn1}, so one can in principle recover the same result from \ac{sn1} alone.\footnote{Indeed, restricting the CF data to \ac{sn1}, one confirms the large bulk flow, but large errors make this consistent with \lcdm\ expectations. We thank Rick Watkins for private communication.} Here it is worth noting that \ac{sn1} redshifts are typically corrected for peculiar velocities of the host galaxy, so it is imperative to make sure that anomalously large bulk flows are not simply swept under the rug to reinforce an isotropic expansion (see criticism of \ac{sn1} peculiar velocity corrections in Refs.~\cite{Rameez:2019nrd, Rameez:2019wdt}). Another cautionary note on making peculiar velocities corrections regards the non-linearities that arise on local scales. This is an extremely delicate subject, especially given that the boundary scales where linear and non-linear regimes are applicable, together with the regime where perturbation theory is no longer valid, are still under investigation. In order to avoid this point, one must go to high redshift sources.

The other way to test the CP model independently is through aberration and Doppler effects with a large sample of sources. Although the Ellis-Baldwin test \cite{1984MNRAS.206..377E} was outlined last century, the test only became feasible this century \cite{Blake:2002gx, Singal:2011dy, Gibelyou:2012ri, Rubart:2013tx, Colin:2017juj, Bengaly:2017slg, Secrest:2020has, Siewert:2020krp,Secrest:2022uvx, Singal:2023lqm, Singal:2023wni, Wagenveld:2023kvi, Oayda:2024hnu}. While there may be outliers \cite{Gibelyou:2012ri, Cheng:2023eba, daSilveiraFerreira:2024ddn, Darling:2022jxt, Mittal:2023xub, Wagenveld:2024qhn}, there is a growing consensus that radio galaxy and \ac{qso} data sets do not inhabit the same rest frame as the \ac{cmb} \cite{Singal:2011dy, Rubart:2013tx, Colin:2017juj, Bengaly:2017slg, Secrest:2020has, Secrest:2022uvx, Singal:2023lqm, Singal:2023wni,  Wagenveld:2023kvi, Oayda:2024hnu}. Nevertheless, these are challenging studies, both technically and sociologically, thus it is always conceivable that observational \cite{Tiwari:2015tba, Mittal:2023xub, Abghari:2024eja, Wagenveld:2024qhn} and or theoretical systematics \cite{Dalang:2021ruy, Guandalin:2022tyl} (however see Ref.~\cite{vonHausegger:2024jan}) are at play. The earliest studies have typically employed frequentist statistics, but later papers have provided a Bayesian interpretation \cite{Dam:2022wwh, Mittal:2023xub, Oayda:2024hnu}. Taken at face value, the disagreement between \ac{cmb} and radio galaxies/\ac{qso}s on the cosmic dipole points to a breakdown of the CP -- and, therefore, of \ac{flrw}. The problem is interpreting this result. At what scales or redshifts does this happen? How badly is \ac{flrw} broken? None of this would be obvious without corroborating \ac{flrw} anomalies in the local Universe.  

So where are we now on \ac{flrw}? Both in the local Universe $z \lesssim 0.2$ and large samples of radio galaxies/\ac{qso}s with assumed median redshifts of $z \sim 1$ we see hints of a violation of \ac{flrw}. Given the rich structures in the local Universe \cite{Hoffman:2017ako, Giani:2023aor, Boehringer:2025uxi}, variations in $H_0$, or alternatively bulk flows, are expected. See Fig.~\ref{fig:H0var_on_sky}.  The important question is whether these effects deviate from the expectations of the \lcdm\ model. On the other hand, a mismatch in the cosmic dipole, could easily be explained by a relatively local effect and not an \ac{flrw} violation at large scales. Thus, it is of interest to study the Hubble diagram out to $z \sim 1$ in order to check if one recovers an \ac{flrw} universe. One could adopt the \lcdm\ model, e.g., see Refs.~\cite{Javanmardi:2015sfa, Krishnan:2021jmh, Zhai:2022zif,Luongo:2021nqh}, and check that variations in $H_0$ become less significant as effective redshifts of \ac{sn1} samples, etc, increase, but one can also tackle this problem model independently through cosmography \cite{Visser:2003vq} up to $z=1$, the radius of convergence of the expansion \cite{Cattoen:2007sk}. See Refs.~\cite{Bengaly:2024ree, Hu:2024qnx} for preliminary studies in this direction. Note, cosmography introduces additional parameters, especially if one wants to describe models beyond \ac{flrw} \cite{Heinesen:2020bej}, so one will require large data sets to conduct these studies. Nevertheless, the motivation should be clear. Current hints of \ac{flrw} violation at low and high redshift need to be connected in a common framework.    
The big question that needs to be addressed is how good is \ac{flrw} as an approximation to the physical Universe? While few scientists outside the field of cosmology, especially relativists, would challenge the assertion that \ac{flrw} is an approximate symmetry, since there is no argument that it is fundamental, this may be cold comfort for precision cosmology. $\sim 9 \%$ variations in $H_0$ \cite{Migkas:2020fza, Migkas:2021zdo, Pandya:2024jqg}, suggest that \ac{flrw} is a good approximation to $\lesssim 10 \%$. This is easy to square with the perceived isotropy of the \ac{cmb}, when one factors in residual asymmetries in \ac{cmb} data or \ac{cmb} anomalies \cite{Schwarz:2015cma}. A subset of the latter has persisted through three independent satellite experiments, namely COBE \cite{COBE:1992syq}, \ac{wmap} \cite{WMAP:2012nax} and {\it Planck} \cite{Planck:2018vyg}, allowing one to make the case that the Universe is unlikely to be statistically isotropic based on \ac{cmb} data alone \cite{Jones:2023ncn}. Unsurprisingly, $H_0$ can vary by up to $\sim 10\%$ along the axis of the hemispherical power asymmetry \cite{Fosalba:2020gls, Yeung:2022smn}, a recognized \ac{cmb} anomaly, and attempts to recover the \ac{cmb} dipole from aberration and Doppler boosts are contaminated at lower multipoles (larger scales) by this anomaly (see Fig. 3 of Ref.~\cite{Planck:2013kqc}).\footnote{For this reason, it is customary to remove large scales, e.g. see Refs.\cite{Ferreira:2021omv, Saha:2021bay}.} In addition, it is evident that \ac{wmap}/{\textit{Planck}} \ac{cmb} data prefers a \textit{phenomenological} Bianchi model over \lcdm\ \cite{Jaffe:2005pw, Jaffe:2005gu, Bridges:2006mt, Planck:2013okc},\footnote{These phenomenological Bianchi models cannot be seen as deformations of the \lcdm\ model since the cosmological parameters adopt different values.} which is only possible if there is a residual anisotropy in the data. Nevertheless, it is not clear if beyond \ac{flrw} models, especially the models in Sec.~\ref{sec:beyondFLRW}, can rival \ac{flrw} cosmology. This is not a statement about fits to data, this is the statement that at a purely technical level one would need a complete understanding of cosmological perturbation theory in these models. Given the anomalies in this white paper, it is hard not to contemplate that precision cosmology, defined through models where fitting parameters are constrained at the $\sim 1 \%$ error, may be at a crossroads. See Ref.~\cite{Aluri:2022hzs} for a recent overview of observations and the key claim that we have exceeded the precision where the approximate nature of the \ac{flrw} assumption is evident.

\subsubsection{Moving beyond FLRW} \label{sec:beyondFLRW}

In light of these \ac{flrw} anomalies, there is a growing interest in exploring more general metrics that relax the assumptions of homogeneity and isotropy. In this respect, if the CP is relaxed so that the homogeneity holds but not isotropy, then the corresponding exact solutions of Einstein's field equations are given by the class of cosmological models called Bianchi universes, named after the Italian mathematician Luigi Bianchi who first classified these spaces \cite{Bianchi_class1, Bianchi_class2,Akarsu:2019pwn,Akarsu:2021max}. Homogeneity in space implies that Einstein's field equations reduce from partial to ordinary differential equations in time, making Bianchi models exact solutions. A complete list of all such solutions for all Bianchi cosmological models from type I to type IX and for perfect fluid is given by Ref.~\cite{Stephani:2003tm}. Bianchi universes also contain, as a subclass, the standard isotropic \ac{flrw} universes. Although Bianchi models were previously considered to be inconsistent with observations, recent studies highlighted above challenge the isotropy assumption and revive interest in these models, which therefore remain widely studied \cite{Pontzen:2007ii,Russell:2013oda}. Ref.~\cite{Akarsu:2019pwn,Akarsu:2021max} presents an observational analysis of Bianchi type-I, -V, and -XI spacetime extensions of the \lcdm\ model. This section discusses some of the alternative metrics that have been proposed to address these issues. 

\paragraph{Lemaître-Tolman-Bondi (LTB) metric} 

The LTB metric \cite{Lemaitre:1933gd,Tolman:1934za,Bondi:1947fta} describes a spherically symmetric but radially inhomogeneous universe. It is given by
\begin{equation}
    ds^2 = -c^2 dt^2 + \frac{R'(r,t)^2}{1 + 2E(r)} dr^2 + R(r,t)^2 (d\theta^2 + \sin^2 \theta \, d\phi^2)\,,
\end{equation}
where \(R(r,t)\) is the radius of the spherical shell at time \(t\) and radial coordinate \(r\), and \(E(r)\) is an arbitrary function related to the energy of the shells. The LTB metric allows for variations in the density and expansion rate along different radial directions, providing a more flexible framework to model inhomogeneities in the Universe \cite{Lemaitre:1933gd,Tolman:1934za,Bondi:1947fta}.
The series of $N$-body simulations discussed in Ref.~\cite{Marra:2022ixf} explores the development of large-scale structures against an LTB background that incorporates a cosmological constant, specifically within the $\Lambda$LTB model framework.

Spherical inhomogeneities in LTB models can emerge either due to matter voids or due to Hubble scale spherical inhomogeneities of \ac{de}. In the latter case, such inhomogeneities can be supported by topological considerations, particularly in the context of recently formed global monopoles. These topological defects arise when the vacuum manifold of the quintessence scalar field has non-trivial \(\pi_2\) topological properties \cite{Barriola:1989hx,Perivolaropoulos:2008ud}. This scenario corresponds to the ``Topological Quintessence'' class of models, which may support deviations from the \ac{flrw} metric on cosmological scales.

In the context of topological quintessence, a global monopole formed during a recent phase transition with a core size comparable to the present Hubble scale could induce the observed accelerated expansion of the Universe. The monopole's scalar field is trapped near a local maximum of its potential in a cosmologically large region of space. This setup leads to an inhomogeneous but isotropic \ac{de} distribution, where the core of the monopole exhibits an effective cosmological constant-like behavior, while the outer regions revert to a matter-dominated Einstein-de Sitter universe.

An off-center observer within such a topological quintessence framework would naturally observe cosmic dipoles due to the asymmetric positioning relative to the monopole core. This asymmetry can manifest as dipoles in various cosmological observations, including the \ac{cmb} and large-scale velocity flows \cite{Sanchez:2010ng,Perivolaropoulos:2014lua, Maartens:2023tib}. 

Recent numerical simulations have explored the dynamics of such global monopoles minimally coupled to gravity in an expanding universe with initially homogeneous matter. These studies show that when the energy density of the monopole core starts dominating the background density, the spacetime in the core begins to accelerate its expansion in accordance with a \lcdm\ model with an effective inhomogeneous spherical \ac{de} density parameter \(\Omega_\Lambda(r)\) \cite{BuenoSanchez:2011wr}. Away from the core, the Universe appears as an Einstein-de Sitter Universe, while near the core, \(\Omega_\Lambda(r)\) reaches a maximum.

These findings suggest that topological quintessence models could provide viable explanations for certain cosmological observations and anomalies. The key is that the presence of a large-scale topological defect could naturally introduce the observed anisotropies without requiring exotic modifications of gravity or the introduction of multiple new fields.

\paragraph{Bianchi metrics} 

Bianchi models \cite{Bianchi_class1,Ellis:2006ba,Schucker:2014wca,Valent:2009htd,King:1991jd,Ringstrom:2000mk,Ashtekar:1991wa} generalize the \ac{flrw} metric by allowing for anisotropies while maintaining homogeneity. They are classified into nine different types based on their symmetry properties. For example, the Bianchi I metric is given by
\begin{equation}
    ds^2 = -c^2 dt^2 + a(t)^2 dx^2 + b(t)^2 dy^2 + c(t)^2 dz^2\,,
\end{equation}
where \(a(t)\), \(b(t)\), and \(c(t)\) are scale factors along the \(x\), \(y\), and \(z\) axes, respectively. These models can describe universes that have directional dependencies in their expansion rates, which might help explain certain observed anisotropies \cite{Ellis:1968vb,Cea:2014gga,Koivisto:2007bp,Aluri:2022hzs}. See Ref.~\cite{Akarsu:2019pwn,Akarsu:2021max} for observational analysis of Bianchi type-I extension of the \lcdm\ model.

\subsubsection{Mc-Vittie spacetime} 
The McVittie metric is the exact solution of Einstein's field equations describing a black hole or a massive object immersed in an expanding cosmological spacetime~\cite{McVittie:1933zz,Kaloper:2010ec}. The metric can be written by
\begin{equation}
    ds^2 =-(1-\Phi_N)\,dt^{2}+a\left(t\right)^2 \left[\frac{dr^{2}}{1-\Phi_N} + r^{2} \,d\Omega^2\right]\,,
\end{equation}
where $\Phi_N = 2 G M/(rc^2)$. In the low-energy limit ($\Phi_N \ll 1$ and $r H \ll 1$) the equation of motion reads~\cite{Sereno:2007tt,Faraoni:2007es,Nandra:2011ug}: $\ddot{r}/r = - G M/r^3 + \ddot{a}/a \;$. For the de Sitter–Schwarzschild metric the $\ddot{a}/a$ changes to $\Lambda c^2/3$. For $z \approx 0.67$ the $\ddot{a}/a$ acts as an attractive force for early cosmic times while for the de Sitter–Schwarzschild metric only the constant term $\Lambda c^2/3$ remains. Galaxy groups in the local Universe are bounds on the interplay between the expansion force and the Newtonian attraction~\cite{Benisty:2024tlv}. This gives another setup to determine the Hubble constant~\cite{Peirani:2005ti,Peirani:2008qs,Karachentsev:2008st,Penarrubia:2014oda,Teerikorpi:2010zz,DelPopolo:2021hkz,DelPopolo:2022sev}.

\paragraph{Inhomogeneous general relativity solutions} 

Beyond the specific cases of LTB and Bianchi metrics, there are more general inhomogeneous solutions in \ac{gr} that do not assume any specific symmetry. These solutions can be constructed using various techniques, such as analytic or perturbative methods or numerical relativity. One notable example of an exact solution of the \ac{gr} field equations, obtained with a dust gravitational source, is the Szekeres model, which generalizes the matter (+ cosmological constant) dominated \ac{flrw} solution by dropping any symmetry in the metric
\begin{equation}
    ds^2 = -c^2 dt^2 + e^{2\alpha} dx^2 + e^{2\beta} (dy^2 + dz^2)\,,
\end{equation}
where $\alpha$ and $\beta$ are functions of both spatial and temporal coordinates. These models allow for full inhomogeneity and anisotropy, providing a more general framework to describe the Universe \cite{Szekeres:1975dx}.
The Szekeres models can be divided into three classes: quasi-spherical,  quasi-plane and quasi-hyperbolic. The quasi-spherical Szekeres class of solutions, which includes the LTB model as a subclass, can therefore be viewed as a generalization of this LTB model where the spheres of constant $\{t,x\}$ coordinates are non-concentric and exhibit each a given mass dipole whose direction is rotated from one sphere to the other.
However, the surfaces $x = \text{const}$ within a space $t = \text{const}$ might be quasi-spherical in regions of the space and quasi-hyperbolic elsewhere, with a zero curvature at the boundaries. Every class possesses the \ac{flrw} solutions as a limit for particular forms of the functions defining its metric which can be obtained at some large scale for any of the classes, quasi-spherical, quasi-plane and quasi-hyperbolic. Hence, the possibility to use the Szekeres model for representing the inhomogeneous late Universe with a transition to homogeneity (\ac{flrw}) at some larger scale. These models have been used to study various cosmological phenomena, including the formation of large-scale structures and the impact of inhomogeneities on the \ac{cmb}. Now, in the era of precision cosmology, their application to the analysis of the largest cosmological data sets will have to be performed \cite{Celerier:2024dvs}.

\paragraph{Fully model independent cosmographic approach} 

Another possibility to test the CP in a fully covariant, model independent way without imposing any a priori metric, is using the general cosmographic approach developed by Kristian and Sachs \cite{Kristian:1965sz} (see also Refs.~\cite{Heinesen:2020bej, Clarkson:1999zq, Maartens:2023tib}). 
Its only assumption is that the galaxies can be described in terms of a pressureless dust fluid following flow lines that 
%movements of galaxies 
can be described in terms of a unique congruence -- 
an assumption made by all of the other theoretical approaches as well. The idea then lies on the fact that the matter flow, and its kinematical parameters -- the expansion $\Theta$, the shear $\sigma_{ab}$ and the vorticity $\omega_{ab}$ --
are a direct probe to the expansion of the Universe. One then can Taylor expand the generalized luminosity distance in redshift and obtain the generalized Hubble and deceleration parameters, which are now naturally directional dependent quantities. For an observer boosted with respect to the matter frame (one must always assume this is the case and directly measure the boost velocity), the generalized Hubble parameter has been proven to contain a monopole, a dipole induced only by the boost velocity, and a quadrupole component \cite{Maartens:2023tib}. This approach also provides a fully covariant non-perturbative generalization of the perturbative result obtained by Ref.~\cite{Nadolny:2021hti} in which they develop a way to disentangle the intrinsic dipole from a kinematically originated one \cite{Maartens:2023tib}, a very important need in order to test the CP.

\paragraph{Timescape cosmology} 
A new study of the Pantheon+ catalog has now claimed that an inhomogeneous alternative \cite{Wiltshire:2007jk,Wiltshire:2009db,Duley:2013ksw,Wiltshire:2013wta}  to the models discussed above fits better than \lcdm\ with strong to very strong Bayesian evidence ($\ln B\sim3$--$5$) \cite{Seifert:2024bqr}. Positive evidence ($\ln B\sim1$--$2$) remains even when restricted to \ac{sn1} with $z>z_{\rm min}=0.06$. In contrast to models based on a single metric, timescape combines small scale \ac{flrw} geometries---regionally valid on scales of $\sim$\,$3$--$30$ Mpc---via a Buchert average \cite{Buchert:1999er,Buchert:2001sa,Buchert:2019mvq}. A closure condition is needed to supplement the Buchert equations, without which its physical interpretation is open to debate \cite{Ishibashi:2005sj,Buchert:2015iva}. Timescape extends the Strong Equivalence Principle to a Cosmological Equivalence Principle to apply on small scales over which average isotropic motion in empty space is operationally indistinguishable from average isotropic expansion in nonempty space \cite{Wiltshire:2008sg}, leading to the quasilocal Hubble expansion condition as a closure relation for Buchert averages. The predicted variance in local Hubble expansion can be calibrated \cite{Duley:2013ksw} and tested observationally, leading to a potential resolution of the Hubble tension \cite{Lane:2023ndt}, and a natural framework for resolving dipole anomalies in terms of small scale non-kinematic differential expansion \cite{Bolejko:2015gmk}. This appears to be consistent with new analysis of void statistics in numerical relativity simulations using the full Einstein equations \cite{Williams:2024vlv}.

The timescape and \lcdm\ expansion histories differ by $\sim\,1$--$3$\% over small redshift ranges, but can be distinguished with a long enough lever arm. Independent projections made for the Euclid mission in 2014 \cite{Sapone:2014nna} show that with Euclid \ac{bao}s plus 1000 independent \ac{sn1} distances, the \ac{flrw} expansion history and the timescape alternative can be definitively tested via the Clarkson--Bassett-Lu (CBL) test \cite{Clarkson:2007pz}. The new Pantheon+ results for \ac{sn1} \cite{Seifert:2024bqr} have been independently confirmed independently with the \ac{des} survey \cite{DES:2024fdw}, where it was found that for events with $z>z_{\rm min}=0.033$ timespace is preferred over \lcdm\ with $\ln B=1.7$. Based on simple geometric scaling arguments, the \ac{des} analysis \cite{DES:2024fdw} finds that \ac{bao}s strongly favor \lcdm\ over timescape. However, in timescape the \ac{bao} must be extracted directly from raw galaxy clustering data \cite{Heinesen:2018hnh}, and independently recalibrated from the \ac{cmb} anisotropy spectrum and the sound speed in the primordial plasma. The nonbaryonic to baryonic matter densities ratio for timescape still has large uncertainties \cite{Duley:2013ksw}. Thus reanalysis of the matter matter model in the early Universe in conjunction with constraints from a variety of forthcoming datasets is an urgent goal for implementing the CBL test to definitively decide between \ac{flrw} and timescape by 2030.

\paragraph{Constraints from observations} 

Detailed cosmological observations have imposed constraints on some of the alternative models described above. For instance, the Generalized LTB model with inhomogeneous isotropic \ac{de} has been studied to understand its consistency with observations such as the Union2 \ac{sn1} data and the \ac{cmb} multipoles \cite{Grande:2011hm}. More recently, Ref.~\cite{Camarena:2022iae} confronted $\Lambda$LTB models to a host of data sets finding that the models could not resolve Hubble tension. It has been shown that for such models to be consistent with observations, the size of the inhomogeneity must be large, typically on the order of a few Gpc. Additionally, the observer must be located relatively close to the center of the inhomogeneity to avoid large dipole anisotropies. 

The exploration of these alternative metrics is crucial for addressing the current tensions in cosmology. By considering more general spacetimes, we can test the robustness of the CP and potentially uncover new physics that could reconcile discrepancies between observations and the standard model. As high-precision data continue to pour in, the development and testing of these models will be an essential part of the future of cosmology.

\subsubsection{What needs to happen going forward}

Even though a more precise definition of the CP took a little while to be developed \cite{1930MNRAS..90..668E, 1930MNRAS..90..668E} following the pioneering works of Vesto Slipher \cite{1915PA.....23...21S}, Henrietta Swan Leavitt \cite{Leavitt:1908vb}, Georges Lema\^{i}tre \cite{Lemaitre:1927zz} and Edwin Hubble \cite{Hubble:1929ig}, it was formulated in a time when, more than a principle, it was a \emph{need} for cosmology to go forward. Its historical importance and scientific contribution to the advancement of cosmology is undeniable. But, given the amount of present and future data, continuing to apply it without a proper investigation of its validity and limitations is no longer a matter of science, but faith.

Here we develop a brief list of the necessary requirements in order to properly test the CP:
\begin{itemize}
    \item Robust data with large sky coverage, in order to test for possible directional dependent effects;
    \item Model independent techniques for cleaning up possible foreground contamination which does not assume (or imposes) the CP;
    \item Model independent analysis and parameter estimation;
    \item The practice of clearly and forthrightly stating all of the (explicit and implicit) assumptions made throughout the analysis;
    \item A robust model independent theoretical method connected to observations, in a way that allows for the interpretation of the data.
\end{itemize}

Note that angular variations of $H_0$ on the sky and the cosmic dipole can both be studied independently. The former will benefit from large \ac{sn1} samples with excellent sky coverage from \ac{ztf}, Vera Rubin observatory, and Romans space telescope. The \ac{ska} will provide large radio galaxy samples allowing a definitive conclusion on the cosmic dipole anomaly \cite{Schwarz:2015pqa, Bengaly:2018ykb}.

\bigskip
\subsection{Quantum gravity phenomenology \label{sec:QG}}

\noindent \textbf{Coordinator:} Giulia Gubitosi\\
\noindent \textbf{Contributors:} Christian Pfeifer, Elias C. Vagenas, Gaetano Lambiase, Manuel Hohmann, Nikolaos E. Mavromatos, Saurya Das, and Vasiliki A. Mitsou
\\

\noindent One might reasonably conjecture that at least some of the \ac{mg} models discussed in the previous sections of this review might emerge in an appropriate limit from quantum gravity. In fact, for some of the models the connection is more apparent, as is the case, for example, of Ho\v{r}ava-Lifshitz gravity \cite{Horava:2009uw}, that introduces anisotropic scaling between space and time at high energies. This leads to a power-counting renormalizable theory of gravity that deviates from \ac{gr} at short distances. This model has been proposed by Ho\v{r}ava in Ref.~\cite{Horava:2009uw}, where an effective Quantum Gravity approach not requiring the Lorentz invariance at fundamental ultraviolet scales has been formulated. This invariance, however, emerges at large distances. It mainly aims to solve the high-energy issues suffered by \ac{gr} through a spacetime foliation capable of reproducing the causal structure out of the quantum regime. Basic foundations and applications of this approach can be found e.g., in Refs.~\cite{Horava:2009if, Lu:2009em, Calcagni:2009ar, Charmousis:2009tc, Brandenberger:2009yt, Sotiriou:2009bx, Cai:2009pe, Panotopoulos:2020uvq, Vernieri:2019vlh, Sotiriou:2014gna,Leon:2019mbo,Postolak:2024xtm}. This model is potentially capable of addressing the $H_0$ tension, as demonstrated in Ref.~\cite{DiValentino:2022eot}. Specifically, the authors obtain a positive result on the cosmic tensions between the Hubble constant $H_0$ and the cosmic shear $S_8$ due to a shift of $H_0$ towards a higher value. Moreover, in Ref.~\cite{Nilsson:2019bxv} the authors show that up to 36\% of the Hubble tension can be explained by Lorentz-violating effects in a Ho\v{r}ava–Lifshitz scenario. This, in fact, is a common feature of theories involving gravitational Lorentz violation, where local $G$ no longer equals cosmological $G$ \cite{Carroll:2004ai}.

However, a full-fledged fundamental theory of quantum gravity is still elusive. This motivates the adoption of a bottom-up approach, that is complementary to the top-down approach attempting to formulate fundamental theories and then working out their predictions in specific limits. In this bottom-up approach,  possible features of the quantum interaction and dynamics of gravity as well as particles and fields propagating on a quantum spacetime are described at an effective level via phenomenological models. This field of research, called quantum gravity phenomenology, successfully leads to observable predictions whose confirmation or constraints serve as guidelines for building a fundamental theory of quantum gravity, as it is explained in detail in the review in Ref.~\cite{Addazi:2021xuf} and white papers in Refs.~\cite{AlvesBatista:2023wqm,AlvesBatista:2021eeu}.

In the following, we will outline how phenomenological models of quantum gravity can affect the understanding of cosmological tensions and how cosmological observations may lead to the discovery of quantum aspects of gravity.

\subsubsection{QG modified gravitational dynamics}

\paragraph{Hubble tension and generalized uncertainty principle}

The \ac{gup} was originally motivated by considerations coming from string theory \cite{Amati:1988tn} and from the analysis of the relation between gravitational and quantum theories in black hole physics \cite{Maggiore:1993rv}. Subsequent work found relations to noncommutative geometry \cite{Quesne:2006is, Quesne:2006fs}. Recently, in Ref.~\cite{Lambiase:2017adh} it was shown that the deformation parameter entering the \ac{gup} can be related to the coefficients of the standard model extension (SME) in the gravity sector \cite{Bailey:2006fd} and in this case, stringent bounds on the \ac{gup} parameter can be inferred from SME parameters. 

The idea of \ac{gup} was suggested as a possible explanation of the Hubble tension in Refs.~\cite{Capozziello:2020nyq,Moradpour:2022oxr}. \ac{gup} introduces Planck-scale corrections to the standard relations between canonical variables, due to the interplay between the quantum theory and general relativity. Therefore, it is expected that such \ac{gup} corrections will be relevant during the very early/Planck epochs of cosmology, and leave their fingerprints on the quantum fluctuations. As these  \ac{gup}-modified quantum fluctuations propagate in a cosmological spacetime, they affect primordial fluctuations during cosmic inflation. Since these fluctuations are encoded in the \ac{cmb} anisotropies \cite{Abazajian:2013vfg,Kaya:2021oih}, one expects to read the \ac{gup} corrections in the \ac{cmb} power spectrum. In particular, one can select the cosmological spacetime to be the isotropic and homogeneous \ac{flrw} Universe with Hamiltonian, in natural units, of the form \cite{Rasouli:2014dba,Alvarenga:2001nm}
\begin{equation}
    \mathcal{H}_{\rm FLRW}(p_{a},a)=N\frac{p_a^{2}}{24a}+6Nka-N\rho a^{3}+\kappa\Pi\,,
\end{equation}
where $a$ and $p_{a}$ are, respectively, the scale factor which plays the role of the generalized coordinate operator and the generalized canonical momentum conjugate to the scale factor. Note that $N$ is the lapse function and $\Pi$ is its conjugate momentum, while $\kappa$ is its corresponding coefficient. Then, one introduces the \ac{gup}-modified canonical momentum \cite{Kempf:1994su,Das:2008kaa,Ali:2009zq,Das:2020ujn}
\begin{equation}
    P_{a}=p_{a}\Big(1+\lambda_{1} p_{a}+\lambda_{2}p_{a}^{2}+\mathcal{O}(p_a^{3}) \Big)\,,
\end{equation}
and substitutes it in the Hamiltonian $\mathcal{H}_{\rm FLRW}(p_{a},a)$, in order to obtain the \ac{gup}-modified Hamiltonian (keeping terms up to 2nd order in momentum)
\begin{equation}
    \mathcal{H}_{\rm FLRW}^{\rm GUP}(p_{a}, a) = \frac{1}{24}\frac{p_{a}^{2}(1+2\lambda_{1} p_{a}+2\lambda_{2}p_{a}^{2}+\lambda_{1}^{2}p_{a}^{2})}{a}+6ka -\rho a^{3}+\kappa\Pi\,.
\end{equation}

By combining  the \ac{gup}-modified Hamiltonian equations, one finds the \ac{gup}-modified Hubble function to be \cite{Aghababaei:2021gxe}
\begin{equation}
    H_{\rm GUP} = H \Big[1+48\lambda_{1}a^{2}H+\left(864\lambda_{2}
+\,576\lambda_{1}^{2}\right)a^{4}H^{2}\Big]^{1/2}\,,
\end{equation}
where $H$ is the standard Hubble function.

As already mentioned, \ac{cmb} will include signatures of
the \ac{gup} corrections, and thus, the fingerprints of quantum gravity. Therefore, on the one hand, the \ac{gup}-modified Hubble parameter, i.e., $H_{\rm GUP}$, can represent the one obtained by the Planck collaboration which utilizes the \ac{cmb} data \cite{Planck:2018vyg}, $H_{CMB}$. On the other hand, the unmodified Hubble parameter $H$, can be assumed to be the one obtained from the \ac{hst} which utilizes the \ac{sn1} data \cite{Riess:2019cxk}. In this case, we dub it $H_{\rm SN}$. 
 Based on this, the above expression for the \ac{gup}-modified Hubble parameter becomes
\begin{eqnarray}
H_{CMB}&=&H_{\rm SN} \Big[1+48\lambda_{1}a^{2}H_{\rm SN}+\left(864\lambda_{2}
+\,576\lambda_{1}^{2}\right)a^{4}H^{2}_{\rm SN}\Big]^{1/2}\,.
\end{eqnarray}
It is evident that the \ac{gup} can, in principle,
provide at least a partial explanation of the Hubble tension problem. Detailed analyses are currently ongoing.

\subsubsection{Quantum gravity effects on the physics of particles and fields}

\paragraph{Propagation of particles and fields on quantum spacetime }

One intensively studied aspect of the interaction between quantum spacetime and particles and fields is a modification of their propagation properties. Usually, this is studied in terms of a modified dispersion relation (MDR) of the particles, which encodes a modified light cone and mass shell structure. One distinguishes two cases: the Lorentz invariance violating (LIV) case \cite{Jacobson:2005bg}, where just an MDR is considered, and the deformed relativity (DSR) case \cite{Amelino-Camelia:2008aez, Arzano:2022har}, where the MDR is supplemented by a modified energy-momentum conservation law and deformed (most often non-linear) Lorentz transformations between observers which leave the MDR invariant and transform the deformed energy-momentum conservation in a covariant way.

In the context of cosmology, the most interesting consequence of an MDR is a different time of arrival of photons of different energy, when they are emitted simultaneously at the same spacetime event at redshift $z$~\cite{Amelino-Camelia:1997ieq}, such as \ac{grb}s~\cite{Ellis:1999sd,Ellis:2002in} or \ac{agn}s~\cite{MAGIC:2007etg}. Such a time delay can be thought of analogously as to what happens when electromagnetic radiation propagates through an optical non-trivial medium and photons of different energy are affected differently by the medium.

In general, such an MDR for photons can be parametrized in the form \cite{Pfeifer:2018pty}
\begin{align}\label{eq:mdrGeneral}
    E^2 = p^2 (1 + f(E,\boldsymbol{\mathrm{p}}, E_{\rm QG},z))\,,
\end{align}
where $f$ is a function that parametrizes the Planck scale modification in terms of the energy $E$ of the photon at emission, its comoving momentum $\boldsymbol{\mathrm{p}} = P/a(t)$, the quantum gravity energy scale $E_{\rm QG}$ (which might or might not be the Planck scale) and possibly the redshift. For two different photons, emitted with an energy $E_1$ and $E_2$, to leading non-vanishing order in $E_{\rm QG}$, this leads to a time delay of the form
\begin{equation}
    \label{eq:timedelay}
    \Delta t_{\rm QG} = t_2 - t_1 = \frac{1}{H_0}\frac{E_2^n-E_1^n}{E_{\rm QG}^n} \kappa(z)\,,
\end{equation}
where $\kappa(z)$ is the redshift distance function to the source at redshift $z$. It depends on the choice of the MDR model, i.e., in the choice of the function $f$ and on the choice of the cosmological model and its parameters. Some examples are
\begin{itemize}
    \item non-critical string-inspired models with redshift dependent $E_{\rm QG}$~\cite{Ellis:1999uh,Ellis:2008gg, Ellis:2009vq} or the Jacob-Piran model \cite{Jacob:2008bw} with constant $E_{\rm QG}$, with $\kappa(z) = \int_0^z \tfrac{(1+z^\prime)^n}{E_{\rm QG}^n(z') H(z')} dz'$.
    \item For $n=1$ in Eq.~\eqref{eq:timedelay}, DSR models like $\kappa$-Poincare in the bicrossproduct basis \cite{Barcaroli:2016yrl}, where the redshift distance function reads  $\kappa(z)=\int_0^z dz' \tfrac{1}{(z'+1)H(z')}$ or in more  general DSR models in \ac{flrw} spacetime, where the redshift distance function is described by the three-parameter $(\eta_1, \eta_2, \eta_3)$ model \cite{Amelino-Camelia:2023srg}
    
    $\kappa(z)=\int_0^z dz'\left(\tfrac{(z'+1)}{H(z')} 
     \left[\eta_1+\eta_2 \left(1-\left(1-\tfrac{H(z') I(z')}{z'+1}\right)^2\right)+\eta_3 \left(1-\left(1-\tfrac{H(z') I(z')}{z'+1}\right)^4\right)\right]\right)$, $I(z') = \int_0^{z'} \tfrac{dz''}{H(z'')}$.
\end{itemize} 

Taking into account these time delay effects in the analysis of the Hubble tension might lead to an alleviation of the present tensions~\cite{Mavromatos:2010nk,Mavromatos:2007zh,Basilakos:2011kc}. More detailed analyses remain a future prospect.

\paragraph{Interactions of particles and fields on quantum spacetime}

Another relevant feature related to quantum gravity effects on particles concerns modifications of relativistic interactions, that might induce changes to the predictions of the particle content of the Universe. 

In LIV models, the combination of MDR with the expected standard conservation of energy and momentum gives rise to strong modifications to interaction thresholds \cite{Jacobson:2005bg}, with effects for example on the opacity of the Universe to high-energy particles \cite{Protheroe:2000hp, Amelino-Camelia:2000ono}.

Going beyond LIV, an important ingredient for self-consistent DSR models is a modified energy momentum conservation. In a process in which two particles with 4-momenta $p$ and $q$ collide, the center of mass energy is not given by the simple sum of their momenta, since this would not be invariant under modified Lorentz transformations, but given by a modified addition law, to first order of the form \cite{Arzano:2022har}
\begin{align}
    (p \oplus q)_\mu = p_\mu + q_\mu + \frac{1}{E_{\rm QG}} f(p,q)_{\mu}\,,
\end{align}
where the precise expression of $f(p,q)_{\mu}$ depends on the model under consideration. In general, it is a non-linear function of $p$ and $q$.

This modified energy-momentum conservation law leads as well to modifications in relativistic threshold reactions, however they would be much weaker than in the LIV case, typically only relevant for particles of Planck-scale energy. While these effects have no direct link to cosmological tensions, their concomitant presence with the propagation effects discussed above might help constrain the specific form of MDR modification that is compatible with observations. 

\bigskip
\subsection{Varying fundamental constants and their role in the Hubble tension \label{sec:Var_fun_const}}

\noindent \textbf{Coordinator:} Jens Chluba\\
\noindent \textbf{Contributors:} Catarina Marques, Dan Grin, Gabriel Lynch, Leo Vacher, Nils Sch\"oneberg, Ruchika, and Vitor da Fonseca
\\

\noindent Fundamental physical constants need not be constant, neither spatially nor temporally. This seemingly simple statement has profound implications for a wide range of physical processes and interactions, and can be probed through a number of observations, see Ref.~\cite{Uzan:2002vq,Uzan:2010pm,Martins:2017yxk} for a broad review. Studies of fundamental constants (FCs) and their possible temporal and spatial variations are thus of utmost importance, and could provide a glimpse at physics beyond the standard model, possibly shedding light on the presence of additional scalar fields and their couplings to the standard sector, e.g., see Ref.~\cite{Bekenstein:1982eu, Martins:2015dqa}.

In the cosmological context, the fine-structure constant, $\aEM$, and electron rest mass, $\me$, are the most interesting, although variations of Newton's constant have also been considered \cite{Rich:2015jla, Galli:2011dsa, Lamine:2024xno} subject to some
recent theoretical and observational constraints \citep{VANPUTTEN2024107425, Banik:2024yzi}. The former can be directly probed with measurements of the \ac{cmb} temperature and polarization anisotropies, e.g., see Ref.~\cite{Kaplinghat:1998ry, Avelino:2000ea, Battye:2000ds, Avelino:2001nr, Rocha:2003gc, Martins:2003pe, Scoccola:2009xtv, Menegoni:2012tq} through their effect on the cosmological recombination history and photon scattering rate. A detailed description of individual physical effects on the \ac{cmb} power spectra is given in Ref.~\cite{Hart:2017ndk}, with calculations of the cosmological recombination history carried out using {\tt CosmoRec} \cite{Chluba:2010ca}. In short, {\it increasing} $\aEM$ and/or $\me$ leads to {\it earlier} recombination. This is primarily driven by the changes to the atomic energy levels, which scale as $E\propto \aEM^2 \me$, thereby enforcing a higher temperature for recombination to occur. Beyond this leading dependence, several subtle effects are encountered leading to differences in how $\aEM$ and $\me$ variations affect the \ac{cmb} signals. Crucially, the effect of $\aEM$ and $\me$ on the Thomson scattering rate, $\sigma_{\rm T}\propto \aEM^2/\me^2$, has to be carefully taken into account to yield consistent constraints \cite{Hart:2017ndk, Hart:2019dxi}. For additional discussion of the effects on recombination see Ref.~\cite{Chluba:2023xqj}.

Analyzing {\it Planck} 2013 data, the values of $\aEM$ and $\me$ around recombination were proven to coincide with those obtained in the lab to within $\simeq 0.4\%$ for $\aEM$ and $\simeq 1\%-6\%$ for $\me$ \cite{Planck:2014ylh}. These limits are $\simeq 2-3$ orders of magnitude weaker than constraints obtained from other ``local'' measurements \cite{Bize:2002vap,Rosenband:2008qgq, Bonifacio:2013vfa, Kotus:2016xxb} and those from \ac{bbn} \cite{Seto:2022xgx}; however, the \ac{cmb} is sensitive to very different phases in the history of the Universe, centered around the time of last scattering some $380,000$ years after the Big Bang, thereby complementing these measurements. In addition, \ac{cmb} measurements can be used to probe spatial variations of the FCs at cosmological distances \cite{Smith:2018rnu, Lucca:2023cdl}, opening yet another avenue for exploration.

\begin{figure}
    \centering
    \includegraphics[width=0.48\columnwidth]{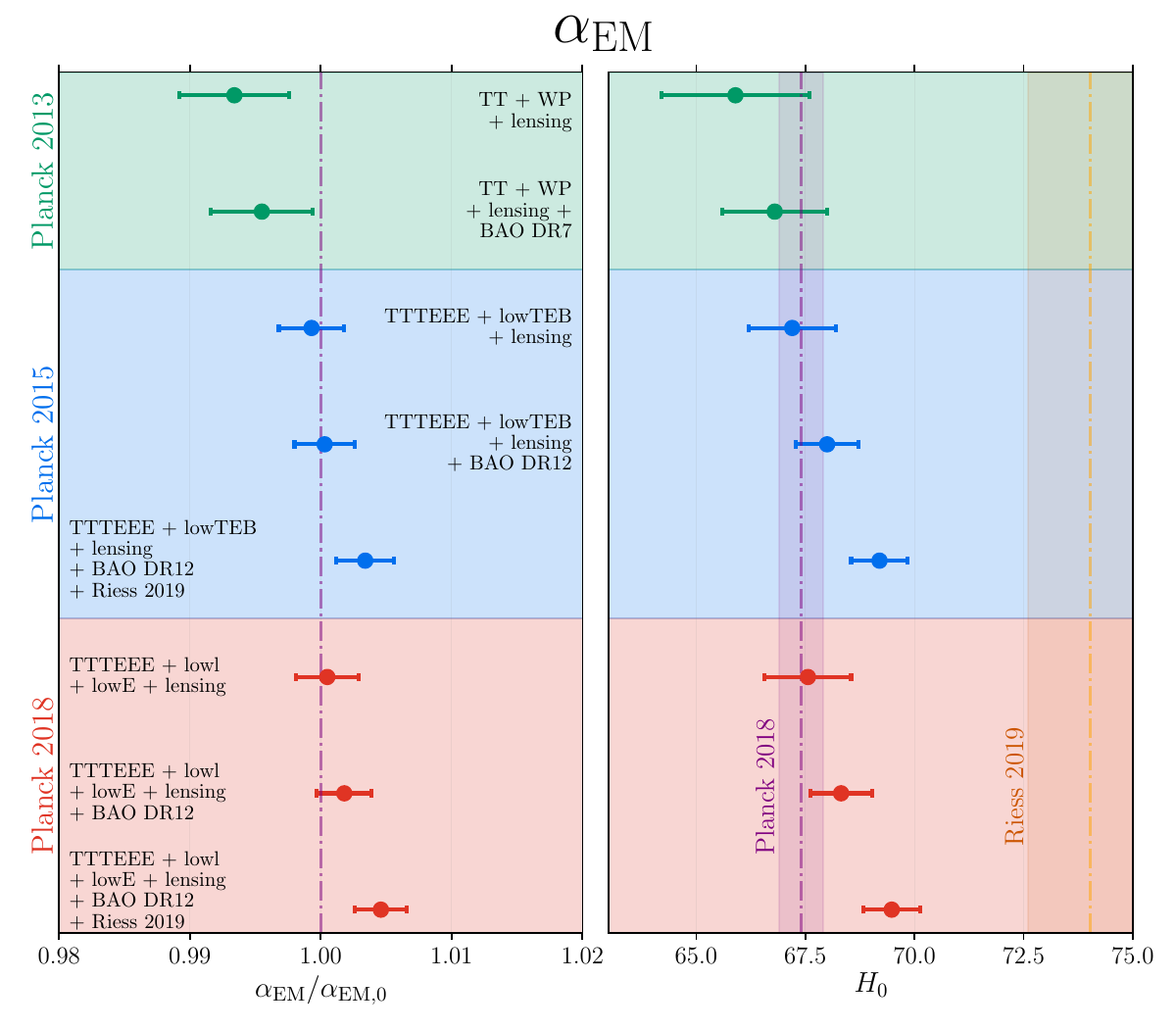}
    \hspace{2mm}
    \includegraphics[width=0.48\columnwidth]{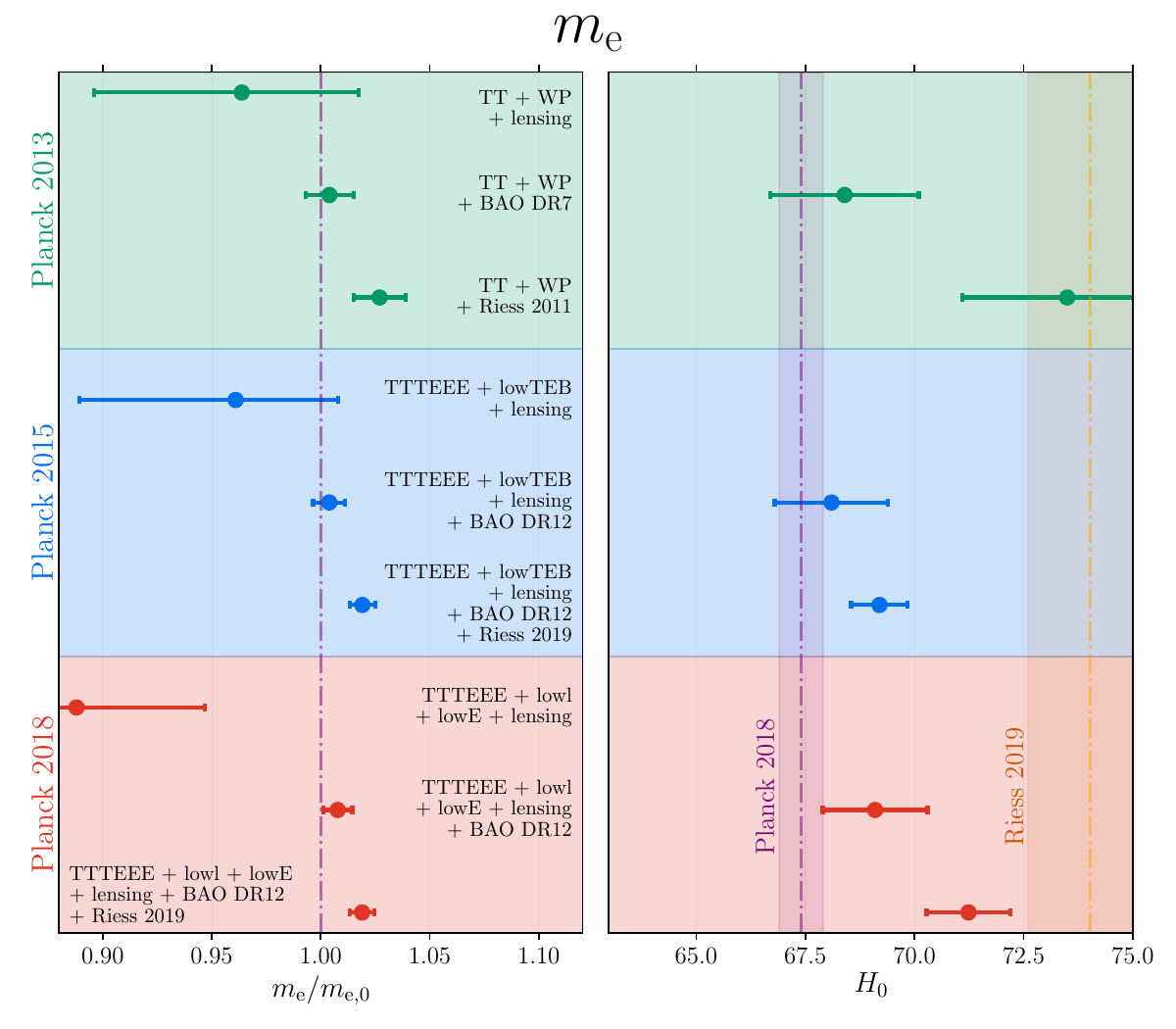}
    \caption{Constraints on the fundamental constants using various combinations of {\it Planck} data together with their $H_0$ values and errors. 
     {\it Left:} results from the fine structure constant $\aEM$. {\it Right:} similar results but for the effective electron mass $\me$. Here, we have redacted the constraint for $H_0$ from \ac{cmb} data only because the error bars are so large. For the $\me$ \ac{mcmc} analysis, we have widened the prior on the Hubble constant such that $H_0>20$\kms. The figure is from Ref.~\cite{Hart:2019dxi}, which illustrated that varying $\me$ can alleviate the Hubble tension.}
    \label{fig:me_alpha}
\end{figure}

With the {\it Planck} 2013 results in mind, no significant surprises were expected from the analysis of improved \ac{cmb} data of the {\it Planck} 2015 and 2018 releases. It, however, turned out that when considering models with varying $\me$, the geometric degeneracy becomes significant and can accommodate shifts in the value of the Hubble parameter when multiple probes are combined \cite{Hart:2019dxi}. The same geometric freedom is {\it not} encountered when varying $\aEM$ due to the modified dependence of the visibility function on this parameter. Indeed, when allowing $\me$ to depart from the standard (local) value, a non-standard value of $\dme=1.0191 \pm 0.0059$ ($\simeq 3.2\sigma$ significance) can be traded for a reduction of the Hubble tension, suggesting that new physics may be at work (see Fig.~\ref{fig:me_alpha} for illustration). The addition of extra degrees of freedom that influence the post-recombination Universe such as $\Omega_k$ or $w_0/w_a$ to variations of the electron mass allows to further ease the Hubble tension with a $\sim 3.6\sigma$ preference, robust to \ac{sn} data and \ac{bbn} constraints \cite{Schoneberg:2021qvd,Khalife:2023qbu,Schoneberg:2024ynd}.

This finding has spurred an increased interest in studying VFCs in this context, with scenarios that allow for varying $\me$, e.g., see Ref.~\cite{Hart:2019dxi, Sekiguchi:2020teg, Hart:2021kad, Lee:2022gzh, Hoshiya:2022ady} for additional discussion, ranking high in model comparisons \cite{Schoneberg:2021qvd}. Since VFCs can be caused by the presence of scalar fields, e.g., see Ref.~\cite{Bekenstein:1982eu, Damour:1994zq, Chiba:2006xx}, a natural question is whether the same scalar field could also be causing effects relating to \ac{ede}, possibly indicating a `two sides of the same coin' interplay. Constraints using the synergy of cosmological and local data have been put on a variety of well-motivated scalar fields models allowing for physical changes in $\aEM$ \cite{daFonseca:2022qdf,Barros:2022kpo,Vacher:2022sro,Tohfa:2023zip,Vacher:2023gnp}. However, local data, such as atomic clocks, are putting very strong constraints on such models such that they are mostly unable to produce significant variations of $\aEM$ during the recombination era \cite{Vacher:2024qiq}, while such constraints on varying $\me$ models are less restrictive \cite{Schoneberg:2024ynd}. In addition, VFCs could play a role in solving the Hubble tension even in light of the recent \ac{desi} measurements \cite{Lynch:2024gmp, Seto:2024cgo, Lynch:2024hzh, Schoneberg:2024ynd, Toda:2024ncp}, although a general mechanism causing early recombination could simply be the main cause of the tensions \cite{Lynch:2024hzh}.

Given the state of affairs, it will be extremely important to ask how different cosmological probes can be combined to shed light on the physical origin of the Hubble tension. One important avenue forward is to directly constrain the electron recombination history, given its crucial role in the formation of the \ac{cmb} anisotropies \cite{Sunyaev:1970plh, Peebles:1970ag, Hu:1994uz, Hu:1995fqa, Chluba:2005uz, Lewis:2006ym, Shaw:2011ez}. This has recently been achieved using a non-perturbative and model-independent approach \cite{Lynch:2024gmp}, providing a clear target for theoretical exploration. Indeed, the cosmological data prefers early recombination over the standard recombination history obtained using {\tt CosmoRec}, very much like what is caused by varying $\me$ or models with early structure formation \cite{Lynch:2024gmp}. This data-driven result may therefore indicate new physics in the redshift one thousand Universe, but at this point cannot distinguish the physical cause of this finding.

One way of directly probing the recombination history is through measurements of the cosmological recombination radiation (CRR) \cite{Chluba:2007zz, Sunyaev:2009qn, Hart:2020voa, Hart:2022agu, Lucca:2023cdl}. This tiny spectral distortion signal is created by photons emitted in the hydrogen and helium recombination eras, and thus directly depends on the time and duration of the recombination process. It was shown 
in Refs.~\cite{Hart:2022agu, Lucca:2023cdl} 
that various extensions to \lcdm\ can in principle be distinguished with future \ac{cmb} spectrometer measurements as envisioned for the \ac{esa} Voyage 2050 program \cite{Chluba:2019nxa}. Should the Hubble tension persist, then this will provide the ultimate test for various theoretical models. In addition, one can expect the observational uncertainties in the cosmological recombination history to hamper our ability to answer questions about extensions to \lcdm. A measurement of the CRR is therefore highly motivated even beyond questions about the Hubble tension, and provides a direct probe of one of the main pillars in our interpretation of CMB data.
\bigskip
\subsection{Local New physics solutions to the Hubble and growth tensions \label{sec:Local_sols}}

\noindent \textbf{Coordinator:} Leandros Perivolaropoulos\\
\noindent \textbf{Contributors:} Bhuvnesh Jain, Harry Desmond, Indranil Banik, Jeremy Sakstein, Nick Samaras, and Ruchika
\\

A recent comprehensive analysis provides new insights into the Hubble tension, suggesting that the core of the problem may lie in the distance ladder measurements rather than in conflicts between early and late Universe observations \cite{Perivolaropoulos:2024yxv}. This study compiled and analyzed two distinct groups of $H_0$ measurements: those based on the distance ladder approach, and those derived from one-step methods independent of both the distance ladder and the sound horizon scale. The analysis revealed a significant discrepancy between these two groups. Distance ladder-based measurements yielded a best-fit $H_0 = 72.8 \pm 0.5$\kms, while one-step measurements resulted in $H_0 = 69.0 \pm 0.48$\kms. Notably, when two outlier measurements were removed from the one-step sample, the best-fit value reduced to $H_0 = 68.3 \pm 0.5$\kms, which is fully consistent with sound horizon-based measurements like those from {\it Planck} \ac{cmb} observations. A Kolmogorov-Smirnov test yielded a $p$-value of 0.0001, indicating that the two samples are not drawn from the same underlying distribution.

Since the distance ladder is the only method for measuring $H_0$ that is based on local physics (via calibrators of \ac{sn1}), these findings lend support to the hypothesis of local physics solutions to the Hubble tension. They suggest that the discrepancy may not be between early and late-time measurements, but rather between distance ladder measurements and all other $H_0$ determinations. This points to either a systematic effect influencing all distance ladder measurements or new physics differentially impacting rungs of the ladder.

Such an intriguing class of potential solutions to the Hubble tension involves new physics acting in the local part of the distance ladder (first and second rungs). This idea was first developed by Ref.~\cite{Desmond:2019ygn}, who identified that screened fifth forces (for reviews see Refs.~\cite{Burrage:2016bwy,Burrage:2017qrf,Baker:2019gxo,Sakstein:2018fwz,Brax:2021wcv}) may have a differential impact on different rungs of the ladder such that their neglect could bias the inference of $H_0$. Ref.~\cite{Desmond:2019ygn} developed a range of phenomenological screening models, in which the degree of screening (and hence the effective value of Newton's constant, $G_\text{eff}$) is set by various gravitational properties of the galaxies used to calibrate \ac{sn1} and their environments. This included a novel screening mechanism governed by the local \ac{dm} density arising from the viable \ac{de} candidate of baryon--\ac{dm} interactions~\cite{Sakstein:2019qgn, Berezhiani:2016dne}. Ref.~\cite{Desmond:2019ygn} determined ranges for the screening proxies such that the anchor galaxies of the Cepheid PLR (the \ac{mw}, \ac{lmc}, and N4258) are screened while some of the galaxies used to infer the \ac{sn1} absolute magnitude are not. Since Cepheid pulsation periods are reduced and luminosities increased by an unscreened fifth force, the standard analysis neglecting the fifth force would then underestimate the distance to the \ac{sn1}-calibrator galaxies, which at fixed redshift would overestimate $H_0$.

\cite{Wojtak:2022bct, Wojtak:2024mgg} present another perspective by analyzing the extinction models used for SnIa in the calibration galaxies. Their study suggests that the standard extinction model developed by \citep{Popovic:2021yuo} for the Pantheon+ SnIa compilation underestimates the brightness of reddened supernovae in high stellar-mass calibration galaxies. They propose a modified extinction model that assumes a Milky Way-like distribution of the total-to-selective extinction coefficient $R_B$ in all calibration galaxies, which is also consistent with the extinction corrections employed for Cepheids, and a modified shape of the reddening distribution. This approach yields a lower value of the derived Hubble constant, reducing the tension with the Planck measurement assuming a flat $\Lambda$CDM cosmology from 5.2$\sigma$ to 2.8$\sigma$. This result highlights the importance of accurately modeling local astrophysical effects, such as dust extinction, to resolve the Hubble tension. Different extinction properties in the calibration galaxies do not necessitate violations of the Copernican principle, but they may likely result from the fact that the calibration galaxies are not a representative sample of galaxies in the Hubble flow (the former are solely late-type galaxies, whereas the latter include all morphological types). This selection bias can propagate to the \ac{sn1} sector via environment-dependent properties, whose physical origin is currently the subject of intense research.

A detailed modification to the SH0ES analysis pipeline allowing for this effect revealed that the Hubble tension could be eliminated with a fifth-force strength $(G_\text{eff}-G)/G \approx 0.1$ in some of the screening models. However, an important constraint derives from the consistency of Cepheid and \ac{trgb} distances to galaxies where both can be measured. The fifth force affects \ac{trgb} distances oppositely to Cepheid distances~\citep{Desmond:2019ygn}, so the fact that these distances agree under \ac{gr} implies that they will disagree under a strong fifth force. This requires $(G_\text{eff}-G)/G \lesssim 0.05$ in the most effective models, preventing this theory from reducing the $H_0$ tension below the $2\sigma$ level. Other constraints were studied, but found not to yield such a strong bound. Nevertheless, in conjunction with some other effects (of which a myriad are presented in this white paper), the tension could be resolved.

Ref.~\cite{Desmond:2019ygn} only addressed the Cepheid-calibrated distance ladder. The principal alternative is \ac{trgb}, which yields a slightly smaller $H_0$ value of $70-72$\kms under the assumption of GR. The model was extended to the \ac{trgb} calibration in Ref.~\cite{Desmond:2020wep}, where it was shown to be fully effective at solving the Hubble tension in that case. This requires the \ac{lmc} (the anchor galaxy of the \ac{trgb} absolute magnitude) to be less screened than the \ac{sn1} calibrators. As the \ac{lmc} is the least massive of the SH0ES anchors -- and \ac{trgb} stars tend to be found in higher-mass galaxies than Cepheids, which are young stars found in star-forming disks -- significant regions of the \ac{mg} parameter space can solve both the \ac{trgb} tension and reduce the Cepheid tension to $\sim 2.5\sigma$. Note that the \ac{sn1} themselves are assumed screened in all these analyses, as would typically be expected in viable \ac{mg} models due to their dense environments -- discussion of possible effects on \ac{sn1} may be found in Appendix B of Ref.~\cite{Desmond:2019ygn}. Variants of the model are studied in Ref.~\cite{Hogas:2023vim, Hogas:2023pjz}.

An alternative to screening is the more recent idea that new physical phenomena or transitions in the form of physical laws occur locally, specifically at distances $\lesssim 40$ Mpc or redshifts $z\lesssim 0.01$, affecting the connection between the distance ladder's second rung (\ac{sn1} calibration by Cepheids or \ac{trgb}) and third rung (Hubble flow \ac{sn1}). These Local Physics Transitions (LPTs) are assumed to involve abrupt changes of fundamental constants like the gravitational constant $G$ or the fine structure constant. They can affect the calibrators and \ac{sn1} used in the first and second rungs of the distance ladder, leading to a different behavior compared to that in the Hubble flow (at larger scales).

The simplest class of this paradigm involves an abrupt transition in the value of $G$ while the other fundamental constants remain fixed. Refs.~\cite{Marra:2021fvf} and \cite{Perivolaropoulos:2022txg} propose that a rapid transition in $G$ at a redshift $z_t \approx 0.01$ could resolve the Hubble tension. This $G$-step model (GSM) could imply that \ac{sn1} have lower luminosity $L_<$ at local scales (second rung of distance ladder) than the luminosity $L_>$ of distant \ac{sn1} in the Hubble flow (third rung). The ratio that would be required to solve the Hubble tension is $L_>/L_<=1.15$, leading to a higher inferred value of $H_0$ due to the degeneracy between the \ac{sn1} $L_>$ and $H_0$ in the Hubble flow.

The LPT paradigm was studied in Ref.~\cite{Perivolaropoulos:2022txg} through a reanalysis of the SH0ES data. By allowing for a transition in the absolute magnitude of \ac{sn1} at a distance of about 50~Mpc, they find that the best-fit value of $H_0$ drops significantly from $H_0 = 73.04 \pm 1.04$\kms to $H_0 = 67.32 \pm 4.64$\kms, in full consistency with the {\it Planck} value. This model also shows a substantial improvement in the fit to the data when an additional constraint from the inverse distance ladder is included. Another study further tested and confirmed that the Cepheid and \ac{sn1} datasets do not disfavor such a transition \cite{Ruchika:2023ugh}.

Assuming $L-A\propto G^\gamma$ (where $A$, $\gamma$ are constants and $L$ is the \ac{sn1} absolute luminosity), an exponent $\gamma<0$ ($\gamma > 0$) would mean that a lower (higher) value of $G$ is required at $z \gtrsim 0.01$ to solve the Hubble tension, leading also to a potential resolution (exacerbation) of the growth rate tension \cite{Marra:2021fvf}.

Recent constraints on $G$ variation between the present time and recombination suggest that $\left| \Delta G/G \right| <0.05$ at $2\sigma$, where $\Delta G \equiv G_{>}-G_{<}$ \cite{Lamine:2024xno}. These studies assume \lcdm\ expansion and that all other fundamental constants remain unchanged; relaxing these could weaken the constraints. Using the strong constraints of Ref.~\cite{Lamine:2024xno}, requiring $L_>/L_<=1.15$ with $A=0$ requires $\gamma \notin \left[ -4.5, 2.8 \right]$. 

Early studies assumed that $L\propto M_\text{Ch}$ (where $M_\text{Ch}$ is the Chandrasekhar mass), which implies $\gamma = -3/2$ \cite{Amendola:1999vu, Garcia-Berro:1999cwy, Gaztanaga:2001fh}. However, this does not account for the standardization procedure required for turning \ac{sn1} into standard candles; building a semi-analytic model to account for this, more recent studies use a simplified \ac{sn1} standardization procedure (no full use of light curve stretch and no use of color) and approximate semi-analytical arguments (no hydrodynamical simulation) to suggest that $\gamma \simeq 1.46$ with $A=0$ \citep{Wright:2017rsu, Zhao:2018gwk, Desmond:2019ygn}. Both approaches make significant simplifications to the underlying physics; a fully reliable analysis should use full 3D hydrodynamical simulations of \ac{sn1} with a range of properties and obtain their light curves, properly treat \ac{mg} effects on white dwarf structure, and adopt a standardization protocol identical to that employed empirically, also producing uncertainties on $\gamma$ and $A$.

It is thus possible that $\gamma \notin [-4.5,2.8]$ could be viable in the GSM. $\gamma<0$ is however also strongly constrained: a sharp rise in $G$ up to $150~$Myr ago ($z\lesssim 0.01$) would have dramatic unobserved effects on neutron stars by causing them to contract, thereby releasing vast amounts of energy \citep{Goldman:2024kot}. This is however also based on significant simplifying assumptions that may not hold in practice. In particular, roughly constant neutron star and binary formation rate over time is assumed, all galaxies are assumed similar to the \ac{mw}, higher black hole formation at early-times due to more massive stars and low metallicity is ignored, etc.

Further strong challenges to the GSM are presented in Ref.~\cite{Banik:2024yzi}. Their main criticisms relate to the expansion of the Earth's orbit around the Sun coupled with lower $G$ leading to a 10\% increase in the number of days per year that is not seen in the geochronometric and cyclostratigraphic records, a higher helioseismic age of the Sun due to higher $G$ over the vast majority of its history creating a mismatch with the ages of the oldest meteorite samples from the early Solar System, and a sharp drop in Solar insolation on the Earth in the geologically recent past due to a drop in $G$, leading to a large temperature drop.

Even if $\gamma \notin [-4.5,2.8]$ were to be conclusively ruled out or \ac{sn1} modeling found to require $\gamma<0$ which is then ruled out by neutron stars, or the tests in Ref.~\cite{Banik:2024yzi} were verified, the possibility of a simultaneous transition of $G$ with other fundamental constants \cite{Gupta:2023lwi,Gupta:2023jcm,Gupta:2021tma} like the fine structure constant would remain, keeping the general LPT paradigm alive. Transitions of the type envisaged by the paradigm could be produced by theories such as Dilaton and Kaluza-Klein theories \cite{Damour:1994zq}, TeVeS-like theories \cite{Moffat:2005si}, or varying $\alpha$ theories \cite{Barrow:1998df}. Note that the above constraints do not apply to the screening models discussed at the start of this section because screening naturally hides the fifth force locally and there is no universal change to fundamental constants like $G$.

In conclusion, the proposed local physics modifications discussed here, including both the screened fifth force model and transitions in the effective gravitational constant, appear to constitute a promising approach to the resolution of the Hubble tension. These results underscore the importance of further investigation of new physics in the local Universe that could affect the calibration of the cosmic distance ladder, and of constraints on such physics from other sources. \newpage
%%%%%%%%%%%%Section_5:
\section{Discussion and future opportunities}\label{sec:conclusion}
\noindent The problem of cosmic tensions poses several challenges and a number of opportunities in more deeply assessing possible systematics in observational surveys, developing new data analysis tool kits to refine our treatment of these data products, and developing more robust physical models that can build on the successes of the concordance model and pose new questions and give new solutions to the next decade of cosmology. To meet these challenges, a community-wide effort will be needed involving advances in each separate field, but also strongly and more interconnected relationships between these disciplines.

The coming decade will bring a number of pivotal surveys and observatories that aim to confront the central questions posed by the last decades of concordance physics. This may bring additional unforeseen discoveries in the cosmic history of the Universe, as well as new tools from the increasingly dominant spectrum of \ac{ml} and statistical physics toolkits, and possibly a new paradigm in our perspective of fundamental physics as an explanation of the driver of cosmic evolution. In Sec.~\ref{sec:fut_survey_prospects} the most significant observatories and survey prospects are discussed in the context of the coming decade of observational cosmology. In Sec.~\ref{sec:recomms} several key areas of development are identified which will be crucial to meeting the growing observational, statistical, and fundamental physics challenges of the upcoming decade of cosmology. Finally, we close in Sec.~\ref{sec:conclu} with a summary of the central themes of this work and an outlook perspective on the impact of cosmic tensions over the next few years.

\subsection{Future survey prospects \label{sec:fut_survey_prospects}}

\noindent Cosmology has undergone astounding advancements in the last few decades, due in part to the rapid progress of robust surveys, including unprecedented instrument precision and impressive new approaches to statistical analyses. These advancements have enabled a wider spectrum of tests of theoretical models, as well as unforeseen discoveries such as the accelerating expansion of the Universe in the late 1990s and the unexpectedly high number of galaxies imaged in the Hubble Deep Field. The next decade is set to extend this discovery potential with a swath of exciting missions nearing completion.

The dynamics of future surveys have pivoted to extreme precision surveys with vastly expanded mission objectives. These include measurements reaching much closer to the early Universe, such as the potential detection of the first luminous objects in the Universe through the 21-cm hydrogen line, the possible detection of signatures of inflation in the \ac{cmb}, and significant reductions in the uncertainty of current \ac{cmb} measurements. There are also a host of surveys that will track the spatial distribution of galaxies in the Universe to unprecedented levels in terms of the number of galaxies and the precision of these measurements. These surveys will facilitate even deeper investigations into the large-scale structure of the cosmos, providing more information on the nature and evolution of \ac{de} and \ac{dm}.

There is a vast array of planned observational surveys and measurements for the coming decade, which will add to the already considerable number of active survey collaborations. Below, we describe some of the major planned and ongoing missions and survey analyses.

\subsubsection{Cosmic microwave background} 

\noindent A key focus of the next generation of cosmological surveys is the study of the \ac{cmb}. Building on decades of groundbreaking discoveries, future \ac{cmb} experiments are poised to address some of the most pressing challenges in modern cosmology, including the persistent tensions in the Hubble constant and the amplitude of matter fluctuations ($S_8$). Upcoming \ac{cmb} missions aim to achieve unprecedented precision in measurements of the \ac{cmb}'s temperature and polarization anisotropies, offering unique insights into the physics of the early Universe and its evolution.

The $H_0$ tension stems from a significant discrepancy between the value of the Hubble constant inferred from early Universe observations, such as the \ac{cmb}, and local measurements using the cosmic distance ladder. Future \ac{cmb} experiments will refine measurements of the sound horizon at the baryon drag epoch—a critical calibration scale for early-time $H_0$ estimates. These high-precision data will test proposed solutions to the $H_0$ tension, including \ac{ede} models, extra relativistic species, or modifications to pre-recombination physics, by either validating or ruling out scenarios that alter the expansion history before recombination.

The $S_8$ tension, which reflects a discrepancy in the amplitude of matter clustering, is another key focus of upcoming \ac{cmb} experiments. By producing high-resolution maps of the \ac{cmb} lensing power spectrum, they will directly probe the distribution of \ac{dm} and the growth of large-scale structures. These measurements will be critical for cross-checking \ac{wl} and galaxy clustering observations, helping to clarify whether the $S_8$ tension arises from unknown systematics in late-time surveys or signals a breakdown in the standard cosmological model.

Additionally, \ac{cmb} experiments will probe primordial B-mode polarization, providing a unique window into the inflationary era and the energy scale of the early Universe. These observations could also inform models that connect the physics of inflation to current cosmological tensions, such as scenarios involving new scalar fields or \ac{mg}.

Future \ac{cmb} missions will also contribute to our understanding of \ac{de} and neutrino physics. Improved measurements of secondary anisotropies, such as the Sunyaev–Zel’dovich effect, will enable detailed studies of galaxy clusters and baryonic feedback processes, while precise constraints on the sum of neutrino masses and extra relativistic species will help refine models of the early Universe. By combining these capabilities with cross-correlations to other cosmological probes, such as \ac{bao} and \ac{wl} surveys, \ac{cmb} experiments will play a pivotal role in addressing cosmological tensions and advancing our understanding of the fundamental properties of the Universe.

Below, we highlight some of the major planned and ongoing \ac{cmb}-focused missions and their scientific goals:

\begin{itemize}
    \item \textbf{South Pole Telescope -- 3rd Generation (SPT-3G)}: The \ac{spt3g} \cite{2022SPIE12190E..03A} is the latest upgrade to the \ac{spt}, focusing on high-resolution studies of the \ac{cmb} from the exceptional observational site at the South Pole. This third-generation camera, featuring over 16,000 detectors, has significantly improved sensitivity, enabling deeper and more precise measurements of \ac{cmb} temperature and polarization anisotropies. What sets \ac{spt3g} apart is its ability to perform high-resolution surveys over small sky patches, optimized for studies of galaxy clusters via the Sunyaev–Zel’dovich effect, as well as gravitational lensing of the \ac{cmb}. These measurements are critical for understanding the distribution of \ac{dm} and large-scale structure formation. Additionally, \ac{spt3g} contributes to constraints on the sum of neutrino masses and potential new physics beyond the standard cosmological model. The compact field-of-view and deep integration capabilities make \ac{spt3g} uniquely suited for detecting subtle signals, such as \ac{cmb} lensing, and studying small-scale anisotropies.

    \item \textbf{Atacama Cosmology Telescope (ACT)}: Located in the Atacama Desert of Chile, \ac{act} \cite{ACT:2023wcq} is a ground-based observatory dedicated to high-resolution studies of the \ac{cmb}. The telescope has been instrumental in mapping \ac{cmb} temperature and polarization anisotropies and probing large-scale structure. The \ac{act} Data Release 4 (DR4) provided high-sensitivity maps of the \ac{cmb}, including measurements of temperature, E-mode polarization, and cross-correlations, covering over 17,000 square degrees of the sky. DR4 data significantly refined estimates of cosmological parameters, including constraints on $H_0$, $\sigma_8$, and the sum of neutrino masses. It also included improved measurements of the lensing power spectrum, advancing our understanding of the distribution of matter in the Universe. The forthcoming \ac{act} Data Release 6 (DR6) is expected to feature even more precise maps with reduced noise levels and expanded sky coverage. DR6 aims to provide improved constraints on primordial B-mode polarization, enhancing our ability to test inflationary physics. Additionally, it will enable more detailed studies of secondary anisotropies, such as the Sunyaev–Zel’dovich effect, and cross-correlations with galaxy surveys.

    \item \textbf{PolarBear and Simons Array}: PolarBear, along with its successor, the Simons Array \cite{POLARBEAR:2015ixw}, is focused on high-resolution measurements of the \ac{cmb}'s polarization anisotropies. Located in the Atacama Desert, Chile, these experiments aim to detect B-mode polarization caused by gravitational lensing and primordial \ac{gw}s. The Simons Array, an upgraded version of PolarBear, consists of multiple telescopes with advanced detectors for improved sensitivity. These measurements will refine constraints on the tensor-to-scalar ratio $r$, test inflationary models, and map the lensing power spectrum, shedding light on the distribution of \ac{dm} and the evolution of large-scale structure.

    \item \textbf{SPIDER}: SPIDER \cite{SPIDER:2021ncy} is a balloon-borne experiment designed to detect the large-scale polarization of the \ac{cmb}, focusing on B-modes linked to primordial \ac{gw}s. By operating above the atmosphere, SPIDER achieves reduced contamination from ground-based noise and atmospheric effects. Its payload consists of multiple telescopes equipped with cryogenic polarimeters, optimized for observing large angular scales. SPIDER has already completed successful flights, with future missions planned to improve sensitivity and expand sky coverage.

    \item \textbf{Ali Cosmic Polarization Telescope (AliCPT)}: AliCPT \cite{Li:2017drr} is a ground-based \ac{cmb} experiment located at the high-altitude Ali Observatory in Tibet. AliCPT focuses on measuring the polarization of the \ac{cmb} at large angular scales, particularly the B-modes associated with primordial \ac{gw}s. Its high-altitude location minimizes atmospheric contamination, allowing for precise measurements.

    \item \textbf{Q\&U Bolometric Interferometer for Cosmology (QUBIC)}: QUBIC \cite{QUBIC:2020kvy} is an international collaboration designed to measure the B-mode polarization of the \ac{cmb} using bolometric interferometry. It uniquely combines the sensitivity of bolometric detectors with the spatial filtering capabilities of interferometry, allowing for precise measurements of polarization anisotropies. Located at high altitude in Argentina, QUBIC targets primordial B-modes associated with inflation, as well as lensing-induced B-modes, providing critical constraints on the tensor-to-scalar ratio $r$. The first module began operations in 2022.

    \item \textbf{The Simons Observatory (SO)}: The Simons Observatory \cite{SimonsObservatory:2018koc} is a next-generation ground-based experiment located in the Atacama Desert, Chile. It will consist of three small-aperture telescopes (SATs) and one large-aperture telescope (LAT), all equipped with advanced cryogenic detector arrays. The primary goal of the observatory is to measure the \ac{cmb} temperature and polarization anisotropies with high precision across a wide range of angular scales. SO will target the primordial B-mode polarization of the \ac{cmb}, providing critical tests of inflationary physics and insights into the early Universe's energy scale. It will also probe secondary anisotropies, such as the Sunyaev–Zel’dovich effect, to map the growth of large-scale structure. The observatory will place constraints on the sum of neutrino masses and search for new physics beyond the standard model. Additionally, SO will refine measurements of cosmological parameters like $H_0$ and $S_8$, addressing key tensions in the \lcdm\ model. The Simons Observatory is expected to begin operations in the mid-2020s. 

    \item \textbf{Cosmology Large Angular Scale Surveyor (CLASS)}: The CLASS experiment \cite{CLASS:2023fvz} is designed to study the polarization of the \ac{cmb} on large angular scales. Located in the Atacama Desert, Chile, CLASS employs a series of telescopes equipped with cryogenic detectors to measure the faint polarization signals from the early Universe. The experiment focuses on detecting the reionization and recombination bumps in the \ac{cmb} polarization power spectrum, aiming to constrain primordial \ac{gw}s and the optical depth of reionization. CLASS's innovative strategy includes observing large sky patches with rapid rotation to minimize systematic errors caused by atmospheric and instrumental noise.

    \item \textbf{GroundBIRD}: GroundBIRD \cite{GroundBIRD:2020wax} is a ground-based experiment designed to measure the polarization of the \ac{cmb} on large angular scales. Located at the Teide Observatory in Spain, it uses fast rotation and superconducting detectors to reduce atmospheric noise and systematic errors. The experiment focuses on detecting B-mode polarization signals, particularly those associated with primordial \ac{gw}s. GroundBIRD's innovative observing strategy, which involves continuous scanning with a rotating cryostat, enhances its sensitivity to large-scale polarization while minimizing contamination.

    \item \textbf{BICEP/Keck Array and BICEP Array}: The BICEP/Keck Array and its successor, the BICEP Array \cite{Hui:2018cvg}, are focused on detecting B-mode polarization of the \ac{cmb} with high sensitivity. Operating from the South Pole, these experiments target large-scale polarization signals to constrain the tensor-to-scalar ratio $r$ and probe the inflationary epoch. The BICEP Array incorporates advanced detector technology and expanded frequency coverage to improve sensitivity and reduce systematic uncertainties. 
    
    \item \textbf{The Lite satellite for the study of B-mode polarization and Inflation from cosmic background Radiation Detection (LiteBIRD)}: The LiteBIRD probe \cite{LiteBIRD:2023aov,LiteBIRD:2024twk,LiteBIRD:2024dbi,LiteBIRD:2023zmo,LiteBIRD:2025mvy} aims to measure the unique imprints of B-mode polarization in \ac{cmb} photons, which are related to primordial \ac{gw}s and inflation. It will succeed the Planck mission in conducting full-sky surveys. Expected to launch in late 2029, LiteBIRD is a Japanese initiative by JAXA involving collaborations with agencies in North America and Europe. Positioned at the Lagrange point L2 in the Sun-Earth system, it will conduct a three-year survey. The relic \ac{cmb} radiation features E-mode polarization, linked to scalar perturbations, and B-mode polarization, associated with tensor perturbations. While the Cosmic Background Imager provided the first detailed E-mode polarization map, the B-mode signal remains undetected. LiteBIRD will take cosmic-variance-limited measurements of E-mode polarization to study large-scale correlations, shedding light on initial conditions for cosmological perturbations and providing insights into physics beyond the standard model. Regarding B-mode polarization, LiteBIRD will aim for a tensor-to-scalar ratio limit of $r < 0.001$. A positive detection would have a profound impact on fundamental physics, offering insights into inflationary physics, parity violation, and primordial cosmological magnetism.

    \item \textbf{CMB—Stage 4 (CMB-S4)}: The \ac{cmbs4} observatory~\cite{CMB-S4:2016ple} will operate telescopes at both the South Pole and the Atacama Desert in Chile, enabling deep microwave observations over small and large sky fields. It will employ advanced superconducting detector array technologies to reduce galactic foreground contamination. Expected to start operations in the late 2020s, though currently on hold, \ac{cmbs4} aims to measure primordial \ac{gw}s associated with the early rapid expansion of density fluctuations, characterized by the B-mode polarization of the \ac{cmb}. This would provide critical information about cosmic inflation and fundamental physics, targeting a tensor-to-scalar ratio of $r < 0.002$, an order of magnitude improvement over current limits. \ac{cmbs4} will constrain the sum of neutrino masses, critical for understanding the sterile neutrino theory and the inverted neutrino mass hierarchy. Beyond neutrino physics, it will impose stringent limits on possible light particles beyond the standard model, addressing extra effective degrees of freedom. On \ac{dm}, \ac{cmbs4} will probe \ac{cmb} anisotropies for potential signals of WIMP annihilation, placing constraints on their masses, while exploring non-thermal \ac{dm} effects on lensing power spectra. Another key goal is probing \ac{de}, with precision measurements of \lcdm\ cosmological parameters such as $H_0$ and $S_8$, enabling comparisons with other early-time probes. Additionally, \ac{cmbs4} will generate higher-resolution maps of matter distribution in the Universe by measuring distortions in \ac{cmb} photons caused by gravitational lensing from the surface of last scattering. This will significantly enhance future galaxy surveys.
    
    \item \textbf{Cosmic Microwave Background - High Definition (CMB-HD)}: CMB-HD~\cite{Sehgal:2019ewc,CMB-HD:2022bsz} is a proposed next-generation ground-based observatory that aims to revolutionize cosmology with its unparalleled resolution and sensitivity. Unlike other \ac{cmb} experiments, CMB-HD focuses on small-scale anisotropies, achieving an angular resolution of 0.5 arcminutes and surveying over 50\% of the sky. This capability allows it to probe previously inaccessible signals and extend our understanding of the Universe to finer detail. CMB-HD’s primary science goals include precise measurements of the Sunyaev–Zel’dovich effects to map galaxy clusters, their gas content, and baryonic physics. It will also produce detailed maps of the small-scale \ac{cmb} lensing power spectrum, offering new insights into the distribution of \ac{dm} and the growth of large-scale structure. By detecting primordial B-mode polarization, CMB-HD will constrain the tensor-to-scalar ratio $r$ to test inflationary physics at energy scales far beyond current limits. Additionally, CMB-HD will place stringent constraints on the sum of neutrino masses, light relics, and possible deviations from the \lcdm\ model.

    \item \textbf{Probe of Inflation and Cosmic Origins (PICO)}: PICO~\cite{2019arXiv190210541H} is a proposed next-generation satellite mission designed to provide a comprehensive, all-sky measurement of the \ac{cmb}. Operating from space, PICO will avoid the challenges of atmospheric contamination and cross-calibration issues that affect ground-based experiments. Its observations will span a broad frequency range from 20 GHz to 800 GHz, enabling highly accurate foreground removal and calibration consistency across the entire sky. A key advantage of PICO’s space-based platform is its ability to precisely measure the optical depth to reionization, $\tau$. This measurement, critical for understanding the early history of star formation and the growth of cosmic structures, is less accessible to ground-based telescopes due to the need for absolute calibration. By combining this capability with its unparalleled sensitivity to B-mode polarization, PICO will constrain the tensor-to-scalar ratio $r$ to levels below $0.001$, providing definitive tests of inflationary models. In addition to probing inflation, PICO will deliver high-resolution maps of the E-mode polarization and \ac{cmb} lensing power spectrum, offering insights into the distribution of \ac{dm} and large-scale structure. It will also place stringent constraints on the sum of neutrino masses, detect potential signals from light relics, and test extensions to the \lcdm\ model. If approved, PICO will complement ground-based experiments by addressing systematic challenges and providing the calibration accuracy and sensitivity only possible from space.

    \item \textbf{Polarized Radiation Imaging and Spectroscopy Mission (PRISM)}: PRISM~\cite{PRISM:2013fvg} is a proposed satellite mission designed to provide extremely sensitive, all-sky measurements of the \ac{cmb}'s temperature and polarization anisotropies. With its advanced instrumentation and broad frequency coverage, PRISM aims to enhance our understanding of both early and late-time cosmology. A key goal of the mission is to detect primordial B-mode polarization, which would provide direct evidence of inflation and constrain the energy scale of the early Universe. PRISM's ability to refine measurements of the \ac{cmb} power spectrum and polarization will enable tighter constraints on the sound horizon at the baryon drag epoch, directly addressing the $H_0$ tension by testing \ac{ede} models and other modifications to pre-recombination physics. Additionally, PRISM's high sensitivity to secondary anisotropies, such as \ac{cmb} lensing, will improve measurements of the lensing power spectrum, aiding investigations into the $S_8$ tension and the growth of large-scale structure. If approved, PRISM would complement ground-based and balloon-borne experiments by avoiding atmospheric contamination, offering a clean and highly detailed dataset for cosmological studies.
    
    \end{itemize}

\subsubsection{Baryon acoustic oscillations} 

\noindent The study of \ac{bao} has become a cornerstone of modern cosmology, providing a powerful method for measuring the expansion history of the Universe. \ac{bao} are the relic imprints of sound waves that propagated in the early Universe, leaving a characteristic scale in the large-scale distribution of galaxies and matter. This standard ruler has proven invaluable for calibrating cosmic distances and constraining key cosmological parameters, including the Hubble constant and the \ac{de} equation of state.

As the precision of cosmological measurements continues to improve, future \ac{bao} experiments are poised to deliver new insights into the late-time Universe. By mapping the three-dimensional distribution of galaxies, \ac{qso}s, and other tracers across an extensive redshift range, these experiments will probe the dynamics of cosmic expansion and the growth of large-scale structure. This enhanced precision will enable a deeper exploration of the $H_0$ and $S_8$ tensions, testing whether they arise from unaccounted systematics or new physics beyond the \lcdm\ model.

In addition to refining our understanding of \ac{de} and its influence on cosmic acceleration, upcoming \ac{bao} surveys will provide complementary constraints on modifications to gravity and potential extensions to standard cosmology. Synergies with other probes, such as \ac{wl} and \ac{cmb} lensing, will further bolster the ability to test fundamental physics and the consistency of the cosmological model.

Below, we describe some of the major planned and ongoing \ac{bao}-focused missions and their scientific objectives:

\begin{itemize}
    \item \textbf{Dark Energy Spectroscopic Instrument (DESI)}: \ac{desi}~\cite{DESI:2016fyo} is a state-of-the-art spectroscopic survey currently operating on the Mayall 4-meter telescope at Kitt Peak National Observatory. Its primary goal is to create the most detailed three-dimensional map of the Universe by observing tens of millions of galaxies and \ac{qso}s across a wide redshift range. \ac{desi}'s high-precision \ac{bao} measurements will provide critical constraints on the cosmic distance scale and the expansion history of the Universe, directly addressing the $H_0$ tension. Additionally, \ac{rsd} analyses from \ac{desi} will help probe the growth rate of large-scale structures, contributing to our understanding of the $S_8$ tension and testing potential modifications to gravity.

    \item \textbf{Canadian Hydrogen Intensity Mapping Experiment (CHIME)}: CHIME~\cite{CHIME:2022dwe} is a revolutionary radio interferometer located at the Dominion Radio Astrophysical Observatory in Canada. Designed to operate in the 400--800 MHz band, it maps the large-scale structure of the Universe using the 21 cm hydrogen line, covering a redshift range of $0.8 < z < 2.5$. By using intensity mapping techniques, CHIME provides precise measurements of the \ac{bao} scale, complementing optical surveys like \ac{desi} and HIRAX. The experiment's wide field-of-view and innovative cylindrical reflector design enable it to conduct highly efficient surveys, making CHIME a key player in refining the cosmic distance scale. .

    \item \textbf{BAO from Integrated Neutral Gas Observations (BINGO)}: BINGO~\cite{Abdalla:2021nyj} is a 21 cm intensity mapping experiment specifically designed to measure \ac{bao} in the redshift range $0.13 < z < 0.48$. Located in Brazil, it utilizes a compact array of radio antennas to map the large-scale distribution of neutral hydrogen. BINGO is optimized to minimize instrumental noise and systematics, providing precise constraints on the expansion history of the Universe and \ac{de}. The project is currently under construction and expected to begin operations in the mid-2020s.

    \item \textbf{Euclid}: The Euclid mission~\cite{Amendola:2016saw}, led by \ac{esa}, is designed to create a detailed three-dimensional map of the Universe, enabling precise measurements of the \ac{bao} scale across a wide redshift range ($0.7 < z < 2.0$). Its near-infrared spectroscopic survey will provide accurate redshifts for tens of millions of galaxies, establishing a robust standard ruler for cosmological distances. By focusing on high-redshift galaxy clustering, Euclid will refine the expansion history and \ac{de} equation of state. Its spectroscopic data will also complement optical \ac{bao} studies, bridging gaps in redshift coverage and enhancing multi-probe analyses with other cosmological surveys.

    \item \textbf{Prime Focus Spectrograph (PFS):} The PFS survey~\cite{2014PASJ...66R...1T,Tamura:2016wsg} is a spectroscopic survey operating on the 8-meter Subaru Telescope. It is designed to observe 2,400 objects simultaneously within a 1.2 deg$^2$ field of view, covering wavelengths from the near-ultraviolet to the near-infrared. With its wide field and spectral coverage, PFS focuses on three primary science programs: cosmology, galaxy evolution, and galactic archaeology. For cosmology, PFS will observe approximately 4 million emission-line galaxies (ELGs) over 1,200 deg$^2$, covering redshifts from $0.8$ to $2.4$. Unlike other surveys, it will uniquely map ELGs at $2.0 < z < 2.4$, complementing \ac{desi}. PFS will provide high-precision measurements of BAO and the Alcock--Paczy\'nski (AP) effect, enabling constraints on the Hubble expansion history and testing potential evolution in dark energy out to $z = 2.4$. By combining these results with lower-redshift BAO constraints, PFS will determine the dark energy density to approximately $7\%$ accuracy per redshift bin. Additionally, \ac{rsd} measurements will reconstruct the growth rate of cosmic structure, $f\sigma_8(z)$, with $6\%$ accuracy up to $z = 2.4$, allowing for a precise determination of the sum of neutrino masses, with an uncertainty of $\sigma(\sum m_\nu) = 0.02$.

    \item \textbf{Roman Space Telescope (formerly WFIRST)}: The Roman Space Telescope~\cite{Spergel:2015sza}, led by NASA, is a highly ambitious mission designed to address key questions about \ac{de}, exoplanets, and the structure of the Universe. Through its High Latitude Survey, Roman will map the large-scale distribution of galaxies and perform \ac{wl} and \ac{bao} analyses with unprecedented precision. This mission will provide robust constraints on the expansion history, \ac{de} dynamics, and the growth of cosmic structures.

    \item \textbf{Rubin Observatory's Legacy Survey of Space and Time (LSST)}: The Rubin Observatory~\cite{Blum:2022dxi}, in its final construction phase in northern Chile, will conduct the \ac{lsst}, a 10-year survey of $18,000\,\text{deg}^2$ of the sky across six wavelength bands. Public data release is expected approximately two years after first light in 2025. \ac{lsst} will address a broad range of fundamental questions, including: (1) Is \ac{de} dynamical, as characterized by the $w_0$-$w_a$ parametrization? (2) Can the expansion history and large-scale structure help distinguish between exotic energy densities and \ac{mg}? (3) What are the properties of \ac{dm}, as revealed by microlensing searches? (4) Do \ac{dm} halos exist without hosting galaxies? (5) Are matter fluctuations in the late Universe consistent with \ac{cmb}-derived constraints? \ac{lsst} is expected to profoundly impact our understanding of \ac{de}, \ac{dm}, and fundamental physics.
    
    \item \textbf{SKAO (Square Kilometer Array Observatory)}: \ac{ska}~\cite{Braun:2015B3} is a next-generation radio telescope array under development, with sites in South Africa and Australia. Its unprecedented sensitivity and angular resolution will enable precise 21-cm intensity mapping, allowing for detailed \ac{bao} measurements across a wide redshift range. These data will be instrumental in tracing the evolution of cosmic expansion and structure growth, providing crucial information for resolving the $H_0$ and $S_8$ tensions and testing the consistency of the \lcdm\ model. (see Sec.~\ref{21cm})

    \item \textbf{The Hydrogen Intensity and Real-time Analysis eXperiment (HIRAX)}: HIRAX~\cite{Crichton:2021hlc} is a radio interferometer array under construction in South Africa, optimized for intensity mapping of the 21-cm hydrogen line. HIRAX will measure \ac{bao} across a redshift range of $0.8 < z < 2.5$, directly probing the expansion history during the period of cosmic acceleration driven by \ac{de}. These measurements will complement optical \ac{bao} surveys, refining constraints on \ac{de} models and addressing potential new physics.

    \item \textbf{Dark Energy Spectroscopic Instrument - Phase II (DESI-II)}: \ac{desi}-II~\cite{DESI:2022lza} is a proposed extension of the highly successful \ac{desi} survey, aiming to build on its existing infrastructure to further refine cosmological measurements. By expanding its redshift coverage and increasing the volume of observed galaxies and \ac{qso}s, \ac{desi}-II will provide even more precise measurements of the \ac{bao} and \ac{rsd}. The extended survey will target fainter galaxies and higher redshift objects, allowing for a deeper exploration of the late-time Universe's expansion history and the growth of cosmic structures.
    
    \item \textbf{The 4-metre Multi-Object Spectroscopic Telescope (4MOST)}: 4MOST~\cite{2012SPIE.8446E..0TD}, based at the ESO’s VISTA telescope in Chile, is designed to conduct massive spectroscopic surveys of galaxies and \ac{qso}s over large sky areas. Its multi-object spectrograph will enable precise \ac{bao} measurements and \ac{rsd} analyses, improving constraints on the cosmic expansion rate and structure growth. 4MOST will work synergistically with imaging surveys like \ac{lsst} to provide redshift information critical for cosmological studies, including cross-correlation analyses to probe the $S_8$ tension.

    \item \textbf{Spectro-Photometer for the History of the Universe, Epoch of Reionization, and Ices Explorer (SPHEREx)}: SPHEREx~\cite{SPHEREx:2014bgr} is a NASA mission designed to perform an all-sky spectral survey. By mapping galaxies across a wide range of redshifts, SPHEREx will measure \ac{bao} and \ac{rsd} signals, enabling precise constraints on the expansion history and the growth of cosmic structure. Its unique spectral coverage will complement optical surveys and provide new insights into the physics.
    
    \item \textbf{CO Mapping Array Project (COMAP)}: COMAP~\cite{Cleary:2021dsp} is a pioneering experiment focused on intensity mapping of carbon monoxide (CO) lines at high redshift. COMAP will trace the large-scale structure of the Universe during the epoch of galaxy formation, providing a complementary approach to \ac{bao} measurements. These data will help refine constraints on the expansion history and test models of \ac{de} and \ac{mg}.

    \item \textbf{Packed Ultra-Wideband Mapping Array (PUMA)}: PUMA~\cite{PUMA:2019jwd} is a proposed next-generation radio interferometer optimized for detecting \ac{bao} and measuring \ac{rsd} over a wide redshift range of $2 < z < 6$. Using 21 cm intensity mapping, PUMA will map the large-scale distribution of neutral hydrogen, providing unprecedented insights into the Universe’s expansion history and the growth of cosmic structure. Its innovative design, featuring a densely packed array of antennas, will enable high sensitivity and wide bandwidth, making it a critical experiment for studying \ac{de}, testing modifications to gravity, and resolving tensions in cosmological parameters. Expected to begin operations in the 2030s, PUMA will complement optical \ac{bao} surveys and push the boundaries of cosmological research.

\end{itemize}

\subsubsection{Weak lensing experiments} 

\noindent \ac{wl} is a cornerstone tool for investigating cosmological tensions, particularly the $S_8$ discrepancy, which reflects a persistent difference between the amplitude of matter fluctuations inferred from early- and late-Universe observations. \ac{wl} measures the subtle distortions in galaxy shapes caused by the gravitational lensing effect of intervening mass distributions, providing a direct probe of the growth of cosmic structures. By combining the clustering amplitude $\sigma_8$ with the matter density parameter $\Omega_{\rm m,0}$, the derived parameter $S_8 = \sigma_8 \sqrt{\Omega_{\rm m,0}/0.3}$ becomes a critical test of the \lcdm\ model.

In addition to its crucial role in addressing the $S_8$ tension, \ac{wl} data also contribute to resolving the $H_0$ tension. Synergistic analyses that combine \ac{wl} with other cosmological probes, such as \ac{cmb} lensing and \ac{bao} measurements, enable a multi-probe approach to jointly constrain both early- and late-Universe parameters. This integrated strategy improves the robustness of results and provides deeper insights into the underlying physics driving these tensions.

Below, we highlight planned and ongoing \ac{wl} experiments and their contributions to addressing the $S_8$ tension:

\begin{itemize}
    \item \textbf{Kilo-Degree Survey (KiDS)}: \ac{kids}~\cite{KiDS:2020suj} provided high-precision \ac{wl} and photometric redshift measurements across $1,350\,\mathrm{deg^2}$, with a strong focus on controlling systematics such as shear calibration, photometric redshifts, and intrinsic galaxy alignments. Using high-quality imaging from the Very Large Telescope (VLT) and tomographic redshift binning, \ac{kids} delivered some of the most precise $S_8$ constraints, highlighting a persistent tension with \lcdm\ predictions. Its integration with external datasets, such as galaxy clustering and \ac{cmb} lensing maps, enabled multi-probe analyses and established a strong foundation for addressing the $S_8$ tension in future surveys.

    \item \textbf{Dark Energy Survey (DES)}: \ac{des}~\cite{DES:2016jjg} observed $5,000\,\mathrm{deg^2}$ of the southern sky, combining \ac{wl} and galaxy clustering to provide robust $S_8$ constraints. With precise photometric redshift calibration and careful control of systematics, \ac{des} advanced studies of large-scale structure growth. Its cross-correlations with \ac{cmb} lensing maps and tomographic analyses tested \lcdm\ predictions and explored extensions like evolving \ac{de}. \ac{des}’s extensive data and methodological innovations set a high standard for future \ac{wl} surveys.

    \item \textbf{Hyper Suprime-Cam (HSC)}: \ac{hsc}~\cite{Aihara:2017paw}, conducted on the Subaru Telescope, delivered \ac{wl} data over $1,400\,\mathrm{deg^2}$ with exceptional resolution and depth, enabling precise studies of cosmic shear. Its high-resolution imaging allowed detailed analyses of smaller-scale structures and provided key insights into $S_8$, revealing persistent tensions with \lcdm\ predictions. The survey's innovative techniques for photometric redshift estimation and systematic error control ensured high accuracy in its results. \ac{hsc} also contributed significantly to cross-correlation studies with galaxy clustering and \ac{cmb} lensing, further refining constraints on the growth of structure and testing extensions to the standard cosmological model.

    \item \textbf{Super-pressure Balloon-borne Imaging Telescope (SuperBIT)}: SuperBIT~\cite{2018SPIE10702E..0RR} is a stratospheric, balloon-borne telescope designed for high-resolution, wide-field imaging, enabling precise \ac{wl} measurements of galaxy clusters. By operating above most of Earth’s atmosphere, it minimizes atmospheric distortions, delivering exceptional data quality. SuperBIT’s observations are critical for mapping the distribution of \ac{dm} and studying large-scale structure formation, providing complementary insights to ground- and space-based surveys. Its unique capabilities contribute to addressing the $S_8$ tension by offering an independent probe of structure growth.

    \item \textbf{Euclid}: Euclid’s high-resolution optical and near-infrared imaging capabilities~\cite{Amendola:2016saw} are designed to map the weak gravitational lensing of billions of galaxies over $15,000\,\mathrm{deg^2}$ of the sky. Its ability to probe redshifts up to $z \sim 2.5$ makes it uniquely suited for tomographic studies of structure growth, providing tight constraints on the $S_8$ parameter. By achieving unparalleled depth and resolution in \ac{wl}, Euclid will play a crucial role in testing modifications to \lcdm, such as evolving \ac{de} models and \ac{mg} theories. Its combination of imaging and spectroscopic data will enable synergy with Rubin \ac{lsst} and \ac{cmb} lensing maps, further enhancing our understanding of cosmic structure formation.

    \item \textbf{Rubin Observatory's Legacy Survey of Space and Time (LSST)}: The Rubin Observatory~\cite{Blum:2022dxi} will survey $18,000\,\mathrm{deg^2}$ of the sky over a 10-year period, providing deep, multi-band imaging across six optical filters. Rubin's \ac{lsst} high precision in \ac{wl} and galaxy clustering measurements will allow detailed tomographic studies of structure growth and yield tighter constraints on $S_8$. Its ability to detect millions of faint galaxies at high redshifts will improve our understanding of the evolution of cosmic structures and test potential extensions to \lcdm. Rubin's \ac{lsst} rich dataset will also facilitate cross-correlations with \ac{cmb} and other \ac{wl} surveys, strengthening the multi-probe approach to resolving cosmological tensions.

    \item \textbf{Roman Space Telescope (formerly WFIRST)}: The Roman Space Telescope’s High Latitude Survey~\cite{Spergel:2015sza} will deliver high-resolution \ac{wl} data over $2,000\,\mathrm{deg^2}$, focusing on the distribution of \ac{dm} and the growth of cosmic structures. Its combination of near-infrared imaging and spectroscopy enables precise photometric redshift estimation, essential for tomographic studies of \ac{wl}. Roman’s unparalleled sensitivity at high redshifts ($z \sim 2$) will refine measurements of the $S_8$ parameter, providing stringent tests of \lcdm\ and its alternatives, such as evolving \ac{de} and \ac{mg} models. Its synergy with Euclid and Rubin's \ac{lsst} will enhance cross-calibration efforts and improve constraints on cosmic structure formation across a wide range of scales and epochs.

    \item \textbf{Square Kilometer Array Observatory (SKAO)}: \ac{ska}~\cite{Braun:2015B3}, utilizing 21-cm intensity mapping and \ac{wl}, will provide unique and independent constraints on $S_8$ by probing the distribution of \ac{dm} and the growth of structures. Its radio-based approach will extend \ac{wl} studies to higher redshifts and larger scales, offering a complementary perspective to optical surveys. By addressing potential systematics in traditional probes, \ac{ska} will play a crucial role in testing deviations from \lcdm\ and enhancing multi-probe strategies to resolve cosmological tensions.

    \item \textbf{Einstein Telescope (ET) and Cosmic Explorer (CE)}: These proposed third-generation \ac{gw} observatories~\cite{Sathyaprakash:2012jk,Evans:2021gyd,Abac:2025saz,Punturo:2010zz}, expected to begin operations in the mid to late 2030s, will achieve unprecedented sensitivity in detecting \ac{gw}s. In addition to their primary focus on \ac{gw} astrophysics, they will enable measurements of \ac{wl} effects on \ac{gw} signals. This innovative approach provides a novel and independent method to probe the $S_8$ parameter, offering insights into the growth of cosmic structures and testing extensions to the standard cosmological model. Their unique capabilities will complement traditional \ac{wl} surveys, further enriching multi-probe cosmological studies. (see Sec.~\ref{GWs})

\end{itemize}

\subsubsection{Gravitational waves as probes of cosmological tensions} \label{GWs}

\noindent \ac{gw}s offer a revolutionary perspective in addressing cosmological tensions, functioning as ``standard sirens'' that enable the measurement of cosmic distances independently of electromagnetic calibrations. By directly determining the luminosity distance to \ac{gw} events, particularly those accompanied by electromagnetic counterparts (e.g., binary neutron star mergers), this method bypasses systematics associated with traditional distance ladder techniques. For a recent review on the capabilities of \ac{gw} observatories see \cite{Chen:2024gdn}. 

In addition to these ``bright sirens,'' \ac{gw}s from events lacking identifiable electromagnetic counterparts, termed ``dark sirens,'' can also contribute to cosmology. By correlating the \ac{gw} signal with galaxy catalogs to infer host redshifts, dark sirens expand the scope of \ac{gw} cosmology, providing complementary constraints on the Hubble constant and other cosmological parameters.
Furthermore, advancements in spectroscopic observations of host galaxies have introduced ``spectroscopic sirens,'' which refine redshift measurements associated with \ac{gw} events. These high-precision techniques enhance the reliability of constraints on the expansion history and reduce uncertainties linked to host galaxy identification.

Beyond $H_0$, \ac{gw}s contribute to understanding large-scale structure formation through \ac{wl} of \ac{gw}s by intervening matter. These lensing effects provide unique insights into the amplitude of matter fluctuations, addressing the $S_8$ tension. Additionally, third-generation observatories such as the \ac{et} and the Cosmic Explorer will extend \ac{gw} observations to higher redshifts ($z \sim 10$), probing the early Universe's expansion history and potential deviations from the \lcdm\ model.
The integration of \ac{gw} observations with traditional cosmological probes, such as \ac{bao}, \ac{cmb}, and \ac{wl}, opens new pathways for multi-messenger cosmology. Together, these complementary approaches promise to resolve key tensions and deepen our understanding of the fundamental physics of the Universe.

Below, we highlight ongoing and planned \ac{gw} experiments and their contributions to addressing cosmological tensions.

\begin{itemize}
    \item \textbf{LIGO-Virgo-KAGRA Network}: The collaboration between the \ac{ligo}, Virgo, and KAGRA interferometers~\cite{KAGRA:2013rdx} has already achieved a remarkable number of significant detections, including the merger of a neutron star with an unknown compact object. This success is attributed to substantial sensitivity improvements, stemming from upgrades to individual detectors and the synergistic effect of combined observations. The fourth observation run (O4) is scheduled to conclude in the summer of 2025, after which further upgrades will be implemented. These upgrades will include advancements such as reduced thermal noise, new test mass mirrors, and the installation of a larger beamsplitter in the \ac{ligo} detectors. These enhancements will significantly increase the network's overall sensitivity, enabling a larger detection volume and improved precision in measurement. As the number of detections grows, the uncertainty in key cosmological parameters like $H_0$ will decrease. Specifically, for neutron star-neutron star (NS-NS) mergers, the uncertainty in $H_0$ estimates is expected to improve by approximately $15\%/\sqrt{N}$, where $N$ is the number of detections. This progress underscores the pivotal role of the \ac{ligo}-Virgo-KAGRA network in refining our understanding of the Universe.
        
    \item \textbf{Laser Interferometer Space Antenna (LISA)}: \ac{lisa}~\cite{LISA:2017pwj} will be the first space-based \ac{gw} detector, consisting of three spacecraft in a triangular configuration, separated by millions of kilometers, and following a heliocentric orbit. Scheduled for launch in the mid-2030s, \ac{lisa} will primarily focus on astrophysical phenomena, including high-redshift mergers, extreme mass ratio inspirals, galactic binaries, and planetary objects. However, its contributions to cosmology will be equally groundbreaking. One of \ac{lisa}'s key cosmological objectives is to dramatically expand the catalog of dark sirens. These \ac{gw} events, devoid of electromagnetic counterparts, serve as alternative standard candles, independent of the traditional cosmological distance ladder. By leveraging statistical methods with galaxy catalogs or identifying electromagnetic counterparts where possible, \ac{lisa} will test the distance-redshift relation on cosmological scales, providing new constraints on \ac{de} models. With sensitivity extending to redshifts as high as $z \sim 10$, \ac{lisa} will open a window to the Universe's distant past, offering unprecedented opportunities to explore the nature of \ac{de} at scales inaccessible to current methods. Furthermore, \ac{lisa}'s observations will enable precise and independent measurements of $H_0$, contributing to resolving the persistent cosmological tensions.

    \item \textbf{Einstein Telescope (ET)}: \ac{et}~\cite{Sathyaprakash:2012jk,Abac:2025saz,Punturo:2010zz} is a proposed next-generation \ac{gw} observatory designed to achieve unprecedented sensitivity through advancements in quantum noise suppression, cryogenics, and interferometry. With a triangular configuration and a planned underground location to minimize seismic noise, the \ac{et} represents a significant leap forward from current ground-based detectors. A pathfinder prototype \cite{Utina:2022qqb}, established at Maastricht University in 2021, has already demonstrated promising results in key enabling technologies, paving the way for the full-scale project. Like \ac{lisa}, the \ac{et} will test the distance-redshift relation on cosmological scales and provide independent measurements of $H_0$, free from the assumptions of the traditional distance ladder \cite{Chen:2024gdn}. This will primarily be achieved through the detection of binary compact object coalescences, which serve as robust standard sirens.  Beyond $H_0$, the \ac{et} will be uniquely equipped to detect \ac{sgwb}s of cosmological origin, offering a direct window into the early Universe. These observations could shed light on fundamental processes such as inflation, the formation of primordial black holes, phase transitions in the early Universe, and potential topological defects. The \ac{et}’s ability to probe these phenomena would significantly deepen our understanding of the Universe's origins and evolution, making it a cornerstone of future \ac{gw} astronomy and cosmology.

    \item \textbf{Cosmic Explorer (CE)}: The proposed CE observatory~\cite{Evans:2021gyd,Evans:2023euw} will build on the design principles of \ac{ligo}, but with significantly enhanced capabilities. Its arms, each spanning $40\,\mathrm{km}$, will dramatically improve sensitivity, particularly in the low-frequency range. This enhanced sensitivity will enable the detection of black hole-black hole mergers at unprecedented distances and increase the detection rate to as many as $10^5$ events per year. With its ability to observe \ac{gw}s in the $5$–$4000\,\mathrm{Hz}$ frequency band, CE will produce an unparalleled map of the \ac{gw} sky, reaching back to high redshifts of $z \sim 20$ \cite{Borhanian:2022czq,Gupta:2023lga}. By probing this deep into the Universe's history, CE will offer complementary insights into binary mergers, tracing their origins to the first stars and providing critical information on the star formation rate and galaxy evolution over cosmic time. In addition to astrophysical discoveries, CE will play a pivotal role in cosmology. By observing a vast catalog of dark sirens, CE will enable precise inferences of $H_0$ and provide a valuable independent check on the expansion history of the Universe. This makes CE an essential component of next-generation \ac{gw} astronomy and its intersection with cosmology.

    \item \textbf{TianQin}: TianQin~\cite{TianQin:2020hid,Luo:2025ewp} is a proposed Chinese space-based \ac{gw} observatory designed to detect \ac{gw}s in the millihertz frequency range. It will consist of three spacecraft in a geocentric orbit, forming an equilateral triangle with arm lengths of approximately $10^5$ km. TianQin's primary scientific objectives include the detection of signals from supermassive black hole mergers, extreme mass ratio inspirals (EMRIs), and \ac{sgwb}s. One of TianQin's unique features is its orbit near the Earth, which facilitates precise laser interferometry while minimizing challenges associated with deep-space communication. Similar to \ac{lisa}, TianQin will employ advanced laser metrology and drag-free control to achieve exceptional sensitivity to low-frequency \ac{gw}s. In the context of cosmology, TianQin will contribute significantly to resolving cosmological tensions. By expanding the catalog of both bright and dark sirens, TianQin will enable independent measurements of $H_0$ and test the distance-redshift relation on cosmological scales. Additionally, its observations will probe the early Universe, offering insights into the \ac{sgwb} and potential new physics beyond the standard model. TianQin is expected to launch in the 2030s, complementing other space-based observatories and advancing the era of precision \ac{gw} cosmology.

    \item \textbf{Taiji}: Taiji~\cite{Ren:2023yec} is a proposed Chinese space-based \ac{gw} observatory designed to detect \ac{gw}s in the millihertz frequency band. Unlike TianQin, which operates in a geocentric orbit, Taiji will be positioned at the Sun-Earth Lagrange point $L_2$, providing a quieter observational environment and allowing for longer baselines. Its design includes three spacecraft forming an equilateral triangle with arm lengths of $3 \times 10^6$ km, optimized for detecting low-frequency \ac{gw}s. Taiji’s scientific goals include probing the evolution of the Universe at redshifts as high as $z \sim 20$ and detecting \ac{gw}s from sources such as massive black hole binaries, intermediate-mass black hole mergers, and the \ac{sgwb}. Its high sensitivity will also enable the study of rare cosmic phenomena inaccessible to other observatories, offering an unparalleled view of early-Universe processes. In cosmology, Taiji will contribute to addressing key tensions, including the $H_0$ discrepancy, by expanding the catalog of dark sirens and providing precise measurements of the distance-redshift relation. Additionally, Taiji will enhance our understanding of \ac{de} and inflation by exploring signals from primordial \ac{gw}s, phase transitions, and primordial black holes. Scheduled for launch in the 2030s, Taiji complements the capabilities of both TianQin and \ac{lisa}, focusing on high-redshift phenomena and deepening our understanding of the early Universe while advancing China’s leadership in \ac{gw} astronomy.

     \item \textbf{DECi-hertz Interferometer Gravitational Wave Observatory (DECIGO)}: DECIGO~\cite{Kawamura:2020pcg} is a proposed Japanese space-based \ac{gw} observatory designed to fill the frequency gap between ground-based detectors like \ac{ligo} and Virgo and low-frequency observatories like \ac{lisa}. With a target frequency range centered around decihertz ($0.1$–$10$ Hz), DECIGO will enable the study of intermediate-mass black hole binaries, early Universe \ac{gw} backgrounds, and other phenomena inaccessible to existing detectors. The DECIGO mission will consist of three spacecraft forming a triangular configuration with arm lengths of $1,000$ km, operating in a heliocentric orbit. The observatory will use highly sensitive laser interferometry and drag-free control technologies to achieve unprecedented precision in \ac{gw} detection. A prototype mission, B-DECIGO, is planned to demonstrate key technologies in Earth orbit before the full-scale DECIGO mission launches. In cosmology, DECIGO will provide unique insights by detecting \ac{gw}s from primordial sources such as phase transitions, cosmic strings, and inflation. Its sensitivity to the \ac{sgwb} will allow for constraints on the physics of the early Universe, offering a direct window into energy scales far beyond those probed by \ac{cmb}. DECIGO will also expand the catalog of standard sirens, enabling precise measurements of $H_0$ and the distance-redshift relation, which are crucial for resolving cosmological tensions. DECIGO’s ability to study \ac{gw}s from $z \sim 1000$ to the present will bridge the gap between early-Universe physics and late-time structure formation, providing complementary data to missions like \ac{lisa}, Taiji, and TianQin. Expected to launch in the mid-2030s, DECIGO represents a major advancement in \ac{gw} astronomy, with profound implications for cosmology and fundamental physics.

    \item \textbf{Pulsar Timing Arrays (PTAs)}: Pulsar Timing Arrays are an innovative approach to detecting low-frequency \ac{gw}s in the nanohertz frequency band. Unlike ground- or space-based interferometers, PTAs utilize the precise timing of millisecond pulsars as natural clocks to measure distortions in spacetime caused by passing \ac{gw}s. International collaborations, such as the North American Nanohertz Observatory for Gravitational Waves (NANOGrav)~\cite{NANOGrav:2023gor}, the European Pulsar Timing Array (EPTA), the Parkes Pulsar Timing Array (PPTA)~\cite{2013PASA...30...17M}, and the International Pulsar Timing Array (IPTA)~\cite{2013CQGra..30v4010M}, form a global network to maximize sensitivity and coverage. PTAs are particularly sensitive to \ac{gw}s produced by supermassive black hole binaries (SMBHBs), which emit at nanohertz frequencies during their inspiral phase. Additionally, PTAs can probe the \ac{sgwb} arising from the superposition of signals from numerous SMBHBs or from cosmological sources, such as cosmic strings, phase transitions, or inflation. In cosmology, PTAs provide unique insights by constraining the evolution of structure and the formation of massive galaxies. These measurements complement other \ac{gw} observatories by covering a distinct frequency range, extending the spectrum of observable \ac{gw}s. PTAs can also indirectly contribute to addressing the $H_0$ tension by improving our understanding of galaxy mergers and structure growth. Recent breakthroughs, such as the NANOGrav 15-year data release, have hinted at the first detection of a \ac{sgwb}. These results, if confirmed, would open a new window into the Universe's evolution and provide evidence for processes occurring at very high energy scales, far beyond the reach of current particle accelerators. PTAs represent an essential component of the global \ac{gw} detection strategy, providing a unique and complementary perspective to ground- and space-based observatories. With the advent of next-generation radio telescopes like \ac{ska}, the sensitivity of PTAs is expected to dramatically improve, enabling more precise measurements and extending the range of detectable sources.

\end{itemize}

\subsubsection{21 cm Cosmology} \label{21cm}

\noindent The 21 cm line, arising from the hyperfine transition of neutral hydrogen (HI), provides a powerful and versatile tool for probing the Universe's structure and evolution across a wide range of redshifts. This signal serves as a unique tracer of the \ac{igm} and large-scale structure, enabling the study of key epochs such as the cosmic dawn, the \ac{eor}, and the post-reionization era. Unlike traditional probes, 21 cm cosmology captures three-dimensional information, offering tomographic insights into the distribution of matter and its interaction with radiation fields over cosmic time.
The potential of the 21 cm line extends beyond mapping large-scale structure. It offers a novel avenue to address key cosmological tensions, including constraints on $H_0$ and the amplitude of matter fluctuations $S_8$. Moreover, the sensitivity of 21 cm surveys to high-redshift phenomena makes them an invaluable complement to other cosmological probes such as \ac{cmb}, \ac{bao}, and \ac{wl}.

In this part, we outline the current and future 21 cm experiments, their capabilities, and their expected contributions to cosmology. Special emphasis is placed on their role in addressing fundamental questions about the early Universe, the formation of cosmic structures, and the underlying physics driving cosmic tensions:

\begin{itemize}

    \item \textbf{Low-Frequency Array (LOFAR)}: LOFAR~\cite{LOFAR:2013jil} is a cutting-edge radio interferometer designed to observe the Universe at low radio frequencies, operating primarily in the range of $10\,\text{MHz}$ to $240\,\text{MHz}$. With its core located in the Netherlands and stations distributed across Europe, LOFAR provides high-resolution imaging and wide-field observations. A key focus of LOFAR is probing the $21\,\text{cm}$ hydrogen signal from the \ac{eor}, offering insights into the period when the first stars and galaxies ionized the \ac{igm}. By mapping the structure of neutral hydrogen during this era, LOFAR aims to uncover the processes governing cosmic reionization and the emergence of the first luminous sources. Additionally, LOFAR’s sensitivity to low-frequency signals makes it a valuable tool for studying other phenomena, including cosmic magnetism, \ac{dm}, and the large-scale structure of the Universe. The array's modular and scalable design allows it to complement ongoing and future surveys.

    \item \textbf{Murchison Widefield Array (MWA)}: MWA~\cite{2019PASA...36...50B} is a cutting-edge low-frequency radio interferometer located in Western Australia, designed to observe the Universe at frequencies between $80\,\text{MHz}$ and $300\,\text{MHz}$. With its compact configuration of 4,096 dipole antennas spread across 1.5 kilometers, the MWA excels at capturing wide-field, low-frequency observations. A primary objective of the MWA is to detect the $21\,\text{cm}$ hydrogen signal from the \ac{eor}, enabling a detailed study of the processes that ionized the early Universe and the formation of the first stars and galaxies. Additionally, the MWA probes large-scale cosmic structures, providing insights into the distribution of matter and the role of \ac{de} in cosmic evolution. The array is a precursor instrument to the \ac{ska} and serves as a testbed for developing advanced techniques in calibration, foreground removal, and signal extraction. Its location at the radio-quiet Murchison Radio-astronomy Observatory ensures minimal interference, enhancing its sensitivity to faint cosmological signals. Beyond cosmology, the MWA is also used to study solar physics, ionospheric science, and transient phenomena, making it a versatile instrument for advancing astrophysical research.

    \item \textbf{Hydrogen Epoch of Reionization Array (HERA)}: HERA~\cite{DeBoer:2016tnn} is a dedicated radio interferometer designed specifically to study the $21\,\text{cm}$ hydrogen signal from the \ac{eor}. Located in the radio-quiet Karoo region of South Africa, HERA consists of 350 closely packed parabolic dishes, each $14\,\text{m}$ in diameter, optimized for high sensitivity to faint cosmological signals at redshifts corresponding to the \ac{eor} ($6 < z < 12$). HERA’s primary goal is to provide a detailed characterization of the \ac{igm} during the \ac{eor}, tracing the formation and evolution of the first luminous sources—stars, galaxies, and black holes—and their impact on the ionization state of the Universe. By mapping fluctuations in the $21\,\text{cm}$ emission, HERA will constrain the timing and progression of reionization, offering insights into the astrophysical processes driving it. The array's design prioritizes simplicity and precision calibration, which are critical for mitigating systematics and isolating the faint \ac{eor} signal from overwhelming foreground contamination. As a successor to PAPER, HERA incorporates lessons learned from earlier experiments, significantly improving sensitivity and resolution. HERA is expected to complement other 21 cm experiments like LOFAR, MWA, and \ac{ska} by focusing specifically on the \ac{eor}. Its measurements will also provide valuable data for testing models of cosmic structure formation, the properties of the first galaxies, and potential new physics beyond the standard cosmological model.

    \item \textbf{The Radio Experiment for the Analysis of Cosmic Hydrogen (REACH)}: REACH~\cite{Cumner:2021wmh} is an innovative experiment designed to detect the globally averaged 21 cm signal from neutral hydrogen, targeting the cosmic dawn and the \ac{eor}. What sets REACH apart is its novel calibration techniques and an advanced antenna design aimed at mitigating the impact of instrumental systematics and foreground contamination, long-standing challenges in 21 cm cosmology. By achieving unprecedented precision in isolating the faint cosmological signal, REACH opens a new window into the thermal and ionization history of the early Universe. Beyond its methodological advancements, REACH has the potential to address broader cosmological tensions. Its ability to constrain the timing and physics of reionization provides indirect insights into the nature of \ac{dm} and \ac{de}. Additionally, deviations in the 21 cm signal from standard theoretical predictions could indicate new physics, such as interactions between baryons and \ac{dm} or non-standard early heating mechanisms. These contributions place REACH at the forefront of innovative probes for resolving foundational challenges in cosmology.

    \item \textbf{Canadian Hydrogen Intensity Mapping Experiment (CHIME)}: Located at the Dominion Radio Astrophysical Observatory in Canada, CHIME~\cite{CHIME:2022dwe} operates in the $400$–$800\,\mathrm{MHz}$ band, corresponding to a redshift range of $0.8$ to $2.5$. Its primary goal is to map neutral hydrogen intensity across this range to measure the expansion history of the Universe. By producing the largest three-dimensional map of hydrogen intensity in this critical epoch, CHIME provides unprecedented insights into the large-scale structure of the Universe and probes cosmic evolution during the dark ages through the $21\,\mathrm{cm}$ transition line. In addition to tracing the evolution of structure, CHIME offers statistical constraints on \ac{bao} using the Hydrogen Intensity Mapping technique. These measurements enable synergistic analyses with other cosmological surveys, enhancing the precision of constraints on fundamental parameters like $H_0$ and the amplitude of matter fluctuations $S_8$. Since beginning operations in 2018, CHIME has already yielded significant data, with planned upgrades set to further improve the precision and resolution of intensity maps.

    \item \textbf{Giant Metrewave Radio Telescope (GMRT)}: Located near Pune, India, the GMRT~\cite{Intema:2016jhx} is an array of 30 fully steerable parabolic dishes, each $45\,\text{m}$ in diameter, operating in the frequency range of $50\,\text{MHz}$ to $1.5\,\text{GHz}$. It is one of the most sensitive radio telescopes for low-frequency observations and plays a significant role in $21\,\text{cm}$ cosmology. GMRT has been used extensively to study the $21\,\text{cm}$ hydrogen signal from the Cosmic Dawn and the \ac{eor}. Its ability to probe neutral hydrogen at different redshifts provides critical insights into the thermal and ionization history of the \ac{igm} and the formation of the first luminous sources. In addition to \ac{eor} studies, GMRT contributes to understanding large-scale structures, galaxy formation, and the properties of \ac{de} and \ac{dm}. As an established facility, GMRT continues to produce high-impact results and serves as a vital complement to ongoing and upcoming experiments such as LOFAR, HERA, and \ac{ska}.

    \item \textbf{Five-hundred-meter Aperture Spherical Telescope (FAST)}: FAST~\cite{2011IJMPD..20..989N}, located in Guizhou, China, is the world's largest single-dish radio telescope, with an aperture of $500\,\text{m}$. Operating in the frequency range of $70\,\text{MHz}$ to $3\,\text{GHz}$, FAST is a highly versatile instrument capable of a broad range of astrophysical studies, including $21\,\text{cm}$ observations. Its massive collecting area provides unparalleled sensitivity, making it an invaluable tool for studying neutral hydrogen in the Universe. FAST contributes to 21 cm cosmology by probing the large-scale distribution of hydrogen gas, investigating the \ac{eor}, and tracing the evolution of cosmic structures during the Cosmic Dawn. Its ability to perform both single-dish and interferometric measurements allows it to complement other instruments in the global effort to constrain models of cosmic evolution. Beyond 21 cm studies, FAST is used for pulsar searches, extragalactic surveys, and studies of \ac{ism} dynamics. With its extraordinary sensitivity and high precision, FAST is poised to make new contributions to our understanding of the early Universe and fundamental physics.

    \item \textbf{Hydrogen Intensity and Real-time Analysis eXperiment (HIRAX)}: Building on the technology of CHIME, HIRAX~\cite{Crichton:2021hlc} is located near the \ac{ska} site in South Africa and is designed to achieve similar scientific objectives in the southern sky. Operating in the $400$–$800\,\mathrm{MHz}$ frequency band, HIRAX aims to map neutral hydrogen intensity over the redshift range of $0.8$ to $2.5$, providing critical insights into the evolution of the Universe during this epoch. The array consists of $1,000$ closely packed $6\,\mathrm{m}$ radio dishes, which will create a high-resolution hydrogen intensity map of the southern sky using the $21\,\mathrm{cm}$ emission line. Initially conceived as a demonstration of \ac{ska} technology, HIRAX has evolved into an ambitious project, poised to deliver a detailed three-dimensional map of large-scale structure in this redshift range. A primary science goal of HIRAX is to measure \ac{bao} signature in the matter distribution, enabling precise constraints on the expansion history of the Universe. HIRAX’s southern sky coverage complements CHIME’s northern sky observations, offering an opportunity for joint analyses to refine measurements of cosmological parameters, including $H_0$ and $S_8$.
     
    \item \textbf{Square Kilometre Array Observatory (SKAO)}: The \ac{ska}~\cite{Braun:2015B3} is expected to begin operations for Phase 1 in the late 2020s, with sites located in Australia (covering low-frequency observations) and South Africa (mid-frequency observations). Together, these will form the world's largest radio telescope system, with a total collecting area of approximately $1\,\text{km}^2$, enabling unprecedented sensitivity and resolution. Operating across a frequency range of $50\,\text{MHz}$ to $14\,\text{GHz}$, the \ac{ska} will span a physical radius of roughly $3,000\,\text{km}$, hosting hundreds of radio dishes and over 100,000 antennas. By producing the most comprehensive $21\,\text{cm}$ hydrogen intensity maps to date, the \ac{ska} will probe the large-scale structure of the Universe, reaching back to the first galaxies and stars. This ambitious project will provide new insights into the \ac{eor}, the emergence of the first luminous sources, and the role of \ac{de} in shaping cosmic evolution. Additionally, it will complement other observational probes, offering synergies with \ac{wl}, \ac{bao}, and \ac{cmb} experiments to refine our understanding of cosmological tensions.

\end{itemize}

\subsubsection{Type Ia supernovae and distance ladder}

\noindent \ac{sn1} play a pivotal role in modern cosmology as highly reliable standard candles for measuring cosmic distances. By leveraging the uniform intrinsic luminosity of these \ac{sn1}, calibrated through the cosmic distance ladder, precise determinations of $H_0$ are possible. The cosmic distance ladder comprises several hierarchical steps: (1) geometric distance measurements using Cepheid variables, \ac{trgb} stars, masers, etc.; (2) calibration of \ac{sn1} luminosities based on these measurements; and (3) application of \ac{sn1} to measure $H_0$ in the Hubble flow, where cosmic expansion dominates.
Recent advancements in distance ladder measurements have refined the determination of $H_0$, notably through Cepheid-calibrated \ac{sn1} in the SH0ES project, which reports a value significantly higher than early Universe estimates from the \ac{cmb}. This discrepancy, known as the Hubble tension, remains one of the most pressing challenges in cosmology. Complementary approaches, including \ac{trgb}, Mira variables, and \ac{sbf} methods, offer alternative calibrations and independent cross-checks, helping to ensure robustness and reduce systematic biases. However, all these methods consistently yield higher $H_0$ values than those derived from early Universe probes, highlighting the tension.

Future surveys and advanced instrumentation promise to transform \ac{sn1} measurements and distance ladder studies. \ac{jwst}’s unprecedented resolution has already begun improving Cepheid observations in crowded fields, while upcoming missions aim to reduce systematic uncertainties further. New facilities will also increase the number and diversity of \ac{sn1} observations, extending measurements to higher redshifts and improving statistical power.
Below, we highlight ongoing and planned \ac{sn} and distance ladder experiments and their contributions to addressing cosmological tensions.

\begin{itemize}

    \item \textbf{Dark Energy Survey (DES)}: \ac{des}~\cite{DES:2016jjg}, conducted from 2013 to 2019, used the 4-meter Blanco Telescope at Cerro Tololo Inter-American Observatory (CTIO) to observe approximately $5,000 \,\text{deg}^2$ of the southern sky. \ac{des} contributed significantly to supernova cosmology by discovering and monitoring thousands of \ac{sn1} across a broad redshift range ($z \sim 0.01$ to $1.2$). Its high-quality photometry enabled precise constraints on $H_0$ and \ac{de} properties. \ac{des} complements other distance ladder experiments by providing a statistically robust dataset of \ac{sn1} at intermediate redshifts, improving constraints on the cosmic expansion history and addressing potential systematics in supernova cosmology.
    
    \item \textbf{The Gaia mission}: Gaia~\cite{Gaia:2016zol}, operating since 2013, has revolutionized the cosmic distance ladder by providing precise parallaxes for millions of stars, including Cepheid variables and \ac{trgb} stars. These highly accurate measurements are critical for calibrating the absolute luminosities of \ac{sn1} and improving the determination of the Hubble constant. Gaia’s unparalleled astrometric precision significantly reduces uncertainties in the local distance scale and mitigates systematic biases in \ac{sn1} calibrations. By anchoring the first rung of the distance ladder, Gaia plays a key role in addressing the Hubble tension and refining our understanding of cosmic expansion.

    \item \textbf{Zwicky Transient Facility (ZTF)}: The \ac{ztf}~\cite{Graham:2019qsw}, operating since 2018 at the Palomar Observatory, utilizes a 48-inch Schmidt Telescope with a wide field of view ($47 \,\text{deg}^2$) for high-cadence transient surveys. \ac{ztf} has been instrumental in discovering and monitoring nearby \ac{sn1}, providing critical data for calibrating the distance ladder. By targeting local \ac{sn1}, \ac{ztf} supports precise measurements of $H_0$ and cross-validates calibrations derived from other standard candles, such as Cepheids and \ac{trgb} stars. Its extensive dataset complements space-based missions like Gaia and \ac{jwst} by increasing the sample size of well-observed \ac{sn1}.

    \item \textbf{Nearby Supernova Factory (SNfactory)}: The SNfactory~\cite{Wood-Vasey:2004thg}, an international collaboration, focuses on obtaining detailed spectrophotometry of low-redshift \ac{sn1}. Using the SuperNova Integral Field Spectrograph (SNIFS) on the University of Hawaii’s 2.2-meter telescope, SNfactory has been essential in characterizing the diversity of \ac{sn1} and improving standardization techniques. SNfactory’s precise spectrophotometric data reduce systematic uncertainties in \ac{sn1} luminosities, refining the calibration of the distance ladder and improving constraints on $H_0$.

    \item \textbf{Asteroid Terrestrial-impact Last Alert System (ATLAS)}: Although primarily designed for asteroid detection, ATLAS~\cite{Wang:2019jig}, operating on multiple telescopes in Hawaii, South Africa, and Chile, has made substantial contributions to supernova cosmology. Its wide field of view and high-cadence observations are well-suited for identifying and monitoring nearby \ac{sn1}. ATLAS’s discoveries provide valuable data for calibrating the distance ladder, especially at low redshifts, complementing ground-based and space-based missions in improving the accuracy of $H_0$ measurements and addressing cosmological tensions.

    \item \textbf{James Webb Space Telescope (JWST)}: \ac{jwst}~\cite{Gardner:2006ky}, launched in 2021, provides unparalleled near-infrared imaging and spectroscopy, enabling precise observations of Cepheid variables and \ac{trgb} stars in crowded fields. These measurements significantly enhance the calibration of \ac{sn1} and improve the accuracy of $H_0$. \ac{jwst}’s ability to observe faint objects at high resolution allows it to extend the cosmic distance ladder to greater distances and reduce systematic uncertainties. By complementing ground- and space-based surveys, \ac{jwst} plays a crucial role in addressing the Hubble tension and probing potential new physics beyond \lcdm.

    \item \textbf{Rubin Observatory's Legacy Survey of Space and Time (LSST)}: The Rubin Observatory~\cite{Blum:2022dxi}, currently under construction in Chile, will conduct the \ac{lsst}, a ten-year survey covering approximately $18,000 \,\text{deg}^2$ of the southern sky. With its 8.4-meter aperture and a 9.6-square-degree field of view, Rubin's \ac{lsst} will discover and monitor hundreds of thousands of \ac{sn1} across a broad redshift range ($z \sim 0.01$ to $1.2$). Rubin's \ac{lsst} will revolutionize supernova cosmology by significantly increasing the sample size and improving statistical precision in measuring $H_0$. Its high-cadence, multi-band photometry will mitigate systematics such as dust extinction and host-galaxy effects, and its synergy with spectroscopic follow-up surveys will ensure robust classifications and redshift measurements. 

    \item \textbf{Roman Space Telescope (formerly WFIRST)}: The Roman Space Telescope~\cite{Spergel:2015sza}, scheduled for launch in the late 2020s, will perform a High Latitude Survey to observe thousands of \ac{sn1} across a broad redshift range ($z \sim 0.1$ to $2.0$). With its wide field of view and high-resolution imaging in the near-infrared, Roman will extend the cosmic distance ladder, improving the precision and accuracy of the Hubble constant. Roman’s deep multi-band photometry will reduce systematics like dust extinction and enhance the calibrations of standard candles. By complementing ground-based surveys such as Rubin's \ac{lsst} and incorporating follow-up spectroscopy, Roman will play a critical role in resolving the Hubble tension.

    \item \textbf{Extremely Large Telescope (ELT)}: The \ac{elt}~\cite{2009ASSP....9..225H}, currently under construction in Chile, will leverage its 39-meter aperture and advanced adaptive optics to achieve unprecedented precision in \ac{sn1} observations. By providing high-resolution imaging and spectroscopy, the \ac{elt} will play a vital role in calibrating the distance ladder through Cepheids, \ac{trgb} stars, and direct measurements of \ac{sn1} in the local Universe. The \ac{elt}’s ability to observe \ac{sn1} at higher redshifts will complement space-based surveys like Rubin's \ac{lsst} and Roman, reducing systematic uncertainties and improving the determination of the Hubble constant.

    \item \textbf{Future Extremely Large Telescopes (GMT and TMT)}: Beyond the \ac{elt}, upcoming ground-based observatories such as the Giant Magellan Telescope (GMT)~\cite{2018arXiv180905804C} and the Thirty Meter Telescope (TMT)~\cite{TMTInternationalScienceDevelopmentTeamsTMTScienceAdvisoryCommittee:2015pvw} will significantly enhance \ac{sn1} and distance ladder measurements. With their large apertures of 24.5 meters (GMT) and 30 meters (TMT), these telescopes will provide unparalleled resolution and sensitivity, enabling precise observations of Cepheids, \ac{trgb} stars, and \ac{sn1} in both the local and high-redshift Universe. These telescopes will extend the reach of the distance ladder by detecting fainter Cepheids and \ac{sn1} at greater distances, improving calibration precision and reducing systematic uncertainties. Their advanced instrumentation will complement space-based missions like \ac{jwst} and Roman by providing high-resolution follow-up spectroscopy and imaging, further refining $H_0$ and contributing to resolving the Hubble tension.
    
\end{itemize}

\subsubsection{Time delay cosmography}

\noindent The strong lensing of \ac{qso}s and \ac{sn} offers a unique opportunity to measure the expansion rate of the Universe across different redshifts. This approach leverages the variable nature of these objects, enabling the use of the time-delay method—an independent high-redshift cosmological probe that does not rely on the distance ladder. The technique involves observing multiple images of a strongly lensed source, such as a \ac{qso} or \ac{sn}, produced by an intermediary lensing galaxy. The mass profile of this galaxy, combined with the respective distances between the observer, lens, and source, determines the offsets of the lensed images and is crucial for obtaining precise estimates of cosmological parameters.

This method relies on the time delays between light rays traveling along different paths, caused by both geometric differences in the light geodesics and the distribution of matter within the lensing galaxy. By accurately measuring these delays, the time-delay method provides an estimate of the Hubble constant, independent of large-scale structure parameters.

Strong lensing is inherently challenging due to the specific alignment required between the observer, lens, and source. Moreover, the rarity of suitable \ac{qso}s or \ac{sn} reduces the probability of detection. Despite these challenges, over 300 strongly lensed \ac{qso}s have been identified to date, typically in the redshift range $\sim 1 - 3$, with lensing galaxies found at redshifts $\sim 0.2 - 0.8$. The lensed images are usually separated by a few arcseconds, well within the resolution capabilities of modern ground-based telescopes. Time delays ranging from a few days to a year are sufficient for accurate measurements, making the time-delay method a powerful tool for cosmology.

Below, we highlight ongoing and planned experiments leveraging the time-delay method and their contributions to measuring $H_0$ and addressing cosmological tensions.

\begin{itemize}

    \item \textbf{Very Large Telescope Interferometer (VLTI)}: The VLTI~\cite{VLT1998}, operated by the European Southern Observatory (ESO), provides ultra-high angular resolution imaging through its interferometric array of telescopes. With its ability to resolve fine details in strongly lensed \ac{qso}s, VLTI plays a critical role in time-delay cosmography by enabling precise measurements of the lensing geometry and time delays. Its high spatial resolution is particularly useful for modeling lensing galaxies and reducing systematic uncertainties in the determination of $H_0$.
    
    \item \textbf{Atacama Large Millimeter/submillimeter Array (ALMA)}: ALMA~\cite{2014PASP..126.1126B}, located in Chile, offers unparalleled sensitivity and resolution for observing lensed submillimeter galaxies and \ac{qso}s. By measuring time delays in radio and submillimeter wavelengths, ALMA provides a complementary approach to optical time-delay cosmography, reducing systematic biases and refining constraints on the Hubble constant.

    \item \textbf{Keck Observatory}: The Keck Observatory~\cite{Davis:2002aq}, located on Maunakea, Hawaii, features two 10-meter telescopes equipped with advanced adaptive optics systems. These capabilities allow Keck to obtain high-resolution imaging and spectroscopy of strongly lensed \ac{qso}s and \ac{sn}, essential for precise time-delay measurements. Keck’s sensitivity and angular resolution enable detailed modeling of lensing galaxies and accurate determination of light path geometries, contributing to independent constraints on $H_0$. Its role as a follow-up instrument for time-delay cosmography complements surveys like Rubin's \ac{lsst} and Euclid, providing critical data for refining cosmological models.

    \item \textbf{Subaru Telescope}: The Subaru Telescope~\cite{2014PASJ...66R...1T}, a 8.2-meter optical-infrared telescope located on Maunakea, Hawaii, plays a key role in observing strong lensing systems for time-delay cosmography. Equipped with instruments such as the \ac{hsc}, Subaru excels at wide-field imaging and detailed follow-up of lensed systems. Subaru’s capabilities allow it to identify and resolve lensed \ac{qso} systems and \ac{sn}, particularly in crowded fields, enabling precise time-delay measurements. Its contributions to modeling lensing mass distributions and improving light path geometries are vital for reducing systematic uncertainties in the determination of $H_0$. Subaru serves as a valuable complement to space-based missions like Roman and ground-based facilities like Keck.
    
    \item \textbf{Euclid}: The Euclid mission~\cite{Amendola:2016saw}, covering a $14,000\,\text{deg}^2$ area of the sky, is expected to observe approximately $170,000$ strong lensing events during its lifetime. These events will provide valuable insights into the properties of \ac{dm} halos in dwarf galaxies, the mass distribution within galaxies, and galaxy cluster structures in the redshift range $0 < z < 2$.  Strong lensing events are inherently rare and require robust detection methods. To address this challenge, Euclid employs \ac{cnn}s, which are trained on extensive simulated datasets to extract image features and identify strong lensing properties. These advanced pipelines allow for precise identification of lensing phenomena, including sharp Einstein rings, which are expected to be detectable down to an angular resolution of $0.5$ arcseconds. Additionally, Euclid is anticipated to detect up to $\sim2300$ strongly lensed \ac{qso} sources, enabling precise measurements of time delays. The time-delay cosmography enabled by Euclid’s high-resolution imaging will place stringent constraints on $H_0$, making it one of the most powerful tools for addressing cosmological tensions.

    \item \textbf{Rubin Observatory's Legacy Survey of Space and Time (LSST)}: The Rubin Observatory’s~\cite{Blum:2022dxi} unparalleled depth and sky coverage will enable the discovery of up to ten million \ac{qso}s over its operational lifetime, spanning a redshift range of $z \sim 0 - 7$. This extensive dataset will have profound implications for \ac{qso}-related physics, including strong lensing and time-delay cosmography. Rubin's \ac{lsst} is expected to detect approximately $\sim 2600$ time-delayed lensing systems, assuming optimistic projections. These systems will play a pivotal role in refining measurements of $H_0$ and probing the dynamical properties of \ac{de} at high redshifts. With its deep and wide survey capabilities, Rubin's \ac{lsst} will significantly enhance the statistical precision of time-delay cosmography, providing critical insights into the expansion history of the Universe.

    \item \textbf{Roman Space Telescope (formerly WFIRST)}: The Roman Space Telescope~\cite{Spergel:2015sza}, set to launch in the late 2020s, will make significant contributions to time-delay cosmography through its deep, high-resolution near-infrared imaging and wide field of view. Roman is expected to detect thousands of strongly lensed \ac{qso}s and \ac{sn}, enabling precise measurements of time delays across a broad redshift range. With its exceptional sensitivity, Roman will identify and characterize lensing systems with high precision, including detailed modeling of lensing galaxies. These observations will reduce uncertainties in the Hubble constant $H_0$ and provide crucial constraints on the properties of \ac{de}. Roman’s synergy with other surveys, such as Rubin's \ac{lsst}, will further enhance the statistical power and accuracy of time-delay cosmography.

    \item \textbf{Square Kilometre Array Observatory (SKAO)}: The \ac{ska}~\cite{Braun:2015B3}, with its radio arrays in South Africa and Australia, will enable high-resolution observations of strongly lensed radio sources. By detecting and monitoring time-delayed radio-lensed systems, \ac{ska} will provide independent measurements of $H_0$. Its sensitivity to faint radio signals and ability to resolve sub-arcsecond structures make it a powerful tool for time-delay cosmography.

    \item \textbf{Extremely Large Telescope (ELT)}: The \ac{elt}~\cite{2009ASSP....9..225H}, currently under construction in Chile, will leverage its 39-meter aperture to achieve unparalleled angular resolution and sensitivity. This will allow for detailed imaging and spectroscopy of strongly lensed \ac{qso}s and supernovae, facilitating precise time-delay measurements. The \ac{elt}'s advanced instrumentation will help model lensing galaxies with greater accuracy, reducing uncertainties in the determination of the Hubble constant.

    \item \textbf{Thirty Meter Telescope (TMT)}: The TMT~\cite{TMTInternationalScienceDevelopmentTeamsTMTScienceAdvisoryCommittee:2015pvw}, planned for construction in Hawaii or the Canary Islands, will provide high-resolution imaging and spectroscopic capabilities, enabling detailed observations of strongly lensed systems. By resolving lensed images with extreme precision, the TMT will contribute to time-delay cosmography by improving measurements of $H_0$ and probing the dynamics of \ac{de}.

    \item \textbf{Giant Magellan Telescope (GMT)}: The GMT~\cite{2018arXiv180905804C}, under construction in Chile, will utilize its segmented 24.5-meter primary mirror to achieve exceptional imaging and spectroscopic capabilities. Its high resolution and sensitivity will support time-delay cosmography by enabling precise modeling of lensed systems and accurate measurements of $H_0$, particularly at intermediate redshifts.

\end{itemize}

\subsubsection{Fast Radio Bursts (FRBs)}

\noindent \ac{frb}s are extremely bright, short-duration radio transients lasting only a few milliseconds, with the potential to serve as powerful tools for probing cosmology. Their extragalactic origin is supported by their large dispersion measures, which are observed to evolve with redshift. A significant number of \ac{frb}s have been localized to host galaxies, allowing their redshifts to be determined and their origins, such as possible magnetar emissions, to be studied.

Localized \ac{frb}s provide valuable insights into several cosmological phenomena, including the baryon fraction in the \ac{igm}, strong lensing events, the equivalence principle, the expansion history of the Universe, and the cosmic reionization history. The interaction of \ac{frb}s with the \ac{igm} makes them excellent probes of baryonic matter in these regions, as the interaction of radio signals with gas distributions offers a new perspective on the ``missing baryon problem''. Regarding the Hubble constant $H_0$, \ac{frb}s alone face degeneracies that limit their ability to directly constrain $H_0$. However, combining \ac{frb} data with complementary datasets such as \ac{cmb}, \ac{bbn}, or \ac{gw} measurements breaks these degeneracies and enables precise cosmological constraints.

Below, we highlight the list of ongoing and planned experiments leveraging \ac{frb}s to address cosmological tensions and improve our understanding of the Universe's evolution.

\begin{itemize}

    \item \textbf{Canadian Hydrogen Intensity Mapping Experiment (CHIME)}: CHIME~\cite{CHIME:2022dwe}, situated in British Columbia, Canada, is a pioneering radio telescope designed to monitor transient radio signals, including \ac{frb}s. Its fixed-cylinder design and wide field of view allow CHIME to observe the sky continuously, detecting thousands of \ac{frb}s annually across a range of redshifts. This high detection rate enables the identification of rare \ac{frb}s suitable for cosmological studies, such as those with high dispersion measures or unique signatures that indicate strong lensing events. CHIME’s advanced time-domain sensitivity is optimized for detecting short-duration radio transients, making it a valuable tool for measuring the arrival times of \ac{frb}s with high precision. These capabilities allow CHIME to constrain propagation effects, such as delays caused by intervening structures, which can be cross-referenced with \ac{gw} studies to extract cosmological information. Its unparalleled \ac{frb} detection volume provides a complementary dataset for investigating the evolution of the cosmic expansion when paired with optical and \ac{gw} experiments.

    \item \textbf{Australian Square Kilometre Array Pathfinder (ASKAP)}: ASKAP~\cite{Johnston:2008hp}, located in Western Australia, is a cutting-edge radio telescope optimized for detecting and localizing \ac{frb}s. Its unique combination of wide field-of-view and interferometric imaging capabilities allows ASKAP to efficiently detect and pinpoint \ac{frb}s to sub-arcsecond accuracy. This precise localization is critical for identifying host galaxies and determining redshifts of \ac{frb} sources. ASKAP's advanced beamforming technology enables the detection of high-redshift \ac{frb}s, providing insights into the \ac{igm} and the evolution of the cosmic expansion. Its ability to associate \ac{frb}s with host galaxies allows for independent constraints on $H_0$ when combined with other cosmological probes.

    \item \textbf{Upgrade of the Molonglo Observatory Synthesis Telescope (UTMOST)}: UTMOST~\cite{2017PASA...34...45B}, located in Australia, is a reconfigured radio telescope optimized for detecting \ac{frb}s. Its high-cadence observing strategy makes it ideal for the real-time discovery and timing of \ac{frb}s, particularly at low redshifts. Although less sensitive than newer facilities, UTMOST’s rapid response capability enables timely follow-up studies and statistical surveys of \ac{frb}s. By characterizing nearby \ac{frb}s, UTMOST complements high-sensitivity instruments such as ASKAP and FAST, contributing to our understanding of the local \ac{frb} population and its cosmological implications.

    \item \textbf{MeerKAT}: MeerKAT~\cite{2016mks..confE...1J}, situated in South Africa, is a highly sensitive radio telescope designed for high-resolution imaging and precision timing, making it an invaluable tool for studying \ac{frb}s. Its wide frequency coverage and long baselines enable MeerKAT to detect faint and distant \ac{frb}s, particularly those at high redshifts, offering insights into the early Universe. MeerKAT excels in pinpointing \ac{frb} host galaxies and measuring dispersion measures with high precision. This allows for a detailed study of the large-scale structure of the Universe and the evolution of baryonic matter over cosmic time. MeerKAT also plays a complementary role to \ac{ska} by serving as a pathfinder and refining techniques for future large-scale \ac{frb} surveys. Its contributions to the Hubble constant determination and reionization studies position MeerKAT as a crucial instrument for resolving cosmological tensions.

    \item \textbf{Five-hundred-meter Aperture Spherical Telescope (FAST)}: FAST~\cite{2011IJMPD..20..989N}, located in Guizhou, China, is the world’s largest single-dish radio telescope, offering unparalleled sensitivity for detecting faint and distant \ac{frb}s. Its ability to detect \ac{frb}s with extremely high signal-to-noise ratios enables precise measurements of dispersion measures and localization for follow-up studies. FAST contributes significantly to cosmology by detecting high-redshift \ac{frb}s and analyzing their interaction with the \ac{igm}. Its sensitivity allows it to probe the cosmic expansion and test reionization models, complementing large-scale surveys such as \ac{ska} and CHIME. FAST’s unique capabilities make it a critical instrument for advancing \ac{frb}-based cosmological studies.

    \item \textbf{Deep Synoptic Array (DSA-110)}: The DSA-110~\cite{Law:2023ibd}, located in California, is a dedicated radio array designed exclusively for detecting and localizing \ac{frb}s. With 110 antennas operating in the $1.28 \,\text{GHz}$ band, the array achieves sub-arcsecond localization accuracy, enabling precise identification of \ac{frb} host galaxies. DSA-110’s focus on real-time detection and localization provides critical insights into the redshift distribution of \ac{frb}s and their cosmological applications. By linking \ac{frb}s to their host environments, DSA-110 contributes to breaking degeneracies in cosmological parameter estimation, particularly when combined with other datasets such as \ac{gw} observations and optical surveys.

    \item \textbf{Giant Metrewave Radio Telescope (GMRT)}: The GMRT~\cite{Intema:2016jhx}, located in Pune, India, has been instrumental in studying \ac{frb}s at low radio frequencies. Its unique sensitivity and wide frequency coverage make it a valuable tool for characterizing the spectral properties of \ac{frb}s and their interaction with the \ac{igm}. GMRT has been used for follow-up observations of \ac{frb}s detected by other instruments, particularly at frequencies below $800\,\text{MHz}$. These observations complement higher-frequency detections by telescopes like CHIME and ASKAP, providing new insights into \ac{frb} emission mechanisms and their propagation through cosmic structures. While not a dedicated \ac{frb} facility, GMRT’s versatility and precision continue to support advancements in \ac{frb} cosmology.

    \item \textbf{Very Large Array Low-band Ionosphere and Transient Experiment (VLITE)}: VLITE~\cite{2018AAS...23135411C}, operating on the Very Large Array (VLA), focuses on low-frequency transient detection, including \ac{frb}s. Its ability to monitor the sky continuously allows it to identify faint \ac{frb}s and investigate their spectral properties at lower frequencies.

    \item \textbf{Square Kilometre Array Observatory (SKAO)}: The \ac{ska}~\cite{Braun:2015B3}, under construction in South Africa and Australia, is poised to become the largest and most sensitive radio telescope array in the world. Its exceptional capabilities will enable precise localization of \ac{frb}s at sub-arcsecond resolution and direct measurement of their redshifts, significantly advancing \ac{frb} cosmology. By detecting and studying \ac{frb}s across a broad redshift range, \ac{ska} will provide insights into the evolution of the \ac{igm}, reionization history, and cosmic expansion. Its ability to map dispersion measures over cosmic time will complement other datasets, such as \ac{gw} observations and optical surveys, to break degeneracies in cosmological parameters, including $H_0$. While its primary focus is on large-scale structure studies through 21 cm intensity mapping, \ac{ska}’s \ac{frb}-specific contributions, including redshift localization and host galaxy characterization, make it a new instrument for addressing cosmological tensions.

    \item \textbf{Next Generation Deep Synoptic Array (DSA-2000)}: DSA-2000~\cite{2023AAS...24123907H}, a planned expansion of the DSA-110, will feature 2,000 antennas, dramatically increasing sensitivity and detection rates for \ac{frb}s. This next-generation array will provide precise localization for tens of thousands of \ac{frb}s across a wide redshift range. With its enhanced capabilities, DSA-2000 will enable detailed studies of the cosmic expansion, baryon distribution, and reionization history.

    \item \textbf{Low-Frequency Array (LOFAR2.0)}: LOFAR2.0~\cite{Nakoneczny:2023nlt}, an upgraded version of the Low-Frequency Array, will provide enhanced sensitivity and temporal resolution for detecting \ac{frb}s at low frequencies. Its broad frequency coverage and improved capabilities will allow it to probe previously inaccessible redshift ranges for \ac{frb}s, making it a unique tool for studying the early Universe. LOFAR2.0’s ability to detect low-frequency \ac{frb}s will contribute to understanding their emission mechanisms and their interaction with the \ac{igm}.

    \item \textbf{APERTure Tile In Focus (Apertif)}: Apertif~\cite{2022A&C....3800514A}, an upgrade to the Westerbork Synthesis Radio Telescope (WSRT) in the Netherlands, provides wide-field imaging for detecting \ac{frb}s. Its combination of sensitivity and survey speed enables it to identify transient events, contributing to population studies of \ac{frb}s and complementing other instruments in the field.

    \item \textbf{Burst Transient Telescope (BURSTT)}: BURSTT is a planned experiment designed to search for \ac{frb}s with enhanced detection rates and localization capabilities. Its wide field of view will make it particularly effective in identifying high-redshift \ac{frb}s and improving constraints on their cosmological applications.

    \item \textbf{Canadian Hydrogen Observatory and Radio-transient Detector (CHORD)}: CHORD~\cite{2019clrp.2020...28V}, a planned successor to CHIME, will combine higher sensitivity with improved localization capabilities for \ac{frb}s. CHORD is expected to detect tens of thousands of \ac{frb}s across a broad redshift range, providing critical insights into the evolution of cosmic structures and addressing cosmological tensions.

    \item \textbf{Pathfinder for Real-time Extragalactic Cosmic Intensity Survey Experiment (PRECISE)}: PRECISE is a dedicated pathfinder for \ac{frb} studies, focusing on real-time detection and precise localization of \ac{frb}s. It will enhance the study of their host galaxies and the \ac{igm}, making it a valuable addition to the next generation of \ac{frb} experiments.

\end{itemize}

\subsubsection{Line-intensity mapping}

\noindent The large-scale structure of the Universe can be surveyed through the radiation intensity emitted by gas clouds, in a manner similar to galaxy surveys, but without resolving individual sources. This approach significantly accelerates the survey process. While many surveys focus exclusively on the $21\,\mathrm{cm}$ line associated with the hyperfine transition of neutral hydrogen, line-intensity mapping (LIM) extends this technique to include additional spectral lines, such as carbon monoxide (CO) and ionized carbon (CII), which provide complementary tracers of large-scale structure and galaxy evolution. The intensity distribution maps derived from LIM can be translated into matter fluctuation power spectra and enable studies across a broader range of cosmic epochs.

Unlike traditional $21\,\mathrm{cm}$ experiments, which focus specifically on the neutral hydrogen distribution, LIM captures the integrated emission of multiple spectral lines, making it a more versatile approach. This broader capability allows LIM to probe different phases of the \ac{ism} and the \ac{eor}, as well as to reduce parameter degeneracies by spanning a wider range of redshifts and tracers. 

This innovative technique employs angular resolution observations to trace the spatial distribution of matter, even at high redshifts where traditional galaxy surveys face challenges. If foreground contamination can be effectively mitigated, line-intensity mapping has the potential to (1) provide wide redshift coverage, averaging out cosmological parameter degeneracies across different redshifts, (2) yield large-scale information that could reveal signatures of inflation and gravitational effects, and (3) access matter density modes that are less affected by non-linear physics.
By complementing $21\,\mathrm{cm}$ experiments, LIM plays a critical role in addressing cosmological tensions, such as the $H_0$ and $S_8$ discrepancies. Its ability to combine data from multiple spectral lines and explore different epochs of cosmic evolution allows it to test new models of \ac{de}, modifications to gravity, and other potential extensions to the standard cosmological model.

The further development of this technique will require advancements in millimeter-wave detectors and the refinement of data analysis methods to mitigate astrophysical and modeling systematics. Below, we identify the key ongoing and planned experiments leveraging line-intensity mapping to study cosmic large-scale structures and the \ac{eor} using this method.

\begin{itemize}

    \item \textbf{The Hobby-Eberly Telescope Dark Energy Experiment (HETDEX)}: HETDEX~\cite{Hill:2008mv} is a groundbreaking survey leveraging line-intensity mapping of Lyman-$\alpha$ emitters to study the large-scale structure of the Universe. By mapping the intensity distribution of Lyman-$\alpha$ emission over the redshift range $1.9 < z < 3.5$, HETDEX provides insights into the matter power spectrum and the evolution of cosmic structures during a pivotal epoch in the Universe’s history. HETDEX’s line-intensity mapping measurements play a crucial role in addressing cosmological tensions by offering independent constraints on large-scale structure and its growth, as well as the expansion history of the Universe. Additionally, HETDEX’s use of intensity mapping reduces reliance on resolving individual sources, enabling efficient surveys of faint, distant cosmic structures.
    
    \item \textbf{CO Mapping Array Project (COMAP)}: COMAP~\cite{Cleary:2021dsp} is a pioneering experiment designed to map the large-scale structure of the Universe through intensity mapping of the CO rotational transition lines. By targeting the $z \sim 2.4 - 3.4$ redshift range, COMAP traces the distribution of molecular gas in galaxies during the peak of star formation activity, offering insights into galaxy evolution and the assembly of cosmic structures. COMAP plays a crucial role in addressing cosmological tensions by providing independent constraints on the large-scale matter distribution and the expansion history.

    \item \textbf{CarbON CII line in post-rEionization and ReionizaTiOn epoch (CONCERTO)}: CONCERTO~\cite{Fasano:2023udw} is a state-of-the-art experiment designed to study the large-scale structure of the Universe using line-intensity mapping of the [CII] emission line. Operating in the redshift range $4.5 < z < 8.5$, CONCERTO aims to trace the distribution of ionized carbon during the \ac{eor} and the early stages of galaxy formation. By capturing the integrated [CII] signal from unresolved sources, it provides a unique probe of the \ac{ism} and star formation in high-redshift galaxies. CONCERTO’s sensitivity to high-redshift structures complements other intensity mapping experiments and traditional surveys, enabling a more detailed understanding of reionization-era physics and its connection to late-time cosmological parameters. Furthermore, by probing large-scale modes, CONCERTO helps refine models of \ac{de}, inflation, and modifications to gravity, providing new insights into the underlying causes of cosmological tensions.
    
    \item \textbf{Spectro-Photometer for the History of the Universe, Epoch of Reionization, and Ices Explorer (SPHEREx)}: This proposed all-sky survey aims to provide spatial distribution data using various tracers to explore different cosmic eras~\cite{SPHEREx:2014bgr}. The survey will utilize the H$\alpha$, H$\beta$, and [OIII] lines to probe $0.1 < z < 5$, $0.5 < z < 2$, and $0.5 < z < 3$, respectively, mapping low-redshift gas clouds. The higher redshift range, spanning $5.2 < z < 8$, will be investigated using the Lyman-$\alpha$ line, providing a direct probe of the \ac{eor}. By spanning a wide range of redshifts and utilizing multiple spectral lines, SPHEREx can help address cosmological tensions, such as the $H_0$ and $S_8$ discrepancies, by providing independent measurements of large-scale structure growth and the expansion history of the Universe.

    \item \textbf{Tomographic Ionized-Carbon Mapping Experiment (TIME)}: TIME~\cite{2024SPIE13102E..2GB} is designed to probe the \ac{eor} by mapping the redshifted 157.7~$\mu$m emission line of singly ionized carbon ([CII]) over the redshift range $5 \lesssim z \lesssim 9$. In addition, it observes carbon monoxide (CO) rotational transitions from galaxies at intermediate redshifts $0.5 \lesssim z \lesssim 2$. These measurements aim to trace the early stages of galaxy formation and star formation history, providing a complementary view of large-scale structure during key epochs in cosmic evolution.

    \item \textbf{The Cosmic Dawn Intensity Mapper (CDIM)}: CDIM~\cite{2019BAAS...51g..23C} is a proposed space-based mission designed to explore the large-scale structure of the Universe through line-intensity mapping of multiple spectral lines. By targeting key tracers such as hydrogen Lyman-$\alpha$, H$\alpha$, H$\beta$, and [OIII], CDIM will probe a wide redshift range, spanning from the \ac{eor} ($6 < z < 10$) to the era of peak star formation ($z \sim 2$). These measurements will provide a comprehensive view of the distribution of ionized and molecular gas, as well as star formation processes. CDIM’s ability to map the matter distribution across a wide range of redshifts and tracers reduces parameter degeneracies and systematic uncertainties. By bridging the gap between early- and late-time probes, CDIM provides complementary data to galaxy surveys and \ac{cmb} experiments. Additionally, its sensitivity to faint, diffuse emission makes it a powerful tool for studying large-scale modes that are less accessible with traditional surveys.

    \item \textbf{Fred Young Submillimeter Telescope (CCAT-prime)}: CCAT-prime~\cite{Stacey:2018yqe}, equipped with the advanced Prime-Cam instrument, is designed to perform line-intensity mapping of emission lines such as [CII] at redshifts relevant to the \ac{eor} and cosmic dawn. By tracing the [CII] line, along with other potential tracers, CCAT-prime aims to explore the star formation history, the \ac{ism}, and the role of galaxies in reionizing the early Universe. Its high-altitude site and state-of-the-art instrumentation provide exceptional sensitivity for mapping large-scale structure during key phases of cosmic evolution.

\end{itemize}

\subsubsection{Potential new probes}

\noindent As our understanding of the Universe deepens, new observational techniques and theoretical frameworks are being developed to challenge existing paradigms and explore unresolved tensions in cosmology. These potential new probes leverage advanced technologies and novel methodologies to address persistent discrepancies in key cosmological parameters, such as the Hubble constant $H_0$, the amplitude of matter fluctuations $S_8$, etc. By pushing the boundaries of precision measurements and theoretical modeling, these probes aim to uncover new physics or systematic effects that may underlie these tensions. 

This section explores these innovative probes, highlighting their potential contributions to resolving the current challenges in cosmology and paving the way for new discoveries about the Universe.

\begin{itemize}

    \item \textbf{Visible and Infrared Survey Telescope for Astronomy (VISTA)}: VISTA~\cite{2015A&A...575A..25S} is a ground-based facility dedicated to wide-field surveys in the visible and near-infrared wavelengths. Equipped with an advanced infrared camera, VISTA is uniquely suited to mapping the large-scale structure of the Universe by detecting galaxies and \ac{qso}s at high redshifts. Its unprecedented sensitivity and wide-field capabilities allow it to probe the cosmic web, trace the distribution of matter, and explore the evolution of structures over cosmic time. By providing precise measurements of galaxy clustering and \ac{wl} signals, VISTA contributes to addressing key cosmological tensions, such as those related to the amplitude of matter fluctuations $S_8$ and the Hubble constant $H_0$. Furthermore, its ability to survey large areas of the sky with high resolution offers unique opportunities to identify deviations from standard cosmological models, potentially unveiling signatures of new physics.

    \item \textbf{Transient High-Energy Sky and Early Universe Surveyor (THESEUS)}: THESEUS~\cite{THESEUS:2021uox} is a proposed space-based mission designed to explore the high-energy transient sky and study the early Universe. By detecting \ac{grb}s from the first generation of stars (Pop III stars) and other energetic phenomena, THESEUS provides a unique probe into the era of cosmic reionization and early star formation. THESEUS combines wide-field monitoring and precise localization of \ac{grb}s with follow-up spectroscopy, enabling measurements of the cosmic star formation rate and the chemical enrichment history of the Universe. Its ability to observe the high-energy sky at unprecedented sensitivity offers a complementary approach to resolving cosmological tensions, such as refining constraints on the matter density parameter ($\Omega_{\rm m,0}$) and testing models of \ac{de}. By bridging the gap between high-energy astrophysics and cosmology, THESEUS represents a transformative tool for uncovering new physics and deepening our understanding of the early Universe.

    \item \textbf{Quasars as Geometric Rulers}: \ac{qso}s, the luminous cores of \ac{agn}, offer a novel approach to cosmological measurements by serving as geometric rulers. Their variability over time, combined with precise measurements of time delays in gravitationally lensed \ac{qso}s, enables the determination of absolute distances in the Universe. This method bypasses many of the traditional assumptions in standard candles or rulers, providing an independent means of measuring $H_0$ and testing the expansion history of the Universe. In addition to their role in time-delay cosmography, \ac{qso}s also probe the large-scale structure of the Universe through clustering and absorption-line studies. This dual utility allows \ac{qso}s to contribute to resolving tensions in cosmological parameters, such as $H_0$ and $S_8$. Their utility as geometric rulers and tracers of structure represents a powerful avenue for exploring deviations from the standard cosmological model.

    \item \textbf{Redshift Drift with ANDES (ArmazoNes High Dispersion Echelle Spectrograph)}: ANDES, a planned instrument for the \ac{elt}, offers a unique opportunity to measure the redshift drift—a direct, model-independent observation of the Universe's expansion history. By observing spectral lines of distant \ac{qso}s and tracking their subtle changes over decades, ANDES aims to detect the tiny redshift drift predicted by the standard cosmological model. This groundbreaking method provides a direct probe of the dynamics of the Universe, bypassing the need for distance ladder calibrations or assumptions about standard candles or rulers. By directly measuring the acceleration or deceleration of cosmic expansion, ANDES will place new constraints on \ac{de}, modifications to gravity, and the fundamental properties of the Universe. Its unparalleled precision makes it a new tool for addressing unresolved tensions in cosmological parameters.

    \item \textbf{CMB Spectral Distortions}: Deviations from a perfect blackbody spectrum of the \ac{cmb} encode information about energy injections in the early Universe. These distortions arise due to processes such as Silk damping of primordial perturbations, decaying or annihilating particles, and early structure formation. The two primary types of distortions, $\mu$-distortions (sensitive to energy injections at $z \gtrsim 10^5$) and $y$-distortions (from later energy injections), provide a complementary approach to testing the standard \lcdm\ model. Measurements of spectral distortions could reveal signatures of new physics, including primordial non-Gaussianity, modifications to the recombination history, or interactions between \ac{dm} and radiation. Future space missions such as \textit{PIXIE} and \textit{PRISM} aim to achieve the required sensitivity to detect these distortions, opening new avenues for cosmological exploration.

    \item \textbf{Lunar Astronomy}: Lunar astronomy~\cite{Silk:2020bsr}, leveraging the unique environment of the Moon, presents a groundbreaking opportunity to probe the cosmos in ways that are challenging or impossible from Earth. The Moon's lack of atmosphere eliminates atmospheric distortions, enabling precise observations across a wide range of wavelengths, including radio, infrared, and X-rays. Furthermore, the Moon's far side provides an unparalleled radio-quiet environment, ideal for low-frequency observations of the early Universe, including the cosmic dawn and the \ac{eor}. Lunar observatories have the potential to address cosmological tensions by providing high-precision measurements of large-scale structure, the \ac{cmb}, and the 21 cm signal from neutral hydrogen. Additionally, their stable environment offers the possibility of ultra-long-term observations, enabling direct measurements of cosmological parameters such as the Hubble constant and constraints on the nature of \ac{de}. By extending observational capabilities beyond Earth-based and space-based platforms, lunar astronomy could transform our understanding of the Universe.

\end{itemize}

\bigskip
\subsection{Recommendations \label{sec:recomms}}

\noindent The open question of \textbf{cosmic tensions presents a unique opportunity} to reassess how observational measurements in cosmology are reconciled with physical models through diverse data analysis approaches. This may signal a turning point in our understanding of how to interpret measurements, particularly in the presence of potential systematics at a fundamental level. Such a systematic effect would need to manifest across all phenomena in local distance ladder measurements or be ubiquitous in early-time measurements. Another possibility is the emergence of new, independent measurements that do not rely on either the distance ladder or early Universe constraints. In particular, advances in \ac{gw} astronomy, where standard sirens provide a novel approach to constructing distance scales, hold great promise. While this method remains in its early stages and has yet to reach the precision required to compete with traditional approaches, its continued development—particularly with upcoming detectors such as \ac{lisa} and third-generation ground-based observatories—could play a crucial role in resolving cosmic tensions in the future.

A key tool in analyzing observational data and potential systematics, alongside the exploration of fundamental physics theories, is the application of data analysis methods. Traditionally, data analysis has relied on Bayesian statistical techniques, both for interpreting observational data and for comparing physical models. However, there is growing interest in developing frequentist approaches more broadly, given their reduced dependence on prior information. At the same time, there is strong motivation to advance statistical methods that leverage the broad range of \ac{ml} techniques. These include supervised learning approaches such as \ac{gp}, as well as unsupervised methods like neural networks. The development of toolkits based on these statistical frameworks holds significant potential, both in enhancing the extraction of information from observational data and in refining model-building strategies. In particular, such tools could provide valuable insights into fundamental physics by systematically addressing the open issues surrounding cosmic tensions. It will be essential to integrate these methodologies into standard analysis pipelines for upcoming large-scale surveys.

A plethora of physically motivated fundamental physics theories have been studied in recent years, with particular attention given to their connection to resolving the open problem of cosmic tensions and associated questions within the standard model of cosmology. These efforts have focused on further developing geometric gravity, exploring potential physics beyond the standard model of particle physics, and investigating emergent phenomenology and possible quantum gravity effects. On the other hand, many practical theories in these approaches were originally motivated by theoretical or semi-physical considerations. The further integration of the growing body of observational evidence into the development of new classes of physical theories will drive the next generation of cosmological models, aligning them with the most pressing questions from observational surveys. Additionally, the use and advancement of model-building toolkits will contribute to the creation of physically motivated theoretical models that address cosmic tensions while preserving the successes of the current standard model of cosmology.

The breadth of observational surveys has been reviewed in Sec.~\ref{sec:obs}, where their potential systematics are discussed in detail. In Sec.~\ref{sec:data_ana}, recent developments in traditional data analysis approaches are examined alongside discussions of emerging statistical tools that are expected to become competitive in the coming years, including those based on \ac{ml} methods. The development of new physically motivated fundamental theories is explored in Sec.~\ref{sec:fun_phys}, with a focus on their potential to resolve cosmological tensions while preserving the successful elements of the current standard model of cosmology. The future of the field will depend not only on the advancement of each of these areas individually but also on the ability of the community to integrate these diverse approaches into a coherent framework to address the growing challenges posed by cosmic tensions. This effort will require close collaboration between observational cosmologists, data analysts, and fundamental physicists. In the coming decade, it will be essential to:
\begin{itemize}
    \item Leverage next-generation observational surveys to improve the precision of key cosmological parameters and systematically test for hidden systematics.
    \item Incorporate advanced statistical techniques, including \ac{ml}-based inference methods and profile likelihood approaches, into standard cosmological pipelines.
    \item Continue developing theoretical models that remain consistent with observational constraints while addressing the root causes of cosmological tensions.
    \item Encourage interdisciplinary collaboration between observational, computational, and theoretical communities to ensure that different approaches are synthesized into a cohesive framework.
\end{itemize}

Resolving the cosmic tensions represents a defining challenge for the cosmology community. Success in this endeavor will require a concerted effort across multiple fronts, balancing innovative new techniques with a rigorous assessment of existing methodologies. By systematically integrating new data, analysis techniques, and theoretical advancements, the next decade of cosmological research has the potential to transform our understanding of the Universe’s fundamental properties.
\bigskip

\noindent \textbf{Challenges in observational cosmology:}
\begin{itemize}
    \item \textbf{Cepheid variables} -- The next generation of ground- and space-based optical surveys will provide deep time-series observations of distant galaxies, expanding the volume over which the universality of the Period-Luminosity relation can be tested. These observations will also improve our understanding of the impact of metallicity in different environments. As new Cepheids are discovered and observational precision increases, reducing uncertainties in Cepheid measurements will require better constraints on dust reddening, zero-point corrections in Cepheid parallaxes, and new geometric anchors in the distance ladder.

    \item \textbf{Masers} -- Systematics in megamaser-hosting galaxies, such as non-circular orbits and non-gravitational forces, do not significantly impact constraints on cosmological parameters. However, the primary challenge for maser-based techniques lies in discovering new high-quality megamaser-hosting galaxies suitable for precise distance measurements.

    \item \textbf{Tip of the Red Giant Branch (TRGB)} -- This method relies on a statistical approach to standardizing \ac{trgb} stars, making the calibration process a critical aspect of the analysis. The main challenges in applying the \ac{trgb} method include (1) potential imbalances between \ac{trgb} and calibrating galaxy brightness intensities, (2) the need for accurate calibration of the target field, and (3) effective outlier detection methods for poorly sampled \ac{lf}s.

    \item \textbf{Mira variables} -- Oxygen-rich Mira stars can serve as anchors in the first rung of the distance ladder for calibrating absolute magnitudes. However, their luminosity, color, large angular size, and surface variations introduce significant uncertainties in parallax measurements. On the second rung, the number of known Mira stars remains low, presenting an opportunity for future observations to expand their use in distance calibration.
    
    \item \textbf{Type Ia supernovae} -- \ac{sn1} primarily appear in the second and third rungs of the three-rung distance ladder. Their systematics are among the most extensively studied and best mitigated. However, the effects of (1) calibration systematics, (2) redshift-dependent biases in the third rung, and (3) stellar physics and sample selection may each contribute to potential changes in the estimates of certain cosmological parameters. Another challenge in this sector is the potential expansion of the distance ladder framework to include higher or lower rungs.

    \item \textbf{J-regions of the asymptotic giant branch (JAGB)} -- This is a relatively new but promising method for extending the distance ladder. The primary challenges for establishing this approach as a robust distance indicator include a better understanding of the effects of metallicity, atmospheric influences on photometry, and variations in pulsation across different stellar populations.

    \item \textbf{Type II supernovae} -- \ac{sn2} offer an independent path to circumvent \ac{sn1} systematics while introducing different systematic considerations. This method has the potential to achieve greater precision with the addition of further calibrators and a better understanding of how spectroscopic data can be leveraged to improve the standardization procedure.

    \item \textbf{Surface brightness fluctuations (SBF) methods} -- \ac{sbf}-based methods have shown promise in constraining distances to very distant galaxies, relying primarily on imaging data. The next generation of optical telescopes will provide opportunities to refine these techniques further. Given the statistical nature of the approach, a deeper understanding of stellar population studies and peculiar evolutionary stages will help reduce systematic uncertainties. Addressing these factors will enable \ac{sbf} to continue improving as a robust method for constraining cosmological parameters.
    
    \item \textbf{Cosmic chronometers} -- This method has the advantage of being theoretically economical in terms of its underlying assumptions while providing a direct numerical measurement of the Hubble parameter in redshift space. Beyond the need to increase the number of measurements using this approach, some systematic challenges include (1) uncertainties in stellar metallicity estimates, (2) errors in star formation history modeling and underlying assumptions, and (3) contamination by young or star-forming outliers.

    \item \textbf{HII galaxies} -- These galaxies are characterized by bursts of massive star formation, and their strong emission in specific spectral lines makes them valuable probes for high-redshift cosmology. However, for this method to become more robust, several systematics need to be addressed, including (1) improved physical modeling of emission line width profiles, (2) better understanding of stellar age effects, (3) corrections for extinction-induced deformations in emission lines, and (4) refined metallicity distribution calibrations.

    \item \textbf{Baryonic Tully-Fisher relation methods} -- The Tully-Fisher and \ac{btfr} can be applied to cosmology to constrain the Hubble constant, given an underlying cosmological model. While this method offers good control over systematics, the relatively small number of observed galactic objects currently limits its competitiveness compared to other techniques. This limitation will be addressed with upcoming large-scale surveys. On the modeling side, further development may help reduce model dependence, for instance, by incorporating \ac{ml} techniques into the analysis.

    \item \textbf{Strong lensing and time-delay cosmography} -- Strong lensing provides a distance-ladder-independent method for measuring $H_0$ through time-delay cosmography. However, several systematics currently limit its precision, including uncertainties in lens mass modeling, the impact of line-of-sight structures, and degeneracies in the inferred time delays. Additionally, assumptions about mass profile parametrizations can introduce biases in cosmological constraints. Improving this technique will require higher-resolution imaging, better characterizations of lens environments, and larger statistical samples of well-measured lensing systems to reduce both statistical and systematic uncertainties.

    \item \textbf{Gravitational wave astronomy} -- Free from electromagnetic radiation, \ac{gw} cosmology provides uniquely independent methods to constrain cosmological parameters, separate from all other approaches. Standard sirens offer the additional advantage of serving as absolute distance indicators, making them independent of a cosmological model in determining the Hubble constant. However, an electromagnetic counterpart is required to identify the host galaxy, which is then used to infer certain galactic properties; alternatively, statistical methods utilizing galaxy catalogs can be employed. On the other hand, dark sirens suffer from potential systematics related to assumptions about galactic properties.

    \item \textbf{Cosmic microwave background (CMB) radiation} -- Like other observatories, \ac{cmb} experiments are subject to instrumentation systematics and internal tensions, which can be more pronounced given the precision of the measurements and the vast volume of data collected. Ensuring consistency between high- and low-multipole analyses in cosmological parameter estimation remains a critical challenge for future \ac{cmb} surveys. The next generation of \ac{cmb} experiments will aim to minimize instrumental and astrophysical systematics, improving constraints on fundamental physics. By reducing these contamination effects, it may even become possible to robustly measure B-mode polarization, which remains a crucial target for detecting primordial \ac{gw}s and probing inflationary physics.
    
    \item \textbf{Baryon acoustic oscillations (BAO)} -- This method enables robust cosmological analyses based on millions of extragalactic objects. However, its statistical nature introduces a number of systematics that must be carefully accounted for. One key challenge is that the source density field is displaced due to peculiar motions, particularly in low-redshift regions, which is corrected using density field reconstruction techniques. As more galaxies are observed with higher precision, these reconstruction methods must continue to improve. Another important aspect of \ac{bao} surveys is the reliance on an underlying fiducial cosmology to transform between coordinate systems. While the influence of this model is relatively small, developing techniques to further reduce this dependence remains a priority. Additionally, \ac{bao} measurements are taken relative to the sound horizon, introducing another model-dependent feature that should be further refined to minimize its impact.

    \item \textbf{Galaxy clustering} -- The growth of large-scale structures can be inferred from the statistical clustering of galaxies in redshift space, which depends on estimates of correlation functions and power spectra. The statistical nature of these measurements introduces several systematics, including those originating from measurement techniques (such as atmospheric and extragalactic extinction corrections), galaxy survey systematics (such as calibration errors and foreground contamination), and uncertainties in nonlinear modeling for redshift-space galaxy clustering.

    \item \textbf{Large-scale structure and higher-order statistics} -- Beyond two-point statistics such as the galaxy correlation function and power spectrum, higher-order statistics and cosmic web topology provide additional tools to probe cosmology. Measurements of cosmic voids, filament structures, and the bispectrum offer sensitivity to non-Gaussianities and potential modifications to gravity. The main challenges in this sector include modeling nonlinear structure formation, correcting for observational systematics, and integrating these statistics into existing cosmological frameworks.

    \item \textbf{Weak lensing} -- Weak gravitational lensing surveys rely on statistically significant samples of galaxies, introducing several challenges that must be addressed. These include uncertainties in individual galaxy redshift estimates due to the inherent limitations of photometric redshift determinations, systematic biases in shear and shape ellipticity measurements, and difficulties in accurately modeling nonlinearities in the matter power spectrum. Improving these aspects will be essential for maximizing the precision of \ac{wl} constraints on cosmological parameters.

    \item \textbf{Galaxy cluster counts} -- Galaxy clusters are the largest gravitationally bound structures in the Universe and are highly sensitive to cosmic evolution and the underlying cosmological parameters. Aside from the challenge of distinguishing how different cosmological models affect cluster evolution, a major issue in this method is the mass calibration problem. This refers to the relationship between the total halo mass and observable cluster properties, which is crucial for using cluster counts as a cosmological probe. The primary method for determining halo masses is \ac{wl}, which itself introduces additional systematic uncertainties. As survey data become more precise, the mass calibration problem will increasingly become a limiting factor in the effectiveness of this approach.

    \item \textbf{Quasar and GRB distance ladder} -- Both \ac{qso}s and \ac{grb}s have shown strong potential as probes of the high-redshift Universe, providing constraints across multiple cosmological parameters. However, their precision remains limited compared to other methods, partly due to systematics related to the physical nature of the sources, which remain an active area of discussion in the literature. Resolving these open questions would significantly enhance the statistical power of cosmological parameter constraints when these data are used in combination with other survey observations.

    \item \textbf{Fast radio bursts (FRBs)} -- The relatively recent discovery of \ac{frb}s provides a powerful new probe of cosmological models. These events have shown promising potential to independently constrain the Hubble constant, as well as other cosmological parameters, particularly offering insights into the matter density parameter within \lcdm. However, systematics related to the modeling of \ac{dm} contributions and the \ac{igm} remain crucial open questions for this technique. Refinements in these areas will be necessary to improve its precision.

    \item \textbf{Radio background excess} -- The discrepancy between the expected emission from known astrophysical and cosmological sources and the observed brightness in the electromagnetic spectrum presents an anomaly in the radio segment of the spectrum. Resolving this requires a careful balance in accounting for infrared radiation production in early galaxies, X-ray contributions affecting inverse Compton scattering of \ac{cmb} photons, the depth of the $21\,$cm signal, and constraints from the cross-correlated angular power spectrum with optical sources. The next generation of observatories has the potential to refine and test possible explanations for this excess signal.
    
    \item \textbf{Bulk flow measurements} -- The average peculiar velocity of a statistically significant sample of galaxies should diminish at sufficiently high redshifts. However, measurements of bulk flows face several challenges, including large uncertainties in peculiar velocity estimates and biases in the spatial distribution of observed galaxies. Recent studies have suggested that bulk flow magnitudes increase with distance, a result that appears counterintuitive. Addressing this challenge will require the continued development of bulk flow measurement techniques, as well as improved observational data quality at higher redshifts.

    \item \textbf{Cosmic dipole and anisotropies} -- The observed \ac{cmb} dipole is expected to be consistent with the dipole measured in other astrophysical tracers. However, discrepancies in dipole measurements using different observational datasets suggest the possibility of systematic effects or new physics. Additionally, potential anisotropies in cosmological parameters such as $H_0$ and $S_8$ challenge the assumption of statistical isotropy. Addressing these issues requires improved methodologies and cross-correlations between independent datasets to test for underlying systematics or signatures of new physics.

    \item \textbf{Cross-correlations between cosmological probes} -- Combining different observational techniques provides a means to reduce systematics and improve cosmological parameter constraints. Key cross-correlations include (1) \ac{wl} and galaxy clustering to calibrate galaxy bias and structure formation models, (2) \ac{cmb} lensing and large-scale structure to refine \ac{dm} distribution models, and (3) standard sirens and \ac{bao} measurements to independently test the cosmic expansion history. As observational precision improves, cross-correlations will become increasingly important in resolving cosmological tensions.

    \item \textbf{21 cm intensity mapping} -- The neutral hydrogen 21 cm signal provides a powerful method to trace the large-scale structure of the Universe across a wide range of redshifts. This technique has the potential to probe the cosmic dawn, reionization, and the post-reionization epochs, offering insights into the distribution of matter over cosmic time. However, several challenges must be addressed, including foreground contamination from astrophysical sources, calibration uncertainties, and the need for precise theoretical modeling of the signal's evolution.

    \item \textbf{Line-intensity mapping (LIM)} -- LIM provides a promising method to trace large-scale structure by measuring integrated emission from spectral lines such as CO and Lyman-$\alpha$ rather than relying on resolved galaxy surveys. This method has the potential to map cosmic structure at high redshifts, but faces challenges such as foreground contamination and the need for precise theoretical modeling.

    \item \textbf{Sunyaev-Zel’dovich (SZ) and X-ray surveys} -- Measurements of galaxy clusters using the \ac{sz} effect and X-ray emission offer valuable constraints on cosmic structure formation and expansion. However, systematic uncertainties related to gas physics, temperature profiles, and mass calibration remain key challenges.

    \item \textbf{Multi-messenger cosmology} -- Observations across multiple cosmic messengers, including electromagnetic signals, \ac{gw}s, and high-energy particles, provide new opportunities for constraining cosmological parameters and testing fundamental physics. By combining information from different observational channels, multi-messenger approaches can help break degeneracies inherent in individual methods. However, challenges remain, including the low detection rates of some sources, difficulties in precise localization, and uncertainties in the modeling of extreme astrophysical environments.

    \item \textbf{Neutrino Cosmology} -- Neutrinos remain among the least understood fundamental particles, and their extremely weak interactions make direct measurements of their properties exceptionally difficult. A key open question is whether these particles have a nonzero mass. Moreover, terrestrial neutrino oscillation experiments provide constraints that appear to contrast with indirect cosmological measurements, potentially hinting at physics beyond the Standard Model in the neutrino sector. Reconciling these discrepancies will require further advancements in both laboratory-based and cosmological neutrino studies.
\end{itemize}
\bigskip

\noindent \textbf{Challenges in the development of numerical and data analysis tools:}
\begin{itemize}

    \item \textbf{Model constraints in conjunction with Bayesian statistics} -- Bayesian statistical methods have been widely used to constrain cosmological model parameters against observational data, and they now feature a broad suite of tools designed to meet these challenges. However, as survey data volumes increase and cosmological models become more complex, the computational demands of traditional Bayesian approaches are becoming increasingly problematic. This motivates the development of more efficient algorithms and novel implementations of Bayesian inference techniques to maintain accuracy while improving computational feasibility.

    \item \textbf{Frequentist approaches} -- Profile likelihoods provide a standard statistical method for analyzing cosmological models with high-dimensional parameter spaces, particularly when the likelihood surfaces exhibit non-Gaussian features. This approach involves running multiple traditional \ac{mcmc} analyses for fixed parameters to construct a profile likelihood. Several community tools already implement this method, some incorporating \ac{ml} techniques to enhance performance. However, due to the computationally intensive nature of this approach, further work on accelerating the sampling process will be crucial for making it more practical for a wider range of applications.
    
    \item \textbf{Hybrid Bayesian-Frequentist approaches} -- While Bayesian inference dominates cosmological data analysis, frequentist methods such as profile likelihoods and bootstrapping offer complementary insights, particularly in scenarios where priors strongly influence results. Hybrid approaches that combine these frameworks can provide more robust parameter constraints, reduce sensitivity to prior assumptions, and improve statistical interpretability in high-dimensional cosmological models.

    \item \textbf{Fast approximate inference techniques} -- Traditional Bayesian inference methods such as \ac{mcmc} can be computationally prohibitive, especially for high-dimensional cosmological models. Alternative approaches such as Hamiltonian Monte Carlo, Variational Inference, and Neural Density Estimation offer promising routes to accelerate parameter estimation while maintaining accuracy. Developing and benchmarking these techniques for cosmological applications is an open challenge.

    \item \textbf{Cosmology simulator inference techniques} -- The vast amount of information contained in the large-scale structure of the Universe presents a significant challenge for traditional simulation and analysis methods. Extracting cosmological parameters from these simulations is particularly difficult, as non-Gaussian and higher-order effects may become increasingly relevant for improving accuracy. Recent advances, including neural network architectures and decision tree learning approaches, have shown promise in addressing these challenges. However, the availability of community tools and a robust comparison of different inference techniques remain open issues that need further exploration.

    \item \textbf{Computational challenges in large-scale simulations} -- Cosmological simulations are essential for understanding structure formation and model testing. However, the increasing size of observational datasets and complexity of simulations present major computational challenges. Key issues include (1) optimizing numerical algorithms for high-performance computing architectures, (2) improving resolution and scalability in N-body and hydrodynamical simulations, and (3) reducing computational costs while maintaining accuracy in statistical inference from simulations.

    \item \textbf{Reconstruction techniques} -- The use of observational survey data to reconstruct physical models is still in its early stages; however, powerful techniques have already been developed and made publicly available. The development of more mature and statistically robust reconstruction methods, including those leveraging \ac{ml} approaches, will be an important aspect of progress in this direction. A major challenge for these techniques lies in effectively utilizing the wide variety of available datasets, which may include background, perturbative, and other physical aspects of cosmological models.

    \item \textbf{Genetic algorithm model selection} -- \ac{ga}s provide a mechanism to explore an otherwise infinite model space within a finite time by applying selection criteria. These methods primarily rely on Bayesian constraint analyses, but practical toolkits for their implementation are not yet publicly available in the literature. The development of such toolkits is essential for advancing this technique into a more comprehensive classification framework. Additionally, combining \ac{ga}s with other, faster constraint methods may enhance their suitability for practical model selection analyses.

    \item \textbf{Machine learning for theoretical model selection} -- The increasing complexity of cosmological models necessitates efficient methods for exploring large parameter spaces. \ac{ml} techniques, such as deep learning classifiers and \ac{gp} regression, are being developed to identify viable theoretical models by learning patterns in high-dimensional likelihood landscapes. However, challenges remain in ensuring interpretability, robustness, and generalization across different datasets.

    \item \textbf{Calibration and validation of machine learning in cosmology} -- \ac{ml} algorithms are increasingly used for inference, parameter estimation, and anomaly detection in cosmological datasets. However, ensuring that these models generalize well across different observational conditions, remain unbiased, and do not overfit to specific datasets is a key challenge. Benchmarking \ac{ml} models against traditional methods and developing interpretable AI approaches will be crucial for their broader adoption in cosmology.
    
    \item \textbf{Dark siren cosmology} -- The number of standard sirens without an associated electromagnetic counterpart continues to grow. These sources can also be used to constrain cosmological parameters, albeit through a statistical reconstruction of their population properties using galaxy catalogs. However, the simultaneous modeling of cosmology and mass distributions introduces computational challenges that must be addressed to extract robust constraints.

    \item \textbf{Spectroscopical sirens} -- The use of spectroscopic data from host galaxies associated with \ac{gw} events offers a promising avenue for improving the precision of standard siren measurements. By obtaining redshift information through spectroscopic observations, it is possible to reduce uncertainties in distance measurements and break degeneracies in cosmological parameter estimation. However, challenges remain in accurately identifying the host galaxy among multiple candidates, mitigating observational systematics in spectroscopic measurements, and optimizing survey strategies to maximize the detection of viable spectroscopical sirens.

    \item \textbf{Systematics mitigation in statistical inference} -- The increasing precision of cosmological surveys requires robust methods to identify and mitigate systematics in data analysis. Key challenges include (1) the impact of prior choices on Bayesian inference, (2) biases introduced by incomplete or inhomogeneous datasets, and (3) uncertainties in the modeling of astrophysical and instrumental effects. The development of statistical techniques that can cross-validate results across different datasets while minimizing these biases remains an important open problem.

    \item \textbf{Big Bang Nucleosynthesis numerical estimates} -- The precision of \ac{bbn} predictions for primordial element abundances has significantly improved in recent years. Advanced numerical codes have revealed tensions in some species ratios that were not apparent in previous lower-order calculations. Future numerical developments incorporating these higher-order effects will be crucial for reassessing these tensions and refining \ac{bbn} constraints on early Universe physics.

    \item \textbf{Cross-correlations and multi-probe statistical methods} -- Combining multiple cosmological probes is essential for breaking degeneracies and improving parameter constraints. However, this requires sophisticated statistical frameworks to account for systematics, survey-specific biases, and correlated uncertainties across datasets. Developing robust cross-correlation pipelines that integrate \ac{wl}, large-scale structure, and \ac{cmb} measurements is a key challenge for upcoming surveys.

    \item \textbf{Data-driven approaches for systematics mitigation} -- Systematic uncertainties remain a major challenge in cosmology, particularly in photometric redshift estimation, instrumental biases, and survey calibration. \ac{ml}, \ac{gp}, and hybrid statistical methods can provide real-time corrections for systematics, but require careful validation to ensure robustness across different datasets.

    \item \textbf{High-volume and high-dimensional data processing} -- The increasing size of cosmological datasets from next-generation surveys requires the development of scalable numerical methods and high-performance computing techniques. Challenges include managing high-dimensional parameter spaces, optimizing data compression without loss of information, and implementing real-time data processing pipelines for massive observational datasets.

\end{itemize}
\bigskip

\noindent \textbf{Challenges in fundamental physics:}
\begin{itemize}

    \item \textbf{Early-time physics} -- The early Universe provides a compelling regime for introducing exotic physics that remains shielded by recombination. Mechanisms such as early scalar fields, additional neutrino species, or modified expansion histories primarily affect the sound horizon at recombination, which in turn impacts inferences of the Hubble constant. These early modifications also present opportunities for testing inflationary scenarios. However, a significant challenge remains in explaining the evolution of large-scale structure, ensuring that any proposed early-time physics can simultaneously produce the correct growth rate to remain consistent with all cosmological observations.
    
    \item \textbf{Inflation} -- Phenomenological models successfully provide a spectrum of scenarios that can alleviate some tensions even under the assumption of Gaussian fields. However, it remains crucial to construct physically realistic explicit models that respect fundamental physical principles such as locality, causality, and unitarity, while also approximating key features of these phenomenological models. As future galaxy surveys improve precision on possible non-Gaussian features associated with primordial conditions, it will become increasingly important to develop accurate numerical simulations and theoretical descriptions of the nonlinear aspects of the evolution of these fields.

    \item \textbf{Big Bang Nucleosynthesis (BBN) tensions} -- \ac{bbn} predictions generally exhibit strong agreement with \lcdm\ in the production rates of primordial species. However, discrepancies remain, particularly for lithium isotopes, and increasingly for helium and deuterium, when compared with their relative abundances inferred from observations of the oldest stars. These discrepancies may stem from stellar astrophysics rather than cosmological considerations, but they could also signal challenges for early Universe dynamics in non-standard cosmological models. Resolving these tensions requires improved modeling of stellar evolution, nuclear reaction rates, and alternative \ac{bbn} scenarios.

    \item \textbf{Primordial magnetic fields} -- The possible existence of non-negligible \ac{pmf}s could have influenced \ac{cmb} anisotropies and small-scale inhomogeneities in the large-scale structure of the Universe. Various realizations of these fields have shown promise in addressing aspects of cosmic tensions, particularly in modifying early-Universe physics. However, more precise small-scale measurements of \ac{cmb} anisotropies are needed to robustly test these scenarios and establish whether \ac{pmf}s play a significant cosmological role.

    \item \textbf{Quantum decoherence in the early Universe} -- The classical nature of primordial perturbations is assumed in standard cosmology, yet their quantum mechanical origin requires a consistent explanation of decoherence processes. Understanding how quantum fluctuations transitioned to classical density perturbations and testing for potential residual quantum signatures remains an unresolved issue in fundamental physics.

    \item \textbf{Parity violations in cosmology} -- Parity-violating interactions, such as those arising in extensions to general relativity or axion-like particles coupled to gravity, could leave unique imprints on the \ac{cmb} polarization or large-scale structure. The detection of a nonzero cosmic birefringence signal, or deviations in E/B-mode power spectra, could offer evidence for such interactions. However, current observational constraints remain inconclusive, and refining the observational tests for parity violation remains an open challenge.

    \item \textbf{Nonflat spatial curvature} -- The cosmological curvature parameter is a fundamental aspect of the \ac{flrw} geometric background and is predicted to be zero in most inflationary models. However, any possible nonzero value remains highly degenerate with other cosmological parameters in standard analyses. Methods to break this degeneracy, such as \ac{bao} measurements or other observational techniques, could provide important insights into potential departures from the standard model and their role in addressing cosmological tensions.

    \item \textbf{Testing the Cosmological Principle} -- The standard model of cosmology assumes large-scale homogeneity and isotropy. However, several observations, including cosmic dipoles, bulk flows, and peculiar velocity anomalies, suggest that these assumptions may need to be tested more rigorously. Developing more precise observational tests and statistical methodologies to assess the validity of the cosmological principle remains an important challenge in the field.

    \item \textbf{Nonvanishing cosmic dipoles} -- The cosmic dipole represents the largest expected deviation from isotropy if the cosmological principle is broken at any level. Potential dipole signals have been investigated in various datasets, including the \ac{cmb}, \ac{qso} distributions, fine-structure constant measurements, \ac{sn1}, radio galaxies, \ac{grb}s, and bulk flow studies. Recent observations have hinted at slight nonvanishing dipoles in some of these probes. To assess the validity and implications of such findings, it is necessary to develop more robust cosmological backgrounds that can accommodate or rule out these potential departures from isotropy.

    \item \textbf{CMB anisotropy anomalies} -- Several features in the \ac{cmb} anisotropic power spectrum exhibit mild tensions with the predictions of the \lcdm\ model. Early Universe modifications that explain or incorporate these features more naturally may offer new insights into fundamental physics beyond standard cosmology. These anomalies include:  
  (1) a stronger-than-expected \ac{isw} effect in certain analyses,  
  (2) non-Gaussian features in the \ac{cmb}, such as the Cold Spot,  
  (3) a slight hemispherical asymmetry in the \ac{cmb} power spectrum,  
  (4) an unexpected lack of large-scale \ac{cmb} temperature correlations,  
  (5) a higher-than-predicted lensing parameter, and  
  (6) a moderate inconsistency between parameter values derived from high- and low-multipole regions of the \ac{cmb} power spectrum.
  Understanding whether these features stem from new physics or residual systematics remains an important challenge for upcoming surveys.

    \item \textbf{Anomalous Integrated Sachs-Wolfe (ISW) effect} -- The \ac{isw} effect arises from the interaction of \ac{cmb} photons with evolving gravitational potentials associated with large-scale structure. Observations suggest a discrepancy between the predicted \ac{isw} signal in \lcdm\ and its measured imprint in cosmic superstructures and supervoids, both in amplitude and sign. This may indicate a more complex growth history than expected in the concordance model. Understanding whether this excess signal can be reconciled within \lcdm, or if it points to new physics, remains an open question.
    
    \item \textbf{The nature of dark matter} -- The connection between the fundamental properties of \ac{dm} and cosmic evolution has significant implications both for the early Universe, through thermal and non-thermal relic physics, and for the late Universe, influencing the formation of large-scale structure. Modeling how the thermal history and other fundamental characteristics of \ac{dm} may lead to deviations from standard cosmology via solutions of the Boltzmann equation will be crucial in assessing whether \ac{dm} plays a role in alleviating cosmological tensions.

    \item \textbf{Axion-like particles and ultralight fields} -- The existence of ultralight scalar fields, such as axion-like particles, could impact the evolution of the Universe through novel interactions with \ac{dm}, \ac{de}, or radiation. These fields have been proposed to address cosmic tensions by altering the expansion rate or structure growth. Developing observational tests for their presence and impact remains a crucial task.

    \item \textbf{Complex dark sector interactions} -- Many extensions to \lcdm\ propose that \ac{dm} and \ac{de} may not be independent, but instead part of a larger dark sector with self-interactions or couplings to other fields. These models introduce new forces in the dark sector that could impact structure formation, cosmic expansion, and small-scale clustering. However, constraining these interactions remains difficult, as current probes (e.g., \ac{cmb}, \ac{bao}, \ac{wl}) primarily test gravitational effects rather than direct couplings.

    \item \textbf{Modified gravity} -- Extensions beyond \lcdm\ may involve modifications to gravity, ranging from mild to radical departures from General Relativity. The literature is rich with competing formulations of \ac{mg} theories, but many are disfavored at smaller scales and exhibit strong degeneracies both among themselves and with scalar-tensor theories. Additionally, \ac{mg} theories often introduce challenges such as unwanted fifth-force effects. On the other hand, these theories provide an avenue for constructing alternative cosmological models independent of \lcdm, which could yield departures from concordance cosmology in both early- and late-time evolution.

    \item \textbf{Late-time physics} -- Measurements in the local Universe place strict constraints on the cosmic expansion profile, leaving little room for significant deviations from \lcdm. However, recent results from large-scale surveys suggest possible hints of dynamical properties in \ac{de} at late-times. While numerous competing theories focus on late-time modifications, many of their observational properties remain highly degenerate. To address this issue, the development of more distinct and testable theoretical models, as well as more precise predictions from specific scenarios, would be beneficial.

    \item \textbf{Exotic dark energy models} -- The majority of \ac{de} models assume a scalar-field-driven component, but alternative formulations involving vector or tensor fields have been proposed. These include vector-tensor theories, emergent gravity models, and Lorentz-violating \ac{de} frameworks. Understanding the observational signatures and constraints on these exotic models is an open challenge.

    \item \textbf{Variation of fundamental constants} -- The possible variation of fundamental constants could arise from various exotic physics scenarios, including non-Standard Model particle interactions, \ac{mg}, or coupled scalar fields, among others. Temporal or spatial variations in fundamental constants would have profound consequences for fundamental physics, potentially impacting atomic structure, nuclear reactions, and cosmological evolution. Observational efforts have primarily focused on detecting variations in the fine-structure constant and the electron mass, as these have shown promise in providing an alternative explanation for the Hubble tension. However, current \ac{cmb} observations lack the precision to distinguish between models that allow for varying fundamental constants. Addressing this challenge will require next-generation observatories with significantly improved sensitivity to such effects.

    \item \textbf{Higher-dimensional cosmological models} -- Theories incorporating extra spatial dimensions, such as brane-world scenarios and string-inspired cosmologies, propose modifications to gravity that could influence cosmic expansion, the dark sector, or structure formation. Developing testable predictions for these theories and determining their viability in explaining cosmic tensions remains a key theoretical challenge.

    \item \textbf{Cosmological quantum gravity} -- It is expected that classical gravity breaks down at high energy scales, but the impact of UV-complete quantum gravity theories on cosmological observables remains unclear. In particular, it is uncertain whether such effects could resolve the Hubble tension or other cosmic anomalies. Possible quantum gravitational effects, such as the \ac{gup}, Lorentz invariance violations, or nonlinear modifications, could provide a new perspective on cosmic tensions as manifestations of non-classical physics. However, fully developed quantum gravity models that remain consistent with small-scale physics while simultaneously addressing these challenges are still in early stages of development.

    \item \textbf{Testing the equivalence principle on cosmological scales} -- The equivalence principle, a cornerstone of general relativity, has been tested at small scales with high precision. However, cosmological tests remain limited. Anomalies in the behavior of standard sirens, \ac{wl}, or cosmic dipoles could indicate potential violations. Future large-scale surveys will play a crucial role in refining these tests.

    \item \textbf{Relic neutrinos and direct detection} -- The cosmic neutrino background (C$\nu$B) is a fundamental prediction of Big Bang cosmology, yet it has never been directly detected. Cosmological observations place strong constraints on its properties, but upcoming laboratory experiments and indirect probes may provide the first evidence for relic neutrinos. If detected, these neutrinos could offer crucial insights into early-Universe physics, including possible deviations from the Standard Model.
\end{itemize}

By confronting these challenges, the cosmology community will be poised to revolutionize our understanding of the Universe, ensuring that future cosmological models remain consistent with increasingly precise observations while harnessing the full power of advanced statistical and computational methodologies. \bigskip
\subsection{Conclusions \label{sec:conclu}}

\noindent The issue of cosmic tensions has emerged over the past decade as a significant challenge to the \lcdm\ concordance model, which also faces underlying consistency issues and additional anomalies. While this is not unexpected for a phenomenological framework, the mounting observational discrepancies suggest that a new concordance model may ultimately be required to resolve these tensions. Despite the long-standing success of \lcdm, the breadth and persistence of cosmic tensions represent the most serious and potentially insurmountable test for the standard cosmological paradigm.

Among these challenges, the continued statistical significance of the Hubble constant tension has solidified its status as a defining question in cosmology. This tension not only tests the robustness of the concordance model but also probes the ability of the cosmology community to reconcile local (late-time) measurements with global (early-time) constraints. The latter, being model-dependent, introduces additional complexities, while the former is subject to astrophysical systematics that may exert only a limited influence on the magnitude of the discrepancy. However, the widespread and consistent detection of the Hubble tension across independent measurement techniques casts doubt on the plausibility of a single cross-survey systematic effect that could fully account for the discrepancy.
In Sec.~\ref{sec:obs_H0}, we review the diverse range of methods used to infer $H_0$. Many of these techniques rely on the distance ladder to measure the Hubble flow, including refined \ac{sn1} analyses, Cepheid variable calibrations, \ac{trgb} and \ac{jagb} methods, \ac{sn2} measurements, Mira variables, and the \ac{sbf} technique. Additionally, \ac{cc}s provide a direct means of measuring the Hubble parameter to comparable redshifts. Collectively, these approaches yield a consistently high value of $H_0$, a result reinforced by distance-ladder-independent methods such as maser-driven constraints and estimates based on the \ac{btfr}. Furthermore, higher-redshift techniques—including HII galaxy distance indicators, time-delay strong lensing, \ac{frb}s,  \ac{qso}s and \ac{grb}s  Hubble diagrams—further substantiate the discrepancy. 
Another promising avenue for constraining $H_0$ lies in \ac{gw}-based measurements using standard sirens, which provide an independent approach free from electromagnetic systematics. This method continues to improve and has already delivered encouraging results, potentially adding new insights into the nature and expression of the Hubble tension. On the other hand, every \ac{cmb}-based survey—including Planck, \ac{act}, and \ac{spt}—consistently yields a significantly lower value of the Hubble constant, assuming a \lcdm\ cosmological framework. This lower range of $H_0$ values is further reinforced through the combination of baryon acoustic oscillation data and constraints from \ac{bbn}. The core of the Hubble tension thus lies in the stark contrast between low-redshift measurements, which suggest a higher $H_0$, and early-Universe estimates derived from the \ac{cmb} and similarly high-redshift probes, which favor a lower value.

The Hubble tension serves as a critical test of \lcdm—or any alternative cosmological model—spanning from the primordial Universe to the present day. A closely related parameter that traces the expansion and structure formation history of the Universe is $S_8$, which encapsulates both the total matter density and the amplitude of matter fluctuations on scales of $8 \, h^{-1}\,$Mpc. This parameter is widely regarded as an effective measure of cosmic structure growth and is examined in Sec.~\ref{sec:S8tension_2.2}. Similar to the Hubble tension, a statistical discrepancy emerges in the observed values of $S_8$ when comparing \ac{cmb}-based primary anisotropy measurements with local probes such as \ac{wl}, galaxy clustering, and galaxy cluster abundance studies. The $S_8$ parameter is also closely connected to the $f\sigma_8$ parameter, which is constrained through \ac{rsd}, as well as other structure growth parameters, including the growth index, growth parameter, and growth rate.
\ac{cmb}-based measurements yield a relatively high and stable estimate of $S_8$ across multiple experiments, including different data subsets from the Planck mission’s legacy release. In contrast, independent analyses from \ac{wl} surveys, galaxy clustering, \ac{rsd}, and galaxy count-based approaches predominantly indicate a lower $S_8$ value at a statistically significant level. When considered alongside the Hubble tension, the potential $S_8$ tension presents an even greater challenge to \lcdm\ and alternative cosmological models, further testing their predictive power and ability to fully reconcile current observational data.

Additionally, several anomalies and challenges arising from the confrontation between \lcdm\ and observational data have become increasingly apparent. These issues, described in Sec.~\ref{sec:data_ana}, include the $A_{\rm lens}$ anomaly, which suggests an excess lensing amplitude in the Planck data and a potential deviation from \lcdm\ predictions. This is further compounded by a slight preference for a closed universe in spatial curvature parameter constraints within the \lcdm\ framework. Other \ac{cmb} anisotropy anomalies include the low quadrupole moment and its alignment with the octupole, hemispherical asymmetry, and the Cold Spot anomaly. Beyond the \ac{cmb}, additional unresolved questions concern the exact nature of cosmological neutrino physics, potential deviations from the cosmological principle due to large-scale bulk flows, and anomalies in the primordial abundances of certain species in the early Universe. More broadly, discrepancies have emerged in Lyman-$\alpha$ measurements, the \ac{isw} effect, radio synchrotron background observations, and the possible existence of large cosmic voids.

The vast amount of observational data related to cosmic expansion, large-scale structure formation, and the evolution of fundamental parameters presents a formidable challenge for the data analysis and theoretical modeling community. Traditional implementations of \ac{mcmc} methods are increasingly struggling to distinguish between competing cosmological scenarios due to the complexity of their parameter dependencies and the high dimensionality of their parameter spaces. In Sec.~\ref{sec:data_ana}, novel approaches to MCMC methods are discussed alongside alternative data analysis techniques. Frequentist parameter inference through profile likelihoods offers an intriguing pathway to mitigating prior volume effects that can dominate posterior constraints in Bayesian approaches. However, this method is even more computationally demanding than standard \ac{mcmc}, making it impractical for certain classes of models beyond \lcdm.
Similarly, extracting cosmological information from large-scale simulations has become computationally intractable using conventional methods, necessitating the development of novel tools. \ac{ml} techniques, particularly deep generative models, have shown significant promise in constraining cosmological parameters with improved efficiency. More broadly, \ac{ml} inference frameworks have begun replacing traditional \ac{mcmc} techniques, leveraging various neural network architectures to refine parameter estimation. In parallel, model selection methodologies have incorporated \ac{ga}s as a systematic approach for exploring complex model spaces against observational constraints. These rapidly evolving techniques are proving essential in addressing the challenges posed by next-generation observatories, which are producing unprecedented volumes of data and introducing a vast array of potential new physics scenarios.

The increasingly robust expression of cosmic tensions—spanning the Hubble constant, large-scale structure growth, and various other anomalies—necessitates a reevaluation of potential new physics within both current and future observational surveys. The cosmology community has proposed a wide range of possible directions for addressing these challenges, which are reviewed in Sec.~\ref{sec:fun_phys}. Early-Universe modifications have the advantage of introducing exotic dynamics prior to recombination, beyond the reach of direct electromagnetic observations. However, while altering the sound horizon can help reconcile certain tensions, these modifications may introduce inconsistencies with the early growth of large-scale structure. Addressing this issue requires exploring different realizations of new physics in this sector alongside novel data analysis techniques, such as frequentist parameter inference. 
On the other hand, late-time modifications to cosmology have also shown promise in addressing cosmic tensions through diverse mechanisms, including additional fields, modifications to gravitational physics, and other extensions to the standard paradigm. However, the significantly more detailed and structured nature of late-Universe observations places stringent constraints on the evolution of such models, restricting the amplitude of any new physics in this regime. An alternative to modifying gravity in the field equations is to reconsider the behavior of the matter sector, such as through interacting or decaying \ac{dm} scenarios, or more intricate physics governing \ac{dm} properties.
Beyond standard modifications to cosmic evolution, alternative explanations have been explored, including the role of large cosmic voids, the influence of \ac{pmf}s, and variations in the dynamics of cosmic inflation. The severity of these tensions has also revived interest in non-standard cosmological geometries that challenge aspects of the cosmological principle, though such proposals remain in early stages of development. Similarly, while quantum gravity theories have not yet demonstrated a direct resolution to cosmic tensions, the potential variation of fundamental constants has been shown to meaningfully impact both the inferred Hubble constant and the growth of large-scale structures, with broader implications across multiple areas of fundamental physics.
Ultimately, reconciling early- and late-time cosmological observations, alongside non-standard phenomenology, may necessitate a combination of modifications to the concordance model. However, it remains crucial to break the strong degeneracy among different classes of modified cosmologies and ensure that proposed models not only fit existing data but also yield new, testable predictions while preserving the well-established successes of the concordance framework across both astrophysical and cosmological scales.

The coming years will see an explosion of high-precision observational data from upcoming surveys, as reviewed in Sec.~\ref{sec:fut_survey_prospects}. New \ac{cmb} experiments, such as LiteBIRD and \ac{cmbs4}, have the potential to detect primordial \ac{gw}s and refine measurements of \ac{cmb} anisotropies, which may provide key insights into fundamental physics. In parallel, large-scale structure surveys—including those conducted by Euclid, the Roman Space Telescope, and the Rubin Observatory—will probe the evolution of cosmic structure and the nature of \ac{de} in unprecedented detail. \ac{wl} studies, \ac{rsd}, and clustering analyses will provide complementary information on cosmic structure formation, while 21 cm intensity mapping from \ac{ska}, FAST, and CHIME will open a new observational window into the cosmic dawn. Additionally, radio astronomy efforts, such as \ac{frb} surveys with UTMOST and MeerKAT, will offer independent constraints on cosmic expansion and structure evolution.
Beyond electromagnetic probes, the next generation of \ac{gw} observatories—including the \ac{ligo}-Virgo-KAGRA network, the \ac{et}, Cosmic Explorer, and the space-based \ac{lisa} mission—will provide direct insight into gravitational radiation, potentially offering novel constraints on early-Universe physics and cosmic expansion. Additionally, advances in standard siren measurements may provide an independent means of resolving the Hubble tension. Meanwhile, improvements in the cosmic distance ladder, enabled by \ac{jwst}, \ac{elt}, and other facilities, will refine local measurements of $H_0$ with increasing precision.
As the volume of observational data continues to grow exponentially, computational advancements will play an increasingly critical role in extracting cosmological information. The limitations of traditional \ac{mcmc} methods necessitate the development of alternative approaches, such as frequentist inference through profile likelihoods, deep generative models, and other \ac{ml} techniques. These innovations have already shown promise in improving parameter estimation and accelerating statistical inference, while model selection algorithms, including \ac{ga}s, offer systematic strategies for exploring vast cosmological parameter spaces. The integration of artificial intelligence and high-performance computing will be essential in processing and analyzing the unprecedented data streams expected from next-generation observatories.

The comprehensive review presented in this White Paper, along with the collaborative efforts that shaped it, reflects a growing consensus within the cosmology community on the key directions that must be pursued in the coming years. As detailed in Sec.~\ref{sec:recomms}, these efforts span observational challenges, advancements in data analysis techniques, and the theoretical development of new physics models. Crucially, resolving cosmic tensions will require a synergistic approach, combining observational precision, computational innovation, and theoretical creativity. The next decade of cosmology will be pivotal in shaping our understanding of the Universe, with the potential to refine the current paradigm or uncover new physics that reshapes our fundamental model of cosmic evolution.

\bigskip

%%%%%%%%%%%% Conventions
%%%%%%%%%%%%%%%%%%%%%%%%%%%%%%%%%%
\clearpage
\section{Conventions \label{sec:conventions}}

The White Paper spans research communities in observational, data analysis, and fundamental physics areas which observe vastly different notational traditions. A big effort was made to homogenize this notation across the breadth of communities in the separate sections of this project. These notation conventions are defined in Table~\ref{tab:notation}. Any deviations from this notation convection are noted in the specific subsections in which they occur.

\begin{table*}[ht]
\begin{center} 
\begin{tabular}{|c|l|}
\hline
Definition & Meaning \\
\hline
$\hbar=c=k_B=1$ & Natural units\\
$\kappa^2\equiv 8\pi G= M_{\rm Pl}^{-2}$ & Gravitational constant\\
$\ln := \log_{\rm e}, \,\,\, \log:=\log_{10}$ & Logarithm conventions\\
$(-\,+\,+\,+)$ & Metric signature\\
  $g_{\mu\nu}$ & Metric tensor\\
   $G_{\mu \nu} \equiv R_{\mu \nu} - \frac{1}{2} g_{\mu \nu} R$ & Einstein tensor\\
$\Lambda$ & Cosmological constant\\
  ${\rm d}s^2 = -{\rm d}t^2 + a^2 (t) \left[ \frac{{\rm d}r^2}{1-k r^2} + r^2  \left(  {\rm d} \theta^2 + \sin^2 \theta {\rm d} \phi^2 \right)\right]$ & 
Friedmann-Lema\^{i}tre-Robertson-Walker (FLRW) spacetime metric\\
$a(t)$ & Scale factor \\
$a_0=1$ & Scale factor today (set to unity) \\
$t$ & Cosmic (proper) time\\
$\tau(t) = \int_{0}^{t}\frac{{\rm d}t'}{a(t')}$ & Conformal time\\
$\dot{\;} \equiv \frac{\rm d}{{\rm d}t}$ & Cosmic time derivative\\
$\;^{\prime} \equiv \frac{\rm d}{{\rm d}\tau}$ & Conformal time derivative\\
$T^{\mu \nu} = \frac{2}{\sqrt{-g}} \frac{\delta \mathcal{L}_m}{\delta
  g_{\mu \nu}}$ & Energy-momentum tensor of the Lagrangian density $\mathcal{L}$\\
$z=-1+\frac{1}{a}$ & Cosmological redshift\\
$H(z)=\frac{\dot a}{a}$ & Hubble parameter\\
$H_0$ & Hubble constant\\
$h\equiv H_0/100\, {\rm km\,s}^{-1}{\rm Mpc}^{-1}$ & Dimensionless reduced Hubble constant \\
$\rho_m$, $\rho_b$, $\rho_r$ & Energy density of total matter, baryonic matter, and radiation\\
$\rho_{\rm DM}$, $\rho_{\rm DE}$ & Energy density of dark matter and dark energy \\
$\Omega_m$ & Present-day matter density parameter \\
$\Omega_r=2.469\times10^{-5}h^{-2}(1+0.2271N_{\rm eff})$ & Present-day radiation density parameter \\
$\Omega_{\rm DM}$, $\Omega_{\rm DE}$ & Present-day density parameters of dark matter and dark energy\\
$\Omega_{\rm CDM}$ & Present-day density parameters of cold dark matter\\
$\Omega_m(z)=\frac{\kappa^2\rho_m}{3H^2}$ & Matter density parameter \\
$\Omega_r(z)=\frac{\kappa^2\rho_r}{3H^2}$ & Relativistic content density parameter \\
$\Omega_{\rm DE}(z)=\frac{\kappa^2\rho_{\rm DE}}{3H^2}$ & Dark energy density parameter \\
$w\equiv\frac{p}{\rho}$ & Equation of state (EoS) parameter \\
$c_s$ & Sound speed\\
$r_s\equiv\int_{0}^{\tau}c_s(\tau') {\rm d}\tau'$ & Sound horizon\\
$r_d \equiv r_s(\tau_d)$ & Sound horizon at drag epoch\\
$\sigma_8$ & Amplitude of mass fluctuations on scales of $8\, h^{-1}$ Mpc \\
$S_8=\sigma_8 \sqrt{\Omega_m/0.3}$ & Weighted amplitude of matter fluctuations \\
\hline
\end{tabular}
\end{center}
\caption{List of notation conventions used in the White Paper (unless otherwise stated).}
\label{tab:notation}
\end{table*}

\clearpage
%%%%%%%%%%%%%%%%%%%%%%%%%%%%%%%%%%

\section{List of Acronyms} \label{sec:acronyms}
\begin{acronym}[ICANN]
    \acro{act}[ACT]{Atacama Cosmology Telescope}
    \acro{agn}[AGN]{Active Galactic Nucleus}
    \acro{agb}[AGB]{Asymptotic Giant Branch}
    \acro{ann}[ANN]{Artificial Neural Network}
    \acro{bao}[BAO]{Baryon Acoustic Oscillations}
    \acro{btfr}[BTFR]{Baryonic Tully-Fisher Relation}
    \acro{bnn}[BNN]{Bayesian Neural Network}
    \acro{bbn}[BBN]{Big Bang Nucleosynthesis}
    \acro{boss}[BOSS]{Baryon Oscillation Spectroscopic Survey}
    \acro{cc}[CC]{Cosmic chronometers}
    \acro{cdm}[CDM]{Cold Dark Matter}
    \acro{cmb}[CMB]{Cosmic Microwave Background Radiation}
    \acro{cnn}[CNN]{Convolutional Neural Network}
    \acro{cpl}[CPL]{Chevallier-Polarski-Linder}
    \acro{dde}[DDE]{Dynamical Dark Energy}
    \acro{de}[DE]{Dark Energy}
    \acro{dm}[DM]{Dark Matter}
    \acro{eboss}[eBOSS]{Extended Baryon Oscillation Spectroscopic Survey}
    \acro{ede}[EDE]{Early Dark Energy}
    \acro{eftoflss}[EFTofLSS]{Effective Field Theory of Large Scale Structure}
    \acro{elt}[ELT]{Extremely Large Telescope}
    \acro{eor}[EoR]{Epoch of Reionization}
    \acro{eos}[EoS]{Equation of State}
    \acro{esa}[ESA]{European Space Agency}
    \acro{et}[ET]{Einstein Telescope}
    \acro{des}[DES]{Dark Energy Survey}
    \acro{desi}[DESI]{The Dark Energy Spectroscopic Instrument}
    \acro{flrw}[FLRW]{Friedmann-Lema\^{i}tre-Robertson-Walker}
    \acro{frb}[FRB]{Fast Radio Burst}
    \acro{cmbs4}[CMB-S4]{CMB—Stage 4}
    \acro{ga}[GA]{Genetic Algorithm}
    \acro{gp}[GP]{Gaussian processes}
    \acro{gr}[GR]{General relativity}
    \acro{grb}[GRB]{Gamma-ray burst}
    \acro{gup}[GUP]{Generalized Uncertainty Principle}
    \acro{gw}[GW]{Gravitational Wave}
    \acro{hst}[HST]{Hubble Space Telescope}
    \acro{hsc}[HSC]{Hyper Suprime-Cam}
    \acro{ide}[IDE]{Interacting Dark Energy}
    \acro{idm}[IDM]{Interacting Dark Matter}
    \acro{igm}[IGM]{Intergalactic Medium}
    \acro{ism}[ISM]{Interstellar Medium}
    \acro{isw}[ISW]{Integrated Sachs–Wolfe}
    \acro{lde}[LDE]{Late Dark Energy}
    \acro{lf}[LF]{Luminosity Function}
    \acro{lisa}[LISA]{Laser Interferometer Space Antenna}
    \acro{lmc}[LMC]{Large Magellanic Clouds}
    \acro{jagb}[JAGB]{J-region Asymptotic Giant Branch}
    \acro{jwst}[JWST]{James Webb Space Telescope}
    \acro{kids}[KiDS]{Kilo-Degree Survey}  
    \acro{ligo}[LIGO]{Laser Interferometer Gravitational Wave Observatory}
    \acro{lss}[LSS]{Large Scale Structure}
    \acro{lsst}[LSST]{Legacy Survey of Space and Time}
    \acro{mg}[MG]{Modified Gravity}
    \acro{mcmc}[MCMC]{Markov chain Monte Carlo}
    \acro{ml}[ML]{Machine Learning}
    \acro{mw}[MW]{Milky Way galaxy}
    \acro{mond}[MOND]{Modified Newtonian dynamics}
    \acro{nede}[NEDE]{New Early Dark Energy}
    \acro{pmf}[PMF]{Primordial Magnetic Field}
    \acro{pdf}[PDF]{Probability Density function}
    \acro{rgb}[RGB]{Red Giant Branch}
    \acro{qso}[QSO]{QSO}
    \acro{rsd}[RSD]{Redshift-Space Distortions}
    \acro{rvm}[RVM]{Running Vacuum Model}
    \acro{sbf}[SBF]{Surface Brightness Fluctuations}
    \acro{sgwb}[SGWB]{Stochastic Gravitational Wave Background}
    \acro{ska}[SKAO]{Square Kilometer Array Observatory}
    \acro{smc}[SMC]{Small Magellanic Clouds}
    \acro{sn}[SN]{Supernovae}
    \acro{sn1}[SNIa]{Type Ia supernovae}
    \acro{sn2}[SNII]{Type II supernovae}
    \acro{spt}[SPT]{South Pole Telescope}
    \acro{spt3g}[SPT-3G]{South Pole Telescope - 3rd Generation}
    \acro{sz}[SZ]{Sunyaev–Zeldovich}
    \acro{trgb}[TRGB]{Tip of the Red Giant Branch}
    \acro{wdm}[WDM]{Warm Dark Matter}
    \acro{wl}[WL]{Weak Lensing}
    \acro{wmap}[WMAP]{Wilkinson Microwave Anisotropy Probe}
    \acro{ztf}[ZTF]{Zwicky Transient Facility}
\end{acronym}

\newpage

\section*{Acknowledgements}\label{sec:acknowledgements}
{\small

This paper is based upon work from COST Action CA21136 {\it Addressing observational tensions in cosmology with systematics and fundamental physics} (CosmoVerse) supported by COST (European Cooperation in Science and Technology).
EDV is supported by a Royal Society Dorothy Hodgkin Research Fellowship.
JLS would also like to acknowledge funding from ``Xjenza Malta'' as part of the ``Technology Development Programme'' DTP-2024-014 (CosmicLearning) Project and the ``XM-TÜBİTAK R\&I Programme'' BridgingCosmology project.
AGV is funded by “la Caixa” Foundation (ID 100010434) and the European Union’s Horizon 2020 research and innovation programme under the Marie Sklodowska-Curie grant agreement No 847648, with fellowship code LCF/BQ/PI23/11970027.
AP acknowledges support from the Polish National Science Centre through the grant 2023/50/A/ST9/00579.
AP is supported by NSF Grant No. 2308193.
CvdB is supported by the Lancaster–Sheffield Consortium for Fundamental Physics under STFC grant: ST/X000621/1.
CU was supported by UKRI STFC ST/W001020/1 and European Union ERC StG, LSS\_BeyondAverage, 101075919.
DE acknowledges support from the Swiss National Science Foundation (SNSF) under grant agreement 200021\_212576.
EMT is supported by funding from the European Research Council (ERC) under the European Union’s HORIZON-ERC-2022 (grant agreement no. 101076865).
ENS acknowledges the contribution of the LISA CosWG.
This project has received funding from the European Research Council under the European Union’s Horizon 2020 research and innovation programme (grant agreement 853291). FB acknowledges the support of the Royal Society through the University Research Fellowship.
The work of FN is supported by VR Starting Grant 2022-03160 of the Swedish Research Council.
GB is supported by the Spanish grants CIPROM/2021/054 (Generalitat Valenciana) and  PID2023-151418NB-I00 funded by MCIU/AEI/10.13039/501100011033.
IS acknowledges NASA grants N4-ADAP24-0021 and 24-ADAP24-0074, and this research was supported in part by grant NSF PHY-2309135 to the Kavli Institute for Theoretical Physics (KITP).
JT is supported by a Ram\'{o}n y Cajal contract by the Spanish Ministry for Science, Innovation and Universities with Ref.\ RYC2023-045660-I.
JM would also like to acknowledge funding from ``Xjenza Malta'' as part of the ``Technology Development Programme'' DTP-2024-014 (CosmicLearning) Project.
KS acknowledges support from the Australian Government through the Australian Research Council Centre of Excellence for Gravitational Wave Discovery (OzGrav), through project number CE230100016.
K.F.D. was supported by the PNRR-III-C9-2022–I9 call, with project number 760016/27.01.2023, and funding from ``Xjenza Malta'' as part of the ``XM-TÜBİTAK R\&I Programme'' BridgingCosmology project.
LZ is supported by the NAWA Ulam fellowship (No. BPN/ULM/2023/1/00107/U/00001) and the National Science Centre, Poland (research grant No. 2021/42/E/ST2/00031).
LG acknowledges financial support from AGAUR, CSIC, MCIN and AEI 10.13039/501100011033 under projects PID2023-151307NB-I00, PIE 20215AT016, CEX2020-001058-M, ILINK23001, COOPB2304, and 2021-SGR-01270.
LV acknowledges support by the National Natural Science Foundation of China (NSFC) through the grant No. 12350610240 ``Astrophysical Axion Laboratories'', and also thanks INFN through the ``QGSKY'' Iniziativa Specifica project.
The work of LAA is supported by the US National Science Foundation Grant PHY-2412679.
MA acknowledges the UK Science and Technology Facilities Council (STFC) under grant number ST/Y002652/1 and the Royal Society under grant numbers RGSR2222268 and ICAR1231094.
MG acknowledges support from the European Union (ERC, RELiCS, project number 101116027) and the PRIN (Progetti di ricerca di Rilevante Interesse Nazionale) number 2022WJ9J33.
MF is funded by the PRIN (Progetti di ricerca di Rilevante Interesse Nazionale) number 2022WJ9J33.
MC, GR, and RH acknowledge support from the project “INAF-EDGE” (Large Grant 12-2022, P.I. L. Hunt), from the ASI-INAF agreement “Scientific
Activity for the Euclid Mission” (n.2024-10-HH.0; WP8420) and from the INAF “Astrofisica Fondamentale” GO-grant 2024 (PI M. Cantiello).
MM acknowledges support from MIUR, PRIN 2022 (grant 2022NY2ZRS 001) and from the grant ASI n. 2024-10-HH.0 “Attività scientifiche per la missione Euclid – fase E”.
RCN thanks the financial support from the CNPq under the project No. 304306/2022-3 and FAPERGS under the project No. 23/2551-0000848-3.
RCB is supported by an appointment to the JRG Program at the APCTP through the Science and Technology Promotion Fund and Lottery Fund of the Korean Government, and was also supported by the Korean Local Governments in Gyeongsangbuk-do Province and Pohang City.
RC acknowledges support from the CONAHCYT research grant CF2022-320152.
This project has received funding from the European Research Council (ERC) under the European Union's Horizon 2020 research and innovation programme (Grant Agreement No. 947660). RIA is funded by the SNSF Eccellenza Professorial Fellowship PCEFP2\_194638.
SK acknowledges funding by the National Center for Science, Poland, grant no. 2023/49/B/ST9/02777.
SL is supported by the National Science Foundation Graduate Research Fellowship Program under grant No. DGE2139757.
SV acknowledges support from the Istituto Nazionale di Fisica Nucleare (INFN) through the Commissione Scientifica Nazionale 4 (CSN4) Iniziativa Specifica ``Quantum Fields in Gravity, Cosmology and Black Holes'' (FLAG), and from the University of Trento and the Provincia Autonoma di Trento (PAT, Autonomous Province of Trento) through the UniTrento Internal Call for Research 2023 grant “Searching for Dark Energy off the beaten track” (DARKTRACK, grant agreement no. E63C22000500003).
SP acknowledges the financial support from the Department of Science and Technology (DST), Govt. of India under the Scheme  ``Fund for Improvement of S\&T Infrastructure (FIST)'' (File No. SR/FST/MS-I/2019/41).
TT acknowledges support from the National Science Foundation, the National Areonautics and Space Administration, and the Gordon and Betty Moore Foundation.
VP is supported by funding from the European Research Council (ERC) under the European Union's HORIZON-ERC-2022 (grant agreement no.~101076865) and from the European Union's Horizon 2020 research and innovation program under the Marie Sk{\l}odowska-Curie grant agreement no.~860881-HIDDeN.
AD acknowledges the support of the European Union’s Horizon 2021 research and innovation programme under the Marie Sklodowska-Curie grant agreement No. 101068013 (QGRANT).
AK has been supported by a \emph{Lend\"ulet} excellence grant by the Hungarian Academy of Sciences (MTA). This project has received funding from the European Union’s Horizon Europe research and innovation programme under the Marie Skłodowska-Curie grant agreement number 101130774. Funding for this project was also available in part through the Hungarian Ministry of Innovation and Technology NRDI Office grant OTKA NN147550.
ARL was supported by FCT through the Investigador FCT Contract CEECIND/02854/2017 and the research project PTDC/FIS-AST/0054/2021.
AP is grateful for the support of Vicerrectoría de Investigación y Desarrollo Tecnológico (Vridt) at Universidad Católica del Norte through Núcleo de Investigación Geometría Diferencial y Aplicaciones, Resolución Vridt No - 096/2022 and Resolución Vridt No - 098/2022. AP was supported by the Proyecto Fondecyt Regular 2024, Folio 1240514, Etapa 2024.
AB acknowledges support from the National Science Center, project no. UMO-2022/45/B/ST2/01067.
ABR is supported by the Funda\c{c}\~{a}o Carlos Chagas Filho de Amparo \`{a} Pesquisa do Estado do Rio de Janeiro (FAPERJ), Grant No E-26/200.149/2025 \'{e} 200.150/2025 (304809)
AJS received support from NASA through STScI grants HST-GO-16773 and JWST-GO-2974.
AM acknowledges the financial support by Conacyt-Mexico through the Post- doc Project I1200/311/2023.
AB acknowledges the fellowship of the Brazilian research agency CNPq.
AH is funded by the Carlsberg foundation.
BSS acknowledges National Science Foundation awards AST-2307147, PHY-2207638, PHY-2308886 and PHY-2309064.
B.~L. acknowledges the support of the National Research Foundation of Korea (NRF-2022R1F1A1076338), the Partenariat Hubert Curien STAR  with the National Research Foundation of Korea (RS-2023-00259422), and of the Korea Institute for Advanced Study (KIAS) grant funded by the government of Korea.
The research activities of BT is supported in part by the U.S.\ National Science Foundation under Grant PHY-2014104.
CGB is supported by the Spanish Grant PID2023-149016NB-I00 (MINECO/AEI/FEDER, UE) and the Basque government Grant No. IT1628-22 (Spain).
CM is supported by an FCT fellowship, grant number 2023.03984.
CZ is supported by the China Scholarship Council for 1 year study at SISSA.
CP acknowledges the financial support by the excellence cluster QuantumFrontiers of the German Research Foundation (Deutsche Forschungsgemeinschaft, DFG) under Germany's Excellence Strategy -- EXC-2123 QuantumFrontiers -- 390837967 and was funded by the Deutsche Forschungsgemeinschaft (DFG, German Research Foundation) - Project Number 420243324.
D.B. acknowledges support from projects PID2021-122938NB-I00 funded by the Spanish “Ministerio de Ciencia e Innovación” and FEDER “A way of making Europe”, PID2022-139841NB-I00 funded by the Spanish “Ministerio de Ciencia e Innovación” and SA097P24 funded by Junta de Castilla y León.
DS acknowledges support from Bulgarian National Science Fund grant number KP-06-N58/5.
DRG acknowledges support from grant PID2022-138607NBI00, funded by MCIN/AEI/10.13039/501100011033.
DB acknowledges funding from the Ministry of Science, Technological Development and Innovations of the Republic of Serbia, Project contract No. 451-03-136/2025-03/200017.
The work of EG was supported in part by grant NSF PHY-2309135 to the Kavli Institute for Theoretical Physics (KITP).
EF is supported by ``Theoretical Astroparticle Physics'' (TAsP), iniziativa specifica INFN and by the research grant number 2022E2J4RK “PANTHEON: Perspectives in Astroparticle and Neutrino THEory with Old and New messengers”under the program PRIN 2022 funded by the Italian Ministero dell’Università e della Ricerca (MUR).
EDMF is supported by World Premier International Research Center Initiative (WPI Initiative), MEXT, Japan.
FA thanks CNPq and Fundação Carlos Chagas Filho de Amparo à Pesquisa do Estado do Rio de Janeiro (FAPERJ), Processo SEI 260003/014913/2023 for financial support.
The work of FB was supported by the postdoctoral grant CIAPOS/2021/169.
FP acknowledges partial support from the INFN grant InDark and from the Italian Ministry of University and Research (\textsc{mur}), PRIN 2022 `EXSKALIBUR – Euclid-Cross-SKA: Likelihood Inference Building for Universe's Research', Grant No.\ 20222BBYB9, CUP C53D2300131 0006, and from the European Union -- Next Generation EU. FP also acknowledges support from the FCT project ``BEYLA -- BEYond LAmbda'' with ref. number PTDC/FIS-AST/0054/2021.
FSNL acknowledges support from the Funda\c{c}\~{a}o para a Ci\^{e}ncia e a Tecnologia (FCT) Scientific Employment Stimulus contract with reference CEECINST/00032/2018, and funding through the research grants UIDB/04434/2020, UIDP/04434/2020 and PTDC/FIS-AST/0054/2021.
GL was funded by Agencia Nacional de Investigación y Desarrollo (ANID) through Proyecto Fondecyt Regular 2024,  Folio 1240514, Etapa 2024. He also thanks Vicerrectoría de Investigación y Desarrollo Tecnológico (VRIDT) at Universidad Católica del Norte for support through Núcleo de Investigación Geometría Diferencial y Aplicaciones (Resolución VRIDT N°096/2022).
GG acknowledges the financial support from the COSMOS network (www.cosmosnet.it) through the ASI (Italian Space Agency) Grants 2016-24-H.0 and 2016-24-H.1- 2018.
GDS acknowledges support from INAF-ASTROFIT fellowship and Istituto Nazionale di Fisica Nucleare (INFN), Naples Section, for specific initiatives QGSKY and Moonlight2, as well as GAIA DPAC funds from INAF and ASI (PI: M.Lattanzi).
GJO acknowledges financial support from the Spanish Grants  PID2020-116567GB-C21, PID2023-149560NB-C21 funded by MCIN/AEI/10.13039/501100011033, by CEX2023-001292-S funded by MCIU/AEI, and by i-COOP23096 funded by CSIC.
GSDj acknowledges the support by the Ministry of Science, Technological Development and Innovation of the Republic of Serbia under contract 451-03-137/2025-03/ 200124 and support by the ICTP - SEENET-MTP NT03 Project TECOM-GRASP.
GCH acknowledges support through the ESA research fellowship programm.
HAF was partially supported by NSF grant AST-1907404.
The work of IDG was supported by the Estonian Research Council grants MOB3JD1202, RVTT3,  RVTT7, and by the CoE program TK202 ``Fundamental Universe''.
The work of IAM is supported by the Basque Government Grant IT1628-22, by Grant PID2021-123226NB-I00 (funded by MCIN/AEI/10.13039/501100011033 and by “ERDF A way of making Europe”).
IDM acknowledges support from the grant PID2021-122938NB-I00 funded by MCIN/AEI/10.13039/501100011033 and from the grant SA097P24 funded by Junta de Castilla y Le\'{o}n and by “ERDF A way of making Europe”.
JA acknowledges support from the Diputaci\'on General de Arag\'on-Fondo Social Europeo (DGA-FSE) Grant No. 2020-E21-17R of the Aragon Government.
JG and BK have been supported in part by the Polish National Science Center (NCN) under grant 2020/37/B/ST2/02371.
JR is supported by a Ram\'on y Cajal contract of the Spanish Ministry of Science and Innovation with Ref.~RYC2020-028870-I. This research was further supported by the project PID2022-139841NB-I00 of MICIU/AEI/10.13039/501100011033 and FEDER, UE.
JS is supported by the Taiwan National Science \& Technology Council.
JARC is supported by the project PID2022-139841NB-I00 funded by MICIU/AEI/10.13039/501100011033 and by ERDF/EU.
JJ acknowledge partial support from STScI under grants HST-GO-16262 and JWST-GO-03055.
JGB acknowledges support from the Spanish Research Project PID2021-123012NB-C43 [MICINN-FEDER], and the Centro de Excelencia Severo Ochoa Program, Spain CEX2020-001007-S at IFT.
KL is a recipient of the John and Pat Hume Scholarship and acknowledges support from the Friedrich Naumann Foundation for Freedom and from the Swiss Study Foundation.
The research activities of KRD are supported in part by the U.S.\ National Science Foundation through its employee IR/D program as well as by the U.S.\ Department of Energy under Grant DE-FG02-13ER41976 / DE-SC0009913.
LLG acknowledges support from Conselho Nacional de Desenvolvimento Cientıfico e Tecnologico (CNPq), Grant No. 307636/2023-2 and from the Fundacao Carlos Chagas Filho de Amparo a Pesquisa do Estado do Rio de Janeiro (FAPERJ) Grant No. E-26/204.598/2024.
LY acknowledges support from YST Program of APCTP and Natural Science Foundation of Shanghai 24ZR1424600.
LC acknowledges support from FCT - Fundação para a Ciência e a Tecnologia through the projects with DOI identifiers 10.54499/2023.11681.PEX and 10.54499/2024.00249.CERN.
MB is supported by the Polish National Science Center through grants no. 2020/38/E/ST9/00395 and 2020/39/B/ST9/03494.
MGE acknowledges the financial support of FONDECYT de Postdoctorado, N° 3230801.
MdC acknowledges support from INFN iniziativa specifica GeoSymQFT.
The work of MR is supported by the European Union – Next Generation EU and by the Italian Ministry of University and Research (MUR) via the PRIN 2022 project n. 20228WHTYC.
ME was supported by the Estonian Ministry of Education and Research (grant TK202, ``Foundations of the Universe''), by Estonian Research Council grant PRG2172, and by the European Union's Horizon Europe research and innovation programme (EXCOSM, grant No. 101159513).
The work of MBL, CGB \& PM is supported by the Spanish Grant PID2023-149016NB-I00 (MINECO/AEI/FEDER, UE). This work is also supported by the Basque government Grant No. IT1628-22 (Spain).
MC was supported by FCT through the Investigador FCT Contract No. CEECIND/02581/2018 and the research project PTDC/FIS-AST/0054/2021.
MM acknowledges funding by the Agenzia Spaziale Italiana (\textsc{asi}) under agreement no. 2018-23-HH.0 and support from INFN/Euclid Sezione di Roma.
MASP acknowledges support from the FCT through the Fellowship UI/BD/154479/2022, through the Research Grants UIDB/04434/2020 and UIDP/04434/2020, and through the project with reference PTDC/FIS-AST/0054/2021 (``BEYond LAmbda''). MASP also acknowledges support from the MICINU through the project with reference PID2023-149560NB-C21.
MAS received support from the CAPES scholarship.
MS acknowledges the support of ANSEF 24AN:PS-astroth-3095 and Higher Education and Science Committee of RA (24WS-1C001 project)
MK was funded through  SASPRO2 project AGE of Gravity: Alternative Geometries of Gravity, which has received funding from the European Union's Horizon 2020 research and innovation programme under the Marie Skłodowska-Curie grant agreement No. 945478. 
MM acknowledges the support by the Ministry of Science, Technological Development and Innovation of the Republic of Serbia under contract 451-03-137/2025-03/ 200124.
MI acknowledges that this material is based upon work supported in part by the Department of Energy, Office of Science, under Award Number DE-SC0022184 and also in part by the U.S. National Science Foundation under grant AST2327245.
NF acknowledge support from the Research grant TAsP (Theoretical Astroparticle Physics) funded by INFN. NF is supported by the European Union – Next Generation EU and by the Italian Ministry of University and Research (MUR) via the PRIN 2022 project n. 20228WHTYC.
The Work of NEM. is supported in part by the UK Science and Technology Facilities research Council (STFC) under the research grant ST/X000753/1.
NAN was financed by IBS under the project code IBS-R018-D3, and acknowledges support from PSL/Observatoire de Paris.
NS is supported by the Charles University Grant Agency (GAUK) - 94224.
PS is supported by Science and Technology Facilities Council (STFC) training grant ST/X508287/1.
PJ acknowledges funding from the Ministry of Science, Technological Development and Innovations of the Republic of Serbia, Project contract No. 451-03-136/2025-03/200002.
RG acknowledges the support from the SNF 200020\_175751 and 200020\_207379 ``Cosmology with 3D Maps of the Universe'' research grant.
RH work has been supported by the Spanish grants FPU19/03348 of MU, MCIN/AEI/10.13039/501100011033 grants PID2020-113644GB-I00, PID2023-148162NB-C22, by the European Union’s Horizon 2020 research and innovation program under the Marie Sklodowska-Curie grants HORIZON-MSCA-2021-SE-01/101086085-ASYMMETRY and H2020-MSCA-ITN-2019/860881-HIDDeN and by the Generalitat Valenciana grants PROMETEO/2019/083 and CIPROM/2022/69.
RH acknowledges funding from the Italian National Institute of Astrophysics (INAF) through large grant PRIN 12-2022 “INAF-EDGE” (PI L. Hunt).
RR acknowledges financial support from the STFC Consolidated Grant ST/X000583/1.
RBN acknowledges funding by the Royal Society through the University Research Fellowship Renewal URF\ R\ 221005.
RK is supported by Project SA097P24 funded by Junta de Castilla y Leon.
SB is supported by “Agencia Nacional de Investigación y Desarrollo” (ANID), Grant “Becas Chile postdoctorado al extranjero” No. 74220006.
ST is supported by the National Science Centre, Poland (research grant No. 2021/42/E/ST2/00031).
SSC acknowledges support from the Istituto Nazionale di Fisica Nucleare (INFN) through the Commissione Scientifica Nazionale 4 (CSN4) Iniziativa Specifica ``Quantum Fields in Gravity, Cosmology and Black Holes'' (FLAG) and from the Fondazione Cassa di Risparmio di Trento e Rovereto (CARITRO Foundation) through a Caritro Fellowship (project ``Inflation and dark sector physics in light of next-generation cosmological surveys'').
SJL is supported by grant PIP 11220200100729CO CONICET and grant 20020170100129BA UBACYT.
SP is supported by the international Gemini Observatory, a program of NSF NOIRLab, which is managed by the Association of Universities for Research in Astronomy (AURA) under a cooperative agreement with the U.S. National Science Foundation, on behalf of the Gemini partnership of Argentina, Brazil, Canada, Chile, the Republic of Korea, and the United States of America. SP acknowledges the financial support of the Conselho Nacional de Desenvolvimento Científico e Tecnológico (CNPq) Fellowships 300936/2023-0 and 301628/2024-6.
TP acknowledges the financial support from the Slovenian Research Agency (grants I0-0033, P1-0031, J1-8136, J1-2460 and Z1-1853).
TB was supported by ICSC -- Centro Nazionale di Ricerca in High Performance Computing, Big Data and Quantum Computing, funded by European Union -- NextGenerationEU.
TS is supported by the Della Riccia foundation grant 2025 and the Galileo Galilei Institute Boost Fellowship 2024.
TK was supported by the Estonian Research Council grants CoE TK202 “Foundations of the Universe” and PRG2608 “Space - Time  - Matter”.
VM thanks CNPq (Brazil) and FAPES (Brazil) for partial financial support.
VAM is supported by Generalitat Valenciana via the Excellence Grant CIPROM/2021/073, by the Spanish MICIN/AEI/10.13039/501100011033 and the EU/FEDER via grant PID2021-122134NB-C21, and by the Spanish MCIU/AEI via the Severo Ochoa project CEX2023-001292-S.
VP acknowledges the support by the Ministry of Education and Science of the Federation of Bosnia and Herzegovina, under Project Number 05-35-2467-1/23.
VK acknowledges financial support from Research Ireland Grant 21/PATH-S/9475 (MOREHIGGS) under the SFI-IRC Pathway Programme. Report number: DIAS-STP-25-05.
VM acknowledges support from ANID FONDECYT Regular grant number 1231418, Millennium Science Initiative AIM23-0001, and Centro de Astrof\'{\i}sica de Valpara\'{\i}so CIDI N21.
VBJ acknowledges funding from the Ministry of Science, Technological Development and Innovations of the Republic of Serbia, Project contract No. 451-03-136/2025-03/200017.
WY has been supported by the National Natural Science Foundation of China under Grant No. 12175096, and Liaoning Revitalization Talents Program under Grant no. XLYC1907098.
WY is supported via the research projects `COLAB'  funded by the National Science Center, Poland, under agreement number UMO-2020/39/B/ST9/03494.
``YH is funded by NSFC under the grant No.12347137 and the China Postdoctoral Science Foundation under Grant No. 2024M753076''.
}
\newline

\noindent \textit{Afternote:} The opinions and conclusions expressed herein are those of the authors, and do not represent any funding agencies.

\newpage
\bibliographystyle{utphys}
\bibliography{references,references2}

%Author information
\let\author\myauthor
\let\affiliation\myaffiliation
\let\maketitle\mymaketitle
\title{White Paper Authors}

%%% Editors:
\author{Eleonora Di Valentino}
\affiliation{School of Mathematical and Physical Sciences, University of Sheffield, Hounsfield Road, Sheffield S3 7RH, United Kingdom}

\author{Jackson Levi Said}
\affiliation{Institute of Space Sciences and Astronomy, University of Malta, Malta}
\affiliation{Department of Physics, University of Malta, Malta}

%%% Forward Writers:
\author{Adam Riess}
\affiliation{Department of Physics and Astronomy, Johns Hopkins University, Baltimore, MD 21218, USA}
\affiliation{Space Telescope Science Institute, 3700 San Martin Drive, Baltimore, MD 21218, USA}

\author{Agnieszka Pollo}
\affiliation{National Centre for Nuclear Research, Pasteura 7, 02-093, Warszawa, Poland}

\author{Vivian Poulin}
\affiliation{Laboratoire Univers \& Particules de Montpellier (LUPM), CNRS \& Universit\'e de Montpellier (UMR-5299),Place Eug\`ene Bataillon, F-34095 Montpellier Cedex 05, France}

%%% Coordinators:
\author{Adri\`a G\'{o}mez-Valent}
\affiliation{Departament de F\'isica Qu\`antica i Astrof\'isica and Institut de Ci\`{e}ncies del Cosmos, Universitat de Barcelona, Av. Diagonal 647, E-08020 Barcelona, Catalonia, Spain}

\author{Amanda Weltman}
\affiliation{High Energy Physics, Cosmology and Astrophysics Theory (HEPCAT) Group, Department of Mathematics and Applied Mathematics, University of Cape Town, Cape Town 7700, South Africa}

\author{Antonella Palmese}
\affiliation{McWilliams Center for Cosmology and Astrophysics, Department of Physics, Carnegie Mellon University, 5000 Forbes Avenue, Pittsburgh, PA 15213, USA}

\author{Caroline D. Huang}
\affiliation{Harvard-Smithsonian Center for Astrophysics, 60 Garden St., Cambridge, MA 02138, USA}

\author{Carsten van de Bruck}
\affiliation{School of Mathematical and Physical Sciences, The University of Sheffield, Hounsfield Road, S3 7RH Sheffield, United Kingdom}

% ...

\newpage
\maketitle

\end{document}